\documentclass[manyauthors,nocleardouble]{cernphprep} 
\usepackage{colortbl}
\usepackage{tablefootnote}
\usepackage{wrapfig}
\usepackage{amssymb}
\usepackage{amsfonts}
\usepackage{amsmath}
\usepackage{epigraph}
\usepackage{longtable}
\usepackage{eurosym}
\usepackage{lscape}
\usepackage[T2A]{fontenc}
\usepackage[utf8x]{inputenc}
\usepackage[english,russian]{babel}
\usepackage{wrapfig}
\usepackage{amssymb}
\usepackage{amsfonts}
\usepackage{authblk}
\usepackage{amsmath}
\usepackage[titletoc]{appendix}
\usepackage{lineno}
\usepackage{color}
\usepackage{xcolor}
\usepackage{multirow}
\usepackage{longtable}
\usepackage{lscape}
\usepackage{hyperref}
\usepackage{bigstrut}
\usepackage{isotope}
\usepackage{soul}
\usepackage{textcomp}

\usepackage{titlesec} 

\usepackage{cleveref}
\labelformat{section}{\thechapter.#1}

\originalTeX

\usepackage[dvipdfmx]{graphicx} 
\usepackage{bmpsize}

\DeclareGraphicsExtensions{.png,.pdf,.jpg,.mps,.eps}
\graphicspath{{./fig/}}
\usepackage{xspace} 

\usepackage{subfig}
\usepackage{comment}
\usepackage{fix2col}
\usepackage[sort&compress,square,comma,numbers,merge]{natbib}
\usepackage{amsmath,amssymb}
\usepackage{amsfonts}
\usepackage{upgreek} 
\usepackage{hepnames}
\usepackage{units}
\usepackage{import}
\usepackage{multirow}
\usepackage{booktabs}
\usepackage{color}
\usepackage{tikz}
\usepackage{placeins}

\usepackage{academicons}
\usepackage{orcidlink}
\usepackage{tikz}
\usepackage{hyperref}

\input ./sty/mymathphys.sty
\input ./sty/myunits.sty
\usepackage[textsize=tiny]{todonotes}
\newcommand{\inmmode}[1]{{\ifmmode{#1}\else{$#1$}\fi}}

\labelformat{section}{\thechapter.\thesection}
\begin{document}
\selectlanguage{english}


\hypersetup{pageanchor=false}   
\begin{titlepage}
\PHdate{\today}

\title{Technical Design Report  of the Spin Physics Detector at NICA}
\begin{center}
For the SPD collaboration \\
\end{center}
\begin{center}
Version 2.01 \\
\vspace{2cm}

\begin{center}
\end{center}
\vspace{5cm}
Internal draft version 2024.088 \\
\end{center}
\end{titlepage}

\hypersetup{pageanchor=true}


\tableofcontents

\chapter*{Preface}
According to astrophysical and cosmological data, about 5\% of the mass of the Universe consists of visible baryonic matter, the properties of which are determined by strong and electromagnetic interactions. 
With respect to the two other components, dark matter and dark energy, baryonic matter seems to be a well-studied subject. In fact, despite the great advances in quantum chromodynamics made in describing the interaction of quarks and gluons within the framework of the perturbative approach, the question of why nucleons are exactly like we see them, remains open. Understanding the structure and fundamental properties of the nucleon directly from the dynamics of its quarks and gluons based on first principles is one of the main unsolved problems of QCD.

The nucleon behaves like a spinning top with a spin of $\hslash/2$. This spin is responsible for such fundamental properties of Nature as the nucleon magnetic moment, different phases of matter at low temperatures, the properties of neutron stars, and the stability of the known Universe. That is why the study of the spin structure of the nucleon is of particular importance. The naive quark model has successfully  predicted most of the gross properties of hadrons, such as charge, parity, isospin, and symmetry properties and their relations. Some of the dynamics of particle interactions can  be qualitatively understood in terms of this model as well. However, the model falls short of explaining the spin properties of hadrons in terms of their constituents.
Since the famous ''spin crisis'' that began in 1987, the problem of the nucleon spin structure remains one of the most intriguing puzzles in contemporary high-energy physics. 
The main  question, which for many years has attracted enormous theoretical and experimental efforts, is how the spin of the nucleon is built up from spins and orbital momenta of its constituents -- the valence and sea quarks, as well as gluons. 
The three-dimensional description of constituents is often performed in terms of the so-called (polarized or unpolarized) transverse-momentum dependent parton distribution functions (TMD PDFs).

Over the last 25 years, both polarized deep inelastic scattering experiments (CERN, DESY, JLab, SLAC) and high-energy polarized proton-proton collisions (RHIC at BNL)  have provided the majority of information about spin-dependent structure functions of the nucleon. Nevertheless, our knowledge of the internal structure of the nucleon is still limitedespecially about its gluon content. New facilities for spin physics, such as the Electron-Ion Collider at BNL and the fixed-target experiments at CERN LHC are planned to be built in the near future to obtain the missing information.

The Spin Physics Detector, a universal facility for studying the nucleon spin structure and other spin-related phenomena with polarized proton and deuteron beams, is proposed to be placed in one of the two interaction points of the NICA collider that is under construction at the Joint Institute for Nuclear Research (Dubna, Russia). At the heart of the project is extensive experience with polarized beams at JINR.
 The main objective of the proposed experiment is the comprehensive study of the unpolarized and polarized gluon content of the nucleon.  
Spin measurements at the Spin Physics Detector at the NICA collider will make a unique contribution and will challenge our understanding
 of the spin structure of the nucleon. 

The Conceptual Design Report of the Spin Physics Detector was presented at the 54th meeting of the JINR Program Advisory Committee for Particle Physics in January 2021 and approved at the 56th meeting in January 2022, based on the report of the international SPD Detector Advisory Committee. 

In this document, the technical design of the Spin Physics Detector is presented.

\chapter{Executive summary}

The Spin Physics Detector collaboration proposes to install a universal detector in the second interaction point of the NICA collider under construction (JINR, Dubna)  to study the spin structure of the proton and deuteron  and other spin-related phenomena using a unique possibility to operate with polarized proton and deuteron beams at a collision energy up to 27 GeV and a luminosity up to $10^{32}$ cm$^{-2}$ s$^{-1}$.  As the main goal, the experiment aims to provide access to the gluon TMD PDFs in the proton and deuteron, as well as the gluon transversity distribution and tensor PDFs in the deuteron, via the measurement of specific single and double spin asymmetries using different complementary probes such as charmonia, open charm, and prompt photon production processes.  Other polarized and unpolarized physics is possible, especially at the first stage of NICA operation with reduced luminosity and collision energy of the proton and ion beams. The SPD physics program is described in detail in \cite{SPD:2021hnm, Arbuzov:2020cqg, Abramov:2021vtu}.  This document is dedicated exclusively to technical issues of the SPD setup construction.

 \begin{figure}[!h]
  \begin{center}
    \includegraphics[width=1.\textwidth]{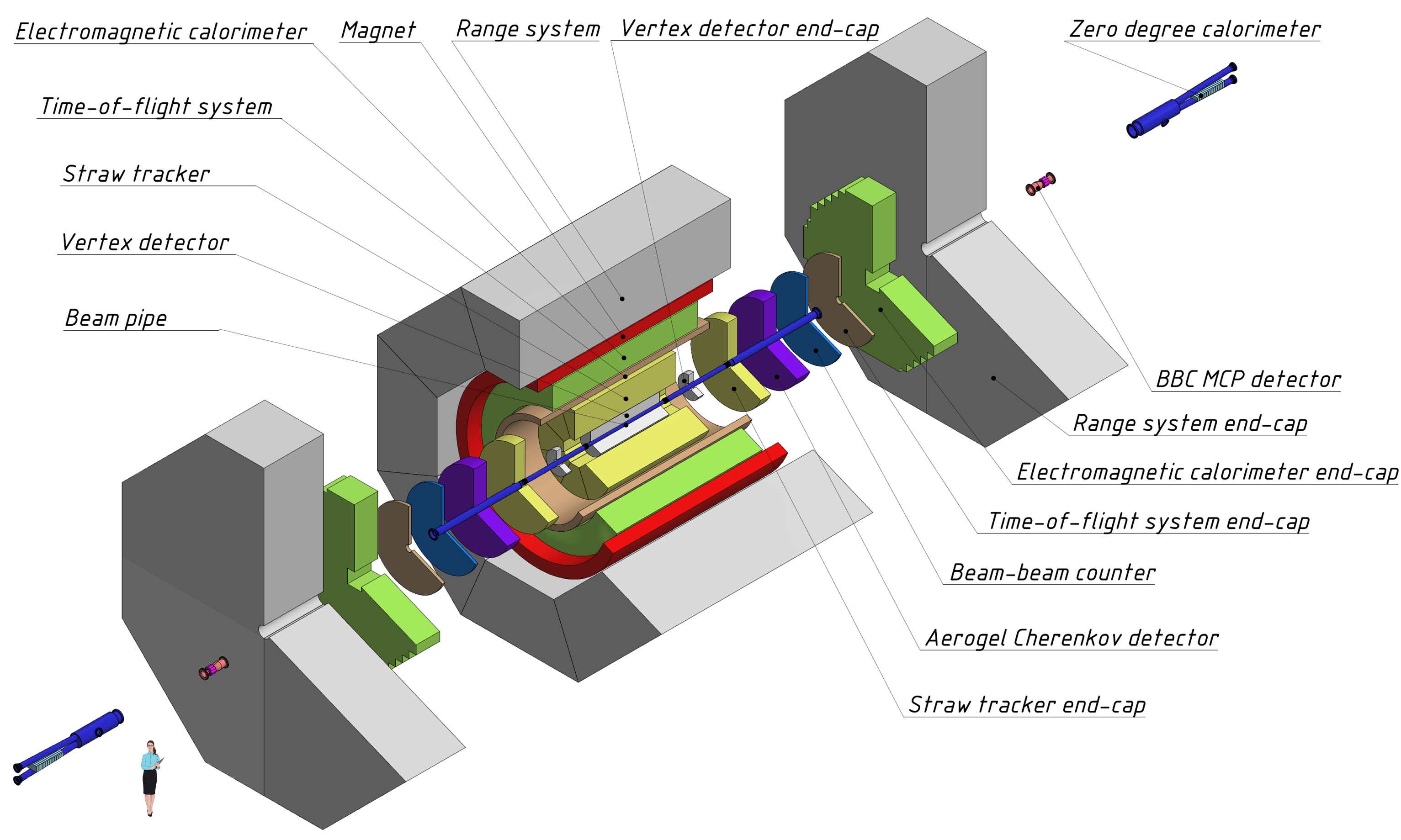}
  \end{center}
  \caption{General layout of the SPD setup.}
  \label{SPD-LA_EX}
\end{figure}
    The SPD experimental setup is designed as a universal $4\pi$ detector with advanced tracking and particle identification capabilities based on modern technologies that will be installed in the SPD experimental hall of the NICA collider. The Silicon Vertex Detector (VD) will provide  resolution for the vertex position on the level of below 100 $\mu$m needed for the reconstruction of secondary vertices of $D$-meson decays. The Straw tube-based Tracking system (ST) placed within a solenoidal magnetic field of up to 1 T at the detector axis should provide the transverse momentum resolution $\sigma_{p_T}/p_T\approx 2\%$ for a particle momentum of 1 GeV/$c$. The Time-of-Flight system (TOF) with a time resolution of about 60 ps will provide $3\sigma$ $\pi/K$ and $K/p$ separation of up to about 1.2 GeV/$c$ and 2.2 GeV/$c$, respectively. Possible use of the Focusing Aerogel Ring-Imaging Cherenkov detector (FARICH) in the end-caps will extend this range significantly. Detection of photons will be provided by the sampling Electromagnetic Calorimeter (ECal) with the energy resolution $\sim 5\%/\sqrt{E}\oplus 1\% $. To minimize multiple scattering and photon conversion effects for photons, the detector material will be kept to a minimum throughout the internal part of the detector. The Range (muon) System (RS) is planned for muon identification. It can also act as a rough hadron calorimeter. The pair of Beam-Beam Counters (BBC) and Zero Degree Calorimeters (ZDC) will be responsible for the local polarimetry and luminosity control.  To minimize possible systematic effects, the SPD will be equipped with a free-running (triggerless) DAQ system. A high collision rate (up to 4 MHz) and a few hundred thousand detector channels pose a significant challenge to the DAQ,  online monitoring, offline computing system, and data processing software.
    
The SPD operation should start a few years after the collider starts operating using the possibilities of polarized $p$-$p$ and $d$-$d$ collisions at $\sqrt{s}<9.4$ GeV and $\sqrt{s}<4.5$ GeV/nucleon, respectively, as well as $A$-$A$ collisions. The starting configuration should consist of the Range System, solenoidal superconducting magnet, Straw tube-based Tracker,  a pair of Zero Degree Calorimeters, and a pair of Beam-Beam Counters. A simple Micromegas-based Central Tracker (MCT) will be installed in the central region instead of the sophisticated silicone vertex detector to keep a reasonable momentum resolution. Partial installation of other detectors (ECal, TOF, etc.) is also possible in the first stage.

The proposed design of the Spin Physics Detector has potential for future upgrades.

 The estimated cost of the Spin Physics Detector in its full configuration is about 110.6 M\$ at current prices ({\bf Nov 2023, 1 Euro = 1.08\$, 1 \$ = 92 RUB}). 50.0 M\$ of this amount is the cost of the first stage. 
 Any expenses related to the development and construction of the infrastructure for polarized beams at NICA are out of this estimation.

\chapter{General concept of the SPD experiment}

NICA, a new research facility aimed at the study of the properties of the strong interaction, is under construction at JINR and should be put into operation in 2025. At the first stage of operation,  the study of hot and dense baryonic matter will be performed in heavy-ion collisions with the MultiPurpose Detector (MPD) placed at the first interaction point of the collider. This study should shed light on the in-medium properties of hadrons and the nuclear matter equation of state, the onset of deconfinement,  chiral symmetry restoration, phase transition, existence of the mixed phase, the critical end-point, etc. The collider also provides the ability to operate with polarized proton and deuteron beams of high intensity that will be used for the study of the polarized structure of proton and deuteron with the Spin Physics Detector (SPD) installed in the second interaction point. 

The main goal of the SPD experiment is to get information about the gluon Transverse Momentum-Dependent Parton Distribution Functions (TMD PDFs) in the proton and deuteron, as well as the gluon transversity distribution and tensor PDFs in the deuteron, via the measurement of specific single and double spin asymmetries using such complementary probes as charmonia, open charm, and prompt photon production processes \cite{Arbuzov:2020cqg}.  This physics task  imposes general requirements on the concept of the experimental setup.  

Unlike the case of high-energy collisions where the collision energy $\sqrt{s}$ is a few orders of magnitude higher than a typical hard scale $Q$
of the studied reactions, at the SPD energies for all the probes planned to be used to access the gluon content of the colliding particles $Q\sim M_{J\psi} \sim 2M_{D}\sim p_{T\gamma~min}$ is just a few times less than $\sqrt{s}/2$.
Therefore, one should expect quite a uniform  distribution of all signal particles (muons from the $J/\psi$ decay, prompt photons, products of $D$-mesons decay, etc.) over the kinematic range. In other words, there is no preferable range in rapidity, which could be specified for each probe for optimal overall performance. Together with the relatively small cross-sections of the discussed probes, this fact  leads to a requirement of $\sim4\pi$ coverage of the SPD setup.

The Spin Physics Detector must have sufficient tracking capabilities and a magnetic system for spectrometric purposes  for the majority of the addressed physics tasks. It has to be equipped with a  muon system thick enough for effective separation of muons and hadrons to make it possible to deal with the decay $J/\psi\to \mu^+\mu^-$. A precision vertex detector  is needed for the recovery of the secondary vertices from the decays of $D^{\pm/0}$ mesons and other short-lived particles. An electromagnetic calorimeter  ensures the capability to detect signal and background photons. A low material budget and general transparency of the setup should also provide favorable conditions for photon physics. Hadron identification capability  is needed for any physics task with protons and/or kaons in the final state, in particular,  to enforce a signal-to-background ratio for $D$-mesons selection, and also to improve tracking at low momenta.  Since tiny effects are intended to be investigated, a triggerless DAQ system is planned in order to minimize possible systematic uncertainties of the measurements. Table \ref{Tab_physics} brings together the elements of the SPD physics program and the requirements for the experimental setup.

\begin{table}[htp]
\caption{Required setup configuration for each  point of the SPD physics program. (++) --  needed, (+) -- useful.}
\begin{tabular}{>{\raggedright}p{7cm}|ccccccc}
\hline\hline
 Physics goal & SVD & ST+ & TOF+ & ECal & RS & BBC & ZDC\tabularnewline
 &  & MCT  & FARICH & &  & & \tabularnewline
\hline 
\hline 
Study of polarized gluon content in proton and deuteron with:&            &           &               &                       &                     &       &    \\
~~~~~~-- charmonia &     + &    ++       &  +           &    ++               & ++                  &     &    \\   
~~~~~~-- open charm&   ++ &  ++       & ++          &   +                  & +                  &     &   \\
~~~~~~-- prompt photons&   & +         &              &   ++               &                   &       &  \\
\hline
Elastic $p$-$p$ and $d$-$d$ scattering && ++ & + & &  & ++ & +\tabularnewline
\hline 
Single-spin physics && ++ & ++ & &  & ++ & +\tabularnewline
\hline 
Vector light and charmed meson production && ++ & ++ & & ++ &  & \tabularnewline
\hline 
Scaling behavior of exclusive reactions with lightest nuclei and spin
observables && ++ & + & & & ++ & ++\tabularnewline
\hline 
Multiquark correlations and exotic hadron state production && ++ & ++ & &  &  & \tabularnewline
\hline 
Exclusive processes in $d$-$d$ collisions && ++ & + & &  & ++ & ++\tabularnewline
\hline 
Search for deconfinement in $p$-$p$ and $d$-$d$ central collisions && ++ & ++ & &  &  & \tabularnewline
\hline 
Search for dibaryons && ++ & + &  & & &  \tabularnewline
\hline 
Search for lightest neutral hypernuclei with strangeness $-1$ and $-2$ && ++ & ++ & &  &  & \tabularnewline
\hline 
Problems of soft $p$-$p$ interactions && ++ & ++ & & &  & \tabularnewline
\hline 
Measuring antiproton production cross-section for dark matter search && ++ & ++ & &  &  & +\tabularnewline
\hline 
Hadron formation effects in heavy ion collisions && ++ & ++ & & & ++ & + \tabularnewline
\hline 
Polarization of hyperons && ++ & + & & &  & \tabularnewline
\hline 
Central diffractive production &+& ++ & + & +& +&  & \tabularnewline
\hline\hline
\end{tabular}
\label{Tab_physics}
\end{table}%

Strict limitations to the SPD detector layout arise from external conditions, such as the maximal possible load to the floor of the SPD experimental hall (1500 tons together with the lodgement and the detector moving system).  Together with the requirement to have the overall thickness of the muon system not less than 4 nuclear interaction lengths ($\Lambda_I$), this limits the outer size of the SPD detector and the size of the inner part. The location of the collider infrastructure, in particular, of the focusing elements, also defines the size of the SPD setup along the beam axis. 

Although $p$-$p$ and $d$-$d$ collisions can be studied at MPD as a reference, and SPD setup has limited capabilities to operate with 
ion collisions, each of the detectors is optimized for its own tasks. MPD with the TPC-based tracking system can disentangle hundreds of charged tracks, but can not operate at luminosity above $10^{29}$ cm$^{-2}$ s$^{-1}$. The design of the MPD detector does not provide for the construction of a full-fledged muon system. Conversely, the SPD is optimized for operation at high luminosity (up to $10^{32}$ cm$^{-2}$ s$^{-1}$) but low track multiplicity. Optimum physics analysis in these two neighbouring experiments requires different luminosities. Therefore they have to take turns in recording data at NICA.
Based on the plans for putting into operation and commissioning the accelerator complex, development of the NICA polarized infrastructure, the plans of the MPD collaboration for technical and physics runs, as well as on realistic assumptions of the SPD funding, we propose the following stages of the SPD project implementation.
 \begin{figure}[!h]
  \begin{center}
    \includegraphics[width=1.\textwidth]{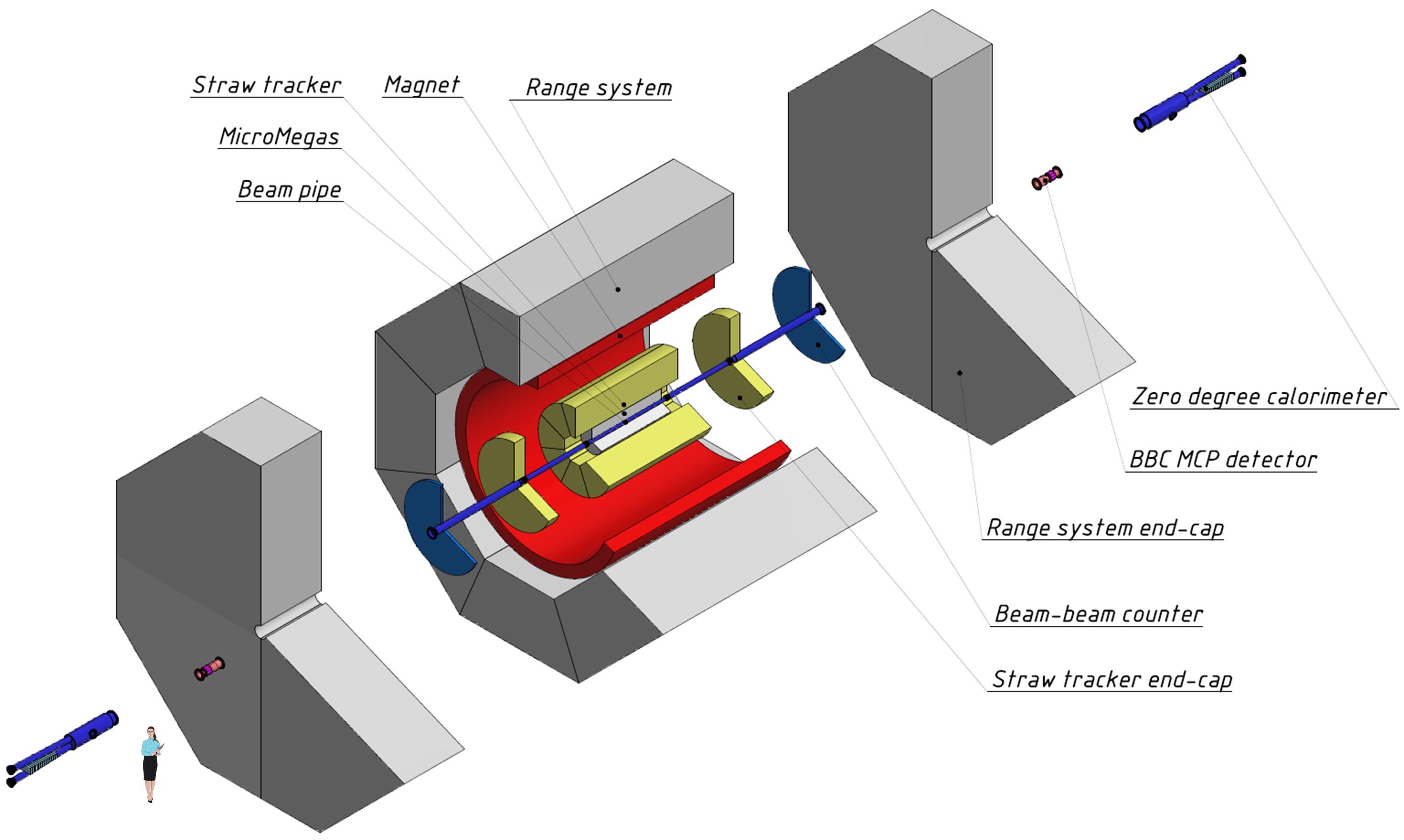}
  \end{center}
  \caption{General layout of the SPD setup at the first stage.}
  \label{SPD-start}
\end{figure}
The {\bf first stage} of the SPD experiment will be devoted to the study of polarized and non-polarized phenomena at low energies and reduced luminosity using heavy ion (up to Ca) and polarized proton and deuteron beams such as polarized phenomena in elastic $p$-$p$ and $d$-$d$ scattering and other exclusive reactions, spin effects in hyperon production, production of dibaryon resonances and hypernuclei, near-threshold charmonia production, etc. 
The duration of the first stage can be up to two years \footnote{here and hereafter, a year of the data taking implies 10$^7$ seconds of NICA operation at the nominal luminosity.}.
It implies the construction of a minimum setup configuration that should include (see Fig. \ref{SPD-start}):
\begin{itemize}
\item a Range System (RS) -- supporting structure of the entire installation and a magnet yoke, muon identification;
\item a Superconducting Solenoid (SS)  -- charged particle momentum reconstruction;
\item a Straw-based Tracking system (ST)  -- charged particle momentum reconstruction, PID via $dE/dx$ measurement;
\item a simple Micromegas-based Central Tracker (MCT) -- to improve charged particle momentum reconstruction;
\item  a system of two Beam-Beam Counters (BBC), inner and outer -- local polarimetry, luminosity control, and, possibly, timing;
\item  a system of two Zero Degree Calorimeters (ZDC) -- local polarimetry and luminosity control).
\end{itemize}
It  could  also include some elements of an Electromagnetic Calorimeter (ECal) for physics with photons and a local $\pi^0$-based polarimetry. 

We expect that for the first stage, the collider will be able to operate with polarized protons and deuterons in the spin transparency mode. The absolute value of the beam polarization should be not less than 0.5 for protons and 0.6 for deuterons.  Collisions of longitudinally and transversely polarized particles, $p$-$p$ and $d$-$d$ (in all combinations: LL, TT, TL, and LT), will be available at energy up to $\sqrt{s}=9.4$ GeV for protons and  $\sqrt{s}=4.5$ GeV/nucleon for deuterons. The corresponding luminosity is up to about $10^{31}$ cm$^{-2}$ s$^{-1}$ and $10^{30}$ cm$^{-2}$ s$^{-1}$, respectively. Both vertical and radial directions will be possible for the transverse polarization. Tensor polarization for deuteron will also be available. Absolute polarimetry is expected to be provided with accuracy not worse than 5\% for both vertical and radial directions. We also expect that it will be possible to operate in the mode of heavy-ion collision.
 \begin{figure}[!h]
  \begin{center}
    \includegraphics[width=1.\textwidth]{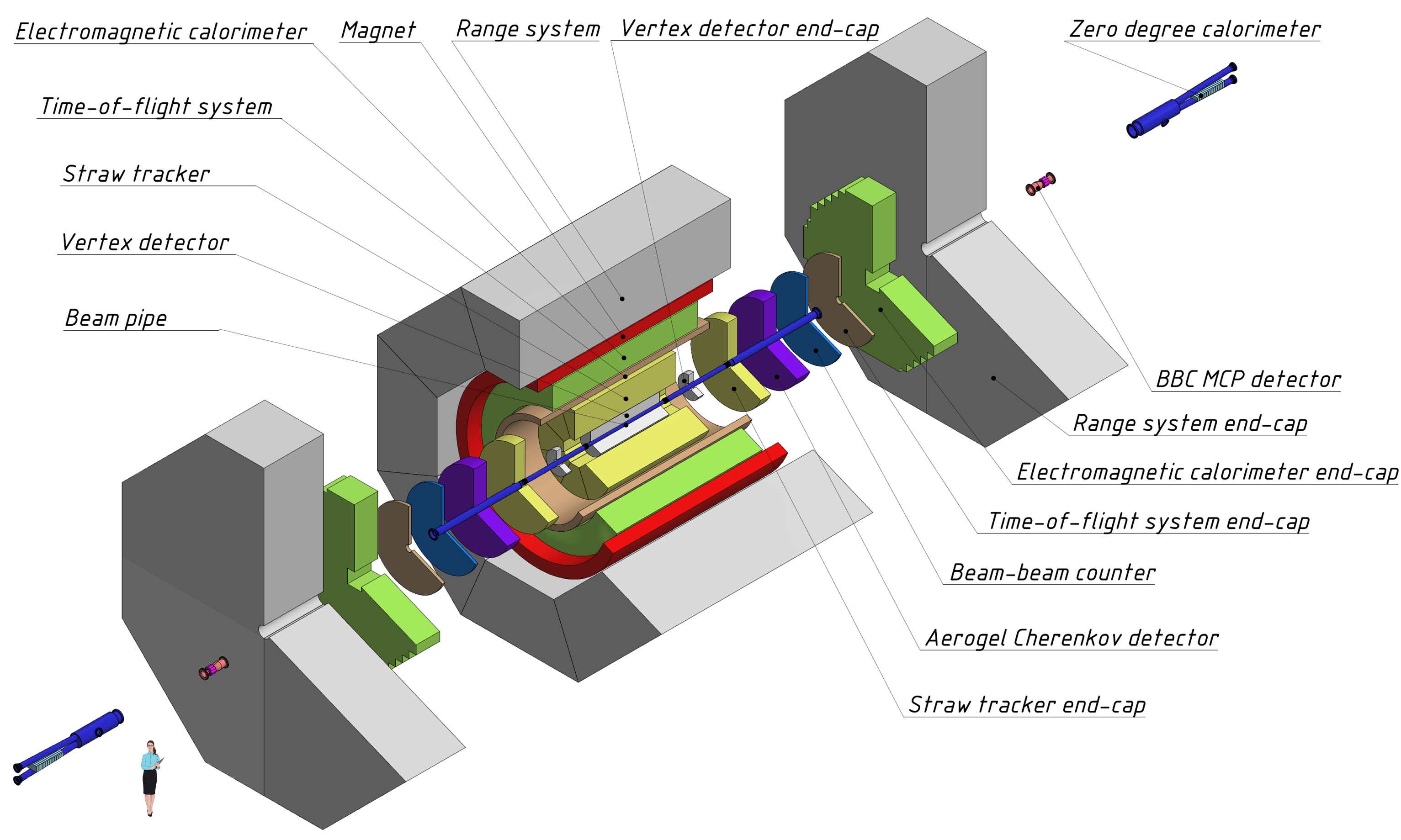}
  \end{center}
  \caption{Final layout of the SPD setup.}
  \label{SPD-full}
\end{figure}

 \begin{figure}[!h]
  \begin{center}
    \includegraphics[width=1.\textwidth]{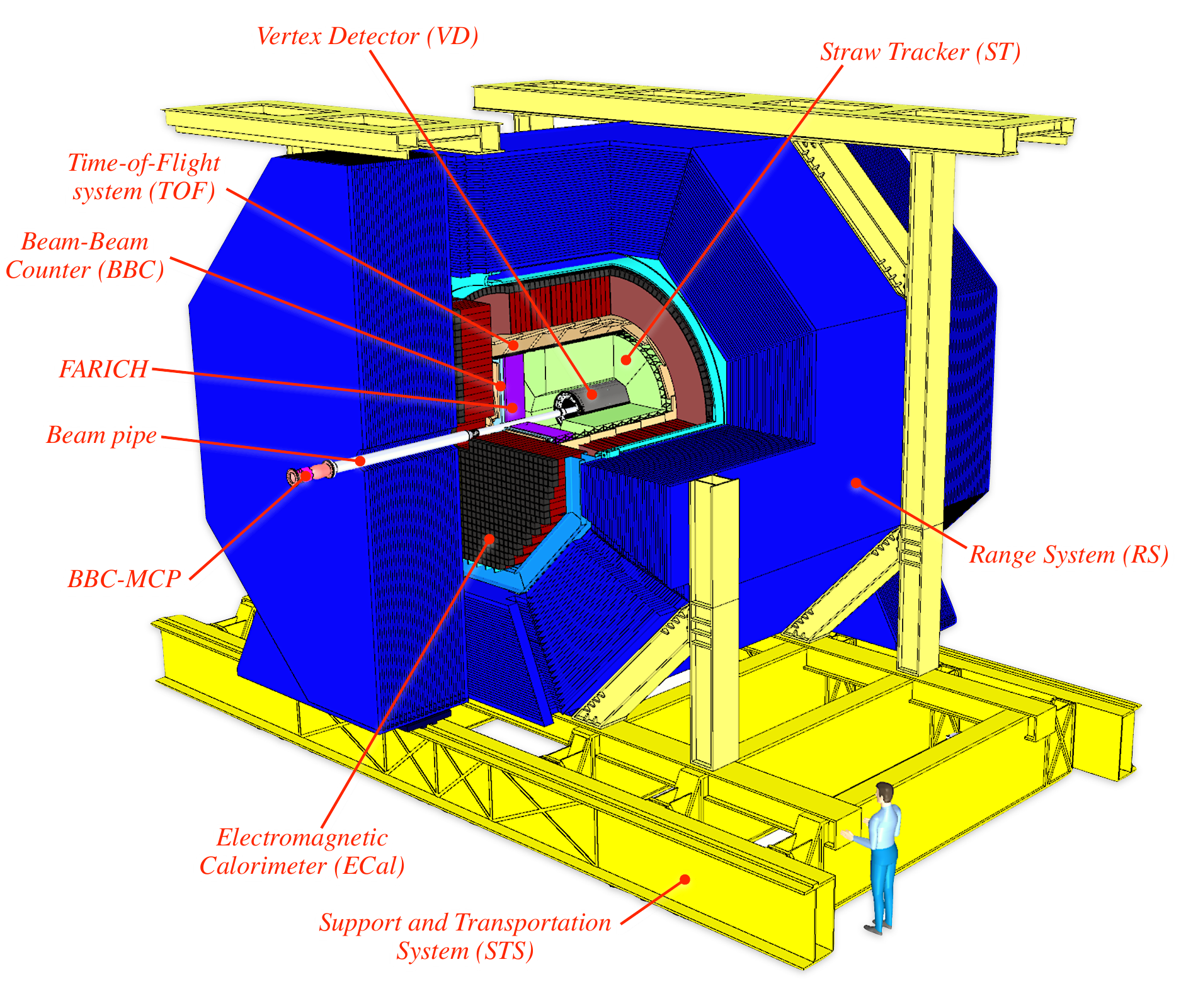}
  \end{center}
  \caption{SPD setup and the Support and Transportation System.}
  \label{SPD-full_STS}
\end{figure}

\begin{figure}[!h]
  \begin{center}
    \includegraphics[width=1.\textwidth]{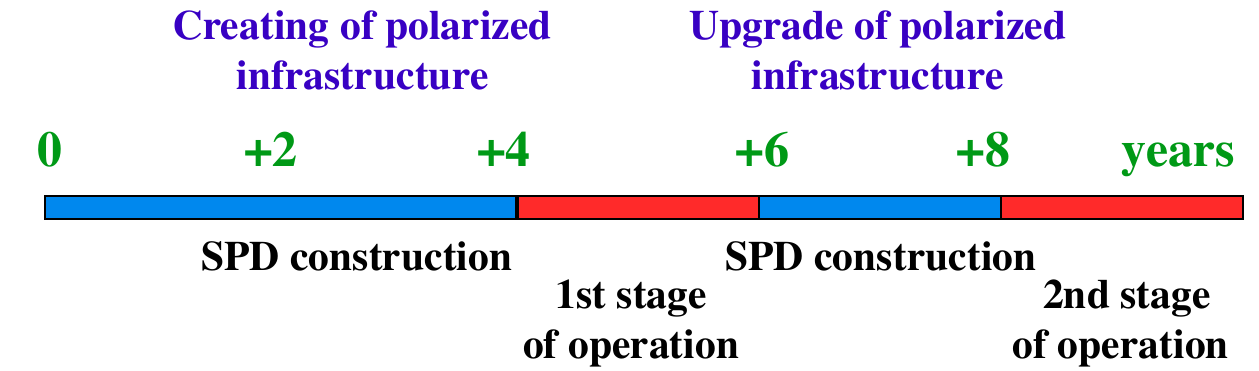}
  \end{center}
  \caption{Tentative operating plan of the SPD project.}
  \label{SPD-plans}
\end{figure}

The main part of the SPD physics program of the experiment, the study of the polarized gluon content in proton and deuteron,   is planned to be implemented during the {\bf second stage} with the full setup (see Fig. \ref{SPD-full}).  For this purpose, the SPD setup must be supplemented with a full-scale Electromagnetic Calorimeter (ECal), a Time-of-Flight system (TOF), and an aerogel-based FARICH detector for particle identification. The Micromegas-based Central Tracker will be replaced by a Silicon Vertex Detector (SVD). Figure \ref{SPD-full_STS} shows the SPD setup with the Support and Transportation System.
This stage should last not less than 4 years. By then we expect the accelerator to be able to deal with polarised protons and deuterons up to energies of 27 and 13.5 GeV/nucleon  in the center of mass system and luminosities of $10^{32}$ cm$^{-2}$ s$^{-1}$ and $10^{31}$ cm$^{-2}$ s$^{-1}$, respectively. The transverse polarisation for protons will be available for all energies, while the longitudinal polarisation will be available at spin resonance points with a beam energy step of 0.51 GeV. The tentative operating plan of the SPD project is presented in Fig. \ref{SPD-plans}.

Taking into account the high degree of integration of the detector subsystems, we decided to present the SPD Technical Design Report as a single document. The general subsystems as well as the subsystems that are assumed to be part of the first stage are highly elaborated. The subsystems of the second stage are described in a more schematic way.

\chapter{Detector summary}

To describe the SPD setup, we use a global coordinate system, where the $z$-axis is oriented along the nominal beam direction, the $y$-axis is vertical, and the $x$-axis is perpendicular to them and is directed toward the center of the collider ring. The origin of the coordinate system is the nominal center of the setup. We often refer to it as the Interaction Point (IP), although the real beam-crossing region has a gaussian shape along $z$-axis with $\sigma_z$ of about 30 cm (zero beam-crossing angle) \cite{Igamkulov:2019gdu}. As for timing, we suppose that in the $p$-$p$ mode we will have a  bunch crossing every 76 ns. 

\begin{figure}[h]
  \begin{center}
    \includegraphics[width=1\textwidth]{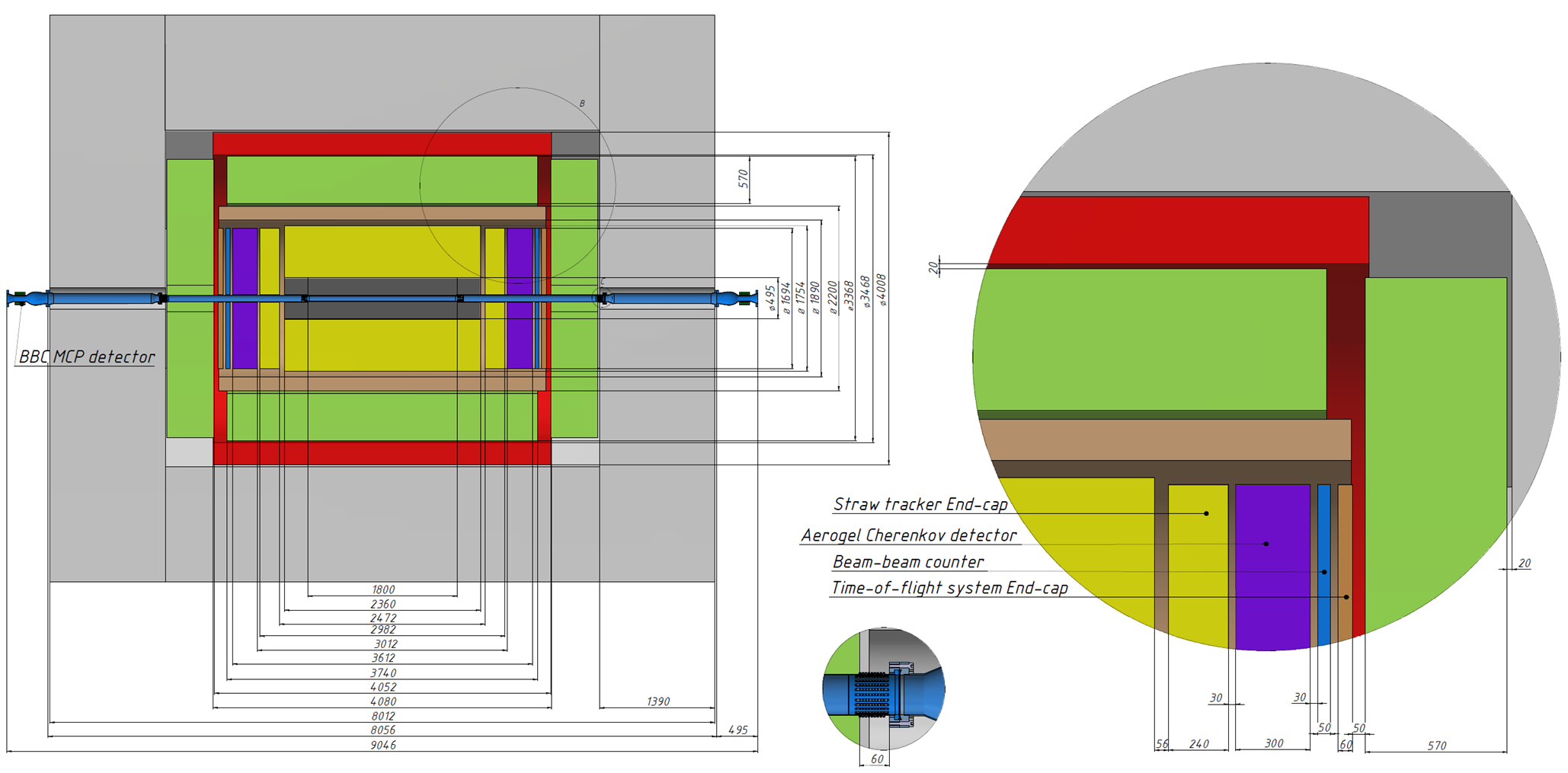}
  \end{center}
  \caption{Schematic side sectional view of the SPD detector with dimensions specified in millimeters.}
  \label{dims1}
\end{figure}

\begin{figure}[h]
  \begin{center}
   \includegraphics[width=0.5\linewidth]{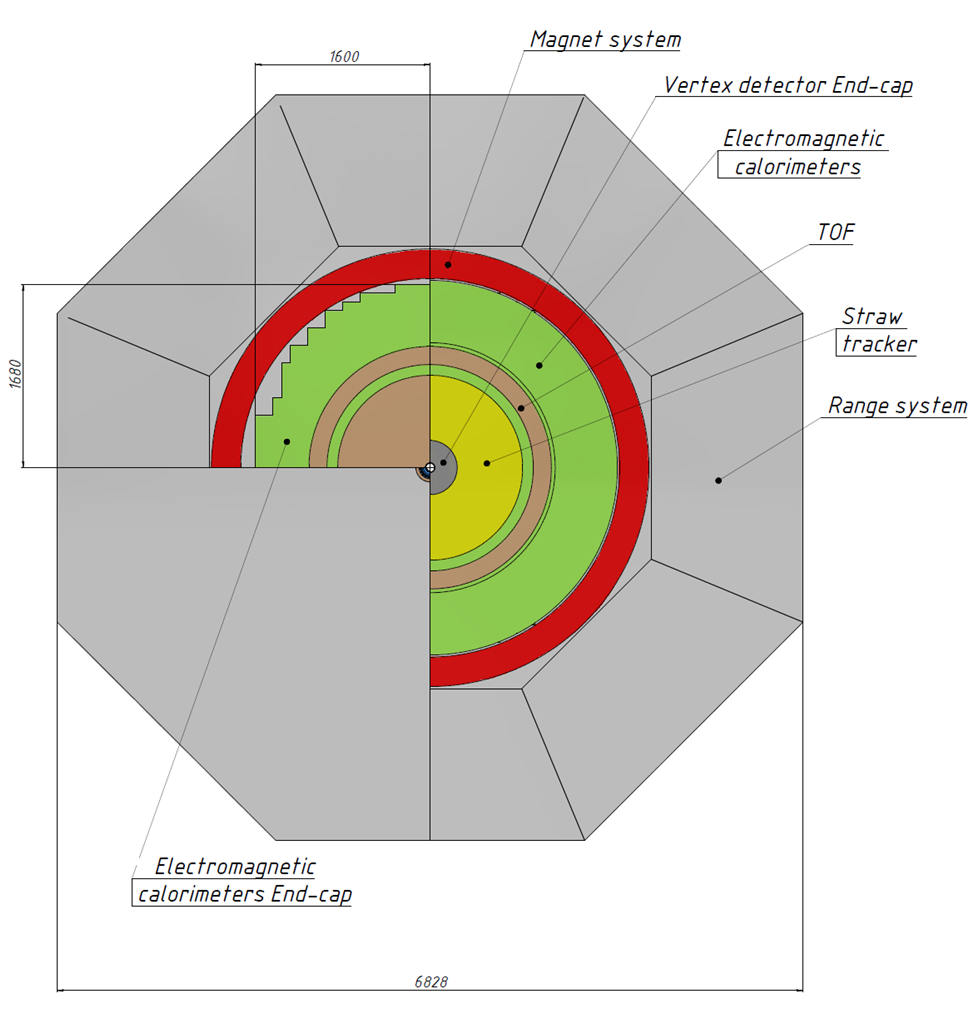}
  \end{center}
  \caption{Schematic cross-sectional view of the SPD detector from the end with dimensions specified in millimeters.}
\label{dims2}
\end{figure}

The detector consists of a barrel part and two end-caps. A pair of Zero Degree Calorimeters should be installed far enough from the interaction point to be surrounded by a beam-line equipment. A scheme of the detector in the full configuration with basic dimensions is shown in Figs. \ref{dims1} and \ref{dims2}.

\begin{table}[h]
\caption{Main parameters of the SPD setup.}
\begin{center}
\begin{tabular}{|l|c|c|}
\hline
                            &  Stage I  & Stage II \\
\hline  
Maximum luminosity, $10^{32}$ cm$^{-2}$ s$^{-2}$ & up to 0.1 & 1 \\
Interaction rate, MHz & up to 0.4 & 4 \\
Magnetic field at IP, T  &  up to 1.0  & 1.0 \\
Maximum stored magnetic field energy, MJ & 21 & 21 \\
Track momentum resolution $\frac{\delta p}{p}$ at 1 GeV/c, \% &  $\sim$1.7    &  $\sim$1.0\\
Photon energy resolution, \%  &  & 5/$\sqrt{E}\oplus 1$ \\
$D^0\to K\pi$ vertex spatial resolution, $\mu$m  &   &  60 for MAPS  \\
                                                          &         &   80 for DSSD \\  
PID capabilities & $dE/dx$, RS &$dE/dx$, ECal, RS, TOF, FARICH \\
Number of channels, $10^3$ & 177  & 280 for MAPS SVD \\
                                               &         & + 6.2$\times$10$^9$ pixels \\
                                                     &         &   608 for DSSD SVD\\  
Raw data flow, GB/s      &  up to 1  & up to 20   \\
Total weight, t         &  1249$^*$     & 1253      \\   
Power consumption$^{**}$, kW &  75  & 109  for MAPS  SVD\\
                                                     &        &  89 for DSSD SVD\\  
\hline
\end{tabular}
\end{center}
{\small *ECal mock-up of similar weight will be used for the first stage\\
**Without ZDC because it is infrastructurally weakly connected to the main setup.\\}
\label{tab_main}
\end{table}%
\newpage
The main parameters of the SPD setup as a whole, expected resolutions, and the parameters of each subsystem are listed in Tables.\ref{tab_main},  \ref{tab_sub_res} and   \ref{tab_sub}, respectively.

\begin{table}[h]
\caption{Spatial, time and energy resolution, as well as, signal length for each of the subsystem.}
\begin{center}
\begin{tabular}{|l|c|c|c|c|c|}
\hline
Detector  & Spatial resolution & Time resolution & Energy resolution & Signal length \\
 \hline
 RS          &    3 mm (wires), 1 cm (strips)  &     150 ns          &      90\%/$\sqrt{E}$ (p, n) & 250$\div$500 ns \\
 ECal     &     5 mm ($\gamma$, 1 GeV)  &  1 ns  &  5\%/$\sqrt{E}\oplus 1\%$ & \\
 TOF       &    10 cm                  &     50 ps            &     -- & \\
 FARICH  &           &       $<$1 ns             & $d\beta/\beta<10^{-3}$     & 10 ns \\
 Straw     &    150 $\mu$m      &       $\lesssim$100 ns         &     $8.5\% (dE/dx)$   &  120 ns\\
 SVD MAPS &   5 $\mu$m       &        --                 &     --  &\\
 SVD DSSD &   27.4 $\mu$m ($\phi$) &        --                &    --   & \\
                &   81.3 $\mu$m  ($z$)      &   &  & \\
MCT & 150   $\mu$m &      10 ns           &    --  &  $\sim$ 300 ns\\
 BBC    &   $\sim$ 10 cm  &    400 ps        &      --   & \\
  BBC MCP  &   1.5 mm     &       50 ps        &     --   & \\
 ZDC             &    $\sim$  1 cm   &     150 ps at 0.4 GeV        &    50\%/$\sqrt{E}\oplus 30\%$ ($n$) & \\
                     &                            &                       &    20\%/$\sqrt{E}\oplus 9\%$ ($\gamma$) & \\ 
 \hline                           
\end{tabular}
\end{center}
\label{tab_sub_res}
\end{table}%

\begin{table}[h]
\caption{Main parameters of the SPD setup subsystems.}
\begin{center}
\begin{tabular}{|c|c|c|c|c|c|c|c|}
\hline
Subsystem  & Stage & Main task & Active element & Weight, & Power, & Channels, \\
                    &           &                  &                         &       t          &   kW    & 10$^3$   \\
\hline
\hline
{\normalsize Range  System  (RS)}       & I+II  & {\normalsize $\mu$-ID}  & {\normalsize mini drift tubes} & 927 & 47 & 137.6\\
             &        &                  & {\normalsize Ar:CO$_{2} 70:30$} &        &       &      \\
\hline
{\normalsize Electromagnetic  }    & II     &{\normalsize  $\gamma$ detection}& {\normalsize Pb/scintillator} & 82 & 8 & 27.2 \\
{\normalsize Calorimeter (ECal) }     &   &   & shashlyk  &  &  & \\
\hline
{\normalsize Time-of-Flight   }    &  II     &{\normalsize  PID}       & {\normalsize RPC chambers} & 4 & 4 & 12.2\\
{\normalsize  system (TOF) }            &        &                  & {\footnotesize C$_2$H$_2$F$_4$:C$_4$H$_{10}$:SF$_6$ }&        &       &      \\
                                                &  &    & 90:5:5  &  &  & \\
\hline
{\normalsize Focusing Aerogel } &  II    &  {\normalsize PID }      & {\normalsize aerogel }   &  0.3 &  0.8  & 70\\
{\normalsize RICH (FARICH)}  &   &   &  &  &  & \\
\hline
{\normalsize Straw Tracker (ST) }   & I+II     &{\normalsize  tracking, PID} & {\normalsize straw tubes} & 0.2 & 4 & 30.5\\
               &        &                  & {\normalsize Ar:CO$_{2}$ 70:30} &        &       &      \\
\hline
{\normalsize Silicon Vertex}  &   &   &  &  &  & \\
{\normalsize Detector (SVD)}  &   &   &  &  &  & \\
{\normalsize ~~~~~ -- MAPS }   & II    & {\normalsize vertex, tracking} &{\normalsize  Si pixels } & $<0.1$ &22 & 6.2$\times$10$^9$ \\    
                                                   &                                                          &                                      &             & &     & pixels \\
{\normalsize ~~~~~ -- DSSD}  & II    & {\normalsize vertex, tracking} &{\normalsize  Si strips}  & $<0.1$ &2 & 327.7\\ 
\hline
{\normalsize Micromegas-based} & I &{\normalsize  tracking} & {\normalsize gas chambers} & $<0.1$& 1 & 5.4 \\
{\normalsize Central Tracker }              &        &                  & {\normalsize Ar:C$_4$H$_{10}$, } &        &       &      \\
         (MCT)      &        &                  & {\normalsize  90:10 }&        &       &      \\               
\hline
{\normalsize Beam-Beam }  &  I+II  & {\small local polarimetry,}   &  &  &  & \\
{\normalsize  Counter (BBC)}   &  & {\small luminosity control,} & {\normalsize scintillator} & 0.1 & 0.5 & 1 \\
{\normalsize }      &   & {\small  timing}  &  &  &  & \\
\hline
{\normalsize BBC MCP}  & I+II  & {\small local polarimetry}  &  &  &  & \\
{\normalsize }  &   & {\small  luminosity control,}  & {\normalsize MCP} & $\ll 0.1$ & $\ll 1$ &  0.1 \\
{\normalsize }      &   & {\small  timing}  &  &  &  & \\
\hline
{\normalsize Zero Degree}   &  I+II  & {\normalsize $n$,$\gamma$} & {\normalsize W/scint.} & 0.3 & 2&  2 \\
{\normalsize Calorimeter (ZDC)}     &  &  {\normalsize detection}  &  &  &  & \\
\hline
\hline
{\normalsize Magnet} &  I+II     &                       &         &  20  &   23 &   \\
\hline
\hline
{\normalsize Support  } &   I+II    &                       &         &  80.3   &    &   \\
{\normalsize and Transportation}  &   &   &  &  &  & \\
{\normalsize System (STS)}  &   &   &  &  &  & \\
\hline
\hline
{\small Top platform (loaded) } &   I+II    &                       &         &  40   &    &   \\
{\small Side platform (loaded) } &   I+II    &                       &         &  100   &    &   \\
\hline
\end{tabular}
\end{center}
\label{tab_sub}
\end{table}%

\newpage
\section{Operation conditions}
\begin{figure}[!h]
  \begin{center}
    \includegraphics[width=0.86\textwidth]{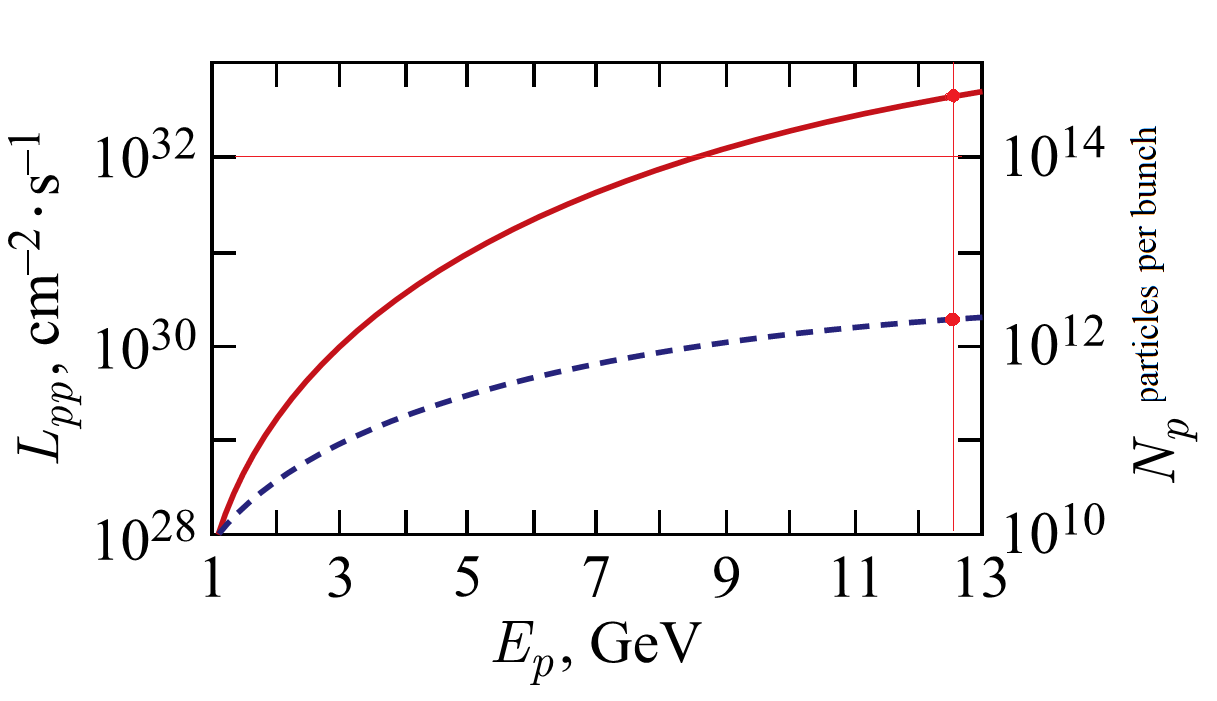}
  \end{center}
  \caption{Normalized dependence of the  luminosity $L_{pp}$ (the red curve and the left scale) and the beam intensity $N_p$ (the blue curve and the right scale) on the proton kinetic energy in the $p$-$p$ collision \cite{Meshkov:2019utc}..}
  \label{Lum}
\end{figure}

\begin{figure}[!h]
  \begin{center}
    \includegraphics[width=0.86\textwidth]{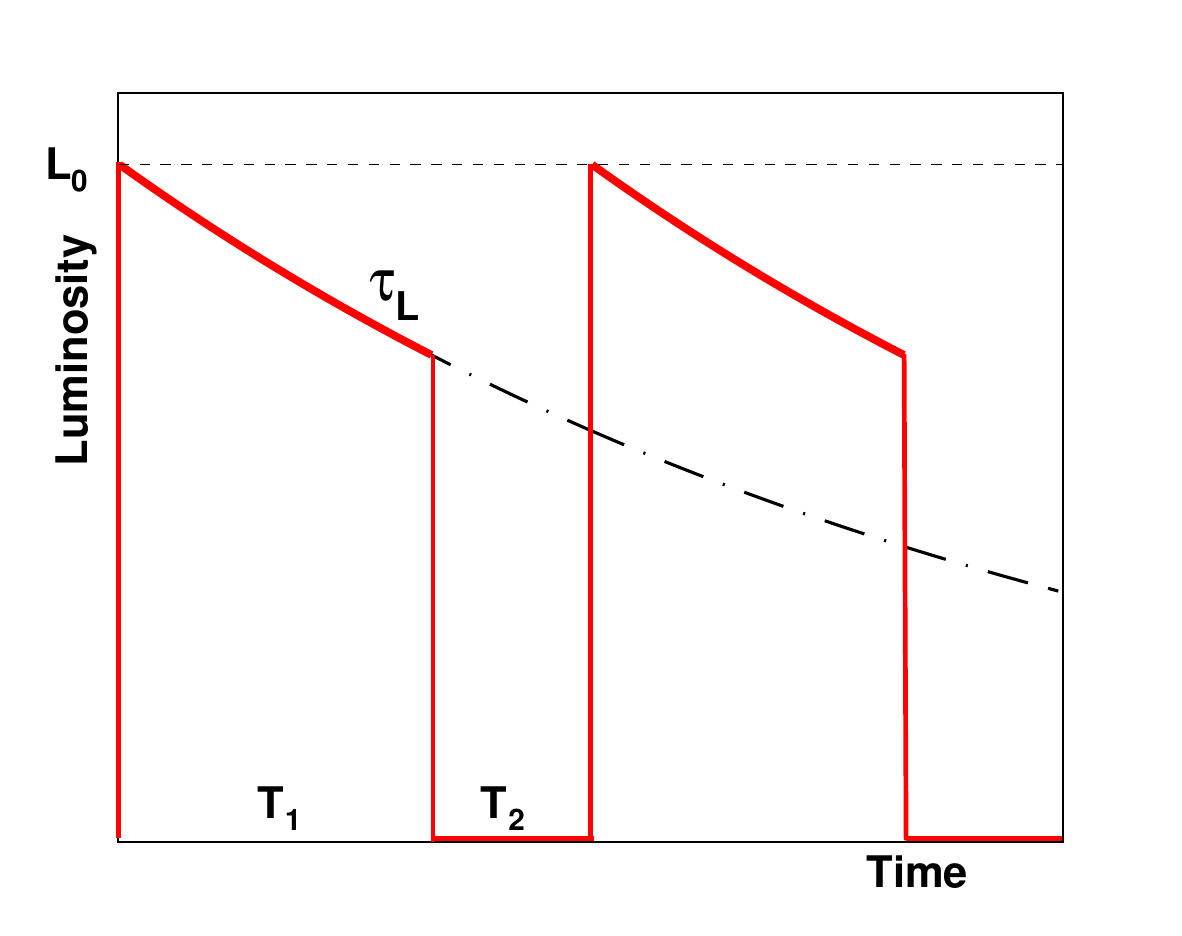}
  \end{center}
  \caption{Cycle of the collider operation.}
  \label{dims3}
\end{figure}

The dependence of the $p$-$p$-collision maximal luminosity and the intensity of proton beams on the CMS energy is presented in Fig. \ref{Lum} \cite{Meshkov:2019utc}.
The expected cycle of the collider operation is presended in Fig. \ref{dims3}.  After the beginning of the beam collisions at a maximal luminosity $L_0$, the data taking continues for a period $T_1$ that is much lower than the luminosity decay time $\tau_L$ (which, in turn, is much longer than the polarization decay time $\tau_P$). During this time the RF system of the collider provides an acceptable longitudinal size of the colliding bunches. After that, the existing beams are dumped and the accelerator complex spends time $T_2$ to accelerate and store a new portion of particles. Assuming $T_1=2$ hours, $T_2=1$ hour and  $\tau_L=6$ hours, the effective luminosity $L_{eff}$ averaged over the cycle is about $0.6\times L_0$.

\chapter{Range (muon) System}
\section{General description}
The Range System (RS) of the SPD detector serves the following purposes:
(i) identification of muons in the presence of a significant hadronic background and 
(ii) estimation of hadronic energy (coarse hadron calorimetry).
It is important to stress that the Range System is the only subdetector of the main part of the SPD setup, which can identify neutrons (by combining their signals with the electromagnetic calorimeter and the inner trackers). Muon identification (PID) is performed via muonic pattern recognition and further matching of the track segments to the tracks inside the magnets. The precise muon momentum definition is performed by the inner trackers in the magnetic field. The Mini Drift Tubes (MDT) \cite{Abazov2010, Abazov20101} are used in the Range System as tracking detectors, providing two-coordinate readout (wires and strips running perpendicularly). Such readout is mostly needed for the events with high track multiplicity and also for the reconstruction of the neutron spatial angle.   

One of the main physics goals of the Range System is the identification of muons from
$J/\psi\to\mu^+\mu^-$ decays contaminated by misidentified pions and their decays.

As for the design and construction of the present system, we hope to capitalize on the experience gained by the JINR group in the development of the PANDA (FAIR, Darmstadt) Muon System \cite{MUON_TDR}. These two systems (PANDA and SPD), dealing with muons of comparable momentum ranges and solving the same PID tasks, should look very similar in their design and instrumentation.

\section{System layout}
\begin{figure}[!h]
  \begin{minipage}[ht]{0.49\linewidth}
   \center{\includegraphics[width=0.8\linewidth]{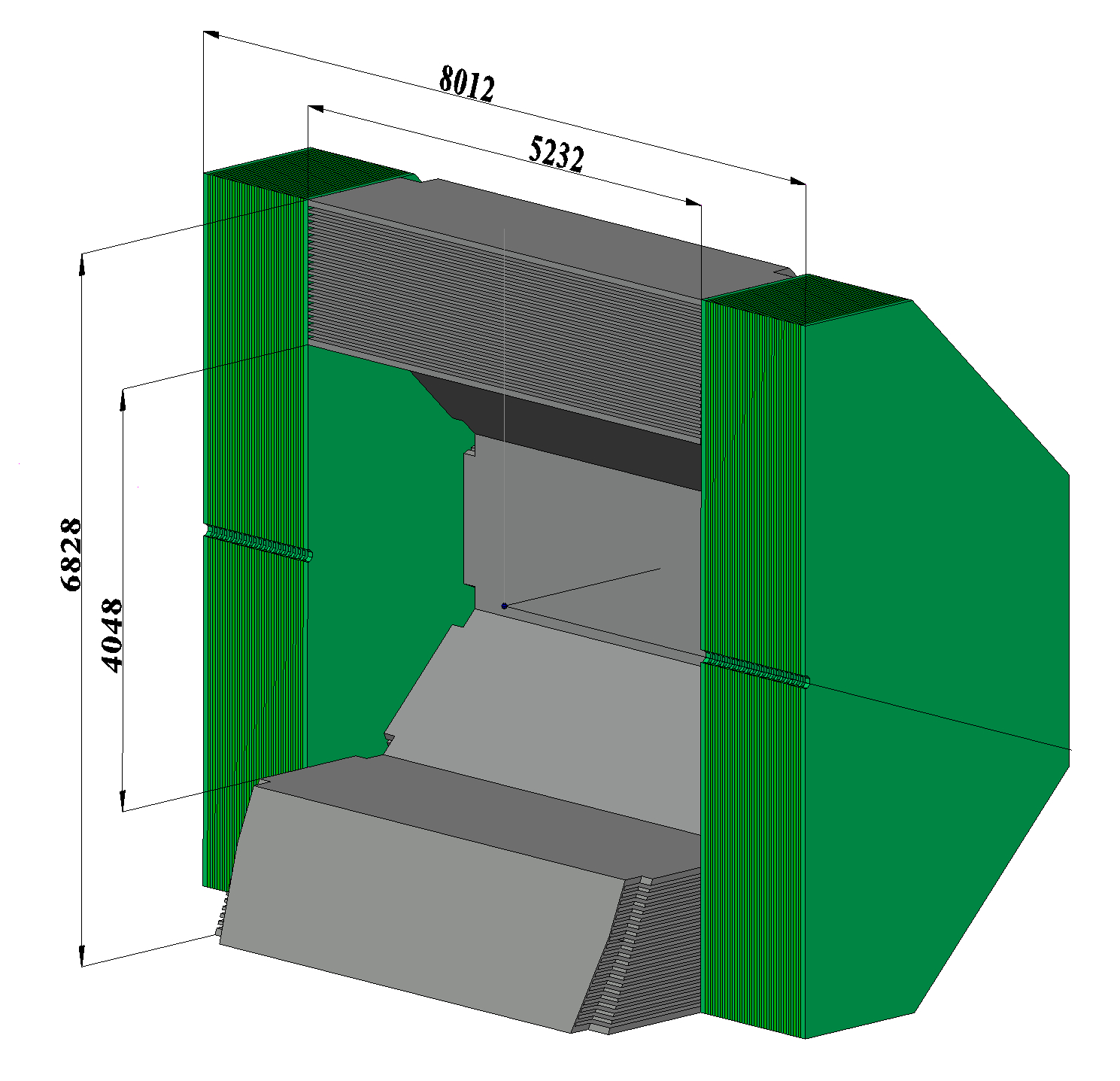} \\ (a)}
  \end{minipage}
  \hfill
  \begin{minipage}[ht]{0.49\linewidth}
    \center{\includegraphics[width=1.0\linewidth]{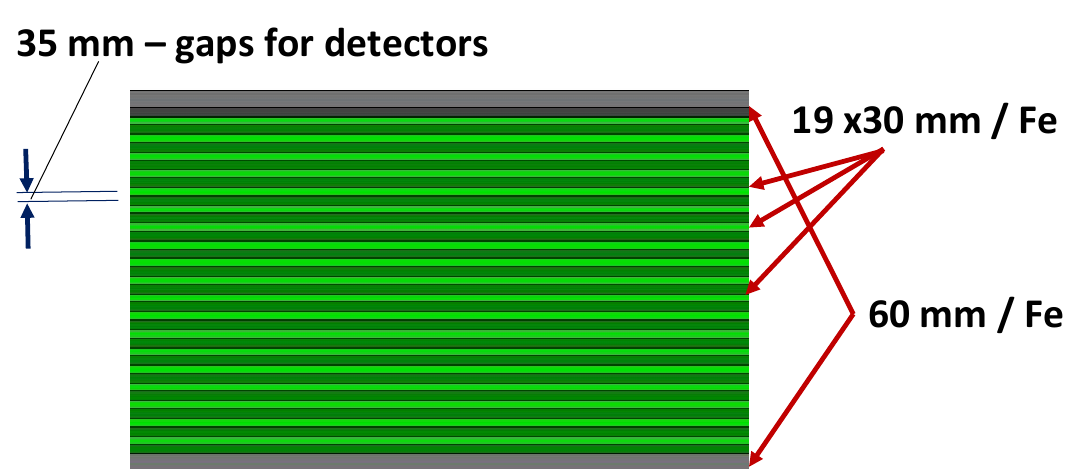} \\ (b)}    
  \end{minipage}
  \caption{3D view (half cut)  and sizes (in mm) of the Range System: (a) Barrel is shown in grey, End Cap disks $-$ in green; (b) absorber structure.}
  \label{fig:RS_layout} 
\end{figure}

The Range System serves as an absorber for hadrons and a ''filter'' for muons. It also forms the magnet yoke. It consists of a Barrel and two End-Caps (ECs). The schematic 3D view of the system and its main  sizes are shown in Fig~.\ref{fig:RS_layout} (a). 
The absorber structure is shown in Fig. \ref{fig:RS_layout} (b). 
The two outer 60-mm steel layers are used for bolting the modules together, and for attachment of internal detectors (inside) and external service devices (outside). The 30-mm thickness of the main absorber plates (nineteen in total) is selected as comparable with muon straggling in steel, thus giving the best possible muon-to-pion separation, and also providing a rather good sampling for hadron calorimetry. 
The interlayer gaps of 35 mm are taken for the reliable mounting of the detecting layers (twenty in total) comprising of the MDTs, the stripboards, and the front-end electronic boards  on top of them, and corresponding cables.

The Barrel consists of eight modules, and each  End Cap consists of two halves (sashes) divided vertically, each sash in its turn is also divided into two equal parts in a vertical plane running perpendicular to the beam direction. Such subdivision of the
system (16 parts in total) is chosen to optimize its further assembly and to satisfy the construction
requirements of the SPD experimental hall (80-ton crane capacity). 
The total weight of the system is about 1006.4 tons, including 37 tons of detectors. The total number of MDT detectors is about 10000 units. The MDTs are deployed in the following way: along the beam direction in the Barrel, and perpendicular to the beam (horizontally) in the End Caps. 
 
The absorption thicknesses of the Barrel and End Caps are selected to be equal to 4 nuclear interaction lengths ($\lambda_I$) each. It provides a rather uniform muon filtering in all directions. Together with the thickness of the electromagnetic calorimeter ($\sim$ 0.5 $\lambda_I$), the total thickness of the SPD setup is about 4.5 $\lambda_I$.

\section{Mechanical design simulation}
The calculations were made to evaluate the stress and displacement of the SPD detector 
(the Barrel and two End Caps with all internal detectors/systems),
together with the Support and Transportation System (STS), as well as to choose a rational design for STS, 
 wherein the Range System is fully included in the model (force diagram/scheme) of the entire design of the SPD detector. 
Ultimately, the total weight of the installation and its distribution over 3 pairs of support carts and additional supports in the case of open End Caps were determined.

The following cases were considered: the full SPD installation  in the operating position - the End Caps are closed, and the case with the EC sashes as far apart as possible. 
For control purposes, the cases with intermediate positions of the EC sashes were also considered.
To reduce the order of the calculation task, a double symmetry of the entire setup was used, that is, a quarter of the installation was considered.
The pictures demonstrating the results, which are presented below, were obtained using this double symmetry of the task (a quarter of the setup).

It was assumed that: 
\begin{itemize}
\item in the working position and while moving along the rails, the installation is based on three pairs of support carts, the weight of  one cart was assumed to be $\sim$680 kg;
\item the weight of all internal detectors and systems is about 100 tons in total, which is applied as a distributed load over the entire internal surface of the Barrel;
\item the weight of the upper platform with the equipment (cryogenics and electronics racks) located on it was taken as 40 tons, which are applied as a distributed load over the surface of the upper load-bearing elements of the SPD setup model;
\item in the case of the EC open sashes, the consoles of the STS power beams were additionally loaded (supported) on 2 pairs of supports (except for the carts).
\end{itemize}
\begin{figure}[!b]
  \begin{minipage}{0.99\linewidth}
    \center{\includegraphics[width=0.54\linewidth]{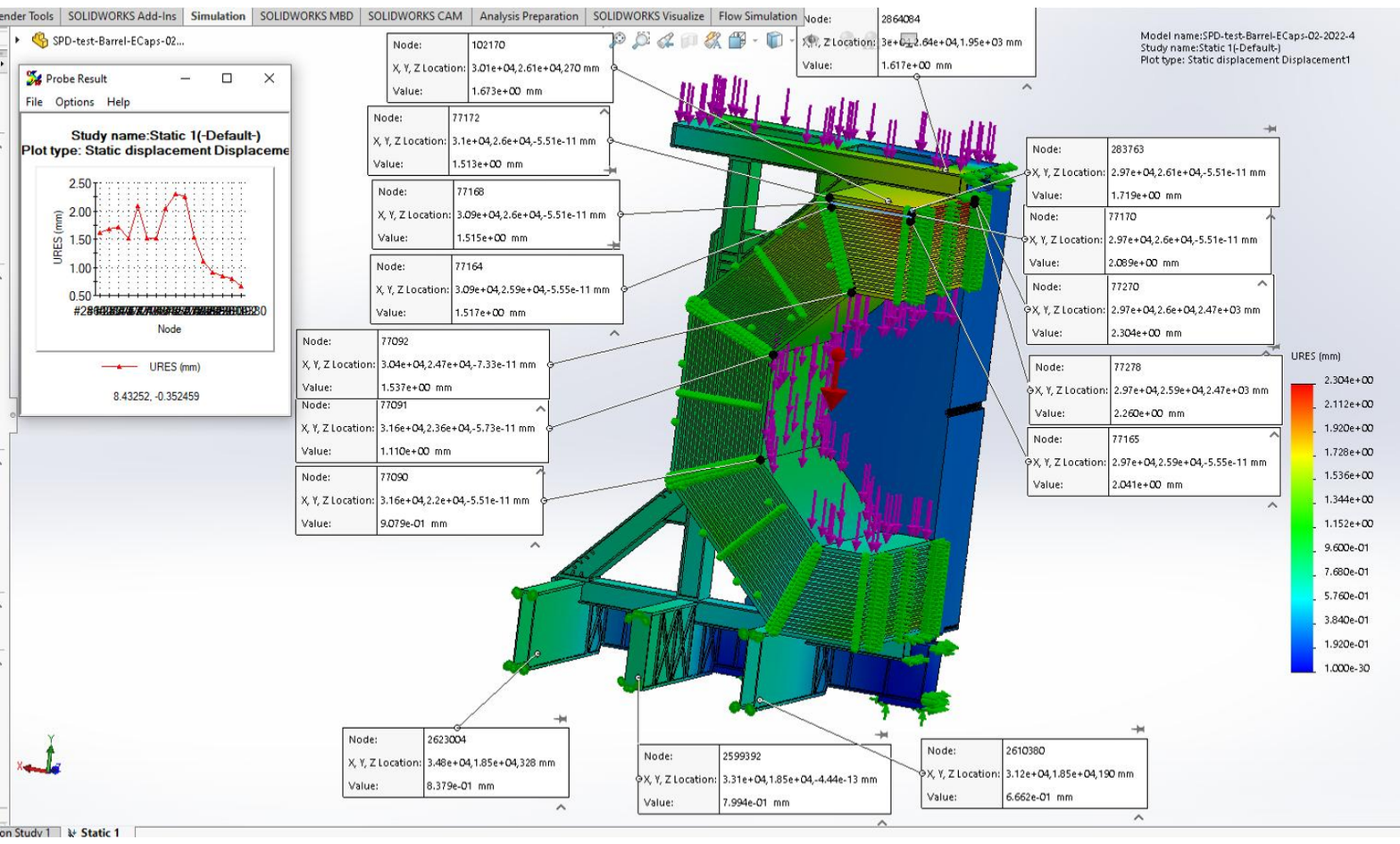} (a)}
    \center{\includegraphics[width=0.54\linewidth]{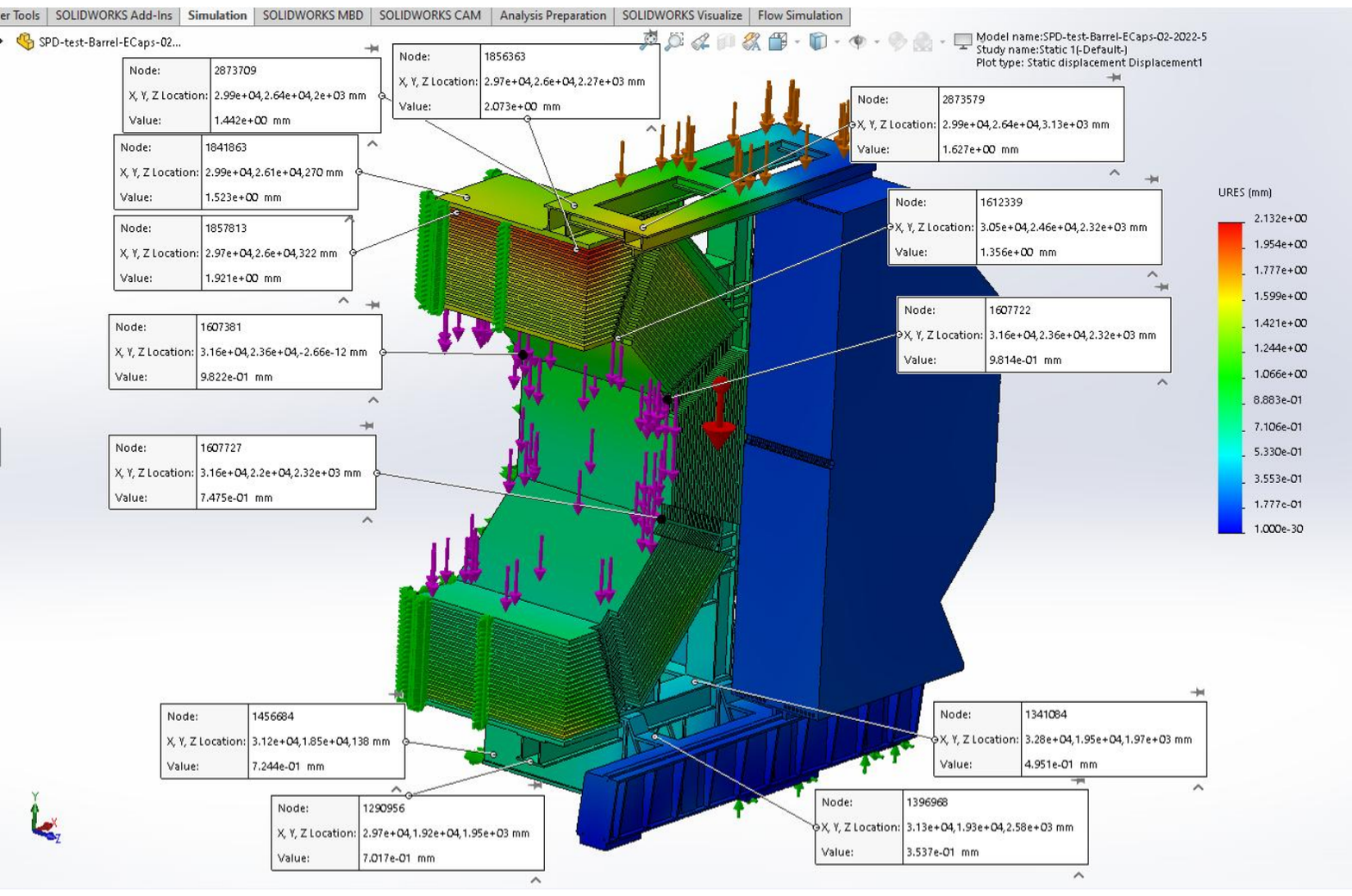} (b)}
    \center{\includegraphics[width=0.54\linewidth]{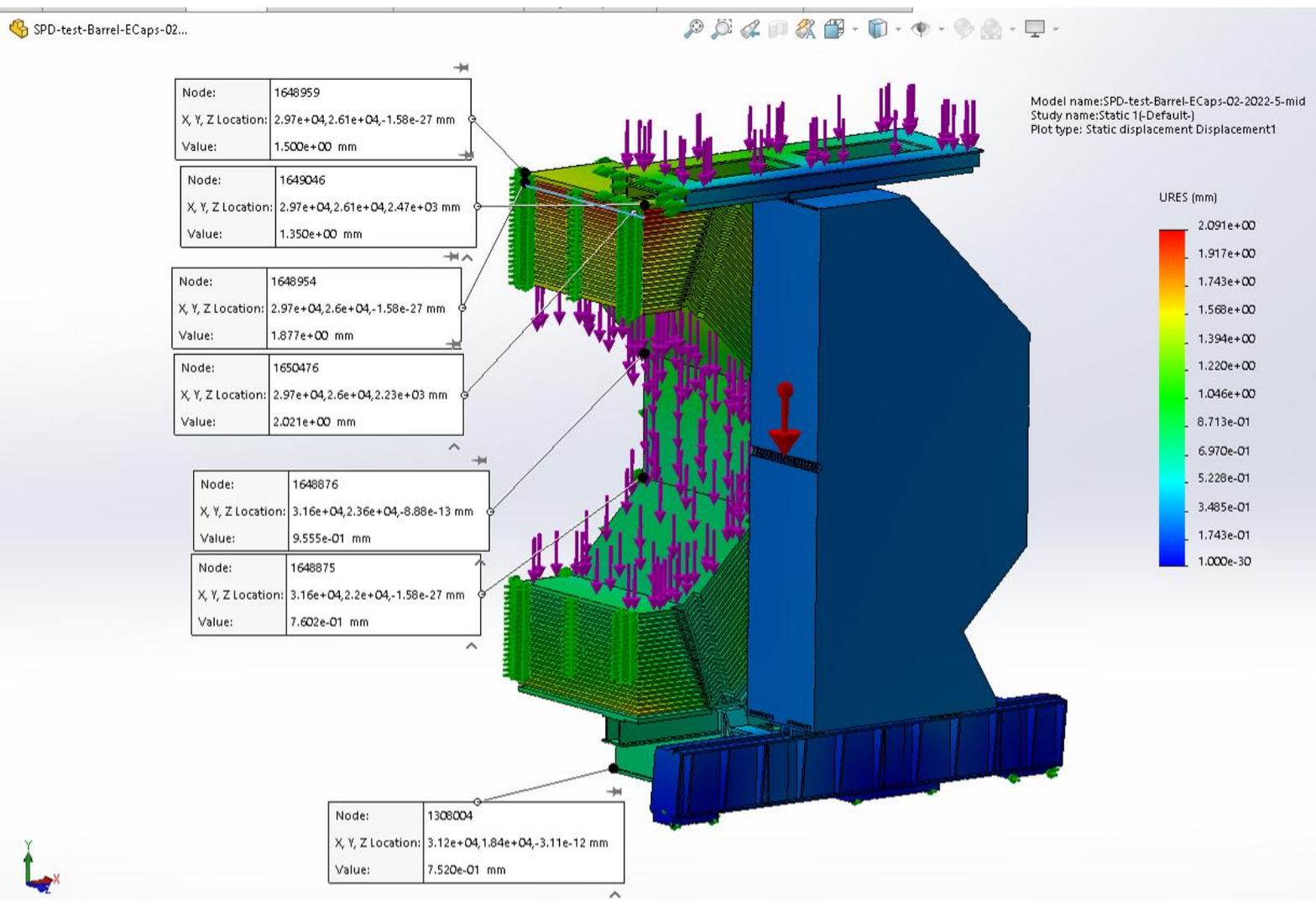} (c)}
  \end{minipage}
  \caption{Displacements of system elements for the case of fully assembled SPD detector (including Range System, all inner detectors, solenoid magnet, and the weight of the top platform with corresponding equipment) loaded on support and transportation system (six carts) (a), fully opened End Caps (b) and partially opened End Caps (c).}
  \label{fig:RS_Displace_1}  
\end{figure}
This simulation of the mechanical design takes into account only the forces of gravity. 

The maximum displacements (see Fig.~\ref{fig:RS_Displace_1}) are observed in the elements of the upper Barrel module in the region of the largest plates of the Barrel module and do not exceed 1.8 $\div$ 2 mm, which at this stage is considered as an acceptable result. 
In the elements of End Caps, displacements are insignificant,  due to the vertical arrangement of the metal plates of the structure.

\begin{figure}[!htb]
   \begin{minipage}{0.48\textwidth}
     \centering
     \includegraphics[width=0.96\linewidth]{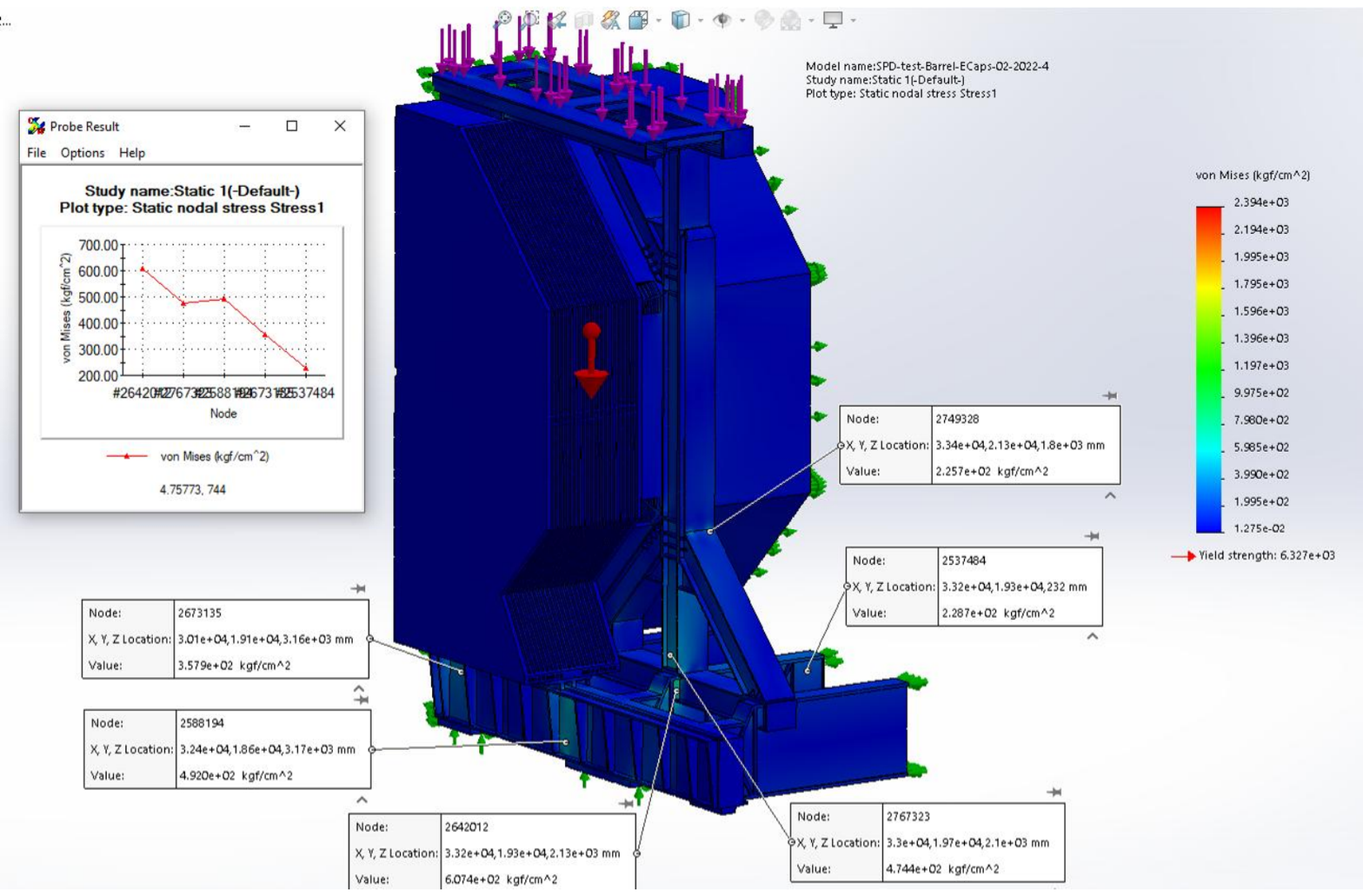} \\ (a)
   \end{minipage}\hfill
   \begin{minipage}{0.48\textwidth}
     \centering
     \includegraphics[width=1.04\linewidth]{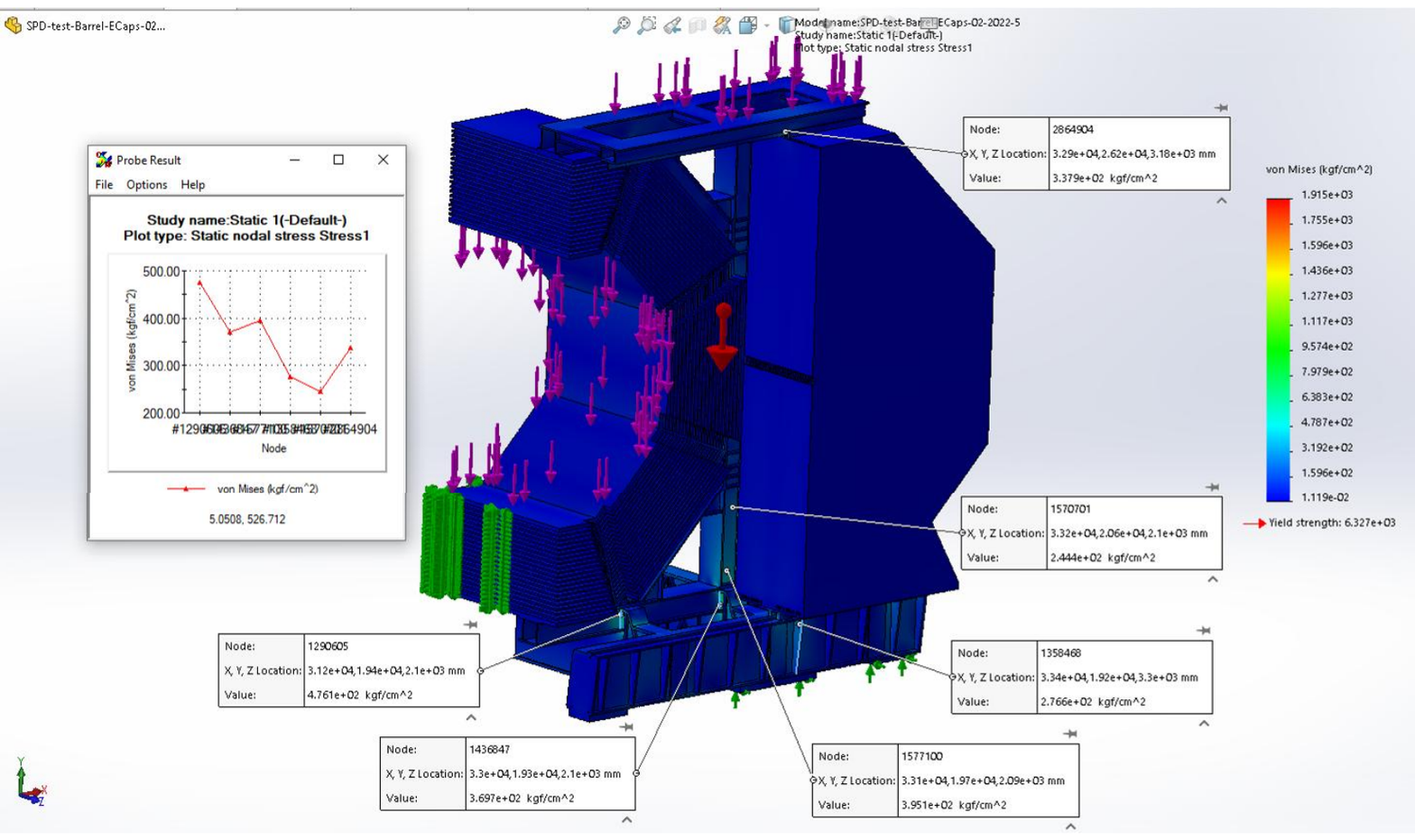} \\ (b)
   \end{minipage}
     \caption{Stress map of fully assembled SPD detector (a) and SPD detector with fully opened End Caps (b) Blue color demonstrates an acceptable level of stress.}
      \label{fig:RS_Stress_2}
\end{figure}

It should be mentioned that the stresses in all considered cases (see Fig. ~\ref{fig:RS_Stress_2}) correspond to low or medium values for the main material of the structure (usual construction steel), and do not exceed 600$\div$800 kg/cm$^{2}$.

A rational arrangement of the support cart pairs has been selected, which provides an approximately uniform distribution of the total weight of the SPD on these cart pairs. 
In the case of the assembly in the working position, the weight distribution is 36\% for the central pair and 32\% for each of the outer pairs.

In other cases, the maximum load on any pair of carts or additional support does not exceed 33.5\%.

The RS and STS weights (excluding (100 + 40) tons mentioned above) are shown in Table~\ref{RS:RSSTSweight}.

\begin{table}[!h]
\caption{The RS and STS weight.}
\begin{center}
\begin{tabular}{|l|c|c|c|c|c|c|c|}
\hline
SPD elements: & Barrel module & Barrel total & 1/2 EC & EC & 2 $\times$ EC & 6  carts & STS \\
\hline 
\hline
Weight, tons & 66.9 & 535.2 & 117.8 & 235.6 & 471.2 & 4.08 & 121  \\
\hline
\end{tabular}
\end{center}
\label{RS:RSSTSweight}
\end{table}

The total weight of the RS is about 1006.4 tons. The total weight of the fully constructed SPD detector: RS with all internal systems (electromagnetic calorimeter, solenoid, PID systems, inner tracker, etc.) together 
with top platform (including the corresponding equipment) and support and transportation systems equals to about 1271.5 tons.

\begin{figure}[!b]
  \begin{minipage}{0.99\linewidth}
    \center{\includegraphics[width=0.5\linewidth]{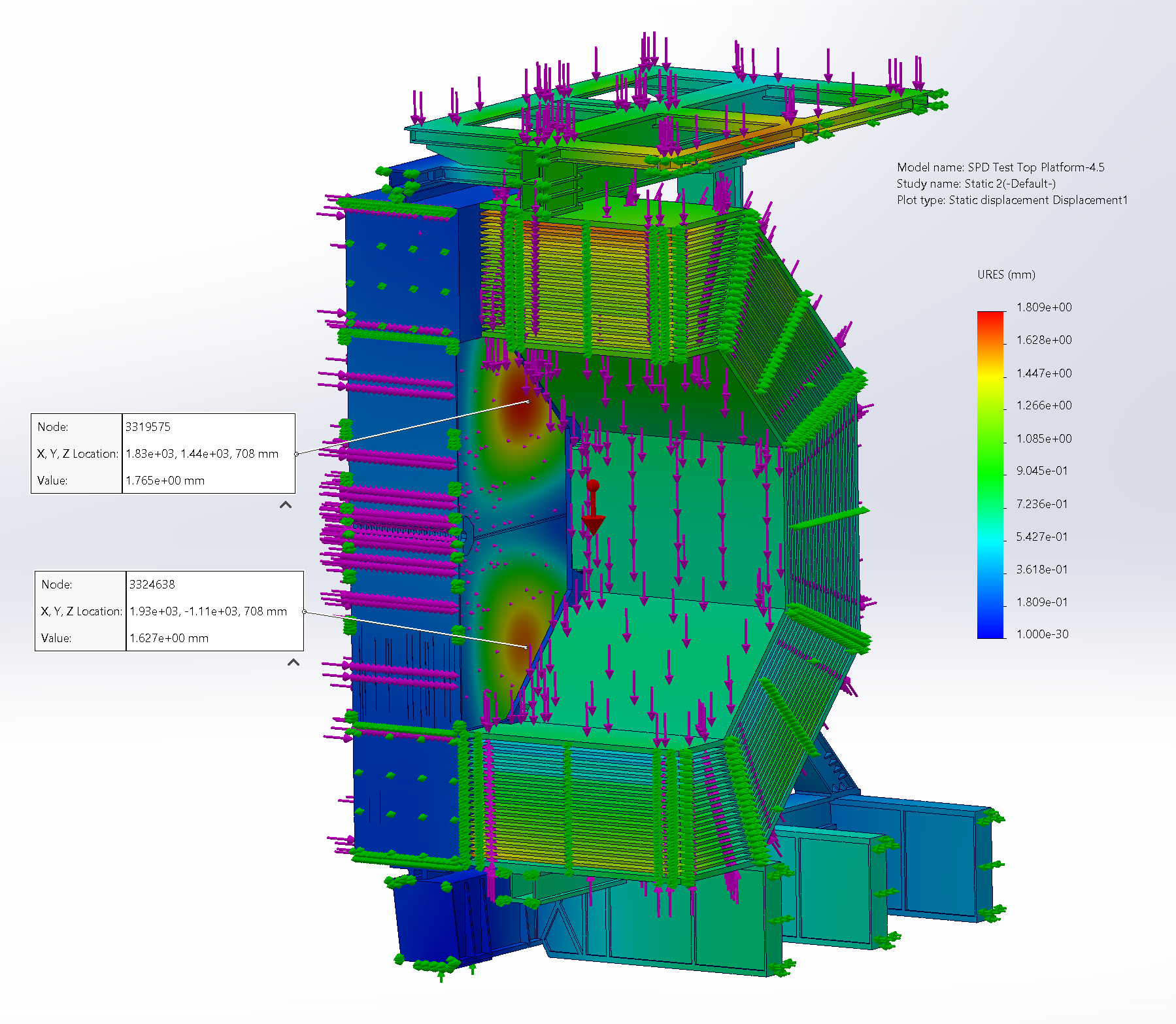} (a)}
    \center{\includegraphics[width=0.5\linewidth]{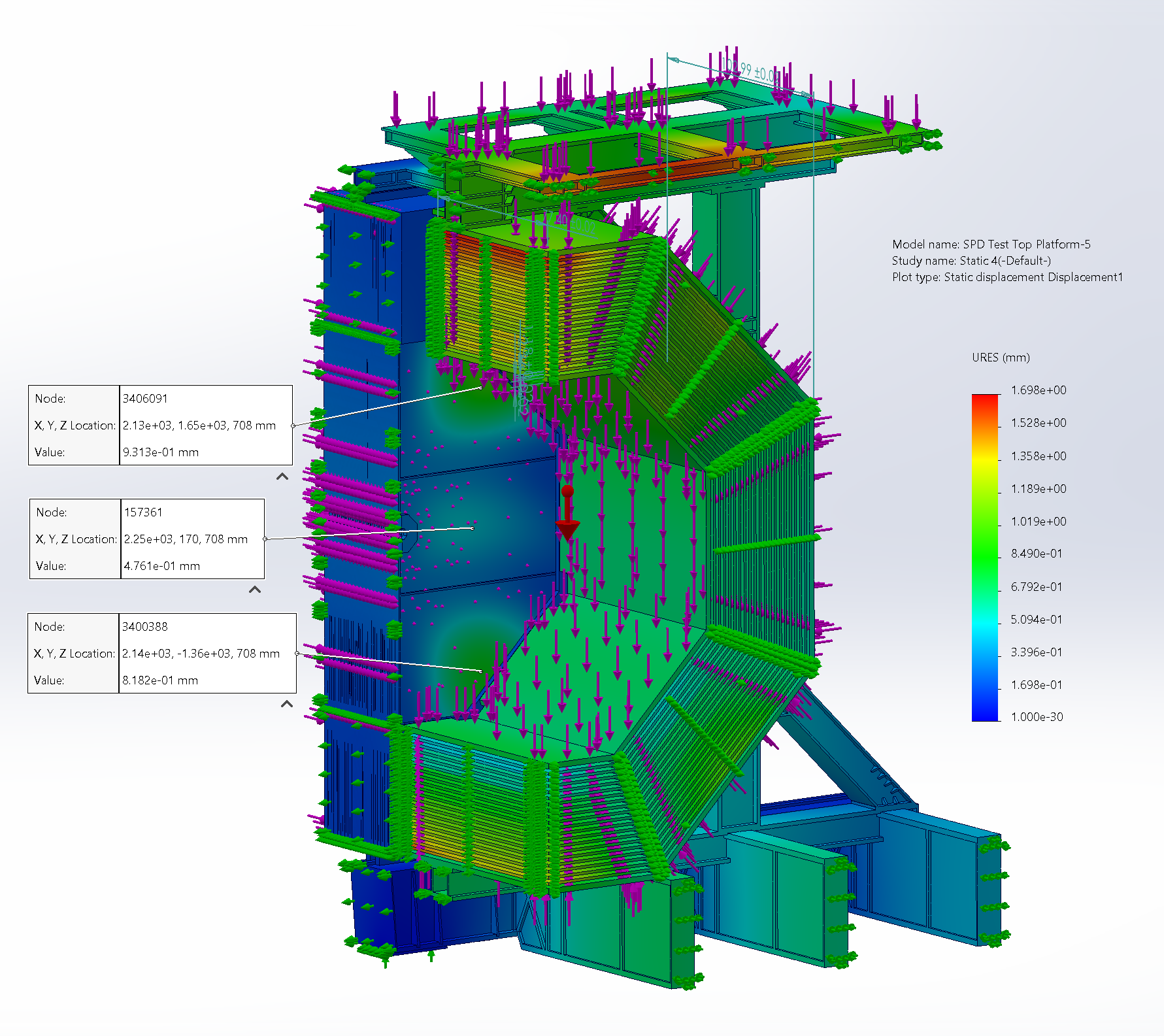} (b)}
    \center{\includegraphics[width=0.5\linewidth]{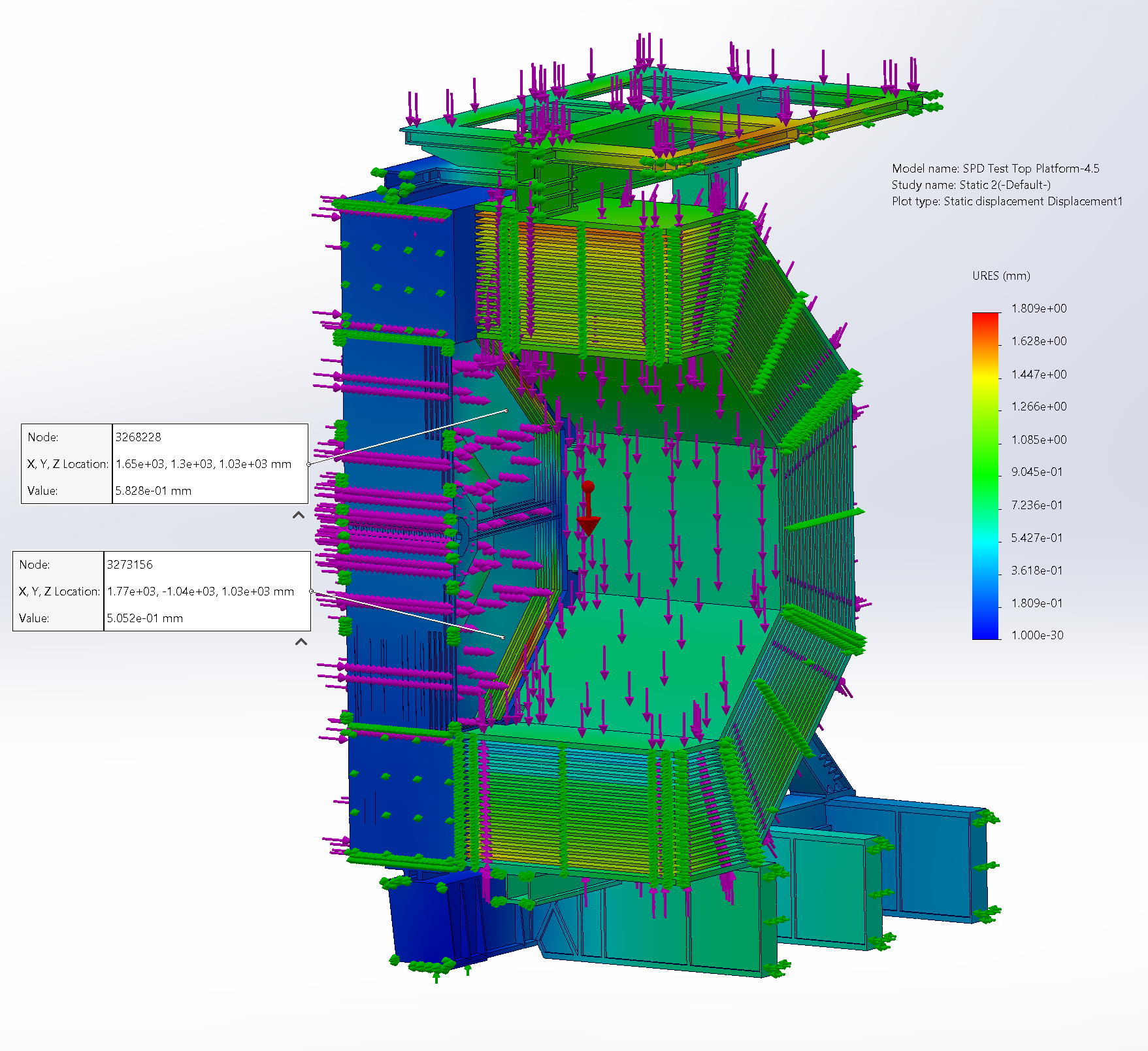} (c)}
  \end{minipage}
  \caption{Effect of magnetic field forces.}
  \label{fig:RS_MFF}  
\end{figure}

When the magnetic field of the SPD setup is turned on, the elements of the Range System will experience additional magnetic forces, since the Range System is the yoke of the Superconducting Solenoid of the assembly. 
Magnetic forces do not change the influence of gravity in any way and do not modify the results of the mechanical finite element analysis (FEA) given above. 
But these forces will have the greatest impact on the central (axial) part of the End Caps -- the field will push it inward (see Fig. ~\ref{fig:RS_MFF}). 
According to preliminary estimates, it will be possible to keep displacements within acceptable limits by simple design solutions presented below. 

The FEA  of magnetic forces (acting always in direction to the center of the magnet) is conducted with the map of these forces provided by the BINP (Novosibirsk) group at the field 1.2 Tesla. 
Figure~\ref{fig:RS_MFF} demonstrates the main results of this analysis. 
Additional mechanical stresses together with gravitational ones do not represent a challenge -- they are well within the elastic limit of the steel. 
But displacements of the steel plates of End Caps towards the center of the setup require special attention.

The initial design of the RS End Caps halves used two ribs (one vertical and one horizontal) crossed at the center of End Caps near the beam pipe. 
These ribs were composed of steel bars (35~mm by 20~mm in cross-section) welded in between the steel absorber plates. 
The rigidity of the planes composed of these ribs acts against magnetic forces (withstand/minimize the displacements). 
Unfortunately, the FEA of displacements under magnetic forces has shown rather big values at two spots (see Fig.~\ref{fig:RS_MFF} (a)). 
The sagitta of these two "bubbles" is about 2~mm. The picture shows the sagitta of the first thin (30 mm) steel layer. 
The first thick layer (60 mm) is removed from the picture to let see the next thin layer's behavior, its displacement is just about 0.3 mm. 
So, the gap width for mounting the detecting plane decreased by $\sim +0.3 - 2 = - 1.7$~mm! It is almost twice larger than our present limit of 1~mm (comparable with expected mechanical accuracy after manufacturing and assembly of RS modules) adopted as tolerance criteria for the gap size.

That is why, at the present level of understanding of the RS End Caps structure behavior under additional magnetic forces, it is decided to use two horizontal ribs instead of one to get a significant improvement of the construction stiffness and to preserve the gap width for detecting planes. 
Figure~\ref{fig:RS_MFF} (b) demonstrates the result of corresponding FEA calculations. It is seen that the sagittas of three spots are in the range of 0.5$\div$1~mm only, which is acceptable for our mechanical design.

But the design with two stiffness ribs adopted for the present TDR has also remarkable negative feedback in terms of the increased number of strips in the RS End Caps system. 
To minimize this effect in the future/final RS design we need to explore peculiarities of magnetic field distribution in our yoke. 
Its thickness is selected under the physical requirement of reliable muon-to-pion separation, which leads to 4~$\lambda_I$ (0.69 m of steel absorber, in our design it is limited by weight and size of SPD setup) rather than the technical requirement to return the magnetic flux of solenoid at 1.2~Tesla. 
This flux is being returned by just a few inner layers (for both, Barrel and End Caps). 
Figure~\ref{fig:RS_MFF} (c) demonstrates this effect: for the previous configuration of two crossed ribs (see Fig.~\ref{fig:RS_MFF} (a)) with the first 6 layers removed from the picture we have very small sagittas of the "bubbles" $\sim$ 0.5~mm, what indicates that field penetrates only in a few layers.

So, using the peculiarity of the magnetic field in the End Caps, it is possible to combine 1-rib and 2-ribs solutions to minimize the increase in a number of strip readout channels: the first few planes (out of 20 in total) should have 2 ribs and the rest -- just 1 rib. 
The optimal solution (depth of 2 ribs and distance between them) requires additional calculations with more detailed mapping of magnetic forces (distributed towards the solenoid). 
Worth mentioning that for the Barrel design influence of the field on the gap size may be neglected, but displacement of the inner 60 mm steel plates towards the magnet is remarkable ($\sim$ 2 mm for the uppermost Barrel module), which should be taken into account in the design of solenoid suspension system.


\section{Assembly of Range System}
\begin{figure}[!htb]
   \begin{minipage}{0.48\textwidth}
     \centering
     \includegraphics[width=1.\linewidth]{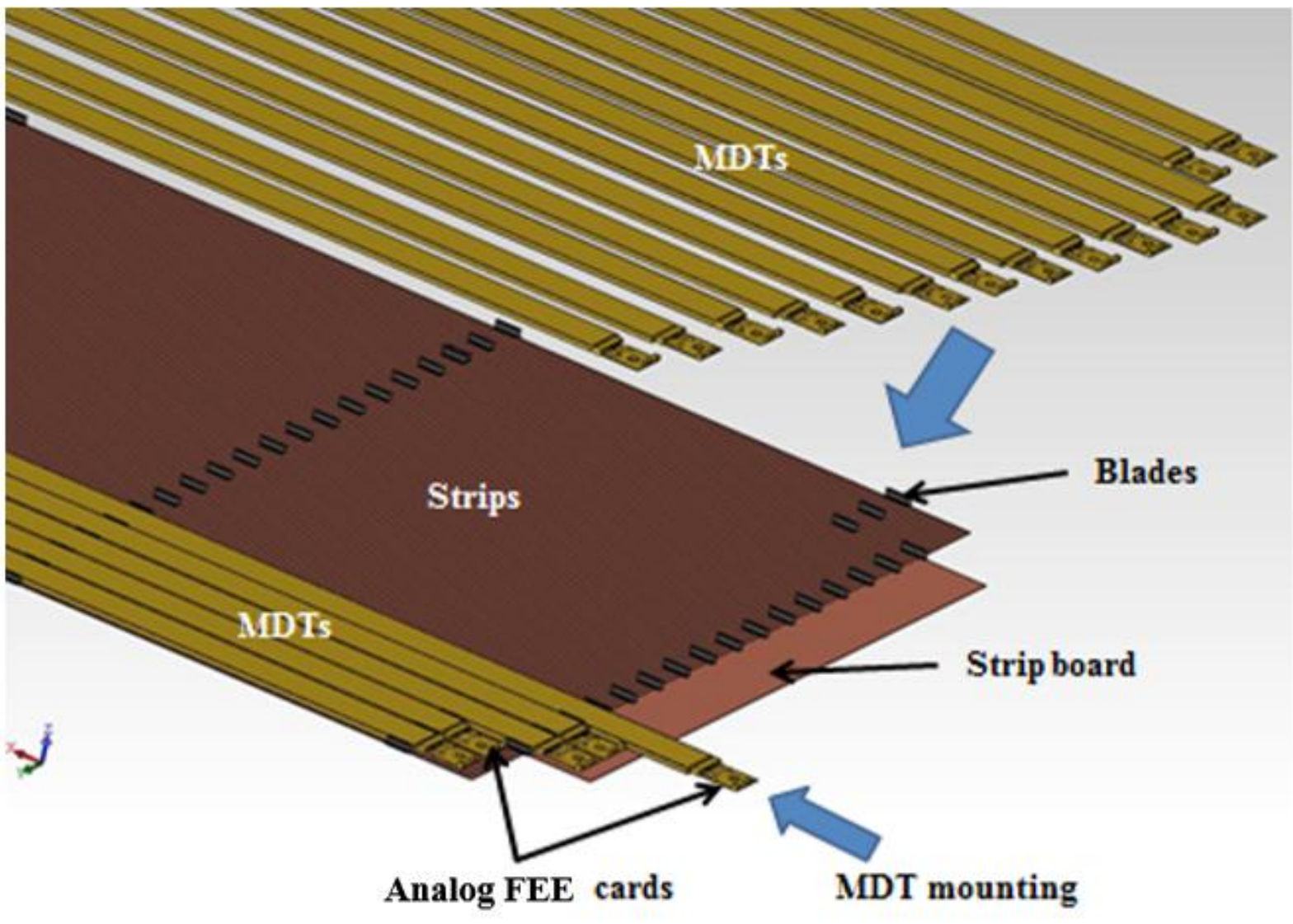} \\ (a)
   \end{minipage}\hfill
   \begin{minipage}{0.48\textwidth}
     \centering
     \includegraphics[width=1.\linewidth]{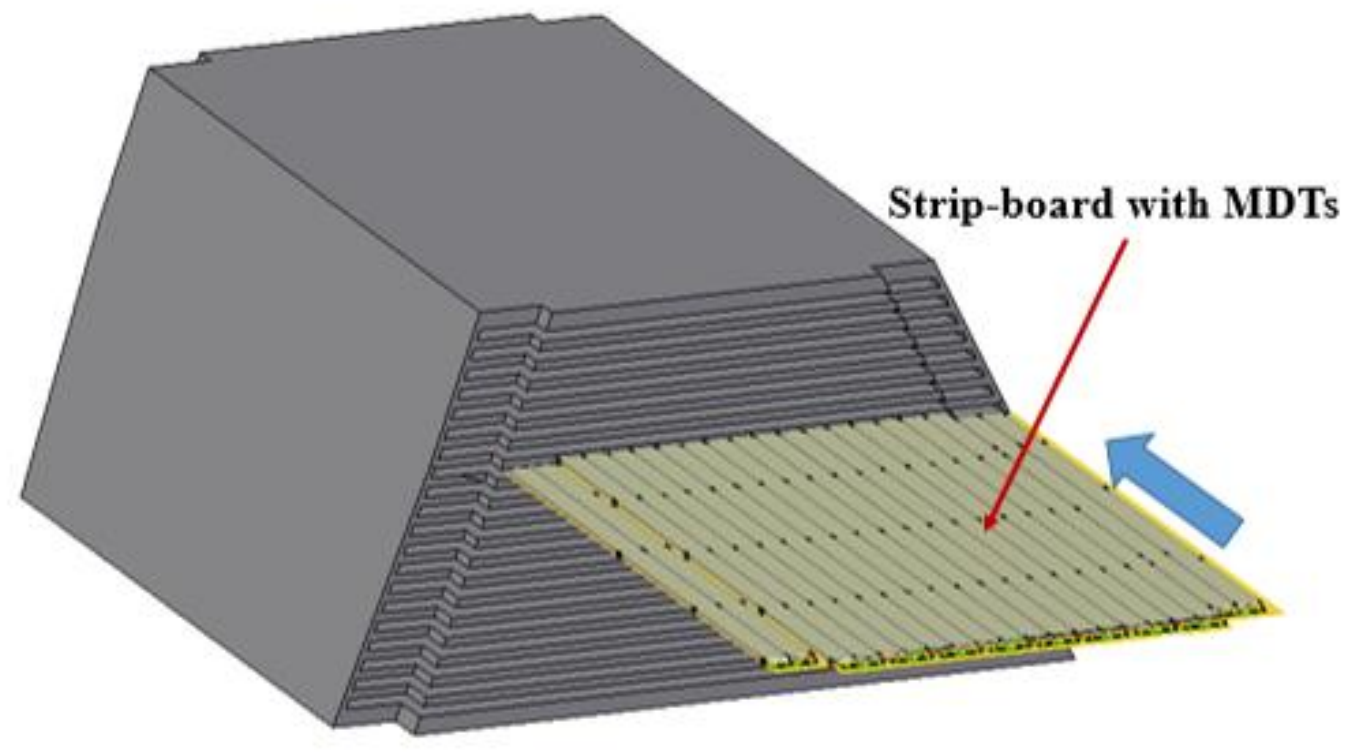} \\ (b)
   \end{minipage}    
    \caption{ (a) Mounting of MDT detectors on the stripboard, forming the detecting plane. 
    (b) Assembly of Barrel module with detecting layers (MDTs with stripboard and corresponding electronics).}
      \label{fig:RS_Assembly}   
\end{figure}

\begin{figure}[!htb]
    \center{\includegraphics[width=0.48\linewidth]{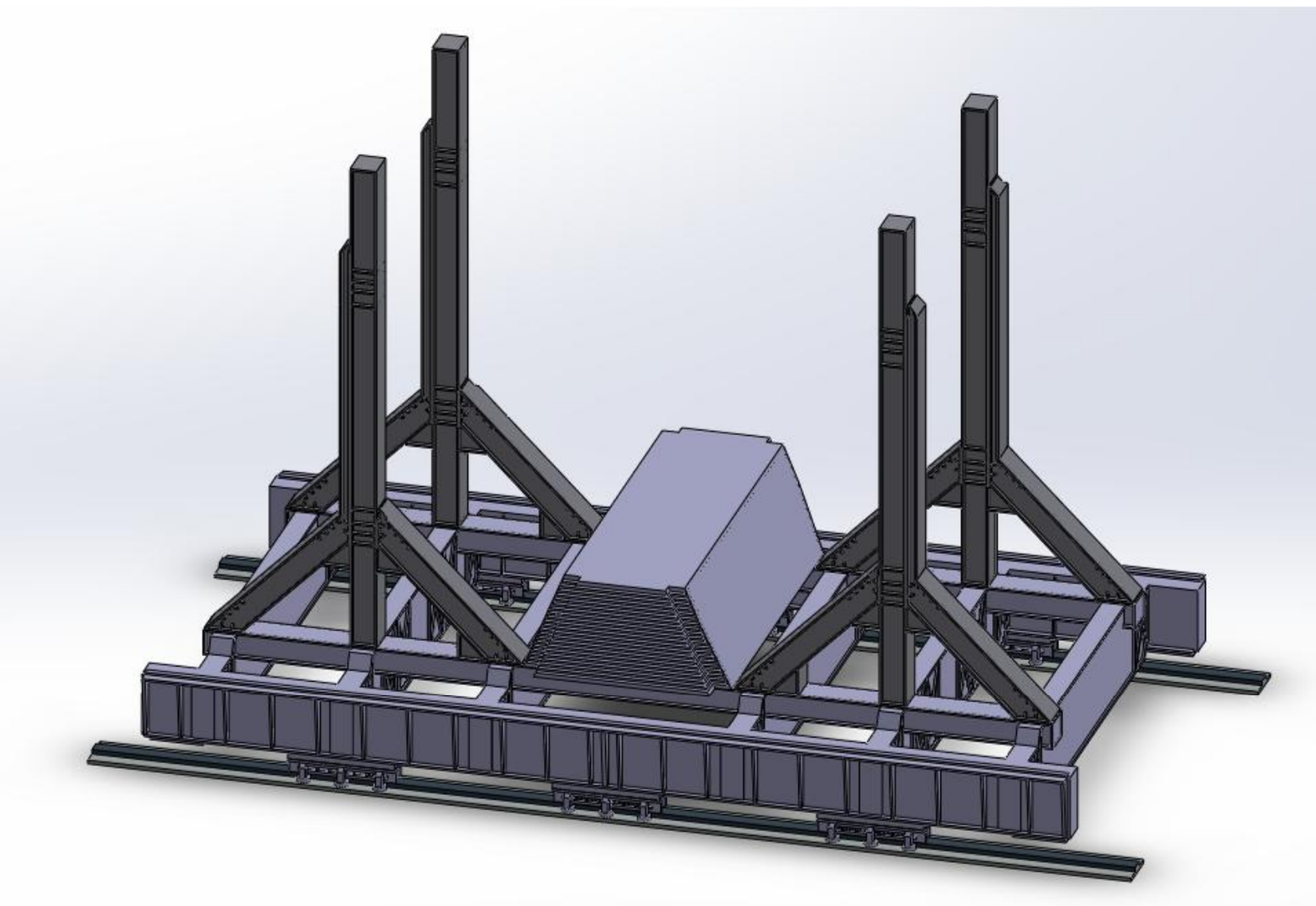}}
     \caption{Support structure of SPD detector with first Barrel module positioned on the transportation system.}
       \label{fig:RS_Support}
\end{figure}

\begin{figure}[!htb]
   \begin{minipage}{0.48\textwidth}
     \centering
     \includegraphics[width=1.\linewidth]{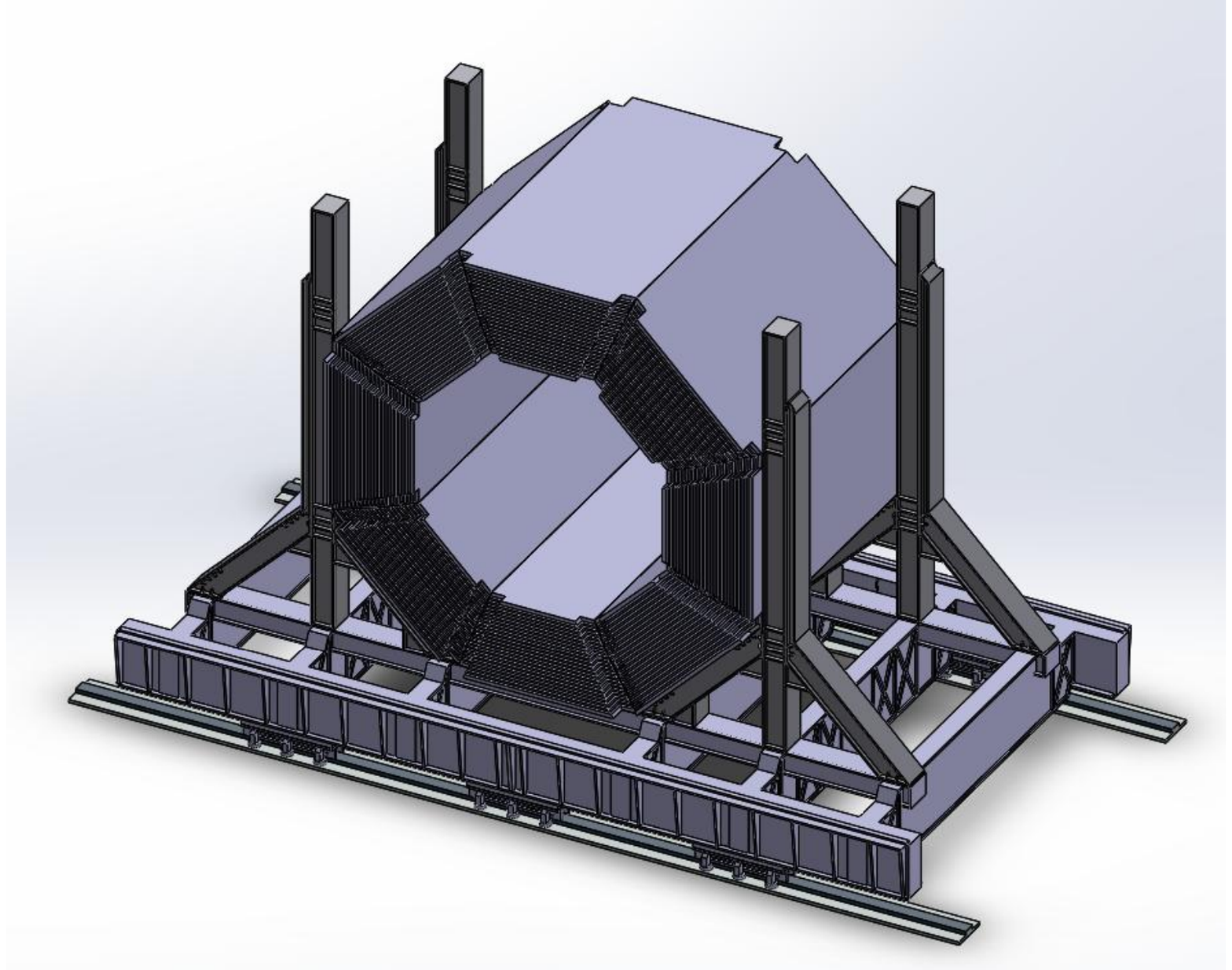}
     \caption{Fully assembled RS Barrel.}
      \label{fig:RS_Assembled_1}
   \end{minipage}\hfill
   \begin{minipage}{0.48\textwidth}
     \centering
     \includegraphics[width=1.\linewidth]{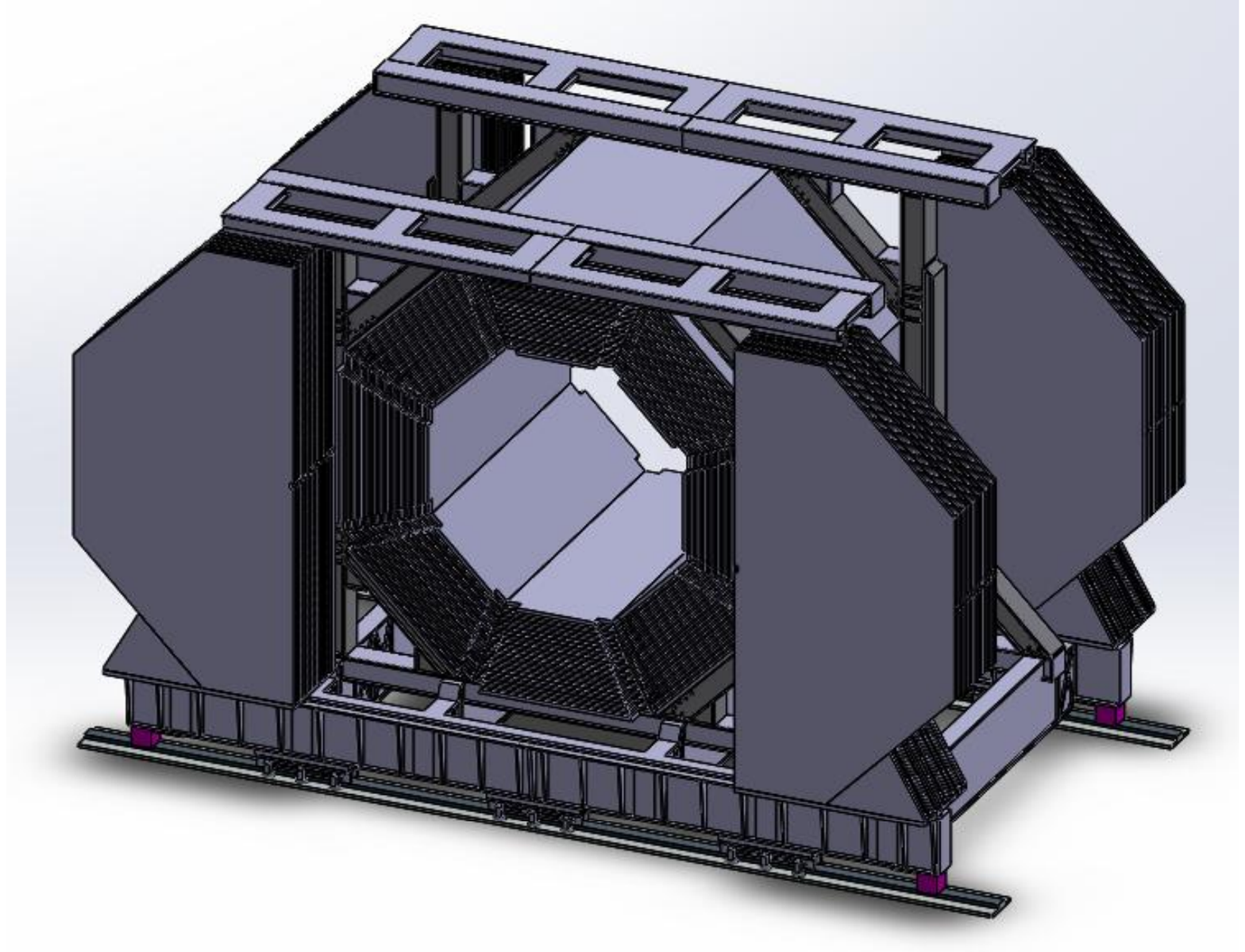}
     \caption{Fully assembled Range System with opened End Caps.}
      \label{fig:RS_Assembled_2}
   \end{minipage}
\end{figure}

The RS should be assembled in the following sequence. Pre-assembled detector layers (each consisting of MDT detectors attached to a stripboard and equipped with FEE electronic cards) are placed in the assembly hall in a horizontal position (Fig.~\ref{fig:RS_Assembly} (a)). 
Each Barrel module, also placed in a horizontal position in the assembly hall, is equipped with these detector layers (Fig.~\ref{fig:RS_Assembly} (b)).
The module is lifted by a crane using a special traverse and moved on to the support and transportation system for assembling (Fig.~\ref{fig:RS_Support}). 
The modules are mounted in pairs, maintaining symmetry (left-right) about the longitudinal axis of the installation. 
The assembly of the Barrel is completed by the last upper closing module (Fig.~\ref{fig:RS_Assembled_1}). After that, the STS upper load-bearing elements are mounted.

Installation of the End Caps is carried out in a similar way, first equipping them with the detector layers of MDTs in a horizontal position. 
Since the weight of one EC sash exceeds 80 tons (maximum crane capability), the design of the sash involves structurally dividing it into two parts along the beam. 
The final view of the RS before mounting the internal SPD detectors and the upper platform is shown in Fig.~\ref{fig:RS_Assembled_2}.

\section{Mini drift tubes detector}
The Mini Drift Tubes detector was initially developed and produced at JINR for the Muon System of the D0 experiment at FNAL \cite{Abazov:2005uk}. Later on, an MDT-based muon system was also produced  for the COMPASS experiment at CERN \cite{Abbon:2007pq}.  The two-coordinate readout modification of the MDT with open cathode geometry and external pickup electrodes was  developed, proposed to, and accepted by the PANDA collaboration at FAIR for the muon system of their experimental setup. This new version of the MDT is proposed for the SPD project, as it has all the necessary features -- radiation hardness, coordinate resolution and accuracy, time resolution, robustness, as well as the advanced level of already conducted R\&D within the PANDA project. 

\begin{figure}[!h]
    \center{\includegraphics[width=1.0\linewidth]{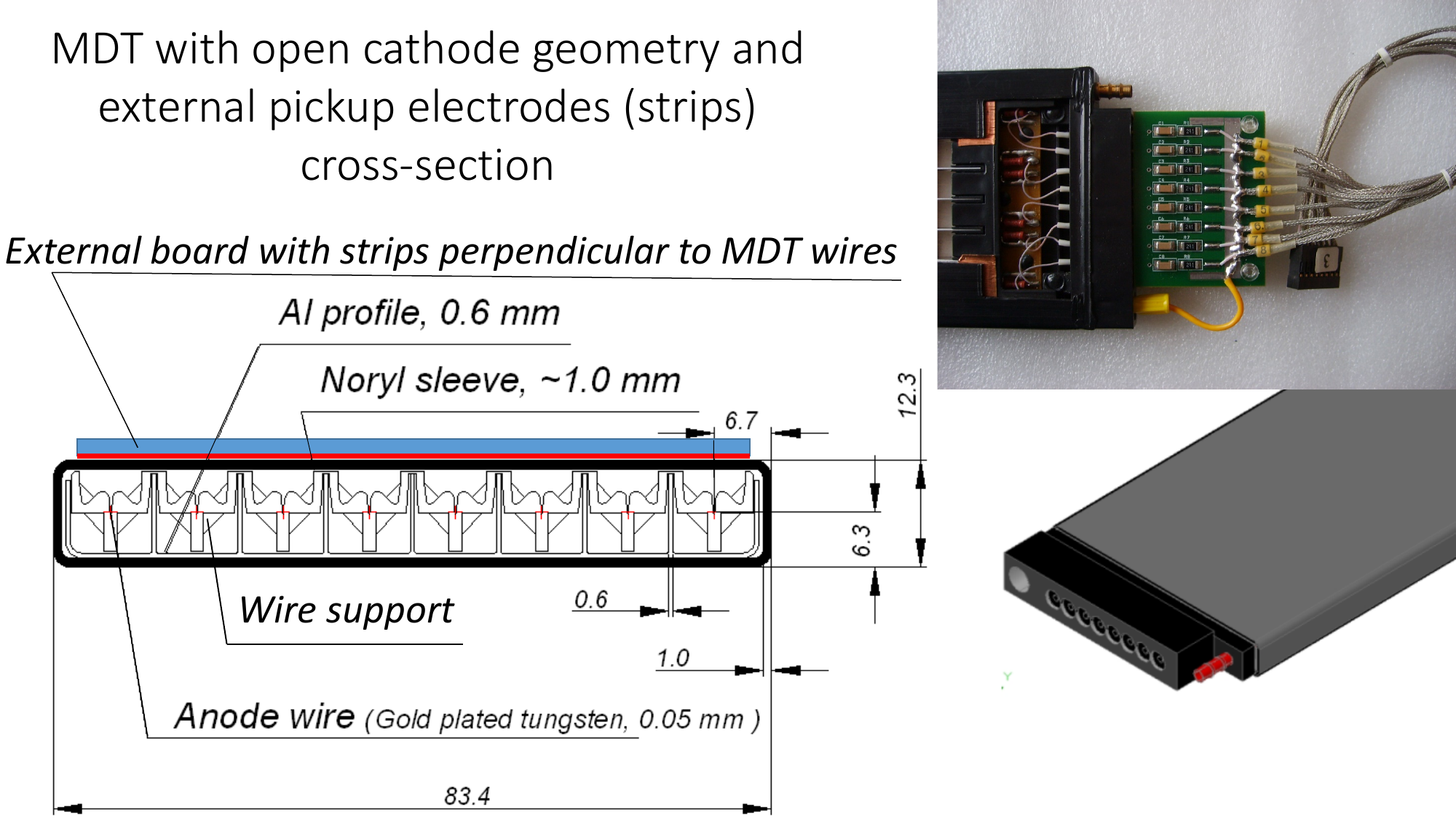}}
     \caption{Mini Drift Tube with open cathode geometry cross-section (left) and layout (right).}
       \label{fig:RS_MDT}  
\end{figure}

The cross-section and  layout of the MDT with open cathode geometry are shown in Fig.~\ref{fig:RS_MDT}. 
The detector consists of a metallic cathode (aluminum extruded comb-like 8-cell profile), anode wires with plastic supports, and a Noryl envelope for gas tightness. 

\begin{figure}[!h]
    \center{\includegraphics[width=0.5\linewidth]{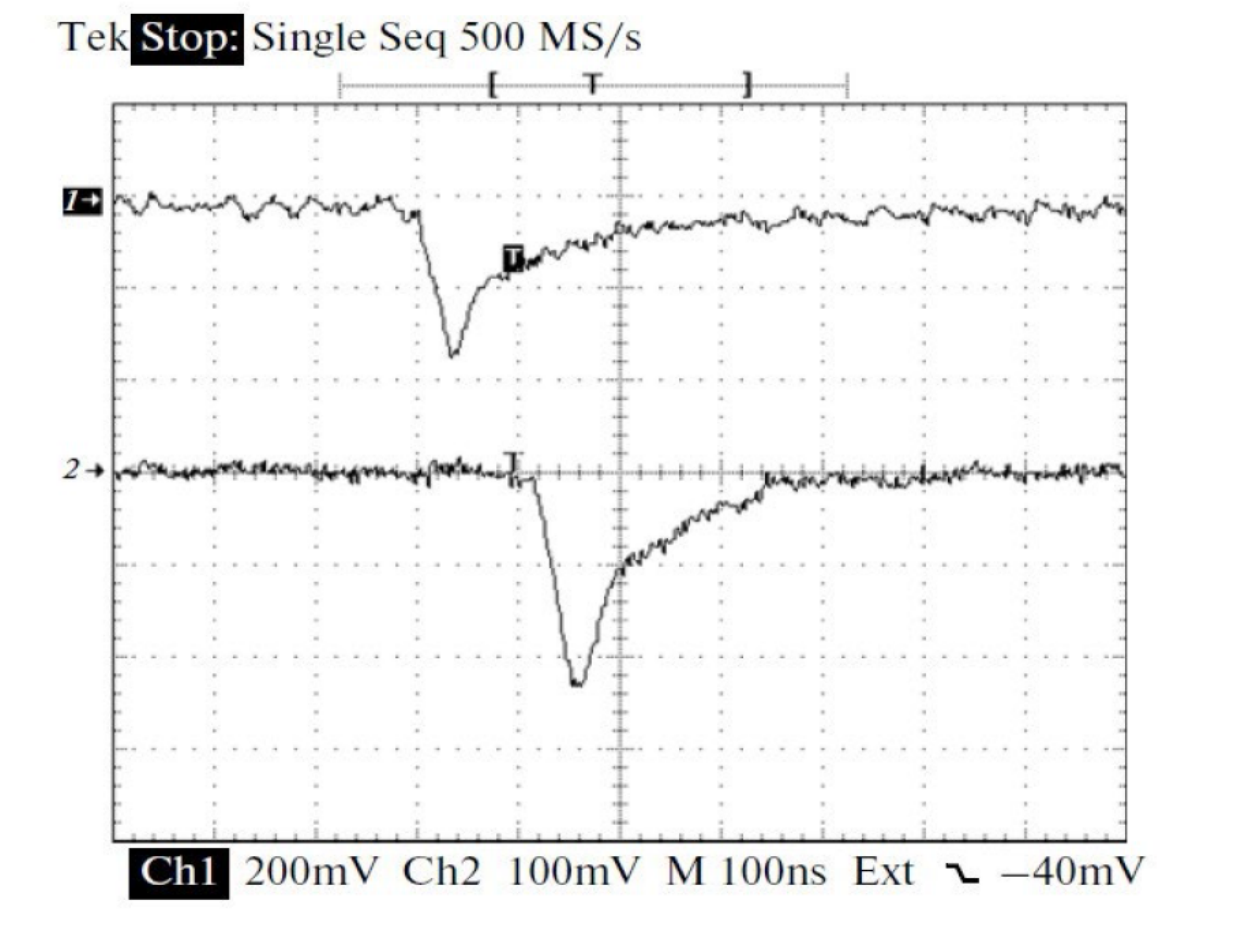}}
     \caption{Oscillograms of single signals: from the anode wire (1) and the strip (2, inverted); the conversion factors are 60 and 480 mV/$\mu$A, respectively.}
       \label{fig:RS_Shape}  
\end{figure}

The comb-like profile of the cathode provides each wire with an opening left uncovered to induce wire signals on the external electrodes (strips) perpendicular to the wires. The strips are applied to obtain the second coordinate readout. The shape of the induced signal repeats the initial one, having the opposite polarity, but the amplitude is about 15$\%$ of the wire signal (see Fig.~\ref{fig:RS_Shape}). Thus, the strip signal readout requires higher signal amplification and proper electromagnetic shielding.

\begin{figure}[!h]
    \center{\includegraphics[width=1.0\linewidth]{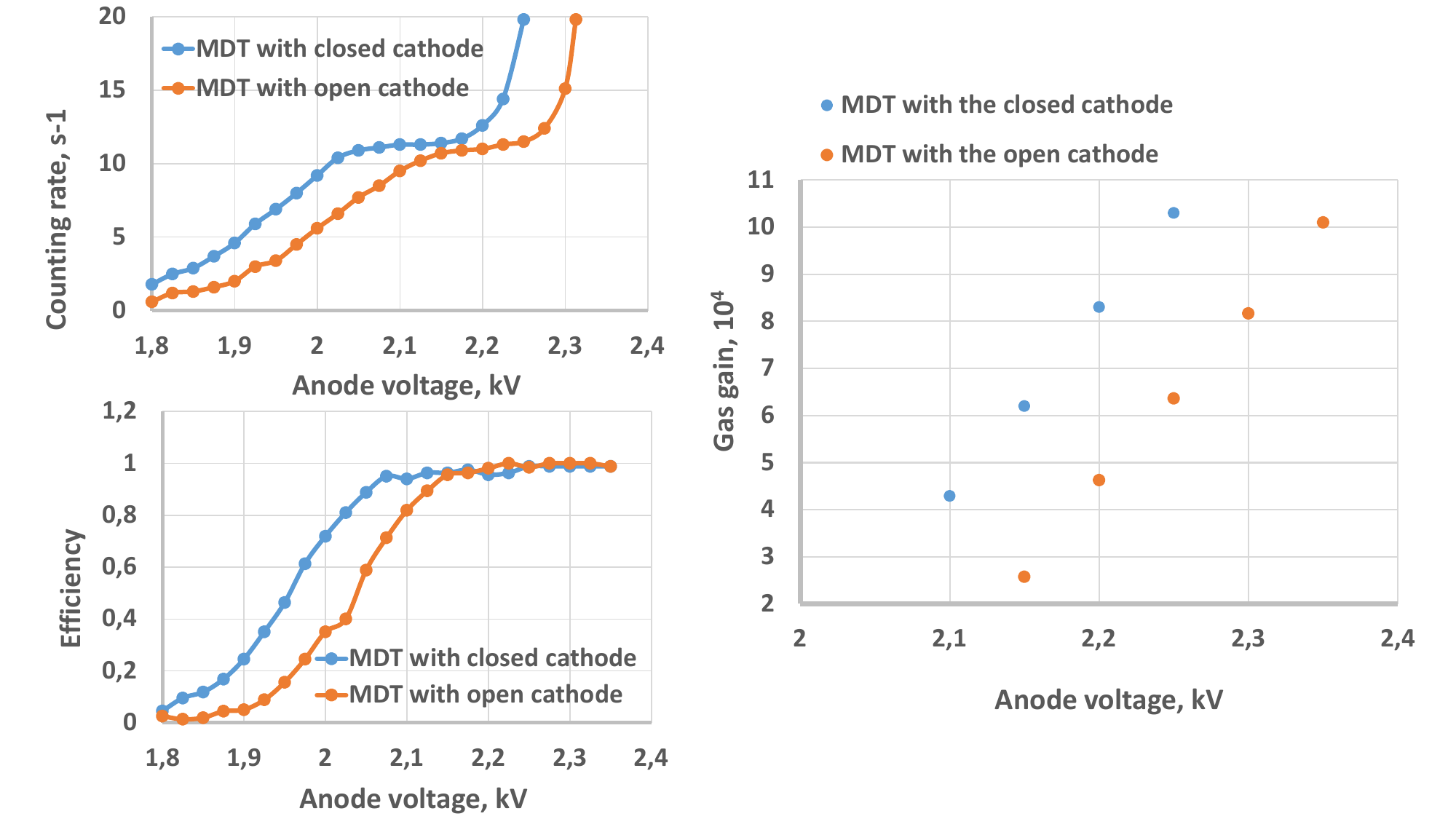}}
     \caption{Comparative plots  for counting rate, efficiency, and gas gain versus the supply voltage for the MDT with closed and open cathode geometry.}
       \label{fig:RS_Rate}  
\end{figure}

Application of an open cathode leads to the loss of the electric field symmetry in each of the 8 detector cells, resulting in a lower gas gain for the applied voltage compared to the standard MDT (cathode openings closed with stainless steel lid). The conducted R$\&$D proved that the MDT with open cathode geometry easily achieves the parameters of the one with a closed cathode at higher voltages. The comparative plots  for counting rate, efficiency, and gas gain for both detector types (see Fig.~\ref{fig:RS_Rate}) show that the MDT with open cathode geometry repeats the standard MDT performance at a high voltage shift of +100 V. The drift time and the amplitude spectra of both detector variants also match, if we set this voltage shift between their operating points.

\begin{figure}[!hb]
    \center{\includegraphics[width=1.0\linewidth]{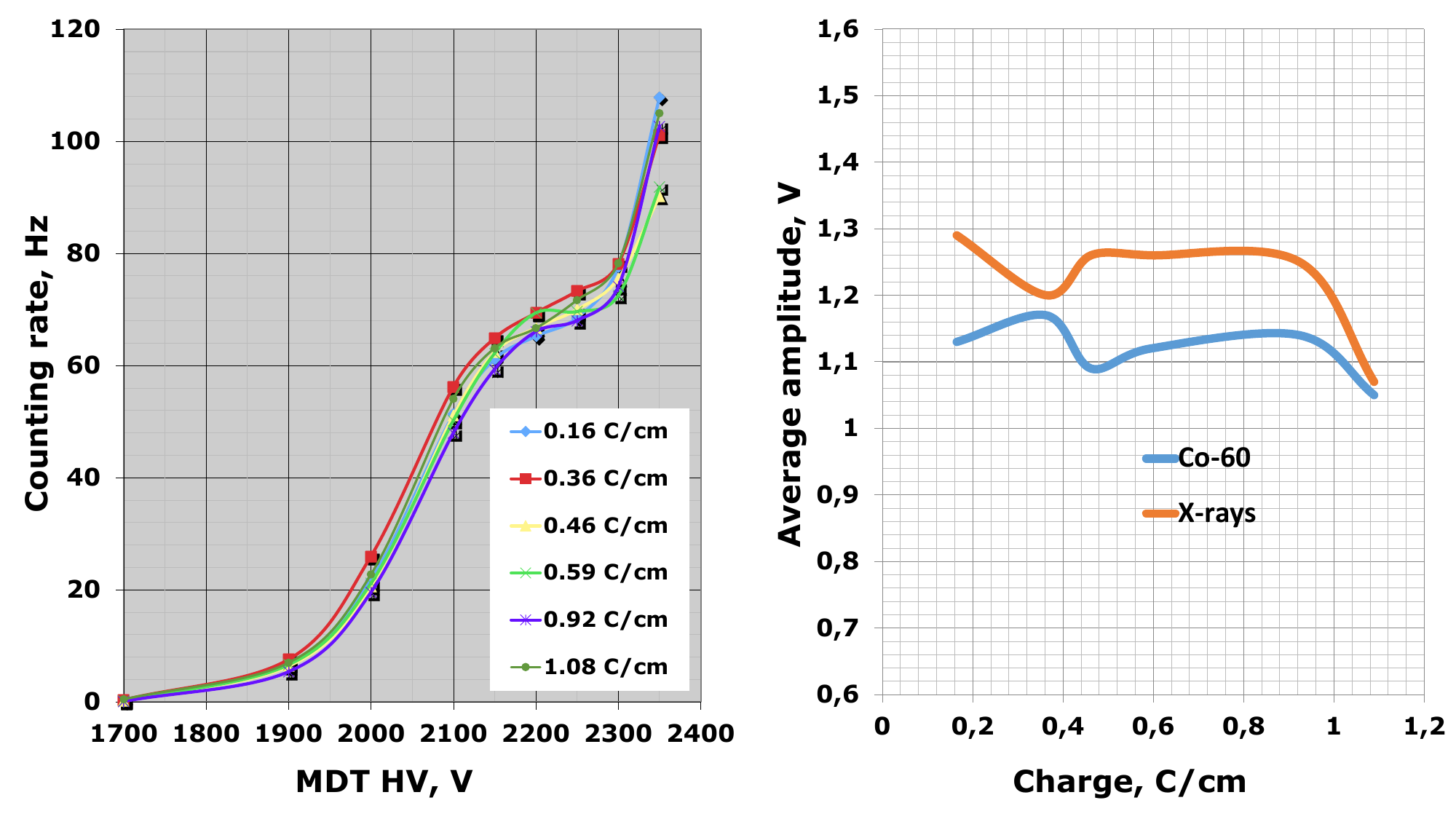}}
     \caption{Counting rate curves for different accumulated charges (0.16$\div$1.08 C/cm) (left); average wire signal amplitudes vs accumulated charge for Co-60 and X-ray sources (right), the peculiar view of the curves (generally demonstrating the stability of the amplitude within few
percent) is due to the long duration of the monitoring  (several months)} and the absence of amplitude correction for
the atmospheric conditions (pressure and temperature).
       \label{fig:RS_Curves}  
\end{figure}

According to the results of the MDT (open cathode geometry) aging tests, the accumulation of  a 1 C/cm total charge does not produce any significant effect on the detector performance. To monitor the aging effects, measurements of the counting rate curves (Co-60 source) together with oscilloscopic observations of the MDT average signals (256 events) for Co-60 and X-rays were made twice a week over the whole period of intense irradiation (see Fig.~\ref{fig:RS_Curves}). Later on,  these measurements (with X-rays) were repeated for up to 3.5 C/cm of irradiation without any visible degradation of the MDT performance.  This should ensure a stable MDT performance for the lifetime of the SPD project.

All R$\&$D studies were made with a gas mixture of 70$\%$ Ar + 30$\%$ CO$_2$ at atmospheric pressure, the one to be used in the proposed SPD Muon System. It is non-flammable, radiation hard, and fast enough (150$\div$200~ns drift time, depending on the incoming angle).
The wire pitch in the present design equals 1 cm, and a 3-cm strip width is selected for the second coordinate. These spatial parameters provide the Range System with coordinate accuracy well enough for the identification of muons and give the system the features of a digital hadron calorimeter.

\section{Gas system}
\begin{figure}[!h]
    \center{\includegraphics[width=0.7\linewidth]{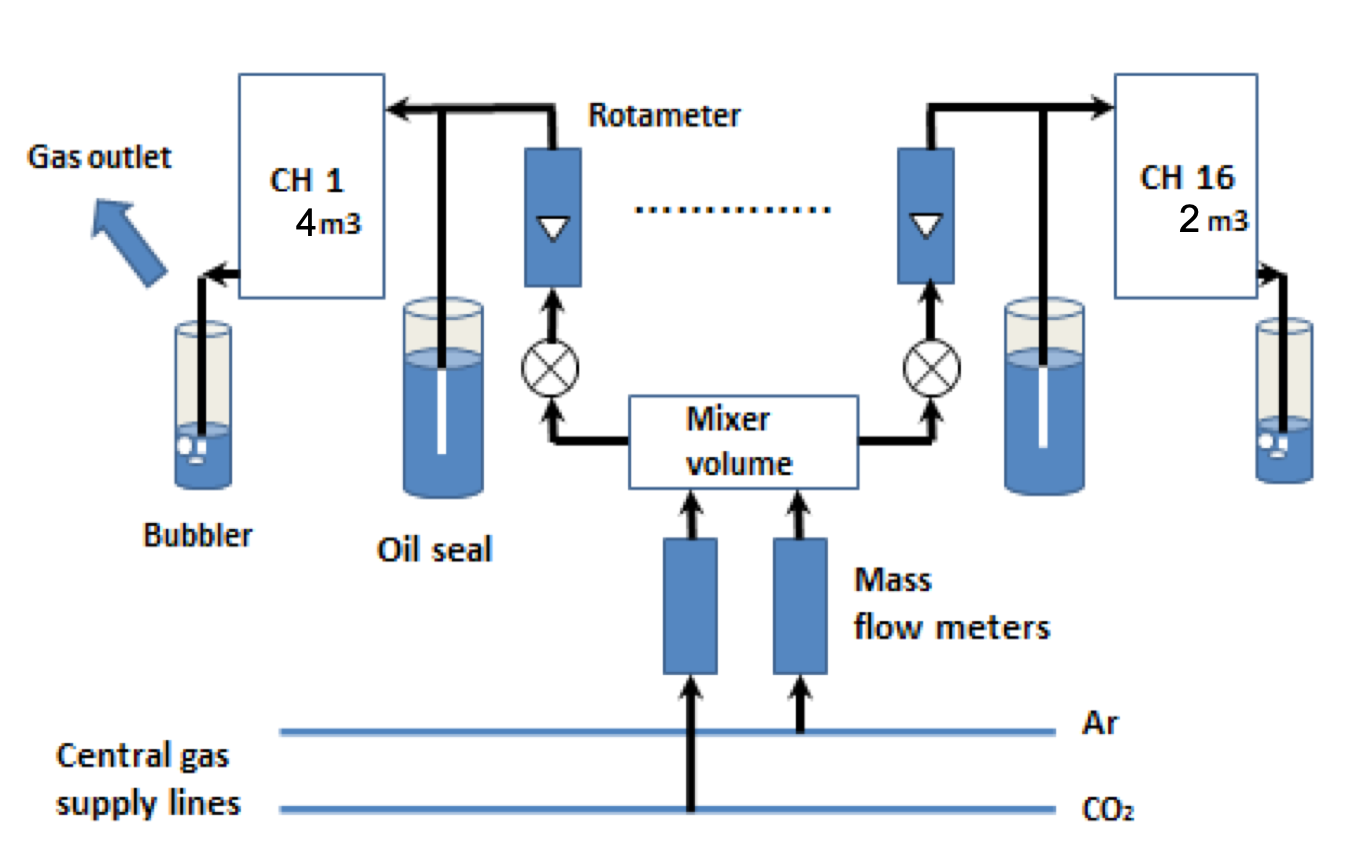}}
     \caption{Schematic diagram of the gas system.}
       \label{fig:RS_GasSystem}  
\end{figure}

The Range System will use quite a simple gas system (see Fig.~\ref{fig:RS_GasSystem}). 
It should prepare and distribute over the whole detector volume the non-flammable, non-explosive, and cheap gas mixture of argon and carbon dioxide: Ar:CO$_2$ = 70:30. 
Such a mixture, being cheap, does not require a closed circuit for recuperation of the outgoing gas flow. 
This mixture had been successfully used for years in the operation of the Muon Wall 1 and Rich Wall subsystems of the COMPASS experiment at CERN, which have practically the same MDT detectors. 
It is supposed now that the mixture composition will be prepared by the mass flowmeters. 
The required accuracy is about 1\% (relative value with respect to the smaller component, so CO$_2$), and purity is of the standard technical value. 
The main components - Ar and CO$_2$ - should be centrally supplied by a standard gas distribution system of the SPD setup to the inputs of mass flowmeters. 
The two outputs of mass flowmeters feed a mixer volume. 
From this volume, the mixture is distributed over the Range System channels. 
The distribution is done simply by using mechanical flowmeters (rotameters). 

Presently, we assume to have 16 main channels for the control of the incoming gas flow (rotameters) and their outlets (simple bubblers). 
These 16 channels are comprised of the following parts: 8 octant modules of the Barrel and 8 half-sashes of the End Caps. The gas volumes of these channels are the following: about 4 cubic meters for each Barrel module and 2 cubic meters for each half-sash of ECs. In each channel/module, the detecting planes are connected in parallel, and the MDT detectors of each plane are connected in series.

Technical details which reflect the features of the gas system are the following:
\begin{itemize} 
    \item The total volume of the Range System is about 50 cubic meters.
    \item The gas exchange rate, which defines the flow, is about (1/2$\div$1)$\times$V$_0$/day, where V$_0$ is the 
nominal system volume ($\sim$50 m$^3$). This corresponds to the total fluxes in the range of 1000$\div$2000 liters per hour. For the normal system operation (long-term stable running) the flux of 1000~l/h is assumed.
    \item For fast gas exchange, the flushing mode should provide for 2$\times$V$_0$/day, which corresponds to
4000~l/h. This large value occurs on very rare occasions and should not be taken into consideration in the overall total gas balance of the SPD setup.
\end{itemize} 

\section{Analog front-end electronics}
As the SPD Muon System inherits all technical solutions developed for the PANDA Muon System, 
for the analog front-end electronics (FEE)  we also will use general ideas of wire and strip signals readout, 
already developed and well-tested using the PANDA Range System prototype. 
Thus, we are planning  to use the analog front-end electronics solutions (with minor modifications) 
developed for the D0/FNAL and the COMPASS/CERN experiments and also accepted by the PANDA/FAIR. 

\begin{figure}[!h]
\centering
  \begin{minipage}[ht]{0.49\linewidth}
    \center{\includegraphics[width=1\linewidth]{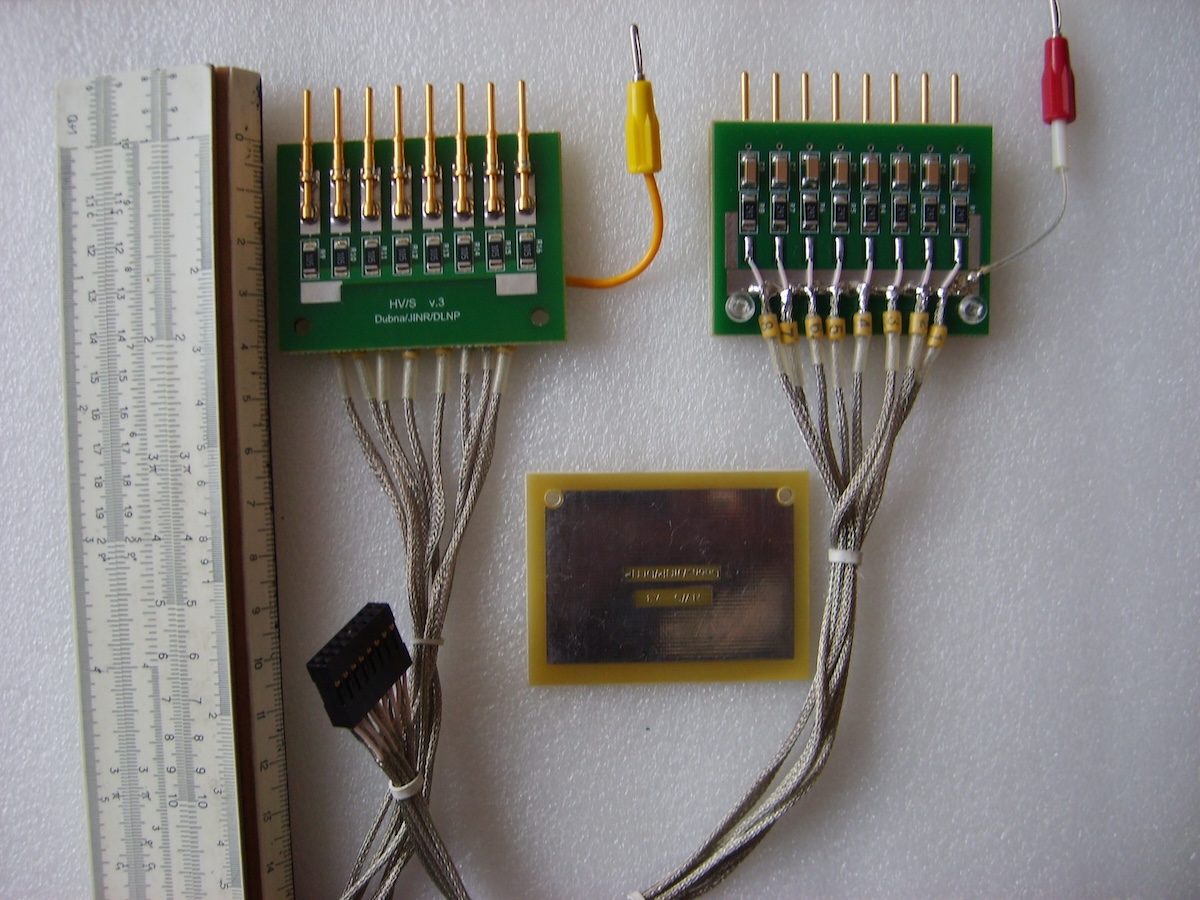} \\ (a)}
  \end{minipage}
  \hfill
  \begin{minipage}[ht]{0.49\linewidth}
    \center{\includegraphics[width=1\linewidth]{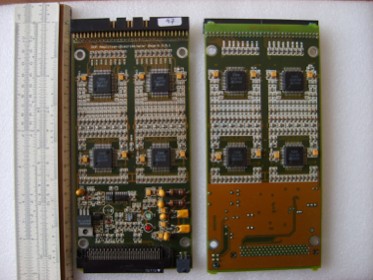} \\ (b)}
  \end{minipage}
  \begin{minipage}[ht]{0.49\linewidth}
    \center{\includegraphics[width=1\linewidth]{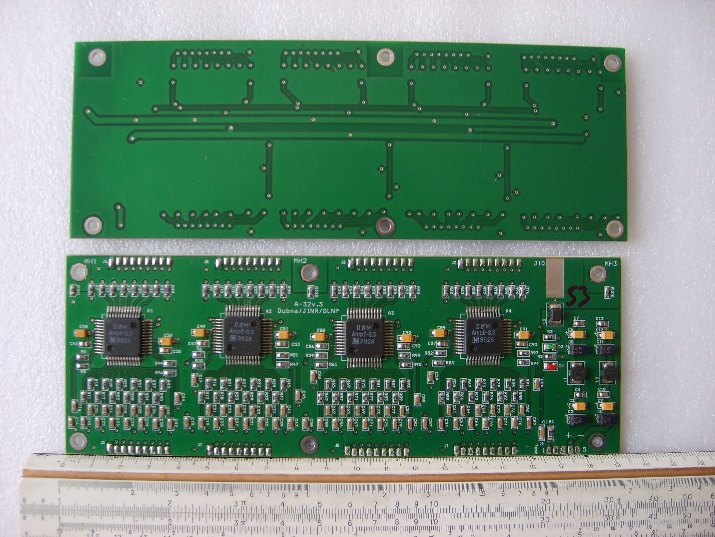} \\ (c)}
  \end{minipage}
   \hfill
  \begin{minipage}[ht]{0.49\linewidth}
    \center{\includegraphics[width=1\linewidth]{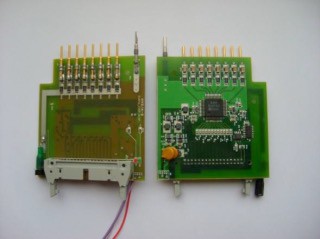} \\ (d)}
  \end{minipage} 
  \caption{Front-end analog electronics cards: HV/S-8 (a), ADB-32 (b), A-32 (c), and HVS/A-8 (d).} 
  \label{fig:RS_FE}  
\end{figure}

The signal readout is based on two ASIC chips: 8-channel amplifier Ampl-8.3 \cite{Alekseev:2001xn} and 
8-channel comparator/discriminator Disc-8.3 \cite{ALEXEEV1999157}.
To arrange wire readout, we need a high voltage distributor and a signal connector circuit followed by an amplifier -- discriminator circuit with low-voltage differential signaling (LVDS) outputs. 
Combination of HV/S-8 and ADB-32 cards ~\cite{Alekseev:2001cg} is the most common among 
the existing variants (see Fig.~\ref{fig:RS_FE} (a), (b)). 
For the strip signal readout, we use a circuit with two stages of Ampl-8.3 followed by the discriminator with the LVDS output. 
Combination of preamplifier cards A-32 (Fig.~\ref{fig:RS_FE} (c)) and ADB-32 (Fig.~\ref{fig:RS_FE} (b)) fulfills this task. 
The mentioned electronic cards without any modifications can be applied for the wire and strip signals readout of the End Caps.
The view of the basic FEE cards is shown in Fig.~\ref{fig:RS_FE}. 

For the Barrel part, we use the same general idea of the readout circuitry, but its implementation will differ 
from electronics mentioned above. 
This is connected to the fact that the Barrel analog FEE is a constructive part of the detector layer and 
will be buried inaccessible.

By now, we assume to use 8-channel electronic cards for both wire and strip readout in the Barrel with all needed circuitries implemented in a single card.
For the wire readout it will be a modification of $\rm{HVS/A-8}$ (Fig.~\ref{fig:RS_FE} (d)) designed for the PANDA Barrel -- 
the new board will contain high voltage distribution circuit, amplifier and discriminator with all related circuitry. 
Strip analog FEE will also be an 8-channel board containing a preamplifier, amplifier, and discriminator with all related circuitry. 
To match 8-channel electronics with 32-channel inputs of the digital FEE part, we are planning to use 
special ''patch panel'' cards that will be placed on the edge of the detector layer and stay accessible. 
These cards will also act as LVDS drivers and LV and threshold distributors with a manual LV adjustment.

Analog FEE cards are located mostly inside the slots of the RS iron absorber -- fully (for the wires and strips) in the Barrel and partially (for the strips only) in the End Caps. 
Corresponding power dissipation (see Table ~\ref{RS:Power}), despite looking quite large, does not impose any practical limitations, as it is ''immersed'' into the massive iron ''radiator'' of the RS absorber. 
So, we do not plan to use additional ventilation apart  from the natural convective cooling.

\begin{table}[!h]
\caption{Analog FEE power dissipation.}
\begin{center}
\begin{tabular}{|l|c|c|c|}
\hline
 & Barrel, kW & End-Caps, kW & Total, kW \\
\hline 
\hline
Wires & 8.9 & 13.6 & 22.5  \\
\hline
Strips & 9.5 & 8.6 & 18.1  \\
\hline
Total & 18.4 & 22.2 & 40.6  \\
\hline
\end{tabular}
\end{center}
\label{RS:Power}
\end{table}

We still continue  our R\&D work intended to improve strip board performance, results of this work will influence the detector layer design. 
In collaboration with the Institute of Physics of the National Academy of Science of Belarus 
we also continue R\&D devoted to Ampl-8.3 modification that will improve its applicability for the strip signal readout and reliability of long-term operation. 
The results of these R\&D works can significantly influence the PCB design of the Barrel analog FEE, 
that is why the development of these particular cards is postponed. 

In total, the Range System has 137600 readout channels (83600 of which are wires and 54000 strips), see  Table ~\ref{RS:FEEchannels}. 

\begin{table}[!h]
\caption{FEE channels.}
\begin{center}
\begin{tabular}{|l|c|c|c|}
\hline
 & Barrel& End-Caps& Total \\
\hline 
\hline
Wires & 32400 & 51200 & 83600  \\
\hline
Strips & 27800 & 26200 & 54000  \\
\hline
Total & 60200 & 77400 & 137600  \\
\hline
\end{tabular}
\end{center}
\label{RS:FEEchannels}
\end{table}

\subsection{Ampl-8.51 -- low input impedance amplifier for the Muon System wire and strip readout}
As mentioned above, all present analog FEE designed for the Muon System uses Ampl-8.3 to read both wire and strip signals. The amplifier ASIC was specially designed for MDT but also managed to cope with strip signals when used in the two-stage circuit. Generally, Ampl-8.3 meets the strip readout requirements well, but not perfectly. In terms of the set stripboard parameters, the value of the strip wave impedance varies in the range of $R_z$ $\sim$10$\div$30 $\Ohm$, but Ampl-8.3 has a 50 $\Ohm$ input impedance. 
The present impedance mismatch does not seem significant, but anyway, matching is preferred, especially for the long strip signal readout. Moreover, it would be much better to have an input impedance less than 10 $\Ohm$, so that after matching the wave impedance of the strip (or wire) with the input impedance of the amplifier, carried out using a series-connected resistor $R_d$, the maximum value of the high-voltage surge is reduced by the resulting voltage divider in $(R_d + R_{in})/R_{in}$ times, increasing the amplifier's resistance to high-voltage breakdown, increasing its lifetime and reliability. In addition, the low input impedance of the amplifier reduces external pickup at its input, which increases its immunity to self-excitation and external interference. Low or absent accessibility of strip and wire readout electronics of the RS Barrel makes a real demand for increased reliability of the amplifiers. Following the given considerations, it was decided to develop a low input impedance modification of Ampl-8.3 that would keep all the  proven virtues of its predecessor.

The new amplifier chip design was carried out under the following guidance:
\begin{enumerate}
\item the amplifier general parameters should correspond to the ones of Ampl-8.3 (or similar);
\item an input stage should be made on a common base NPN transistor to provide low input resistance;
\item amplifier inputs should be protected by the diodes against HV pulses (both negative and positive);
\item input Rush amplifier circuit should be followed by differential amplifying stages to minimize cross-talks;
\item the amplifier should belong to a trans-resistance type; it should not have a common feedback to exclude the possibility of parasitic generation by excluding stray capacitances and resistors of the feedback integrated capacitors and resistors between the input and output of the amplifier;
\item maximum layout symmetry to minimize DC voltage shift at the amplifier's differential outputs at a quiescent mode.
\end{enumerate}

As a result, an Ampl-8.51 circuit has been developed to meet all requested parameters. Amplifier circuit design together with spice simulations have been done using Cadence 6 CAD software. General parameters of the Ampl-8.51 spice model are presented in Table~\ref{RS:ampl851}. All parameters are given taking into account the technological spread of the circuit elements' values. For resistors, it is $\pm$ 20\% of the nominal values. The Rush cascade cannot be implemented in an integral design with a small scatter of parameters, since they are determined not only by the ratio of the resistors but also by their nominal values. The spread of the main parameters of the ASIC Ampl-8.51 will repeat the technological spread of the resistor values.

\begin{table}[htp]
\caption{Ampl-8.51 general parameters (spice simulation).}
\begin{center}
\begin{tabular}{|l|c|}
\hline
Parameter & Value \\
\hline 
\hline
Input impedance for signal bandwidth: & \\
1 MHz, $\Ohm$  & 0.25   \\
10 MHz, $\Ohm$  & 0.9$\div$2.2   \\
30 MHz, $\Ohm$  & 3.32$\div$8.7   \\
\hline
Input signal polarity & $\pm$ \\
\hline
Input $\pm$ overvoltage protection: & Yes \\
\hline
Differential output & Yes \\
\hline
Gain, mV/µA & 100$\div$150 \\
\hline
DC output offset voltage, V & $<$ 1.0 \\
\hline
Output load, W & 1000 \\
\hline
Output signal leading/trailing edge (0.1, 0.9), ns & 8$\div$12 \\
\hline
$I_{noise}$ for detector capacitance: & \\
CD = 0, r.m.s., nA & 63$\div$110 \\
CD = 40 pF, r.m.s., nA & 96 \\
CD = 1800 pF, r.m.s., nA & 315 \\
CD = 5000 pF, r.m.s., nA & 572 \\
\hline
Dynamic range for signals of both polarities, dB & 48 \\
\hline
Channel-to-channel cross-talk, dB & $<$-40 \\
\hline
Voltage supply, V & $\pm$ (2.8$\div$3.2) \\
\hline
Dissipated power at $\pm$3 V supply, mW/channel & 64 \\
\hline
Channels per chip & 8 \\
\hline
\end{tabular}
\end{center}
\label{RS:ampl851}
\end{table}%

\begin{figure}[!h]
    \center{\includegraphics[width=0.8\linewidth]{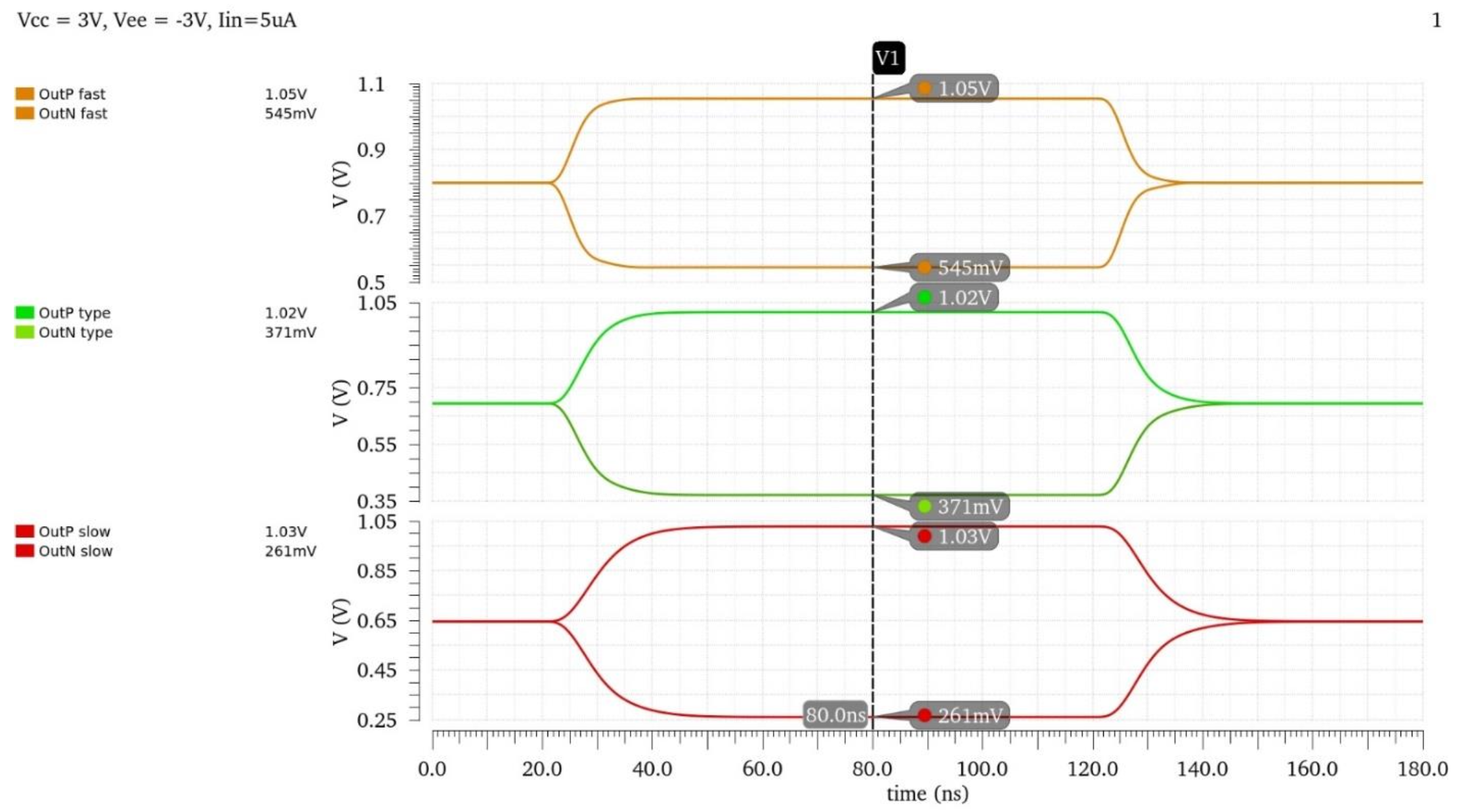}}
\caption{Ampl-8.51 output signals, Iin $=$ 5 $\mu$A.}
  \label{fig:RS_amplOutput}  
\end{figure}

The oscillograms of the Ampl-8.51 output signals within the boundaries of the technological spread of the circuit elements parameters are shown in Fig.~\ref{fig:RS_amplOutput} -- the simulation is performed for the so-called fast, typical, and slow models of the circuit elements. In these figures, fast models are represented by brown lines, typical models - by green lines, and slow models - by red lines.

\begin{figure}[!h]
    \center{\includegraphics[width=0.8\linewidth]{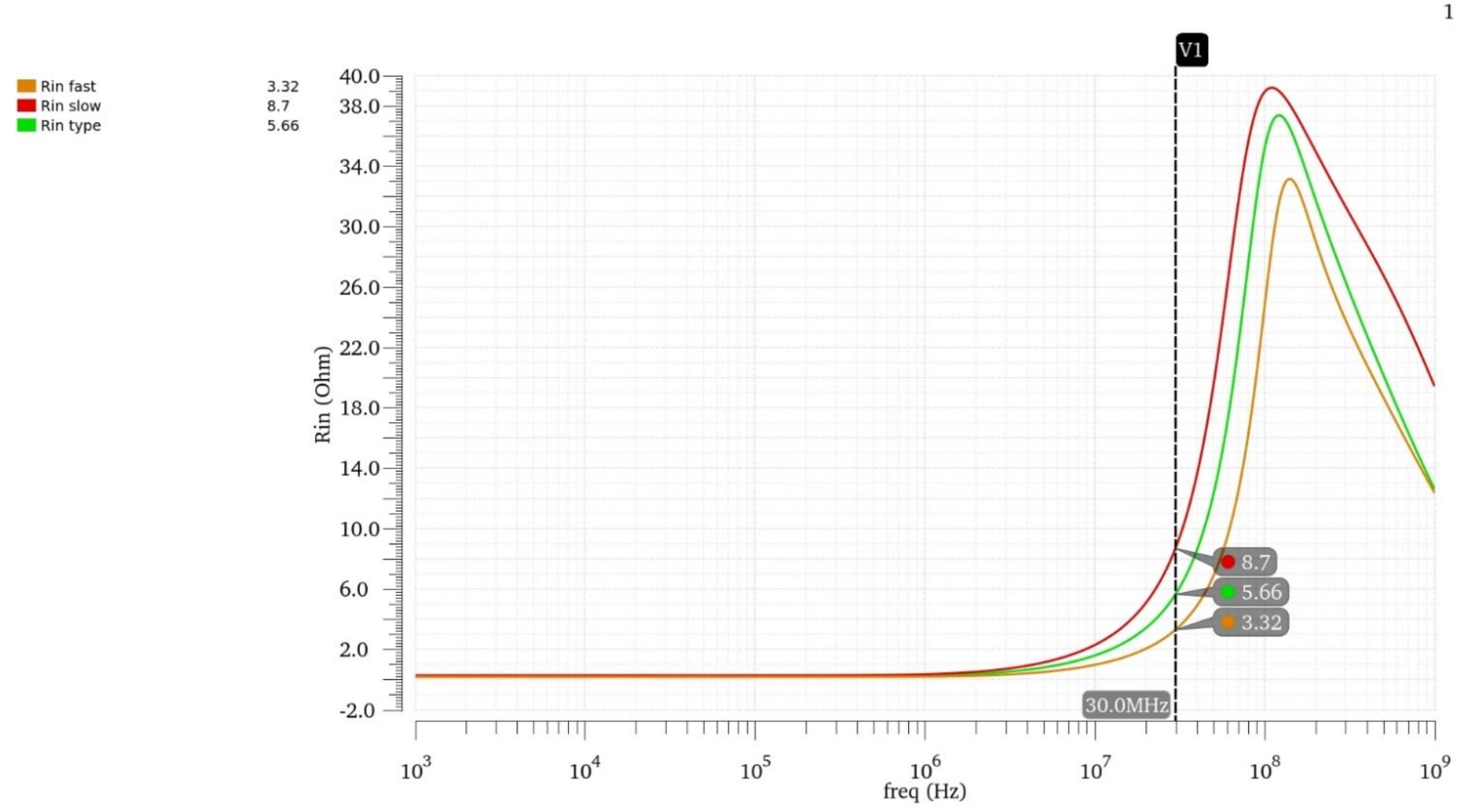}}
\caption{Ampl-8.51 input impedance frequency response.}
  \label{fig:RS_amplInput}  
\end{figure}

Figure~\ref{fig:RS_amplInput}  presents the plots of  the Ampl-8.51 input impedance frequency response. At a frequency of 30~MHz, the value of the input resistance will vary within 3.32 $\div$8.7~$\Ohm$ due to technological spread in the values of the circuit elements. 
At a frequency of 10~MHz, the value of the input impedance remains within 0.9$\div$2.2~$\Ohm$, and at low frequencies it does not exceed 0.25~$\Ohm$.

The presented simulation results of the Ampl-8.51 with the Rush cascade show that the main parameters of the developed integrated circuits correspond to the requirements of the technical specifications for typical values of the parameters of the circuit elements and retain the operating values within the technological spreads of the circuit elements parameters.

 The topology of the Ampl-8.51 and the AISIC have been made using BJT-JFET technology by the research and production corporation ''Integral'' (Minsk, Belarus).

The first tests of the manufactured chip showed that this iteration has two general problems -- low operating stability (self-excitations) and imbalance of  the differential output offset. According to preliminary examination, these problems are due to the following mistakes in microchip topology (circuit peculiarities were not taken into account):

\begin{itemize}
\item inadequate power bus and traces width (R$\neq$0, V drop);
\item series power connections (not star type) in the circuit parts most sensitive to power;
\item complimentary parts of the differential circuit powered from different power bus points (operation point floats);
\item the same power bus for input current amplifier and following  differential amplifier, power traces are mixed (input senses current jerks of the output circuit, resulting in self-excitations);
\item common input ground (input amplifier circuits sense neighbors, resulting in self-excitations).
\end{itemize}

\begin{figure}[!h]
    \center{\includegraphics[width=0.9\linewidth]{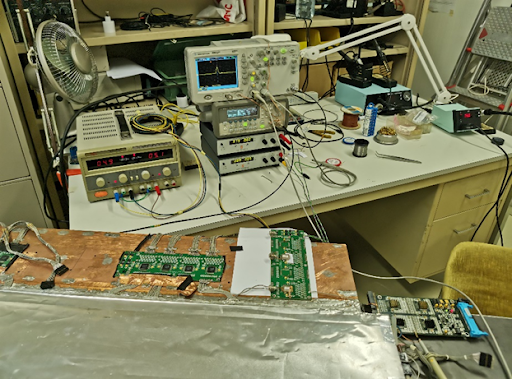}}
\caption{Ampl-8.51 single-channel tests.}
  \label{fig:RS_amplTest}  
\end{figure}

To minimize described negative effects 1-st and 8-th amplifier channels were disconnected from common ground and power bus (applied to 4 chips).  Separated channels restore the output baseline and operate without self-excitations, giving a chance to check general amplifier parameters.  General parameters of Ampl-8.51 individual channels were measured, and its ability to provide a strip signal readout was checked, the test stand is shown in Fig.~\ref{fig:RS_amplTest}.

Measured single-channel parameters (see Table~\ref{RS:ampl851Channel}) perfectly matched the technical requirements.

\begin{table}[htp]
\caption{Ampl-8.51 single-channel parameters.}
\begin{center}
\begin{tabular}{|l|c|}
\hline
Parameter & Value \\
\hline 
\hline
Input impedance (30 MHz), $\Ohm$ & 5.4  \\
\hline
Gain, mV/µA & 131   \\
\hline
Rise time/fall time, ns & 9.2  \\
\hline
$I_{noise}$ (CD = 1.8 nF), nA  &  550  \\
\hline
Dynamic range (CD = 0 nF), dB & 48  \\
\hline
Dissipated power, mW/channel & 63  \\
\hline
\end{tabular}
\end{center}
\label{RS:ampl851Channel}
\end{table}

 As it stands, we still need to achieve a combined 8-channel stable operation of the amplifier. We will try to apply all possible and necessary corrections to the metal layers of existing chips (with unfinished metallization) to fulfill this task.

\section{Digital front-end electronics}
\begin{figure}[!h]
    \center{\includegraphics[width=1.\linewidth]{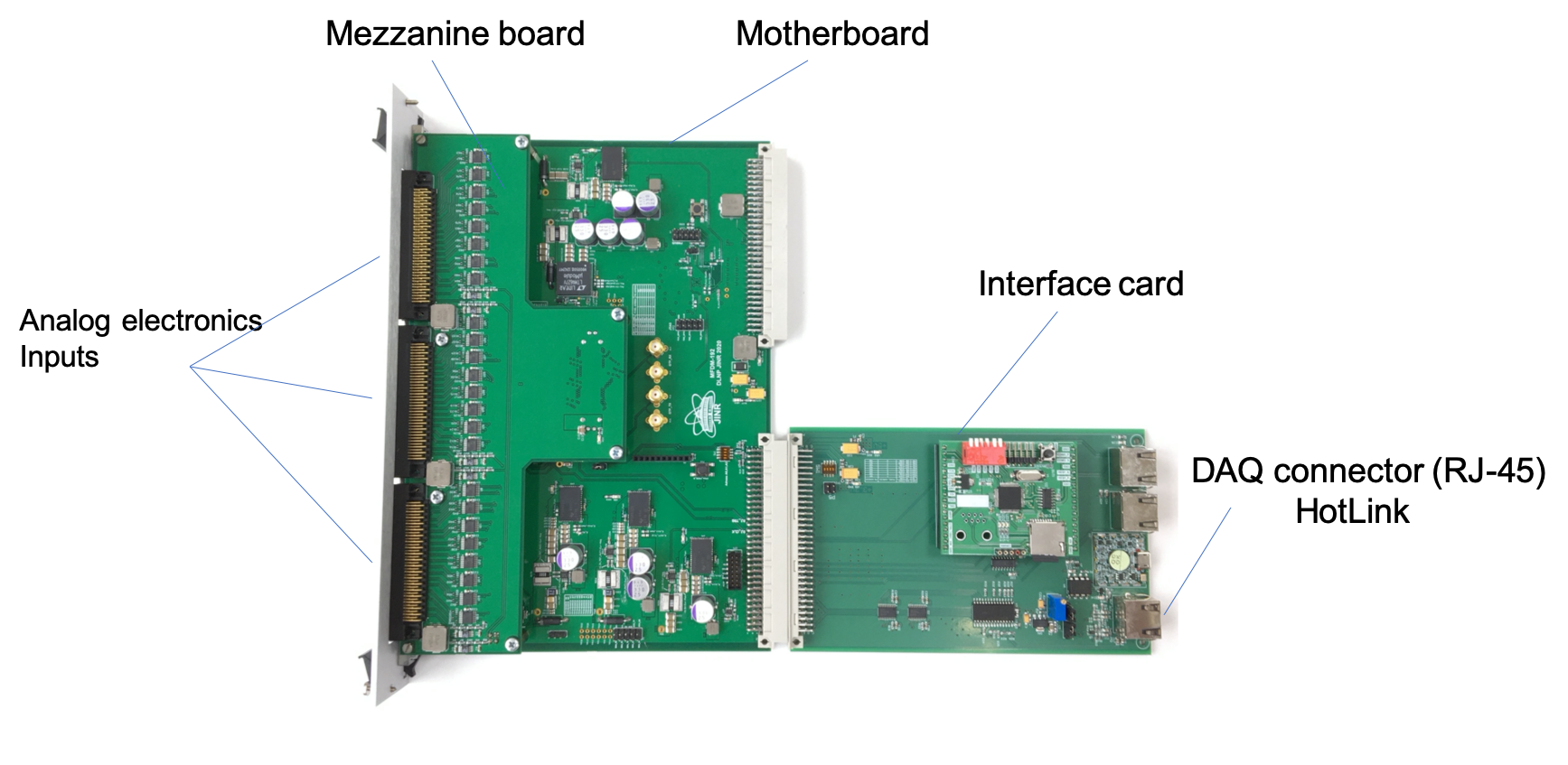}}
\caption{MFDM readout unit.}
  \label{fig:RS_dMFDM}  
\end{figure}

The digital electronics being created for the Muon System is based on the use of FPGA chips. 
The prototype of the digital 192-channel MFDM module (see Fig.~\ref{fig:RS_dMFDM}) (Muon FPGA Digital Module) that we have developed includes an XC7A200T chip of the Xilinx Artix 7 family. 
In terms of functionality, mechanics, data format, and DAQ interface this unit is compatible with the previously developed MWDB (Muon Wall Digital Board) unit \cite{Alexeev2005} made based on the TDC F1 ASIC and  successfully used for data  readout from the Muon System of the COMPASS experiment (CERN). 
This approach allows both types of units to be used in the same readout system, thus making it possible for the new MFDM cards to be tested under actual operating conditions.

The MFDM unit is made in the VME standard (6U, 2M) with the following technical parameters:
\begin{itemize}
\item the number of readout channels - 192;
\item input signals levels - LVDS;
\item range of the threshold voltage adjustment (for the analog FEE) is 0 $\div$ +3V;
\item the unit operates in the latch mode;
\item the maximum time interval for input signals digitizing $\sim$ 9 $\mu$s;
\item discreteness of the time interval digitizing (bin size) -- (4$\div$5) ns;
\item power consumption per unit is about 24 W.
\end{itemize}

\begin{figure}[!h]
    \center{\includegraphics[width=0.9\linewidth]{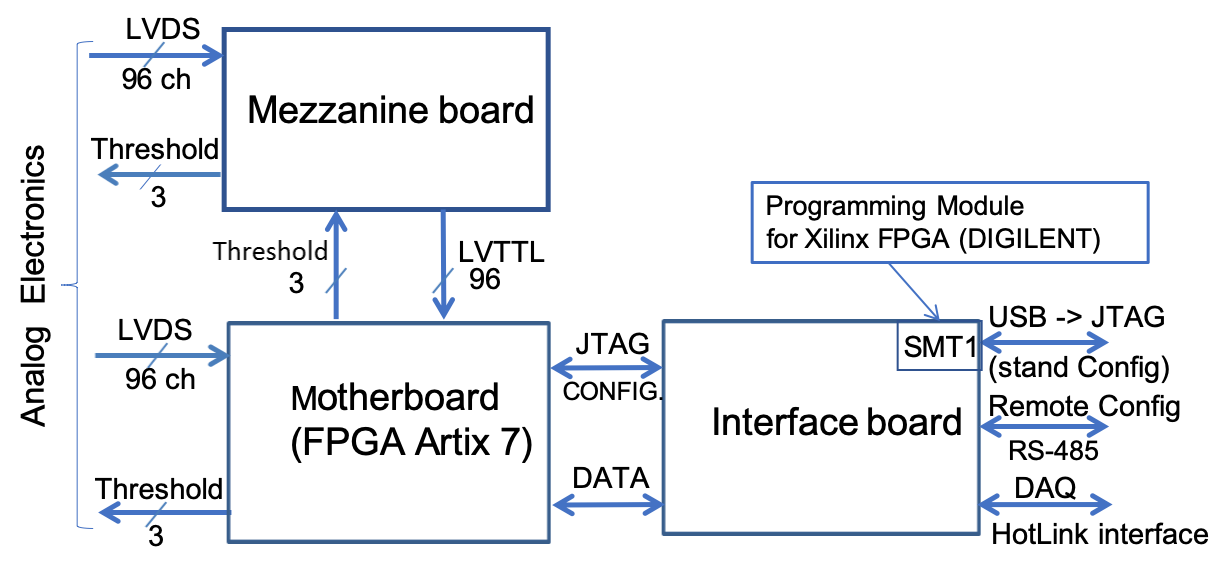}}
\caption{Block-diagram of the MFDM module.}
  \label{fig:RS_MFDM}  
\end{figure}

The unit includes three electronic boards (Fig.~\ref{fig:RS_MFDM}): motherboard, mezzanine card, and interface card.

The motherboard accepts 96 LVDS signals from the analog electronics through 3 high-density connectors, converts them to LVTTL levels, writes to the FPGA, and also communicates with two other boards. 
The mezzanine card accepts the remaining 96 LVDS signals through 3 connectors, converts them to LVTTL levels, and transmits through the 120-pin board-to-board connector to the motherboard. 
The interface card is designed to perform communication of the MFDM module with the DAQ via the HotLink interface (RJ45 connector), together with FPGA firmware  downloading both from a local computer via the JTAG interface and a remote terminal via the RS-485 interface (RJ45 connector).
Interface card is installed on the backside of the VME crate in the P2 connector.

\begin{figure}[!h]
    \center{\includegraphics[width=0.9\linewidth]{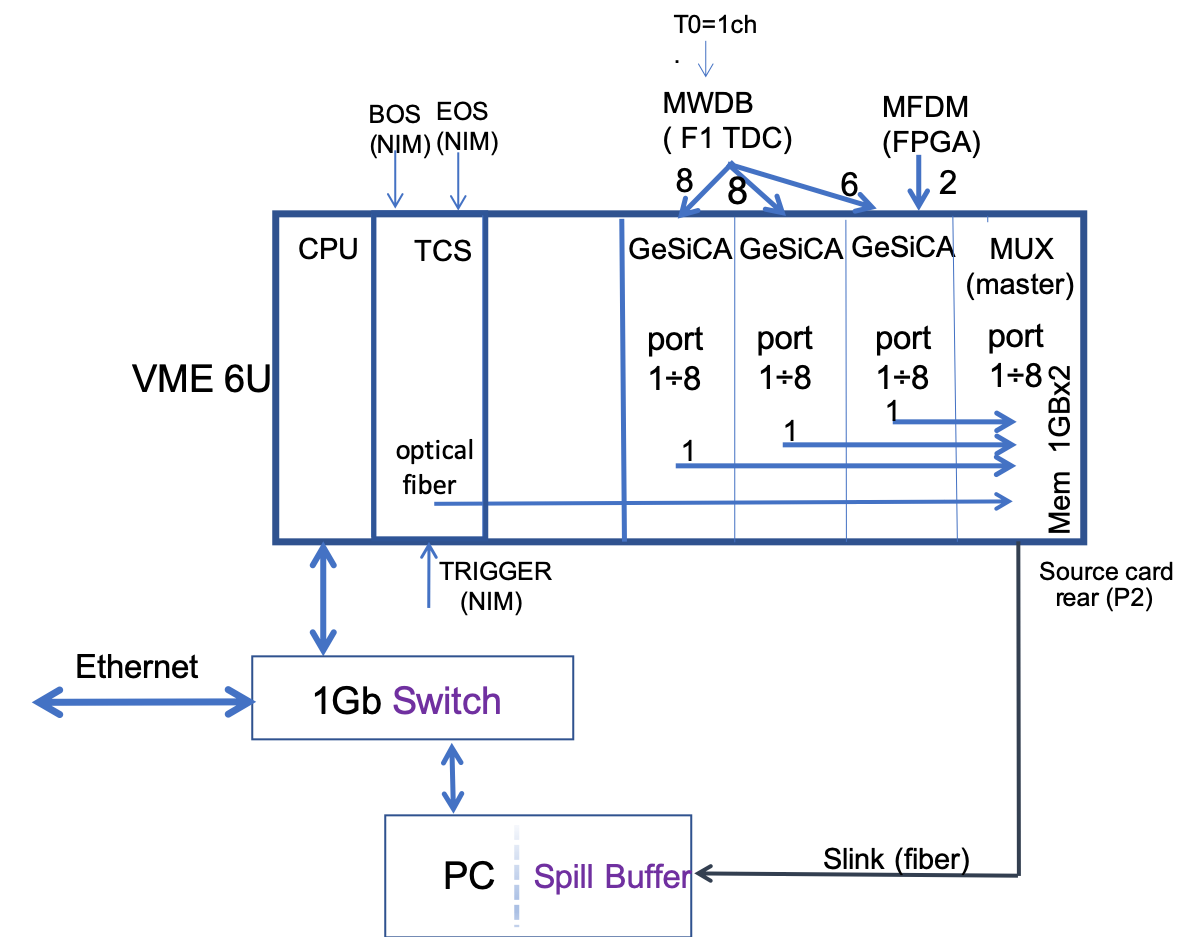}}
\caption{Stand-alone DAQ block-diagram.}
  \label{fig:RS_MFDM_DAQ}  
\end{figure}

To treat the total information from 137600 channels (wires and strips) we need 760 digital
modules (housed in 85 VME 6U crates) with total power consumption of about 18.2 kW.

Tests of the MFDM unit performed at CERN with the Muon System prototype (250 MDT detectors, 4000 readout channels) on cosmic  rays gave encouraging results. 
Figure~\ref{fig:RS_MFDM_DAQ} shows the structure of a stand-alone data acquisition system, based on GeSIca and a multiplexer (MUX) units from COMPASS DAQ, in which two MWDBs (F1 TDC / ASIC) were replaced by two MFDM modules (FPGA) in order to verify their performance.

\begin{figure}[!h]
    \center{\includegraphics[width=0.85\linewidth]{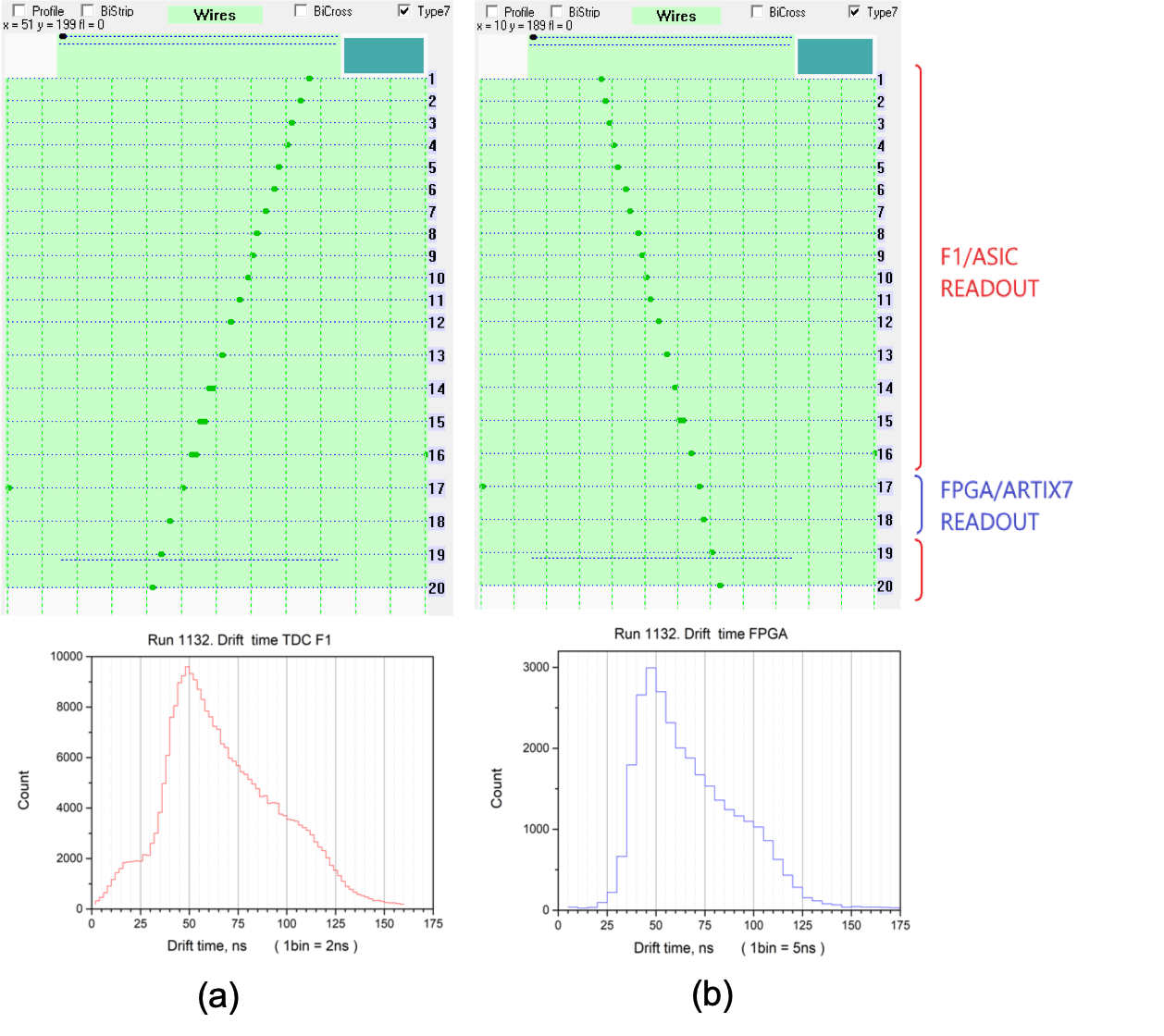}}
\caption{FPGA digital readout test.}
  \label{fig:RS_FPGA_test}  
\end{figure}

Figure~\ref{fig:RS_FPGA_test} displays the tracks and the time spectra obtained from the tests, using only MWDB (F1 TDC) units (a) in one case, and two MFDMs (FPGA) (b) instead of two MWDBs (planes No. 17, 18) in the other. 
In both cases, the obtained results were the same -- in planes No. 17 and 18 the triggered wires are in the correct position on the track, and the time spectra also coincide with small deviations  due to the peculiarities of TDC F1 operation in the latch mode. 
The maximal drift time of MDT ($\sim$150 ns) defines the time resolution of the Muon System providing the time interval / window for assembly of the event (association of the hits from different detectors in one event) by event builder.

The final tests of the MFDM unit will be carried out with the SPD Range System prototype (1300
channels of wire and strip readout) in the SPD test beam zone at Nuclotron. 
The described stand-alone readout system will be used only with the prototype.
But for the SPD Muon System readout, we have developed an FDM module (FPGA Digital Module)
similar to the MFDM card described above, but with a different board form factor and interface for communication with DAQ.

The connection between the FEE (FDM) of the Muon System and the DAQ first-level concentrator L1 is
supposed to be carried out through serial electrical links (e-links), including the following communication signals:
\begin{itemize}
\item reset (LVTTL) from DAQ;
\item reference Clock (LVDS/SLVS) from DAQ;
\item synchronization signals Start of Slice, Start of Frame, and Set next Frame (LVDS/SLVS) from DAQ; 
\item data/clock for FEE initialization (I2C interface, bidirectional, LVTTL);
\item data link FEE (LVDS/SLVS).
\end{itemize}

\begin{figure}[!h]
    \center{\includegraphics[width=0.8\linewidth]{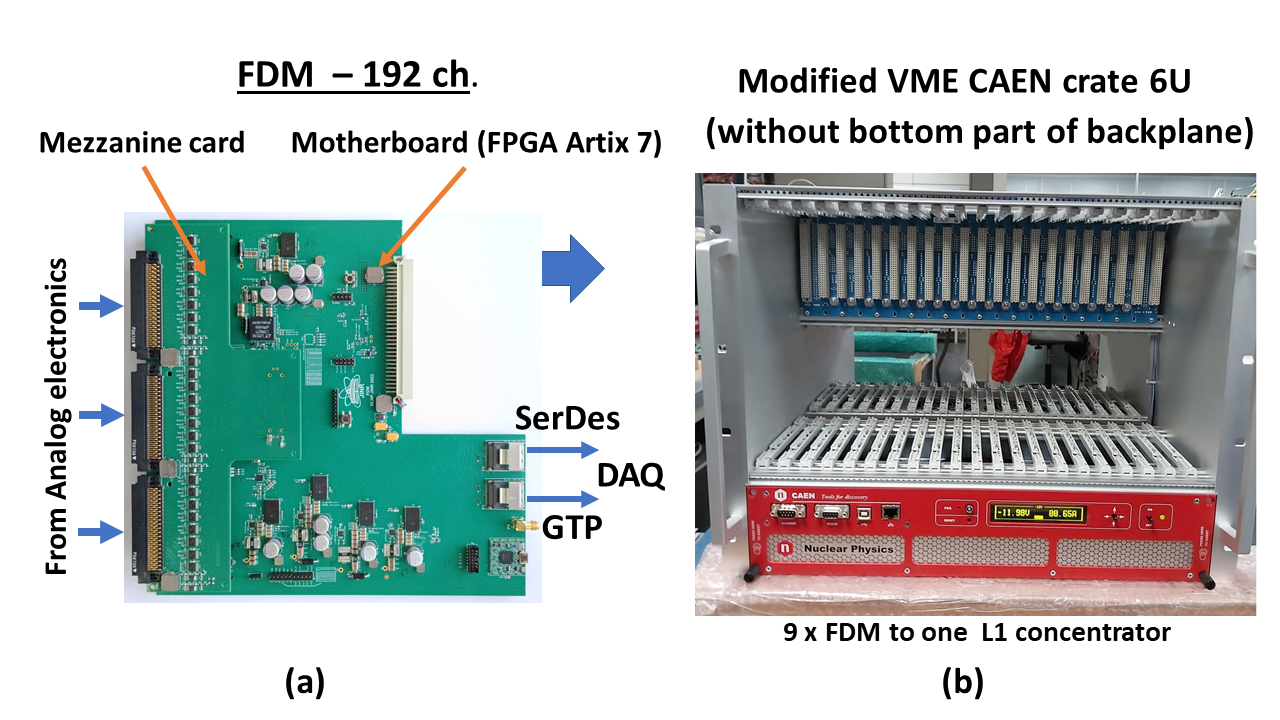}}
\caption{"Boot" shaped FDM module (a) and corresponding VME crate with modified backplane~(b).}
  \label{fig:RS_FDM_module}  
\end{figure}

The FDM module consists of a motherboard and a mezzanine board (Fig.~\ref{fig:RS_FDM_module} (a)). 
The motherboard is made in the shape of a ''boot'', so to accommodate such modules in the crate, its backplane must be modified. 
The modification implies that the lower part of the backplane with P2 connectors is removed (Fig.~\ref{fig:RS_FDM_module} (b)). Otherwise, the crate parameters are the same as in the basic configuration.

Due to the technical capabilities of the Artix 7 FPGA two selectable communication channels with the DAQ are provided in the FDM module:
\begin{itemize}
\item high-speed serial connectivity with built-in multi-gigabit transceivers (GTP channel);
\item SerDes (serialization/deserialization) channel.
\end{itemize}
 
\begin{figure}[!h]
    \center{\includegraphics[width=1\linewidth]{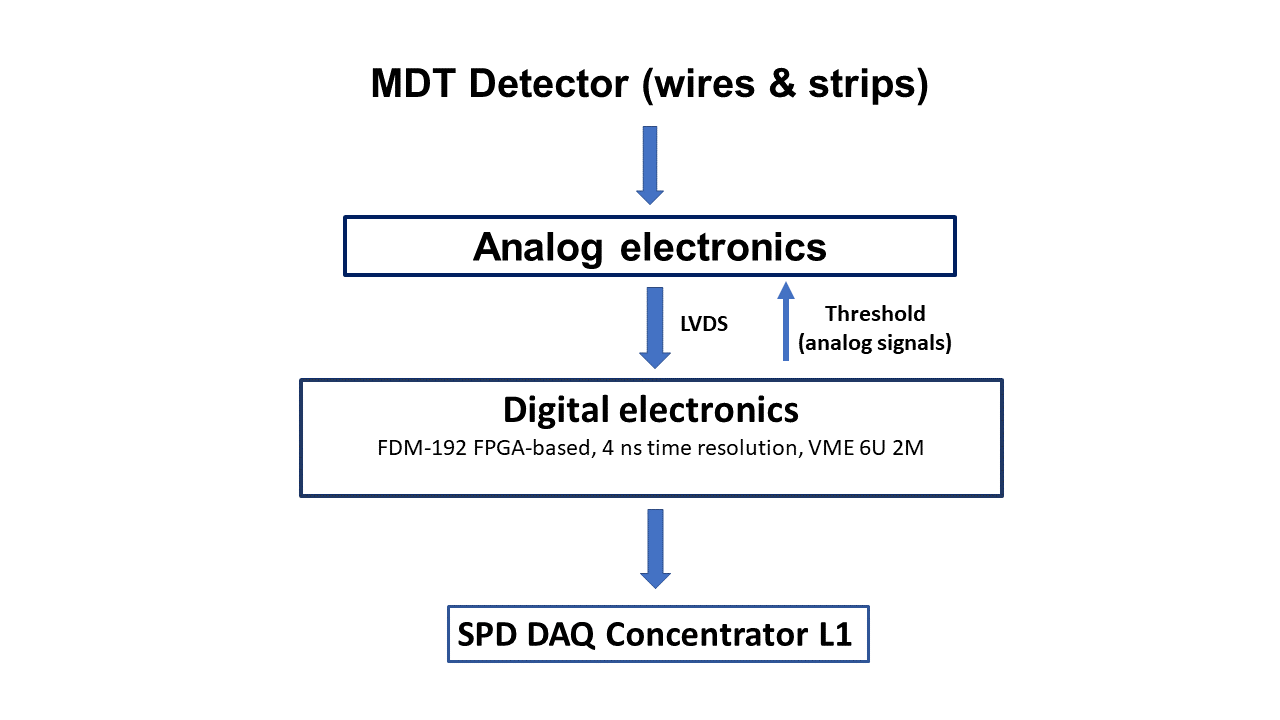}}
     \caption{Data flow diagram -- from the RS to the DAQ. }
       \label{fig:RS_DATA_FLOW}  
\end{figure}

Each communication channels use an individual Mini-SAS connector (36 pins).
The final choice of communication channel will be made after studying the operation of the FDM module
with the first-level concentrator L1.
A general view of the data flow structure for the Muon System is shown in Fig.~\ref{fig:RS_DATA_FLOW}.

\section{Prototyping}
\label{sec:prot}
\begin{figure}[!h]
    \center{\includegraphics[width=0.9\linewidth]{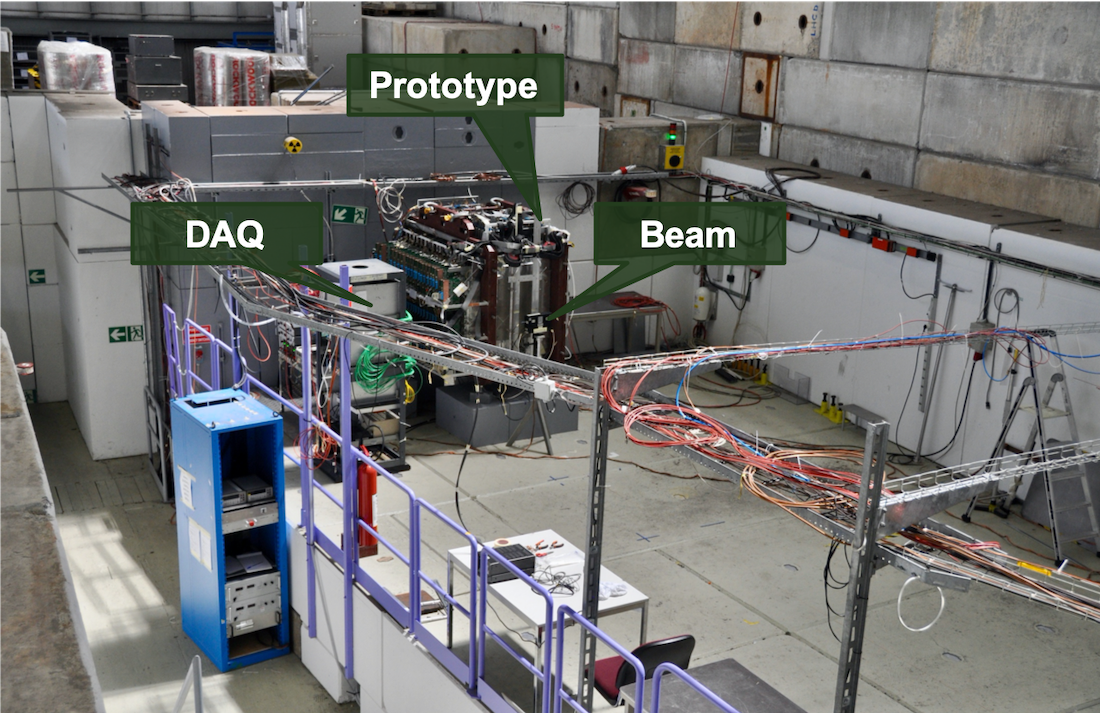}}
     \caption{Range System prototype (10 tons, 4000 readout channels) at CERN.
       \label{fig:RS_proto}}  
\end{figure}

The evaluation of the main parameters of the proposed Range System is being performed with a big prototype installed at CERN within the PANDA program. The prototype (see Fig.~\ref{fig:RS_proto}) has a total weight of about 10 tons (steel absorber and detectors with electronics) and comprises 250 MDT detectors with 4000 readout channels (2000 for the wires and 2000 for the strips, 1 cm wide; finally, the 3 cm wide strips were adopted for the SPD Range System). 
It has both samplings (3 cm and 6 cm) present in the system (PANDA Barrel and End Caps) and provides an opportunity for direct calibration of the response to muons, pions, protons, and neutrons. 

\begin{figure}[!hbt]
  \begin{minipage}[ht]{0.32\linewidth}
 \center{\includegraphics[width=1\linewidth]{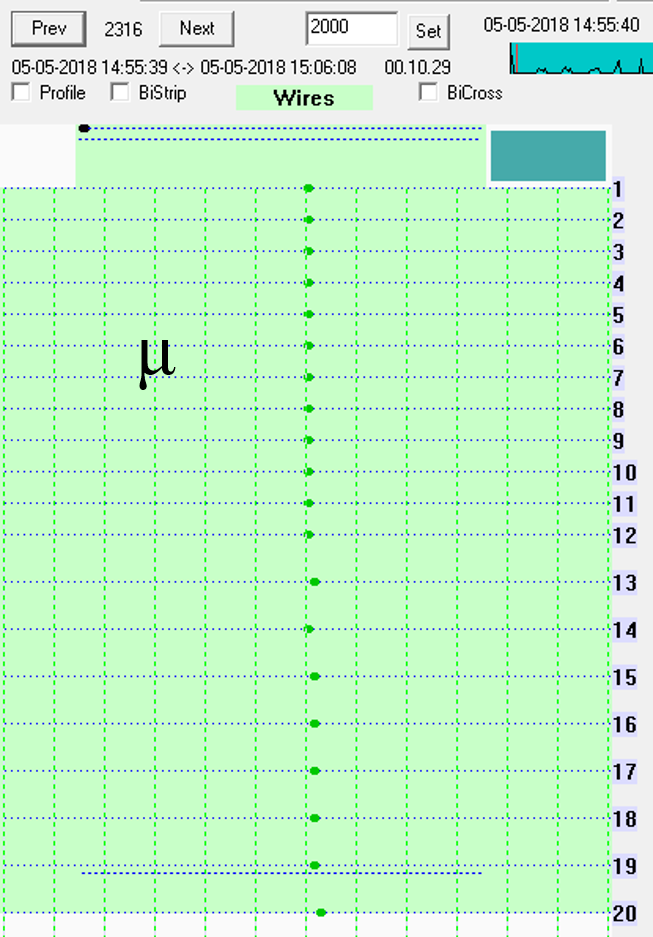} \\ (a)}
 \end{minipage}  
 \hfill
  \begin{minipage}[ht]{0.32\linewidth}
    \center{\includegraphics[width=1\linewidth]{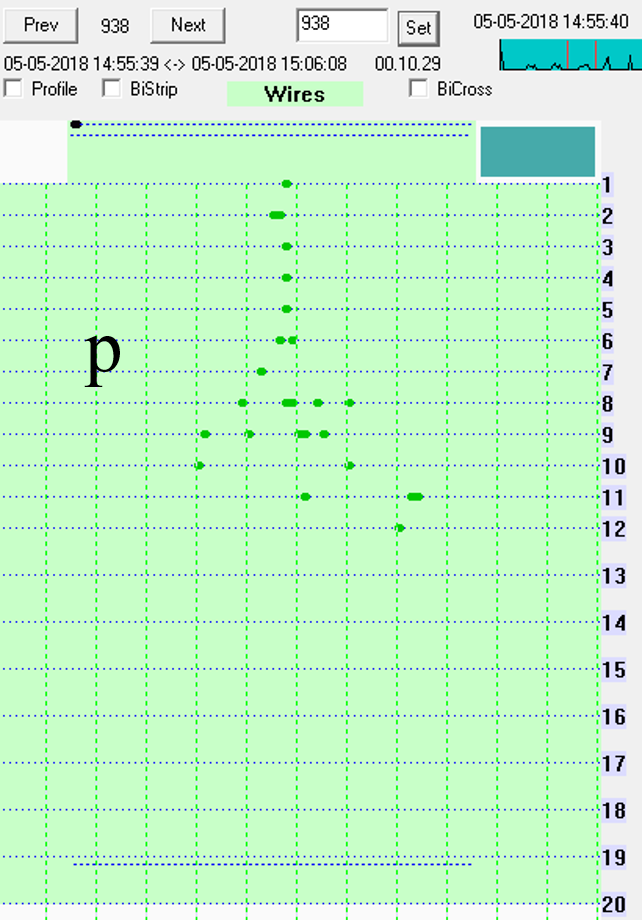} \\ (b)}
  \end{minipage}
   \hfill
  \begin{minipage}[h]{0.32\linewidth}
    \center{\includegraphics[width=1\linewidth]{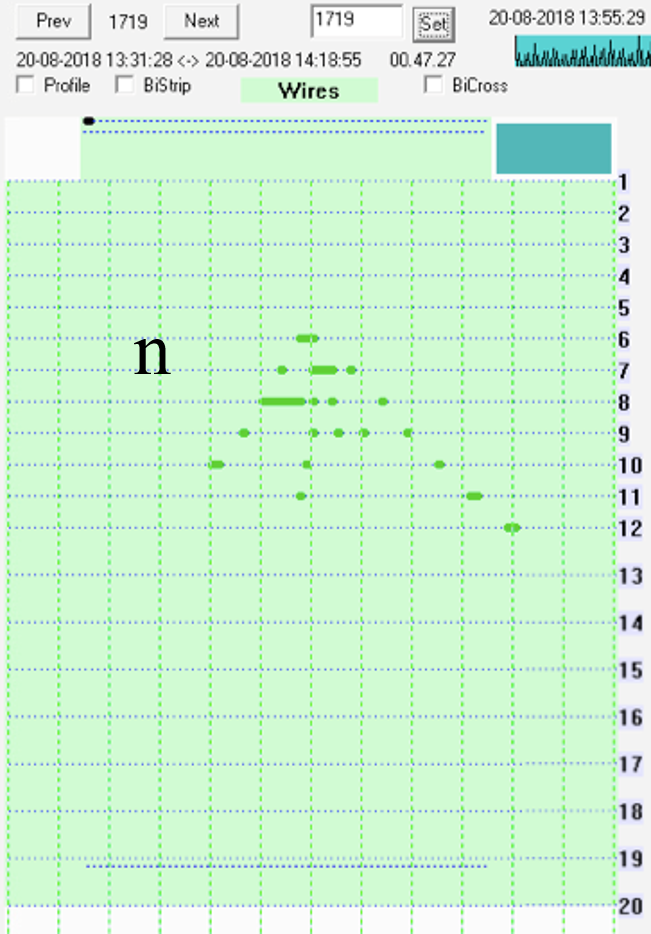} \\ (c)}
  \end{minipage} 
  \caption{Demonstration of PID abilities: patterns for (a) muon, (b) proton, and (c) neutron. }
  \label{fig:RS_PID}  
\end{figure}

Figure~\ref{fig:RS_PID} gives examples of the prototype response to different particles. The patterns demonstrate excellent PID abilities of the Range System. The data were taken during the May and August runs  in 2018 at the T9/PS/CERN test beam. The beam particles hit the prototype from the top of the picture. The beam momentum for all the particles is 5.0 GeV/c. Neutrons were generated by a proton beam on a carbon target placed in the very vicinity of the first detecting layer. The points on the pictures represent hit wires, thus giving the impression of a typical device response with an accuracy of $\sim$ 1~cm.

\begin{figure}[!hbt]
  \begin{minipage}[ht]{0.5\linewidth}
 \center{\includegraphics[width=1.0\linewidth]{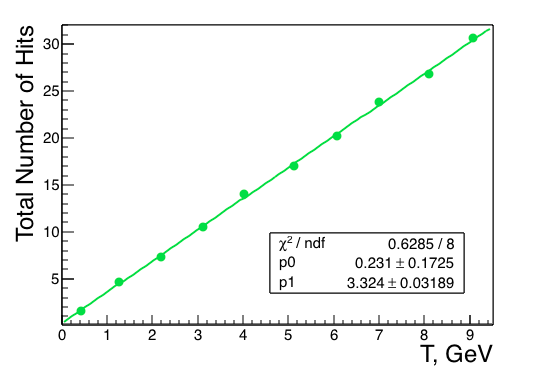} \\ (a)}
 \end{minipage}  
 \hfill
  \begin{minipage}[ht]{0.5\linewidth}
    \center{\includegraphics[width=1.0\linewidth]{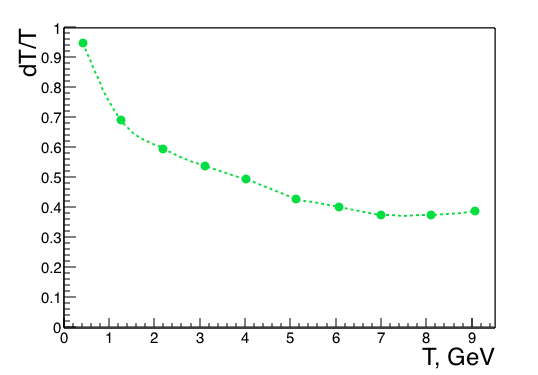} \\ (b)}
  \end{minipage}
  \caption{Calorimetric signal (a) and energy resolution (b) vs kinetic energy (T) of hadrons.}
  \label{fig:RS_Calo/Reso}  
\end{figure}

In addition to good PID quality, the Range System also has  rather good features as a measuring device -- in calorimetry and tracking. 
Figure~\ref{fig:RS_Calo/Reso} demonstrates the linearity of the RS response to the kinetic energy of the hadrons and modest resolution (due to the sampling). 
Such calorimetric signals may potentially help in data analysis in combination with the signals from the electromagnetic calorimeter and tracking system (used for precise momentum definition). 
 Most importantly, this calorimetric signal is the only one to register the neutrons and roughly estimate their energy.

\begin{figure}[!hbt]
  \begin{minipage}[ht]{0.5\linewidth}
 \center{\includegraphics[width=1.0\linewidth]{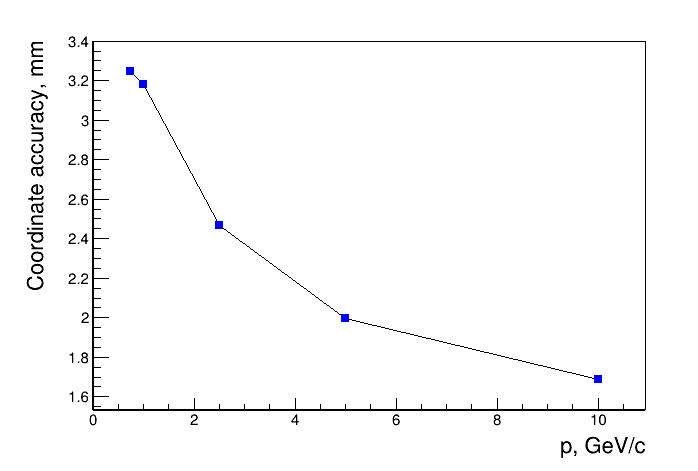} \\ (a)}
 \end{minipage}  
 \hfill
  \begin{minipage}[ht]{0.5\linewidth}
    \center{\includegraphics[width=1.0\linewidth]{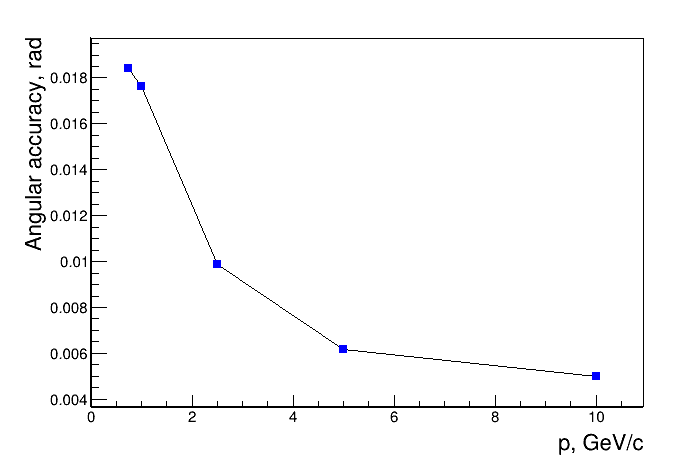} \\ (b)}
  \end{minipage}
  \caption{Coordinate (a) and angular (b) accuracy of the Range System vs muon momentum.}
  \label{fig:RS_Coordinate/Accuracy}  
\end{figure}

 The Range System may also provide additional space information -- space angle for track segment or shower, and entry point to the system. 
Figure~\ref{fig:RS_Coordinate/Accuracy} shows  an example of track
and angle resolution obtained for the muons in ($R, \varphi$) plane. (It depends on the accuracy of about 1 cm provided by the wire readout and smeared by the multiple scattering in the massive absorber.) 
The accuracy of the reconstruction for the entry space point may be estimated as $\sim 3$ cm and is limited by the strip width.

\section{Simulation and performance}
The performance of the Range (Muon) System was evaluated using Monte Carlo simulations of the detector within the SpdRoot framework as well as the R\&D program conducted with the RS prototype on test beams (see Section~\ref{sec:prot}). 
The studies included system response to the passage of particles with different momenta, the efficiency of muon/hadron separation, testing  pattern recognition algorithms, and tuning the parameters of Monte Carlo simulation using experimental data. 

The full simulation of the RS detector is based on the geometrical model of the SPD Range System, incorporated within the SpdRoot framework from engineering drawings.
The model of the RS prototype was also implemented within the SpdRoot package, which made possible a direct comparison of the detector response to traversing particles.

\subsection{Detector model}
The preliminary Range System model was created and simulated using Geant4 and encapsulated within the SpdRoot framework. It consists of a Barrel and two End Caps with equal thickness 4$\lambda_I$. MC-point is created by Geant4 in a sensitive detector volume when a particle traverses through it.

Digitization is used while converting MC-points into hits (detector signal). The procedure is based on at least one of two requirements: 
\begin{enumerate}
\item minimum passage in active volume $\ge$ 1~cm;
\item minimum energy deposited in active volume $\ge$ 1~keV.
\end{enumerate}

\begin{figure}[!hbt]
  \begin{minipage}[ht]{0.33\linewidth}
 \center{\includegraphics[width=1\linewidth]{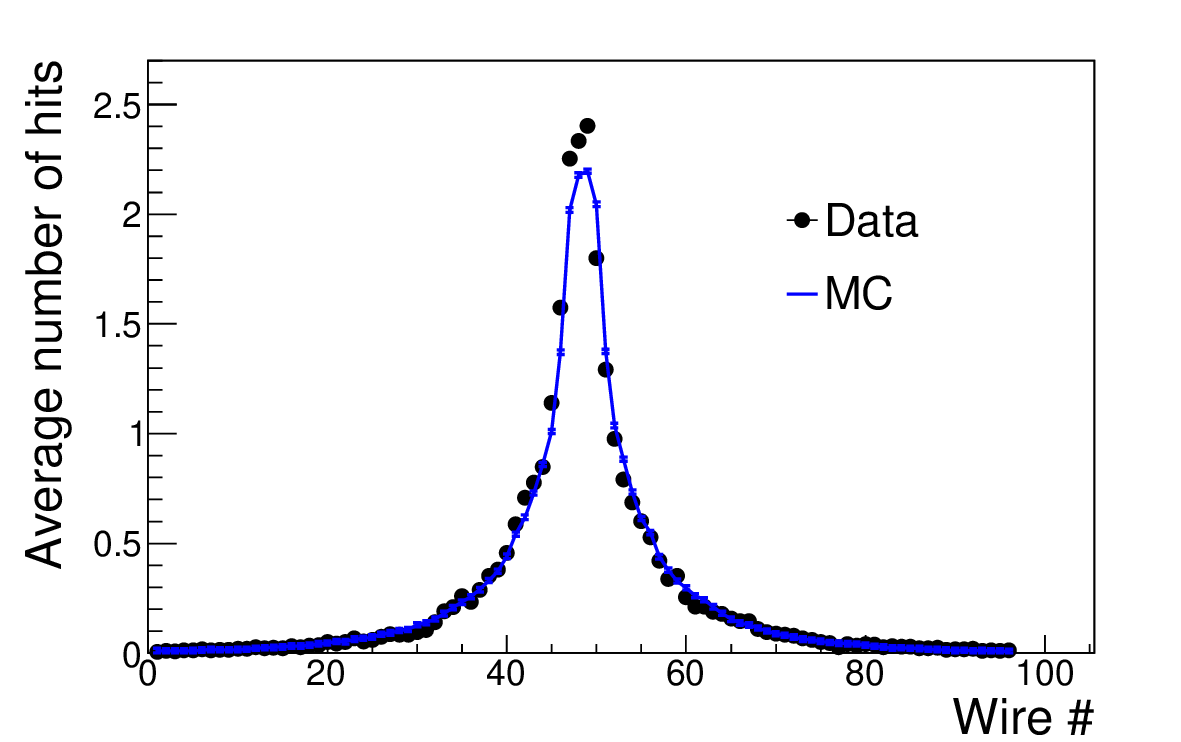} \\ (a)}
 \end{minipage}  
  \begin{minipage}[ht]{0.33\linewidth}
    \center{\includegraphics[width=1\linewidth]{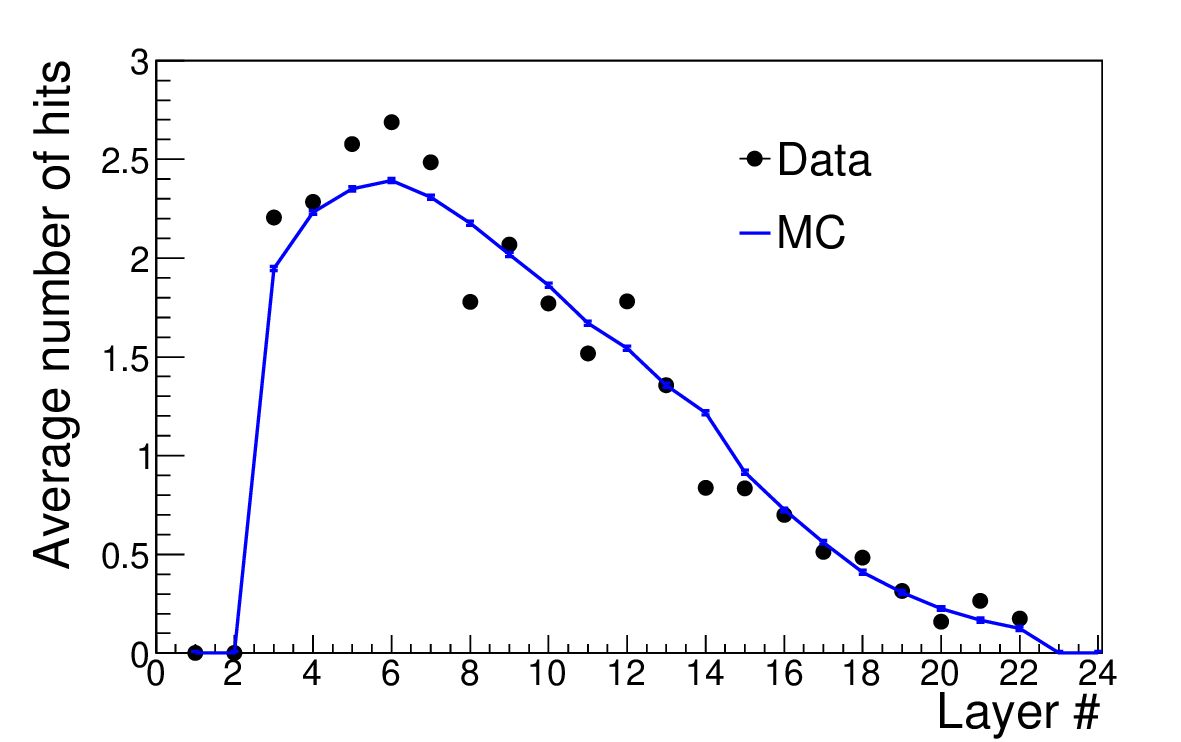} \\ (b)}
  \end{minipage}
  \begin{minipage}[h]{0.33\linewidth}
    \center{\includegraphics[width=1\linewidth]{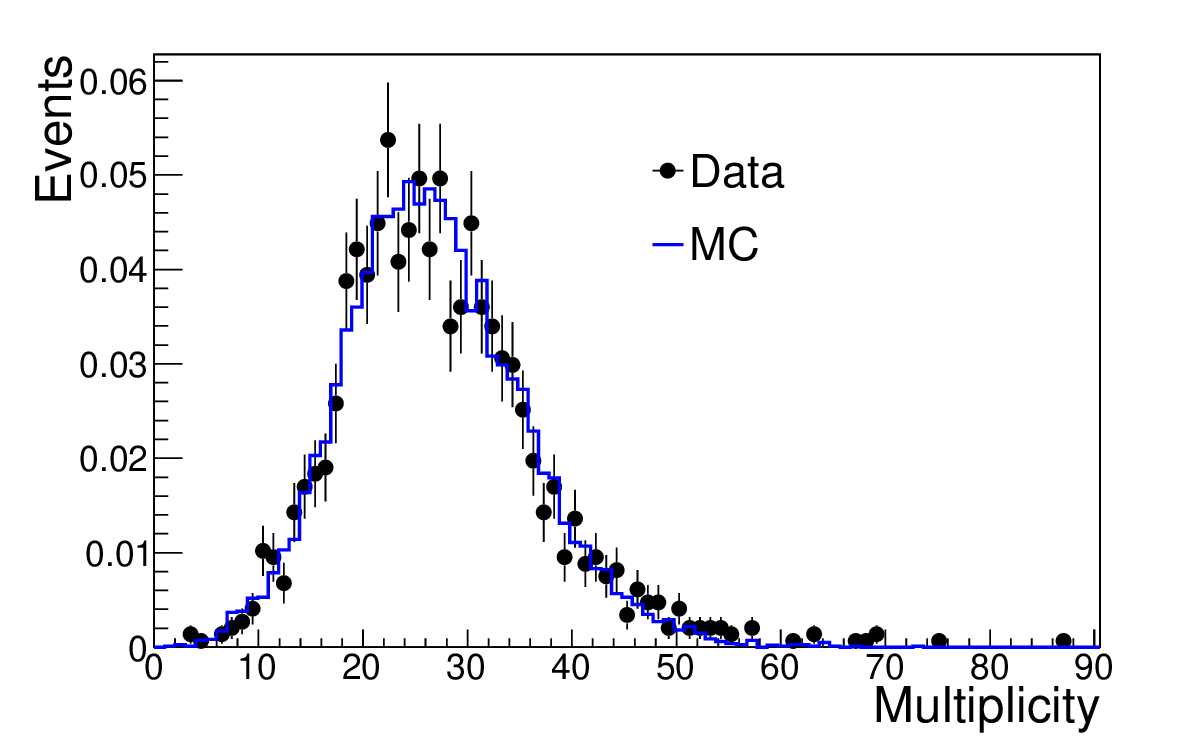} \\ (c)}
  \end{minipage} 
  \caption{Calorimetric signal/response to the protons with momentum 6 GeV/$c$: (a) average hits per wire (all layers), (b) average hits per layer, and (c) normalized hit multiplicity in an event. }
  \label{fig:RS_Response}  
\end{figure}

Figure~\ref{fig:RS_Response} shows the main reference signal/response distributions:
average hits per wire (all layers), average hits per layer, and normalized hit multiplicity in an event.
It shows good agreement between data and MC with the digitization procedure applied.

\subsection{PID algorithms}
One of the primary purposes of the Range System is the reconstruction of charged particles traversing the RS and the identification of muons and hadrons. 
Charged particles produce hits in the MDT detectors forming a hit profile in a layered structure of the Range System. 
Two approaches may be applied: a Kalman propagation algorithm for fitting tracks and algorithms based on machine learning (ML) techniques. 
The second  approach can be represented as a pipeline of algorithms that process the resulting hit profile in two steps: (a) forming groups of hits by clustering algorithms and (b) assigning the obtained clusters to muons or hadrons by classification algorithms. 
The main advantage of the described procedure is a relatively high processing speed, compared to traditional track-reconstruction algorithms, since it exploits only the signals from the Range System alone with no need to assign tracks to those reconstructed in the Straw Tracker.

\subsection{Clustering}
Clustering is one of the unsupervised machine learning tasks and
only hard clustering algorithms (each hit belongs to a cluster or not)  are considered.
The most basic class of clustering algorithms is the centroid-based K-Means algorithm.
Each data point is classified by computing the distance between that point and each group center, 
and then classifying the point to be in the group whose center is closest to it.
Then the group center is recomputed by taking the mean of all the vectors in the group.
These steps are repeated for a chosen number of iterations or until the group centers do not change much between iterations.
Despite the speed and linear complexity, the application of the algorithm is found to be nearly impossible for our aims,
since the number of clusters should be known in advance.
As a result, it splits hadron showers and muon tracks into segments.

A possible solution could be the application of density-based algorithms that do not require a pre-set number of clusters at all.
The DBSCAN algorithm (Density-Based Spatial Clustering Application with Noise) considers clusters as areas of high density 
separated by areas of low density.
The algorithm identifies outliers as noise, associates hits to clusters of arbitrary shapes, naturally exploits 3D coordinates of hits (wire/strips), and has only two hyper-parameters.
The main downside is that it doesn't perform as well as others when the clusters are of varying density.

\begin{figure}[!h]
    \center{\includegraphics[width=0.8\linewidth]{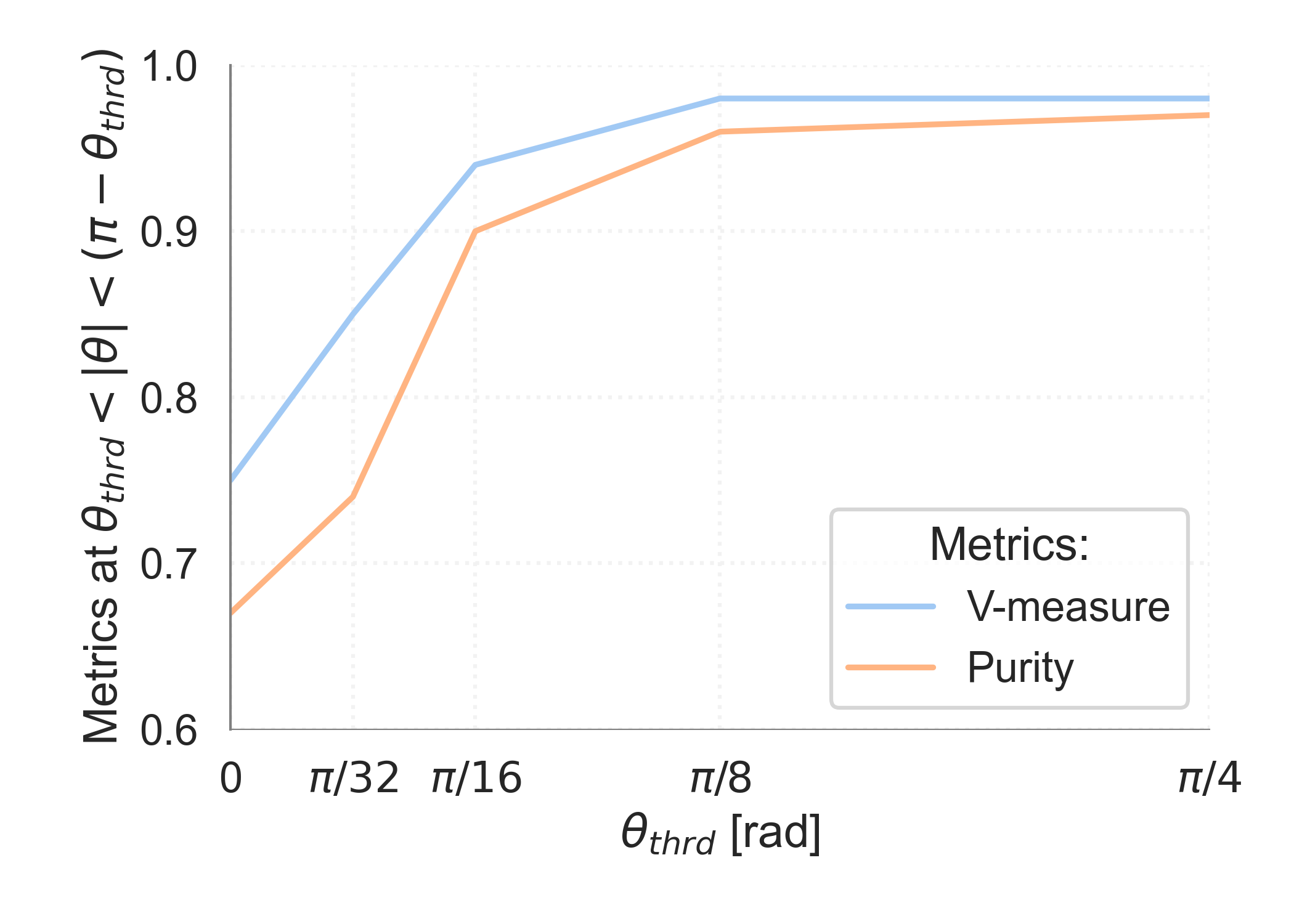}}
     \caption{The DBSCAN evaluation metrics as a function of polar angle threshold. }
       \label{fig:RS_DBSCAN_Metrics}
\end{figure}

To estimate the quality of the algorithm, we use two evaluation metrics: clustering purity and V-measure.
Clustering purity ($P$) is defined as a ratio of the sum of correctly defined hits over the total number of hits
$P = \sum_i N^{correct}_{i, hits}/ N^{total}_{hits}$.
It is found to be a very straightforward and transparent metrics, but it increases as the number of clusters increases.
Another metrics is defined by the values of \textit{homogeneity}, where each cluster contains only members of a single class, and 
\textit{completeness}, where all members of a given class are assigned to the same cluster.
The V-measure is defined as a harmonic mean between homogeneity and completeness.
The advantages of this metrics is that it is normalized between 0 and 1, and it can be used to compare 
different clustering models that have different numbers of clusters.
The downside is that the random labeling will not yield zero scores, especially when the number of clusters is large.
Figure~\ref{fig:RS_DBSCAN_Metrics} shows the DBSCAN evaluation metrics as a function of the polar angle threshold
$\theta_{thrd} < |\theta| < (\pi - \theta_{thrd})$.
The performance of the clustering procedure looks reasonable, starting from the $\pi/16<\theta<15\pi/16$
that approximately corresponds to pseudorapidity requirement $|\eta|<2.4$, quite regularly used by $J/\psi$ physics analyses. 

\subsection{Muon/hadron separation}

\begin{figure}[!hbt]
  \begin{minipage}[ht]{1\linewidth}
 \center{\includegraphics[width=0.8\linewidth]{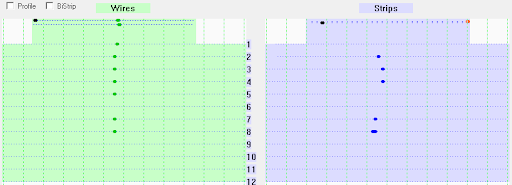} \\ (a)}
 \end{minipage}  
 \hfill
  \begin{minipage}[ht]{1\linewidth}
    \center{\includegraphics[width=0.8\linewidth]{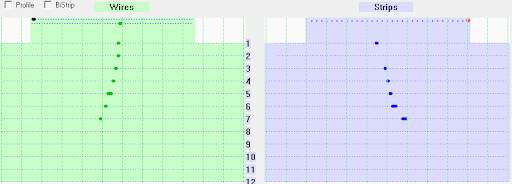} \\ (b)}
  \end{minipage}
  \caption{Comparison of wire (green) and strip (purple) patterns in the RS Prototype for (a) muons and (b) pions with the same momentum 0.5 GeV/$c$. }
  \label{fig:RS_MuPi05}  
\end{figure}

\begin{figure}[!hbt]
  \begin{minipage}[ht]{1\linewidth}
 \center{\includegraphics[width=0.8\linewidth]{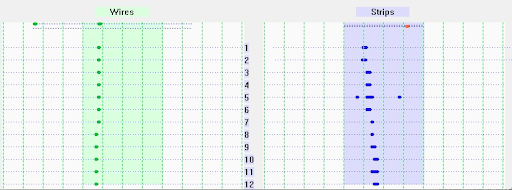} \\ (a)}
 \end{minipage}  
 \hfill
  \begin{minipage}[ht]{1\linewidth}
    \center{\includegraphics[width=0.8\linewidth]{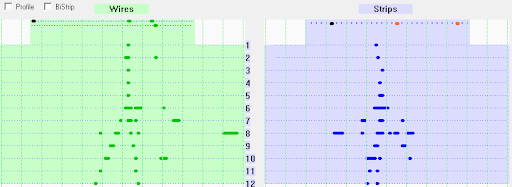} \\ (b)}
  \end{minipage}
  \caption{Comparison of wire (green) and strip (purple) patterns in the RS Prototype for (a) muons and (b) pions with the same momentum 10 GeV/$c$. }
  \label{fig:RS_MuPi10}  
\end{figure}

The second stage of the reconstruction is used to separate between muon and hadron clusters found in the clustering stage by using ML classification algorithms.
A hit profile in the Range System corresponding to a particular kind of particle with a certain momentum has a specific pattern. Low-momentum hadrons (p $<$ 1 GeV/$c$)  are almost indistinguishable from muons with the same momentum (see Fig.~\ref{fig:RS_MuPi05}). The increasing energy of hadrons significantly changes the profile of hits, forming a cascade of secondary particles for hadrons with momentum up to 10 GeV/$c$ (see Fig.~\ref{fig:RS_MuPi10}).

Finding variables sensitive to differences in such patterns is directly related to the possibility of separation between muons and hadrons. It can be used as an input to various Machine Learning techniques.

\begin{figure}[!hbt]
  \begin{minipage}[ht]{0.5\linewidth}
 \center{\includegraphics[width=1.0\linewidth]{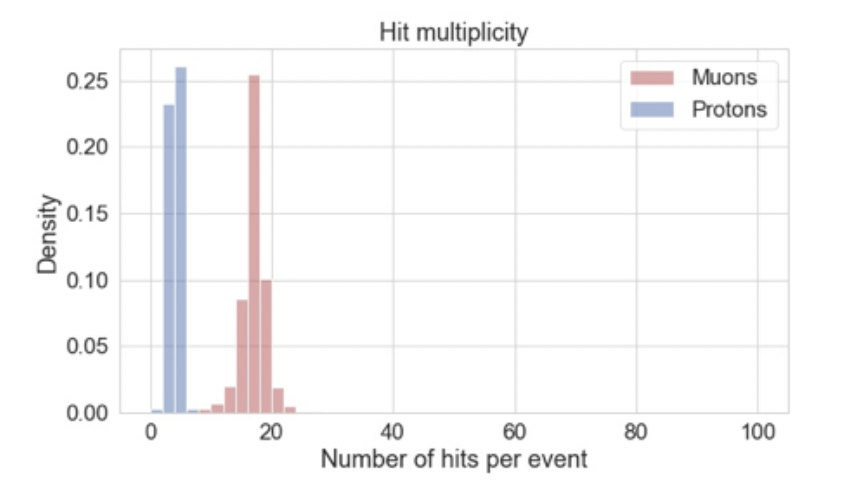} \\ (a)}
 \end{minipage}  
 \hfill
  \begin{minipage}[ht]{0.5\linewidth}
    \center{\includegraphics[width=1.0\linewidth]{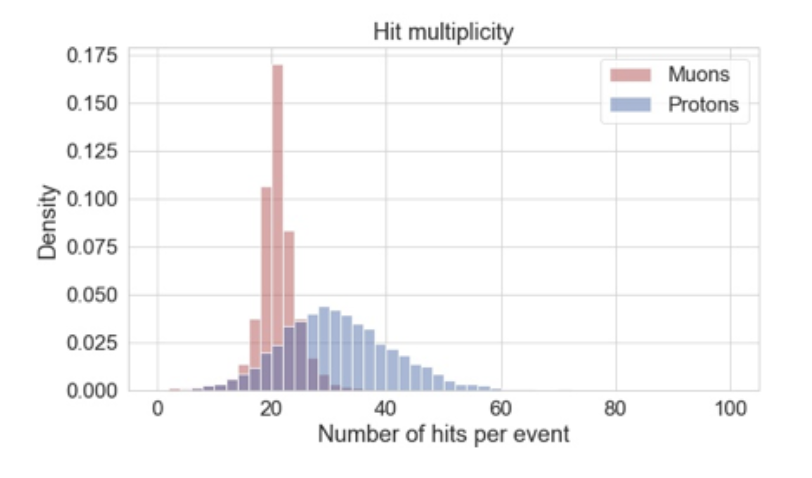} \\ (b)}
  \end{minipage}
  \caption{Normalized distribution of hit multiplicity per event for muons compared to protons with momentum  (a) p = 1 GeV/$c$ and (b) p = 6 GeV/$c$ in SPD Range System (MC). }
  \label{fig:RS_HitsMulti1and6}  
\end{figure}

\begin{figure}[!hbt]
  \begin{minipage}[ht]{0.5\linewidth}
 \center{\includegraphics[width=1.0\linewidth]{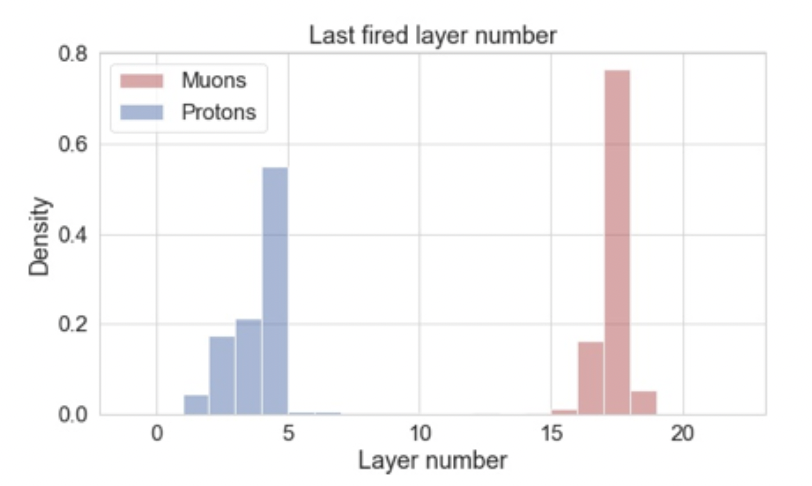} \\ (a)}
 \end{minipage}  
 \hfill
  \begin{minipage}[ht]{0.5\linewidth}
    \center{\includegraphics[width=1.0\linewidth]{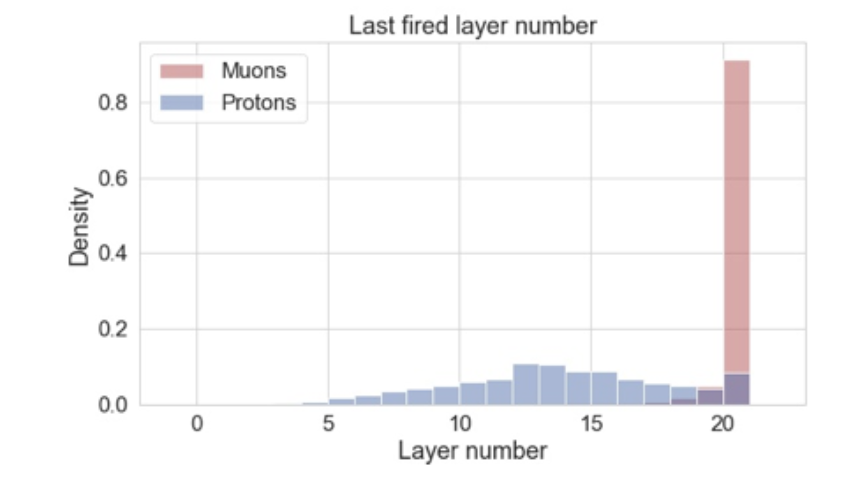} \\ (b)}
  \end{minipage}
  \caption{Normalized distribution of last fired layer number for muons compared to protons with momentum  (a) p = 1 GeV/$c$ and (b) p = 6 GeV/$c$ in SPD Range System (MC). }
  \label{fig:RS_LastLayer1and6}  
\end{figure}

As a starting point, a Decision Tree classification algorithm is used to separate between signal (muon) and background (protons and pions) samples in Data and MC. The chosen technique is a well-known solution for binary classification (DecisionTreeClassifier from the scikit-learn library). The main advantages of the algorithm are its robustness, transparency, and limited number of hyper-parameters for optimization. Events in both samples were labeled using time-of-flight detectors only. Later, the Cherenkov counter tags will be used to fix the muons. The following variables are chosen as  input to the Decision Tree: hit multiplicity in an event, first and last fired layers, splitting layer number (the lowest number of a layer having $\ge$ 2 hits), and number of hits in the last layer. The normalized distributions of hit multiplicity per event and last fired layer number for muons compared to protons with momentum p =~1~GeV/$c$ and p =~6~GeV/$c$ obtained with the SPD Range System (MC) are shown in Fig.~\ref{fig:RS_HitsMulti1and6} and Fig.~\ref{fig:RS_LastLayer1and6}.

\begin{figure}[!h]
   \begin{minipage}{0.48\textwidth}
     \centering
     \includegraphics[width=0.98\linewidth]{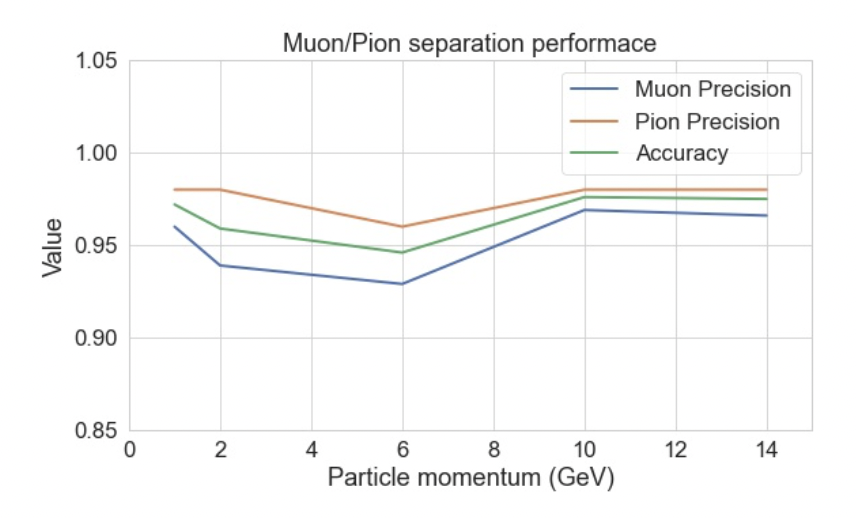} \\ (a)
   \end{minipage}\hfill
   \begin{minipage}{0.48\textwidth}
     \centering
     \includegraphics[width=1.\linewidth]{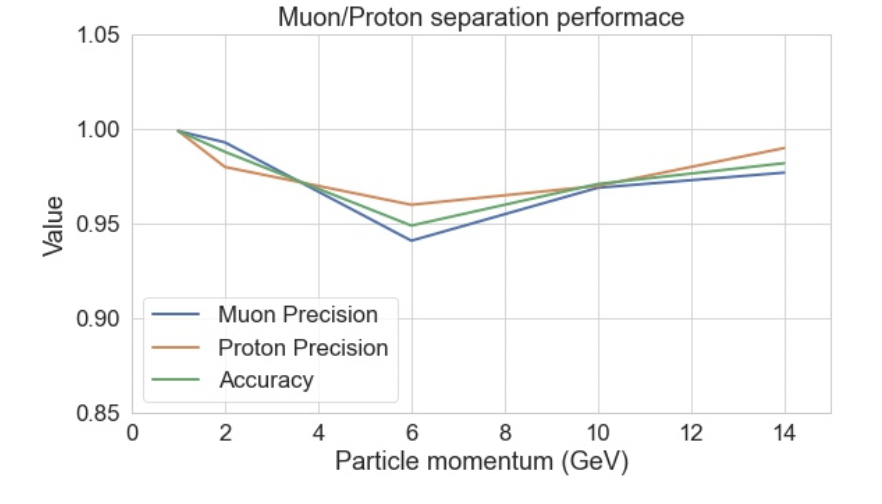} \\ (b)
   \end{minipage}
     \caption{Muon/pion (a) and muon/proton (b) separation performance as a function of particle momentum.}
      \label{fig:RS_Separation}
\end{figure}

The hit multiplicity is found to be the most powerful discriminative variable while the last fired layer number also showed high importance for muon/hadron separation. The applied technique allows to differentiate between the classes with 95$\div$97\% accuracy for muons/pions (see Fig.~\ref{fig:RS_Separation}~(a)) and 96$\div$99\% for muons/protons (see Fig.~\ref{fig:RS_Separation}~(b)). The analogous method was applied to RSP dataset and showed less accuracy for muons/protons separation (93\%), mainly due to the impurity of the signal  in the muon sample, since the events in data were labeled using TOF detectors only. Later, the Cherenkov counter tags will be used to fix the muon data sample.

The combined algorithms based on the ML-technics and Kalman propagation can be expected to be more effective. 
The preliminary MC-simulations show that the pion suppression rate below 1\% can be achieved, but it
has not been confirmed yet in the analysis of data from the RS prototype.

\clearpage

\section{Cost estimate}
The preliminary cost estimate for the Range System is presented in Table \ref{RS:cost}.

\begin{table}[htp]
\caption{Cost estimate for the RS.}
\begin{center}
\begin{tabular}{|l|c|}
\hline
 & Cost, M\$ \\
\hline 
\hline
MDT detectors with stripboards  & 3.03   \\
Analog front-end electronics (including cabling) & 3.72 \\
Digital front-end electronics (including VME crates and racks) & 4.32 \\
Yoke/absorber (without support and transportation system) & 6.18 \\
\hline
Total & 17.25 \\
\hline
\end{tabular}
\end{center}
\label{RS:cost}
\end{table}%

\chapter{Magnetic system}
\section{SPD superconductive solenoid}
A superconducting magnet is one of the most important parts of the SPD setup. Together with the tracking system, it provides the measurement of charged particle momenta with an accuracy of about 2\%. The steel elements of the Range System serve as a yoke of this magnet. 
\subsection{General performance requirements}
\subsubsection{Magnetic field}
The magnetic field along the solenoid axis should be  1.0 T.
\begin{figure}[!hb]
    \center{\includegraphics[width=1.\linewidth]{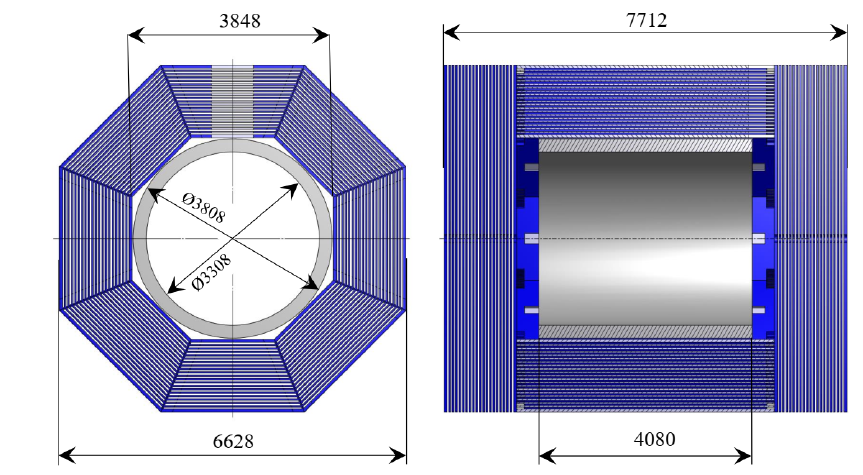}}
  \caption{ Main dimensions of the cryostat and the yoke.
	}
\label{lb_mag_1}
\end{figure}
\subsubsection{Main dimensions and parameters of the magnet}
The cryostat of the magnet with the coils, cold mass, and thermal shields is located inside the yoke. A distribution box named Control Dewar is located on the upper octant of the yoke.
The overall dimensions are driven by the space planned for the detectors of SPD and the magnet field parameters. The outside diameter of the cryostat is 3808 mm and the gap between the yoke and the cryostat is about 20 mm. Radially a free diameter of 3308 mm is left for the SPD detectors. The length of the magnet is 3800 mm and the magnet should be installed symmetrically inside the yoke (see Fig.  \ref{lb_mag_1}). 

\subsection{Technical specification of components}
\subsubsection{Magnetic analysis: field homogeneity optimization}
The SPD solenoid is designed to provide a magnetic field of 1.0 or 1.25 T over a length of about 4 m in a bore of 3.2 m. 
The field homogeneity constraints at the center of the solenoid ($|z|<1.5$~m, $r<0.9$~m) uniformity is better than $\pm$2\%. To meet the above requirements, the cold mass design involved splitting the solenoid into three interconnected windings (Upstream/Downstream and Center coils, respectively). The study has been carried out for diffrent number of turns in each coil and operating current from  4.4 to 5.2 kA. The final magnet configuration corresponds to an operating current of 5 kA and  the number of turns 312 -- 156 -- 312 for the Upstream, Center and Downstream coils. 

\begin{figure}[!ht]
  \begin{minipage}[h]{0.5\linewidth}
    \center{\includegraphics[width=1\linewidth]{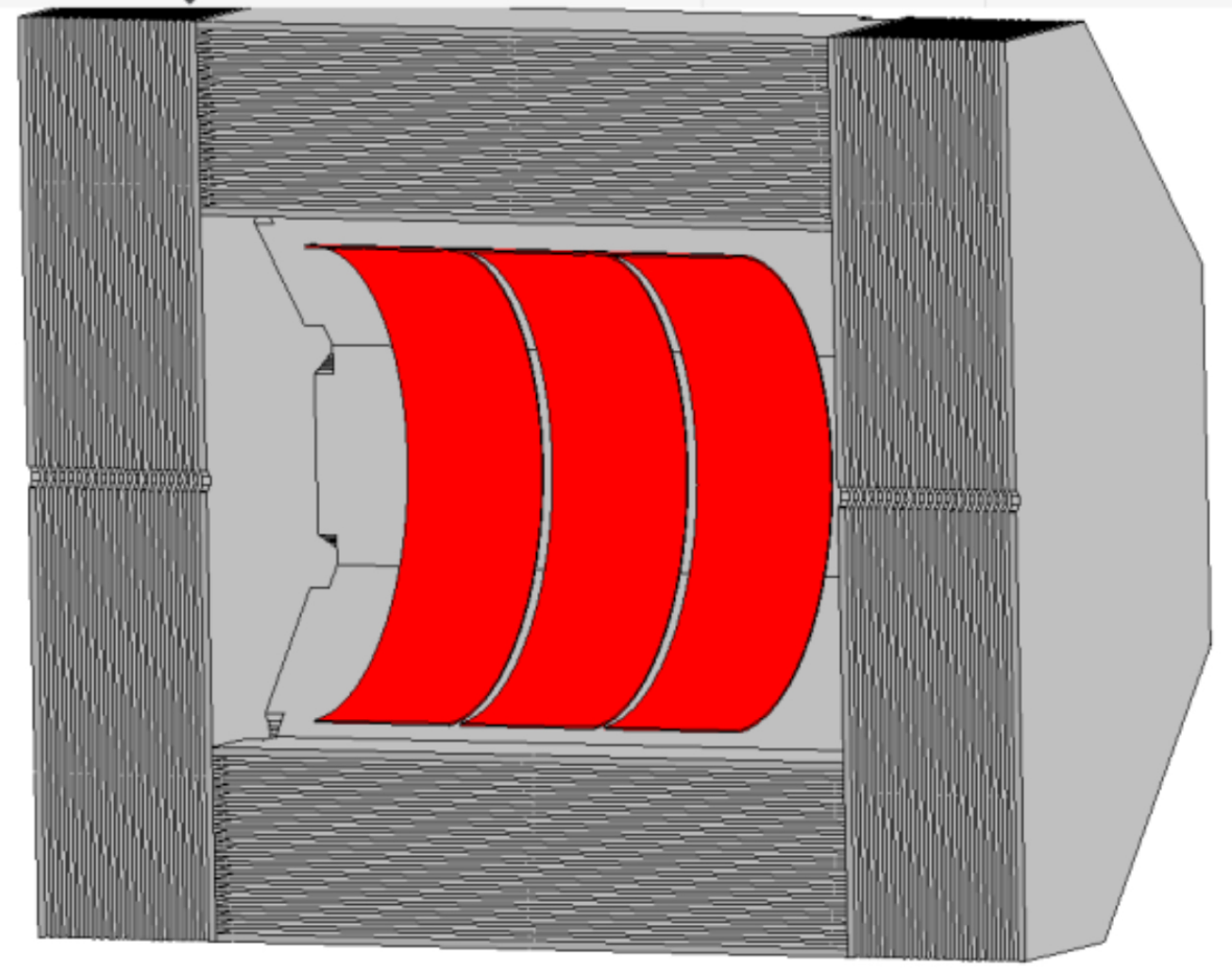} \\ (a)}
  \end{minipage}
  \hfill
  \begin{minipage}[h]{0.5\linewidth}
    \center{\includegraphics[width=1\linewidth]{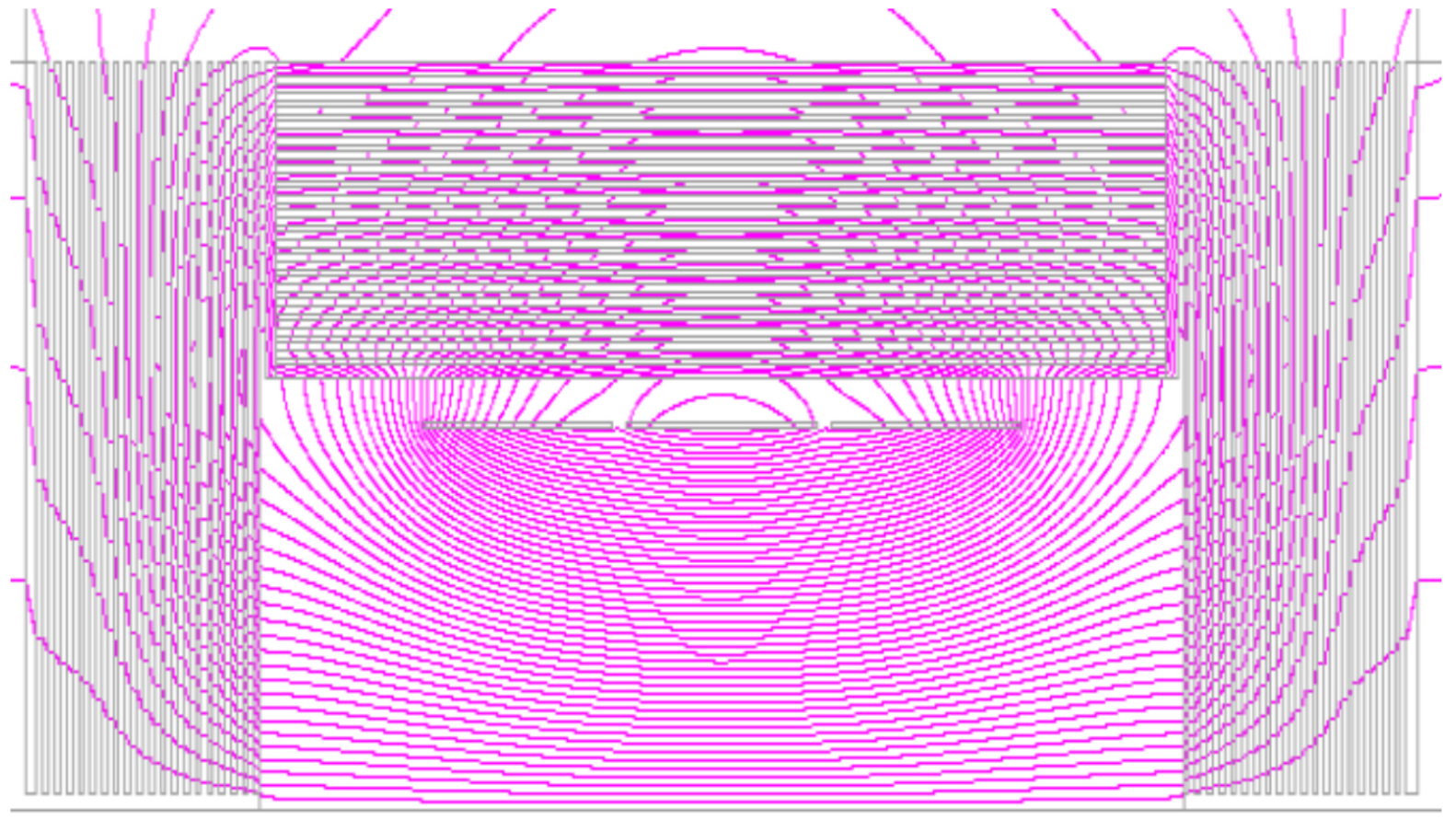} \\ (b)}
  \end{minipage} 
  \caption{(a) Genegal layout of the solenoid with three coils. (b) Configuration of the magnetic field.}
\label{M_lb_mag_2}
\end{figure}

\paragraph{2D simulation: axially symmetric case}
We considered the original version with three identical coils of 250 turns each and a current of 6300 A. The arrangement for coils was 250-250-250. The general layout of the solenoid and the corresponding magnetic field configuration are shown in Fig. \ref{M_lb_mag_2}. 
As well as the same arrangement of the side coils was considered, but with half the number of turns in the central coil. That is the arrangement for coils as 250-125-250. 

\begin{figure}[!h]
    \center{\includegraphics[width=1.\linewidth]{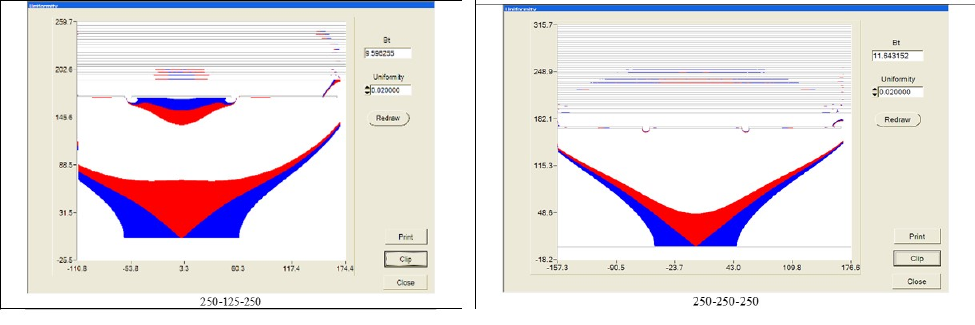}}
  \caption{ Uniformity of the $B_z$ component within the range $\pm$ 2\%  for two arrangements of coils: 250-125-250 (left) and 250-250-250 (right).
	}
\label{M_lb_mag_3}
\end{figure}

\begin{figure}[!h]
    \center{\includegraphics[width=1.\linewidth]{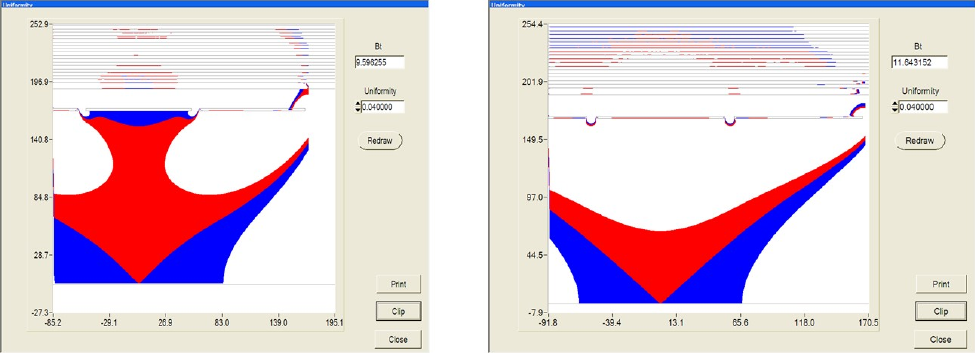}}
  \caption{Uniformity of the $B_z$ component within the range $\pm$ 4\%  for two arrangements of coils: 250-125-250 (left) and 250-250-250 (right).
	}
\label{M_lb_mag_4}
\end{figure}

\begin{figure}[!h]
    \center{\includegraphics[width=1.\linewidth]{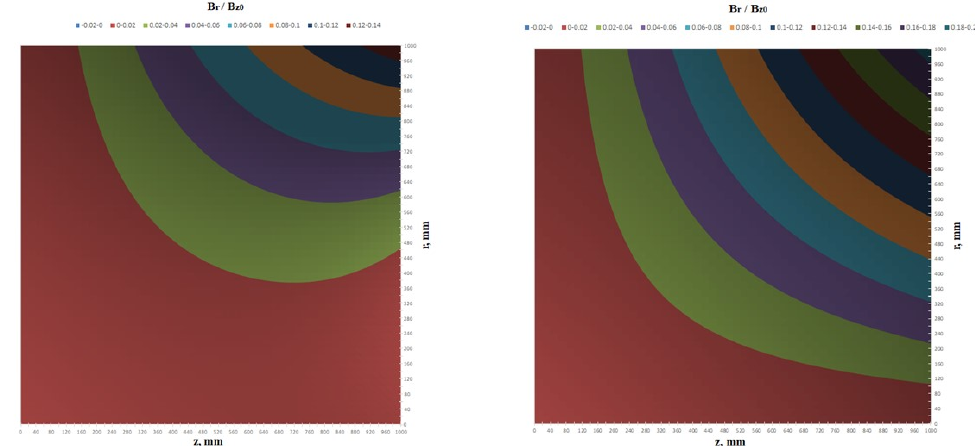}}
  \caption{ Uniformity of $B_r/B_z$ for two arrangements of coils: 250-125-250 (left) and 250-250-250 (right).
	}
\label{M_lb_mag_5}
\end{figure}

The uniformity of the $B_z$ component within the ranges $\pm$ 2\%  and $\pm$ 4\% for two arrangements of coils is presented in Fig. \ref{M_lb_mag_3} and Fig. \ref{M_lb_mag_4}, respectively. The uniformity of the $B_r/B_z$ value is shown in Fig. \ref{M_lb_mag_5}. As one can see, the uniformity region for the $B_z$ component and the $B_r/B_z$ value  for the arrangement scheme 250-125-250 is about 2$\div$2.5 times larger than for the 250-250-250 case.

\begin{figure}[!h]
    \center{\includegraphics[width=1.\linewidth]{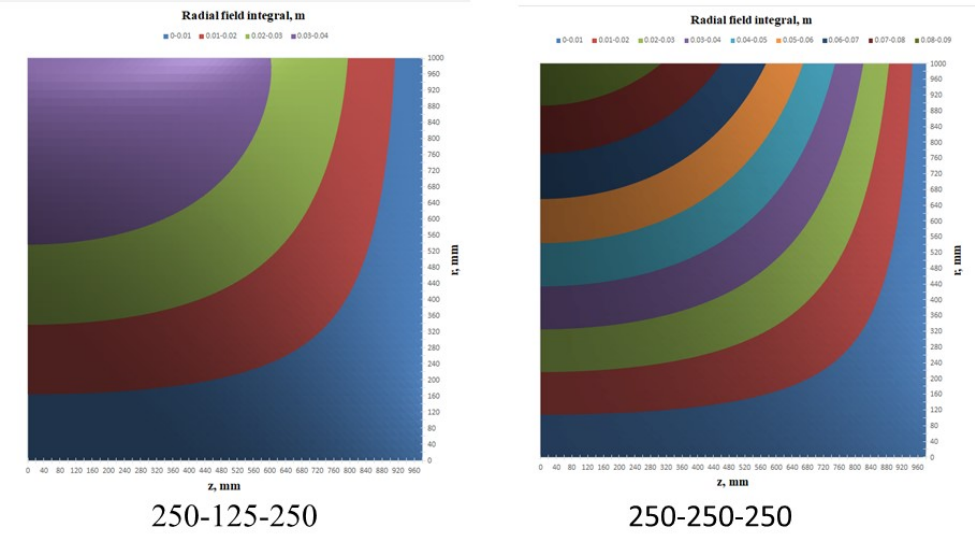}}
  \caption{ Radial integral for two arrangements of coils: 250-125-250 (left) and 250-250-250 (right).
	}
\label{M_lb_mag_6}
\end{figure}

The radial integral is calculated for different radii $r$ as:  
 \begin{equation}
Int(r) = |\int_{z}^{100~cm}\frac{B_r(r,z)}{B_z(r,z)}dz|
\end{equation}
 is presented in Fig. \ref{M_lb_mag_6}. The maximum value of the radial integral decreased from 88 mm to 30 mm for the 200 cm ($z$) $\times$ 100 cm ($r$) region. 
At the same time, the value of the central magnetic field was reduced from 1.16 T for the 250-250-250 case to 0.96 T for the 250-125-250 case. 

To provide the value of the central field above 1 T, it is necessary to increase the number of turns in the side coils to 300 and in the central coil to 150. For a current of 6300 A, the value of the central magnetic field of 1.12 T is obtained. 
In conclusion, it should be noted that reducing the number of ampere-turns in the central coil twice as compared to the number of ampere-turns in the side coils has a positive effect on the quality of the magnetic field in terms of increasing the zone of uniformity of the field component $B_z$, the ratio $B_r/B_z$, as well as minimizing the radial integral.
At the same time, the reduction of the central field by 20\% should be compensated by increasing the turns in all coils, also by 20\%, respectively, to ensure a central field of at least 1.1 T. 
The above reasoning was performed for the two-dimensional axial case. After analyzing the coil design, we came up with a more technologically advanced version of it, which reflects all the considerations of this section.
This new version has been considered in the 3D calculations of the next paragraph.

\paragraph{3D simulation}

\begin{figure}[!h]
    \center{\includegraphics[width=1.\linewidth]{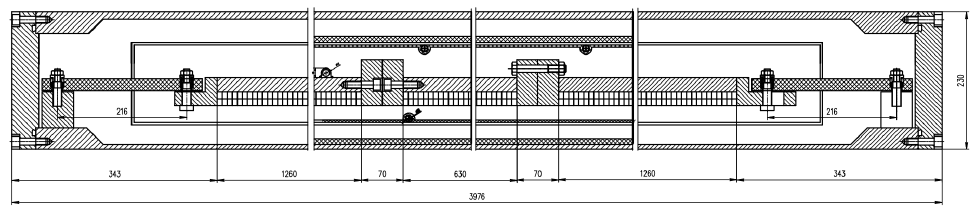}}
  \caption{ 
  Cross-sectional view of the magnet demonstrating the arrangement of superconducting coils for the basic option with three main coils and without correction coils.
	}
\label{fig:coils}
\end{figure}

The following conditions were applied for the 3D simulation:
the configuration of the superconducting coil and magnetic design according to the latest STEP model with three coils. The coil arrangement is shown in Fig. \ref{fig:coils}.
The current of the coil is 5200 A. The conductor cross-section is 8 mm $\times$ 12 mm. The length of  the side coils is 2$\times$1260 mm, while the length of the central coil is 630 mm. 

\begin{figure}[!h]
    \center{\includegraphics[width=1.\linewidth]{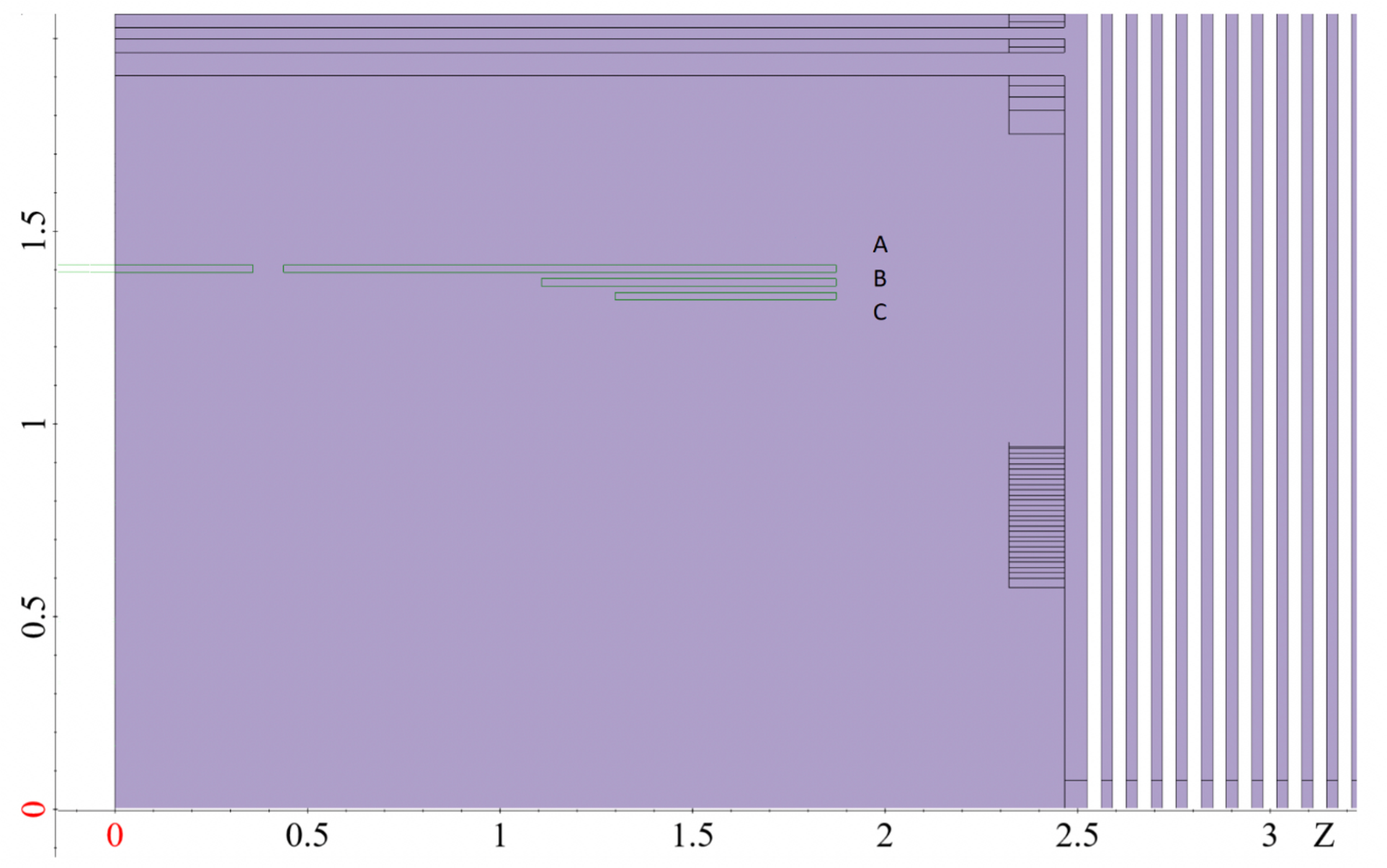}}
  \caption{ 
  Schematic cross-section view of the yoke and coils for the configuration of the solenoid with correction coils.
	}
\label{fig:coils2}
\end{figure}
The calculations were carried out using the MASTAC software package.
Three options were considered:
\begin{enumerate}
\item the basic version with three coils;
\item the correction coils added to the basic version (see Table \ref{tab:coils1} );
\item the central coil is absent to increase the field uniformity.
\end{enumerate}
Figure \ref{fig:coils2} shows the identification of the superconducting coils. The ''A'' configuration is a basic three coils arrangement,
 ''B'' and ''C'' are the types of correction coils added. Table \ref{tab:coils1} shows the correction coil designs in terms of the number of layers and ampere-turns. 

\begin{table}[h!]
\caption{Design of the correction coils.}
\begin{center}
\begin{tabular}{|c|c|c|}
\hline
Configuration & Coil ''B'' & Coil ''C''   \\
\hline
1   &  2 layers of 60 turns & 2 layers of 60 turns \\
2   &  2 layers of 20 turns & 2 layers of 50 turns \\
3   &  2 layers of 50 turns & 2 layers of 50 turns \\
4   &  2 layers of 40 turns & 2 layers of 40 turns \\
\hline
\end{tabular}
\end{center}
\label{tab:coils1}
\end{table}%

\begin{table}[h!]
\caption{Maximum values of deviations of $B_z(z)$ values from $B_z(0)$ in \% at different vertical Y positions.}
\begin{center}
\begin{tabular}{|c|c|c|c|}
 \hline
 & Y=0 m & Y=0.5 m & Y=1.0 m \\
 \hline
 65$\times$2 & 12.2 & 10.8 & 3.7 \\
  60$\times$2  +   60$\times$2    &  5.7  & 2.9 & 13.7 \\
  20$\times$2  +   50$\times$2    &  9.7  & 8.0 & 3.9 \\                      
  50$\times$2  +   50$\times$2    &  6.6  & 4.2 & 10.1 \\
    40$\times$2  +   40$\times$2    &  8.3  & 6.3 & 5.8 \\
    80$\times$2  +   30$\times$2 +   30$\times$2    &  3.9  & 3.4 & 15.8 \\
\hline
\end{tabular}
\end{center}
\label{tab:coils2}
\end{table}%

\begin{figure}[!h]
  \begin{minipage}[ht]{0.5\linewidth}
    \center{\includegraphics[width=1\linewidth]{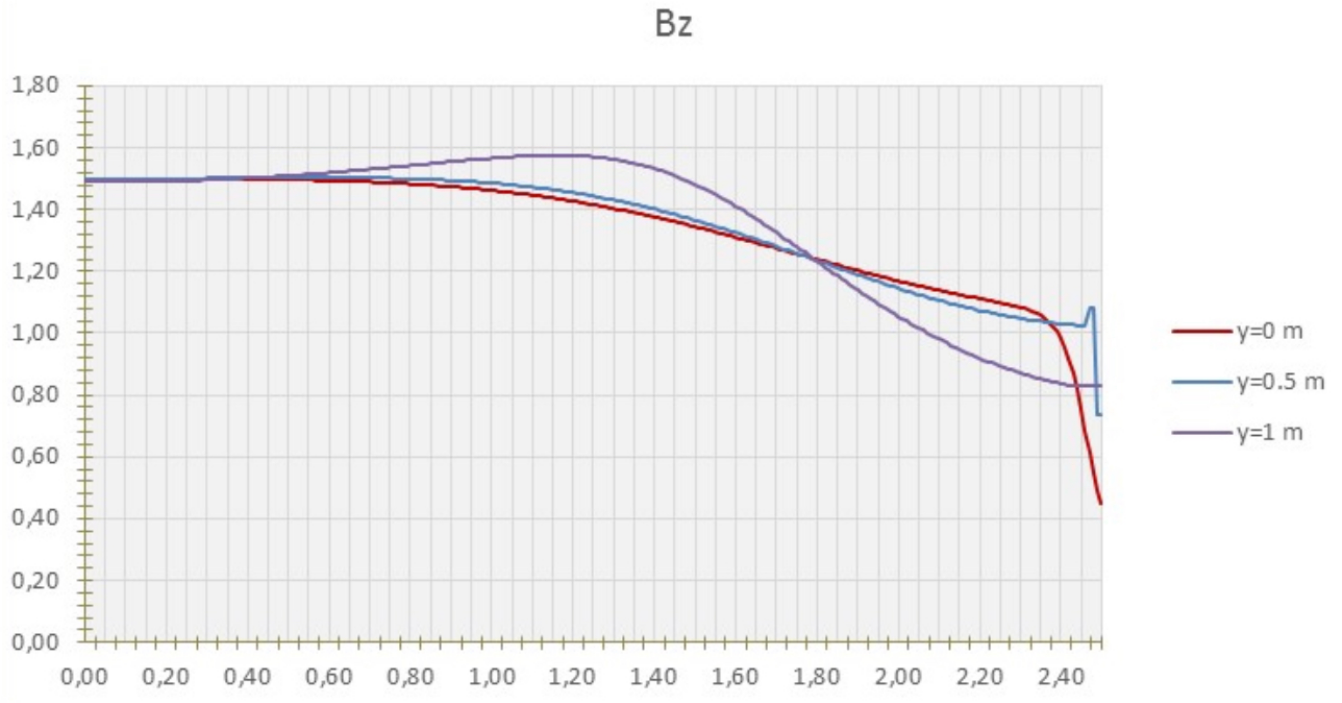} \\ (a)}
  \end{minipage}
  \hfill
  \begin{minipage}[ht]{0.5\linewidth}
    \center{\includegraphics[width=1\linewidth]{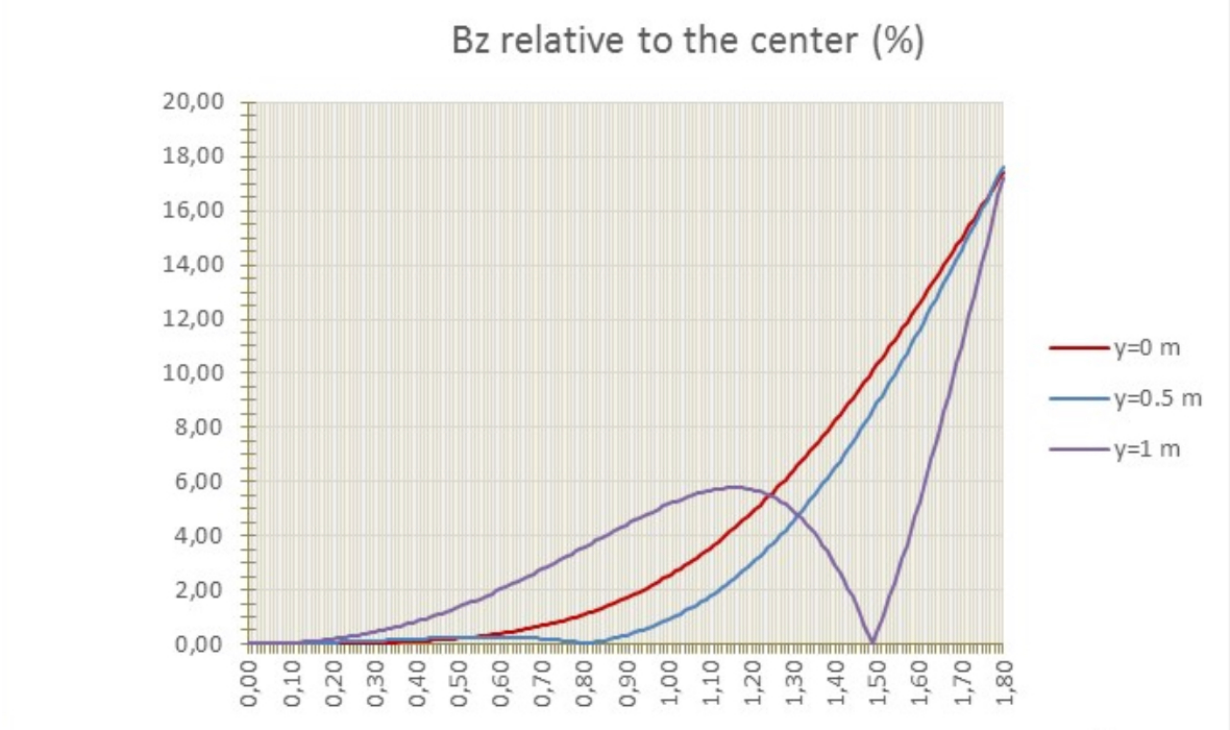} \\ (b)}
  \end{minipage} 
  \caption{
  Distribution of the axial component of the magnetic field $B_z$ along the $z$ axis for the magnet configuration with correction coils $40 \times 2 + 40 \times 2$ and the current of 5200~A. (a) shows the absolute value of $B_z$ and (b) shows $B_z$ normalized to its value at $z=0$. Three curves correspond to the displacements $y=0$~m, $y=0.5$~m, $y=1$~m at $x=0$~m.
	}
\label{fig:field_2}
\end{figure}

\begin{figure}[!h]
    \center{\includegraphics[width=1.\linewidth]{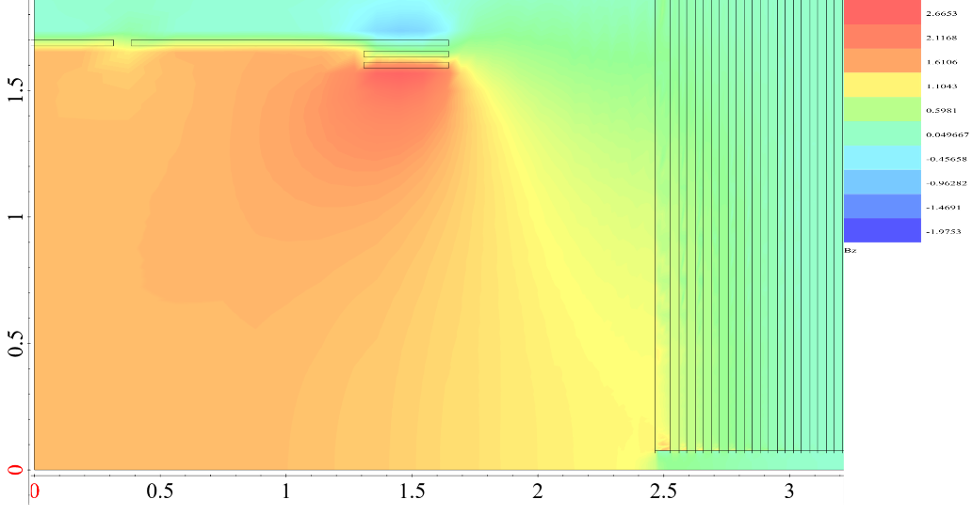}}
  \caption{ 
  Distribution of the $B_z$ component of the magnetic field in the $YZ$ section at $X = 0$ for correcting coils configuration $40 \times 2 + 40 \times 2$.
	}
\label{fig:field_3}
\end{figure}

Table \ref{tab:coils2} presents the uniformity for the simulation of cases described in Table \ref{tab:coils1}. From Table \ref{tab:coils2}, one can see that for coils 40$\times$2 + 40$\times$2 the results are optimal. 
This field map can be considered as a reference for estimating the rate of reconstruction in a homogeneous field.
The distribution and uniformity of $B_z$ field component in the section $YZ$ at $X = 0$ for coils 40$\times$2 + 40$\times$2 are shown in Fig. \ref{fig:field_2} (a) and (b), respectively. The color plot for the $B_z$ component is presented in Fig. \ref{fig:field_3}. 

\begin{figure}[!h]
    \center{\includegraphics[width=1.\linewidth]{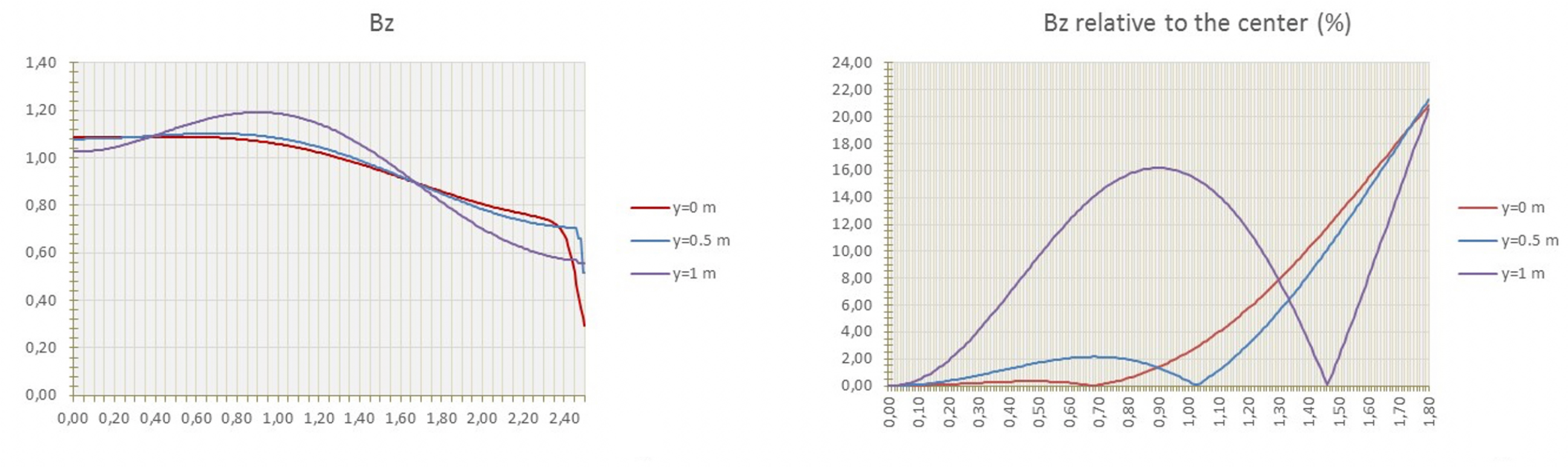}\\ (a)~~~~~~~~~~~~~~~~~~~~~~~~~~~~~~~~~~~~~~~~~~~~~~~~~~~~~~~~~~~~~~~~~~~~~~~~~~~~~~~~(b)}
  \caption{
  Distribution of the axial component of the magnetic field $B_z$ along the $z$ axis for the magnet configuration without the main central coil and the current of 6500~A. (a) shows the absolute value of $B_z$ and (b) shows $B_z$ normalized to its value at $Z=0$. Three curves correspond to the displacements $Y=0$~m, $Y=0.5$~m, $Y=1$~m at $X=0$~m.
	}
\label{fig:field_4}
\end{figure}

\begin{figure}[!h]
    \center{\includegraphics[width=1.\linewidth]{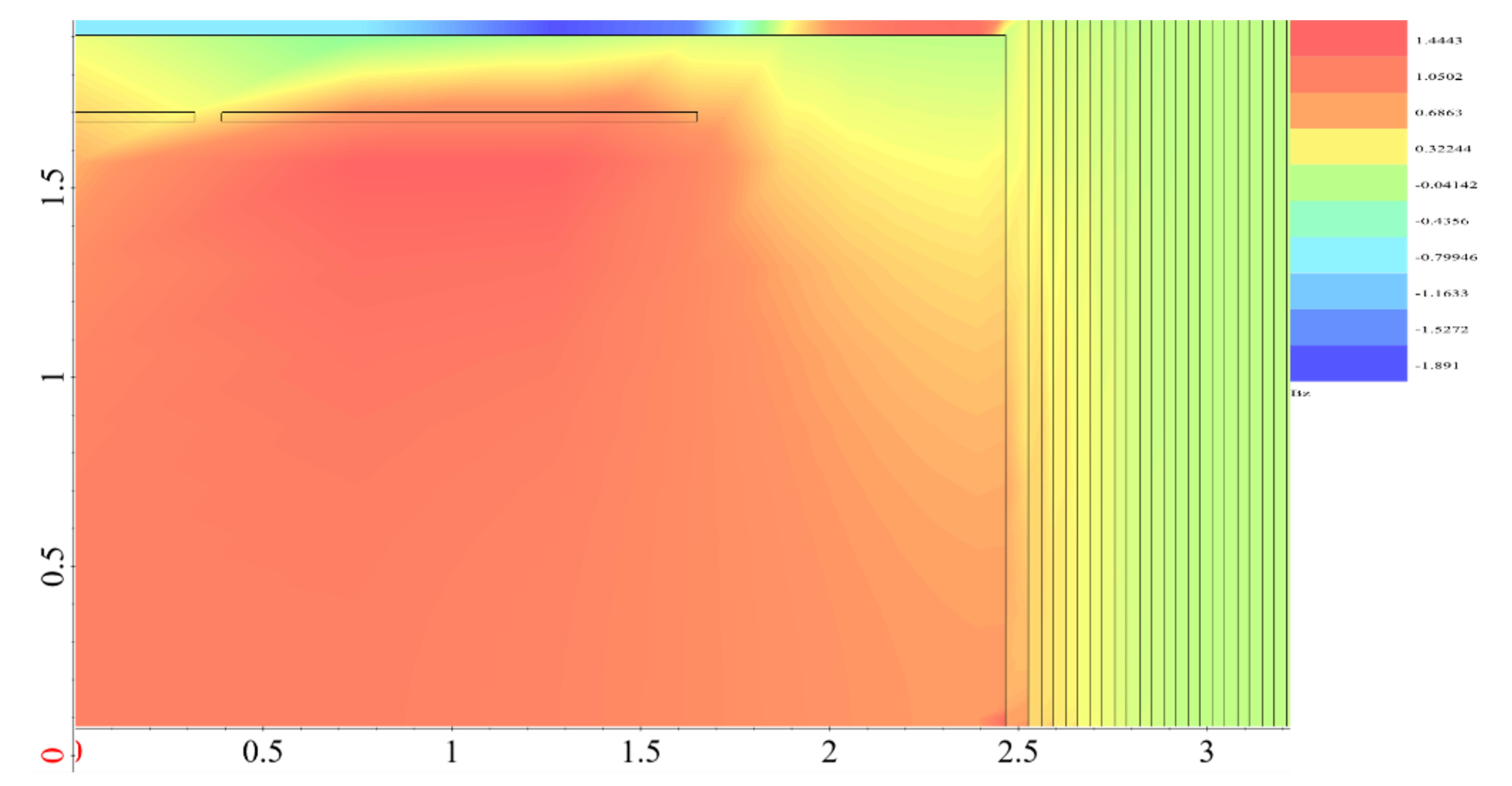}}
  \caption{ $B_z$ component of the magnetic field in the $YZ$ section at $X = 0$ for the central coil switched off and current 6500 A.
	}
\label{fig:field_5}
\end{figure}

\begin{figure}[!hp]
    \center{\includegraphics[width=1.\linewidth]{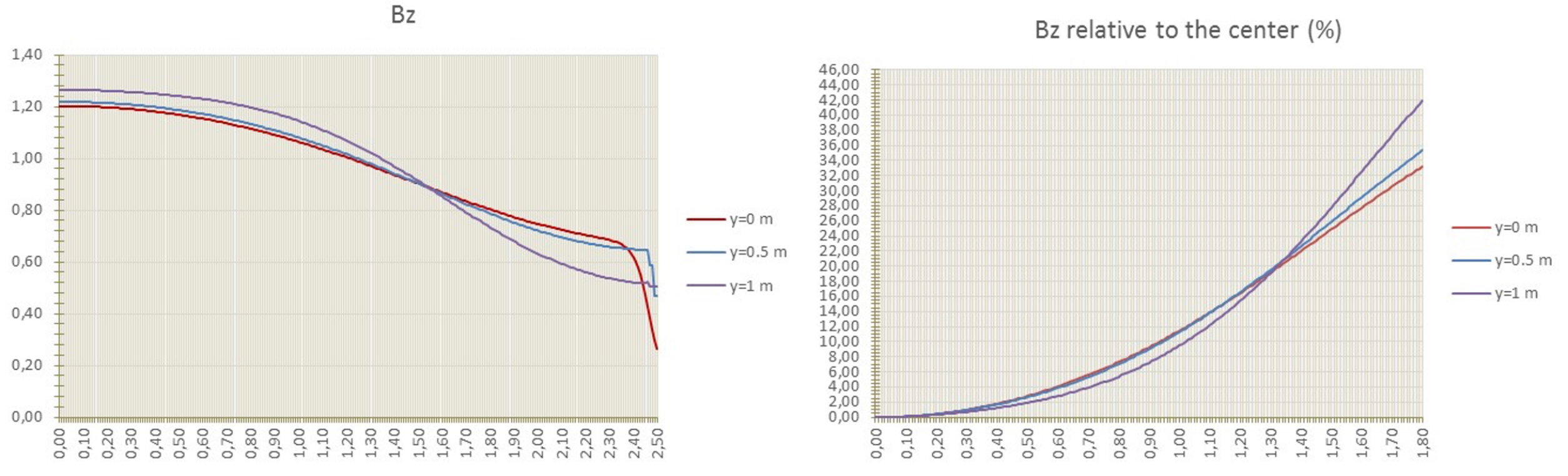}\\ (a)~~~~~~~~~~~~~~~~~~~~~~~~~~~~~~~~~~~~~~~~~~~~~~~~~~~~~~~~~~~~~~~~~~~~~~~~~~~~~~~~(b)}
  \caption{ 
  Distribution of the axial component of the magnetic field $B_z$ along the $Z$ axis for the basic magnet configuration with three main coils and a current of 5200~A. (a) shows the absolute value of $B_z$ and (b) shows $B_z$ normalized to its value at $Z=0$. Three curves correspond to the displacements $Y=0$~m, $Y=0.5$~m, $Y=1$~m at $X=0$~m.
	}
\label{fig:field_6}
\end{figure}

A potential option for improving the field uniformity without the use of correction SC coils is considered to switch off the central coil in order to reduce its contribution to the central region of the solenoid. In this case, the uniformity improves up to 10\% (see Figure \ref{fig:field_4} and \ref{fig:field_5}).
The disadvantage of this method of improving uniformity is the decrease in the solenoid field and, as a consequence, the need to increase the current from 5200 A to 6500 A to achieve 1.1 T at the center of the solenoid.
The distribution and uniformity of $B_z$ in the section $YZ$ at $X = 0$ for the central coil switched off are shown in Fig. \ref{fig:field_6} (a) and (b). 
Therefore, the three coils version of Fig. \ref{fig:coils} is the basic variant. 


For the final optimization, three-dimensional calculations were carried out using the TELMA software package. 
In the final coil configuration the current is reduced to 5000 A and  the number of turns is increased to 312 – 156 – 312. 

\begin{figure}[!hp]
  \begin{minipage}[ht]{1\linewidth}
    \center{\includegraphics[width=1\linewidth]{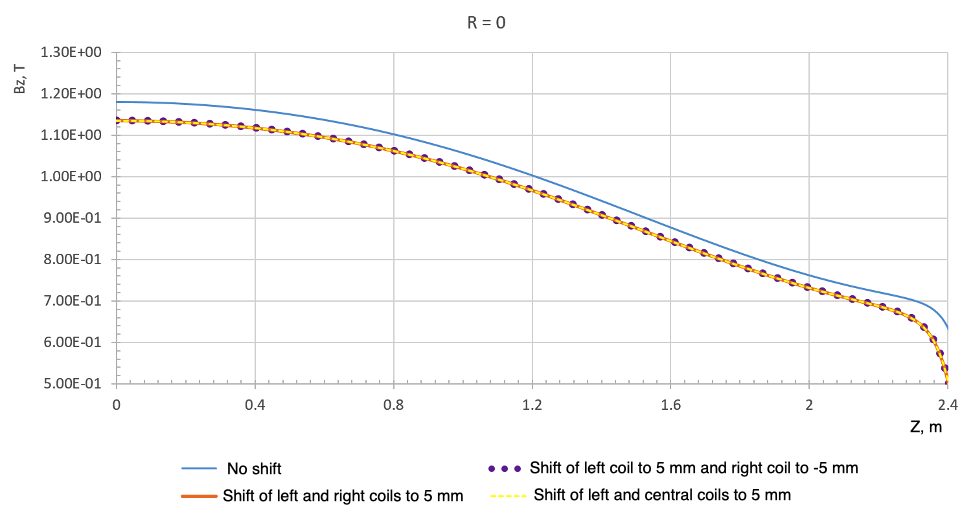} \\ (a)}
  \end{minipage}
  \hfill
  \begin{minipage}[ht]{1\linewidth}
    \center{\includegraphics[width=1\linewidth]{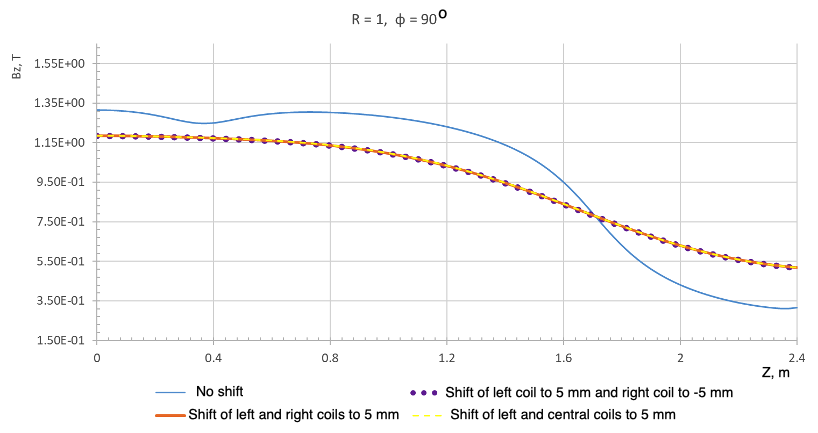} \\ (b)}
  \end{minipage} 
  \caption{Distribution of the $B_z$ field component along $Z$ axis for R=0 m (a) and R=1 m (b) for the nominal position and radial 5-mm shifts of the coils.}
\label{fig:field_7}
\end{figure}
\begin{figure}[!hp]
    \center{\includegraphics[width=1.\linewidth]{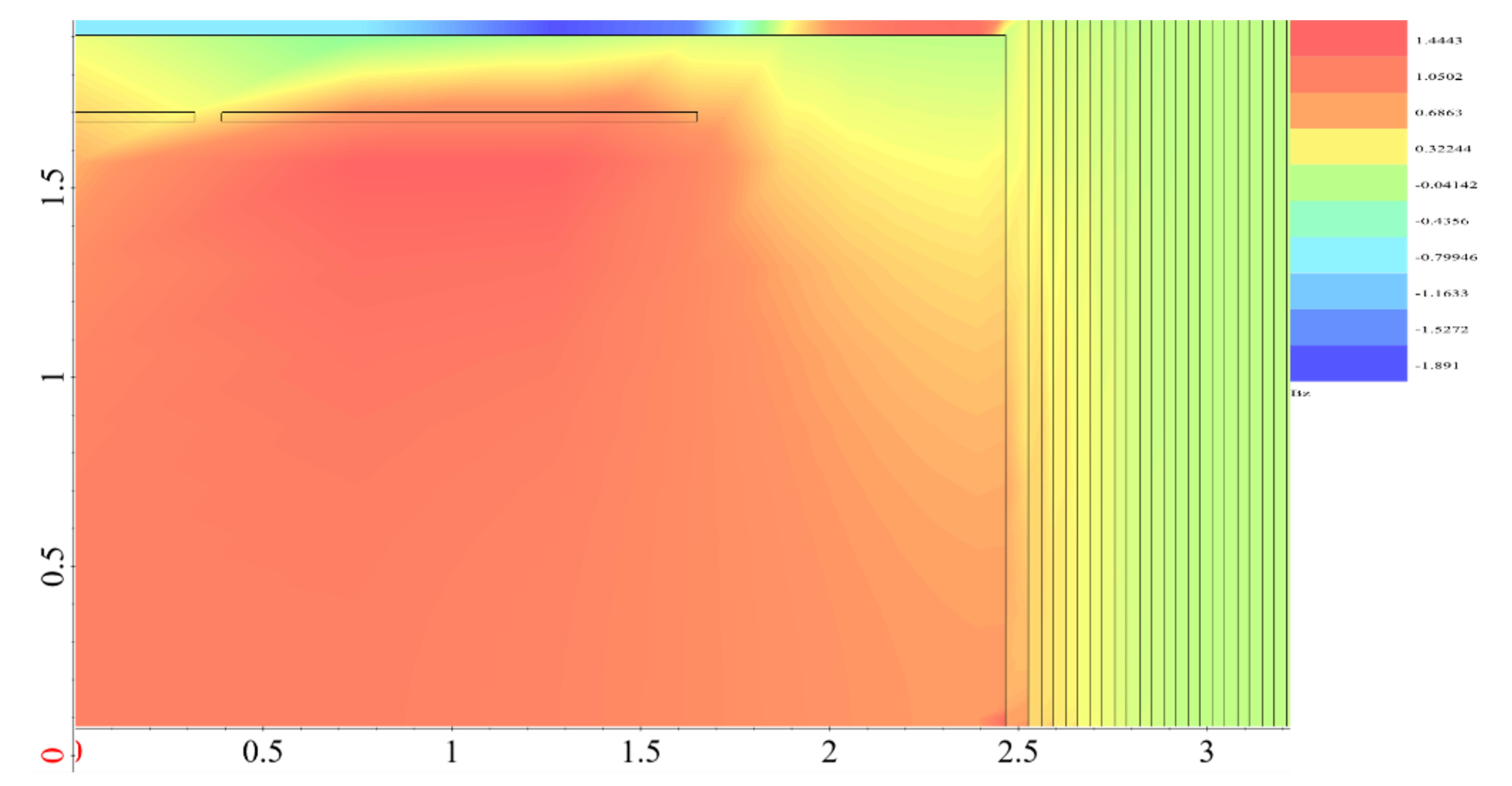}}
  \caption{ Distribution of the field magnitude $|B|$ in the $ZY$ plane for the final coils and yoke geometry. 
	}
\label{fig:field_8}
\end{figure}

The field component $B_z$ distributions along the solenoid axis at R=0 m and at R=1 m are shown in Fig. \ref{fig:field_7} (a) and (b), respectively.  The results for the different 5-mm displacements of the coils are also shown. The field change due to different displacements is about 4-5\%  which remains within the specified field homogeneity range. 
The color field map for the final magnet configuration is presented  in Fig. \ref{fig:field_8}.

\begin{figure}[!hp]
    \center{\includegraphics[width=1.\linewidth]{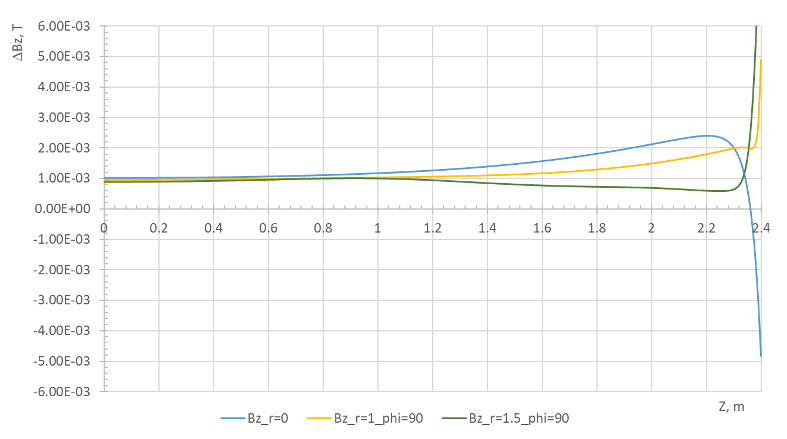}}
  \caption{  Impact of the 5-mm gap between the barrel and the end cap on the field distribution.
	}
\label{fig:field_81}
\end{figure}

Impact of the 5-mm gap between the barrel and the end cap on the field distribution is illustrated by Fig. \ref{fig:field_81}.

\begin{figure}[!hp]
    \center{\includegraphics[width=1.\linewidth]{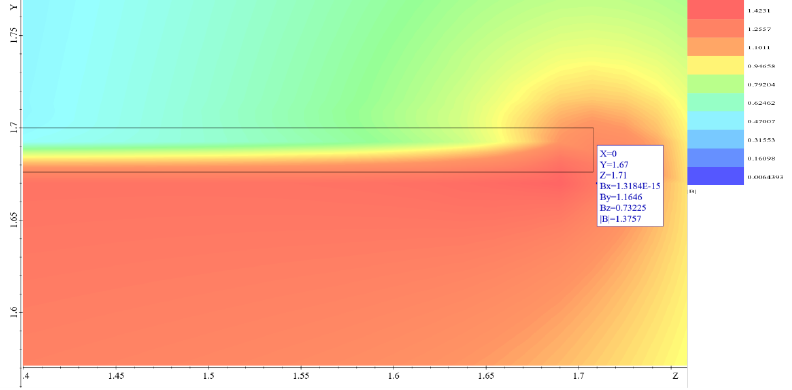}}
  \caption{  Parameters of the maximum magnetic field on the SC cable at the current of 5000 A.
	}
\label{fig:field_82}
\end{figure}

The maximum field on the superconducting cable at maximum current is indicated in Fig. \ref{fig:field_82}.
Integration of the specific magnetic energy over the solenoid volume at maximum current shows that the value of the stored total magnetic field energy is 21 MJ.

\FloatBarrier
\paragraph{Magnetic forces}

\begin{figure}[!h]
    \center{\includegraphics[width=1.\linewidth]{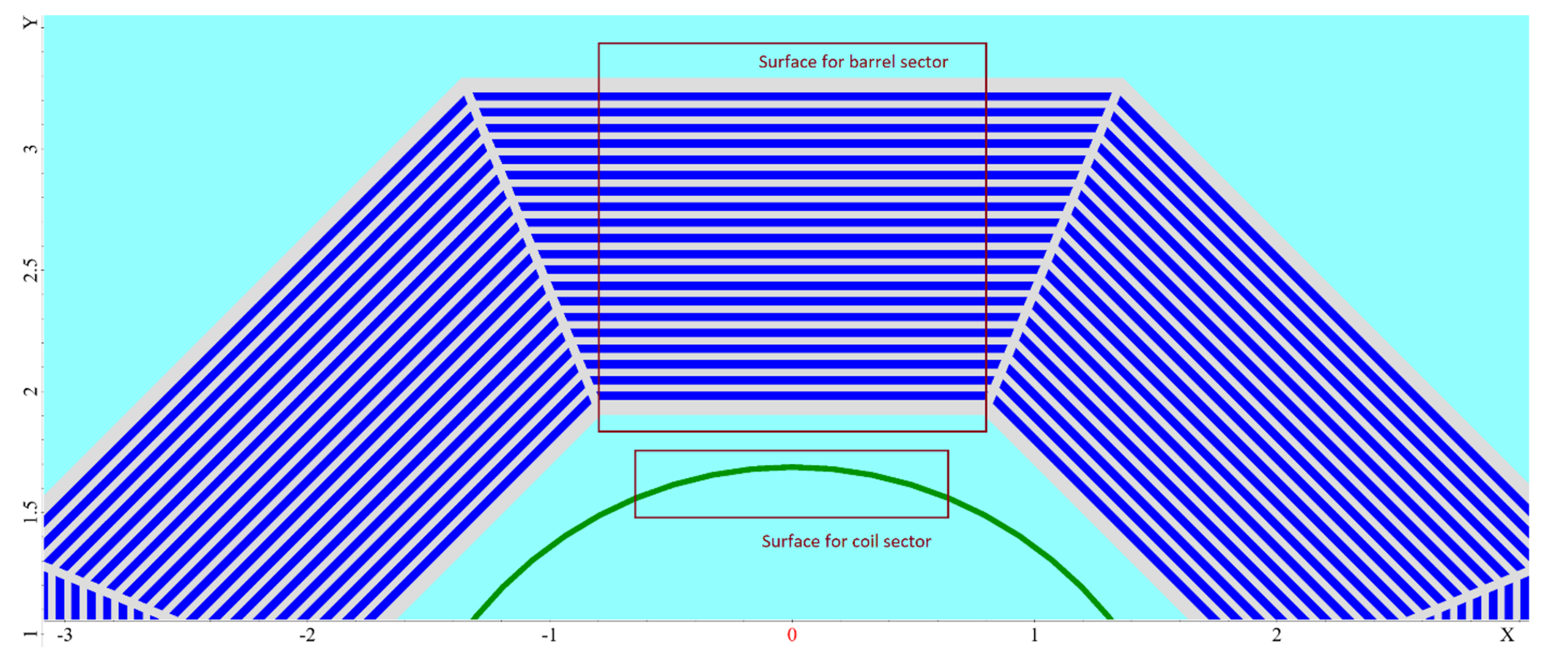}}
  \caption{ Schematic view of contour surfaces for force calculation.
	}
\label{fig:field_9}
\end{figure}

Simulation of the magnetic forces was performed with the Maxwell tensor method using 3D magnetic field data. The results of the simulation are shown in Table \ref{tab:mag_forces}. The elements on which the force acts are shown in Figure \ref{fig:field_9}.

\begin{figure}[!h]
    \center{\includegraphics[width=0.7\linewidth]{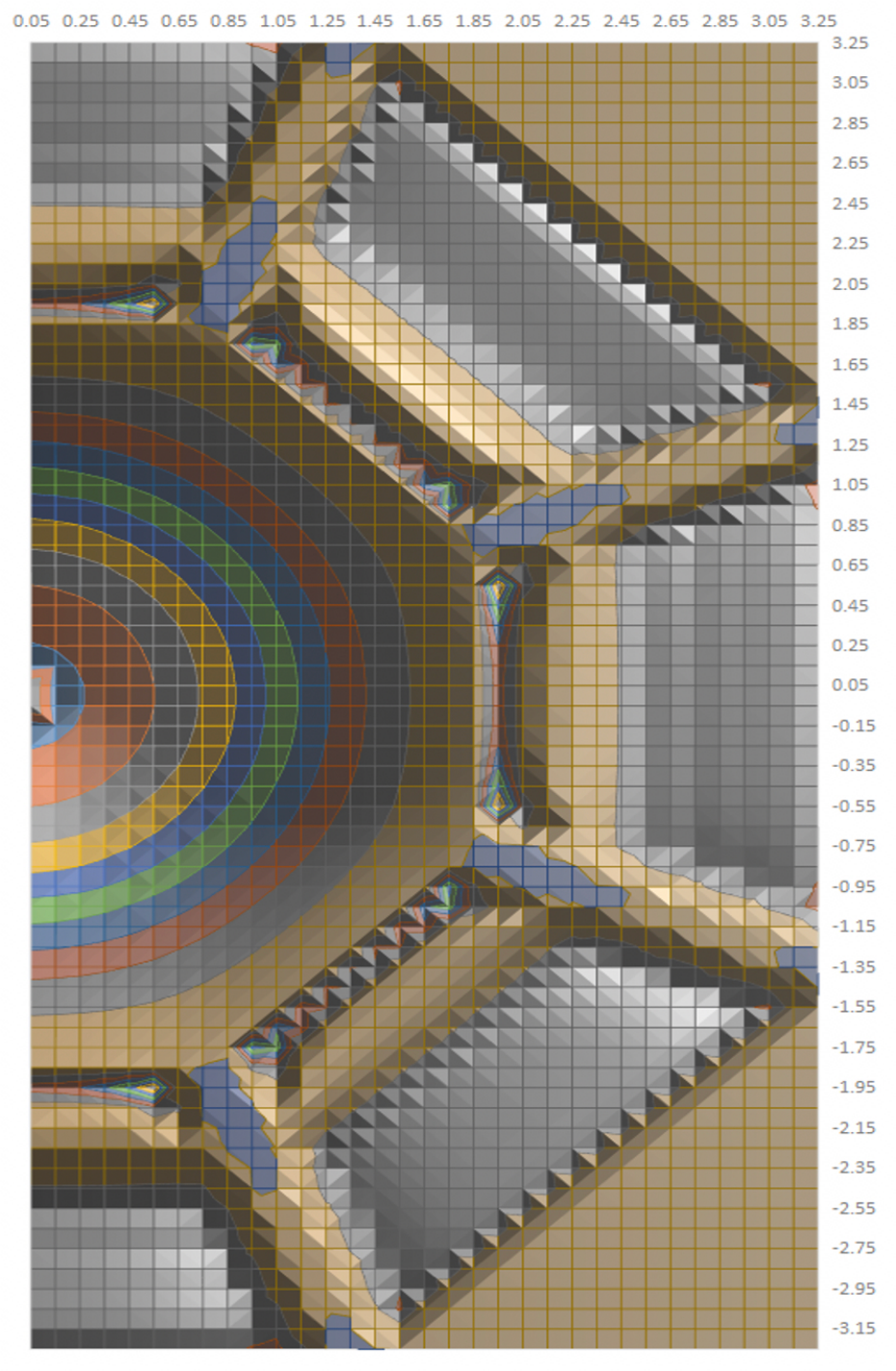}}
  \caption{ Magnetic forces map for one of the door wings.
	}
\label{fig:field_10}
\end{figure}

Figure \ref{fig:field_10} shows a map of the distribution of magnetic forces acting on one of the door wings. The force values are calculated for the centers of the corresponding grid elements 0.1 m $\times$0.1 m and displayed in color ranges of 200 newtons. The results of the calculation for magnetic forces acting on the coils during their displacement are shown in Table \ref{tab:mag_forces2}. 

\begin{table}[h!]
\caption{Magnetic forces in kN acting on the elements of the solenoid (1 mm air gap before end-caps).
Calculations were performed for the basic configuration of the magnet with three coils.}
\begin{center}
\begin{tabular}{|c|c|c|c|c|c|c|}
\hline
 Force  com- & Barrel top sector & Barrel top  & End-caps & End- & Central coil (1/8 & Side coil (1/8   \\
ponent & (half along Z) &  sector   &  (half)  &  caps & of top sector)  & of top sector) \\
\hline
X & 0 & 0 & -31.6 & 0 & 0 & 0 \\
Y & -158.2 & -304.3 & 0 & 0 & 671.7 &1134.2 \\
Z & 73.4 & 0 & -672.4 & -1412.7 & 0 & -608.3 \\
\hline
\end{tabular}
\end{center}
\label{tab:mag_forces}
\end{table}%


\begin{table}[h!]
\caption{Magnetic forces in kN acting on the displaced coils.}
\begin{center}
\begin{tabular}{|c|c|c|c|c|}
\hline
Displacement, mm & Force comp. & Left coil & Central coil & Right coil \\
\hline
\hline
0 & X & 0 & 0 & 0 \\
0 & Y & 0 & 0 & 0 \\
0 & Z & 5172.3 & 0 & -5175.2 \\
\hline
Z+5 mm & X & 0 & 0 & 0 \\
Z+5 mm & Y & 0 & 0 & 0 \\
Z+5 mm & Z & 4860.0 & -86.3 & -5099.2 \\
\hline
Y+5 mm & X & 0 & 0 & 0\\
Y+5 mm & Y & -2288.6 &	-1132.1	&-2234.9 \\
Y+5 mm & Z & 4967.4 & 0.0 & -4852.6 \\
\hline
\end{tabular}
\end{center}
\label{tab:mag_forces2}
\end{table}%

\subsubsection{Cold mass with conductor and coil}
The SPD solenoid is designed to operate at a current of 5.0 kA, that is 27\% of the critical current at 4.5 K, and 2.05 T peak magnetic field in the coil and 1.0 T magnetic field in the axis of the solenoid. 

The conductor is a superconducting NbTi/Cu wire-based Rutherford cable, co-extruded with a high-purity aluminum-stabilizing matrix. The insulated conductor dimensions at 4.5 K are 10.90 mm in width and 7.90 mm in height. The Rutherford cable is composed of 8 strands with a diameter of 1.40 mm and a Cu/SC ratio of 1.0. The critical current density of the superconductor at 4.2 K and 5 T shall be larger than 2800 A/mm$^2$ to ensure a temperature margin for a quench well above 2.0 K. The same type of conductor is produced in Russia and used for the PANDA solenoid (FAIR, Darmstadt).

The SPD solenoid consists of three 2-layer coil modules to be wound on a collapsible mandrel. After curing, aluminum alloy support cylinders are placed over each module and the assemblies are epoxy-bonded. The pre-assembled modules are then removed from the mandrel to be electrically connected and bolted together to form a single cold mass assembly before installation in the cryostat.

The magnet is indirectly cooled by a network of aluminum alloy cooling tubes, securely bonded to the outer surface of the aluminum alloy support cylinder, ensuring proper heat conduction. During normal operation, two-phase helium is circulated through the cooling tubes by natural convection.

The magnet quench protection system relies on the continuous monitoring of the coil's voltages and the extraction of the energy to an external dump resistor, once a quench is detected. The maximum coil winding temperature and electrical voltage across the coil terminals during a quench has to be limited to 100 K and 500 V, ensuring a robust and extremely low-risk operation, as appropriate for the unique particle detector magnets. Quench propagation across the coil winding is accelerated by using the quench-back effect of the support cylinder, as well as the aluminum heat drains.

The cold mass of the SPD solenoid consists of three epoxy resin-impregnated coils, reinforced by shells made of structural aluminum. 

The upstream and downstream winding packs are identical and feature 2 layers of 150 turns. The center coil, instead, is smaller, featuring 2 layers of 75 turns. The conductor is wound around the aperture with tension. The tension of the conductor is set with tensioning rollers weighing about 50 kg. The coils should be prepared and impregnated in a vacuum according to the standard BINP technological scheme TTS4 STO 103-2011.

\begin{figure}[!ht]
    \center{\includegraphics[width=1.\linewidth]{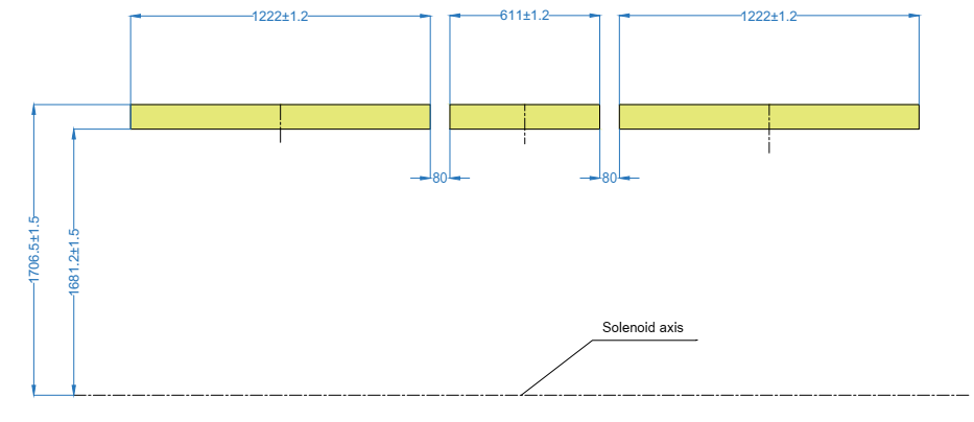}}
  \caption{ Room temperature dimensions in mm of the coil envelopes, including cable and ground insulation.
	}
\label{lb_mag_2}
\end{figure}

Figure \ref{lb_mag_2} shows the dimensions of the coil envelopes at room temperature, as computed taking into account thermal shrinkage, shrink fit, and the magnetic field. Note that the dimensions include both cable insulation and ground insulation. 

\begin{figure}[!h]
  \begin{minipage}[ht]{0.65\linewidth}
    \center{\includegraphics[width=1\linewidth]{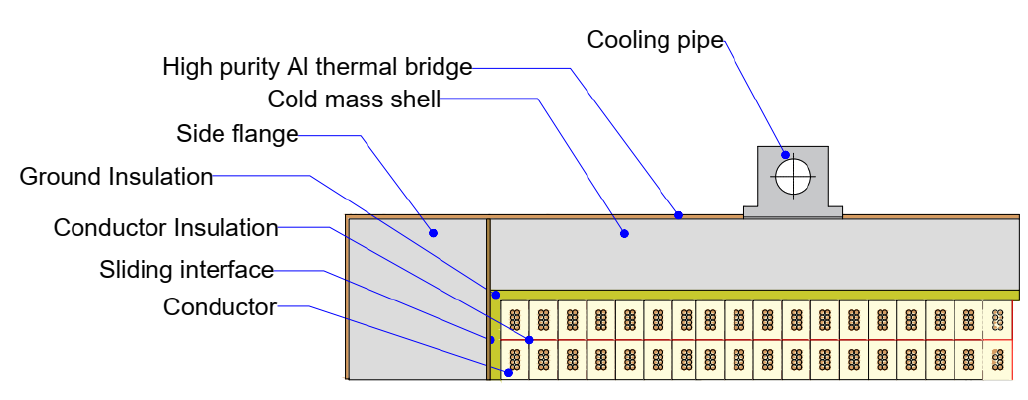} \\ (a)}
  \end{minipage}
  \hfill
  \begin{minipage}[ht]{0.35\linewidth}
    \center{\includegraphics[width=1\linewidth]{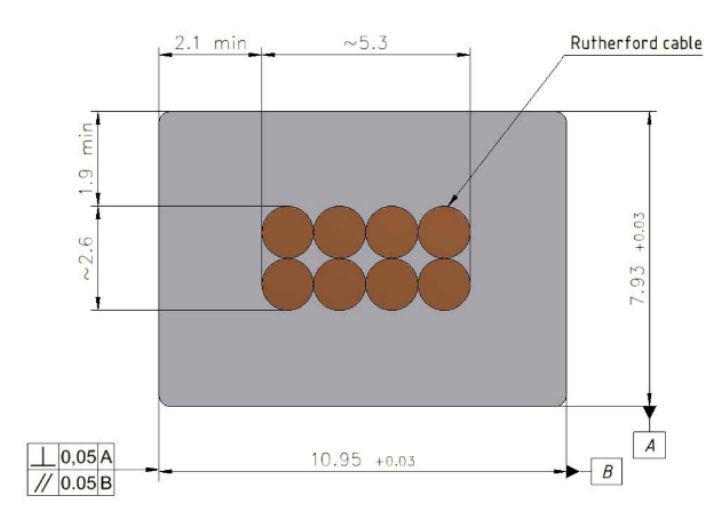} \\ (b)}
  \end{minipage} 
  \caption{(a) Part of the cold mass cross-section, showing the layout at the end flange. (b)  Cross-section of the solenoid conductor, showing the 8 strands of Rutherford cable enclosed by the Al matrix, including sizes and their tolerances.
	}
\label{lb_mag_3}
\end{figure}

The cold mass components are shown in Fig. \ref{lb_mag_3}~(a).

\subsubsection{Conductor}
Al stabilized conductors for detector magnets are selected for the following reasons:
\begin{itemize}
\item simplicity of conduction cooling; affordable, since no dynamic operation is needed; quasistationary;
\item simplicity and reliability of electrical connection;
\item high-purity Al stabilized, the residual-resistance ratio (RRR)  $\geq$ 1000, maximum MPZ (m), much larger $\lambda/\rho$ than copper, where $\lambda$ is the mean free path of conduction electrons and $\rho$ is resistivity;
\item particle transparency for minimum particle scattering.
\end{itemize}
This type of conductor was used in the production of larger detectors, such as CELLO, CDF, TOPAZ, VENUS, ALEPH, DELPHI, CLEO, SDC, BELLE, ATLAS CS, ATLAS ECTs, ATLAS BT, CMS, PANDA and Mu2e solenoids.
The same conductor, as was used for the PANDA solenoid is supposed to be used in SPD. 
The solenoid conductor consists of three components: 
\begin{itemize}
\item superconducting NbTi/Cu multi-filamentary strand with 1.4 mm diameter;
\item Rutherford-type flat cable with 8 strands;
\item high purity aluminum stabilizer of 10.95 mm $\times$ 7.93 mm cross-section, clad by a co-extrusion or plating process (or another qualified process).
\end{itemize}

\begin{table}[]
\centering
\caption{NbTi/Cu strand mechanical and electrical specifications.}
\label{tab:ma_1}
\begin{tabular}{|l|c|c|c|}
\hline
Parameter                           & Unit  & Value  & Tolerance  \\
\hline
Diameter filament                   & $\mu$m    & $<20$   & --           \\
Diameter strand                     & mm    & 1.400  & $\pm$ 0.005     \\
Cu/SC ratio                         & --     & 1.00   & $\pm$ 0.05      \\
Surface coating                     & --     & none   & --           \\
NbTi Jc (at 4.2 K, 4.05 T)          & A/mm$^2$ & $>$ 2800 & --          \\
Critical current (at 4.2 K, 4.05 T) & A     & $>$ 2600 & --          \\
n-value (at 4.2 K, 4.05  T)         & --     & $>$ 30   & --           \\
Conductor RRR                       & --     & $>$ 100  & --           \\
Twist direction                     & --     & left   & --           \\
Twist pitch                         & mm    & 25     & $\pm$ 5        \\
\hline
\end{tabular}
\end{table}

\begin{table}[]
\centering
\caption{Rutherford cable mechanical and electrical specifications.}
\label{tab:ma_2}
\begin{tabular}{|l|c|c|c|}
\hline
Parameter                                                                                                       & Unit     & Value   & Tolerance \\
\hline
Number of strands                                                                                               & --        & 8       & --         \\
Cable width                                                                                                     & mm       & 5.3     & $\pm$ 0.1     \\
Cable thickness                                                                                                 & mm       & 2.60    & $\pm$ 0.05    \\
Transposition angle                                                                                             & degree   & 20.0    & $\pm$ 0.5     \\
Twist direction                                                                                                 & --       & right   & --         \\
Critical current (at 4.2 K, 4.05T)                                                                              & A        & $>$ 20800 & --         \\
\begin{tabular}[c]{@{}l@{}}Critical current degradation of\\   extracted strand (at 4.2 K, 4.05 T)\end{tabular} & \%        & $<$ 5     & --         \\
RRR of extracted strands                                                                                        & --        & $>$ 100   & --         \\
Residual twist on 1 m of cable                                                                                  & degree/m & $<$ 45    & -- \\        
\hline
\end{tabular}
\end{table}

\begin{table}[]
\centering
\caption{Solenoid conductor mechanical and electrical specifications.}
\label{tab:ma_3}
\begin{tabular}{|l|c|c|c|}
\hline
Parameter                                                           & Unit & Value        & Tolerance \\
\hline
Width (after cold work) at 300 K                                    & mm   & 10.95        & $\pm$ 0.03    \\
Thickness (after cold work) at 300 K                                & mm   & 7.93         & $\pm$ 0.03    \\
Width (after cold work) at 4.5 K                                    & mm   & 10.90        & $\pm$ 0.03    \\
Thickness (after cold work) at 4.5 K                                & mm   & 7.90         & $\pm$ 0.03    \\
Minimum Al layer thickness in height                                & mm   & $>$ 1.9        & --         \\
Minimum Al layer thickness in width                                 & mm   & $>$ 2.1        & --         \\
Surface roughness                                                   & --    & Ra $<$ 3.2     & --         \\
Critical current (at 4.2 K, 4.05  T)                                & A    & $>$ 20800      & --         \\
Critical current (at 4.5 K, 3.6 T)                                  & A    & $>$ 23400      & --         \\
Critical current degradation of extracted strand w.r.t. virgin wire & \%    & $<$ 15         & --         \\
Overall Al/Cu/superconductor ratio                                              & --    & 10.5/1.0/1.0 & --         \\
Extracted strand RRR                                                & --    & $>$ 100        & --         \\
Aluminum RRR (at 4.2 K, 0 T)                                        & --    & $>$ 600        & --         \\
Al 0.2\% yield strength at 300 K                                     & MPa  & $>$ 30         & --         \\
Al 0.2\% yield strength at 4.2 K                                     & MPa  & $>$ 40         & --         \\
Shear strength Al to strands                                        & MPa  & $>$ 20         & --         \\
Unit length for upstream \& downstream coils                         & m    & 2 $\times$ 3203     & --         \\
Unit length for center coil                                         & m    & 1604         & --         \\
Total length of conductor                                           & m    & 8010         & --         \\
Minimum bending radius during production and on spools              & mm   & $>$1000       &  --\\    
\hline   
\end{tabular}
\end{table}

Pure aluminum features a very high electrical and thermal conductivity at low temperatures, providing the best possible stability in terms of maximum MPZ (Minimum Propagation Zone), making the coil least sensitive to mechanical disturbances that can cause coil training during commissioning and quenching during normal detector operation. Furthermore, aluminum-stabilized superconductors can be made by co-extrusion of several kilometers long. Precise rectangular conductor shapes can be obtained, allowing for high accuracy in the coil winding. The cross-section of the conductor is shown in Fig. \ref{lb_mag_3}~(b).

Tables \ref{tab:ma_1}, \ref{tab:ma_2}, and \ref{tab:ma_3} present a summary of the main mechanical and electrical properties of the strands, Rutherford cable, and Al-stabilized conductor. 

The conductor has a temperature margin of about 2,4 K. The shape of the conductor allows to easily wind the coil from both the wide and the narrow sides.

\subsubsection{Insulation}
The main requirement for coil insulation is to provide the required electrical strength in terms of break-down voltage while maximizing heat conduction, in order to reduce the peak temperature in the conductor in the 2-layer windings.
The conductor insulation used in PANDA is proposed for the SPD solenoid. The insulation scheme relies on the use of fiberglass 3$\div$5 mm thick in the first and the last layers, and  a nominal conductor insulation between layers.
The nominal cable insulation thickness is 0.200 mm per side. It is made of two half-overlapped layers of 100 $\mu$m thick fiberglass tape. The tape is wrapped around the surface of the conductor. BINP technology of a vacuum impregnation provides  reliable electrical strength and  good resistance to coil winding imperfections, especially when using  two layers of the conductor, in the SPD case.

Ground insulation is added at the outer surface and sides of the coil modules. No ground insulation is applied at the inner diameter, where the detector vacuum already provides the necessary dielectric strength. At the outer diameter (3$\div$5 mm thick), the first layer is wound, which is then machined, after curing of the coil module to obtain precise fitting to the supporting cylinder-to-coil interface. The precise thickness of ground insulation after machining depends on the control of the radial dimensions of the coil during winding. A minimum thickness of 2.0 mm is recommended. 

The envisaged thickness of ground insulation at the coil sides is 2.80 mm for the upstream and downstream coils and 1.50 mm for the central coil. The thickness of the axial ground insulation can be adjusted to correct winding tolerances since dielectric strength is always ensured by the mylar layer, located at the sliding interface. At the side flange, used as the base for coil winding, the axial ground insulation is provided in the form of epoxy shims, precisely machined, in order to accommodate winding tolerances. At the opposite side, axial ground insulation is provided by filling the empty spaces between the winding pack and the flange with epoxy resin after the flange assembly.

\begin{figure}[!h]
  \begin{minipage}[ht]{0.5\linewidth}
    \center{\includegraphics[width=1\linewidth]{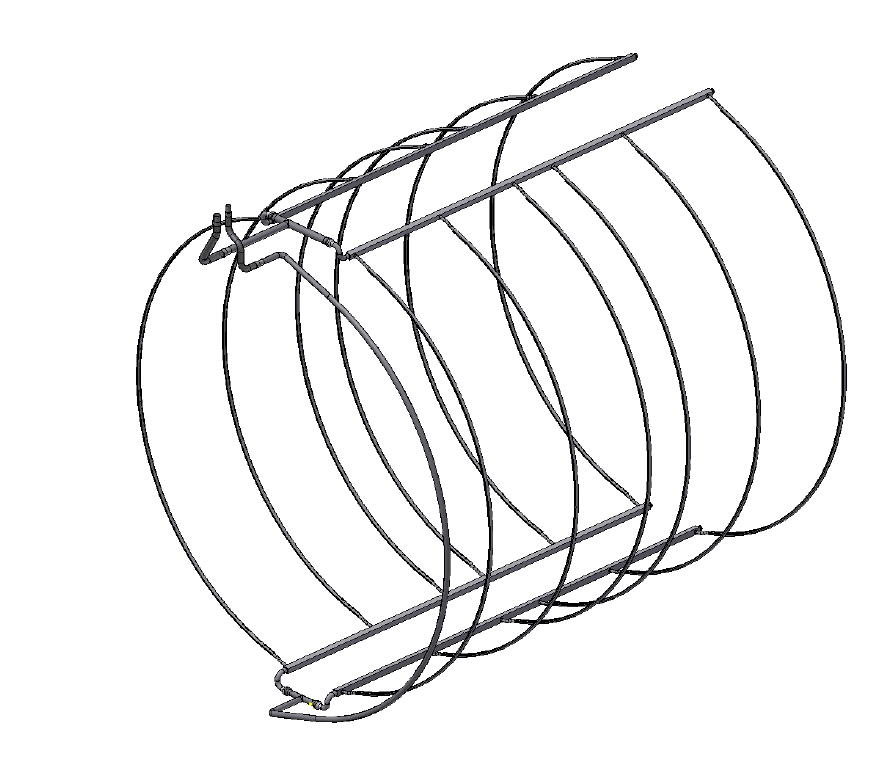} \\ (a)}
  \end{minipage}
  \hfill
  \begin{minipage}[ht]{0.5\linewidth}
    \center{\includegraphics[width=1\linewidth]{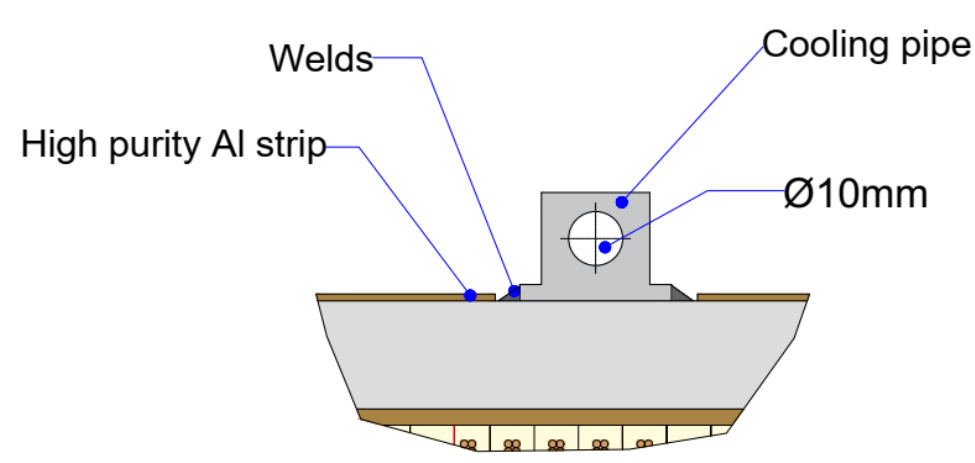} \\ (b)}
  \end{minipage} 
  \caption{(a) Thermosyphon circuit. (b) Connection of the cooling tubes and aluminum strip on the outer surface of the aluminum alloy support cylinder. 
	}
\label{lb_mag_4}
\end{figure}

\subsubsection{Thermosyphon cooling circuit}
The cold mass of the SPD solenoid is indirectly cooled by circulating two-phase helium by natural convection. The homogenous model, used for the preliminary study for CMS \cite{Bredy:2006iy} and then BINP, is also used for cooling of the PANDA detector solenoid and the  CBM detector dipole magnet.

The main advantages of a thermosyphon system are its simplicity and reliability, as it does not require any moving parts, such as cold pumps and high pressure for piping of the coils. On top of this, thermosyphon cooling ensures constant and uniform temperature, since the helium flow automatically adjusts to the heat load distribution.

The thermosyphon circuit consists of two manifolds, at the top and bottom of the cold mass, connected by two sets of six syphon tubes, as illustrated in Fig. \ref{lb_mag_4}~(a).  

The SPD coils feature six cooling tubes with two tubes for each.

The material selected for the thermosyphon circuit is Al-1100 due to its high thermal conductivity. The extruded tubes have a diameter of 10 mm and are welded to the support cylinder -- a technically proven solution to ensure reliable thermal contact, as shown in Fig. \ref{lb_mag_4}~(b).

The thermosyphon circuit is assembled as two halves consisting of six cooling tubes, welded to the top and bottom manifolds. The division in two halves is on purpose, as then each half-circuit can be fully tested for leaks, prior to installation on the cold mass, including the bimetallic (steel-aluminum) joints at the connectors, interfacing the manifolds and the cryogenic lines to the cold box.


\subsubsection{Sliding interface}
Due to the presence of NbTi/Cu strands in the Al-stabilized conductor, the thermal contraction of the coil is less than that of the aluminum alloy, used for the support cylinder and flanges. On top of this, small gaps tend to develop at the coil winding-to-flange interfaces under the bending action of the Lorentz force.

To prevent the risk of stress accumulation at the coil winding-to-flange interfaces that could result in cracks in the insulation, triggering magnet quenches, the coil is allowed to slide with low friction and, if needed, separate from the flange. Therefore, sliding interfaces are created between coil windings and flanges by placing a 0.1 mm mylar foil and 0.4 mm of kapton\textregistered.

\subsubsection{Support cylinder}
To limit coil deformation and stress, due to the Lorentz force, support cylinders are placed around the coil windings. The cylinders are made of graded aluminum alloy Al-5083-O, exhibiting good mechanical strength at cryogenic temperature. Being the main structural elements of the cold mass, the shells must be produced as single pieces with uniform material properties.

The assembly formed by coil winding and support cylinder is referred to as a coil module. The assembly operation is performed via shrink fit with interference in order to ensure radial pre-stress. The required radial interference between coil windings and shells is 0.70 mm at room temperature. For this purpose, the support cylinders are produced with oversized dimensions to allow for machining of the inner surface. The thickness of the cylinders after machining shall not be less than 20~mm.

The coils are permanently bonded to the support cylinders using epoxy resin or grease (for example Apiezon N), suitable for cryogenic applications. The shells are connected together using bolts passing through the flanges, as described in the next section.

\subsubsection{Flanges, bolts, spacers, and venting holes}
The flanges are disks made of aluminum alloy Al-5083-O that cover the coil winding ends (outer flanges) or separate the coil modules (inner flanges). Similar to the requirements for the support cylinders, the flanges have to be produced as single pieces. Joining several flange pieces by welding is not allowed. 

The bolt calculations assume a safety factor of 1.5 and are based on Eurocode 3, which is the standard adopted at CERN/DESY/FAIR. 
All bolted interfaces should be designed to hold the weight of the cold mass in tension during assembly and for all operating conditions. 
All bolted interfaces should be designed to be slip-critical under any condition. As such, the friction force under each bolt shall be higher than the shear force. Therefore, the bolts need to be sufficiently pre-loaded during the assembly. The recommended pre-load at room temperature amounts to 60\% of the tensile strength of the bolt. Since the thermal contraction of the material SS A4 is lower than that of Al-5083-O, the pre-load will reduce to 35\% of the bolt tensile strength after a cool-down. Compressive forces during the magnet operation will also reduce the  bolt pre-tension  to some extent. 

Bolt tightening shall be performed in steps with a torque wrench, and each bolt pre-load should be measured using an ultrasonic stress meter. Bolts shall be tightened uniformly, with an acceptable pre-load variation of $\pm$10\%. It is recommended to re-measure the bolt pre-load and adjust if necessary, prior to the installation of the cold mass in the cryostat to correct possible relaxation, due to a local yielding and creep in the material.

\begin{figure}[!ht]
    \center{\includegraphics[width=1.\linewidth]{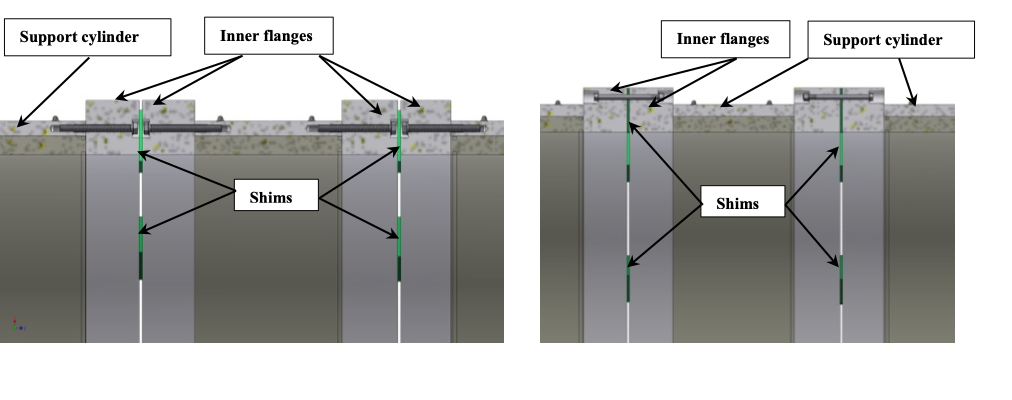}}
  \caption{ Connections "inner flanges -- support cylinder" (left) and "inner flange -- inner flange" (right).
	}
\label{lb_mag_5}
\end{figure}

Gaps of 8 mm are envisaged between the inner flanges that are to be shimmed with aluminum spacers. The gap allows for adjusting the axial position of the coil modules, to compensate for winding and assembly tolerances and obtain the correct magnetic field profile (Fig. \ref{lb_mag_5}). The axial dimensions of the spacers are to be determined after the dimensions of the coil modules have been precisely measured. 

In order to reduce the work associated with shimming and not interfere with the thermal bridges, spacers will be placed only at the locations of the bolts. Shimming of the entire surface of the flange is not required: based on the maximum axial force on the coil modules, the Al 5083-O yield strength (145 MPa), and, assuming a sufficient safety factor, the minimum shim area is 0.04 m$^2$. The value is much smaller than the surface of the inner flanges ($\sim$1 m$^2$).

When the thermal bridges are placed in correspondence with the position of the bolts, and hence of the spacers, attention shall be paid to limit the stress on the soft aluminum, so that the thermal conduction of the strip is not degraded.

To reduce the air evacuation time of the cryostat during pumping, it is recommended that all closed volumes should have a hole to vent to the cryostat vacuum (minimum diameter  of 1 mm). 

Venting holes are recommended at the location of the bolt connections, although at present they are not included in the technical drawings of the cold mass. When the threaded hole is filled with epoxy, venting holes allow for an exit of excess resin, which can be regarded as an indicator of good bond quality. 

\subsubsection{	Cold mass thermalization}

\begin{figure}[!h]
    \center{\includegraphics[width=0.5\linewidth]{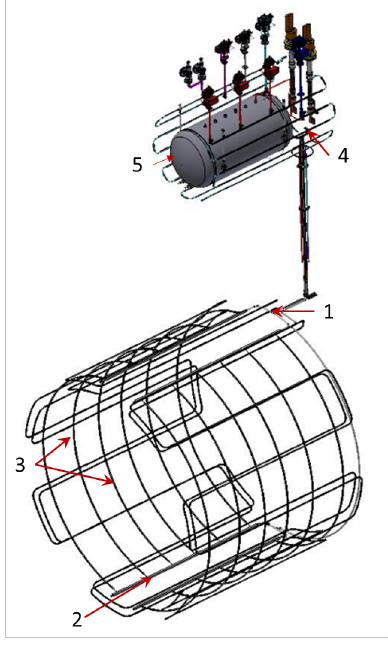}}
  \caption{ Design of the cooling cold mass circuit: 1 -- upper manifold; 2 -- lower manifold; 3 -- syphon tubes; 4 -- Al-SS bi-metal adapters; 5 -- vessel with liquid helium.
	}
\label{lb_mag_9}
\end{figure}

The cryogenic scheme of the SPD solenoid relies on indirect cooling of the cold mass by circulating saturated helium at 4.5 K by natural convection. A thermosyphon circuit consists of a bottom and top manifold connected by 12 parallel syphon tubes, attached to the outer surface of the support cylinder, as in Fig. \ref{lb_mag_9}.

The cooling method chosen for the SPD magnet's superconducting coil is based on the natural convection of liquid helium flow. It is a self-regulating thermosyphon circulation flow system. The natural circulation loop works on the principle that a heat load on the channels of the heat exchanger produces a two-phase flow that is, on average, less dense than the liquid phase. 

\begin{figure}[!h]
    \center{\includegraphics[width=1.\linewidth]{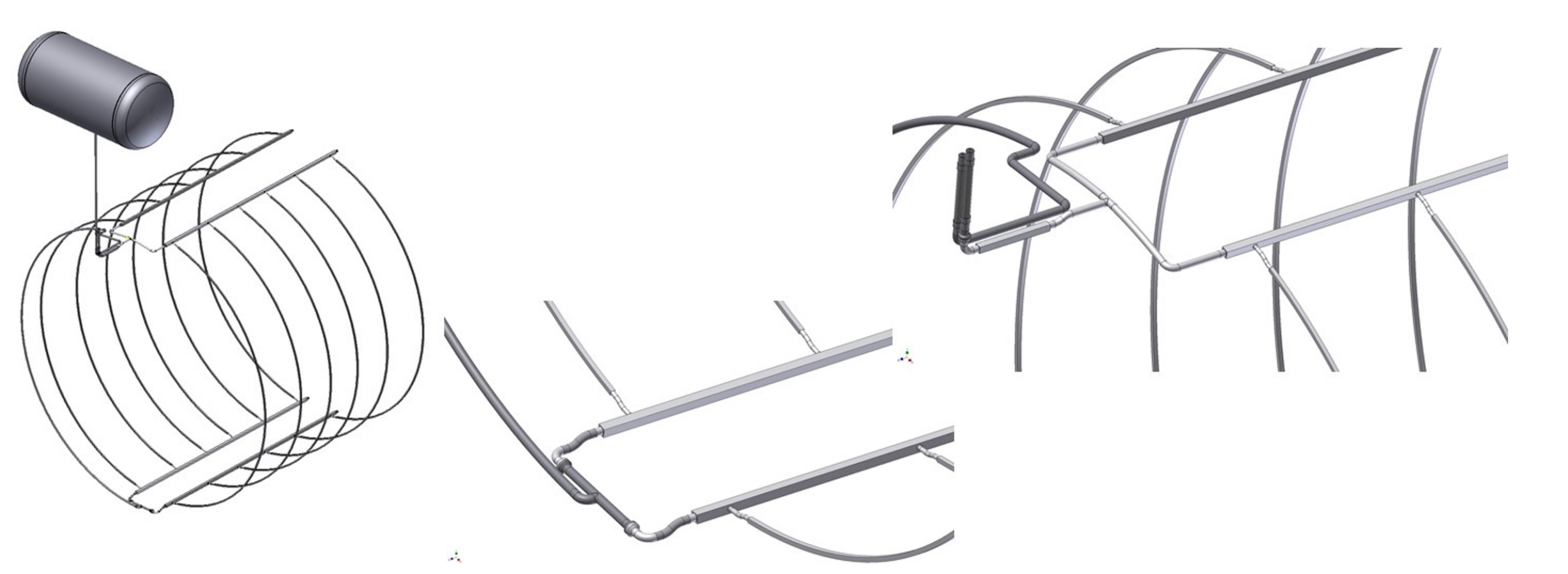}}
  \caption{ Thermosyphon circuit of the SPD solenoid showing the two-halves assembly of the cold mass, as well as the height of the chimney and position of the control dewar.
	}
\label{lb_mag_10}
\end{figure}
The liquid from the helium vessel of the control dewar will be fed through the front pipeline and manifolds at the bottom of the support cylinder. Then, the liquid will be heated in the tubes of the heat exchanger (a rib cage configuration) on the surface of the support cylinder. The two-phase helium from the top manifold will return back through the reverse pipeline to the upper part of the helium vessel.

The design of the thermosyphon cooling circuit includes the definition of the optimal size for the syphon tubes and their position on the cold mass support cylinder. PED and AD2000 pressure vessel rules have to be applied for the construction of the cryogenic system.

In order to verify the feasibility of thermosyphon cooling, the expected mass flow rate and vapor fraction have been assessed with the homogenous model, described in Ref. \cite{Dhanaraj_2015}. In the model, the required  quantities are determined by balancing the driving hydrostatic pressure difference, which results in the thermally induced density gradient between the hot and cold sides of the loop, with the fluid acceleration and friction pressure drops in the supply and return tubes. 
The reported diameters correspond to the inner diameters of the pipes. It must be noted that the current technical drawings may contain out-of-date values for the dimensions of the pipes of the thermosyphon circuit.

The model thus confirms that the height of the control dewar installed on the upper platform of the SPD yoke with respect to the bottom of the magnet is sufficient to guarantee the driving hydrostatic pressure. The thermosyphon tubes and supply/return lines are also properly dimensioned to guarantee minimal pressure drop along the cooling circuit (Fig. \ref{lb_mag_10}).
The main advantages of a thermosyphon system are its reliability since it does not include any moving part, and its ability to maintain a uniform temperature, as the helium flow automatically adjusts to the heat load distribution and variations.

\begin{figure}[!ht]
    \center{\includegraphics[width=1.\linewidth]{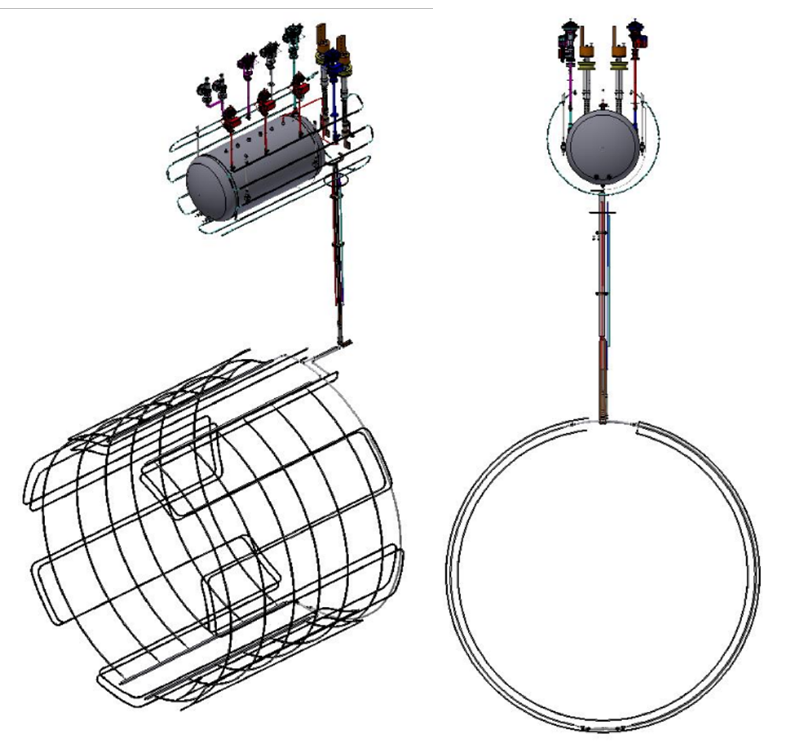}}
  \caption{ General view of the SPD cryostat with the control dewar.
	}
\label{lb_mag_110}
\end{figure}

\subsubsection{	Cryostat and control dewar}

All materials and components used in the construction of the cryostat and control dewar must be suitable for the use for which they are intended, in particular for use at low temperatures. The choice of materials has to be in line with the directive 2014/68/EU PED and must be made according to the regulation prescribed by $\Gamma$OCT P 52630-2012 and the Technical Regulations TP TC 032/2013.

The general view of the SPD cryostat with the control dewar is presented in Fig. \ref{lb_mag_110}.
The cryostat comprises of the cold mass with the superconducting coil, surrounded by a thermal screen and suspended on low heat conduction triangle supports. The cryostat should be mounted on the supports connected to the magnet yoke. The design of the cryostat supports will be defined later.

The control dewar is a functional unit of the cryogenic system and is intended to maintain the required parameters of liquid and gaseous helium (flow rate, pressure, temperature) and to provide cooling of the current leads. In the process flow diagram of the SPD cryogenic system, it is located between the cryostat and the helium liquefier. 

Physically, the control dewar is located on the top of the yoke. The vessel for liquid helium, current leads, piping, control, and relief valves, and temperature sensors are placed in the horizontal vacuum tank of the control dewar. 

Cryogenic and current supplies will be connected from the control dewar to the cryostat through the service chimney in a special slit at the backward end of the yoke barrel. In the bottom part of the chimney, an interface box is located, which is intended to accommodate connections of the cryogenic pipelines and current bus bars from the cryostat and the control dewar.

Based on the experience of creating a similar superconducting magnet, when assessing the cyclic strength of the cryostat, it was estimated that the solenoid parts should allow for 1000 cooldown cycles from room temperature to the operating temperature and back to room temperature, and 2500 energization cycles to the nominal current or any fraction of it. 

The cryogenic system of the cryostat and the control dewar will be designed to handle the loads resulting from all operation scenarios. The design pressure for all pipelines and the helium vessel is 19 bar absolute (bar-a).

\begin{figure}[!h]
    \center{\includegraphics[width=0.9\linewidth]{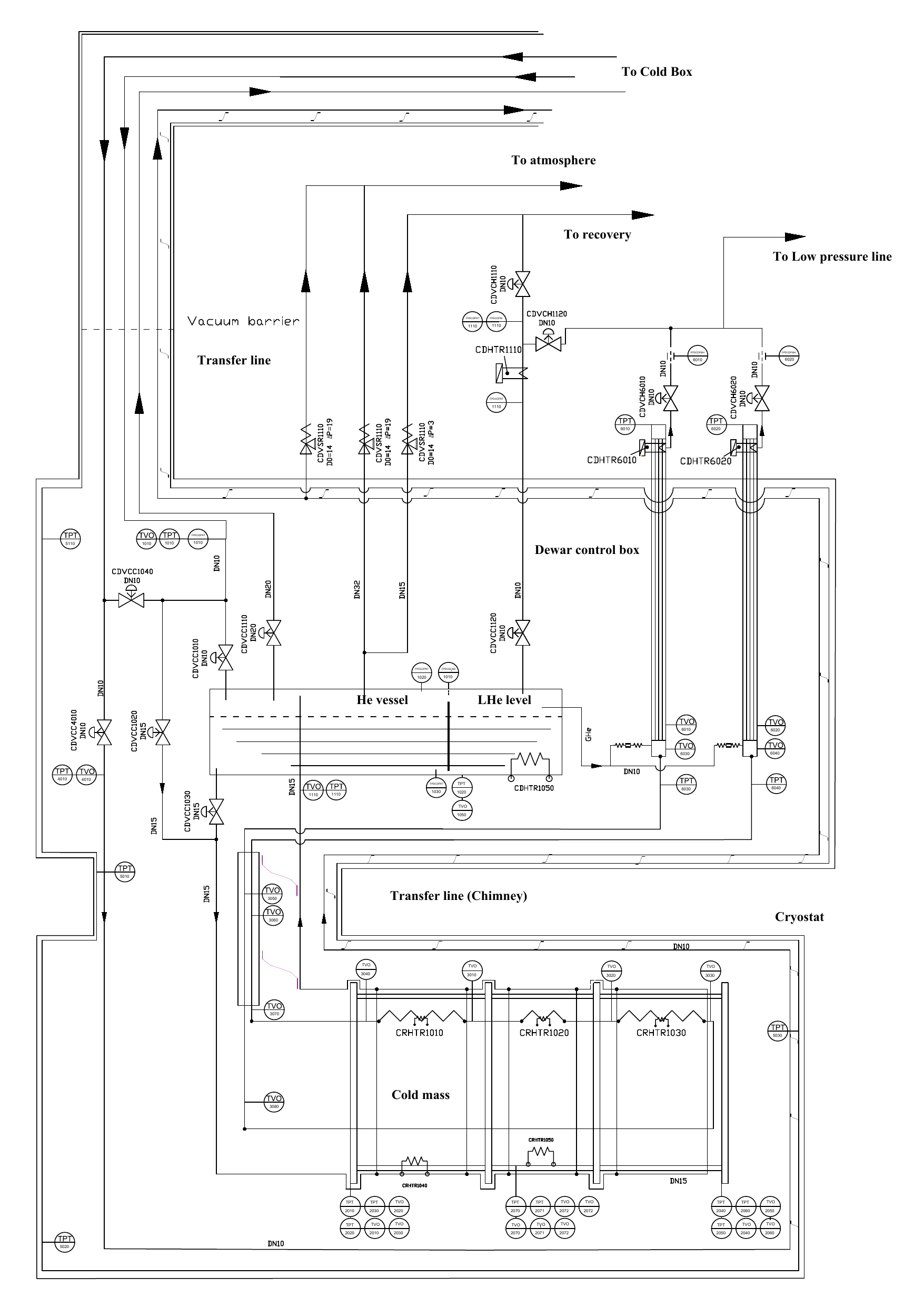}}
  \caption{ Process flow diagram of the SPD cryogenic system.
	}
\label{lb_mag_1222}
\end{figure}

Liquid helium that is used for cooling the cold mass is fed from the liquefier at 4.5 K. Thermal shields of the solenoid surround the cold internal parts and are cooled by gaseous helium, which passes through the pipes of the heat exchanger of serpentine type. The helium flow leaves the liquefier at a temperature level of 40 or 50 K and returns to the liquefier after passing the thermal shields at a temperature level of about 80 K.
The process flow diagram of the SPD cryogenic system has the same principle as for CMS solenoid (CERN) and PANDA (FAIR) and is shown in Fig \ref{lb_mag_1222}.

The main components of the cryogenic system are: 
\begin{enumerate}
\item the cryostat, which includes:
\begin{itemize}
\item the superconducting coil cooled indirectly via the thermal contact with the aluminum support cylinder,
\item thermal shields cooled by gaseous helium,
\item a vacuum vessel;
\end{itemize}
\item the control dewar, which includes:
\begin{itemize}
\item a vessel for liquid helium,
\item current leads,
\item thermal shields cooled by gaseous helium,
\item valves, gauges, and other instrumentation,
\item a horizontal vacuum tank with a pumping system;
\end{itemize}
\item a transfer line (chimney) connecting the vacuum shell of the cryostat and the control dewar;
\item the helium liquefier;
\item the transfer lines for helium supply connecting the refrigerator and the control dewar.
\end{enumerate}

\subsubsection{	Thermal loads of cryostat and control dewar}
Estimated thermal loads of the control dewar, cold mass, and thermal shields of the SPD solenoid are summarized in Table \ref{tab:mag_5}. 
\begin{table}[]
\centering
\caption{Heat loads of the control dewar, transfer line, and cryostat. }
\label{tab:mag_5}
\begin{tabular}{|l|c|c|c|}
\hline
T=4.5K                               & \multicolumn{3}{l}{Heat loads, W}                           \\
\hline
                                     & Normal condition & Without magnetic field & Current ramping \\
\hline                                   
\multicolumn{4}{l}{Cryostat}                                                                       \\
\hline  
radiation                            & 7.8              & 7.8                    & 7.8             \\
supports                             & 5*               & 5*                     & 5*              \\
eddy current loss in cold mass       & -                & -                      & 11.50**         \\
eddy current loss in conductor       & -                & -                      & 0.09**          \\
current leads, 6.5kA B=1.25T         & 15               & 9                      & 9               \\
\hline  
\multicolumn{4}{l}{Control dewar}                                                                  \\
\hline  
radiation                            & 0.45             & 0.45                   & 0.45            \\
supports                             & 0.26             & 0.26                   & 0.26            \\
cold valves                          & 1.05             & 1.05                   & 1.05            \\
safety relief valves                 & 3,22             & 3,22                   & 3,22            \\
vacuum barrier                       & 0.35             & 0.35                   & 0.35            \\
\hline  
\multicolumn{4}{l}{Transfer line}                                                                  \\
\hline  
radiation                            & 0.06             & 0.06                   & 0.06            \\
supports                             & 0.20             & 0.20                   & 0.20            \\
\hline  
Total                                & 33,39            & 27,39                  & 44,98           \\
\hline
\hline
T=60K                                & \multicolumn{3}{l}{Heat loads, W}                           \\
\hline
                                     & Normal condition & Without magnetic field & Current ramping \\
\hline
\multicolumn{4}{l}{Cryostat}                                                                       \\
\hline  
radiation                            & 160              & 160                    & 160             \\
supports thermal shields             & 12.00            & 12.00                  & 12.00           \\
eddy current loss in thermal shields & -                & -                      & 47.00**         \\
\hline  
\multicolumn{4}{l}{Control dewar}                                                                  \\
\hline  
radiation                            & 10.66            & 10.66                  & 10.66           \\
supports thermal shields             & 6.30             & 6.30                   & 6.30            \\
supports helium vessel               & 16.50            & 16.50                  & 16.50           \\
cold valves                          & 22.50            & 22.50                  & 22.50           \\
safety relief valves                 & 1.07             & 1.07                   & 1.07            \\
vacuum barrier                       & 1.18             & 1.18                   & 1.18            \\
\hline  
\multicolumn{4}{l}{Transfer line}                                                                  \\
\hline  
radiation                            & 0.99             & 0.99                   & 0.99            \\
supports                             & 2.00             & 2.00                   & 2.00            \\
Total                                & 233,2            & 233,2                  & 280.2           \\
\hline
\multicolumn{4}{l}{*ATLAS data;  ** PANDA data}                                                   
\end{tabular}
\end{table}

\subsubsection{Cryostat vacuum vessel}
\begin{figure}[!ht]
    \center{\includegraphics[width=1.\linewidth]{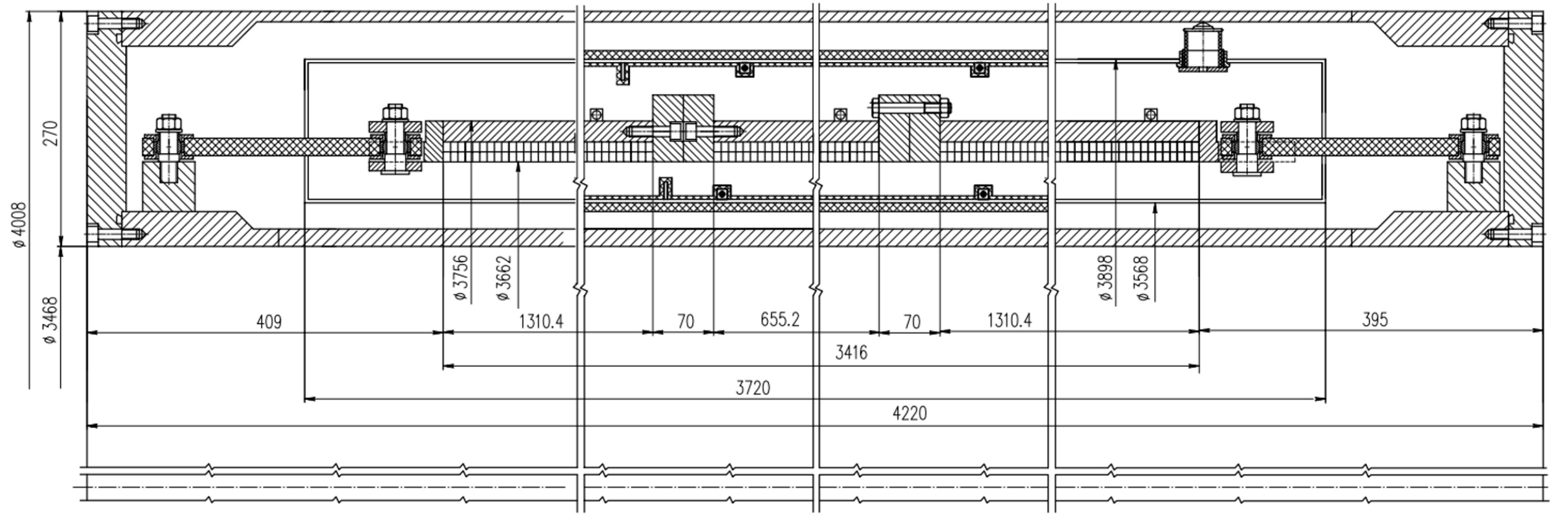}}
  \caption{ Cross-section of the cryostat.
	}
\label{lb_mag_11}
\end{figure}

The cryostat vacuum vessel is designed as two concentric shells with thick annular end plates, all made of stainless steel, according to AD2000 W10; its basic parameters are given in Table \ref{tab:mag_5a}. A cross-section of the cryostat is shown in Fig. \ref{lb_mag_11}.

\begin{table}[]
\centering
\caption{Vacuum vessel dimensions and operating conditions.}
\label{tab:mag_5a}
\begin{tabular}{|l|c|}
\hline
Envelope             &                  Dimension                                                                             \\
\hline
\begin{tabular}[c]{@{}l@{}}Inner radius
     \\ Outer radius
      \\ Length 
      \\ Inner cylinder thickness \\ Outer cylinder thickness \\ End plates thickness\end{tabular} & \begin{tabular}[c]{@{}l@{}}1604 mm\\ 1834 mm\\ 3800 mm
      \\ 8 mm
      \\ 12 mm
      \\ 45 mm\end{tabular} \\
Operating conditions                                                                                                                                           &                                                                                               \\
Operating pressure inside/outside                                                                                                                              & 0/1 bar-a                                                                                    \\
Designed inside overpressure against 1 bar-a                                                                                                                   & 1.45 bar-a      \\
\hline                                                                              
\end{tabular}
\end{table}
The nominal wall thickness is 12 mm for the outer cryostat shell and 16 mm for the inner shell. The thickness of the flanges is 45 mm. The thickness of the shells allows to minimize displacement of the cold mass under the action of the magnetic forces, relative to its nominal position. The flanges are attached to the vacuum vessel shells by 96+96 stainless steel bolts, according to the $\Gamma$OCT 11738-84 (DIN912) M10$\times$1.5 (L = 60 mm).

The installation of the barrel part of  ECal, weighing 53 tons, requires the cryostat to be optimized, taking into account additional loads. It may be possible to increase the thickness of the inner shell. The calorimeter is proposed to be installed on two horizontal rails, welded to the inner shell.

\begin{figure}[!ht]
    \center{\includegraphics[width=1.\linewidth]{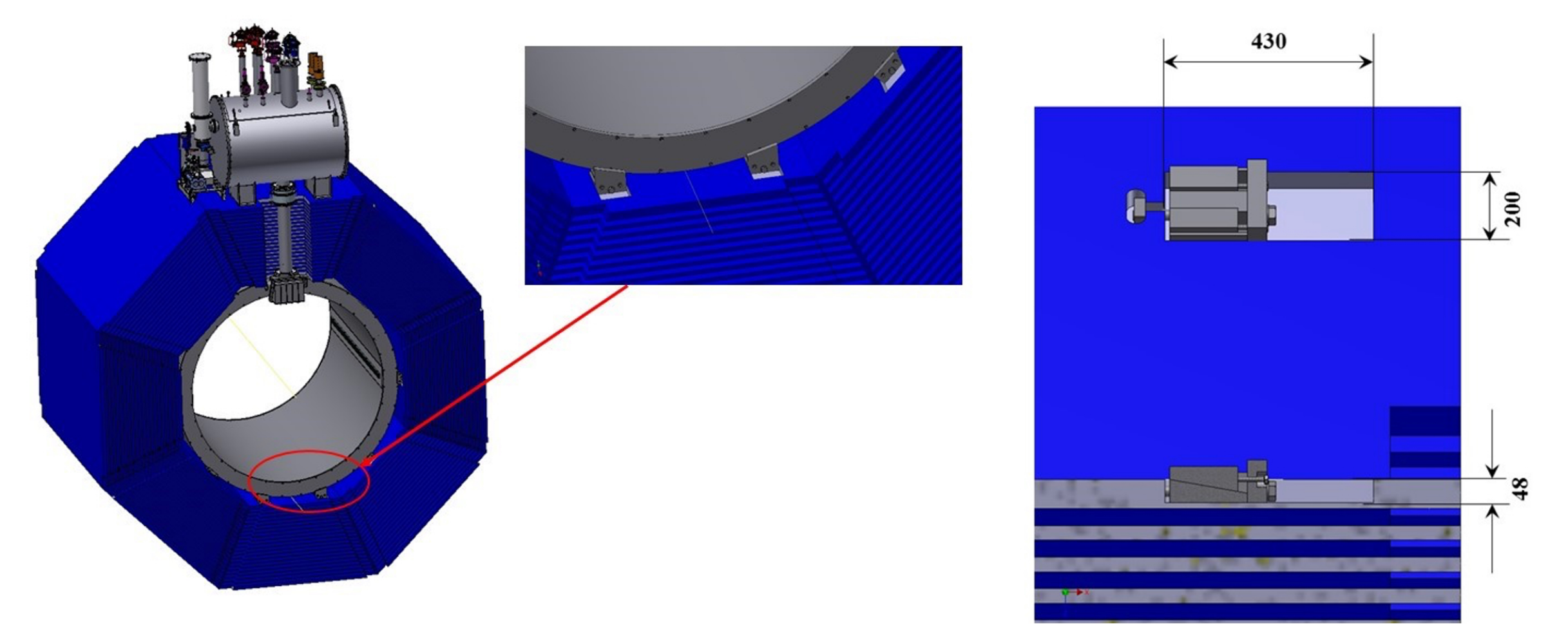}}
  \caption{ Location of the cryostat in the yoke and arrangement of the cryostat supports.
	}
\label{lb_mag_111}
\end{figure}
\begin{figure}[!ht]
    \center{\includegraphics[width=1.\linewidth]{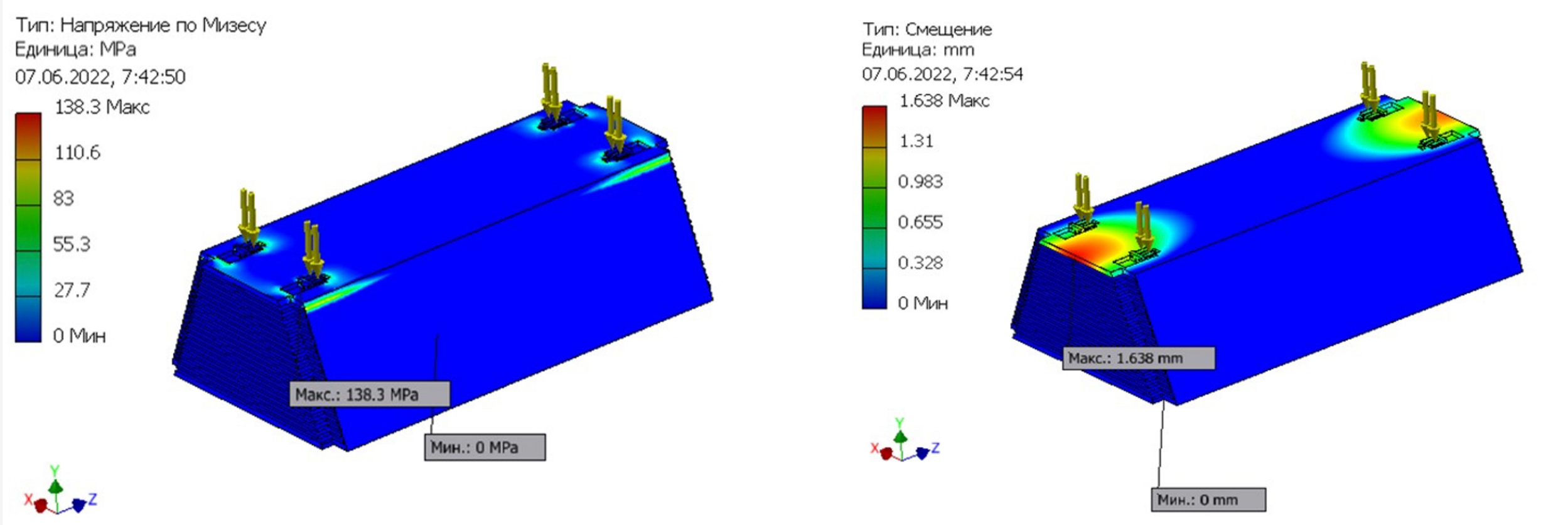}}
  \caption{ Calculation of mechanical and magnetic loads on the bottom octant.
	}
\label{lb_mag_112}
\end{figure}

 The cryostat supports (to be designed later) will provide a high rigidity of the construction for all structural loads, including the action of the magnetic forces, and will take into account the minimal dimensions between the outer shell of the cryostat and the yoke (see Fig. \ref{lb_mag_111}). The 18 supports will be installed, 9 pieces for each side. The maximal load (about 64 tons) should be on the bottom octant of the yoke. To distribute this load, four supports are installed on the top plate of the octant next to the edges, whereas the biggest part of the load goes on the side plate of the octant. The mechanical stress analysis shows that the maximum dimensional deformation is 1.6 mm for the top octant plate and the maximum stress is about 140 MPa (Fig. \ref{lb_mag_112}). Also, these supports make it possible to align the solenoid axis with the beam axis.


\begin{figure}[!ht]
    \center{\includegraphics[width=1.\linewidth]{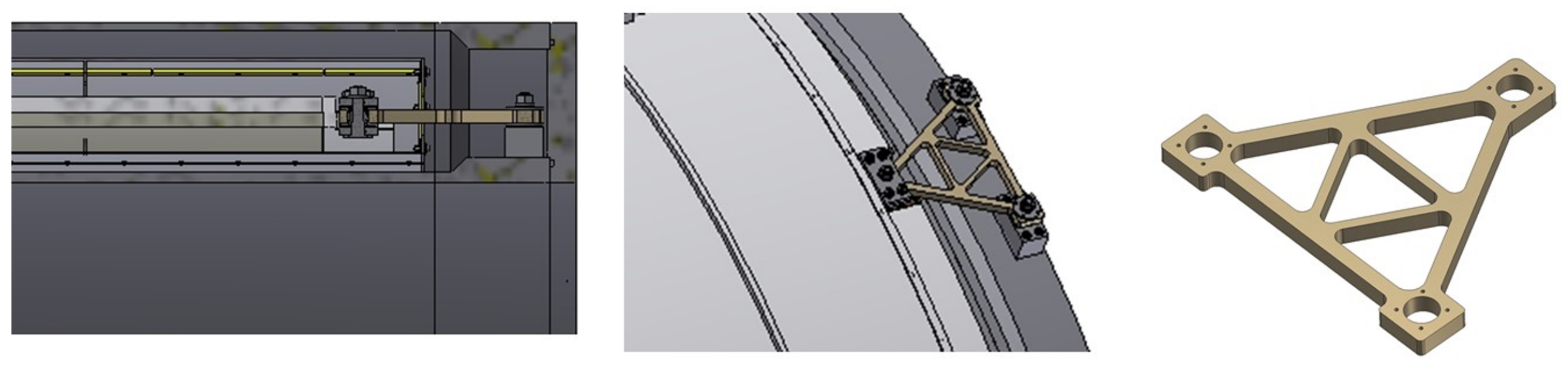}}
  \caption{ Design of the triangle support of the SPD cold mass.
	}
\label{lb_mag_113}
\end{figure}

\begin{figure}[!ht]
    \center{\includegraphics[width=1.\linewidth]{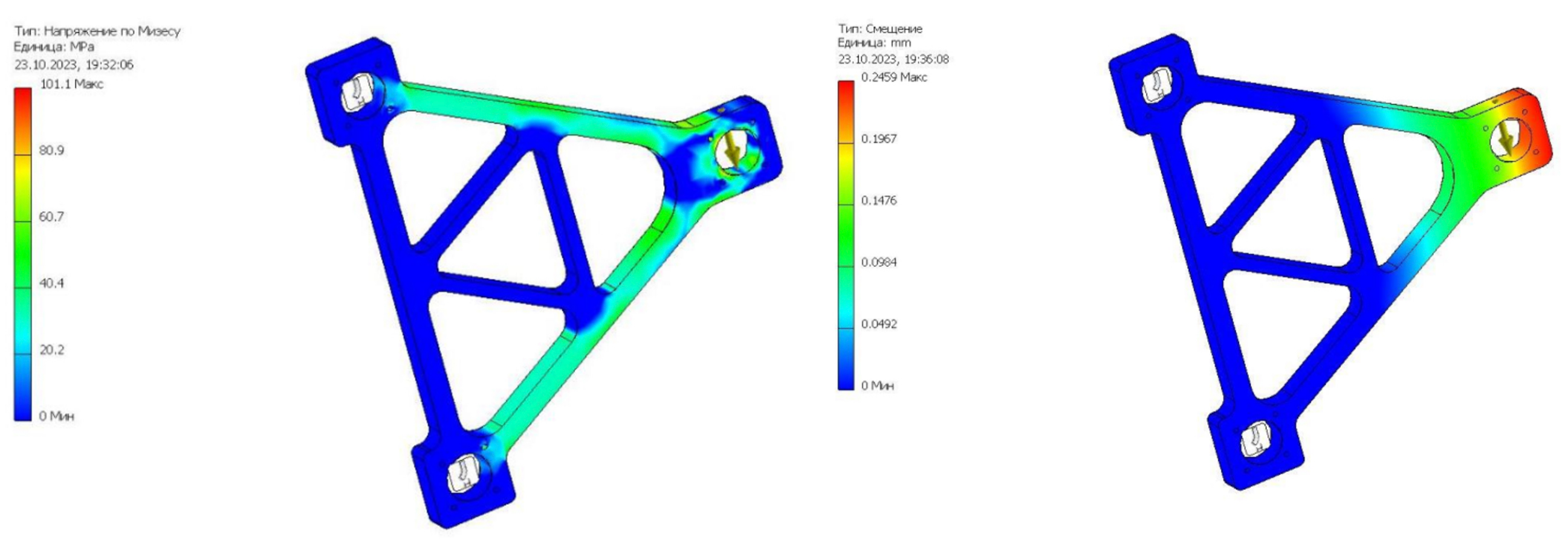}}
  \caption{Calculated mechanical loads on the triangle support.
	}
\label{lb_mag_114}
\end{figure}

The axial and radial gaps between the cryostat vacuum shell and the thermal shield, as well as between the shield and the cold mass in warm and cold states, are indicated in Fig. \ref{lb_mag_11}. They are not sufficient for the cryostat assembly and for avoiding thermal contacts between the shells after cooling. A certain amount of space is needed for adjusting the optimal mutual positions of the holes for the triangle supports in the support flanges on the inner cryostat shell and in the brackets or holes for connecting the triangle supports to the cold mass. 
There are 12 triangle supports on each side. The wide side of a triangular support is attached to a part of the vacuum shell by a special flange with two M16 bolts. The apex of the triangle is attached to the cold mass. The shape of the support takes into account the mechanical and thermal loads, as a compromise.
To compensate for the radial changes in lengths of the cold mass and the vacuum shell  (more than 12 mm), spherical washers are installed in all attachment points. To minimize the change in the position of the cold mass, we assume the pre-setting of the cold mass to be at + 5$\div$6 mm from the plane of the support flange, and after cooling down to 4.5 K -- at -5$\div$6 mm from the plane of the support flange. To compensate for the longitudinal changes in length (more than 12 mm), a slot is proposed in the cold mass, at the point of attachment of the triangle apex. The design of the  triangular support is shown in Fig. \ref{lb_mag_113} and the mechanical analysis in Fig. \ref{lb_mag_114}. The estimated loads on the triangular supports are given in Table \ref{tab:load_to_support}. 
\begin{figure}[!ht]
    \center{\includegraphics[width=1.\linewidth]{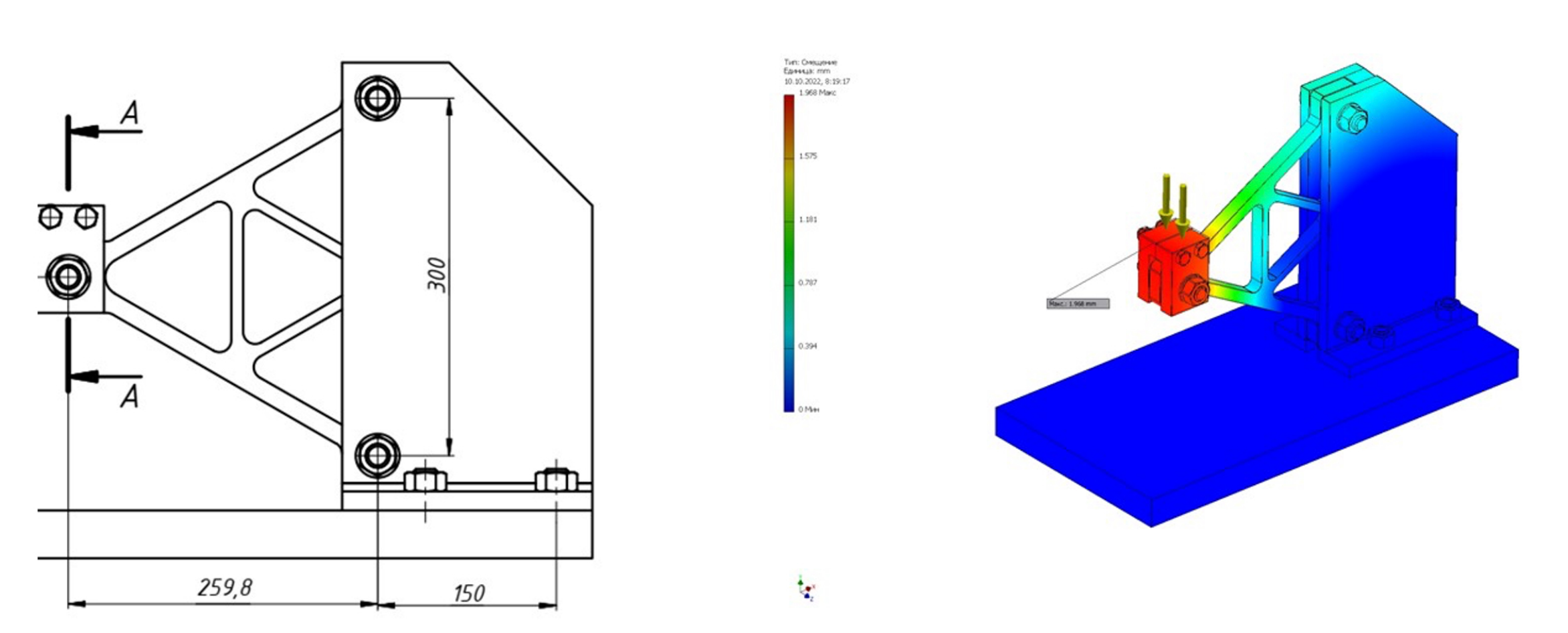}}
  \caption{Triangle support test device.
	}
\label{lb_mag_115}
\end{figure}

\begin{table}[htp]
\caption{Loads on the triangular support.}
\begin{center}
\begin{tabular}{|l|c|c|}
\hline
Load conditions & Load, kN & Maximum load to support, kN \\
\hline
Cold mass weight & 42 & 4.0 \\
Decentering force (5 mm) & 47.5 & 4.47 \\
\hline
Total load  & 89.5 & 8.47 \\
\hline
\end{tabular}
\end{center}
\label{tab:load_to_support}
\end{table}%

As a material for the supports, it is proposed to use glass-textolite STEF1 or some type of carbon fiber. Mechanical calculations were made for the support with a thickness of 20 mm and a load of 10 kN (Fig. \ref{lb_mag_114}). Calculations indicate an acceptable stress level of 96 MPa, with a limit stress of STEF1 -- 132 MPa and with a maximal deformation of 0.9 mm.
Testing of the triangular supports at room and cryogenic temperatures is planned. The proposed scheme of the tests is shown in Fig. \ref{lb_mag_115}.

The exact location of the cold mass inside the cryostat after its assembly should be defined with maximum precision. This is necessary to determine the optimal position of the superconducting coil inside the yoke aperture. After the cryostat assembly, any correction of the cold mass position inside the cryostat is impossible. A correction of the coil position, relative to the yoke aperture, can only be provided by means of fitting spacers in the cryostat support legs. This correction can be implemented in accordance with the results of field measurements in the tracker area and measurements of forces in the suspension rods.

In the normal operative condition mode, the inner cryostat volume is evacuated, which leads to general membrane compressive stresses in the outer cryostat shell, due to the action of the outer atmospheric pressure. In the process of magnet operation, an emergency situation may happen, when an overpressure up to 0.05 MPa appears in the inner volume. This leads to general membrane compressive stresses in the inner cryostat shell.

\begin{figure}[!h]
    \center{\includegraphics[width=1.\linewidth]{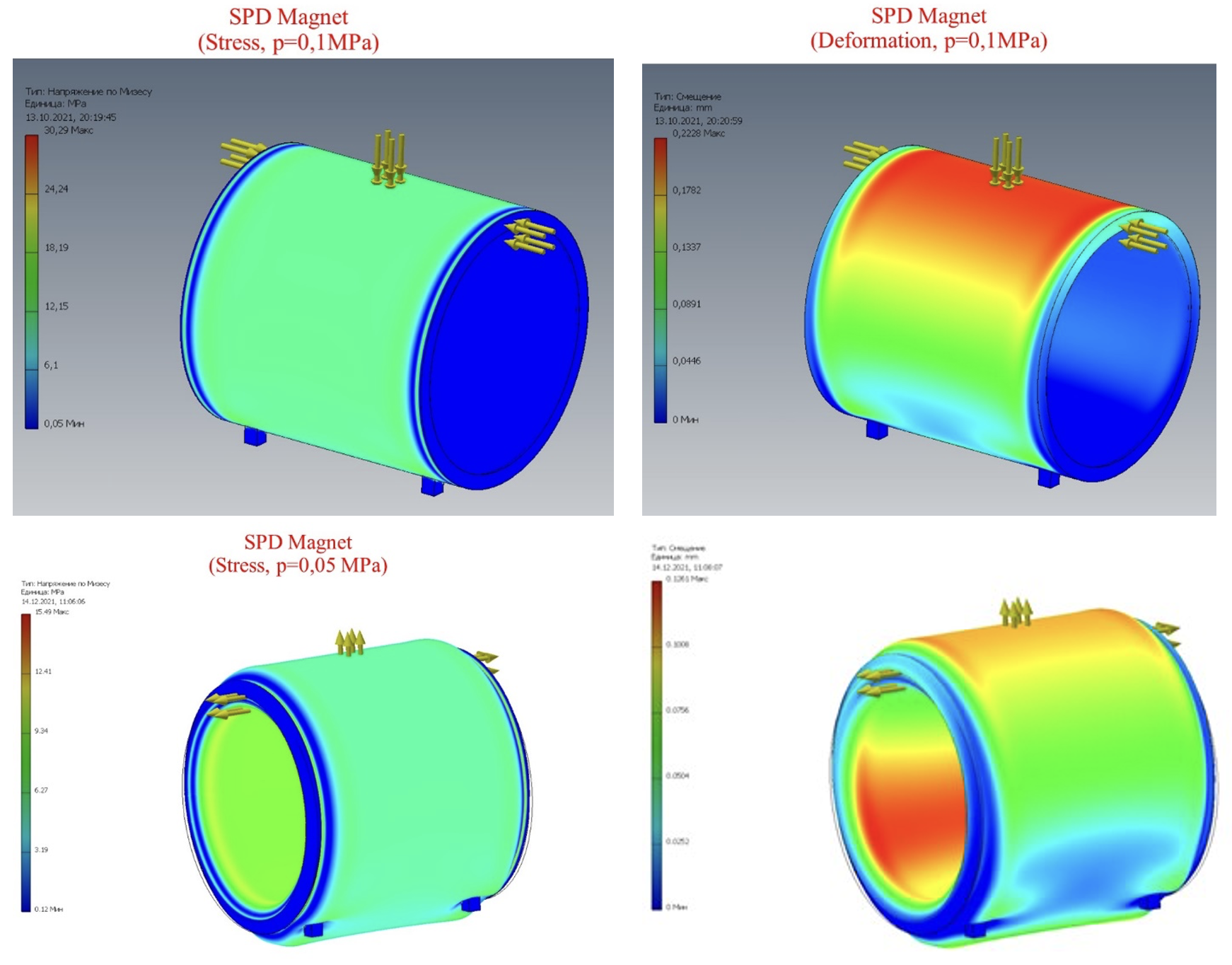}}
  \caption{ Calculations of the cryostat stresses and deviations. The load of the calorimeter is not included.
	}
\label{lb_mag_12}
\end{figure}

The buckling analysis of the cryostat shells is based on an FE model. The considered load cases are the following:
\begin{itemize}
\item	 cryostat weight;
\item	 weight load of the cold mass and thermal screen octants, applied to the outer cryostat shell;
\item	weight loads of the thermal screen octants and of the inner detectors, applied to the inner cryostat shell;
\item	 outer overpressure of 0.1 MPa (first variant);
\item	 inner overpressure of 0.05 MPa (second variant).
\end{itemize}
According to computations, the stability of the cryostat shell under the action of an outer overpressure of 0.1 MPa is ensured with a safety margin of 8.5. The stability of the cryostat shell from the action of an inner overpressure of 0.05 MPa is ensured with a safety margin of 17 (Fig. \ref{lb_mag_12}).

\subsubsection{Control dewar}
\begin{figure}[!h]
    \center{\includegraphics[width=1.\linewidth]{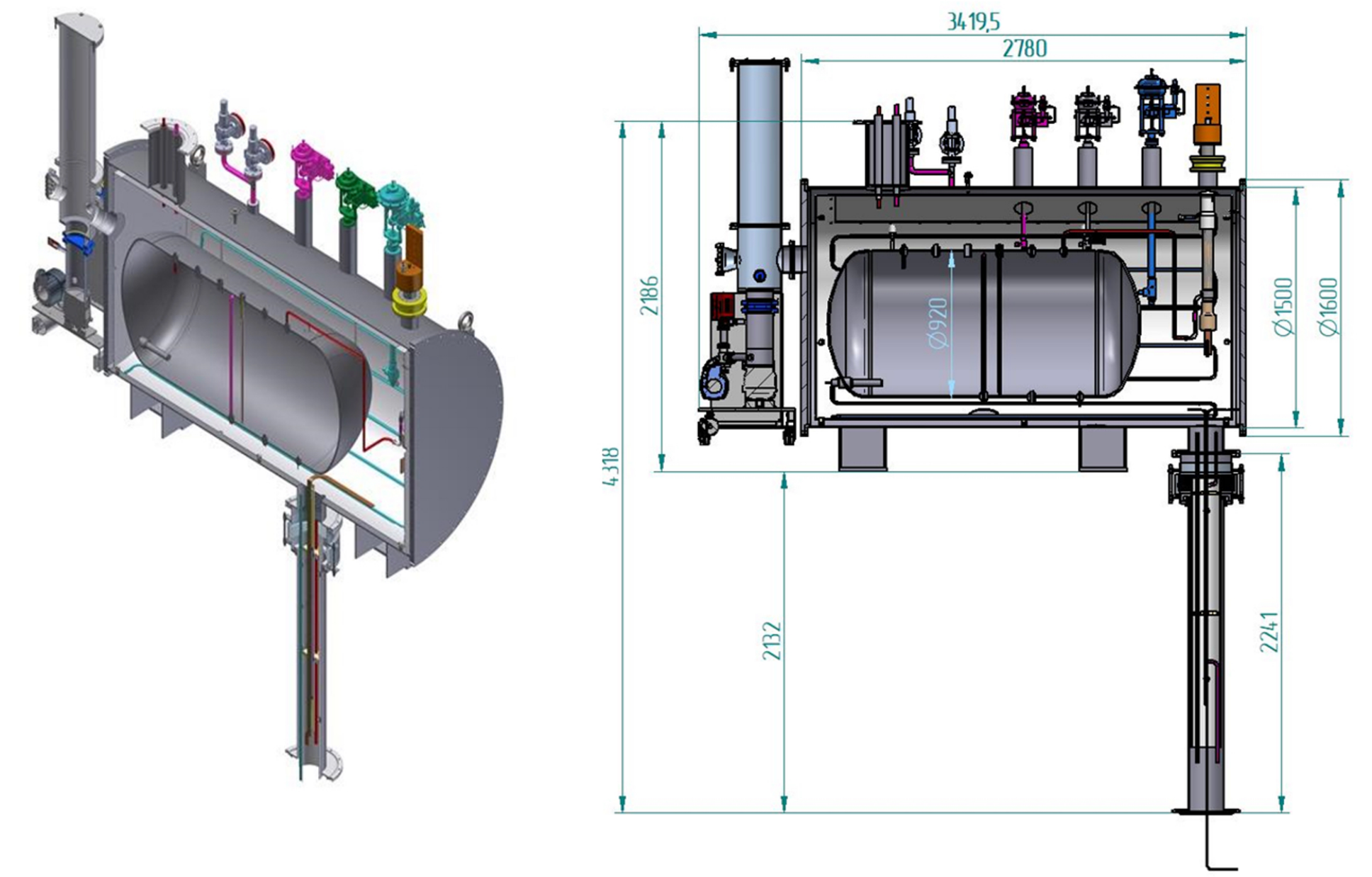}}
  \caption{ Layout of the control dewar. 
	}
\label{lb_mag_13}
\end{figure}

\paragraph{Vacuum vessel}
The outer vacuum housing of the control dewar encloses the stainless steel vessel for liquid helium, control valves, current leads, and process piping, surrounded by a thermal shield, located between the outer housing and the interior of the control dewar (Fig. \ref{lb_mag_13}). The vacuum vessel of the control dewar has a horizontal barrel shape and is made of stainless steel.

The thickness of the outer walls of the vessel is 8 mm and the weight of the assembled control dewar is about 1600 kg. The main dimensions of the control dewar are shown in Fig. \ref{lb_mag_13}. The diameter of the vacuum shell is 1500 mm, the height with valves is about 2500 mm, the width of the control dewar is 2780 mm, and with the vacuum pump system -- 3425 mm. The flanges of the control dewar lids are tightened by 2$\times$48 studs M12.

The vapor-cooled current leads to the solenoid and the cryogenic relief valves are mounted on the top plate of the control dewar housing. 

Vacuum vessels, shells of transfer lines and the cryostat must handle the loads from all possible operating scenarios. The design pressure of all vacuum vessels is 0.45 bar(g).

\paragraph{Valves of the control dewar}
Chosen types of the control dewar valves are listed in Table \ref{tab:mag_6}. The regimes of the flow through the heat exchangers of the cold mass and the thermal shield will be defined in detail later. The positions of valves of the control dewar for various magnet operation regimes are established in accordance with Table \ref{tab:mag_7}, where the positions of controlled valves are indicated for each regime.

\begin{table}[]
\centering
\caption{Control dewar valves.}
\label{tab:mag_6}
\begin{tabular}{|l|c|c|c|l|}
\hline
Part                                & Nominal       & Rated  & Rated & Valve type \\
                                       &diameter,  mm & pressure,  bar & temperature, K & \\
\hline
CV1 & 10                                                & 25                                               & 4.5$\div$300                                           & pneumatically-actuated control valve \\
        &   &   &   & with digital positioner (0...100\%)\\
CV2& 10                                                & 25                                               & 4.5$\div$300                                           & pneumatically-actuated control valve  \\
        &   &   &   & with digital positioner (0...100\%)\\
CV3& 25                                                & 25                                               & 4.5$\div$300                                           & pneumatically-actuated control valve  \\
        &   &   &   & with digital positioner (0...100\%)\\
CV4& 10                                                & 25                                               & 4.5$\div$300                                           & pneumatically-actuated control valve  \\
        &   &   &   & with digital positioner (0...100\%)\\
CV5& 10                                                & 25                                               & 4.5$\div$300                                           & pneumatically-actuated control valve \\
        &   &   &   & with digital positioner (0...100\%)\\
CV6& 8                                                 & 25                                               & 80$\div$300                                            & pneumatically-actuated control valve \\
        &   &   &   & with digital positioner (0...100\%)\\
CV7& 8                                                 & 25                                               & 80$\div$300                                            & pneumatically-actuated control valve  \\
        &   &   &   & with digital positioner (0...100\%)\\
CV8& 8                                                 & 25                                               & 40$\div$300                                            & pneumatically-actuated control valve  \\
        &   &   &   & with digital positioner (0...100\%)\\
CV9& 8                                                 & 25                                               & 40$\div$300                                            & pneumatically-actuated control valve  \\
        &   &   &   & with digital positioner (0...100\%)\\
SV1& 15                                                & 25                                               & 80$\div$300                                            & pneumatically-actuated shut-off valve                   \\
        &   &   &   &(open/closed)  \\
SV2& 20                                                & 25                                               & 80$\div$300                                            & pneumatically-actuated shut-off valve                \\
        &   &   &   &  (open/closed)    \\
RV1& 20                                                & 20                                               & 4.5$\div$300                                           & safety relief valve                                                   \\
RV2& 20                                                & 20                                               & 4.5$\div$300                                           & safety relief valve     \\                                             
\hline                                             
\end{tabular}
\end{table}

\begin{table}[]
\centering
\caption{Positions of the control dewar valves for various SPD magnet operation regimes. C: valve closed, O: valve open, R: valve regulated.}
\label{tab:mag_7}
\begin{tabular}{|c|c|c|c|c|c|c|c|c|c|c|c|}
\hline
{\tiny Valve} & {\tiny  Initial } & {\tiny Cool down}  & { \tiny Steady-state} & {\tiny  Warm-up} & {\tiny  Cooling of }  & {\tiny  Emergency} & {\tiny  Cooling} & {\tiny Emergency} & {\tiny Emergency} & {\tiny Emergency}  & {\tiny Emergency}  \\
          &{\tiny  setting }& \tiny {from} &  &  &  { \tiny the thermal} & {\tiny (coil}  & {\tiny  after coil } & {\tiny (refrigerator}  & {\tiny (power}  & {\tiny (current lead} & {\tiny (loss of} \\
          &           & \tiny {300 to 4.5 K} &  &  & {\tiny shields} & {\tiny quenching)} & {\tiny quenching} &{\tiny failure)} & {\tiny failure)} &  {\tiny voltage rise above} &{\tiny vacuum)}\\
          &            &     & & & & & & & &{\tiny allowable level)} & \\
\hline
CV1                    & C               & R                                                               & C            & R       & C                              & C                          & R                            & C                                                                          & C                                                                   & R                                                                                                          & C                                                                    \\
CV2                    &   C            &     R                                                         &R            & C       & C                              & C                          &R                            & C                                                                          & C                                                                   & C                                                                                                          & C                                                                    \\
CV3                    & C               & C                                                              & O            & C       & C                              & C                          & C                            & O                                                                          & O                                                                   & C                                                                                                          & C                                                                    \\
CV4                    & C               & O                                                               & O            & R       & C                              & C                          & O                            & C                                                                          & C                                                                   & R                                                                                                          & C                                                                    \\
CV5                    & C               & C                                                               & C            & R       & C                              & O                          & R                            & R                                                                          & R                                                                   & R                                                                                                          & O                                                                    \\
CV6                    & C               & R                                                               & R            & R       & C                              & C                          & R                            & R                                                                          & R                                                                   & R                                                                                                          & C                                                                    \\
CV7                    & C               & R                                                               & R            & R       & C                              & C                          & R                            & R                                                                          & R                                                                   & R                                                                                                          & C                                                                    \\
CV8                    & C               & R                                                               & R            & R       & R                              & R                          & R                            & C                                                                          & C                                                                   & R                                                                                                          & C                                                                    \\
CV9                    & C               & O                                                               & O            & O       & O                              & O                          & O                            & C                                                                          & C                                                                   & O                                                                                                          & C                                                                    \\
SV1                    &C              &C                                                              & C            & O       & C                              & C                          & C                            & C                                                                          & C                                                                   & O                                                                                                          & C                                                                    \\
SV2                    &C             &O                                                           & O            & C       & C                              & O                          & O                            & O                                                                          & O                                                                   & C                                                                                                          & O            \\                                                       
\hline
\end{tabular}
\end{table}

\paragraph{Vessel for liquid helium }

A stainless steel vessel for liquid helium, required to maintain the operating conditions of the solenoid, is placed inside the control dewar. The hydraulic volume of the helium bath of the vessel is chosen to provide stable operation of the magnet with heat inflows to the control dewar and the superconducting coil during the period necessary for safe de-energizing of the superconducting coil when the liquefier is stopped. The heat flow into the vessel and to the superconducting coil, including heat losses in current leads, amounts to 50 W, which corresponds to a liquid helium flow rate of 71 l/h. Since it takes more than 60 minutes for safe de-energization of the coil, the volume of the helium vessel is chosen to be 300 liters, to have a sufficient safety margin. The outer surface of the vessel is covered by a multilayer screen-vacuum insulation to reduce heat inflows.

\begin{figure}[!h]
    \center{\includegraphics[width=1.\linewidth]{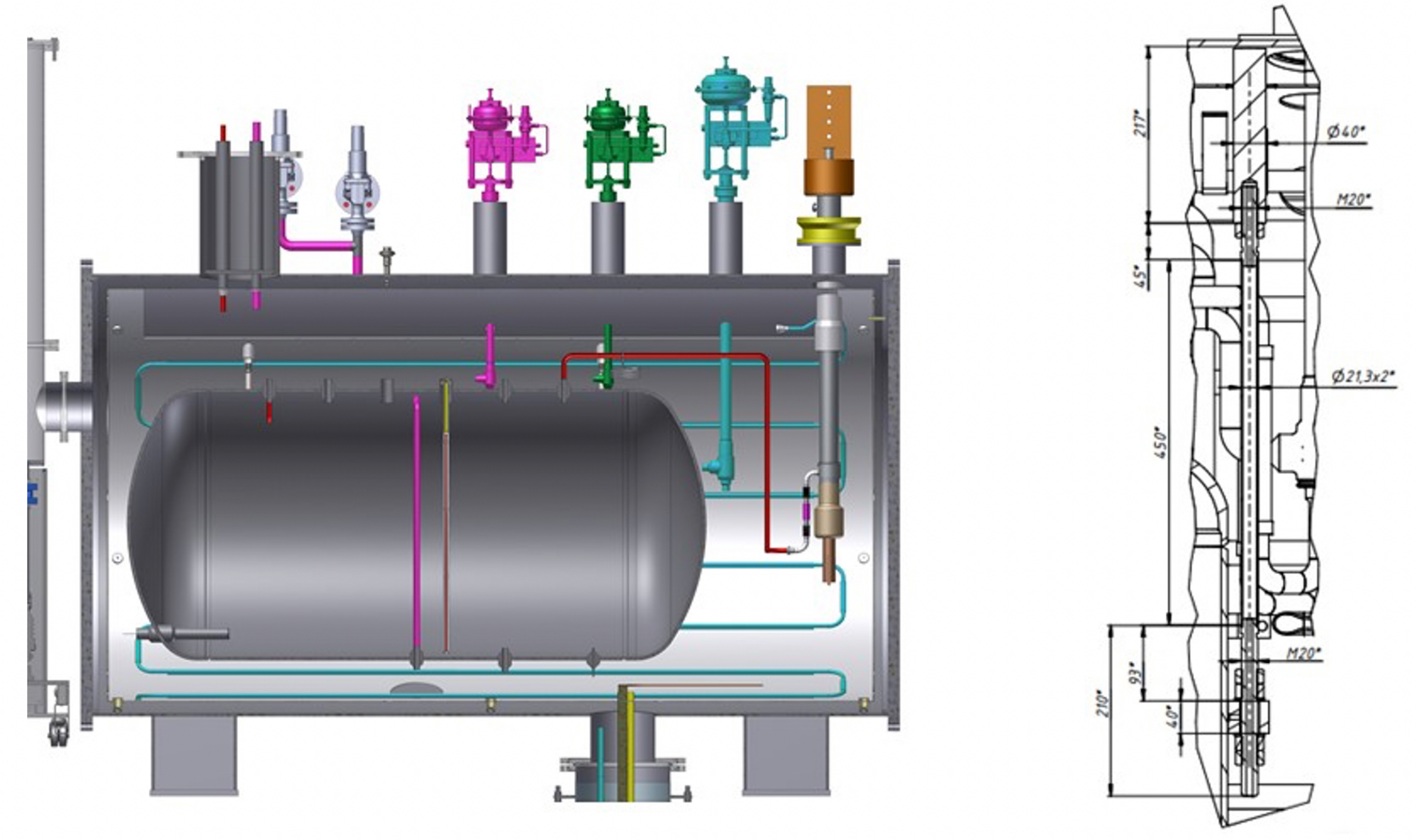}}
  \caption{ Arrangement of the helium vessel into the control dewar and a sketch of the suspension of the helium vessel.
	}
\label{lb_mag_15}
\end{figure}

The volume and the outer surface area of the vessel for liquid helium are 300 l and 2.5 m$^2$, respectively. It is possible to increase 
the volume of the helium vessel,
because the space on the upper yoke platform allows a placement of the control dewar with bigger dimensions.
The helium vessel suspensions are made by combining a pipe and a rod (Fig. \ref{lb_mag_15}). The heat load on the suspension (0.09 W)  is calculated with the following parameters: the temperature of the shell -- 300 K, of the helium vessel -- 4.5 K, and of the thermal shields -- 60 K. The components of the suspension are:
\begin{itemize}
\item a stainless steel rod 217 mm long with a diameter of 40 mm;
\item a stainless steel  pipes with a diameter of 21.3, thickness of 2, and length of  588 mm;
\item a rod 200 mm long, stainless steel M12.
\end{itemize}

The stability of the helium vessel, when it is exposed to external pressure of $-1$ bar(g) (vacuum inside the vessel), is provided with a sufficient safety margin. The conditions of static and cyclic strength and the stability condition will be fulfilled in this case as well. 

An electrical heater is installed in the helium bath to maintain the required level of liquid helium and to control the rate of the return helium flow to the helium liquefier (maintenance of thermal balance). The heater power of about 400 W corresponds to the maximum cooling capacity of the liquefier. This heater is also used for rapid evaporation of liquid helium from the bath. The power capacity of the heater may vary from zero to the maximum value, determined by the cooling power of the liquefier at a temperature level of 4.5 K. Two identical liquid helium level-meters (one active, one spare) will be installed in the helium vessel. 

It is also suggested to install a differential pressure sensor. The first point of pressure measurement is the upper part of the helium vessel with gas medium; the second capillary should be installed at the bottom of the helium vessel.  The value of the $\Delta P$ provides a measurement of the liquid helium level in the vessel. The control of the liquid helium level  in the vessel will be included in the control system of the magnet. 

\subsubsection{	Chimney and interface}
\begin{figure}[!t]
    \center{\includegraphics[width=1.\linewidth]{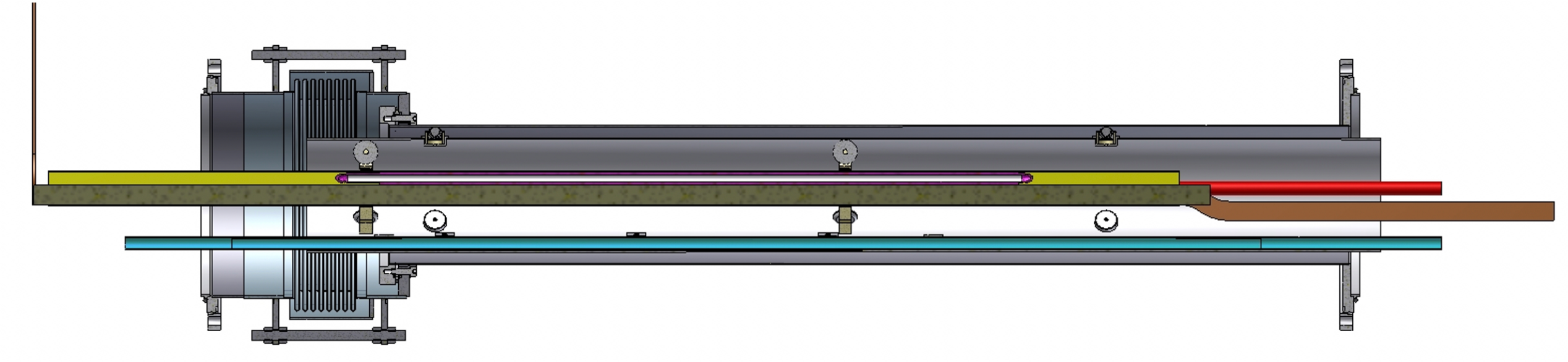}}
  \caption{ Section of the chimney with the control dewar.
	}
\label{lb_mag_16}
\end{figure}

The vacuum pipe (chimney) connects the vacuum volumes of the cryostat and the control dewar. The outer diameter of the chimney is 219.1 mm, and the wall thickness is 2 mm. It encloses the superconducting bus-bars, direct and return pipes for gaseous and liquid helium flow, and measurement wiring, all of which are surrounded by a thermal shield (Fig. \ref{lb_mag_16}).  Its bottom part is supplied by a weld ring flange DN 250 ISO-K with rotatable bolt ring DN 250 ISO-K (both made of stainless steel 304/1.4301). This flange is tightened to the flange on top of the interface box by 12 bolts M10. The rotatable bolt ring facilitates the connection procedure of the flanges at the stage of the cryostat and control dewar assembly.

\begin{figure}[!b]
    \center{\includegraphics[width=1.\linewidth]{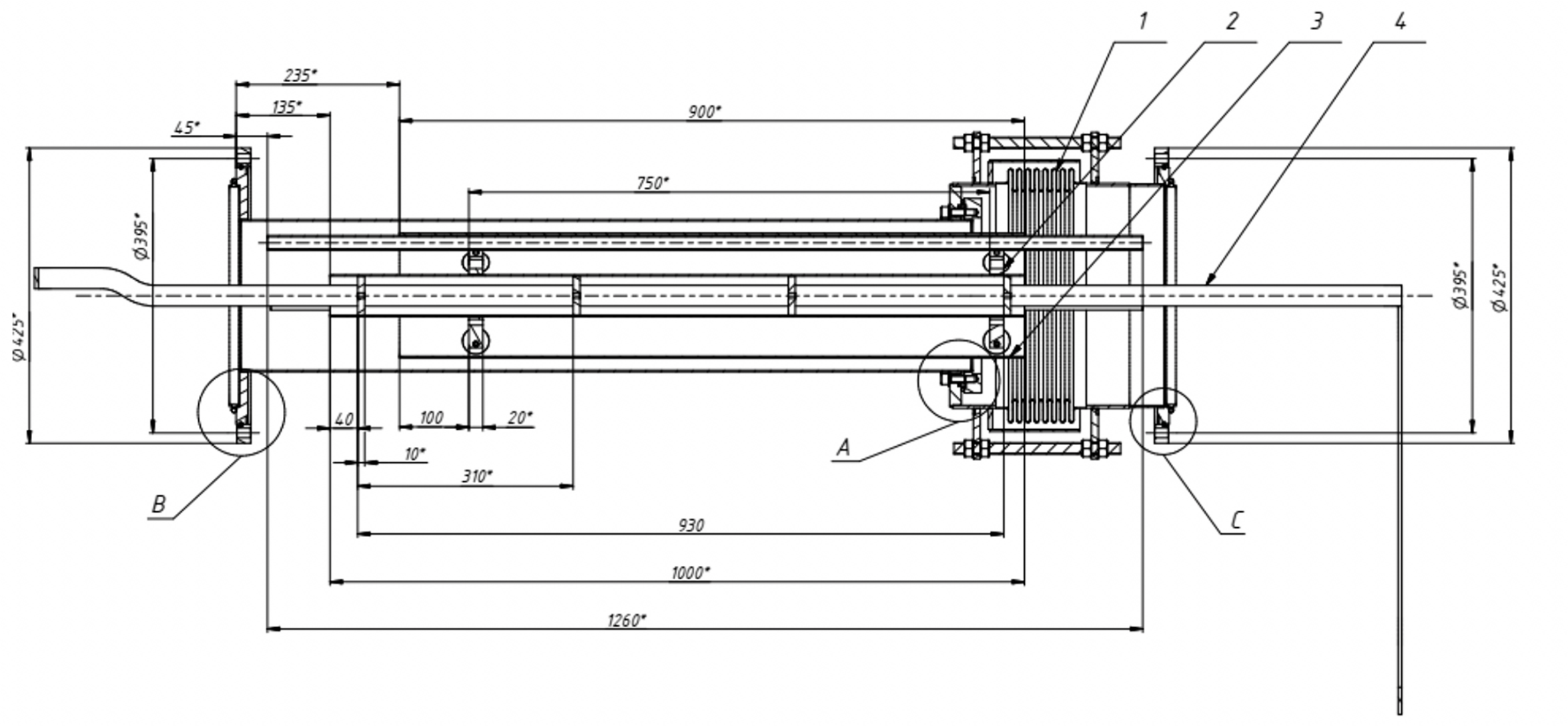}}
  \caption{ The main dimension of the chimney. 1 -- bellows; 2 -- G10 spacer; 3 -- thermal shield; 4 -- bus-bar.
	}
\label{lb_mag_17}
\end{figure}

A decoupling spring bellow (made of stainless steel 1.4541), supplied with two flanges DN 250 ISO-K (made of stainless steel 304/ 1.4301), is placed at the top of the chimney, in order to compensate for mutual shifts of the control dewar and the cryostat during assembly, reloading of the magnet from stationary supports to roller skates and back, and during magnet movement along the rail track. The upper flange of the decoupling spring below is connected to the flange at the bottom of the control dewar. 
The main parts of the chimney are shown in Figure \ref{lb_mag_17}.

\begin{figure}[!h]
    \center{\includegraphics[width=1.\linewidth]{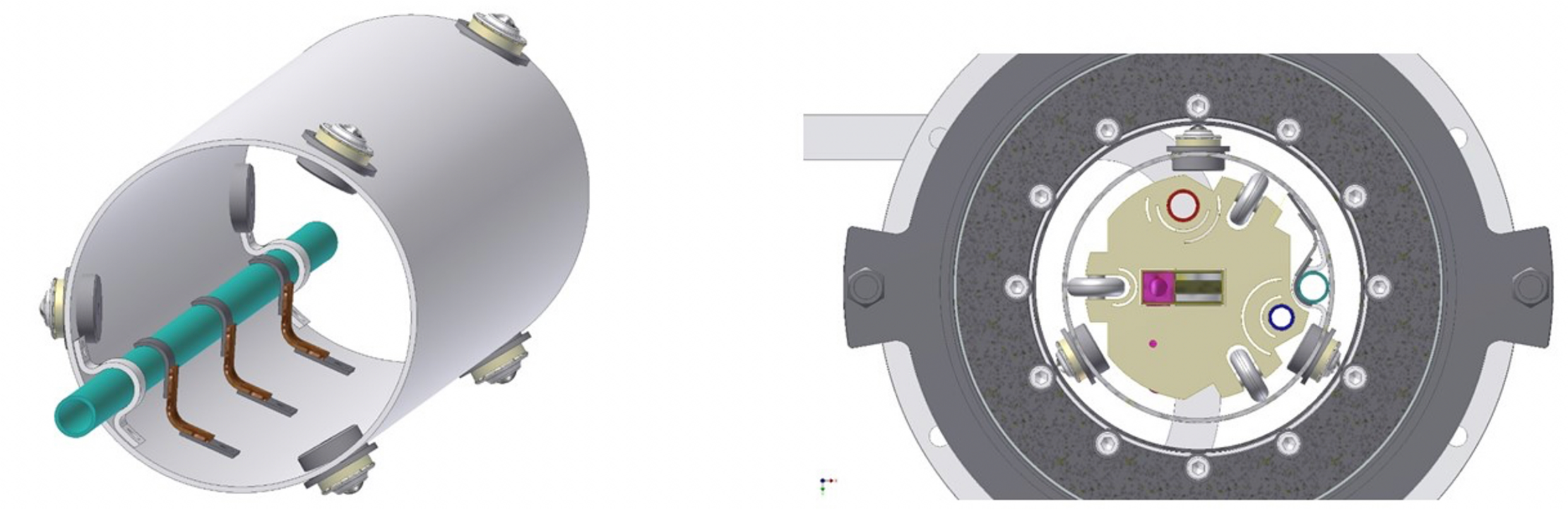}}
  \caption{ Views of the chimney's thermal shield with thermal bridges and a cross-section of the chimney.
	}
\label{lb_mag_18}
\end{figure}

The current leads and process piping surrounded by a thermal shield are located between the outer shell and cold part with temperature 4.5 $\div$ 50 K. The 80-K return pipe should be fixed to the chimney thermal shield and have the copper thermal bridges installed every 350$\div$400 mm. To fix the location of the thermal shield in the vacuum shell, a few ball supports are installed (see Fig. \ref{lb_mag_18}). Also, these supports allow the movement of the thermal shield, relative to the vacuum shell, after cooling and shrinkage of the thermal shield. The material of the thermal shield is stainless steel 304/ 1.4301. One edge of the copper thermal bridges is brazed to the stainless steel 80-K return pipe, and another edge is connected to the thermal shield by screw rivets. Apiezon N grease provides a reliable heat contact of the connection.

\begin{figure}[!h]
    \center{\includegraphics[width=0.8\linewidth]{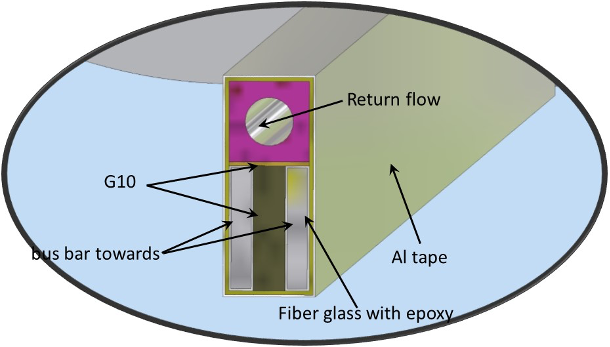}}
  \caption{ Cross-section of a bus-bar assembly.
	}
\label{lb_mag_19}
\end{figure}

\begin{figure}[!h]
    \center{\includegraphics[width=0.82\linewidth]{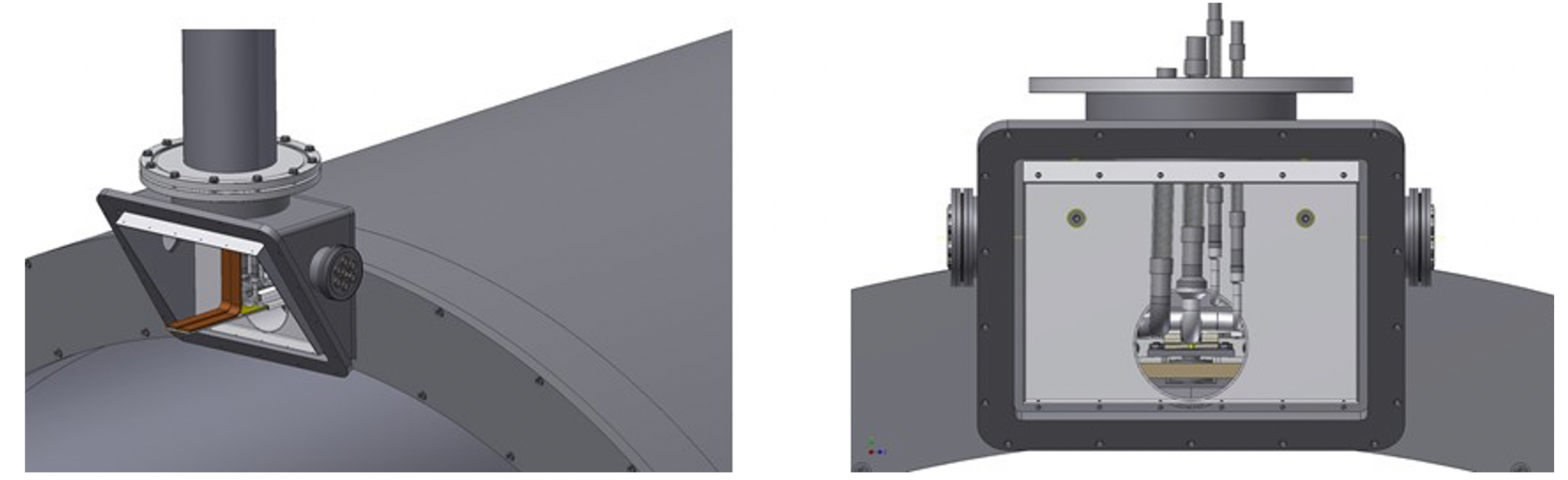}}
  \caption{ Connecting bus-bars and process pipes into the Interface box. 
	}
\label{lb_mag_20}
\end{figure}

The positions of the 4.5-K forward and return pipes, the 50-K forward pipe, and bus-bars into the chimney are fixed by two G10 spacers. The spacers will be equipped with two or three G10 wheels, to compensate for the changing dimensions of this assembly, relative to the thermal shield during all operating conditions. The 50-K return pipe has a rigid fixation to the spacers, while other pipes and bus-bars have a slippery fixation. To avoid any risk of heating the bus-bars and high risk of quenching or even overheating of the superconductor, the bus-bars are connected to an assembly with a 4.5-K return pipe (Fig. \ref{lb_mag_19}).
\paragraph{Interface box}
The interface box is located in the bottom part of the chimney and is fixed on the flange of the cryostat. The box accommodates connections of the pipelines, as well as joints of the superconducting bus-bars of the cryostat and the control dewar. The flanges with connectors for sensors are located on the side cover of the interface box.
Testing of the solenoid welds of superconducting cables in the interface box will be performed using extensions of the bus-bars (Fig. \ref{lb_mag_20}). The joints will be made by edge welding of Al to Al stabilizer, up to three times. These extensions can be cut to allow the dismounting of the solenoid a few times.

\begin{figure}[!h]
    \center{\includegraphics[width=1\linewidth]{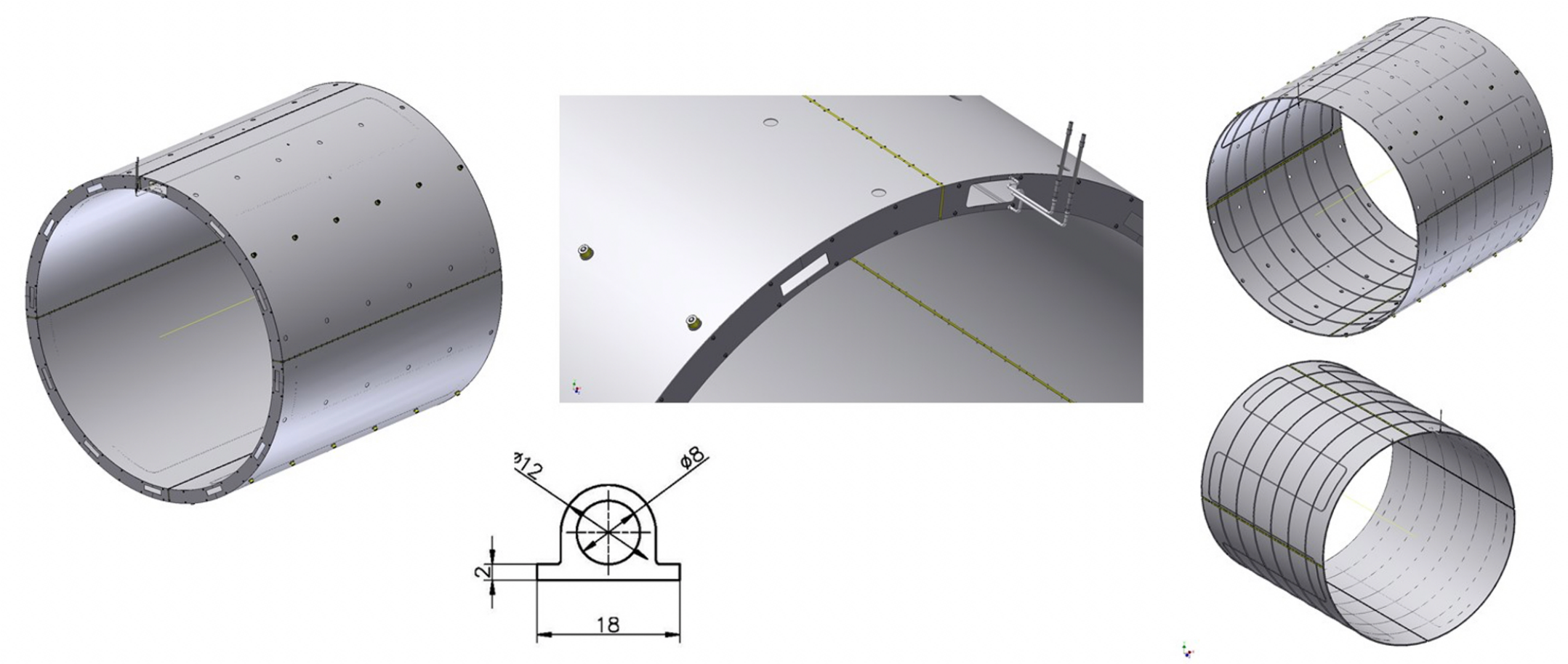}}
  \caption{ Cryostat thermal shield.  
	}
\label{lb_mag_21}
\end{figure}

\paragraph{Thermal shields}
Thermal shields of the solenoid surround the cold internal parts and are cooled by gaseous helium, which flows through the pipes of the heat exchangers. The design of the cooling circuit of the screen provides its cooling independently from the cold mass cooling. The helium flow leaves the liquefier at a temperature of 50 K and passes the heat exchangers of the thermal shields of the cryostat, chimney, and control dewar successively. After that, it returns to the liquefier at a temperature of 80 K. Heat gains to the thermal shields of the cryostat, chimney, and control dewar, as well as the temperature of the cooling gas at the inlet and outlet of the heat exchangers of each circuit element in the steady-state regime, are indicated in the flow scheme. The total length of the cooling tube of the heat exchangers is about 76.3 m (cryostat, control dewar, and chimney).

\subsubsection{Thermal shield of the cryostat }
The cryostat thermal shield surrounds the cold mass with the coil. It consists of two cylindrical sectors and two flanges. The heat exchanger of the cryostat thermal screen consists of Al alloy pipes with a diameter of 12 and thickness of 2 mm fixed on the outer and inner shells of the quadrants and connected by intermittent welding. The cooling pipes were optimized for 50 K helium gas, but it is possible also to use liquid nitrogen.

The shield envelope is split up into electrically separated parts (quadrants) (Fig. \ref{lb_mag_21}), to avoid damage and deformations of the screen, when the coil undergoes an emergency discharge. The quadrant shells are separated by gaps of about 14 mm, covered by sheets of insulating material G10. In addition, there are radial flange cuts of 10 mm. As can be seen from these pictures, there are four different types of quadrant configurations  (depending on the shape of the heat exchanger tube and on the positions of the attachments for the temperature interceptions of the cold mass suspension). 

The inner and outer shells of the octants are supported by the stiffening ribs (angled profile 30 $\times$ 20 mm) and interconnected by the flanges at the ends. Shells, flanges, and stiffening ribs of the octants are made of aluminum alloy AL 5083 (AMg5). The thickness of the sheets of the shells and flanges is 4 mm. The mass of the outer shell of a quadrant is about 60 kg, and the mass of the inner shell is about 52 kg. These masses are convenient for assembling. The total area of the screen surface of the cryostat is 78 m$^2$. The total mass of the cryostat screen is 880 kg.

The shield is positioned into the cryostat with the help of the  ball support units with a G10 ring. On the outer and inner surface of the shield,  30 layers of superinsulation will be placed. The total heat inflow through the shield supports is about 15 W. 
Appropriately oriented slots, made in the octant shells, are intended to reduce the lateral loads applied to the studs, caused by thermal shrinkage of the shield material.

\subsection{Electrical systems}
\subsubsection{	Electrical connections of the coils}
The SPD solenoid relies on the following electrical joints:
\begin{itemize}
\item two layer-to-layer joints in the upstream and downstream coil modules;
\item two coil-to-coil joints connecting the center coil module to the upstream and downstream coil modules;
\item two terminal joints connecting the solenoid to the bus-bars.
\end{itemize}

\begin{figure}[!h]
    \center{\includegraphics[width=0.8\linewidth]{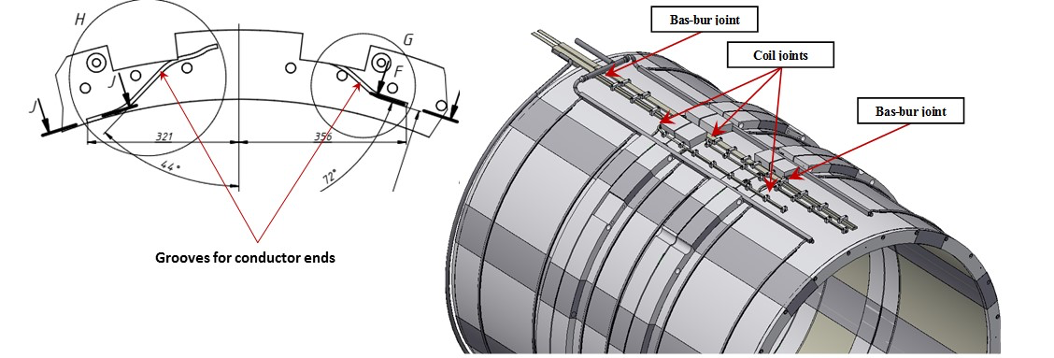}}
  \caption{ Position of coil-to-coil and terminal joints on the PANDA solenoid cold mass as an example.
	}
\label{lb_mag_24}
\end{figure}

\begin{figure}[!h]
  \begin{minipage}[ht]{0.5\linewidth}
    \center{\includegraphics[width=1\linewidth]{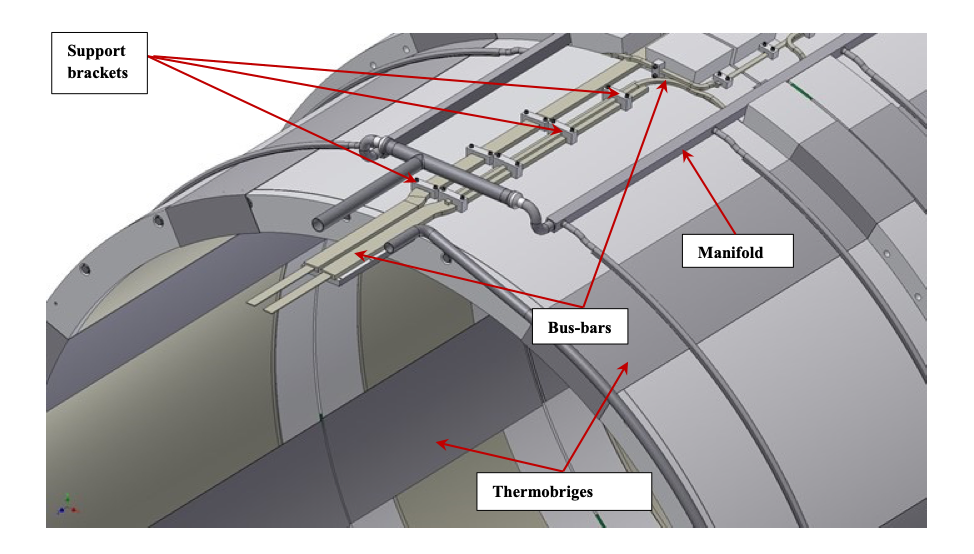} \\ (a)}
  \end{minipage}
  \hfill
  \begin{minipage}[ht]{0.5\linewidth}
    \center{\includegraphics[width=1\linewidth]{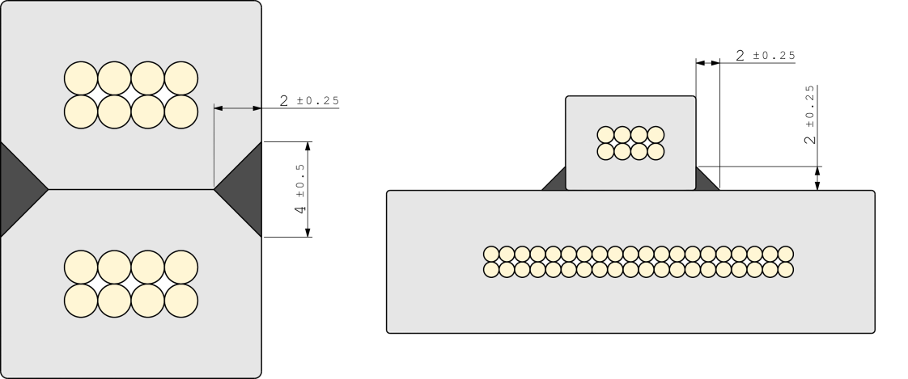} \\ (b)}
  \end{minipage} 
  \caption{(a) Upstream end of the cold mass showing the bus-bars and their support bracket. The area is crowded due to the presence of radial rod supports and a cooling line manifold connecting the thermosyphon circuit to the cold box. (b) Layer-to-layer and coil-to-coil (left) and terminal (right) joint layouts. }
\label{lb_mag_25}
\end{figure}

The routing of the coil-to-coil joints and terminal joints on the cold mass is displayed in Fig. \ref{lb_mag_24}. The joints are aligned with the magnetic field lines to avoid forces. The joints are clamped and glued to the cold mass, resulting in good thermal contact with the cold surface of the support cylinder. No thermalization of the joints to the cooling circuit is envisaged in this reference design, due to the expected low heat generation. A support bracket, shown in Fig. \ref{lb_mag_25}~(a), holds the weight of the bus-bars at the upstream end of the cold mass. The bracket is designed to sustain a total weight of 800 N, while the expected weight of the bus-bars is about 300 N.

Although the current routing of the bus-bars and joints is such as not to interfere with the other components installed on the cold mass, it is advisable to further optimize the layout of the upstream end of the cold mass. The present design of the area is crowded, due to the necessity to route both the electrical and cooling lines through the cryogenic chimney.

In Fig. \ref{lb_mag_25}~(a) it is possible to distinguish the grooves in the inner flanges, allowing routing of the conductor from the coil to the outer surface of the cold mass. The cross-section of the coil conductors is 8$\times$11 mm$^2$.

The area is crowded due to the presence of the radial rod supports and cooling line manifold, connecting the thermosyphon circuit to the cold box.

The design of the joints between the SPD solenoid conductors is essentially the same as applied in the ATLAS magnets. The configuration of the layer-to-layer and coil-to-coil joints is shown in Fig. \ref{lb_mag_25}~(b) left.  Terminal joints are instead illustrated in Fig. \ref{lb_mag_25}~(b) right. The use of the left-over conductor, developed for the ATLAS end-cap toroids, is intended in this reference design for the bus-bars of the SPD solenoid.

All conductor joints inside the cryostat are made without removing the aluminum stabilizer. The joints are made by TIG welding of the adjoining cables. Al-1050 is recommended as the filler material, as used for the conductor in the ATLAS toroids. The use of different aluminum alloys is also possible, depending on the expertise of the firm, building the SPD solenoid. After welding and cleaning, the conductor insulation is applied around the layer-to-layer joints. Ground insulation, instead, has to be wrapped around the coil-to-coil and terminal joints.

The electrical resistance of the joint shall not exceed 1.0 n$\Ohm$ at 4.2 K and 3 T. The joint length resulting in the envisaged resistance is determined experimentally by welding and testing the conductor samples. In this reference design, a minimum joint length of 700 mm is assumed for all joints (note that some joints in the drawings of this document may have erroneously a shorter length). Cooling and interruption of welding must be applied to the joints during welding, in order to limit the duration of the superconducting cables inside the conductors at temperatures above 600 K, to avoid critical current degradation. The critical current degradation in the superconducting cables due to the joint welding shall not exceed 5\%.

\subsubsection{Power supply}
\begin{figure}[!h]
  \begin{minipage}[ht]{0.6\linewidth}
    \center{\includegraphics[width=1\linewidth]{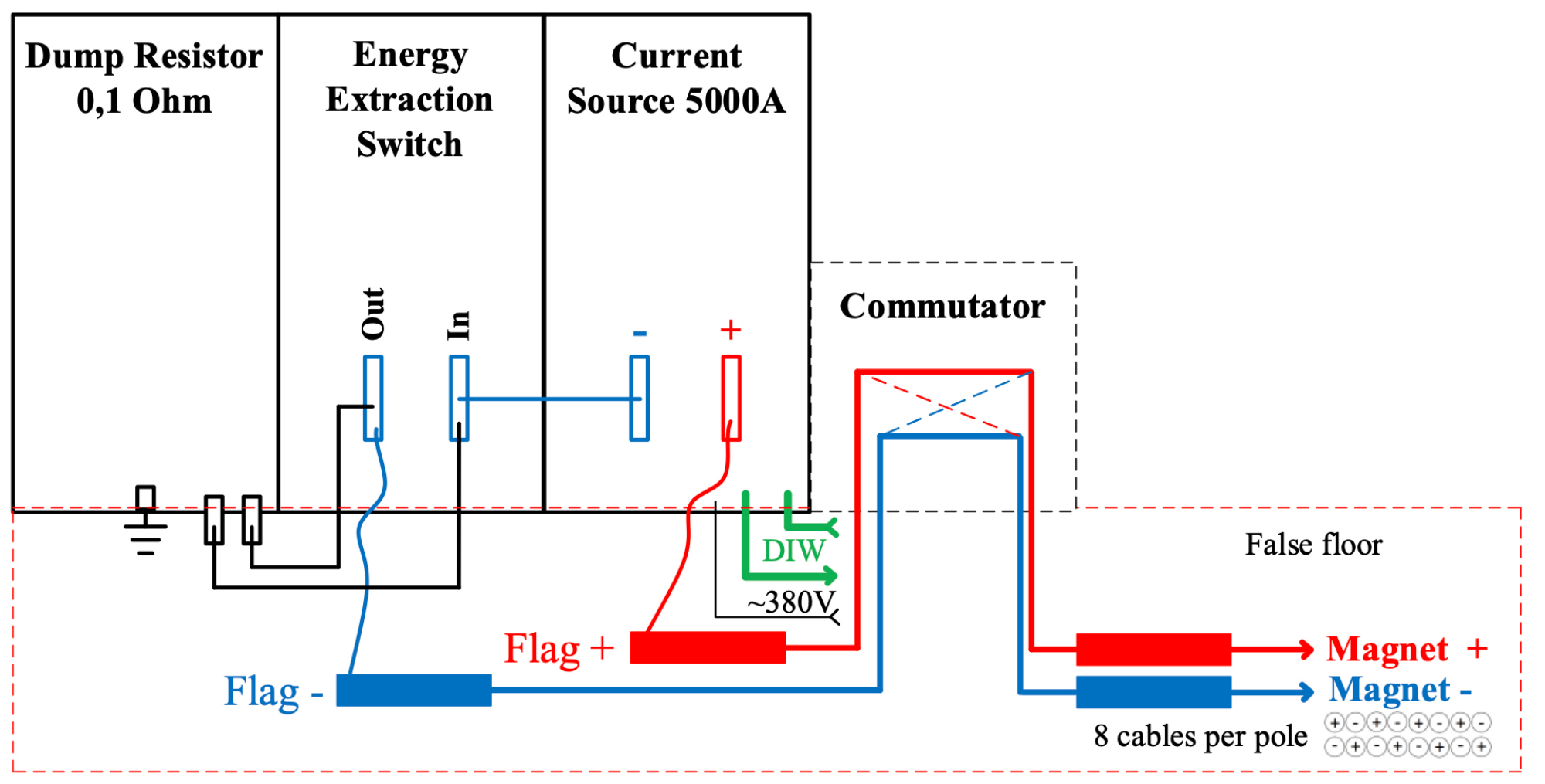} \\ (a)}
  \end{minipage}
  \hfill
  \begin{minipage}[ht]{0.4\linewidth}
    \center{\includegraphics[width=1\linewidth]{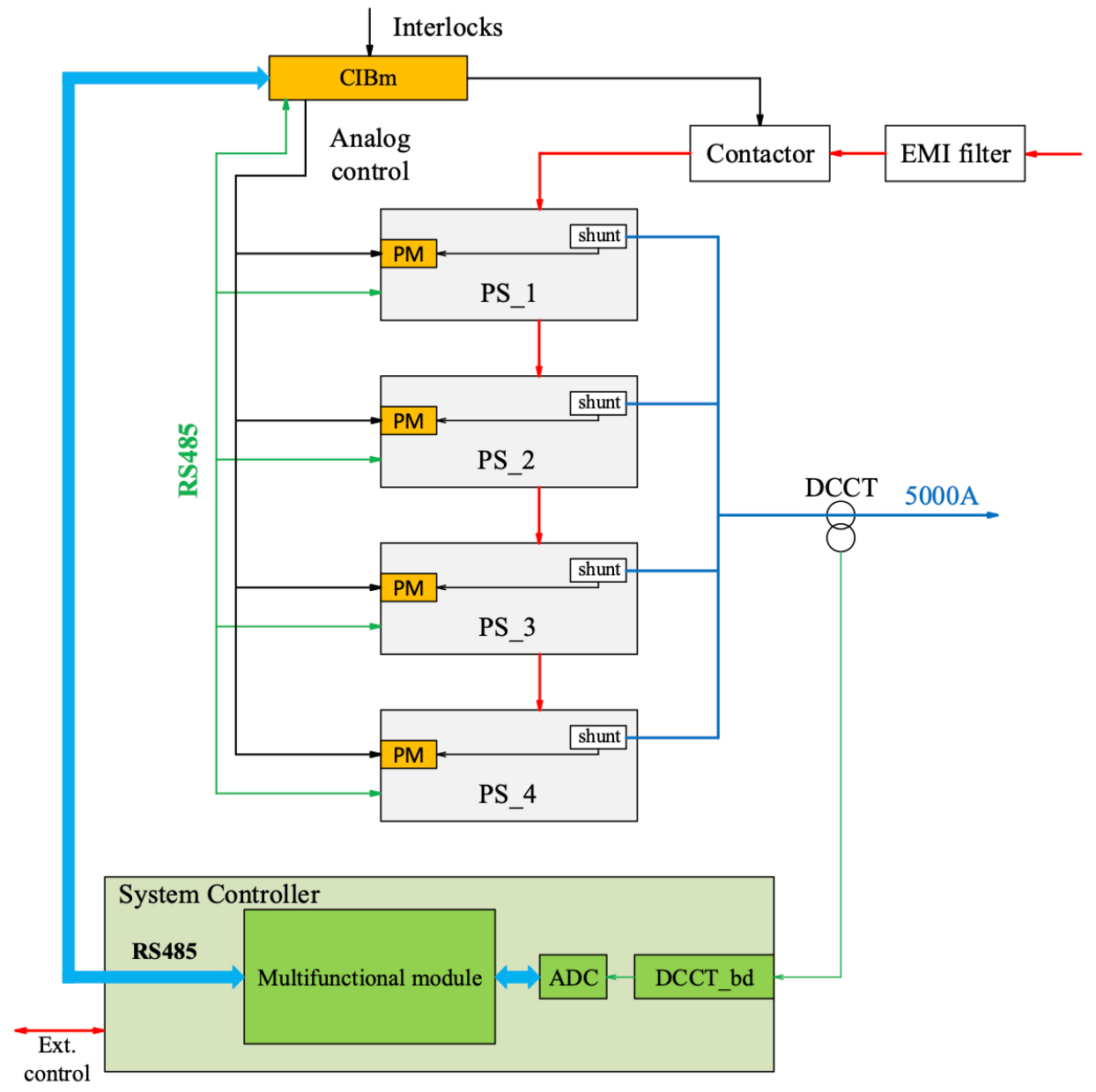} \\ (b)}
  \end{minipage} 
  \caption{(a) Racks and cabling diagram of the powering scheme. (b) General design of four paralleled power supplies with the external feedback system and DCCT.}
\label{lb_mag_26}
\end{figure}

The magnet is powered by a DC power supply, providing a nominal current of 5 kA. 
 The bus-bars connecting to the power supply are equipped with circuit breakers. The bus-bars resistance is negligible, compared to the circuit resistance during slow and fast dumps.
The fast discharge of the magnet is achieved by opening the circuit breakers, disconnecting the magnet from the power supply, and introducing the dump resistor to the circuit. In this case, the current circuit is looped via two freewheeling diodes, the dump resistor, and the magnet itself, so the energy will be extracted with the time constant $L / R_{dump}$. In case of slow energy extraction the circuit breakers remain closed but the current source is switched off. In this case, the current circuit is looped via two freewheeling diodes, circuit breakers, and the magnet itself, so the energy will be extracted with the time constant determined by the voltage drops over the cables and diodes. A discharge time in the case of slow energy extraction will take about 2000 s.
A block scheme of the electrical racks is shown in Figure \ref{lb_mag_26}~(a).

\begin{figure}[!h]
    \center{\includegraphics[width=0.8\linewidth]{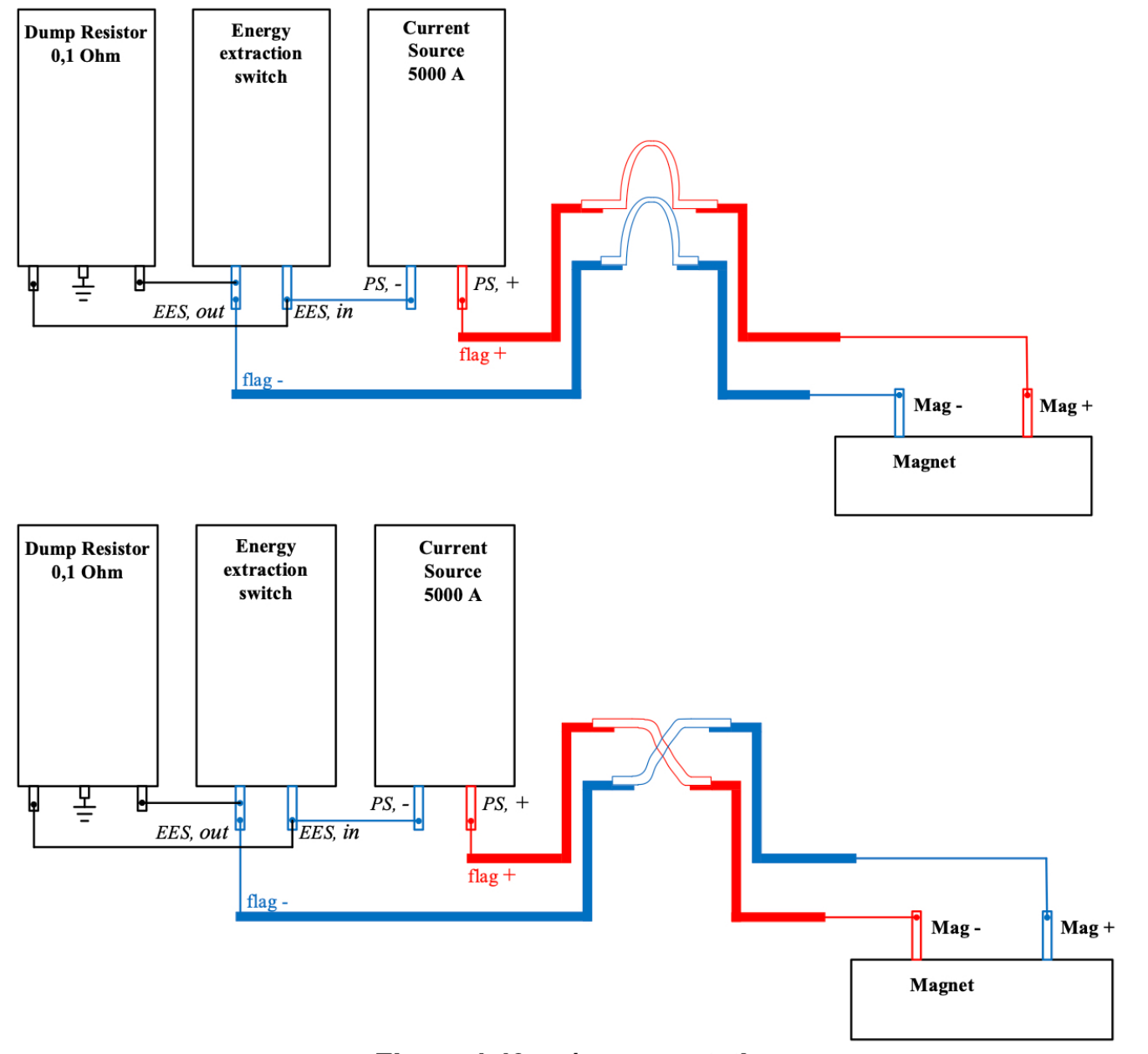}}
  \caption{ Polarity commutation scheme.
	}
\label{lb_mag_266}
\end{figure}

The power supply assembly design is shown In Fig. \ref{lb_mag_26}~(b). It is proposed to use the TDK-Lambda type current source. One current source has a value of the current up to 1500 A. Four power supplies are connected in parallel and looped with the external feedback system and DC current transformer (DCCT). 
\paragraph{Commutator.}
To reverse  polarity of the power supply which
might be necessary to eliminate systematics in the spin asymmeries measurement with polarized beams, a comutation scheme is  provided for like it is shown in Fig. \ref{lb_mag_266}.

\subsubsection{Quench protection}
\begin{figure}[!h]
  \begin{minipage}[ht]{0.6\linewidth}
    \center{\includegraphics[width=1\linewidth]{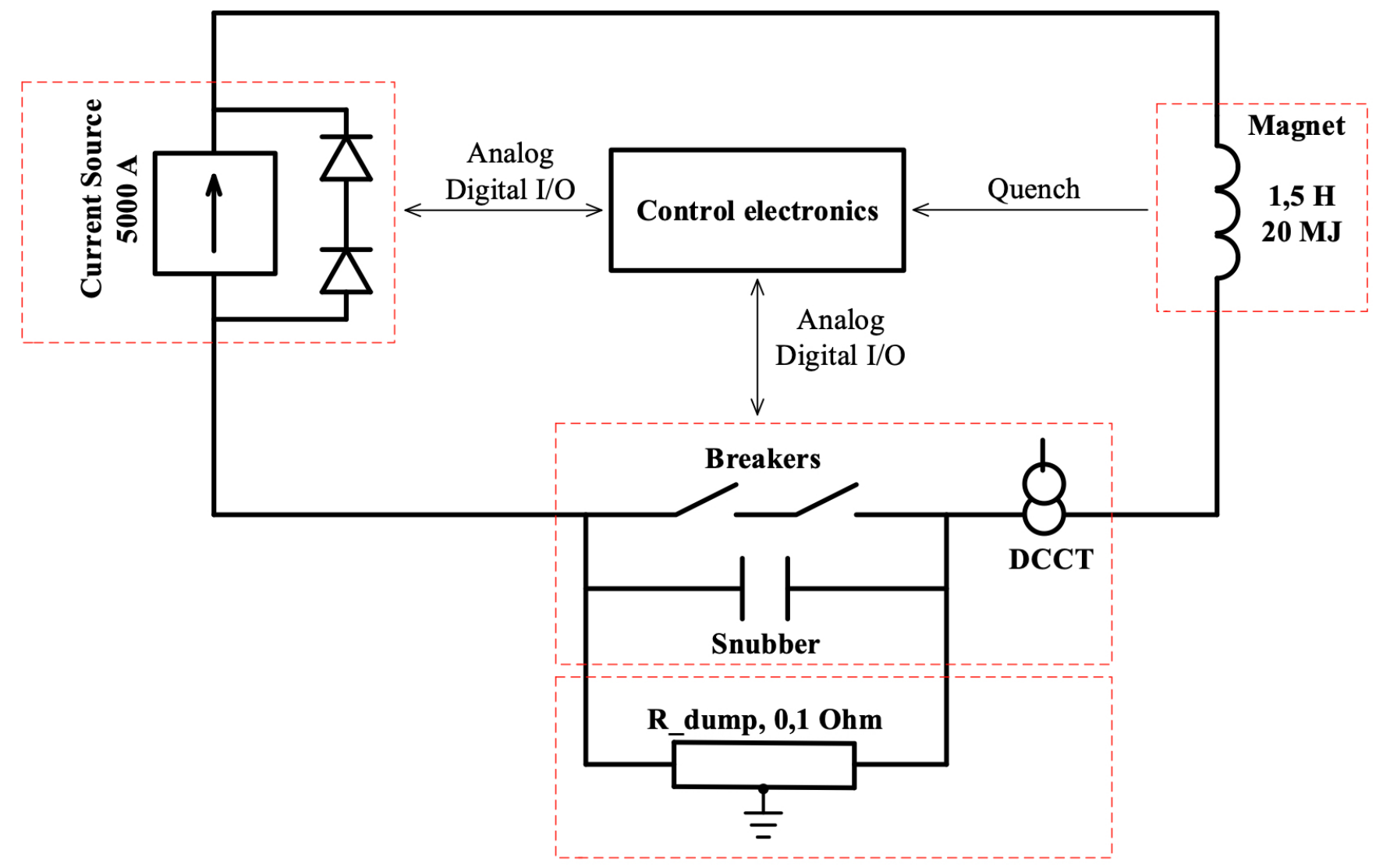} \\ (a)}
  \end{minipage}
  \hfill
  \begin{minipage}[ht]{0.4\linewidth}
    \center{\includegraphics[width=1\linewidth]{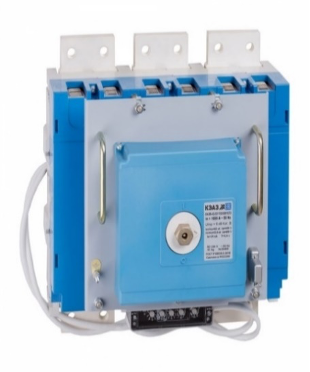} \\ (b)}
  \end{minipage} 
  \caption{ Electrical quench protection scheme of the solenoid (a). Electromechanical circuit breaker VA55-43  (b).
}
\label{lb_mag_27}
\end{figure}

A quench protection system (QPS) is designed to limit the peak temperature on the coil windings to 80 K and the peak voltage to ground to 500 V during any normally protected quench. The system relies on the monitoring of the resistive voltages, associated with the normal zone spread, and the quench development via the voltage taps, installed on the coil modules, as well as on the joints and bus-bars. A schematic view of the magnet electrical circuit is shown in Fig. \ref{lb_mag_27}~(a).

The quench protection system continuously monitors the magnet voltage during operation. In case a resistive voltage exceeding the threshold of 0.5 V is detected, the QPS generates the opening signal for the energy extraction system and interlock to the current source. In this case, the energy stored in the magnet will be extracted to the dump resistor with the time constant $L/R_{dump}$. The value of the resistor is set to 0,1 $\Ohm$ to compromise the maximal voltage during the energy extraction and the extraction rate. The middle point of the dump resistor is grounded to decrease twice the maximal voltage of the coils to ground.

\paragraph{Energy extraction system.}
The energy extraction system (EES) consists of:
\begin{itemize}
\item	two (for redundancy) electro-mechanical breakers;
\item	0.1 $\Ohm$ Dump Resistor ($R1_{dump}$ and $R2_{dump}$) with the grounded middle point;
\item	snubber capacitor to prevent the arc during the opening of the main contacts;
\item	equalizing resistors; 
\item	electronic boards forming the signals for internal coils and motor drives of the breakers;
\item	measurement board for the analog signals: total EES voltage, currents through the phases of the breakers.
\end{itemize}
Their main parameters are presented in Tab. \ref{EES}.
\begin{table}[htp]
\caption{Parameters of EES.}
\begin{center}
\begin{tabular}{|l|c|}
\hline
Paramener  & Value \\
\hline
Nominal current				&	5000 A \\
Maximal extracted energy			&	20.8 MJ\\
Current polarity				&	any\\
Inductance in a circuit			&	1.5 H\\
Dump resistor value			&	0.1$ \div$ 5\% $\Ohm$ \\
$R_dump$ maximal overtemperature	&	60 K\\
Time constant for the energy extraction	&	15 s \\
\hline
\end{tabular}
\end{center}
\label{EES}
\end{table}%

\paragraph{Breakers and snubber}
As an electro-mechanical switch VA55-43 (KEAZ, Russia) was chosen (Fig. \ref{lb_mag_27}(b)). It is three-phase AC/DC breaker with a phase current up to 2000A. The previous R\&D and twenty years of working experience with the breakers of that family have shown excellent reliability.

\begin{figure}[!h]
    \center{\includegraphics[width=0.8\linewidth]{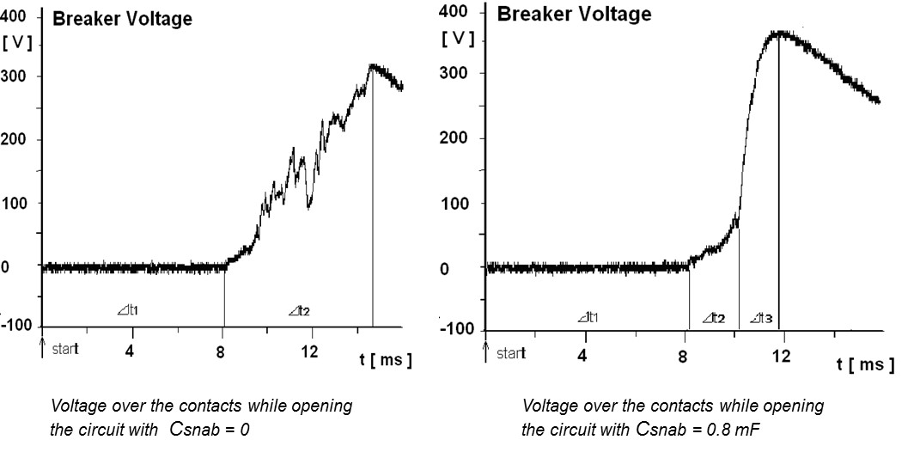}}
  \caption{ Use of the snubber against arc.
	}
\label{lb_mag_28}
\end{figure}

To avoid an arc between the power contacts the extra-low impedance capacitors are used as the snubber. One can see the difference between opening the DC circuit without snubber (Fig. \ref{lb_mag_28}~(left)) and with the snubber (Fig. \ref{lb_mag_28}~(right)). Here: $\Delta t_1$ is the mechanical time delay of the breaker, $\Delta t_2$ is the arcing time interval, and $\Delta t_3$ is charging the snubber capacitor time interval. Therefore, using the snubber significantly decreases the opening time, minimizes the arc, and prevents the burning damage to the main contacts. To equalize the phase currents the equalizing resistors (20 $\mu \Ohm$) are introduced to each phase. 

The total energy $E$ stored in the magnet is
\begin{equation}
E=\frac{1}{2\mu_0} B_0^2\times V = 20.8 ~MJ.
\end{equation}
With the presented layout of the quench protection system, 58\% of the initial magnetic field energy is dissipated in the dump resistor, 26\% as eddy current loss in the coils and casing, and the remaining 16\% in the diode unit.

The results of the quench analysis show that proper operation of the QPS limits the peak coil temperature and voltage well below the threshold values. The maximum temperature reached by the coils is here re-assessed in the case of failure of the QPS, which results in the entire stored energy being dissipated in the winding. Under the assumption that the quench-back effect of the support cylinder causes a rapid propagation of the normal zone to the entire coil without extracting energy from the system, the magnet would heat up to 80 K. Even in the unrealistic case that the energy is entirely dumped in only one of the sub-coils, the temperature would reach 112 K and 161 K for the upstream/downstream and center coils, respectively. Although unwanted from the cooling point of view, these temperature values do not pose risks to the integrity of the solenoid.

\subsection{Vacuum system of the SPD magnet}
\begin{figure}[!h]
  \begin{minipage}[ht]{0.45\linewidth}
    \center{\includegraphics[width=1\linewidth]{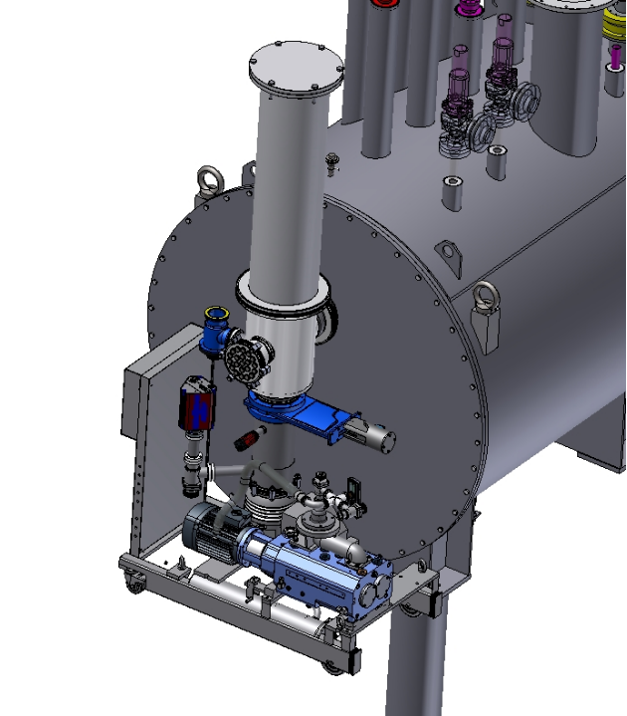} \\ (a)}
  \end{minipage}
  \hfill
  \begin{minipage}[ht]{0.55\linewidth}
    \center{\includegraphics[width=1\linewidth]{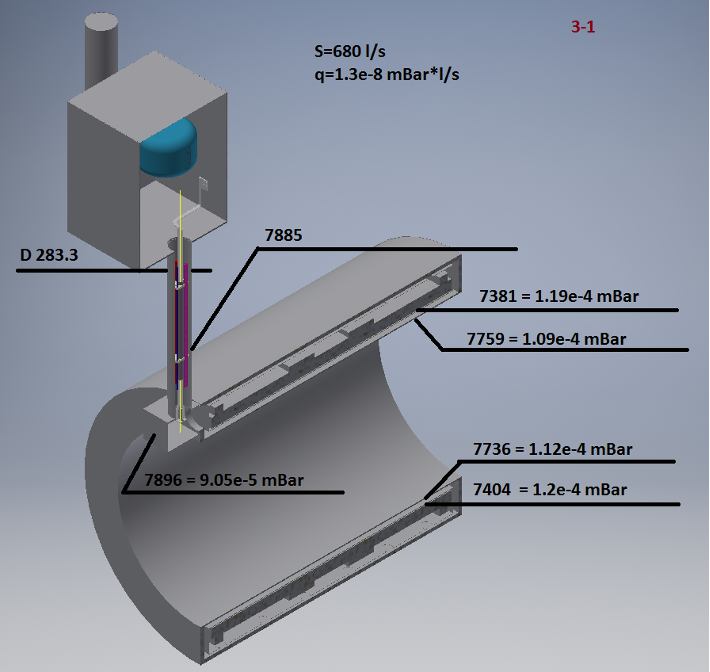} \\ (b)}
  \end{minipage} 
  \caption{(a) Vacuum equipment. (b) Distribution of the pressure inside of the vacuum system.
}
\label{lb_mag_29}
\end{figure}

The control dewar, chimney, and cryostat with superconducting coil have a common vacuum volume. A pumping station with rotary and turbo-molecular pumps performs the vacuum pumping. The pressure to be obtained by vacuum pumping before cooling and cryostatic temperature regulation is 10$^{-5}$ mbar. Further pumping will be performed in the process of cooling by cryogenic panels, mounted on the outer surface of the support cylinder. The vacuum system is shown in Fig. \ref{lb_mag_29}~(a). The pumping station is placed on the upper platform of the magnet. Two pressure sensors are fitted on the vacuum casing of the control dewar to check the pressure. 
For pumping of the vacuum volumes of the control dewar, chimney, and superconducting coil cryostat up to $10^{-1}$ mbar with a safety factor of 2, a rotary pump could be employed with a pumping speed of 15$\div$18 l/s. For pumping down to 10$^{-5}$ mbar, a turbo pump with a pumping speed of 500$\div$680 l/s can be chosen. The distribution of the pressure inside the vacuum system is presented in Figure \ref{lb_mag_29}~(b). The calculation was done by software ''MOLFLOW+ 2.6.62''.  

The cryostat and control dewar of the SPD solenoid will be built according to AD2000. The worst-case scenario concerning the safety of the cryostat would arise if a leak evolved in the part of the liquid helium (LHe) cooling circuit that runs through the vacuum vessel. Only in this case, the pressure balance could actually be reversed, and the pressure inside the cryostat could exceed the atmospheric pressure.

The cryostat vacuum vessel, control dewar, and all connections have to be verified to resist up to a pressure of 1.50 bar (absolute pressure) from the inside. This provides a good safety margin, as the coolant circuit will be working at an absolute pressure of 1.30 bar, and we foresee safety flange DN250. For the outer cryostat shell, the stability against a pressure reverse is easily guaranteed, since it is designed to work with 1.0 bar from the outside. This is naturally different for the inner wall of the cryostat, which is designed to keep, under normal operation, the atmospheric pressure from inside. However, the inner wall of the cryostat is designed to keep safely the full load of the detector, mainly the weight of the ECal with minimum deformation. The extra thickness needed to fulfill this requirement is more than enough to guarantee  against buckling of the inner wall of the cryostat in case of a pressure reversal, due to a leak in the helium circuit. 

A system of safety valves has been foreseen to prevent an excessive pressure rise in the vacuum vessel of the cryostat, control dewar, and chimney. A relief flange opening at 0.3 bar is foreseen. In case of LHe vaporization, the relief valve would open, venting the helium gas to the atmosphere and preventing damage to the cryostat vessel. 
\begin{table}[]
\centering
\caption{Sensor instrumentation summary for the cold mass of the SPD solenoid.}
\label{tab:mag_8}
\begin{tabular}{|l|c|c|c|c|}
\hline
\multicolumn{1}{c}{\textbf{Function}} & \multicolumn{1}{c}{\textbf{Location}} & \multicolumn{1}{c}{\textbf{Quantity}} & \multicolumn{1}{c}{\textbf{Sensor type}} & \multicolumn{1}{c}{\textbf{Notes}} \\
\hline
\multirow{6}{*}{Quench detection}     & Coil  modules                         & 6                                     &                                          & 3 Primary/Redundant pairs          \\
                                      & Coil-to-coil joints                   & 4                                     &                                          & 2 Primary/Redundant pairs          \\
                                      & Terminal joints                       & 4                                     & Voltage                                  & 2 Primary/Redundant pairs          \\
                                      & Bus-bar joints                        & 4                                     & taps                                     & 2 Primary/Redundant pairs          \\
                                      & Vertical bus-bars                     & 4                                     &                                          & 2 Primary/Redundant pairs          \\
                                      & Current leads                         & 4                                     &                                          & 2 Primary/Redundant pairs          \\
\hline                                      
\multirow{3}{*}{Cryogenic monitoring} & Coil modules                          & 6                                     & Cryogenic                                & 1 Top/Bottom                       \\
                                      & Bus-bars                              & 4                                     & temperature                              & 2 Primary/Redundant sensors        \\
                                      & Cryogenic lines                       & 2                                     & sensor                                   & 1 Supply/Return line   \\
\hline                                                 
\end{tabular}
\end{table}
\begin{table}[]
\centering
\caption{List of purchasing equipment for control dewar cryogenic system.}
\label{tab:mag_9}
\begin{tabular}{|c|l|l|}
\hline
\multicolumn{1}{c}{No.} & \multicolumn{1}{c}{Type device/ instrument} & \multicolumn{1}{c}{Producer}                                                       \\
\hline
1                       & Cold Valves                                 & \begin{tabular}[c]{@{}l@{}}Velan\\ Weka\\ Stoehr\\ Flowserve\end{tabular}          \\
2                       & Relief valves                               & Leser                                                                              \\
3                       & Warm valves                                 & \begin{tabular}[c]{@{}l@{}}Velan\\ Weka\\ Stoehr\\ Flowserve\end{tabular}          \\
4                       & Pressure transmitters                       & Emerson/ Siemens/ WIKA                                                             \\
5                       & Pressure indicators                         & Swegelok                                                                           \\
6                       & T platinum sensors                          & Heraeus                                                                            \\
7                       & Low-temperature thermometers                       & \begin{tabular}[c]{@{}l@{}}Business Logistix Ltd\\ OXFORD/ JINR/ BINP\end{tabular} \\
8                       & Warm flow meters                            & BROOKS Smart series/ BRONKHORST                                                    \\
9                       & Heaters                                     & many companies                                                                     \\
10                      & Level meters                                & AMI                                                                                \\
11                      & Connectors, 2 bars                          & Fischer                                                                            \\
12                      & Connectors, 2 bars for heaters              & Fischer                                                                            \\
13                      & Connectors, 40 bars                         & Serie\_Hermetique\_Jager                                                             \\
14                      & Pumps, preliminary and TMP                  & Edwards/ Laybold or equal                                                          \\
15                      & Gate valve DN150                            & VAT                                                                                \\
16                      & Vacuum angle valves                         & many companies                                                                     \\
17                      & Fittings                                    & Swegelok                                                                           \\
18                      & MLI                                         & Groupe HUTCHINSON ex- JEHIER/ RUAG                                                 \\
19                      & Ball supports                               & Bosch Rexroth            \\                                                         
\hline
\end{tabular}
\end{table}

\subsection{Cryogenics safety}
The different parts composing the cryostat should be designed, fabricated, and inspected according to the intent of the 
actual norms including corresponding documentation, certificates, and used standards. 

The choice of indirect cooling of the solenoid will greatly reduce the amount of liquid helium in the cryogenic system. The pipes, guaranteeing the coil refrigeration, will contain $\sim$10 l of refrigerant. This includes the two manifolds placed on the top and on the bottom of the cold mass for the thermosyphon circulation. Another  $\sim$300 l of liquid helium will be stored in the helium vessel, housed in the control dewar.

The operating pressure of the SPD coil cryogenic system will be 130 kPa (absolute pressure), corresponding to a coolant temperature of 4.5 K during cooldown and warmup. The two phases of liquid helium, used by the cooling system, will be fed to the helium vessel of the control dewar through a transfer line from the helium liquefier during normal operation. Liquid helium from this vessel will give the coil cryogenic system the capability to operate for up to 60 min. for safe de-energization of the coil.

The transfer line will decouple the refrigerator from the solenoid cryogenics, in case of refrigerator failure. In the worst case, the gas would be transferred into the helium vessel at the full pressure of the running compressor of the liquefier ($\sim$16 bar). In this case, it would be vented to the air through the relief valve and safety flange DN250. Furthermore, it will help that the liquefier and the control dewar will be connected through a long transfer line. This line will hydraulically decouple them in the scenario of a liquefier failure with the 16-bar helium gas, at the temperature of the last heat exchanger, transferred into the system.

As an additional safety measure, all the cooling loops and distribution lines inside the cryostat (including the intermediate radiation shields) are designed for a nominal operating pressure of 19 bar. The helium vessel of the control dewar, operating at 1.3 bar, is designed for an overpressure of 19 bar and will be equipped with a relief valve, venting the vessel into the atmosphere for the pressure p = 1.8 bar in case of helium boiling off, due to a quench or vacuum failure.

\subsection{Cold mass instrumentation}
The instrumentation of the cold mass of the SPD solenoid, taken as an example, falls under two groups: quench protection and cryogenic monitoring. Table \ref{tab:mag_8} provides a summary of the sensors to be installed.

The quench protection system monitors the coil modules and superconducting bus-bars for resistance growth and provides a trigger for the current breaker to isolate the power converter from the solenoid, in order to limit the rise of conductor temperature, due to Joule heating. The quench detection system relies on voltage taps. Voltage taps are critical sensors and, therefore, must have redundancy. The voltage taps, installed on the joints, also provide the capability to monitor the joint resistance. Similarly, voltage taps are installed on the current leads to monitor the lead resistance.

Cryogenic monitoring is based on temperature sensors to control cooldown, warmup, steady state operation, and changes in response to a heat load, such as a quench. Temperature sensors are installed on the cold mass and bus-bars, as well as on the supply and return cryogenic lines.

Appropriate instruments and tools are required to control cryogenic processes. There are cryogenic cold and warm control valves with transmitters and actuators, flowmeters, pressure transmitters and manometers, thermosensors, level meters, and related accessories. The minimal list of needed instruments and tools is presented in Table \ref{tab:mag_9}.

\section{Cost estimate}
An estimate of different contributions to the cost of the magnetic system  is presented in the Table \ref{Mag_cost}. To the indicated  values, we must add the cost of the system of detailed measurement and control of the magnetic field of 600 k\$. So, the total cost of the SPD magnet is 9.4 M\$.

\begin{table}[htp]
\caption{Cost estimate.}
\begin{center}
\begin{tabular}{|c|c|c|}
\hline
Milestone & Description                 &   kEuro   \\
\hline
1 & Project management and testing  &  \bf{2070} \\
	& Technical project                      &    800 \\
	& 	Plan review                         &    170 \\
	& 	Preliminary design review  &    150  \\
	& 	SAT full solenoid                &     950  \\
\hline
2	& Conductor                                                    &  \bf{1970} \\
	& 	Contract conductor with ext. firm           &    460        \\
	& 	FDR conductor                                        &   270  \\
	& 	Start of conductor production at ext. firm&   880   \\ 
	& 	FAT conductor                                         &   360  \\
\hline
3	& 	Cryostat and cold mass      &  \bf{1260}  \\
	& 	Cryostat design                   &   160 \\
	& 	FDR cryostat                      &   160 \\
	& 	Procurement  components  &   410 \\
	& 	FAT cryostat                        &   420 \\
	& 	SAT cryostat                       &    110 \\
\hline	
4	& 	Cryogenics                            & \bf{865} \\
	& 	Control dewar design                         & 100 \\
	& 	Control dewar vacuum equipment     &  150 \\
	& 	FDR                           &  195 \\
	& 	FAT                             &  320 \\
	& 	SAT                           &  100 \\
\hline
5	& 	Electrical components               &  \bf{946}\\
	& 	Contract el. components           &  140 \\
	& 	FDR el. components                  &  200\\
	& 	Procurement of el. components&  440 \\
	& 	FAT el. components                   &  166 \\
\hline	
6	& 	Magnet alarm safety system	              &   \bf{140}  \\
	& 	FDR safety system                                 &   40 \\
	& 	Tender of components for safety system&   80 \\
	& 	FAT safety system                                  &   20 \\
\hline	
7	& 	Coil winding                                    &   \bf{717} \\
	& 	Design coil winding                        &    50 \\
	& 	Tooling design                                &    40 \\
	& 	FDR coil                                         &    90 \\
	& 	Procurement of coil components   &   220 \\
	& 	FAT coil winding                             &   252 \\
	& 	Cold mass integration                    &    65 \\
\hline
	& 	Total  (including installation and commissioning) &   \bf{7968} \\
\hline

\end{tabular}
\end{center}
\label{Mag_cost}
\end{table}%

\section{Cryogenic system}
The magnetic field is formed by a superconducting (SC) coil. The NbTi superconductor, which has an operating temperature of 4.5 to 9.0 K, has been selected for the SC coil of the detector SPD. Keeping the SC coil in the SC state implies the use of a non-standard cryogenic system operating at 4.5 K. The working substance is liquid helium.

The SC coil presented by a working group from the Institute of Nuclear Physics (INP) in Novosibirsk was selected for the SPD magnetic system project. The concept of production SC coil involves the use of a Rutherford SC cable. The thermosiphon method is used to cool this SC coil, which means the use of a cryogenic liquefaction plant.


The cryogenic helium plant with a control cryostat will be placed on a technical platform, which is moved together with the SPD detector and is located on top of the detector. The cryogenic system consists of three main subsystems: helium and nitrogen systems, as well as auxiliary systems. The helium system includes a cryogenic plant of liquefaction type, high- and low-pressure helium supply pipelines, as well as a cryogenic flexible pipeline for liquid and vaporous helium between the plant and a control cryostat. The nitrogen system consists of two cryogenic tanks for liquid nitrogen storage, one cryogenic tank each with liquid argon and carbon dioxide, cryogenic pipelines, and a nitrogen vapor pipeline. Auxiliary systems include  vacuum, measuring, cryogenic monitoring and control system, gas supply system, and water-cooling system.

\subsection{Helium cryogenic system  \label{sec:Cryogenic_He_system}}
\subsubsection{Cryogenic plant}
The SC coil requires the use of a cryogenic helium liquefaction plant, which allows to cooling of the SC coil and heat shields with helium vapor (50 K), the maximum cooling rate is 1 K/h. This concept uses the thermosiphon method of SC coil cooling. A similar cooling principle and SC coil design is used in such projects of MPD, PANDA, etc. detectors. The main operating parameters of the cryogenic plant are summarized in Tab \ref{tab:cryo_2}

\begin{table}[]
\centering
\caption{Main operating parameters of the cryogenic plant.}
\label{tab:cryo_2}
\begin{tabular}{|l|c|}
\hline 
 Operating parameters                            & Value                                                                \\
\hline 
 Cooling capacity                                & 100$\div$130 l/h                                                             \\
 Mass flow rate                                   &  6$\div$8  g/s          \\
 Temperature of outlet flow from the SPD         & 4.3 K (1.05 bar)                                                    \\
 Temperature of inlet flow from the SPD &          4.4 K (1.15 bar)                                     \\
 Hydraulic resistance of the SC coil             & 0.1 bar                                                             \\
 Cold mass                                     & 5000 kg                                                             \\
Maximum pressure in coil                        & 5 MPa                                                               \\
Maximum heat load    (for He 4.5 K)                           & 60$\div$80 W                                                                \\
Maximum heat load    (for He 50 K)                           & 200$\div$300 W                                                                \\
 Equipment requirement                           & Maximum reliability,  \\
     & energy efficiency, \\
      & compactness, automatic mode \\
 Interval of repair/regulatory work of the plant & Not more than once a year      \\         
\hline                    
\end{tabular}
\end{table}

The plant must have an additional outlet of 50 K helium flow with a smooth adjustable temperature from 297 K to 50 K, to cool the heat shield of the detector SC coil. The heat load of the thermal shield is about 300 W. 

The plant includes flexible lines to connect to the control cryostat for the transportation of liquid and gaseous helium flows. The maximum heat load for flexible lines of helium communications is 3 W/m. All cryogenic flexible lines and pipelines are vacuum insulated, condensate formation on external surfaces of communications is not allowed. 

The power consumption of the cryogenic plant is 2 kW.

\begin{figure}[!h]
  \begin{minipage}[ht]{1.\linewidth}
    \center{\includegraphics[width=1\linewidth]{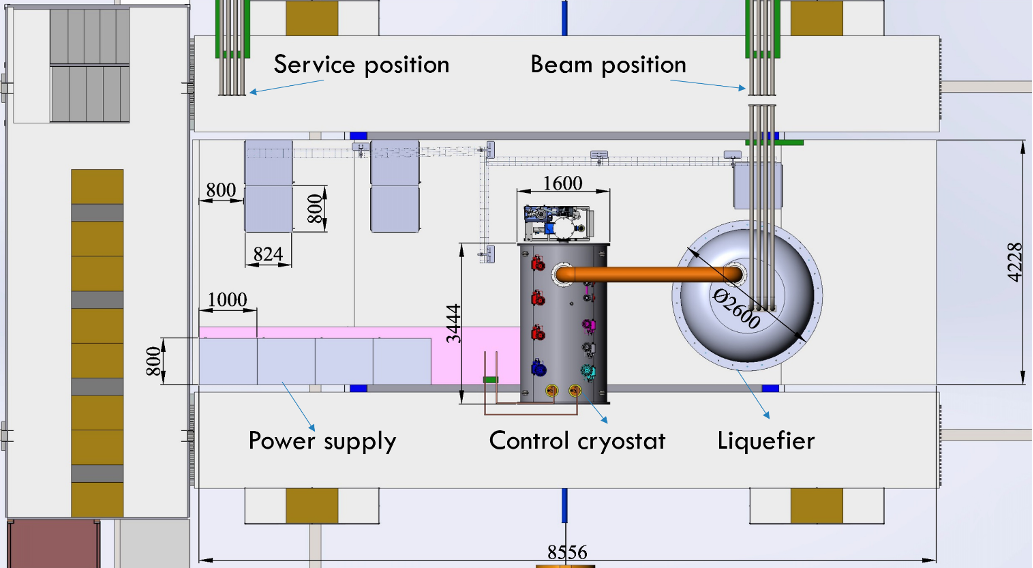}}
  \end{minipage}
  \caption{Location of main elements of the SPD cryogenic system.
	}
\label{lb_cryo_022}
\end{figure}

Figure \ref{lb_cryo_022} shows the location of the main equipment of the cryogenic system: the liquefier, which is connected by a pipeline to the control cryostat, power supply cabinets, just above there are cabinets for control and measurement of various parameters of the solenoid and liquefier operation. The equipment is concentrated in the center of the technical platform and does not occupy the side platforms.  This is due to the necessity to provide mounting and dismounting of the detector doors, as well as repair works on the magnet with its dismantling. The pipelines have two connection points with the rigid pipelines of the technical platform for gaseous helium and liquid nitrogen.

\subsubsection{Helium pipelines}

\begin{figure}[!h]
  \begin{minipage}[h]{1.\linewidth}
    \center{\includegraphics[width=1\linewidth]{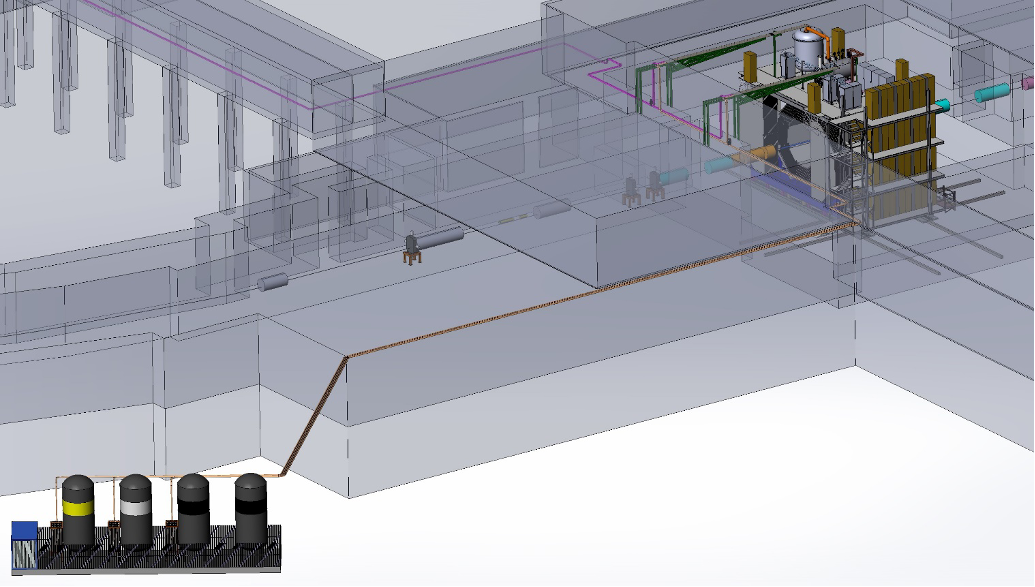}}
  \end{minipage}
  \caption{Tracing pipelines of high- and low-pressure helium flows.
	}
\label{lb_cryo_02}
\end{figure}

The cryogenic system requires pipeline installation along the route from the central cryogenic station to the location of the SPD detector. Figure \ref{lb_cryo_02} shows the routing of the high- and low-pressure helium flows. The total length of the traces is about 220 meters. Pipelines of high-pressure helium flow have a 60 mm diameter and low-pressure helium flow has a 100 mm diameter. The material of the pipelines is stainless steel 12X18H10T or analog. Tracing has the following route. From Room 177 the route goes piping with other communications in one route to Room 109/1 (about 120 meters of traces). From Room 109/1 the route goes to the courtyard of Building 17 and on the ramp comes to Room 140/2 (the SPD experimental site, see Chapter \ref{lbl:servises}).

\begin{figure}[!h]
  \begin{minipage}[h]{1.\linewidth}
    \center{\includegraphics[width=0.8\linewidth]{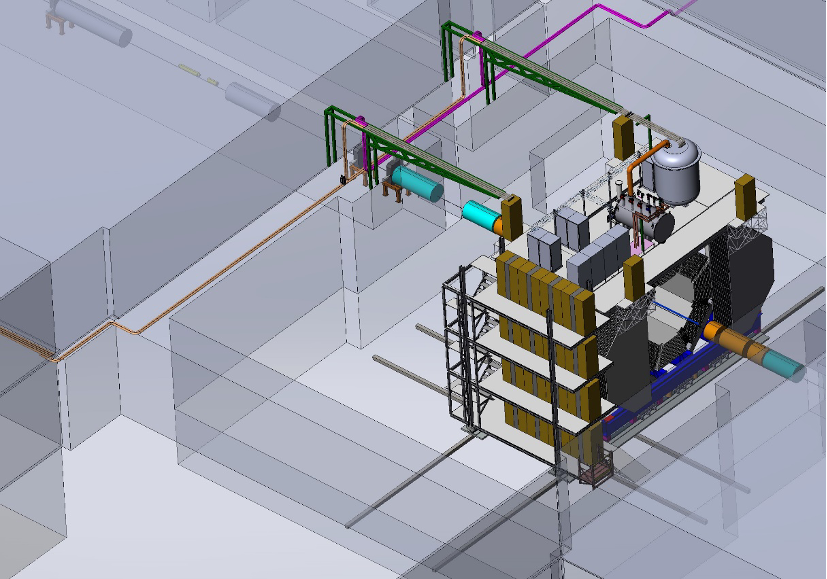} \\ (a)}
  \end{minipage}
  \hfill
  \begin{minipage}[h]{1.\linewidth}
    \center{\includegraphics[width=0.8\linewidth]{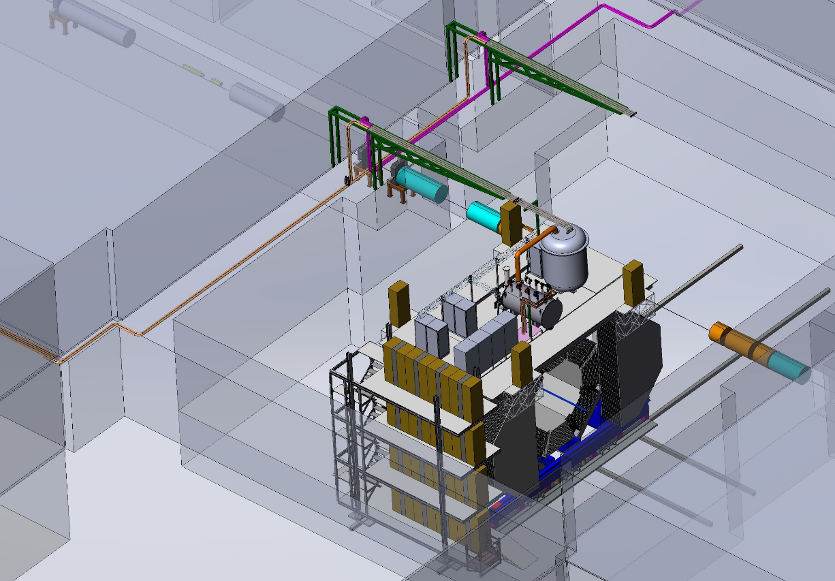} \\ (b)}
  \end{minipage} 
  \caption{Rigid lines in the room 140/2 when  SPD is in the ''beam'' (a) and ''service'' (b) position.
	}
\label{lb_cryo_3}
\end{figure}

The pipeline section from the entrance to Room 140/2 to the cryogenic plant is made of flexible lines. Figure \ref{lb_cryo_3} shows the location of the cryogenic pipelines and two positions ''beam'' and ''service''. The pipelines have a system of safety devices, to provide safe operation.

\subsection{	Nitrogen system \label{sec:Cryogenic_N2_system}}
\begin{figure}[!h]
  \begin{minipage}[ht]{1.\linewidth}
    \center{\includegraphics[width=0.8\linewidth]{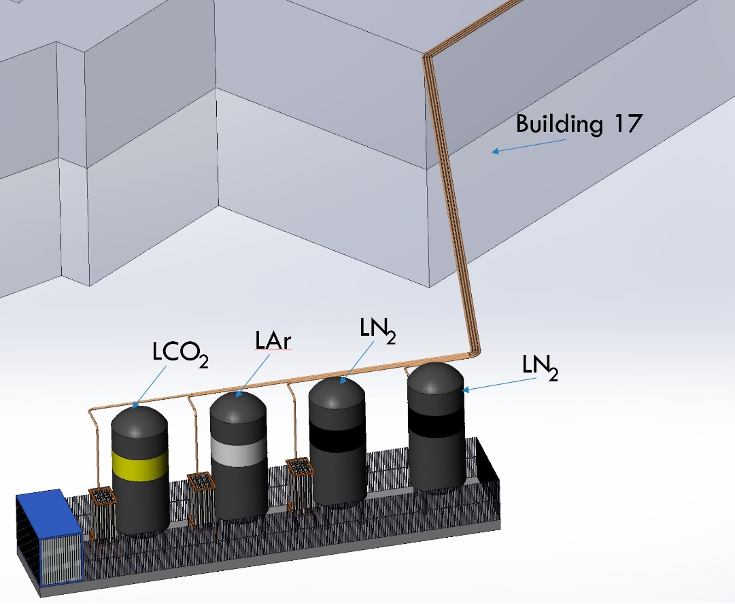}}
  \end{minipage}
  \caption{Special foundation with four cryogenic tanks.
	}
\label{lb_cryo_04}
\end{figure}

To provide operation of the cryogenic helium plant it is necessary to create a nitrogen system. The nitrogen system consists of four cryogenic tanks of 10 m$^{3}$ volume each: two to store liquid nitrogen and one each for liquid argon and liquid carbon dioxide, as well as cryogenic pipelines (Fig. \ref{lb_cryo_04}). 
Their main technical parameters are presented in Table \ref{tab:cryo_3}.

\begin{figure}[!h]
  \begin{minipage}[ht]{1.\linewidth}
    \center{\includegraphics[width=0.8\linewidth]{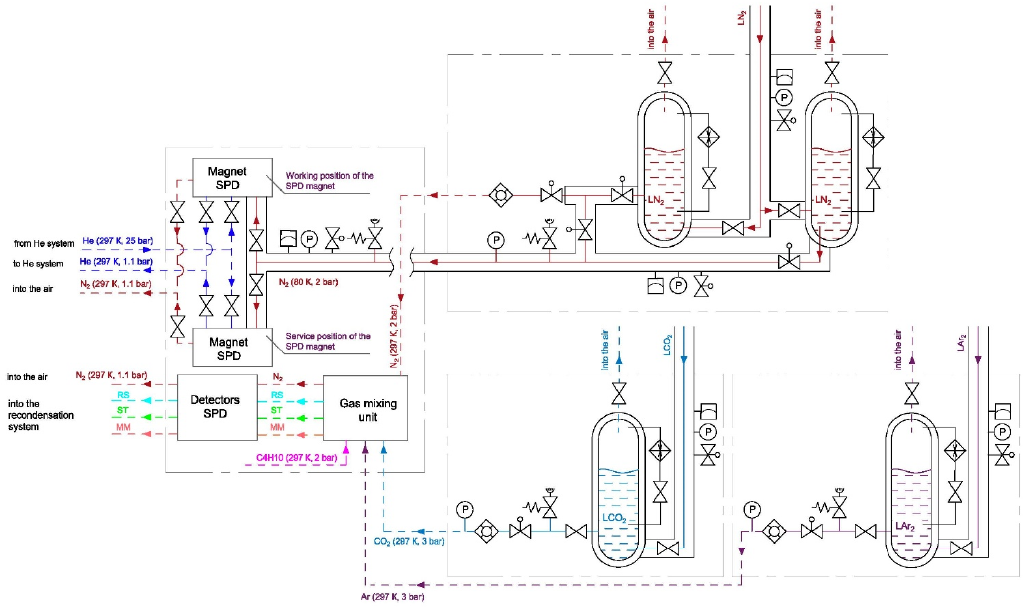}}
  \end{minipage}
  \caption{Technological scheme of the nitrogen system.
	}
\label{lb_cryo_05}
\end{figure}

The cryogenic storage tanks are located on a special foundation in Building 17 on the side of Room 140/1 (the SPD production site, see Chapter \ref{lbl:servises}). 
Liquid nitrogen is supplied through a cryogenic pipeline by raising the pressure to 4.0 bar in the cryogenic storage tank. When the minimum liquid nitrogen level in the first cryogenic storage tank is reached, the control system automatically switches the liquid nitrogen supply from the first cryogenic storage tank to the second cryogenic storage tank.  The cryogenic pipeline is vacuum insulated and condensate formation on external surfaces of communications is not allowed. The total length of the route is about 150 meters. The pipeline is made on a tube-in-tube type, the inner pipeline is 30 mm in diameter, and the outer pipeline is 70 mm in diameter. The material of the pipeline is stainless steel 12X18H10T or analog. The route has the following: from the tanks, the pipeline rises to a height of 6 meters for the passage over the road, then along the wall of Building 17, Room 140/1, enter the room and bring it to the cryogenic plant. The route has a system of safety devices, to ensure safe operation. Figure \ref{lb_cryo_05} shows the technological scheme of the nitrogen system.

The discharge of gaseous nitrogen is carried out according to the principle of the shortest route with exit to the atmosphere. The discharge pipelines's diameter is 70 mm.  The energy consumption of the system is 2 kW.

\begin{table}[b]
\centering
\caption{Main technical parameters of the cryogenic storage tank.}
\label{tab:cryo_3}
\begin{tabular}{|l|c|}
\hline
Operating parameters                                                                       & Value                        \\
\hline
Overall volume                                                                             & 10 000 l                    \\
Maximum daily evaporation rate depending & 0.17 \%                      \\
     on pressure 100 kPa and ambient temperature15 $^{\circ}$ C       & \\
Maximum outlet flow rate (nitrogen)                                                        & 600 Nm$^3$ / h                 \\
Maximum operating pressure                                                                 & 1.8 MPa                     \\
Operating pressure range                                                                   & From 1.2 bar to 4.0 bar     \\
Minimum permissible wall temperature (inner vessel)  & 77 K \\
Maximum pressure in vacuum space at T=293 K & 10 Pa \\
Interval of repair / regulatory work of the plant & Not more than once a 7 years. \\
\hline
\end{tabular}
\end{table}%

\begin{figure}[!h]
    \center{\includegraphics[width=0.8\linewidth]{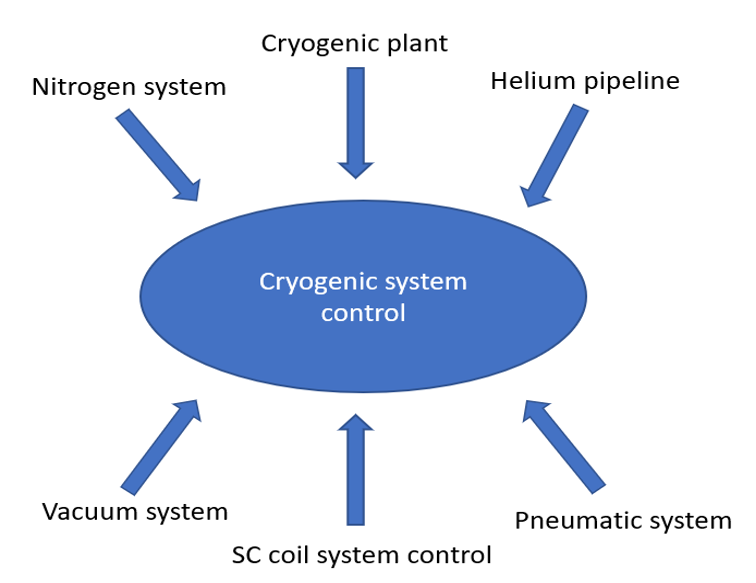}}
  \caption{ Block diagram of the cryogenic control system.
	}
\label{lb_cryo_5}
\end{figure}
\subsection{Auxiliary systems}

The vacuum subsystem provides maintenance of discharged medium in isolation volumes of the cryogenic plant, input cryostat, and SC coil cryostat. The maximum pressure value in the isolation volumes at 293 K (before the start of cooling) is $1\times10^{-1}$ Pa, maximum pressure value at 4.5 K (after cooling) for the cryogenic plant is $1\times10^{-3}$ Pa, for inlet cryostat and SC coil cryostat is $1\times10^{-5}$ Pa.

The subsystem includes several types of vacuum pumps: rotary vane pumps, Roots pumps, diffusion vapor pumps, and turbomolecular pumps. The subsystem equipment is located on the technological platform of the detector SPD. The total power consumption of the subsystem is 14 kW.

Pneumatic valves and valves are used to control the technological processes of the cryogenic system, their stable operation is ensured by the ordering of gas. The ordered gas subsystem has the following characteristics: working pressure 6 bar, working gas -- dry air. The system includes an air screw compressor, receiver, moisture separator, and various fittings (valves and reducers). The power consumption of the subsystem is 3 kW.

The cryogenic system is controlled by means of data acquisition from all related subsystems (cryogenic unit, helium pipelines, nitrogen system, pneumatic system, vacuum system, and data from SC coil control system). Figure \ref{lb_cryo_5} presents a block diagram of the cryogenic control system. The control system has automatic circuits with which the following steps are carried out: preparation, cooling of the SC coil, cryostatting at workloads, and heating.

\subsection{Cost estimate}
The cost of the equipment described above and the required capacity are presented in Tab. \ref{tab:cryo_4}.

\begin{table}[b]
\centering
\caption{Costs and required equipment capacity}
\label{tab:cryo_4}
\begin{tabular}{|l|c|c|}
\hline
 Name of system                           & Cost, k\$ & Power consumption, kW \\
\hline
 Cryogenic plant                            & 3 330     & 2         \\
 Helium pipeline                          & 500       & -         \\
 Nitrogen system                          & 1 500     & 2         \\
 Vacuum subsystem                         & 500       & 14        \\
 Pneumatic subsystem                      & 50        & 3         \\
  Cryogenic system control                 & 500       & 2         \\
\hline
  Total               & 6 380     & 23      \\ 
\hline
\end{tabular}
\end{table}

\chapter{Electromagnetic Calorimeter}
\newcolumntype{$}{>{\global\let\currentrowstyle\relax}}
\newcolumntype{^}{>{\currentrowstyle}}
\newcommand{\rowstyle}[1]{\gdef\currentrowstyle{#1}%
  #1\ignorespaces
}

\section{General concept}
\label{sec:ecal:concept}

The electromagnetic calorimeter (further calorimeter) should meet the criteria imposed by the
physical goals of the SPD \cite{ecal:spd_cdr} experiment of various nature and importance. The most important
criteria arise from the physical requirements  for the measurement accuracy of energies, impact positions,
and timing of photons and electrons.  the technological possibilities of modern experimental physics
should be taken into account when choosing the calorimeter setup. Cost factors should be
considered also to ensure the feasibility of the project. The high multiplicity of secondary particles
leads to  the requirements of high segmentation and high density of the absorber's medium with a small
Moliere radius. It is required in order to have  a sufficient spatial resolution and a capability to
resolve overlapping showers. The transverse size of the calorimeter cell should be of the order of
the Moliere radius. A reliable reconstruction of photons and neutral pions is possible only for
small  a overlap of showers. Occupancy should not exceed 5\%,  to make possible an efficient
photons reconstruction with high precision. The SPD experiment imposes the following
requirements on the calorimeter characteristics:
\begin{itemize}
 \item reconstruction of photons and electrons in the energy range from 50 MeV to 10 GeV;
\item  energy resolution for the above-mentioned particles: $ \sim 5\%/\sqrt{E} [\textrm{GeV}] $; 
\item  good separation of two-particle showers; 
\item  operation in the magnetic field;
\item  long-term stability: 2$\div$3\% in a six-month period of data taking.
\end{itemize}

The energy range requirement follows from the kinematic range of secondary particles, which
are produced in a collision of protons with energy up to 27 GeV and emitted into $4\pi$ solid angle.
Good energy resolution is required for  the identification and quantitative measurement of single
photon and neutral pion energies. Good two-particle separation is needed to separate photon
showers from the $\pi^0$ decay in order to suppress background events in measurements with prompt
photons. Long-term stability is necessary for polarization measurements featuring $\pi^0$
reconstruction in the calorimeter, especially in the end-caps. Calorimeter instability may result in
false asymmetry values. While it is essential to meet the physics requirements imposed on the
calorimeter design, one should also take into account  the cost estimate and  the technical feasibility
when choosing its granularity,  because a larger number of cells leads to higher costs of manufacturing and readout electronics.

\begin{figure}
\centerline{\includegraphics[width=0.9\linewidth]{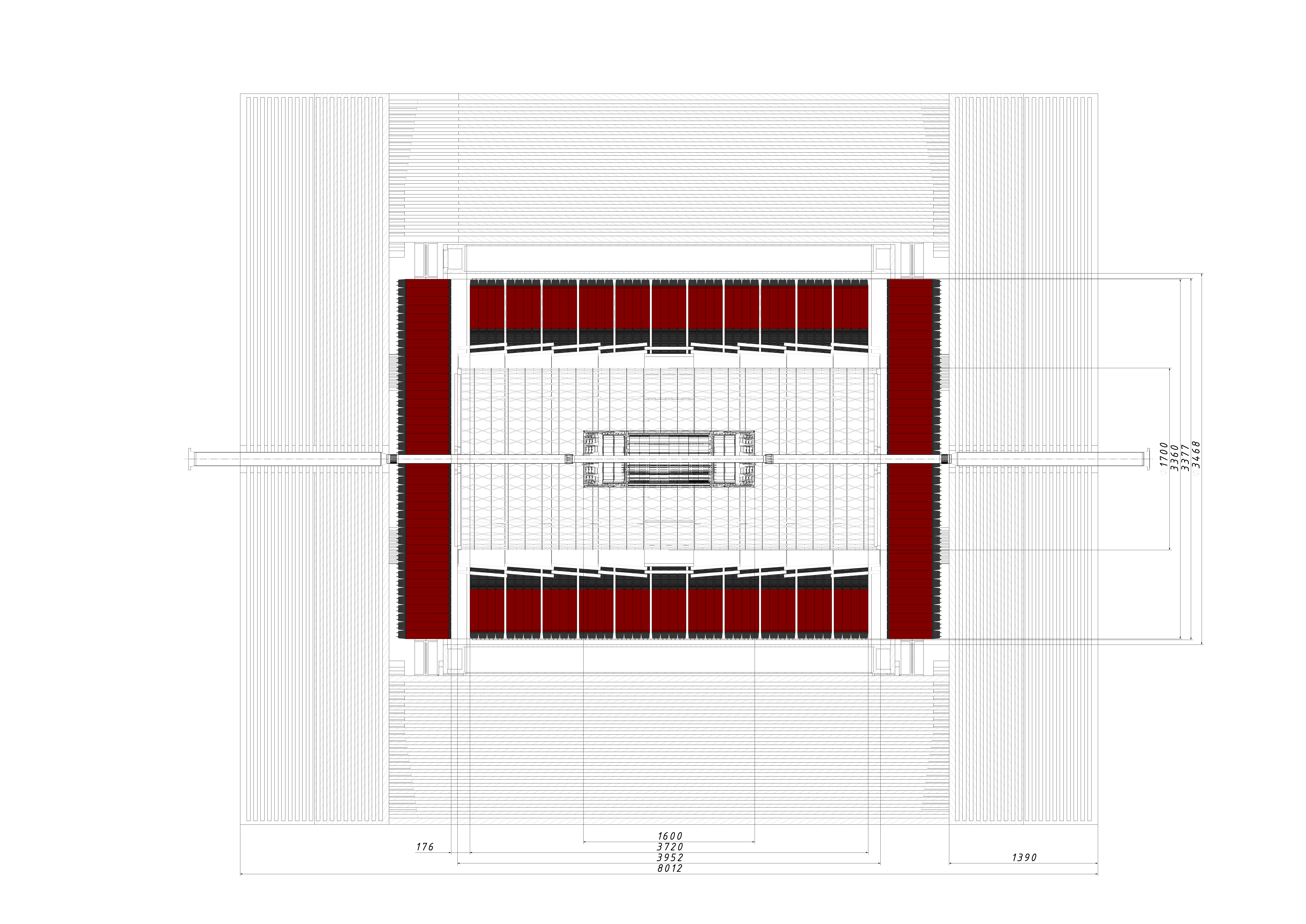}}
    \caption{Side view of the electromagnetic calorimeter. The barrel part and the end-caps are shown in red.
}
\label{fig:ecal:rz_view}
\end{figure}

\section{Overview of the SPD calorimeter}
\label{sec:ecal:overview}

The SPD electromagnetic calorimeter is placed between the cryostat with magnet coils and the
Particles Identification Detector (PID). The calorimeter consists of a barrel
and two end-caps, covering a $4\pi$ solid angle. It is shown  in Fig.~\ref{fig:ecal:rz_view} in red (barrel part) and in blue (end-caps).  The outer dimensions of the calorimeter are defined by the inner size of the cryostat. The thickness of the calorimeter is dictated by
the required thickness of the active part and the size of the readout block consisting of  a
photodiode and  the amplifier boards, as well as by the size of the flexible part of the fibers. For  an
efficient absorption of electrons and photons with energies up to 10 GeV, the calorimeter
thickness, which is defined by the number of sampling layers, should be at least 17$\div$20 $X_0$ in
terms of radiation lengths $X_0$.

For the sampling structure of 190 layers of 1.5-mm polystyrene scintillator and 0.5-mm lead,
the length of the active part is 380 mm, which corresponds to 17.6 $X_0$. The flexible parts of the fibers take
up to 8 cm. The transverse size of the calorimeter cell should be  on the order of the effective
Moliere radius of the calorimeter medium, which is, in turn, defined by the
scintillator-to-lead sampling ratio. The selected structure has a Moliere radius of 5.8 cm. The
separation efficiency of two photons with energies from 200 MeV to 500 MeV depends on the
cell size and reaches a plateau at a cell size of 40 mm, as it was determined in the MC
simulation. Therefore, we have selected $\sim$40 mm cell granularity for both  the barrel and  the end-caps.
 The cells in the barrel part of the calorimeter have  a trapezoidal shape in the azimuthal direction to
minimize the gaps between the modules. The vertex angle of the trapezoid equals 1.87$^\circ$.

\subsection{Barrel}

A schematic drawing of the calorimeter barrel part is shown in Fig.~\ref{fig:ecal:rz_view}. Its transverse size is
limited by the cryostat and PID. The barrel part is composed of the cells of trapezoidal shape in the
azimuthal direction with a vertex angle of 1.87$^\circ$. The front size of one cell is equal to 34 mm. In
the direction along the beam axis, the cell size is equal to 48 mm. A schematic drawing of the
calorimeter barrel part is shown in Fig.~\ref{fig:ecal:xy_view_assembly} (a), an isomeric calorimeter view is shown in Fig.~\ref{fig:ecal:xy_view_assembly}~(b).
The barrel consists of 44 annular layers, located one after another along the beam axis. Each
layer of the calorimeter's barrel is composed along the $\phi$ angle of 96 modules (4 cells per module). The design of the
modules is described in more detail in Section ~\ref{sec:ecal:modules}.  Every four layers of  the calorimeter barrel are
grouped, forming a ring containing 384 modules  weighing 4 tons. The
ring is made up of two half-rings to facilitate  moving and handling at installation. 
 Four layers are combined into a single block  using 2 mm thick stainless steel sheets,
which finally determine the gap between the layers. Subsequently, the half-rings are combined
into a single ring and mounted on two rails fixed to the cryostat inner side, as shown in Fig.~\ref{fig:ecal:rails_support}. The
calorimeter ring is divided into 24 sectors by the angle $\phi$, thus forming a cluster of 64 cells
which are read by one ADC, as shown in Fig.~\ref{fig:ecal:xy_view_assembly} (b). The calorimeter barrel is composed  of 11 rings, as shown in Fig.~\ref{fig:ecal:rails_support}. In total, there are 16896 cells with a weight of 52.9 tons.

\begin{figure}[!h]
    \begin{minipage}[b][][b]{0.60\linewidth}
    \center{\includegraphics[width=1\linewidth]{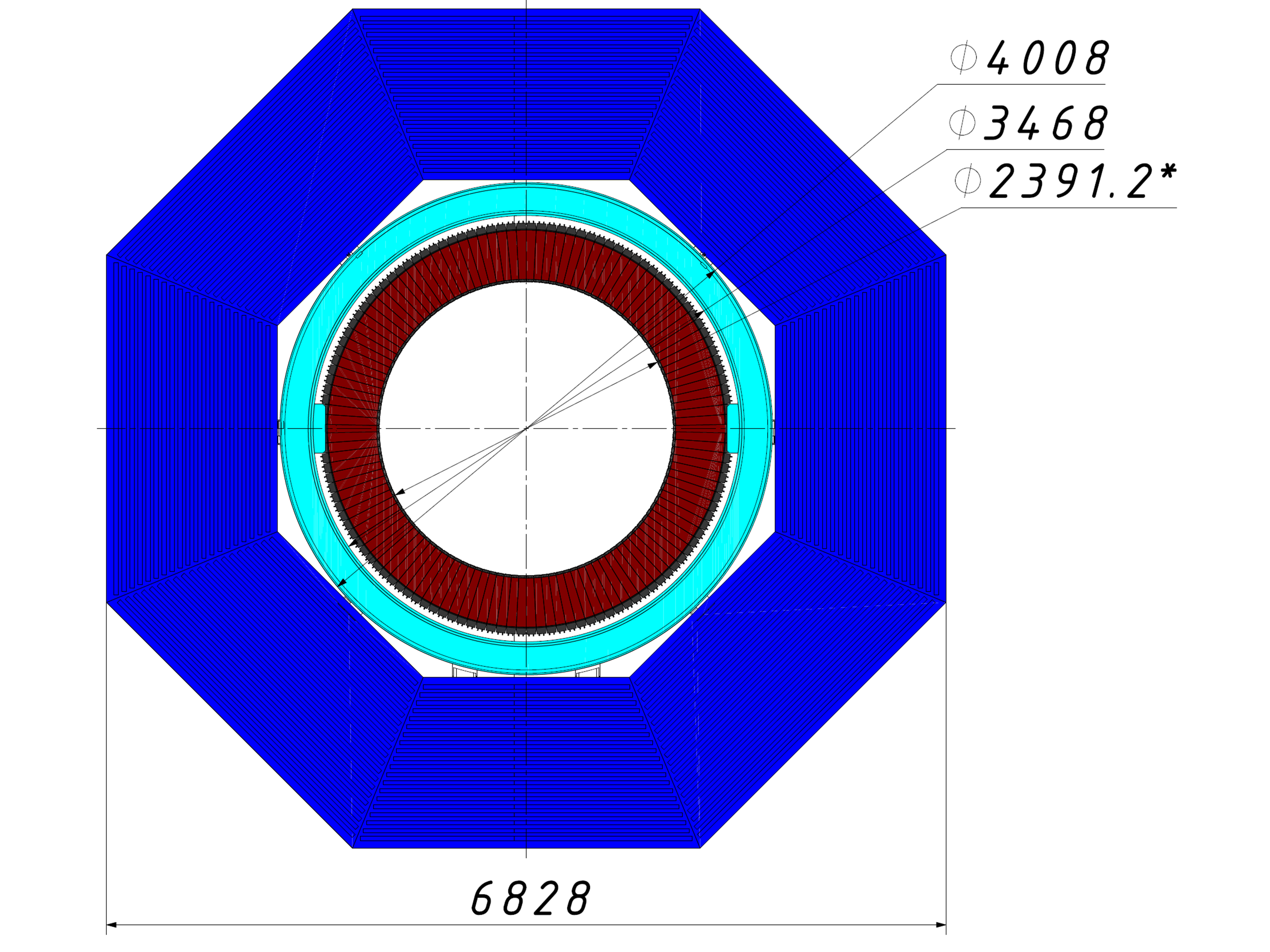}}
    \captionsetup{labelformat=empty}
    \caption{(a)}
    \addtocounter{figure}{-1}
  \end{minipage}
  \hfill
    \begin{minipage}[b][][b]{0.40\linewidth}
    \center{\includegraphics[width=1.\linewidth]{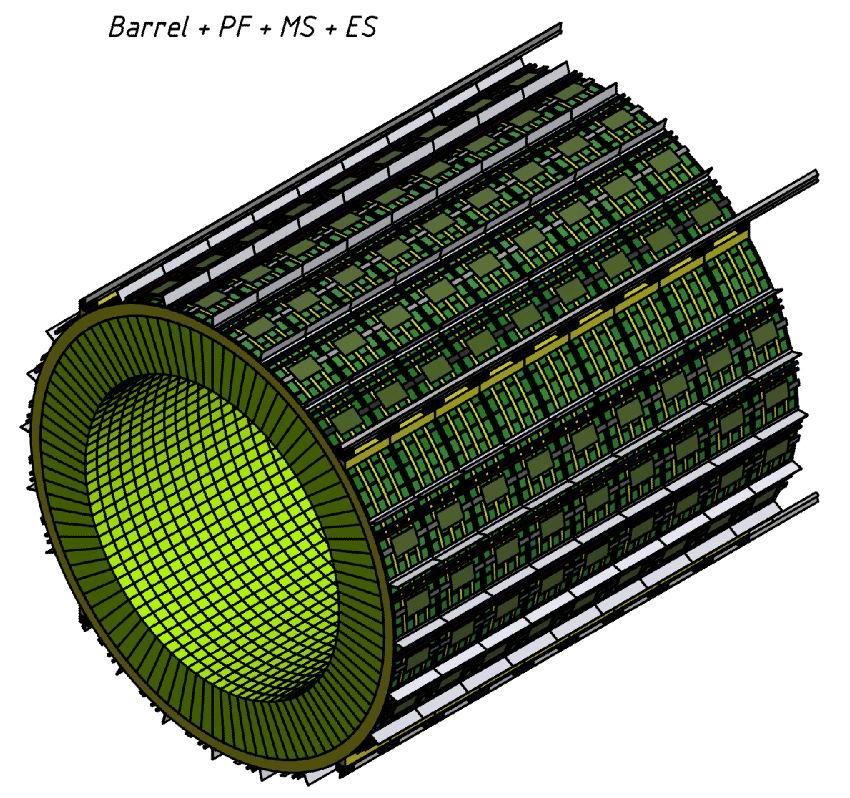}}
    \captionsetup{labelformat=empty}
    \caption{(b)}
    \addtocounter{figure}{-1}
  \end{minipage}
    \caption{Schematic drawing (a) of a cross-section of the barrel part of the calorimeter. The isometric view (b) shows  an assembled barrel part.}
  \label{fig:ecal:xy_view_assembly} 
\end{figure}

\begin{figure}
\centerline{\includegraphics[width=1.0\linewidth]{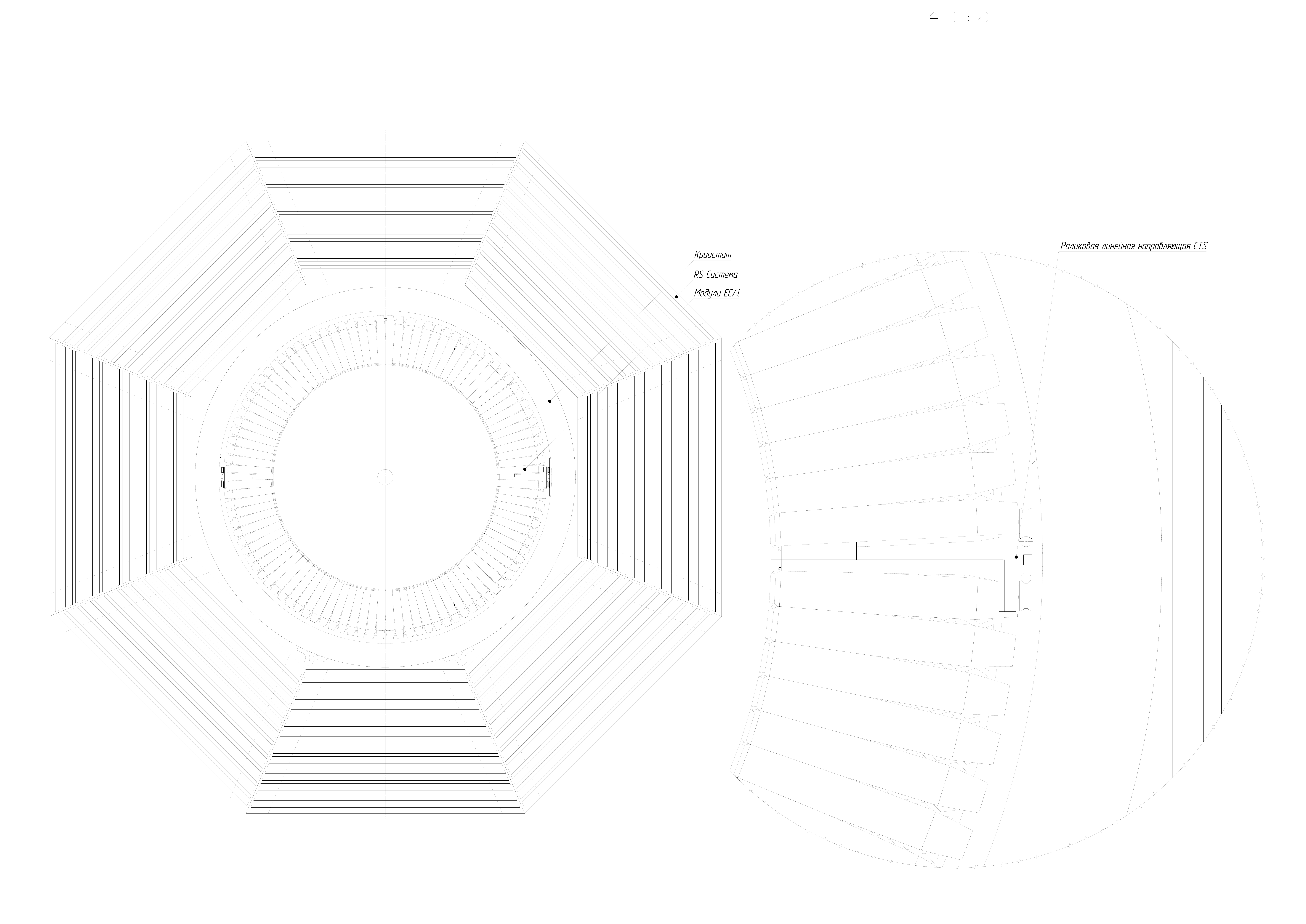}}
    \caption{The rails to ECal support and fastening to the cryostat inner side. Two roller trays
    support each calorimeter ring. There are 9 such rings installed inside of cryostat. One ring
     weights about 4 tons.}
\label{fig:ecal:rails_support}
\end{figure}

Bias voltage for all 1536 cells of one ring is provided by 96 front end-cards (FE-boards), and data
acquisition from all these cells is provided by 24 pcs. of  64-channel ADC boards. All of 96 pcs.
of FE-boards, intended for one ring bias voltage, are controlled and  powered by one HV
module. See more details about related electronics below in Sections ~\ref{sec:ecal:mppc}, ~\ref{sec:ecal:readout_hv}. The HV control box is
located outside of the barrel. It can be remotely placed at a distance of up to 100 meters from the
calorimeter. All 96 pcs. of FE-boards of one ring are connected to one HV control box by one
flat cable of 5-pair twisted wires. FE-cards are mounted directly on the calorimeter module together with scintillator light sensors MPPCs. The power supply concept of ADCs and
FE-boards is shown in Table ~\ref{tab:ecal:power}.

\begin{figure}[!h]
    \begin{minipage}[b][][b]{0.50\linewidth}
    \center{\includegraphics[width=1\linewidth]{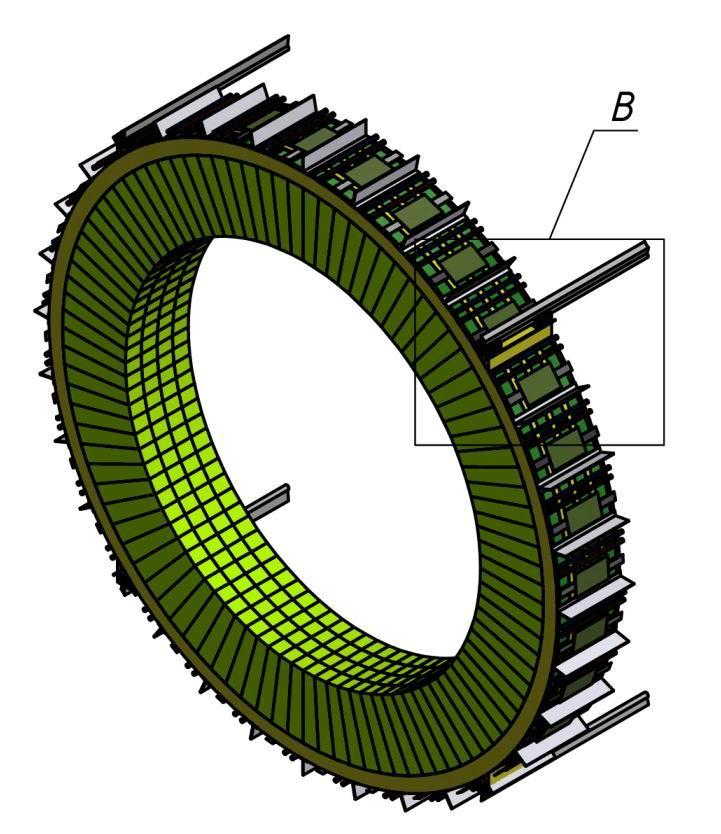}}
    \captionsetup{labelformat=empty}
    \caption{(a)}
    \addtocounter{figure}{-1}
  \end{minipage}
  \hfill
    \begin{minipage}[b][][b]{0.50\linewidth}
    \center{\includegraphics[width=1.\linewidth]{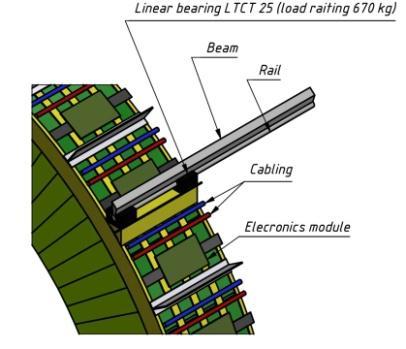}}
    \captionsetup{labelformat=empty}
    \caption{(b)}
    \addtocounter{figure}{-1}
  \end{minipage}
    \caption{The isometric view of one ECAL barrel ring (a) and  supporting rails are
    shown (b). Electronics (ADC) are located according to calorimeter clusters.}
  \label{fig:ecal:z_ring} 
\end{figure}

ADC and cable traces are located along the surface of the ring, as shown in Fig.~\ref{fig:ecal:z_ring}. A free
space of 20 mm between the cryostat and the top of the calorimeter should be provided for  the
placement of ADC and cable traces. This space is also required for the airflow that cools the
ADCs. Airflow must be forced from outside of the calorimeter to avoid overheating. The
rails for fastening the ECal's barrel inside of the cryostat are mounted in this gap also.

\subsection{End-caps}

Each end-cap (one is shown  in Fig.~\ref{fig:ecal:endcap}) consists of 5136 cells, grouped by 64, forming 80
clusters. All 64 cells of each cluster are connected to 4 pcs. of 16-channel FE-boards for
MPPC's bias voltage control and their readout is provided by one ADC. A complete list of ECal's
main components -- ADC, FE, HV, MPPC, etc. is presented in Table ~\ref{tab:ecal:number_components}. The cell cross-section is
40$\times$40 $\text{mm}^2$ and the length is 500 mm along  the beam. The end-cap has the absorber length equal to
17.6 $X_0$. The weight of one end-cap is 14.4 tons and for two parts 28.8 tons, respectively, 
as it is shown in Table ~\ref{tab:ecal:weight}. In total, there are 9152 cells in both end-caps.
There is a hole for the beam pipe in the center of each end-cap. The hole has a size of 320$\times$320 $\text{mm}^2$, which is equivalent to 64 cells.

\begin{figure}
\centerline{\includegraphics[width=0.7\linewidth]{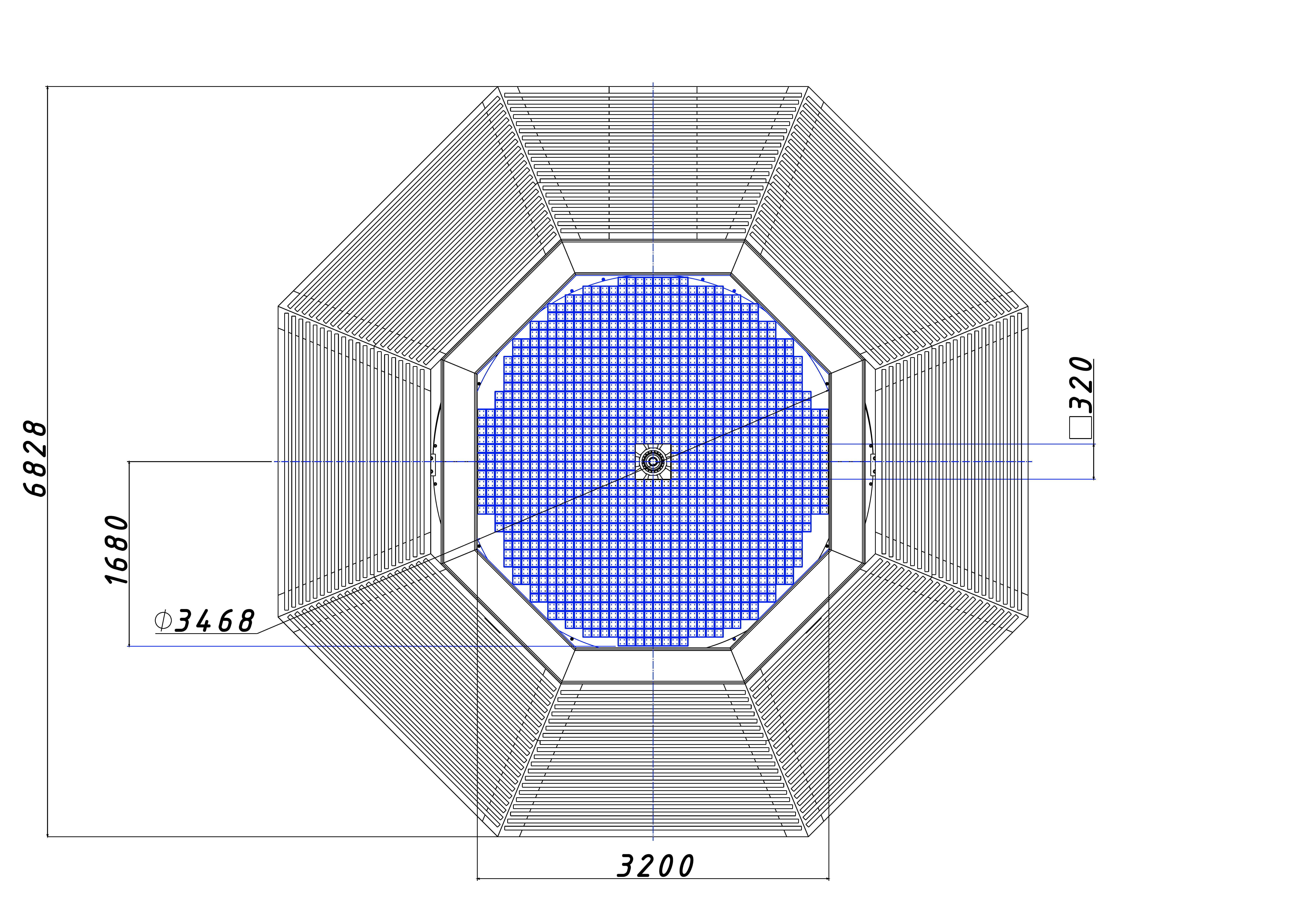}}
\caption{The end-cap part of the calorimeter in the frame, installed in RS. Each end-cap consists of
5136  cells with a total weight of 14.4 tons. The holes of size 320$\times$320 mm$^2$ for the beam pipe can be
seen in the centers of the end-caps. All dimensions are in millimeters.
}
\label{fig:ecal:endcap}
\end{figure}

The end-cap is mounted in the frame that supports it and shapes its geometry. The frame is
installed directly on the barrel RS, as shown  in Fig. ~\ref{fig:ecal:endcap}. There is a gap about 6 cm between the end-cap of the RS and the calorimeter's end-cap for the ADC's placement and cable routes (Fig.~\ref{fig:ecal:rz_view}). This
gap is also necessary  for air circulation of the ADC cooling.

\section{Design of the calorimeter module prototype}
\label{sec:ecal:modules}

The first version of the module, which was described in CDR of SPD in 2020 and published in
\cite{ecal:moliere_studies}, had a sampling structure of 220 layers of 1.5 mm and 0.3 mm scintillator and lead
respectively. This version had the absorber length of 12.6$X_0$ and  a Moliere radius equal to
7.6 cm. The second (new) version of the module, which was made for testing purposes, consisted
of 190 alternating layers of polystyrene scintillator and lead with a thickness of 1.5 mm and 0.5
mm, respectively. The new version of the modules with 190 layers of scintillator and lead has a
shorter length along  the absorbed particles path (48.8 cm instead of 58 cm) but has a greater
absorption quality of 17.6$X_0$ and a smaller Moliere radius of 5.8 cm. The lead plates are
intended to absorb the particle energy and develop an electromagnetic shower, whereas the
scintillator plates produce an amount of light proportional to the energy of the particles. The
properties of the absorber and the scintillator define the Moliere radius, which is equal to 5.8 cm
for the selected structure \cite{ecal:ecal_spd}. The energy resolution for 1 GeV photons is assumed to depend on
the calorimeter sampling fraction and is expected to be $\sim$5\% \cite{ecal:energy_resolution_combinations}. The scintillator plates are made
of polystyrene beads with an added luminophore admixture of 1.5\% p-Terphenyl and 0.05\%
POPOP ($\text{C}_{24}\text{H}_{16}\text{N}_2\text{O}_2$) \cite{ecal:p_terphenil, ecal:popop}. It has a scintillation time of about 2.5 ns and a light output of 60\% of
anthracene, which are good results. The radiation hardness of the scintillator is sufficient for
radiation doses up to about 10 Mrad ($10^5$ Gy), which is important for operating the
calorimeter in the radiation field of secondary particles in the vicinity of the interaction point.
The luminophore admixtures re-emit the energy of excitations in polystyrene in the form of
visible light. The first admixture (p-Terphenyl) emits light with a wavelength of maximum
emission at 340 nm. This light is absorbed by the second admixture (POPOP) and is re-emitted
into a spectrum with a wavelength of maximum emission of 420 nm, which is seen as a light of
blue glow. The light from the scintillator plates is gathered using wavelength-shifting fibers
(WLS) \cite{ecal:wls}. Fibers of type Y-11 manufactured by KURARAY are used.

\begin{figure}[!h]
    \begin{minipage}[b][][b]{0.50\linewidth}
    \center{\includegraphics[width=1\linewidth]{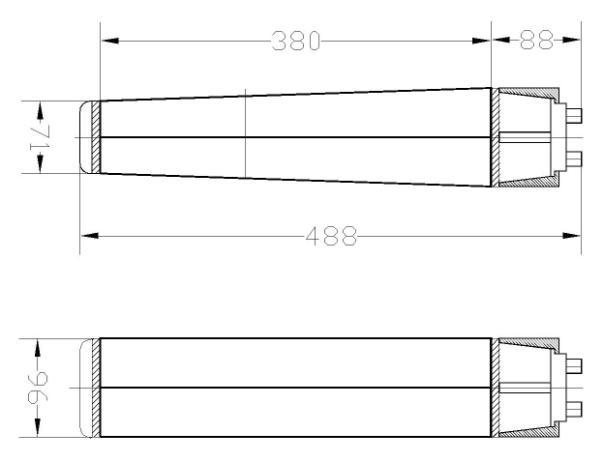}}
    \captionsetup{labelformat=empty}
    \caption{(a)}
    \addtocounter{figure}{-1}
  \end{minipage}
  \hfill
    \begin{minipage}[b][][b]{0.50\linewidth}
    \center{\includegraphics[width=1.\linewidth]{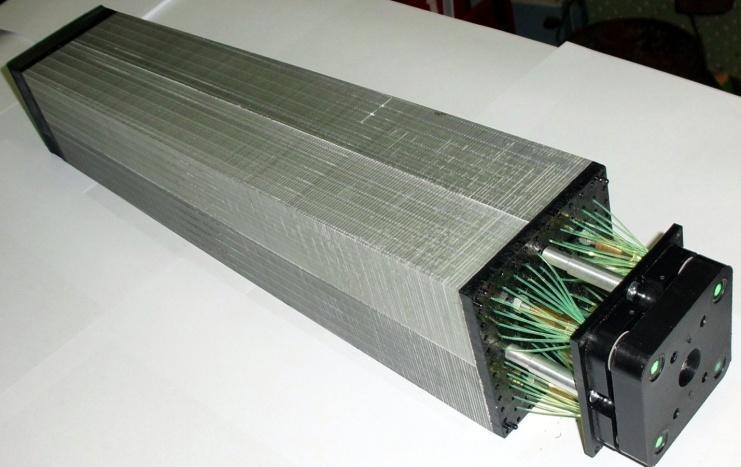}}
    \captionsetup{labelformat=empty}
    \caption{(b)}
    \addtocounter{figure}{-1}
  \end{minipage}
    \caption{ECal module drawing (a) and photo (b) without light shielding cover.}
  \label{fig:ecal:module} 
\end{figure}

\begin{figure}
\centerline{\includegraphics[width=0.7\linewidth]{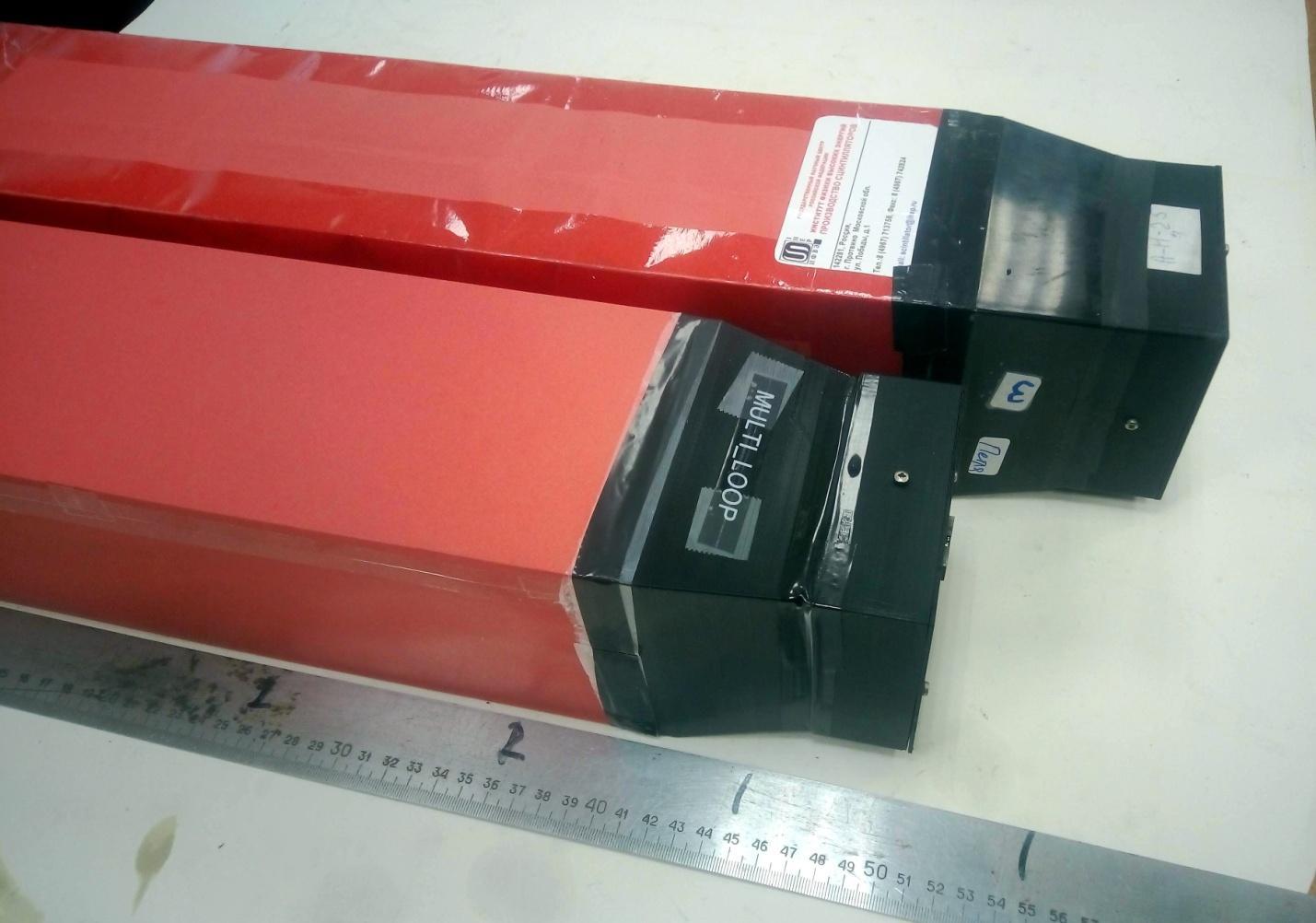}}
    \caption{Two calorimeter modules covered in light isolation paper are shown in this photo.
    The modules have different lengths due to  a different number of layers (190 and 220) and a different
    absorber thickness of 0.5 and 0.3 mm, respectively.}
\label{fig:ecal:module_photo}
\end{figure}

The fibers absorb the light from the POPOP and re-emit it into a spectrum with a wavelength of
maximum emission of 490 nm. Thirty six WLS fibers go along each cell, gather in one bundle,
and transmit light to one multi-pixel 6$\times$6 $\text{mm}^2$ photodiode (multi-pixel photon counter, or
MPPC). The schematic module drawing is shown in Fig.~\ref{fig:ecal:module} (a). The active part of the module is
380 mm and  the total module length is $\sim$490 mm without MPPC board. A single module consists of
4 cells with 190 layers of the scintillator and the absorber with a thickness of 1.5 mm and
0.5 mm, respectively. The period of the structure
is set to 2 mm in order to avoid optical contact between the lead and the scintillator, and because
 the connection technique involves special ''Lego'' spikes. 
Four bundles of fibers for guiding the light to the MPPC can be seen  on the
photo in Fig.~\ref{fig:ecal:module} (b). Two modules with different sampling structure are shown at Fig.~\ref{fig:ecal:module_photo}. These
modules have different lengths due to  differing numbers of layers -- the first version with 220 layers
and the second version with 190 layers, and  a different absorber thickness  of 0.3 mm and 0.5 mm,
respectively.  The length of the shortest one (second version) does not exceed 
500 mm.

\section{Scintillator production}
\label{sec:ecal:scint_production}

\subsection{Injection molding technology}

As it was mentioned in detail in Section~\ref{sec:ecal:modules}, the calorimeter employs polystyrene scintillator plates
from Polystyrene with an added luminophore admixture of 1.5\% p-Terphenyl and 0.05\% POPOP
(C24H16N2O2) \cite{ecal:p_terphenil, ecal:popop}. Scintillation plates for calorimeter prototypes were manufactured by
injection molding technology at the pilot plant of IHEP \cite{ecal:scint_ihep}. Scintillation material of this type
has been successfully used in calorimeters for the last 20 years. It has a high radiation hardness
(about $10^7$ rad = $10^5$ Gy), a good light output (60\% of anthracene), a fast decay time (1.2 ns), a
high transparency for blue light (420 nm) with attenuation length $\sim$60 cm. The injection molding
technology requires a special machine (Fig.~\ref{fig:ecal:molding_machine}) and  a matrix form (Fig.~\ref{fig:ecal:scint_press}) for scintillation plates
production with given dimensions. Thermo plastic injection molding machines are a standard
series of injection molding machines designed to perform most typical tasks that do not require
the use of special materials or particularly high requirements for molded products.  These 
injection molding machines have incorporated the most reliable, time-tested and effective
solutions. The production procedure is automatic and allows  a production rate of about one cycle per
minute (4 tiles). Granulated Polystyrene PSM-115 \cite{ecal:psm_115} with dopants (1.5\% Pt-Terphenyle \cite{ecal:p_terphenil}
and 0.05\% POPOP \cite{ecal:popop}) is used for scintillator production. All used components are pre-dried and
then mixed in special mixers. Then the prepared mixture is fed into the receiver of the injection
molding machine.

\begin{figure}
\centerline{\includegraphics[width=0.7\linewidth]{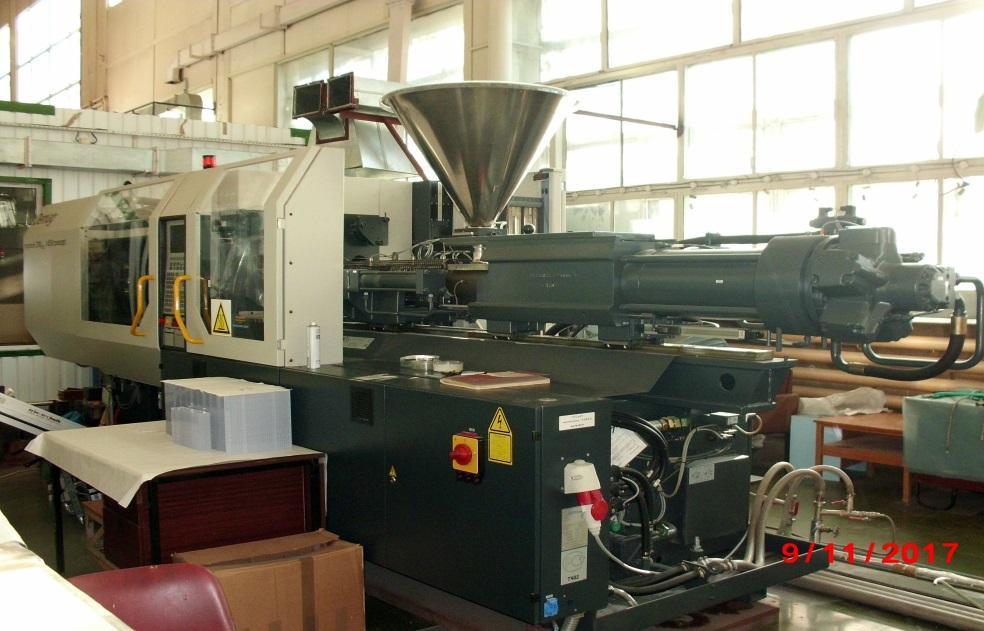}}
    \caption{Thermo plastic injection molding machine, general view.}
\label{fig:ecal:molding_machine}
\end{figure}

\subsection{Matrix form}
One of the important points is the molding press form for the scintillator production. It should be
built of high-precision components, made of high-quality materials (see Fig.~\ref{fig:ecal:scint_press}~(a)). The number of molding cycles should be  in  the
order of one million to produce 4 million plates required for the SPD calorimeter. Figure ~\ref{fig:ecal:scint_press}
shows 4 pcs. of scintillator plates coming out of the injection machine. Each injection cycle for
4 plates  takes one minute. The total number of SPD calorimeter scintillator plates is about
4,000,000 pieces with a total weight of about 15 tons.

\subsection{Time  estimate for calorimeter modules production}

Taking into account that one injection molding machine cycle is one minute and with two molds
 running 24 hours a day, a total of 583 days are needed to produce the required amount of the
scintillator and absorber tiles. Considering 273 working days per year, the production process of the
tiles will take about 2 years. Assembly of  the calorimeter modules at the rate of 10 modules per day
 allows to complete this process within 2.4 years, as it is shown in Table ~\ref{tab:ecal:manufacture_time}. In order to produce  the SPD calorimeter by 2030, it is necessary to manufacture components and
assemble modules in parallel.

\begin{figure}[!h]
    \begin{minipage}[b][][b]{0.50\linewidth}
    \center{\includegraphics[width=1\linewidth]{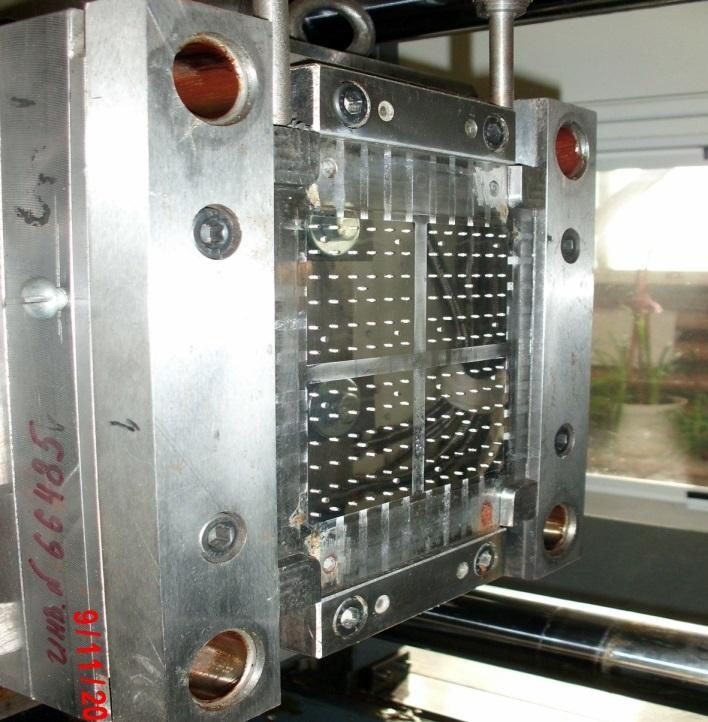}}
    \captionsetup{labelformat=empty}
    \caption{(a)}
    \addtocounter{figure}{-1}
  \end{minipage}
  \hfill
    \begin{minipage}[b][][b]{0.50\linewidth}
    \center{\includegraphics[width=1.\linewidth]{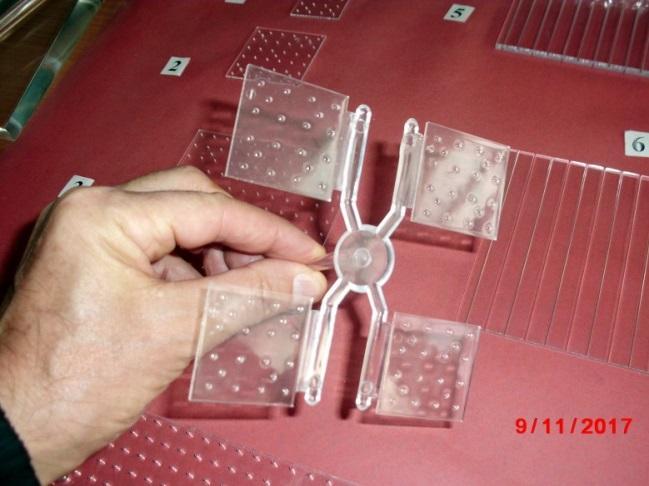}}
    \captionsetup{labelformat=empty}
    \caption{(b)}
    \addtocounter{figure}{-1}
  \end{minipage}
    \caption{a) 4-set matrix for scintillator production; b) The scintillation plates released from
    the molding machine, details of the molding system are visible.}
  \label{fig:ecal:scint_press} 
\end{figure}

\section{Multi-pixel photodiodes}
\label{sec:ecal:mppc}

All of the MPPCs  that are used in this prototype have the same size of 6$\times$6 $\text{mm}^2$. The
S13360-6025 \cite{ecal:hamamatsu_mppc} series has the best response speed, low capacitance and  a large number of pixels,
but the largest temperature coefficient of $K_t$ $\sim$0.054 V/$^\circ$C. The temperature coefficient
represents a linear dependence of breakdown voltage on temperature. Temperature dependence
of MPPC's breakdown voltage leads to MPPC's gain variation, if this temperature dependence
is not compensated. To achieve a calorimeter's stability of about 2\%, one needs to ensure
temperature stability of the surrounding environment, or use the breakdown voltage
compensation scheme:

\begin{equation}
U_{OP}(T) = U_{BR} + \Delta U + (K_t\times\Delta T),
\end{equation}

where: $U_{OP}$\ and $U_{BR}$\ are MPPC bias voltage at temperature T and MPPC's breakdown voltage
at device characterization temperature, respectively. $\Delta U$ is a given over-voltage and $\Delta T = T - 20^\circ C$
is a deviation of the current temperature ($T$) from temperature selected by the compensation
program, typically 20 $^\circ$C. MPPCs of  the S14160-6050 series have  a higher photodetector efficiency 
but fewer pixels, which is worse in terms of the dynamic range. This series has a small
temperature coefficient. An optimal solution would be the usage of MPPCs with properties similar
to  the S13360-6025 or S14160-6050 series but with a smaller pixel size of 15 $\div$ 20 $\mu \text{m}$, i.e. larger
amount of pixels, which would make them more suitable for the calorimeter. Four MPPCs are
surface-mounted on a circuit board, as it is shown in Fig.~\ref{fig:ecal:circuit_board}. A thermistor is also installed on the
board to measure the photodiode's temperature. The circuit board is attached to a module in such
a way that the photodiodes are located exactly opposite the ends of fiber bundles. There is no
optical contact between the photodiode and the WLS, there is an air gap of about 0.1 mm
instead. Optical grease is not used in order to avoid instability in the conditions of light guiding.
A light insulating basket made of black plastic is installed above the circuit board.

\begin{figure}[!h]
    \begin{minipage}[b][][b]{0.50\linewidth}
    \center{\includegraphics[width=1\linewidth]{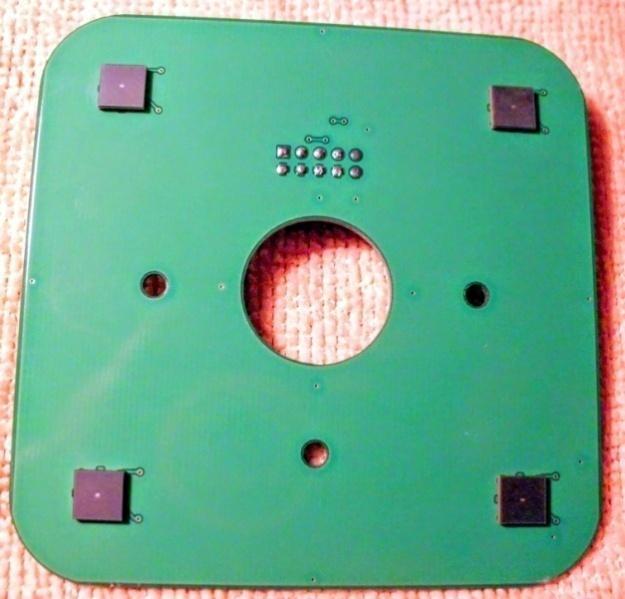}}
    \captionsetup{labelformat=empty}
    \caption{(a)}
    \addtocounter{figure}{-1}
  \end{minipage}
  \hfill
    \begin{minipage}[b][][b]{0.50\linewidth}
    \center{\includegraphics[width=1.\linewidth]{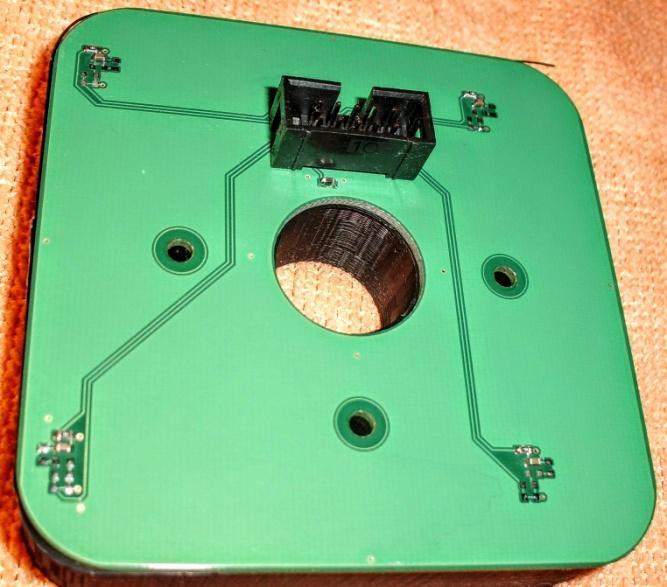}}
    \captionsetup{labelformat=empty}
    \caption{(b)}
    \addtocounter{figure}{-1}
  \end{minipage}
    \caption{Printed circuit board with 4 MPPC diodes: front (a) and back (b) sides.}
  \label{fig:ecal:circuit_board} 
\end{figure}

The MPPC boards are connected to the amplifier / bias control board (FE-board) (Fig.~\ref{fig:ecal:adc} (a))
with up to 1-meter of flat 5-pair twisted-pair cable. Four wires transmit signals from four
MPPCs to the amplifiers of FE-board. One wire is used to provide a regulated common bias voltage
to all MPPCs on the board (up to $\sim$55 V) and another one to connect the thermistor for local
temperature measurement of MPPCs. Output signals of each MPPC are transmitted via twisted
pairs of wires (signal/GND). The small regulated individual reverse bias voltage can be applied by
FE-board on signal wires of each MPPC in  the range from 0.0 to $\sim$3.0 V for individual adjusting of
bias voltage.  This way, the bias voltage of each MPPC can be individually and precisely  trimmed in
a small range with  a 10-bit precision (i.e. $\sim$3 mV step) in order to take into account and
compensate for the possible variations of individual parameters of each MPPC. FE-board provides
hardware control of regulated MPPC's bias voltage and also amplifies MPPC's signals and
converts them to  a differential type, which is required by  the ADC board. Calculations required for  the bias
voltage compensation with temperature are performed at a software level by  the control PC, taking
into account the local temperature measured by the thermistor installed on  the MPPC's circuit board.
This approach allows calorimeter operation without special measures for MPPC's temperature
stabilization. Signal stability  on the order of 1\% was achieved during measurement over an
extended period of time with  the usage of such a technique.

\section{MPPC readout and high voltage control}
\label{sec:ecal:readout_hv}

\subsection{Analog-to-digital converter (ADC)}
\label{sec:ecal:adc}

The readout electronics consists of a 64-channel ADC board -- an analog-to-digital converter
ADC-64 (Fig.~\ref{fig:ecal:adc} (b)). The ADC continuously performs simultaneous samples on all 64 input
channels with a fixed frequency and provides  a full digital representation of  an input signal time-shape. 
Samples are done at a 64-MHz frequency, which corresponds to the sampling period of
15.625 ns. Each sample is digitized with a 12-bit conversion. At present, there is an
ADC-64ECal modification, which improves digitization up to 14-bit and significantly extends
the dynamic range of the measured amplitudes. The new ADC-64ECal \cite{ecal:afi} modification also
allows operation in strong magnetic fields, which is necessary for experiments at the NICA
accelerator complex.

\begin{figure}[!h]
    \begin{minipage}[b][][b]{0.50\linewidth}
    \center{\includegraphics[width=1\linewidth]{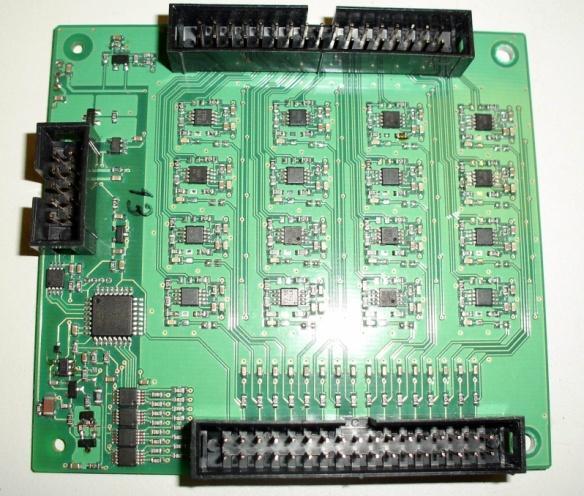}}
    \captionsetup{labelformat=empty}
    \caption{(a)}
    \addtocounter{figure}{-1}
  \end{minipage}
  \hfill
    \begin{minipage}[b][][b]{0.50\linewidth}
    \center{\includegraphics[width=1.\linewidth]{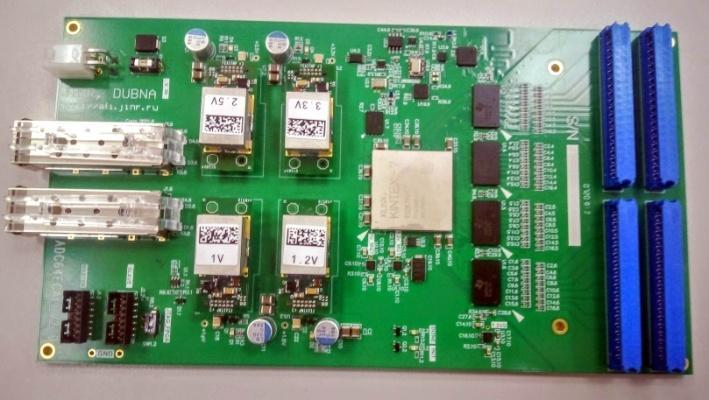}}
    \captionsetup{labelformat=empty}
    \caption{(b)}
    \addtocounter{figure}{-1}
  \end{minipage}
    \caption{(a) 16-channel amplifier / bias control board (FE-board).
   (b) ADC-64ECal digitizer.
    The ADC-64ECal power consumption is about 13 W per one 64-channel board.}
  \label{fig:ecal:adc} 
\end{figure}

The ADC-64ECal power consumption  is about 13 $W$ per one 64-channel board, i.e. $\sim$200 mW/ch.
 ADC's total power consumption is shown in Table ~\ref{tab:ecal:power}. It is equal  to 2.8 kW for  the barrel and 0.9 kW for  the calorimeter end-cap.  the ADC card will be located directly on the calorimeter to reduce signal
noise and reduce the cabling bundles between ADC and data collectors. ADC provides two Ethernet
interfaces (shown in Fig.~\ref{fig:ecal:adc} (b)): one for data transfer to  a higher level of data
acquisition system, and another one for time synchronization (White Rabbit technology is
employed), which provides sub-nanosecond accuracy. 50 $\Ohm$ coaxial input can be used as an
external trigger for readout synchronization. The ADC can also operate in  the streamer mode due to  a
dedicated firmware.

\subsection{Front-end amplifier}

 the Front-end (FE) electronics will be located directly on the calorimeter. A small path of  the analog
signal to FE means less signal distortion and pickup noise. FE electronics card with an
16-channels amplifier / bias control is shown in Fig.~\ref{fig:ecal:adc} (a). It is used to:

\begin{itemize}
    \item control the MPPC's bias,
    \item amplify MPPC's output signals,
    \item convert them to  a differential type and transmit them to  the analog inputs of ADCs (Fig.~\ref{fig:ecal:adc}~(b)).
\end{itemize}

Each FE-board can service up to 4 pcs. of MPPC boards (Fig.~\ref{fig:ecal:circuit_board}). The FE-board and the MPPCs
must be located close to each other, therefore they will be placed on the same printed circuit
board. The FE-board has  a power consumption of about 33 mW, i.e. $\sim$2 mW/ch., and  their total
 power consumption is $\sim$1.4 kW for the whole calorimeter.

\subsection{High voltage system}

\begin{figure}
\centerline{\includegraphics[width=0.7\linewidth]{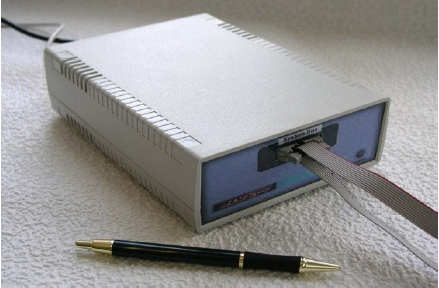}}
    \caption{Control power box (HV) for MPPC's bias voltage.}
\label{fig:ecal:hv_control}
\end{figure}

All required power supply voltages for FE-boards (High voltage, required for MPPC's bias, and
Low voltage, required for internal circuits of FE-boards) are supplied from the specially
designed HV control power box \cite{ecal:hvsys_small}, which is shown  in Fig.~\ref{fig:ecal:hv_control}. The HV control box can provide  the
required power supply for up to 127 pcs. of FE-boards, which are connected in parallel on a 10-wire flat cable. The HV control box has Ethernet control from the outside. The HV control box
communicates with FE-boards by  the RS-485 interface for MPPC's bias voltage control and
temperature feedback. The flat cable length can be over 100 meters. Therefore, this box can be
installed outside of the calorimeter at a distance of up to 100 meters. The consumed power of this
unit is up to 100 watts. Taking into account that 20 such units are needed for the calorimeter,
their total consumption is up to 2 kW (Table ~\ref{tab:ecal:power}).

\subsection{LED generator}

To make a precise calibration of each ECal channel and for the continuous monitoring of
channel amplification  a LED generator has been designed \cite{ecal:hvsys} and built. The main requirement for this
device is its stability. Light intensity variation should be below 1\%. To reach such stability and
to reduce residual variations of LED light intensity, the feedback loop has been implemented  on
the device. A pin diode has been located on the LED back side. Light from the LED back side
is detected by the pin diode and analyzed by control circuits to organize a feedback control and
improve the LED's stability. Light from  the LED front side  has a high luminosity and can be
distributed by  the optical fiber light guides to a large number of calorimeter cells. One such generator
can supply light  for up to 100 channels with  the intensity of about 1000 photoelectrons, which is enough
to provide MPPC's gain control and sensitivity monitoring. This LED control system was used in  the
test setup, shown  in Fig.~\ref{fig:ecal:test_setup}, for monitoring long-term stability during data-taking on cosmic rays.

\subsection{Slow-control system}

The main objectives of  the slow-control system are:

\begin{itemize}
    \item operating equipment parameters control;
    \item monitoring of low and high-voltage power supplies;
    \item recording of slow-control commands and data;
    \item notification about problems (alarms);
    \item recording of  LED signals for the  long-term stability check.
\end{itemize}

High voltage (HV) has been designed especially for the MAPD-based devices and is part of the
slow-control system. Specific points of HV are:

\begin{itemize}
    \item multichannel ($\sim$30 000 channels);
    \item very precise voltage setting for each channel;
    \item need for voltage correction depending on temperature of the MAPD;
    \item temperature control for  the ADC boards.
\end{itemize}

\section{Cosmic ray test results}
\label{sec:ecal:cosmic_results}

\subsection{Energy resolution}

A test setup was made from four calorimeter modules consisting of 16 cells with a cross-section
of 55$\times$55 $\text{mm}^2$ and  then tested on cosmic rays. In this prototype, light detectors were based on
MPPC types S13360-6025 with 25-micron pixel pitch. For testing on cosmic rays, a small setup
 of 4 modules (each 11$\times$11 $\text{cm}^2$ ) with a total cross-section of 22$\times$22 $\text{cm}^2$ was used (Fig.~\ref{fig:ecal:test_setup}). The
cells, 55$\times$55 $\text{mm}^2$ each, are assembled in a 4$\times$4 setup. The modules are placed vertically, while
the direction of the registered cosmic rays is determined by trigger counters. Trigger counters are
scintillator detectors based on multi-pixel photodiodes of FC6035 type and size 6$\times$6 $\text{mm}^2$. 
 All photodiodes of the counters were connected to   a coincidence circuit to make   a trigger for the ADC for
cosmic rays events selection. Auxiliary trigger signals from the generator which controls the
LEDs were also logically added (by the ''or'' function) to   the cosmic rays trigger signals. LEDs were
used for control and calibration of   the calorimeter cells, estimations of the light yield, and for   the
long-term stability check. Data acquisition was conducted by the 64-channel ADC board,
similar to the one described  in Section ~\ref{sec:ecal:adc} with intended software usage from the ADC developer \cite{ecal:afi}. During a
data-taking period of 5$\div$6 days, statistics  on the order of a million events was obtained. The setup
allows one to measure energy depositions and trajectories of cosmic ray particles. Relativistic
muons with energy above 250 MeV pierce through the calorimeter and form a peak in the
deposited energy spectrum. In order to select straight tracks of   the particles, which pass vertically
through one module, only those events  that have a number of hits equal to 1 are selected to
avoid side tracks.

\begin{figure}
\centerline{\includegraphics[width=0.7\linewidth]{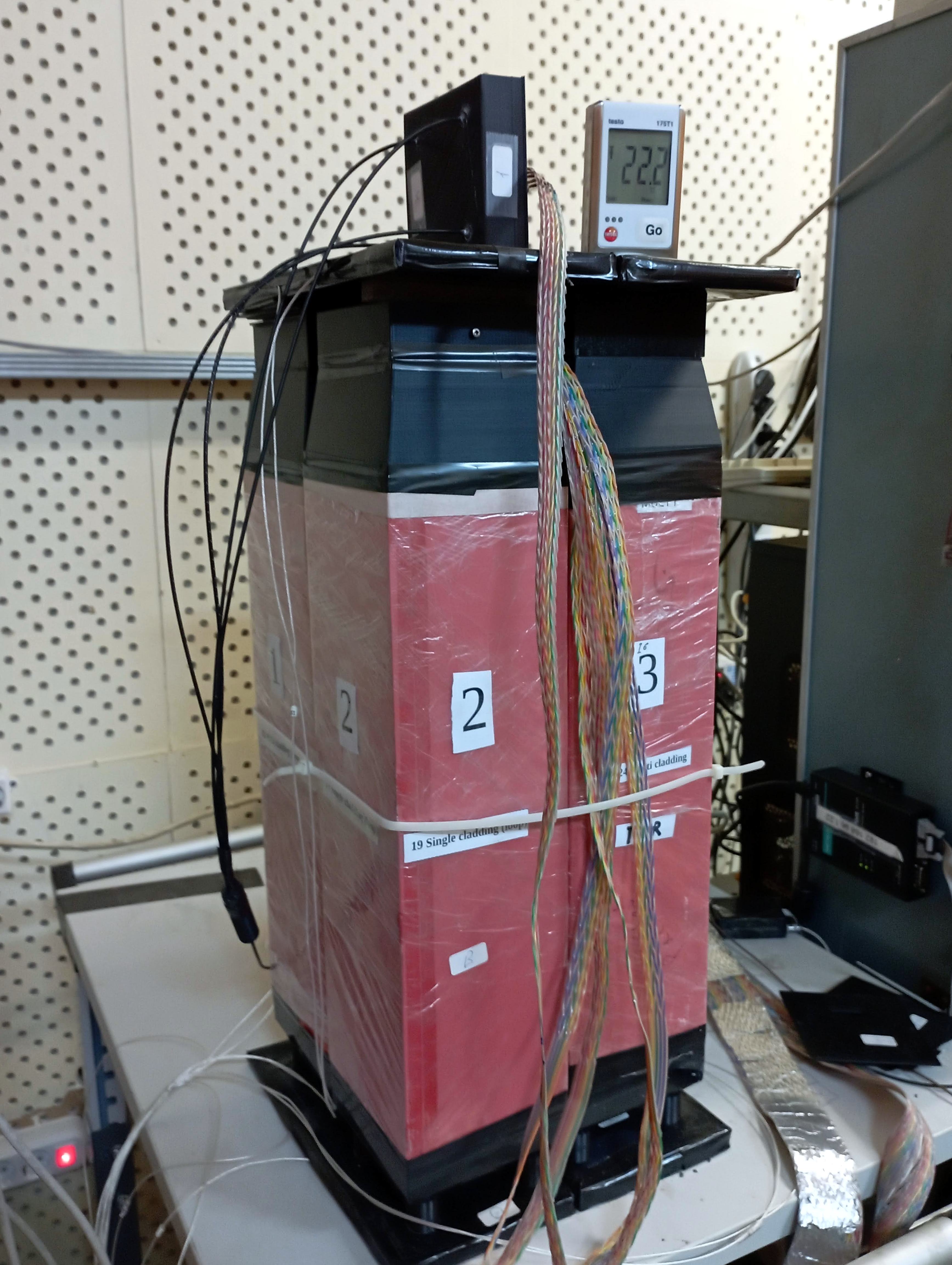}}
    \caption{Photo of the calorimeter test setup consisting of 4 modules of the size 11$\times$11 $\text{cm}^2$ ,
    with the total cross-section of 22$\times$22 $\text{cm}^2$. The environment temperature was measured by
    thermistors, provided for each  of the 4 MPPC boards, and by   a digital thermometer as well. One of the
    boards in the light protection cover and the digital thermometer  were placed on top of the setup.
    The optical fibers (visible in this photo) distributed the light from the LED generator.}
\label{fig:ecal:test_setup}
\end{figure}

\begin{figure}[!h]
    \begin{minipage}[b][][b]{0.50\linewidth}
    \center{\includegraphics[width=1\linewidth]{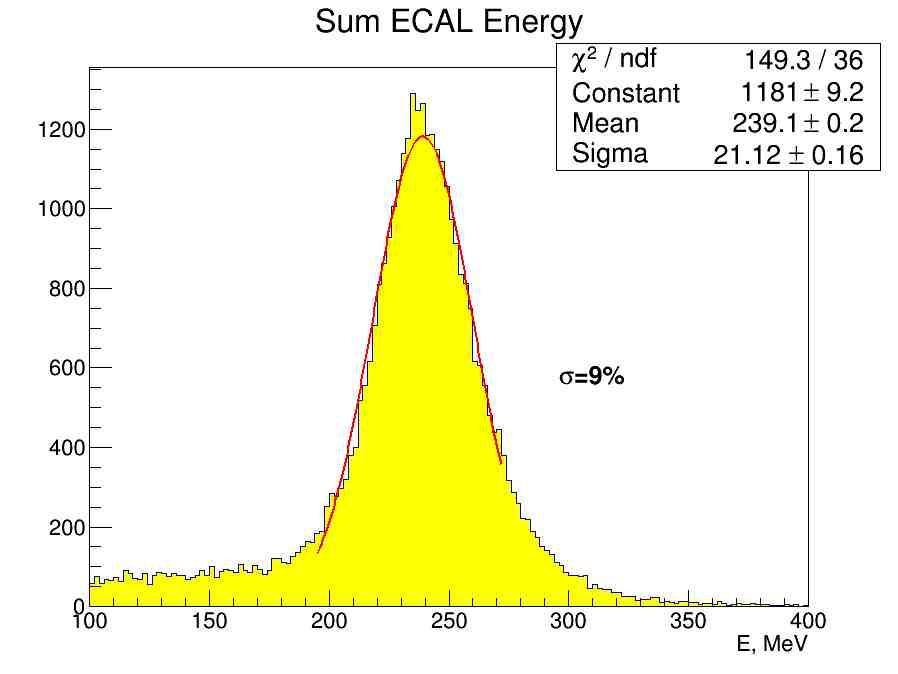}}
    \captionsetup{labelformat=empty}
    \caption{(a)}
    \addtocounter{figure}{-1}
  \end{minipage}
  \hfill
    \begin{minipage}[b][][b]{0.50\linewidth}
    \center{\includegraphics[width=1.\linewidth]{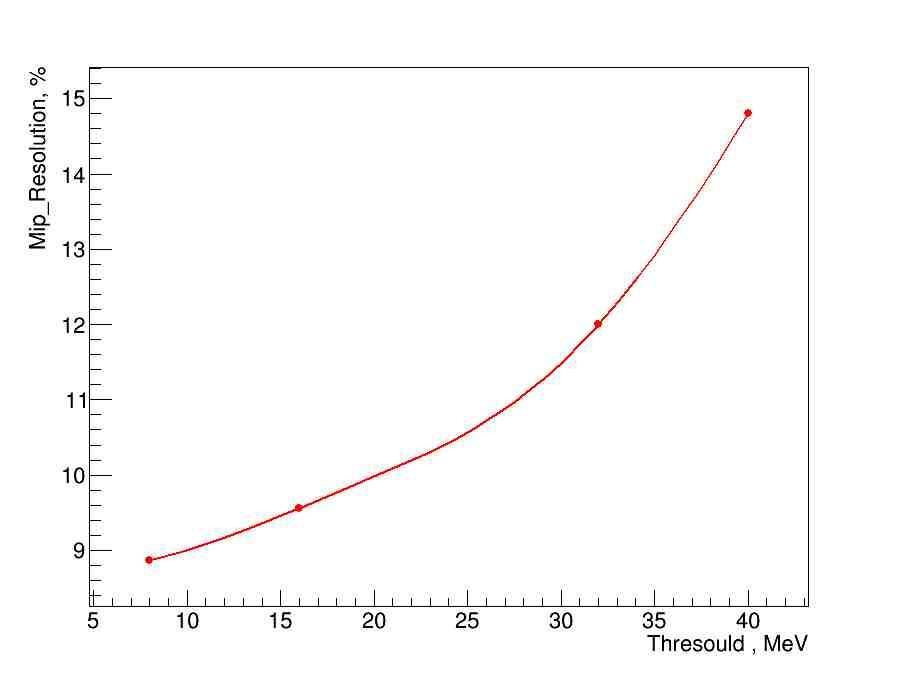}}
    \captionsetup{labelformat=empty}
    \caption{(b)}
    \addtocounter{figure}{-1}
  \end{minipage}
    \caption{(a) Energy spectrum from the calorimeter for vertical cosmic ray particles with
    energy resolution of 9\%. This point corresponds to the ADC threshold $\sim$8 MeV.
   (b) Energy resolution dependence on the ADC threshold in MeV.}
  \label{fig:ecal:test_energy_spectrum} 
\end{figure}

Signals obtained on cosmic muons were used for amplitude alignment and calorimeter energy
calibration. Only events with exactly one cell hit were selected. The bordering cells have more
events with smaller amplitudes due to angled tracks. We perform calorimeter calibration using
only vertical tracks. Each maximum position in terms of ADC units is mapped to the
corresponding energy deposition. The energy scale is determined from the Monte Carlo
simulation as the scale factor between the energy deposition in scintillator plates for the given
structure of an electron with 1-GeV energy and a relativistic muon with energy above 1 GeV.
From this proportion, we estimate the MIP signal in this calorimeter to be 240 MeV. This value
divided by the position of the muon's peak maximum is used as a calibration coefficient for each
cell. This calibration procedure involving the MIP energy deposition is not absolute or
conclusive. Primarily, it aligns the gain coefficients in each cell to ensure   an equal response for each
cell. The measured electron or photon energy can be further revised by reconstructing neutral
pions or calibrating the calorimeter using electron or photon beams of known energy. The
electromagnetic calorimeter measures electron or photon energy by summing up signals from all
16 cells. Each cell can only contain a fraction of energy deposited by the particle in the
calorimeter (if the particle is not a relativistic muon or a MIP). The energy resolution of the
calorimeter for vertical cosmic ray particles is 9.0\% (Fig.~\ref{fig:ecal:test_energy_spectrum} (a)). This value corresponds to
energy deposition of 240 MeV. Assuming the resolution depends on the energy as $\sqrt{E}$, the
energy resolution at 1 GeV is estimated to be 5\%. The energy resolution dependence on the
ADC's threshold is shown in Fig.~\ref{fig:ecal:test_energy_spectrum} (b).

\subsection{Long-term stability}

Temperature dependence of calorimeter stability was estimated with the usage of daily temperature
variations in the range of 18-26 $^\circ$C. Signals from cosmic rays particles as well as signals from
LEDs of a 1-Hz frequency were captured during the measurement over 10 days. The photodiode's
temperature was continuously monitored by FE-board through MPPC's board thermistor. The
bias voltage on photodiodes was corrected during data capture in accordance with a measured
temperature using a linear dependence:

\begin{equation}
U_{bias}(T) = U_{base} + K_t \times (T-20),
\end{equation}

where T is the MPPC board temperature, ($^\circ$C), 
\newline $U_{bias}(T$ is the total bias voltage applied to MPPC at temperature T, (V), 
\newline $U_{base}$ is the MPPC base voltage from power supply \cite{ecal:hvsys_small} at 20 $^\circ$C (V).

The temperature coefficient $K_t$ = 0.054 V/$^\circ$C that was used for temperature compensation of bias
voltage is shown in Fig.~\ref{fig:ecal:mppc_breakdown} (a, b), and depends on MPPC type.

\begin{figure}[!h]
    \begin{minipage}[b][][b]{0.50\linewidth}
    \center{\includegraphics[width=1\linewidth]{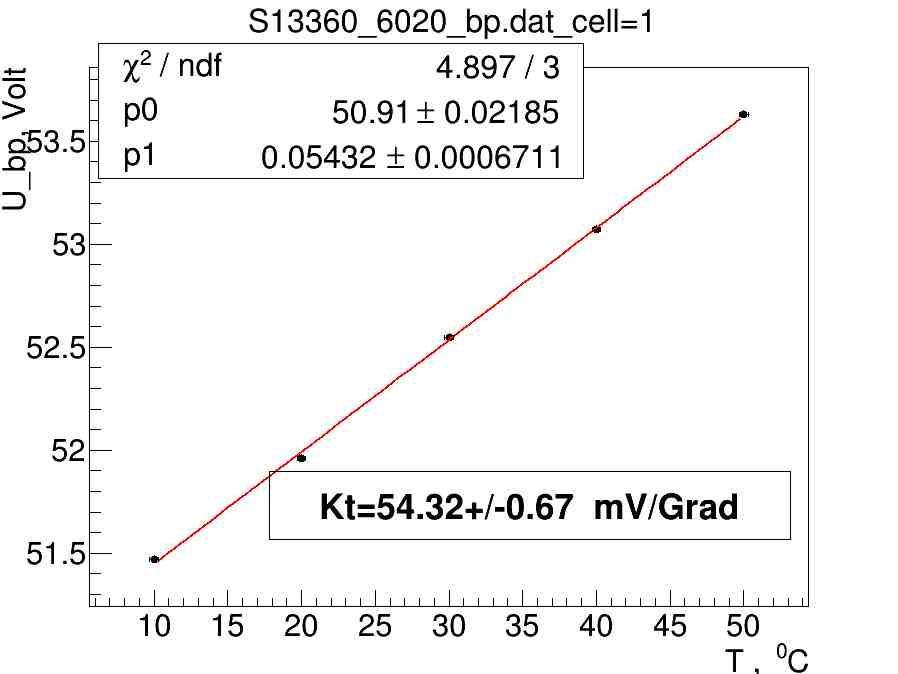}}
    \captionsetup{labelformat=empty}
    \caption{(a)}
    \addtocounter{figure}{-1}
  \end{minipage}
  \hfill
    \begin{minipage}[b][][b]{0.50\linewidth}
    \center{\includegraphics[width=1.\linewidth]{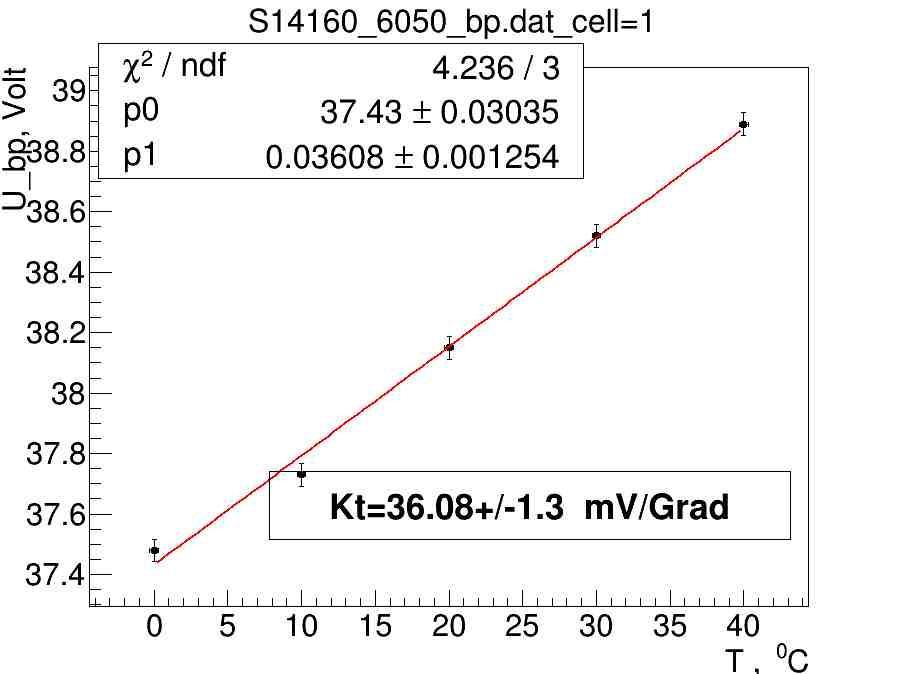}}
    \captionsetup{labelformat=empty}
    \caption{(b)}
    \addtocounter{figure}{-1}
  \end{minipage}
  \caption{Dependencies of MPPC breakdown voltage on temperature for different MPPC
    types: (a) for S13360-6025, (b) for S14160-6050.}
  \label{fig:ecal:mppc_breakdown} 
\end{figure}

The lower temperature dependence typical for diodes with a low breakdown voltage as
S14160-6050. But this type of MPPC has few pixel numbers ($\sim$14000) and therefore recently new
type S14160-6015 was developed with 160000 pixels of 15 $\mu$m pixel pitch and it will be
tested soon.

\begin{figure}[!h]
    \begin{minipage}[b][][b]{0.50\linewidth}
    \center{\includegraphics[width=1\linewidth]{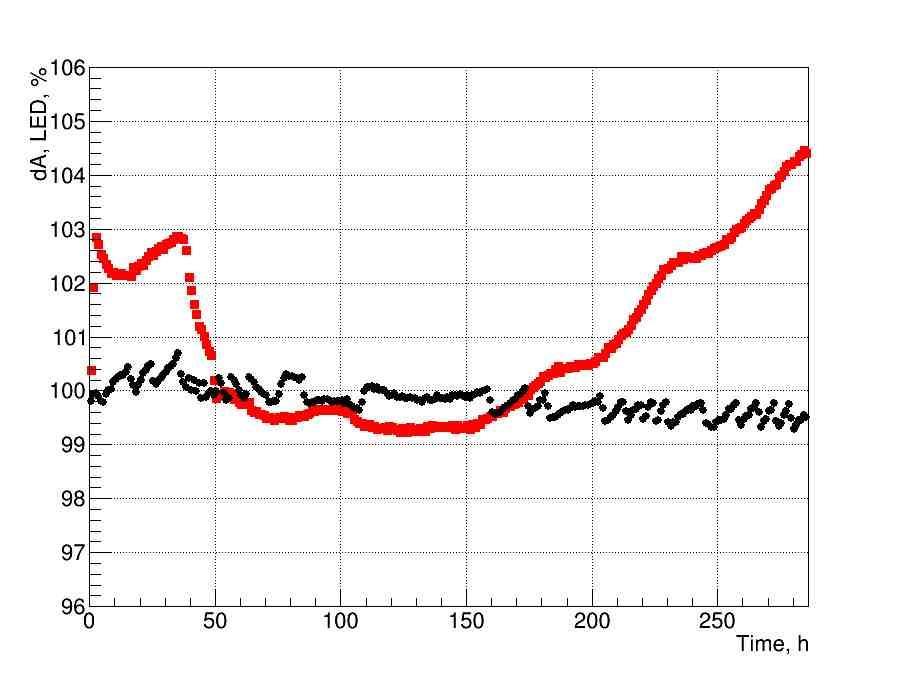}}
    \captionsetup{labelformat=empty}
    \caption{(a)}
    \addtocounter{figure}{-1}
  \end{minipage}
  \hfill
    \begin{minipage}[b][][b]{0.50\linewidth}
    \center{\includegraphics[width=1.\linewidth]{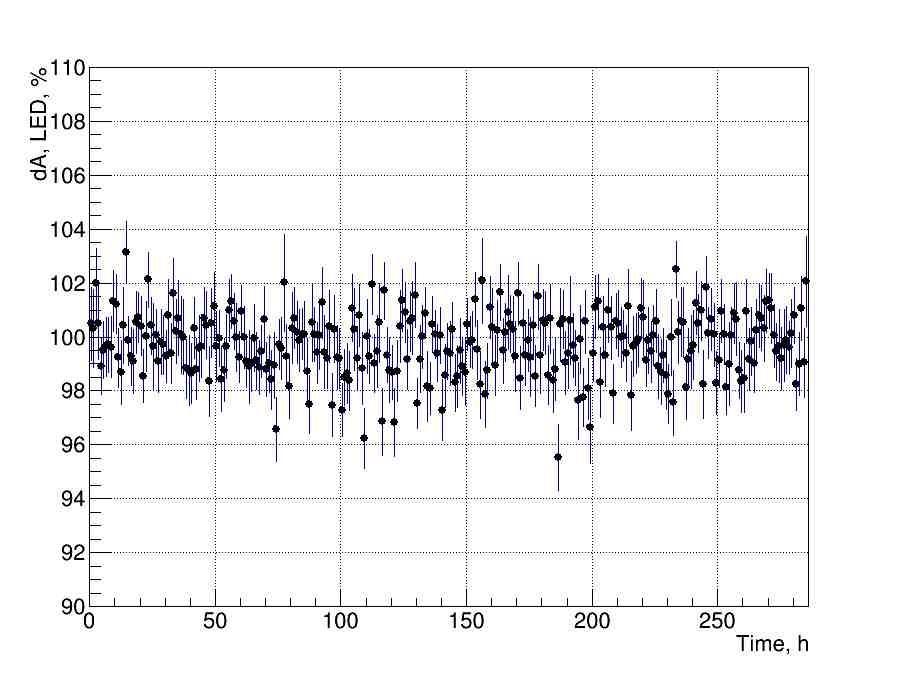}}
    \captionsetup{labelformat=empty}
    \caption{(b)}
    \addtocounter{figure}{-1}
  \end{minipage}
    \caption{Dependences of the sum (average value) of MPPC signals on   the time of measurement
    (in hours). (a) LED signals with temperature-dependent bias voltage compensation (black) and
    ones without compensation (red). (b) MIP signals with temperature-dependent bias voltage.}
  \label{fig:ecal:voltage_compensation} 
\end{figure}

Dependences of the calorimeter signal on the measurement time are shown   in Fig.~\ref{fig:ecal:voltage_compensation} (a, b).
These data are presented in \% with respect to the first 5 minutes of the measurement period for
normalization. The temperature variations during the measurement were about $\pm 4 ^\circ$C. After
compensation was employed, daily variations in the MPPC signal amplitude were constrained
within $\pm$0.5\%. The calorimeter can operate with a stability of $\sim$1\% during the time over 10 days if
the temperature compensation of the operating voltage is maintained, as it can be seen from these
results and is shown in Fig.~\ref{fig:ecal:voltage_compensation} (a, b).



\section{Cost estimate and the time scale}
\label{sec:ecal:cost_time}

The cost of the calorimeter is proportional to the number of channels. Mechanical assembly of a
calorimeter cell from the scintillator and the lead plates costs 200\$ per channel. Another
expensive element is the wavelength-shifting fibers. For a 40$\times$40 $\text{mm}^2$ cell, 16 fibers of the total
length of 8 m are used. Assuming an average price of 10\$/m, the price per channel amounts to 96\$.
The cost of photodiodes depends on the quantity. For purchases of tens of thousands of
units, their price is about 33\$ per unit. The electronics also contributes significantly to the total
cost, especially the ADC with a price of 45\$ per channel. 
The cost of the front-end (FE) and voltage (HV) control systems is 38\$ per channel.
The total cost of a calorimeter cell is about 442\$. 
Thus, the total cost of a 27168-cell calorimeter is 12.0 M\$, as it is shown in Table ~\ref{tab:ecal:cost}. 
We need to prepare  the technical documentation with
drawings of the calorimeter components. Then we can start production operations. The readout front-end
electronics should correspond to   the data acquisition system. It should have low power
consumption and a high rate of data acquisition. For these,   a new electronic base should be used to
achieve these goals.

\begin{table}[h]
\centering
    \caption{\label{tab:ecal:power} Power consumption of the ECal in kW for: 2 -- ADC \cite{afi_VME},  3 -- FE-boards \cite{ecal:hvsys},
    4 -- HV power units, 5 -- total in kW.}
    \begin{tabular}{|c | c | c | c | c|} 
 \hline
    1 & 2 & 3 & 4 & 5 \\ 
 \hline
    Item & ADC & FE & HV & Total \\
 \hline
    mW/ch & 200 & 63 & 1.2 & 264 \\

    Barrel & 3.4 & 1.1 & 2.4 & 6.8 \\

    End-cap-1 & 1.0 & 0.3 & 0.4 & 1.7 \\

    End-cap-2 & 1.0 & 0.3 & 0.4 & 1.7 \\
 \hline
    Total & 5.4 & 1.7 & 3.2 & 10.3 \\ 
 \hline
\end{tabular}
\end{table}

\begin{table}[h]
\centering
    \caption{\label{tab:ecal:number_components} The number of   the main ECal components: 2 -- the number of calorimeter cells, 3 -- the number of
    64-channel ADC boards, 4 -- the number of 16-channel front-end boards (FE-boards), 5 -- the number of
    HV power control units, 6 -- the number of MPPC , 7 -- the total length of WLS fiber in meters.}
    \begin{tabular}{|c |c |c |c |c |c |c|} 
 \hline
    1 & 2 & 3 & 4 & 5 & 6 & 7 \\  
 \hline
    Item & N cells & ADC64E & FE & HV & MPPC & Fiber, m \\ 
 \hline    
    Barrel & 16896 & 264 & 1056 & 24 & 16896 & 162202 \\
    EC-1 & 5136 & 80 & 321 & 4 & 5136 & 24653 \\
    EC-2 & 5136 & 80 & 321 & 4 & 5136 & 24653 \\
 \hline
    Total & 27168 & 426 & 1698 & 32 & 27168 & 217344 \\ 
 \hline
\end{tabular}
\end{table}

\begin{table}[h]
\centering
    \caption{\label{tab:ecal:weight} The weight distribution for the ECal parts. Supporting structures with a total weight of 11.5 tons are not included in the table.  Columns 2, 3 are related to the absorber (lead) and
    scintillator in   a single cell; 
    4 -- single cell weight in kg;  5, 6 -- lead and scintillator contribution in
    tons. 7 -- the total weight of the calorimeter main parts,   the
    supporting mechanical frame is also taken into account.}
\begin{tabular}{|c|c|c|c|c|c|c|}
\hline
1 & 2    & 3     & 4    & 5     & 6     & 7     \\
\hline
Material             & Lead & Scint & Cell & Lead  & Scint & Total \\
Units                 &   kg   &  kg    &  kg   &  ton   &  ton   &  ton   \\
Barrel                & 2.07 & 0.56  & 2.63 & 44.50 & 8.35  & 52.86 \\
EC-1                 & 1.82 & 0.49  & 2.31 & 11.87  & 2.54  & 14.41 \\
EC-2                 & 1.82 & 0.49  & 2.31 & 11.87  & 2.54  & 14.41 \\
\hline
Total                 & 5.70  & 1.55  & 7.25 & 68.23 & 13.43 & 81.67 \\
\hline
\end{tabular}
\end{table}

\begin{table}[h]
\centering
    \caption{\label{tab:ecal:cost} The calorimeter cost  estimate. EC-1 and EC-2 -- the cost of the end-caps. 
    The cost per channel of various calorimeter components in \$ is shown on the bottom line. 
    In the lines above, the cost in k\$ is shown for the barrel, end-caps, and the entire calorimeter.
    The cost of assembling  the calorimeter modules and manufacturing of the supporting mechanical
    frame is estimated in columns 6 and 7.}
    \begin{tabular}{|c |c |c |c |c |c |c |c |c |c |c |c|} 
 \hline
    1 & 2 & 3 & 4 & 5 & 6 & 7 & 8 & 9 & 10 & 11 & 12 \\ 
 \hline
    Item & N cells & ADC & FE & HV & Assembl. & Frame & MPPC & Fiber & Lead & Scint & Total \\ 
    Barrel & 16896 & 752 & 596 & 54 & 3379 & 300 & 558                & 1622 & 920 & 686 & 7465 \\
    EC-1  & 5136 & 229 & 181    & 9 & 1027 & 100 & 169                   & 493    & 280 & 209 & 2278 \\
    EC-2  & 5136 & 229 & 181    & 9 & 1027 & 100 & 169                 & 493   & 280 & 209 & 2278 \\
\hline
    Total & 27168 & 1209 & 958 & 72 & 5433 & 500 & 897 & 2608   & 1480 & 1103  & 12022 \\ 
 \hline
 \hline
   \$/ch &            & 45     & 35 & 3 & 200  & 18  & 33  &  96 &  54 & 41  & 442 \\ 
 \hline
\end{tabular}
\end{table}

\begin{table}[h]
\centering
    \caption{\label{tab:ecal:manufacture_time} ECal components manufacturing time estimate. 2 -- number of cells; 3*, 4* -- time  for 
    scintillator and lead plates production,   in days; 5 -- time for calorimeter modules assembling, in days,
    at an estimated production rate of 10 modules per day; 6 -- the total time required for the calorimeter's main parts
    production, in years. *) -- tasks in columns 3 and 4 must be carried out in parallel.}
    \begin{tabular}{|c |c |c |c |c |c|} 
 \hline
    1 & 2 & 3* & 4* & 5 & 6 \\ 
 \hline
    Item & N cells & Scintillator & Lead & Assembling & Years \\  
 \hline
    Barrel & 16896 & 279 & 279 & 422 & 1.8 \\
    EC-1 & 5136 & 169 & 169 & 128 & 0.5 \\
    EC-2 & 5136 & 169 & 169 & 128 & 0.5 \\
 \hline
\end{tabular}
\end{table}

\chapter{Time-of-Flight system}
The purpose of the time-of-flight (TOF) system is to distinguish between charged particles of different masses in the momentum range up to a few GeV/$c$. The current configuration of the TOF detector at about 87.7 cm from the collision point implies that to separate pions from kaons up to a momentum of 1.2 GeV/$c$, the time resolution of the TOF system would have to be 70 ps or better. The longitudinal RMS of the collision region is about 30 cm, so, the true collision event time $t_0$ can not be set to the nominal bunch crossing clock time from the accelerator as the uncertainty would be of the order of 1 ns (larger than the required time resolution). Therefore, the TOF system will be used in events with multiple charged tracks to determine both the collision event time $t_0$ and the individual track identification. For details of this analysis, see Section 1.4 of Chapter 9 of the SPD CDR \cite{SPD:2021hnm}. In addition to particle identification, the detector will also provide a start time for the straw drift tubes.

The TOF system will consist of a barrel and two end-cap parts with an overall active area of 22.6 m$^2$. The charged particle rate that the detector will have to withstand is 0.1 kHz/cm$^2$ for the barrel. The rate increases rapidly when moving closer to the beam axis. Thus, for the TOF elements located in the end-caps 20 cm off the beam axis, the rate of about 1 kHz/cm$^2$ is expected (see Fig. \ref{fig:flux} for details). The MRPC technology is being considered for the TOF detector.

\section{General layout}
\begin{figure}[!ht]
  \begin{minipage}[ht]{1\linewidth}
    \center{\includegraphics[width=1\linewidth]{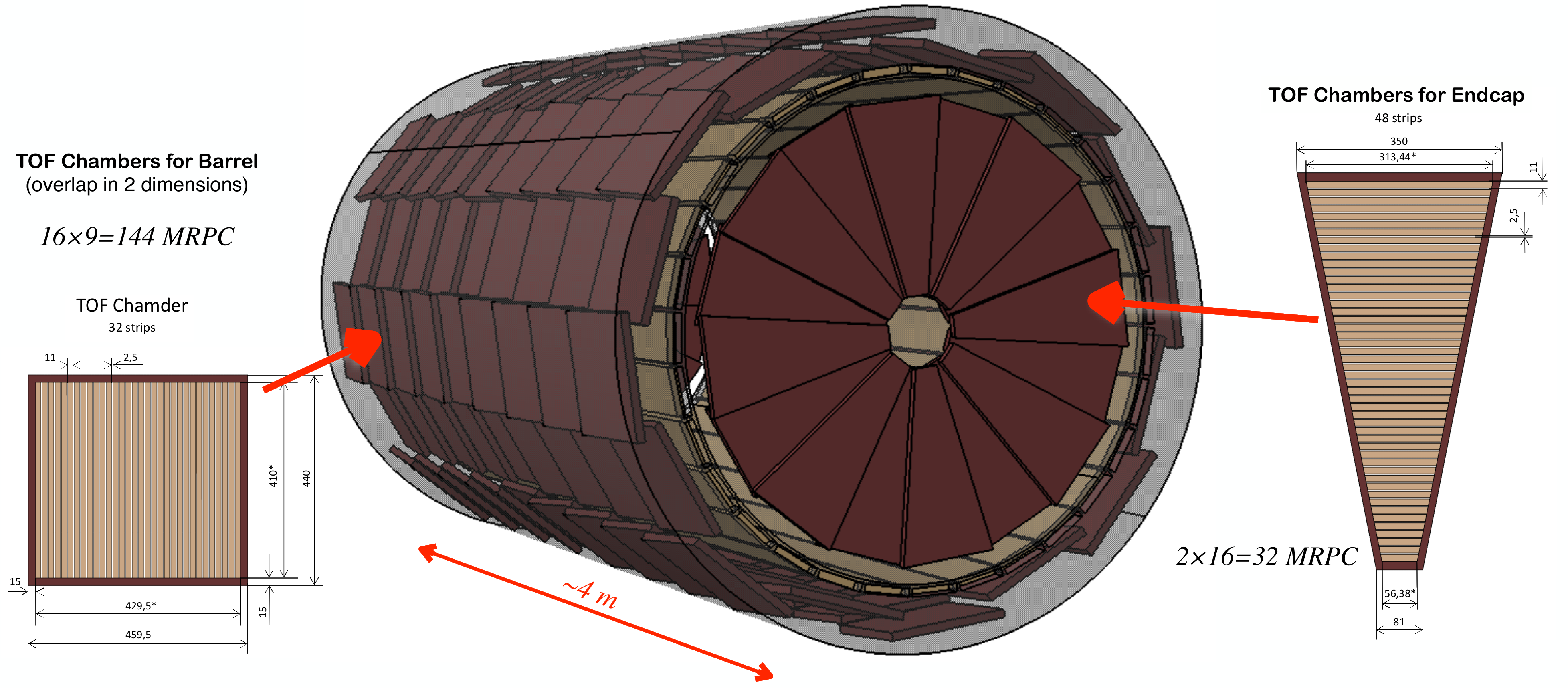}}
  \end{minipage}
    \caption{Multigap Resistive Plate Chambers, MRPCs, are considered for the PID system. Barrel and one of two end-cap parts are shown. }
\label{TOF_general}
\end{figure}

The layout of the SPD TOF detector is shown in Fig. \ref{TOF_general}. 

   The Multigap Resistive Plate Chamber (MRPC) is a stack of resistive glass plates with high voltage applied to external surfaces.  The pickup electrodes are located inside the chamber. A fast signal, induced on the pickup electrodes by an electron avalanche, is further transported to FEE located nearby. To increase the speed of gas exchange and reduce gas consumption, we designed a new sealed MRPC. The schematic cross-section of the MRPC is shown in Fig. \ref{TOF_sealed} \cite{Wang:2021mmx}. In order to achieve a good time resolution of around 50 ps, two MRPC stacks with 10$\times$0.22 mm gaps  can be used in SPD. The number of MRPCs for barrel and end-caps TOF, and the number of the corresponding readout channels (two-sides readout) are shown in Table \ref{TOF_params}. In total, the TOF system consists of 176 MRPCs and 12288 readout channels.
    
    \begin{table}[htp]
\caption{Parameters of the SPD TOF system.}
\begin{center}
\begin{tabular}{|l|c|c|c|}
\hline
  & Number of MRPCs & Number of strips per MRPC & Number of readout channels \\
  \hline
  Barrel       &  144        &  32    & 9216 \\
  End-caps &$2\times16$ & 48    &  3072 \\ 
\hline
 Total         &  176  &     &12228\\ 
\hline
\end{tabular}
\end{center}
\label{TOF_params}
\end{table}%

\section{MRPC-based TOF system}

\begin{figure}[!ht]
  \begin{minipage}[ht]{0.49\linewidth}
    \center{\includegraphics[width=1\linewidth]{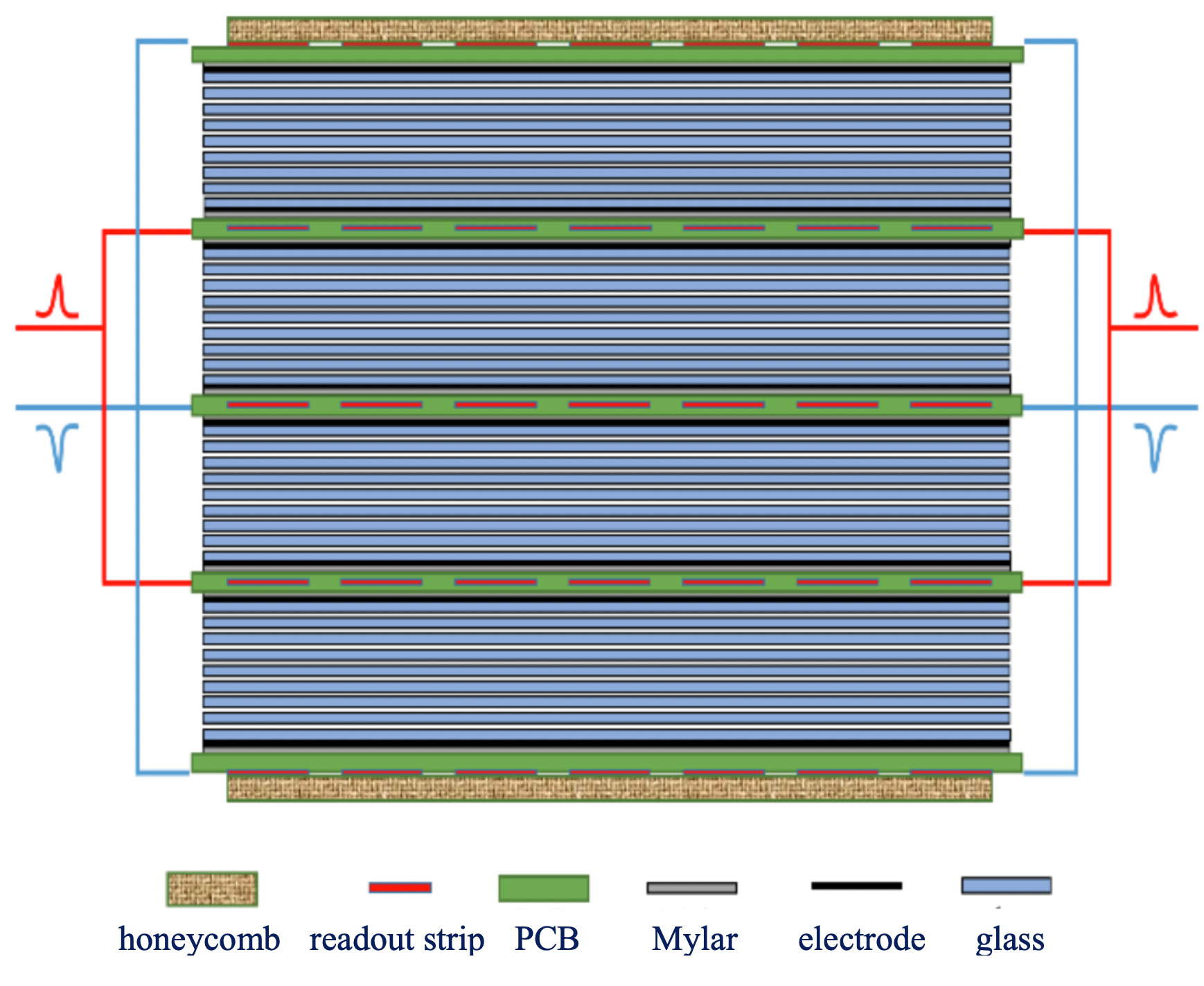} \\ (a)}
  \end{minipage}
  \hfill
  \begin{minipage}[ht]{0.49\linewidth}
    \center{\includegraphics[width=1\linewidth]{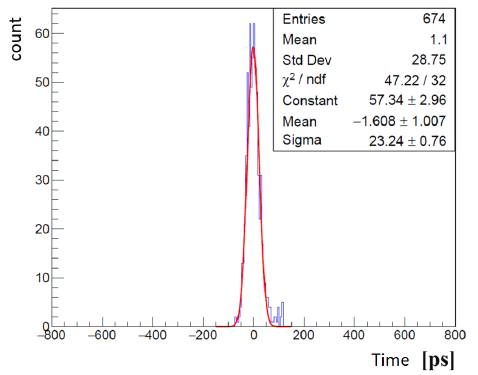} \\ (b)}
  \end{minipage} 
  \caption{(a) Schematic picture of the 32-gap MRPC chamber \cite{Wang:2020iwn}.
  	(b) Time resolution results of the cosmic test.
	}
\label{TOF32}
\end{figure}

The required time resolution for SPD is better than 60 ps, while the efficiency of particle registration at a high rate (few kHz/cm$^2$) should be above 98\%. Based on the experience of building similar systems in such experiments as ALICE \cite{Akindinov:2004fr}, HARP \cite{Ammosov:2009zza}, STAR \cite{STARTOF1}, PHENIX \cite{PHENIXTOF1}, and BM@N \cite{BMN1}, a glass Multigap Resistive Plate Chamber could be used as a base time detector. For example, the TOF-700 wall in the BM@N experiment, placed at a distance of 8 m from the target, provides the $\pi$/$K$ separation of up to 3 GeV/c and $p/K$ separation of up to 5 GeV/$c$, under the assumption that the time resolution of the start timing detector is below 40 ps. As we know, the time jitter of MRPC should be correlated with the width of the gas gap.  In order to study the intrinsic time resolution of MRPC, a 32-gap MRPC with 0.128 mm of gas gap was developed. The structure is shown in Figure \ref{TOF32}~(a) \cite{Wang:2020iwn}. Six readout strips with 1 cm pitch are configured on the PCB sheets. Five PCBs are required in this design. The cathode and anode signals are transmitted through differential cables. During the preliminary cosmic ray test, the high-performance analog front-end electronics and the Lecroy oscilloscope (10 GHz pulse sampling) were used. The crossing time of a signal is determined when setting a fixed threshold, and it is related to the amplitude of the signal. The time spectrum of the difference of the two MRPCs is shown in Figure \ref{TOF32}~(b). It can be seen that the time resolution of each MRPC is 23.24 ps / $\sqrt{2} = $16.4 ps. 

Besides the implementation of the MRPC type described above, it is possible to realize cameras in the
variant with the application of the resistive layer on each of the camera glasses.
This option was created and tested in IHEP (Protvino) \cite{tof_talk}, and it showed the time resolution better than 40 ps.
An additional advantage in terms of reduced power supply voltage, in comparison with the camera described above.

\section{Advantage of sealed MRPC}
\begin{figure}[!ht]
  \begin{minipage}[ht]{1\linewidth}
    \center{\includegraphics[width=1\linewidth]{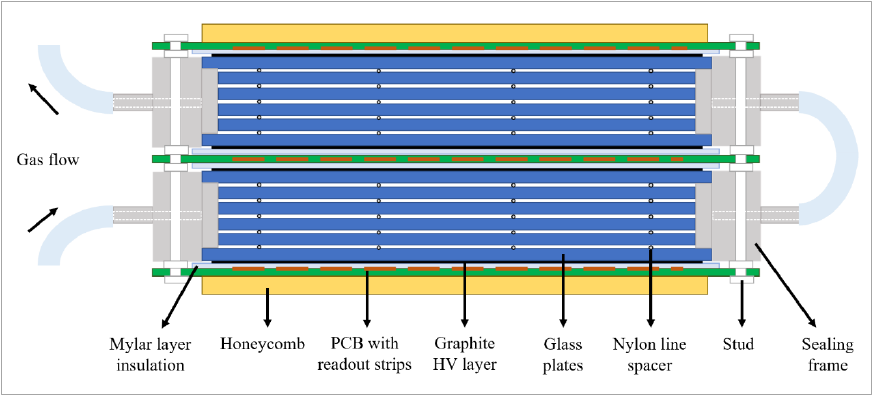}}
  \end{minipage}
    \caption{Schematic view of sealed MRPC \cite{Wang:2021mmx}.}
\label{TOF_sealed}
\end{figure}

The choice of the working gas mixture for MRPCs has always been an important topic. It should allow the MRPC detector to perform successfully and stably for different purposes, and be eco-friendly at the same time. This indicates that the gas mixture should have a low ozone depletion power and global warming potential (GWP). The tetrafluoroethane currently used in MRPCs is ozone-friendly but with a GWP of about 1430 (the reference GWP of CO$_2$ is 1). Therefore, a lot of research has gone into looking for possible replacements. Among the possibilities, HFO-1234ze (1,3,3,3-tetrafluoropropene, C$_3$H$_2$F$_4$)  with a GWP of 6 is one of the most popular candidates, and tests of gas mixtures based on it are ongoing.
Another reasonable approach is to reduce gas consumption or recycle gas. The CSR external target experiment (CEE) in Lanzhou, China, will adopt a sealed technology of MRPC to construct the TOF system. The MRPC detector, shown in Figure \ref{TOF_sealed}, is sealed by gluing an integral 3D-printed frame and the outermost electrodes together. It can operate stably with a gas flux of 4 ml/min, which is extremely low, compared to when MRPCs are placed in a sealed box. 

\subsection{Prototype test results}
\begin{figure}[!ht]
  \begin{minipage}[ht]{0.49\linewidth}
    \center{\includegraphics[width=1\linewidth]{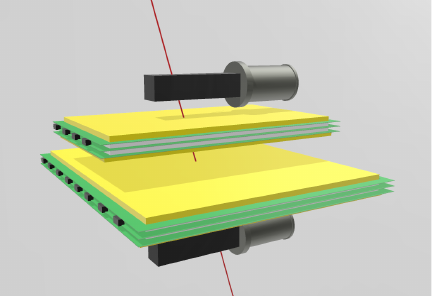} \\ (a)}
  \end{minipage}
  \hfill
  \begin{minipage}[ht]{0.49\linewidth}
    \center{\includegraphics[width=1\linewidth]{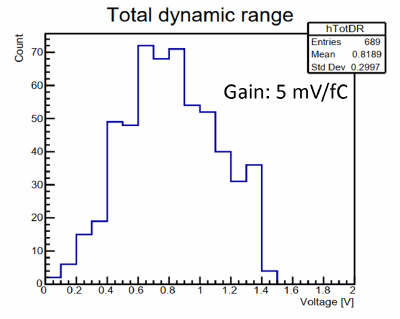} \\ (b)}
  \end{minipage} 
  \begin{minipage}[ht]{0.44\linewidth}
    \center{\includegraphics[width=1\linewidth]{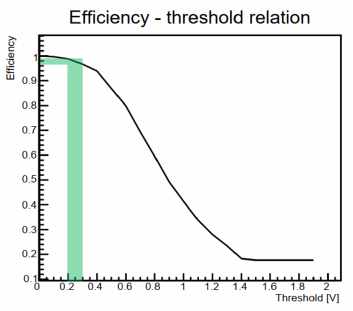} \\ (c)}
  \end{minipage}
  \hfill
  \begin{minipage}[ht]{0.54\linewidth}
    \center{\includegraphics[width=1\linewidth]{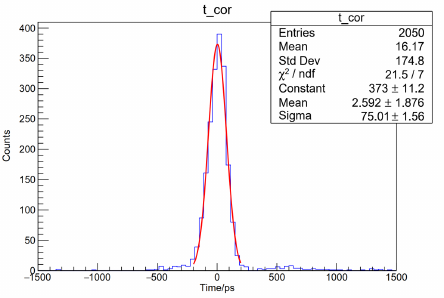} \\ (d)}
  \end{minipage} 
  \caption{(a) Cosmic test layout. 
  	(b) Dynamic range of signal amplitudes of the sealed MRPC.
	(c) Efficiency vs. threshold dependence for the sealed MRPC.
	(d) Flight time distribution of two prototypes.
	}
\label{TOF_cosmic}
\end{figure}

Two sealed MRPC prototypes have been assembled for the testing performance. They were different in number of strips (32 and 16), while the same in geometry parameters: 10 gas gaps of 0.25 mm thickness, 48 $\times$ 1.5 cm$^2$ strip size, and 0.2 cm gap between strips. Working (sensitive) area of 5 cm width was split into a 4-strip interval and signals
were read out from a connector located at both ends of the strips. Under cosmic rays, the signal characteristics and time resolution were examined with different readout methods. Figure \ref{TOF_cosmic}~(a) shows the cosmic test layout, in which both prototypes operate under $\pm$6900 V high voltage. The readout strips are read out at both ends for each counter. We use a fast amplifier (described in Section \ref{tof_el_1}) and a Tektronix MSO58 oscilloscope to process the waveform. In this way, the dynamic range of the MRPC signal is obtained, as shown in Fig. \ref{TOF_cosmic}~(b). It can be translated to a 40-200 fC dynamic range.

The distribution is filled by the amplitudes of the largest signal for each event, so the loss of these signals will lead to the loss of detection. Based on the idea above, the MRPC efficiency in variant electronic thresholds is plotted in Fig. \ref{TOF_cosmic}~(c). Based on the result, the suggested threshold is 40 fC, and the prototype can reach an efficiency of over 97\%.

\begin{figure}[!ht]
  \begin{minipage}[ht]{1\linewidth}
    \center{\includegraphics[width=1\linewidth]{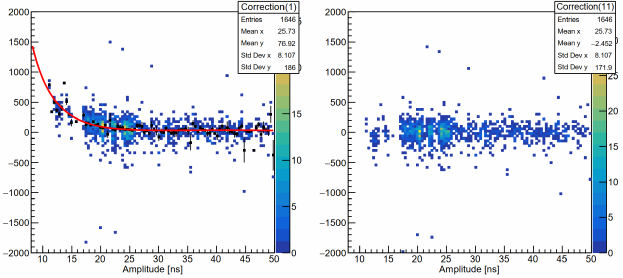}}
  \end{minipage}
    \caption{Slewing correction and the result. }
\label{TOF_corr}
\end{figure}

\begin{figure}[!ht]
  \begin{minipage}[ht]{0.44\linewidth}
    \center{\includegraphics[width=1\linewidth]{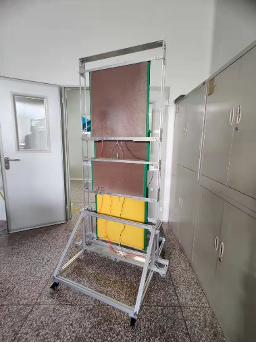} \\ (a)}
  \end{minipage}
  \hfill
  \begin{minipage}[ht]{0.54\linewidth}
    \center{\includegraphics[width=1\linewidth]{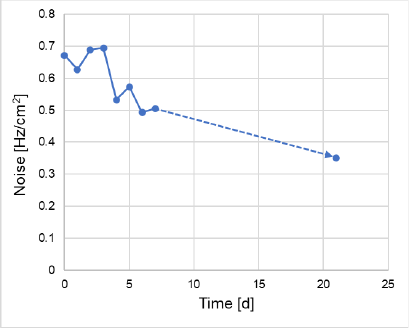} \\ (b)}
  \end{minipage} 
  \caption{(a) The CEE-eTOF supermodule.
  	(b) Noise decay of a brand new sealed MRPC.
	}
\label{TOF_tests}
\end{figure}

The dynamic range is very helpful when testing the prototypes with the NINO-TDC chain since the proper level of threshold for NINO has been determined in advance. Figure \ref{TOF_cosmic}~(d) shows the flight time distribution between two counters after a series of corrections, e.g., gain correction, time delay correction, and slewing correction in Figure \ref{TOF_corr}. For a single time start/stop of counter-electronic combination, the time resolution is $75/\sqrt{2}=53$ ps. Meanwhile, from the number of reconstructed events and that of triggers, we can estimate that the efficiency of both counters is greater than 97.3\%, It compares well with earlier tests.

At last, it is important to note that the two prototypes worked stably during the test period of about 25 days, with a gas flow of less than 5 ml/min. The working gas consumption is decreased by a factor of 10, compared to the operation of traditional MRPCs.

\subsection{Test results on the sealed MRPC constructed TOF supermodule}
The proposed prototype of the TOF system is based on the end-cap Time of Flight (eTOF) wall, developed for the external
target experiment (CEE)\cite{Wang:2020ehx}. It has a sensitive area of 320$\times$160 cm$^2$ divided into 7 supermodules.
Each module includes 3 or 4 sealed MRPCs according to the position of the module. The assembled
module for the test is shown in Fig. \ref{TOF_tests}~(a).
For the gas flow, we follow a strategy of connecting each gas chamber in sequence. All the counters worked stably at nominal high voltage under a 10 mL/min gas flow. During that time, noise level and dark current were recorded to study the detector training procedure. As it is shown in Figure \ref{TOF_tests}~(b), a brand new counter shows an initial noise level around 0.7 Hz/cm$^2$, and after the 3-week training in high voltage the noise dropped to 0.35 Hz/cm$^2$. This will benefit the signal-to-noise ratio in experiment operations. Besides, after the training the summed dark current of the supermodule settled to around 70 nA.

\section{TOF-related electronics}
A very important part of the high-performance time-of-flight system is the readout electronics. For the full exploitation of the excellent timing properties of the Multigap Resistive Plate Chamber, front-end electronics with special characteristics are needed. The signals from MRPCs must be amplified as fast as possible without losses. The leading times of the signal must be digitized and measured with accuracy much better than the time resolution of the detector. The readout electronics for the SPD TOF will consist of the front-end electronics (FEE) and the data acquisition system (DAQ).

To maximize the time performance of MRPC, the FEE should have a  high bandwidth and low time jitter. The time digitization  should also have a low jitter. The readout of MRPC can be done with the technologies described in \cite{LIU201953,5941022}.

\subsection{Option 1: fast amplifier + pulse shape analyzer \label{tof_el_1}}
\begin{figure}[!ht]
  \begin{minipage}[ht]{1\linewidth}
    \center{\includegraphics[width=1\linewidth]{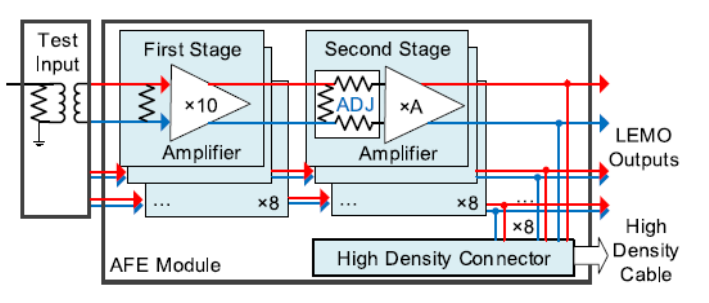}}
  \end{minipage}
    \caption{Schematic diagram of the USTC fast amplifier.}
\label{TOF_FE1}
\end{figure}

\begin{figure}[!ht]
  \begin{minipage}[ht]{1\linewidth}
    \center{\includegraphics[width=1\linewidth]{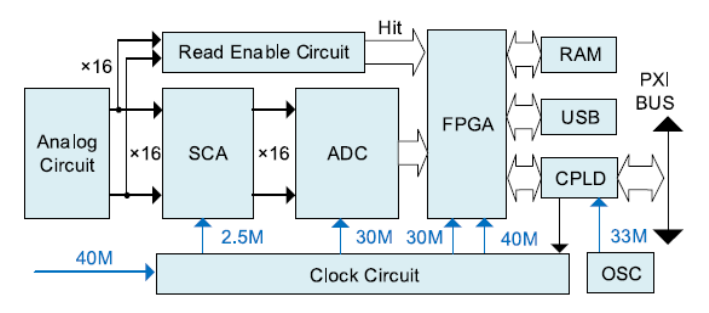}}
  \end{minipage}
    \caption{Schematic diagram of pulse shape digitizer. }
\label{TOF_FE2}
\end{figure}

As we know, the time jitter of a high sampling rate pulse shape analyzer is very low. For example,  the time jitter of DRS4 is usually less than 5 ps. The time jitter of a high bandwidth fast amplifier is also low. The schematic diagram of the USTC fast amplifier is shown in Fig. \ref{TOF_FE1}. It consists of eight channels and its time jitter is around 4 ps. The schematic diagram of the DRS4-based pulse shape digitizer is shown in Fig.~\ref{TOF_FE2}. 


   \begin{figure}[!ht]
  \begin{minipage}[ht]{1\linewidth}
    \center{\includegraphics[width=1\linewidth]{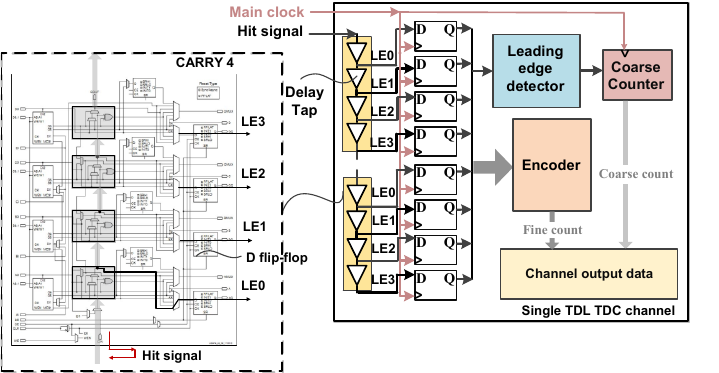}}
  \end{minipage}
    \caption{Schematic diagram of the FPGA TDC.}
\label{TOF_FE3}
\end{figure}
 
 \subsection{Option 2: CFD-based solution}
The analog signals from any detector have a spread in their amplitude and width. Therefore, to determine the time of occurrence of the signal very accurately, it is necessary to take into account the shape parameters of that signal. Traditionally, correction is carried out using three possible solutions: 
\begin{itemize}
\item hardware Constant Fraction Discriminator (CFD);
\item multichannel Analog to Digital Converter (ADC);
\item Time over Threshold method (ToT).
\end{itemize}

The first is a hardware solution, implemented in various applications of the CFD method. 
The commonly used in the past CFD method is currently not utilized  for the following reasons:
\begin{itemize}
\item the cost: cheap and fast ADCs have appeared on the market for some time;
\item the complex structure and specific components of the CFD dramatically increased in cost;
\item the CFD-based systems, known from past experiments, turned out to be ineffective and required constant complex configuration.
\end{itemize}

	The ADC is used in the second method. The method is characterized by high power consumption and price. The price also includes the cost of communication cables, which further reduces the reliability of the system.

	The ToT method has lowered the price for different applications. We are considering two versions of this method. The first and simplest one is implemented in the NINO chip. The second one is implemented in the electronics of the TOF detector of the BM@N experiment. The second one meets our requirements to a greater extent but has higher costs.
	Two methods of constructing the readout electronics with MRPC are being considered today. The first is the long-standing time-over-threshold method. The second one is the  constant fraction (CF) method. Both methods have their own advantages and disadvantages. The FE electronics for the ToT method are quite simple to manufacture and use. The ToT method shows the time resolution of about 45 ps, instead of the desired 30 ps. The CF method requires  more complex FE electronics. However, according to our measurements, this method allows one to receive a better time resolution. To implement the capabilities of the CF method, a TDC with a time resolution of at least 10 ps is required. An attempt is now being made to combine both methods to achieve the utmost time resolution. It should be noted that such amplifiers (ToT and CF), integrated into the microcircuit are not available now. Therefore, it was decided, at the first stage, to use commercial microcircuits of the amplifiers and comparators. The developed amplifiers with the CF method have already shown encouraging results. A time resolution of about 40 ps was obtained.
	The main priority of the readout system is its cost. The main contribution to the cost is made by multi-channel readout systems, including analog ASIC and TDC digitization channels.
	Schematic diagram of the FPGA-based TDC setup proposal presented in Fig. \ref{TOF_FE3}. 
	The main problem is that there are no  commercially available parts, unlike  the DAQ system, which is entirely based on commercial components.
	
\subsubsection{Special analogue ASIC for MRPC readout}
	Systems with a large number of readout channels require the use of special integrated circuits, such as NINO and PADI. We can not yet  get the right number of chips, but we could fix the parameters of these chips as  required to be completely sufficient. The Table  \ref{tab:nino} summarizes the main parameters of NINO and PADI ASICs.
	
	Both chips are specially designed for time-of-flight applications, but they have significant differences. The main difference of PADI is the absence of the ToT function of measuring amplitudes. The absence of the ToT function will require the use of an additional ADC channel to measure the charge. The second problem of PADI is a large Preamplifier Voltage Gain 250, against Gain 30 of the  NINO chip. For this reason, PADI is unstable, especially when building large systems. 
	PADI applications are limited, due to the absence of a pulse stretcher in its structure (pulse duration range is 1$\div$6 ns), the used TDCs have restrictions on the minimum duration of the input pulse, usually 5$\div$10 ns. This is an important parameter that must be considered when choosing both an amplifier-discriminator and TDC types.
	Summing up, we can say that the PADI chip, although developed keeping in mind manufacturing of the NINO chip and produced using faster technology, does not exceed the parameters of the NINO chip due to errors in the task formulation, and in some applications is inferior. For these reasons, when determining the required parameters of the analog part of the electronics in the project, we will use as a prototype the basic parameters of the NINO chip indicated in Table \ref{tab:nino}.

\begin{table}[htp]
\caption{NINO  and PADI ASICs specifications.}
\begin{center}
\begin{tabular}{|c|c|c|}
\hline
Main parameters comparison &	NINO ASIC (ALICE) &	PADI ASIC (GSI) \\
\hline
Channels per chip &	8&	8 \\
Conversion Gain, mV/fC	  &   1080	& 1900\\
PA Bandwidth, MHz &	500 & 410 \\
PA Voltage Gain, V/V &30 &	250 \\
ToT function available & Yes & No\\
Stretch timer &Yes &	No \\
Baseline DC offset, mV & 2 & 1 \\
Equivalent Noise Charge, e RMS &	1750	 & 1150 \\
Input Impedance Range, $\Ohm$ &35 $\div$ 75  &	30 $\div$160 \\
Power consumption, mW/channel &27 &17 \\
Threshold type &	external &	SPI protocol \\
Timing jitter, ps 	& $<5$ &	$<5$ \\
\hline
\end{tabular}
\end{center}
\label{tab:nino}
\end{table}%

 \section{TOF performance}
\begin{figure}[htb]
    \centering
\includegraphics[width=.95\textwidth]{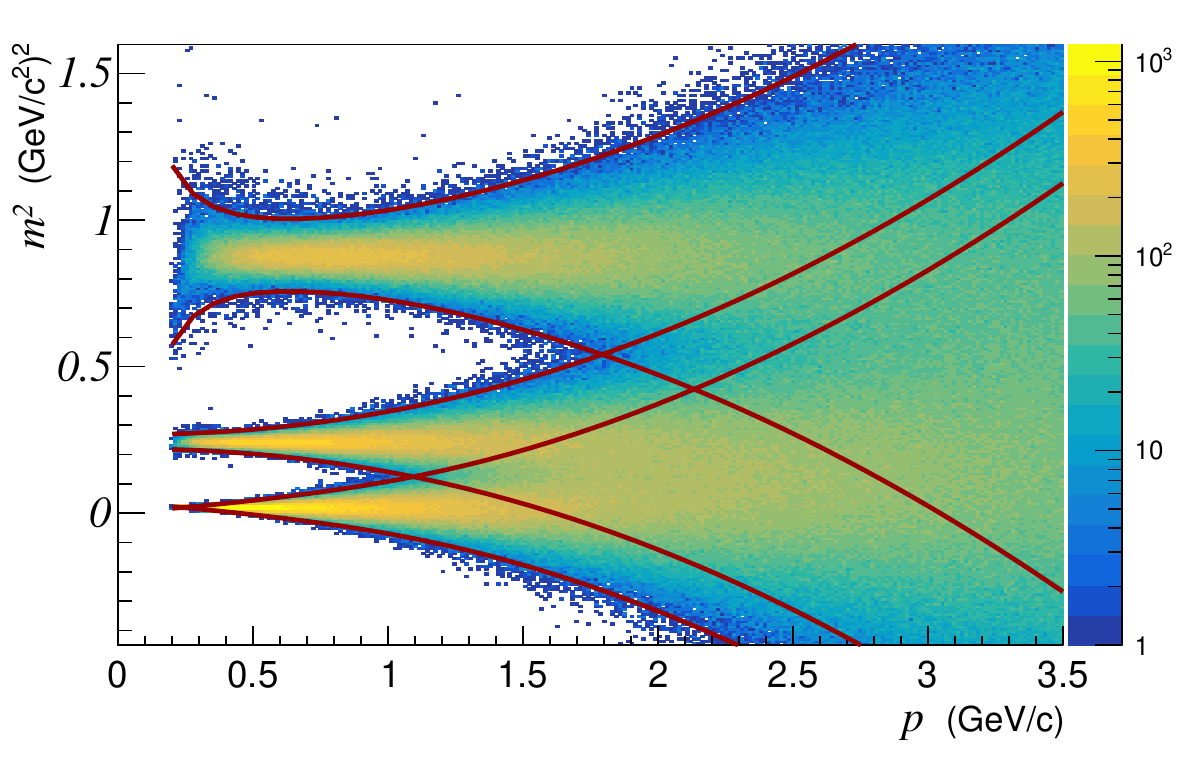}
\includegraphics[width=.95\textwidth]{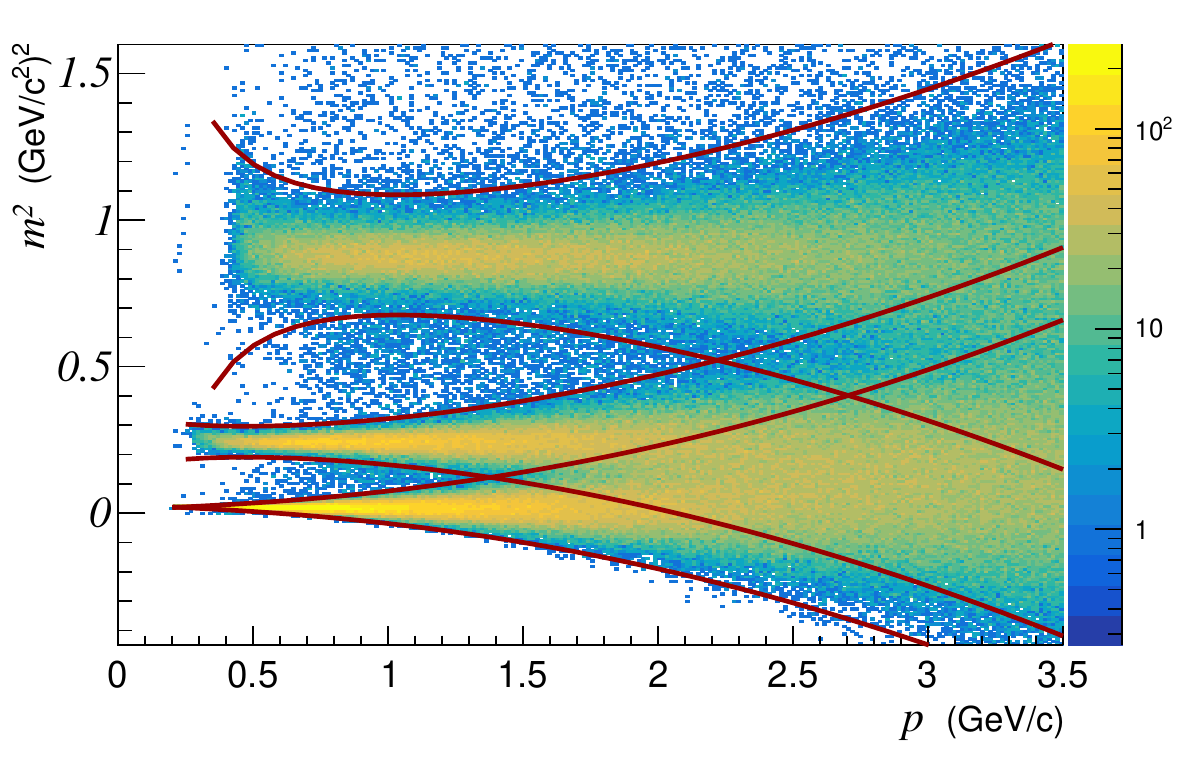}
    \caption{Reconstructed particle $m^2$ as a function of the initial track momentum for pions, kaons, and protons for the barrel (top) and the end-cap (bottom) parts of SPD. For each particle type, the $3\sigma$ intervals are also shown. The figure is obtained for the minimum bias sample of $pp$ collisions at 27~GeV.}
    \label{fig:tof-pid-perfomance}
\end{figure}

The performance of particle identification with TOF is estimated with MC simulation in the SpdRoot framework. For the simulation it assumed that $t_0$ is determined (see CDR~\cite{SPD:2021hnm} for details) and the flight-time resolution $t_{TOF}-t_0$, where $t_{TOF}$ is the TOF measurement) is 60~ps. Given the flight time and reconstructed track, the particle mass squared is reconstructed. In this procedure, the average velocity of each track from the Kalman fit is used.
The performance estimated for minimum bias events in the $p$-$p$ collisions at  27~GeV is shown in Fig.~\ref{fig:tof-pid-perfomance}.

 \section{Cost estimate}
  As it was mentioned above,  the whole TOF consists of 176 MRPCs and 12288 readout channels. At around 3000\$ per MRPC, the cost of 176 MRPCs is about 550 k\$. Assuming that the price of FFE and DAQ is about 120 \$ per channel, the total cost adds up to 1.5 M\$. The cost of ASIC development ($\sim$ 150 k\$) has to be accounted for additionally. So, the total cost estimate is 2.2 M\$.


\chapter{Focusing Aerogel RICH detector}
\begin{figure}[h]
	\begin{center}
		\includegraphics[width=0.8\textwidth]{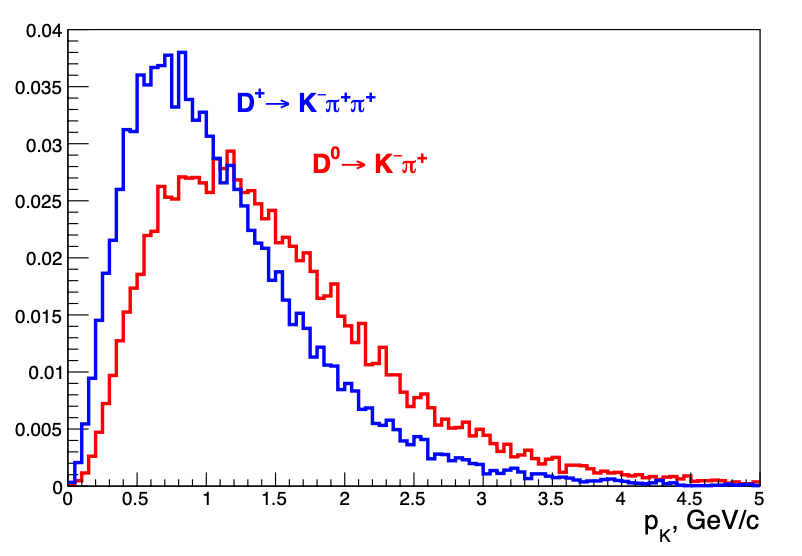}		
		\caption{Simulated kaons momentum from $D$-mesons decay.}
		\label{fig:kaonsmom}
	\end{center}
\end{figure}

According to the physics program of the SPD experiment, the major task of the specialised particle identification (PID) system is to separate pions and kaons in final open charmonia states.  The simulated momentum spectra for kaons from decays of the charged and neutral $D$-mesons are shown in Fig~\ref{fig:kaonsmom}.

\begin{figure}[h]
	\begin{center}
		\includegraphics[width=1.\textwidth]{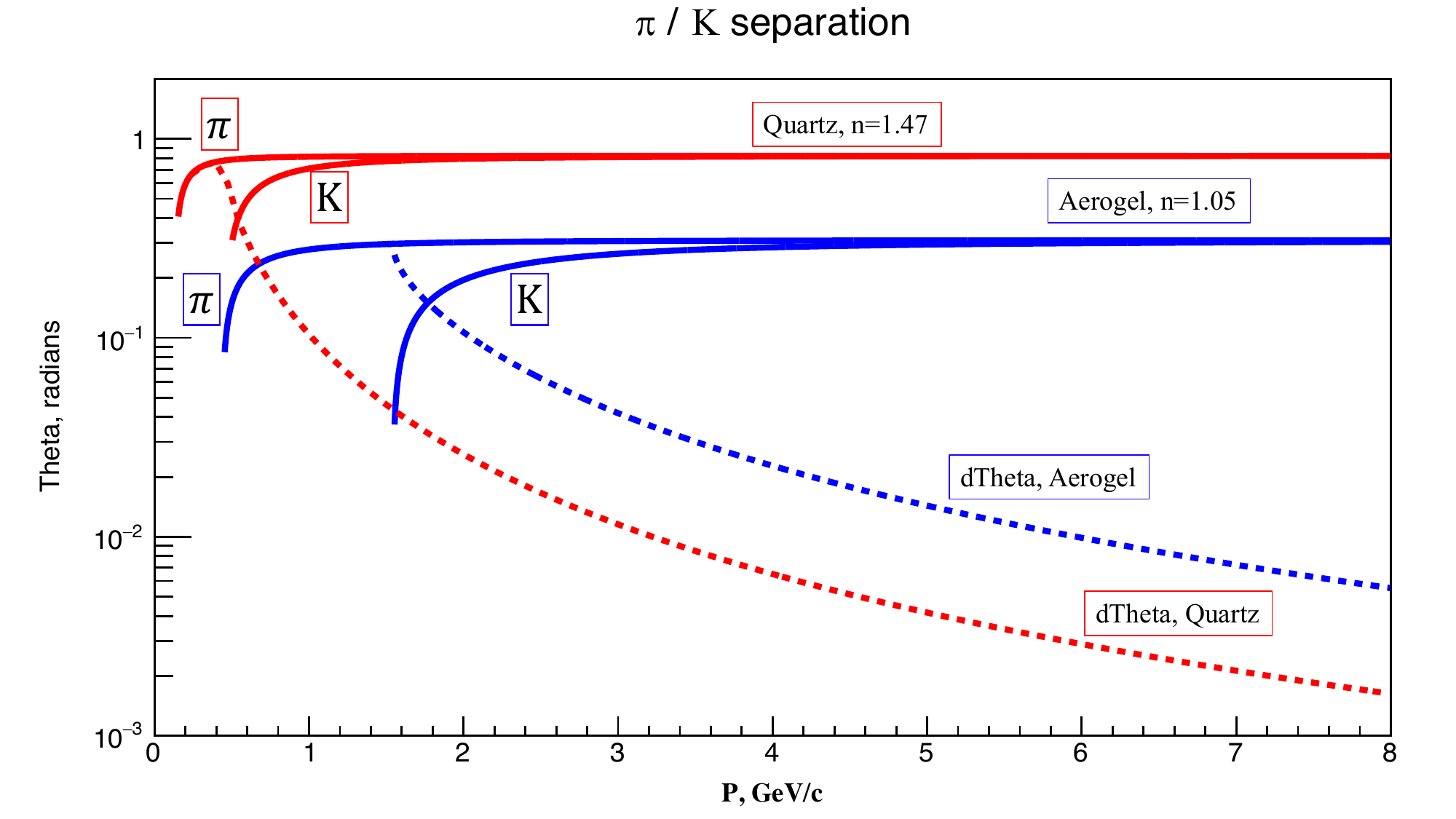} 
		\caption{Dependence of Cherenkov angles and their differences for $\pi$- and $K$-mesons on momenta.}
		\label{fig:dTheta}
	\end{center}
\end{figure}

The RICH detector is the best option to provide reliable $\pi/K$-separation in a wide momentum range (from 0.6 to 5.0 GeV/c). According to simple calculations it is clear that the refractive index of the Cherenkov radiator should be in the range from 1.03 to 1.06.  Only silica-based aerogels have such refractive indices and a rather good transparency in the visible light range. In Fig.~\ref{fig:dTheta} it is shown that the difference of the Cherenkov angles between pions and kaons with momentum $p=5.5$~GeV/c is about 3~mrad for refractive indices $n$=1.48 (quartz) and  about 10~mrad for $n=1.05$ (aerogel).  It means that for reliable separation at the level of three standard deviations (3$\sigma$) it is necessary to provide a Cherenkov angle resolution of about 3~mrad per track for aerogel ($n$=1.05), while for quartz ($n$=1.48) it has to be about 1~mrad, which is a more complicated task from the technical point of view, in case of a limited space for Cherenkov rings expansion, photon detector pixel size and refractive index dispersion. 

Silica aerogel is a porous silicon dioxide with a variable
index of refraction ($1.006\div 1.2$) that is applicable
in Cherenkov detectors for particles with a few GeV$/c$ momentum.
A first RICH detector using silica aerogel as a Cherenkov radiator 
was successfully employed in the HERMES experiment \cite{farich:Akopov2002}.
It was followed by RICH1 of the LHC$b$ detector \cite{farich:LHCbTDR}. Both
RICH detectors use gas as a second radiator and focusing mirrors 
to form Cherenkov ring images on a photon detection plane.

\begin{figure}[h]
	\begin{center}
		\includegraphics[width=0.8\textwidth]{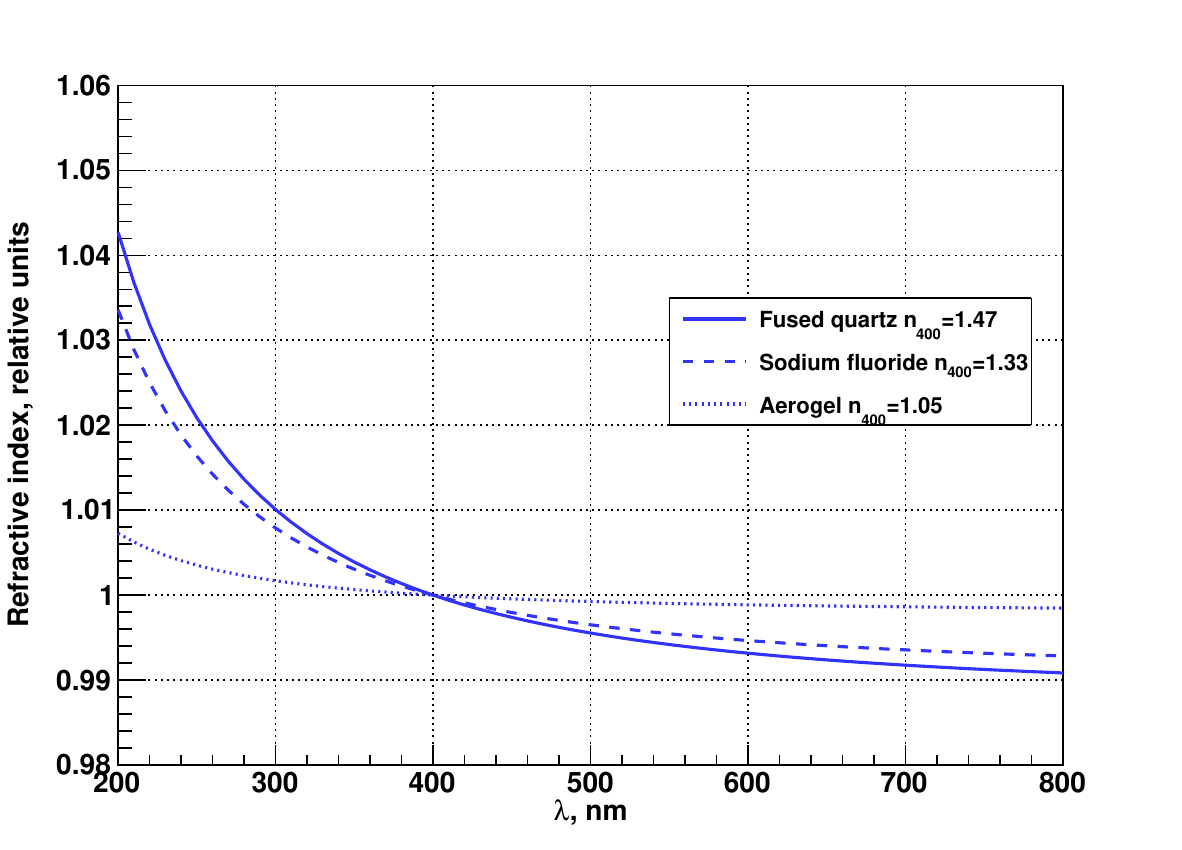} 
		\caption{Dependence of relative refractive indices normalised to $n_{400~nm}$ on wavelength for quartz, NaF and aerogel.}
		\label{fig:dnwl}
	\end{center}
\end{figure}

The major impacts on the Cherenkov angle resolution in RICH detectors are the following:
\begin{itemize}
\item geometry of the photon detector, such as photon sensitive pixel size $\Delta_{pix}$ and the distance from the radiator to the photon detection plane ($L$); 
\item Cherenkov radiator parameters, such as refractive index dispersion ($\sigma_{disp}=\frac{\sigma_n}{n}\frac{1}{\tan{\Theta_C}}$) and thickness ($t$);
\item Number of detected photons ($N_{pe}$), which depends on the photon detection efficiency (PDE) and Cherenkov radiators thickness.
\end{itemize}
In general, the Cherenkov angle resolution could be expressed as follows:
\begin{equation}
\sigma_{track}\approx\sqrt{\left(\frac{\Delta_{pix}\cos{\Theta_C}}{L\cdot\sqrt{12}}\right)^2/N_{pe}+\left(\frac{\sigma_n}{n\cdot\tan{\Theta_C}}\right)^2/N_{pe}+\left(\frac{t\cdot\sin{\Theta_{C}}}{L\cdot\sqrt{12}}\right)^2/N_{pe}}.
\end{equation}
If we assume that $L=300 mm$, $\Delta_{pix}=3$ mm, refractive index for aerogel $n=1.05$ and its relative dispersion $\frac{\sigma_n}{n}=0.001$ (see Fig.~\ref{fig:dnwl} for $\lambda=300\div700nm$), we will get the following dependence for a 1\,cm thick radiator for particles with a velocity very close to speed of light ($\beta\approx1$):
\begin{equation}
\sigma_{track}\approx\frac{1}{\sqrt{N_{pe}}}\cdot\sqrt{\left(0.0027\right)^2+\left(0.0032\right)^2+\left(0.003\cdot t[cm]\right)^2}.
\end{equation}
From this formula it is seen that to provide a Cherenkov angle resolution of about 2.5\,mrad, it is necessary to have 5 detected photons per track for a 1\,cm thickness.  Such a number of detected Cherenkov photons, on the one hand, looks like very optimistic hardly achievable upper limit for 1\,cm thick aerogel radiators. On the other hand, 5 detected photons per track is the minimal number of detected photons required to reconstruct a Chernkov ring in the experimental conditions, when most tracks are not perpendicular to the radiator plane and the rings transform to ellipses.  Therefore, it is highly required to increase the radiator thickness to increase the number of detected Cherenkov photons, but the angle uncertainty will increase $\sigma_{track}\propto\sqrt{t}$. That is why the proximity focusing RICH detector based on a multilayer focusing aerogel radiator  is considered to provide a reliable $\pi/K$-separation up to momentum $p=5\div6~$GeV/$c$. 

A RICH with proximity focusing uses a gap between a layer of the radiator 
medium and the photon detector to make a ring image. This allows one to
construct more compact RICH detectors as compared to RICH with focusing 
mirrors. Such a design was implemented in the RICH with the dual 
aerogel-NaF radiator for the ISS-born experiment 
AMS-02 \cite{farich:Buenerd2002}.

The Focusing Aerogel RICH (FARICH) concept was suggested in 2004~\cite{farich:Barnyakov2005,farich:Belle2005} and today its advantages are exploited in the ARICH system of the Belle II experiment at the SuperKEKb e$^+$e$^-$--\,collider (KEK, Tsukuba, Japan)~\cite{farich:arich}.


\section{FARICH concept}
\begin{figure}[h]    
\begin{minipage}{0.49\linewidth}
       \center{\includegraphics[width=1\linewidth]{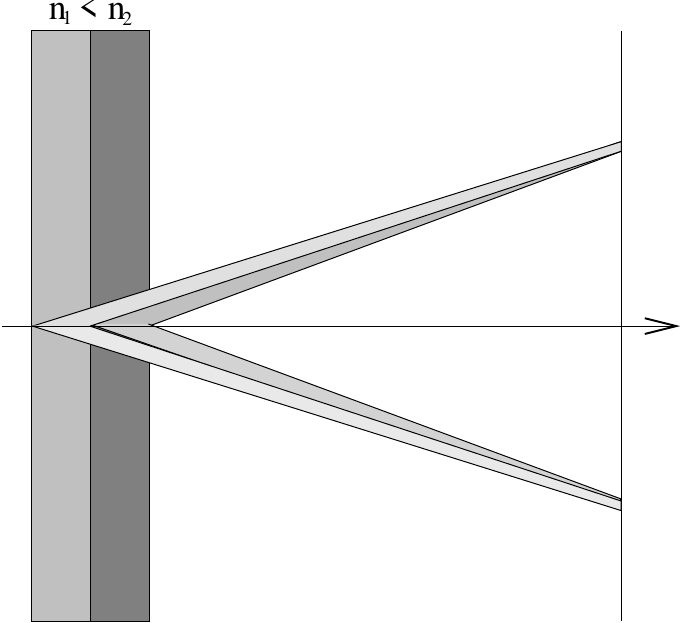} \\ (a)}
  \end{minipage}
  \hfill
\begin{minipage}{0.49\linewidth}
       \center{\includegraphics[width=1\linewidth]{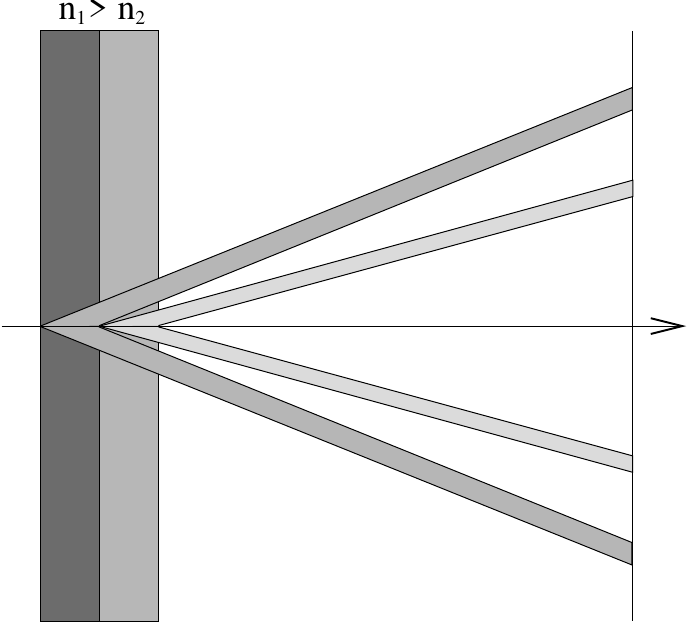} \\ (b)}
  \end{minipage}
   \caption{FARICH in single-ring (a) and multi-ring (b) alternatives.}
   \label{fig:farichconcept}
\end{figure}

In a proximity focusing RICH, one of the main factors limiting Cherenkov
angle resolution is a finite thickness of a radiator. In 
\cite{farich:Barnyakov2005,farich:Belle2005,farich:Barnyakov2008}
it was proposed to use a radiator consisting of several layers of aerogel with 
different refractive indices to overcome this limitation. An index of 
refraction and a thickness of each layer are chosen so that rings from 
all layers coincide on the photon detection plane. Another possibility 
is to have several separate rings 
(Fig.~\ref{fig:farichconcept}). Both options allow one to diminish the photon 
emission point uncertainty. A detector employing this technique is called 
Focusing Aerogel RICH (FARICH).

In 2004 in Novosibirsk the first multilayer aerogel sample was obtained
\cite{farich:Barnyakov2005}. Four layers of that aerogel had indices of refraction 
and thicknesses that match the designed ones with a good precision. At present,
we have several aerogel tiles in a single-ring option with 4 layers with 
dimensions up to $230\times230\times35$~mm$^3$~\cite{farich:TIPP23aer}.

\section{Current status and  progress of FARICH R\&Ds}
Since 2011 a series of beam tests with FARICH prototypes  has been performed by the BINP team. 
In 2011, the first test of FARICH prototype at  the installation ''Extracted beams'' of the VEPP-4M complex was performed~\cite{farich:exbeam,farich:exbeam2}.
32 SiPMs from CPTA (Moscow) with  a sensitive area of 2.1$\times$2.1~mm$^2$, 16-channel discriminator boards with built-in pre-amplifiers, 64-channel multi-hit TDC V1190B (CAEN) with 100\,ps time tick  were used for photon detection.
At the beam test with 1.5~GeV electrons, Cherenkov angle resolution for a single photon with a 4-layer aerogel sample was measured and the focusing effect in comparison with single layer aerogel samples was demonstrated~\cite{farich:farich2014}.

\begin{figure}[ht!]
    \begin{minipage}{0.49\linewidth}
       \center{\includegraphics[width=1\linewidth]{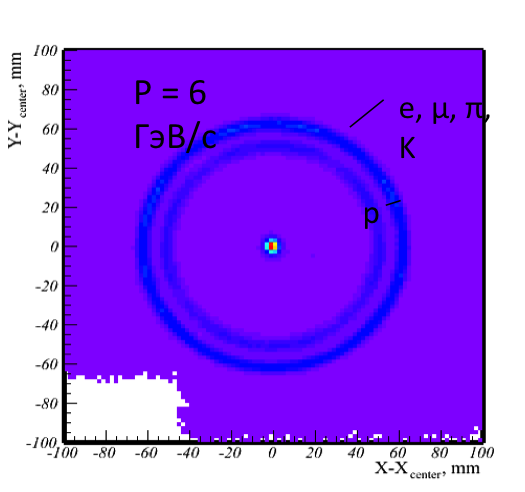} \\ (a)}
  \end{minipage}
  \hfill
      \begin{minipage}{0.49\linewidth}
       \center{\includegraphics[width=1\linewidth]{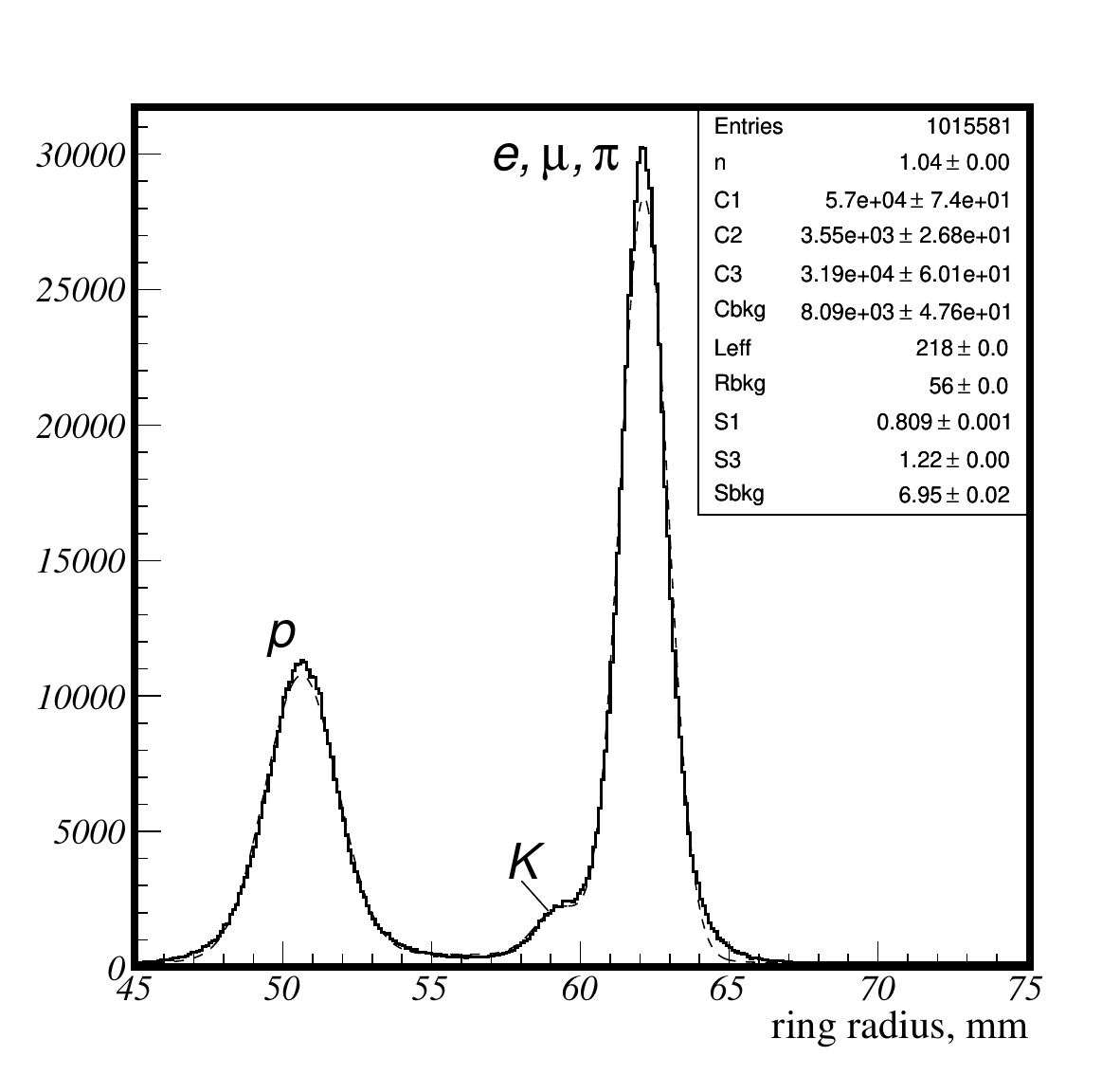} \\ (b)}       
   \end{minipage}
   \begin{minipage}{0.49\linewidth}
       \center{\includegraphics[width=1\linewidth]{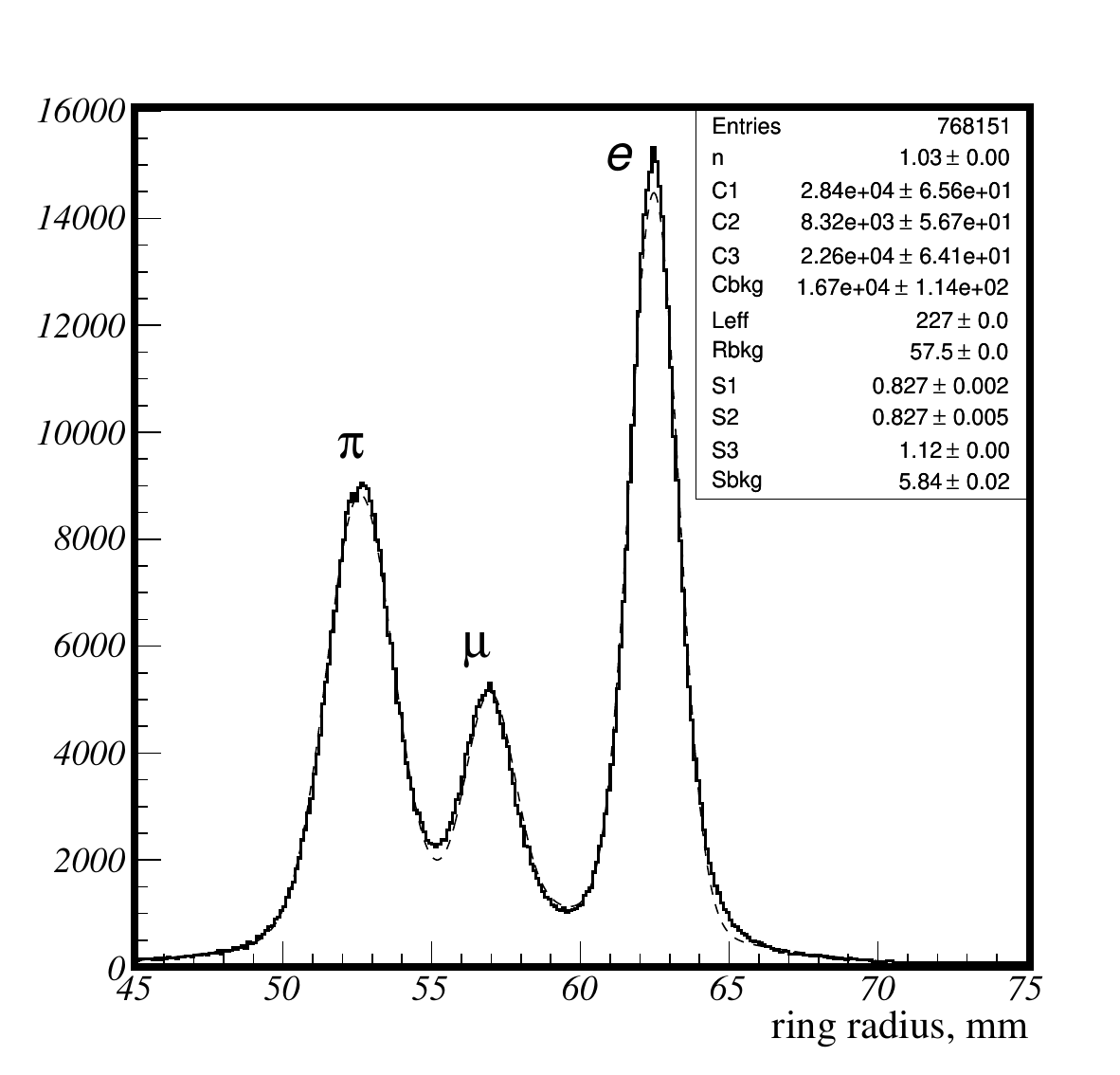} \\ (c)}
     \end{minipage}
  \hfill    
   \begin{minipage}{0.49\linewidth}
       \center{\includegraphics[width=1\linewidth]{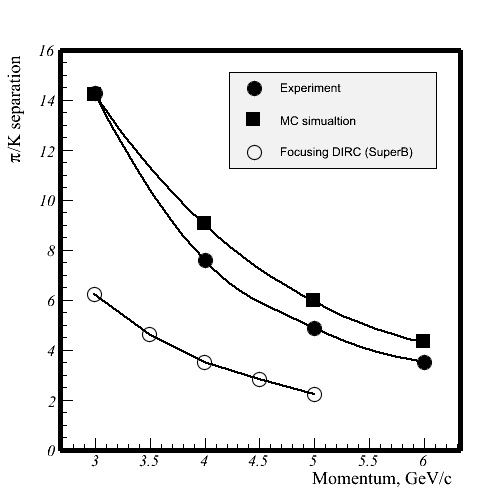} \\ (d)}
  \end{minipage}
 \caption{FARICH beam test results obtained with particles momentum $p=1\div6$~GeV/$c$: Cherenkov rings for particles momentum 6~GeV/$c$ (a), Cherenkov photon distribution on the radius for particles momentum 6~GeV/$c$ (b), Cherenkov photon distribution on the radius for particles momentum 1~GeV/$c$ (c), dependence of $\pi/K$-separation power in $\sigma$ on the momentum for the prototype of FARICH (measurements and simulation) and FDIRC (measurements) (d).
}
 \label{farich:fig:tbeam2}
\end{figure}

Particle identification capability of the FARICH method was tested with hadron beams at CERN in 2012~\cite{farich:farich2014,farich:farich2012}.The prototype was based on a 4-layer aerogel sample with a maximal refractive index $n_{\text{max}}=1.046$ and dimensions 115$\times$115$\times$37.5~mm$^{2}$. The matrix of 24$\times$24 of silicon ``digital'' photosensors DPC3200-22-44 Philips~\cite{farich:pdpc} with overall sizes 200$\times$200~mm$^2$ was used as a photon detector. Each sensor consists of 4 pixels of 3.2$\times$3.9~mm in total, 2304 pixels).
The Cherenkov rings for different particles ($p$, $K^{\pm}$, $\pi^{\pm}$, $\mu^{\pm}$, $e^{\pm}$) with momenta equal to 6~GeV/$c$ are shown in Fig.~\ref{farich:fig:tbeam2}(a). In Fig.~\ref{farich:fig:tbeam2}(b), the distribution of the same events on the radius is presented. Distribution of events on the Cherenkov radius for particles with 1~GeV/$c$ pulses is presented in Fig.~\ref{farich:fig:tbeam2}c. Power of $\mu/\pi$-separation at the level of 5.3$\sigma$ was obtained at the 1~GeV/$c$ momentum, $\pi/K$-separation better than 3.5$\sigma$ until the momentum equals to 6~GeV/c was observed. The obtained results demonstrate a better particle identification FARICH capability in comparison with FDIRC~\cite{farich:DIRC2008,farich:fdirc2015} (see Fig.~\ref{farich:fig:tbeam2}d).

\begin{figure}[h!]
	\begin{center}
		\includegraphics[width=0.8\textwidth]{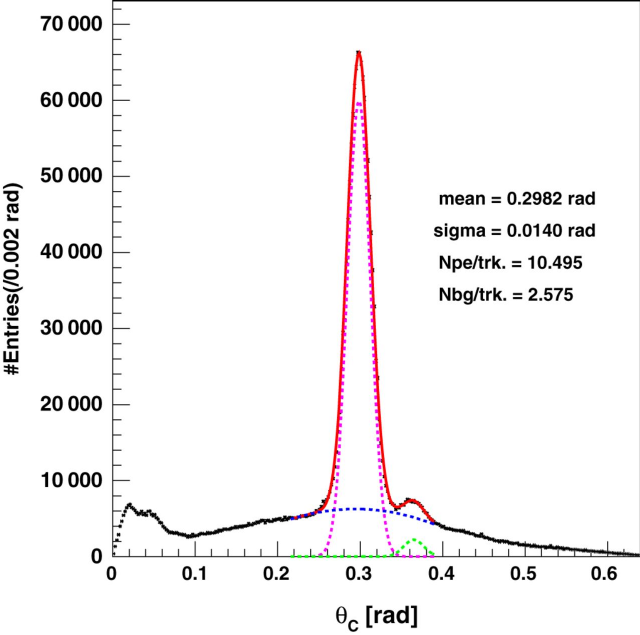}
		\caption{Cherenkov angle distribution obtained with relativistic electrons at the DESY beam test facility for an ARCIH module based on two aerogel tiles with $n_1$=1.045 and $n_2$=1.055 arranged in the focusing mode and an array of HAPD photon detectors.}
		\label{fig:arichtbres}
	\end{center}
\end{figure}

\begin{figure}[h!]
	\begin{center}
		\includegraphics[width=1.\textwidth]{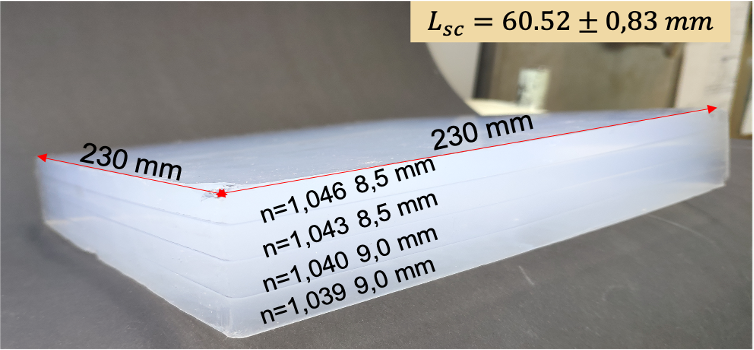}
		\caption{Largest 4-layer focusing aerogel tile produced in Novosibirsk in 2022.}
		\label{fig:4layer}
	\end{center}
\end{figure}

\begin{figure}[h]
    \begin{minipage}{0.49\linewidth}
   \center{\includegraphics[width=1\textwidth]{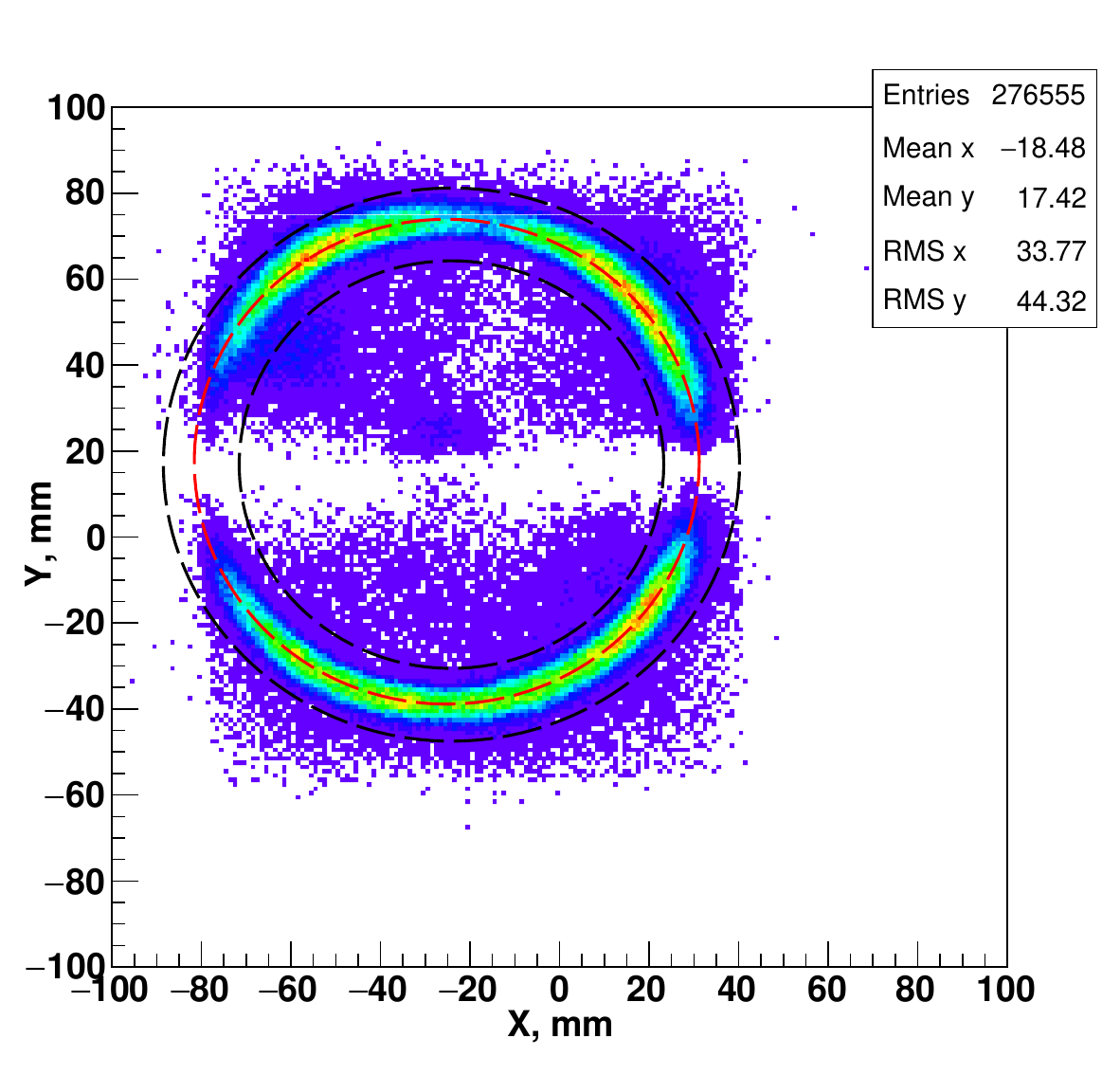}\\(a)}
    \end{minipage}
  \hfill
    \begin{minipage}{0.49\linewidth}
   \center{\includegraphics[width=1\textwidth]{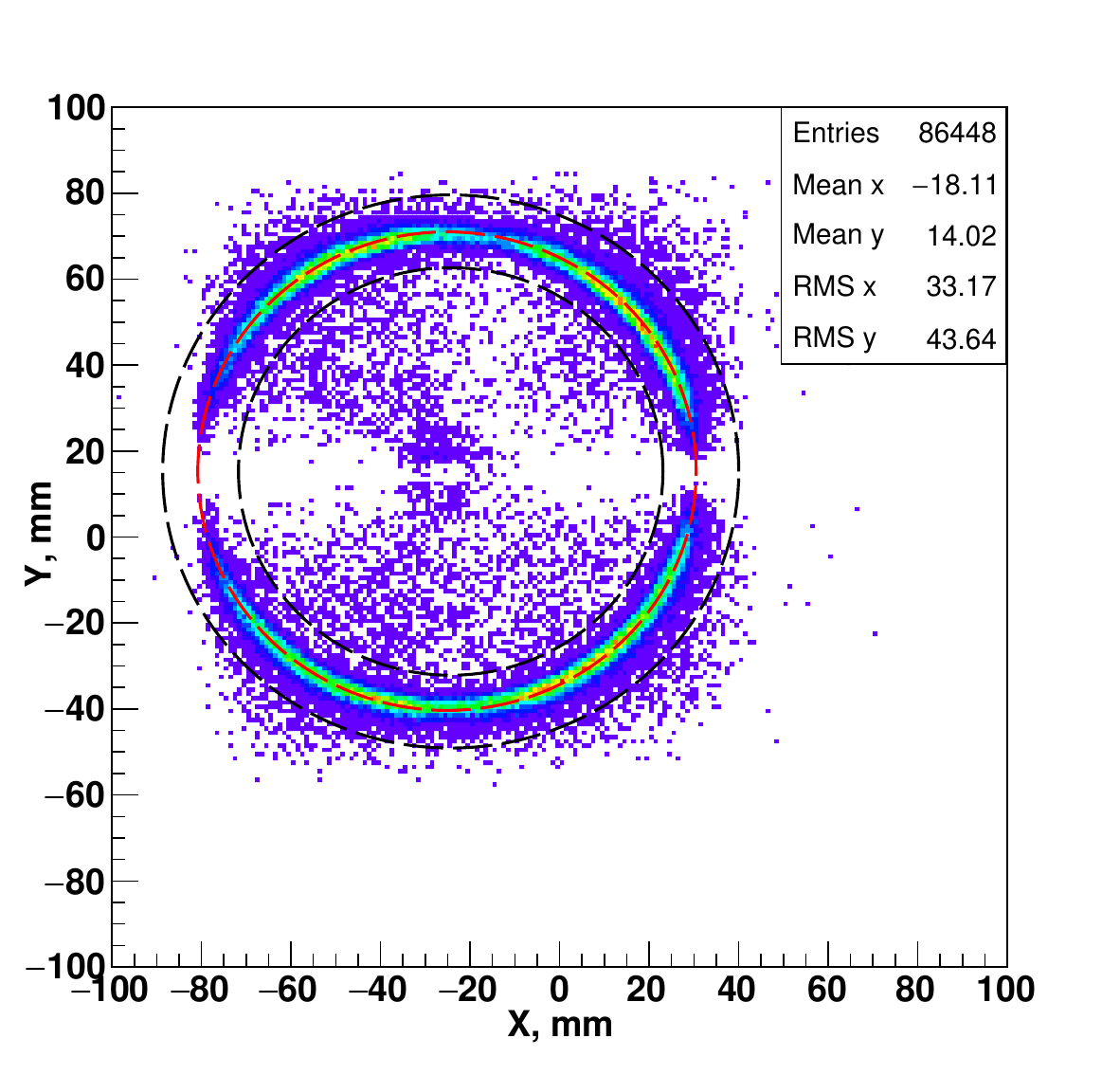}\\(b)}
    \end{minipage}
    \caption{Hitmap in the $XY$ plane accumulated with a FARICH prototype based on 4-layer focusing aerogel produced in 2022 during the beam test with relativistic electrons: photon detector pixel size 6\,mm (a) and 3\,mm (b). }
     \label{fig:TB2023hitmap}
\end{figure}

\begin{figure}[hb!]
    \begin{minipage}{0.46\linewidth}
       \center{\includegraphics[width=1\linewidth]{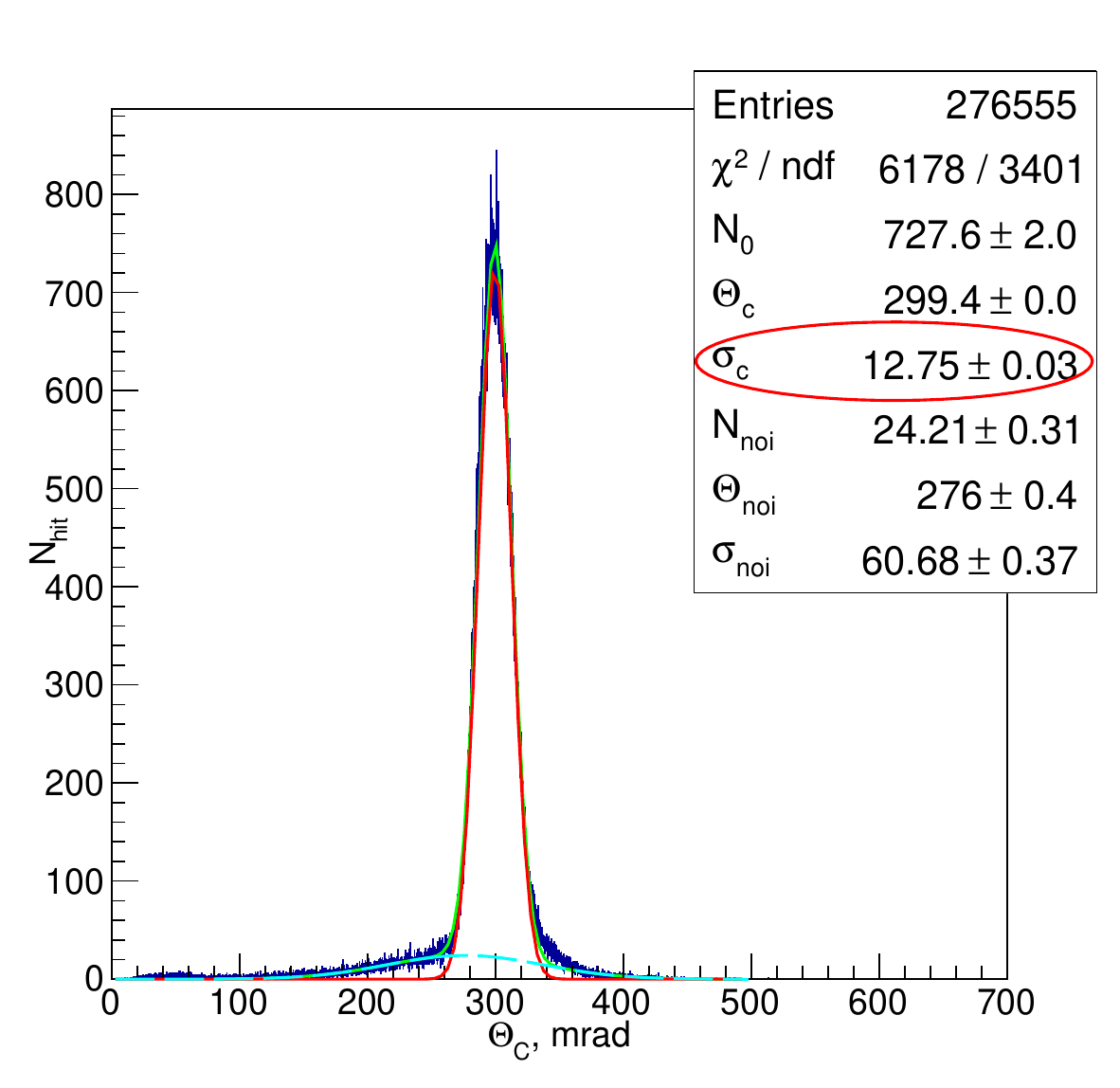} \\ (a)}
  \end{minipage}
  \hfill
      \begin{minipage}{0.52\linewidth}
       \center{\includegraphics[width=1\linewidth]{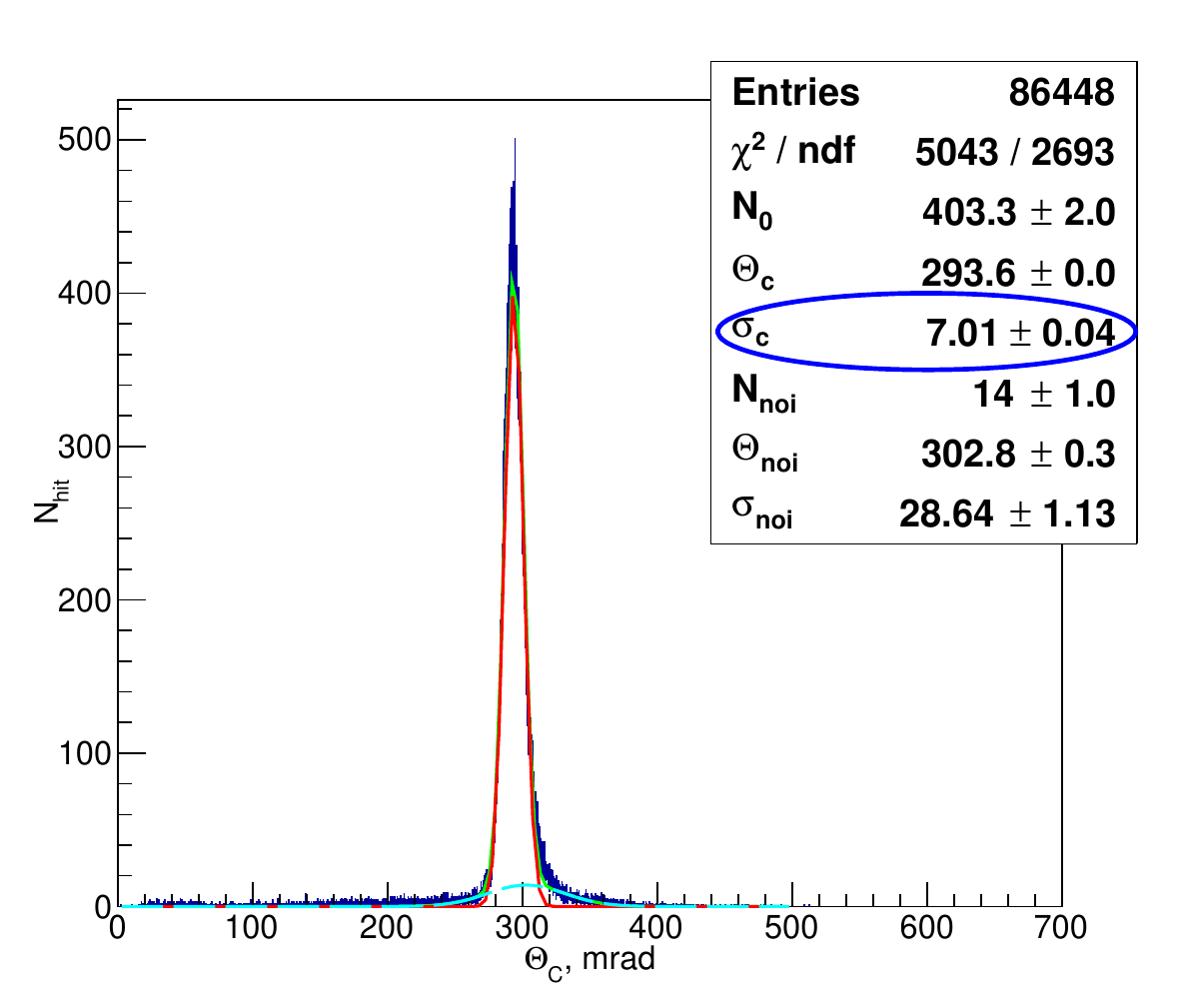} \\ (b)}
  \end{minipage}
		\caption{Cherenkov single-photon angle distribution for a FARICH prototype based on 4-layer focusing aerogel produced in 2022 obtained with relativistic electrons: photon detector pixel size 6\,mm (a) and 3\,mm (b). }
		\label{fig:TB2023theta}
\end{figure}

Since 2017, a RICH detector based on a focusing aerogel radiator has been exploited in the Belle II experiment and it is called ARICH (Aerogel RICH)~\cite{farich:arich}. 
The system is based on two layers of aerogel (20~mm with a refractive index $n=1.045$ plus 20~mm with $n=1.055$) and HAPDs. One segment was tested at the beam. 
For a relativistic particle, a signal of 10.5 photoelectrons and a single photon angle resolution were obtained $\sigma^{\Theta_c}_{\text{1pe}}=14$\,mrad 
(see Fig.~\ref{fig:arichtbres}), which corresponds to  angle resolution per track $\sigma_{\text{track}}=\frac{\sigma^{\Theta_c}_{\text{1pe}}}{\sqrt{N_{\text{pe}}}}=\frac{14.}{\sqrt{10.5}}=4.3\,\text{mrad}$ and $\pi/K$-separation at the level of 5$\sigma$ for momentum $p=$4\,GeV/c~\cite{farich:arich2016}. 
In 2022--2023, the essential progress in production of multilayer focusing aerogel Cherenkov radiators was achieved by the Novosibirsk group (cooperation of Budker Institute of Nuclear Physics and Boreskov Institute of Catalysis researchers). For the first time, several 4-layer focusing aerogel blocks with dimensions 230$\times$230$\times$35\,mm$^3$ were produced~\cite{farich:TIPP23aer} (see Fig.~\ref{fig:4layer}) and tested with relativistic electron beams~\cite{farich:TIPP23}. 
Fig.~\ref{fig:TB2023hitmap} shows the results of the beam test of these blocks with relativistic electrons at the BINP beam test facility: a hitmap of Cherenkov detected photons with a photon detector pixel size of 6$\times$6\,mm$^2$ accumulated for several thousands of tracks (a) and a hitmap with a pixel size of 3$\times$3\,mm$^2$ (b).
From the figure it is seen that the impact of the aerogel on Cherenkov angle resolution is essentially less than the impact of a 6$\times$6\,mm$^2$ pixel of the photon detector. Nevertheless, the obtained single-photon resolutions for both pixel sizes look very attractive and are presented in Fig.~\ref{fig:TB2023theta}.

Cherenkov angle distributions for the case of 6\,mm and 3\,mm pixel  are shown in Fig.~\ref{fig:TB2023theta}(a) and (b), correspondingly. It was found that the angle resolution obtained in the recent experiment with a 6\,mm pixel size is better than the published results of the ARICH module beam tests with a 5\,mm pixel size and the results obtained with a 3\,mm pixel size are in good agreement with the simulation. Taking into account the experimentally measured average number of detected Cherenkov photons per track ($\overline{N_{\textit{pe}}}=16$), it is possible to estimate the particle separation capability of the FARICH detector based on a 4-layer focusing aerogel radiator produced in Novosibirsk in 2022.
\begin{figure}[tbh!]
    \begin{minipage}{0.49\linewidth}
       \center{\includegraphics[width=1\linewidth]{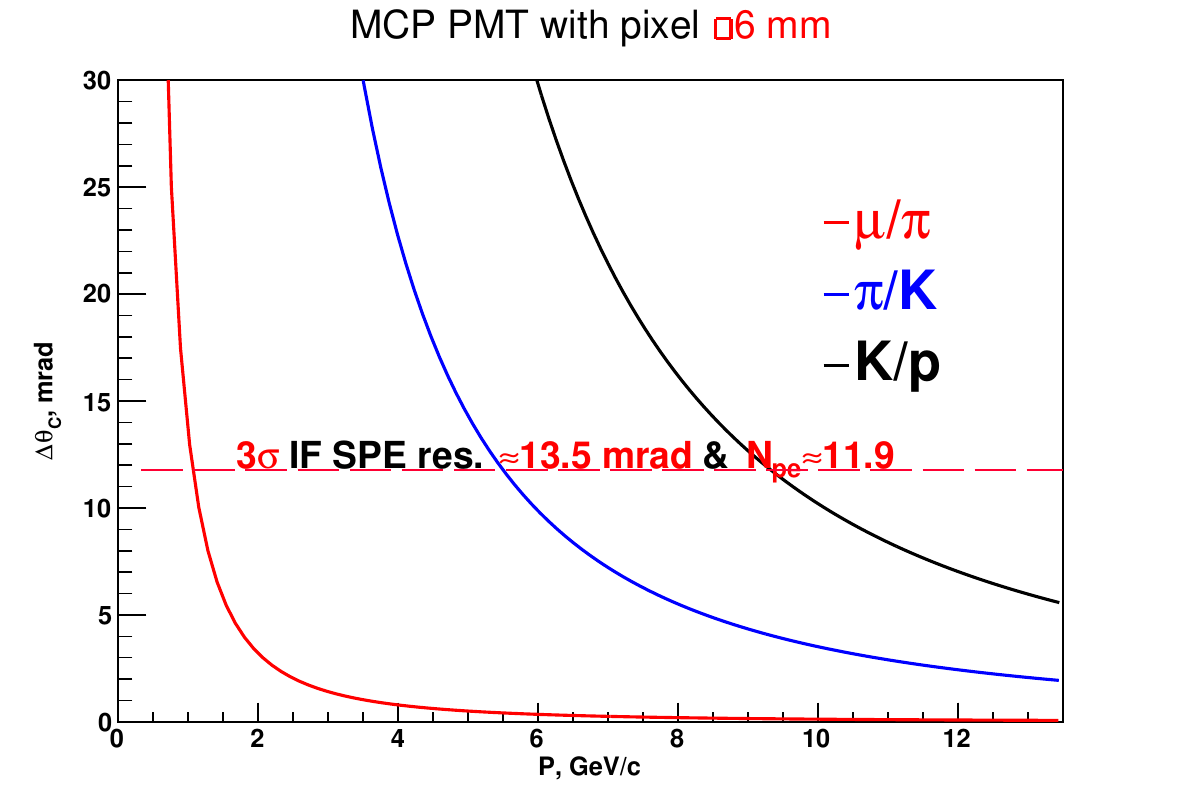} \\ (a)}
     \end{minipage}
  \hfill
      \begin{minipage}{0.49\linewidth}
       \center{\includegraphics[width=1\linewidth]{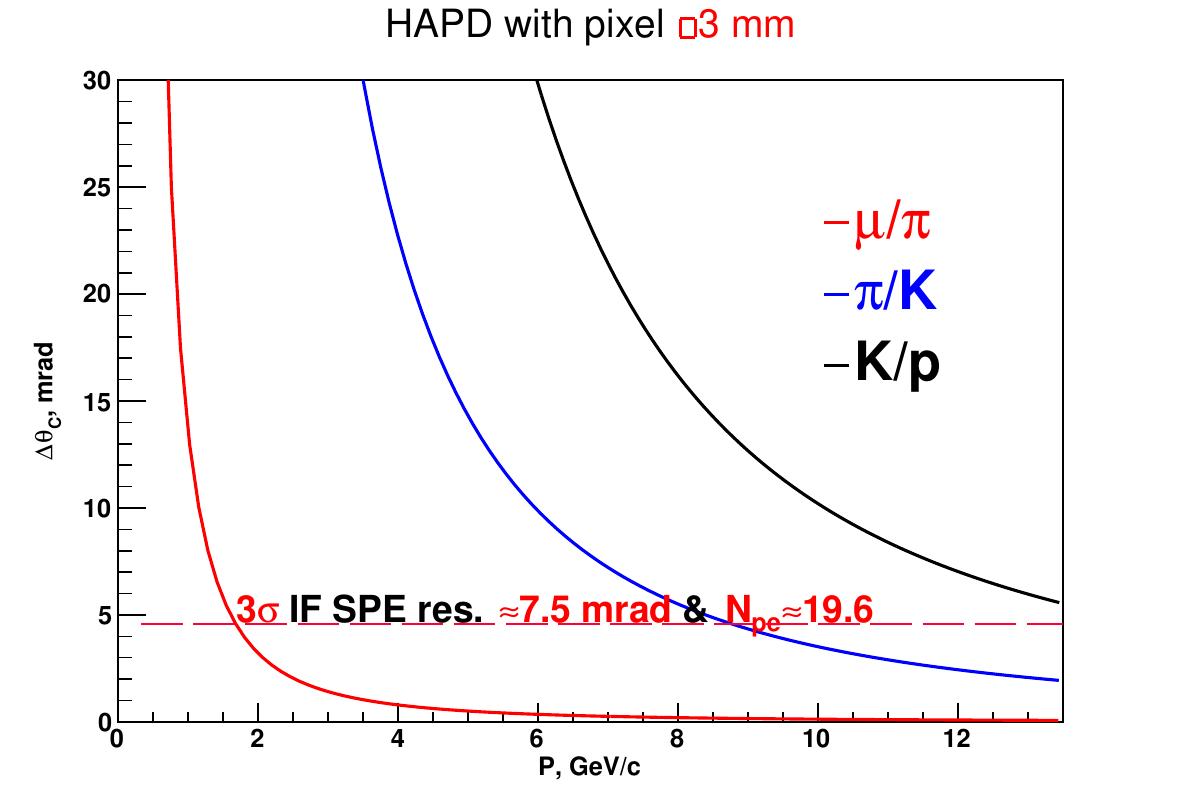} \\ (a)}
     \end{minipage}
		\caption{Cherenkov angle differences for particles pair (\textcolor{red}{$\mu/\pi$} -- \textcolor{red}{red}, \textcolor{blue}{$\pi/K$} -- \textcolor{blue}{blue}, \textcolor{black}{$K/p$} -- \textcolor{black}{black}) and the level of 3$\sigma$ separation: for 6\,mm pixel and $\overline{N_{\textit{pe}}}=12$ (a), for 3\,mm pixel and $\overline{N_{\textit{pe}}}=19$ (b).}
		\label{fig:TB2023separ}
\end{figure}
Fig.~\ref{fig:TB2023separ} (a) shows that it is possible to expect a $\pi/K$-separation better than 3$\sigma$ up to 5.5\,GeV/$c$ if we assume that the pixel size is 6\,mm and the number of detected Cherenkov photons is about $N_{\textit{pe}}=12$. This number was calculated for the MCP PMT option of the photon detector from the recent beam test results $\overline{N_{\textit{pe}}^{TB}}=16$~\cite{farich:TIPP23} as follows: 
\begin{equation}
N_{\text{pe}}=\overline{N_{\textit{pe}}^{TB}}\cdot\frac{GE}{GE^{TB}}\frac{CE_{\textit{MCP PMT}}}{CE_{\textit{H12700}}}\frac{\text{ff}_{\textit{MCP PMT}}}{\text{ff}_{\textit{H12700}}},
\end{equation}
where $GE$ and $GE^{\textit{TB}}$ is the geometrical efficiency of the photon detector plane expected in the SPD experiment and observed in the recent beam test experiment, correspondingly, $GE=1.0$ and $GE^{\textit{TB}}=0.8$. The $CE_{\text{MCP PMT}}=0.58$ is the photoelectron collection efficiency of the MCP PMT and $CE_{\text{H12700}}=0.9$ is the collection efficiency for the flat-panel multi-anode PMT H12700 from Hamamatsu. The $\text{ff}_{\textit{MCP PMT}}=0.81$ is the so-called fill factor for MCP PMT devices, which is determined by its photon sensitive and total areas. Here the values are taken for MCP PMT N6021 from NNVT (China) which have a photosensitive rectangular area of 46$\times$46\,mm$^2$ and the overall size 51$\times$51\,mm$^2$. The same value for H12700 MaPMT is equal to $\text{ff}_{\textit{MCP PMT}}=0.88$. Some improvement of particle separation could be reached by improving the photon detector performance: decreasing the pixel size from 6 to 3\,mm pixel and increasing the photon detection efficiency as for HAPD (Hybrid Avalanche Photon Detector). The estimated particle separation capability for this case is shown in Fig.~\ref{fig:TB2023separ}(b). It is demonstrated that in this case it is possible to provide $\pi/K$-separation at 3$\sigma$ level up to 8.7\,GeV/$c$ and $\mu/\pi$-separation up to 1.7\,GeV/$c$.

\section{FARICH system design}
\subsection{Aerogel}
\begin{figure}[th!]
	\begin{center}
		\includegraphics[width=0.75\textwidth]{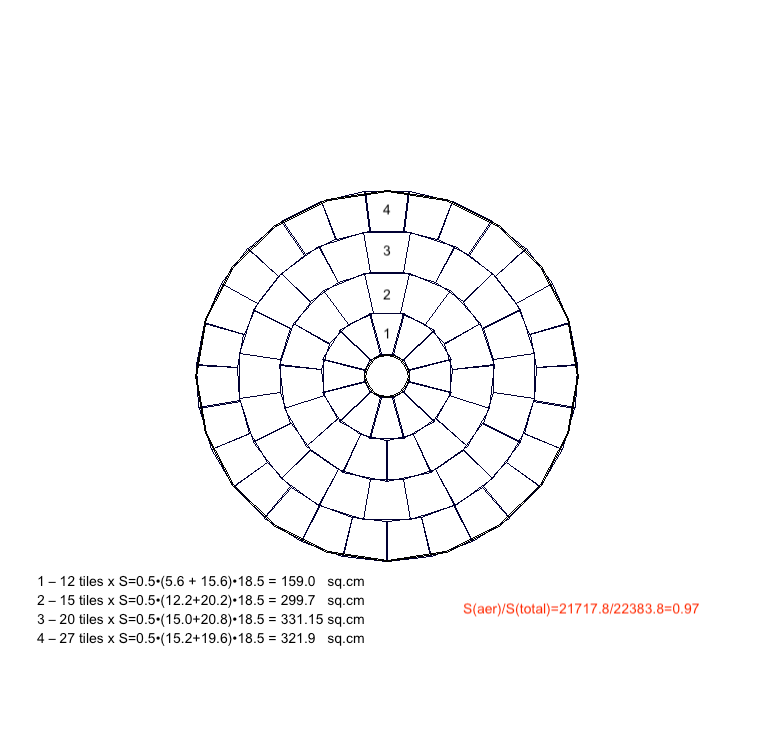}
		\caption{ Layout of aerogel trapezoidal tiles in the endcap regions of the SPD.}
		\label{fig:trapezaer}
	\end{center}
\end{figure}

\begin{figure}[tbh!]
	\begin{center}
		\includegraphics[width=0.75\textwidth]{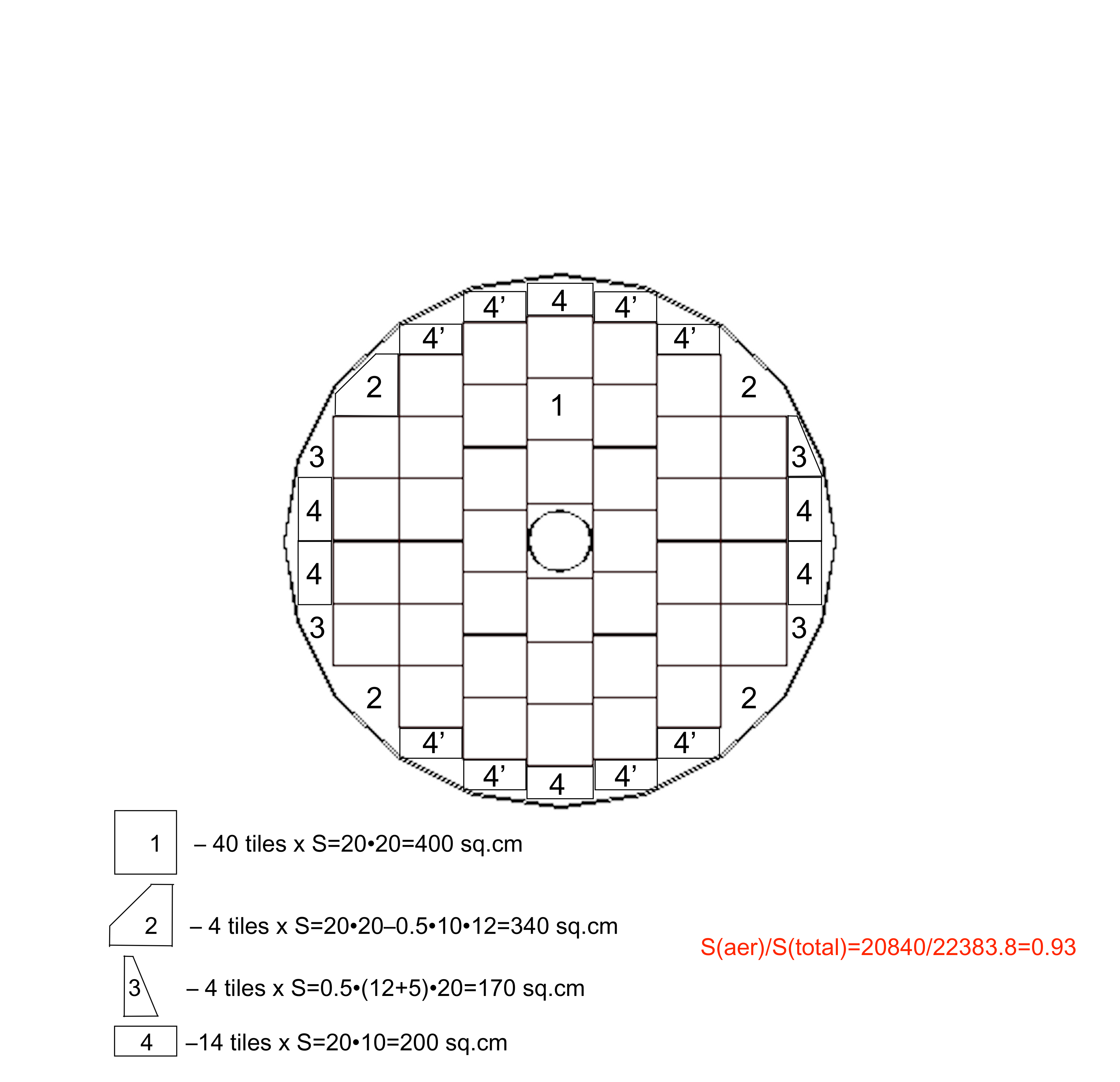}
		\caption{Layout of aerogel rectangular tiles in the endcap regions of the SPD.}
		\label{fig:rectangaer}
	\end{center}
\end{figure}

The FARICH detector for the SPD experiment is considered for endcap regions and the total area of two endcaps will not exceed 4\,m$^2$. To fill the area of the endcap discs by aerogel, two options are being considered: trapezoidal and rectangular. Fig.~\ref{fig:trapezaer} shows the option based on trapezoidal shapes of the aerogel tiles. In this case, there are four form-factors of the tiles and the expected geometrical efficiency (ratio of the active area of the aerogel radiator to the total area of the disc) is equal to 0.97, while for the other option based on four rectangular aerogel tiles (see Fig.~\ref{fig:rectangaer}) the geometrical efficiency could be estimated as 0.93.

\subsection{Photon detectors}
The FARICH will work in the magnetic field of about 1~T that imposes a serious 
limitation on the photon detector choice. Presently, the only three types of visible photon detectors can operate in the axial magnetic field: Silicon Photomultiplier (SiPM), Hybrid Avalanche Photon Detector (HAPD), and Micro Channel Plate (MCP) PMT. SiPM is a multipixel Geiger-mode avalanche 
photodiode (MPGM-APD or SiPM)~\cite{farich:gapd,farich:gapd2}.
SiPM is almost insensitive to magnetic field and has a gain of the order of $10^6$,
a high PDE in the visible region, a low voltage bias, and a size of a few~mm.  The most well-known manufacturers of such devices are Hamamatsu (Japan)~\cite{farich:mppc}, FBK (Italy)~\cite{farich:sipm64fbk}, and SensL (Irland)~\cite{farich:sensl}. Also, there are several less known manufacturers, such as Integral (Belorussia), NDL, and JoinBon (China).
The main parameters (PDE, dark count rate (DCR), probability of optical crosstalks) of the recently designed SiPMs from these manufacturers are very similar. Comparison of these devices is presented in~\cite{farich:sipmchar}.
The main limitation of their usage in the particle collider experiments is their low radiation hardness~\cite{farich:sipmvsn,farich:pdpcvsn,farich:sipmspevsn}. According to several investigations, it was shown that after an accumulated dose of about 10$^{10}n_{\text{eq}}/\text{cm}^2$ the SiPM DCR increased by two times. It could lead to distortion of the Cherenkov ring reconstruction process and additional loads to the data transfer system.

HAPDs were developed for only one project -- ARICH -- in the Belle II experiment by Hamamatsu. The possibility to work with HAPD in a parallel magnetic field of 1.5~T was demonstrated. The radiation hardness tests show that the HAPD are able to work up to an integrated neutron dose of up to 10$^{12}n_{\text{eq}}/\text{cm}^2$~\cite{farich:arich2019}. Today, Hamamatsu does not produce such devices any more. However , potential opportunities to develop  a technology of HAPD production in Russia are considered from time to time.

 The MCP PMT-based position-sensitive photon detector looks more reliable for the purposes of the FARICH option of the SPD experiment. MCP PMTs are able to work in parallel magnetic fields without  any decreasing efficiency. And at a 45$^{\circ}$ tilt  to a magnetic field of 1~T, the collection efficiency of photoelectrons  decreases by two times only~\cite{farich:mcpb,farich:mcpb2,farich:mcpb3}. In comparison with SiPM, they have a good radiation hardness and a low level of intrinsic noises (100\,kcps/cm$^2$). There is a lot of experience accumulated in the field of development and exploitation of MCP PMTs in different labs all over the world. The most well-known producers of position-sensitive MCP PMTs are Photonis-Planacone (USA), Hamamatsu (Japan), Photek (Great Britain), Income (USA). Also, a very interesting progress in the development of MCP-based devices has been demonstrated by NNVT (China). The MCP PMT N602.1 has 8$\times$8 pixels with a size of 5.8$\times$5.8\,mm$^2$ and a lateral size of 51$\times$51\,mm$^2$. It is known  that there are several producers of MCP-based devices, such as ``VTC Baspik'' or ``Cathode'' and ``Ekran FEP'' in Russia, which can produce round-shaped MCP PMTs. Production of square-shaped multi-anode MCP PMTs is being negotiated with the above-mentioned manufacturers. 

It is necessary to note that the option of MCP PMT-based photon detectors has an additional attractive feature: it could be used as a Time-of-Flight detector, if it is supplied with proper readout electronics, as one of the most popular features of MCP-based devices is their recordable time resolution which can reach about 6\,ps per track~\cite{farich:mcp6ps}.

\begin{figure}[tbh!]
	\begin{center}
		\includegraphics[width=1.\textwidth]{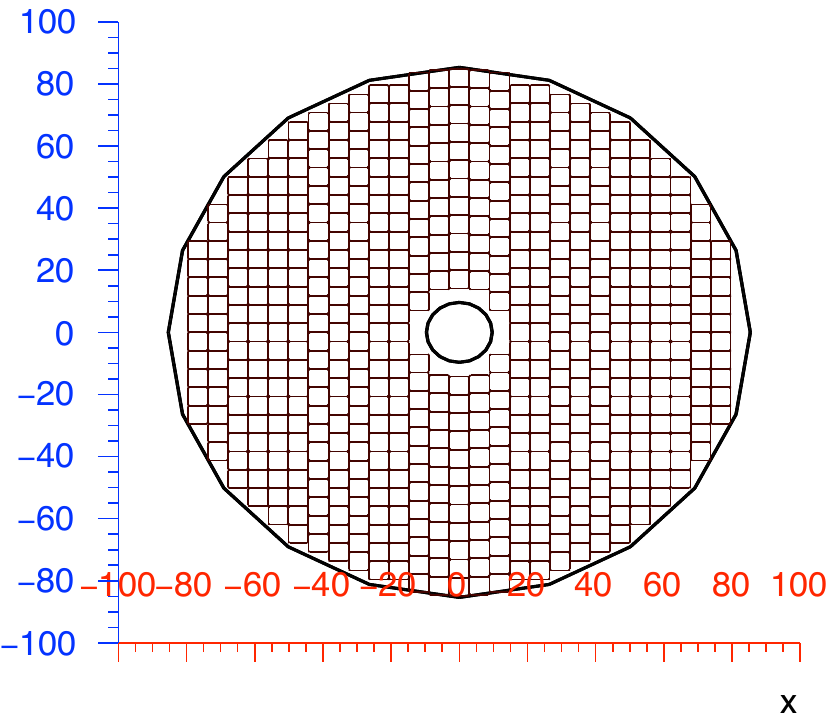}
		\caption{ Layout of MCP PMTs N6021 from NNVT in one endcap of the FARICH system.}
		\label{fig:pmtlayout}
	\end{center}
\end{figure}

It is natural to consider MCP PMTs N6021 from NNVT (China) as a basic option for the FARICH system of the SPD experiment.  In this case, the layout of the photon detector planes will look as in Fig.~\ref{fig:pmtlayout}.
The total amount of PMTs is about 548\,pcs per each endcap and, consequently, 2$\times$548$\times$64=70144 pixels and readout electronics channels in total for the whole system.

\subsection{Electronics}
The  large number of channels and their high density will require
 development of dedicated front-end electronics. The digitizing part (TDC) should be located right in the detector. Today, there are two main approaches to the development of such readout electronics.
\begin{itemize}
\item Field Programmable Gate Array (FPGA), when standard programable chips are used to arrange the desirable configuration of the readout electronics channel and the DAQ system. In this case, some flexibility in the development is present. However stability and performance could be worse in comparison with the other approach. Nevertheless, a very impressive progress in the development of the FPGA-based TDC  is demonstrated by the group of electronics designers from GSI (Germany). Their FPGA-TDC, the so-called DiRICH, developed for RICH detectors in several projects (CBM, HADES, PANDA), is able to provide a time measurement resolution of about 40\,ps and, in addition, measure the signal amplitude with the ToT (Time-over-Threshold) technique ~\cite{farich:dirich}.
\item Application Specialised Integrated Circuit (ASIC), when all necessary electronics elements and circuits are  implemented in one silicon crystal. This approach can provide a very precise and stable performance, but the chip designed in such a way will not be flexible. Any minor change in the readout channels will lead to design and production of a new batch of chips. Therefore, the development process is more expensive in comparison with the FPGA approach. Up to date, a lot of specialised chips (ASIC) for readout arrays of pixelated photon detectors have been developed: NINO~\cite{farich:nino,farich:nino2}, MAROC, SPIROC~\cite{farich:maroc}, PETA~\cite{farich:peta,farich:peta2,farich:peta3}, BASIC~\cite{farich:basic}, VATA64HDR16~\cite{farich:vata} and TOFPET~\cite{farich:tofpet}. Generally, all of them were developed for Positron Emissive Tomography or for another application with fast scintillation crystals (LYSO, LaBr$_3$, etc.). Some of them, such as NINO and VATA64HDR16, are planned  to be used for Cherenkov light registration from quartz. For example, NINO is an amplifier-discriminator which in couple with a fast TDC (i.e. HTDC) provides the registration of signals with a picosecond time resolution. Also, this chip was tested with the MCP PMT as part of the development of the particle identification system TORCH for the LHCb detector at the Large Hadron Collider~\cite{farich:torchmcp}. The other chips mentioned above  are very complex devices able to determine the photon position (pixel number), amplitude, and time of signals.  It takes a lot of time to transfer such information through DAQ). In the FARICH system, only a few detected photons per track are expected and there is no necessity to determine the signal amplitudes in each channel, because the amplitude will be not more than one photoelectron. 
\end{itemize}
For the SPD project, the best way is to design and develop readout electronics based on one of the two approaches. However, ideas and concepts could be adopted, in part or in full, from one of the previously developed chips  mentioned above or from any other chip.

The total thickness of the system for perpendicular tracks is estimated as 14.5\%$X_0$.

\section{Possible design improvements}
\begin{figure}[ht]
	\begin{center}
		\includegraphics[width=1.\textwidth]{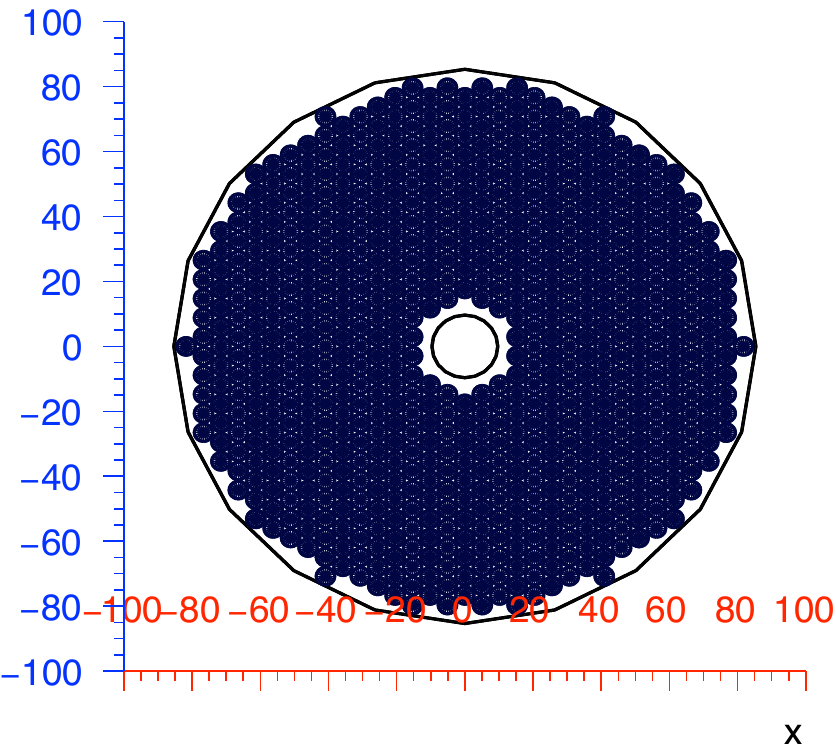}\\
		\caption{Layout of round-shaped MCP PMTs  arranged in hexagonal structure in one of the FARICH system endcaps. Fraction of photocathode diameter to external PMT diameter is equal to $\o$PC/$\o$PMT=50/58=0.86.}
		\label{fig:rpmtlayout}
	\end{center}
\end{figure}

For the moment, there are several obvious possibilities to improve the design of the FARICH system for the SPD experiment. \\
\begin{itemize}
\item Aerogel tiles design (distribution of refractive indices and layers thickness) should be optimised with the help of simulation in accordance with the geometry of the SPD detector, photon sensors PDE, and pixel size. 
\item Optimisation of the photon detectors cost is highly required. Development and organisation of photon detector production in Russia can significantly decrease the cost of the system. Therefore close cooperation with manufacturers in Russia, such as ``VTC Baspik'',  ``Ekran FEP'' are foreseen for the nearest future R\&Ds. For instance,  if due to some technological reasons round-shaped  position-sensitive photon detectors are  cheaper and easy to produce, it  will be possible to optimise  the design of the FARICH system for such devices. Fig.~\ref{fig:rpmtlayout} shows a very preliminary layout of  a hypothetical  round-shaped position-sensitive PMT with  a photocathode diameter $\o$PC=50\,mm and  an external PMT diameter $\o$PMT=58\,mm. 
\end{itemize}
In this case, the total amount of PMTs is equal to 630 and the fill factor for such hexagonal arrangement is less only by 10\% than for the layout of rectangular-shape PMTs with the same dimensions (PC area is 50$\times$50 and total PMT area is 58$\times$58\,mm$^2$). Such a small decrease in the detection efficiency due a decrease in the fill factor could be finally compensated, for instance, by improving the aerogel properties (transparency, thickness or refractive index) or by means of PMT performances (photcathode quantum efficiency, photoelectron collection efficiency).

\section{Cost estimate}
The major inputs to the total cost of the FARICH system are presented in Tab. \ref{FARICH:cost}.
\begin{table}[]
\caption{Cost estimate of the FARICH system.}
\label{FARICH:cost}
\begin{center}
\begin{tabular}{|l|l|c|c|r|}
\hline
Component & Units & Quantity & Price per unit, \$ & Total cost, k\$ \\ \hline\hline
Aerogel & tiles & 136 & 6\,800 & 924.8 \\ \hline
Photon detectors & pcs & 1100 &13\,000 &14\,300.0\\ \hline
Electronics & chann. &70\,144 & 18.5 &1\,297.7\\ \hline
High voltage supplies & chann. & 1100 & 100 & 110.0\\ \hline
Mechanics &      &           &                      &      \\
and assembling & &           &                      & 100.0  \\ \hline
TOTAL: & & & & 16\,732.0\\ \hline
\end{tabular}
\end{center}
\end{table}
All these cost estimations are based on the prices and costs taken from the previous  projects and from the invoices for small batches recently requested from the vendors. For instance, 1 m$^2$ of the aerogel radiator is estimated as the cost of aerogel produced for the CLAS-12 project with some extras due to additional works connected with a more complicated process of 4-layer aerogel synthesis. To estimate the photon detectors cost, the price for four MCP PMTs N6021 from NNVT was taken. The additional costs connected with the Russian distributor fees and custom taxes are included as well. Therefore, this cost of the system could be considered as the upper limit of the cost. Some decrease and detalisation of the costs are expected after carrying out dedicated R\&Ds and direct consultations with vendors and designers. Also, there are two general assumptions in the table: the averaged cost of one High Voltage (HV) supply channel is equal to \euro{100} and the readout electronics channel cost is equal to \euro{17}.

\chapter{Straw Tracker}
 The purpose of the Straw Tracker (ST)  is to reconstruct  tracks of  primary and secondary particles with high efficiency, 
  to measure their momenta with high precision, based on a track curvature in a magnetic field, and to contribute to a particle identification via energy deposition ($dE/dx$) measurements. 
 A spatial resolution of ST is required to be about 150\,\textmu{}m. It has to be achieved when operating in a magnetic field of \mbox{about 1\,T}.
 
  The detector is planned to be built of low-mass straw tubes, similar to those used in many modern experiments, such as NA62 \cite{Sergi:2012twa}, COMET \cite{Nishiguchi:2017gei}, SHiP \cite{Anelli:2015pba}, Mu2e \cite{Lee:2016sdb}, COMPASS \cite{Bychkov:2002xw,Platzer:2005jp}, NA64 \cite{Volkov:2019qhb}, and others. 
  
 The tracker will consist of the barrel part and two end-caps. Two different kinds of straw tubes are considered to be used for them.

\section{Barrel part}
\subsection{Welded straw tubes}
\subsubsection{Material}

The detector is assembled from about 26\,000 straw tubes. 
A single straw tube is manufactured from a thin polyethylene terephthalate (PET) foil, 
which is welded longitudinally by ultrasonic welding to form a tube. 
The straw has an active length from 10~cm to 2.7~m and a nominal inner diameter of 9.8~mm.
Its inner surface has a metal coating (Cu/Au) to provide electrical conductance on the cathode, 
while the electrons, produced in ionization processes, will drift towards the central wire anode.

The choice of the straw material is a compromise between many different requirements, i.e.~radiation length, permeation of gases, mechanical properties, adhesion to metal coating, bonding with epoxy, and the ability to be ultrasonically welded into a tube. A summary of the specifications is given in Table \ref{st_bar_t1}.
\begin{table}[htp]
\caption{Summary of coated material specifications.}
\begin{center}
\begin{tabular}{|c|c|}
\hline
Description  & Specifications \\
\hline
PET film  &  polyethylene terephthalate \\
                & type Hostaphan RNK 2600 of \\
                &(36 $\pm$ 2) $\mu$m thickness\\
Density    &  1.4 g/cm$^3$ \\
Copper layer thickness &   50 nm \\
Gold layer thickness     &   20 nm \\
Resistivity of straw tube (2.1 m) & $\sim$70 $\Ohm$ \\
Permeation of naked film,   & 6 for He, 0.06 for Ar, 1 for CO$_2$ \\
($10^{-12}$ Torr$\times$l$\times$cm/s$\times$cm$^2$$\times$Torr) & \\
\hline
\end{tabular}
\end{center}
\label{st_bar_t1}
\end{table}%

The straw tubes for the tracking detector in the SPD experiment are manufactured from a biaxially oriented coextruded film made of PET. 
One side of the film is chemically pre-treated to improve adhesion.
This side was chosen for the epoxy bonding between the straw tube and the polyetherimide (PEI)  straw fixation plug. 
The non-treated side is coated by a conductive layer of 50 nm of Copper (Cu) followed by a protective layer of 20 nm of gold (Au). 
Once the production of the tube is completed, the conductive layer is on the inside of the tube, while the outside remains uncoated and pre-treated chemically for bonding.
The thickness of the film is 36 $\mu$m.

\subsubsection{Long-term tests}
\begin{figure}[!ht]
  \begin{minipage}[ht]{0.45\linewidth}
    \center{\includegraphics[width=1\linewidth]{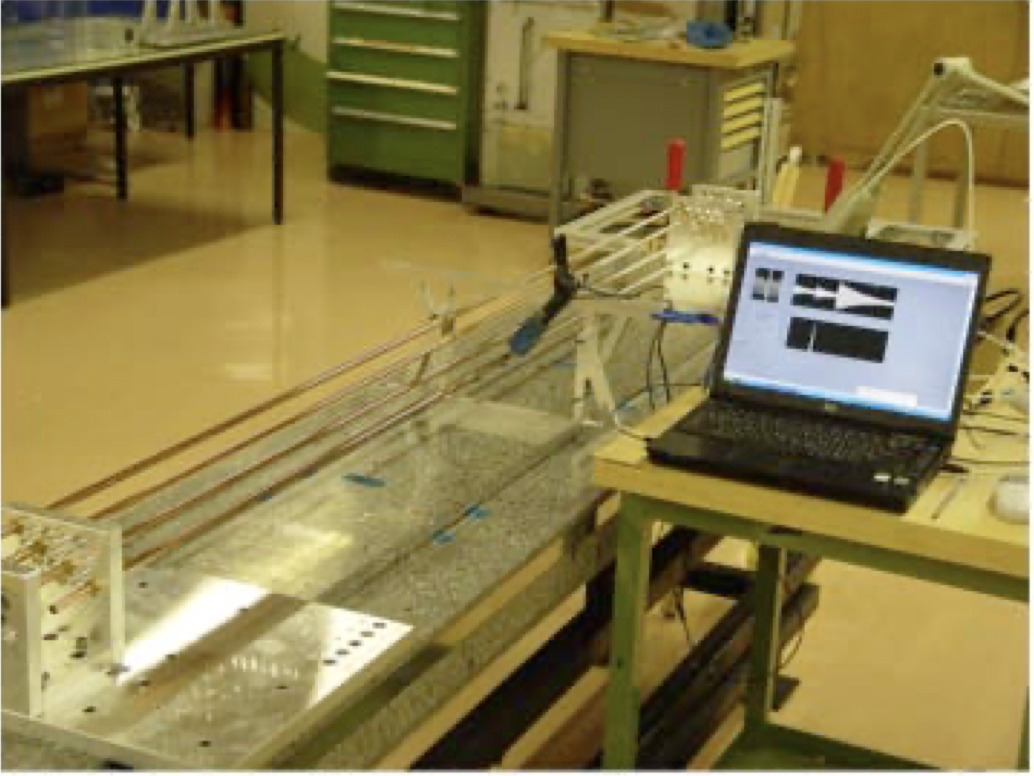} \\ (a)}
  \end{minipage}
  \hfill
  \begin{minipage}[ht]{0.5\linewidth}
    \center{\includegraphics[width=1\linewidth]{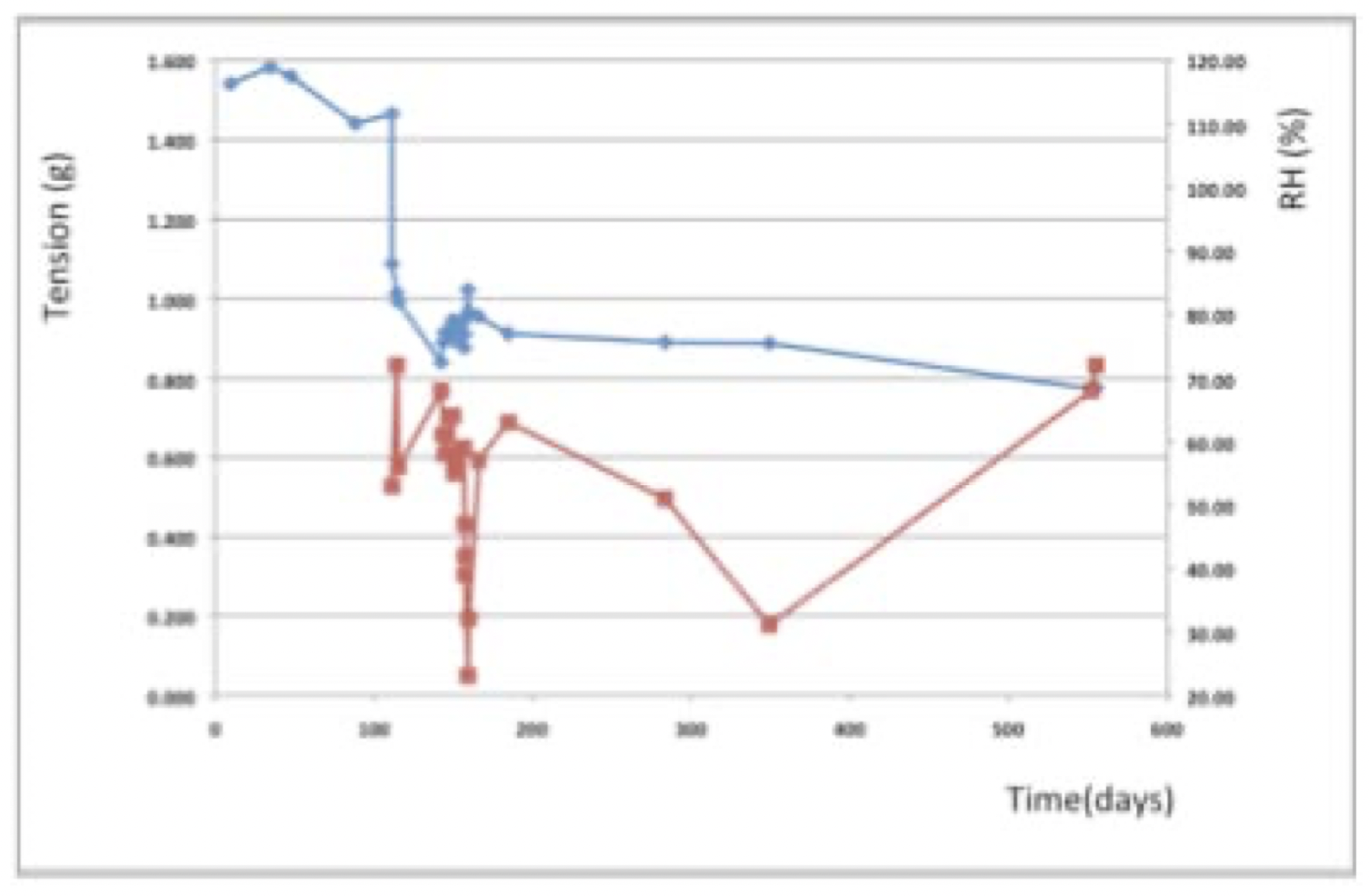} \\ (b)}
  \end{minipage} 
  \caption{(a) View of the long-term test setup. 
  (b) Straw tension as a function of time (blue curve) plotted together with humidity (red curve).
}
\label{st_ba_fig1}
\end{figure}

Long-term tests were carried out with the straws mounted under tension with a load of 15 N applied before the extremities of the tubes were bonded to the PEI supporting parts. The sag and the tension of the tubes were measured during a long-term test lasting 18 months. 
The values are recorded with  the straws under pressure and compared to a reference straw at atmospheric pressure.
The results are shown in Fig.\,\ref{st_ba_fig1}.

\subsubsection{Tensile test of PET samples}

\begin{figure}[!h]
  \begin{minipage}[ht]{0.5\linewidth}
    \center{\includegraphics[width=1\linewidth]{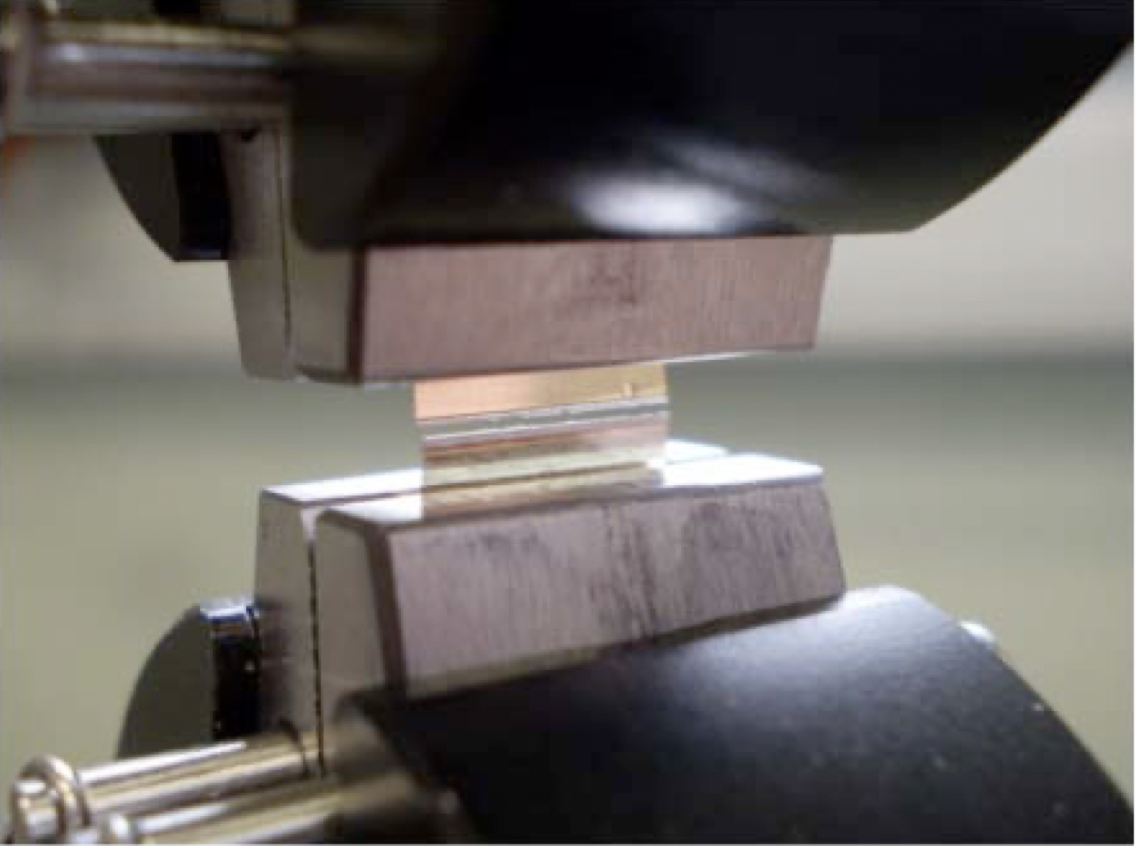} \\ (a)}
  \end{minipage}
  \hfill
  \begin{minipage}[ht]{0.5\linewidth}
    \center{\includegraphics[width=1\linewidth]{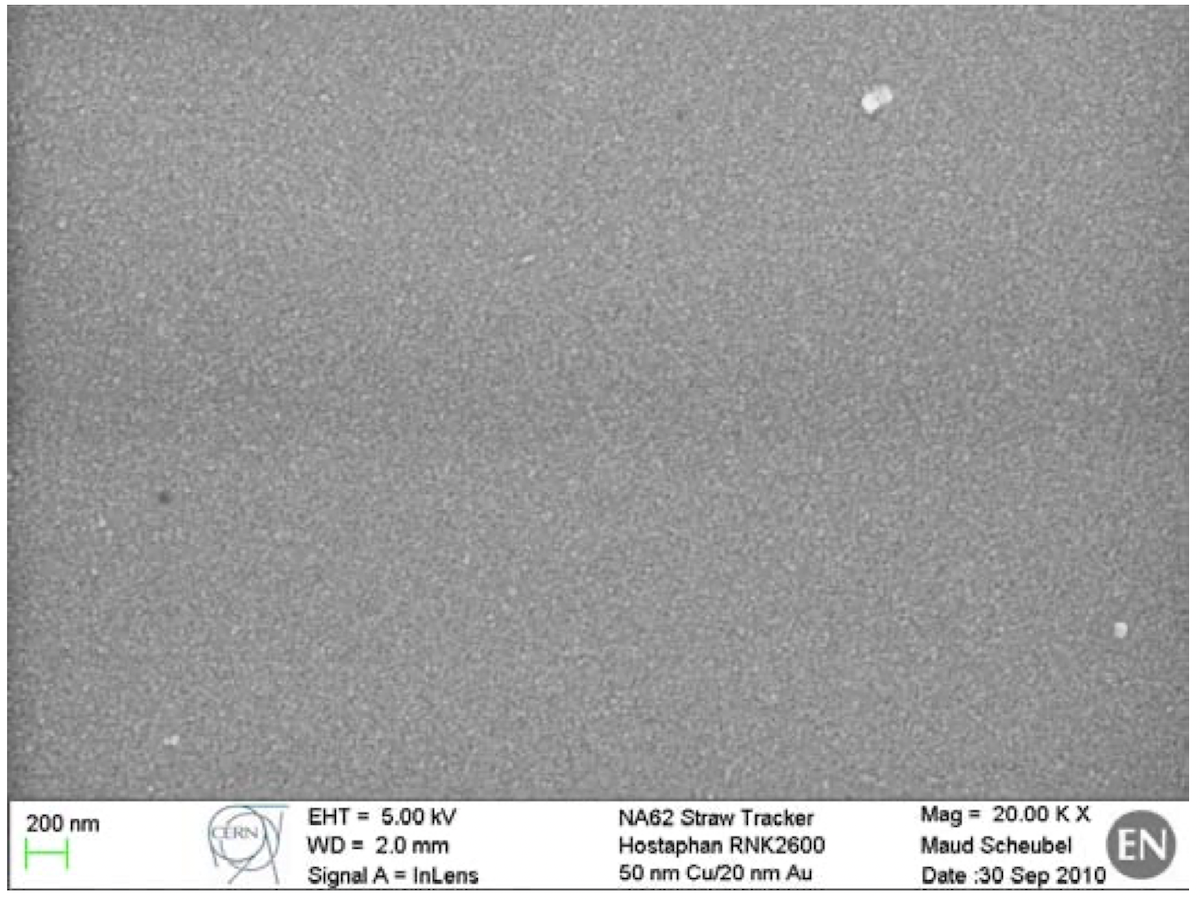} \\ (b)}
  \end{minipage} 
  \caption{(a) Close-up view of the tensile test apparatus with a tested film sample (the weld is in the center). (b) Typical view of a coated PET with a magnification of 20\,000 (RNK 2600).
	}
\label{st_ba_fig2}
\end{figure}

The base material for the straw tubes was tested in CERN using the Instron tensile test apparatus. 
A close-up view of the apparatus with a tested film sample is shown in Fig.~\ref{st_ba_fig2}~(a).
The values observed for welded samples are typically 25\% lower than those of the coated unwelded PET. 
Unwelded PET has a tensile strength close to 130~MPa, whereas welded and coated samples fail at $\sim$100~MPa. 
The stress in the straws due to the pressure difference of 1\,bar is 13\,MPa, which gives sufficient safety margin.

\begin{figure}[!ht]
    \center{\includegraphics[width=0.9\linewidth]{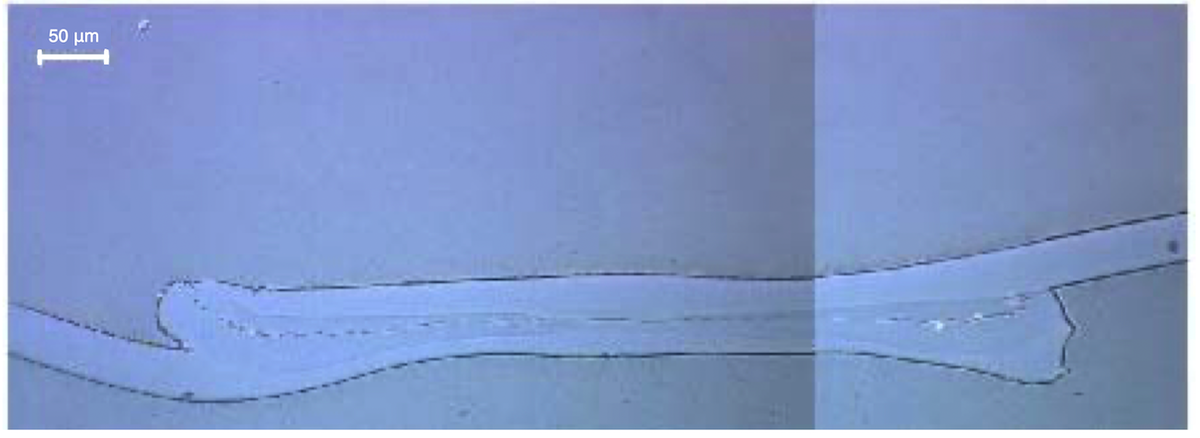}}
  \caption{ Cross-section of a welded seam where the two sides of the strip are fused.
	}
\label{st_ba_fig3}
\end{figure}

\subsubsection{Coating and permeation}

The metal coating is accomplished by a sputtering process.
The metallisation provides electrical conductivity, as well as an improvement in the permeation properties of the PET substrate. 
Measured values of the permeation through a 50 nm Cu coated membrane show a barrier improvement factor of 8.5 for He (helium), 9 for CO$_2$,  and 5 for Ar. The quality of the sputtered coating is measured by a peel test using adhesive tape.

\subsubsection{Studies using a scanning electron microscope}

The coated PET was examined using the Scanning Electron Microscope (SEM) in order to determine the quality of the coating and sputtering on the substrate. The SEM was also used to verify the metal coating thickness, not as a direct measurement, but by using the energy-dispersive X-ray spectroscopy facility and third party software.
The metal coating and weld quality of the seam were checked using both the SEM and the traditional optical microscopy, as shown in Fig.~\ref{st_ba_fig2}\,(b) and Fig.~\ref{st_ba_fig3}.

\subsubsection{Glue bonding test}

The adhesion between the outer side of the straw tube and the PEI straw support is critical. 
The chosen epoxy resin for this operation was TRA-BOND 2115,  fabricated by Tra-Con Inc.,~whose outgassing properties and radiation hardness were studied at CERN. This epoxy has a good resistance to stress and temperature cycles. Its pot life is close to two hours and curing in 24 hours. Samples of PET were peel tested after being bonded to an aluminium plate. The bond of the selected candidate was better than the natural resistance of the material. The selected material was also analyzed by X-ray photoelectron spectroscopy. This allowed us to better understand the chemical treatment made by the PET suppliers and its effects on bonding.

\subsubsection{ Straw manufacturing and welding}

\begin{figure}[!h]
    \center{\includegraphics[width=0.9\linewidth]{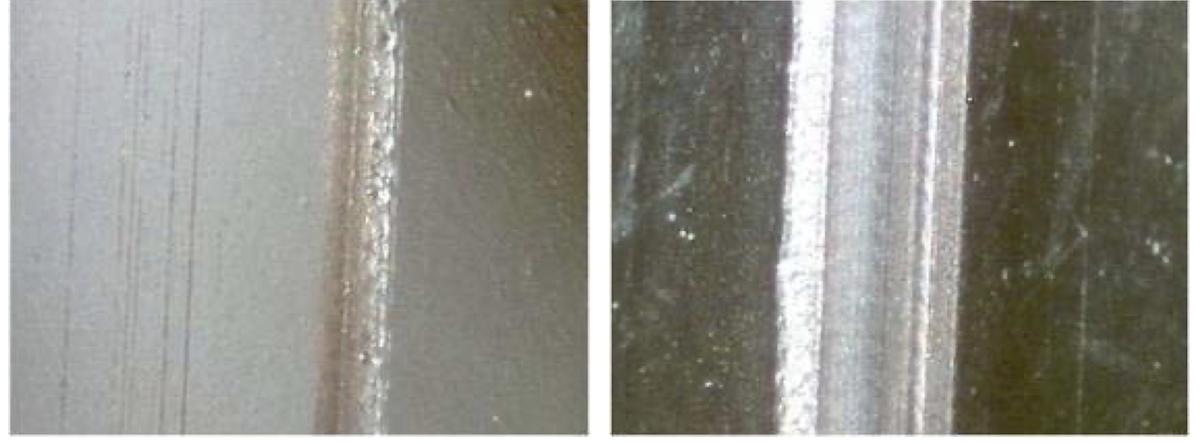}}
  \caption{ Quality of the welding seams of 0.4 mm (left) and 0.85 mm (right) at $\times60$ magnification.
	}
\label{st_ba_fig5}
\end{figure}

\begin{figure}[!h]
    \center{\includegraphics[width=0.7\linewidth]{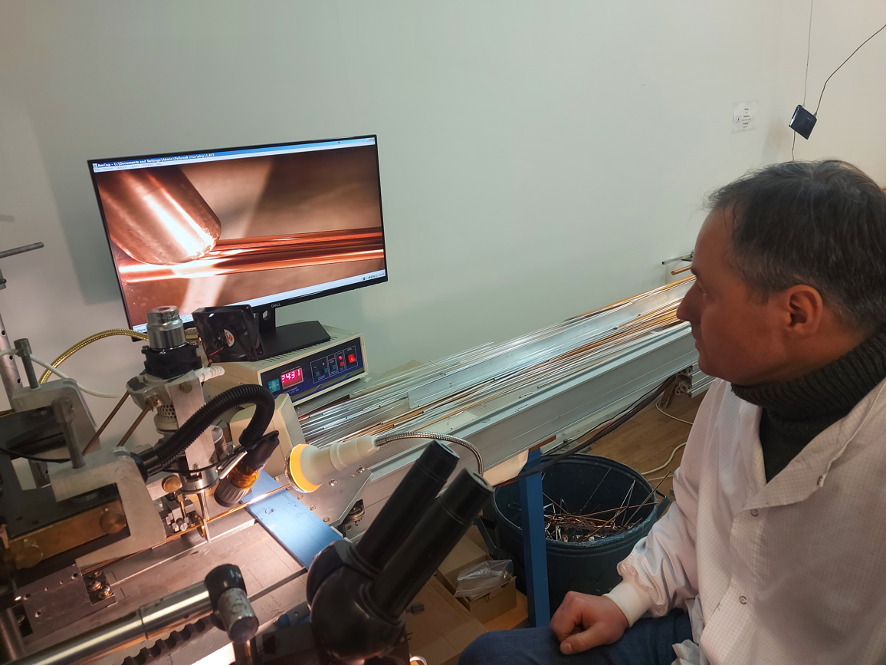}}
  \caption{The straw production setup with an ultrasound head with movable fixation at the frame.
	}
\label{st_ba_fig6}
\end{figure}

\begin{figure}[!h]
  \begin{minipage}[ht]{0.52\linewidth}
    \center{\includegraphics[width=1\linewidth]{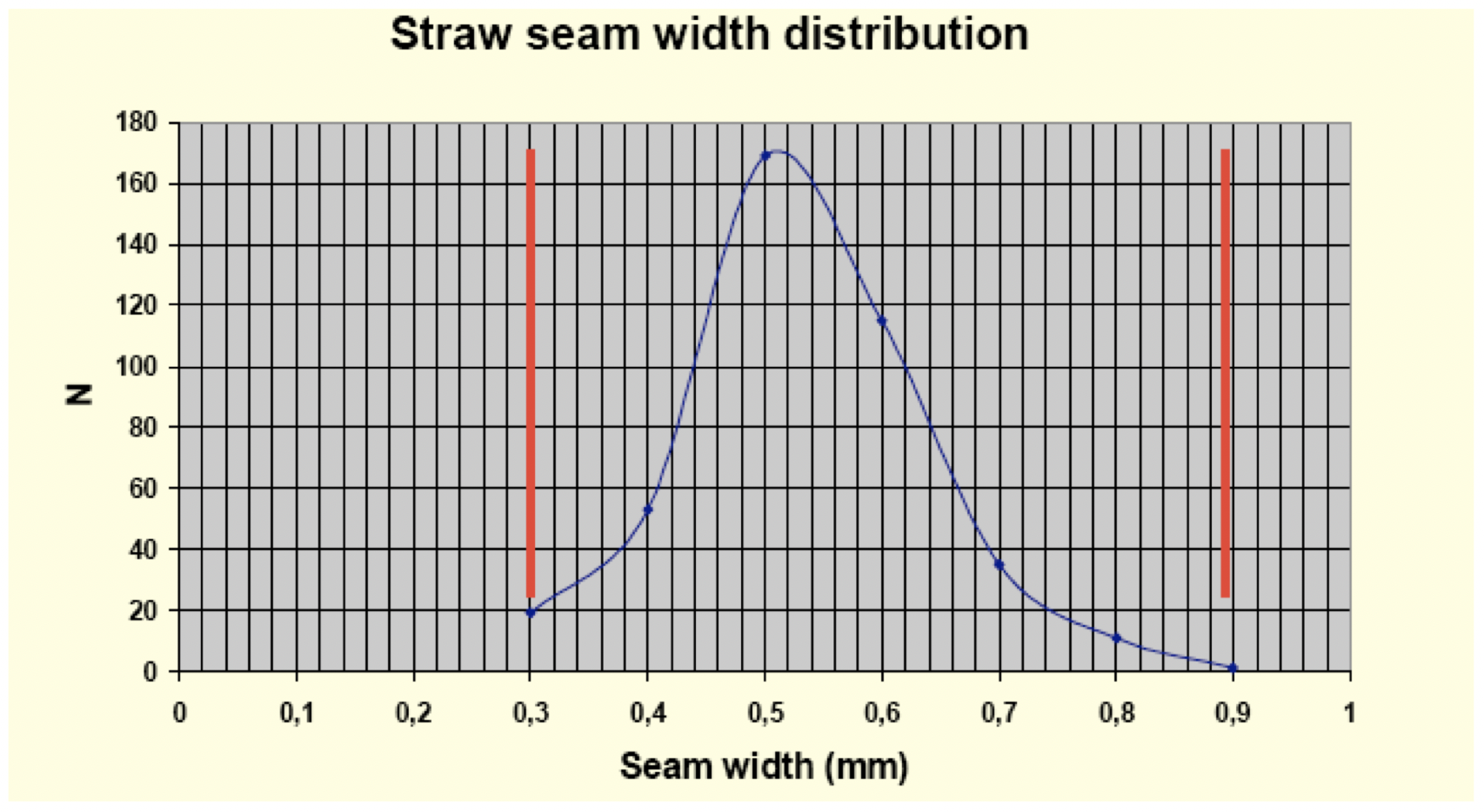} \\ (a)}
  \end{minipage}
  \hfill
  \begin{minipage}[ht]{0.48\linewidth}
    \center{\includegraphics[width=1\linewidth]{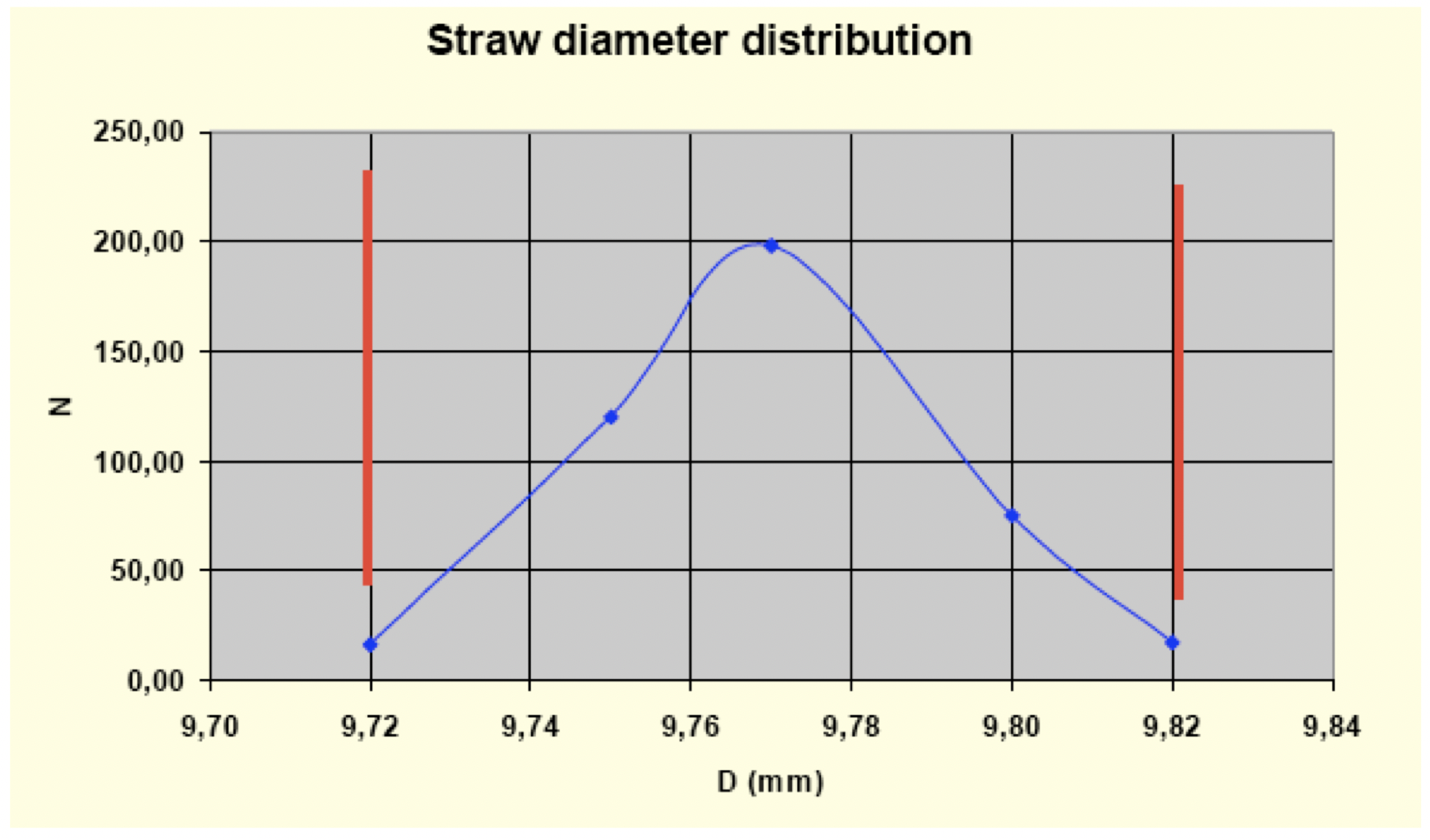} \\ (b)}
  \end{minipage} 
  \caption{Welding seam width (a) and straw diameter (b) distributions.
	}
\label{st_ba_fig7}
\end{figure}

A dedicated machine for the straw production has been designed and constructed to manufacture   
8000 straws for the NA62 experiment (diameter 10~mm, thickness 36~$\mu$m), 
2000 straws for the COMET experiment (diameter 10~mm, thickness 20~$\mu$m), and
100 straws for the SHiP experiment (diameter 20~mm, thickness 36~$\mu$m). 
Straw sets are currently being produced for prototypes intended for the NA62 upgrade and 
the straw tracker for DUNE (diameter 5~mm, thickness 20~$\mu$m). 
An ultrasound head with a movable fixation at the frame and an anvil for the straw welding are shown in Fig.~\ref{st_ba_fig6}.
They were designed and 
constructed to obtain a good quality of the welding seam (see Fig.~\ref{st_ba_fig5}) and 
to minimize differences in the straw diameter down to $\sim$10~$\mu$m. 
The distribution of the straw seam widths and tube diameters, obtained during the production of a large set of straws, are shown in  Fig.~\ref{st_ba_fig7}. 
After manufacturing, each straw is equipped with the glued end-plugs and tested for leaks. 
All measurements are individually recorded in a logbook where all straw characteristics are stored, i.e.~material, manufacturing date, inner diameter, etc. 
A unique serial number is allocated to every straw tube.
Using this number one can access the straw information 
at any moment of the detector assembly.

A few straws with a diameter of 10 mm were used for the dedicated mechanical tests. They were cut into 20 segments about 25~cm long and tested under overpressure until the breaking point. The other straws were cut to 5.3~m and the cut ends were preserved for further analysis. The breaking pressure was found to be 9~bar on average, and not one sample broke under 8.5~bar. The quality control procedure was the same as for NA62 straws. During the ultrasonic welding process,  the seam quality was verified by a digital microscope and recorded to file for each straw. Furthermore, the seam quality was checked by an operator in real time.

\begin{figure}[!h]
    \center{\includegraphics[width=0.6\linewidth]{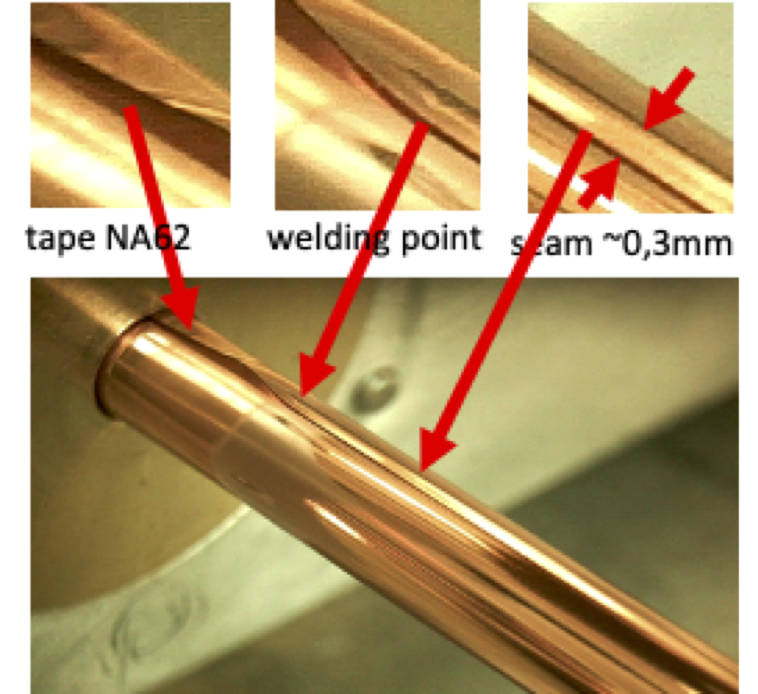}}
  \caption{A single straw tube for the  SPD barrel.
	}
\label{st_ba_fig4}
\end{figure}

Several measurements and tests are performed post-fabrication. The seam width and straw inner diameter are measured by an optical method. The cathode electrical DC resistance is measured. The elongation and breaking force are measured on the test samples (cut straw ends). The straws undergo a long-term overpressure test with temporary end-plugs glued into both ends of each straw. An overpressure test to $\Delta P\approx  2$ bar is performed for a period of about 1~hour. Subsequently, the straw is subjected to a long-term overpressure test at  $\Delta P\approx  2$ for a period of at least 30~days. A gas leak estimation is obtained by measuring the loss of pressure over time. The local straw deformation is measured under an applied weight of 300~g, and the pressure is derived from the calibrated relation between loss of pressure and deformation. The design of an individual straw tube is shown in Fig.~\ref{st_ba_fig4}.

\subsubsection{Straw conditioning}

Once the ultrasonic welding apparatus produces the straws, they need to be carefully stored, prior to installation into the detector. 
The straws need to be stored during a period that varies from several weeks to several months. 
The tube storage has several constraints that need to be addressed.
Namely, the inner diameter of the tube must remain free of dust and of foreign particles. The tube must also have the ability to be manipulated by technicians without buckling. With the above constraints in mind, it was decided that the optimal way to store the tubes would be under pressure. 
This would also have the added value of being able to observe changes in pressure in the pipe over time.
Straws that show gas leakage and have poor weld quality can be removed. 
Taking into account further quality control tests that are performed on the tubes before installation, and the manipulation of the tube itself during its installation into the detector, a dedicated system was developed, in the form of valves.
For storage, the straws are gas-filled with an approximate overpressure of 1~bar. 
The pressure loss over time is measured indirectly by measuring the local deformation of a straw under a 300~g weight and comparing it to the day it was first subjected to pressure.
The first valve must have the ability to let gas into the straw, and then act as a non-return valve, thus keeping the straw under pressure. The valve design is loosely based on the ''Dunlop'' / ''Woods'' valve design, with an additional feature. To prevent leakage through the non-return system, an O ring and  a threaded screw have been added. This is also used to fit the gas connector.

The second valve used at the other end of the straw tube is a simplified version without a non-return mechanism.
To run a leak test, we simply connect the test system to an M4 thread inside the valve and pressurize the straw. 
To simplify measurements of the sagitta of a freely supported tube, both valves have the same weight.
In addition, the ends can be connected to the standard M4 threads. Furthermore, both valves have an outer diameter smaller than the straw tube's outer diameter, which allows the installation of a straw tube under pressure.
This facilitates the handling of the straws and reduces the risk of buckling  the straw wall and the welding seam.

\begin{figure}[!h]
    \center{\includegraphics[width=0.7\linewidth]{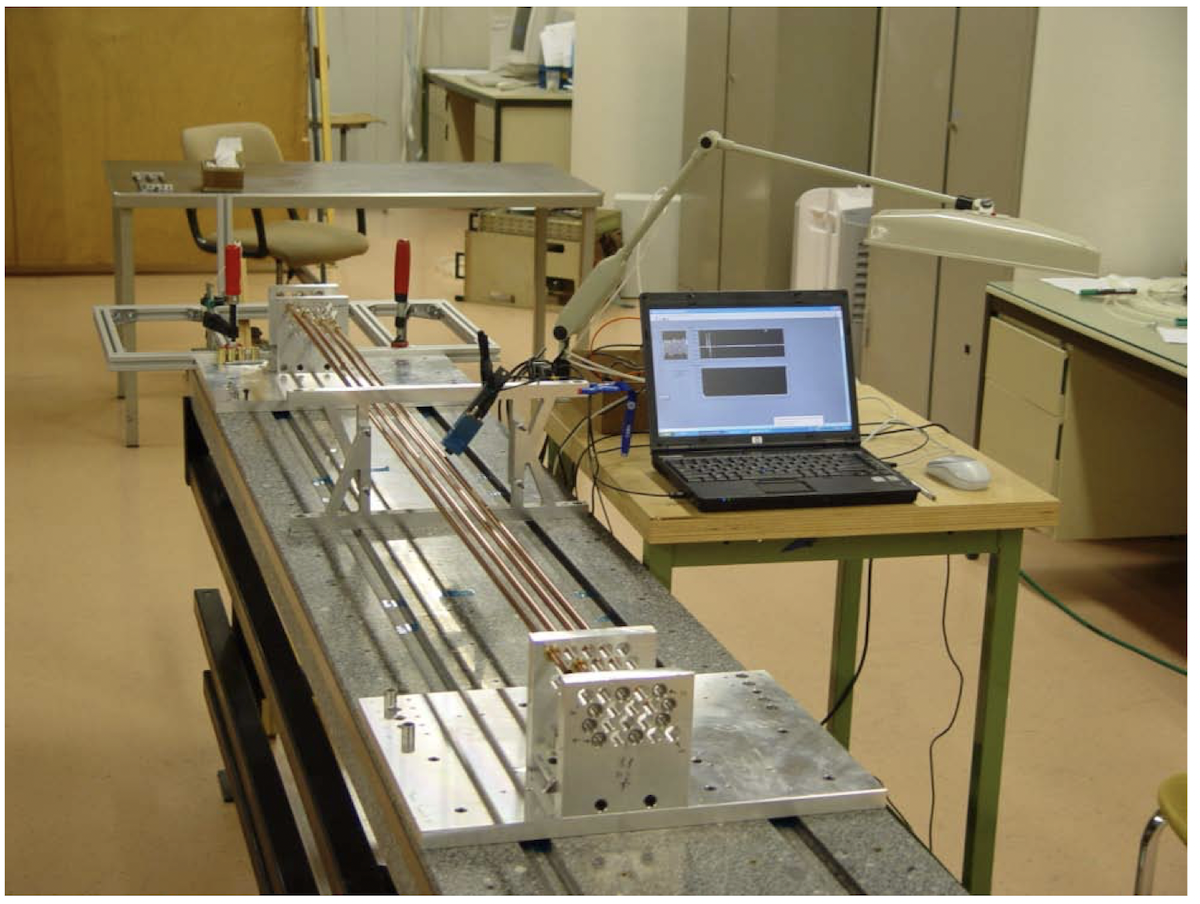}}
  \caption{Setup for the measurement of the straw sag and frequency as a function of different pre-tension and pressure. The computer is connected to an IR emitter and receiver (OPB732) to measure the frequency of the vibrating straw.
	}
\label{st_ba_fig8}
\end{figure}

\subsubsection{Mechanical properties and pre-tension of the straw}

In order to determine the necessary straw pretension, a dedicated setup was developed to measure the straw sag and the vibration frequency of the straws as a function of pressure, as shown in Fig.~\ref{st_ba_fig8}. 
The straw was connected to a gas bottle, and the absolute pressure inside the straws was set to 2 bar. Four straws were glued horizontally into two plates. One straw was equipped with a sliding fixation on one end to allow for a variable straw pre-tension. 
The sliding end of the straw was blocked before measuring the deformation (sag) and frequency. 
The results of the measurements are shown in Fig.~\ref{st_ba_fig9}~(a). 
The increase in sag is clearly visible as the pressure inside the straw decreases.

\subsubsection{ Pressure influence}

\begin{figure}[!h]
  \begin{minipage}[ht]{0.5\linewidth}
    \center{\includegraphics[width=1\linewidth]{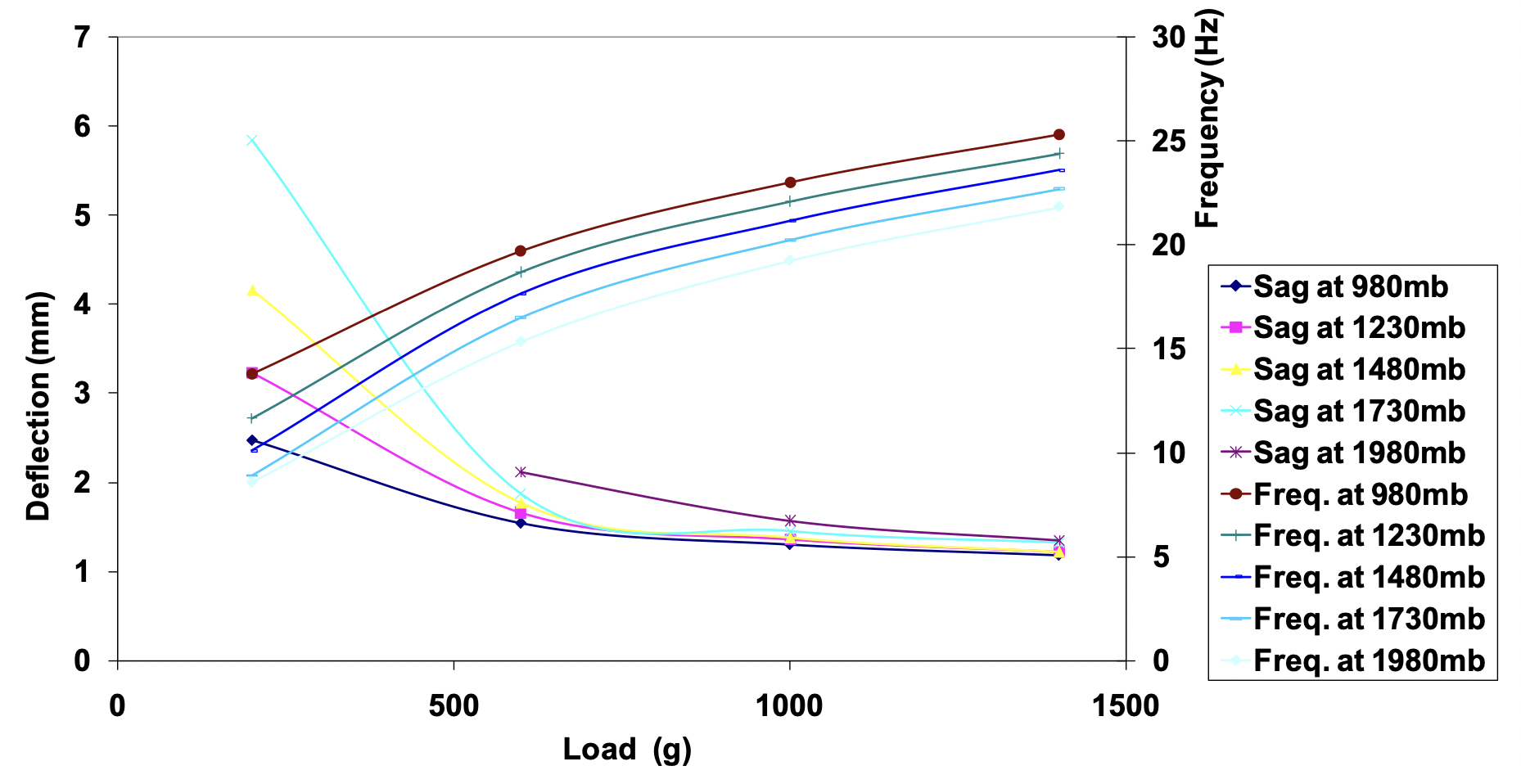} \\ (a)}
  \end{minipage}
  \hfill
  \begin{minipage}[ht]{0.5\linewidth}
    \center{\includegraphics[width=1\linewidth]{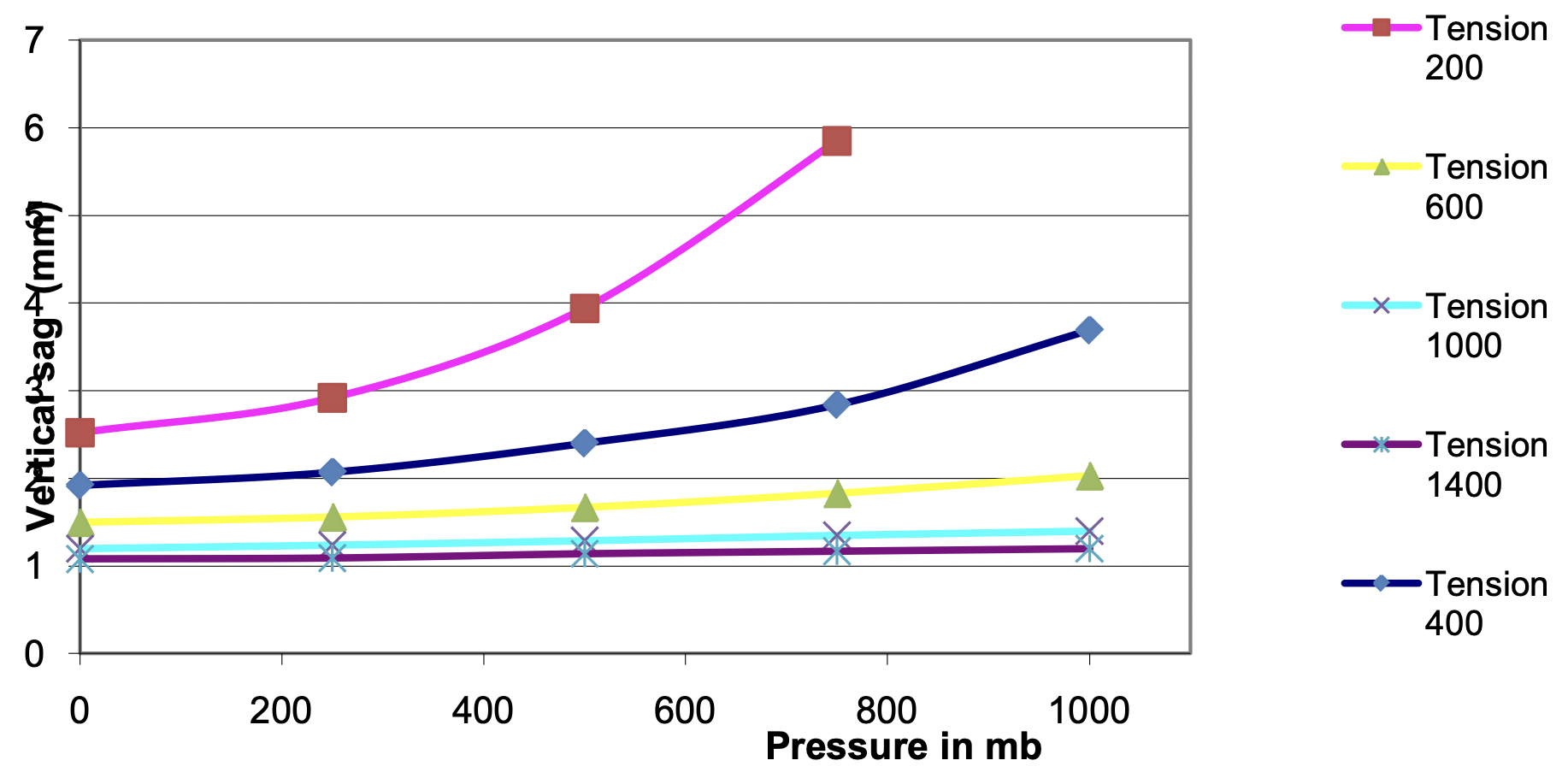} \\ (b)}
  \end{minipage}
  \caption{(a) Measured sag and frequency of the straw as a function of pre-tension at different over-pressures. 
  (b) Straightness as a function of pressure for a 1.85~m long horizontal straw for different values on the pre-tension.
	}
\label{st_ba_fig9}
\end{figure}

A study of the straw sag as a function of overpressure at different pre-tension was carried out to determine the required straw tension. The results are shown in Fig.~\ref{st_ba_fig9}~(b). 
A minimum tension of 10 N is necessary to obtain an acceptable straw straightness. We have decided to apply a force of 15 N during installation,  to allow for some loss of tension over time.

\subsubsection {Wire centering and wire offset}

In order to evaluate the effect of electrostatic forces on the wire, the dependence of the wire displacement on the straw length and high voltage was calculated.
The following formula was applied:
\begin{equation}
y=\frac{C}{k^2}(\frac{1}{\cos (kL/2)} -1),
\end{equation}
where $C=\rho g \sigma$ and $k$ is calculated using the expression
\begin{equation}
k^{2}=2\pi \epsilon_0 E_0^2(b)/T.
\end{equation}
Here $\rho$ is the density of the wire material, $\sigma$ is the cross-section of the wire, $T$ is the straw tension, $E_0$ is the field strength at the straw wall, $L$ is the length of the straw tube, and $\epsilon_0$ is the vacuum permittivity constant.

\begin{figure}[!h]
    \center{\includegraphics[width=0.7\linewidth]{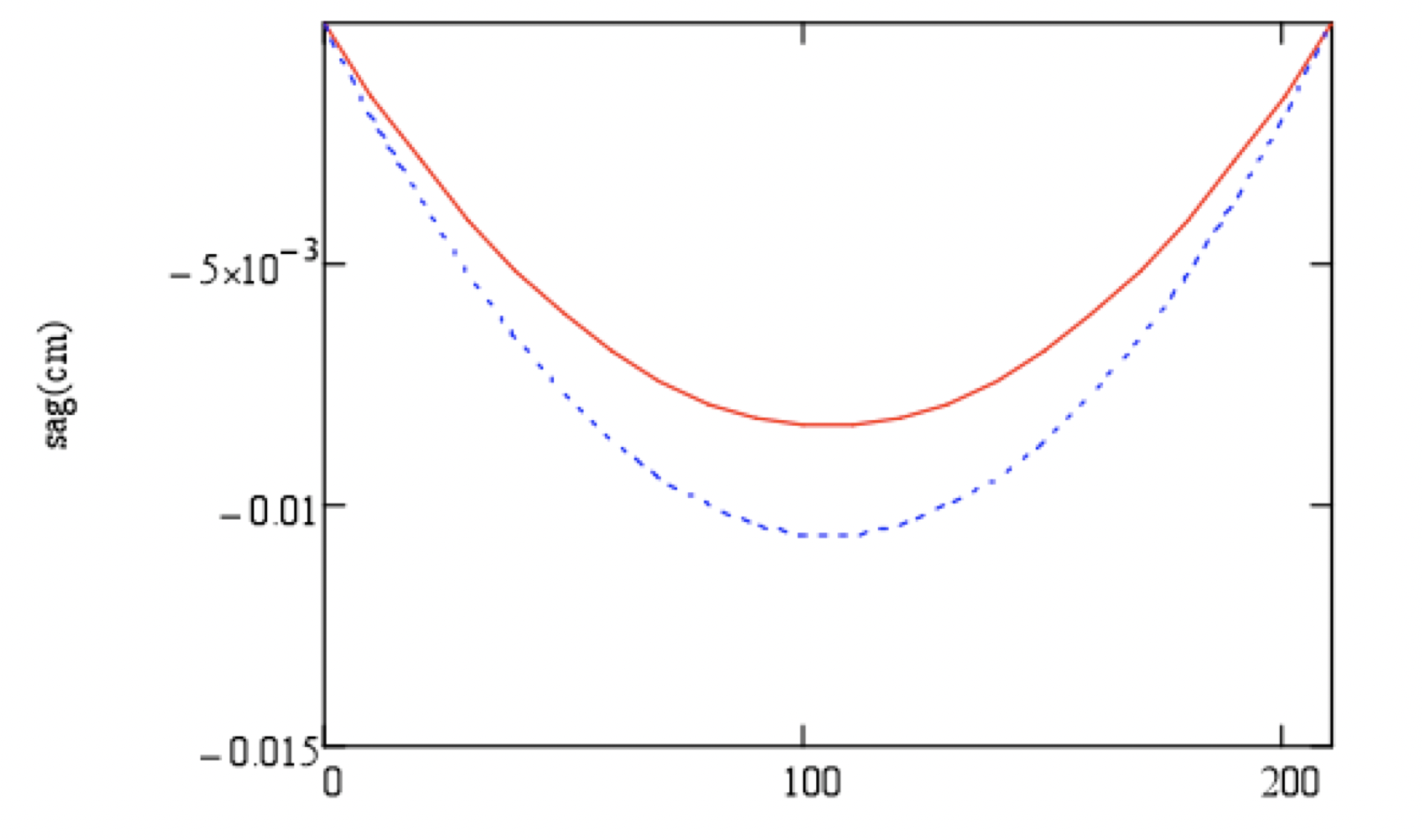}}
  \caption{The wire deflection for a 2.1~m long straw  with (blue dashed line) and without (red solid line)
  	electrostatic deflection at 2.5~kV.
	}
\label{st_ba_fig10}
\end{figure}


The results are presented in Fig.~\ref{st_ba_fig10}.
In the case when the voltage was set to 2.5~kV (slow gas) and the tension in the wire was set to 90~g,
the increase in wire sag due to electrostatic forces is about 27\% concerning the case without an electric field.
For the fast gas with a high voltage of 1.8~kV, the corresponding increase in the wire sag is 9\%.

\subsection{Detector response simulation}
\label{subGarfield}
\subsubsection{Choice of the gas mixture and high voltage operating point}

The initial choice of the gas mixture and high voltage settings was based on the experience of the NA62 experiment \cite{NA62:2017rwk}, where the straw tracker consists of  the straw tubes of identical design. 
Its performance requirements are similar to those of the SPD straw tracker. 
The $Ar(70\%)+CO_2(30\%)$ gas mixture was demonstrated to provide stable straw operation, sufficient spatial resolution, and high detection efficiency. 
Moreover, it does not contain any flammable or non-ecological components. 
The optimal operation voltage was chosen to be 1750 V, which corresponds to a gas gain of about $4.5\times10^4$. In the case of  the NA62 experiment readout electronics,  based on the CARIOCA chip \cite{Moraes:2003rga}, 
such voltage allows for efficient operation with a spatial resolution better than 120 $\mu$m.

\begin{figure}[!h]
  \begin{minipage}[ht]{0.5\linewidth}
    \center{\includegraphics[width=1\linewidth]{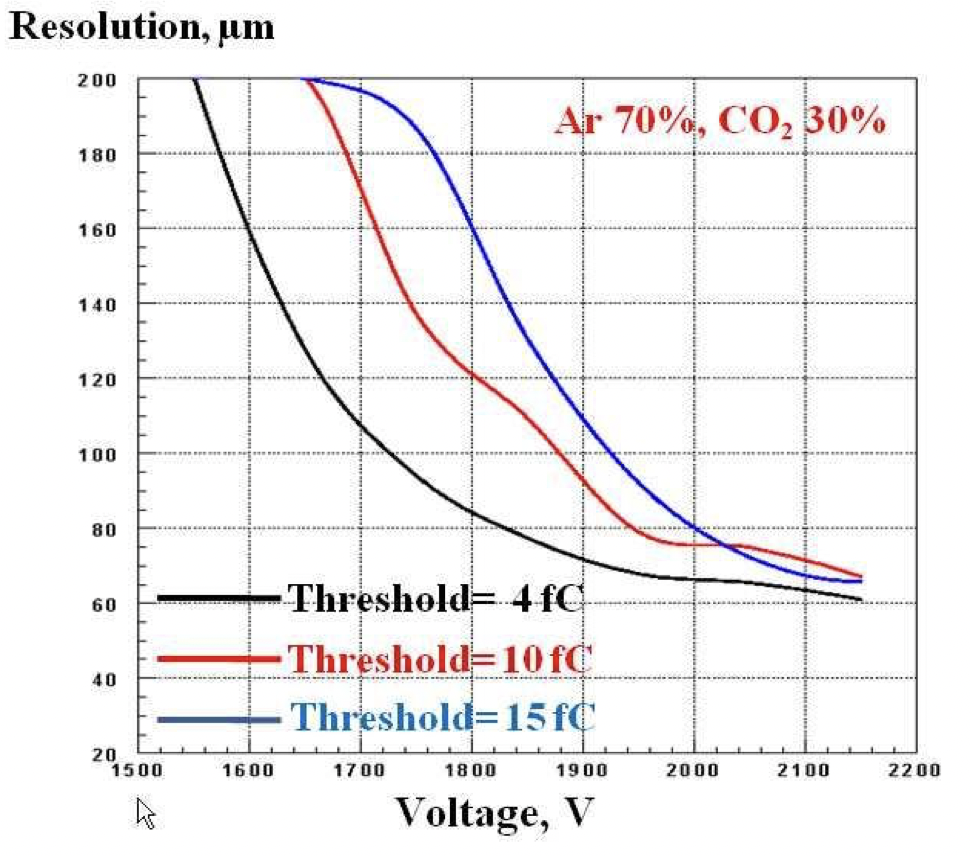} \\ (a)}
  \end{minipage}
  \hfill
  \begin{minipage}[ht]{0.5\linewidth}
    \center{\includegraphics[width=1\linewidth]{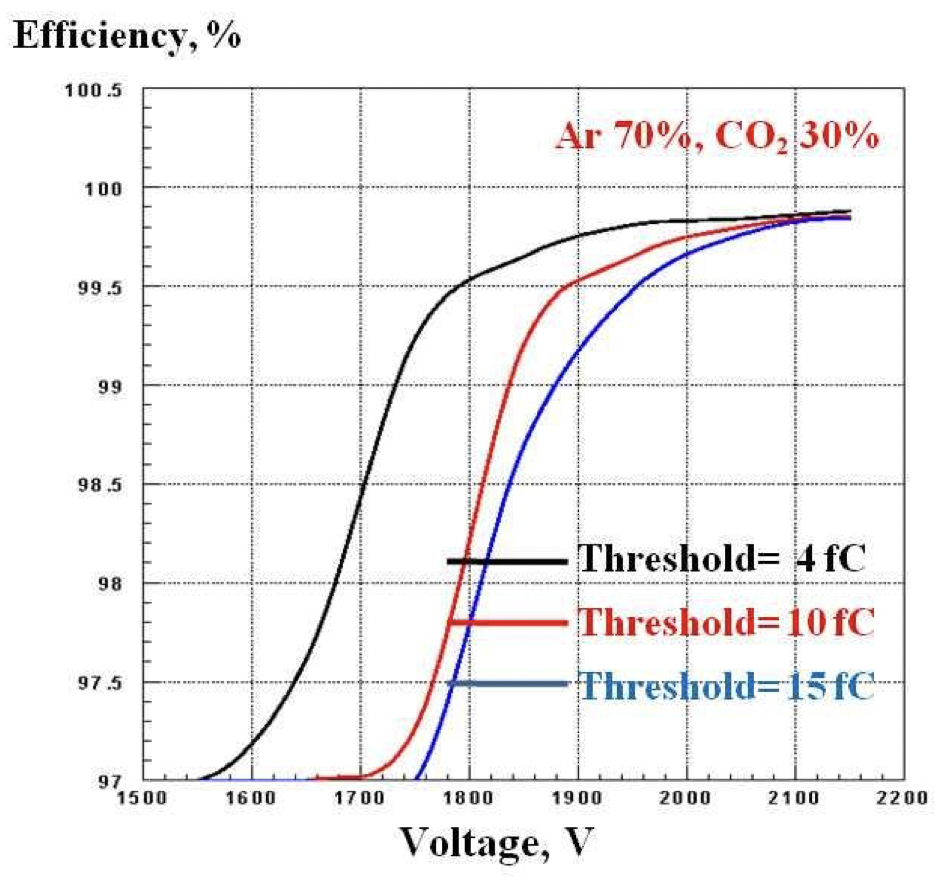} \\ (b)}
  \end{minipage} 
  \caption{GARFIELD simulation results for the straw spatial resolution (a) and detection efficiency (b) as functions of the high voltage and discriminator thresholds for a charged track crossing a straw at the distance of 3 mm from the anode wire \cite{Hahn:1404985}.
	}
\label{st_gar_0}
\end{figure}

The straw spatial resolution and detection efficiency have been studied with the GARFIELD simulation software \cite{Veenhof:1993hz,Veenhof:1998tt} for the NA62 readout electronics models. 
The results as functions of the high voltage and signal discriminator thresholds are presented in Ref.\,\cite{Hahn:1404985} and are also  shown in Fig.~\ref{st_gar_0} . 
The considered discriminator thresholds correspond to 4, 10, and 15 fC of the charge induced on an anode wire. The results are shown for charged tracks crossing the straw at a distance of 3 mm from the anode wire. 

\subsubsection{ GARFIELD simulation of the straw tube response}
\label{subsubGarf}
\begin{figure}[!h]
  \begin{minipage}[ht]{0.5\linewidth}
    \center{\includegraphics[width=1\linewidth]{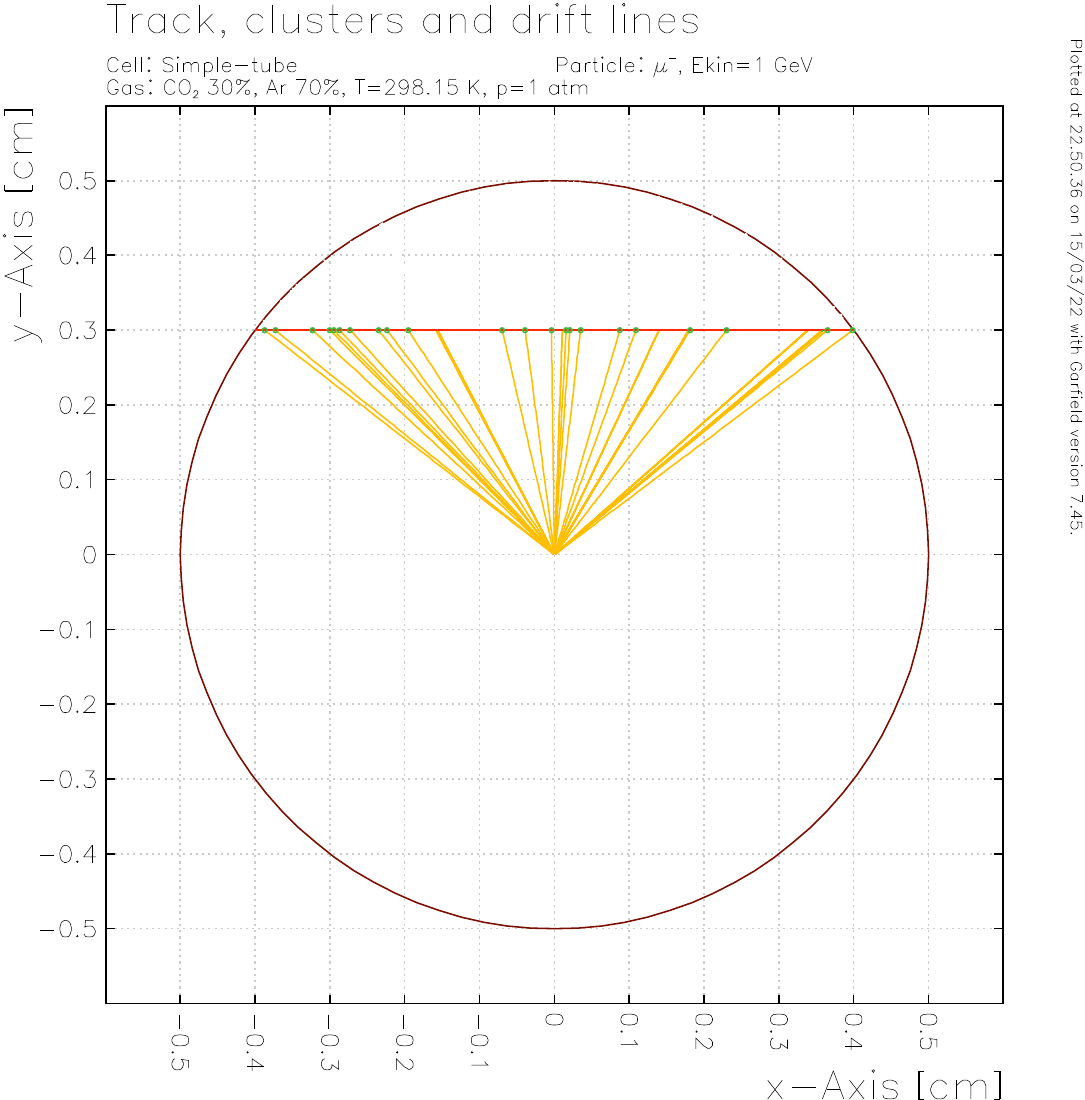} \\ (a)}
  \end{minipage}
  \hfill
  \begin{minipage}[ht]{0.5\linewidth}
    \center{\includegraphics[width=1\linewidth]{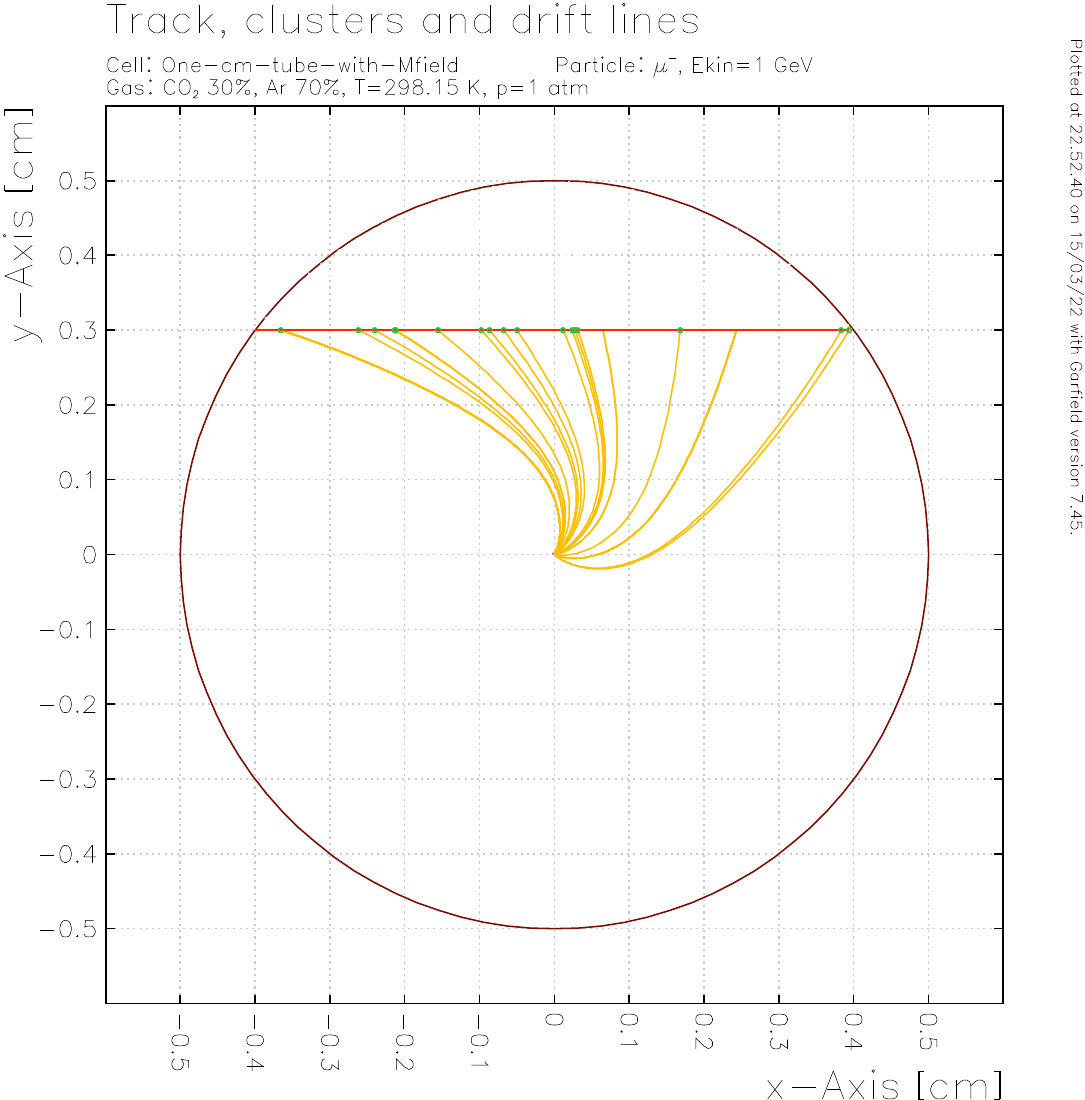} \\ (b)}
  \end{minipage} 
  \caption{Muon track (red), primary ionization clusters (green), and electron drift trajectories (yellow), simulated with GARFIELD for 0 T (a) and 1.5 T (b) magnetic field, aligned with the straw longitudinal axis.
	}
\label{st_gar_1}
\end{figure}

Detailed studies of a straw tube response are ongoing for the SPD straw tracker operation environment. To evaluate the intrinsic performance of a straw in the SPD magnetic field, a GARFIELD simulation of the straw response to a 1 GeV muon has been performed for two extreme cases of 0\,T and 1.5~T field, aligned with the straw longitudinal axis. Muon tracks are generated to be normal to the anode wire at a distance of 3~mm. Drift trajectories of the electrons produced in clusters of primary ionization for 0\,T and 1.5~T magnetic fields
are compared in Fig.~\ref{st_gar_1} (a) and (b), respectively. 

\begin{figure}[!h]
  \begin{minipage}[ht]{0.5\linewidth}
    \center{\includegraphics[width=1\linewidth]{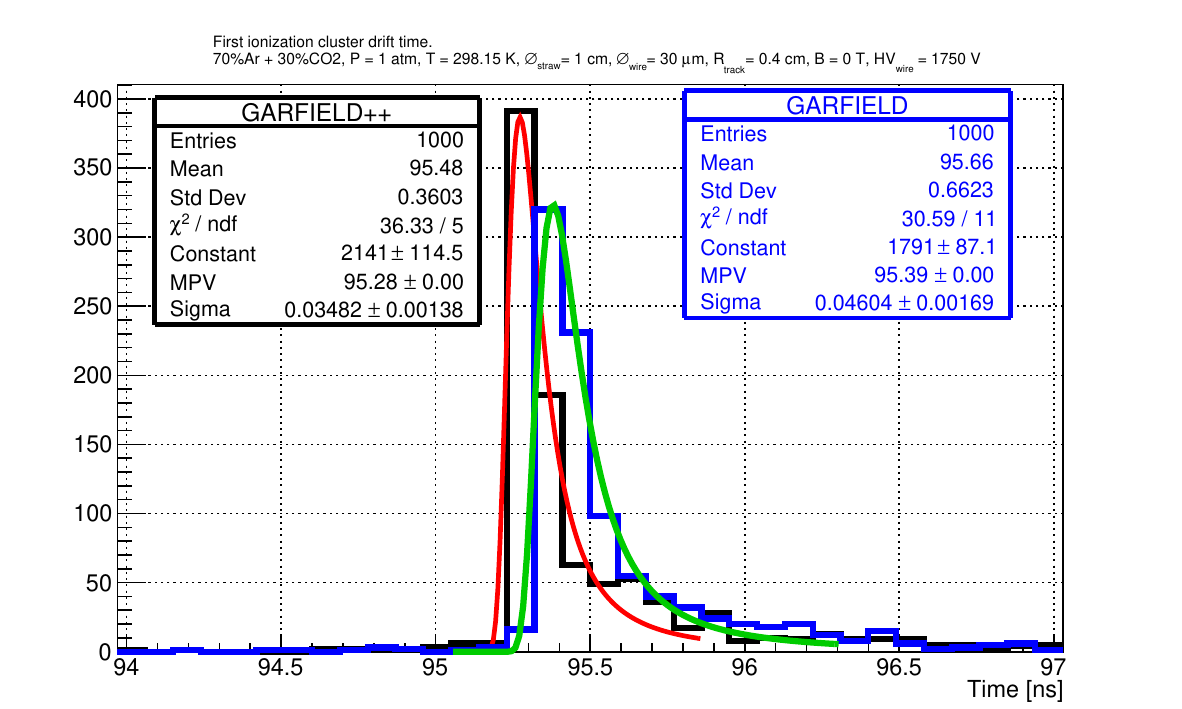} \\ (a)}
  \end{minipage}
  \hfill
  \begin{minipage}[ht]{0.5\linewidth}
    \center{\includegraphics[width=1\linewidth]{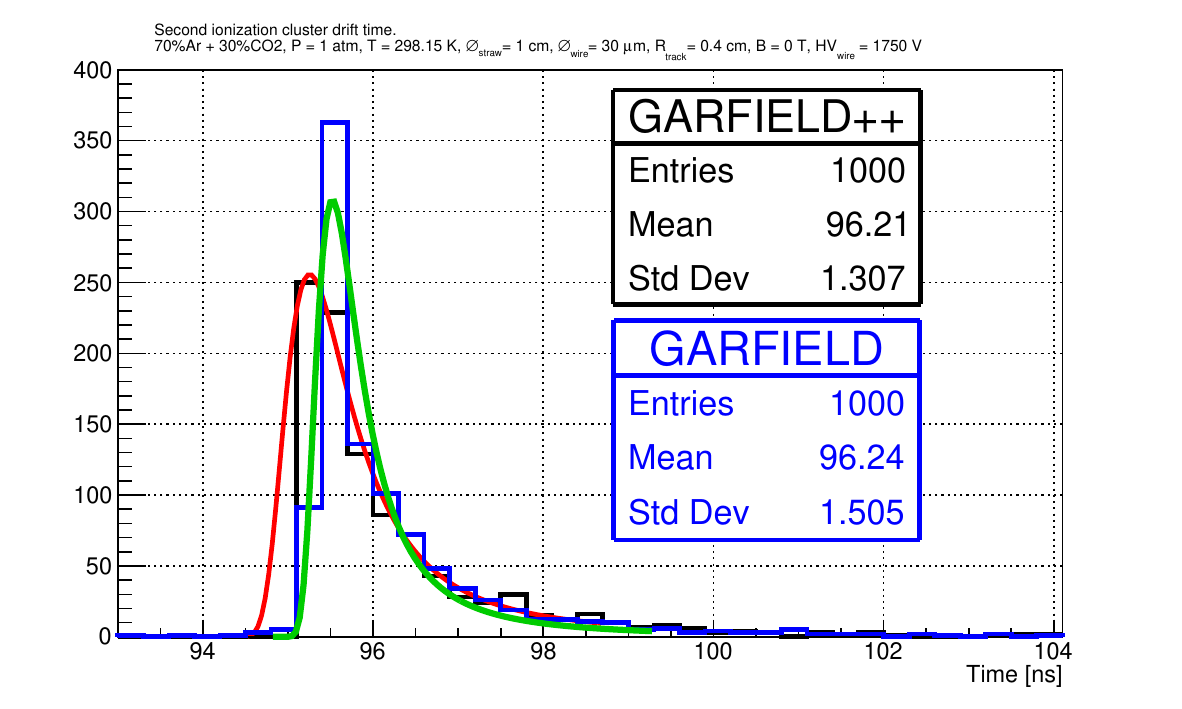} \\ (b)}
  \end{minipage} 
  \begin{minipage}[ht]{0.5\linewidth}
    \center{\includegraphics[width=1\linewidth]{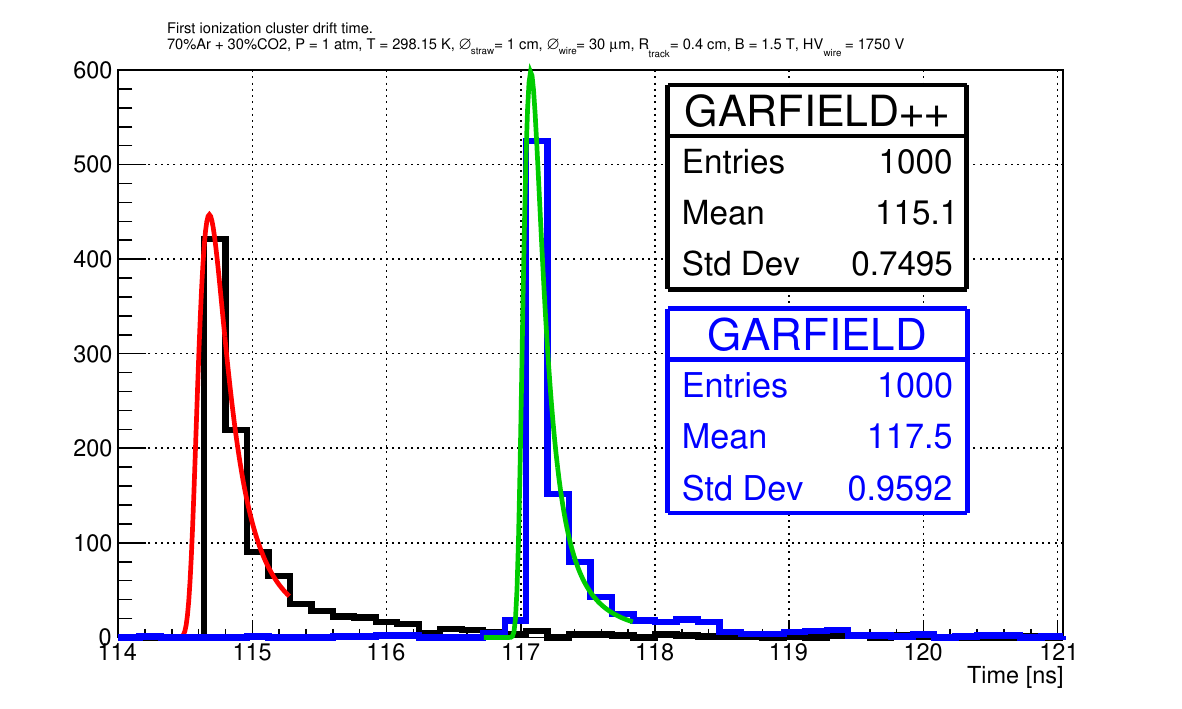} \\ (c)}
  \end{minipage}
  \hfill
  \begin{minipage}[ht]{0.5\linewidth}
    \center{\includegraphics[width=1\linewidth]{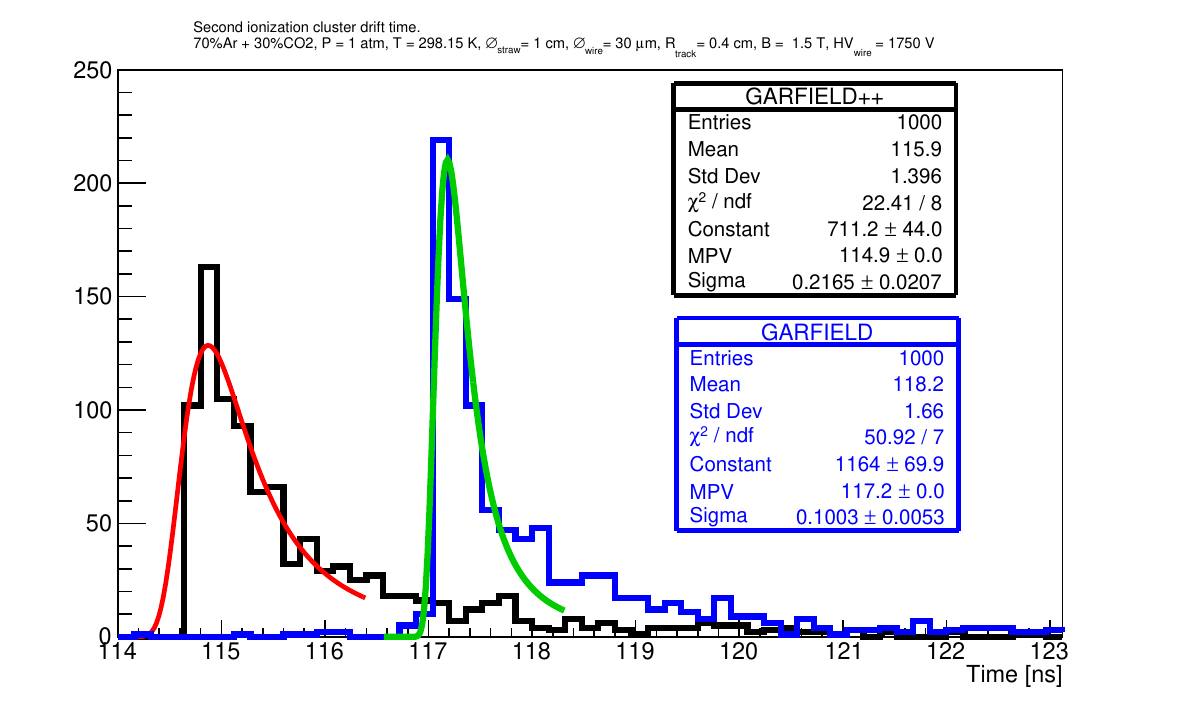} \\ (d)}
  \end{minipage} 
  \caption{ Examples of the obtained drift time distributions for the first (left) and second (right) arriving ionization clusters for a track passing at 4 mm from the anode wire for 0 T (top) and 1.5 T (bottom) magnetic field. To compare the most probable values (MPV) and the distribution widths (RMS), the GARFIELD (blue) and Garfield++ (black) predictions are fitted to a Landau distribution (green and red, correspondingly).	}
\label{st_gar_2}
\end{figure}

The relative coordinate of the primary ionizing particle is reconstructed from the measured electron drift time and is affected by its fluctuations, which, in turn, depend on fluctuations in the primary ionization cluster distribution, together with longitudinal diffusion as well as on characteristics of the readout electronics. To evaluate the intrinsic performance of the straw tubes, the drift time of the first and second ionization clusters arriving at the anode is studied as a function of the distance between the muon track and the anode wire. No transformation of the straw signal by the readout electronics is implemented at that stage. Examples of the obtained time distributions for the first and second arriving clusters,  produced by a muon passing at a distance of 4 mm from the anode wire, are shown in Fig.~\ref{st_gar_2} for a magnetic field of 0\,T  and 1.5~T. Simulation is performed with the old Fortran version of GARFIELD and with the new Garfield++ package. A reasonable agreement between the two predictions was observed for both values of the magnetic field.

The width of obtained distributions does not exceed 0.3 ns even for the magnetic field of 1.5\;T. This points to a negligible contribution of the cluster distribution fluctuations and electron diffusion to the straw time resolution. The influence of the magnetic field on the most probable arrival time is clearly noticeable, while the time distribution widths are not affected by the magnetic field aligned with the anode wire. The dominating influence of the readout electronics on the straw tube time resolution will be demonstrated later in Section \ref{RER}.

Additionally to precise tracking, the Straw Tracker may also be used for particle identification. To evaluate the intrinsic capability of the straw tubes for energy loss measurements, Garfield++ simulation studies of the gas gain fluctuations are performed and the results are compared to estimated values. 

When a charged particle traverses a straw tube, the relative standard deviation of the induced charge $q$ for the given particle energy loss $\Delta E$ can be estimated as 
$$
\frac{\sigma(q)}{q} = \sqrt{\frac{W}{\Delta E}(F+f)}\;,
$$
where $W$ is the mean ionization energy per electron-ion pair in the gas, $F$ is the Fano factor and $f$ is the relative variance of the single-electron multiplication~\cite{Knoll:2000fj}. The Fano factor empirically accounts for correlations in the ionization process which cause a reduction of the statistical fluctuation in the number $N$ of primary electron-ion pairs, $\sigma_N=\sqrt{FN}$, $F<1$~\cite{Fano:425303}. The calculations are performed for the fixed energy loss of 1.5\;keV/cm, which corresponds to the average energy loss of a minimum ionizing particle (MIP), and for the Fano factor of 0.23 as measured in pure argon~\cite{HASHIBA1984305}. 
The average ionization energy and factor $f$ for the Ar/CO$_2$ gas mixture are taken as 28\;eV and 0.8~\cite{Schindler:1500583} respectively.
 
\begin{figure}[!h]
  \begin{minipage}[ht]{0.5\linewidth}
    \center{\includegraphics[width=1\linewidth]{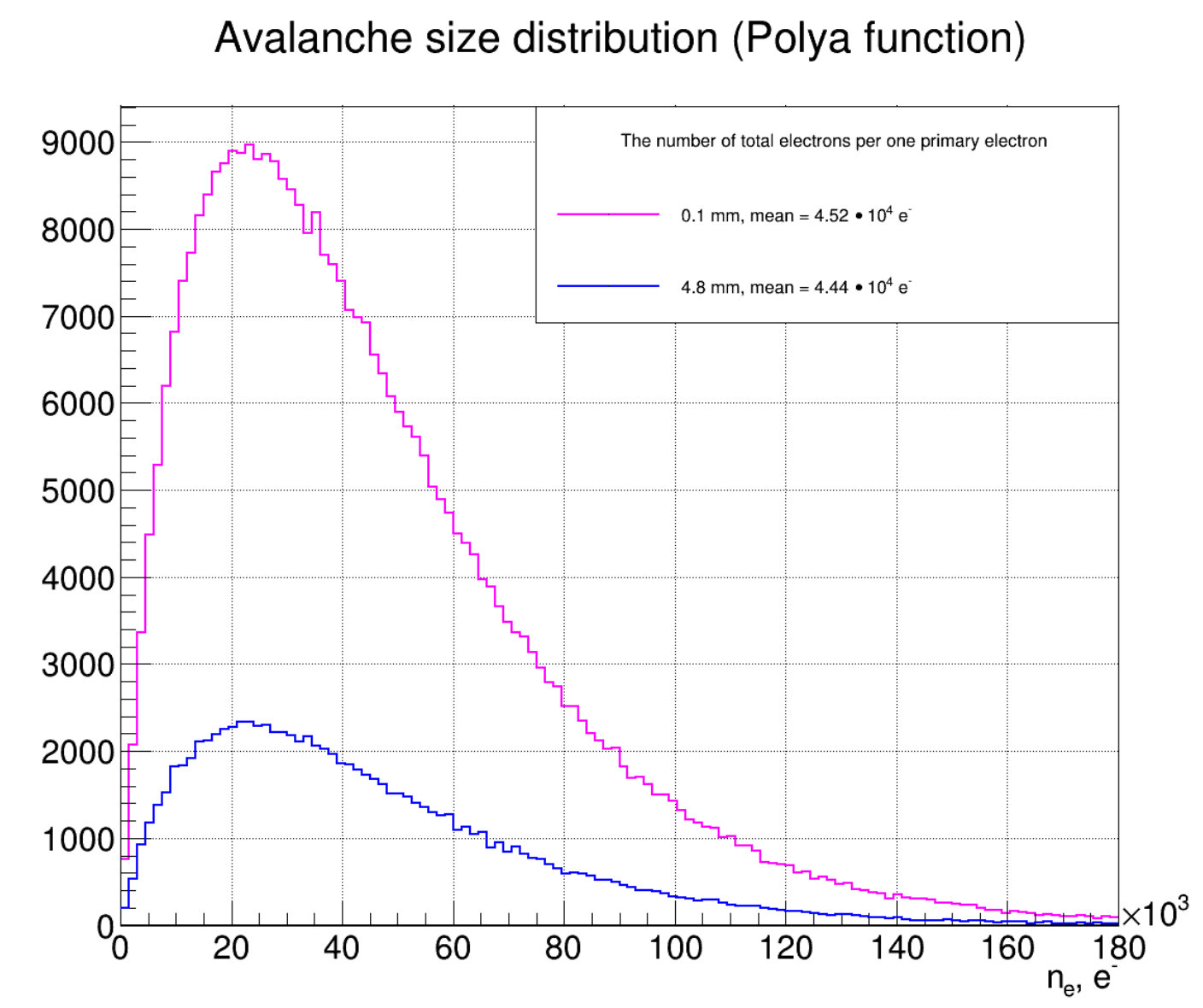} \\ (a)}
  \end{minipage}
  \hfill
  \begin{minipage}[ht]{0.5\linewidth}
    \center{\includegraphics[width=1\linewidth]{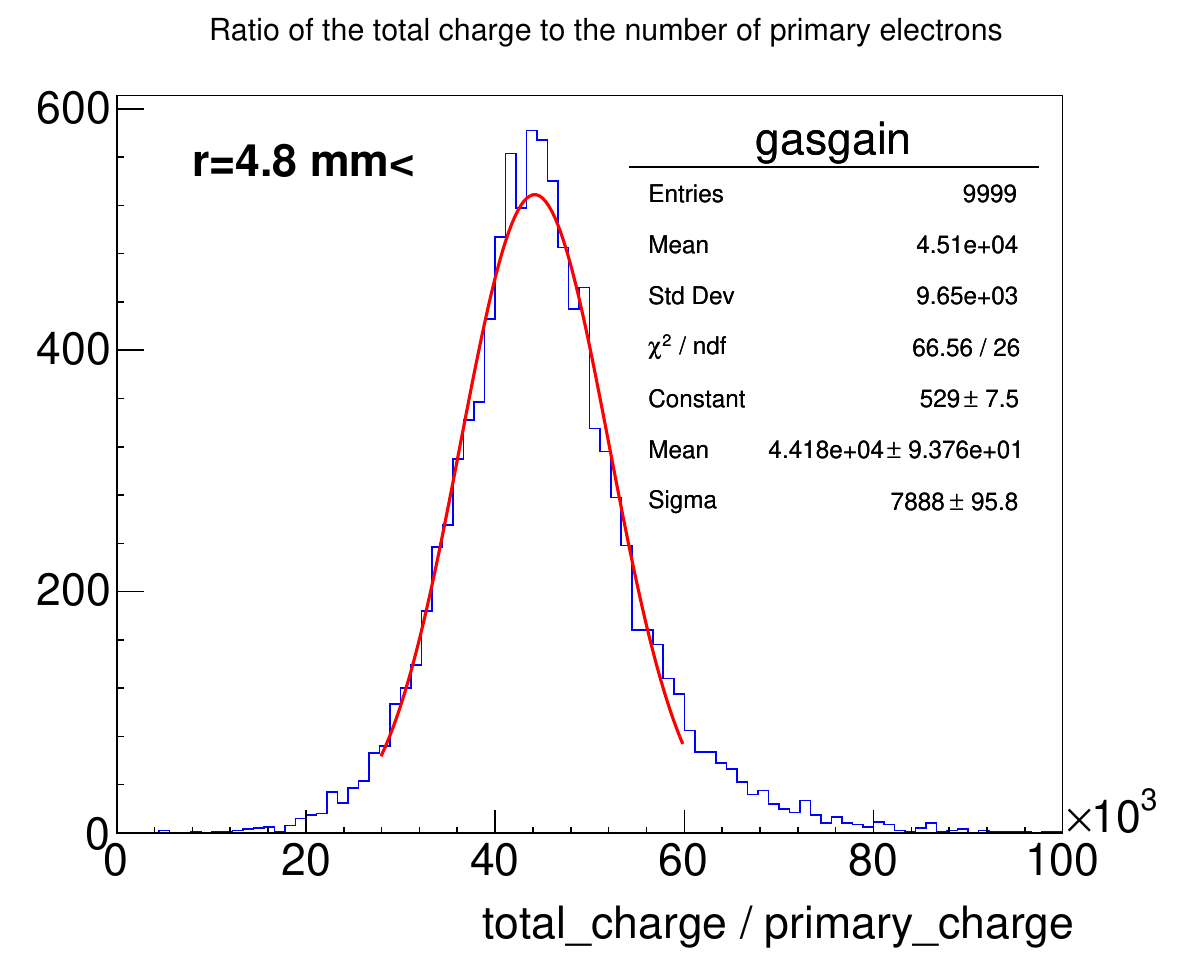} \\ (b)}
  \end{minipage} 
  \caption{Distribution of the single-electron multiplication factor obtained for primary electrons generated at the distances 0.1 and 4.8\;mm from the anode wire (a). Gas gain distribution obtained for muon signals as the ratio of the signal charge to the number of primary electrons created by the muon passing at the distance of 4.8\;mm from the anode (b).}
 \label{garfq}
\end{figure}

For GARFIELD++ simulation studies, two models of the single-electron multiplication factor distribution are considered. Polya distribution~\cite{Schindler:1500583} with the shape parameter $\theta$=1 corresponding to the factor $f=(1+\theta)^{-1}=0.5$ is chosen together with the extreme case of $\theta$=0. Figure~\ref{garfq} shows the distribution of the number of electrons in a single avalanche for $\theta$=1 (a) and the resulting gas gain distribution obtained for muon signals as the ratio of a signal charge to the number of primary electrons created by a 1\;GeV muon (b). 

\begin{figure}[!h]
    \center{\includegraphics[width=0.7\linewidth]{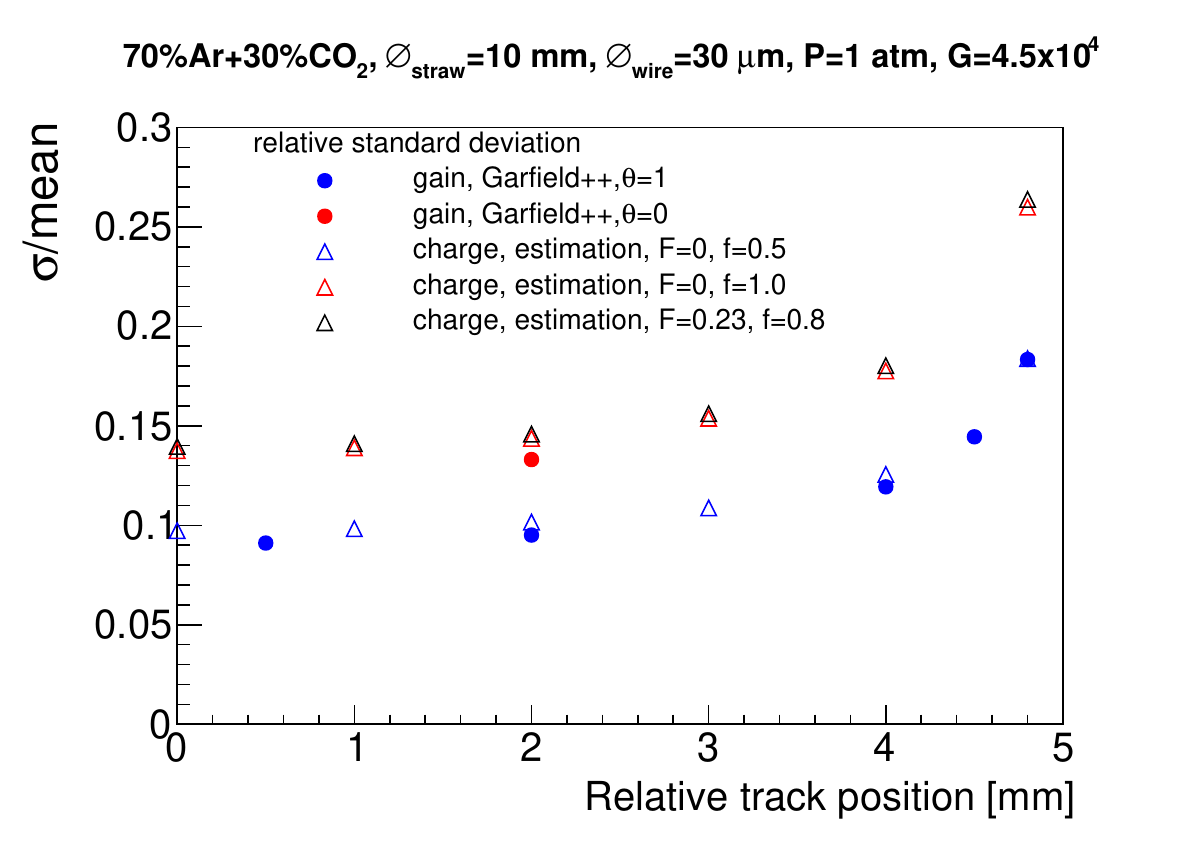}}
  \caption{Relative standard deviation of the gas gain, as obtained in Garfield++ simulation for the Polya function parameters $\theta$ of 0 and 1. The results are in a good agreement with the estimated charge deviation for factors $f$ equal to 1 and 0.5, respectively, neglecting the contribution of the primary charge fluctuation ($F=0$). The complete relative standard deviation of the induced charge for the fixed energy loss of 1.5\;keV/cm is calculated for $f=0.8, F=0.23$.
	}
\label{relSTD}
\end{figure}

The relative deviation of the charge distribution obtained in GARFIELD++ simulation is compared to the calculated values in Figure~\ref{relSTD}. Except for the peripheral straw areas, both the relative variation of the gas gain and the total charge variation for the average MIP energy loss do not exceed 20\%. 

\subsubsection{ Influence of realistic electronics readout on the straw time resolution \label{RER}}

The influence of the readout electronics on the straw spatial resolution is studied using a combination of GARFIELD and LTSpice \cite{LTspice} electronics simulation package. Individual straw signals simulated with GARFIELD are processed with LTSpice for a given electronics chain model. One of the possible SPD ST front-end electronics options, based on a VMM3a ASIC~\cite{Iakovidis_2020}, is considered in the study, and the corresponding model provided by the RD51 Collaboration (CERN) is used. Details of the readout electronics are described in Section~\ref{VMM3a_section}. 
At this stage, electronic noise and the influence of the magnetic field are not taken into account.


\begin{figure}[!h]
    \center{\includegraphics[width=0.7\linewidth]{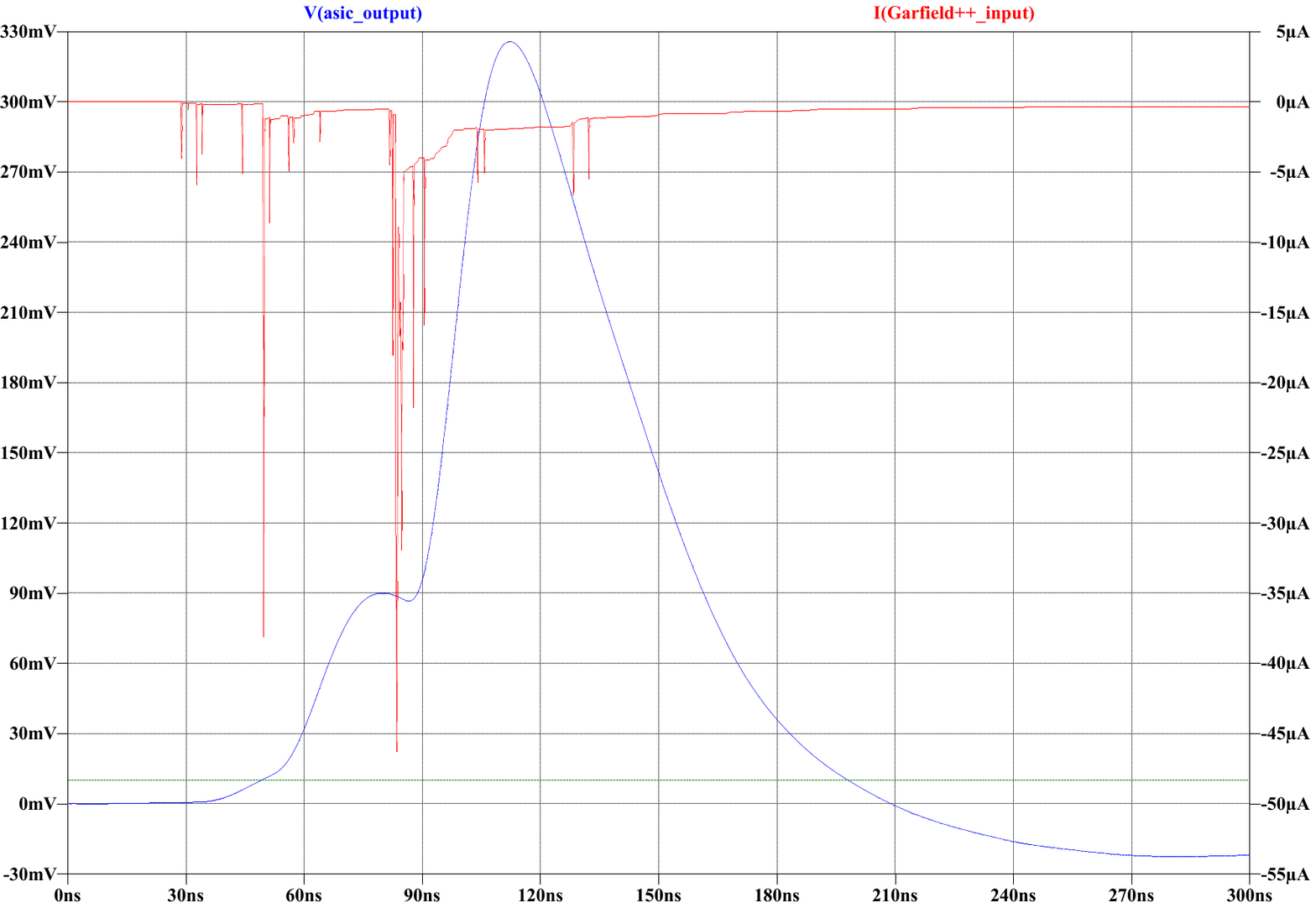}}
  \caption{Signal induced at the straw anode due to an avalanche development as generated by Garfield++ for a muon crossing the straw at the distance of 2\;mm from the anode wire (red, left y-axis) and the corresponding electronics response emulated with LTSpice (blue, right y-axis). The threshold of 10 mV corresponding to 3\;fC of charge for the considered electronics model is shown as a green line.
	}
\label{st_gar_4}
\end{figure}

\begin{figure}[!h]
  \begin{minipage}[ht]{0.32\linewidth}
    \center{\includegraphics[width=1\linewidth]{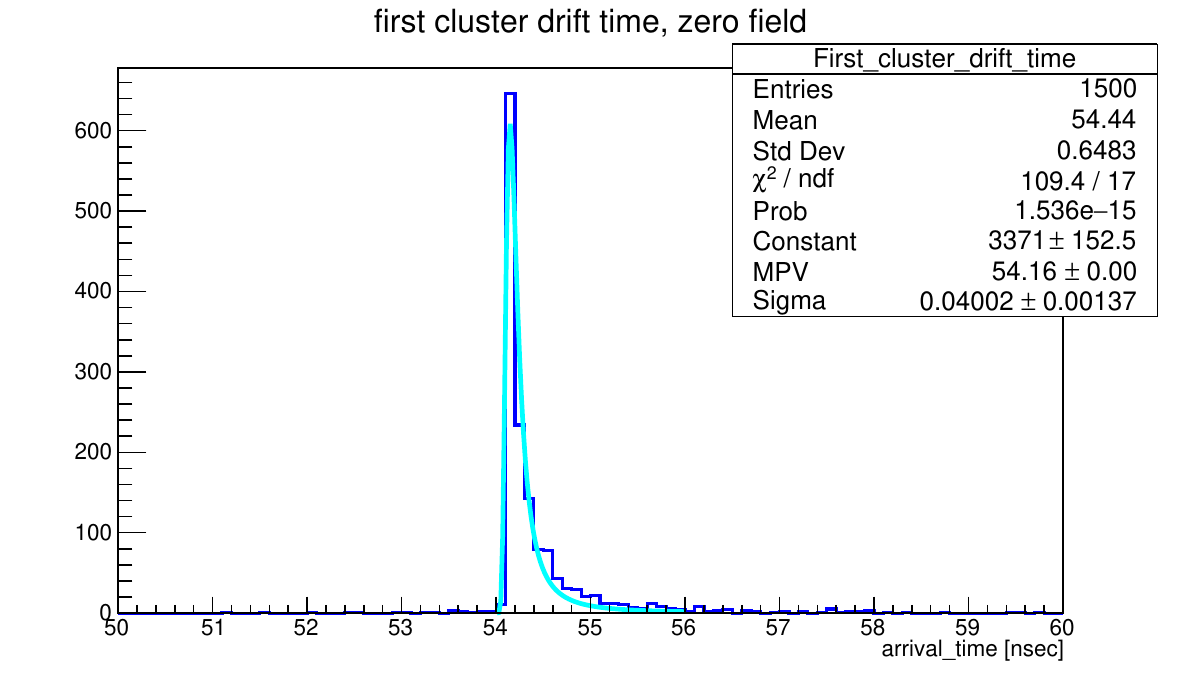} \\ (a)}
  \end{minipage}
  \hfill
  \begin{minipage}[ht]{0.32\linewidth}
    \center{\includegraphics[width=1\linewidth]{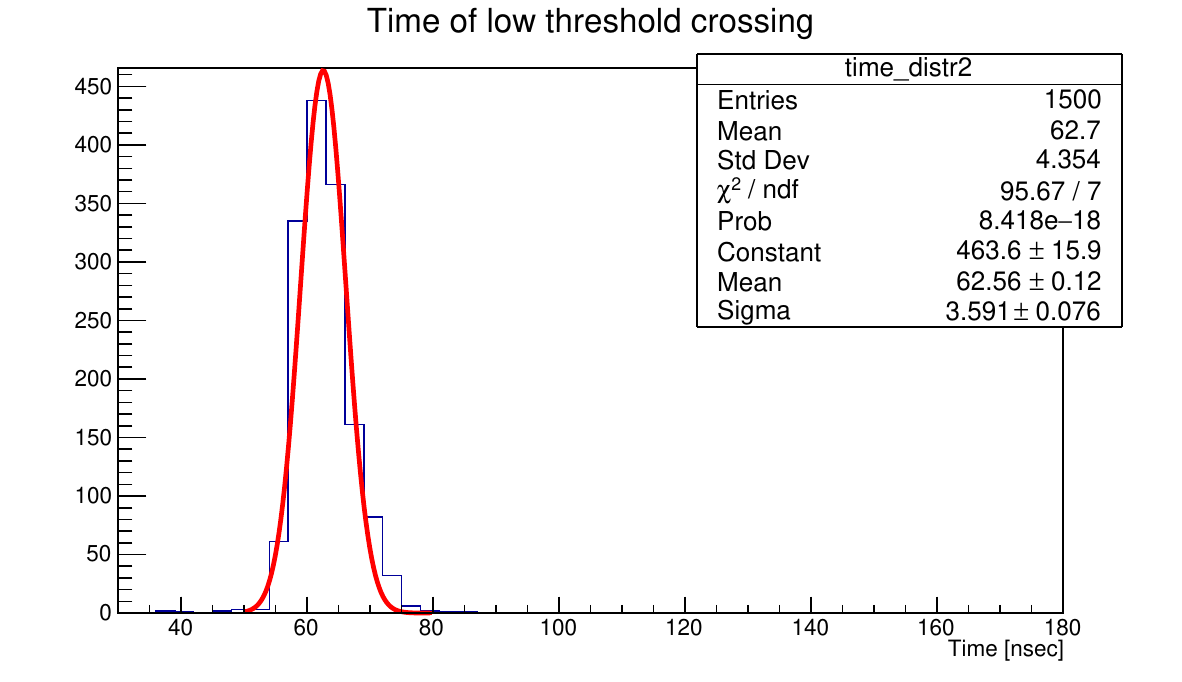} \\ (b)}
  \end{minipage} 
  \hfill
  \begin{minipage}[ht]{0.32\linewidth}
    \center{\includegraphics[width=1\linewidth]{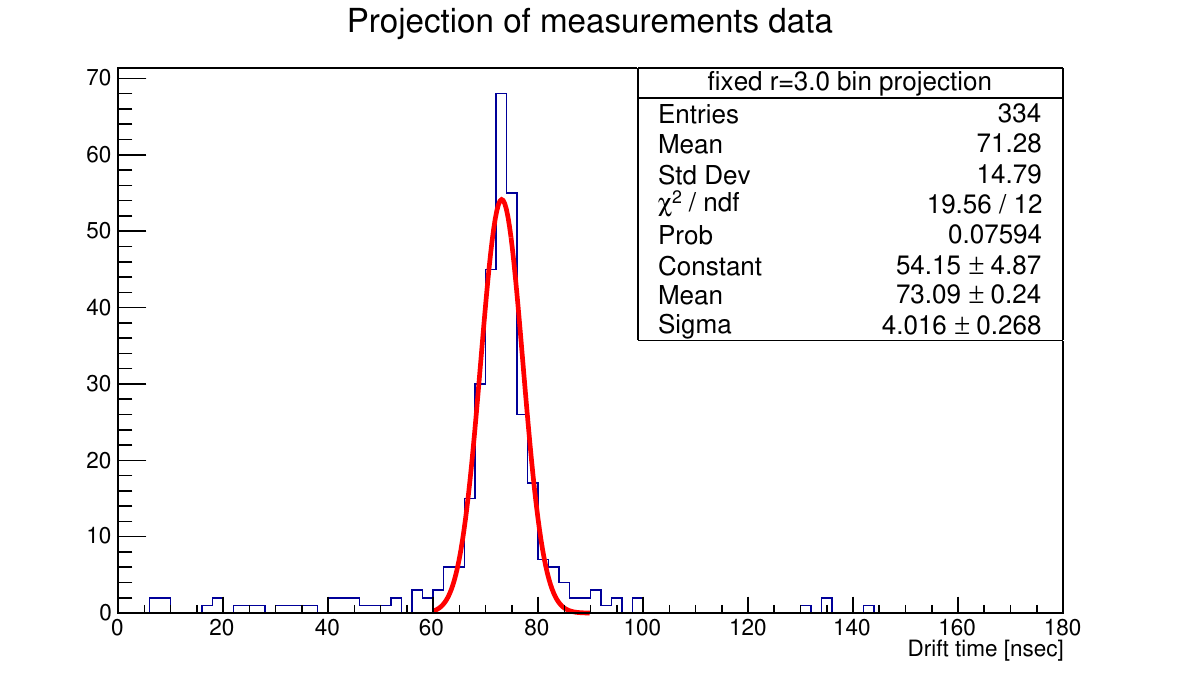} \\ (c)}
  \end{minipage} 
  \caption{Drift time distribution, obtained for first arriving clusters as simulated with GARFIELD together with a Landau fit (a), and the time distribution of the VMM3a-based readout electronics response, modeled for the corresponding signal sample with LTSpice for 25 ns peaking time and 10 mV discriminator threshold together with a Gaussian fit (b). (c) Corresponding distribution obtained with an NA62 straw readout with the original electronics. The time of the discriminator threshold crossing is measured for the relative muon track distances of $3.05\pm0.05$~mm.
	}
\label{st_gar_5}
\end{figure}

A current induced at the straw anode due to an avalanche development upon a muon crossing the tube at the distance of 2 mm from the anode wire as generated with GARFIELD is shown in Fig.~\ref{st_gar_4} together with the corresponding readout electronics signals, emulated with LTSpice. The electronics chain model is set to describe the VMM3a operation mode with the shortest signal peaking time of 25 ns. A 10 mV signal discrimination threshold is chosen to correspond to 3~fC of the input charge. The time distributions of the first arriving cluster and the discriminator threshold crossing are shown in Fig.~\ref{st_gar_5} (a) and (b), respectively. In the case of realistic electronics, the time distribution is significantly wider. For the chosen electronics parameters, the time resolution was found to be 3.6~ns. 
The corresponding distribution, obtained with an NA62 straw readout with the original electronics, is shown  Fig.~\ref{st_gar_5}~(c). 

\begin{figure}[!h]
  \begin{minipage}[ht]{0.5\linewidth}
    \center{\includegraphics[width=1\linewidth]{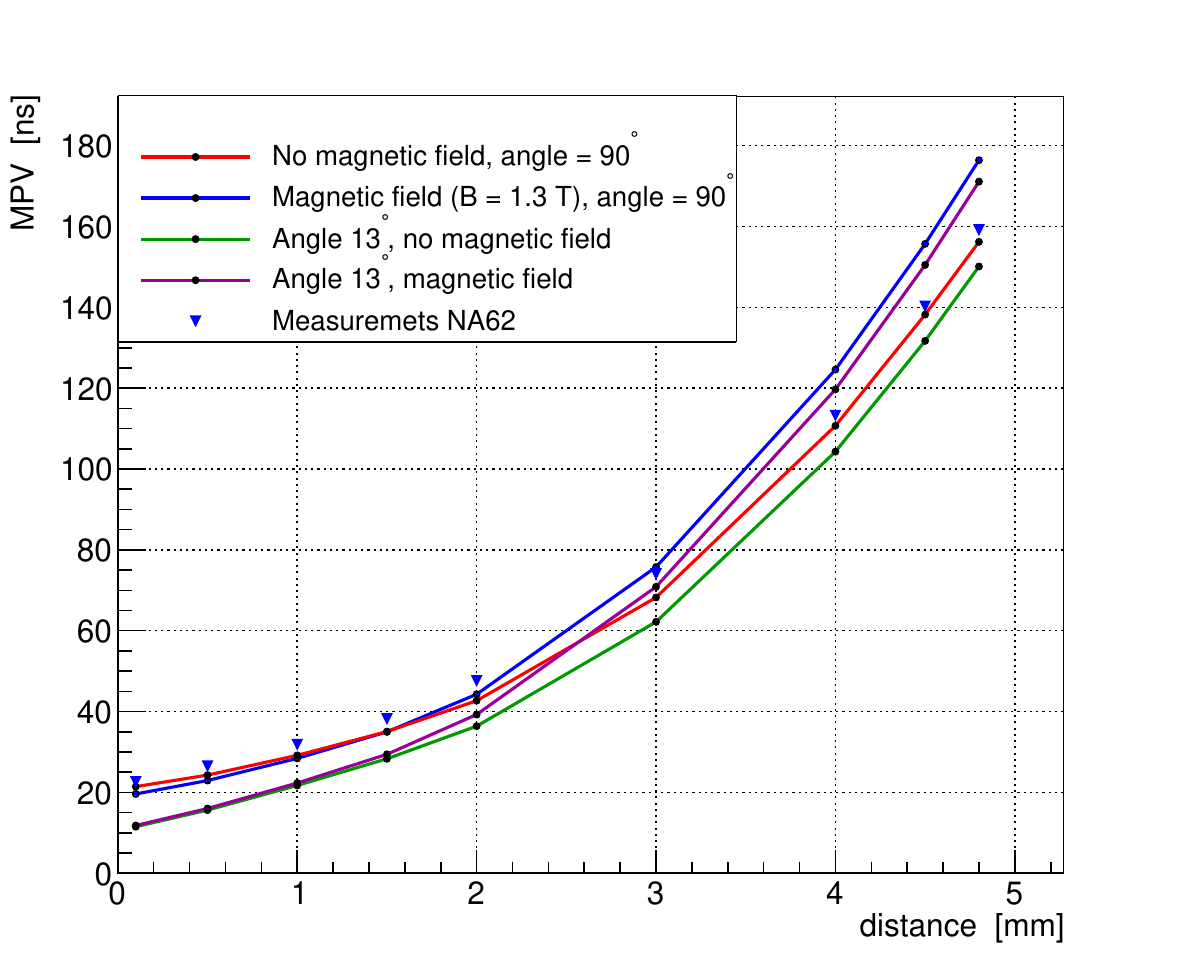} \\ (a)}
  \end{minipage}
  \hfill
  \begin{minipage}[ht]{0.5\linewidth}
    \center{\includegraphics[width=1\linewidth]{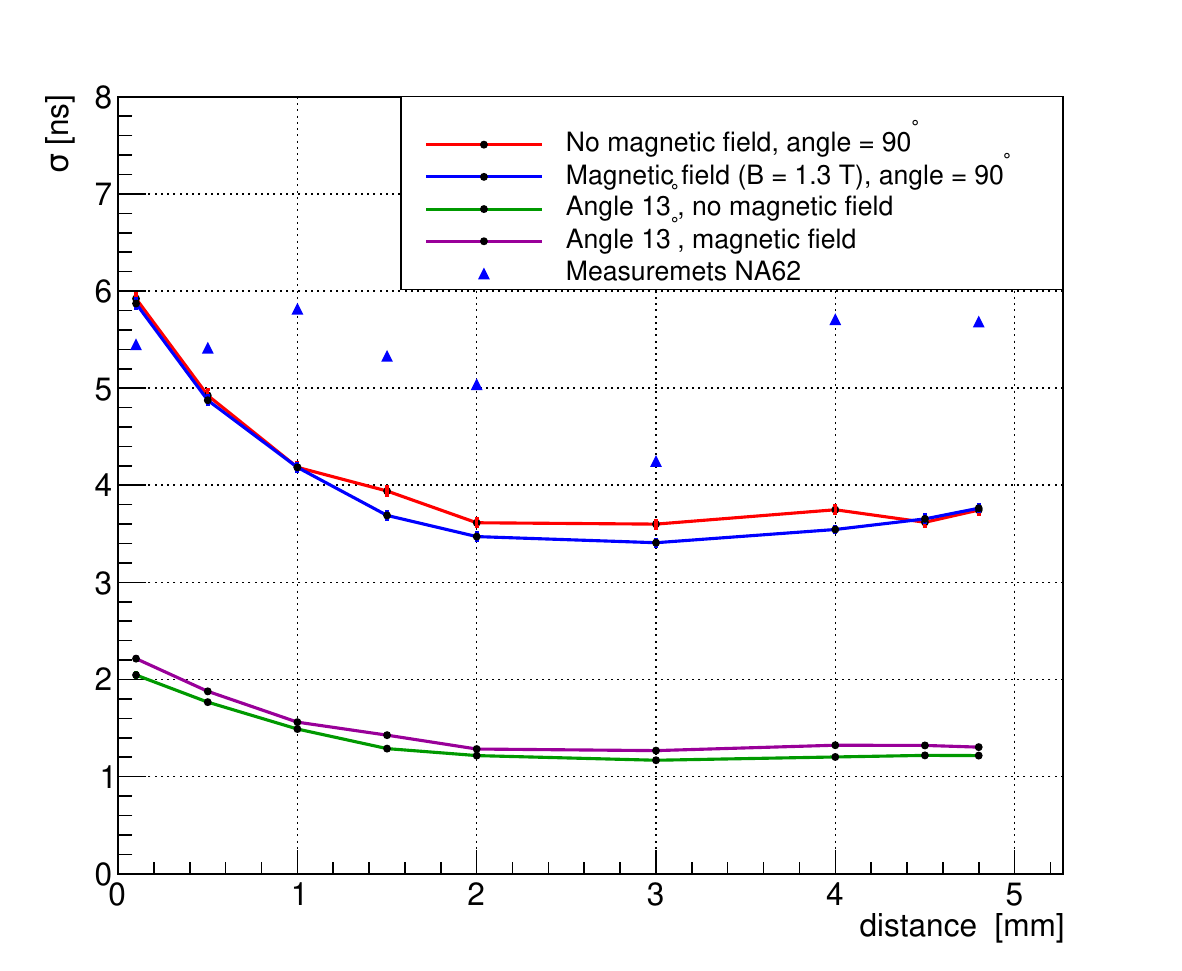} \\ (b)}
  \end{minipage} 
  \caption{Mean (a) and sigma (b) values of the fitted Gaussian function describing the time distribution of the LTSpice output signals crossing the thresholds of 10\;mV as functions of the relative track position obtained with and without magnetic field and compared to the measurements done with NA62 electronics.
  Electronic's noise implemented in the model is of 1500 e$^-$ equivalent noise charge. }
 \label{garft}
\end{figure}

Operation parameters of straw tubes, the influence of readout electronics and the ST operation conditions can not be neglected for precise prediction of the track reconstruction performance. Straw response parametrization accounting for the readout electronics and noise, for the magnetic field and track incidence angle can be included in the Monte Carlo simulation of the SPD setup using results of the GARFIELD++/LTSpice simulation studies. Figure~\ref{garfq} shows the mean (a) and width (b) of the fitted Gaussian function describing the time distribution of the LTSpice output signals crossing the thresholds of 10\;mV. Electronic's noise implemented in the model is of 1500 e$^-$ equivalent noise charge (ENC). The simulation is performed for a 1\;GeV muon crossing a straw tube perpendicular to the straw axis or at the angle of 13 degrees. The results obtained without a magnetic field are compared to the case of 1.3\;T field aligned with the straw axis and to the measurements done with the straw tubes and readout electronics of the NA62 experiment.

\subsection{Chamber design, construction and installation}

\subsubsection{Detector geometry and layout}

\begin{figure}[!h]
    \center{\includegraphics[width=1.\linewidth]{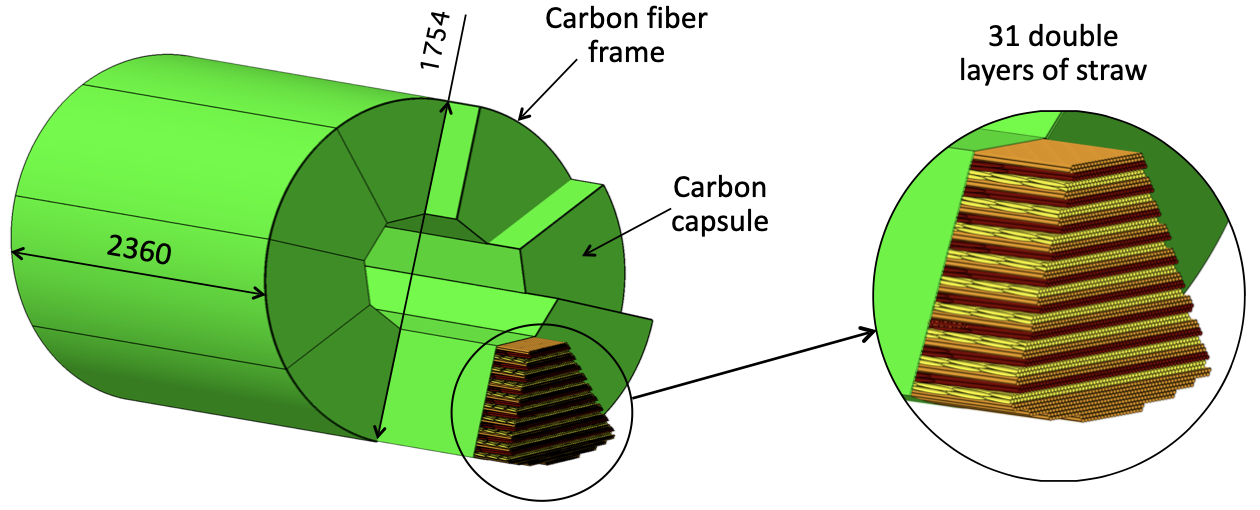}}
  \caption{General layout of the barrel part of ST, which shows 8 modules (octants).
  Each module contains 31 double layers of straw tubes encased in a composite-polymer capsule.
	}
\label{fig:st_4_1}
\end{figure}

\begin{figure}[!h]
  \begin{minipage}[ht]{0.57\linewidth}
    \center{\includegraphics[width=1\linewidth]{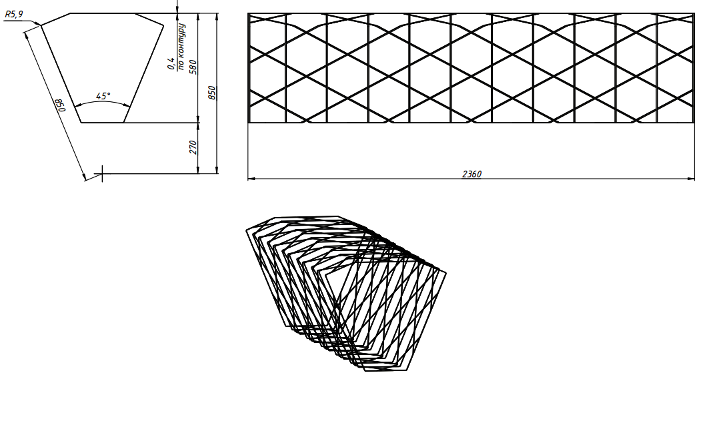} \\ (a)}
  \end{minipage}
  \hfill
  \begin{minipage}[ht]{0.35\linewidth}
    \center{\includegraphics[width=1\linewidth]{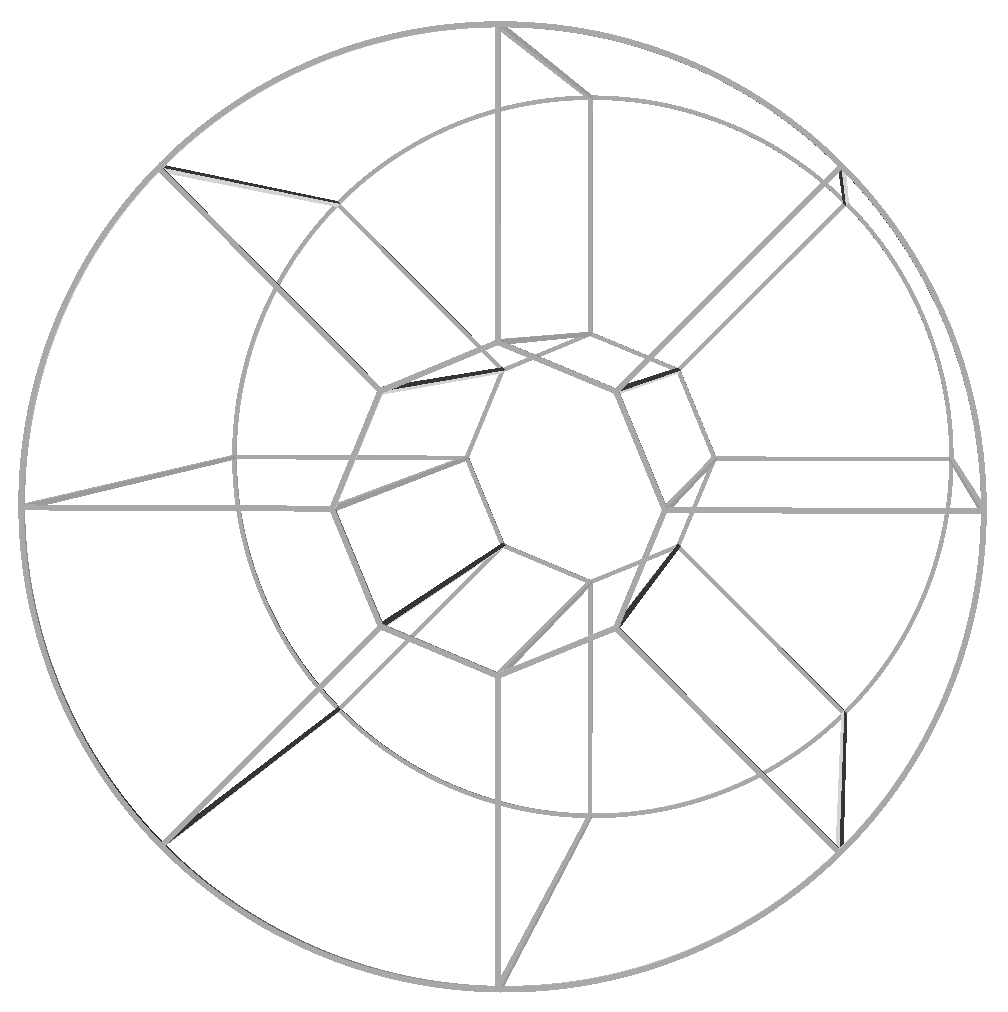} \\ (b)}
  \end{minipage} 
  \caption{(a) Power frame of a single octant. (b) The barrel supporting frame for eight ST octants.
	}
\label{fig:st_4_2}
\end{figure}

\begin{figure}[!h]
    \center{\includegraphics[width=0.9\linewidth]{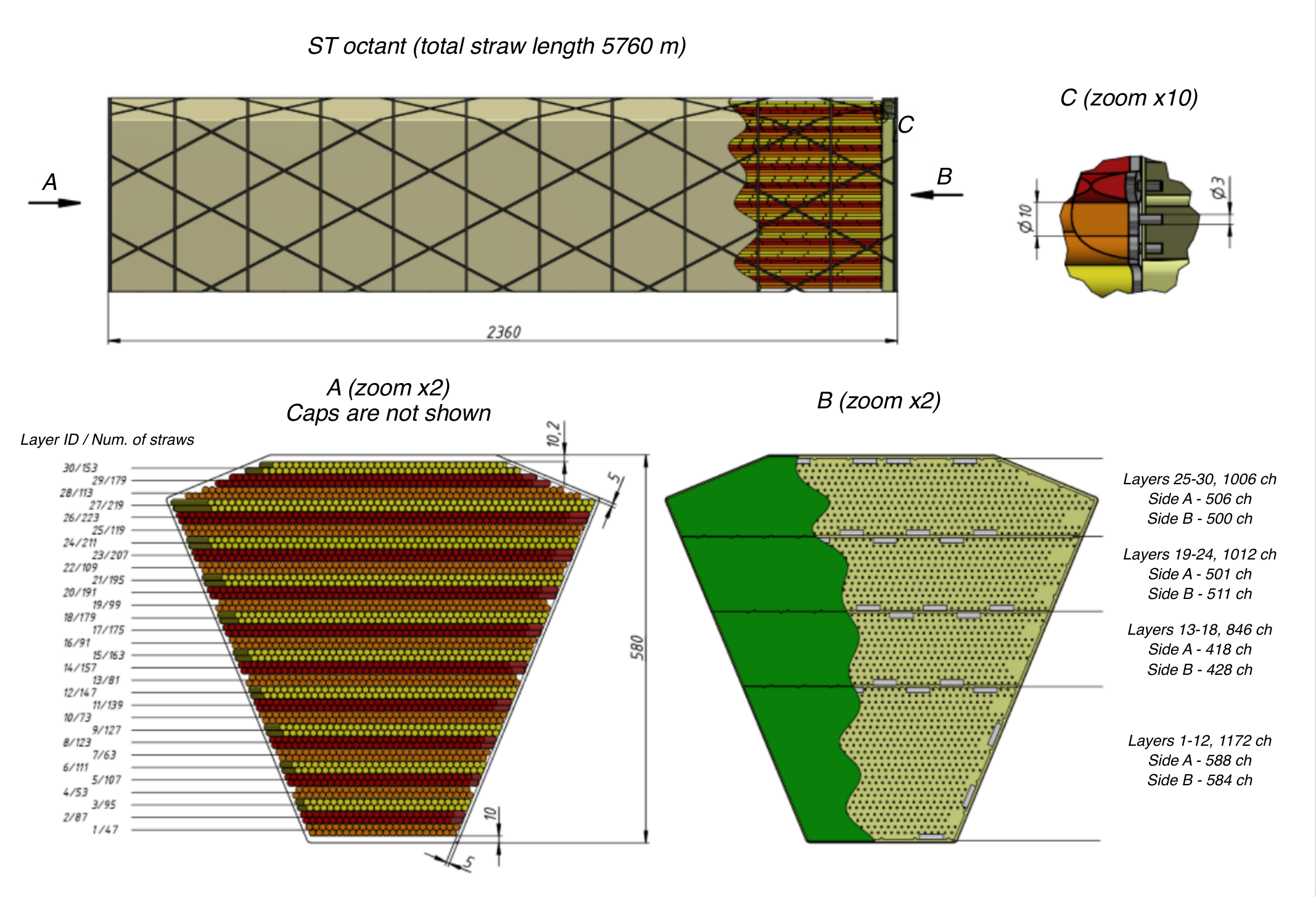}}
  \caption{Sketch of one module (octant) of the barrel part of the straw tracker.
	}
\label{fig:st_4_3}
\end{figure}

The mechanical construction proposed for  the SPD ST is based on engineering solutions, which were already efficiently applied in the ATLAS and PANDA experiments. The straw tracker consists of three parts: a barrel part and two end-caps. The barrel has inner and outer radii of 270 and 867 mm, respectively. It is subdivided azimuthally into 8 modules, with 31 double layers of straw tubes each, as shown in Fig.~\ref{fig:st_4_1}. The layout of a single module with the supporting frame is shown in Fig.~\ref{fig:st_4_2} and  Fig.~\ref{fig:st_4_3}.


\begin{table}[htp]
\caption{Straw tubes in different layers of one octant of the ST barrel part.}
\begin{center}
\begin{tabular}{|c|cccccccccccc|}
\hline
Layer           &1&2&3&4&5&6&7&8&9&10&11&12\\
\hline
Angle, deg  &  0&+5& -5&  0& +5&-5& 0& +5&  -5&  0 & +5 &-5 \\
Straws, pcs.&45&83&85&53&91&95&61&99&103&73 & 111&111 \\
\hline
\hline
Layer          &13&14&15&16&17&18&19&20&21&22&23&24\\
\hline
Angle, deg  &   0&+5&-5&0&+5&-5&0&+5&-5&0&+5&-5\\
Straws, pcs.& 81&119 & 123& 89& 127& 131& 99&139&139&109&147&151 \\
\hline
\hline
Layer          &25&26&27&28&29&30& \multicolumn{1}{c|}{31}  &  \multicolumn{1}{c|}{} &   \multicolumn{4}{c|}{Total} \\
\cline{1-8}
\cline{10-13}
Angle, deg  & 0&+5&-5& 0&0&0& \multicolumn{1}{c|}{0} & \multicolumn{1}{c|}{} & 0 & +5 & -5 & \multicolumn{1}{|c|}{0,$\pm$5} \\
Straws, pcs. &117&155&158&123&103&79&\multicolumn{1}{c|}{39} &  \multicolumn{1}{c|}{} & 1071&1071 & 1096& \multicolumn{1}{|c|}{3238} \\
\cline{1-8}
\cline{10-13}
\end{tabular}
\end{center}
\label{tab:layers}
\end{table}

The main axes of the straw direction are $Z$, $U$, and $V$. The $Z$ axis is along the beam axis. 
The angle between the $U$, $V$ and $Z$ axes is $\pm5$~degree. The value of the angles by layers is shown in Table \ref{tab:layers}.
Each tube has a diameter of 9.8 mm. 
Each module is enclosed in a 400~$\mu$m thick carbon fiber mesh capsule. 
The capsule provides the positioning of the individual straw tubes with a 50 $\mu$m accuracy. One side and two ends of the capsule have 5 mm holes, where straw end-plugs will be fixed. The FE electronic boards, which will be connected to these plugs, will additionally serve as the capsule covers, thereby isolating the inner volume from the external environment. 

The straw tubes will only be read out from one side. 
Each octant contains about 1100 tubes with parallel and about 2200 tubes with oblique angles to the beam direction. 
The total number of electronic channels per octant is 3238. 
Thus, the total number of channels in the barrel part of ST is $3238 \times 8 = 25904$. 
The rigidity of the structure is assured by the low overpressure of gas inside the tubes and their fixation inside the capsule volume. The anode wire positioning accuracy is achieved by the wire fixation in the carbon fiber covers. 
The capsule also serves to thermally stabilize the gas, circulating inside the detector volume, and to protect the straw surface from humidity. 
The sketch of one octant is shown in Fig.~\ref{fig:st_4_3}. 
One meter of straw tube weights of 1.15 g. 
Thus, the weight of all the tubes in one octant with the glued end-caps is about 17~kg.

\subsubsection{Assembling the straw tracker}

Before the octant assembly, the straws of a proper length have to be prepared with the end-plugs glued to both ends of the straw. 
Layers $Z$, $U$, and $V$ will be installed one by one, forming the octant. 
Each straw of the stereo-layers $U$ or $V$ will have one end positioned at one of the octant ends, 
while the opposite end of the straw will be located at the octant side and will be connected to the gas pipes during the assembly.  

\begin{figure}[!h]
    \center{\includegraphics[width=1.\linewidth]{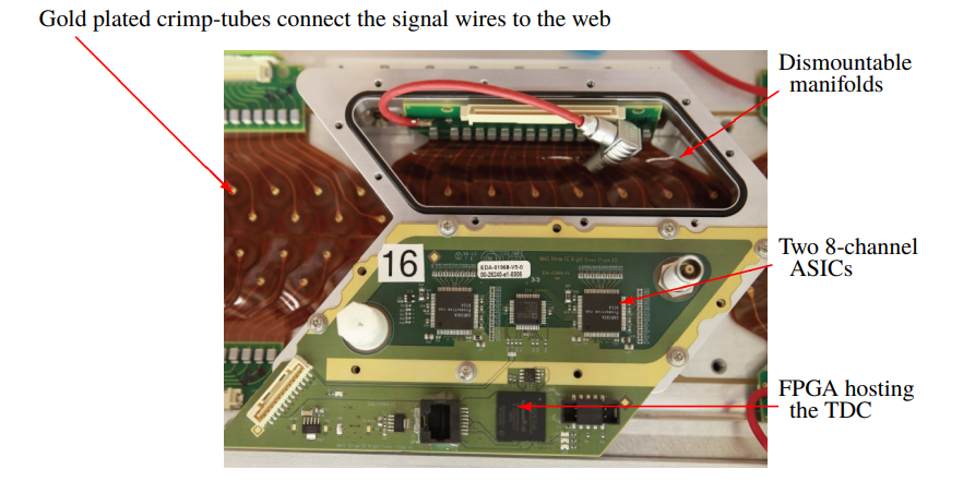}}
  \caption{Picture of the straw end connectivity and gas manifold. A rigid-flex circuit board (web) connects high voltage to the signal wires and transmits the signal to the front-end cover. The cover also houses the high-voltage connector and feedthrough for the gas. 
	}
\label{fig:st_4_4}
\end{figure}

A frame with precise holes for the straw end-plugs will be glued into the octant ends, maintaining the straw alignment. Then the anode wires are inserted into the straws, and their crimp tubes are connected to the flexible webs, as was implemented in the NA62 Straw Tracker and is shown in Fig.~\ref{fig:st_4_4}.  

At each end of the octant,  four front-end covers (one per octant quarter) will be installed,  sealing the gas volume. A 50 $\mu$m thick mylar foil, plated with 50 nm of aluminium,  will be glued over the octant sides. Finally, carbon fibers will be glued onto the foil.

\begin{figure}[!h]
    \center{\includegraphics[width=1.\linewidth]{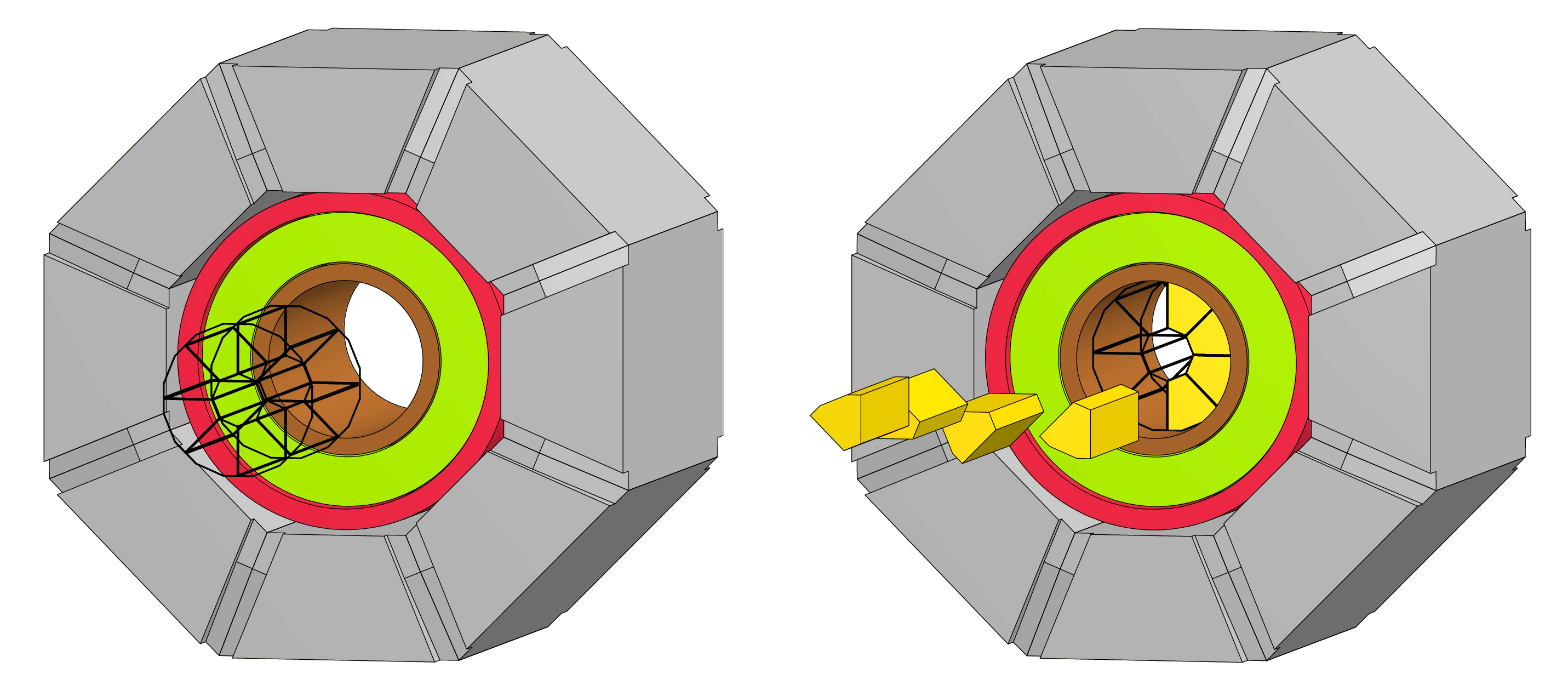}}
  \caption{Straw tracker installation procedure. Left: installation of the carbon frame. Right: installation of the pre-assembled ST octants. 
	}
\label{fig:st_4_5}
\end{figure}

The ST installation procedure is demonstrated in Fig.~\ref{fig:st_4_5}.
Two separate halves of the carbon fiber frame are inserted into the inner volume of the SPD detector before mounting the  octants.
After that, the octants are installed into the frames one by one. 

\subsection{Detector components and assembly principles}
\subsubsection{Active web and wire connection}
\begin{figure}[!h]
  \begin{minipage}[ht]{0.3\linewidth}
    \center{\includegraphics[width=1\linewidth]{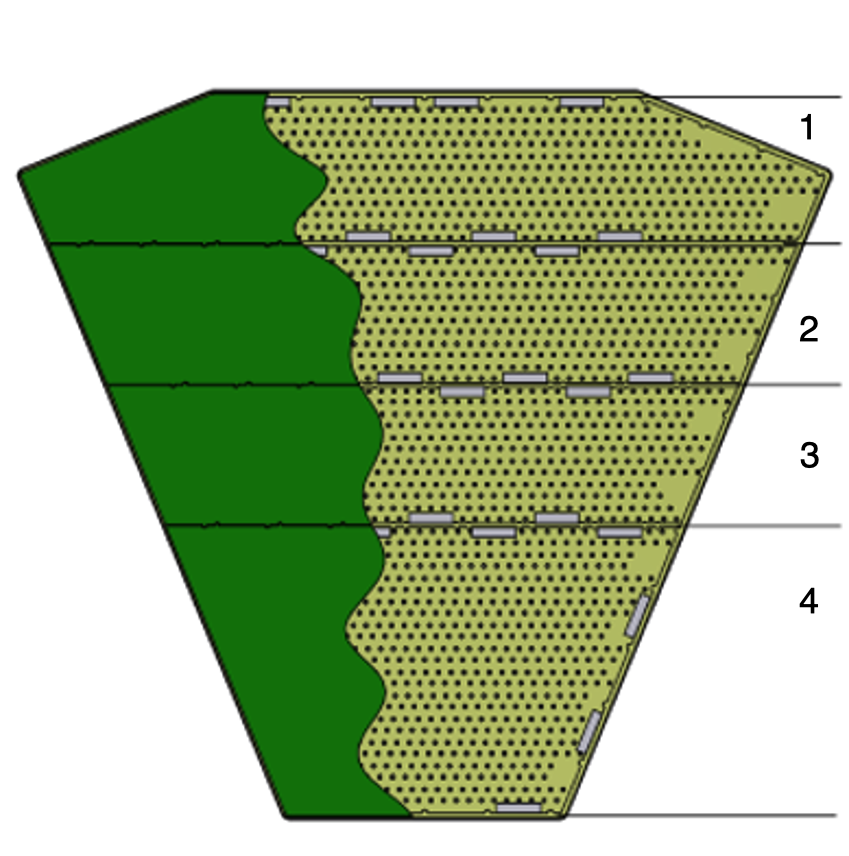} \\ (a)}
  \end{minipage}
  \hfill
  \begin{minipage}[ht]{0.7\linewidth}
    \center{\includegraphics[width=1\linewidth]{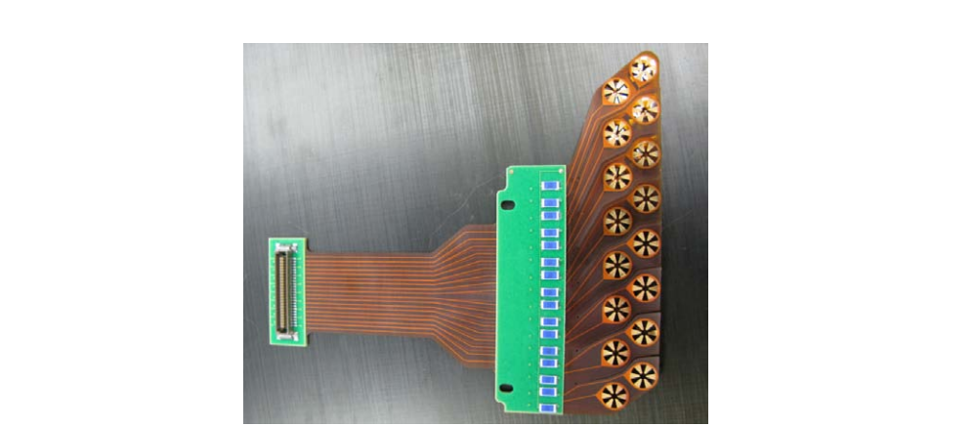} \\ (b)}
  \end{minipage} 
  \caption{(a) Four gas volumes of an octant. (b) Detail of the rigid-flex circuits board (web) that connects to the straws (ground) and the wire (high voltage and signal).
	}
\label{fig:st_4_6}
\end{figure}

A cross-section of the octant end mechanical structure holding the straws is shown in Fig.~\ref{fig:st_4_6}~(a). 
A carbon frame has holes for precise positioning of the straws and wires. 
The modularity in terms of HV, gas supply, and a readout system is represented by four sections containing about an equal number of straws. 
Removable partitions for the gas manifold and a dedicated polyurethane joint for gas tightness were developed. The partitions also serve as supports for the so-called cover, which contains the front-end electronics, high-voltage, and gas connections.
Polyetherimide insulating sockets are glued between the straws and the frame.

High voltage is transferred to the wire by an active web, which also transmits the signal back to the front-end electronics.
The  web is a multi-layer rigid-flex printed circuit board
having two connectors per straw, namely one for HV and one for the signals. 
The web has a connector to be plugged  into the backside of the cover, as shown in Fig.~\ref{fig:st_4_6}~(b). 
The cover does not only provide the leak tightness of the manifold, it also comprises the front-end electronics, high-voltage, and gas connections.

\begin{figure}[!h]
  \begin{minipage}[ht]{0.32\linewidth}
    \center{\includegraphics[width=1\linewidth]{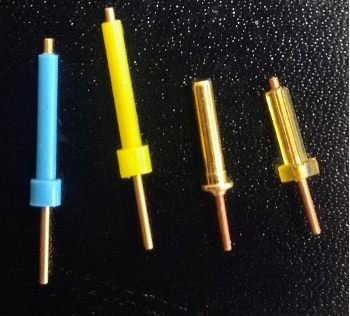} \\ (a)}
  \end{minipage}
  \hfill
  \begin{minipage}[ht]{0.68\linewidth}
    \center{\includegraphics[width=1\linewidth]{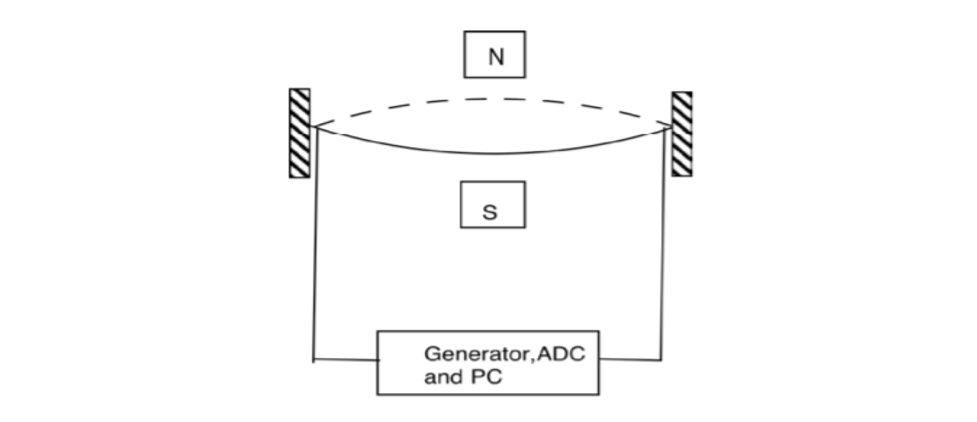} \\ (b)}
  \end{minipage} 
  \caption{(a) Connection plugs.	(b) Schematic view of the wire tension measurement device.}
\label{fig:st_4_7}
\end{figure}

The wire is tensioned and positioned using a copper tube inserted in the so-called connection, shown in Fig.~\ref{fig:st_4_7}. 
The copper tube connects to the second layer of the web, which brings high voltage to the wire and the signal back to the front-end electronics.

\subsubsection{Measurement of the straw straightness}

Once the straws are fixed at both ends in the detector frame, their straightness is measured layer by layer with two light sources and a CCD camera/microscope. The two laser diodes illuminate the straw and the camera/microscope takes a photograph of the straw section. The system is calibrated using a wire stretched from top to bottom, parallel to the straws. A stepper motor moves the camera, and a picture is taken at 10 equidistant positions along the straw. A dedicated LabVIEW\textregistered~program transforms the pictures into a black-and-white image and finds the two straw edges automatically. The program then calculates the position of the straw center (straightness) in the $XY$-plane and the diameter of the straw.

\subsubsection{Wiring}

A gold-plated tungsten wire from Toshiba with a diameter of 30 $\mu$m is chosen for the anode. The limit of elastic deformation has been measured up to 150 g and the rupture occurs around 220 g. The nominal wire tension is set to 90 g. The detector is placed in a horizontal position and the straw resistivity is measured. All straw diameters are measured and recorded in the logbook. The webs are then installed and the ground circuit is fixed to the inside of the straw with the help of the connection plug, which is shown in Fig.~\ref{fig:st_4_7}~(a). A tight fit is necessary to ensure good electrical contact. The ground petals are formed with a special tool to facilitate the insertion of the PET plugs. 
A wire with a diameter of 0.1~mm (''needle'' wire) is blown through the straw.
The 30 $\mu$m wire is fastened to the needle wire and kept under a small tension at all times during installation to avoid kinks. The wire is gently pulled through the straw. From both sides, the 30 $\mu$m wire is inserted in the copper tubes and the copper tubes are then inserted through the electronic circuit into the contact plugs. Once the wire is fixed to the web side, a 90~g weight is suspended to the wire on the opposite side and fixed. The electrical continuity between the two pins is checked, as well as the insulation between the wire and the straw (broken wire). After wiring one layer, the wire tension is measured and a HV test is performed to measure the leak current, which should not exceed 1~or~2~nA per high voltage group at 1600~V.

\subsubsection{Measurement of wire tension}

The full straw tracker contains a total of about 26000 anode wires, and the wire tension needs to be verified during and after installation. The wire resonance frequency is close to 50 Hz for a tension of 80~g and a wire length of 2200 mm. 
Since 50 Hz is close to the frequency of the 220 V voltage supply, in order to avoid this frequency, the nominal wire tension was set to 90 g. The acceptable values of wire tension during the module production are between 85 and 95 g. The upper limit of 95 g is considered to be comfortably below the elastic limit of 150 g, while the lower limit is high enough to allow good operation of the straw. Electrical instability of the wire was found for tensions below 35 g. 
Therefore, to minimize electrostatic deflections on the wire, the lower limit is set to 85 g. 

The principle of measuring wire tension is presented in Fig.~\ref{fig:st_4_7}~(b).
Namely, the wire oscillations are stimulated in an external magnetic field with the help of a current generator, 
and the resonant frequency of the wire is measured. The current induced on the wires will be amplified with a high-input impedance amplifier and an input signal threshold (typically of about 1.5 mV). Signals will be digitized by a PC-based A/D board and the measured frequency will be translated into applied wire tension $T$ (in grams), following the formula:
\begin{equation}
T=\frac{4\mu L^2 \nu^2}{a},
\end{equation}
where $L$ is the free wire length in cm, $\nu$ is the frequency in Hz, $\mu$ is the mass per unit length in g/cm and $a$ is the gravitational acceleration. Uncertainties in the measured tension arise from variations in the wire diameter and the length of the vibrating wire. The minimum uncertainty can be estimated to be about 1\%. The readout electronics, connected to a cell of 16 straws, measures the main resonance vibration frequency of each wire. The magnetic field near the tested wire must be at least 100 G. The operator can modify the parameters of the pulse generator and Fourier analysis on the Labview-based computer panel. The tension of the wires will be measured during the stringing process itself, and a final global wire-tension measurement of all wires in the module will be carried out before the next assembly step. The results of the measurement will be written to the production database. Loss of wire tension can happen between module assembly at the assembly hall and arrival at the SPD hall, so the wire tension will be measured again after delivery to the nominal position. Major attention will be paid to wires, which show a tension loss of more than 10 g. These wires will be investigated carefully and replaced if necessary.

\subsubsection{Gas tightness tests}

The next stage will be to verify the  joints glued between the straws and the frame. It is important to perform this test at this stage because the straws and their  joints  are still easily accessible and repairable in case of a problem. A dedicated gas-tightness setup is to be used to verify the quality of about 8000  joints (per octant), glued between the straws and their end-plugs, and 
between the end-plugs and the straw support frame. The system contains a temperature sensor, pressure gauges, a vacuum stand, and a system of valves and pipes. Straws and gas manifolds will be filled with Ar from a bottle with a pressure of 100~mbar above atmospheric pressure.
The  volume of the octant to be tested is about 1\,m$^3$. 
The leak rate (mbar$\times$ l/sec) will be evaluated by measuring the pressure drop inside the straws as a function of time, once the pump has been stopped. A module will be approved for subsequent assembly, whenever the measured leak rate does not exceed $10^{-2}$ mbar$\times$l/sec. Nevertheless, every effort will be made to get the leak value as low as possible.


\subsection{Gas system}

The gas system consists of two parts. First, the mixer system which delivers quantity, mixing ratio and pressure conditioning to downstream elements. Second, the distribution system, which delivers the gas in well-defined quantities to the individual detector components.

\subsubsection{ Gas system requirements}

The total gas flow chosen corresponds to a normal gas flow rate of ~2$\div$4 cm$^3$/min per straw. The gas modularity is optimized in order to minimize the number of lines between the detector and the distribution racks, and, on the other hand, to minimize the loss of performance in the case of an accidental leak in any module.

\subsubsection{	 Mixer}

The detector shall be supplied with a constant gas mixture of Ar:CO$_2$=70:30 with a precision better than 1\%. The total flow of the mixer will be 1500 l/h. Each primary gas line is equipped with a Digital Mass Flow controller to measure the component flow with appropriate accuracy. An output pressure regulator adjusts the downstream pressure from 0.2 to 2 bar. The control system provides a flow an independent mixing ratio and adequate error handling. The mixer has to automatically follow the demands of the distribution system. 

\subsubsection{	Gas distribution}

The gas from the mixer is distributed to the eight gas distribution racks, one for each octant. The major design criteria of the distribution system is the uniform gas supply to each cell with adequate separation capabilities, in case of pressure loss,  due to a leaking straw. Each octant is divided into four gas volumes, as  it is shown in Fig. \ref{fig:st_4_6}.

\subsection{Aging studies}

In order to validate the different components and materials  in the detector, a dedicated prototype was built. The prototype contains two straws: one with aluminium coating and another with copper/gold. The straw end-pieces were made from ULTEM, and the connection to the straws was made with a section of the web. The wire is the 30 $\mu$m gold-plated Tungsten from Toshiba,  foreseen for the chamber production. Final crimp tubes were used for electrical and mechanical connections of the wire to the web. The prototype was mounted in the CERN aging test facility. A gas mixture of 70\% argon and 30\% CO$_2$ was used. The parameters during the aging test are summarized in Table \ref{tab_aging_test}.
\begin{table}[htp]
\caption{Parameters of setup for aging testing.}
\begin{center}
\begin{tabular}{|c|c|c|c|c|c|c|}
\hline
Gas mixture & Gas flow & Current & Total charge & Irradiation  & High voltage & Fe-55 scan  \\
                    &                 &              &                     & area          &                       &  slit size \\
\hline
Ar (70\%)+ & 0.5 cm$^3$/min & 280 nA& 0.27 C/cm & 10 mm &1700 V & 1 mm  \\
 CO$_2$ (30\%)&                    &            & (500 hours) & & & \\
\hline
\end{tabular}
\end{center}
\label{tab_aging_test}
\end{table}%

\begin{figure}[!h]
    \center{\includegraphics[width=0.9\linewidth]{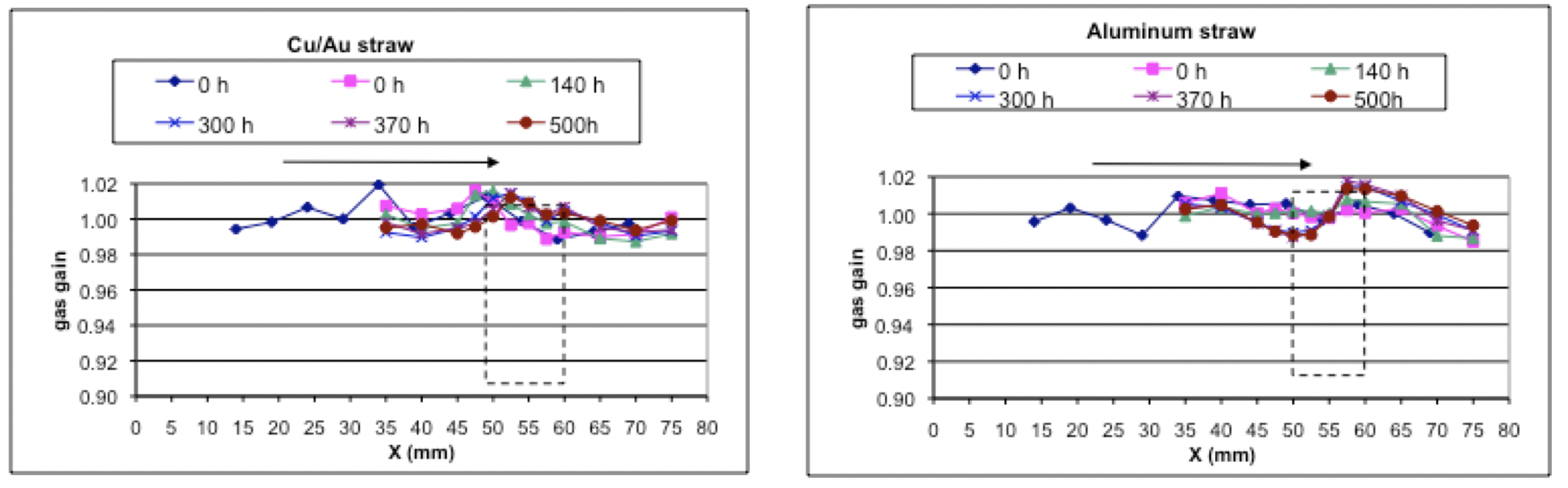}}
  \caption{Scan of signal amplitude along the straw length for straws with  two types of coatings Cu/Au (left) and Aluminium (right). 
  Different graphs correspond to different accumulated charge.
	}
\label{st_ba_fig13}
\end{figure}

A scan with a Fe-55 source is made at intermediate irradiation levels 
along the straw length
from 10~mm to 75~mm. The results are shown in Fig.~\ref{st_ba_fig13}. No change in amplitude can be seen up to a total  charge of 0.27 C/cm. The estimated accumulated charge for the hottest straws in the experiment is 0.04~C/cm. After the initial run with Ar/CO$_2$, it was decided to dismount the prototype and investigate the wire and the inner surface of the straw  around the irradiated region.
\begin{figure}[!h]
    \center{\includegraphics[width=0.9\linewidth]{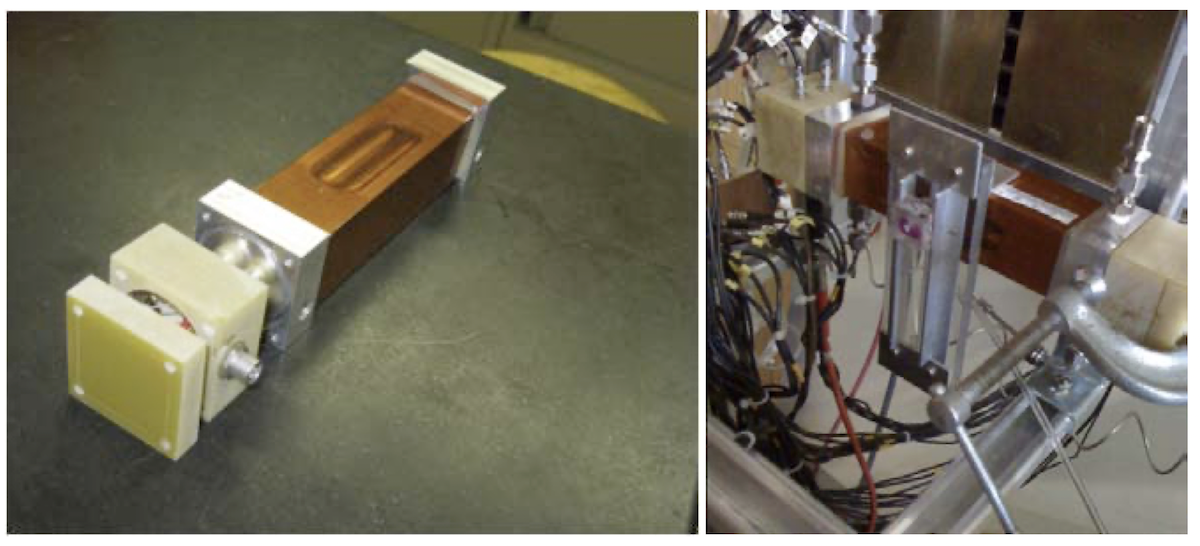}}
  \caption{ Straw prototype for material validation (left), mounted in the setup (right).
	}
\label{st_ba_fig12}
\end{figure}



\section{End-cap part of ST}

The design of the ST end-cap has to satisfy the following criteria.
The detector must have the shape of a disk, have a relatively small central hole, 
which is defined by a vacuum beam pipe, and have a small amount of matter in the sensitive region of the detector. 
The detection layers should be thin, and the number of layers should be sufficient to identify particles via $dE/dx$ measurement.

\begin{figure}[!ht]
    \center{\includegraphics[width=0.7\linewidth]{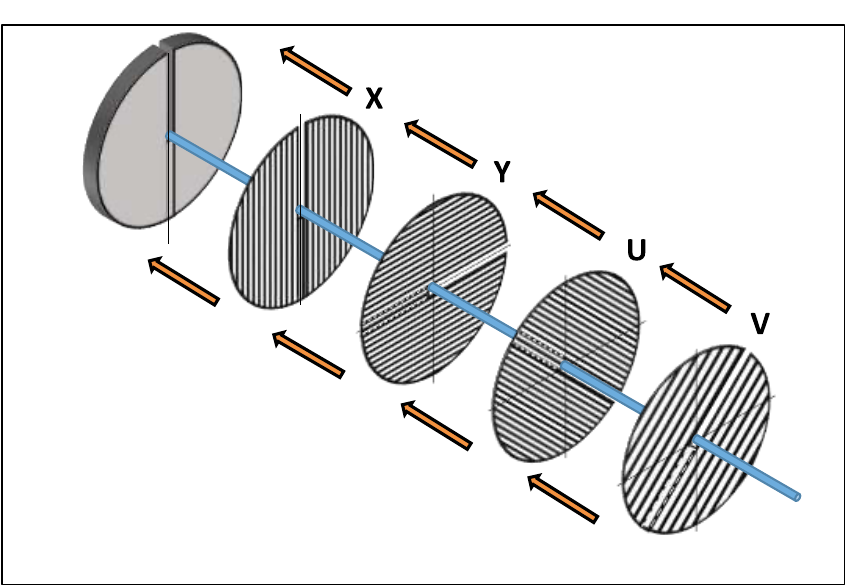}}
  \caption{General layout of the ST end-cap detector with drift tubes.
	}
\label{SEC_1}
\end{figure}

For efficient registration of interactions with a large multiplicity, a detector with 
drift tubes, arranged so that they form an $X$, $Y$, $U$, $V$ coordinate system, is proposed. 
Each coordinate plane consists of two halves separated by a distance for installation on a vacuum tube, as shown in Fig.~\ref{SEC_1}. 
All coordinate planes are mounted sequentially on a solid base that can be attached to another power element of the SPD setup.

\begin{figure}[!ht]
  \begin{minipage}[ht]{0.25\linewidth}
    \center{\includegraphics[width=1\linewidth]{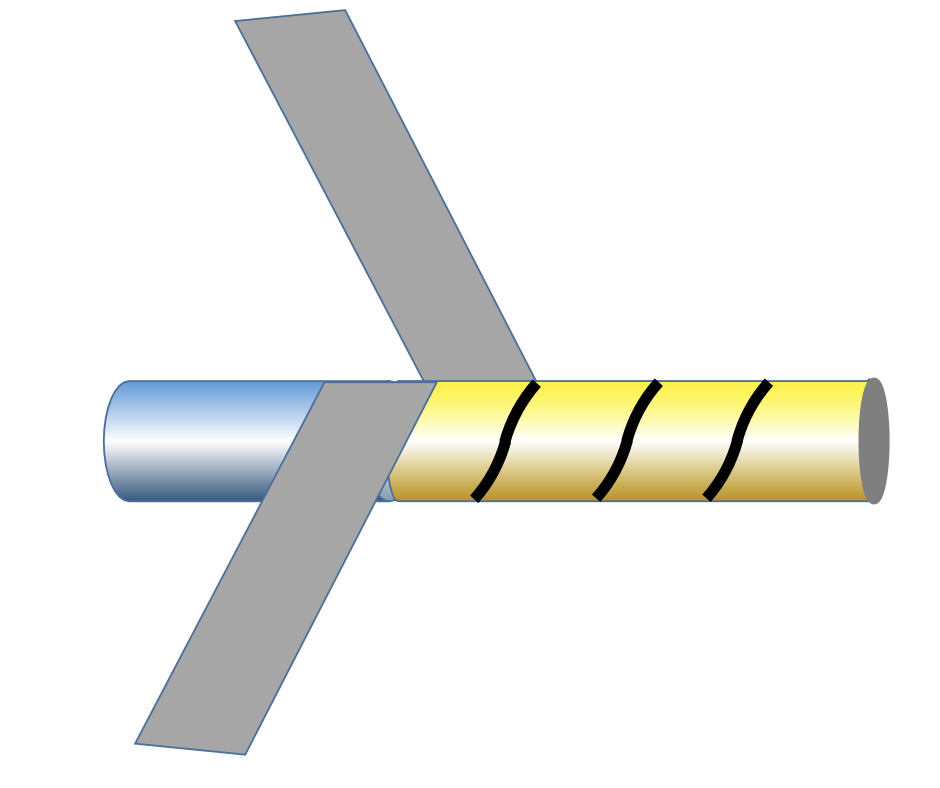} \\ (a)}
  \end{minipage}
  \hfill
  \begin{minipage}[ht]{0.377\linewidth}
    \center{\includegraphics[width=1\linewidth]{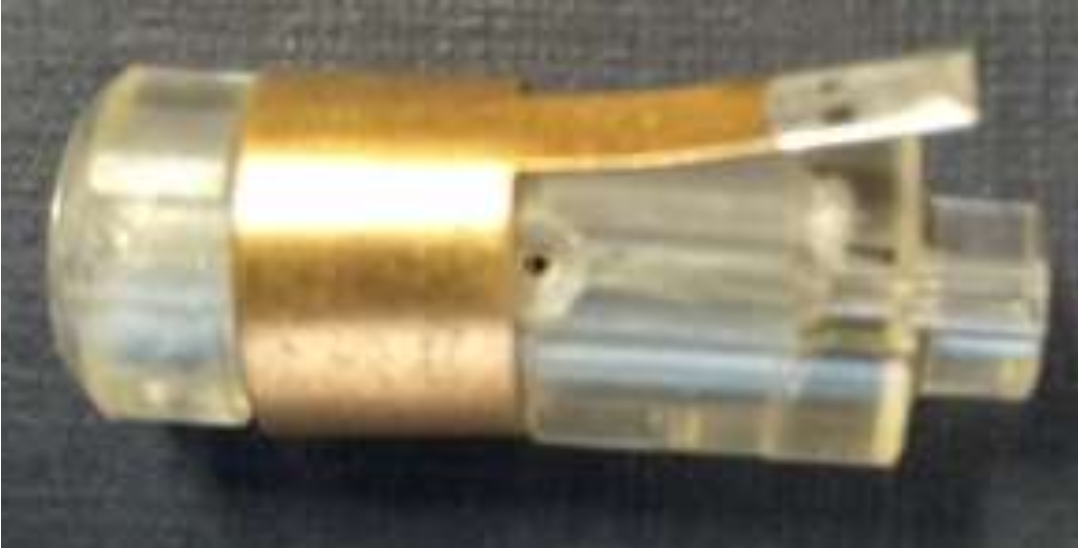} \\ (b)}
  \end{minipage} 
  \begin{minipage}[ht]{0.32\linewidth}
    \center{\includegraphics[width=1\linewidth]{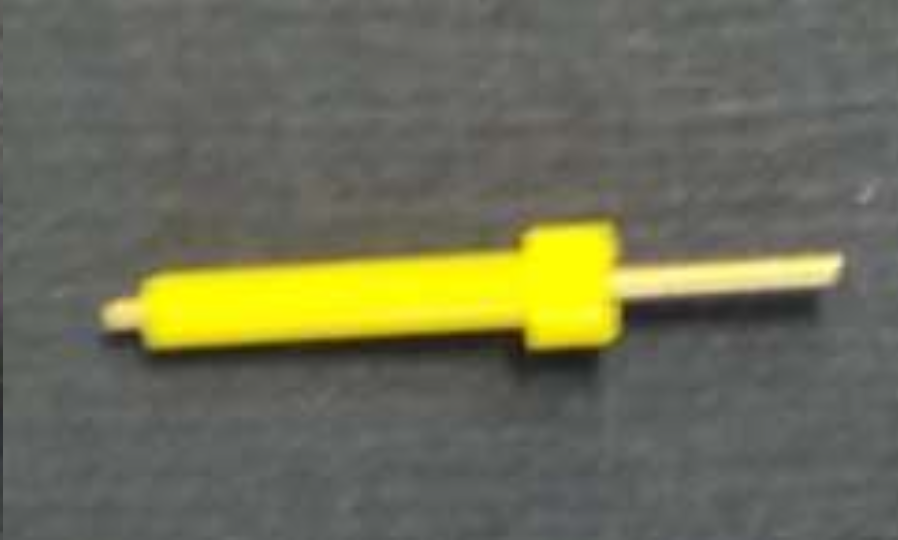} \\ (c)}
  \end{minipage} 
  \caption{(a) Twisted structure of the straw tube. The picture illustrates a winding on a precision rod. The end-plug (b) and the crimping pin (c) are used to fix the anode signal wire.
	}
\label{SEC_2}
\end{figure}

It is possible to implement such a detector based on thin-walled drift tubes, made by winding two "kapton" tapes, as shown in Fig.~\ref{SEC_2}~(a). The technology for manufacturing such detectors is being successfully developed in LHEP JINR. Over two decades, the straw chambers have been created for several large experiments, such as SVD-2 and Thermalisation at Protvino, ATLAS, COMPASS, and NA64 at CERN, CBM in Darmstadt, etc.

It is supposed to use straw tubes with a diameter of 9.56 mm, made of two layers of polyimide film.
The outer layer of the straw tube consists of a polyimide tape with a thickness of 25 microns. A copper coating with a thickness of 100 nm is applied on one side. The surface resistance is $1 \pm 0.1~\Ohm$/square. Polyurethane hot-glue coating with a thickness of 4 $\pm$ 1 $\mu$m is applied on the other side.
The inner layer of the straw tube consists of a similar polyimide tape. The layer of hot glue is applied to one side, and on the other  side - a layer of aluminum 0.2 $\mu$m thick. Graphite with a thickness of 6 $\pm$ 2 $\mu$m is applied on aluminum. The resistivity of this surface is about 10 $\Ohm$/square. The outer and inner tapes are glued together on a calibrated rod, as shown in Fig. \ref{SEC_2}~(a), heated to 170 $^{\circ}$C.

\subsection{Elements of technology for assembling twisted straw tubes}

Before assembling the detector or its parts, each straw tube is filled with precision steel balls with a diameter of $6.010\pm0.003$~mm or $9.525\pm0.003$~mm for the straw tubes of the corresponding diameter. Filling with balls occurs automatically. It takes about 1 minute to fill a 1.8\,m long straw tube. Tubes with balls are assembled into the size of a detector on a precision table and glued together to form a single straw array. Araldite 2011, VK-9, and ALK-5.8 are used for gluing planes and other structural elements. The amount of glue used does not exceed 60\% of the weight of the straw tubes. After removing the straw balls, the tubes are installed in a common plane on a large precision table.

\begin{figure}[!ht]
  \center{\includegraphics[width=0.7\linewidth]{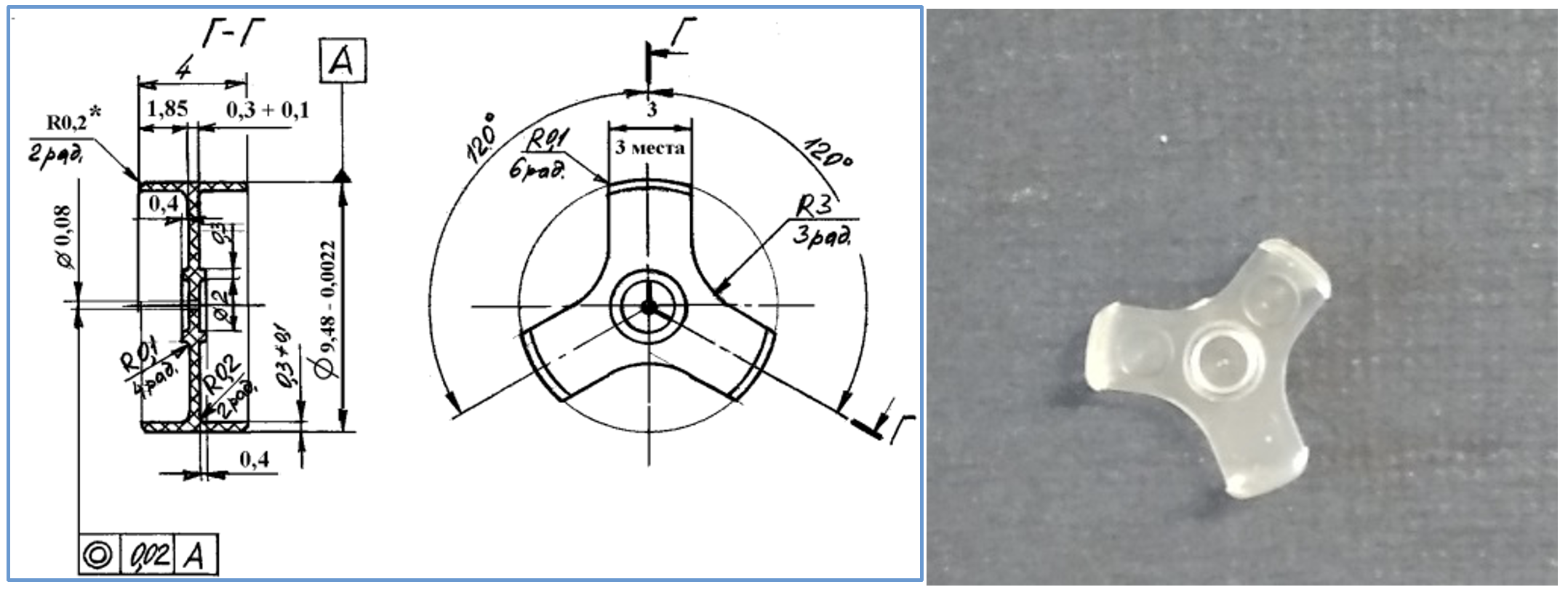}}
  \caption{Technical drawing (left) and photo (right) of the internal backing of the signal spacer wire.
	}
\label{SEC_3}
\end{figure}
\begin{figure}[!ht]
    \center{\includegraphics[width=0.9\linewidth]{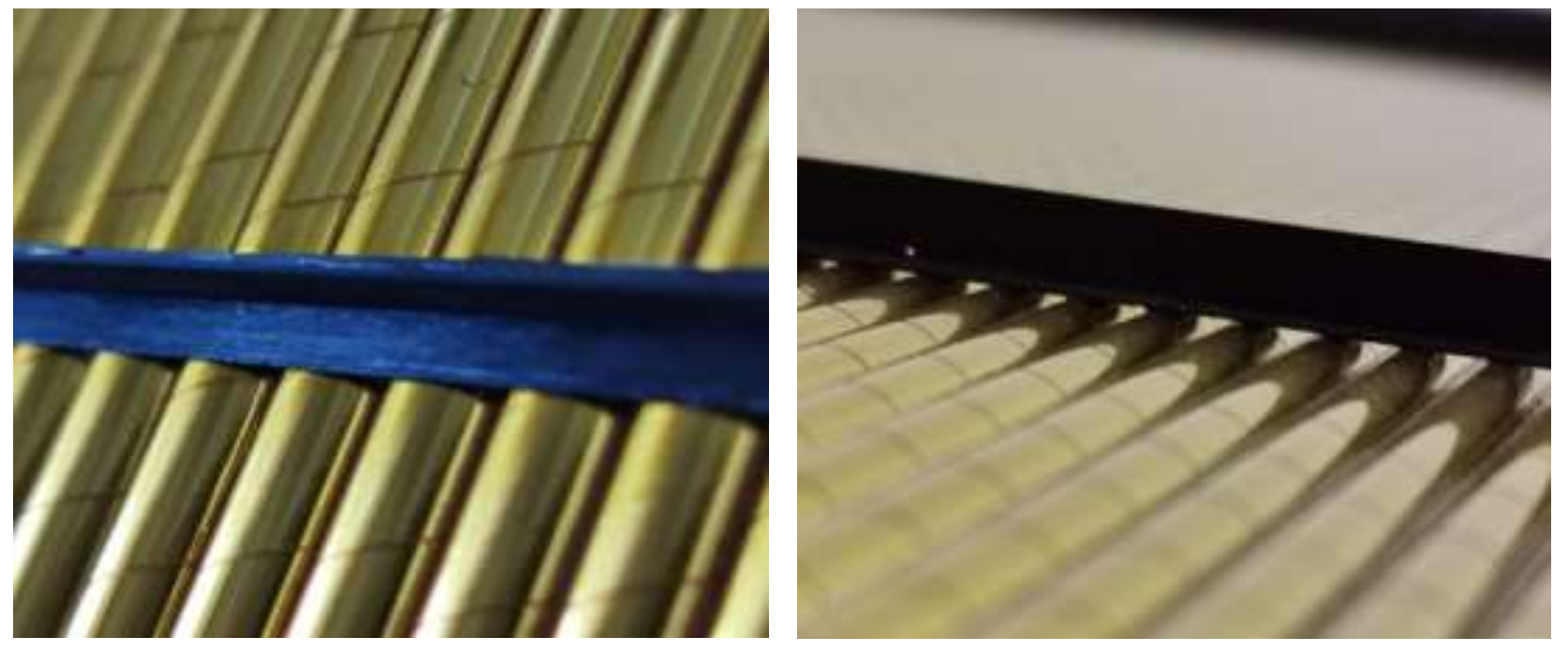}}
  \caption{Example of carbon fiber strips used as external  support for the straw tubes.
	}
\label{SEC_4}
\end{figure}

The end-plug and crimp pins, located at both ends of the straw, are used for the installation and positioning of the anode signal wire. The end-plug and the crimping pin are shown in Fig.~\ref{SEC_2}~(b) and (c), respectively. The positioning accuracy of the signal wire is not worse than 100 $\mu$m. The end-plug has a bronze spring contact that connects the inner conductive layer to the ground. It also has two longitudinal grooved channels for supplying a gas mixture.

Internal supports of the signal wire (spacers) are installed inside the tubes and serve to prevent sagging of the signal wires. 
Spacers are pre-glued to the signal wire in the tubes with a length of more than 800~mm. The drawing and appearance of the spacer are shown in Fig.~\ref{SEC_3}.

External supports serve to reduce the sagging of the straw tubes.
The support elements are made of thin carbon fiber plates, glued to the tubes.
An example of such elements is shown in Fig.~\ref{SEC_4}.
Transverse T-shaped strips of $0.5 \times 9$~mm$^2$ are used to increase the geometric stability of the transverse dimensions of the straw array, which are glued in the direction perpendicular to the straw axis of the tubes.

The positioning accuracy of each signal wire in the straw tubes is determined by the accuracy of the straw manufacturing and 
the accuracy of the plastic end-bushings and pins. 
The general positioning of the tubes in the detector plane is carried out on a precision table using precise rulers, 
as a result of which the accuracy does not exceed 0.1~mm.

\subsection{The main characteristics of  twisted straw  tubes}

The main characteristics of the drift detectors are given in Table \ref{EC_tab1}.
\begin{table}[htp]
\caption{Main characteristics of  the drift straw tubes.}
\begin{center}
\begin{tabular}{|l|c|}
\hline
Drift tube diameter & 6 $\div$10 mm\\
Gas amplification at 1800 V & $2.5\times10^4$ \\
 Operating voltage range &  200 V \\
Electron collection time for $B=0$\,T & 0 $\div$100 ns \\
Electron collection time for $B=2$\,T & +20\% \\
Registration threshold  &  2$\div$3 fC\\
Resolution when measuring drift time &  100$\div$200 $\mu$m \\
Registration efficiency for a two-layer detector at 1 MHz  & 99.7\% \\ 
Gas mixture & Ar:CO$_2$ = 70:30 \\
\hline
\end{tabular}
\end{center}
\label{EC_tab1}
\end{table}%

\subsection{Radiation properties of twisted straw tubes}

Properties of the materials that are usually used for  the straw production are given in Table \ref{EC_tab2}.

\begin{figure}[!h]
    \center{\includegraphics[width=0.9\linewidth]{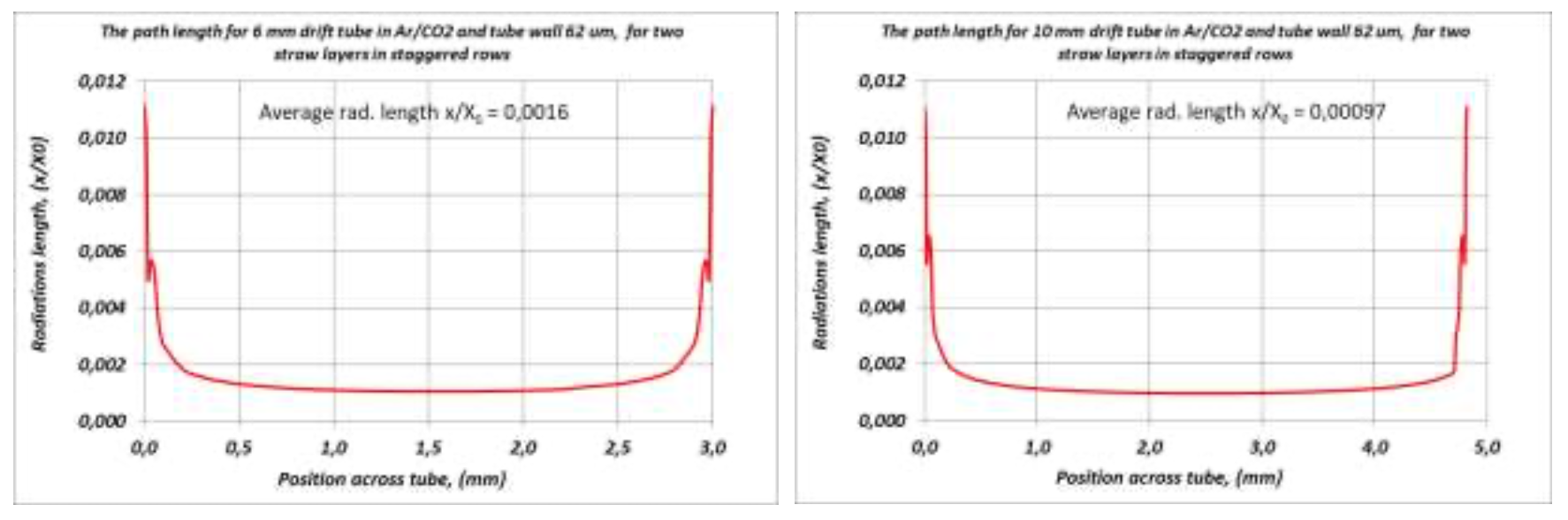}}
  \caption{ Dependence of the thickness of the two-layer straw detectors in radiation units $x/X_0$ on the place of passage of the particle through the two layers of tubes for two diameters of a straw.
	}
\label{SEC_6}
\end{figure}

\begin{figure}[!h]
    \center{\includegraphics[width=0.9\linewidth]{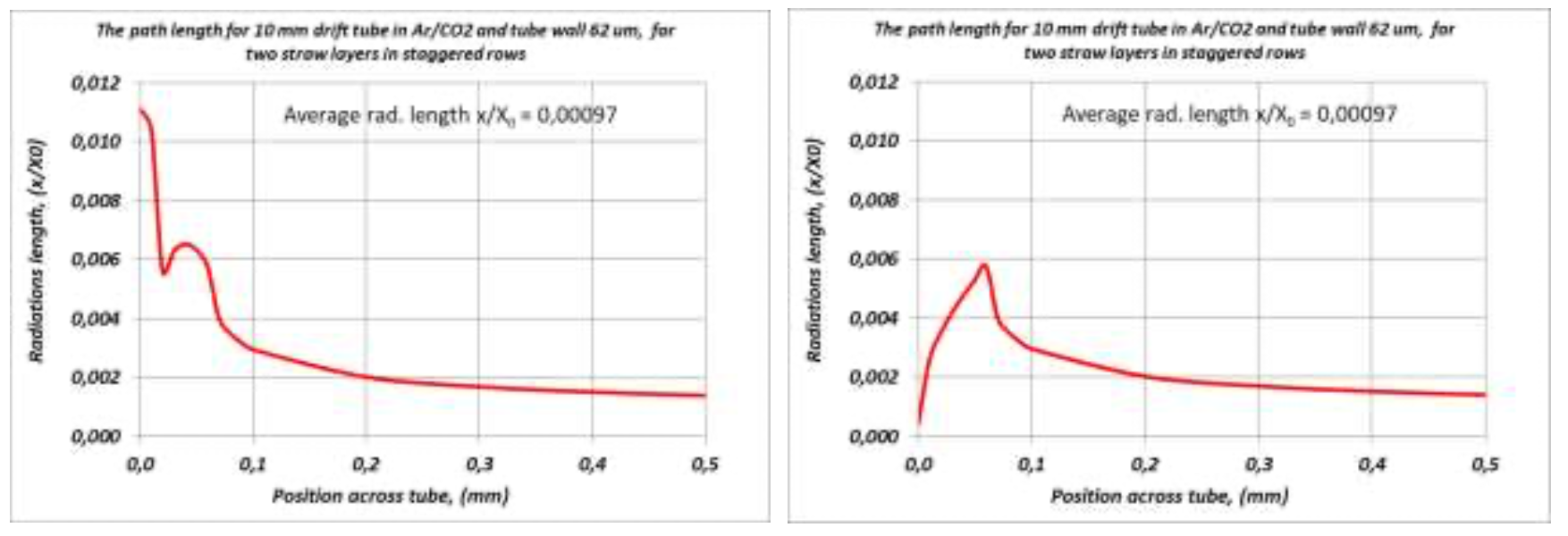}}
  \caption{ More detailed picture for particles passing between tubes of the same layer. The figure on the left takes into account the adhesive layer between the tubes of the first layer and the substance of the signal wire 30 $\mu$m in the second layer. On the right, only the substance of the tube walls in two layers without signal wire and glue is taken into account.
	}
\label{SEC_7}
\end{figure}

Figures  \ref{SEC_6} -  \ref{SEC_7} illustrate the results of calculating the radiation thickness of a straw detector consisting of two layers shifted by half the diameter of  the tubes. The tubes in each layer are glued together without gaps into a single coordinate plane.
\begin{table}[htp]
\caption{Characteristics of the materials used in the construction of the straw tubes.}
\begin{center}
\begin{tabular}{|l|c|c|c|}
\hline
Material & Density, g/cm$^3$ & Rad. length, g/cm$^2$ &  Rad. length, cm \\
\hline 
Tungsten, W     & 19.3          &    6.76   &  0.35  \\
Graphite, C      &    2.2          &   42.70 &  19.4 \\
Polycarbon Lex&   1.2           &   41.50  &   34.6 \\
Kapton             &   1.4           &   39.95  &   28.5 \\
Acrylic (PMMA) &  1.19         &    40.55 &   34.1 \\
Polystyrene      &   1.06        &     43.79 &   41.3 \\
\hline
\end{tabular}
\end{center}
\label{EC_tab2}
\end{table}%

\begin{figure}[!h]
    \center{\includegraphics[width=0.5\linewidth]{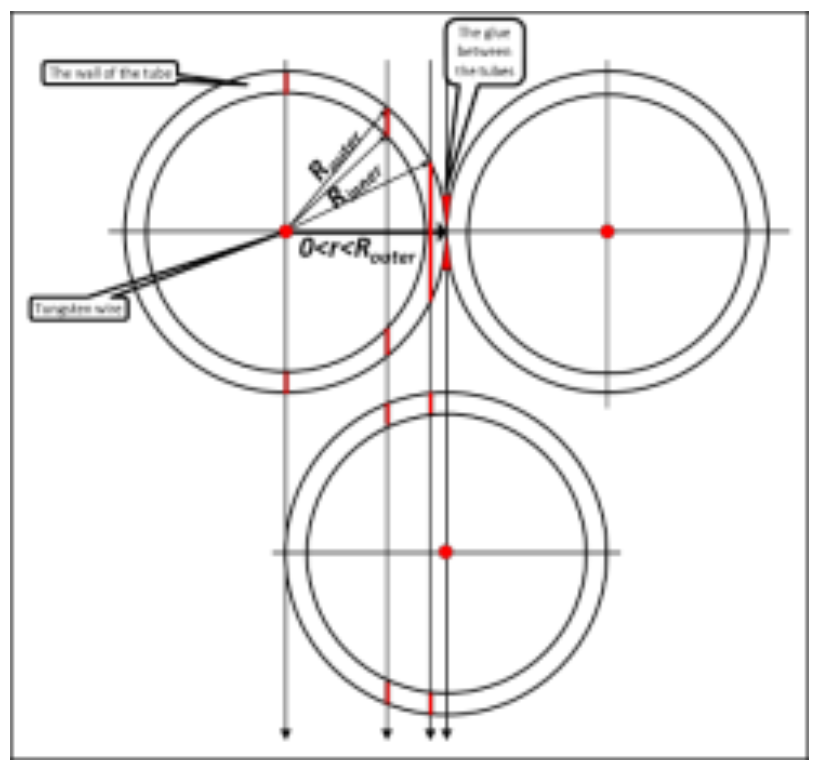}}
  \caption{
  A typical arrangement of tubes in a two-layer detector, where the layers are shifted relative to each other by half the diameter.
	}
\label{SEC_5}
\end{figure}

\subsection{Coulomb scattering in the straw material}

A charged particle passing through the medium is deflected repeatedly on the nuclei of the medium. Most of this deviation is due to Coulomb scattering on the nuclei. For many applications, it is sufficient to use the Gaussian approximation for the projection of the angular distribution with a width of $\theta_0$, for the scattering medium in units of radiation length 
$10^{-3}< x/X_0 <100$:
\begin{equation}
\theta_0 = \frac{13.6~MeV}{p\beta c} z \sqrt{x/X_0} \times (1+ 0.038 ln(x/X_0).
\end{equation}

For 8 double-layer single-coordinate chambers with a diameter of straw 10 mm and a radiation
thickness $x/X_0=9\times 10^{-4}$ and for a single-charged particle with a momentum p = 1 GeV/$c$  the multiple scattering angle $\theta_0$ is $9.4\times 10^{-5}$ that corresponds to a deviation at a 1\,m baseline of 0.094 mm. For the straw tubes of 6 mm ($x/X_0=1.64\times 10^{-3}$) the corresponding numbers are $1.3\times 10^{-4}$ and 0.13 mm. For the detector frame material made of aluminum 0.8 cm, radiation thickness ($x/X_0 = 0.1$) the values $\theta_0$ and spatial deviation are  $3.7\times 10^{-4}$ and 0.37 mm, respectively.

\subsection{Humidity and ambient temperature. Influence on the parameters of the tubes}

\begin{figure}[!h]
    \center{\includegraphics[width=0.7\linewidth]{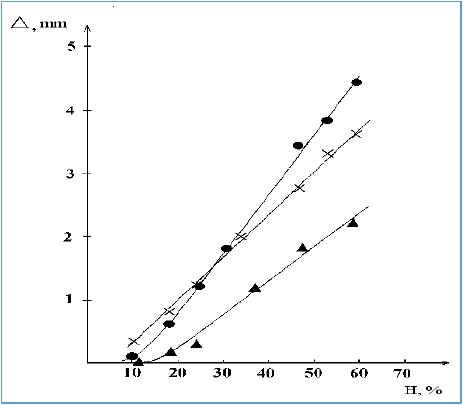}}
  \caption{Dependence of the elongation of single tubes $\varnothing$6~mm and $\varnothing$10~mm with a length of 3.2~m employed in the COMPASS experiment \cite{Bychkov:2006xr}. 
A large spread of data is observed for different tube materials.
	}
\label{SEC_8}
\end{figure}

\begin{figure}[!h]
    \center{\includegraphics[width=0.85\linewidth]{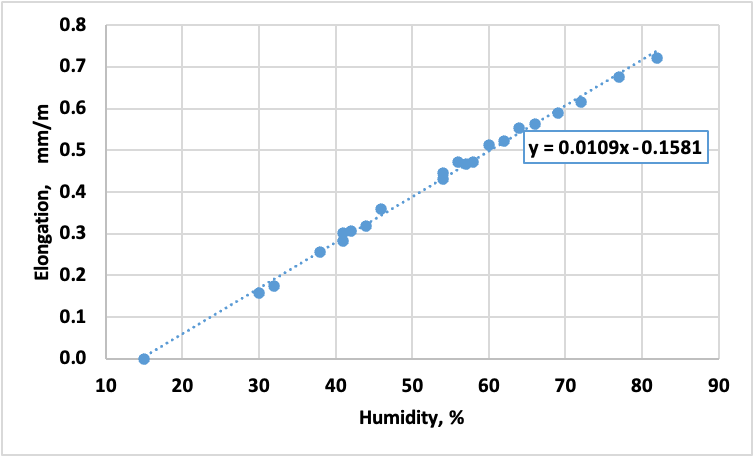}}
  \caption{ Effect of humidity on the elongation of tubes with a diameter of 10 mm and carbon/kapton coating.
	}
\label{SEC_9}
\end{figure}

The straw tubes are made  by winding of kapton tapes on a rod, followed by sintering. It is known that the walls of the tubes change their size depending on the temperature and humidity of the environment. During the development of straw detectors for the COMPASS experiment in 2002, estimates of the elongation value depending on humidity were carried out. 
The elongation of the straw tubes depending on humidity is shown in Figures \ref{SEC_8} and \ref{SEC_9}.
The test showed that the length of the straw tubes is preserved after they are dried with dry air, i.e.~the tubes have a range of elastic deformation.

The same study was carried out when creating straw detectors with copper-coated tubes for the NA64 experiment in 2019. 
As a result, the elongation value of the straw tubes was found to be $1.4 \pm 0.2$~mm/m for copper-coated tubes with diameters of 6~mm and 10 mm for a change in relative humidity from 50\% to 80\%.


The temperature coefficient of expansion of the straw tubes $25.2 \times 10^{-6}$ K$^{-1}$ is mainly determined by the material of the tube walls and does not depend much on the diameter and type of coating. Changing the temperature by 28 degrees elongates the straw tube by 0.7 mm. Elongating the tubes from humidity by 0.7 mm/m is equivalent to changing the temperature of the straw detector from 20 to 48 degrees. In our case, the detectors are in normal conditions at a temperature that will vary within 5$\div$10 $^\circ$C. This may cause a change in the length of the tubes by 0.1 $\div$ 0.2 mm/m. This temperature dependence must be taken into account when developing detectors. For comparison, the coefficients of thermal expansion of some materials that can be used for construction are (in units of 10$^{-6}$ K$^{-1}$): 23 for kapton, 18 for aluminium, and 27 for mylar.

The presented results show that it is advisable to install straw planes in the frame of the chamber at a temperature not higher than the operating temperature of the chamber and humidity not lower than the corresponding
elongation 
will be in the future. 
The effect of different operating and assembling temperature/humidity can be compensated by applying additional compressive forces to the frame elements during the assembling procedure, or by stretching the frame after installing the straw tubes, compensating for the influence of climatic factors.

\subsection{Mechanical properties of the straw tubes}

\begin{figure}[!ht]
  \begin{minipage}[ht]{0.55\linewidth}
    \center{\includegraphics[width=1\linewidth]{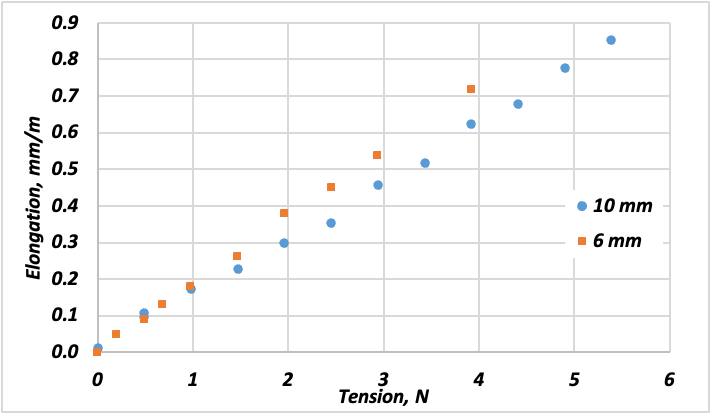} \\ (a)}
  \end{minipage}
  \hfill
  \begin{minipage}[ht]{0.4\linewidth}
    \center{\includegraphics[width=1\linewidth]{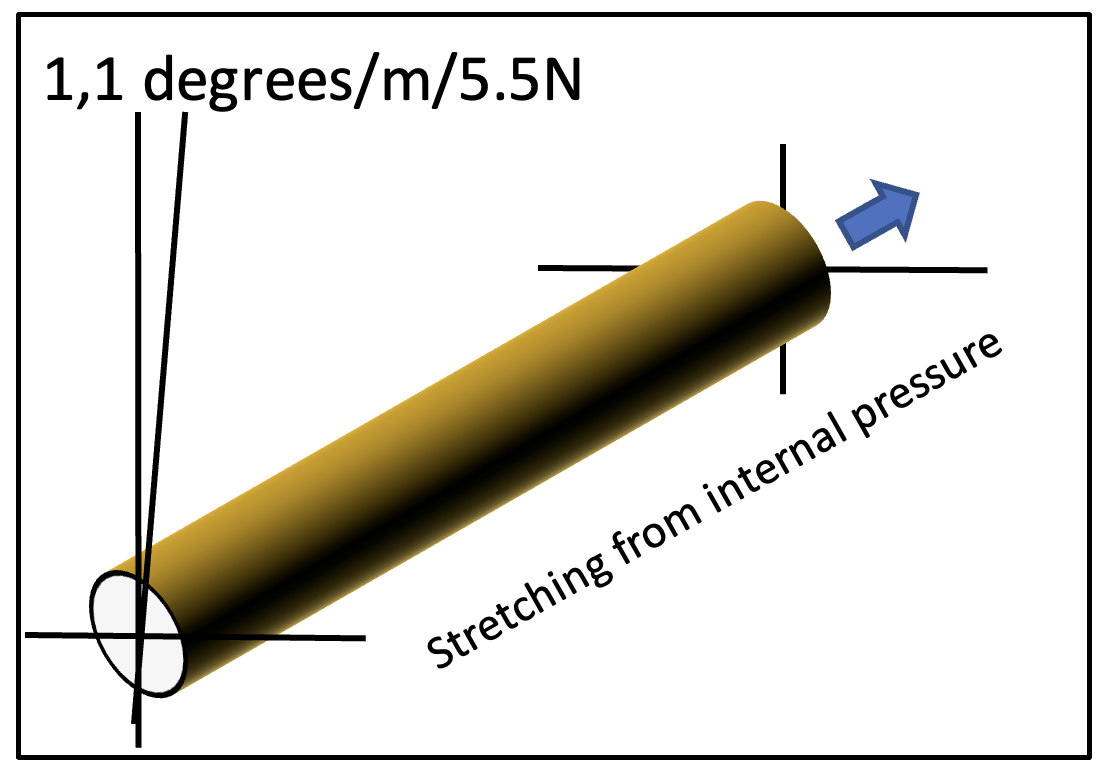} \\ (b)}
  \end{minipage} 
  \caption{(a) Elongation of straw tubes depending on the applied force. Round dots are for tubes with a diameter of 10 mm, while square dots are for 6 mm tubes. (b) 
The twisting angle of a straw tube with a diameter of 6 mm when stretched under pressure. 
The angle is 1.1 degrees for 1 m-long tube and a tensile force of 5.5 N directed along the tube.
	}
\label{SEC_10}
\end{figure}

Mechanical tests were carried out for the straw tubes with a diameter of 6~mm and 10~mm. 
The tubes were subjected to stretching.
The test results are shown in Fig.~\ref{SEC_10}~(a), from which it can be seen that in order to compensate for the influence of humidity, it is enough to stretch the frame of the detector by 0.7~mm per meter of straw length. At the same time, it was shown that the tensile force applied to the tube of about 5.5~N does not cause plastic deformations of the tube for a long period. Measurements were carried out for about 2~years. When the twisted tube is stretched, twisting may occur. The study of this effect was done at high gas pressure inside the tubes. The practical absence of twisting was shown, see Fig.~\ref{SEC_10}~(b). The pressure of 2~atm stretches the tube with a force of 5.5 N (570~g of force).

The tensile force can be created by excessive pressure inside the tubes when gluing the tubes into the frame. 
This pressure changes not only the tube length but also its diameter. 
Thus, at an excessive pressure of 0.5~atmospheres, there is an increase in the diameter of the tube by 0.5~$\mu$m, which leads to a change in the positions of tubes and an increase in the width of the array of 288 tubes by 0.14~mm.

\subsection{End-cap design based on a two-layer array of twisted tubes}

\begin{figure}[!ht]
  \begin{minipage}[ht]{0.5\linewidth}
    \center{\includegraphics[width=1\linewidth]{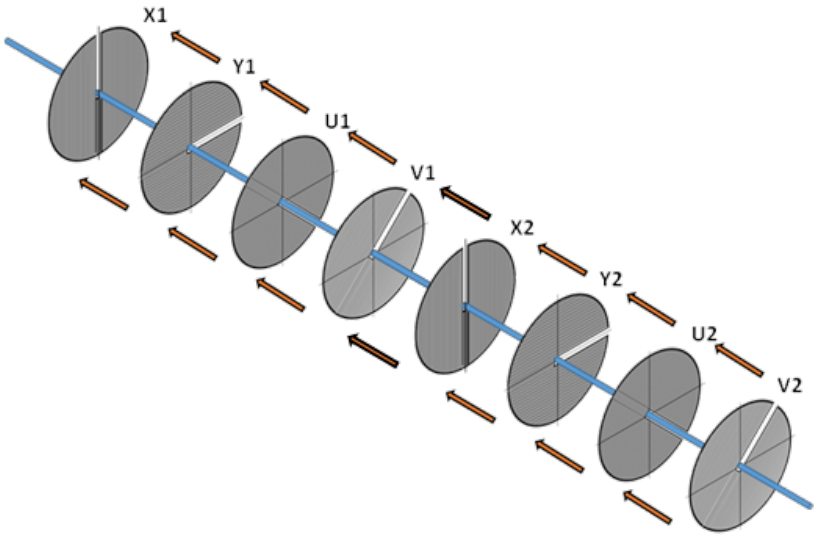} \\ (a)}
  \end{minipage}
  \hfill
  \begin{minipage}[ht]{0.48\linewidth}
    \center{\includegraphics[width=1\linewidth]{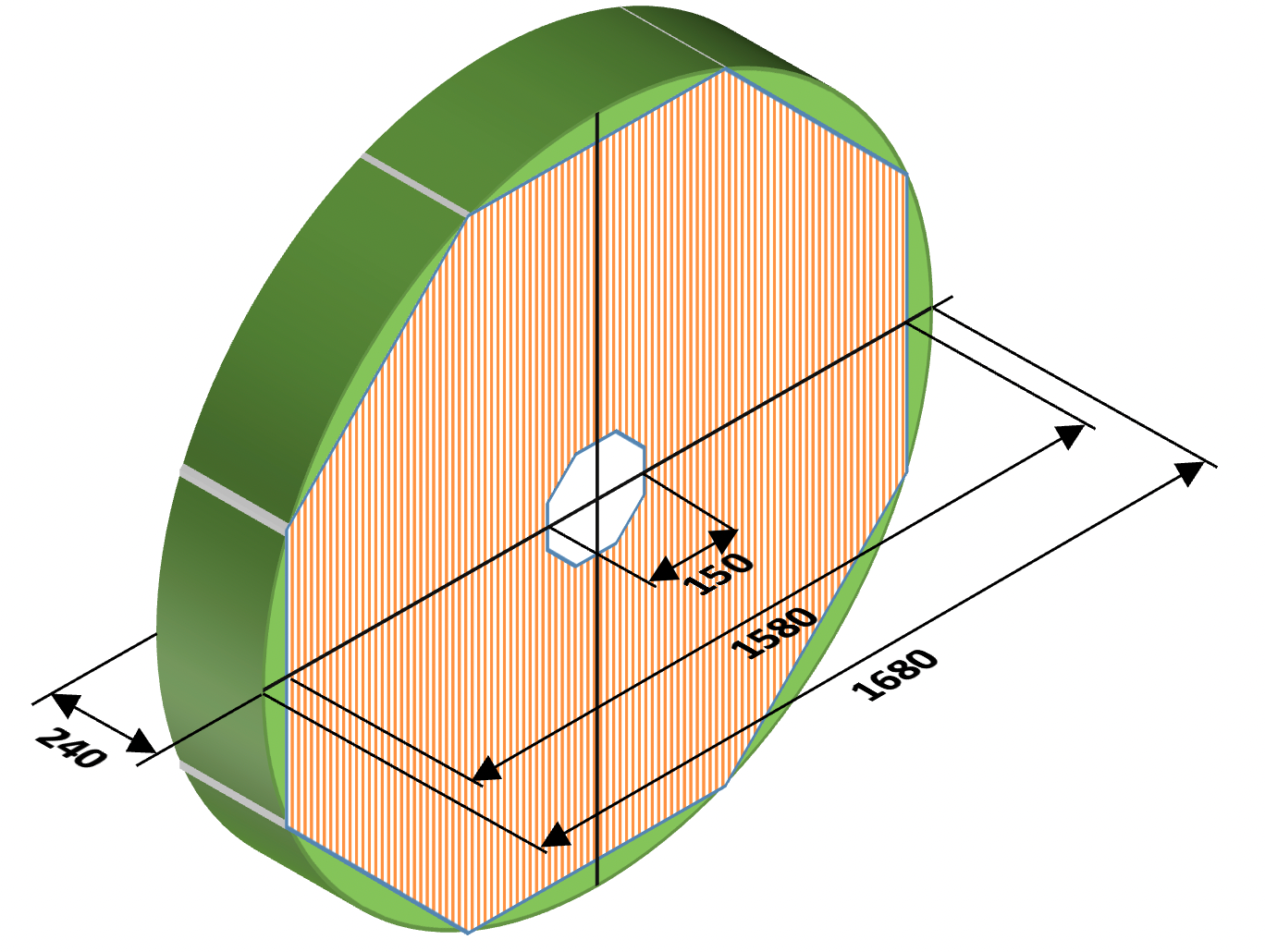} \\ (b)}
  \end{minipage} 
  \caption{(a) ST end-cap consisting of 8 coordinate planes assembled together. (b) Common view and main dimensions.
	}
\label{SEC_11}
\end{figure}

End-cap is proposed with an octagonal arrangement of the drift coordinate planes at an angle of 45~degrees, which form an $X$, $Y$, $U$, $V$ coordinate system, see Fig.~\ref{SEC_11}~(a). In total, 8 coordinate planes are supposed to be used in one end-cap. Each coordinate plane consists of two halves of a disk with an interval for installing a vacuum tube. The thickness of one coordinate plane is 30 mm. Eight coordinate planes are mounted together, forming a rigid block, 240 mm thick. A free octagonal zone 150~mm wide is formed in the center of the block, in which a vacuum pipe of accelerator with a diameter of up to 100~mm is located. The assembled end-cap must be put on the vacuum tube and, together with the rest of the internal detectors, attached to the external part of the SPD installation. The common view of the end-cap is presented in Fig.~\ref{SEC_11}~(b).

\begin{figure}[!h]
    \center{\includegraphics[width=0.7\linewidth]{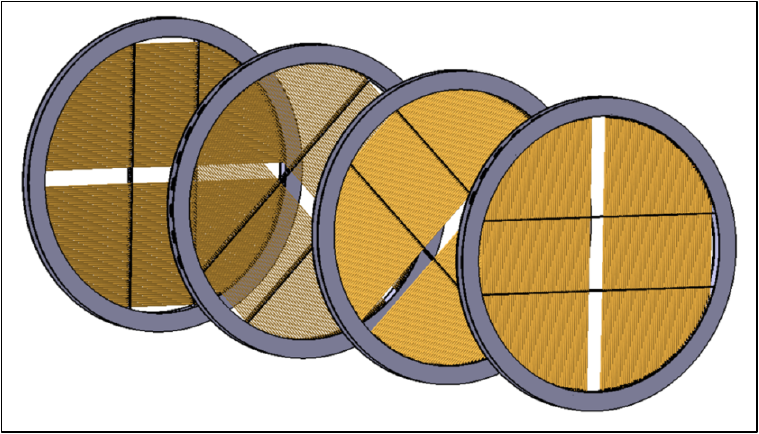}}
  \caption{ 
	Four double-layer planes with a central gap for a vacuum tube are depicted.
	}
\label{SEC_12}
\end{figure}

\begin{figure}[!h]
    \center{\includegraphics[width=0.7\linewidth]{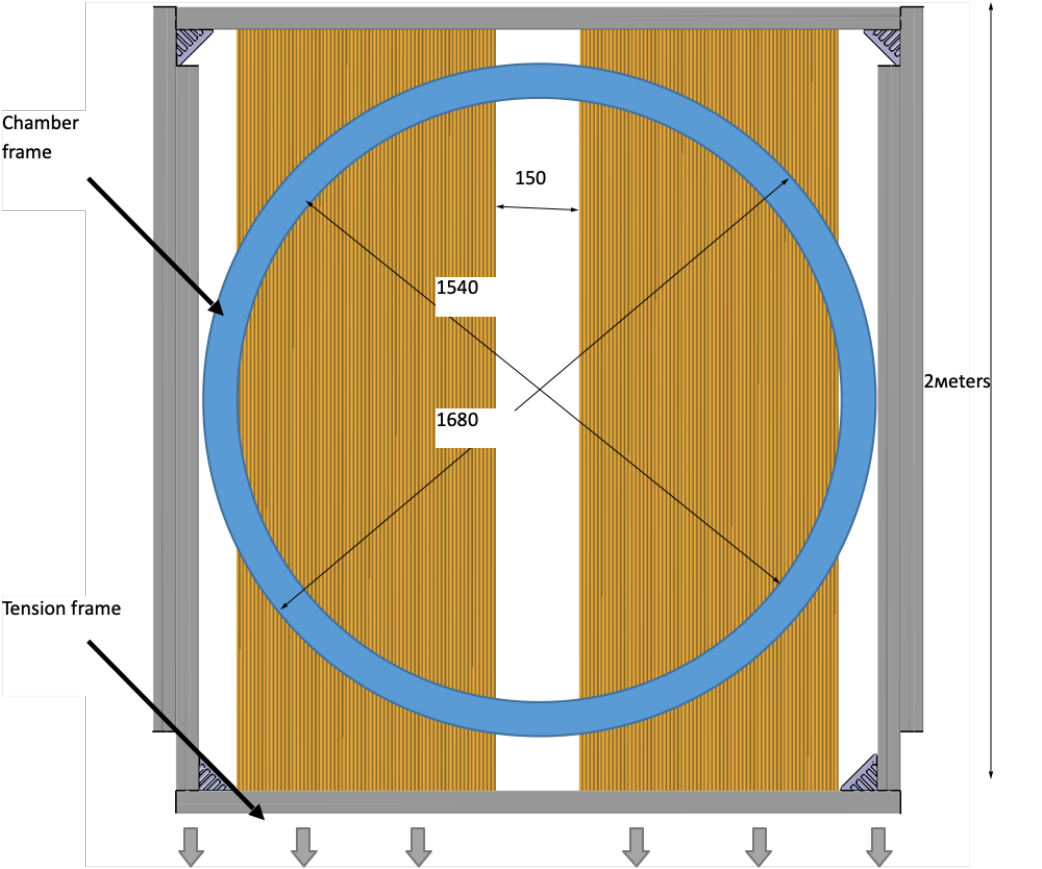}}
  \caption{ Pre-tensioning device for straw arrays.
	}
\label{SEC_13}
\end{figure}

As mentioned above, the tubes in the detector layers must be pre-stretched before gluing into the detector frame. For this purpose, a special technological frame is made, into which all the tubes of the array are glued before the installation of  the signal wires. 
Frame size is $2 \times 2$~m$^2$, see Fig.~\ref{SEC_13}. Then the frame is stretched by mechanical screws together with an array of tubes to the required compensating size. The compensating elongation of the array of tubes should be $0.7 \times 2 = 1.4$~mm for  the tubes 2 meters long. An array of tubes is glued in a stretched state on a precision table into the detector's frame, and each tube is cut along the contour of the frame.
The second array is also glued to the technological frame with a half-shift in the diameter of the tubes, stretched and glued to the chamber frame from the reverse side. Frame thickness is 2 mm of carbon fiber, width is 50 mm.

\begin{figure}[!ht]
  \begin{minipage}[ht]{0.5\linewidth}
    \center{\includegraphics[width=1\linewidth]{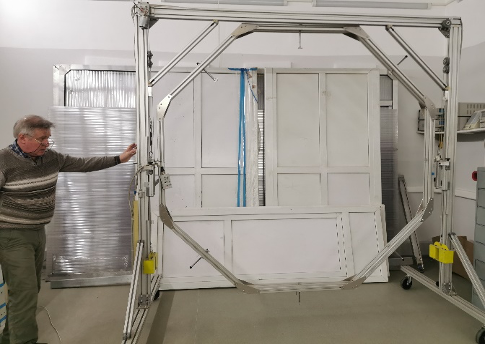} \\ (a)}
  \end{minipage}
  \hfill
  \begin{minipage}[ht]{0.5\linewidth}
    \center{\includegraphics[width=1\linewidth]{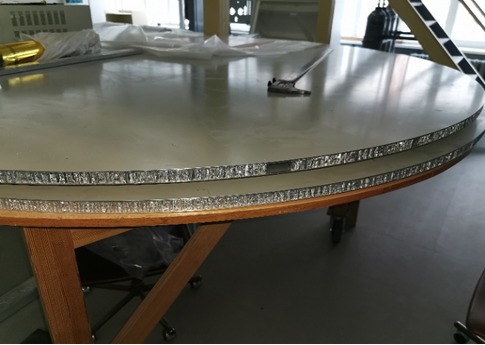} \\ (b)}
  \end{minipage} 
  \caption{(a) Technological equipment for assembly. (b) Octagonal rotary platform and a portable honeycomb board for assembling are shown. The diameter of the portable board is 2300 mm, and the thickness is 50 mm.
	}
\label{SEC_14}
\end{figure}

Each half of the array consists of 72$\times$2 = 144 straw tubes with a diameter of 9.56~mm. The width of the array is 688.3~mm.
There is a total of 288 tubes in  the two layers of one chamber, which must be pre-stretched with a force of 
5N$\times$288 = 1440~N (14.6~kg load). Calculations carried out in the Inventor program show that the deformation of the chamber's frame does not exceed 50~$\mu$m.
Further installation of the signal wires, gas supplies, and matching electronic circuits (motherboards) is carried out according to the proven standard technology.  The straw planes are assembled on a rigid portable honeycomb board on a rotatable octagonal frame shown in Fig. \ref{SEC_14}.

\begin{figure}[!h]
    \center{\includegraphics[width=0.7\linewidth]{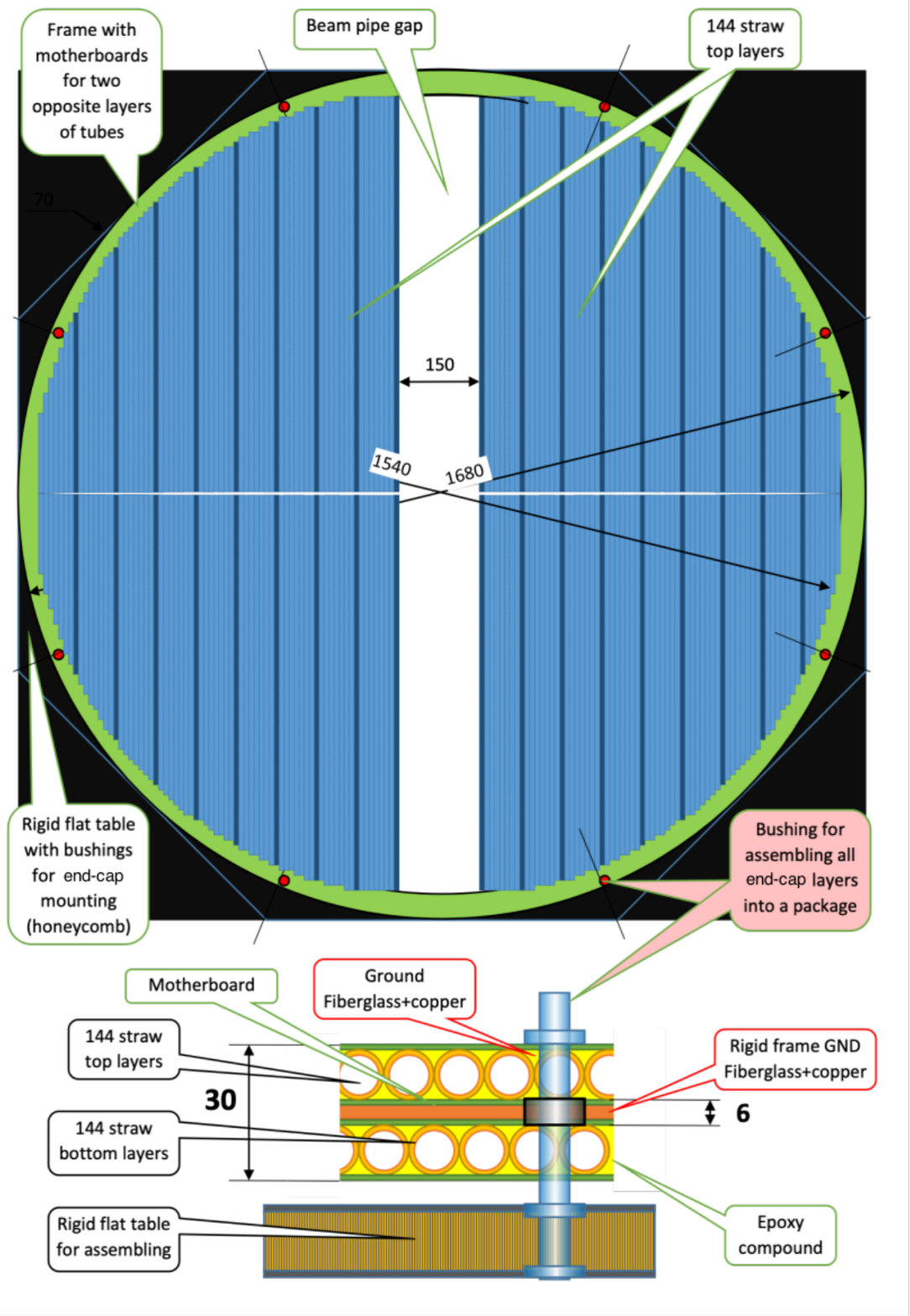}}
  \caption{One of eight coordinate planes with a central gap for the installation of a vacuum tube.
  A technological honeycomb portable board is also shown.
	}
\label{SEC_15}
\end{figure}

\subsection{Readout electronics placement}

\begin{figure}[!h]
    \center{\includegraphics[width=0.8\linewidth]{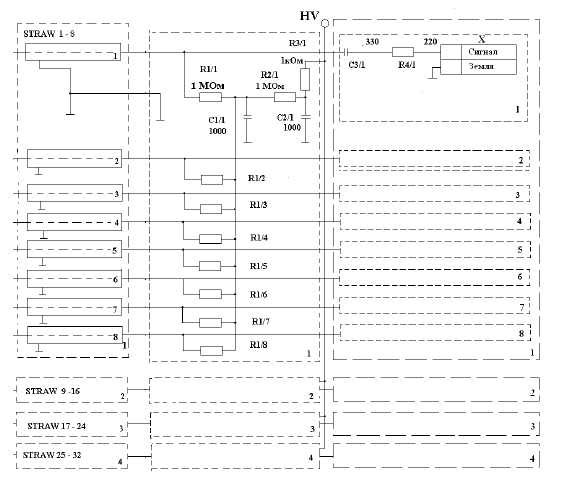}}
  \caption{The scheme of reading signals from one group of 32 straw channels -- ''motherboard''.
	}
\label{SEC_16}
\end{figure}

The electronics, which are located directly on the chambers, perform the function of matching, amplifying, and integrating straw charge signals, as well as providing high voltage to the signal wires of the detector. One straw layer consists of 288 registration channels. 9 ''motherboards'' and 9 amplifiers for 32 channels will be required to read information from one layer. The diagram of the 32-channel motherboard is shown in Fig.~\ref{SEC_16}. In total, for two end-caps with 16 coordinate layers of 4608 registration channels, 144 motherboards, and 144 amplifiers will be required. Subsequent recording electronics are being developed and will be described in Section \ref{VMM3a_section}.

\subsection{End-cap design option with annular cylindrical frame}

\begin{figure}[!h]
    \center{\includegraphics[width=1.0\linewidth]{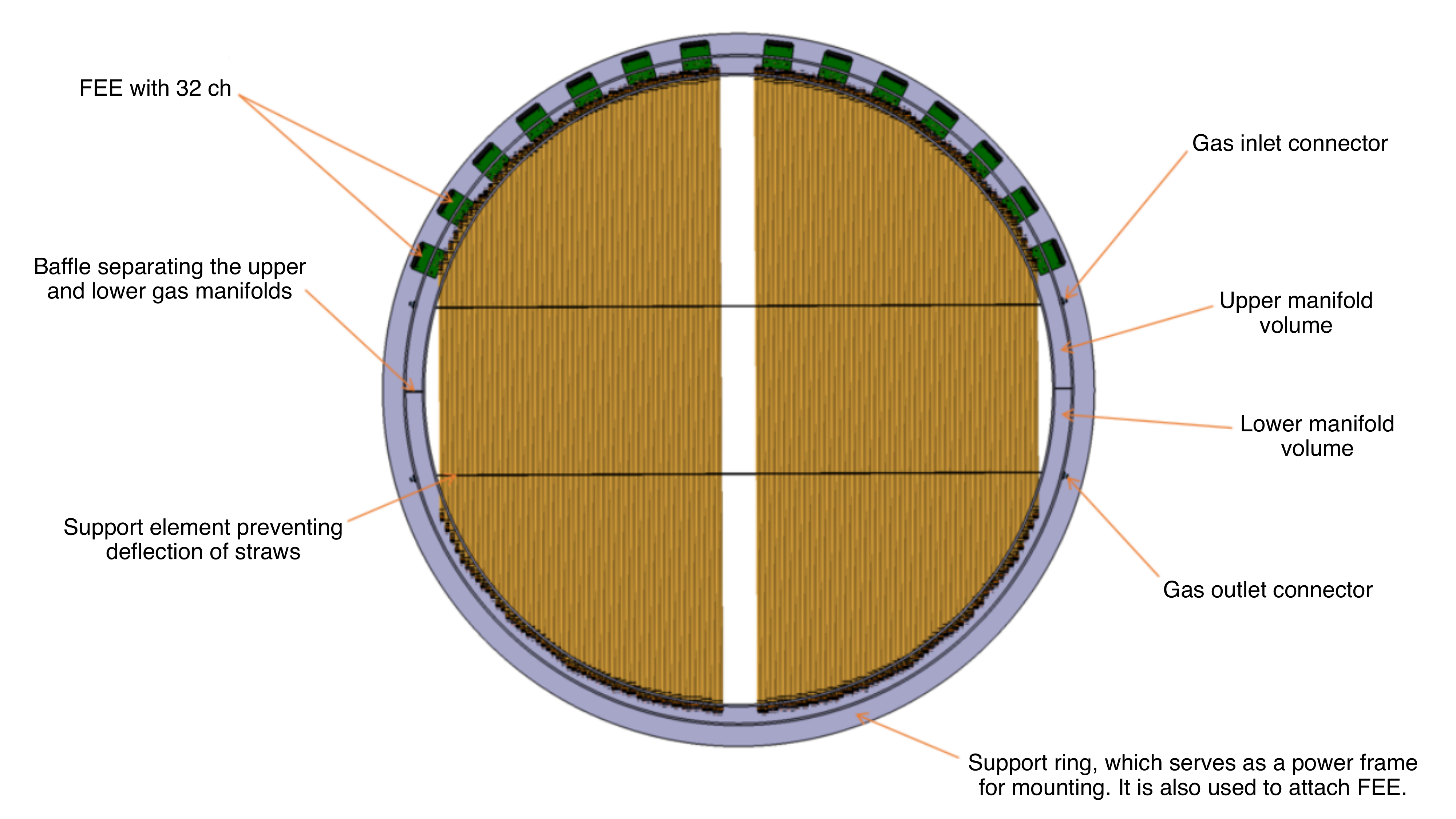}}
  \caption{Cross-section view of one chamber indicating the possible layout of the readout electronics and gas manifolds.
	}
\label{SEC_17}
\end{figure}

\begin{figure}[!h]
    \center{\includegraphics[width=0.7\linewidth]{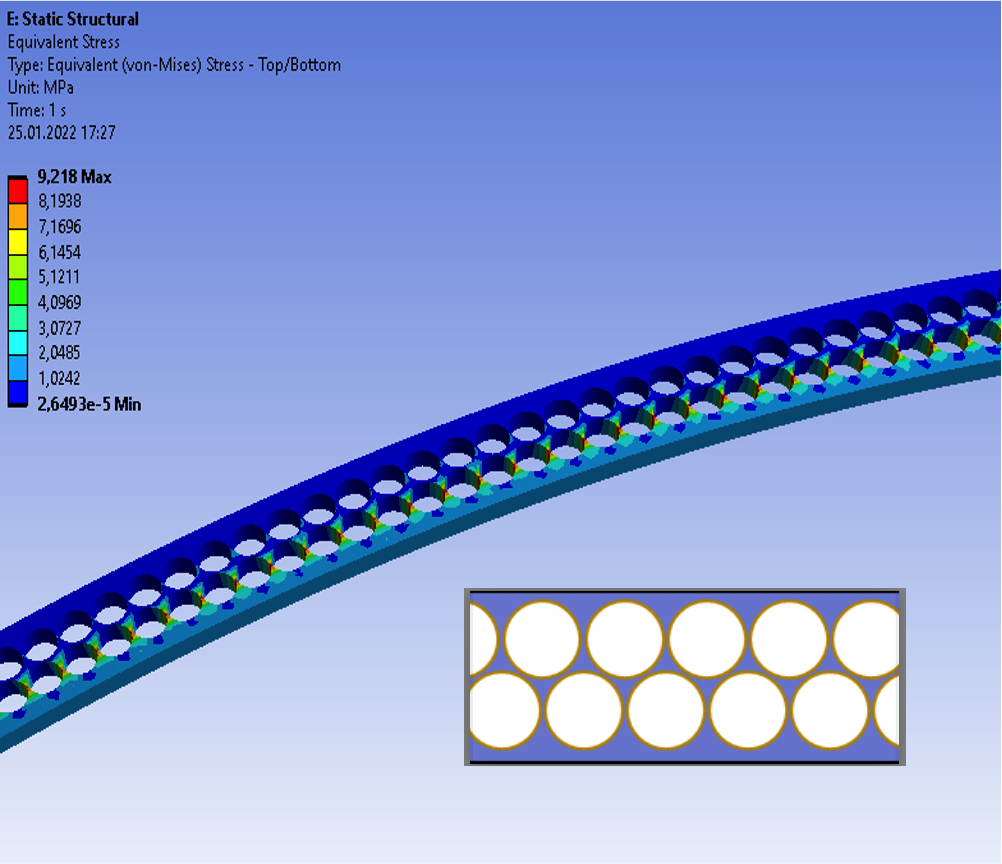}}
  \caption{ Section of the frame with holes in which straw tubes are installed.
	}
\label{SEC_18}
\end{figure}

 An end-cap will consist of eight ring straw chambers, deployed relative to each other. Each chamber consists of two layers of straw mounted in an annular carbon fiber frame with an outer diameter of 1510~mm and a thickness of 5~mm. The total thickness of  one end-cap is 400 mm. The straw layers are shifted relative to each other by the value of the straw radius, see Fig. \ref{SEC_17}. The rows are arranged along the chords of the frame. The working length of the rows in the chamber varies from 360~mm to 1500~mm, depending on the location of the rows. 
A section of the frame with holes, in which straw tubes are installed, is shown in Fig.~\ref{SEC_18}.

The straw tubes are glued individually into the frames under a tension of 500~g each, to avoid sagging of the straw tubes, possible due to high humidity. Also, to avoid sagging of the straw tubes in the chamber, 4 carbon fiber supports are used, shown in Fig.~\ref{SEC_17}, two on each side of the chamber. Each row layer contains 256 rows, 128 on each side relative to the center of the chamber.
After sealing the tubes, anode wires are installed in them under a load of 90 g each. As calculations have shown, the deflection of the frame after that will not exceed 30~$\mu$m.
Motherboards are glued from the end of the frame after installing the anode wires. Motherboards are designed to supply high voltage to the straws and connect the signal from them to the amplifiers. The boards are made in such a way that there are no gaps between the boards after being glued to the frame.
The ground contacts of all lines, as well as the signal pins, are soldered using short insulated wires to the corresponding contacts on the board.
Then, a second carbon fiber ring with a thickness of 5 mm is glued to the boards with an outer diameter of 1570 mm. Gas connectors are installed in the ring to supply the working gas mixture to the straw tubes.

The resulting $\Pi$-shaped box is divided into 4 separate sections to ensure the decoupling of gas flows.
The box is hermetically sealed from above with fiberglass plates.

\section{Front-end electronics}
\label{VMM3a_section}

\subsection{Signal parameters and requirements to the front-end electronics}
As described in Subsection~\ref{subGarfield}, precise tracking is based on the time measurements of  avalanches caused by several closest to the wire primary ionization clusters. This implies the requirement of a small peaking time of the readout electronics (O(10\;ns)) and a low signal discrimination threshold corresponding to several fC. At the same time, good charge measurements have to be done with a reasonably large shaping time to collect charges from all primary electrons created along the charged particle track. Defined by the drift time of electrons created near the straw wall, the signal duration is expected to be about 150\;ns. Another important requirement is defined by the straw tracker occupancy, which is estimated to be up to 200\;kHz per straw in the barrel tracker for Stage2 operation, and about 20\;kHz for Stage1. The front-end electronics should be adjusted to the straw characteristic impedance which is calculated to be about 350~$\Ohm$  at 20~MHz and 1000~$\Ohm$ at 1~MHz, to the capacitance of 23~pF and account for the signal attenuation of 2.3 at 20~MHz.

In order to achieve excellent resolution in terms of coordinates, time, and energy deposition, the following requirements are imposed on the ST for the readout electronics of the straw tubes:
\begin{itemize}
\item possibility to measure both time and deposited charge;
\item  time resolution not worse than 1~ns;
\item low threshold to identify charge from first primary ionization clusters;
\item dynamic range better than 1000 for charge measurement;
\item low power consumption to reduce heating;
\item availability and low cost;
\item sufficient bandwidth.
\end{itemize}

Several possible solutions based on existing Application Specific Integrated Circuits (ASICs) are compared in Section~\ref{subsec:asics}. 

\subsection{External crosstalk, grounding and shielding}

\begin{figure}[!h]
    \center{\includegraphics[width=0.65\linewidth]{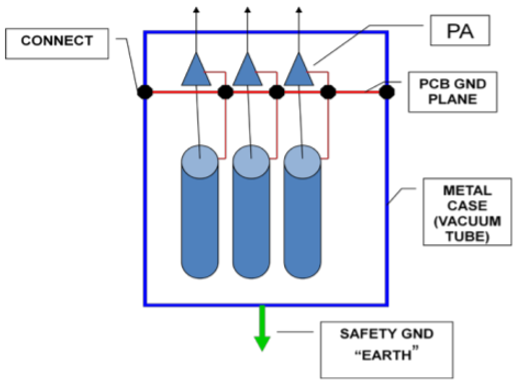}}
  \caption{Grounding and shielding scheme of ST.
	}
\label{st_fee_4}
\end{figure}

Electrically active objects surrounding the detector, such as a crane, a magnet, power supplies of other detectors, wireless connections, etc., can create interference, hereinafter referred to as external crosstalk. Another possible source of problems for the detector electronics is the mismatch in the voltage levels used in front- and back-end electronics. Therefore, it is necessary to ensure that all conductive parts are connected to a safety ground. The proposed grounding and shielding scheme for the ST detector is shown in Fig.~\ref{st_fee_4}. This scheme ensures that signals received from the cathode of the straws are separated and connected close to preamplifier inputs at the front-end board. This scheme was implemented in the 64-straw prototype. 

The ground of the front-end board is connected along its perimeter to a metal case that  surrounds the whole detector, thus forming a tight electromagnetic shield. The conductive layer of the straw tubes is grounded only on one end to prevent external currents flowing through the cathodes. All cable shields are connected on both ends: one side directly to the front-end board, back-end side through damping impedance. 

All metallic pipes for gas supply or cooling of the ST detector will also be electrically connected to the metal frame. The metal case of the station should be connected to the safety ground at one point. The back-end electronics (VME crates) must be connected to the same ground, thus providing both the required safety connection and eliminating potential difference between front-end and back-end electronics. 

\subsection{Available ASIC solutions}
\label{subsec:asics}
The multifunctional ASIC VMM3~\cite{IakovidisVMM} is widely used for readout of MicroPattern Gaseous Detectors (MPGD) and is the base of the advanced VMM3A version~\cite{Iakovidis_2020} developed for the ATLAS New Small Wheel (NSW) readout. Its main benefit is flexible settings for analogue input circuitry. This ASIC is capable for charge and time measurements. Each chip can read out 64 individual channels. Low power consumption of 15\;mW/channel and low cost of around 1\$/channel make it attractive for a compact detector readout. A fast programmable gain preamplifier, a semi-gaussian shaper, a tail cancellation circuitry, a baseline restorer, and a single threshold discriminator are integrated into each channel of the VMM3/3A chip. The chip discriminator thresholds are adjusted in VMM3/3A by a global 10-bit Digital to Analogue Converter (DAC) with additional channel-specific 5-bit trimming DACs. These features enable VMM3/3A to meet the ST requirements of the low threshold level for time measurements. Equivalent noise charge (ENC) of better than 1000 e- can generally be achieved with the input capacitance less than 100\;pF. 

\begin{figure}[htpb]
\centering
\includegraphics[width=0.90\textwidth]{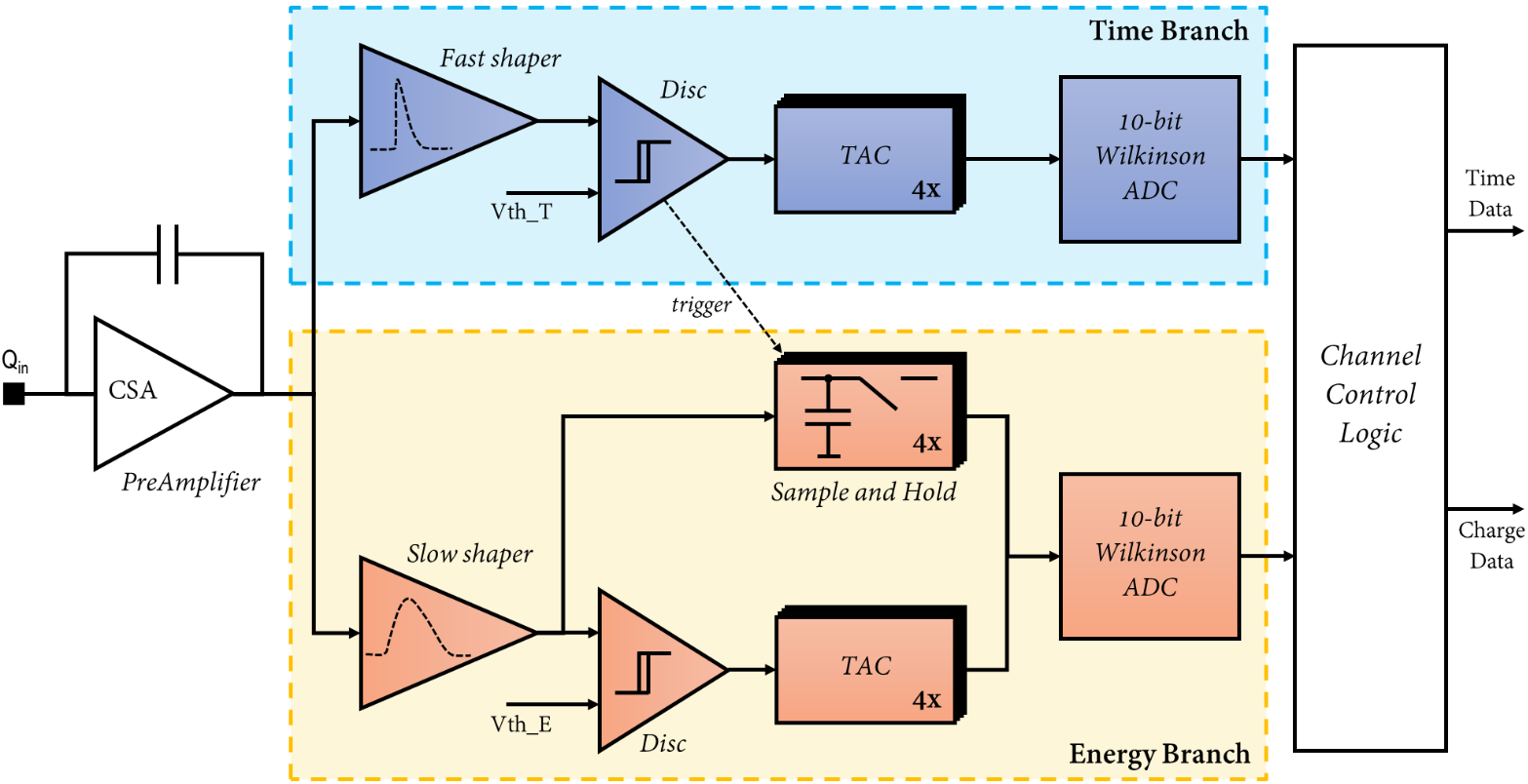}
\caption{ TIGER time and charge digitisation architecture~\cite{TIGER}.}
\label{fig:Architecture}
\end{figure}

The TIGER ASIC~\cite{TIGER} is also developed for MPGD readout and is used in front-end electronics of the CGEM-IT (Cylindrical Gas Electron Multiplier Inner Tracker), the new inner tracker of the BESIII Experiment~\cite{BES3}. The main features of the inner circuitry of the TIGER ASIC are close to those of VMM3/3A, however, TIGER has non-adjustable amplifier gain and shaping time. TIGER processes an amplified signal simultaneously with two independent lines which have different shaping times, as shown in Figure~\ref{fig:Architecture}. This approach is attractive for straw tube readout since the shaping time choice is always a compromise between time and charge measurement precision. 

Table~\ref{tab:asics} compares the main parameters of the VMM3/3A and TIGER ASICs. The time measurement of a straw signal can be performed with the necessary precision with both options and both ASICs can be configured for operation in the triggerless readout mode. However, the dynamic range is too small for charge measurements of low momentum particles even at the lowest amplifier gain. While the straw signal charge of a MIP still can be measured VMM3/3A, the TIGER design adjusted for efficient work with MPGD can not provide charge measurement for straws operated with a reasonable gas gain. Moreover, the VMM3A ASIC has a logic issue influencing operation at the time-over-threshold mode which is necessary for the straw readout and differs from the nominal time-at-peak regime used for the Atlas NSW readout~\cite{Bautin2023aai}. For both ASIC families further development is considered in order to adopt their parameters for efficient readout of straw tubes.

\begin{table}[htpb]
\centering
\begin{tabular}{|c|c|c|}
\hline
                    & VMM3/3A                       & TIGER              \\
\hline
Number of channels  & 64                            & 64                 \\
Clock frequency     & 10...80 MHz                   & 160...200 MHz      \\
Input capacitance   & \textless{}300 pF             & \textless{}100 pF  \\
Dynamic range       & 2000 fC                       & 50 fC              \\
Peaking time        & 25, 50, 100, 200 ns           & 70, 200 ns         \\
Gain                & 0.5, 1, 3, 4.5, 6             & 12 mV/fC           \\
                    & 9, 12, 16 mV/fC               &                    \\
ENC (energy branch) & \textless{}3000 $e^-$         & \textless{}1500 $e^-$ \\
TDC binning         & $\sim$1 ns                    & 50 ps              \\
Maximum event rate  & 140 kHz/ch                    & 60 kHz/ch          \\
Consumption         & 15 mW/ch                      & 12 mW/ch           \\
\hline
\end{tabular}
  \caption{Comparison of the main parameters of the VMM3/3A and TIGER ASICs.}
\label{tab:asics}
\end{table}

\subsubsection{Development of the AST-SPD ASIC}
\begin{figure}[!h]
    \center{\includegraphics[width=0.8\linewidth]{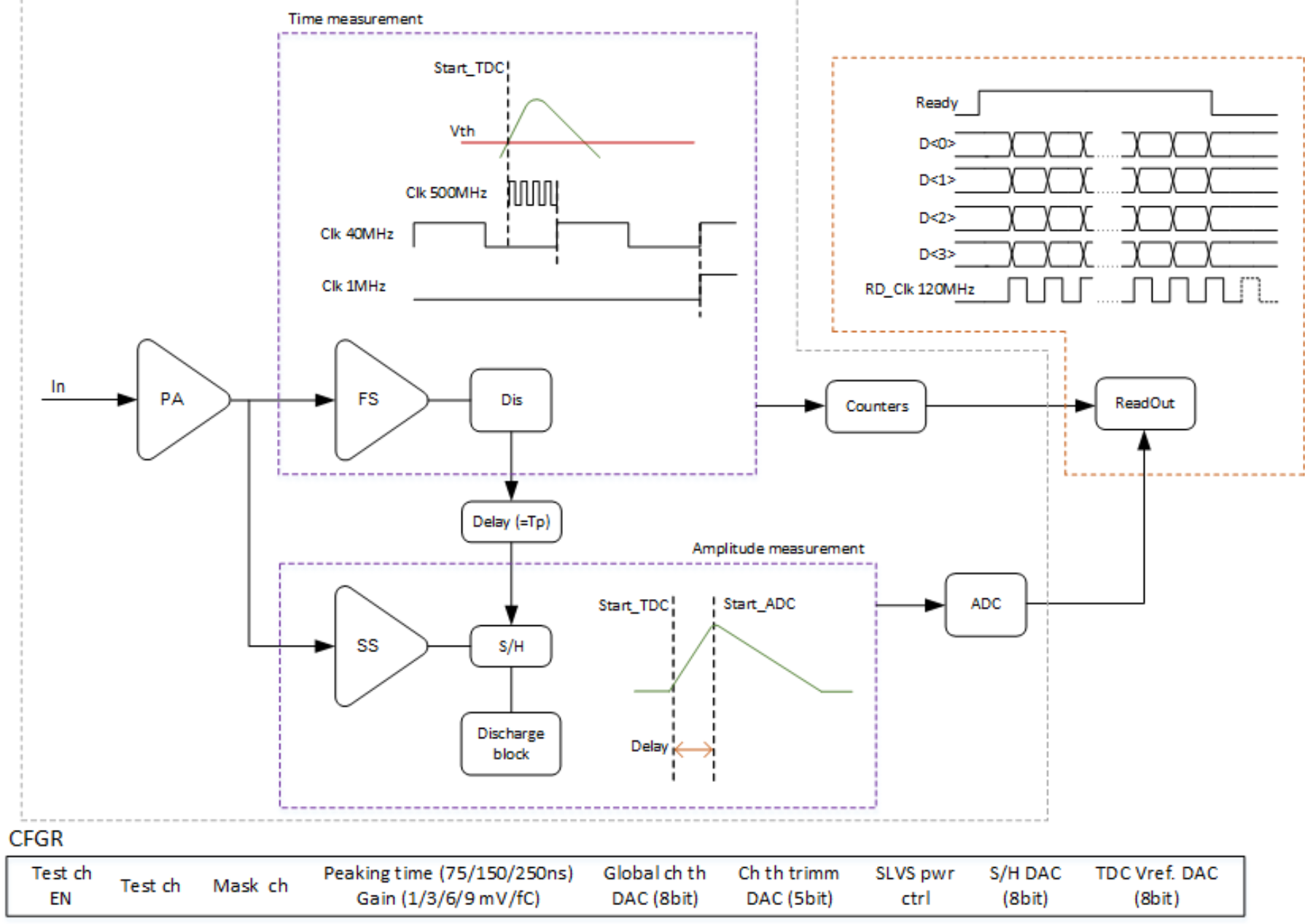}}
  \caption{AST architecture.}
\label{fig:AST}
\end{figure}

Development of an alternative ASIC, AST-SPD, has been started to provide a common readout solution for the ST and the Micro-Megas based Central Tracker of the SPD detector. The AST-SPD design realizes the approach of parallel signal processing similar to that of TIGER. Table~\ref{tab:ast} summarizes the parameter of the developed chip and Figure~\ref{fig:AST} shows its architecture. The preamplifier (PA) output signal is shared between the Fast and Slow signal processing lines. The Fast Shaper has a shaping time of up to 10\;ns and the gain of up to 30\;mV/fC. Those parameters support a good signal-to-noise ratio and the sharp signal rising edge provides the nanosecond precision of the discriminated signal time measurements.

The Slow signal processing line starts with the Slow Shaper which has adjustable shaping time and gain values. This allows reliable operation for different kinds of gaseous detectors and different gas gain values. Signal amplitude digitization is performed with SAR ADC with a conversion time of 250\;ns. The ADC readout is done with the clock frequency of 120\;MHz using four SLVS lines. The size of single hit data including the time and charge values and the channel address is  28 bits. 

\begin{table}[htpb]
\centering
\begin{tabular}{|l|l|}
\hline
\multicolumn{2}{|c|}{Detector Parameter}              \\
\hline
Negative input charge, fC&1000\\
Detector channel capacitance, pF&20-100\\
Loading per channel, kHz&up to 200\\
Working mode& triggerless\\
\hline
\multicolumn{2}{|c|}{Common chip parameters}\\
\hline
Technology & CMOS, 180 nm\\
Number of channels  & 32\\
Supply voltage, V & 1.8 \\
Power dissipation, mW/ch & 10\\
\hline
\multicolumn{2}{|c|}{Fast shaper, time channel}\\
\hline
Shaping time, ns & 6-10\\
Time channel resolution, ns & 1 \\
ENC (r.m.s.), e$^-$ at C$_{det}$=60pF & below 1000\\ 
\hline
\multicolumn{2}{|c|}{Slow shaper, amplitude channel}\\
\hline
Shaping time, ns & 75/150/250\\
Shaper order & 4\\
Gain, mV/fC & 1/3/6/9 \\
ENC (r.m.s.), e$^-$ at C$_{det}$=60pF & below 1000\\ 
ADC, bit & 10 \\
\hline
\end{tabular}
  \caption{Main parameters of the AST chip.}
\label{tab:ast}
\end{table}

\subsection{Studies with a prototype straw readout based on VMM3 and TIGER ASICs}
\begin{figure}[htb]
  \begin{minipage}[ht]{0.5\linewidth}
    \center{\includegraphics[width=1\linewidth]{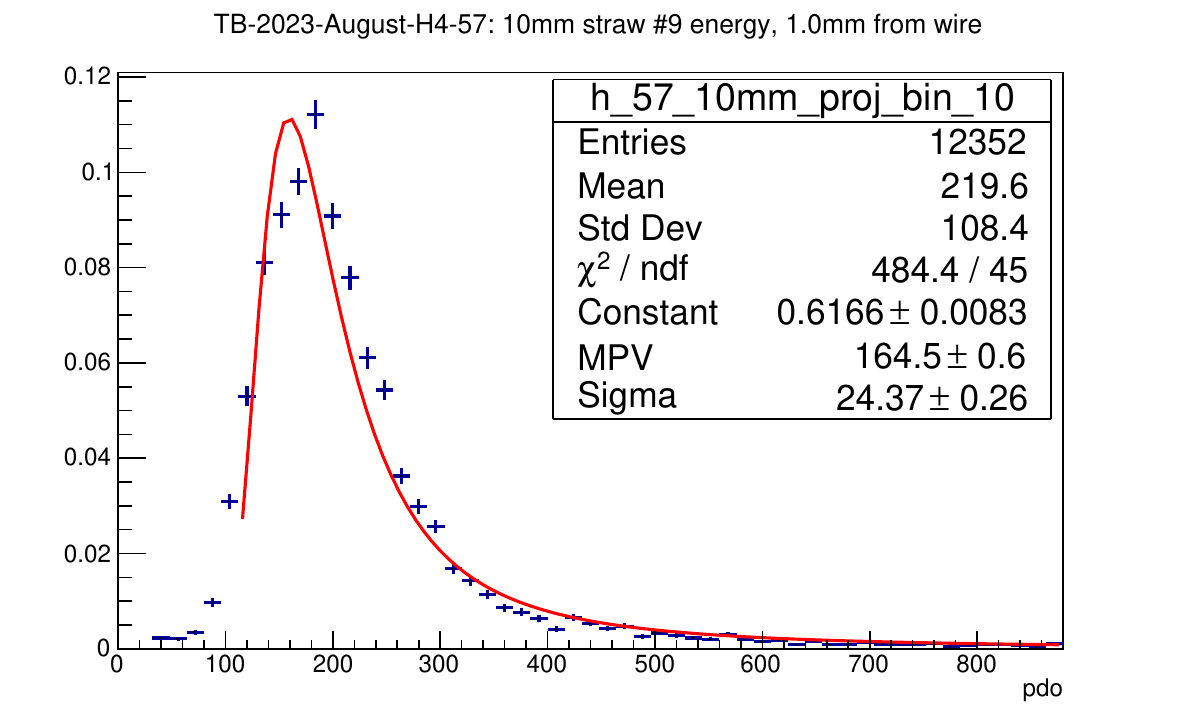} \\ (a)}
  \end{minipage}
  \hfill
  \begin{minipage}[ht]{0.5\linewidth}
    \center{\includegraphics[width=1\linewidth]{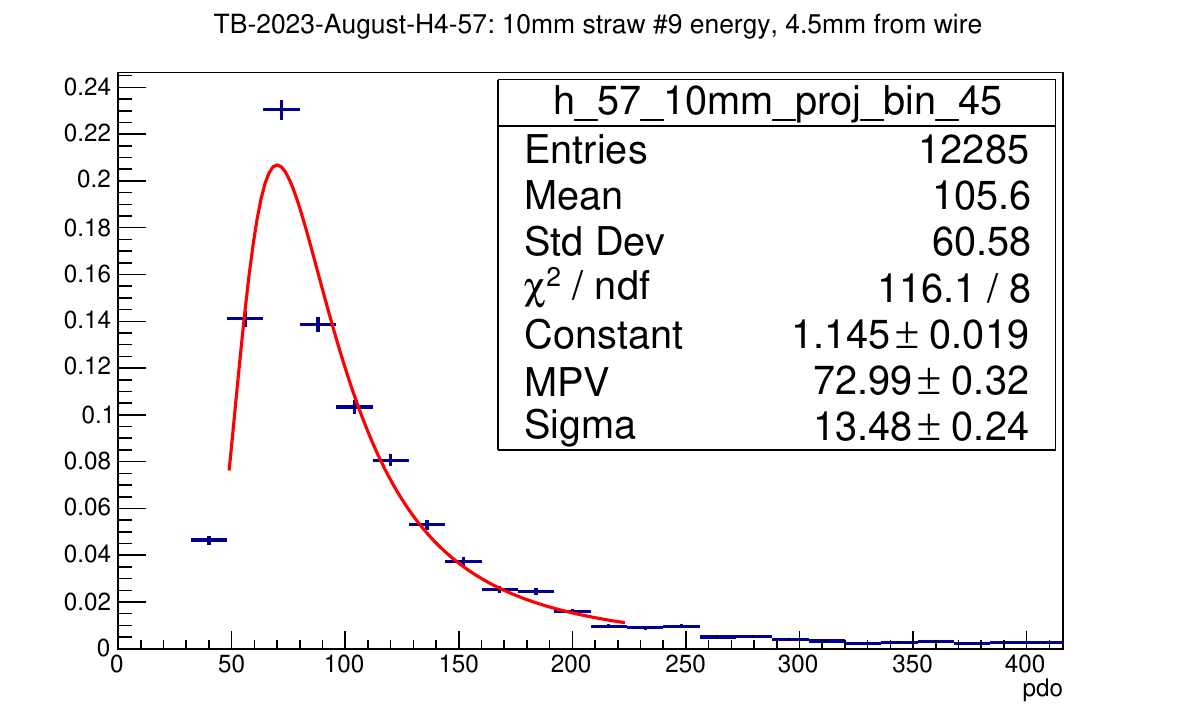} \\ (b)}
  \end{minipage}
    \caption{Signal charge spectra measured with the VMM3-based straw readout in uncalibrated units of ADC counts. The measurements are done for the muon tracks passing at the distance of 1 (a) and 4.5 (b) mm and the straw tubes operated at a high voltage of 1805\;V.}
    \label{fig:resultsQ}
\end{figure}

The VMM3 and TIGER straw readout options have been tested with the SPS muon test beams. The measurement setup  consists of a straw tracker prototype combining straws of 5, 10, and 20\;mm diameter, a reference Micromegas-based tracker, and four scintillator counters used in coincidence for $t_{0}$ time measurements. 
\begin{figure}[hb]
    \centering
    \subfloat[Time resolution]{\includegraphics[width=.4\linewidth]{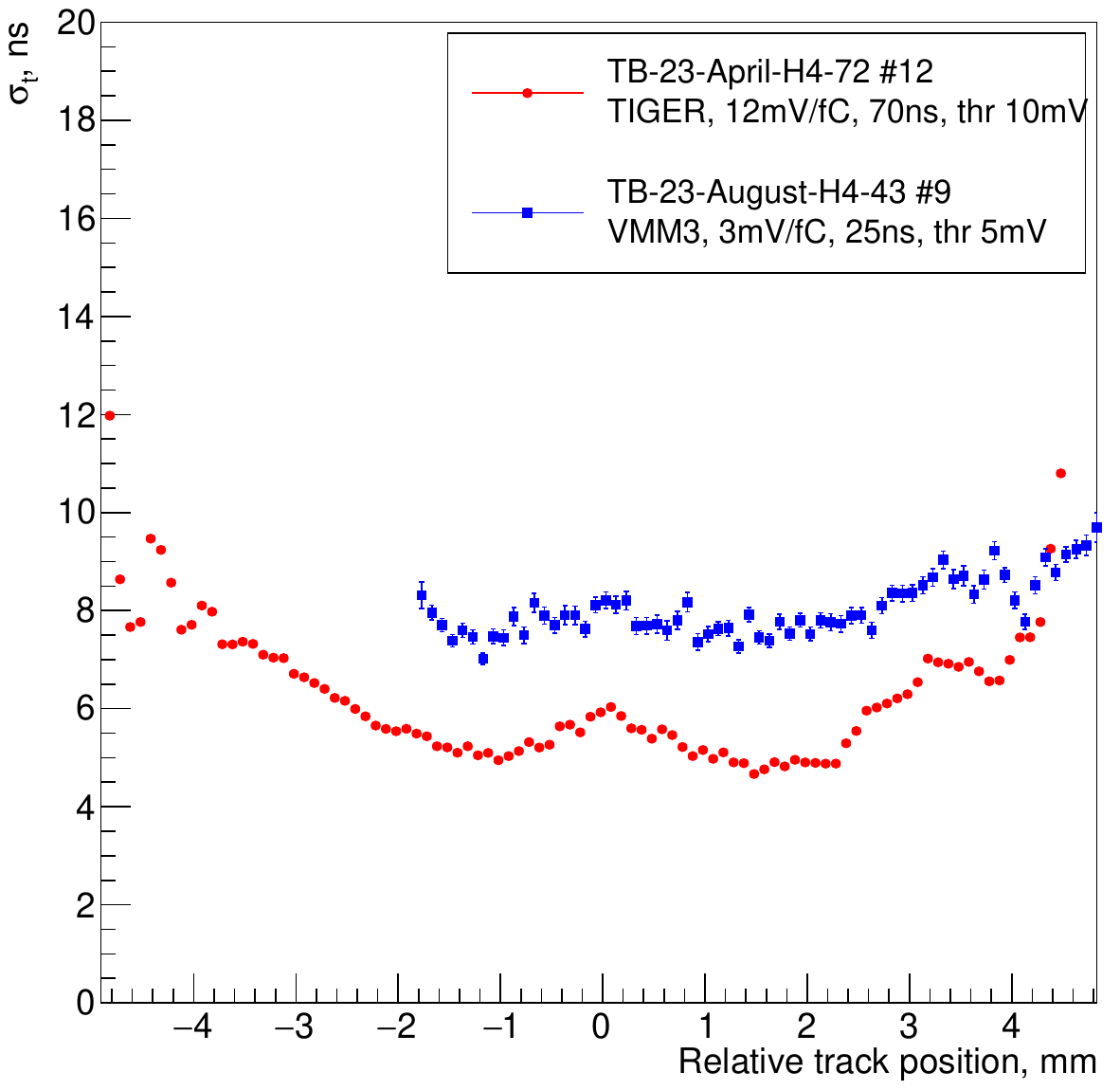}}
    \subfloat[Spatial resolution]{\includegraphics[width=.4\linewidth]{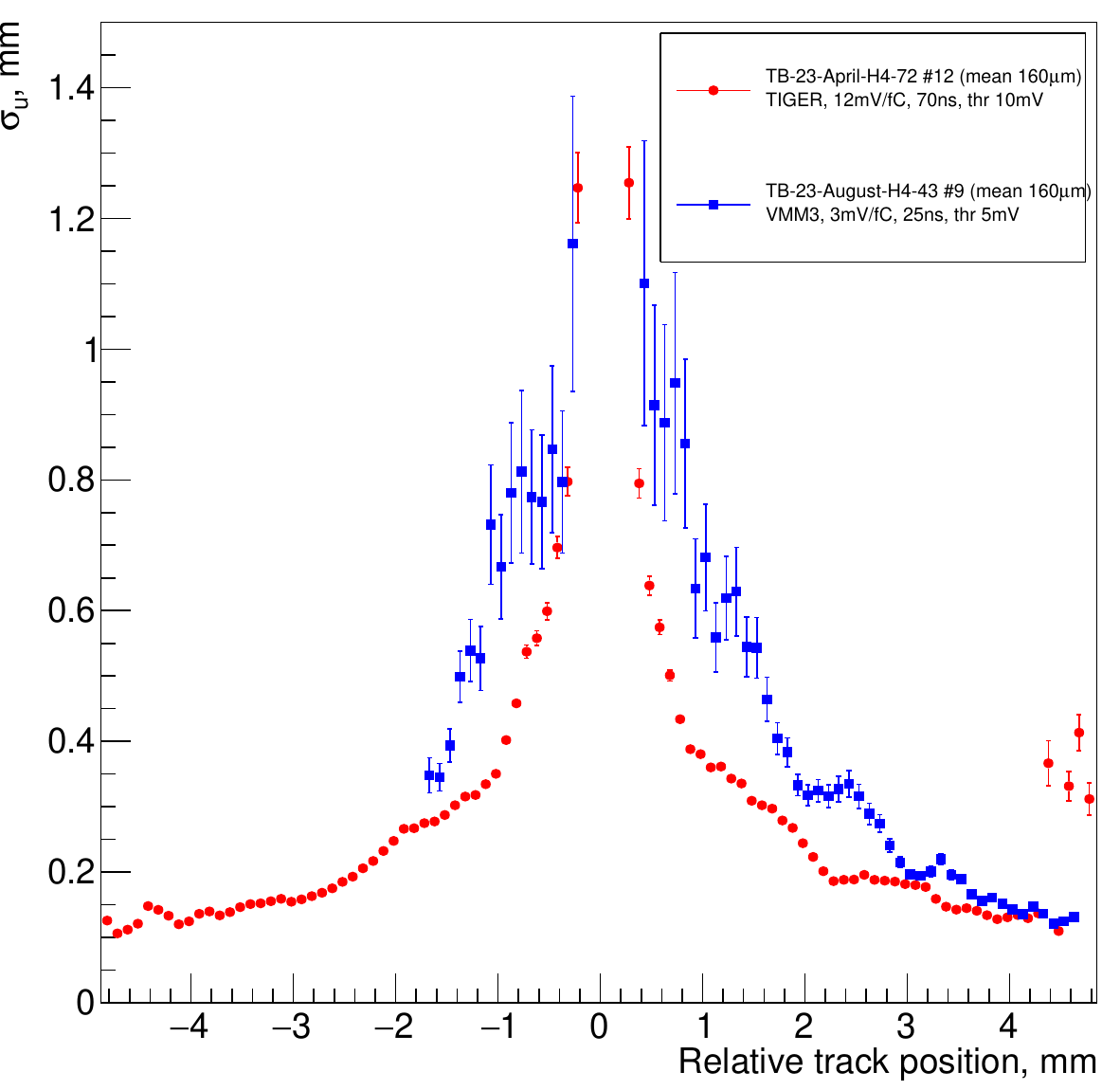}}
\caption{Preliminary results of the VMM3 and TIGER readout comparison for 10\;mm diameter straw tube with ø30um wire, operating with $Ar/CO_{2}$ 70:30 gas mixture at HV = 1750\;V. No correction for the finite resolution of the reference timing and coordinate systems and the electronics noise is applied at this stage.}
\label{fig:results}
\end{figure}
For the VMM3 readout the smallest possible peaking time of 25\;ns and gain of 3\;mV/fC was set during the measurements. The drift time is obtained as a difference of $t_{0}$ and the time when the straw signal exceeds a threshold, both of them are measured with the front-end boards. Figure~\ref{fig:results} shows preliminary results on the time and spatial resolution of the 10\;mm straws as obtained at the initial step of the test beam data analysis without corrections for the finite resolution of the reference timing and coordinate systems and the electronics noise. 
Charge measurements have been done with the VMM3-based readout with the signal peaking time of 200\;ns and gain of 0.5\;mV/fC. Figure~\ref{fig:resultsQ} shows the signal charge spectra for muons passing at the distance of 1 and 4.5\;mm from the anode wire. 

\section{DCS}
\subsection{ DCS architecture}
The Detector Control System (DCS) provides control and monitoring of the detector hardware. In addition, it will perform archiving of the hardware parameters to the database to provide access to this information during the offline data analysis. It has been decided to use the common DCS system, now under development at CERN, which is based on the PVSS II SCADA toolkit and the JCOP framework. The Straw Tracker controlled and monitored equipment includes the following items:
\begin{itemize}
\item low voltage power supplies;
\item high voltage power supplies;
\item gas mixing and distributing systems;
\item low voltage and temperature monitoring system.
\end{itemize}

\subsection{ Low voltage system}
The LV system provides power to the front-end (FE) boards placed on the four straw tracker chambers. Each chamber contains four views and each view contains 30 FE boards. In the present design, one FE board consumes ~ 1 A at 5 volts and, therefore, one module requires the LV power supply to provide a current of about 30A at 5 V. These requirements could be fulfilled by eight Wiener power supplies MPV8008, which have eight channels of 10 A output current each. One view requires four channels and, therefore, two power supplies will provide the required power for one chamber. The MPV8008 could be controlled in a similar  solution used in the LHC experiments. It includes a CAN-bus interface card and the OPC-server providing the interface to PVSS. A second possibility to control the MPV8008 is to use the TCP/IP protocol together with the corresponding TCP/IP PVSS driver, which is part of the PVSS driver package. 

\subsection{ High voltage system}
The Straw Tracker requires a high-voltage (HV) source with a voltage below 2 kV and low current. The present design assumes two HV channels per view, with 32 channels in total. The CAEN HV power supply board A1535, containing 24 channels with 3.5KV/3mA output, should be sufficient. To have 32 HV channels, two boards of this type, housed in the CAEN SY2527LC mainframe, are needed. The mainframe should be controlled via an Ethernet line with TCP/IP protocol and the corresponding CAEN OPC server. The JCOP framework is suitable for the board A1535 and is already available, having been developed for the CMS ME1/1 muon chambers.
	
\subsection{ Gas system controls}
The gas system for the straw tracker should provide and distribute Ar/CO$_2$. Due to the fact that the straws are placed in the inner layers of the detector, the gas system should rapidly close the gas line to a cell, in case of a leak in a straw. Pressure sensors will detect a sudden drop in pressure on the supply lines. In this case, the gas system should also send a signal to the Straw Tracker DCS to switch off the corresponding HV channels or the whole HV system. 
The plan is to use the CERN standard gas mixing/distribution system based on PVSS, with the possibility to control and monitor the gas mixing and gas flow values from the DCS PC. For this purpose, the PVSS distribution manager will be used. Both software and hardware interlock signals prevent HV from turning on if the gas flow is missing or an incorrect gas mixture is distributed.

\subsection{ Thermometry and FE monitoring}
In order to get information about the cover (front-end) temperature at the straw group inlet, one thermo-sensor per FE cover will be mounted. Two voltages (one for the FE boards and  one for the current consumed by a board) will be measured using Embedded Local Monitoring Boards (ELMBs). A chamber view has 30 boards, and four thermo-sensors will be mounted directly on the mechanical structure. In total, one view needs to measure 34 temperature values and 90 voltages. Therefore, to read out all monitoring information from one chamber, 10 ELMBs are needed. The readout should be performed via CAN-bus and OPC server, which is similar to monitoring the LV power supply. It seems reasonable to have one CAN-bus branch per chamber, which gives four CAN-bus branches in total. In case the LV power supplies control protocol is compatible with the ELMB control protocol, the LV power supplies will be connected to the chamber CAN-bus branches. The final decision depends on the final choice of the FE electronics.

\subsection{ Logical trees in DCS and FSM}
To get a convenient way of navigation through the detector elements, the DCS logical tree will have a structure including both tracker nodes and gas system nodes. Due to the fact that the detector has only a few LV and HV channels, the bottom node of the tree will be linked to the corresponding channel of the LV or HV power supply in the hardware tree. The FSM tree should correspond to the Straw Tracker logical tree. The states of FSM should correspond to the states of the whole detector.

\subsection{ 	 DCS development and maintenance}
The plan is to use our own DCS PC running tracker PVSS system and  DCS OPC servers for all detectors on one Windows-based computer. This PC should have CAN-bus adapter cards installed. After the development is finished and to simplify the maintenance of the detector DCS, the plan is to create a JCOP FW component containing all tracker PVSS panels and scripts, and to port it to the SPD central DCS computer. A copy of this component should be stored in the SPD DCS repository. The ST DCS PC should be used to run the DCS servers and to house the CAN-bus adapters. 
To simplify the control of the detector during further hardware development, debugging, and maintenance, the PC to access PVSS panes located at the central DCS PC, will be used. 
For this purpose, the DCS PC will be equipped with the PVSS User Interface software installed.

\section{Cost estimate}
\subsection{ST barrel}

The configuration for the proposed ST is based upon 8 octants in the barrel part. Each octant contains about 3200 straws. 
Assuming one-end reading for each straw, the total number of electronic channels is about 26\,000. 
The core cost is estimated for the construction of the complete ST, using a mixture of updated quotes from vendors, and costs of the similar detector components used in different experiments. 
The risk associated with the project is relatively small. The ST design is based upon well-established technology, using low-mass straws, successfully developed for various modern projects. This does not require serious research and development in the field of detectors.

The corresponding costs for the various items of the ST barrel are summarized in Table \ref{cost_barrel}.
\begin{table}[htp]
\caption{Cost estimate for the barrel part of ST.}
\begin{center}
\begin{tabular}{|l|c|}
\hline
Item   & Cost, k\$ \\
\hline
Straw tubes   & 220\\
Glue      &  20\\
End-plugs & 100\\
Crimping pins & 90 \\
Anode wire  &  60 \\
Other  components & 90 \\ 
\hline
Mechanics \& C-fiber frames & 200 \\
ST tools & 160 \\
Safety equipment \& consumables & 60\\
Gas system  & 300 \\
Cooling system  & 130 \\
\hline
Front-end electronics (ASIC \& boards) & 500\\
Back-end electronics  & 300 \\
HV components & 200 \\
LV components & 200 \\
Cables \& connectors  & 150 \\
\hline
Prototyping  & 320\\
\hline
\hline
Total   & 3100 \\
\hline
\end{tabular}
\end{center}
\label{cost_barrel}
\end{table}%

\subsection{ST end-caps}

The configuration for the proposed ST is based upon 16 double-layer modules in two end-cap parts (8~modules on each side). 
Each module contains 288 straws. 
Assuming one-end reading for each straw, the total number of electronic channels is 4608.

The corresponding costs for the various items of the ST end-cap are summarized in Table \ref{EC_tab3}.
%
\begin{table}[htp]
\caption{Cost estimate for the end-cap part of ST.}
\begin{center}
\begin{tabular}{|l|c|c|c|c|}
\hline
Unit & Quantity & Unit cost, \$ & Total cost, k\$ \\
\hline
Straw tubes  (3m/straw), m & 15000 & 2.5 & 37.5 \\
Anode wire   (4m/straw), m & 20000 & 2    & 40    \\
Pins                                     & 10000 & 3    & 30    \\
End plug                              & 10000 & 2    & 20    \\
Spring contact                     & 10000 & 2    & 20    \\
Spacers                               &  5000  & 1    &  5     \\
Glue, chemical materials, consumables &&&  1     \\
\hline
Straw frame                         &      16 & 3000 & 48 \\
Rigid flat mounting frame    &        2 & 6000 & 12 \\
Covers for frames                &      80 &  100 &   8 \\
Frame for tension                 &       2 & 2000 &  4 \\
Precision rulers for 16 straw, combos & 3000 & 10 & 30 \\
Measuring table for alignment & 1 & 3000 &  3 \\
\hline
Prototype                             & 2 & 25000 & 50 \\
Gas equipment                    & & & 20 \\
Tension control equipment  & 1 & 5000 & 5 \\
Tooling for cutting                & 2 &  500 &  1 \\
\hline
Motherboard plate and frontend & 240 & 1000 & 240 \\
\hline
\hline
 Total & &  & 574.5 \\
\hline
\end{tabular}
\end{center}
\label{EC_tab3}
\end{table}

\section{Identification of particles using energy loss $dE/dx$ in straw tubes}

The ST  will be the only detector able to separate $\pi$/$K$/$p$ during the first stage of the SPD experiment, and its performance is crucial for many physics tasks suggested in Ref.~\cite{Abramov:2021vtu}. Particle identification in the detector is based on measurements of  ionization energy losses. The energy losses follow the Landau distribution and are subjected to strong fluctuations due to rare acts of ionization with large energy loss. 
A truncated mean method is used to achieve a good resolution.
 For this method, a certain fraction (10$\div$50\%) of signals with the highest amplitudes are discarded. The obtained mean value can be compared to different particle hypotheses and the corresponding probability can be assigned for each particle type.

\begin{figure}[htb]
    \centering
    \includegraphics[width=.55\textwidth]{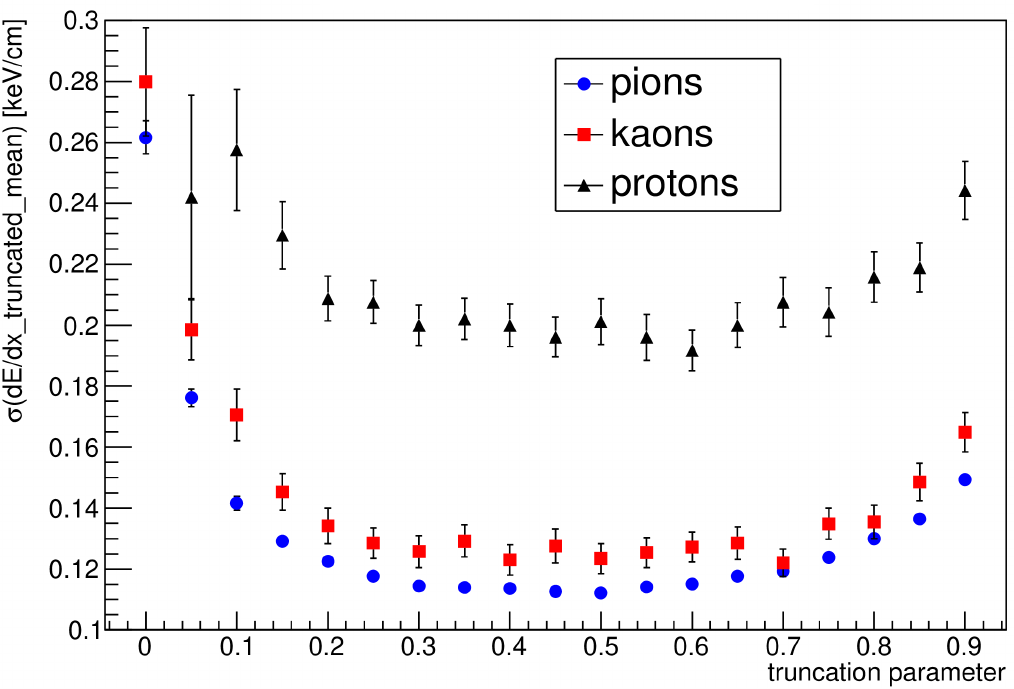}
    \caption{Dependence of the truncated mean resolution on the fraction of discarded signals for tracks, reconstructed in the barrel part of the ST detector.}
    \label{fig:dedx-trancation-par}
\end{figure}

\begin{figure}[htb]
    \centering
   \includegraphics[width=.75\textwidth]{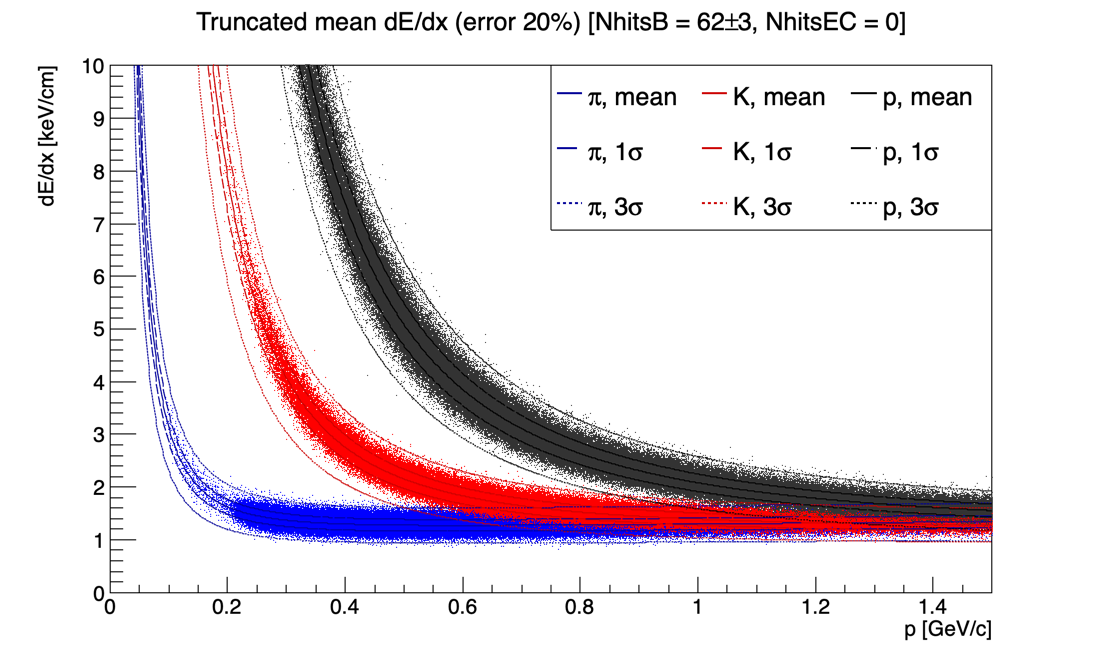}
    \includegraphics[width=.75\textwidth]{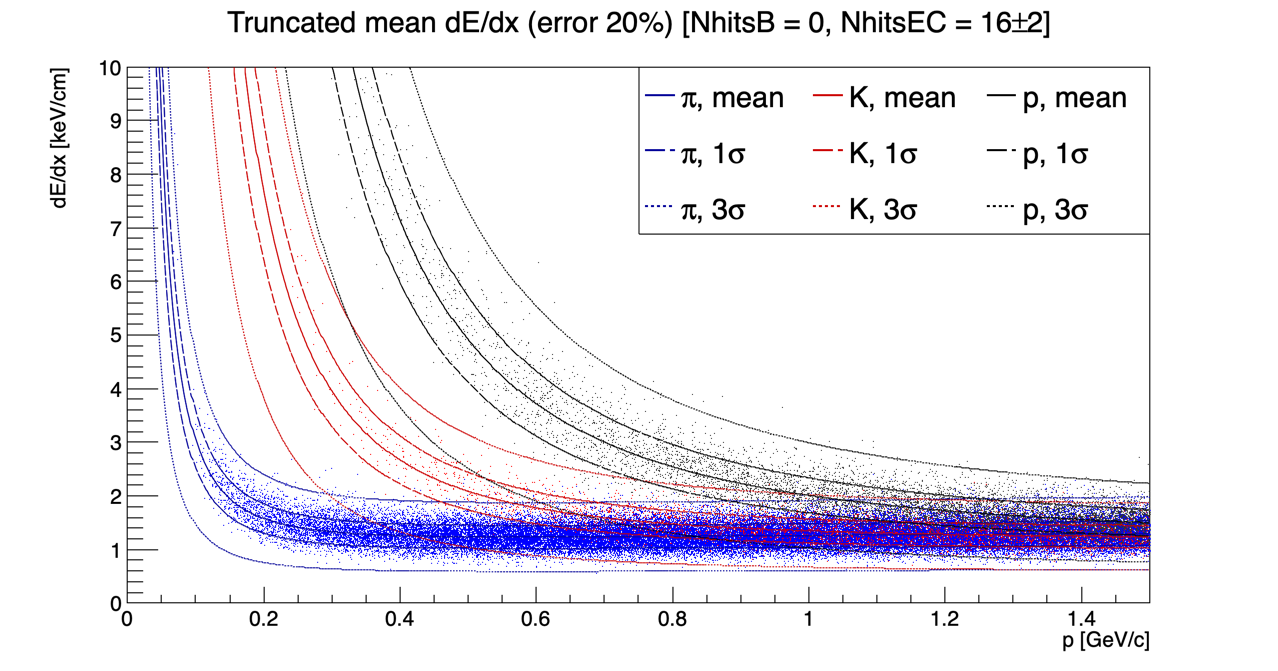}
    \caption{Performance for particle identification with $dE/dx$ in the straw tracker. The dashed and dotted lines indicate $1\sigma$ and $3\sigma$ intervals, respectively. The upper plot shows tracks reconstructed in the ST barrel, the lower~---~in the ST end-caps.}
    \label{fig:dedx-performance}
\end{figure}

The full simulation of PID performance must include a description of the straw tracker, a simulation of the ionization losses and a gas gain in tubes, and the response of the electronics. The simplified simulation accounting for the first two points within the SpdRoot framework has been performed, the impact of the electronics will be discussed later. The resolution dependence on the fraction of discarded signals is shown in Fig.~\ref{fig:dedx-trancation-par}  for the barrel part of the detector.  One can see that it weakly depends on the truncation parameter in the range 0.3$\div$0.6. For the following estimations, this parameter is set to 35\%. 
The truncated mean value of $dE/dx$ calculated per track as a function of momentum follows the Bethe-Bloch distribution
and is shown in Fig.~\ref{fig:dedx-performance} for tracks crossing the ST barrel and the ST end-caps.
The particle separation capability for pions, kaons, and protons can be estimated from curves indicating $1\sigma$ and $3\sigma$ displacement intervals.


\begin{figure}[htb]
    \centering
    \includegraphics[width=1.\textwidth]{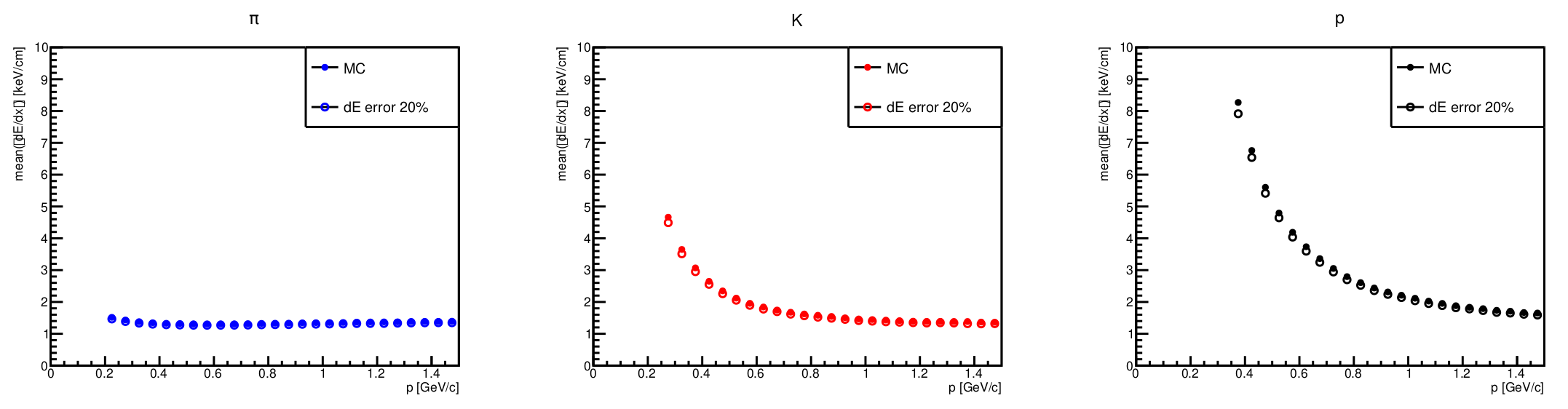}
    \caption{Truncated mean for different particles as functions of their momentum obtained without and with charge smearing of 20\%.}
    \label{fig:dedx-smeared}
\end{figure}

To account for the intrinsic signal charge fluctuations, the study on the particle separation has been repeated with 20\% deviation of the signal charge to account for the straw gas gain fluctuations described in~\ref{subsubGarf}. The results shown in Figure~\ref{fig:dedx-smeared} demonstrate the negligible influence of the intrinsic fluctuations of the straw signal.

In the general case, the signal amplitude is defined by a convolution of the raw signal and the electronic response function. For the long shaping time, the amplitude will depend on the collected charge only. In our case, the shaping time setting for VMM3 is 100 or 200~ns, and the total drift time is about 100~ns for a 10~mm straw tube. The charge collection efficiency in this case will depend on the signal shape, which depends slightly on the track-to-wire distance. The signal duration for tracks coming close to the wire is much larger as compared to tracks passing at large distances. This requires to introduce amplitude, which can be done routinely.

\chapter{Beam-Beam Counters for local polarimetry}

The main goal of the local polarimetry at the SPD is permanent monitoring of the  transverse beam  polarization during data 
taking to reduce the systematic error coming from the beam polarization variation.  
Another task for the local polarimetry at the SPD  is independent from the beam polarization monitoring independent of the 
major polarimeters (CNI and absolute)  and possible usage of this tool to tune
the beam polarization axis.  

The SPD  energy range (below $\sqrt{s}$=27~GeV) is new, therefore, there is a lack of polarization data to find an explicit solution for the local polarimetry.   
One of the tools to control the proton beam polarization is measurements of the azimuthal asymmetry in the inclusive production of charged particles in collisions of transverse polarized proton beams.
Such a method has been successfully adopted  at the STAR detector. Two Beam-Beam Counters (BBCs) are  used for this purpose. Each BBC consists of
two zones, the inner and the outer one, corresponding to a different rapidity range. The inner and outer zones cover 
3.3$<|\eta|<$ 5.0  and 2.1$<|\eta|<$3.3, respectively. BBCs detect
all the charged particles produced in the forward direction within their acceptance. 
The measurements at STAR demonstrated that the BBCs are sensitive to the transverse polarization of the colliding beams.  The value of the effective analyzing power $A_N$ for the inclusive production of charged particles at $\sqrt{s}$= 200~GeV is about $(6\div 7)\times$10$^{-3}$. At NICA energies it should have, in principle, the same value or even a larger one due to a larger analyzing power for the $p$-$p$- elastic scattering. Therefore, on the whole, the BBC can be used for the local polarimetry at the SPD for
the $p$-$p$- scattering. However, a detailed
MC simulation is required to estimate the asymmetries values in the SPD energy range. 

\section{MC simulation results}

An MC simulation has been performed for the simplified 
segmented BBC in order to estimate the expected asymmetries and to optimize the final
BBC design. In the simulation the BBC consisted of 96 scintillation tiles. It was divided into 6 concentric layers with 16 azimuthal sectors each. 
The distance between the tiles was equal to 10 mm. The tile thickness was 5 mm.
The diameter of the BBC was approximately 1700 mm. The distance between each detector and the nominal center of the SPD setup 
was taken as $z$ = 1716 mm.   The uncertainty of the interaction point location is expected to be 
$\Delta z =\pm$300~mm.

\begin{figure}[!h]
  \begin{minipage}[ht]{.9\linewidth}
    \center{\includegraphics[width=0.6\linewidth]{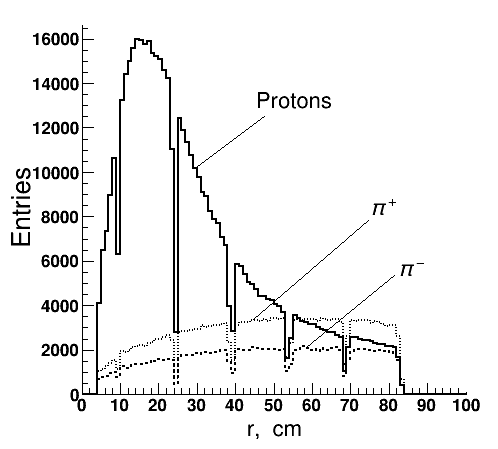}}
  \end{minipage}
  \caption{The events distributions as a function of the BBC plane radius for protons, $\pi^+$ and $\pi^-$ particles are shown by the solid, dotted, and dashed lines, respectively. The results have been obtained using the FTF generator at the total $p$-$p$ energy $\sqrt{s}$ = 6.2 GeV. }
\label{fig:ter1}
\end{figure}

The simulation has been performed at energies $\sqrt{s}$ = 6.2, 10, and 23.5 GeV   using the FRITIOF (FTF) and Pythia8 generators within the SPDroot framework. The events distributions have been obtained in the dependence on radius ($r$) in the BBC plane for protons, $\pi^+$ and $\pi^-$
particles. The $r$-dependencies obtained by means of the FTF events generator at 6.2 GeV are demonstrated in Fig.\ref{fig:ter1}. 

\begin{figure}[!h]
  \begin{minipage}[ht]{.9\linewidth}
    \center{\includegraphics[width=1.0\linewidth]{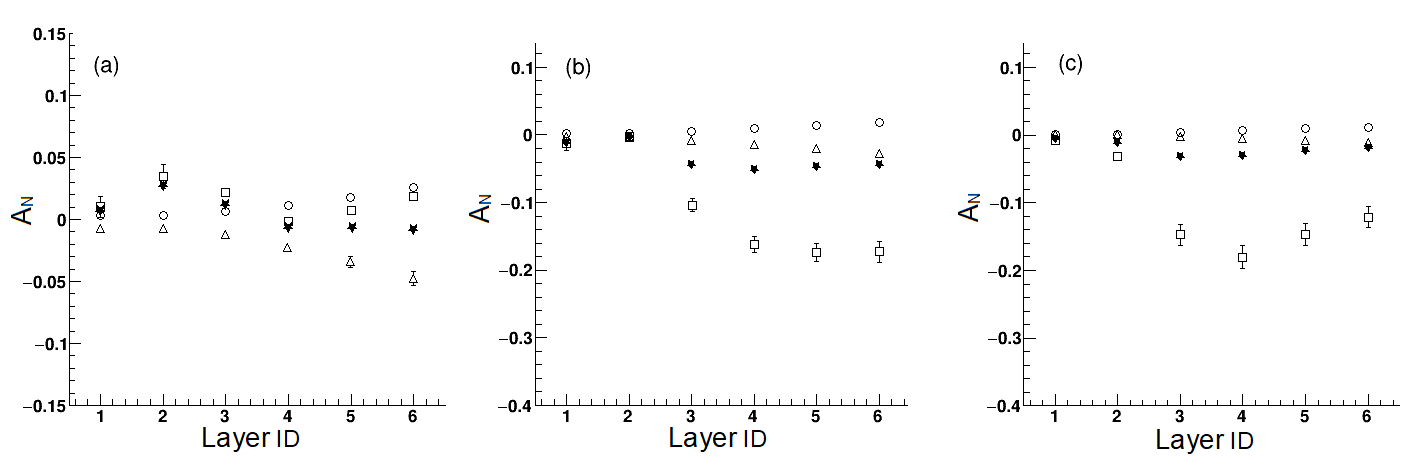}}
  \end{minipage}
  \caption{The analyzing powers $A_N$, and $A^{eff}_N$ obtained within the framework of the model 
CPQ \cite{Abramov:2022dez} using the FTF simulation results. (a), (b), and (c) are the results of the calculations at energies 
$\sqrt{s}$ = 6.2, 10, and 23.5 GeV, respectively. The open squares, triangles, and circles are the 
$A_N$ data for protons,  $\pi^+$ and $\pi^-$, respectively. The solid stars are the $A^{eff}_N$ data.}
\label{fig:ter2}
\end{figure}

The data for the protons, $\pi^+$ and $\pi^-$  particles are shown by the solid, dotted, and dashed lines, respectively. 
The analyzing powers $A_N$ for the inclusive reaction have been calculated within the framework of the phenomenological model for the chromomagnetic polarization of quarks (CPQ) \cite{Abramov:2022dez}.  The efficient analyzing powers $A^{eff}_N$ have been estimated by the formula:
\begin{eqnarray}
A^{eff}_N =\frac{A^p_N\cdot N_p + A^{\pi^+}_N\cdot N_{\pi^+} + A^{\pi^-}_N\cdot N_{\pi^-}}{N_{ch}}.
\end{eqnarray}
Here, $N_{ch}$ is the total number of charged particles, $N_p$, $N_{\pi^+}$ and  $N_{\pi^-}$  are the number of protons, $\pi^+$, and $\pi^-$ in each layer, respectively. The calculated values of the $A_N$ and $A^{eff}_N$ asymmetries using the FTF -- simulation results are demonstrated in Fig.\ref{fig:ter2}. The value of asymmetry $A^{eff}_N$  is small due to the small values of $x_F$ and $p_T$ in this kinematic area. The non-zero value for the second (and third) layer is due to $A^p_N$ (elastic or inelastic scattering). 

The asymmetry estimation has been performed for 10 and 23.5 GeV. At $\sqrt{s}$ $>$ 10 GeV  $A^{eff}_N$ has non-zero values for the layers number $>$2.  The selection of the elastic channel is necessary to estimate its contribution to the behavior of $A^{eff}_N$.  

\begin{figure}[!h]
  \begin{minipage}[ht]{.9\linewidth}
    \center{\includegraphics[width=1.0\linewidth]{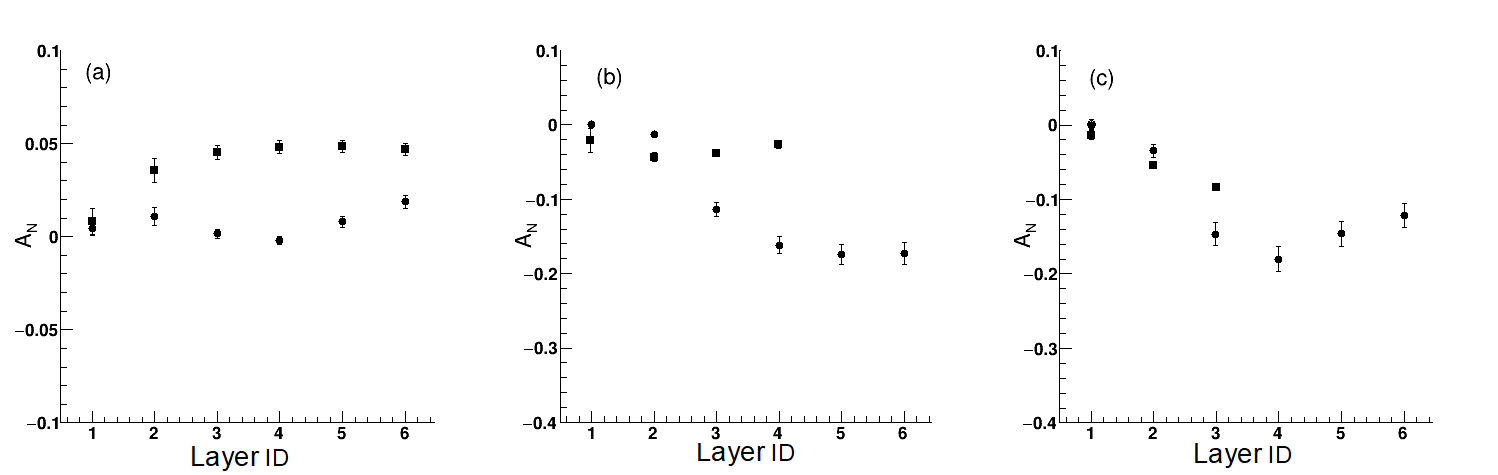}}
  \end{minipage}
  \caption{The analyzing powers $A_N$ have been calculated at the energies 6.2 (a), 10 (b), and 23.5 (c) GeV using the FTF generator. The data for elastic and inelastic scattering are shown by the squares and circles, respectively. }
\label{fig:ter3}
\end{figure}

The analyzing powers $A_N$ have been calculated at three energies: 6.2, 10, and 23.5 GeV, for elastic and inelastic $p$-$p$ scattering (see Fig.\ref{fig:ter3}). The elastic scattering plays a significant role at all layers (except for the first one) at the energy 6.2 GeV. The contribution of the inelastic scattering is close to zero at this energy. The role of the inelastic channel is important starting from the third layer at 10 and 23.5 GeV. The contribution of the elastic scattering is approximately equal to zero and decreases with the energy increasing because of the strong fall of the  differential cross-section of the elastic $p$-$p$ scattering at high energies. Thus, the elastic scattering plays an important role at  $\sqrt{s}<$ 10 GeV. The inelastic $p$-$p$ scattering, on the contrary, is significant at  $\sqrt{s}>$ 10 GeV. 
Therefore, the BBCs design has to be able to select the elastic $p$-$p$ scattering at low energies.
The details of the simulation are given in Ref. \cite{arkady1}.

\section{Beam-Beam Counters}

\begin{figure}[!h]
  \begin{minipage}[ht]{.9\linewidth}
    \center{\includegraphics[width=1.0\linewidth]{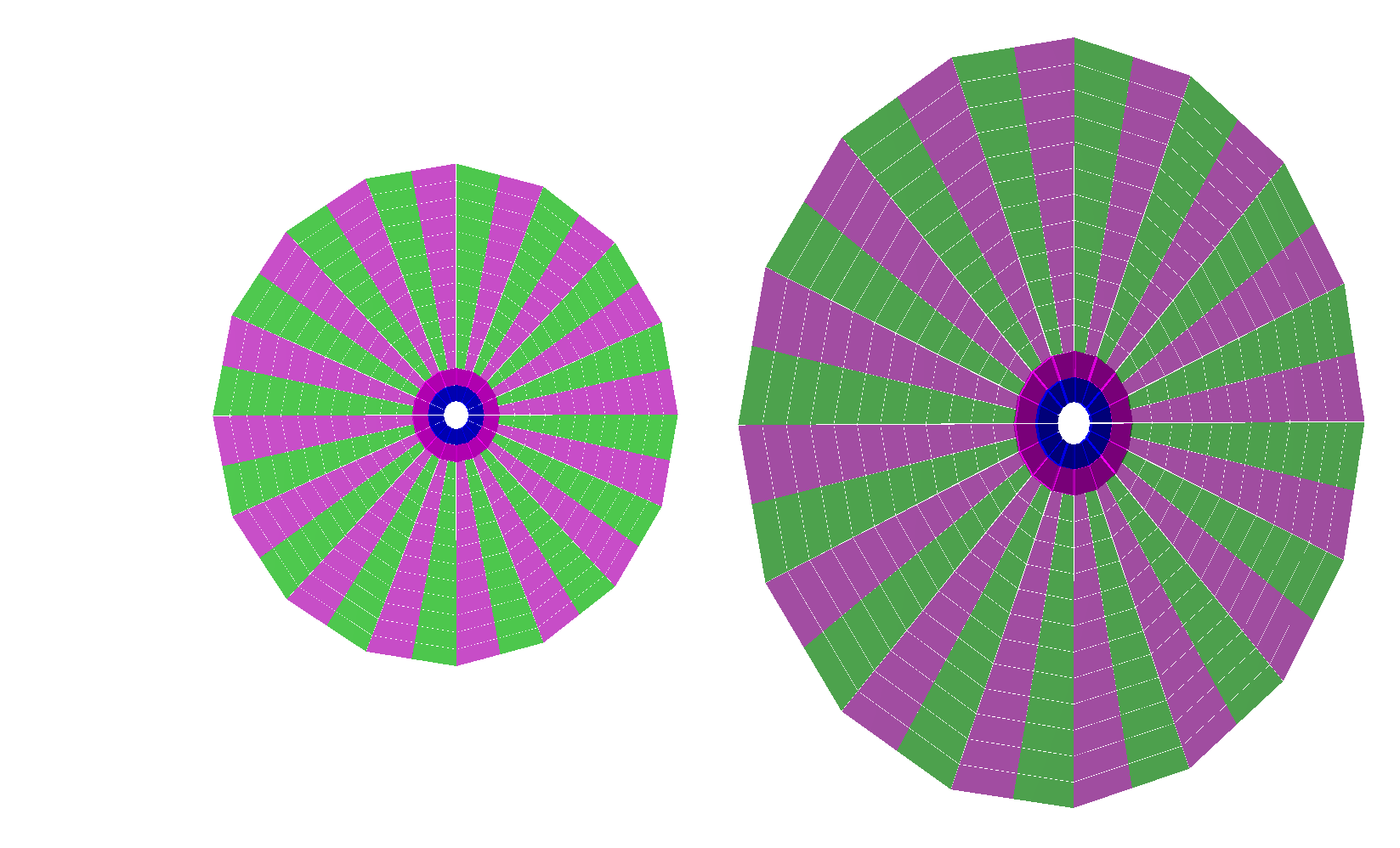}}
  \end{minipage}
  \caption{Sketch of the two BBC detectors.}
\label{BBC_SPD_2}
\end{figure}

Two Beam-Beam Counters (BBCs) are planned to be located in the front of the TOF system in the end-cups of the SPD setup \cite{SPD:2021hnm} (at a distance of $\pm\sim $1.7~m from the detector center, see Fig. \ref{BBC_SPD_2}. 
The detector should consist of two parts: the inner and the outer one, 
which are based on the use of the plastic scintillators.  The inner part of the BBC will
use highly segmented scintillators directly coupled with the silicon photomultiplier
(SiPM), while the BBC outer 
part will be produced from plastic scintillator tiles with the SiPM readout via a wavelengths shifter (WLS).  

The inner part covers the acceptance 30$\div$60 mrad and should be separated into 
4 layers consisting of 32 azimuthal sectors. The outer part covering the polar angles between 60 and 500 mrad will be divided into
13 concentric layers with 16 azimuthal sectors in each of them (Fig. \ref{BBC_SPD_2}).
The final granularity is the matter of further optimization for the entire energy range of collisions at the SPD.    

The main goals of the Beam-Beam Counters are: i) local polarimetry at the SPD based on  the measurements of the azimuthal asymmetries in the inclusive production of charged particles in the collisions of transversely polarized proton beams, ii) monitoring of the beam collisions, iii) participation in the precise determination of the collision time $t_0$ for the events, in which other detectors cannot be used for that purpose (for instance, in case of elastic scattering) and iv) event plane
determination for the global polarizations and flows studies in ion-ion collisions.

Another important goal of the BBCs is fast preselection of different types of events for monitoring purposes. 
The Monte Carlo simulation shows that in the $p$-$p$ collisions at $\sqrt{s}$=27~GeV   
at least one BBC should have a signal in 79\% of events (51\% of events has a signal in both BBCs). However, for hard processes, in 97\% of events only one  
BBC will be hit, while   hits  for both counters could be expected in 68\% of cases.
Therefore, the requirement of the BBC signals allows one to preselect hard processes.

\subsection{Outer part of the BBC: scintillation tiles}

\begin{figure}[!hbt]
  \begin{minipage}[ht]{0.49\linewidth}
    \center{\includegraphics[width=1\linewidth]{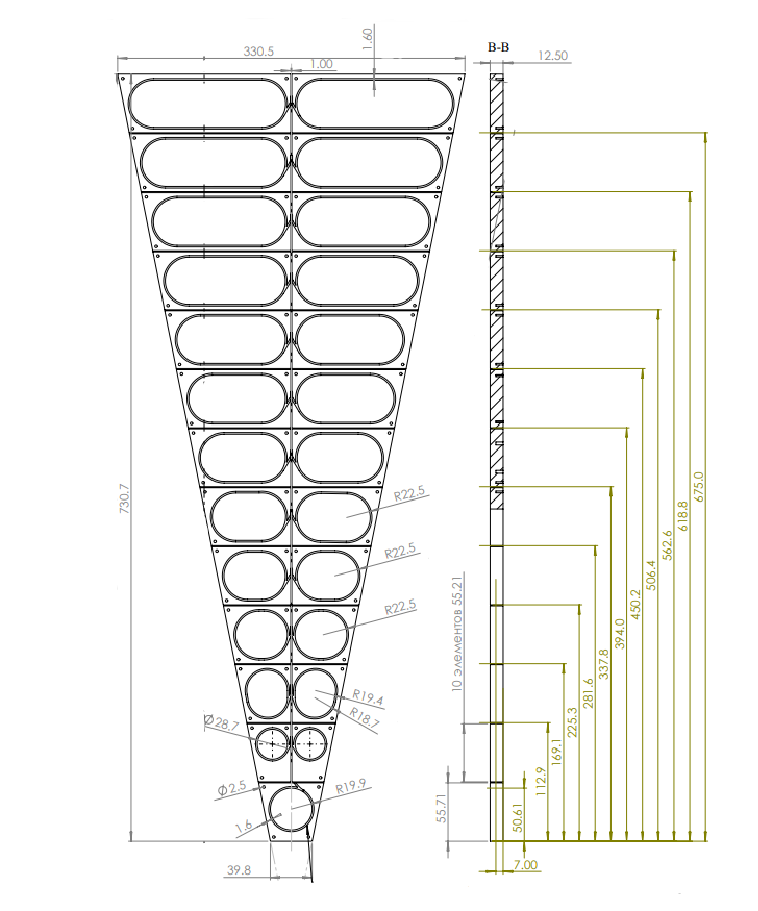} \\ (a)}
  \end{minipage} 
  \hfill
  \begin{minipage}[ht]{0.40\linewidth}
    \center{\includegraphics[width=1\linewidth]{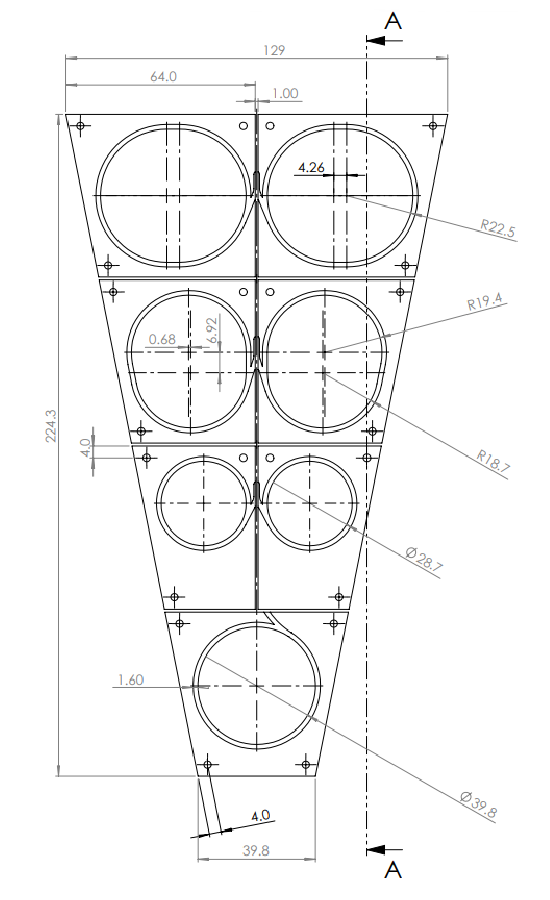} \\ (b)}
  \end{minipage}
  \caption{(a) Geometry of one complete sector of the BBC outer part. (b) Geometry of one sector of the BBC 7 tiles-prototype (the innermost part of the complete sector).}
\label{fig:ars1}
\end{figure}

The SPD BBC geometry has been updated in order to increase radial granularity and is different from the one presented in SPD conceptual design \cite{SPD:2021hnm}. A schematic view of the updated SPD Beam-Beam Counter sector, produced from the plastic scintillator tiles is presented in Fig.~\ref{fig:ars1}(a). The major reasons for the design change are to have the possibility of the event plane reconstruction for the studies of the   global polarizations and flows  in ion-ion collisions similarly to the STAR Event Plane Detector \cite{Adams:2019fpo} and a better selection of  elastic scattering events at low energies for the local polarimetry.

The outer part of the SPD BBC is designed to have 16 sectors with 25 tiles in each sector (Fig.~\ref{fig:ars1}(a)) in one wheel. Each tile is a separate signal source that should be read using a silicon photomultiplier (SiPM) connected to the WLS fiber. The total amount of the readout channels for two outer parts of the BBC is 800, what  leads us to use a CAEN FERS-5200 front-end readout system that was designed for large detector arrays \cite{FERS5200}.

\subsubsection{R\&D studies for BBC outer part}

The R\&D for the BBC outer part has been performed using a prototype consisting of 7 tiles shown in   
Fig.~\ref{fig:ars1}(b). The BBC prototype uses 4 first rows of the tiles (from bottom to top).  We will refer to the bottom one as the central, and the others -- by their position relative to the central tile: first, second, and third rows, correspondingly.  

The R\&D stage is dedicated to the selection of  a material for the detector. The detector prototype is made of polystyrene scintillator tiles with machined grooves for WLS fibers.
In order to force the fiber to collect as much light as it could, the tile's surface should be covered to reflect or scatter light from the coating layer. The gained light will be ''trapped'' inside until it is collected by the fiber or absorbed by absorption centers (defects, traps).
Light collection also depends on optical cement that serves for fiber fixation inside the tile. Also, optical cement is used to get rid of the air gap that spoils the light collection. As soon as the scintillator's light  passes through the interface between the two substances (tile to cement and cement to fiber) and spreads through the cement, the optical parameters of the cement should be such as to minimize any optical effects that make light collection poor. In addition, the WLS parameters should be matched to the scintillator for better light collection.

\begin{figure}[!hbt]
  \begin{minipage}[ht]{0.35\linewidth}
    \center{\includegraphics[width=1\linewidth]{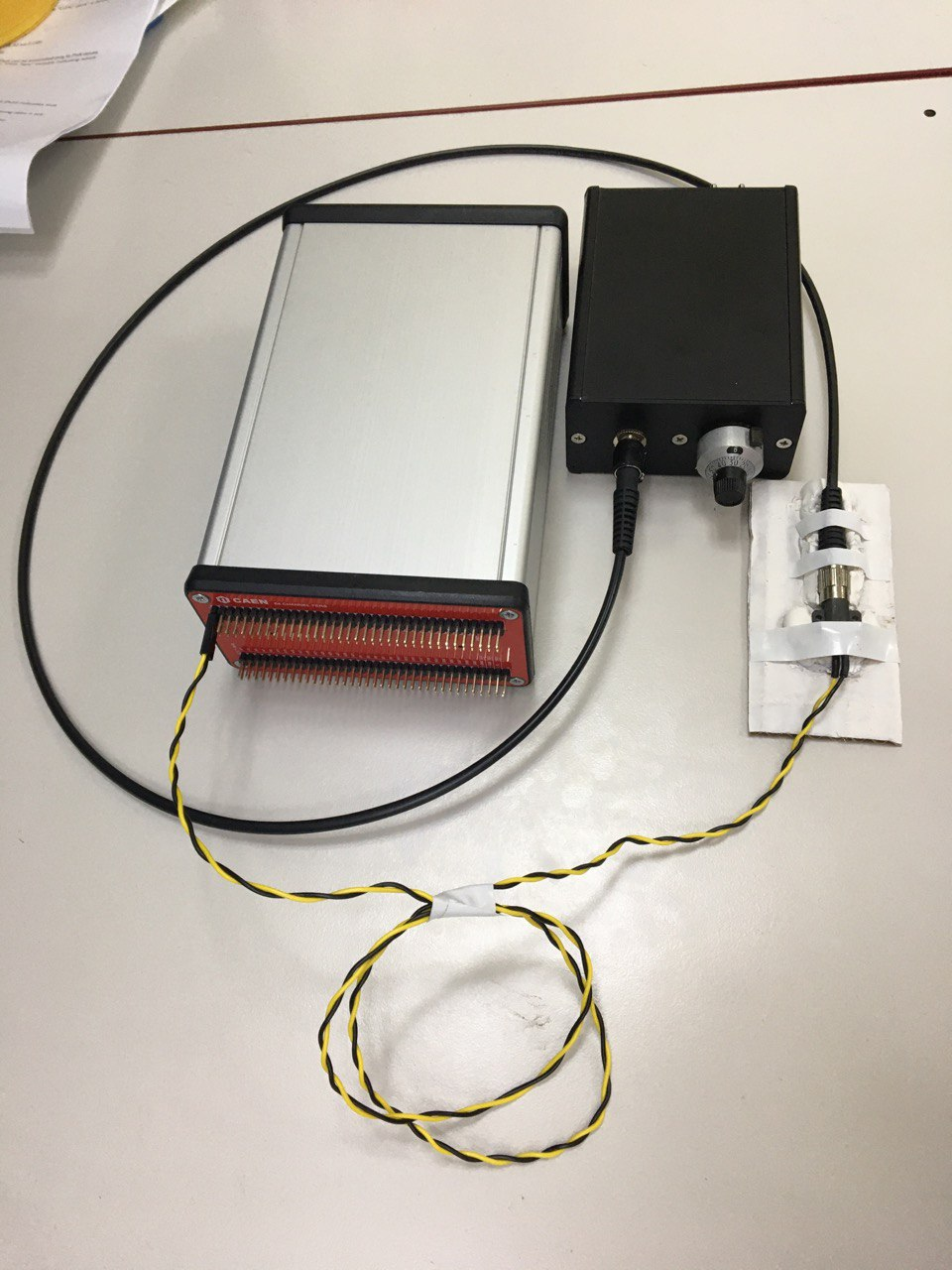} \\ (a)}
  \end{minipage} 
  \hfill
  \begin{minipage}[ht]{0.60\linewidth}
    \center{\includegraphics[width=1\linewidth]{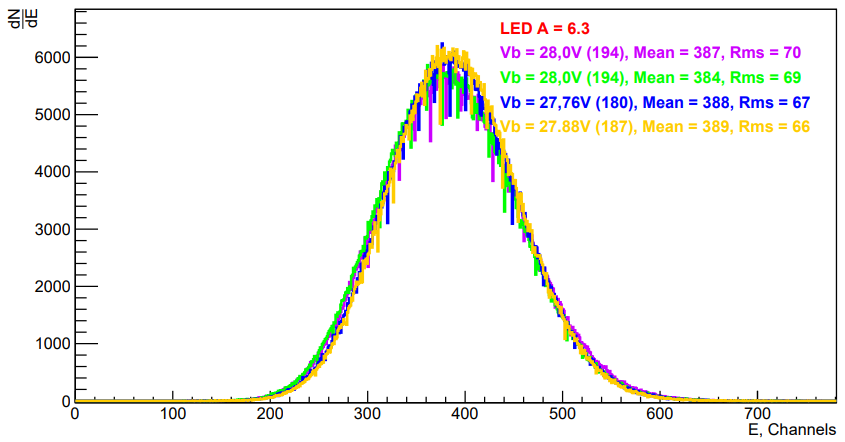} \\ (b)}
  \end{minipage}
  \caption{ (a) Calibration of SiPMs using the DT5202 board and  CAEN 
LED Driver SP5601.   
 (b) Results of the SiPM calibration using ''SPECTROSCOPY" mode of the DT5202 board.}
\label{fig:tish1}
\end{figure}

To start the studies, the SiPM calibration was performed. The CAEN SP5601 led driver was used as a stable signal source. 4 SensL $3\times3 \, mm^2$  or $1\times1 \, mm^2$ SiPMs were connected to the DT5202 board channel by channel. It is important to note that different DT5202 channels  have a different gain, so the signal difference from channel to channel  for various SiPMs is caused not only due to the SiPMs bias voltage difference, but also due to the electronic circuit of FERS-5200 distinctions. Thus, the amplitude of the signal depends on the SiPM and the channel it is connected to, so each SiPM was assigned to its own channel. The experimental stand for the SiPM cabibration and the results of the SiPM calibration using the ''SPECTROSCOPY'' mode of the DT5202 board \cite{FERS5200} are shown in Fig.\ref{fig:tish1}(a) and
Fig.\ref{fig:tish1}(b), respectively. One can see that all four channels have a similar effective gain after the calibration procedure.

\begin{figure}[!h]
  \begin{minipage}[ht]{.9\linewidth}
    \center{\includegraphics[width=1.0\linewidth]{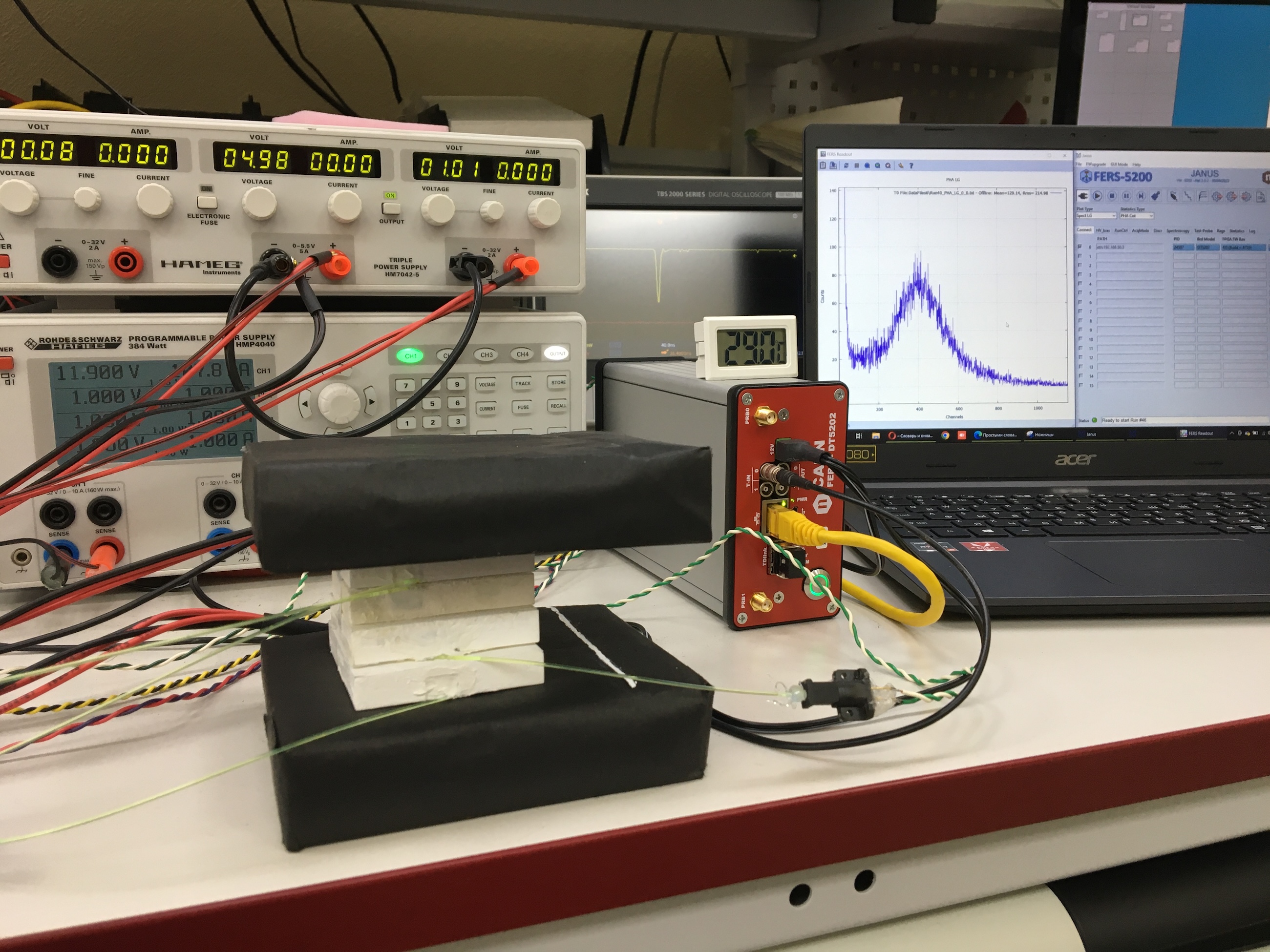}}
  \end{minipage}
  \caption{Picture of the cosmic rays stand for the tiles performance studies with FERS5200.}
\label{fig:tish2}
\end{figure}

The tests have been performed with the cosmic rays. A picture of the cosmic rays stand for the titles performance studies with FERS5200 is shown in Fig.\ref{fig:tish2}. Although the DT5202 board has an internal coincidence circuit (CC), an external trigger has been used for measurements with cosmic rays. Two scintillation counters made of a $100\times 100\times 10\, mm^3$ BC-404 scintillator plate viewed by the  Hamamatsu H10720-110 phototube were used for the trigger purpose.  
The trigger  time resolution was $\sim$650 ps. The trigger counters were located under and over 
the investigated tiles (up to 4).

The light signal from the WLS via the optical connector was fed in the SiPM connected with the DT5202 board.    
This board specifically designed for the readout of SiPMs is one of the main parts of FERS5200 \cite{FERS5200}. The DT5202 is an all-in-one readout system which integrates two Citiroc 1A chips for a total of 64 acquisition channels, an integrated power supply module for the SiPM bias (range of 20$\div$85 V), an FPGA, multiple communication interfaces (USB, Ethernet, and TDlink (optical)), and general purpose I/Os. The bias for the SiPM can be finely tuned channel by channel through an individual 8-bit DAC (2.5V or 4.5V range) in the Citiroc 1A. A temperature feedback circuit is also available to compensate the gain drift.

Each readout channel is composed of a Preamplifier, a Slow Shaper with a pulse height detector, and a Fast Shaper followed by a discriminator. Pulse height values from each Citiroc-1A are converted sequentially by a 13-bit multiplexed ADC to perform energy measurements. The 64 outputs of the discriminators, i.e. the channel self-triggers, are used in the FPGA both to feed the trigger logic (OR, Majority, etc.) and to acquire the timing information (Time Stamp and Time over Threshold) to be used in combination with the amplitude information (Pulse Height Analysis, PHA) or even as an alternative.   
Citiroc-1A outputs the 32-channel triggers with a high timing resolution (better than 100 ps RMS), even though it does not have an internal TAC/TDC to acquire the timing measurement. However, the FPGA of the A5202 is programmed for this purpose and a low resolution TDC (0.5 ns) was implemented to calculate the time distance ($\Delta T$) between a reference signal and the input pulses.

The Janus software on Windows and Linux provided by CAEN \cite{FERS5200} allows one to completely manage the DT5202 module and the data acquisition. The acquired data   can be analyzed either by the Janus software or by software using ROOT libraries.

\begin{figure}[!hbt]
  \begin{minipage}[ht]{0.49\linewidth}
    \center{\includegraphics[width=1\linewidth]{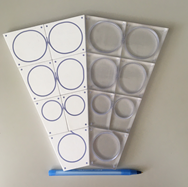} \\ (a)}
  \end{minipage} 
  \hfill
  \begin{minipage}[ht]{0.42\linewidth}
    \center{\includegraphics[width=1\linewidth]{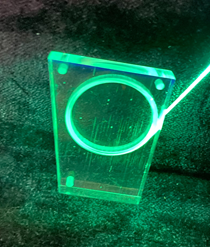} \\ (b)}
  \end{minipage}
  \caption{ (a) "Matted" and polished tiles for two sectors of the BBC 7~tiles-prototype. 
 (b) Polished tile with a WLS.}
\label{fig:ars2}
\end{figure}

The tiles were produced by the Uniplast Enterprise. The scintillator content was 98-98.5\% 
of polystyrene Styrolution 124N, 1.5-2.0\% of p-Terphenyl 
(CAS 92-94-4) and 0.01-0.04\% of POPOP (CAS 1806-34-4). 
A tile covered with white acrylic paint (''matted'') and a polished tile, double covered with a unique non-woven material made from high-density polyethylene continuous filaments (Tyvek) were used for the tests at the first stage of R\&D.
The   "matted" and polished tiles for two sectors of the BBC 7~tiles prototype are shown in Fig.\ref{fig:ars2}(a). The polished tile with a green WLS is demonstrated in Fig.\ref{fig:ars2}(b).

 \begin{figure}[!h]
  \begin{minipage}[ht]{.9\linewidth}
    \center{\includegraphics[width=0.6\linewidth]{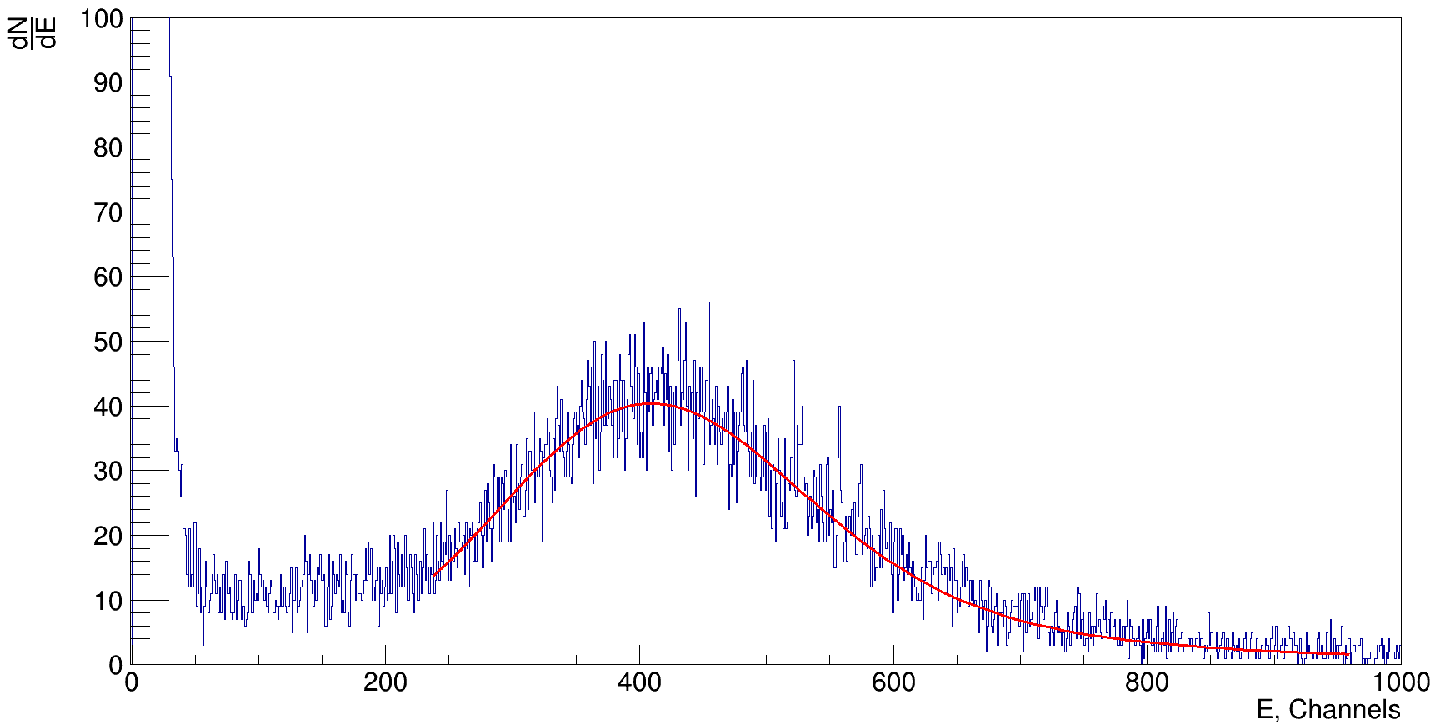}}
  \end{minipage}
  \caption{SiPM amplitude distribution for one tile with convolution of Gaussian and Landau used as the fitting function.}
\label{fig:ars3}
\end{figure} 

The  SiPM amplitude distribution from DT5202 for one tile is shown in Fig.\ref{fig:ars3}.
The line is the results of the fit by the convolution of Gaussian and Landau (the so-called ''langaus'')..

\begin{figure}[!hbt]
  \begin{minipage}[ht]{0.49\linewidth}
    \center{\includegraphics[width=1\linewidth]{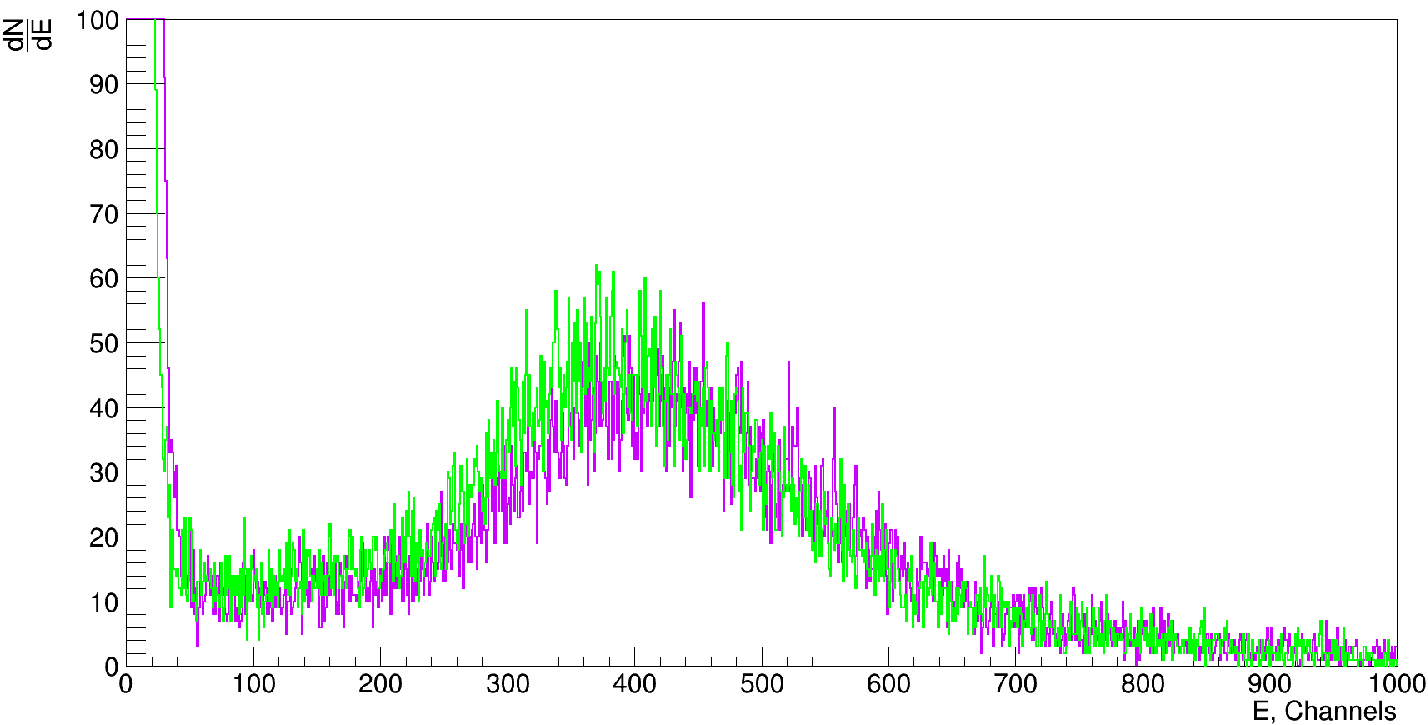} \\ (a)}
  \end{minipage} 
  \hfill
  \begin{minipage}[ht]{0.49\linewidth}
    \center{\includegraphics[width=1\linewidth]{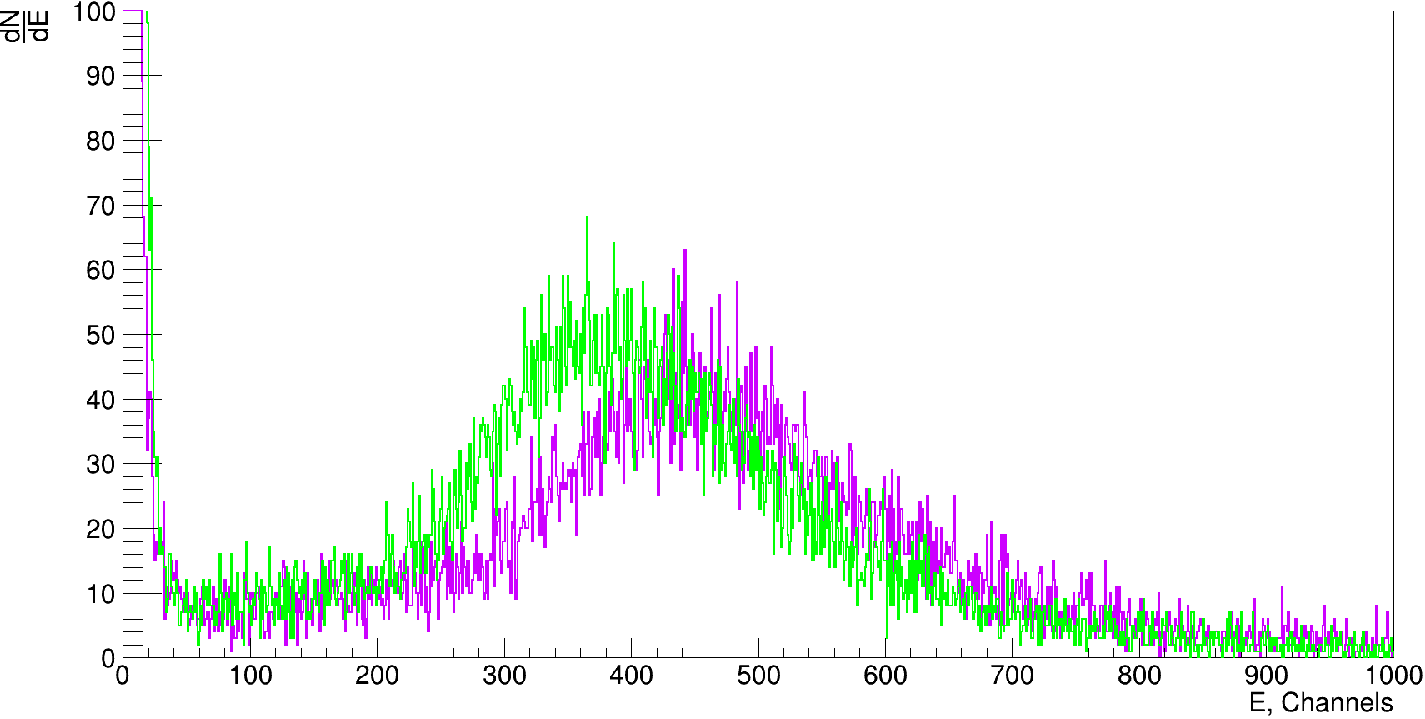} \\ (b)}
  \end{minipage}
  \caption{  Comparison of the matted (purple) and covered with Tyvek (green) tiles  for  row 1  (a)
and row 3 (b).}
\label{fig:ars4}
\end{figure}

The row 1 and 3 geometry tiles, the one covered with Tyvek and the matted one, filled with CKTN MED mark E optical cement and Saint-Gobain Crystals BCF92S WLS fiber were used  for comparison of the
scintillator surface reflection. The results are presented in Fig.\,\ref{fig:ars4}.  
The result on the mean, width and integral value of the fitting area were  obtained 
using the ''langaus'' fit \cite{Arseny2023}. Due to a higher peak position (from 7\% and up to 15\% difference) and technological complexity of mass production for Tyvek covers, the option with the matted one is more appropriate. 
 
\begin{figure}[!h]
  \begin{minipage}[ht]{.9\linewidth}
    \center{\includegraphics[width=0.6\linewidth]{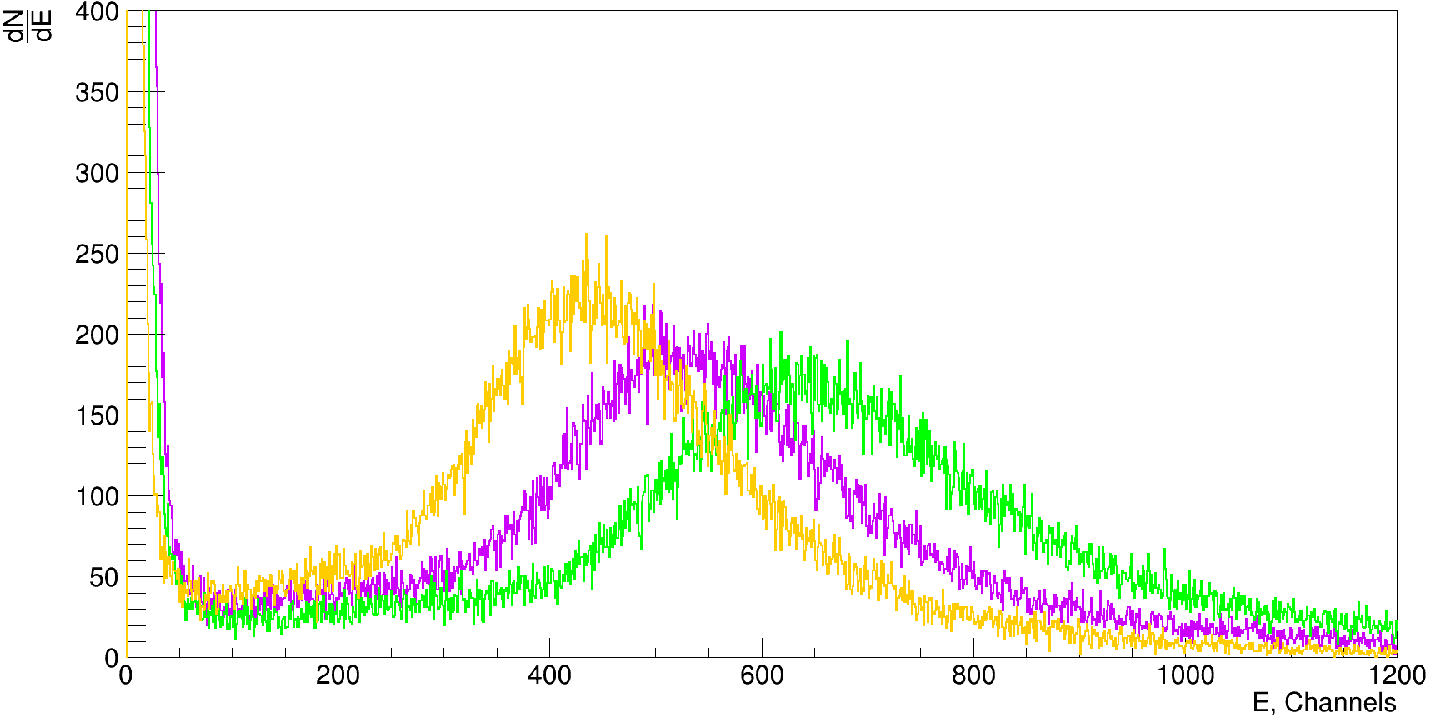}}
  \end{minipage}
  \caption{ Comparison of the Saint-Gobain Crystals BCF91AS (purple), Saint-Gobain Crystals BCF92S (yellow), and Kuraray Y-11 (green) WLS. }
\label{fig:ars5}
\end{figure}

The Saint-Gobain Crystals BCF91AS, BCF92S, and Kuraray Y-11 WLS fibers were compared. All samples with BCF91AS, BCF92S, and Y-11 were made using CKTN MED. The study was performed with row 3 tiles. The distributions are presented in Fig.\,\ref{fig:ars5}.  Due to the fact that the Kuraray Y-11 fiber collects more light, the choice of Y-11 fibers looks more appropriate.

\begin{figure}[!hbt]
  \begin{minipage}[ht]{0.49\linewidth}
    \center{\includegraphics[width=1\linewidth]{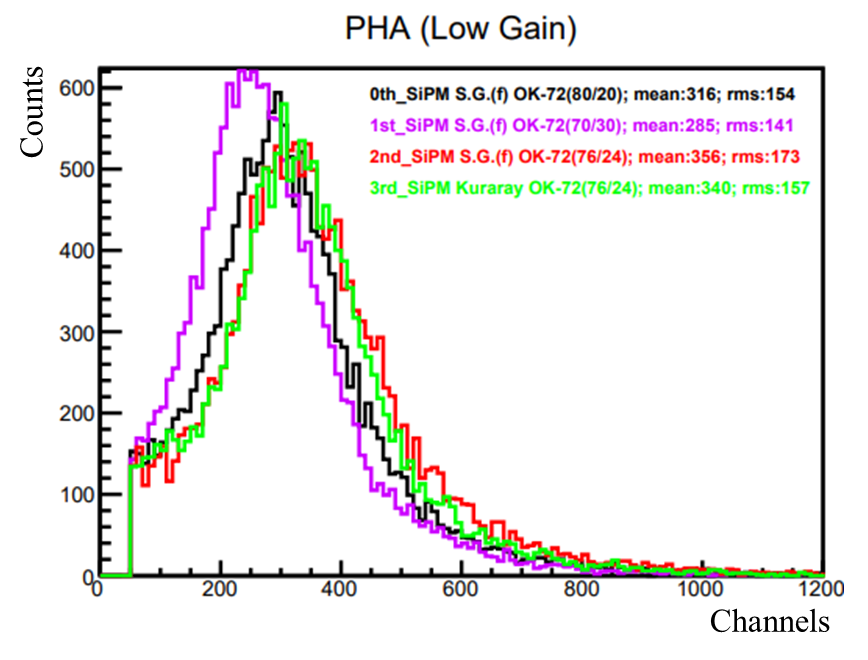} \\ (a)}
  \end{minipage} 
  \hfill
  \begin{minipage}[ht]{0.49\linewidth}
    \center{\includegraphics[width=1\linewidth]{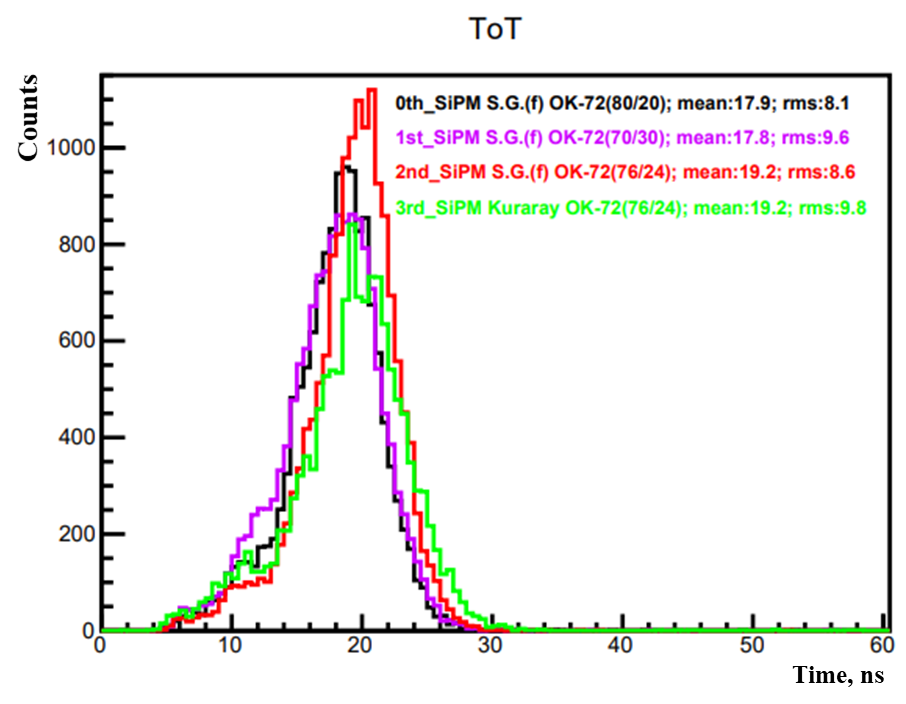} \\ (b)}
  \end{minipage}
  \caption{Comparison of row 3 tiles with different a $A/B$ ratio of OK-72 cement and a different WLS 
for signal amplitude (a), and for ToT (b).}
\label{fig:tish3}
\end{figure}

The use of two optical cements: CKTN MED mark E and OK-72, gave similar results  \cite{Arseny2023}. However, OK-72 is easier to apply due to its low viscosity and extended curing time (in terms of mass production, compared to CKTN mark E). Therefore, further studies were performed 
with the OK-72 cement.

Three different compositions of $A$ to $B$ components for the OK-72 cement were studied for the row 3 tiles
with Saint-Gobain Crystals BCF92S WLS.
The results for an $A/B$ ratio of 80/20, 70/30, and 76/24 are presented in Fig.\ref{fig:tish3}(a) by 
black, purple and red colors, respectively. The results for the tile with the Kuraray Y11 WLS and 
A/B ratio 76/24 is shown by green. One can see that the optimal A/B ratio for the OK-72 cement is 76/24.
On the other hand, the use of WLS Saint-Gobain Crystals BCF92S and Kuraray Y11 provides a very similar result. 
 
DT5202 board can provide only a time-over-threshold (ToT) signal in the free streaming mode.
The results for row 3 tiles with a different $A/B$ ratio of OK-72 cement and a different WLS  are 
presented in Fig.\ref{fig:tish3}(b). Still one can see the difference observed in Fig.\ref{fig:tish3}(a).

Therefore, the R\&D performed gave a preliminary result on the materials for the outer part of the BBC detector:
"matted" scintillator, OK-72 optical cement with an  $A/B$ ratio of 76/24, green WLS Kuraray Y11 
(which can be replaced by the Saint-Gobain Crystals BCF92S).
The final choice will be done after manufacturing and testing  the BBC outer part prototype,
the sector of which is shown in Fig.\ref{fig:ars1}(b).

\subsubsection{R\&D for the BBC inner part}

The BBC inner part cannot be produced using the above-mentioned technology with the WLS due to a smaller size of the 
scintillators. Therefore,  the SiPMs should be attached directly to the scintillator. 
There are two different options for the inner part of the BBC based on the use of high granularity 
scintillators array.

\begin{figure}[!hbt]
  \begin{minipage}[ht]{0.49\linewidth}
    \center{\includegraphics[width=1\linewidth]{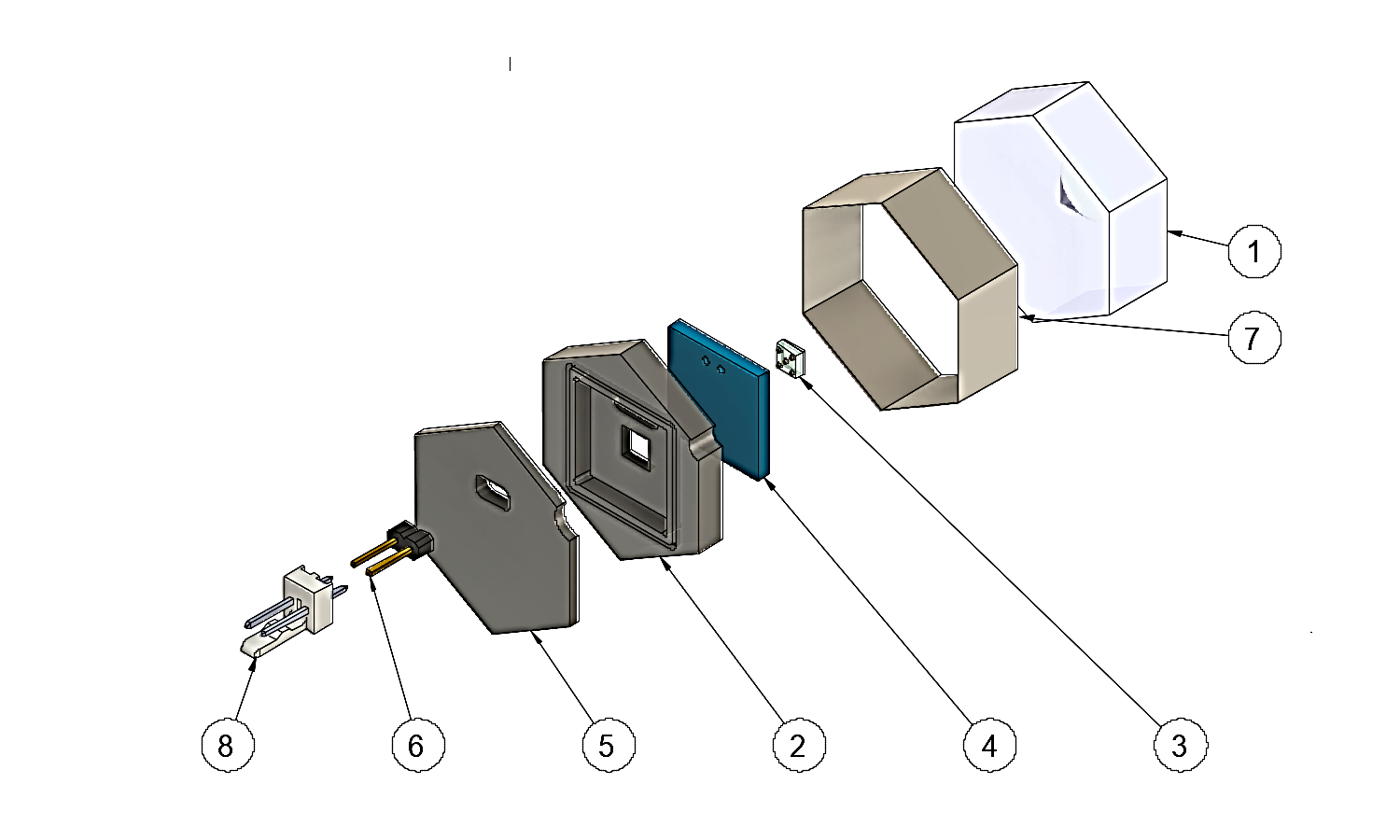} \\ (a)}
  \end{minipage} 
  \hfill
  \begin{minipage}[ht]{0.49\linewidth}
    \center{\includegraphics[width=1\linewidth]{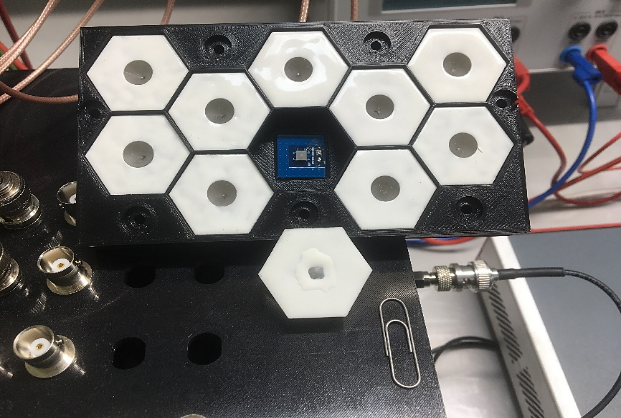} \\ (b)}
  \end{minipage}
  \caption{ (a) Structure of the hexagonal cell: 1-- scintillator hexagon, 2-- housing of the SiPM cavity,
3-- SiPM, 4-- SiPM PCB and support, 5-- SiPM cavity cover, 6-- connector, 7-- mylar or "matted" coating, 8-- male connector. (b) Array of hexagonal ''matted'' cells.   
(b).}
\label{fig:chile1}
\end{figure}

The first option is based on the use of hexagonal ''matted'' scintillators with a  $3\times3 \, mm^2$ SiPM attached to the center of each cell. The structure of one cell 
and  the array of 10 "matted" cells are shown in Fig.\ref{fig:chile1}(a) and Fig.\ref{fig:chile1}(b),
respectively. 

\begin{figure}[!hbt]
  \begin{minipage}[ht]{0.49\linewidth}
    \center{\includegraphics[width=1\linewidth]{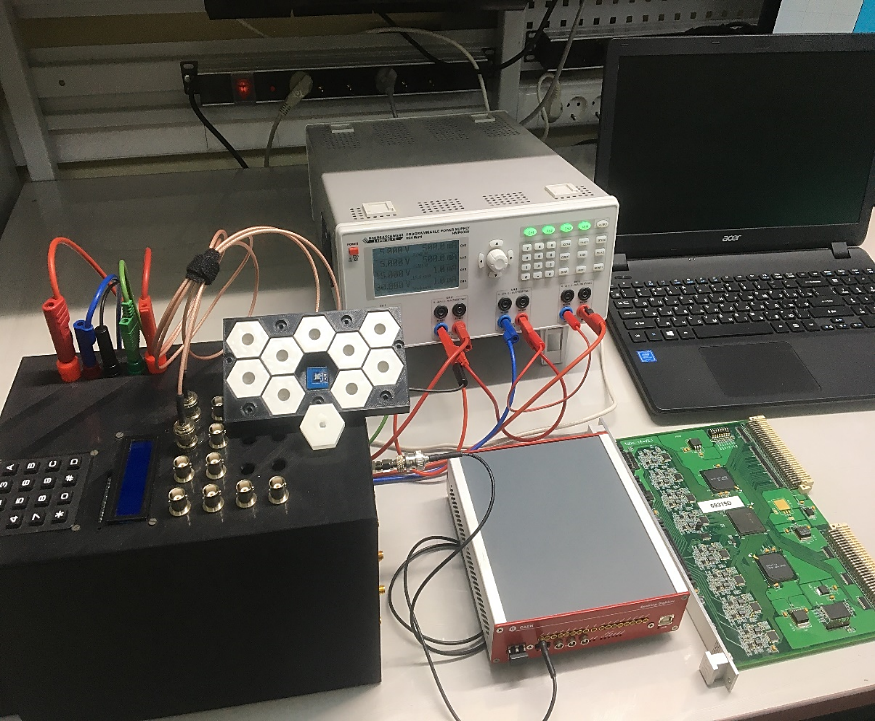} \\ (a)}
  \end{minipage} 
  \hfill
  \begin{minipage}[ht]{0.49\linewidth}
    \center{\includegraphics[width=1\linewidth]{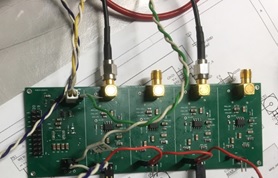} \\ (b)}
  \end{minipage}
  \caption{ (a) Picture of the cosmic rays stand for the hexagonal cells: array of hexagonal "matted" cells with Hamamatsu
SiPM S14160-3050HS, FEE developed at CTEPP-UNAB and  CAEN digitizer DT5742.
(b) The FEE developed at CTEPP-UNAB.}
\label{fig:chile2}
\end{figure}
 
The tests of the hexagonal cells with cosmics were performed using a Hamamatsu 
SiPM S14160-3050HS, FEE developed at CTEPP-UNAB (Chile), and a CAEN digitizer DT5742 
(see Fig.\ref{fig:chile2}(a)). The FEE used is shown   Fig.\ref{fig:chile2}(b). The time resolution achieved was $\sim$400 ps. Further R\&D is required to improve the time resolution.

The second option is based on  the detector proposed for stage 2 of the MPD operation at  NICA
\cite{Torres:2021jgh}.  This detector  consists of
two arrays, each having 80 plastic scintillator cells located 
 symmetrically at the opposite sides of the MPD
interaction point. Each scintillator cell is viewed by 2-4 SiPMs.

\begin{figure}[!h]
  \begin{minipage}[ht]{.9\linewidth}
    \center{\includegraphics[width=1.0\linewidth]{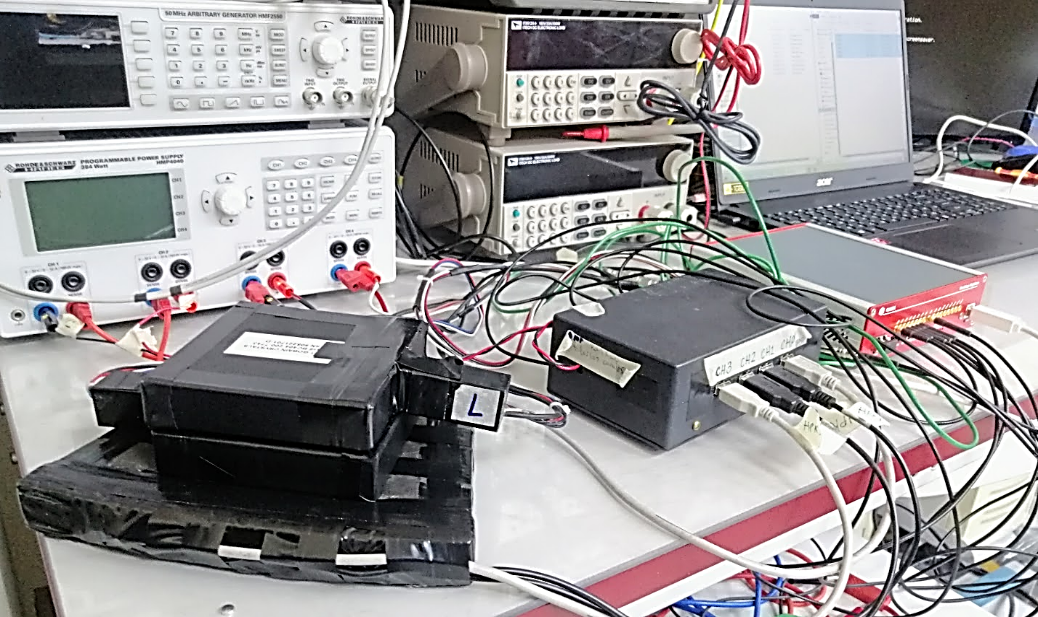}}
  \end{minipage}
  \caption{Picture of the cosmic rays stand for the BeBe scintillator cell \cite{Torres:2021jgh}  studies.}
\label{fig:mexico}
\end{figure}

The picture of the cosmic rays stand at JINR for the BeBe scintillator cell \cite{Torres:2021jgh} studies
is shown in Fig.\ref{fig:mexico}. The scintillator was viewed by $3\times3 \, mm^2$
4 Hamamatsu SiPMs S13360-3050CS attached to the opposite edges of the cell. A CAEN digitizer DT5742 has been used for the data analysis.
The time resolution achieved was $\sim$0.8 ns. The BeBe cell size was quite large compare to the cell size assumed for the inner part of the BBC. Therefore, further R\&D is required for the SPD case.

The inner part of BBC requires time measurement. Several FEE options can be used.
The first option is FEE developed at CTEPP-UNAB (Chile).  The second option is FEE developed at ITEP NRC KI \cite{Tishevskiy:2020sna} and based on the Time-over-Threshold   technique. This technique is a well-known method that allows us to measure the energy deposited in the material by reconstructing the given property of the output current pulse – the total charge collected, the pulse amplitude, etc. The ToT method converts the signal pulse height into a digital value in the early stage of the FEE, which greatly simplifies the system in comparison to analog detectors with serial readout through ADCs. The measurement of the ToT is composed of two measurements of time for the signal going above (leading) and returning below (trailing) the given threshold. The first version of the prototype includes a power supply and electronics  made on a separate PCB  used for a single SiPM. The bias voltage consists of a constant (38/52/67 V) and a part adjustable for each channel (in the range from 0 V to +10 V). 
It is possible to connect up to 40 PCBs simultaneously. The amplifiers used for that do not change the leading edge of the signal. This allows us to get a time stamp of the event. Afterwards, the signal is integrated and transmitted to the comparator.
 
The third FEE option for the BBC detector is the electronics produced by the ITEP group 
for the DANSS experiment \cite{Alekseev:2016llm}.  It has a multi-channel platform created from several printed circuit boards. The first board provides power and communication with the PC via RS-232, and is connected via IDC-34 to a common board on which 15 electronics boards are installed. Power is supplied in a wider range, and consists of a constant (10-65 V) and a bias voltage adjustable for each channel ($\pm$10 V). Each 1.7$\times$1.9 cm$^2$ board contains an offset voltage output for SiPM and an input for the signal. The signal passes through the amplifier and is then output via the IDC-34 connector to the reading electronics.

When comparing the front-end electronics with the ToT function and the DANSS experiment, the time resolution after time-walk correction was approximately 130 ps and 225 ps, respectively \cite{Tishevsky:2022ulx}.

FERS5200 \cite{FERS5200} can be used at the first stage of the BBC inner part operation.
Unfortunately, the use of the DT5202 board provides the time resolution worse than 0.5 ns \cite{Tish2023}.  In this respect, the use of the DT5203 boards based on the use of picoTDC is more attractive. However, it requires to adopt one of the  above-mentioned FEE options.
In future, CAEN is going to develop a new board (DT5204) with a fast amplifier and a picoTDC dizitizer 
for the  SiPM, which can be an option for the FEE upgrade. 

\subsection{BBC DAQ}  

The FEE and dizitizing will be based on the use of FERS5200 \cite{FERS5200}, which   is an extendable system: the same FERS unit can be used as a stand-alone system for prototyping and validation, and then added to a greater network tree for the readout of large arrays. This strategy eliminates waste and maximizes cost efficiency.

\begin{figure}[!h]
  \begin{minipage}[ht]{.9\linewidth}
    \center{\includegraphics[width=0.6\linewidth]{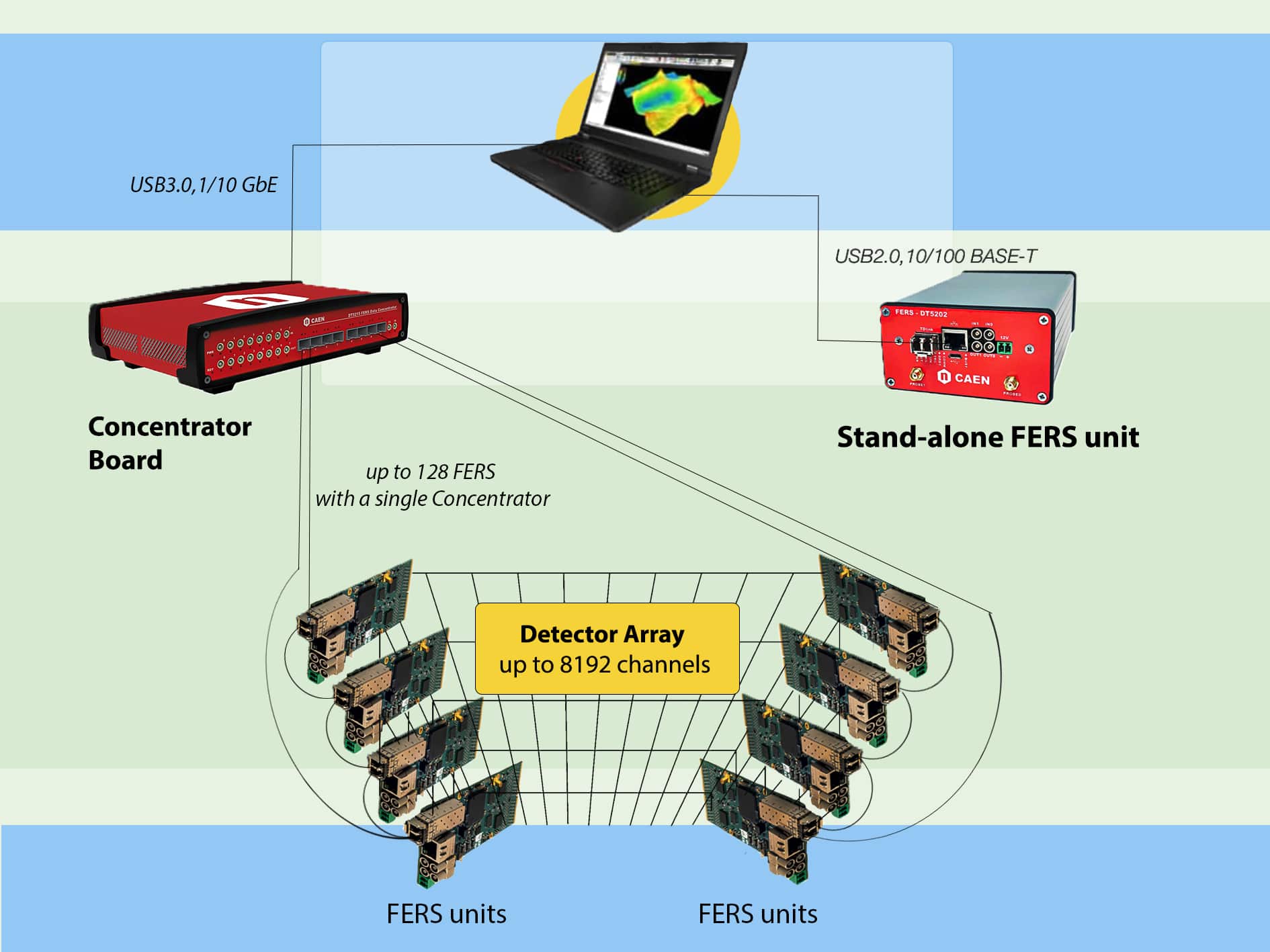}}
  \end{minipage}
  \caption{FERS5200 configuration with DT5215 board.}
\label{fig:fers5200}
\end{figure}

The core of the scalability of the FERS-5200 is the optical TDlink, which manages simultaneously the data stream, unit synchronization, and slow control. The concentrator board DT5215 hosts 8 optical links, each capable of sustaining up to 16 FERS units in the daisy chain. In the case of A5202/A5203, this means 8192 channels managed by a single Concentrator board (see Fig.\ref{fig:fers5200}). The  
multiple concentrator boards can be synchronized to further extend the total number of channels. 
The Concentrator Board manages the data readout from the network of FERS units and the event data building according to the time stamp and/or trigger ID of the event fragments acquired by each unit. Sorted and merged data packets are then stored in the local memory and finally sent to the host computers. 

The outer part of each BBC is covered  by 7 DT5202(A5202) boards, while its inner part 
requires the use of 2-4  DT5202(A5202) or  DT5203(A5203) boards depending on the 
FEE option. 

\newpage

\section{Cost estimate}

The cost estimate of both inner and  outer parts of the BBC is given in Table~\ref{table:BBC}.
The FEE and digitizers costs are estimated based on the option of the FERS5200 use.

\begin{table}[hbt]
\centering
\caption{The cost estimation of the BBC manufacture and assembling.
\label{table:BBC}}  
 
\begin{tabular}{ccc}
\hline\hline
 Item &  number & cost, k\$ \\
\hline
Scintillator & 1200  &  30 \\
\hline
SiPMs  &  1400   &  80 \\
\hline 
WLS (SG or Kuraray) &    &  40 \\
\hline
Clean fibers &   &  40 \\
\hline
Composite &   2  &  20   \\ 
frames & &\\
\hline 
FERS5200   & 25  &   200 \\
\hline
Cabling,   & & 20\\
connectors &  &   \\
\hline
LED boards & 32  & 10 \\
\hline
assembling &  & 30 \\   
\hline 
Total     &  &   470$^*$ \\     
\hline\hline         
\end{tabular}
\end{table}
$^*$ - without the cost of manpower for manufacture and assembling

\chapter{Silicon Vertex Detector}
The SPD Silicon Vertex Detector (SVD) is a silicon-based  part of the spectrometer, responsible for  the precise determination of the primary interaction point and  measurement of the secondary vertices from the decays of short-lived particles (first of all, $D$-mesons). The Vertex Detector is divided into the barrel and two end-cap parts. A version of the vertex detector, based on Monolithic Active Pixel Sensors (MAPS), is discussed as the main one. Nevertheless, we keep in mind  the  possibility of using a vertex detector, based on the Double-Sided Silicon Detectors (DSSD).

\section{MAPS-based vertex detector}
To date, when creating tracking systems, the use of pixel sensors as a basic element of vertex detectors is the most optimal experimental technique in high-energy physics. 
Besides demonstrating great registration efficiency  for charged particles, produced in collisions of beams in modern collider experiments, the modern pixel detectors are resistant to hard radiation, allow to register short-lived particles  with a good spatial resolution, and can detect rare processes and decays \cite{Rossi:2006buh}. Current vertex detector systems use both hybrid pixel detectors and monolithic pixel detectors. Hybrid pixel detectors consist of a separately made sensor matrix and an electronic panel (chip). In the design of hybrid pixel detectors, the chip contact pads are made with the same dimensions as the implemented pixels, that is, an exact match is created between the size of the pixel and the size of the chip contact pad. Thus, a very important characteristic of hybrid detectors is pixel density. In addition, the electronic chip must be very close (10$\div$20 $\mu$m) to the touch panel. 

In monolithic pixel detectors, the electronics and the touch panel are produced in the same process, on the same chip. This has the following advantages, compared to hybrid detectors: there is no need to connect the chip and the sensor array (in hybrid technology, to connect these two parts is quite time-consuming, because of the high pixel density), and the capacity of each pixel is reduced, thus substantially reducing the noise level of the detector. Therefore, this technology of manufacturing pixel detectors is more reliable and less expensive \cite{Rossi:2006buh}. Taking into account the main parameters of the beam collisions at NICA, we can use the existing solutions, implemented in the framework of the ALICE experiment Internal Track System upgrade \cite{Abelevetal:2014dna}. The characteristics of the pixel sensors, developed by the ALICE Collaboration \cite{Abelevetal:2014dna,Mager:2016yvj} meet the requirements for the Silicon Vertex Detector in the SPD setup. It is assumed that this module will consist of 4 cylindrical layers of silicon detectors with a total square of 5.4 m$^2$, based on MAPS technology for high-precision reconstruction of the primary and secondary vertices.

We are also closely following the progress of the MPD-ITS collaboration in developing the technology and building the MAPS-based vertex detector for the MPD facility at NICA \cite{Murin:2021szk,MPDITS:2022yvh,MPDITS:2022ozp,Kondratyev:2022irh}. These developments can also be used in the design of the SPD vertex detector.

\subsection{MAPS technology}
One of the key tasks of the SPD experiment is to study the production of particles, containing heavy quarks ($D$-mesons, $\Lambda_c$, etc.), with small transverse momenta. The main role in these studies will be played by the Silicon Vertex Detector, consisting of four cylindrical layers of pixel detectors, mounted on the ultralight supporting carbon fiber structures with a length of 1500 mm. The physics program of future research imposes strict requirements on the characteristics and parameters of the detectors used. The key factors are to minimize the distance between the collision point and the first detector layer and  reduce the amount of matter along particle tracks. At the same time, the increase in the frequency of beam collisions at the collider calls  for fast detectors with high granularity. Based on the available experience of the ALICE Collaboration \cite{Abelevetal:2014dna,Mager:2016yvj}, and after comprehensive experimental studies of the MAPS characteristics at SPbSU \cite{Zherebchevsky:2016usy,Zherebchevsky:2021fbi}, the following parameters of these detectors are proposed (see Table \ref{maps_tab}).  
\begin{table}[htp]
\caption{Desired parameters of the MAPS-based SVD.}
\begin{center}
\begin{tabular}{|l|c|}
\hline
Parameter & Value \\
\hline
Silicon thickness, $\mu$m & 50 \\
Sensor size, mm$^2$ & 15$\times$30 \\
Detection efficiency, \% & $>$99 \\
Spatial resolution, $\mu$m & 5 \\
Average power consumption, mW/cm$^2$ & 40\\
Operating temperature, $^{\circ}$C & 20$\div$30\\
Pixel noise rate, triggers / (pixel$\times$event) & $<10^{-6}$ \\
Total Ionizing Dose (TID) radiation hardness (safety factor 10), krad & 100 \\
 Non-Ionizing Energy Loss (NIEL) radiation hardness, 1-MeV n$_{eq}$ / cm$^{2}$ & 10$^{12}$\\
 \hline
\end{tabular}
\end{center}
\label{maps_tab}
\end{table}%

\begin{figure}[!t]
    \center{\includegraphics[width=0.9\linewidth]{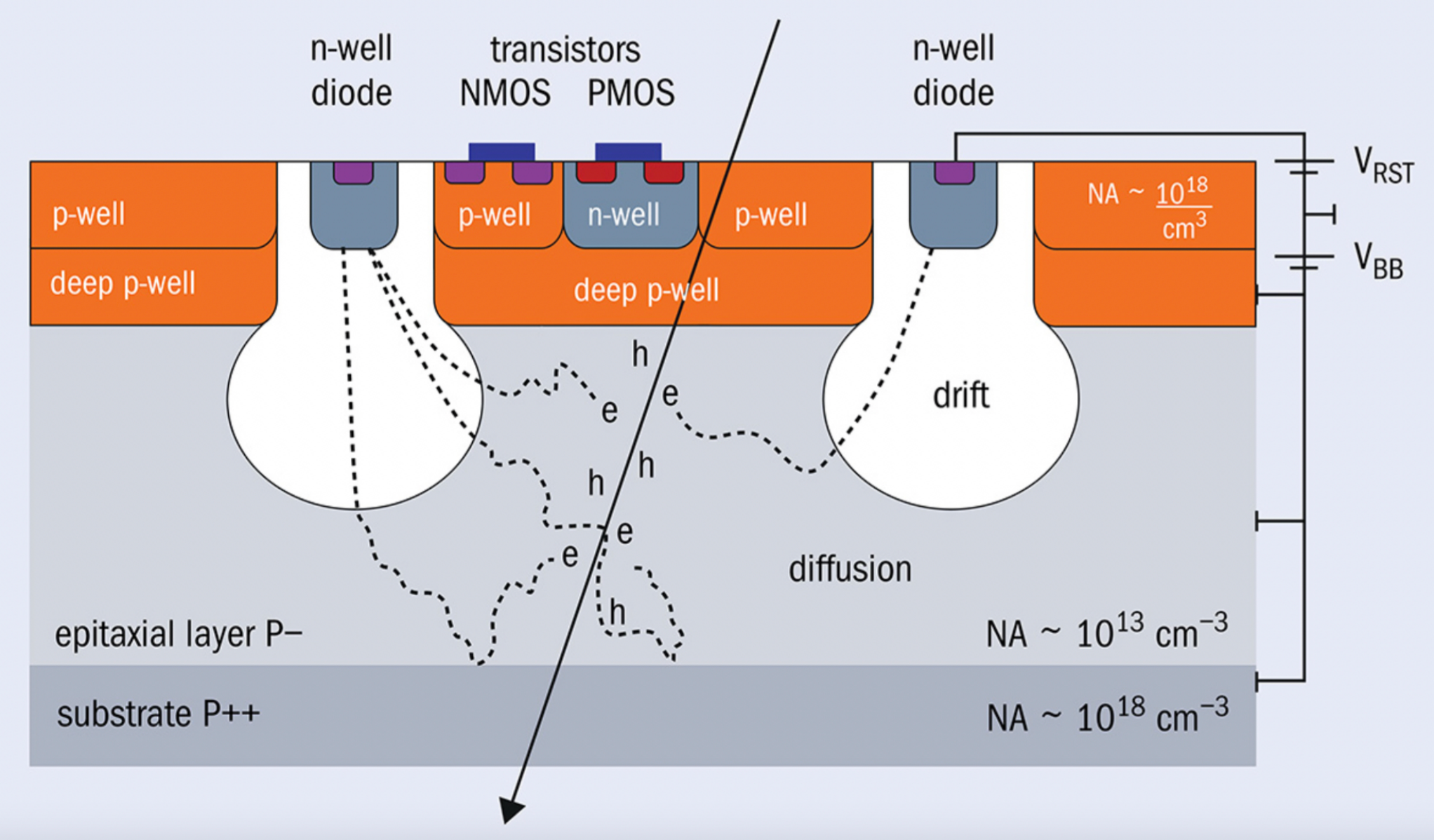}}
  \caption{Scheme of a pixel sensor \cite{CERNcurier}. }
  \label{fig:vd_5maps}  
\end{figure}

To achieve these parameters, it is proposed to use 180 nm CMOS technology.  The main advantages of this technology are good noise conditions, low power consumption, and the radiation hardness of the pixel sensors used. The MAPS sensor consists of a high resistivity epitaxial layer ($\sim$1 k$\Ohm \times$cm) grown on a low resistivity substrate, a matrix of collecting diodes (pixels), and the front-end electronics. The scheme of charge collection by the pixel cell is shown in Fig. \ref{fig:vd_5maps}.
The epitaxial layer is the active volume of the detector (can vary from 20 to 40 $\mu$m). The charge-collecting diode is implemented as a transition between the epitaxial layer and the $n$-type pocket ($n$-well). Due to the small thickness of the epitaxial layer, the collected charge (generated by the minimum ionizing particle) is relatively small for this type of detector: $\sim1000\div1600$ e$^-$. The size of the charge-depleted zone can be changed by applying a VRST bias voltage. To further increase the width of the depleted zone, a reverse bias voltage ($V_{BB}$) can be applied to the detector substrate (Fig. \ref{fig:vd_5maps}). Charges are collected by both drift (in the depleted zone) and diffusion (in the rest of the volume). Since diffusion plays a significant role in the charge collection process, the charge from a single particle (even for tracks perpendicular to the pixel matrix plane) is collected by several neighboring diodes (pixels), forming what is called a cluster. The ratio of the percentage of charge collection by drift and diffusion is determined by the size of the depleted zone, so that the $V_{BB}$ can also influence the size of the cluster. The total detector bias voltage applied to the collecting diode is formed from voltage $V_{RST}$, which fixes the operating point of the input transistor, and the reverse bias voltage $V_{BB}$ applied to the substrate. This greatly increases the depletion zone around the charge-collecting diode and lowers the input pixel capacitance, resulting in more efficient charge collection, reduced detector noise, and improved radiation resistance. To improve detector efficiency, a deep $p$-type pocket ($p$-well) is placed in the front-end electronics area (Fig. \ref{fig:vd_5maps}).  Right above it, there are elements of the detector's electronic circuits, implemented using PMOS structures with the pocket of $n$-well type. Thus, the $p$-type pocket shields the pixel electronics, preventing electrons from collecting on the $n$-well and, thereby, generating additional noise \cite{Abelevetal:2014dna,Mager:2016yvj,Zherebchevsky:2016usy,Reidt:2016bvz}.

As a result, each pixel already contains an amplifier and a signal shaper. The pixel size is 28$\times$28 $\mu$m$^2$, and the pixel matrix size is 15.3$\times$30 mm$^2$. The matrix consists of 512$\times$1024 pixels and is therefore divided into 512 rows and 1024 columns. Analog and digital electronics are formed on the periphery of the pixel matrix. The size of the peripheral circuit is 1.208$\times$30 mm$^2$.

The signal arising from the passage of the particle is read by the Priority Encoder (PE) circuit, implemented according to the  Address-Encoder Reset Decoder (AERD) principle \cite{Yang:2015vpa}. In this case, the entire pixel matrix is divided into 512 double columns, each of which uses an AERD scheme  to read the addresses of triggered pixels only. When the charge collected on the detecting diode exceeds a certain discriminator threshold, the single-bit memory cell (if open) goes to state ''1''. The readout system gives priority to the encoding function for a column of pixels, in which all pixels have occupied the entire digital section. This allows the memory cells of all pixels to be read or zeroed. 

This system enables the reading of signals from pixels in a binary form and storing them in the pixel cell. All analog signals, required to operate the front-end pixel electronics, are generated by fourteen digital-to-analog converters, built into the detector. The readers of each of the 32 regions contain SRAM memory cells for storing the event information. The accumulated data is transferred to a parallel 8-bit port. The high-level control unit provides full access to the control and status registers of the detector, as well as the memory cells of the readers. 

We are also considering using MAPS-based detectors, developed as part of the ARCADIA project \cite{ARCADIA, Neubuser:2020dvr}.
The ARCADIA (Advanced Readout CMOS Architectures with Depleted Integrated sensor Arrays) project is developing FD-MAPS with an innovative sensor design, which uses backside bias to improve the charge-collection efficiency and timing over a wide range of operational and environmental conditions. The sensor design targets very low power consumption, in the order of 20 mW cm$^{-2}$ at 100 MHz cm$^{-2}$ particle flux, to enable air-cooled operation. 

The FD-MAPS architecture, initially embodied in a 512$\times$512 pixel matrix, should enable the scalability of the sensor up to the matrix size of 2048$\times$2048 pixels. Maximizing the active area of a single sensor (10 cm$^{-2}$ or bigger) simplifies and reduces the costs of the detector construction. The total thickness of such detectors could be from 50 to 300 $\mu$m.

\subsection{Supporting structure for MAPS detectors}

\begin{figure}[!t]
    \center{\includegraphics[width=0.9\linewidth]{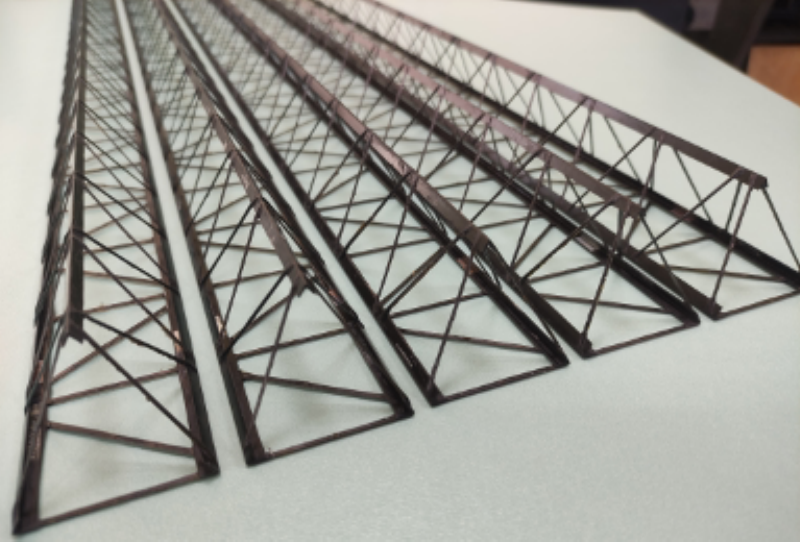}}
  \caption{Examples of ultralight carbon-composite structures to support the cooling panels of the detector modules for SVD SPD.}
  \label{fig:vd_6maps}  
\end{figure}

An important feature of silicon MAPS sensors, planned to be used in the creation of the vertex detector of the SPD experiment, is their high thermal power, which can reach up to 40 mW/cm$^2$. Therefore, with the use of CMOS technology for pixel sensors in the experiments at the NICA collider, the development of an effective cooling system is required. It should be capable  of ensuring  optimal operation of electronics and detectors at temperatures not exceeding +30$^{\circ}$C, with a minimal amount of matter in the particle collision region. The latter leads to the minimization of multiple scattering, i.e. increase in radiation transparency of the whole detector complex. Thus, for the best reconstruction of vertex decays of unstable particles, the radiation thickness of all detector modules and their support structures should not exceed 1\% of the radiation length. That is, the main elements of the vertex detectors (silicon sensors, cable systems, supporting and cooling systems) should have a minimum of material with a small charge number $Z$ \cite{Zherebchevsky:2018erx,Zherebchevsky:2014vra}. At SPbSU, together with the ALICE Collaboration, ultralight carbon fiber structures were developed to support detectors based on MAPS for the new Internal Tracking System of this experiment \cite{Abelevetal:2014dna,Zherebchevsky:2018erx,Zherebchevsky:2014vra}. Similar technologies can be used in the creation of the  SPD SVD. The ultralight radiation-transparent carbon fiber support structures of MAPS with an integrated cooling system consist of trusses, on which thermal panels, made with carbon fiber, are attached.  The cooling system consists of polyamide tubes with a diameter of 1 mm. All ultralight support structures are spatial monolithic structures - trusses (see Fig. \ref{fig:vd_6maps}). The trusses are constructed using prepregs, consisting of high-modulus carbon fiber and impregnated with an appropriate binder. The combination of these components makes it possible to obtain MAPS sensor support structures with the best mechanical and strength properties with minimal substance content. 
The entire detector module (stave) will consist of three main components: (i) pixel detector, (ii) power bus, and (iii) cold plate.
The estimated contributions of one layer stave to the material budget is about 0.81\% $X_0$  \cite{Abelevetal:2014dna}.

\subsection{Cost estimate}
A rough estimates of the cost of a MAPS-based vertex detector are given in Tab. \ref{MAPS_cost_tab}. When estimating the cost of the chips, the possible manufacturing faults are taken into account. The total cost is 11.7 MEuro or 13.5 M\$.

\begin{table}[htp]
\caption{Cost estimates for a MAPS-based vertex detector construction.}
\begin{center}
\begin{tabular}{|l|c|}
\hline
 & Cost, kEuro \\
 \hline
 Pixel chips  (per 1 m$^2$): & \  {\bf 2700} (500)  \\
 ~~~~~~~CMOS wafers + Thinning \& dicing & 1620 (300)  \\
 ~~~~~~~ Series test  & 1080 (200)  \\
 \hline
 Staves:                                          &  {\bf 1600} \\
  ~~~~~~~ FPC construction           &  350 \\
  ~~~~~~~ FPC test                         &  150  \\
   ~~~~~~~ Module assembly \& test   & 400 \\
  ~~~~~~~ Carbon Space Frame  \& Cold Plate construction  & 200 \\
  ~~~~~~~ Carbon Space Frame \& Cold Plate tests & 100 \\
  ~~~~~~~      Stave assembly \& test  & 400 \\
  \hline
  Support and installation:  & {\bf1600} \\
  ~~~~~~~   Layers End-Wheels  & 100 \\
    ~~~~~~~  Barrel Shell &   250  \\
    ~~~~~~~  Service Barrel  & 250 \\
  ~~~~~~~  Installation mechanics \& test & 500 \\
  \hline 
  Readout electronics:  & {\bf 3000} \\
    ~~~~~~~   Development of Readout Unit & 2500 \\
    ~~~~~~~   Test of Readout Unit & 500 \\    
 \hline   
 Power Distribution System: & {\bf 1500} \\
    ~~~~~~~    Power supply  &  1000 \\
     ~~~~~~~   Power regulation &  500 \\
 \hline
 Cooling System:  & {\bf 1000} \\
 \hline
 Detector Control System:    & {\bf 300} \\
 \hline
  TOTAL &  {\bf 11700} \\
\hline   
\end{tabular}
\end{center}
\label{MAPS_cost_tab}
\end{table}%

\section{DSSD option}

\begin{figure}[!h]
  \begin{minipage}[ht]{1.\linewidth}
    \center{\includegraphics[width=0.8\linewidth]{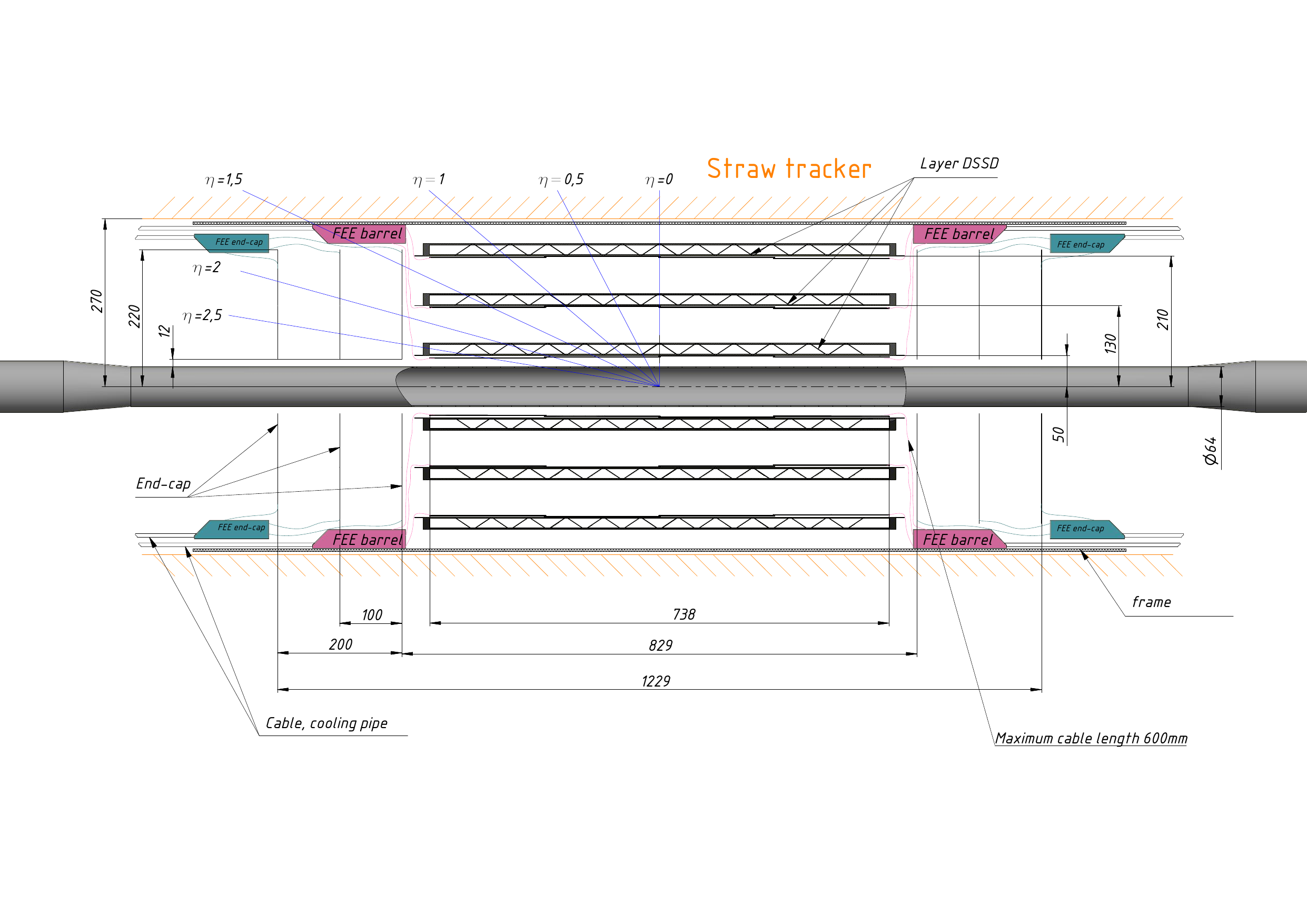} \\ (a)\\}
  \end{minipage}
  \hfill
  \begin{minipage}[ht]{0.99\linewidth}
    \center{\includegraphics[width=0.8\linewidth]{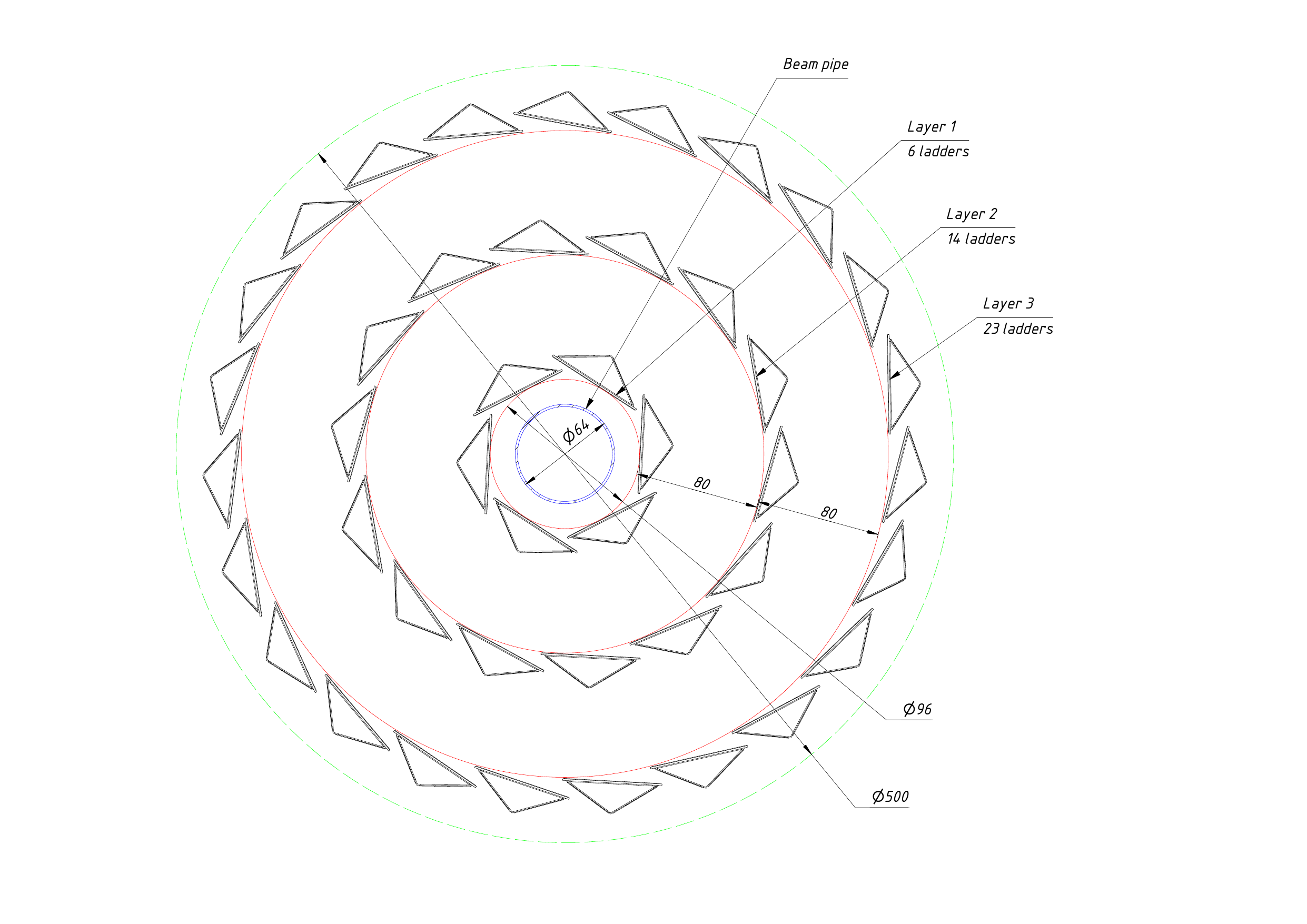} \\ (b)}
  \end{minipage}
  \caption{ Longitudinal (a) and transversal (b) cross-sections of the DSSD Vertex Detector. All dimensions are given in millimeters.
}
  \label{fig:vd_1}  
\end{figure}

The DSSD SVD barrel consists of  three layers, based on the Double-Sided Silicon Detectors (approximately 1.9 m$^2$). The end-cap regions consist of three disks each (approximately 0.22 m$^2$).  
The SVD barrel covers a radius from 32 mm to 250 mm. All three cylindrical layers are equipped with rectangular two-coordinate silicon strip detectors and give information on the coordinates of the tracks ($r$,$\phi$,$z$), which makes it possible  to measure a point in each layer).
The end-cap regions detect particles in the radial region between 32 mm and 250 mm. Each of the three disks is set with a DSSD with concentric ($r$) strips and radial ($\phi$) strips.
The SVD has a length of about 1.2 m and covers the region of pseudo-rapidity up to $|\eta|<2.0$. Each DSSD has a 300-$\mu$m thickness and a strip pitch in the range from 95 $\mu$m to 281.5 $\mu$m. The DSSDs are assembled into detector modules by two detectors per module, forming  18-cm long  strips. The general layout of the DSSD SVD is presented in Fig. \ref{fig:vd_1}.

The detectors and the front-end electronics boards (FEE-PCB) are connected via low-mass polyimide microcables and assembled on the extra-light  carbon fiber mechanical supports with a cooling system in a similar way  as it was done for the ALICE outer  barrel. The relevant numbers for the barrel part  and  the end-caps of the SVD for the DSSD configuration are presented in  Tables \ref{VD_pars} and  \ref{VD_pars_ec}, respectively.


From  the general conditions of the SPD setup, the SVD performance requirements are i) geometry close to $4\pi$; ii) track reconstruction efficiency for muons greater than 99\% at $p\leq 13$ GeV/$c$ (for $0\leq |\eta|\leq 2.5$); iii) low material budget;  iv) coordinates resolutions for vertexing: $\sigma_{r,\phi}< 50~\mu$m, $\sigma_{z} < 100~\mu$m.
The lifetime of the Silicon Vertex Detector is required to be not less than 10 years of NICA operation.

\begin{table}[!h]
\caption{Number of the main components for the barrel consisting of 3 layers of DSSD.}
\begin{center}
\begin{tabular}{|l|c|c|c|c|}
\hline
Parameter & Layer 1 &  Layer 2 &  Layer 3 & Total \\
\hline
\hline
N$_{DSSD}$/module & 2 & 2 & 2 & \\ 
N$_{modules}$/ladder & 4 & 4 & 4  & \\ 
N$_{ladders}$/layer & 6 & 14 & 23 & 43  \\ 
N$_{DSSD}$/layer & 48 & 112 & 184 & 344\\ 
N$_{chip}$/module & 10 & 10 & 10 & \\ 
N$_{chip}$/layer & 240 &  560 & 920 & 1720 \\ 
N$_{channel}$/layer & 30720 & 71680 &  117760 & 220160 \\ 
\hline
\end{tabular}
\end{center}
\label{VD_pars}
\end{table}%

\begin{table}[!h]
\caption{Number of the main components for the end-cap consisting of 3 layers of DSSD-based on trapezoidal modules  (640$\times$640 strips).}
\begin{center}
\begin{tabular}{|l|c|c|c|c|c|c|}
\hline
Parameter & Layer 1 &  Layer 2 &  Layer 3 & End-cap 1& End-cap 2 & Total \\
\hline
\hline
N$_{DSSD}$/module & 2 & 2 & 2 && & \\ 
N$_{ladders}$/layer & 14 & 14 & 14 & 42 & 42 &84  \\ 
N$_{DSSD}$/layer & 28 & 28 & 28 & 84 & 84 & 168\\ 
N$_{chip}$/module & 10 & 10 & 10 & & & \\ 
N$_{chip}$/layer & 140 &  140 & 140 & 420 & 420 & 840\\ 
N$_{channel}$/layer & 17920 & 17920 &  17920 & 53760 & 53760 & 107520 \\ 
\hline
\end{tabular}
\end{center}
\label{VD_pars_ec}
\end{table}%

\begin{figure}[!h]
    \center{\includegraphics[width=0.8\linewidth]{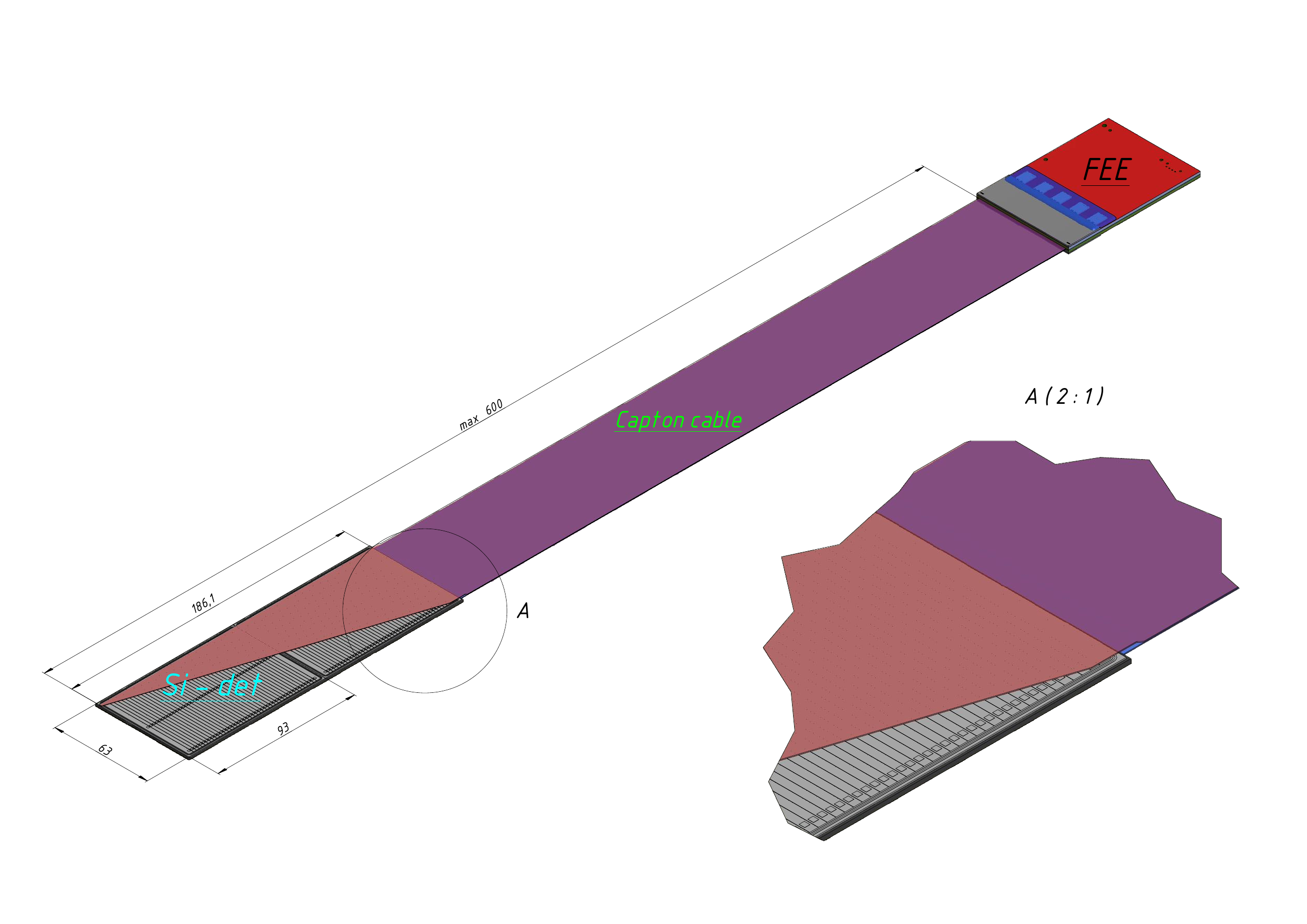}}
  \caption{Concept of the barrel DSSD module. All dimensions are in millimeters.}
  \label{fig:vd_2}  
\end{figure}

The concept of a single DSSD module for the barrel part is shown in Fig. \ref{fig:vd_2}. The module consists of two silicon detectors,  wire-bonded strip-to-strip for the $p+$ side (to reduce the number of readout channels), glued to the plastic frame and connected with two front-end electronics boards via a low-mass polyamide cable. 

The Silicon Vertex Detector will be made using a planar double-sided technology based on the n-type conductivity 6-inch float-zone silicon wafers (produced by ZNTC, Zelenograd, Russia). Its size is 63$\times$93 mm$^2$ and its thickness is 300 $\mu$m.  The pitch for the $p+$ side is 95 $\mu$m and for the $n+$ side 281.5 $\mu$m. The number of strips is 640 and  320 for the $n+$ and $p+$ sides, respectively. The stereo angle between the strips is 90 degrees. The expected spatial resolution for such a detector topology is $pitch_{p(n)+}/\sqrt{12}=$ 27.4 (81.26) $\mu$m for $r-\phi$ and $r-z$ projections, respectively. As mentioned before, the barrel DSSD module contains two DSSDs ($p+$ strips wire bonded strip to strip) and has 640 strips at each side. 

To bring the front-end electronics out of the tracker volume,  two thin polyimide cables with aluminum traces (for each side of the module) will be used. The cable consists of several layers: signal, perforated or solid dielectric (polyimide), and a shielding layer. Cable pins were designed for tape-automated  bonding with the detector and the pitch adapter sides. The maximum cable length is 60 cm, and the total thickness of all cable layers is less than 0.15\% of  $X_0$.  
\begin{figure}[!h]
    \center{\includegraphics[width=0.8\linewidth]{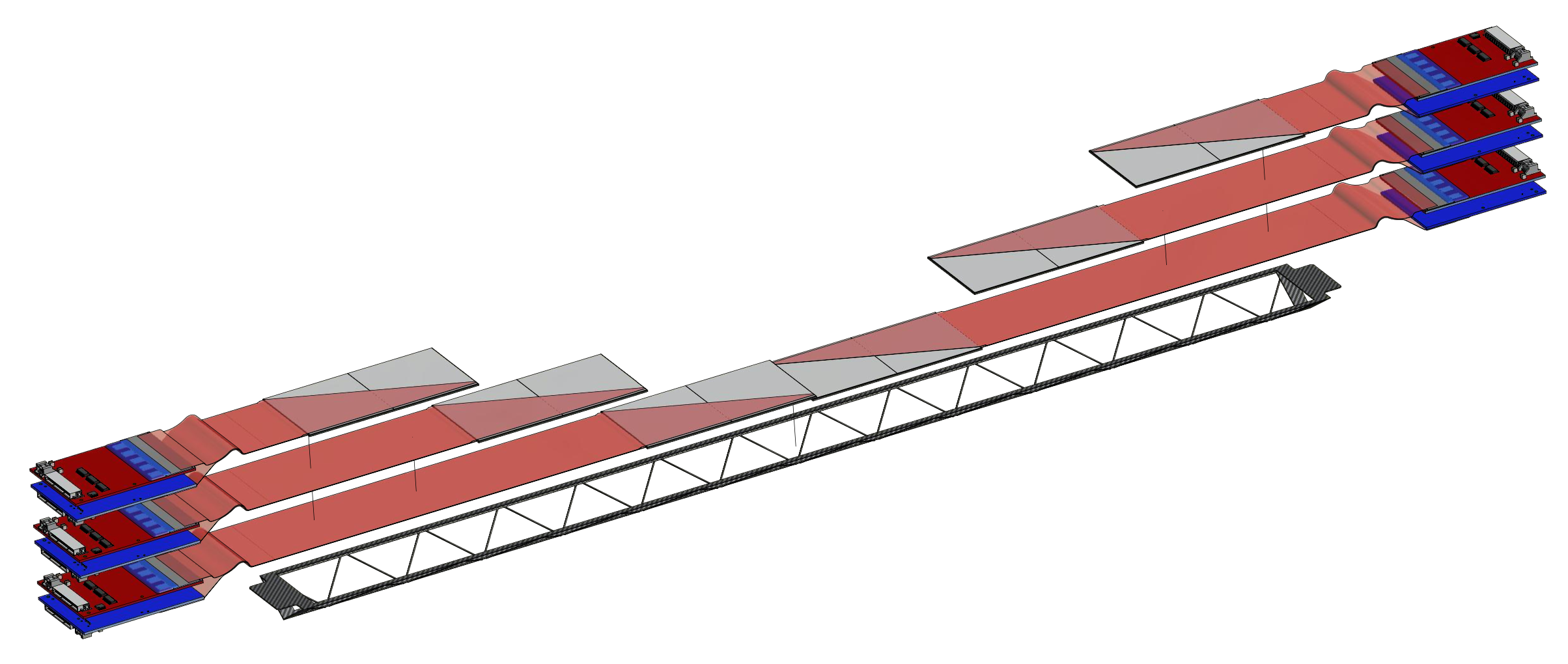}}
  \caption{Sketch of the barrel ladder.}
  \label{fig:vd_3}  
\end{figure}
\begin{figure}[!h]
  \begin{minipage}[ht]{1.\linewidth}
    \center{\includegraphics[width=0.9\linewidth]{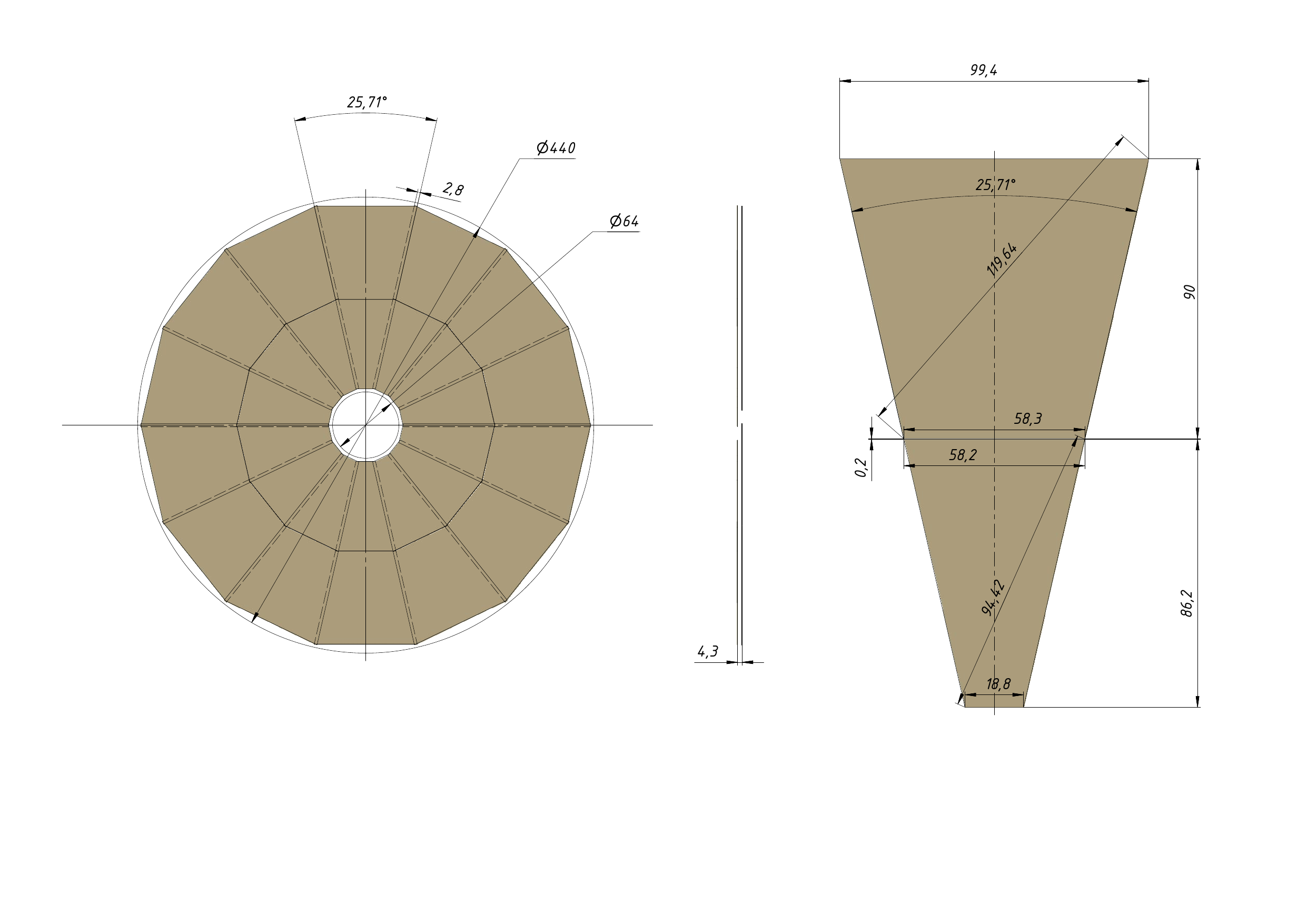} \\ (a)\\}
  \end{minipage}
  \hfill
  \begin{minipage}[ht]{0.99\linewidth}
    \center{\includegraphics[width=0.9\linewidth]{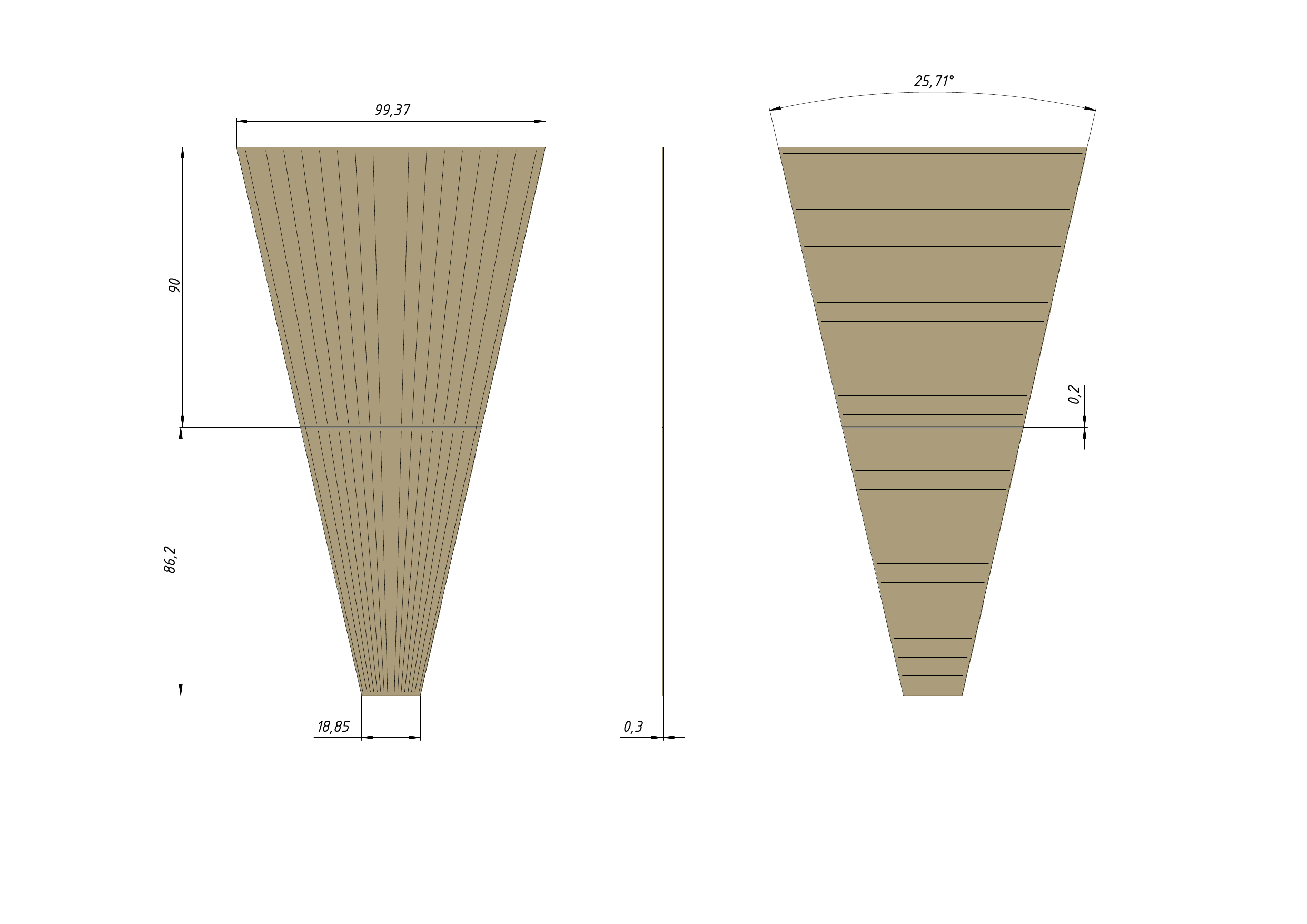}\\ (b)}
  \end{minipage}
  \caption{ (a) Sketch of the SVD end-cap planes consisting of trapezoidal shape DSSD modules. (b) Sketch of a single trapezoidal module.}
  \label{fig:vd_5}  
\end{figure}

Since the DSSDs have a  DC topology, it is necessary to supply bias voltage to the detector and electrically decouple the DC current from the ASIC's electronics inputs. For this purpose, an integrated RC circuit (sapphire plates with Si-epitaxial layer Silicon On Insulator (SOI)) Pitch Adapter (PA) will be used for each side of the module (produced by ZNTC, Zelenograd), designed with different topologies for each side. 
 After the pitch adapter, the detector signal goes to ASIC. Table \ref{Tab:VD} shows a possible ASIC readout solution. The optimal choice should be performed after the ongoing R\&D.

\begin{table}[!t]
\caption{Possible ASIC readout solution for the Vertex Detector.}
\begin{center}
\begin{tabular}{|l|c|c|c|c|}
\hline
ASIC & APV25	& VATAGP7.3	& n-XYTER &	TIGER \\
\hline
\hline
Number of channels & 128 & 128 &128 & 64 (128) \\
Dynamic range         &   -40fC -- 40fC &	-30fC -- 30fC &	Input current 10 nA,   & 1--50fC \\
                                 &                         &                           &  polarity $+$ and $-$ & \\
Gain	                         &25mV/fC	          &20$\mu$A/fC	      &                                  &	10.35mV/fc  \\
Noise	                 &246 e$^{-}$+36 e$^-$/pF	&  70e$^-$+12 e$^-$/pF   &	900 e$^-$  at 30pF &	2000 e$^{-}$  at 100pF \\                        
Peaking time             &	50ns      &	50ns/500ns	         & 30ns/280ns &	60ns/170ns \\
Power consumption  &	1.15mW/ch. &	2.18mW/ch.	&10mW/ch. &	12mW/ch. \\
ADC                        &	No &	No	& 16fC, 5 bit &	10-bit \\
                                &             &         &                  &    Wilkinson ADC  \\          
TDC                        &	No &	No	& &	10-bit  \\
                                &             &         &                  &    Wilkinson ADC  \\    
\hline
\end{tabular}
\end{center}
\label{Tab:VD}
\end{table}%

\subsection{Mechanical structure}

\begin{figure}[!h]
    \center{\includegraphics[width=0.95\linewidth]{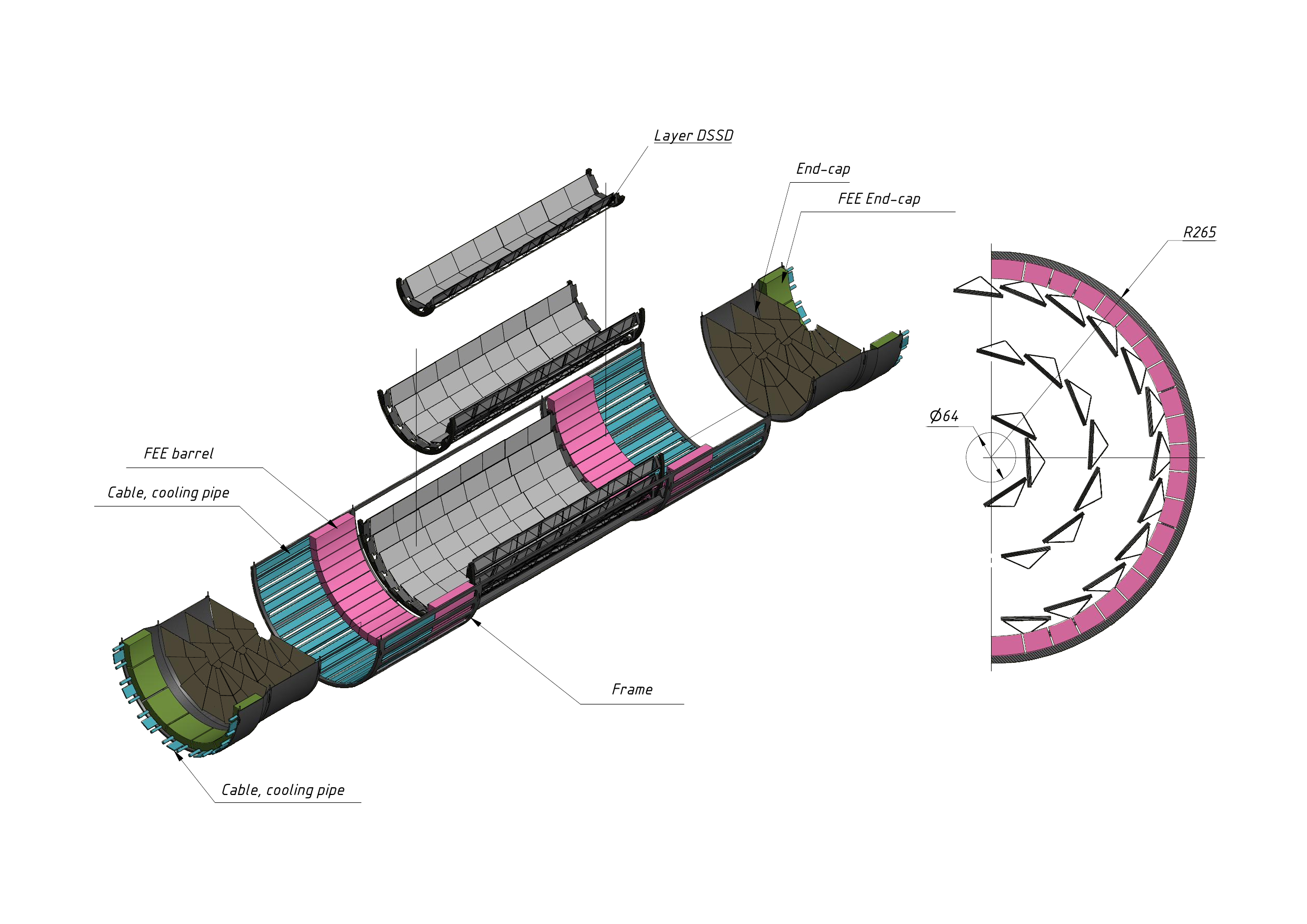}}
  \caption{Sketch of the assembly of the end-cap and the barrel parts of VD.}
  \label{fig:vd_4}  
\end{figure}

\begin{figure}[!h]
    \center{\includegraphics[width=0.95\linewidth]{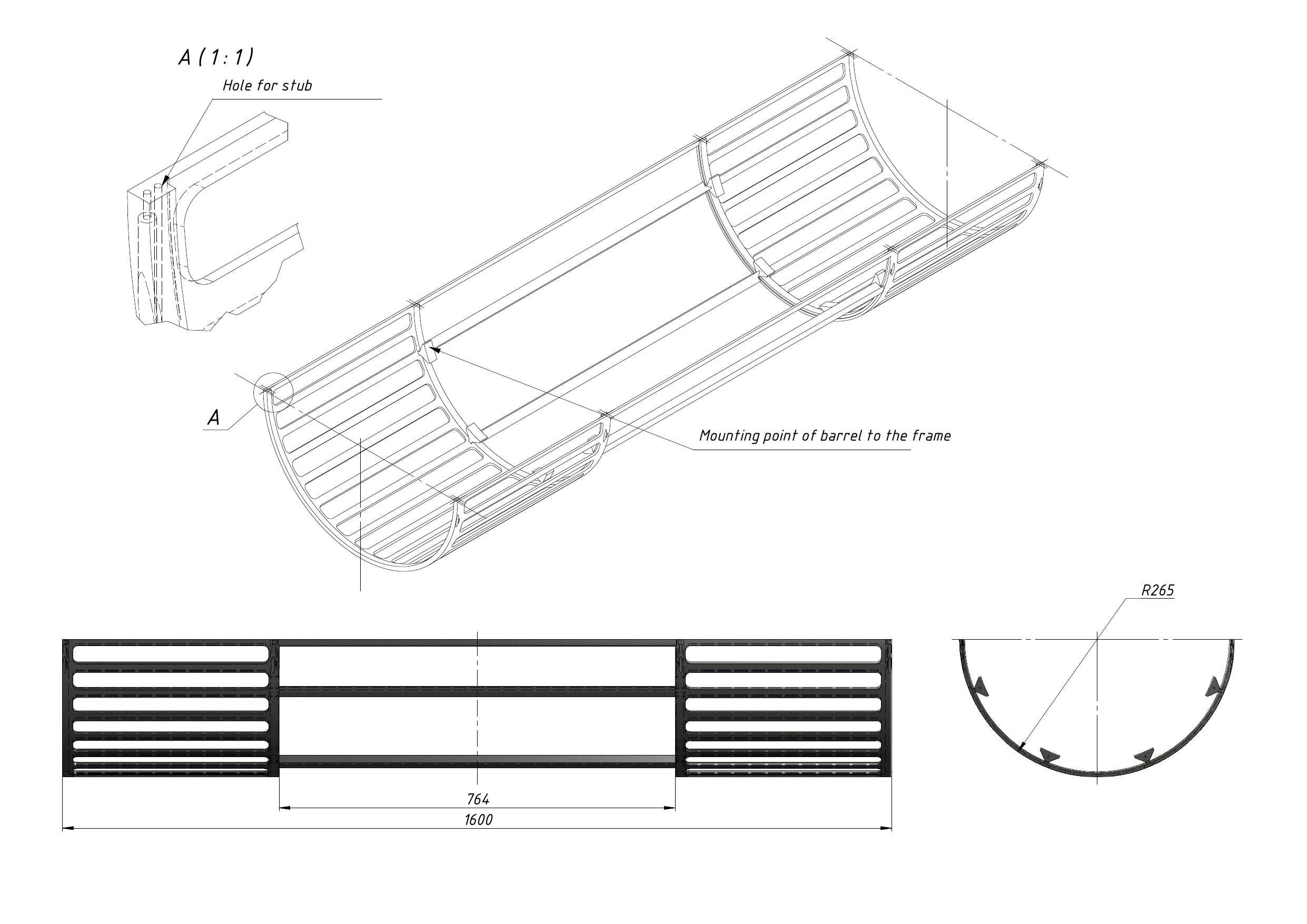}}
  \caption{SVD supporting frame.}
  \label{fig:vd_7}  
\end{figure}

\begin{figure}[!h]
    \center{\includegraphics[width=0.95\linewidth]{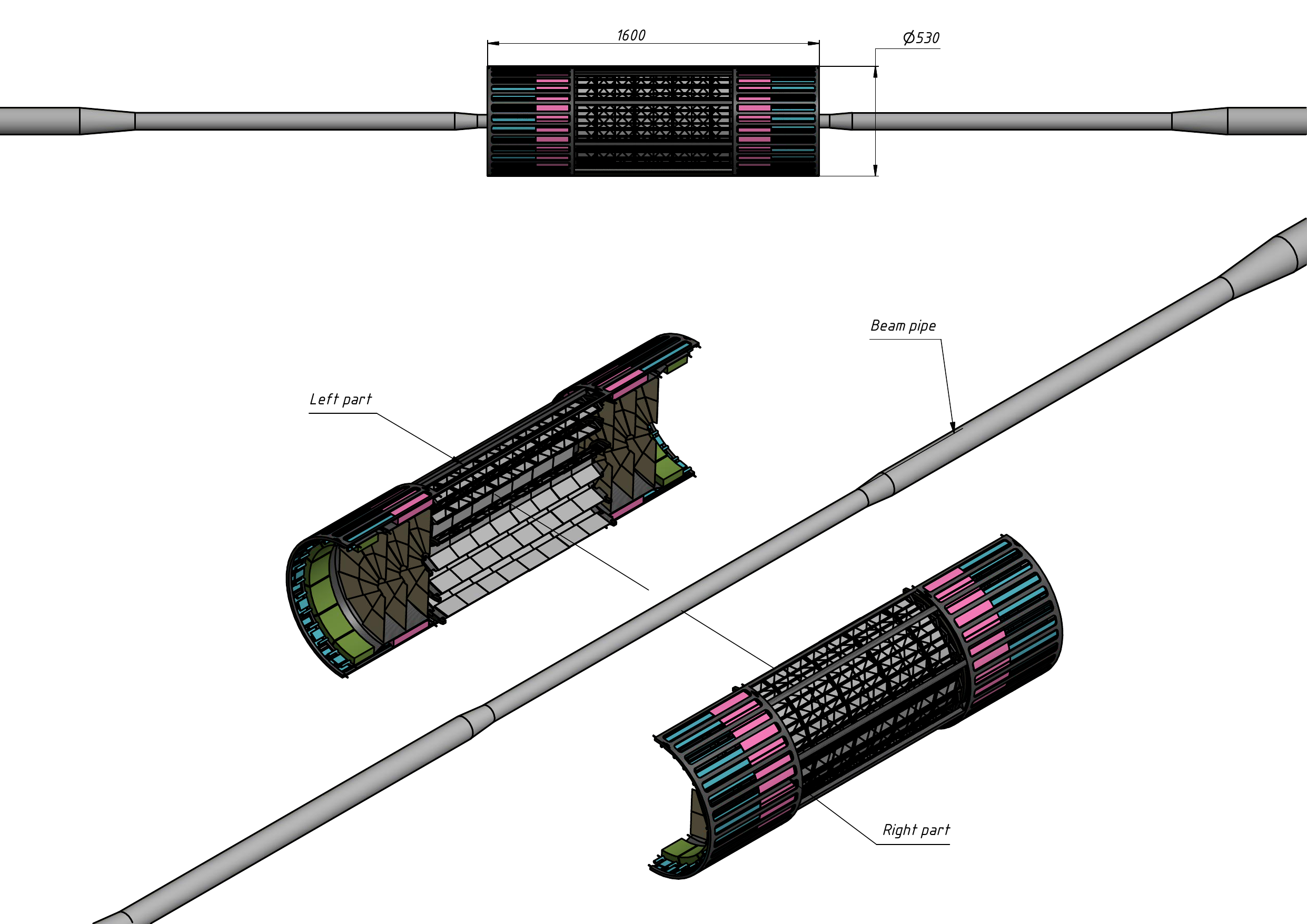}}
  \caption{Sketch of the SVD mechanical support structure with the conical caves for the FEE.}
  \label{fig:vd_6}  
\end{figure}

The concept of the barrel DSSD ladder is shown in Fig. \ref{fig:vd_3}. The silicon modules are laying on a carbon fiber support, from  the center out to the edges. The detectors are connected with the FFE via thin low-mass cables. The front-end electronics is located at the edges of the ladder and is placed in the conical caves to provide a connection to the voltage supply, DAQ,  and the cooling ASIC chips subsystems. A sketch of an end-cap plane, consisting of trapezoidal  modules, is presented in Fig. \ref{fig:vd_5}. The internal structure of the detector, the supporting frame, as well as its placement with respect to the beam pipe  together with communications, are shown in Figs.  \ref{fig:vd_4},  \ref{fig:vd_7}, and  \ref{fig:vd_6}, respectively.

\subsection{Cooling system}
 A liquid cooling system is used to dissipate the thermal power ($\sim$2 kW) from the FEE, located on the two ends of the VD (1 kW each)
to avoid thermal heating of the electronics and surrounding detectors. A water cooling system is chosen to dissipate the generated power. The readout electronics elements are placed on the ends of the detector barrel and on the ends of the end-caps. Using compact aluminum liquid-cooled heat exchangers the heat from the electronics boards will be dissipated. The complete SVD setup will run inside an insulated box. The box temperature should be kept below 20$^{\circ}$ C at all times of operation. 

\begin{figure}[!h]
    \center{\includegraphics[width=0.95\linewidth]{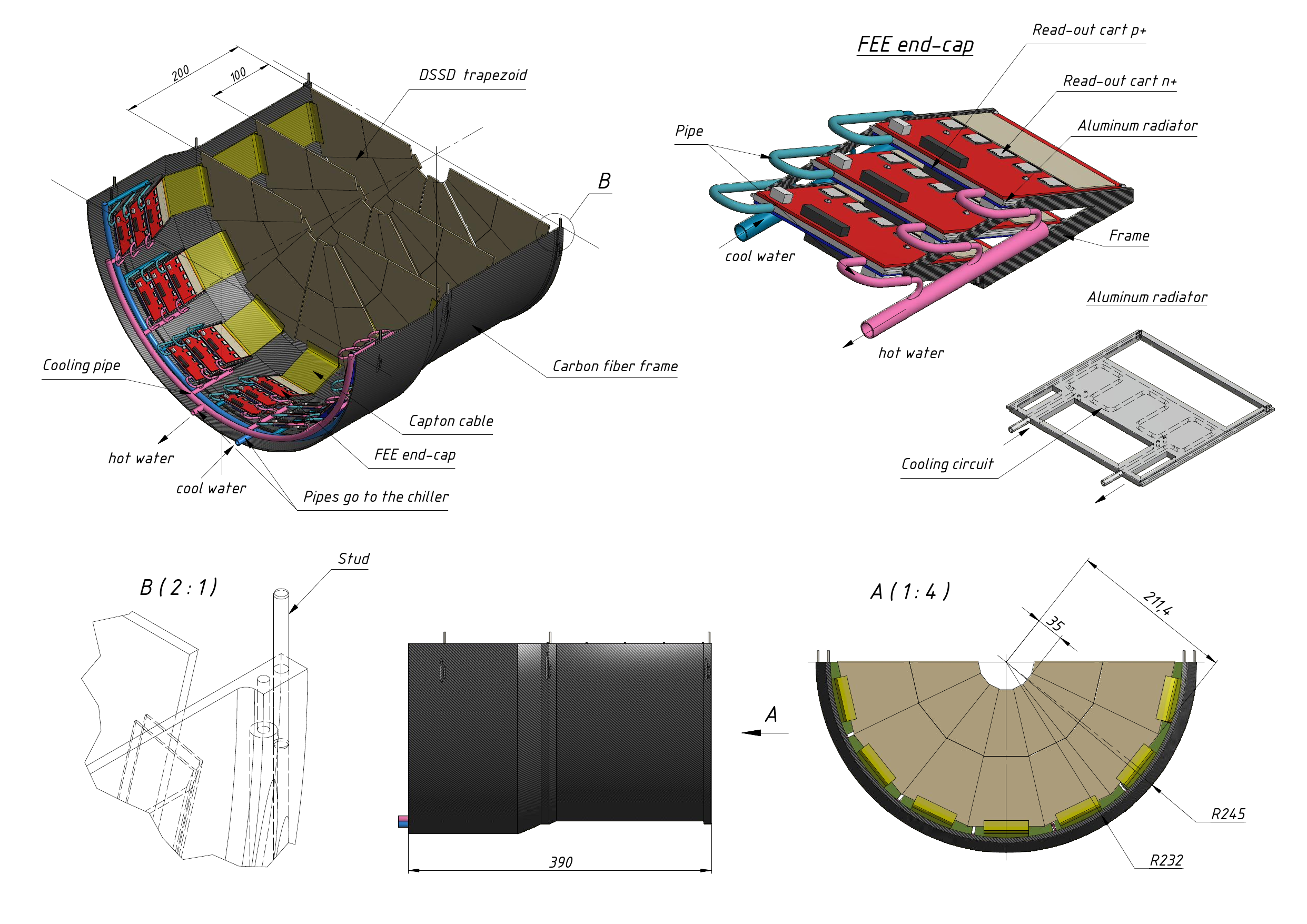}}
  \caption{Sketch of the cooling system.}
  \label{fig:vd_10}  
\end{figure}

The liquid cooling system is based on four Huber Minichillers 900w laboratory chillers with a total cooling capacity of 3.6 kW at 15$^{\circ}$ C coolant temperature. This chiller was chosen because of its high cooling capacity, compact size, and suction capacity of ~18 l/min. The coolant used is deionized water.  Two chillers are used to cool the barrel readout electronics (right and left side), and two more chillers are used to cool the end-cap readout electronics.
The chillers will be located ~25 m away from the vertex detector of the SPD unit. The number of tube connectors and diameters are under development. Fig. \ref{fig:vd_10} shows the principle and variant of heat removal from one of the end-caps, other parts of SVD will be cooled similarly, the only difference is the location of FEE.

\subsection{Cost estimate}
The preliminary cost estimate for the DSSD configuration of the Silicon Vertex Detector is presented in Table \ref{tab:costVD}.
It is based on the following assumptions: 
\begin{itemize}
\item the cost of 1 m$^{2}$ of DSSD Si sensors is 0.6 MEuro;
\item the cost of 1 channel (128 ch./chip) is 7 Euro.
\end{itemize}
The cost of the SVD, which includes also low and high voltage supply, cables, cooling system and  mechanics, is 2.4 MEuro per 1 m$^{2}$ or  2.64 M\$. So, the total cost of the detector is 8.8 M\$.


\begin{table}[htp]
\caption{ Cost estimate for the SVD  DSSD configuration.}
\begin{center}
\begin{tabular}{|c|c|c|c|c|c|c|}
\hline 
Layer  & Number of & Number of  & Barrel              & End-cap              & Cost            & Cost (one) \\
           & sensors     & ASICs         & area, m$^2$    & area, m$^2$       & barrel, MEuro  & end-cap, MEuro \\
\hline 
\hline
1         & 48     & 240       &  0.28      &  0.22 ($\times$2)    &    0.67   &  0.53 \\
2         & 112     & 560        &  0.66      & 0.22  ($\times$2)     &    1.58   &  0.53 \\
3         & 184     & 920       &  1.08     &  0.22  ($\times$2)  &    2.59    & 0.53  \\
\hline
Total    & 344  & 1720       & 2.02     & 1.32  & 4.85 & 3.18 (two end-caps) \\
\hline
\end{tabular}
\end{center}
\label{tab:costVD}
\end{table}%

\section{SVD performance}
\begin{figure}[!h]
    \begin{minipage}{0.49\linewidth}
       \center{\includegraphics[width=1\linewidth]{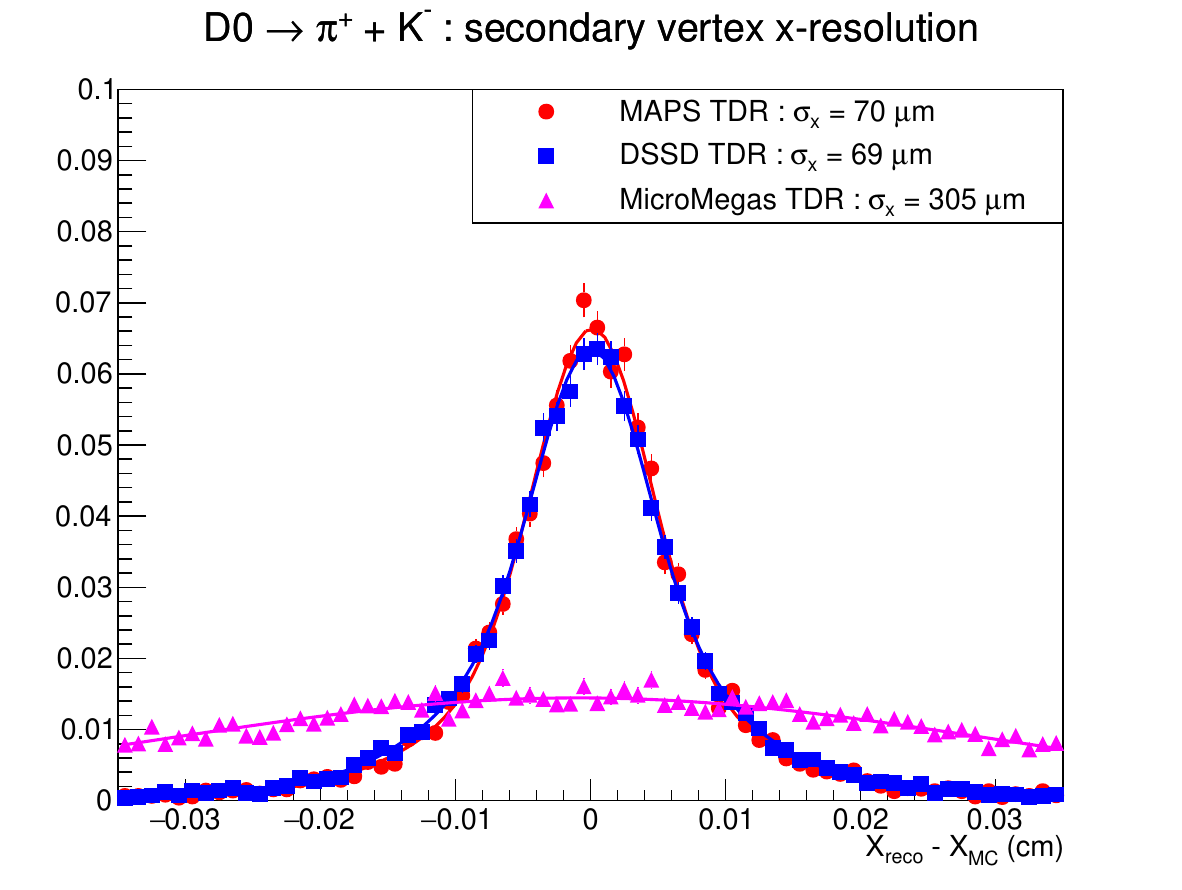} \\ (a)}
  \end{minipage}
  \hfill
  \begin{minipage}{0.49\linewidth}
        \center{\includegraphics[width=1\linewidth]{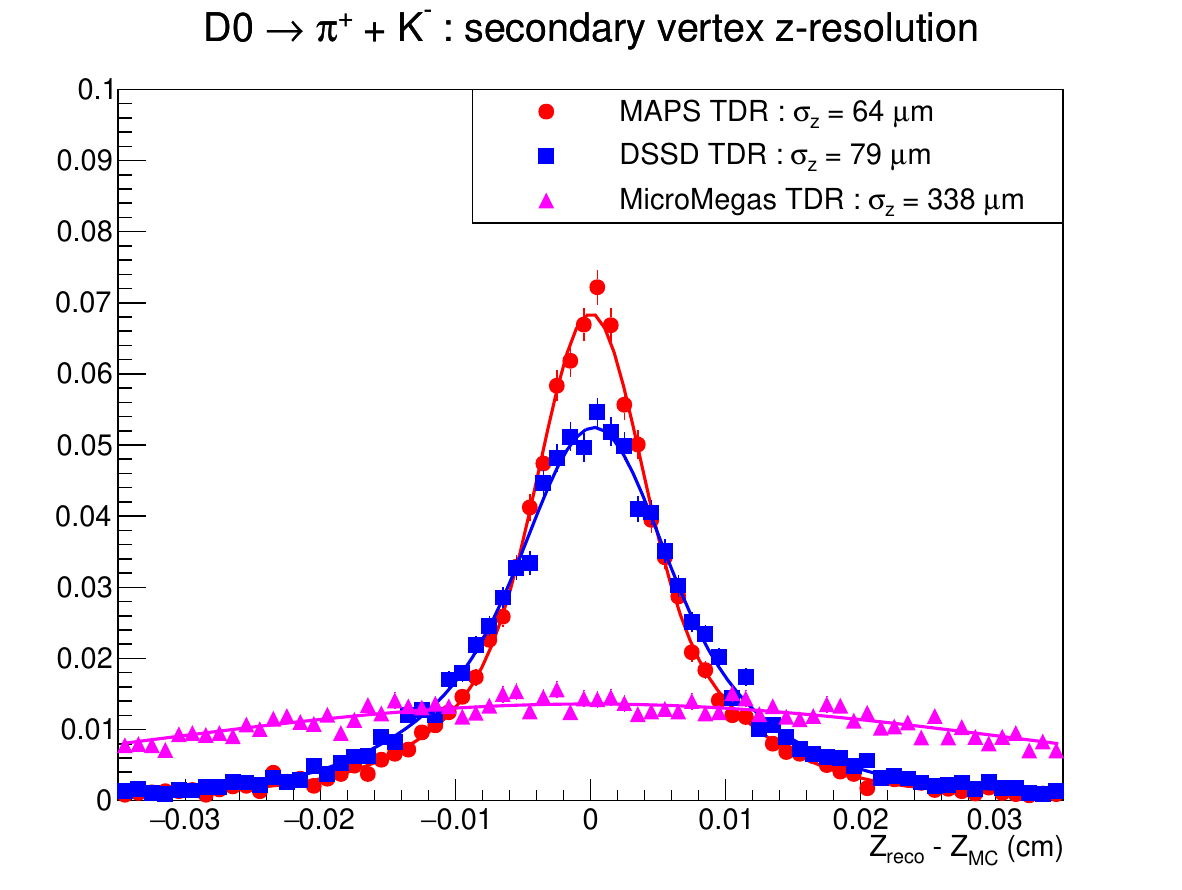} \\ (b)}
  \end{minipage}
  \caption{Spatial resolution along $X$ (a) and $Z$ (b) axes for the  secondary vertex of the $D^0\to K^-\pi^+$ decay for MCT (magenta), 3-layer DSSD SVD (blue), and 4-layer MAPS SVD (red).}  
  \label{fig:3}  
\end{figure}

The Monte Carlo simulation for a VD configuration with 4 equidistant layers of MAPS and 3 layers of DSSD described above was performed. The primary vertex position resolution in $X$ and $Z$ coordinates and the momentum resolution $dp/p$ are presented in Fig. \ref{fig:3}~(a) and (b), respectively. Because of the smaller amount of substance, the MAPS-based detector demonstrates better performance than the DSSD-based one. Such improvement in the momentum resolution should enlarge accordingly  the significance of the $J/\psi$ and $D$ peaks. The choice of a Silicon Vertex Detector design, based on the DSSD technology, can only be motivated by the unavailability of MAPS technology for the SPD Collaboration.

\chapter{ Micromegas-based Central Tracker}

\section{Introduction}
Detection of charged particles and measuring their momenta is important for any
task of the SPD physics program~\cite{Abramov:2021vtu}. Tracking and determining the
 particle momentum for the SPD are performed with a vertex detector and a straw tracker~\cite{SPD:2021hnm}.
The vertex detector based on the MAPS technology (or DSSD
 as a backup option) will not be installed for the first stage of the
SPD operation. Absence of the tracker close to the beam pipe leads to severe worsening of the momentum resolution, and also affects the tracking efficiency and reconstruction of secondary vertices of the long-lived particles. To provide a performance adequate for the physics tasks, it is proposed to install a relatively simple and cheap central tracker based on the Micromegas technology. Such a detector would be used during the first two or three years of the SPD operation.

In brief, the idea of the Micromegas-based Central Tracker (MCT) is to improve the momentum resolution and tracking efficiency   of  the main tracking system during the first period of data taking. The main requirement is that the total cost should not exceed 10\% of the SVD.  The MCT will not provide tracking in the far forward region and precise secondary vertex reconstruction for $D$-tagging, but may be useful for reconstruction of  hyperons and $K^0_s$ decays.

This proposal relies heavily on the Saclay group experience, who developed the Micromegas vertex tracker  for the CLAS12 experiment \cite{Acker:2020qkv}.

\section{Micromesh Gaseous Structure}
\subsection{Principle of operation}
\begin{figure}[htb]
  \begin{minipage}[ht]{0.45\linewidth}
  \center{    \includegraphics[width=1.0\linewidth]{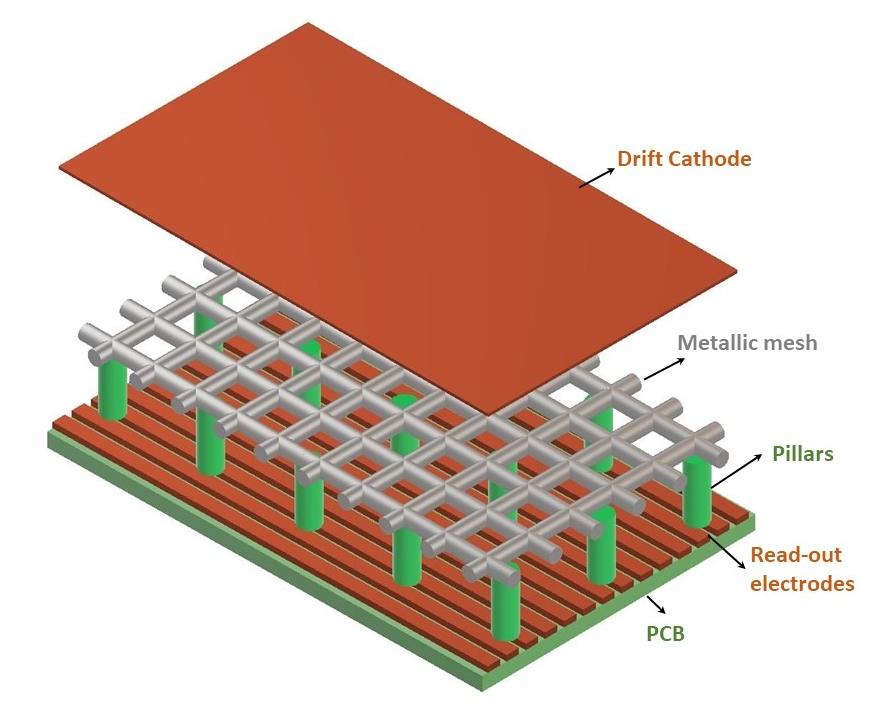}  \\(a)}
  \end{minipage}
  \hfill
  \begin{minipage}[ht]{0.55\linewidth}
  \center{       \includegraphics[width=1\linewidth]{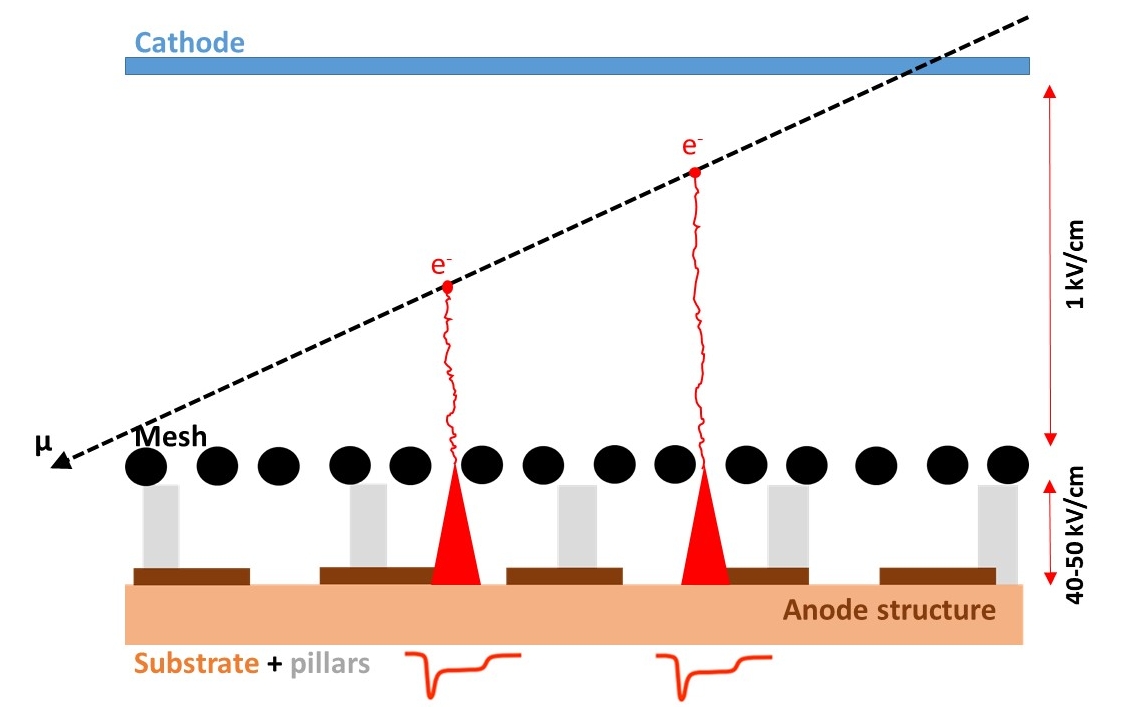}  \\ (b) }     
  \end{minipage}
    \caption{Sketch of the layout and operation principle of a  Micromegas detector.}
    \label{fig:MM_principle}
\end{figure}

 The micromesh gaseous structure (Micromegas, MM-detector) proposed by G. Charpak and  I. Giomataris~\cite{GIOMATARIS199629}, is a parallel plate counter with dedicated ionization and amplification gaps separated by  a fine mesh (Fig.~\ref{fig:MM_principle}).
A typical thickness of  the ionization gap is 3$\div$5 mm with a drift field applied at about 600~V/cm, while  the amplification gap is about 120 $\mu$m thick and the amplification field exceeds 30 kV/cm.  High uniformity of the amplification gap and, hence, amplification field is ensured by the regularly spaced insulating pillars.  High-energy particle crossing detector volume ionizes gas in the conversion gap.  Electrons of primary ionization move toward the mesh and pass through it. Due to a very high field tension difference between the drift and the amplification gaps, the mesh transparency for primary electrons is above 95\%.  In  the amplification gap  an electron starts an avalanche, resulting in  a final signal  of about $10^5$ electrons.  Most of  the electrons and ions in  the avalanche are produced near  the anode, so ions pass almost full amplification voltage before being collected on  the mesh and produce dominant contribution to  the signal. The ion collection time for a single-cluster avalanche is about 150 ns.
	To get  a good space resolution, the anode plane should be segmented. Usually, the readout electrodes are shaped as narrow strips with  a typical pitch 0.35 $\div$ 0.5 mm.   

\subsection {Hit reconstruction and accuracy}
Even  a perpendicular track usually results in signals induced on several neighbor strips,  a ''signal cluster''.  For track angles close to $90^{\circ}$, a standard ''charge centroid'' method works perfectly.  In this method  a hit coordinate is calculated according  to the following formula: $x=\sum{A_ix_i}/\sum{A_i}$, where $A_i$  is  a signal amplitude, $x_i$ is  a strip coordinate, and summation  is done over all strips in  a cluster.   The experience of the COMPASS experiment \cite{COMPASS:2007rjf} demonstrates that space resolution better  than 100 $\mu$m may be routinely achieved for  the perpendicular tracks.  For the inclined tracks, accuracy is much worse. In this case  the so-called ''$\mu$-TPC'' algorithm is much more suitable. 
\begin{figure}[htb]
  \begin{minipage}[ht]{0.5\linewidth}
   \center{    \includegraphics[width=1.0\linewidth]{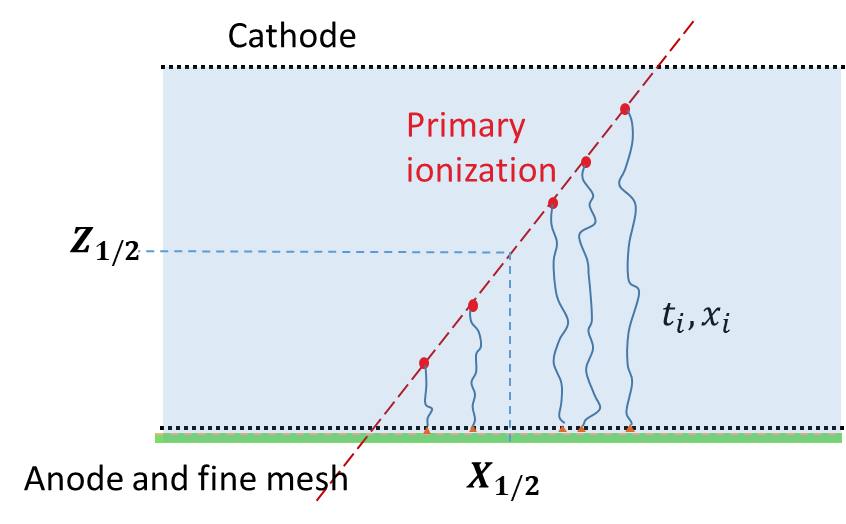}  \\ (a)}
  \end{minipage}
  \hfill
  \begin{minipage}[ht]{0.4\linewidth}
  \center{       \includegraphics[width=1\linewidth]{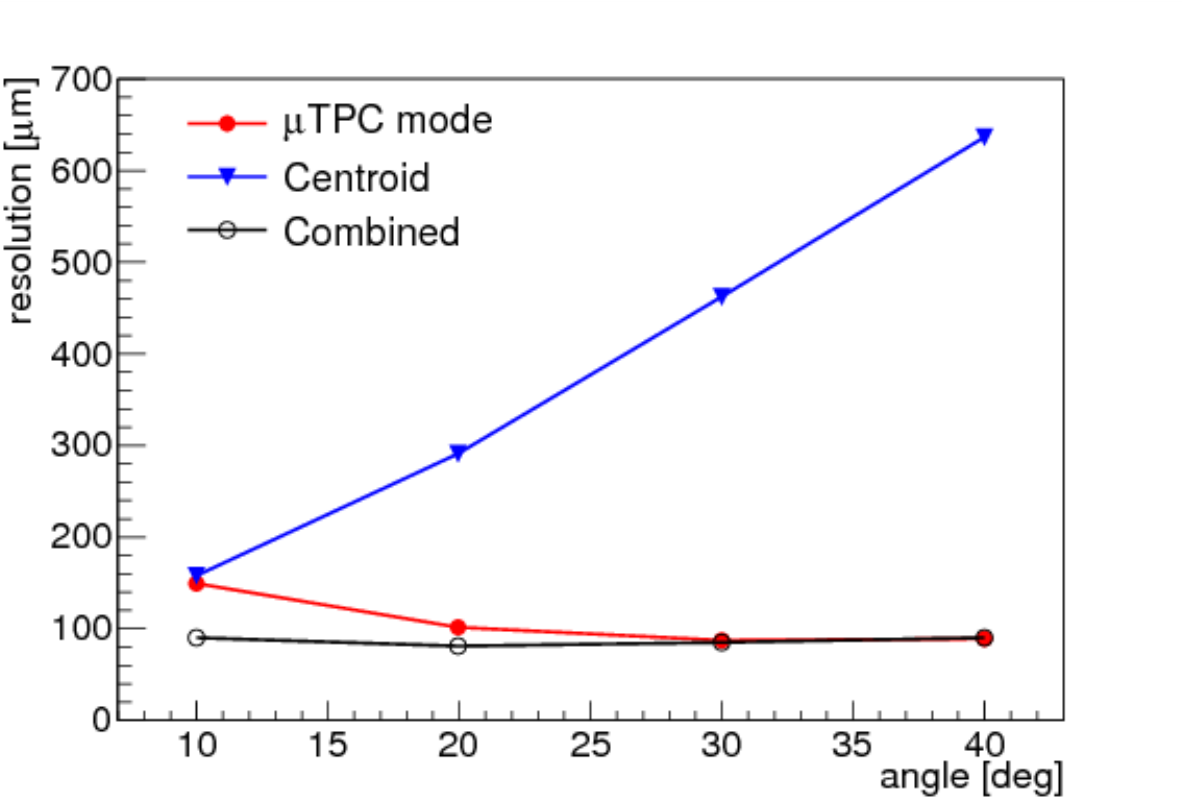}  \\ (b) }     
  \end{minipage}
    \caption{(a) Principle of the Micromegas operation in the $\mu$-TPC mode. (b) Dependence of the space resolution on the incoming track angle for the charge centroid and $\mu$-TPC reconstruction methods, results of the ATLAS Micromegas prototype test. Courtesy of the ATLAS Collaboration.}
    \label{fig:mTPC}
\end{figure}
In this method  a local track segment in  the gas gap is reconstructed using precise timing information for each strip, similar to track building in a common Time Projection Chamber detector:  $x$ coordinate is a strip position, and the second coordinate is defined by  the product of the drift time and electron drift speed (Fig.~\ref{fig:mTPC}~(a)).   The accuracy vs incoming track angle dependence for  the centroid and $\mu$-TPC methods obtained at  the beam test of  the ATLAS MM prototype \cite{Bianco:2016kul} is shown in Fig.~\ref{fig:mTPC}~(b).   As it may be seen,  a combination of  the charge centroid  and $\mu$-TPC methods provides an accuracy  of 100$\div$150~$\mu$m for an incoming track angle from $0^{\circ}$ to approximately $45^\circ$.  The obvious disadvantages of the $\mu$-TPC method are: necessity for the time measurement's accuracy better than 2 ns,  and requirement for a better signal-to-noise ratio and, hence, a higher gas gain.
For  the  SPD experiment, all tracks are almost perpendicular to the detector plane, but  taking into account that the Micromegas Central Tracker operates in the magnetic field, a non-zero Lorentz angle results in the fact that the drift line is inclined in respect to the electric field direction.
\begin{figure}[htb]
    \centering
    \includegraphics[width=.95\textwidth]{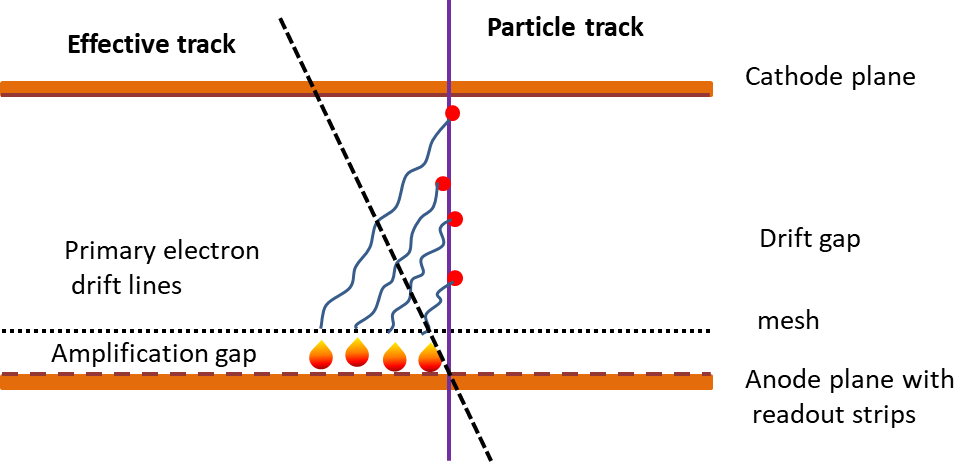}
    \caption{ Micromegas operation in the magnetic field.}
    \label{fig:Lorentz}
\end{figure}
For the hit reconstruction algorithms,  the tracks looks like ''effectively inclined'' as Fig.~\ref{fig:Lorentz} demonstrates. There are several options to get  a good space resolution for the MM operation in a reasonable magnetic field:
\begin{enumerate}

\item use the $\mu$-TPC algorithm for the hit reconstruction. This option requires  a stable MM operation at  a relatively high gas gain $G\approx 10^4$;

\item	operate the MM with an increased drift field to reduce the Lorentz angle. The disadvantage of this solution is  a reduced effective mesh transparency and detector efficiency. Reliability of this approach was demonstrated by the CLAS12 Collaboration \cite{Acker:2020qkv}: the barrel Micromegas tracker operated effectively in a 5 T magnetic field with a drift field increased up to 8 kV/cm. Despite the mesh transparency  being reduced to 60 \%, a final efficiency  of well above 90\% and an accuracy  of about 150 $\mu$m were obtained;   
\item	 choose  a gas mixture with  a very low Lorentz angle. Possible candidates are mixtures like $CO_2-Ar-iC_{4}H_{10}$(70-20-10) with an expected Lorentz angle  of about $5^{\circ}$-$7^{\circ}$. Unfortunately, these mixtures are rather slow in weak drift fields and   have never been used for the MM operation, so additional R\&D is necessary.
\end{enumerate}
 Neither option needs any  hardware modifications; a final decision  will be taken only after additional tests.

\subsection{Spark protection}
\begin{figure}[htb]
    \centering
    \includegraphics[width=.9\textwidth]{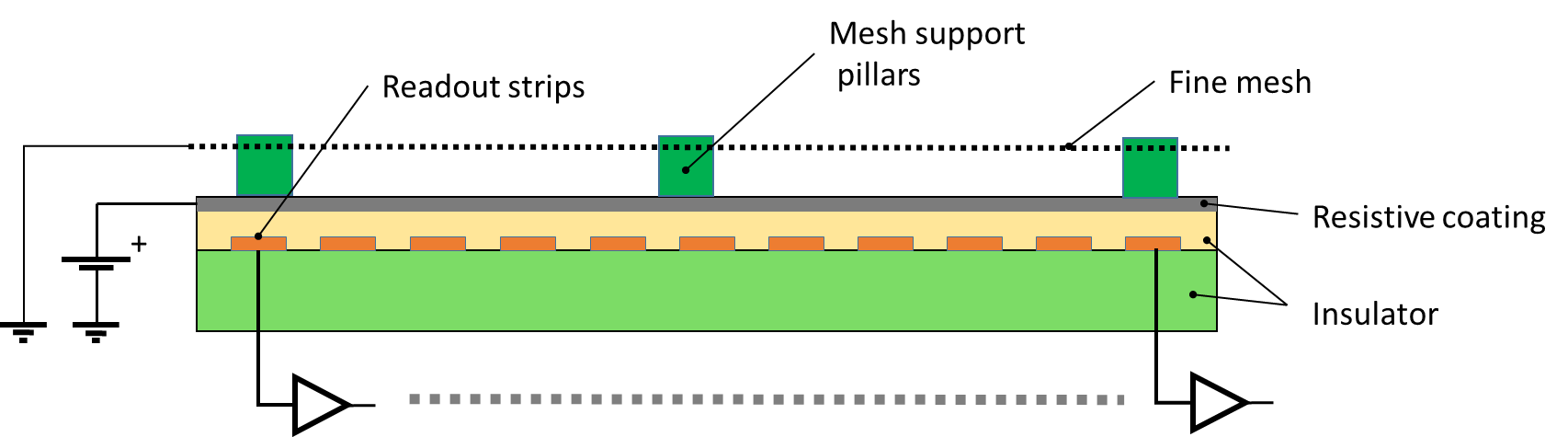} 
    \caption{Sketch of a spark-protected Micromegas: detector cross-section across readout strips}
    \label{fig:SparkProtection}
\end{figure}

As well as any parallel plate counter,  the Micromegas detector is vulnerable to a spark discharge. If  an avalanche charge exceeds  the Raether limit ($10^7 \div 10^8$) electrons, depending on  the gas mixture,  an avalanche transforms to  a streamer followed by  a spark. Very high ionization may be easily produced by  an interaction of a slow proton or neutron in the detector volume.  
 Every discharge results in detector inefficiency for about 1 ms, effectively limiting the maximum flux capacity.   To mitigate this effect,  a special double-layer structure of readout electrodes will be used, as it is shown in Fig.~\ref{fig:SparkProtection}. 
The ''top'' or ''high voltage'' electrodes are made of a high-resistivity material  and serve to create a high-tension electric field in  the amplification gap. The second layer with copper ''pickup'' strips collects the signal initially induced on the HV strip and transports it to  the amplifier. Due to high resistivity, in case of  a spark, only   a small part of  the HV electrode (up to few mm) is discharged, while almost  the entire detector area remains active. The spark protection may be realized with the following techniques:  a strip-patterned resistive layer by screen printing, and uniform DLC (Diamond-Like Carbon). The second method provides better surface planarity and requires no precise resistive and pickup strips alignment. In addition, well-defined charge smearing over $\approx$3 strips~\cite{Iodice:2020jls} improves the accuracy of the charge centroid method. For these reasons we plan to use the DLC option in the MCT design. A slightly higher cost is not important in our case due to small detector size.    

\section{Bulk Micromegas technology}
\begin{figure}[htb]
    \centering
    \includegraphics[width=1.0\textwidth]{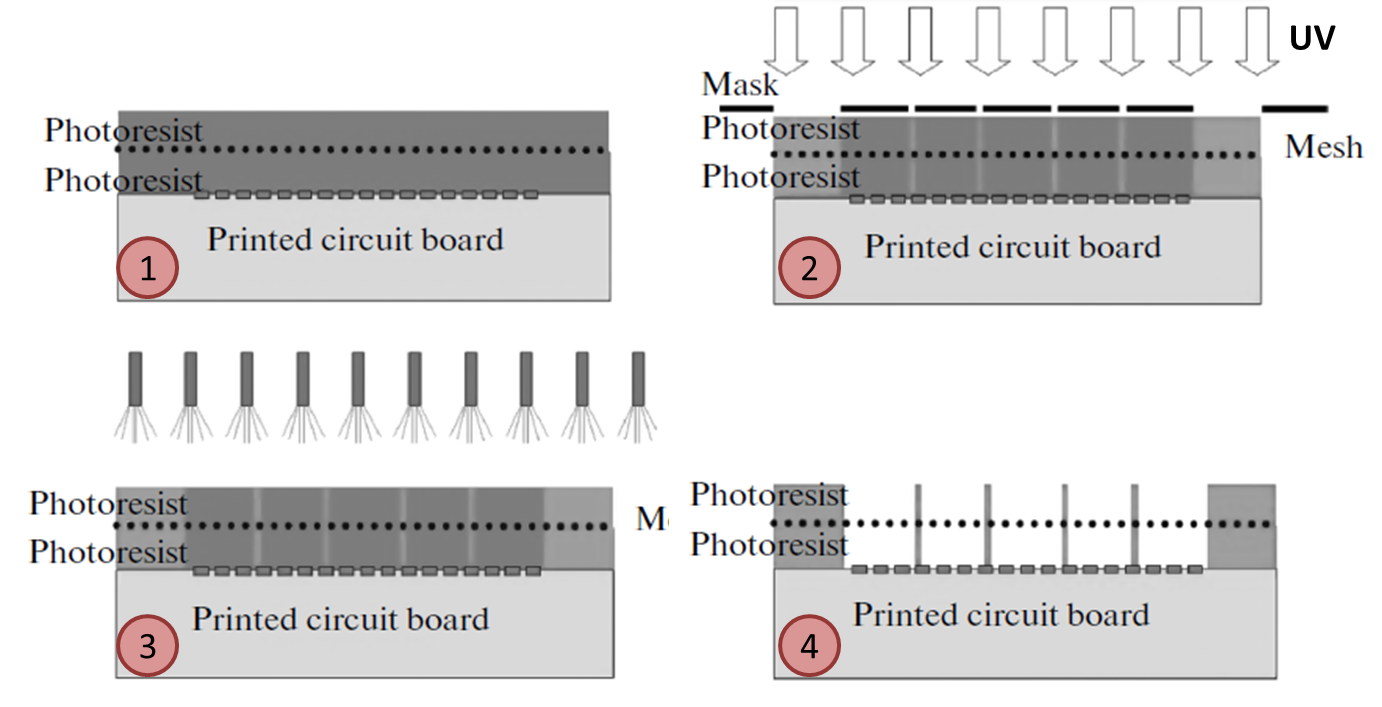}    
    \caption{Main bulk Micromegas production steps: 1 -- readout PCB lamination with photoresist and fixation of a pre-tensioned mesh; 2 -- exposition of photoresist to UV light through a mask corresponding to the desired pillar pattern; 3 -- etching of unexposed photoresist; 4 -- finished bulk Micromegas module (without the anode plane).}
    \label{fig:Bulk}
\end{figure}

There are several methods of building a Micromegas detector. We propose to use the bulk Micromegas technology \cite{Giomataris:2004aa} to build the SPD Micromegas Central Tracker. In this method,  the readout PCB,  the amplification gap, and  the mesh  are produced as an entire module using photo-lithography.  The production procedure includes the following steps:
\begin{enumerate} 
\item  a printed circuit board with readout strips is laminated with  a photoresist material. The thickness of  the photoresist defines the  amplification gap of the Micromegas detector.
\item stainless	steel mesh  pre-tensioned on  a temporary frame is applied over the PCB and  the photoresist  and is fixed by  an additional  photoresist layer.
\item	 the photoresist is exposed  to UV light through  a photomask.
\item	 the unexposed photoresist is removed by chemical etching. The exposed photoresist forms  an edge zone and pillars.
\end{enumerate}
The production procedure is illustrated  in Fig.~\ref{fig:Bulk}.

As a final step,  a cathode plane should be fixed at  a distance of  a few mm from  the mesh, to form  the ionization gap. The bulk technology is simple, reliable and cheap;  it allows one to use commercially available materials and equipment and build rather  large detectors. The PCB with  the mesh may be bent to build  a cylindrical detector, multiple fixation points prevent occurrence of  the mesh waves or folds. The main disadvantage of  this method is the  inability to disassemble the PCB+mesh module to fix possible problems. So here production is  a ''single attempt'' procedure. 

\section{Detector layer layout and production procedure} \label{sec:MM_bent}
 The SPD Micromegas Central Tracker will be produced using the spark-protected bulk technology.
Every single  layer is  an independent,  one-coordinate cylindrically bent detector. As  the MCT inner diameter is smaller than the  vacuum tube flange diameter, it should be finally assembled around  the fixed beam pipe.  For this reason, every detector layer will consist of two independent half-cylindrical parts. The main steps of  the half-cylindrical Micromegas detector production are listed below.
\begin{enumerate}
\item 
 The readout PCB+mesh module is produced using the ''bulk Micromegas'' technology.
\item
The ''Bulk module'' (the readout PCB with a fixed mesh) is bent on  a cylindrical template. Carbon composite long-beames  are glued on  the long edge of the PCB, and half-arcs are glued on  the end-face edge of  the  active area.
\item
The cathode PCB is glued on top of  the long-beams and  the arcs, forming   a drift  gap and closing  the gas volume.
\end{enumerate}
While  the readout and cathode boards themselves are very thin and flexible,  a detector layer after gluing is rigid enough  to be self-supporting due to its shape. Long-beams are empty inside and will be used as gas distribution pipes. The total PCB length is approximately 50 mm bigger than  the active area. This extra space will be used to put  a signal and high voltage connector and for gas communication. To provide  the required durability of the ''communication'' part of the layer, it will be reinforced by 3D printed plastic elements. 
\begin{figure}[htb]
    \centering
    \includegraphics[width=1.0\textwidth]{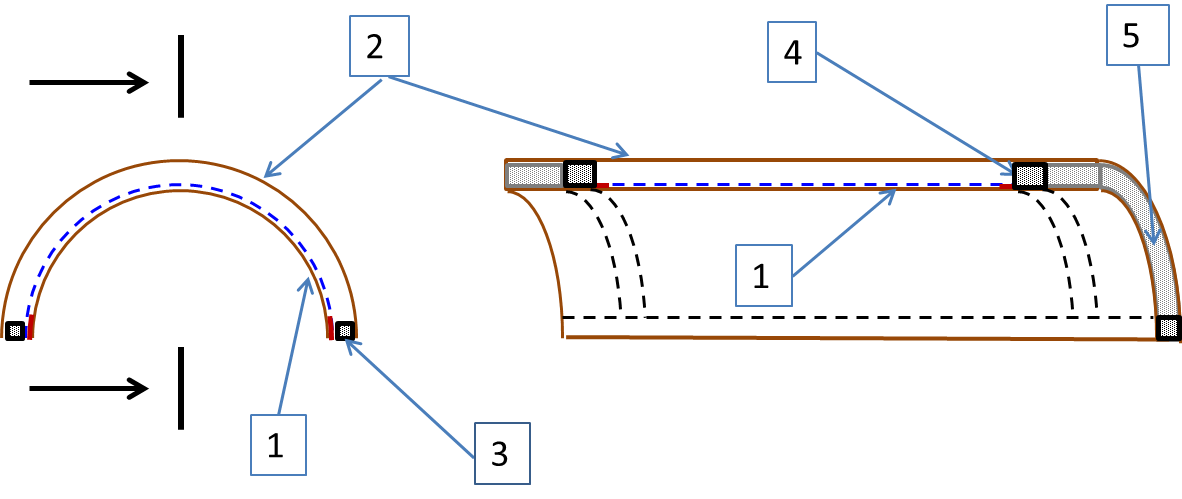}    
    \caption{Simplified sketch of detector half-layer (not to scale). Left: perpendicular cross-section. Right: longitudinal cross-section. 1 - Readout board with mesh and amplification gap; 2 -- cathode board; 3 -- carbon fiber long-beams; 4 - carbon fiber arcs; 5 -- 3D-printed plastic reinforcement. Passivated areas are shown by red color.}
    \label{fig:layer}
\end{figure}

A simplified sketch of  a half-layer is shown in Fig.~\ref{fig:layer}. The minimal layer radius is 50 mm; the total material budget is about 0.4\% of the radiation length per layer.  The micromegas design requires having some dead area near  the detector edges for mesh fixation. In our case, this area  is also used  for reinforcement elements (carbon fiber long-beams). We estimate  a minimal width of  the dead zone as 4$\div$5 mm,  which will result in a geometrical inefficiency  of about 6\% for  the innermost detector layer. 

\section{Front-end electronics} 
The Micromegas operation in  the $\mu$-TPC mode requires precise amplitude and time measurement.  Since one of the primary requirements for  the Micromegas Central Tracker is its moderate cost, FE boards will be based on the existing and available for order VMM3 ASIC \cite{Iakovidis:2018nwp}.

The VMM ASIC was developed to be used with sTGC and the MM detectors of the ATLAS New Small Wheel and are very flexible to be configured to  read out the most types of gaseous detectors. The ASIC provides a peak amplitude and time with respect to the bunch-crossing clock, or any other external signal.
\begin{figure}[htb]
    \centering
    \includegraphics[width=0.85\textwidth]{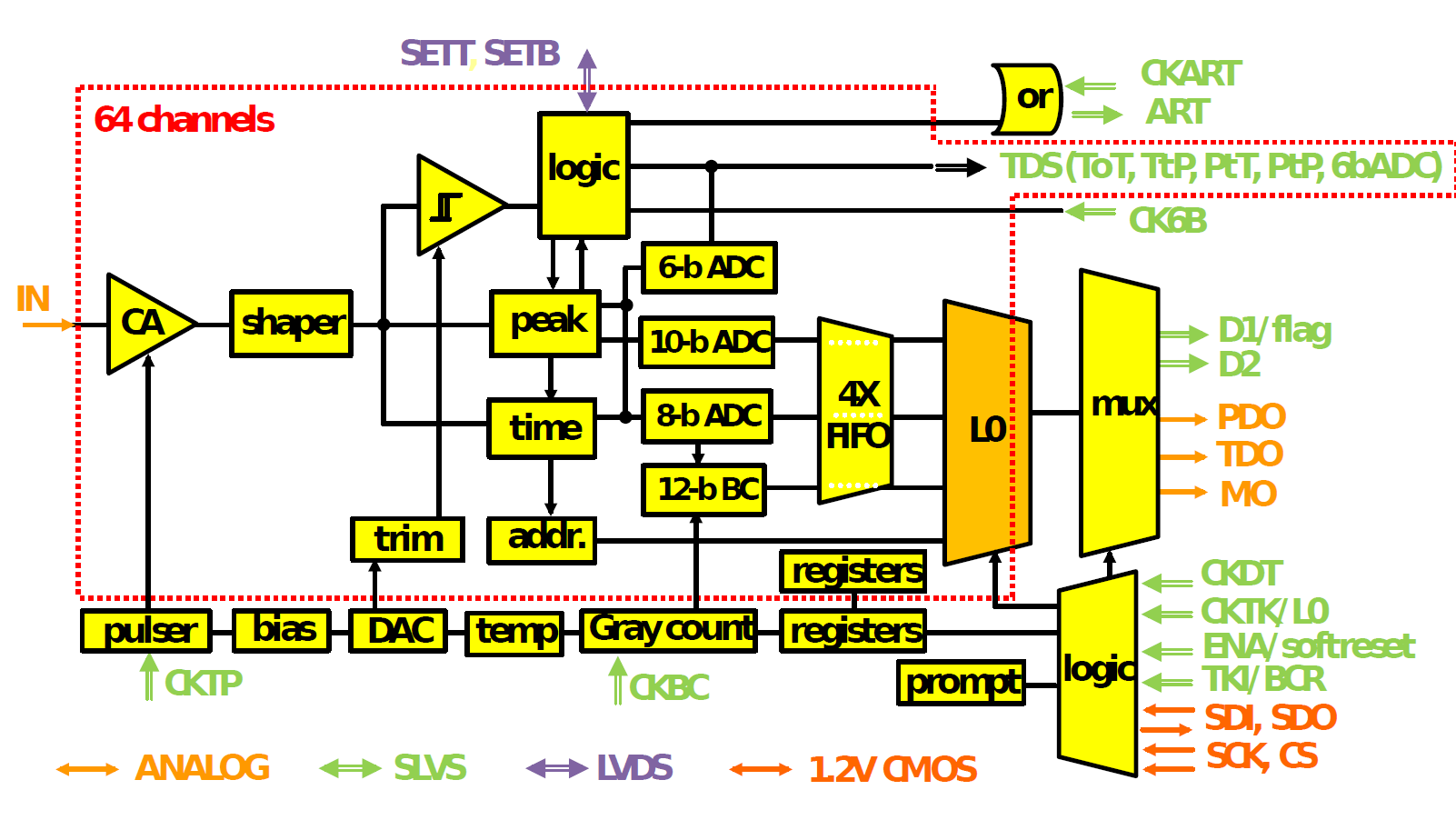}
    \caption{Block diagram of the VMM3 ASIC.}
    \label{fig:vmm}
\end{figure}
A block diagram of one of the identical channels is shown in Fig.~\ref{fig:vmm}. Each channel is equipped with a fast comparator with an individually adjustable threshold. When a signal crosses a set threshold, a peak detection circuit is enabled. Neighbour-enable logic allows the threshold to be set relatively high and yet very small amplitudes  to be recorded. At the peak, a time-to-amplitude converter is started and then stopped by the next bunch-crossing. The two amplitudes are digitized and stored in a de-randomizing buffer and read out serially with a smart token passing scheme that only reads out the amplitude, timing, and addresses of the active channels, thus dramatically reducing the data bandwidth required and resulting in a very simple readout architecture. The so-called ''continuous'' readout mode available for VMM3 allows simultaneous read/write of data, and provides dead-timeless operation that can handle rates up to the maximum of 4 MHz per channel. In addition to the main properties mentioned above, it includes a plethora of features that significantly increase its versatility. These include selectable polarity, gain (0.5, 1.0, 3.0, 9.0 mV/fC) and shaping time (from 25 to 200 ns). The integrated calibration circuit allows precise calibration of amplitude and time measurements on  the channel-by-channel basis. 

 The RD51 Collaboration have developed  a HYBRID128 front-end board based on the VMM3a ASIC, as a part of the Scalable Readout System project. This board is commercially available and fits most of the MCT requirements very well.  Its main limitation is insufficient radiation hardness,  but it is not critical for the SPD project  with quite moderate radiation conditions, see Chapter \ref{ch:rad}. We use  the geometrical characteristics of HYBRID128 (50 $\times$ 80 mm$^2$, 7 mm thickness without the cooling system) to estimate  the space needed to locate  the FE boards near  the detector.  

\section{Detector layout}
\subsection{Preliminary simulation}
To optimize the overall detector layout, preliminary simulations were performed for 3 variants of the detector geometry. Hereafter in the text they are referred to as MCT-1, MCT-2, and MCT-3.   The variants included one, two or three detector superlayers, respectively, each consisting of two or three layers of Micromegas cameras with different orientations of read strips.  The details of the configurations tested are summarized in Table ~\ref{tab:sim-variants}.   
 
\begin{figure}[htb]
    \centering
    \includegraphics[width=0.99\textwidth]{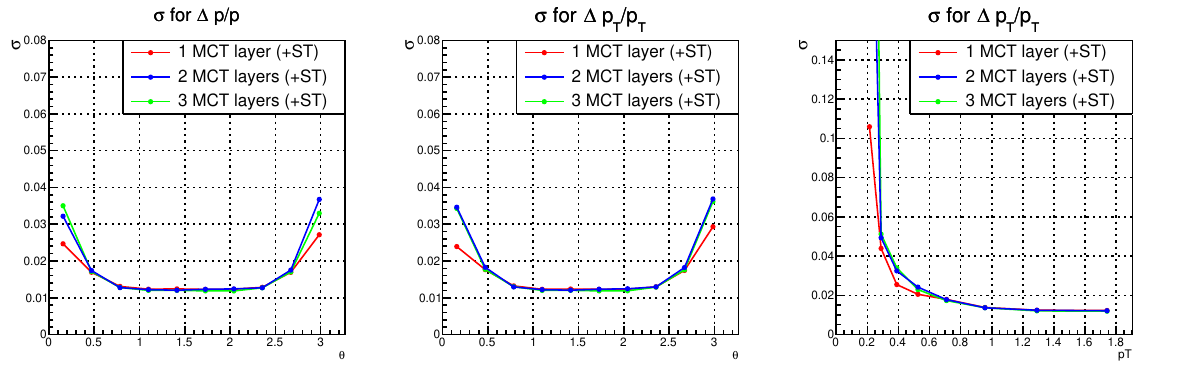}
    \caption{Simulation results: the momentum resolution of SPD setup as a function of the momentum and the polar angle for three variants of the Micromegas Tracker layout. The MCT-1, MCT-2, and MCT-3 configurations are shown in red, blue, and green, respectively.  Layout details are presented in Table \ref{tab:sim-variants}}
    \label{fig:MM_LayoutCompar}
\end{figure}

\begin{table}[htb]
\centering
\caption{Summary of the MCT detector layout variants used in the preliminary MC simulation.}
\label{tab:sim-variants}
\begin{tabular}{|p{1.5cm}|p{1cm}|p{1cm}|p{1cm}|p{1cm}|p{1cm}|p{1cm}|p{1cm}|p{1cm}|p{1cm}|}
 \hline
 \multirow{2}{*}{} & \multicolumn{3}{|c|}{Superlayer 1}& 
\multicolumn{3}{|c|}{Superlayer 2}& \multicolumn{3}{|c|}{Superlayer 3} 
\tabularnewline
\cline{2-10}
   &N layers &Strip angle & Min.R, mm& N layers &Strip angle & Min.R, mm& N layers &Strip angle & Min.R, mm 
\tabularnewline    
\hline
 MCT-1&3& $0^\circ,\pm5^\circ$ & 50 & &  &  &  &  &
\tabularnewline    
\hline
 MCT-2&3& $0^\circ,\pm5^\circ$ & 50 &   &   &   & 3 & $0^\circ,\pm5$ & 190
\tabularnewline    
\hline
MCT-3&3& $0^\circ,\pm5^\circ$ & 50 &  2 &  $\pm5^\circ$ & 120  & 2 & $\pm5^\circ$ & 190
\tabularnewline    
\hline

\end{tabular}
\end{table}
 The spatial resolution of one detector layer was assumed to be 150$\mu$.  The impulse resolution of the tracker system (MCT + Straw tracker) was estimated for a $90^\circ$ and $45^\circ$ 1 GeV track. The simulation results are summarized in Fig.~\ref{fig:MM_LayoutCompar}. 

From the point of view of raw resolution, the configuration with one superlayer located as close to the beam pipe as possible looks preferable. The multiple scattering in the outer layers completely negates the effect of the larger number of measurement points in the MCT-2 and MCT-3 configurations. In addition, the small number of readout channels for the MCT-1 variant allows the detector to be split into two parts along the beam axis.  This will provide a number of important advantages:
\begin{enumerate}
    \item  the occupancy and the count rate per channel of the inner layer will be reduced.
    \item the strip capacitance and, consequently, the noise of the readout electronics will be reduced. This is especially important for the triggerless DAQ system that will be used by the SPD experiment.
    \item the size of printed circuit boards for chamber production will be reduced to 50$\div$55 cm, which is within the standard limitations of the contract manufacturers.
\end{enumerate}
 The dead zone with a width of about 1 cm in the center of the detector is not critical because the length of the beam collision region exceeds 60 cm. Tracks passing through the dead zone (about 2\% of the total track number) will be reconstructed by straw tracker with a slightly worse resolution.

\subsection{General layout}
\begin{figure}[htb]
    \centering
    \includegraphics[width=1.0\textwidth]{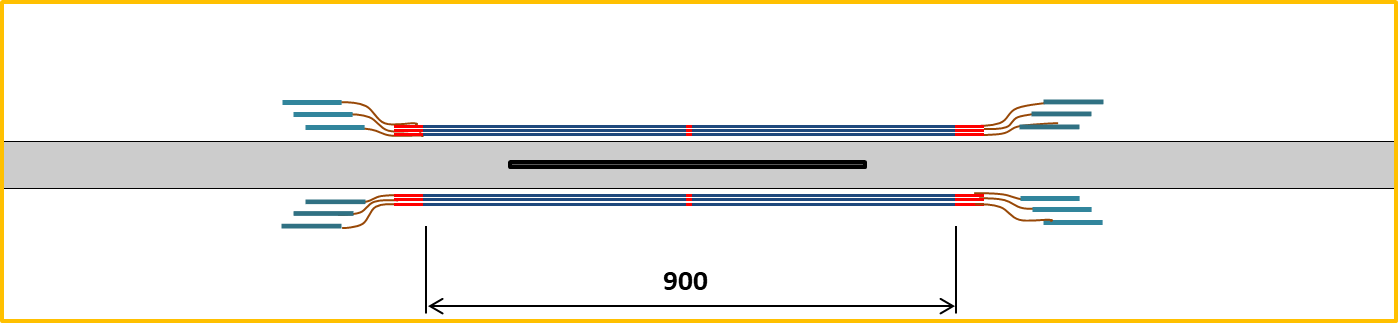} 
    \caption{Layout of the MCT. Blue color marks the active area of the layers. The dead areas are  marked by red and the FE card by turquoise.  The inner border of  the Straw Tracker is shown by a yellow line.  The beam pipe is shaded by gray color, the black area in the middle of the beam pipe is the beam-crossing area.}
    \label{fig:MiVD}
\end{figure}
The general detector layout of the MCT is shown in Fig.~\ref{fig:MiVD}.
Three layers of cylindrical MM cameras are located as close as possible to the beam pipe.  Each detector layer is divided into two independent parts along the beam axis ($Z>0$ and $Z<0$).  The dead zone between the parts is 1 cm. The length of the sensitive area of the detector 2$\times$45 cm is determined by the size of the beam collision area.  Detailed information on the detector geometry is summarized in Table \ref{tab:MM_properties}.
\begin{table}[htb]
    \caption{Main characteristics of the suggested Micromegas Central Tracker.}
    \centering
    \label{tab:MM_properties}
\begin{tabular}{|p{.7cm}|p{1.2cm}|p{0.9cm}|p{1.8cm}|p{1.2cm}|p{1.5cm}|p{1.2cm}|p{2cm}|}
\hline 
Layer  & Radius, cm & Strip angle & Active area length, cm & Pitch, mm & Number of channels & Number of FE boards & Geometrical acceptance \tabularnewline
\hline 
\hline 
1  & 5.0-5.4 & $0^{\circ}$ & $2\times 45$ & 0.4 & $4\times 368$ & 12 & \multirow{3}{*}{$11^{\circ}\!-\!169^{\circ}$}\tabularnewline
\cline{1-8} 
2  & 5.4-5.8 & $+5^{\circ}$ & $2\times 45$ & 0.42 & $4\times474$ & 16 & \tabularnewline
\cline{1-8} 
3  & 5.8-6.2 & $-5^{\circ}$ & $2\times 45$ & 0.42 & $4\times504$ & 16 & \tabularnewline
\hline 
\end{tabular}
\end{table}

 The signal is transmitted from the detector to the readout boards by means of thin flat cables. The readout boards are located $\approx 10$ cm from the detector ($|Z|\approx 60$ cm as close as possible to the beam pipe.
 Due to  the substantial power dissipation,  the FE boards need active cooling. Presently, water cooling is considered as  the main option. The cooling bar will be done as an integrated part of the support structure.  With a minimal cooling bar thickness of 6 mm, the  FE board stack will have  a maximum radius $R\approx 116$ mm. That limits the detector acceptance to $10.8^\circ\leq\theta\leq169.2^\circ$. 

\section{Simulation of detector performance}

A Monte Carlo simulation of the SPD detector with and without the Micromegas Central Tracker was performed to evaluate the detector performance.
The geometrical characteristics of the MCT corresponded to those given in Table \ref{tab:MM_properties}, the coordinate resolution of one detector plane was assumed to be 150 $\mu$.  The distribution of collision points along the beam axis was assumed to be Gaussian with a 60 cm width, and 6 GeV center-of-mass energy minimal bias events were used as the data set.
\begin{figure}[htb]
    \centering
    \begin{tabular}{cc}
    \includegraphics[width=.5\textwidth]{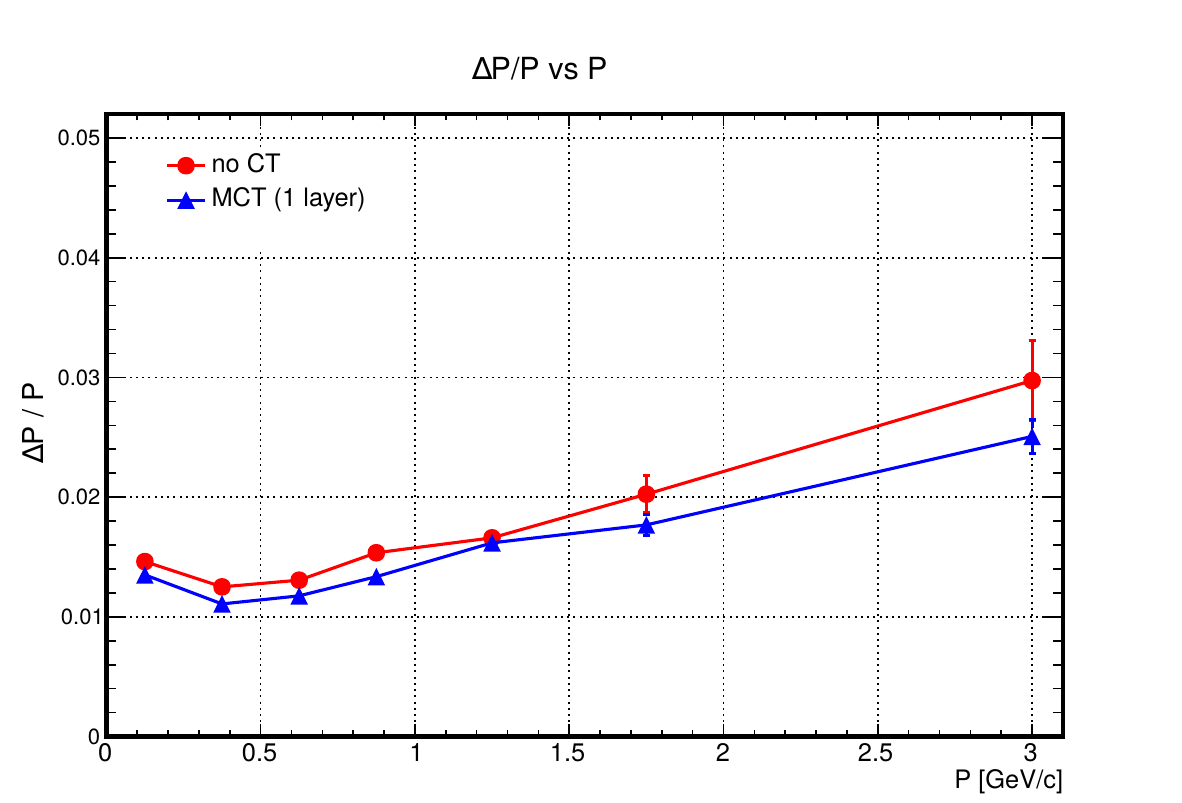} 
    \includegraphics[width=.5\textwidth]{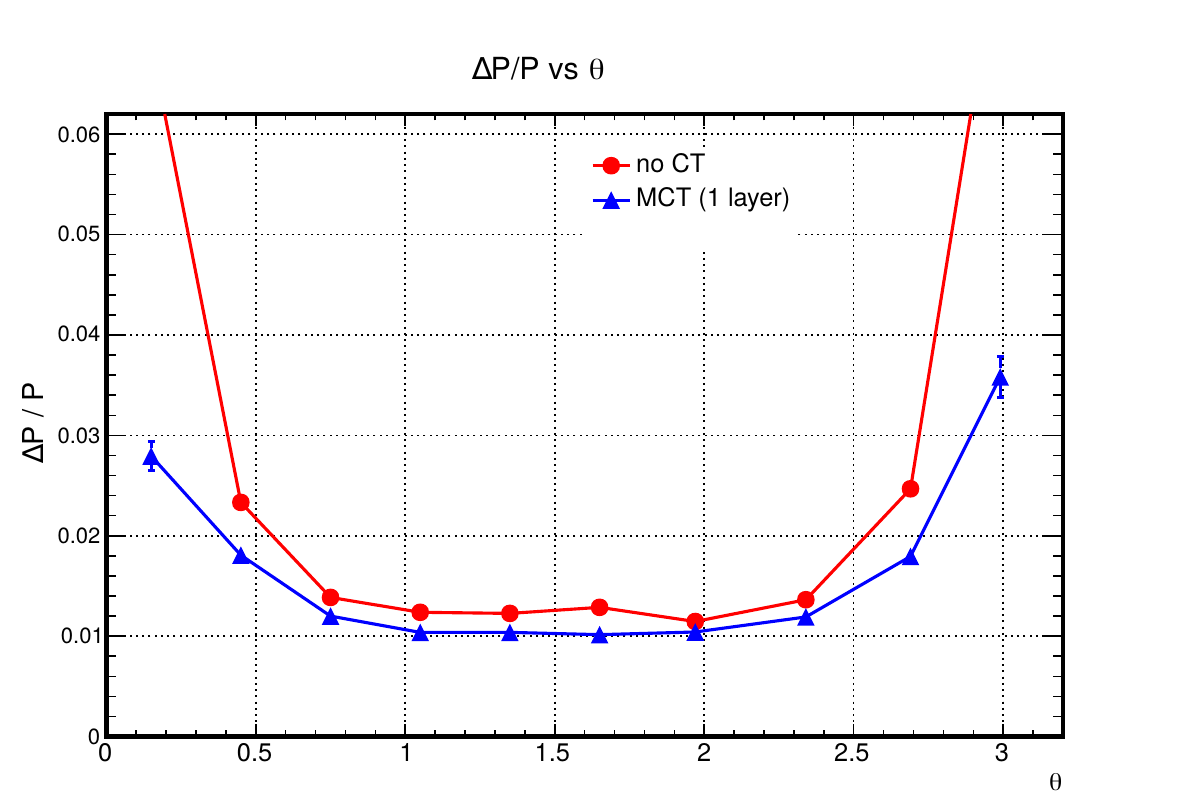} \\    
    \end{tabular}
    \caption{ Simulation results: momentum resolution of the SPD setup with (blue) and without (red) the Micromegas Central Tracker as a function of the momentum (left) and the polar angle (right).}
    \label{fig:MM_SimMomentum}
\end{figure}

\begin{figure}[htb]
    \centering
    \begin{tabular}{cc}
    \includegraphics[width=.5\textwidth]{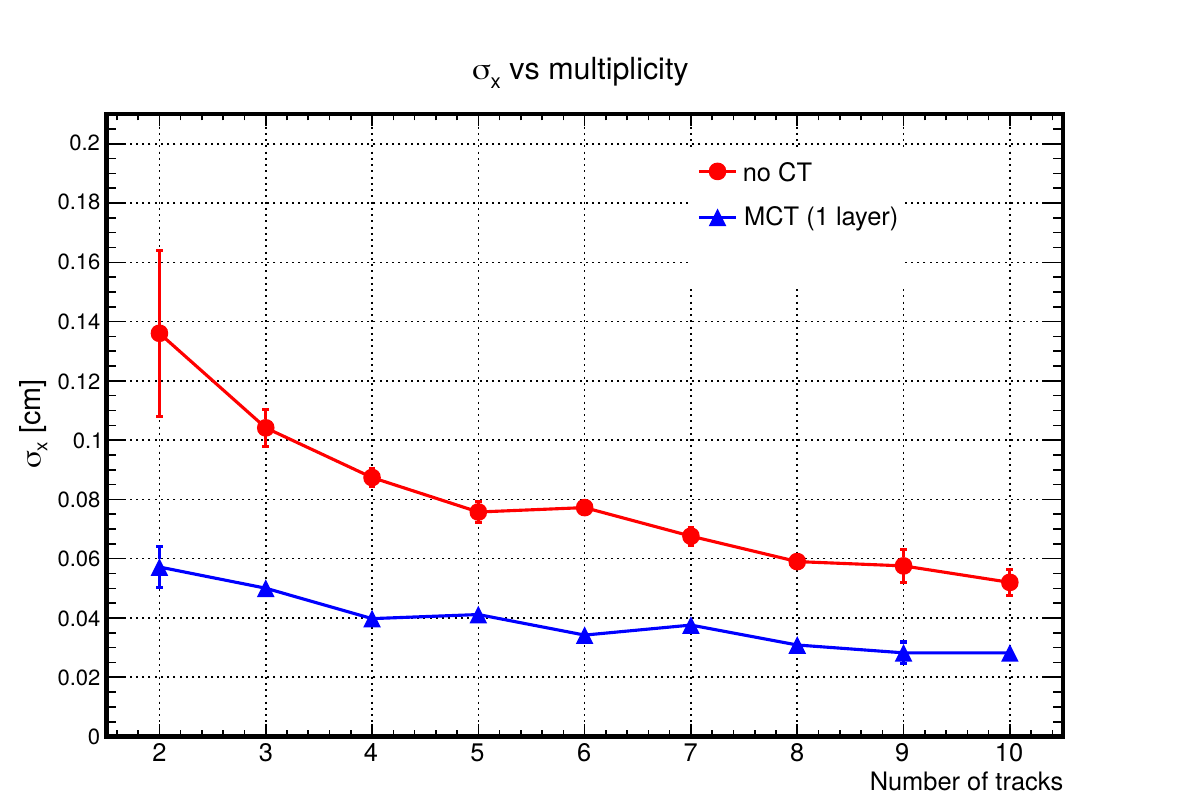} 
    \includegraphics[width=.5\textwidth]{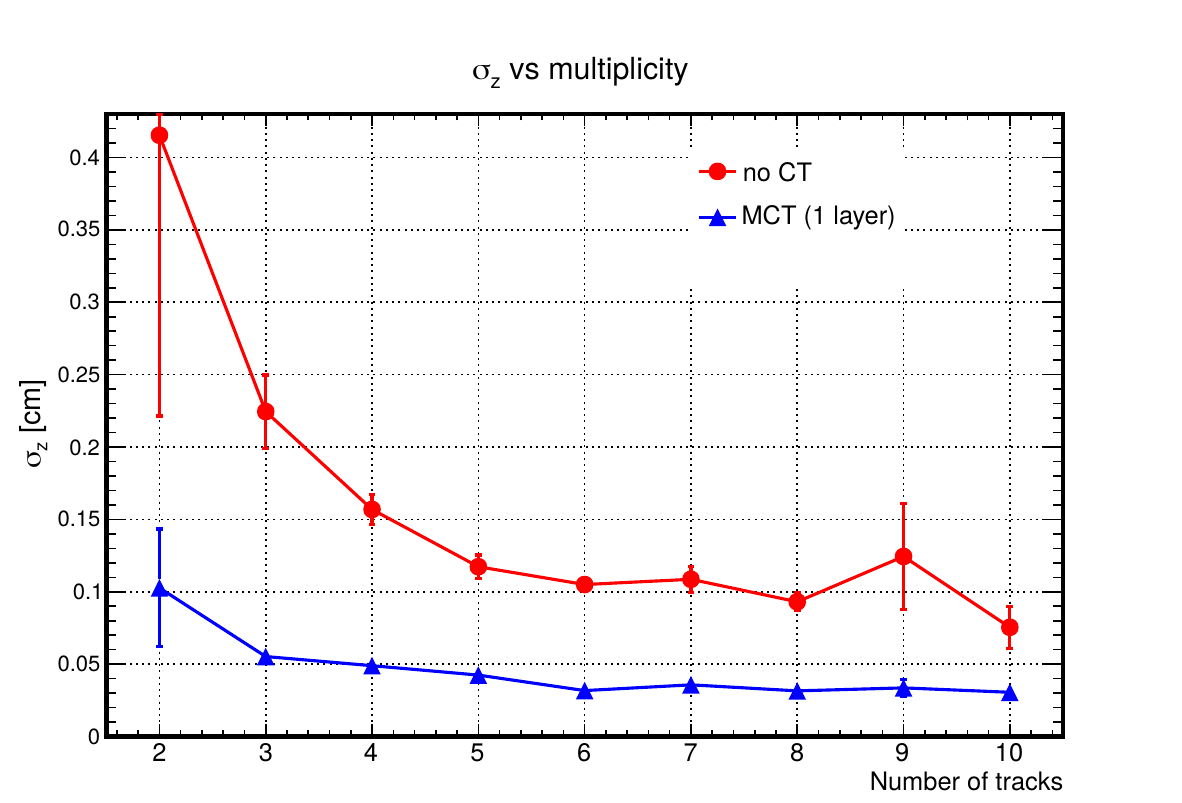} \\     
    \end{tabular}
    \caption{ Simulation results: vertex reconstruction accuracy in $r\phi$ plane (left) and along beam axis(right) with (blue) and without (red) Micromegas Central Tracker }
    \label{fig:MM_SimVertex}
\end{figure}

The dependence of the momentum resolution on the track momentum and the polar angle were evaluated.  The simulation results are shown in Fig.~ \ref{fig:MM_SimMomentum}.  Additionally, the accuracy of the vertex reconstruction as a function of the number of outgoing tracks was evaluated (Fig \ref{fig:MM_SimVertex}). 
Obviously, the MCT does not provide the accuracy necessary to distinguish $D$ meson decays, but it can be useful for the reconstruction of $K_s$ and $\Lambda$.

\section{Water cooling } 
To estimate the total power consumption of the reading electronics, we used real data of the typical consumption for MMFE-8 cards. These cards are used by the ATLAS Collaboration to read signals of Micromegas New Small Wheels cameras. Each card contains 8 64-channel VMM-3A chips, specialized ASICs for implementation of trigger logic and data exchange, and local voltage converters. Typical power consumption of such a card was 30 W. Assuming that the power consumption is proportional to the number of VMM chips and assuming a safety factor of 1.5, we expect that the maximum consumption of one 128-channel card will not exceed 12 W. The total heat dissipation of the MCT readout electronics in this case will be about 1.5 kW, and it depends on the hit rate of the cameras. The drift characteristics and gain of the gas detectors are strongly affected by temperature, so the cooling system must ensure a stable temperature of the FE cards. Water cooling is the easiest and most reliable way to meet these requirements. 

\begin{figure}[htb]
    \centering
    \begin{tabular}{cc}
    \includegraphics[width=.5\textwidth]{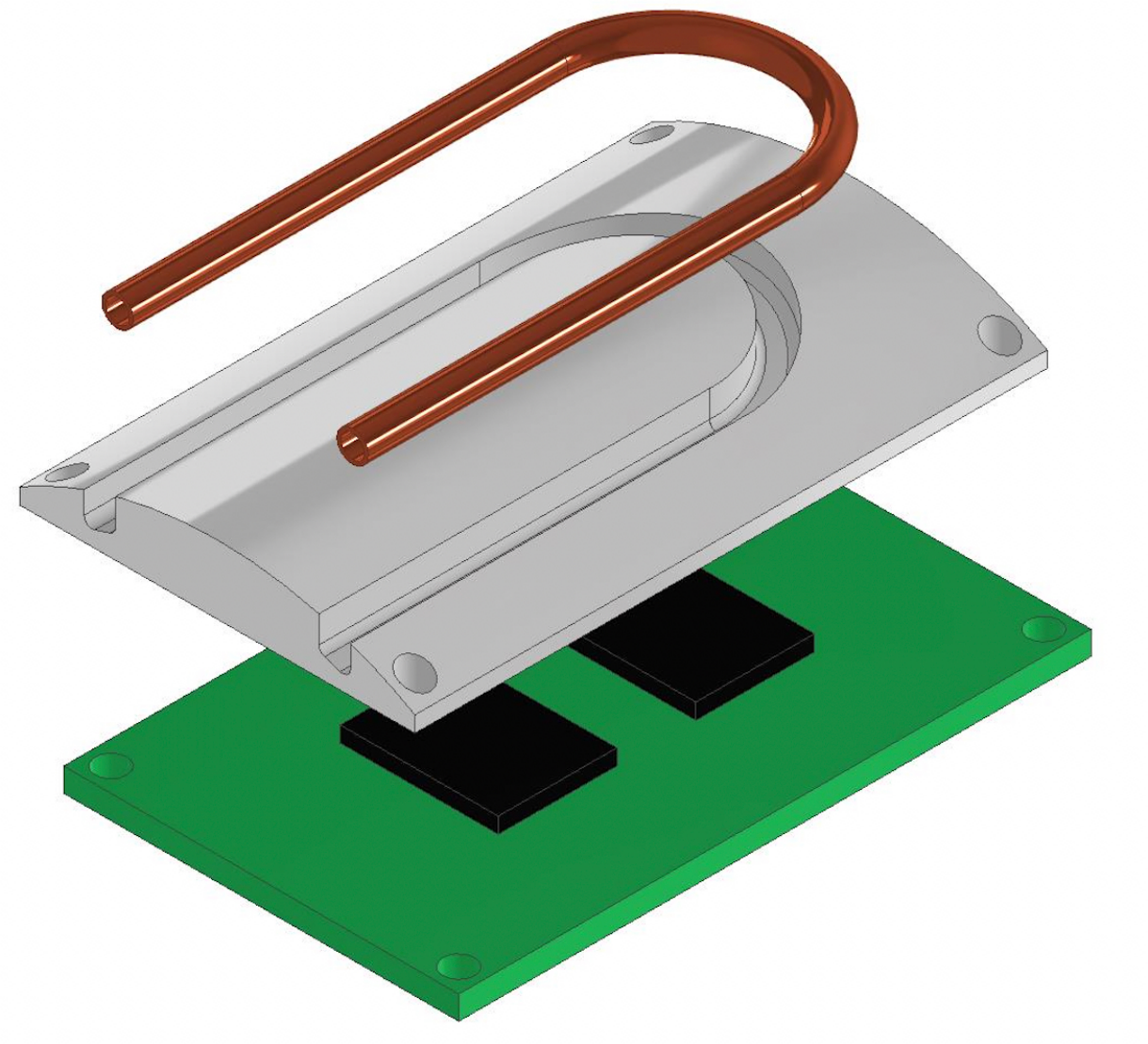} 
    \includegraphics[width=.5\textwidth]{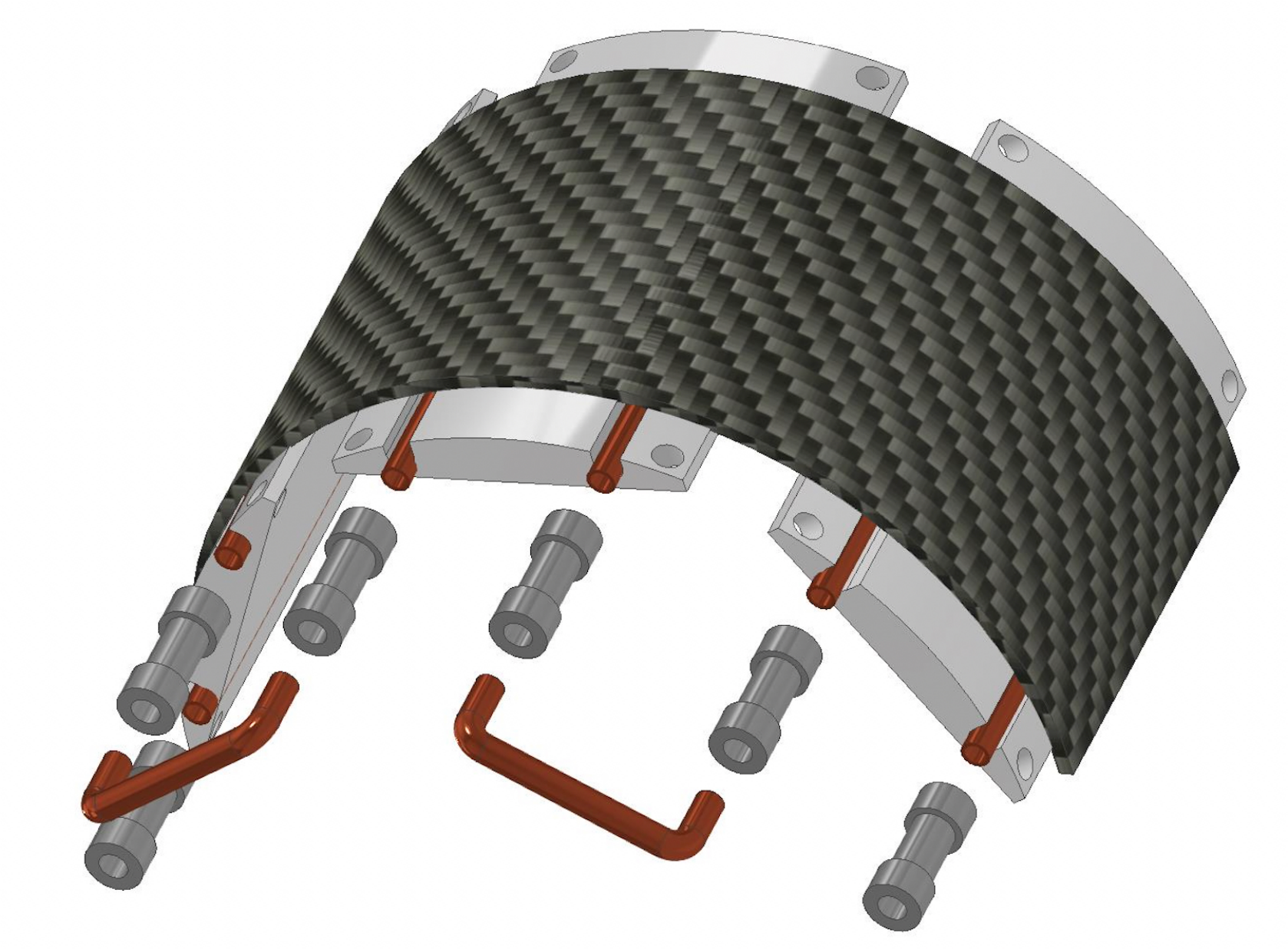} \\     
    \end{tabular}
    \caption{(a) Exploded-view diagram of a single cooling plate with the FE board. (b) Schematic view of the carbon fiber support structure with cooling plates fixed to it.}
    \label{fig:cooling}
\end{figure}

It is proposed that the individual cooling plates be rigidly attached to a supporting carbon fiber structure. The FE boards will be attached to the plates with screws. In this way, the cooling plates will simultaneously act as mounting pads for the readout boards. Thermal contact between the heat-generating elements and the cooling plate will be provided by thermal pads. The proposed design of the radiator is shown in Fig. \ref{fig:cooling}~(a). It consists of an aluminum plate with a U-shaped groove, in which a 4$\times$2.5 mm metal tube is soldered in. A schematic illustration of one layer of the support structure with radiators is shown in Fig. \ref{fig:cooling}~(b). The radiators in one layer are connected in series by copper or polyurethane tubes 4$\times$2.5 mm. The "branches" of the radiators  are connected to a larger opening manifold. The calculations show that at 
a pressure of 1 bar in the cooling system, the  total length of pipes less than 2 m, and a heat dissipation of 120 W, the temperature difference of water between the input and output will not exceed 2.5 degrees. 

\section{Limitation on the spark protection layer resistance}
\begin{figure}[t!]
    \centering
    \includegraphics[width=.7\textwidth]{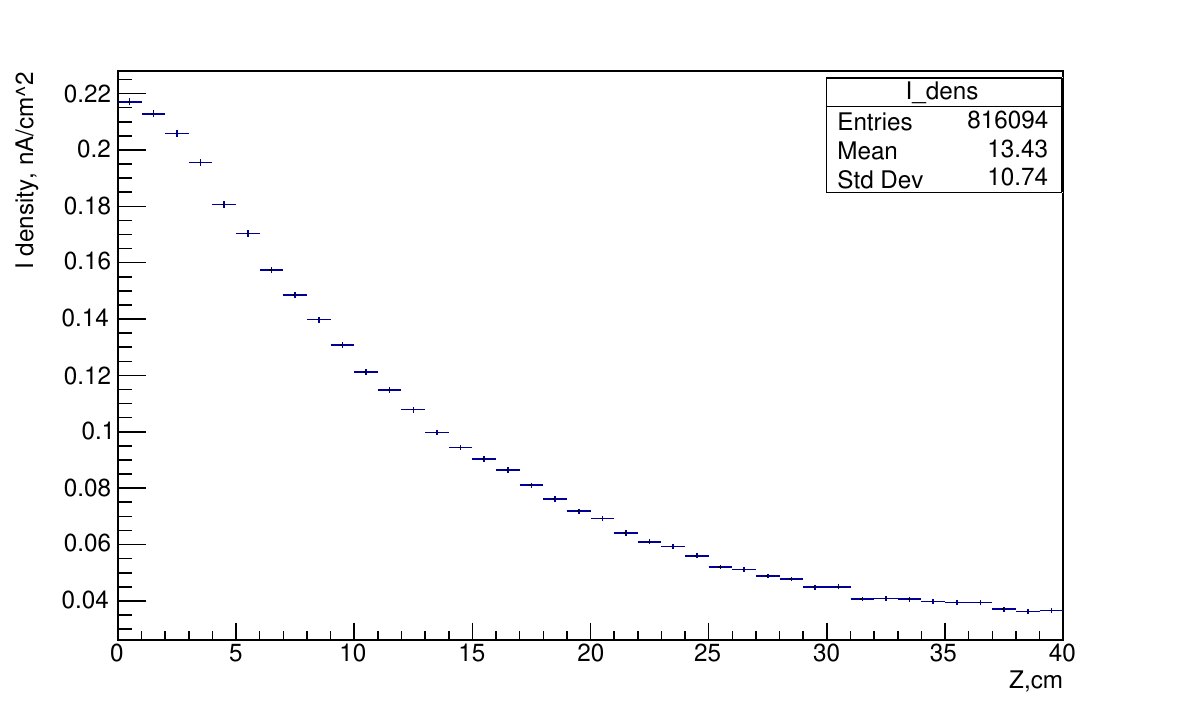} 
    \caption{Expected signal current density of the innermost layer of the MCT detector (simulation result).}
    \label{fig:Idens}
\end{figure}

While the resistive layer mitigates the effect of discharges on the detector efficiency, it also results in some voltage drop in the center of the detector and the following gain reduction. The effect depends on the protective layer sheet resistance $\rho_{sq}$, detector width $W=\pi R$, and the distribution of the signal current density $j,[nA/cm^2]$ over the detector area. Fig. \ref{fig:Idens} shows the current density distribution obtained from the simulation. The primary ionization was defined by the track polar angle, particle momentum and type. The gas gain was assumed to be $10^4$. The maximum signal current density is about $0.22 nA/cm^2$ for the center of the innermost layer of the MCT. 

To be conservative,  we assume that the current density is the same over the detector area and is equal to the maximum one obtained from the simulation. In this case, the voltage drop at the center of the detector can be calculated analytically $ \Delta U[V]=\frac{1}{8}\rho\pi^2R^2j \approx 7\times 10^{-3}\times \rho_{sq}[M\Omega]$ for the first stage of the experiment. For operation at full luminosity, a similar estimate would be $\Delta U[V] \approx 10^{-1}\times \rho_{sq}[M\Omega]$. We consider a 5 V voltage drop as acceptable  and   conclude that spark protective layer resistance  $\rho \leq 10 M\Omega $ is suitable for detector operation with a safety factor 5.

\section{Detector occupancy and gas mixture choice}
\subsection{Requirements}
The nominal CM energy in the SPD experiment is 27 GeV with a luminosity of about $ 10^{32} cm^{-2}s^{-1} $.  These values result in  about 3 MHz  rate of inelastic events and  an average event multiplicity $\approx$10 tracks per a  minimal bias event. 
The Micromegas Central Tracker is planned to be used only at  the first stage of  the detector operation with a beam collision energy below 10 GeV and a luminosity at least an order of magnitude below nominal.  Nevertheless, taking into account  the unknown beam-induced background, we use the nominal luminosity and energy to estimate  the detector working conditions.

\begin{figure}[htb]
    \centering
    \includegraphics[width=0.5\textwidth]{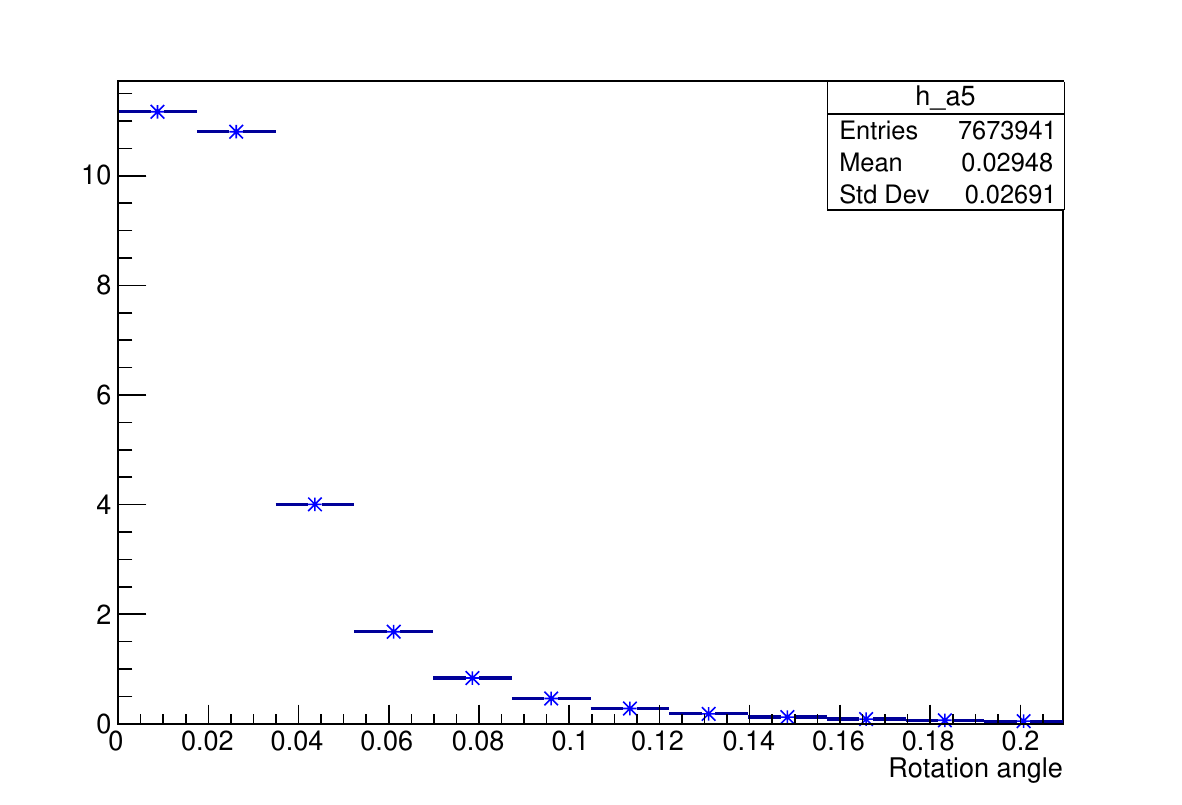} 
    \caption{Track bending angle distribution for $r=5cm$, simulation result.}
    \label{fig:rotation_angle}
\end{figure}

 A single strip count rate depends on  the  track multiplicity and  an average number of hits per cluster $N_{cl}$. For  the  perpendicular tracks without  the magnetic field,  the  last value is defined mainly by diffusion, for  the  common gas mixtures $N_{cl}\approx 2$. In a real experiment,  the   cluster size increased with  the  track bending in the magnetic field, Lorentz angle, and  the  gas gap. For  the  MCT,  the  average track  bending angle is small(Fig.~ \ref{fig:rotation_angle}) and may be neglected.
 
For  a  rough estimate of  the  hit rate, we take $N_{cl}=s\times tan(\Theta_L)/d+2$, where $s$ is  a drift gap   depth, $\Theta_L$ is  a  Lorentz angle, and $d$ is  a  signal strip pitch. For  the  common gas mixtures like $Ne-C_2H_6(10\%)-CF_4(10\%)$, used by  the  COMPASS experiment, or $Ar-CO_2(7\%)$, initially proposed for ATLAS,  the  Lorentz angle is rather large ($\Theta_L \geq\ 40^{\circ}$).  That results in  a  single strip count rate  of about 150 kHz and a single event detector occupancy of about 5$\%$. Despite being manageable, these values are too high for  a  "temporary" detector.  The obvious solution  is  to find  a  combination of gas mixtures and  operation parameters with  a  much smaller Lorentz angle, ideally, below $10^\circ$.

Another two parameters affected by  the  Lorentz angle  are  signal amplitude and space resolution. For  the triggerless operation, noise will not be suppressed by time coincidence, so  the threshold must be rather high. As  the signal amplitude is inversely proportional to  the cluster size,  a small Lorentz angle gives  a clear benefit. A small cluster size ($\approx 1 mm$ or 2$\div$4 strips)  also allows to use  a simple and reliable charge-weighing method for  the coordinate measurement.
Finally, the DAQ and on-line reconstruction form the requirement  to a  maximum electron drift time  and  full signal length  within 300 ns.

Summarising,  a good gas mixture for the MCT must provide:
\begin{itemize}
    \item stable long-term operation at gas gain $G\approx10^4$;
    \item small Lorentz angle;
    \item short enough signal with full length $T_{full} \approx T_{drift} + T_{ion} \simeq 300$ ns. Here, $T_{ion}$ is  the full ion drift time in  the amplification gap and $T_{drift}$ is  the maximum primary electron drift time.
\end{itemize}

\begin{figure}[t!]
    \centering
    \includegraphics[width=1.0\textwidth]{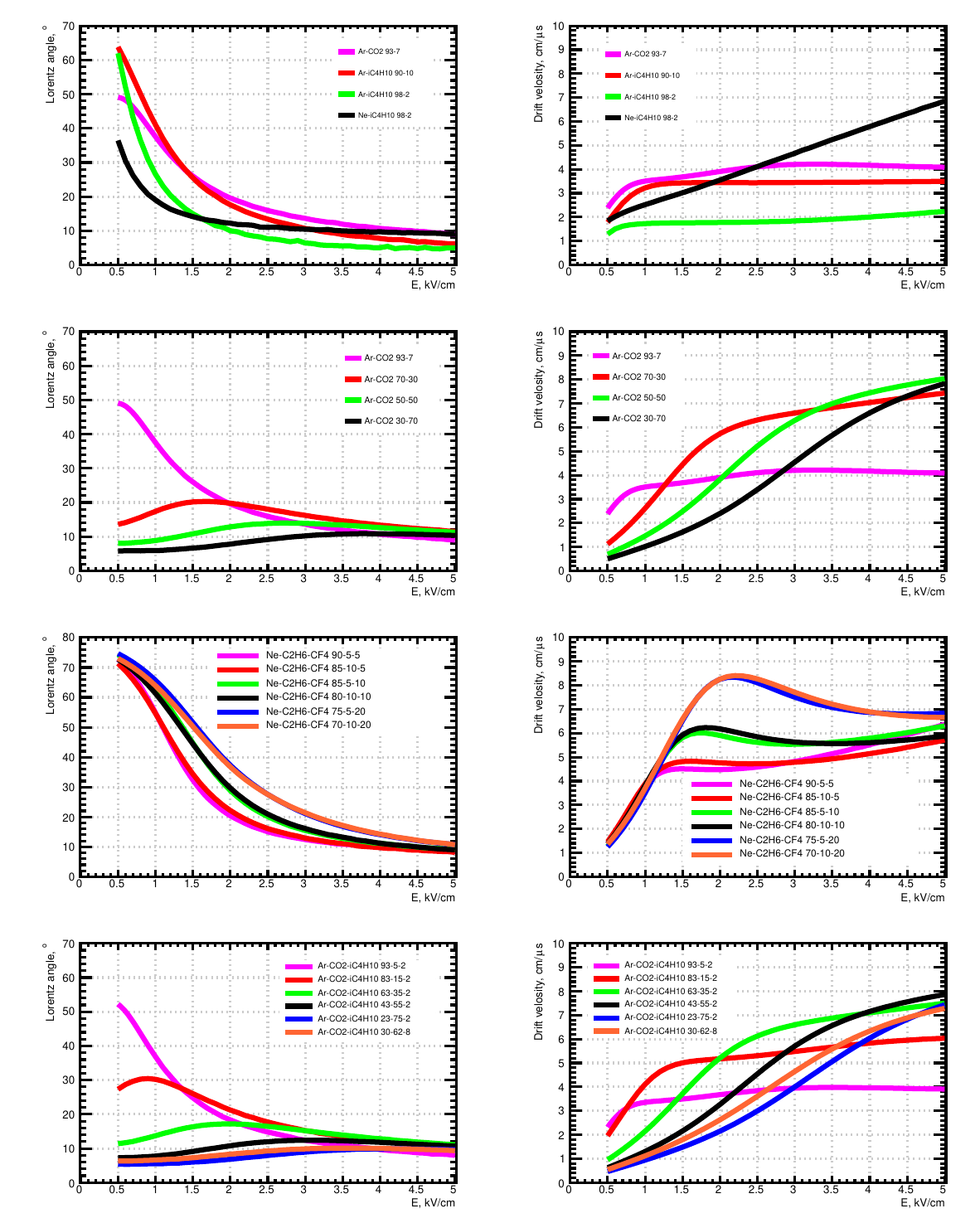} 
    \caption{GARFIELD simulation results for the Lorentz angle (left) and electron drift speed (right) vs the electric field tension (B=1 T).}
    \label{fig:all_gases}
\end{figure}

\subsection{Overview of gas mixtures}
The expected dependence of  a Lorentz angle and electron drift speed on  the electric field tension for several gas mixtures was studied with  the GARFIELD \cite{Veenhof:1993hz,Veenhof:1998tt} simulation.  The magnetic field was set to $B=1$ T.  The result summary  is presented in Fig.~\ref{fig:all_gases}, and a short overview is given below.  

 The $Ne-C_2H_6-CF_4$ mixture was used by  the COMPASS experiment for  the Micromegas operation. The main advantages of  the $Ne$-based mixtures  are a low spark rate and  a high achievable gas gain. Unfortunately, these mixtures have  a too high Lorentz angle in  the magnetic field 1~T and  in the reasonable drift fields. An additional drawback is  a low primary ionization. We do not consider $Ne$-based mixtures as good options for  the  Micromegas Central Tracker operation.

 The $Ar-CO_2$ and $Ar-CO_2-iC_4H_{10}(2\%)$ mixtures with a $CO_2$ fraction of $5-7\%$ were studied by ATLAS in the frame  of  the New Small Wheel project.  For an optimum drift field of about 0.6 kV/cm, these mixtures  have a too large  Lorentz angle , which can be reduced to an acceptable value by increasing the drift field to 3 kV/cm. According to the simulation results, the corresponding reduction in the charge collection efficiency would be $30-40\%$. The drift characteristics of these mixtures are quite similar. In general, a small $2\%$ isobutane admixture  increases the achievable gas gain while keeping the mixture non-flammable.      

 The $Ar-iC_4H_{10}(10\%)$ mixture was used by  the CLAS12 Collaboration for the Micromegas operation in  a  strong magnetic field  of 5 T. A very high drift field of 8 kV/cm was used to reduce  the Lorentz angle below $20^\circ$ and reach $\approx 150$ $\mu$m spatial resolution at the price   of  a substantial loss of the effective mesh transparency  \cite{Konczykowski:2010zz}, \cite{Charles:2013cvy}. Due to  a much smaller magnetic field (1T) in  the SPD spectrometer,  the 2.4 kV/cm drift field will be big enough to get $\Theta_L \approx 15^\circ$ and   $T_{drift}\approx100$ ns with an acceptable amplitude loss of $15\div30$ \%. The only real drawback is  the flammability of this mixture.   
Since in the first phase of the SPD experiment it is not planned to use flammable mixtures for the other subsystems, we consider $Ar-iC_4H_{10}(10\%)$ as very good and tested in a real experiment reserve option, which can provide  a guaranteed result, unless  a better variant is found. 
 
 The $Ar-CO_2$  mixtures with  a high $CO_2$ fraction ($\approx 70\%$) look promising due to a very low Lorentz angle.  This mixture is too slow in  a weak electric field, but  the electron drift speed rises fast with  the field tension, providing both  Lorentz angle $\Theta_L=8^\circ$ and $T_{drift} \approx 100$ ns with a 2.5 kV/cm drift field. The very first test (see next sections) demonstrates that in addition to excellent drift characteristics  these mixtures can provide a stable detector operation with gas gain $G\approx 10^4$.  
According to the simulation results, this gas mixture has the highest density of primary ionization clusters (32 $cm^{-1}$) and provides a charge collection efficiency above $90\%$. This allows us to set a sufficiently high threshold equivalent
to 4 electrons of primary ionization at a registration efficiency of $98\%$.

 Summarizing, this time we consider the $Ar-CO_2$ mixtures with a $CO_2$ fraction of about $70\%$ as a primary option and the $Ar-iC_4H_{10}(10\%)$ mixture as a backup solution. No specific detector feature is needed to operate any of the above-mentioned mixtures, the final decision will be made after additional tests.

\section {Gas mixture tests} 
\begin{figure}[htb]
    \centering
    \includegraphics[width=0.8\textwidth]{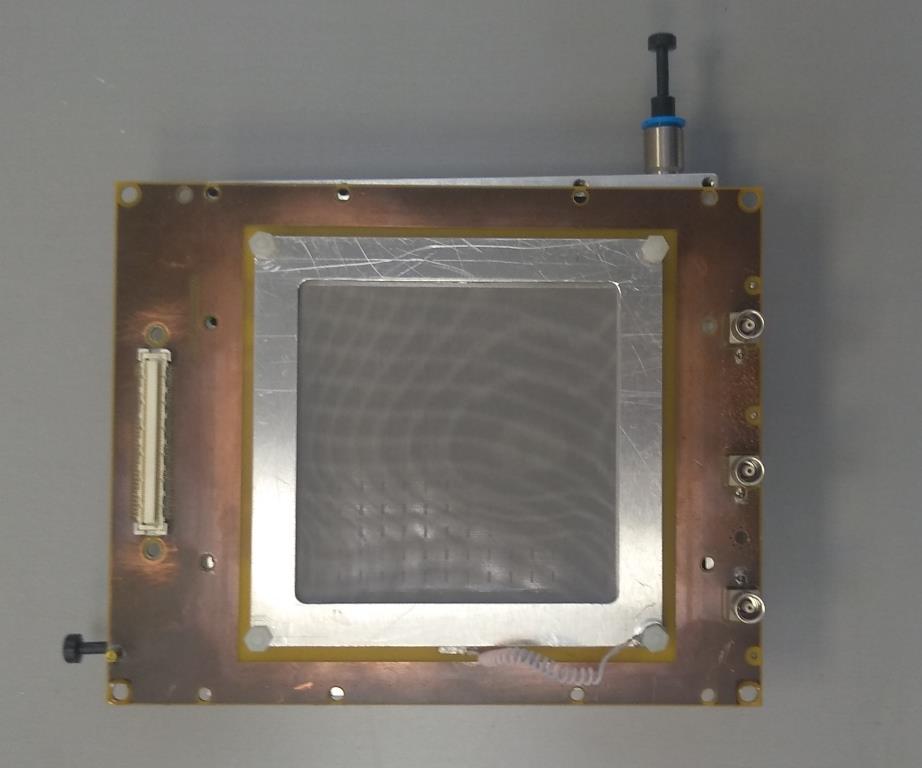} 
    \caption{Micromegas prototype with spark protection DLC layer used for the preliminary gas test.}
    \label{fig:photo}
\end{figure}

A small spark-protected Micromegas prototype with an 8$\times$ 8 cm$^2$ active area and  a DLC resistive layer with a sheet resistance of  about 100 M$\Ohm$ was produced for the very first test. The signal strip pitch was 0.5~mm, amplification gap 0.128 mm, and conversion gap 5 mm. The detector photo is shown in Fig.~\ref{fig:photo}.
All readout strips were connected in parallel to a fast spectrometric amplifier with rise time $\tau_{rise}=7$ ns and decay time $\tau_{decay}=100$ ${\mu}$s. This setup allows one to measure both gas gain with a $^{55}$Fe source and  the MM signal duration for  a single-cluster signal ($^{55}$Fe) and cosmic muons. 
\begin{figure}[htb]
    \centering
    \begin{tabular}{cc}
    \includegraphics[width=.45\textwidth]{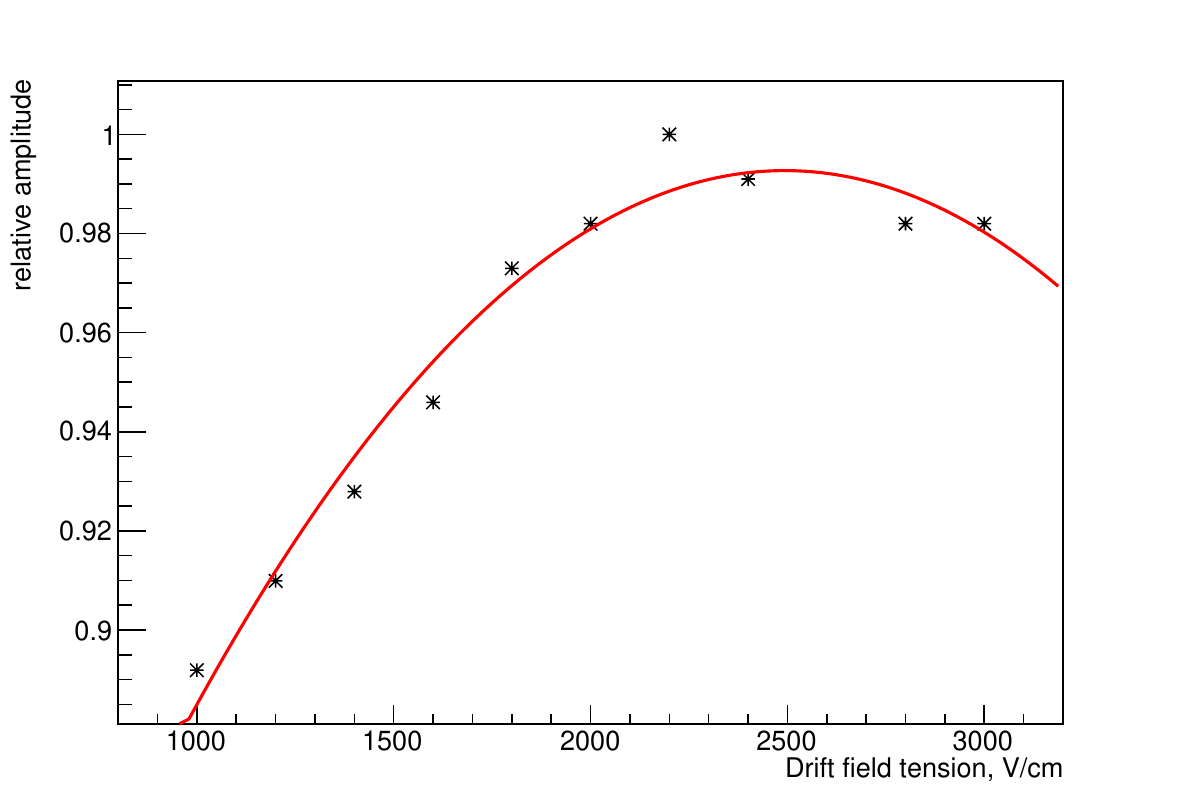} 
    \includegraphics[width=.45\textwidth]{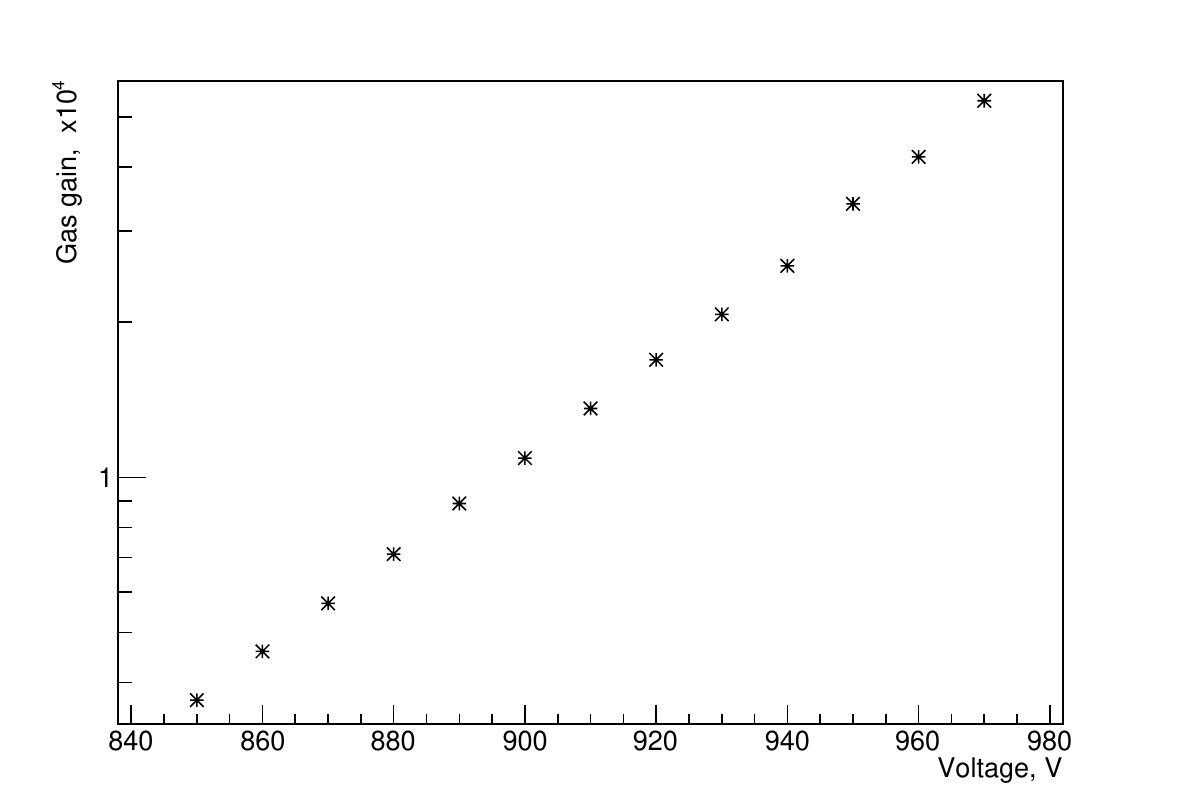} \\     
    \end{tabular}
    \caption{Relative signal amplitude vs. drift voltage (a) and gain vs amplification voltage (b) for the test MM detector with an $Ar~(30\%)-CO_2~(70\%)$ gas mixture.}
    \label{fig:ArCO2_70}
\end{figure}
The prototype was tested with an $Ar-CO_2$ mixture with a $CO_2$ fraction of 7\%, 30\%, 50\%, and 70\%. Stable detector operation was seen for all mixtures and the most important results were  observed for $Ar-CO_2$ (70\%) one. 
The maximum effective gas gain  above $ 4\times 10^4$ was reached with a $^{55}$Fe source at amplification voltage $U=960$ V, while  $U=900$ V was needed for $G=10^4$.  We consider 60 V to be a very solid margin for  a real detector operation. Measurements of  the amplitude vs drift voltage dependence result in  the optimum drift field of about $2.2\div2.5$ kV/cm with  a few percent drop at 3 kV/cm. The single-cluster signal duration was about 130 ns, and  the full signal length for  the MIP tracks  was well within 300 ns, meeting the expectations. The resulting plots are given in Fig.~\ref{fig:ArCO2_70}

\section {Cylindrical chamber prototypes}

\subsection{Mechanical prototypes and geometry test}
To practice the assembly procedure of semi-cylindrical detectors, two mechanical prototypes were fabricated with parameters close to those planned for the final version of the Micromegas-based central tracker: length 40 cm, bend radius 5 cm, drift gap 3 mm. The first prototype was made of 0.3mm FR4, a material with stiffness higher than expected for a real detector. In contrast, the second prototype was made of a very flexible material, 0.2mm kapton foil. In both cases, there were no problems in the gluing process and the strength of the assembled prototypes was sufficient to fabricate self-supporting detectors. After assembly, the shape deviation of the prototypes from the cylindrical shape was measured using a 3D scanner with an optical sensor. The saddle-like deformation of the cathode plane did not exceed 0.15 mm at the center point and the maximum deviation of the long edge of the prototypes from a straight line (barrel-like deformation) was 0.35 mm. Both values are within the permissible limits and should not noticeably affect the operation of the detectors.

\subsection{First working prototype \label{sec:MM_bent}}
It is planned that the ''minimal unit'' of the MCT will be a semi-cylindrical detector with a length of the sensitive region of 40$\div$45 cm and a bending radius of 5.0 $\div$ 6.2 cm. To test the applicability of the developed assembly technique, a first prototype was fabricated with parameters close to the planned MCT parameters, but with the length reduced to 20 cm. 
There are two obvious difficulties in the assembly of cylindrical micromegas detectors:
\begin{enumerate}
    \item 	When a flat bulk-MM module is bent, the mesh tension increases significantly, which can cause deformation of the PCB that is the base of the detector or damage to the mesh itself. To minimize this effect, a new mesh tension machine was fabricated to reduce the initial mesh tension  from the standard value of $7 \div 10 ~ N\cdot cm^{-1}$ to $1~N \cdot cm^{-1}$.
    \item The stretched mesh has a flat shape between the pillars, due to which the distance from the mesh to the cylindrical surface of the anode varies significantly. The magnitude of the variation depends quadratically on the pitch of the pillars. At a pitch of 1 mm, the variation of the amplifying gap decreases to 2.5 $\mu$, which is an acceptable value.  At the same time, the small pitch of the pillars increases the area occupied by them, which can eventually lead to a decrease in efficiency. To mitigate this effect, a special arrangement of pillars was developed, shown in Fig.~\ref{fig:proto_pillars}. 
    \begin{figure}[htb]
    \centering  
    \includegraphics[width=.7\textwidth]{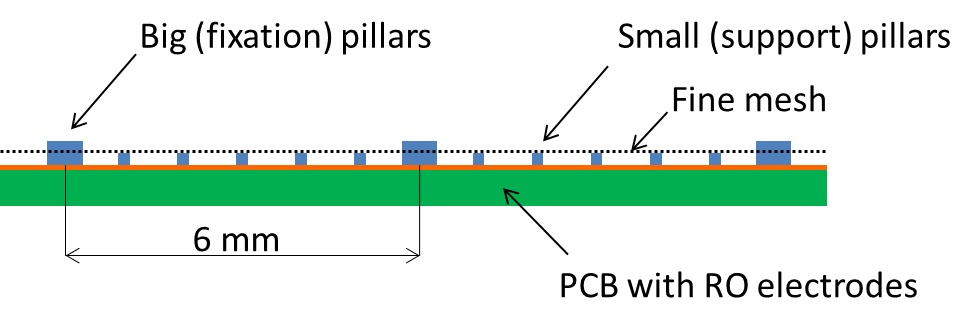} \\     
    \caption{Pillar arrangement of the first prototype of the bent Micromegas chamber}
    \label{fig:proto_pillars}
    \end{figure}
    Small 0.2 mm diameter pillars, which do not fix the mesh but only support it, are placed at a 1 mm spacing. Large fixing 0.75 mm pillars   have a pitch of 6 mm.  The total area occupied by the pillars is less than 4\%. In addition, since the typical value of transverse electron diffusion at drift is 0.1-0.2 mm, and the effective length of the track projection on the anode plane, taking into account the Lorentz angle, is not less than 0.4 mm, we expect that the detector will retain its efficiency even if a particle hits directly into the pillar.
\end{enumerate}

\begin{figure}[htb]
    \centering
    \begin{tabular}{cc}
    \includegraphics[width=.464\textwidth]{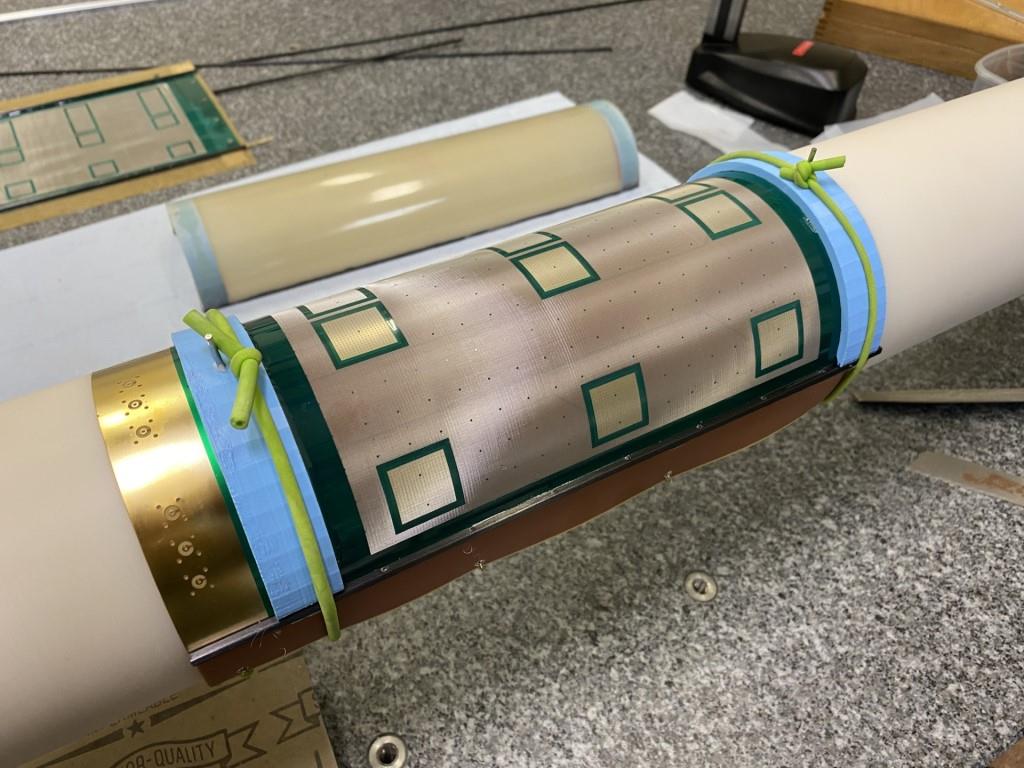} 
    \includegraphics[width=.45\textwidth]{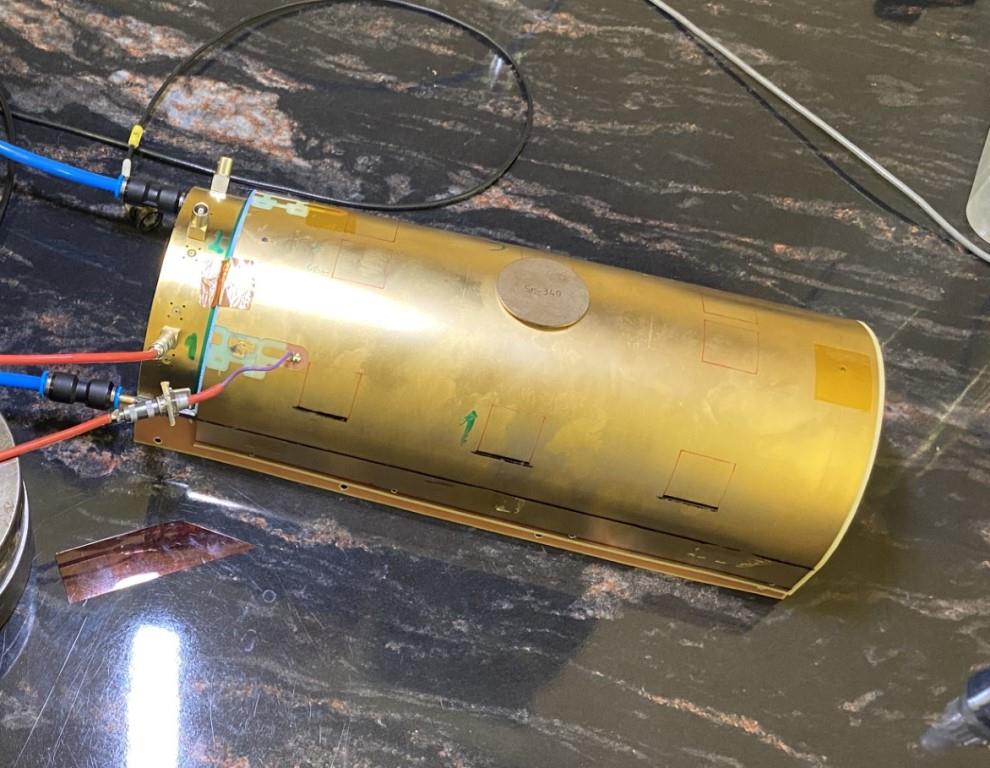} \\     
    \end{tabular}
    \caption{First cylindrical MM prototype. Left: assembling procedure. Bulk MM module bending and fixation on a template for gluing. Right: assembled module prepared for the very first functionality test. }
    \label{foto:bent_proto}
\end{figure}

\begin{figure}[htb]
    \centering
    \begin{tabular}{cc}
    \includegraphics[width=.45\textwidth]{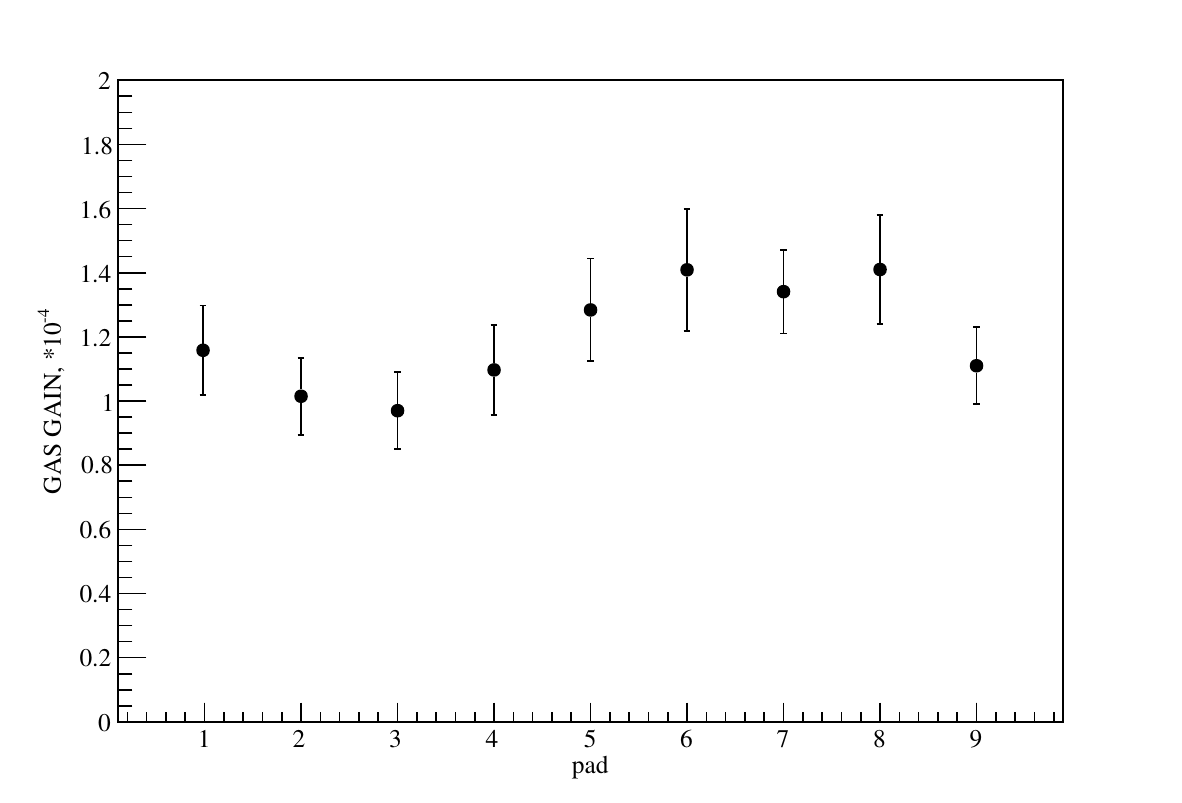} 
    \includegraphics[width=.45\textwidth]{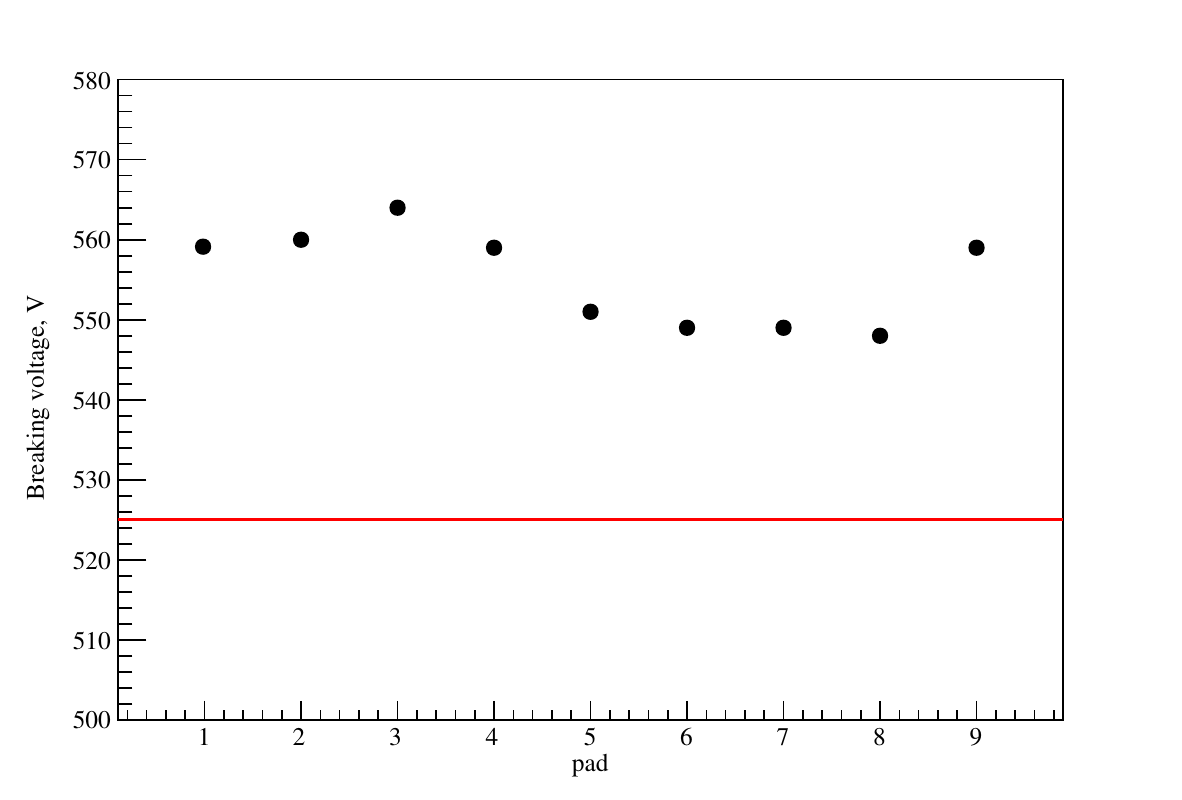} \\     
    \end{tabular}
    \caption{ First half-cylindrical prototype test results. Left: Gas gain vs pad number at fixed gain voltage $U_{gain}=525$ V, error bars represent energy resolution (RMS) measured with a $Fe^{55}$ source. Right: breaking voltage measured separately for all pads. Red line represents working voltage for a gas gain of about $10^4$.} 
    \label{fig:MM_bent_test}
\end{figure}

The most important parameter we want to control for the first prototype is the stability of the gain gap value and hence the gas gain over the detector area. To facilitate such measurements, the signal electrodes are made as $1.5\times1.5$ cm$^2$ and $1.5\times4.5$ cm$^2$ pads. During the production of micromegas, breakages of individual pillars are possible. In order to study the influence of such defects on the detector performance, two small pillars (not neighboring ones) were removed during manufacturing on one pad( $\#6$).
 The basic fabrication sequence is outlined in  Section \ref{sec:MM_bent}. Fig.~\ref{foto:bent_proto}, left,  shows the most critical operation of the prototype assembly - bending of the MM module and its fixation on the cylindrical template. The general view of the assembled prototype prepared for the very first functionality test is shown in Fig.~\ref{foto:bent_proto}, right.
 For the assembled prototype, the breakdown voltage was determined for all pads, and the gas gain and energy resolution were measured at the same operating voltage $U_{gain}=525V$. The results of the tests are summarized in Fig.~\ref{fig:MM_bent_test}. 
 As can be seen, the detector is stable at gas gain $G=10^4$, has a sufficient stability margin over the operating voltage, and has a uniformity of properties good for detectors of this type.It can also be seen that the absence (for pads $\#6$) of individual pillars does not lead to a critical decrease in the breakdown voltage.

\section{Cost estimate}
 The cost estimate  is summarized in Table \ref{tab:cost_MM}.
\begin{table}[htb]
 \caption{Estimate of the Micromegas Central Tracker cost.}
    \centering
    \begin{tabular}{|l| c| c| c|}
    \hline
         & Cost & N units & Cost, k\$ \\
         & per unit, k\$ & & \\
\hline \hline
Detector materials &  & & 220 \\
\hline
FE electronics  & 2.0 & 65 & 130 \\
\hline
Assembling tools &  & & 200\\
\hline
HV PS, HV cables &    &         & 25 \\
\hline
Gas system & & & 20 \\
\hline

Supporting structure, water  cooling & & & 150\\
\hline
Prototype production & & & 20\\
\hline
\hline
TOTAL                 &  & & 765 \\
\hline
    \end{tabular}
    \label{tab:cost_MM}
\end{table}
Materials for detector production include PCBs including 1 spare of each type (120k\$), fine mesh (30 k\$), DLC coating (50 k\$, manufacturer's estimate), photoresist, glue, other consumables (20 k\$). The VMM3 ASICs for the FE card production are available, the cost of materials and production at PRC fabs is estimated at 2k\$ for a 128-channel card. The quoted cost includes 10\% (5 pcs) spare cards.  Level 1 hubs are the part of the DAQ and are not included in the MCT cost calculation.

\chapter{Zero Degree Calorimeter}

\section{General layout}

\begin{figure}[!bt]
\includegraphics[width=\textwidth]{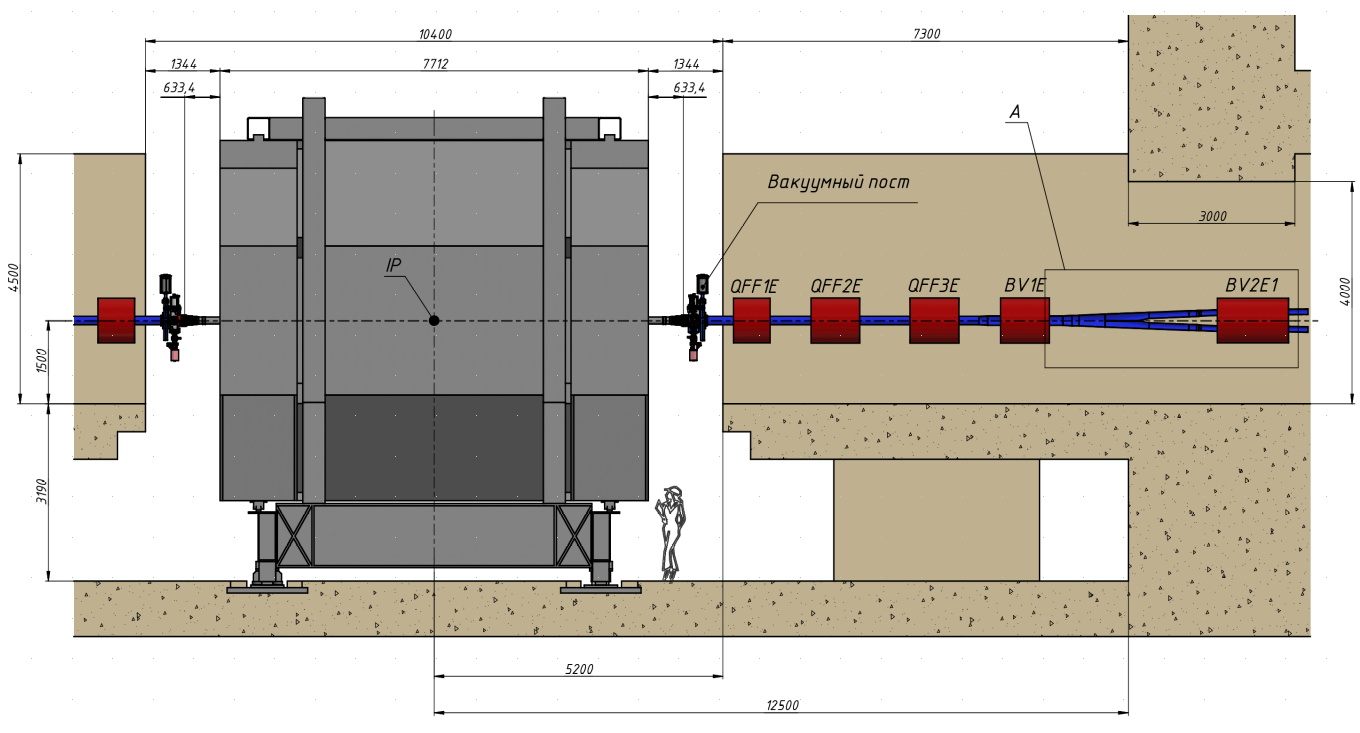}
\includegraphics[width=\textwidth]{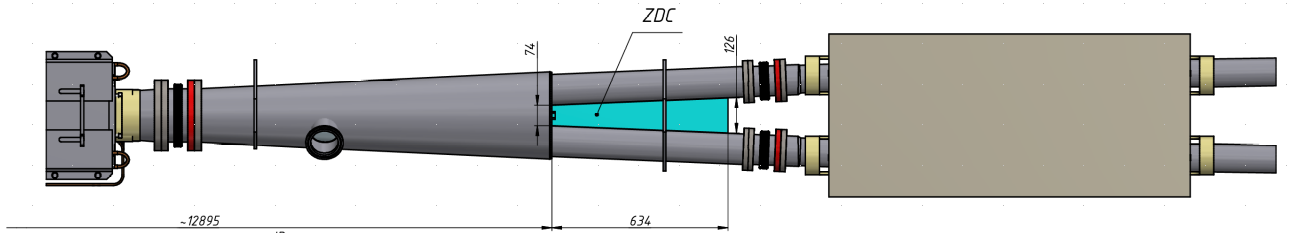}
\caption{\label{zdc:1}ZDC position on the one side from the IP. The top figure shows
the general view while 5 times enlarged area "A" is shown in the bottom. The cryostat
is not shown.}
\end{figure}

A Zero Degree Calorimeter (ZDC) is a standard device for the collider environment. It is
placed in the space between the separation dipole magnet BV1E and the next dipole
magnet BV2E1,  about 13~m from the IP (Fig.~\ref{zdc:1}). The strong magnetic field
before the ZDC efficiently removes all charged particles, allowing clean measurement
of neutrals, so the device can work up to very high luminosities.
Two ZDC devices are supposed to be placed symmetrically
on both sides of the IP. A coincidence between them,  as well as with other detectors, 
will be used.

The ZDC's main tasks are:
\begin{itemize}
\item  luminosity measurement;
\item spectator neutron tagging;
\item time tagging of the events for event selection;
\item local polarimetry with forward neutrons.
\end{itemize}

To accomplish these tasks, the following performance parameters should be met:
\begin{itemize}
\item	time resolution $150\div200$ ps;
\item	energy resolution for neutrons $50\div60\%/\sqrt{E} \oplus 8\div10\%$;
\item	neutron entry point spatial resolution 10 mm.
\end{itemize}

\begin{figure}
\includegraphics[width=\textwidth]{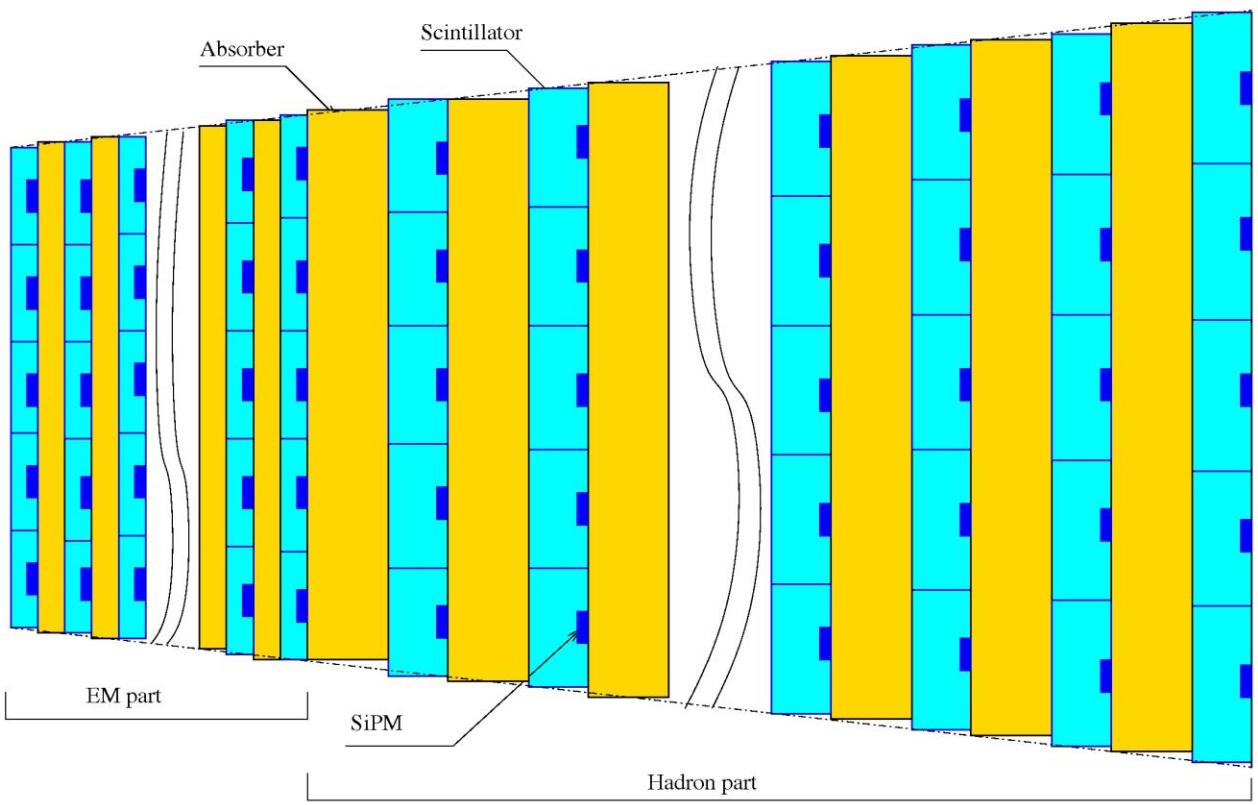}
\caption{\label{zdc:2}A schematic layout of the calorimeter}
\end{figure}

We plan to use a fine-segmented calorimeter based on plastic scintillator active tiles
with the direct SiPM readout and tungsten absorber plates, similar to the calorimeter proposed for the
CALICE \cite{zdc_CALICE}. A schematic view of the calorimeter is shown in Fig.~\ref{zdc:2}.

\section{Detailed description}

\begin{figure}[!bt]
\includegraphics[width=\textwidth]{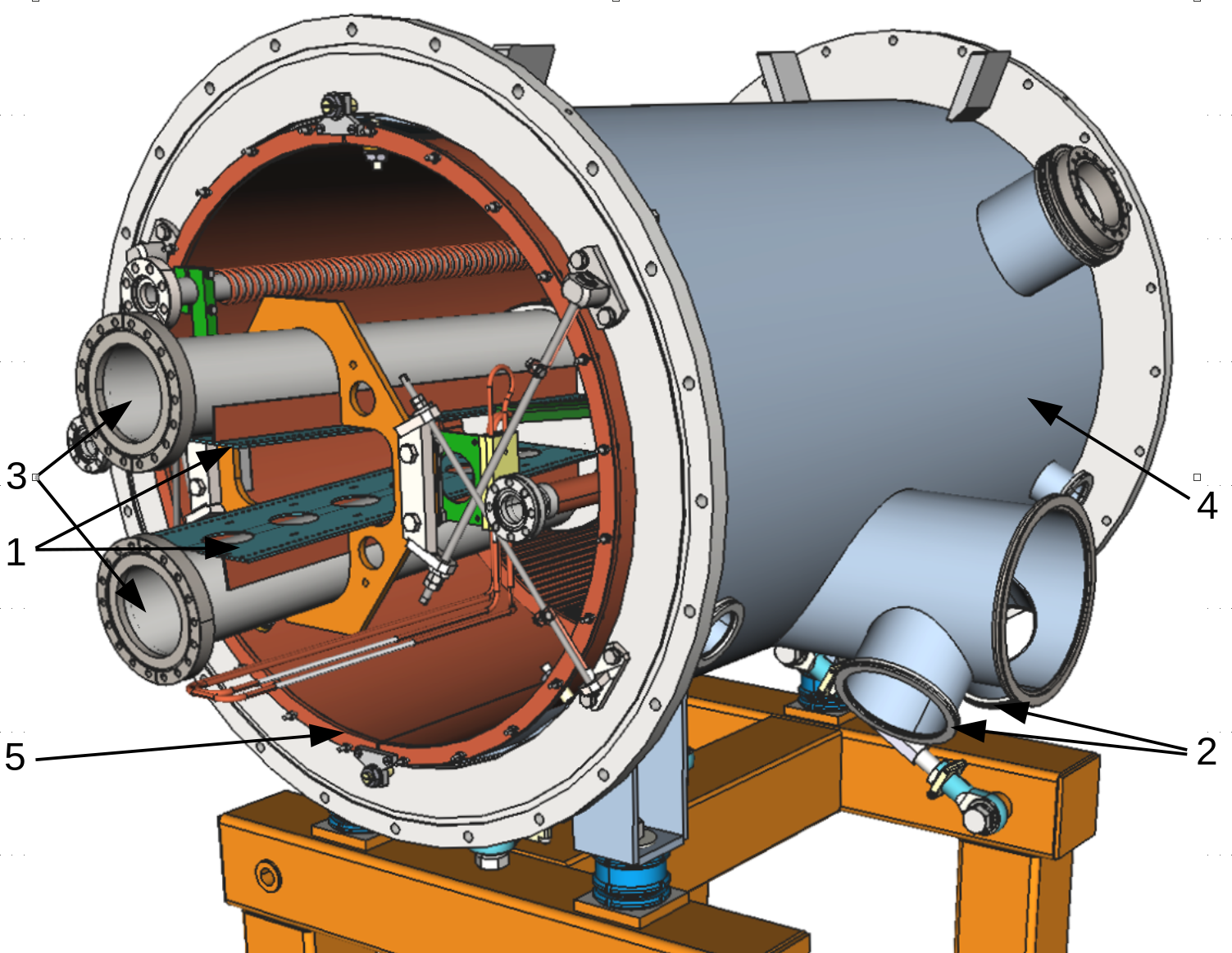}
\caption{\label{zdc:3}ZDC place inside the cryostat. 1 -- the ZDC supporting rails;
2 -- the flanges for the output connectors; 3 -- the beam pipes; 4 -- the cryostat outer shell;
5 -- the copper screen cooled to the liquid nitrogen temperature.}
\end{figure}

The position of the ZDC is inside the cryostat of the NICA magnets (see Fig.~\ref{zdc:3}). This location
provides several challenges:
\begin{itemize}
\item limited space between the two beam pipes;
\item insulation vacuum $\sim 10^{-6}$~Torr;
\item cryogenic temperature $\sim 80$~K;
\item difficult accessibility.
\end{itemize}

\begin{figure}[!bt]
\begin{center}
\includegraphics[width=0.49\textwidth]{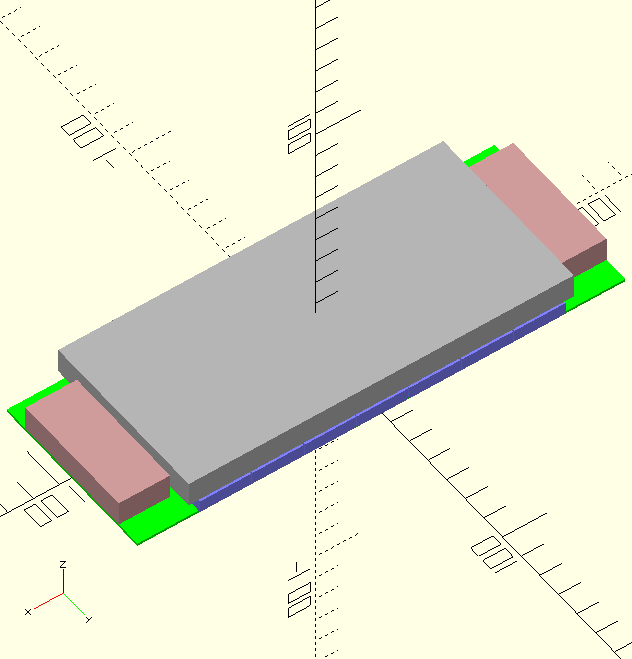}
\includegraphics[width=0.49\textwidth]{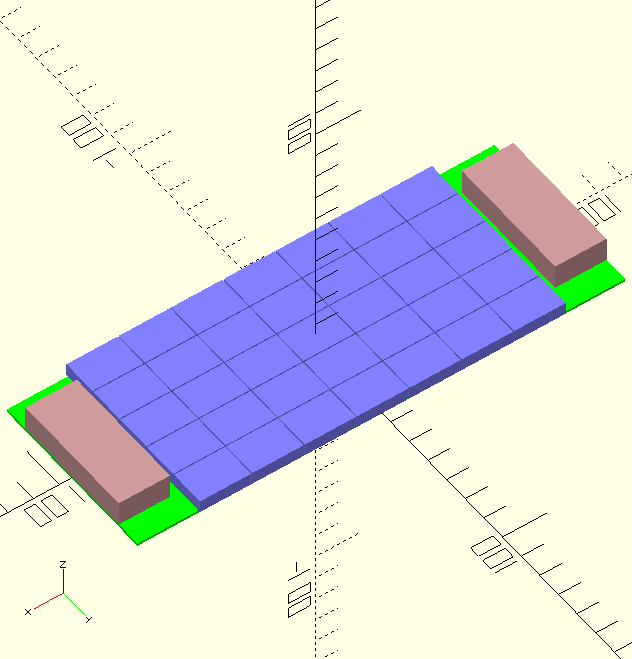}
\end{center}
\caption{\label{zdc:4}Single ZDC module with the absorber (left) and without (right). The elements are shown by
colors: green - printed circuit board, blue -- scintillator tiles, gray -- absorber, magenta -- connector area.}
\end{figure}

The calorimeter is assembled from individual planes, as shown in Fig.~\ref{zdc:4}. Each plane has
a printed circuit board (PCB) with the SiPMs 3$\times$3~mm$^2$ S13360-3050PE produced by Hamamatsu, the scintillator tiles,  and the 
tungsten absorber plate. The SiPMs and  the tiles are organized into a matrix 7$\times$5.
The plane sizes are increasing from the front to the back side of the
calorimeter,  together with the increase of the gap between the beam pipes. Tile sizes are
increasing accordingly. The scintillator tiles for better light collection
are chemically whitened by UNIPLAST (Vladimir, Russia) \cite{zdc_uniplast}.
Then a small polished burrow for the SiPM readout is made.

Each plane is attached to the top and  the bottom rails (position 1 in Fig.~\ref{zdc:3}). These rails are connected by
thermal bridges to the thermal screen (position 5 in Fig.~\ref{zdc:3}), cooled to the liquid nitrogen temperature.
Signals from the SiPMs are lead out by flat cables to the vacuum-tight connectors, placed on four 6-inch flanges
(position 2 in Fig.~\ref{zdc:3}) of the cryostat. We consider
either circular or rectangular connectors 36$\div$40 pins each, 15 connectors per flange. Only SiPMs are
planned to be placed inside the cryostat with a total dissipation power of about 0.1~W. All the electronics,
which includes front-end amplifiers, SiPM power supplies, and signal digitization, 
will be placed outside the cryostat in the racks below and above the beam pipes. The ZDC front-end electronic
will be based on the CAEN FERS-5200 system~\cite{zdc_FERS}. A5202 modules, based on the Citiroc-1A chip produced by Weeroc for SiPM readout,
include all stages necessary for SiPM operation from setting bias voltage and amplification to digitization. 

The calorimeter can be roughly divided into two parts -- the front part and the rear part.
The front part serves mostly as an electromagnetic calorimeter to measure gammas, while the rear part
is responsible for the measurement of neutrons. The preliminary parameters of the layers in each part
are given in Table~\ref{zdc:5}.

\begin{table}
\caption{\label{zdc:5}Preliminary parameters of the calorimeter layers.}
\begin{center}
\begin{tabular}{l|c}
Parameter                       & Value \\
\hline
\multicolumn{2}{c}{Electromagnetic part}\\
\hline
Number of layers                & 8     \\
Scintillator thickness, mm      & 5     \\
Absorber thickness, mm          & 5     \\
PCB thickness, mm               & 1     \\
Total absorber thickness, mm    & 35\footnote{No absorber in front of the first scintillator layer to
tag residual charged particles}         \\
Total part thickness, mm        & 83    \\
Part thickness, X$_0$           & 10    \\
Part thickness, $\lambda_i$     & 0.4   \\
Number of channels              & 280   \\
\hline
\multicolumn{2}{c}{Hadron part}         \\
\hline
Number of layers                & 22    \\
Scintillator thickness, mm      & 10    \\
Absorber thickness, mm          & 13    \\
PCB thickness, mm               & 1     \\
Total absorber thickness, mm    & 286   \\
Total part thickness, mm        & 528   \\
The part thickness, $\lambda_i$ & 3.1   \\
Number of channels              & 770   \\
\hline
\multicolumn{2}{c}{Total}               \\
\hline
Thickness, mm                   & 611   \\
Thickness, $\lambda_i$          & 3.5   \\
Number of channels              & 1050  \\
\end{tabular}
\end{center}
\end{table}

The background conditions in the location of  the ZDC could be found in the
Chapter~\ref{ch:rad},  devoted to NICA ring background estimates.
The operational limit for Hamamatsu SiPM is $10^{11}-10^{12}$~neutrons/cm$^2$.
Beam halo calculations give an estimation of $10^{10}$~neutrons/(cm$^2\times$~year) 
at the ZDC position for the most intensive NICA beams.

\section{Monte Carlo simulation}

\begin{figure}
\includegraphics[width=\textwidth]{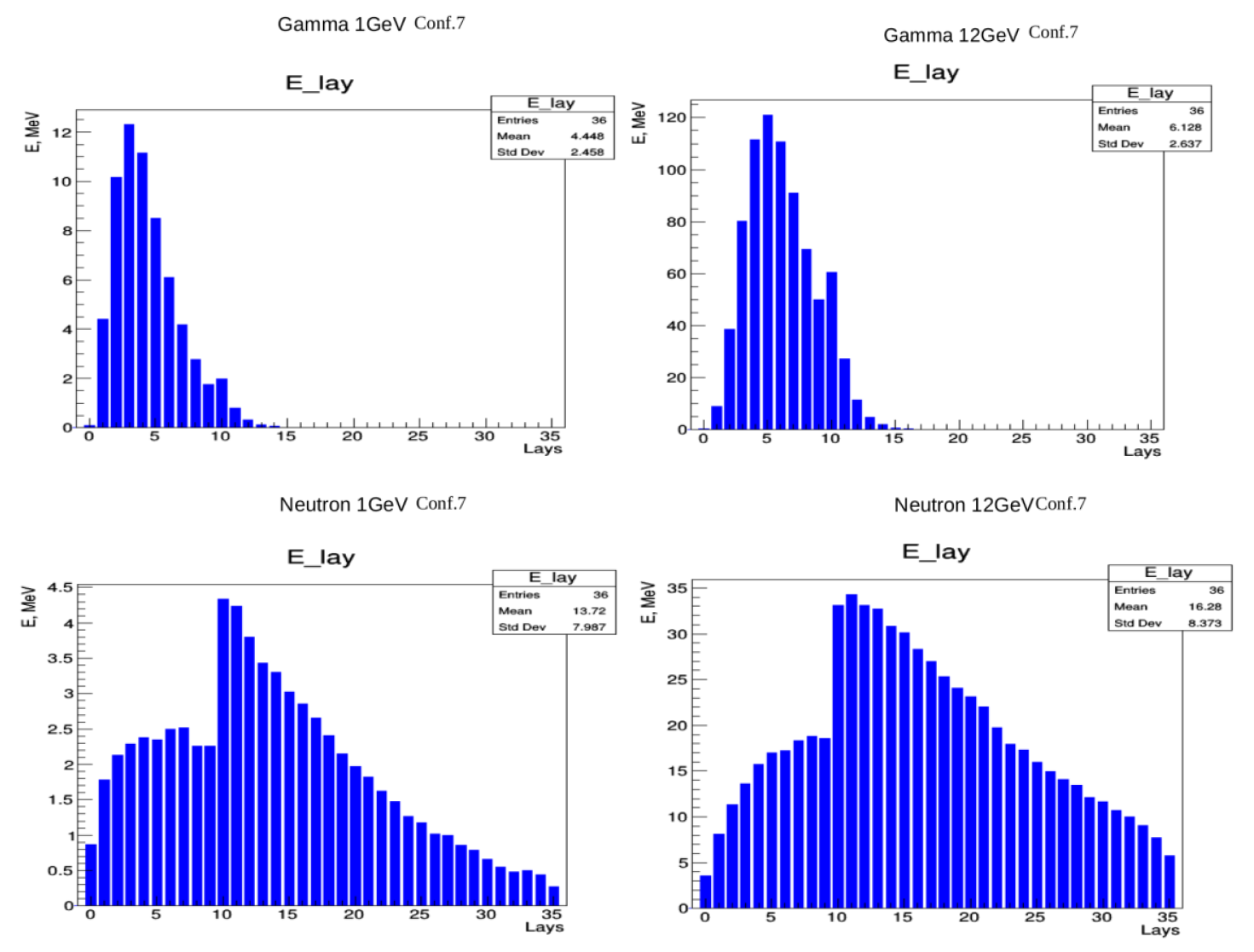}
\caption{\label{MC:1}Photon (up) and neutron (down) longitudinal energy distributions for different particle energy (1 GeV -- left and 12 GeV -- right).}
\end{figure}

The MC model of the ZDC was created to determine the optimal ZDC configuration.
The simulation is done within GEANT4 framework \cite{Agostinelli:2002hh, Allison:2006ve, Allison:2016lfl}. Neutrons and photons of different energies (1, 3, 6, and 12 GeV) are created by a box generator. The particle momentum is parallel to the longitudinal axis of the calorimeter. The interaction point distribution in the frontal transverse plane of the detector is uniform. Energy and space resolution, transverse and longitudinal leakage, and neutron/photon separation have been studied.  Different detector configurations are tested for  neutron and photon identification. For the most longitudinally granulated configuration within available space energy resolution for neutrons is about $50\%/\sqrt{E} \oplus 30\%$  and for photons is about $20\%/\sqrt{E} \oplus 9\%$. The longitudinal energy distributions for 1 GeV and 12 GeV photons and neutrons are shown in Fig.~\ref{MC:1}. One can see from the Figure that the longitudinal energy distributions for photons and neutrons are very different and can be used for neutron/photon separation. A jump at the 11th layer corresponds  to its increased thickness.

\newpage

\begin{figure}[!bt]
\includegraphics[width=7cm,height=7cm]{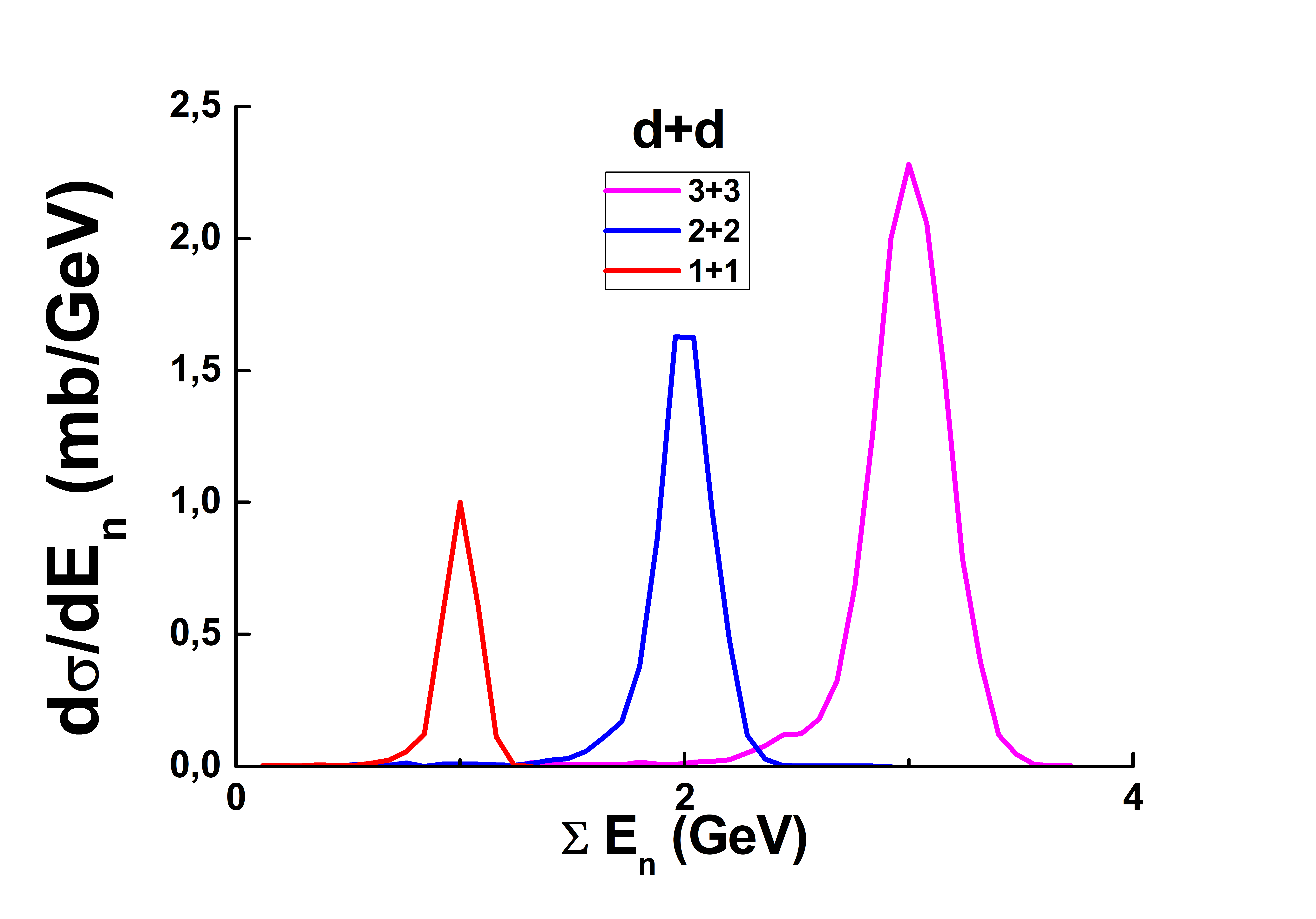}
\includegraphics[width=7cm,height=7cm]{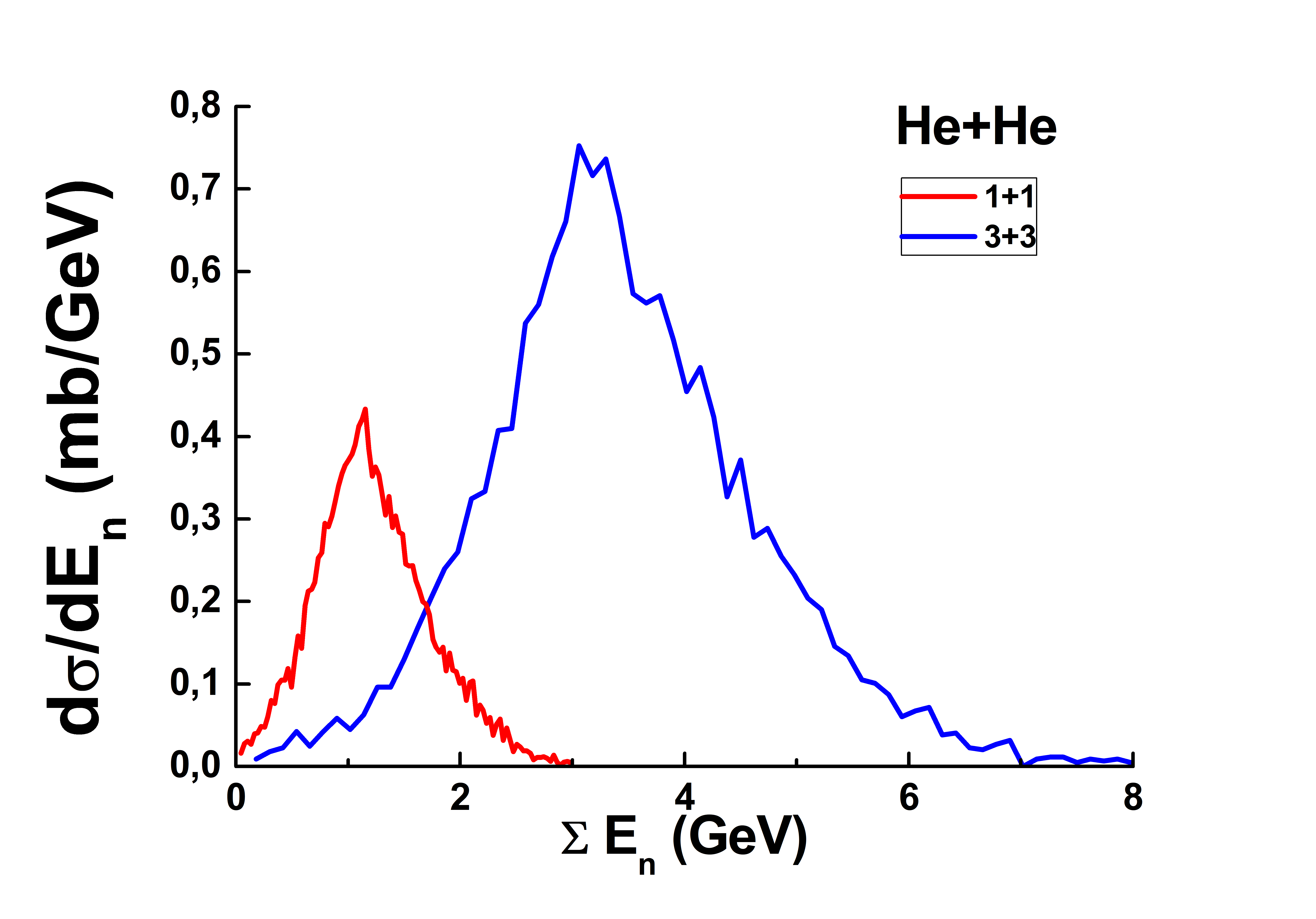}
\includegraphics[width=7cm,height=7cm]{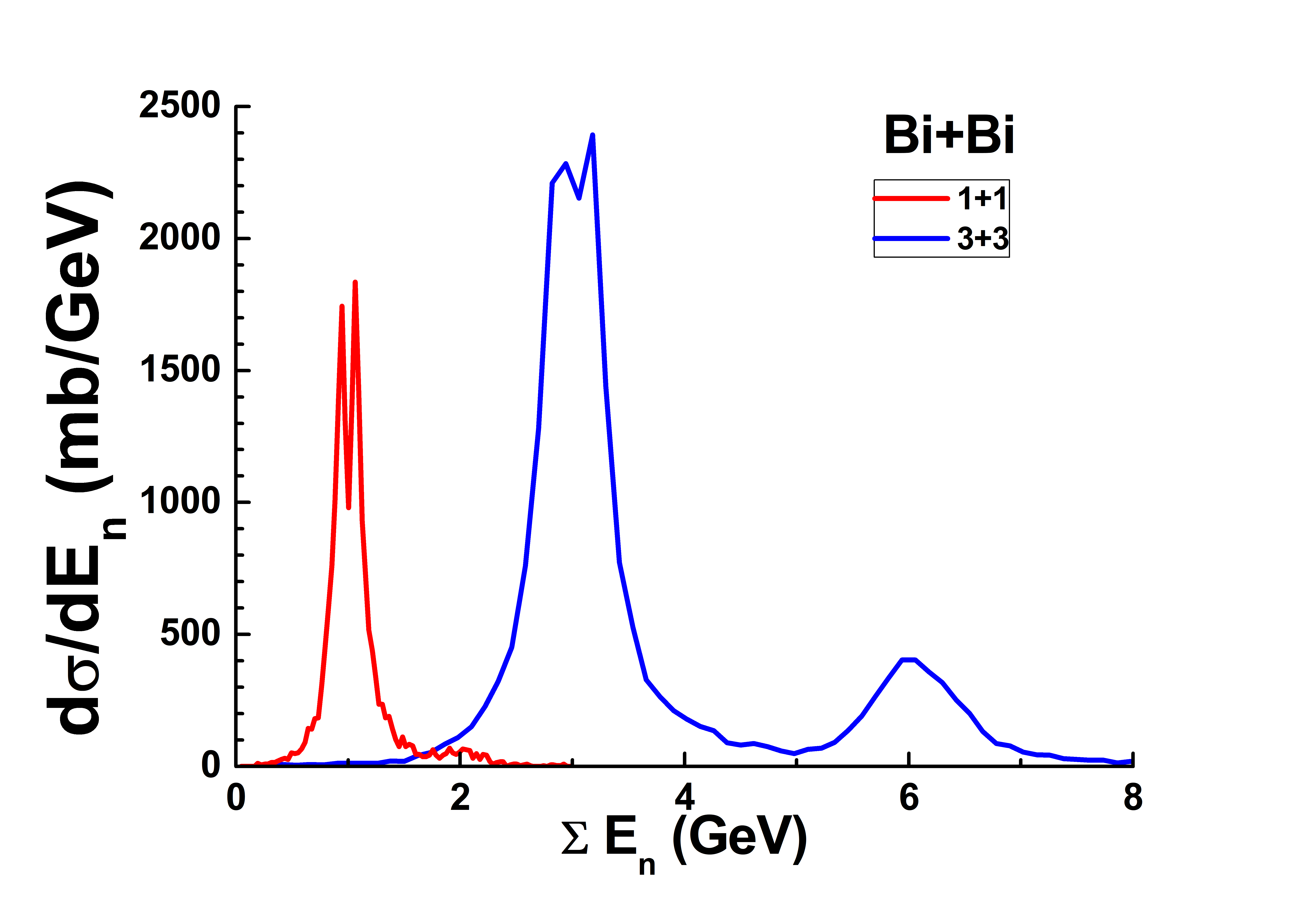}
\caption{\label{MC:2}Distributions for the total energy deposition produced in the ZDC by neutrons  for the single event in the deuteron-deuteron, helium-helium and bismuth-bismuth collisions at different beam energies.}
\end{figure}

To estimate the ZDC  occupancy during operation with various beams,  the simulation was carried out using the GEANT FTF -- the FRITIOF event generator \cite{Allison:2016lfl}. At the initial stage of the NICA accelerator operation,  it is planned to work with beams of light and heavy nuclei.  The distributions of the total energy deposition produced by neutrons in the ZDC for a single event in the deuteron-deuteron, helium-helium,  and bismuth-bismuth collisions at the kinetic energies 1, 2, and 3 GeV per nucleon are shown in Fig.~\ref{MC:2}.

In the spectrum of bismuth-bismuth interactions at 3 GeV, the contribution from the impact of two spectator neutrons is visible. Table~\ref{MC:3} presents the results of the MC analysis for  the number of neutrons detected per second in the ZDC for the different colliding beams and energies at the luminosity $L=10^{26}~cm^{-2} s^{-1}$.  This value of luminosity is planned to be achieved in the NICA operation at the heavy-ions collision mode.

\begin{table}
\caption{\label{MC:3}ZDC neutron rates for the different beams and energies.}
\begin{center}
\begin{tabular}{|c|c|c|c|}
\hline
$L=10^{26}~cm^{-2} s^{-1}$ & {\bf d+d} & {\bf He+He} & {\bf Bi+Bi}\\
\hline
{\bf $T_{beam} (GeV)$} & Neutrons per second & Neutrons per second & Neutrons per second\\
\hline
{\bf 1} & 0.02 & 0.04 & 60 \\
\hline
{\bf 2} & 0.05 & - & - \\
\hline
{\bf 3} & 0.1 & 0.2 & 250 \\
\hline
\end{tabular}
\end{center}
\end{table}

\section{Time resolution measurements}
\begin{figure}[!bt]
\includegraphics[width=\textwidth]{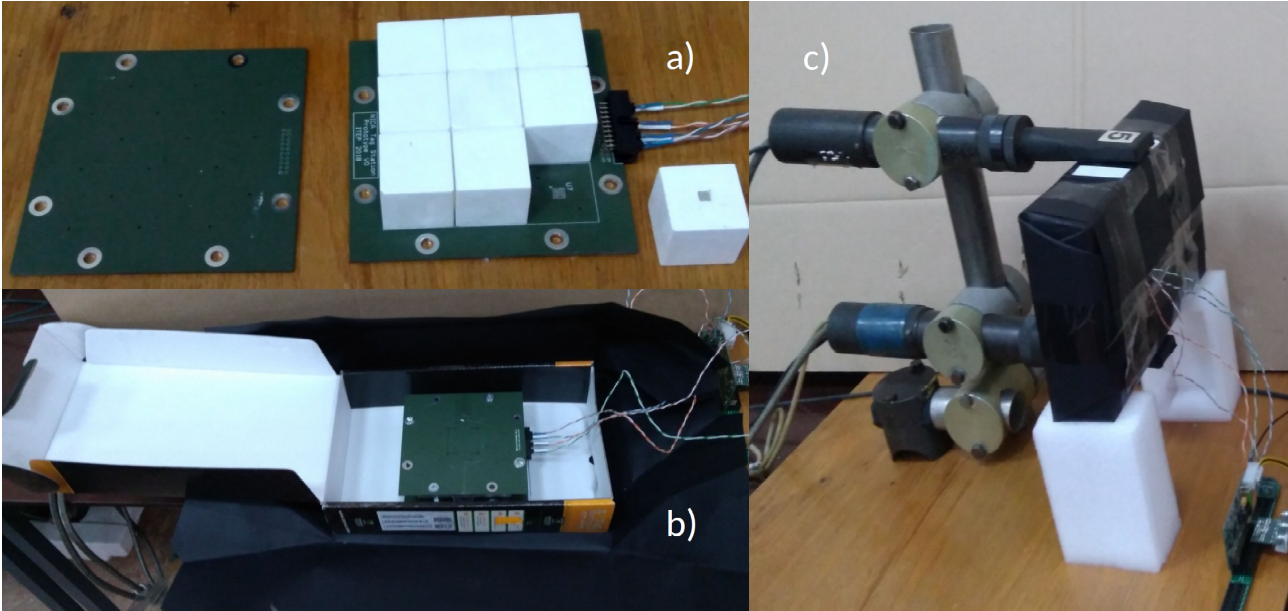}
\caption{\label{zdc:8}One layer prototype: a) 9 cubes of 30$\times$30$\times$30 mm$^3$ plastic, the
SiPM board and the support board; b) the prototype assembly in a box before wrapping in black paper;
c) the box in place for cosmic muon tests.}
\end{figure}

For experimental estimates of the time resolution, an assemblage of 9 plastic cubes, laid
on a printed circuit board with mounted SiPMs and fixed with a support board (see Fig.~\ref{zdc:8}),
was tested with cosmic muons. Each cube was 30$\times$30$\times$30 mm$^3$ in size and  was
chemically covered with a thin, light-reflecting layer. A numerical simulation has shown the
mean number of cells hit in an event is more than 20 for both photons and neutrons, and the
energy deposit in the scintillator for each hit  was in the range of 3$\div$9~MeV. So, the cubes
of the test assemblage, with an average cosmic muon energy deposit of 6~MeV, provide a good
approximation of the real tile response in the calorimeter.

\begin{figure}[!hbt]
\begin{center}
\includegraphics[width=0.6\textwidth]{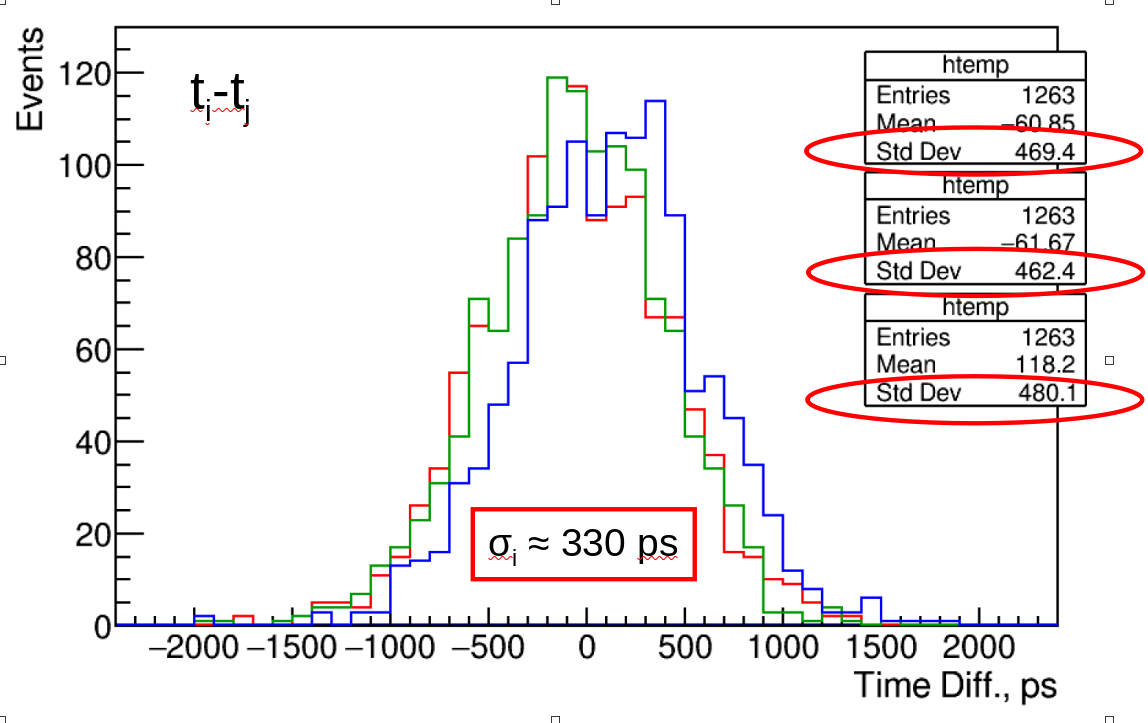}
\end{center}
\caption{\label{zdc:9}Time difference for the three cube pairs. The resolution of
each cube is $1/\sqrt{2}$ of the resolution of difference.}
\end{figure}

The  obtained result for the vertical muons is shown in Fig.~\ref{zdc:9}. It gives a time resolution of
a single cube of 330~ps. Using different cube combinations, we found out that the resolution
did follow  the $1/\sqrt{E}$ law. So the aim of 150~ps will be reached at about 30~MeV energy
deposit in the scintillator. For the approximately 1/12 energy deposit fraction of the scintillator
in the ZDC, this means about 400 MeV energy of the incoming particle.

\section{ZDC for the first NICA run}

The development of the ZDC will consist of two stages. During the initial NICA operation
when no other SPD devices will be installed, we want to test a partial ZDC
with an approximate cost of 20\% of the entire unit. Its main tasks will be the following:
\begin{itemize}
\item test of the device concept technology in the real position, with a special emphasis on the
problems of the high radiation, the cryogenic operation temperature, and the passing of the signals out of
the vacuum;
\item simple measurements with the beam, including beam luminosity and neutron/gamma discrimination efficiency;
\item study of the background conditions in the location of the ZDC;
\item test of the MC simulation and compare the model results with real data;
\item check whether multiple configurations are possible;
\item test of the CAEN FERS-5200 readout system.
\end{itemize}
We assume that all the electronics of the partial ZDC will be later used as a part of the
electronics for the whole ZDC, while the scintillator tiles and the absorber would be replaced.

The exact configuration of the partial ZDC
will be a result of the optimization based on the MC simulation. The preliminary layout
is given in Table~\ref{zdc:6}. It will have only 6 sensitive layers of 10 mm thickness with 186 scintillation
tiles\footnote{We want to abandon corner tiles in this design and have only 31 tiles per plane. This will make a
better fit to the A5202 boards, which have 64 channels, and give space for screws.} and a front scintillator
plane without segmentation to tag charged particles. We are going to use a copper absorber for this design. 

\begin{table}
\caption{\label{zdc:6}Preliminary parameters of the partial calorimeter}
\begin{center}
\begin{tabular}{l|l|c}
Layer & Parameter                       & Value \\
\hline
0     & Layer thickness, mm             & 10    \\
\hline
1     & Absorber thickness, mm          & 15    \\
      & Layer thickness, mm             & 26    \\
      & Layer thickness, X$_0$          & 1.1   \\
      & Layer thickness, $\lambda_i$    & 0.1   \\
\hline
2     & Absorber thickness, mm          & 15    \\
      & Layer thickness, mm             & 26    \\
      & Layer thickness, X$_0$          & 1.1   \\
      & Layer thickness, $\lambda_i$    & 0.1   \\
\hline
3     & Absorber thickness, mm          & 45    \\
      & Layer thickness, mm             & 56    \\
      & Layer thickness, X$_0$          & 3.2   \\
      & Layer thickness, $\lambda_i$    & 0.3   \\
\hline
4     & Absorber thickness, mm          & 45    \\
      & Layer thickness, mm             & 56    \\
      & Layer thickness, X$_0$          & 3.2   \\
      & Layer thickness, $\lambda_i$    & 0.3   \\
\hline
5     & Absorber thickness, mm          & 100   \\
      & Layer thickness, mm             & 111   \\
      & Layer thickness, $\lambda_i$    & 0.7   \\
\hline
6     & Absorber thickness, mm          & 100   \\
      & Layer thickness, mm             & 111   \\
      & Layer thickness, $\lambda_i$    & 0.7   \\
\hline
\multicolumn{3}{c}{Total}                       \\
\hline
      & Number of layers                & 6     \\
      & Thickness, mm                   & 396   \\
      & Thickness, $\lambda_i$          & 2.2   \\
      & Number of channels              & 187   \\
\end{tabular}
\end{center}
\end{table}

\section{Cost estimate}
The cost estimate of a single ZDC device is given in Table~\ref{zdc:7}. The FERS-5200 modules
will be reused in the full design, so only their additional number compared to the partial design is listed.
Some spare modules are also assumed.

\begin{table}
\caption{\label{zdc:7}Cost estimate of a single ZDC detector.}
\begin{center}
\begin{tabular}{l|l|c|c|r|c|r}
                             &       &            &\multicolumn{2}{c}{Partial design} & \multicolumn{2}{c}{Full design}\\
Item                         & Units & Unit price & Amount        & Cost         &  Amount        & Cost          \\
\hline
SiPM S13360-3050PE           & pc.   & 50       & 200           &  10000         &  1050          &  52500        \\
Polystyrene scintillator     & tile  & 5        & 200           &   1000         &  1050          &   5250        \\
Tungsten absorber            & kg    & 160      &               &                &   130          &  20800        \\
Copper absorber              & kg    & 15       &  30           &    450         &                &               \\
SiPM boards                  & pc.   & 50       &   6           &    300         &    30          &   1500        \\
Cables in vacuum             & m     & 15       &  24           &    360         &   120          &   1800        \\
A5202 boards                 & pc.   & 7000     &   7           &  49000         &    25          & 175000        \\
DT5215 concentrator          & pc.   & 7000     &   1           &   7000         &     1          &   7000        \\
\hline
{\bf Total}                  &       &          &               &  68110         &                & 263850        \\
\end{tabular}
\end{center}
\end{table}

\chapter{Beam pipe and  BBC MCP detector \label{Ch_pipe}}
\section{SPD beam pipe}
A beam pipe separates the detector and the high vacuum of the accelerator. It must be mechanically sturdy on the one hand and thin enough, in terms of the number of radiation lengths to minimize multiple scattering and radiation effects, on the other hand. The diameter of the beam pipe is a compromise between the radial size of the particle beams and the requirement to position detectors as close to the 
beam line as possible for better reconstruction of the tracks.
\begin{figure}[bh!]
\centering
\includegraphics[width=1.\textwidth]{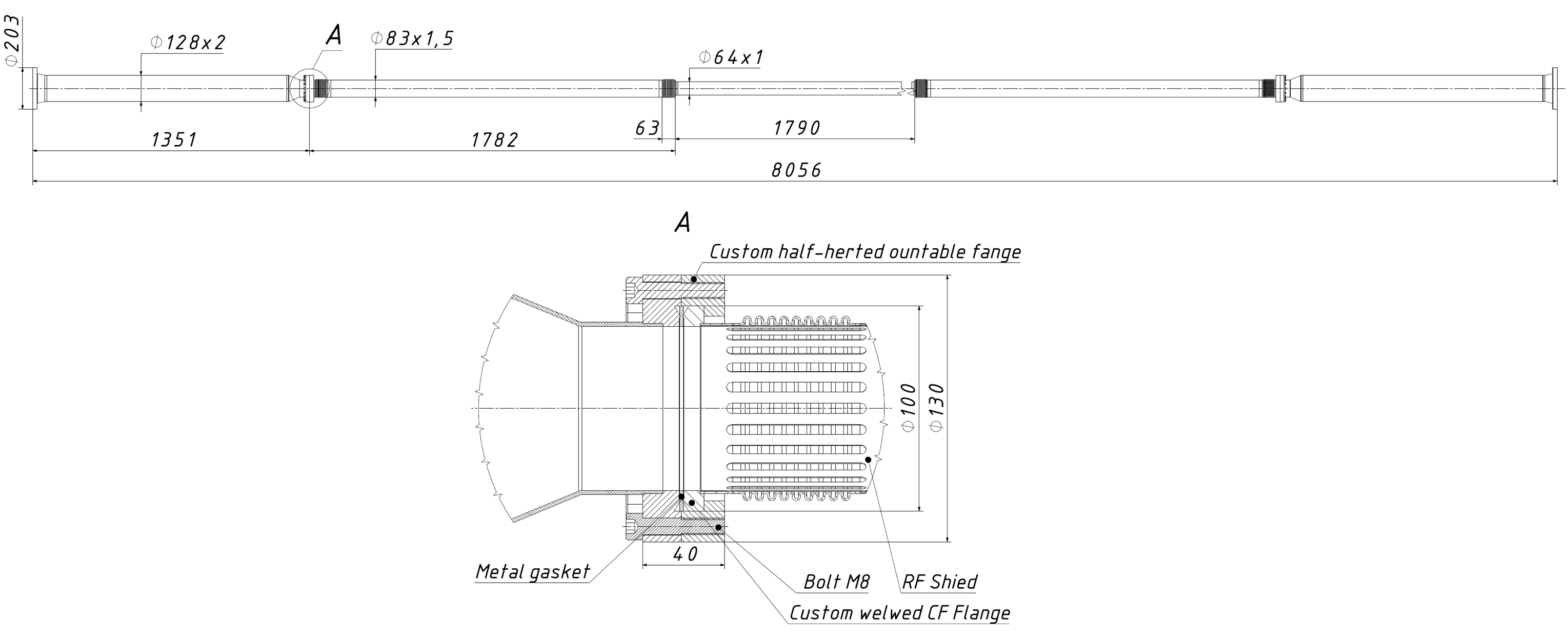}
\caption{Part of the accelerator vacuum beam pipe that will pass through the SPD experimental setup. 
At the first stage of the experiment, the central segment of the pipe will be made of aluminum.
Later it will be replaced with a beryllium segment. 
All dimensions are given in millimeters.}
 \label{fig_beampipe}
\end{figure}
The beam pipe of SPD has similar geometrical parameters as the pipe of MPD and 
will be manufactured using the same technology. 
The schematic view of the  pipe is shown in Fig.\,\ref{fig_beampipe}.
The pipe is almost 9\,m long. It is assembled from 5 segments, the diameter of which expands towards the ends.
At the first stage of the experiment, we plan to start working with an aluminum pipe to gain installation experience.
At a later stage, the material of the central segment will be replaced by beryllium. The thickness of the pipe in both cases will be 1 mm. Thus,  a charged particle, emitted from the IP at 90${^\circ}$, crosses 1.12\% and 0.28\% of $X_{0}$, respectively for aluminium and beryllium pipes.
The end parts of the pipe, providing a connection with the accelerator pipe, will be made removable.
They will be installed at a later stage of the assembly, after the ECal end-caps have been mounted, just before closing   sections of the RS end-cap. 
In order to reduce the diameter of the joint between the end and inner segments,
a conflate sharp edge vacuum connection with a removable bolt flange will be used.
Several bellows will be inserted into the segments to provide the necessary working space in response to thermal expansion and contraction. 
They also provide the necessary tolerance for misalignment.

This assembly is part of the high vacuum system and must be cleaned and tested for leaks according to specification. 
To avoid electron clouds, a  treatment of the inner surface of the beam pipe is required. 
Laser treatment or gettering is used for this purpose.

The cost of the pipe for the first stage of the experiment is estimated at 100 k\$ while replacing the central part with beryllium would cost up to 400 k\$.

\section{ BBC MCP detector}
A pair of the MicroChannel Plate detectors will be installed at a distance of $\sim$4.0 m from the beam interaction point (just to the right and left of the pipe section, shown in Fig. \ref{fig_beampipe}), close to the beam pipe.
These detectors are designed primarily to study a small-angle (mainly elastic) scattering of protons and deuterons at the polar angle of about 15 mrad and will operate together with the scintillator part of BBC.  Due to azimuthal granulation, they can be used for determination and control of the beam polarization via the measurement of $A_N$ asymmetries in the elastic $p$-$p$ and $d$-$d$ scattering. In combination with other detectors, the BBC MPC detector could contribute to $t_0$ reconstruction, which is of special value for the off-line analysis of the time-of-flight spectra and particle identification.

\begin{figure}[th]
\centering
\includegraphics[width=0.7\textwidth]{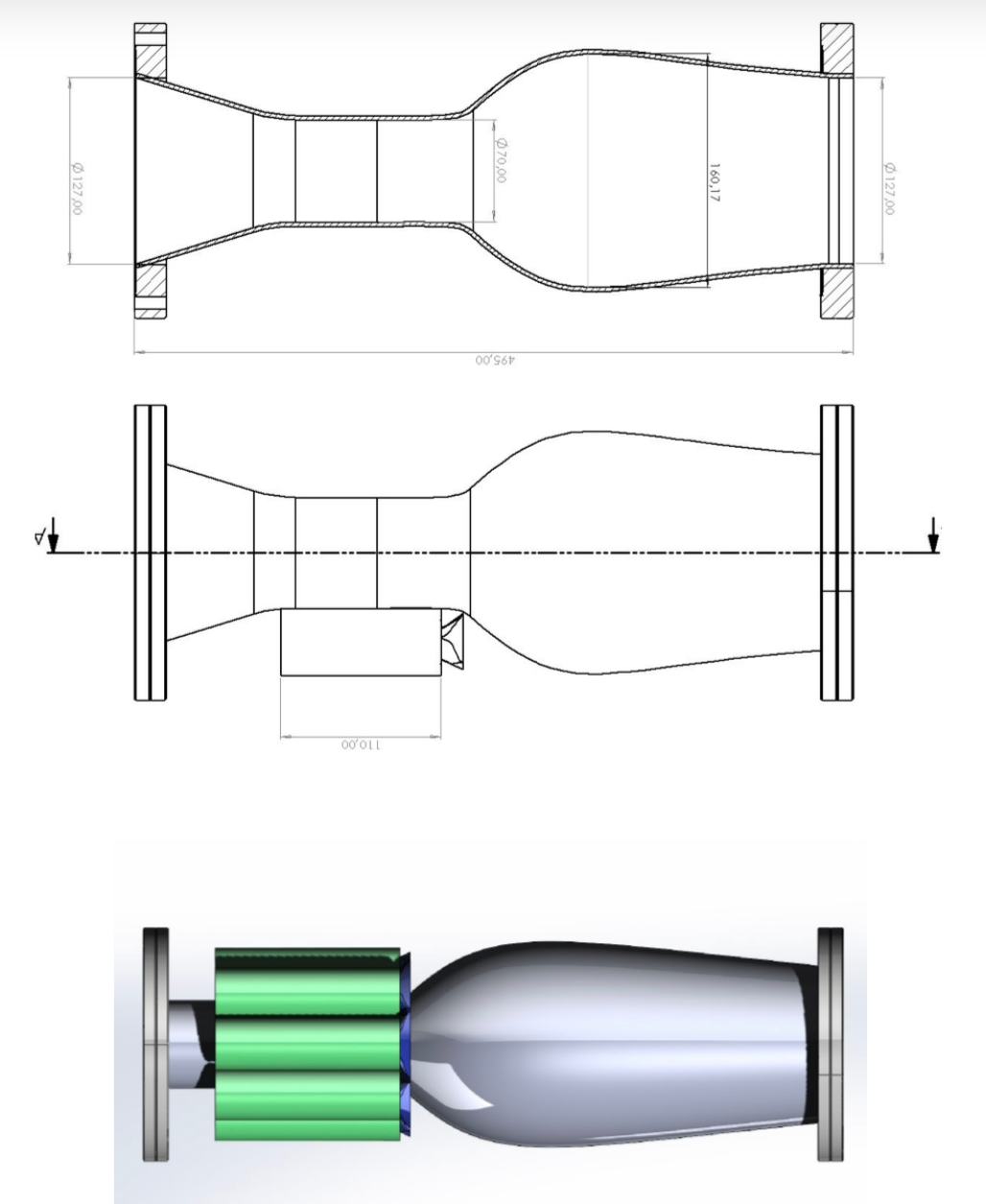}
\caption{View of the BBC MCP detector around the beam line.}
 \label{fig_mcp2}
\end{figure}
\begin{figure}[h]
\centering
\includegraphics[width=0.7\textwidth]{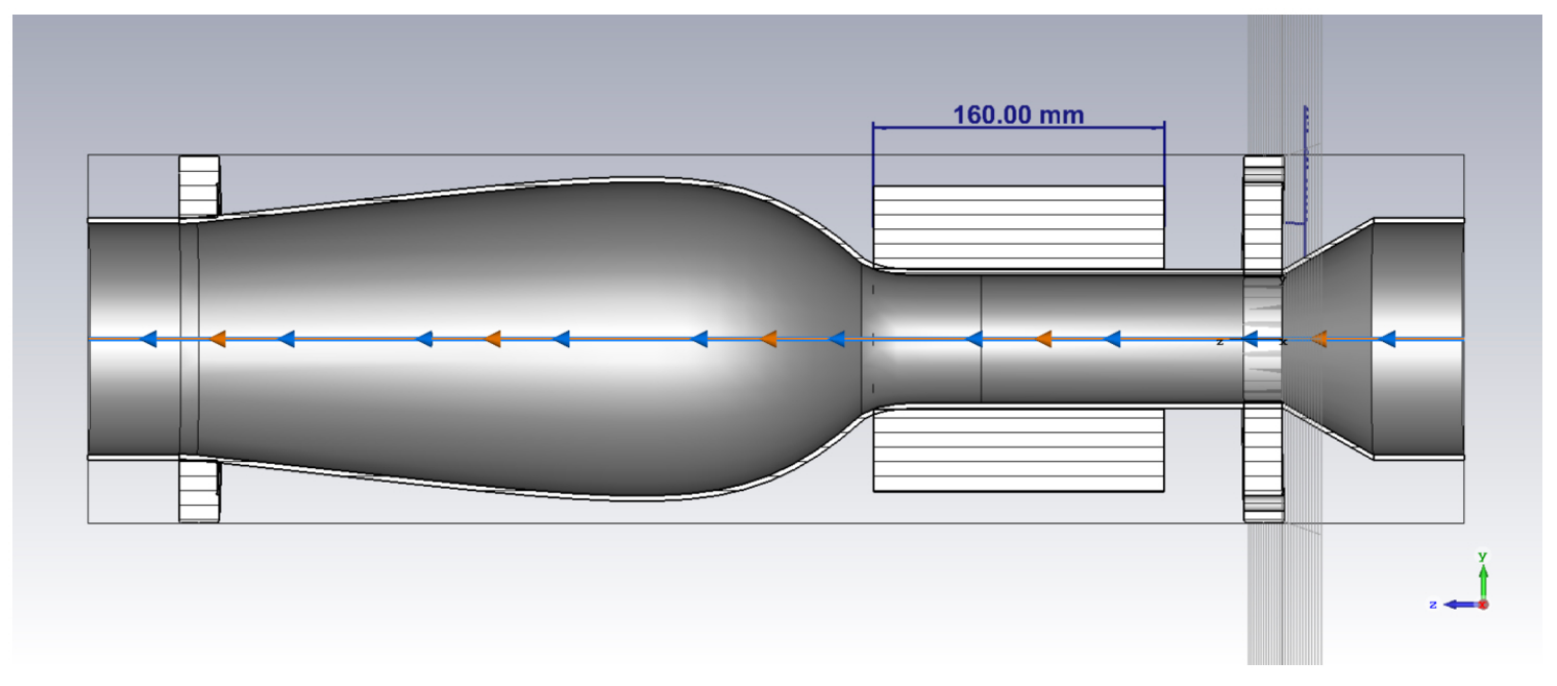}
\caption{Geometry of the vacuum chamber used in simulations.}
 \label{fig_mcp1}
\end{figure}

At maximum luminosity, the average beam current and the peak bunch current in the collider rings are about 0.5 and 10 A. That makes it challenging to obtain a stable circulation of the beams. A standard recipe to minimize problems is: 
\begin{enumerate}
\item to make the vacuum chamber as smooth as possible, avoiding jumps in the vacuum chamber cross-section;
\item to cover all bellows between different elements of the vacuum chamber and vacuum ports by shields, to avoid getting cavities, where electromagnetic oscillations could be supported;
\item all plates, pickups, and other vacuum chamber elements, seen by the beam have to be installed into the same cross-section and properly loaded to minimize their excitation by the beam. 
\end{enumerate}
The quality of the vacuum chamber is characterized by its transverse and longitudinal impedances, which are functions of the frequency.  
 
To minimize scattering and absorption by the vacuum chamber walls for particles produced in the collisions, the MCP  detectors are  suggested to be mounted, as shown in Fig. \ref{fig_mcp1}. Such a choice minimizes the amount of material, which a particle traverses on the way from the interaction point to the detector, and, thus, addresses problems of ions scattering and absorption by the vacuum chamber. However, it creates a cavity, which increases the ring impedances. To minimize the contribution of these cavities, their shape is chosen to have only a smooth transition in the chamber radius, and the maximum radius is chosen to be only slightly above the detector radius. 

\begin{figure}[h]
\centering
\includegraphics[width=1.0\textwidth]{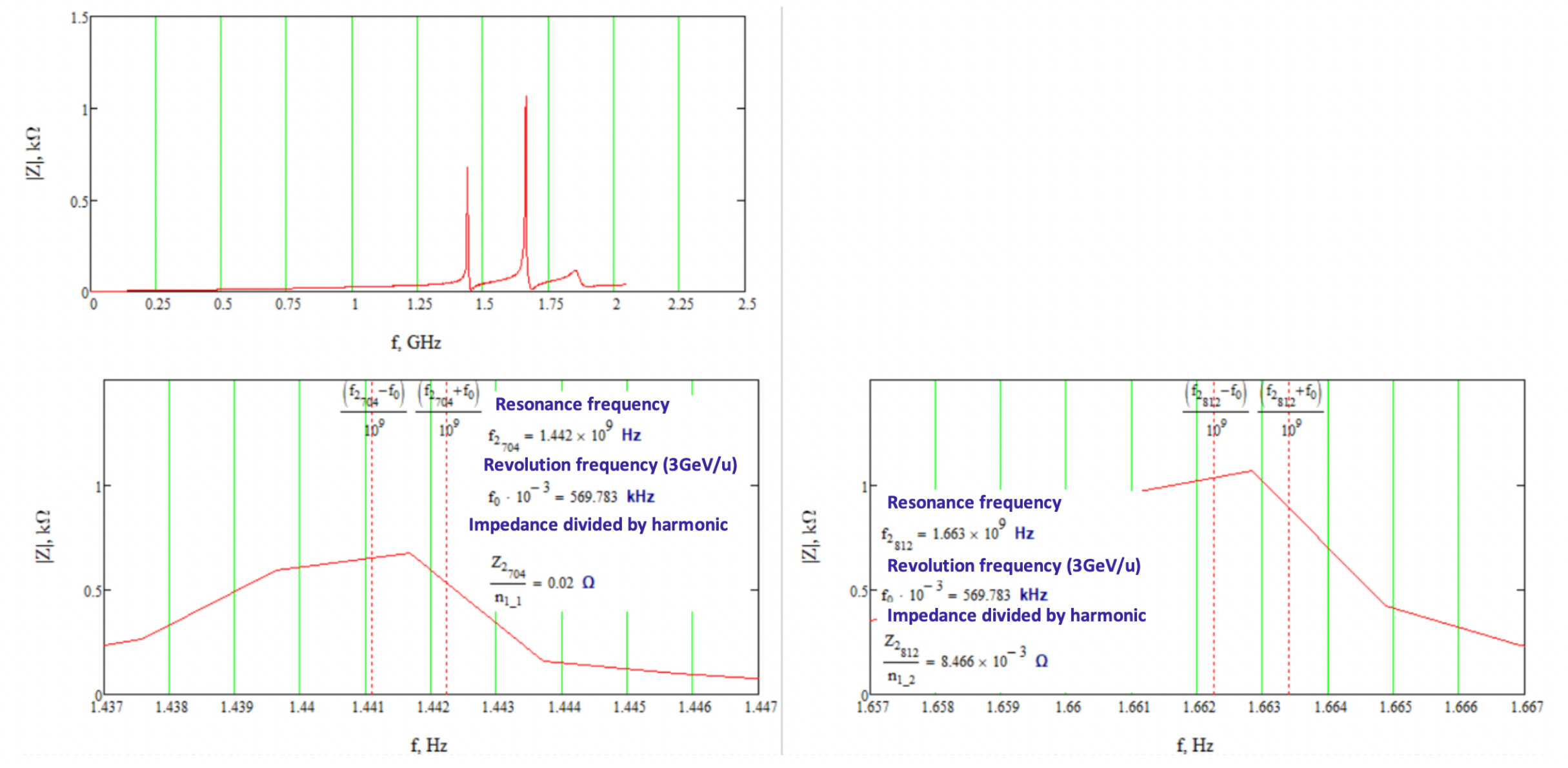}
\caption{Results of simulations: (top) -- longitudinal impedance, (bottom left) -- impedance details near the peak at 1.44 GHz, (bottom right) -- impedance details near peak at 1.66 GHz. }
 \label{fig_mcp3}
\end{figure}

\begin{figure}[h]
\centering
\includegraphics[width=1.0\textwidth]{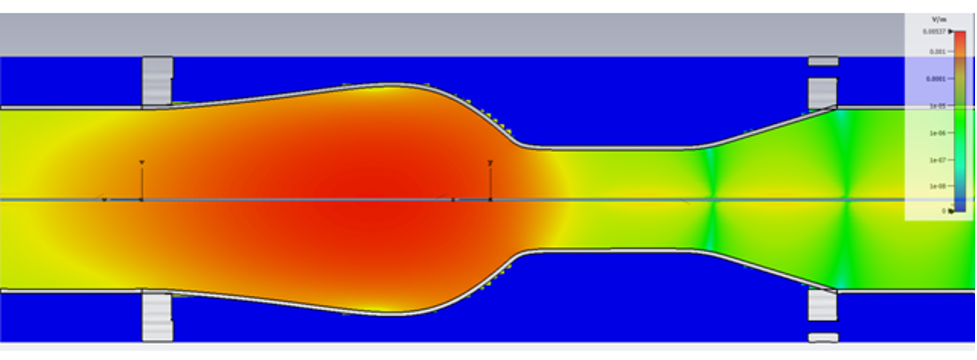}
\caption{Calculated electric field (V/m) for the smoothed beam pipe without RF instabilities.}
 \label{fig_MCPfield}
\end{figure}

Figure \ref{fig_mcp1} shows the cavity geometry, used in the simulations, and Figure \ref{fig_mcp3} presents the computed longitudinal impedance. As one can see, the impedance is significant only at frequencies above 1 GHz, where two resonant peaks can be seen at frequencies of about 1.44 and 1.66 GHz, which are well above the frequency spectrum of NICA bunches. Note also, that the peak values of a single cavity are close to the peak values of a single Beam Position Monitor impedance (800 $\Ohm$, 0.87 GHz). The instability of a continuous beam is driven by $Z_n/n$, where $Z_n$ is the impedance value at $n$-th harmonic of the revolution frequency. For the peaks presented in Figure \ref{fig_mcp3}, $Z_n/n$ is about 0.3 $\Ohm$. This value is well below the space charge impedance 
\begin{equation}
Z_n/n = i \frac{Z_0}{\beta\gamma^2}ln(\frac{a}{r}),
\end{equation}
which sets up the scale of the problem. Here, $\beta$ and $\gamma$ are the relativistic parameters of the beam, $Z_0=377$ $\Ohm$, and $a/r$ is the ratio of the vacuum chamber radius to the beam radius. Thus, we can conclude that an addition of two (or four) cavities, presented in Fig. \ref{fig_mcp2}, should not cause a significant effect on the beam stability in NICA. The calculated electric field distribution for the proposed smoothed beam pipe is shown in Fig. \ref{fig_MCPfield}.
\begin{figure}[h]
\centering
\includegraphics[width=1.0\textwidth]{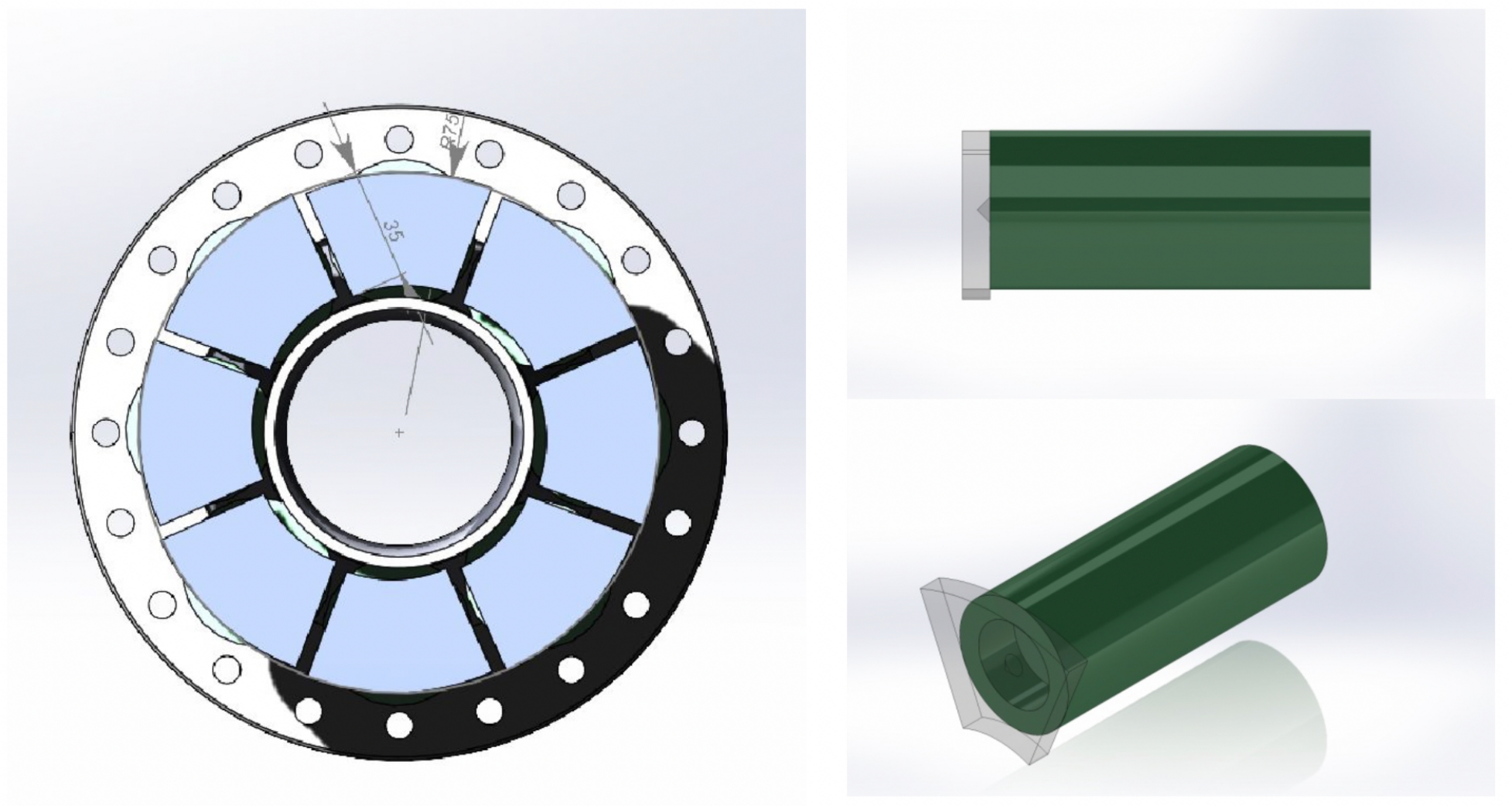}
\caption{View of the photomultiplier casing and fast scintillator attached to its end.}
 \label{fig_mcp4}
\end{figure}

Thus, it was decided to place two 32-channel BBC MCP detectors beyond the vacuum beam line of the accelerator, maximally close to the beam axis. The design of the BBC MCP detector, developed earlier (see SPD CDR \cite{SPD:2021hnm}), is replaced by the one shown below. It was decided to use photomultipliers ''TOPAZ'', based on MCP, manufactured by the BASPIK company (Vladikavkaz) \cite{baspik}. The prototypes of the "TOPAZ"-based detectors (see Fig. \ref{fig_mcp4}) were tested in the laboratory environment with radioactive sources and a picosecond laser, as well as at 200 MeV electron beam of LINAC-200 (DLNP, JINR). A time resolution of about 50 ps was achieved with these detectors. 
This time resolution allows one to find the position of the interaction point within 1.5 cm along the beam line.

It is obviously possible to upgrade these detectors for particle registration at angles larger than 2$^{\circ}$, by increasing the assembly size and the number of detectors. 

\subsection{Readout electronics}
\begin{figure}[!h]
  \begin{minipage}[ht]{0.4\linewidth}
    \center{\includegraphics[width=1\linewidth]{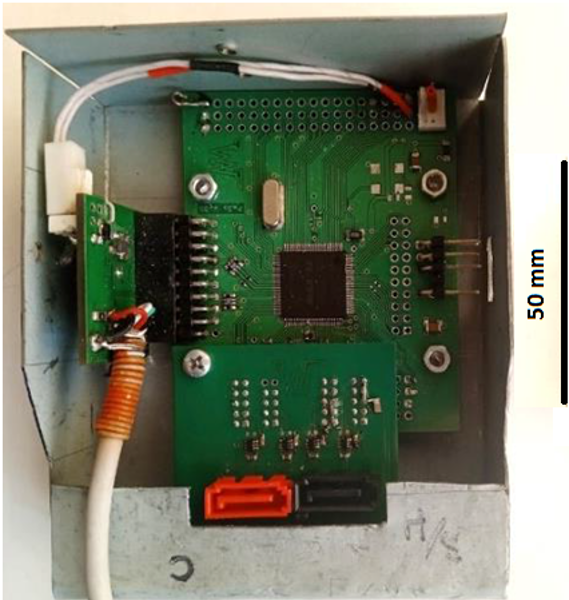} \\ (a)}
  \end{minipage}
  \hfill
  \begin{minipage}[ht]{0.6\linewidth}
    \center{\includegraphics[width=1\linewidth]{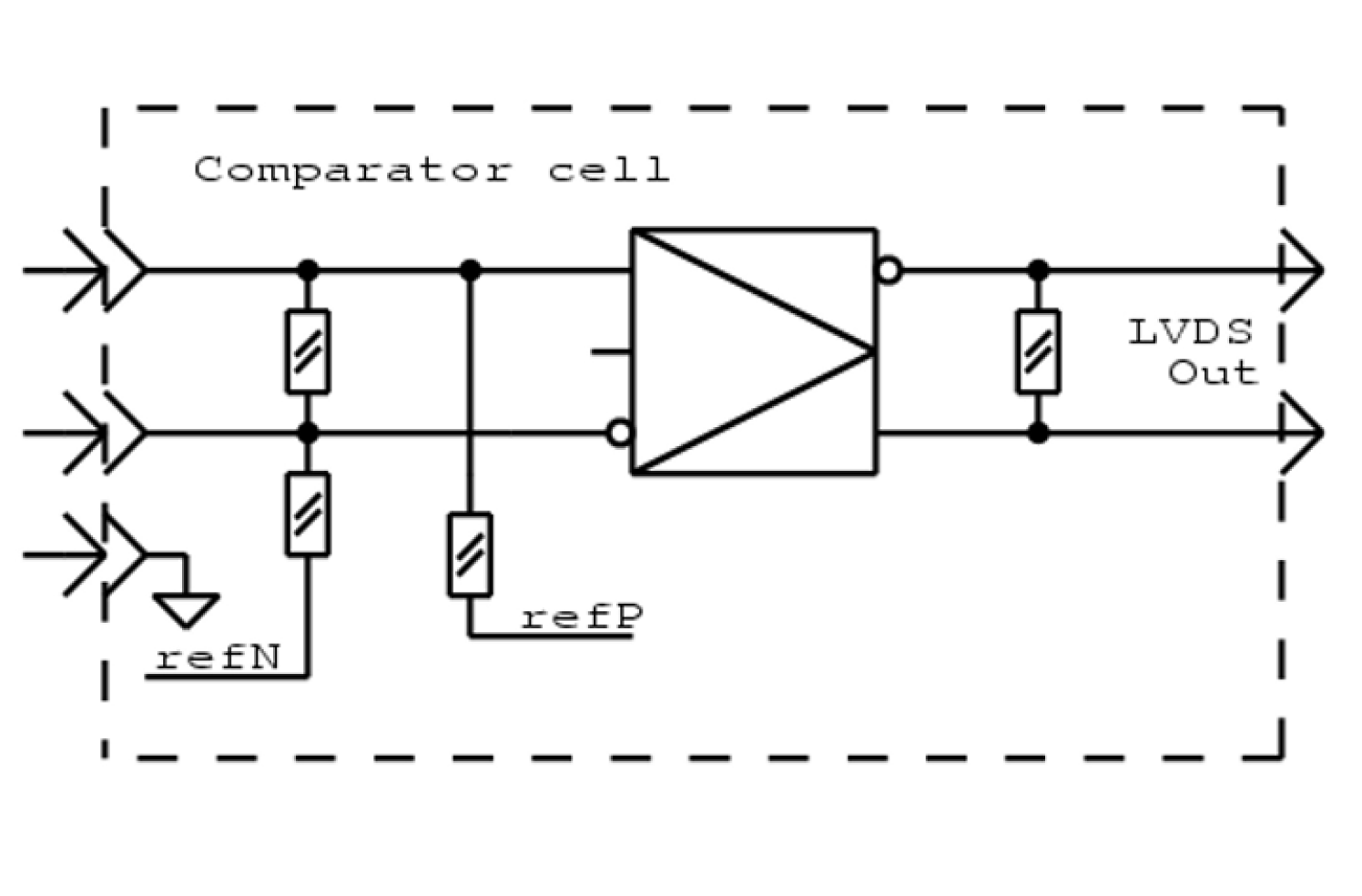} \\ (b)}
  \end{minipage} 
  \caption{ (a) Four-channel electronic module. (b) Block diagram of a cell of the differential comparator. 
	}
\label{lb_mcp_FE_1}
\end{figure}

\begin{figure}[h!]
\centering
\includegraphics[width=1.0\textwidth]{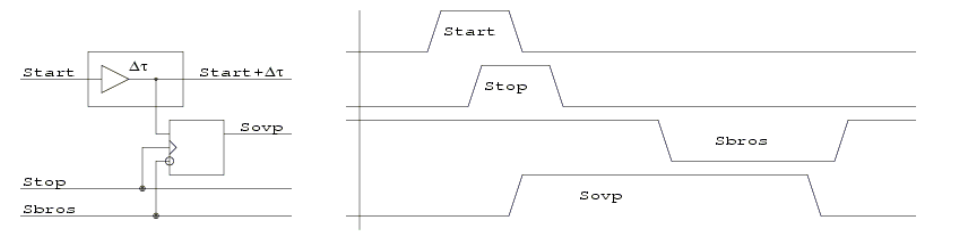}
\caption{Time diagram of a cell.}
 \label{lb_mcp_FE_2}
\end{figure}

The developed prototype of the electronic module for the SPD beam-beam counter is based on the 4-channel module created earlier. The prototype uses a differential comparator (Fig. \ref{lb_mcp_FE_1}(a)). Each channel consists of LVDS-LVCMOS, the TDC, and the output code generator. The block diagram of registration channels of a differential comparator is shown in Fig. \ref{lb_mcp_FE_1}(b). Figure \ref{lb_mcp_FE_2} represents the time diagram of a cell.

\begin{figure}[h!]
\centering
\includegraphics[width=1.0\textwidth]{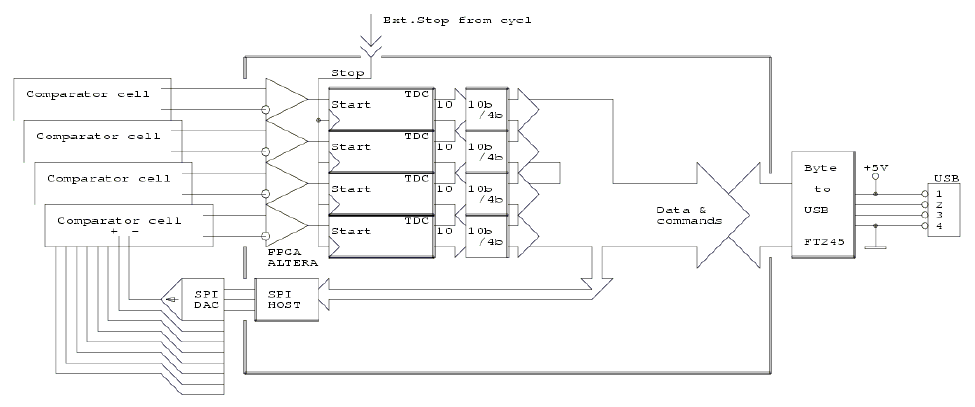}
\caption{Schematic diagram of the 4-channel time-to-code converter.}
 \label{lb_mcp_FE_3}
\end{figure}
\begin{figure}[h!]
\centering
\includegraphics[width=1.0\textwidth]{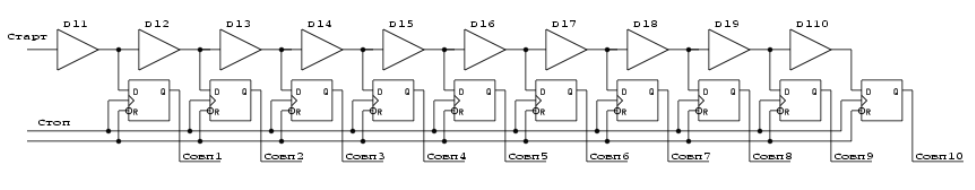}
\caption{Structure of the time-to-code converter.}
 \label{lb_mcp_FE_4}
\end{figure}

To increase the rate of time information processing in the multichannel system, we developed a time-to-code conversion system based on delayed matching. Four time-to-code converter channels with a resolution of 200 ps on an interval of 2 ns were implemented in FPGA EPM240. The interval duration is limited by the structural specific features of the FPGA scheme applied in this prototype. The scheme of the 4-channel time-to-code converter is shown in Fig. \ref{lb_mcp_FE_3}. The structure of the time-to-code converter is shown in Fig. \ref{lb_mcp_FE_4}.

The preliminary design and simulation of the fast readout system for the beam-beam collision counter was performed using the Quartus software.
The modeling in the Timing Simulation mode proved the predicted performance of the readout scheme: not longer than 10 ns per each event for the 4-channel system. The time of conversion from the thermometric to binary code for each channel does not exceed 6 ns, the reset delay does not exceed 3 ns. For the 32-channel coordinate and time registration system using signals from MCP detectors constructed based on the above-mentioned scheme, the total processing time with account of all delays does not exceed 20 ns. The simulation of the system operation showed its capability of determining the particle detection time with a precision (sigma) of at least 100 ps.

\begin{figure}[h!]
\centering
\includegraphics[width=0.7\textwidth]{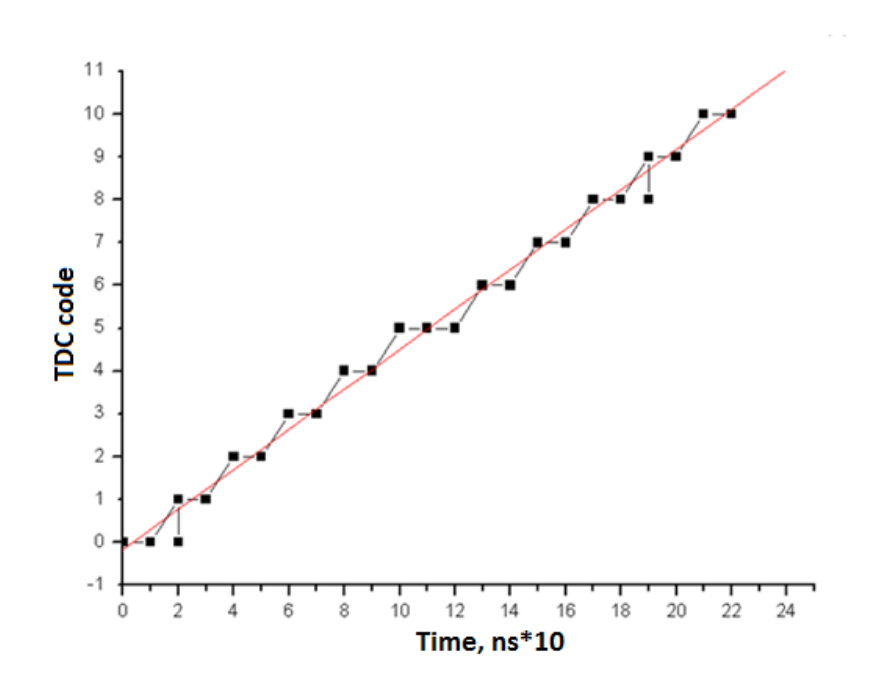}
\caption{TDC code for one of the prototype channels as a function of introduced delay. }
 \label{lb_mcp_FE_5}
\end{figure}

The prototype fast electronic module was tested using a 4-channel generator of nanosecond pulses with a controllable delay between channels and a phase stability of $\pm$25 ps. The delay between channels varied with a step of  100 ps. Output signals of the generator were used as model signals in the tests. 
The obtained data confirm the numerical results. The time resolution of the developed system is 200 ps (see Fig. \ref{lb_mcp_FE_5}). 

Thus, the prototype of the multichannel fast electronic module for the MCP detectors based on the fast discrete comparators and the multichannel time-to-code converter using the FPGA EPM240 was developed and manufactured.
The simulation of the fast readout system performance for the beam-beam collision monitor using Quartus demonstrated the capability of detecting the particle collision time with a precision of at least 200 ps.
The module software was developed. The data obtained using a generator confirm the results of the simulations. The time resolution of the developed system is 200 ps.
Eight electronic modules are capable of digitizing and recording information on the registration time in one event of the number of particles from 32 pads during a time interval of less than 20 ns, which allows one to use this system as a fast trigger.
The final version of the multichannel fast electronic module will be based on a modern FPGA of the Cyclone-10 or MAX-10 series.

\subsection{Cost estimate}
The BBC MCP cost  estimate is based on the cost of building the MCP-based profilers for the booster and NICA rings. For two rings with electronics, high voltage power supply, cables and connectors, it is about 250 k\$. Because of the complex profile of the vacuum chambers, we estimate their fabrication at an additional 100 k\$. The detailed calculation is presented in Tab. \ref{MPC_cost_tab}.

\begin{table}[htp]
\caption{Cost estimate.}
\begin{center}
\begin{tabular}{|l|c|c|}
\hline
 Item & Number & Cost, k\$ \\
 \hline
 MCP detector (1 channel) 	& 64	& 1.7\\
Electronics (1 channel)	& 64 &	0.45\\
HV power supply (1 channel) &	64 &	0.6\\
Cables 	&&	27 \\
FPGA		&& 12.6 \\
Support, mechanics	&&	5 \\
Assembling, tests	&&	28 \\
Vacuum chambers    &&  100 \\
\hline
Total	&& 348.6\\
\hline
\end{tabular}
\end{center}
\label{MPC_cost_tab}
\end{table}%

\chapter{Integration and services \label{lbl:servises}}

\begin{figure}[h!]
\begin{center}
\includegraphics[width=1\textwidth]{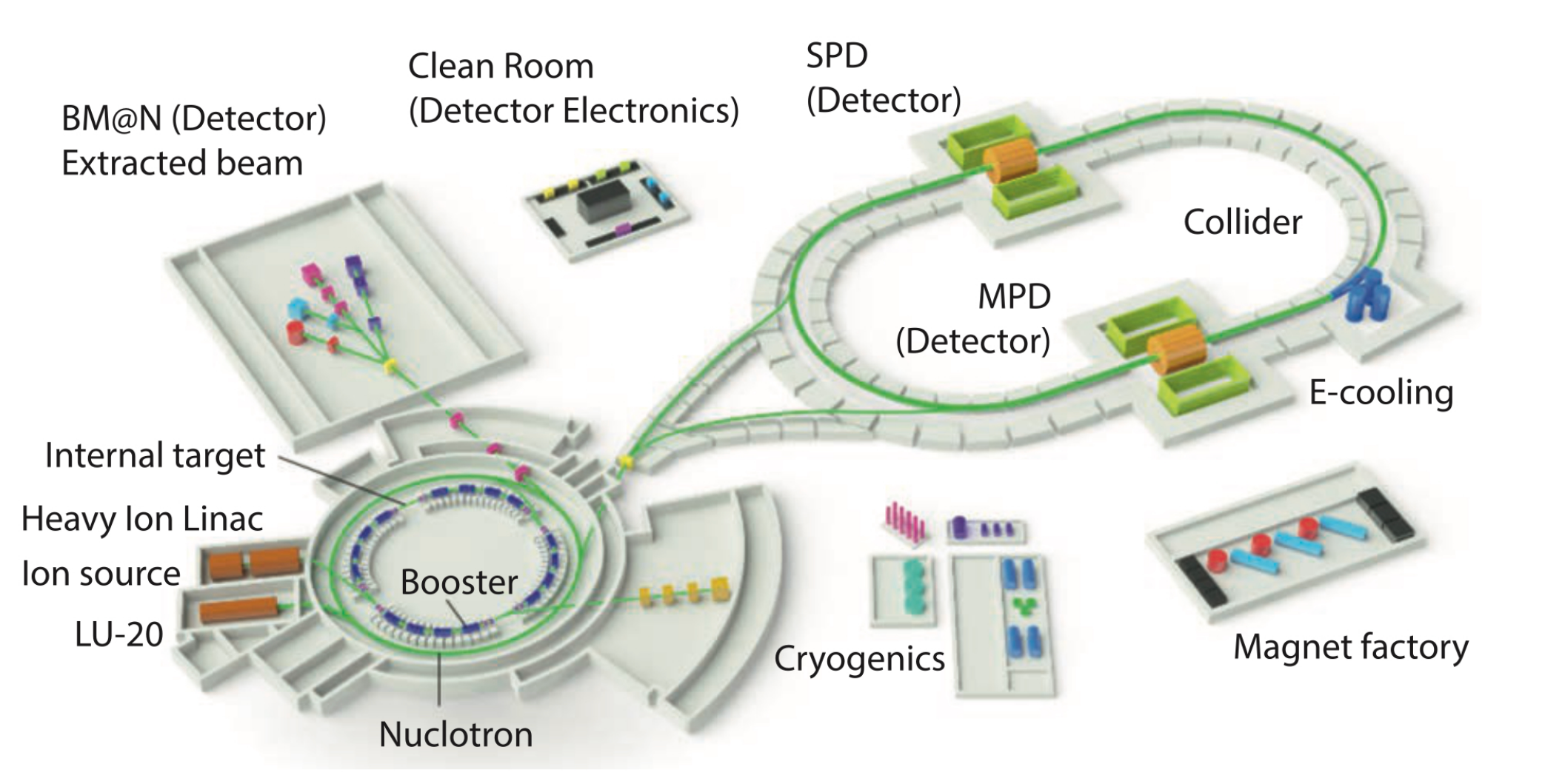}
\caption{Schematic layout of the NICA facility with SPD (southern point) and MPD (northern point).}
\label{fig:nica}
\end{center}
\end{figure}

The NICA complex  is shown in Fig. \ref{fig:nica}.
It includes an injection complex, a booster, an upgraded Nuclotron and two storage rings with two interaction points 
aimed at the MPD and SPD detectors, respectively. 
The SPD area is located in the southern point of beam collisions. 

The layout of the SPD experimental hall is very similar to that of the MPD  \cite{MPDCDR2014}.


\section{SPD experimental hall}

\begin{figure}[t]
\begin{center}
\includegraphics[width=0.7\textwidth]{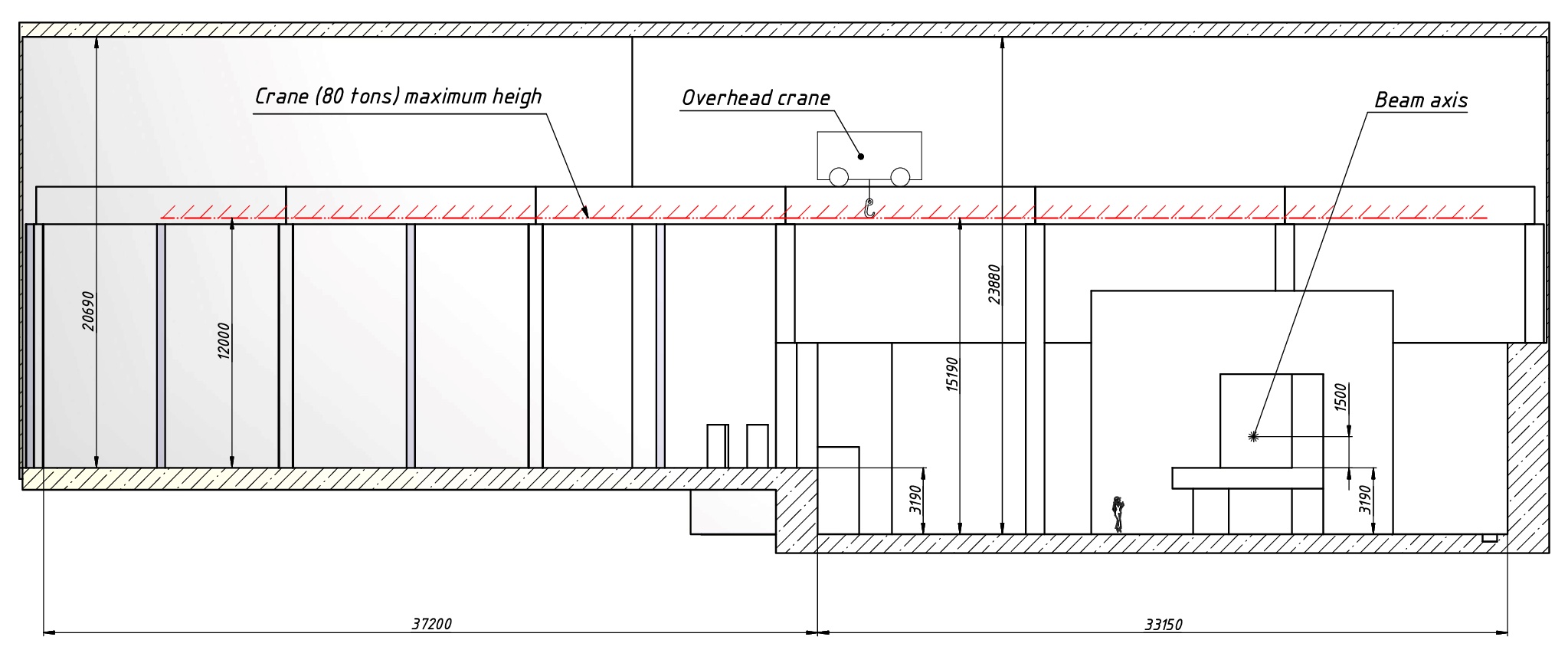}
\caption{Side view of the SPD building.
The production site is one on the left (ground level), and the experimental site is on the right (3.19 m below ground level).
The maximum height accessible to the overhead crane is marked with a red line.}
\label{fig:spd-hall-1}
\vspace{0.5cm}
\includegraphics[width=0.7\textwidth]{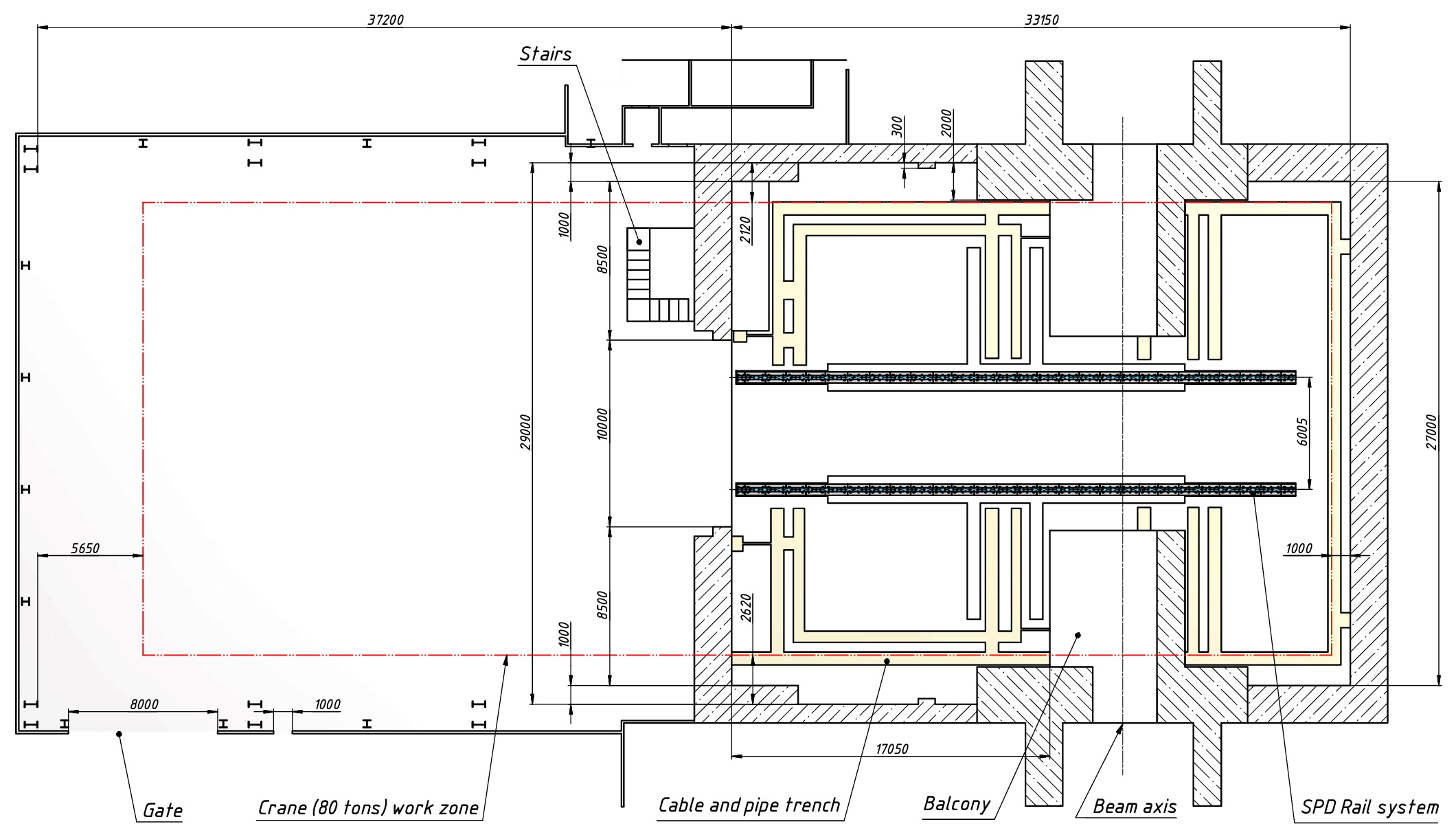}
\caption{View from the top of the SPD building. 
The production site is one on the left, and the experimental site is on the right.
The area accessible to the overhead crane is marked with a red line.
The gutters in the concrete floor used for laying communications are marked in yellow.}
\label{fig:spd-hall-2}
\end{center}
\end{figure}

The SPD experimental facility is accommodated  in  a single-span steel hall (Room 140) with the overall dimensions of $32 \times 72$~m$^2$. 
The top and side views of the SPD building are shown in Figs. \ref{fig:spd-hall-1}, \ref{fig:spd-hall-2}, respectively. 
The building is divided into  {\it production} and {\it experimental} sites (Rooms 140/1 and 140/2, respectively) with  the  total area of about 2000~m$^2$. 

Since the {\it production site} has a gate, the site will be used for unloading materials.
The size of the gate is \mbox{$8\,{\rm m} \times \rm{8\,m}$}, which is large enough for trucks.
The production site will also be used for preparation, testing and maintenance of the detector subsystems.
Elements of the gas system can also be placed on the site, which will distribute, monitor, and supply the necessary gas mixtures to the gaseous detectors of SPD.
The floor level of  the production site coincides with the ground level.

The {\it experimental site} is shown on the right side of Fig. \ref{fig:spd-hall-1}, \ref{fig:spd-hall-2}.
This part of the building is a~reinforced  array made of protective monolithic concrete  
with a wall thickness from 1\,m to 3\,m and height up to 11\,m. 
The dimensions and configuration of the site were designed in accordance with the terms of reference, taking into account the provision of biological protection.
The floor is lowered below the level of the production site by 3.19\,m
in order to match the levels of the beam-line and the symmetry axis of the SPD detector.
There is a 10 m wide mounting opening in the 2\,m thick wall separating the production and experimental sites.
After installing the detector, this opening will be laid with concrete blocks. 
For the passage of personnel and transportation of small-sized equipment, a labyrinth is provided.
It begins in the well at the production site and leads to the level of experimental site.

\begin{figure}[t]
\centering
\includegraphics[width=0.49\textwidth]{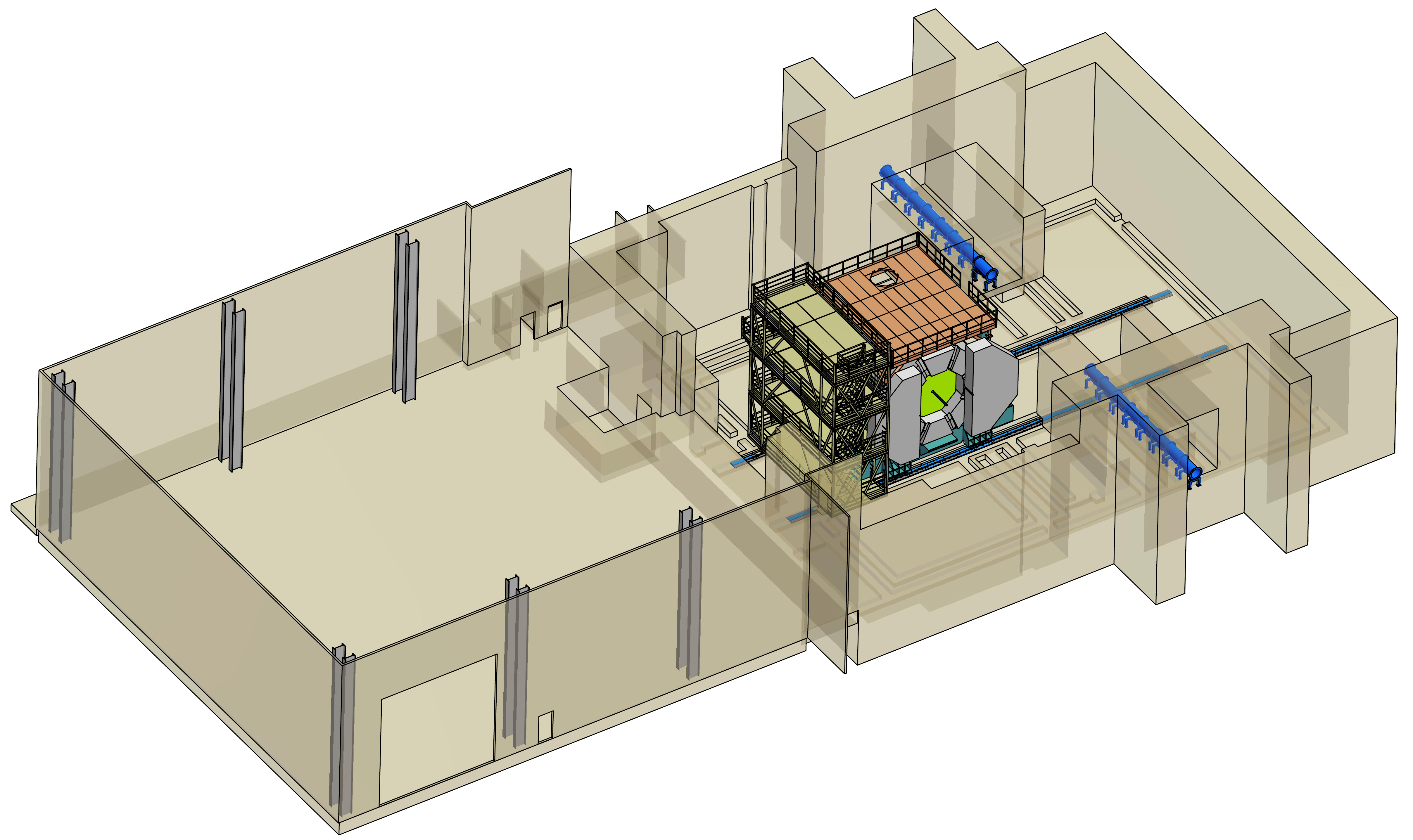}
\hfill
\includegraphics[width=0.49\textwidth]{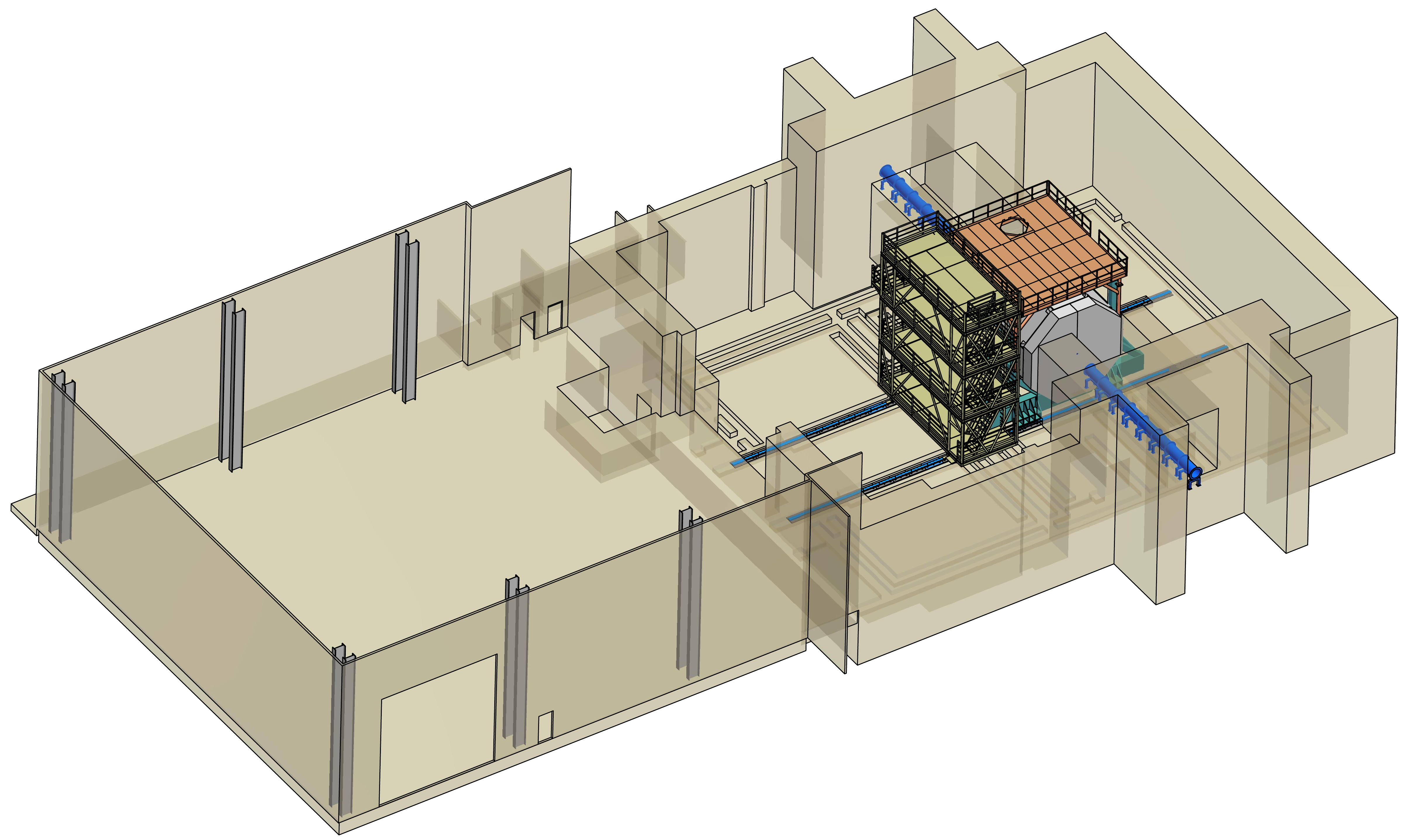}
\caption{3D rendering of the SPD building. 
The detector assembly position is shown in the left figure.
The beam position of the detector is shown in the right figure.
Beam line is shown in blue.}
\label{fig:spd-hall-3D}
\end{figure}

\begin{figure}
 \center{\includegraphics[width=0.8\textwidth]{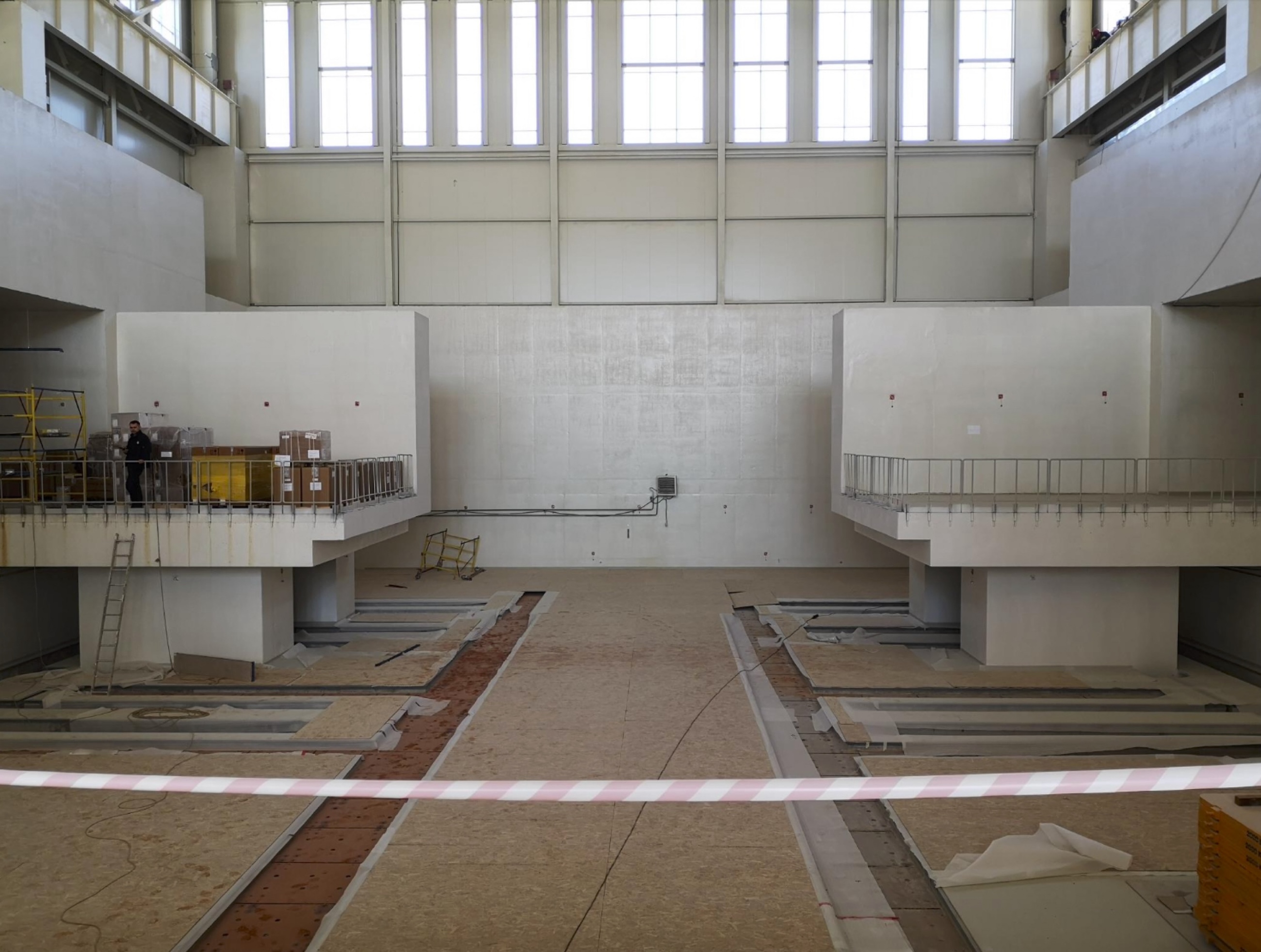}}
\caption{Experimental site of the SPD building (24.03.2022).}
\label{fig:spd-hall-photo}
\end{figure}

Two main rails are installed along the experimental site  to provide transportation of the SPD detector from
the assembly position to the operating (or beam) position.
 Schematic views of the SPD building with the detector in the assembly and operating positions are shown in Fig.~\ref{fig:spd-hall-3D}.
 The view of the experimental site is presented in Fig. \ref{fig:spd-hall-photo}.
Potentially, two pairs of additional rails, located across the main ones, can be installed in the detector assembly area.
They can be used for rail-guided support, in order to insert inner subdetectors.
All rails are electrically connected to the metal reinforcement of the building, which, in turn, 
is connected to an array of rod-electrodes  buried in the earth
by 3\,m 
below the building for grounding.

The concrete solid floor of the experimental site has sufficient load-bearing capacity to allow the assembly and 
operation of the SPD detector, including auxiliary structures.
This capacity will be quite enough:

\begin{itemize}
\item to withstand the weight of the fully assembled detector with all necessary services;
\item to preserve the integrity of the detector during its transportation 
between the assembly position and beam position;
\item to provide a stable position of the detector with high accuracy during operating cycles.
\end{itemize}
The load on the floor of the experimental site was calculated based on the total weight of the experimental setup.
The estimated weight of 1245 tons meets the requirements of safety regulations.
The weight of each detector subsystem contributing to the overall load is presented in Table~\ref{tab:spd-tech-req}.
According to the project documentation, the shrinkage of the earth under the building will not be larger than 3~cm 
for the entire period of maintenance of the building.

An overhead traveling crane with a maximum lifting capacity of 80 tons is installed in the hall of the building.
It will ensure the movement of  the detector parts from the unloading to assembly area.
The crane can be equipped with linear or H-shaped traverses with lifting capacity from few to 75~tons.
The crane service area covers both the production and experimental sites.
The area accessible to the crane is marked with a red line in the drawing of Fig.~\ref{fig:spd-hall-2}.

\section{Gas supply systems}

\begin{figure}[t]
\centering
\includegraphics[width=1.\textwidth]{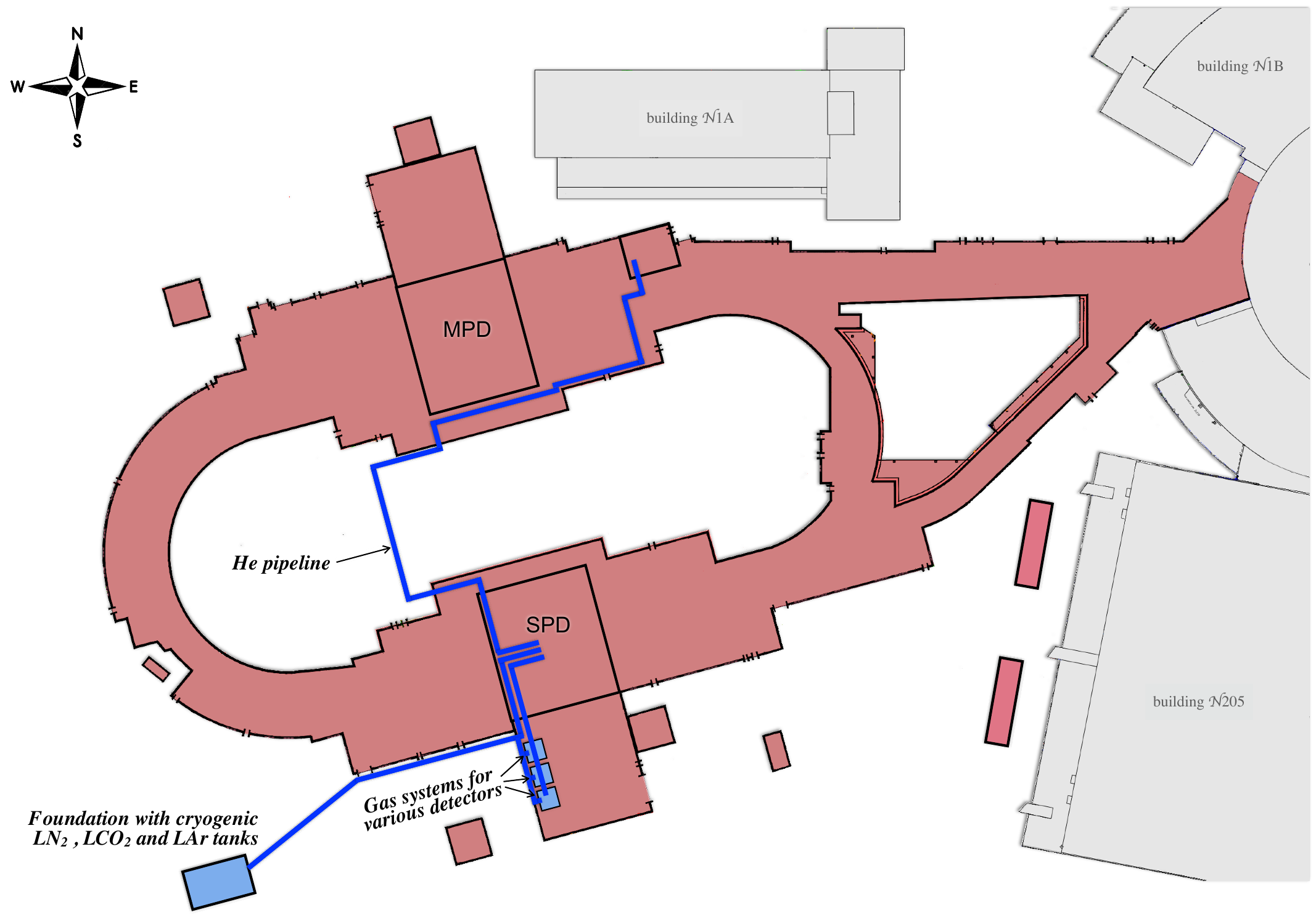}
\caption{Top view of the NICA collider building with indications for elements of the SPD gas systems (shown in blue).}
\label{fig:spd-pipeline}
\end{figure}

The SPD subsystems consuming special gases are listed in Table~\ref{tab:spd-tech-req}.
With regard to gas supply systems, we will follow generally the same strategy as used by MPD.
For safety reasons, as well as due to restricted access to the experimental area during the operation of the accelerator, 
pressurized gas tanks will not be placed on the experimental site.
For those gases that will be used extensively and therefore require large storage tanks, 
the primary gas supply point will be located outside the SPD building.
Vessels for other gases, which are used little and therefore do not require large storage tanks, 
can be placed on the production site.
Those gases that are toxic and harmful to the environment (like freon) can be recycled.
A top view of the NICA collider building with designations of the elements of the SPD gas systems is shown in Fig.~\ref{fig:spd-pipeline}.
A technological scheme of the gas supply system is shown in Fig.~\ref{lb_cryo_05}.

The {\it cryogenic helium plant}
should be close to the detector to provide cryogenic fluids and gas 
to maintain low temperatures for the magnet operation.
A refrigerator will be installed on a platform on top of the detector, as described in Section~\cref{sec:Cryogenic_He_system}.
Helium will be delivered from the central cryogenic station, located on the MPD side of the NICA complex.
A warm (non-cryogenic) pipeline with high and low pressure helium flows will connect the station with the refrigerator. 
The total length of the pipeline is about 220~meters.

A {\it nitrogen system} is required for  operation of the cryogenic helium plant. 
The system will include  two cryogenic storage tanks with liquid nitrogen and  
cryogenic pipeline connecting those tanks with the helium plant, as described in Section~\ref{sec:Cryogenic_N2_system}.
The  storage tanks will be located on a concrete foundation approximately 50 meters from the SPD building.
The connecting pipeline will be raised to a height of 6~meters to pass over the road.
Discharge of gaseous nitrogen into the atmosphere will be carried out according to the principle of the shortest path.

The  gas system for the SPD {\it gas detectors} is planned to be implemented following a modular design, 
so it can be adapted to  the requirements of each  detector,
but still be based on a common architecture.
There are four detectors in SPD which need to be supplied with gas. 
Arranged in descending order of  gas volume, they are RS, ST, TOF, and MCT.
As one can see from Table~\ref{tab:spd-tech-req}, the mixtures used by the detectors contain common components.

For instance, {\it argon} is used in the RS, ST, and MCT detectors,
due to a very low electron attachment coefficient, 
that ensures most of the free electrons to reach the multiplication region without interference.
Depending on the detector, the fraction of Ar in the gas mixture varies from 70\% to 90\% (see Table~\ref{tab:spd-tech-req}).
In order to avoid using multiple Ar vessels for different detectors,
a single tank with argon at cryogenic temperature will be employed.
It will be installed on a concrete foundation that is also used for the nitrogen tanks.
The argon will be warmed and fed into the SPD building via a pipeline, using the same trestle above the road that is used for the nitrogen pipeline.


A key requirement for the {\it TOF gas system} (C$_2$H$_2$F$_4$:C$_4$H$_{10}$:SF$_6$ = 90:5:5) is the possibility 
to recirculate and reuse the gas exiting from the detector. 
The gas mixture can, in fact, be collected after being used, and re-injected into the supply line after purification.
Due to the relatively small volume of MRPC chambers that need to be filled with gas, all components of the system 
can be installed in a specially designated area of the production site of the SPD building.
After mixing in the proper proportion, the gas mixture will be transferred to the detector via a pipeline 
running over the wall, separating the production and experimental sites. 
Another pipeline will return the used gas mixture back to pass through the purification module.
Particular attention will be paid to the monitoring of gas leaks, 
as the mixture used by TOF contains toxic and flammable components.
Moreover, they are quite expensive.
If the leak is detected, the control system will block the gas circulation, and 
an audible alarm will alert people, giving them the opportunity to leave the area.

In order to optimize the space available in the SPD building,  fenced zones,  dedicated to  the racks of the gas system of each detector,
will be allocated on the west side of the building (the far side from the gate), as shown in Fig.~\ref{fig:spd-pipeline}.

\section{Power supply system}

\begin{figure}[t]
\centering
\includegraphics[width=0.99\textwidth]{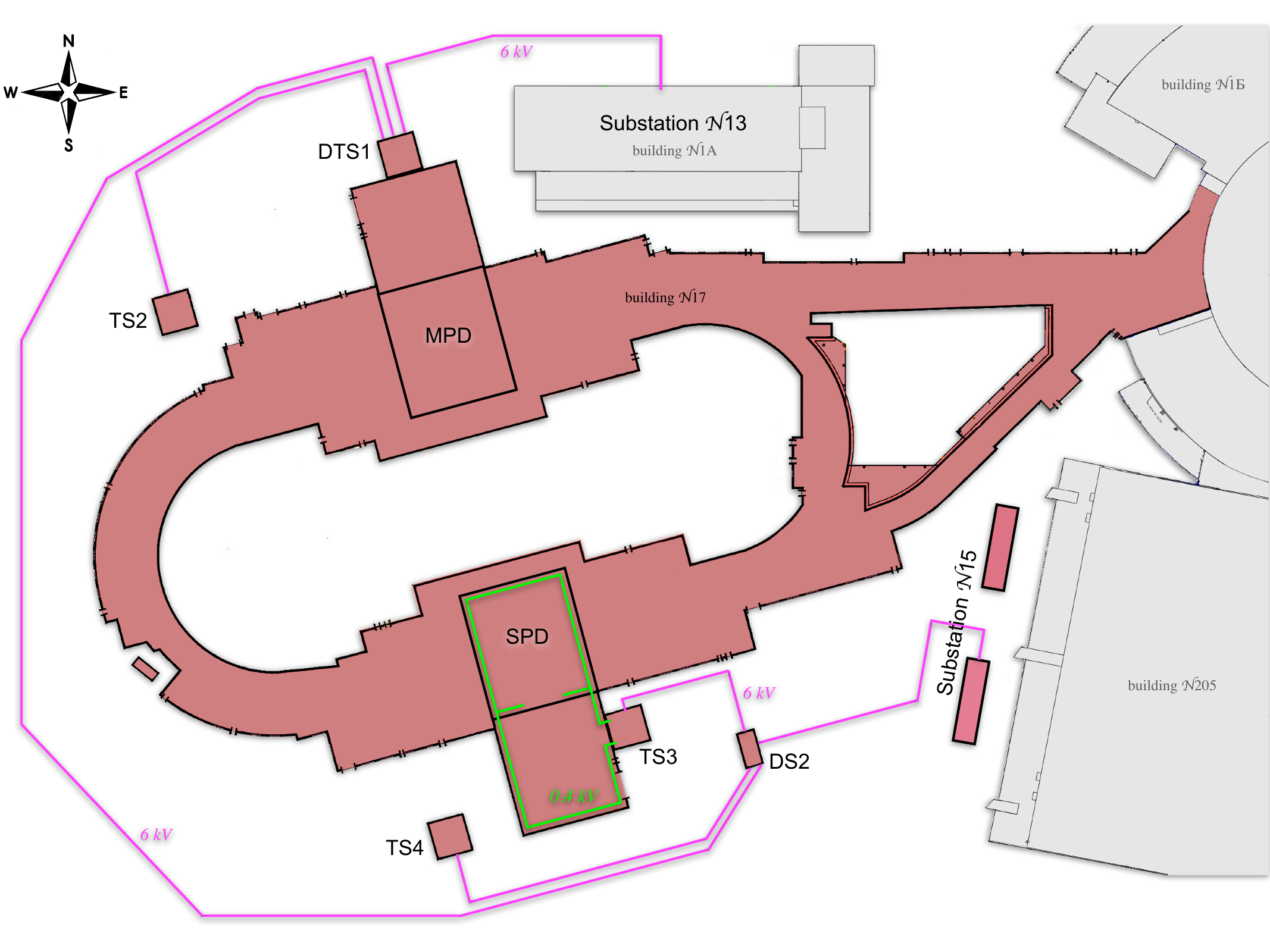}
\caption{Top view of  building \texttt{N}17 (the NICA collider) with indications for elements of the SPD power supply system.
Power lines with an operating voltage of 6~kV are shown in pink. 
Power lines with an operating voltage of 0.4\,kV are shown in green. 
}
\label{fig:spd-electro}
\end{figure}

The power supply system of SPD is schematically presented in Fig.~\ref{fig:spd-electro}. 
The sequence of power substations starting from the principal one to the SPD experimental setup is as follows:
\[
   {\rm PSS1} \rightarrow {\rm S15} \rightarrow {\rm DS2} \rightarrow {\rm TS3} \rightarrow {\rm SPD~setup},
\]
where 
PSS1 is the Principal Step-down Substation 110/6 kV "Dubna" of the VBLHEP site;
S15 is a recently rebuilt transmission Substation \texttt{N}15, which also supplies  power to building \texttt{N}205 
(test  zone for extracted beams of Nuclotron);
DS2 is a Distribution Substation \texttt{N}2; and 
TS3 is a Transformer Substation~\texttt{N}3, which is adjacent to the SPD building.
The collider building has a 6~kV radial power supply network with two distribution points DTS1 (MPD side) and DS2 (SPD side).
This, in general, also makes  it possible to receive the power for SPD from the transmission Substation \texttt{N}13.

A  transformer substation TS3 is a termination of the distribution line intended to supply SPD with electric power.
It contains two three-phase oil-immersed transformers of a sealed version, without oil conservators, 
each with a power of 2.5 MW.
The substation also includes all the necessary devices for control and protection of itself and the lines. 
The transformers step down the voltage from the high transmission level of 6~kV to the low  voltage output of 0.4~kV.
This output from each transformer feeds a complete bus-bar trunking with a rated current of 4 kA,
through which electricity is supplied to the power pannels, located along the walls of the SPD hall.



The power available to the SPD setup is determined by the 2-nd category of reliability of the entire system.
According to the requirements of this category, consumers must be able to switch to an auxiliary power source within a few minutes in case of an emergency situation. 
In practice, this will mean that both 2.5~MW transformers will operate at half power,
and  that all power goes to one of them when the other fails.
The summary of the power reduction factors:

\begin{itemize}

\item due to the 2-nd category of reliability on the 6 kV side, the power of two transformers is reduced down to \mbox{5\,MW / 2 = 2.5\,MW};
\item due to the 2-nd category of reliability on the 0.4 kV side,  the power is further reduced down to  \mbox{2.5 MW / 2 = 1.25 MW}
 (efficiency and own consumption are not taking into account).

\end{itemize}
Thus, as a result, the SPD can receive the power of about 1.2\,MW which can be entirely consumed by the experimental setup.
A tentative summary with the estimated consumption by each detector subsystem is give in Table\,\ref{tab:spd-tech-req}.
The current estimate of the consumption value is approximately 100 kW, which is an order of magnitude lower than the available capacity.

Despite the fact that most of the equipment in the hall belongs to the 2-nd category of power supply reliability,
there is also equipment of the 1-st category. 
Equipment of this category must be able to switch to an auxiliary power source within a second in case of an emergency situation. 
It includes emergency ventilation systems, parts of fire fighting systems, and emergency lighting.
Furthermore, there is a special group of the 1-st category of reliability, which includes technical means of automated control systems, radiation monitoring, and cryogenic system.
All this equipment will be provided with uninterruptible power supplies with built-in rechargeable batteries for 1 hour of autonomous operation.

In order to avoid long cables,
crates with electronics for the data acquisition systems and power supplies 
should be placed in the vicinity of the detector. 
They will be located either on a side platform attached to the detector or on the top platform together with  the cryogenic equipment. 
Both platforms will be electrified and instrumented with an overhead cable carrier to ensure a continuous supply of electricity to the subsystems, even while the entire SPD setup is being moved from the assembly position to the operating position on the beamline.



\begin{table}[h!]
\caption{Technical requirements for the SPD detector subsystems$^*$.}
\renewcommand{\arraystretch}{1.3}
\begin{center}
\begin{tabular}{|l|c|c|c|c|}
\hline
Subsystem  &  Weight,  & Power      &  Gas,  \\
                    &       ton          &      kW      & compositions \\
\hline
\hline
\rowcolor[gray]{0.95}[.9\tabcolsep] SVD (MAPS)   &     $<0.1$  &    22       &       -           \\
\rowcolor[gray]{0.95}[.9\tabcolsep] SVD (DSSD)   &     $<0.1$  &    2       &       -           \\
\rowcolor[gray]{0.95}[.9\tabcolsep] MCT       &     $<0.1$  &    1       &   Ar:C$_4$H$_{10}$ = 90:10               \\
ST               &     0.2        &    4           & Ar:CO$_2$ = 70:30   \\
\rowcolor[gray]{0.95}[.9\tabcolsep] ECal          &   53 + 2$\times$14.5 = 82          &      8       &        -          \\
RS              &  481 + 2$\times$223 = 927          &      47           &  Ar:CO$_2$ = 70:30  \\
\rowcolor[gray]{0.95}[.9\tabcolsep] TOF            &      4          &       4            &   C$_2$H$_2$F$_4$:C$_4$H$_{10}$:SF$_6$ = 90:5:5   \\
\rowcolor[gray]{0.95}[.9\tabcolsep] FARICH         &   0.3         &      0.8       &        -          \\
BBC            &      0.1       &     0.5            &         -        \\
\hline         
\rowcolor[gray]{0.95}[.9\tabcolsep] Magnet       &    20          &                     &   He  \\
\hline    
\rowcolor[gray]{0.95}[.9\tabcolsep] Support \& transporting system &   80.3              &                            &  \\
Top platform       &  40         &                         &             \\
~~~~~ -- including cryogenics  &    5           &         23           &    He+N$_2$\\
\rowcolor[gray]{0.95}[.9\tabcolsep] Side platform  with equipment   &  100         &                         &             \\
\hline
 Total           & 1253      &   89~(109)                  &            \\ 
\hline
\multicolumn{4}{l}{*ZDC is omitted from the table because it is infrastructurally weakly connected to the main setup.}  
\end{tabular}
\end{center}  
\label{tab:spd-tech-req}
\end{table}%

\chapter{Radiation environment \label{ch:rad}}
\section{Radiation background in the detector}
It is well known that the intersection points of hadron collider beams are a powerful source of ionizing radiation because of the relatively large cross-section of hadron interactions and the significant multiplicity of secondary particles. Interaction with the material of the experimental setup leads both to the attenuation of fluxes of some components and to the generation of new particles. Beam halos, as well as induced radiation, also contribute to the radiation environment near the interaction point. However, we believe that their relative contribution is negligibly small. Correct estimation of fluxes and doses is necessary to calculate the loading of different detectors, as well as to take into account the effects of aging of the detector elements and electronics. The total ionizing dose  determines the long-term damage effects on electronics and sensors, while the flux of high-energy hadrons determines the rate of stochastic failures, like single-event upsets.

\begin{figure}[!h]
  \begin{minipage}{0.48\linewidth}
    \center{\includegraphics[width=1\linewidth]{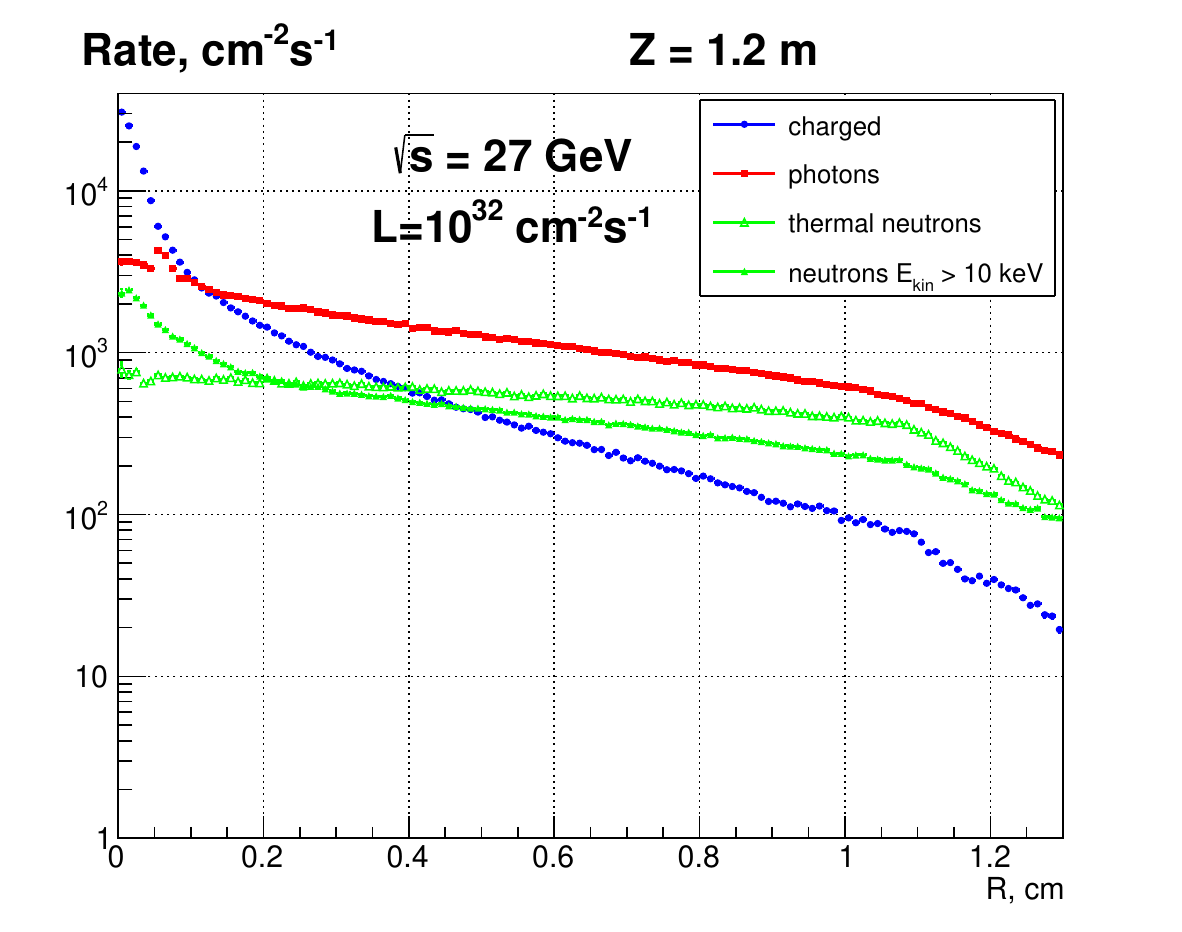} \\ (a)}
  \end{minipage}
  \hfill
  \begin{minipage}{0.50\linewidth}
    \center{\includegraphics[width=1\linewidth]{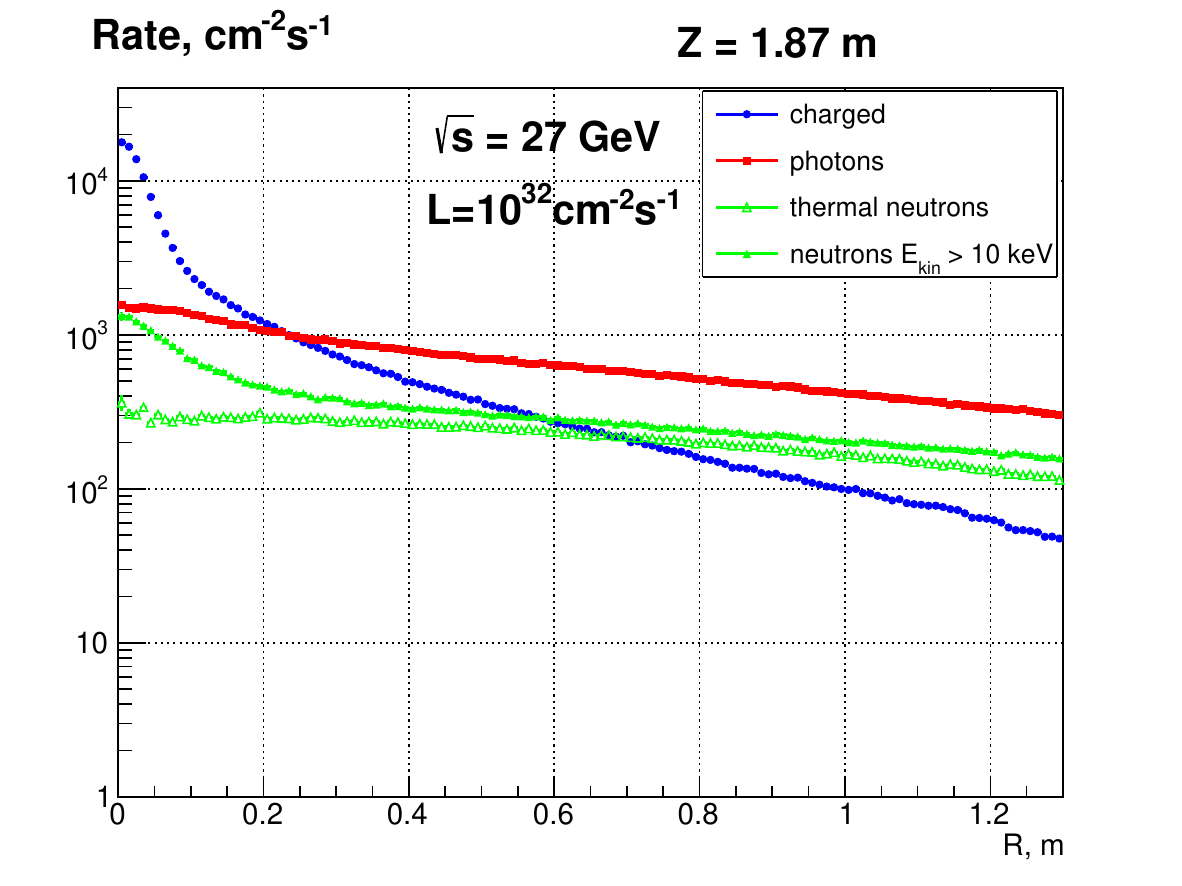} \\ (b)}
  \end{minipage}
    \begin{minipage}{0.47\linewidth}
    \center{\includegraphics[width=0.97\linewidth]{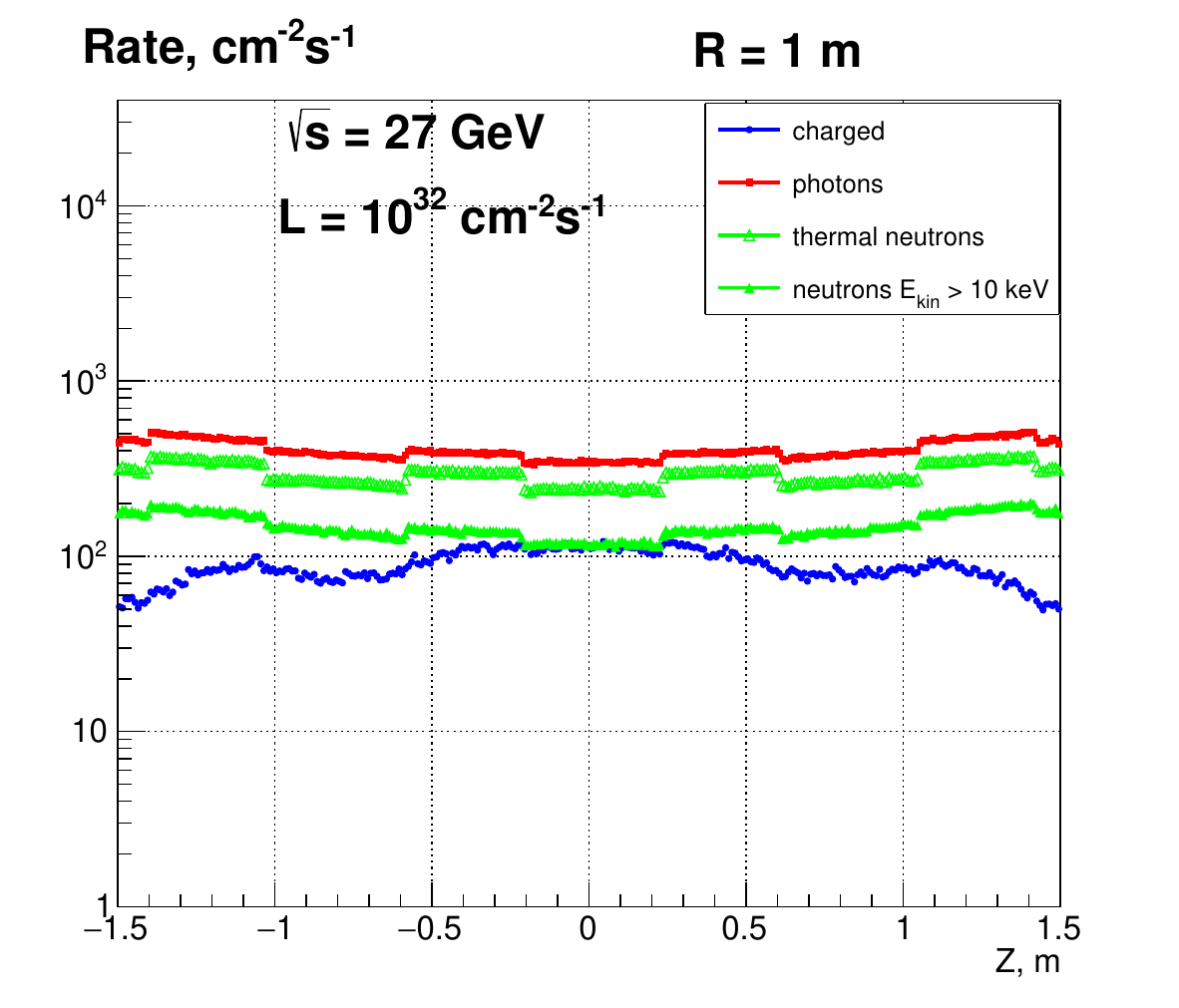} \\ (c)}
  \end{minipage}
  \hfill
  \begin{minipage}{0.52\linewidth}
   \center{\includegraphics[width=1.\linewidth]{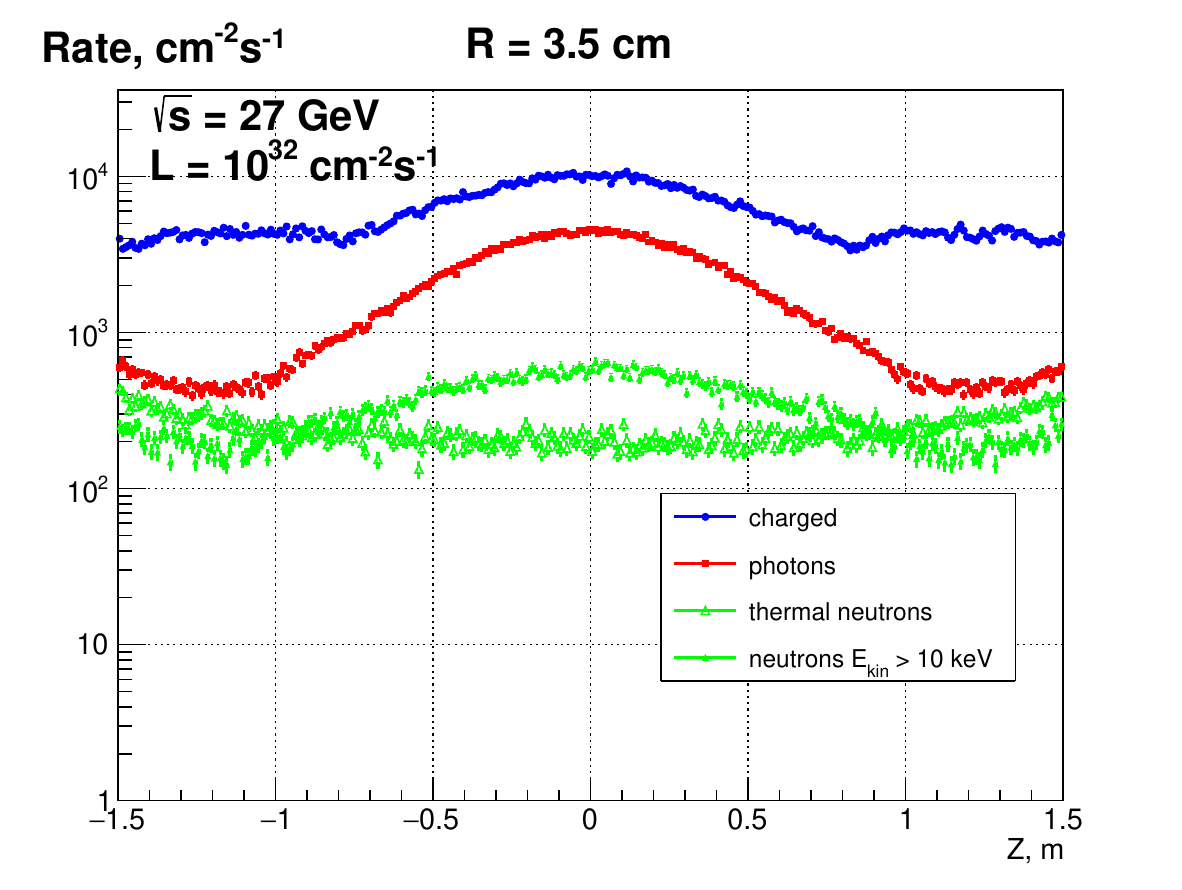} \\ (d)}
  \end{minipage}
  \caption{Flux of charged particles, energetic and thermal neutrons, and photons in the radial direction at (a) Z=1.2 m, (b) Z=1.87 m, and along the beam axis at (c) R=1 m and (d) R=3.5 cm.}
  \label{fig:flux}  
\end{figure}

\begin{figure}[!h]
\center{\includegraphics[width=1\linewidth]{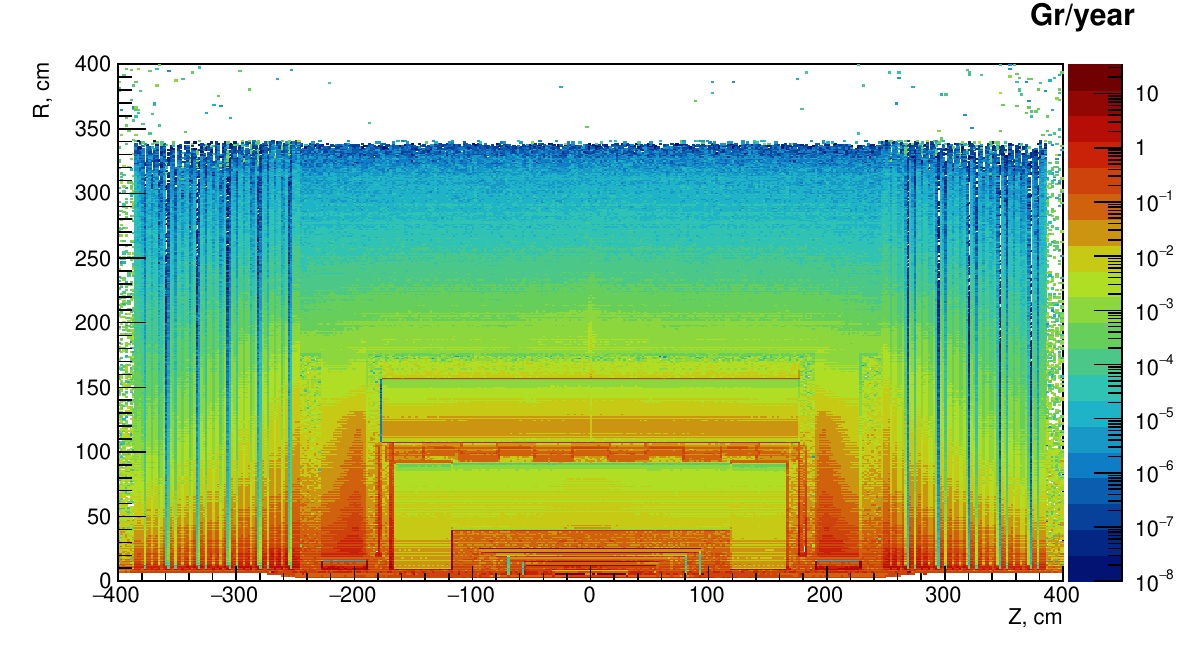}}
  \caption{Dose in Gy after one year of data taking at $\sqrt{s}=27$ GeV and L=10$^{32}$ cm$^{-2}$ s$^{-1}$ (averaged over azimuthal angle $\phi$).}
  \label{fig:dose}  
\end{figure}

The radiation environment in the SPD setup has been simulated  using the Geant4-based SpdRoot software with a recent description of the detector and the LHEP\_HP physics list to take into account the flux of thermal neutrons.  The primary interactions have a Gaussian distribution along the beam axis with respect to the nominal center of the setup with $\sigma_Z=30$ cm. The secondary interactions in the material of the setup, the multiple scattering, the decay of unstable particles, and the influence of the magnetic field modify  the fluxes of particles in the SPD setup significantly. Figure \ref{fig:flux} illustrates the fluxes of the charged particles, energetic and thermal neutrons, and photons at different points of the SPD setup for $p$-$p$ collisions at $\sqrt{s}=27$ GeV and $L=10^{32}$  cm$^{-2}$ s$^{-1}$: Z=1.2 m, in front of the EC ST (a),  Z=1.87 m, in front of the EC ECal (b), R=1 m, in front of the barrel ECal (c), and R=3.5 cm, just after the beryllium beam pipe (d). Figure \ref{fig:dose} shows a map of the absorbed dose distribution over the detector after one conditional year ($10^7$ s) of the data taking at maximum luminosity and energy of $p$-$p$ collisions. As it follows from the calculations, even in the inner part of the SPD facility the annual dose does not exceed a few hundred Grays, while, for example, for the inner layer of the silicon detector of CMS at the LHC this value is up to 1 MGy \cite{Dinardo:2015wra}. So, the expected dose rate at the SPD seems to be quite comfortable for the electronics and the detectors.

In the case of the ion-ion collisions at SPD, despite the large multiplicity of tracks in an individual interaction, the dose rate is about two orders of magnitude lower than in the above-mentioned case of $p$-$p$ collisions because of the low luminosity. However, such collisions will be accompanied by a much larger yield of energetic neutrons.

\section{Radiation background in the SPD experimental hall}

The radiation environment in the pit of the SPD experimental hall is an issue of special importance. The radiation background during the initial period of the heavy-ion beam collider operation is important for the assembly and commissioning of the SPD detector. The background during operation with proton and deuteron beams is essential for the placement and maintenance of the equipment.

 During NICA operation with heavy-ion beams, the main dose-forming component of radiation fields behind the biological shields of the collider and experimental facilities is neutrons of a wide range of energies \cite{RB_4}. The field of leakage neutrons from shields, repeatedly scattered in the air, ground, and surrounding objects (skyshine neutrons) also determines the radiation situation at large distances from the complex accelerators. The working Electron Cooling System (ECS) is also a source of gamma-quanta. Beam collisions in the SPD and MPD detectors are not considered as significant sources of ionizing radiation.
  
\begin{figure}[!h]
    \center{\includegraphics[width=1\linewidth]{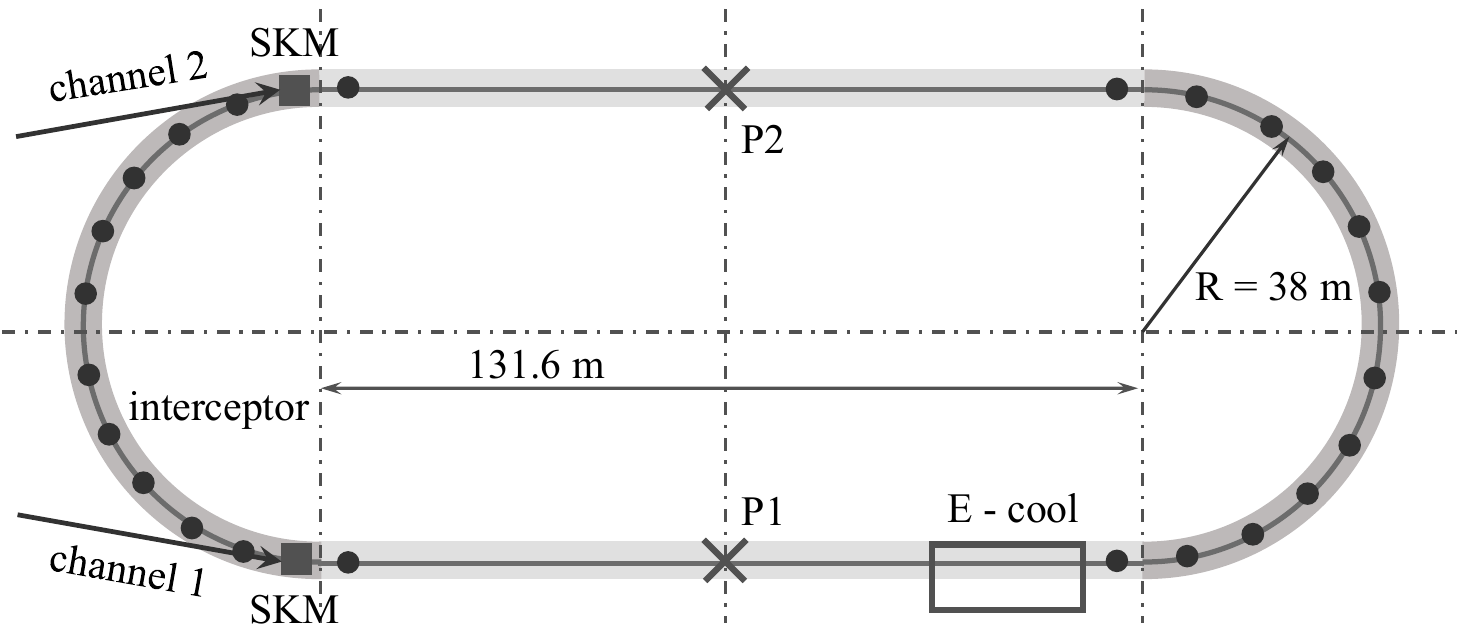}}
  \caption{  Location of the main sources of ion losses in the NICA collider  \cite{RB_2}. Black circles indicate the scrapers while two black bars represent the septum kicker magnets (SKM). The ECS is shown by a rectangle.}
  \label{fig:dose11}  
\end{figure}

 Three sources of ion losses responsible for the radiation environment in adjoining areas are considered in the collider rings: i) ion losses  due to interaction with residual gas in the vacuum chamber uniformly distributed over the ring ($\sim$ 1\% for $^{197}_{79^+}$Au ions of 4.5 GeV/nucleon); ii) beam losses in the scrapers (24 in each ring placed mainly in the arcs of the collider), which serve to intercept ions dropped out of the acceleration or circulation process ($\sim$ 82\%); and iii) recombination of ions with electrons in the ECS located in the vicinity of the MPD setup ($\sim$ 17\%). Additional losses occur during injection in the injection section on the accelerator elements, such as the kicker and septum. The location of  the main sources of ion losses in the collider is shown in Fig. \ref{fig:dose11}. The experimental setups are not considered a significant secondary radiation source on the NICA complex \cite{RB_2}. 
 
\begin{figure}[!h]
    \center{\includegraphics[width=1\linewidth]{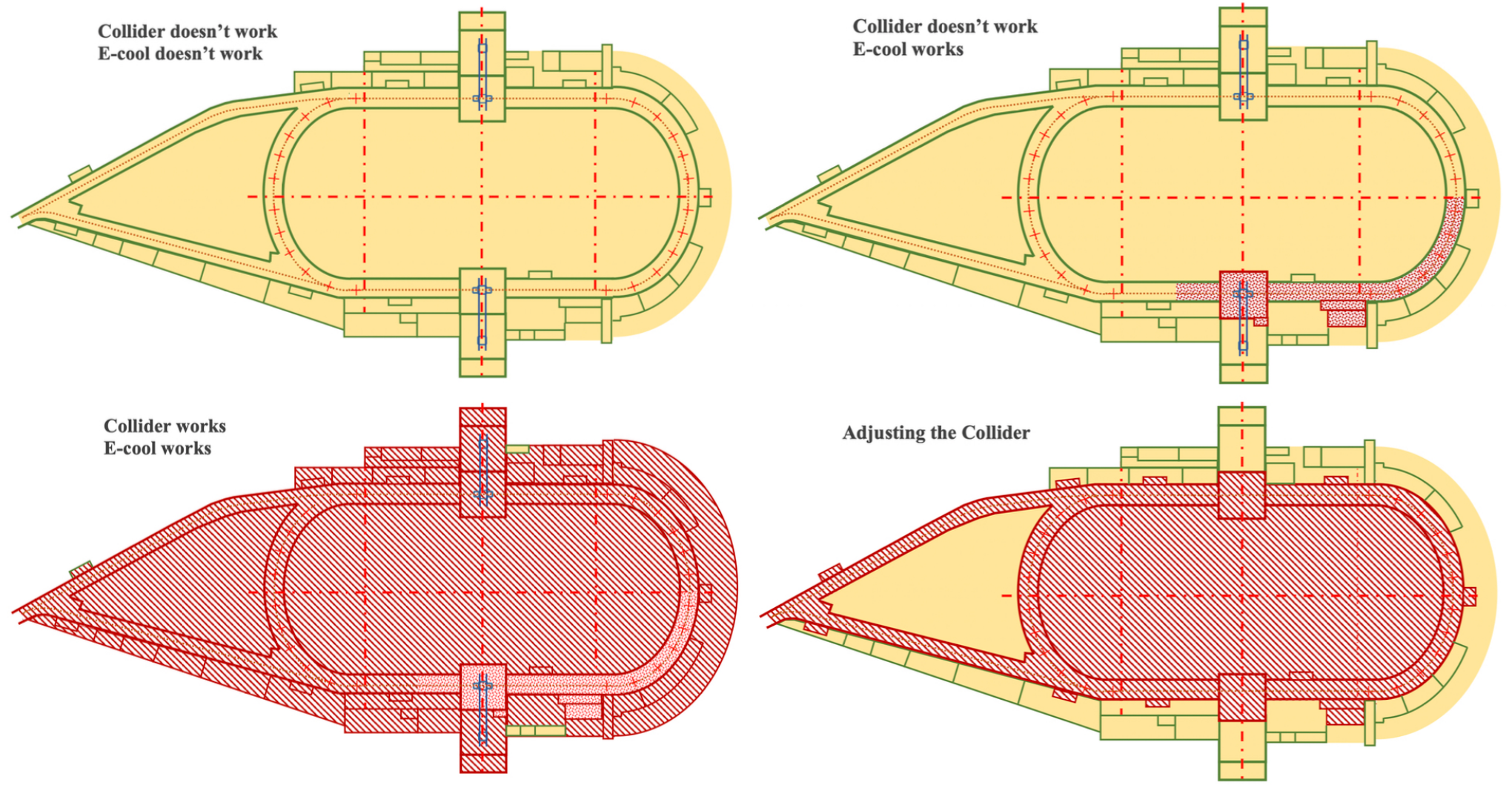}}
  \caption{ Radiation zoning of the collider buildings and areas in different modes of operation \cite{RB_2}.  The restricted access zone is shown in yellow, while the exclusion zone is in red.}
  \label{fig:dose2}  
\end{figure}

  Radiation zoning of the collider buildings and areas in four different modes of operation: i) collider doesn't work, ECS doesn't work; ii) collider doesn't work, ECS works; iii) collider and ECS are both in operation;  iv) collider adjustment; is presented in Fig. \ref{fig:dose2}. The collider spaces are divided into two zones: the restricted access zone, where the level of human exposure may be at the level of 20$\div$200 mSv per year,
   and the exclusion zone, where the level of human exposure may exceed 200 mSv per year \cite{RB_3}.

Despite the protection wall of concrete blocks, separating the main part of the SPD pit from the beam pipe when the collider operates with heavy-ion beams in the absence of collisions at the SPD interaction point, the SPD pit is  considered to be the exclusion zone. The expected dose rate  just behind the protective wall is expected up to 0.7  $\mu Sv/h$ \cite{RB_1}. This means that the assembly of the SPD setup can only be carried out when the collider is not running. This limits us to a guaranteed period of four months  per year.


During SPD operation in the proton-proton collision mode, the dose rate in the pit will  allow the computer equipment of the DAQ and the online filter to be placed there.

\chapter{Detector assembling procedure}

The detector assembling procedure will be carried out at the experimental site of the SPD pavilion.
At this time, the accelerator beam line will be separated from the pavilion space by the concrete blocks and slabs, which will provide  a temporary biological protection. 
Nevertheless, due to the radiation background, the assembly can only take place during the shutdown of the accelerator. 
This limits the time that can be allocated for the assembly, 
and applies a clear requirement that parts of the detectors, which are delivered to the pavilion,
should be maximally integrated modules.

The first step is to install the Range System.
The RS is external to all other detector systems. It  is also the largest and heaviest SPD system, weighting above 900 tons. 
Therefore, it will be mounted first. 
Due to the large weight and dimensions, it can only be assembled using the overhead crane and only on the experimental site of the SPD pavilion.
The RS barrel is divided into eight sectors (octants).
The octants are mounted by sequentially laying  on the lodgement from the bottom up. 
During the installation procedure, a system of internal supports and struts will be required. 

Once the barrel part of the detector is assembled, four sections of  the end-caps are installed. 
Since the lifting capacity of the crane is limited to 80 tons, each of these sections has to be assembled in two parts. 
These parts are connected into a single one when transferred to the working position. 

The RS barrel is the preferred structure for supporting or suspending other SPD systems, i.e.  it can be used to install permanent or temporary mounting structures of other detectors.
In addition, the RS as an external system is convenient  to use as a reference coordinate system of SPD, tied to the coordinate system of the collider.
For other detector systems, only tunnel assembly is available. 
  The RS also serves a magnet yoke. 

\begin{figure}[th]
\centering
\includegraphics[width=0.99\textwidth]{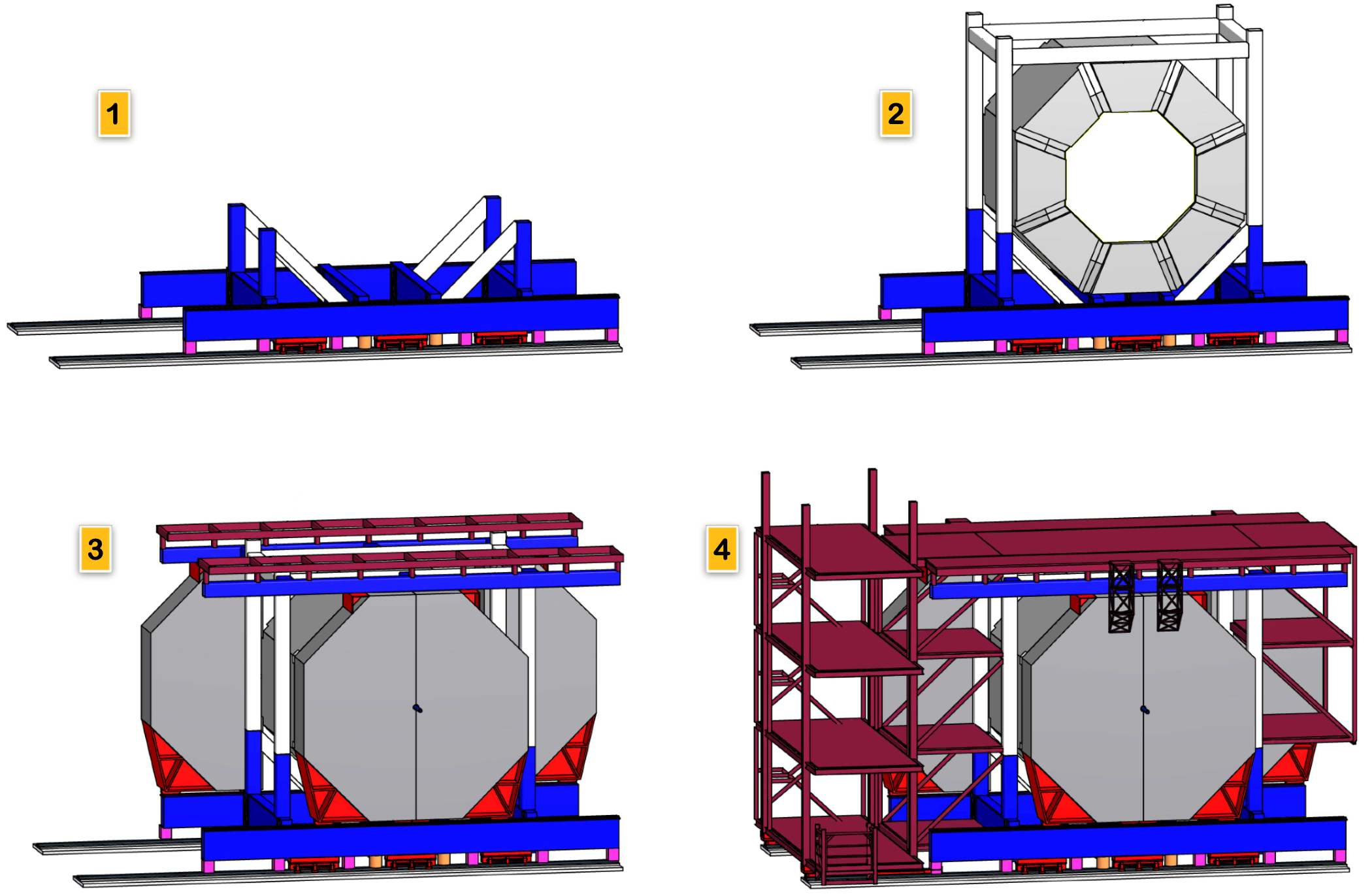}
\caption{
(1) Installation of the lower part of the movable lodgement. 
(2) Installation of the barrel octants of RS with a support structure.
(3) Installation of the RS end-caps.
(4) Installation of top and side platforms for cryogenics and electronics. }
\label{fig_RS_v1}
\end{figure}

\begin{figure}[th]
\centering
\includegraphics[width=0.9\textwidth]{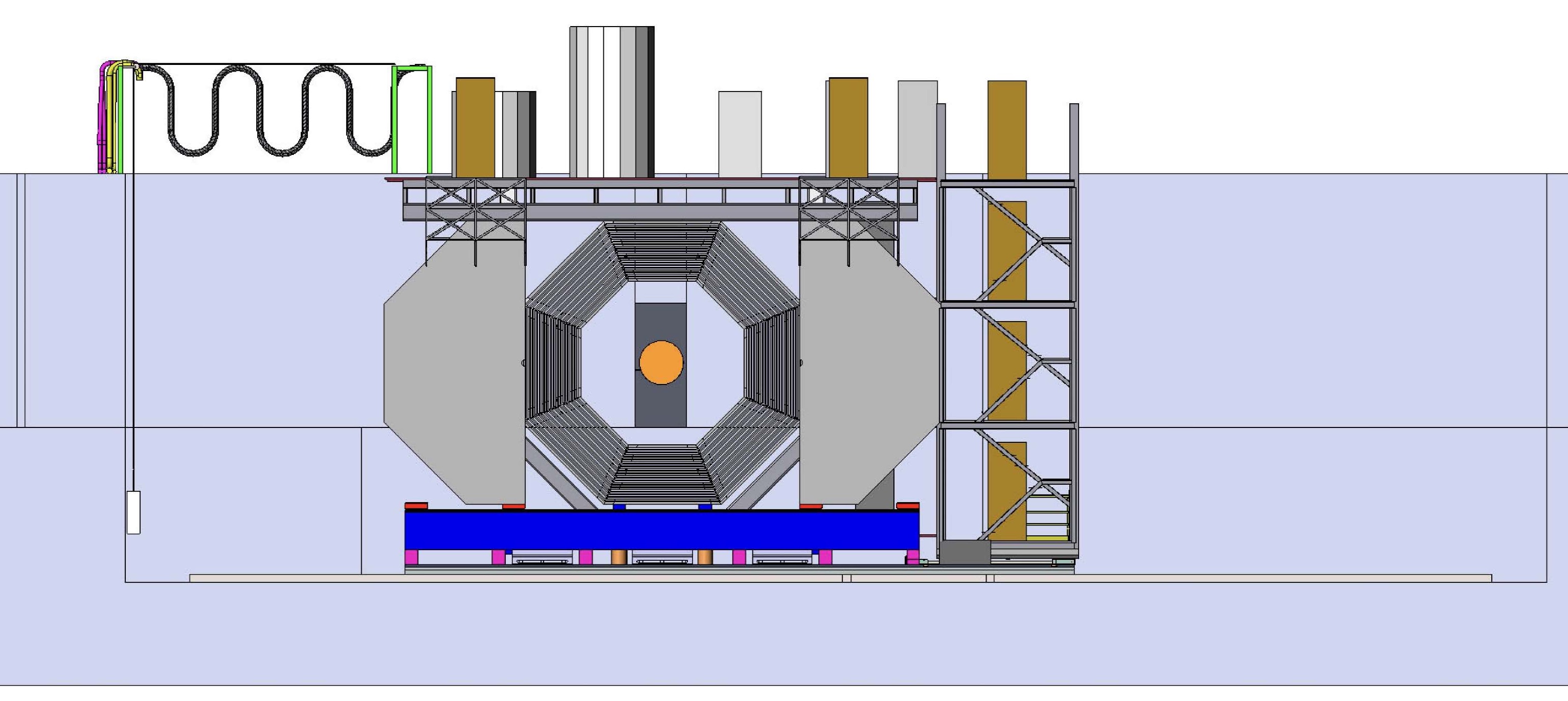}
\caption{Side view of the detector. 
Once the upper and side platforms are ready, one can start installing cryogenic equipment, electronic racks,
set cables and gas pipelines.
This step, however, can be postponed to the moment after the magnet is installed 
(see Figs. \ref{fig_Mag_v1} - \ref{fig_Mag_v3} ).}
 \label{fig_cryogenics}
\end{figure}

 The installation of such structural components as the lower part of the movable lodgement,
 the barrel octants of RS with a supporting structure,
 the RS end-caps, and
 the upper and side platforms for cryogenics and electronics 
are shown in Fig.\,\ref{fig_RS_v1}.
Once the upper and side platforms are in place, one can also start installing cryogenic equipment, electronic racks,
set cables and gas pipelines, as shown in Fig.\,\ref{fig_cryogenics}.

\begin{figure}[th]
\centering
\includegraphics[width=0.55\textwidth]{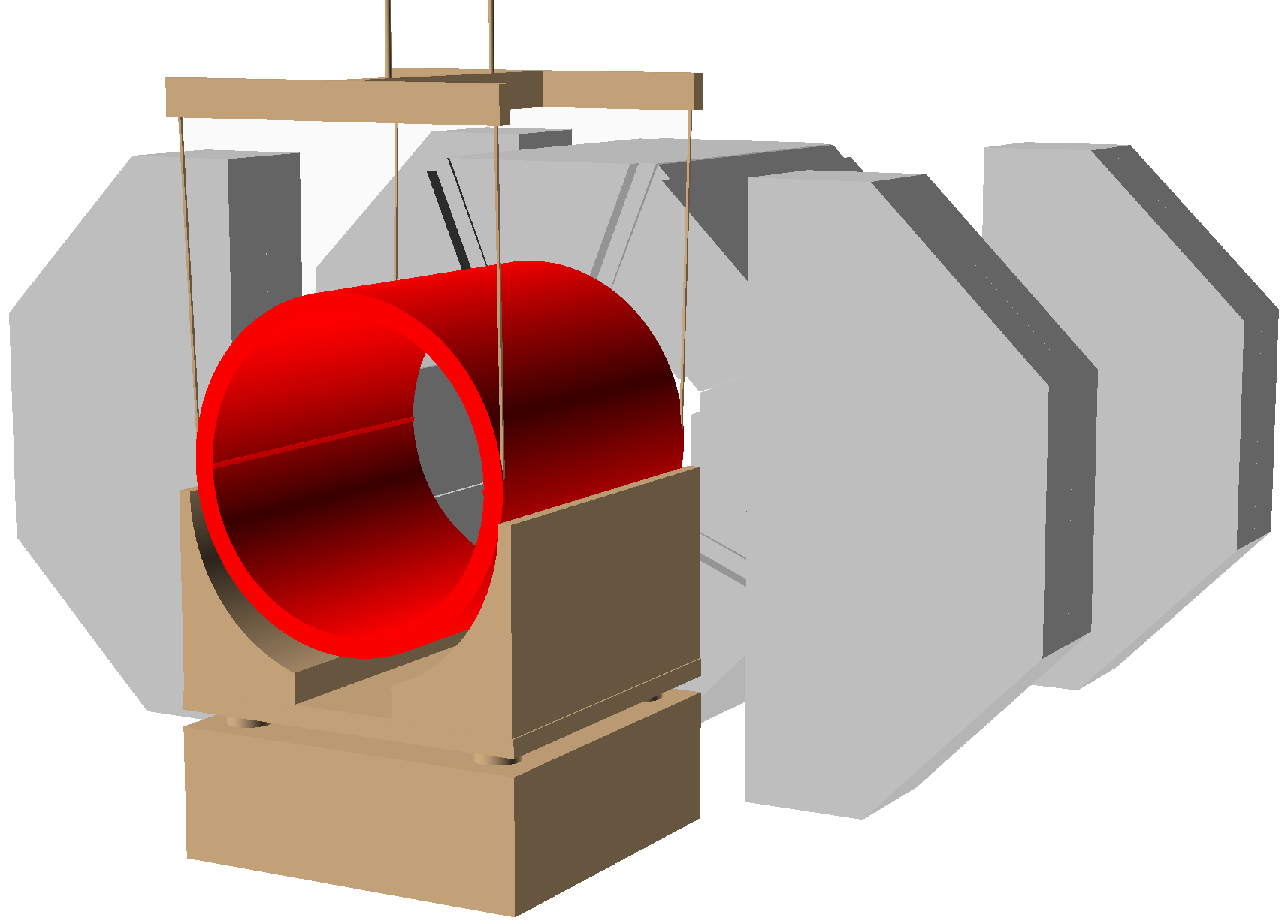}
\caption{(Var-1) The magnet is lowered by a crane into the cradle.
The cradle contains  lifting jacks for adjusting the height and aligning the magnet with the RS barrel.
The magnet is transported in the axial direction to the working position.
The magnet weights 16~tons.}
\label{fig_Mag_v1}
\vspace*{0.5cm}
\includegraphics[width=0.56\textwidth]{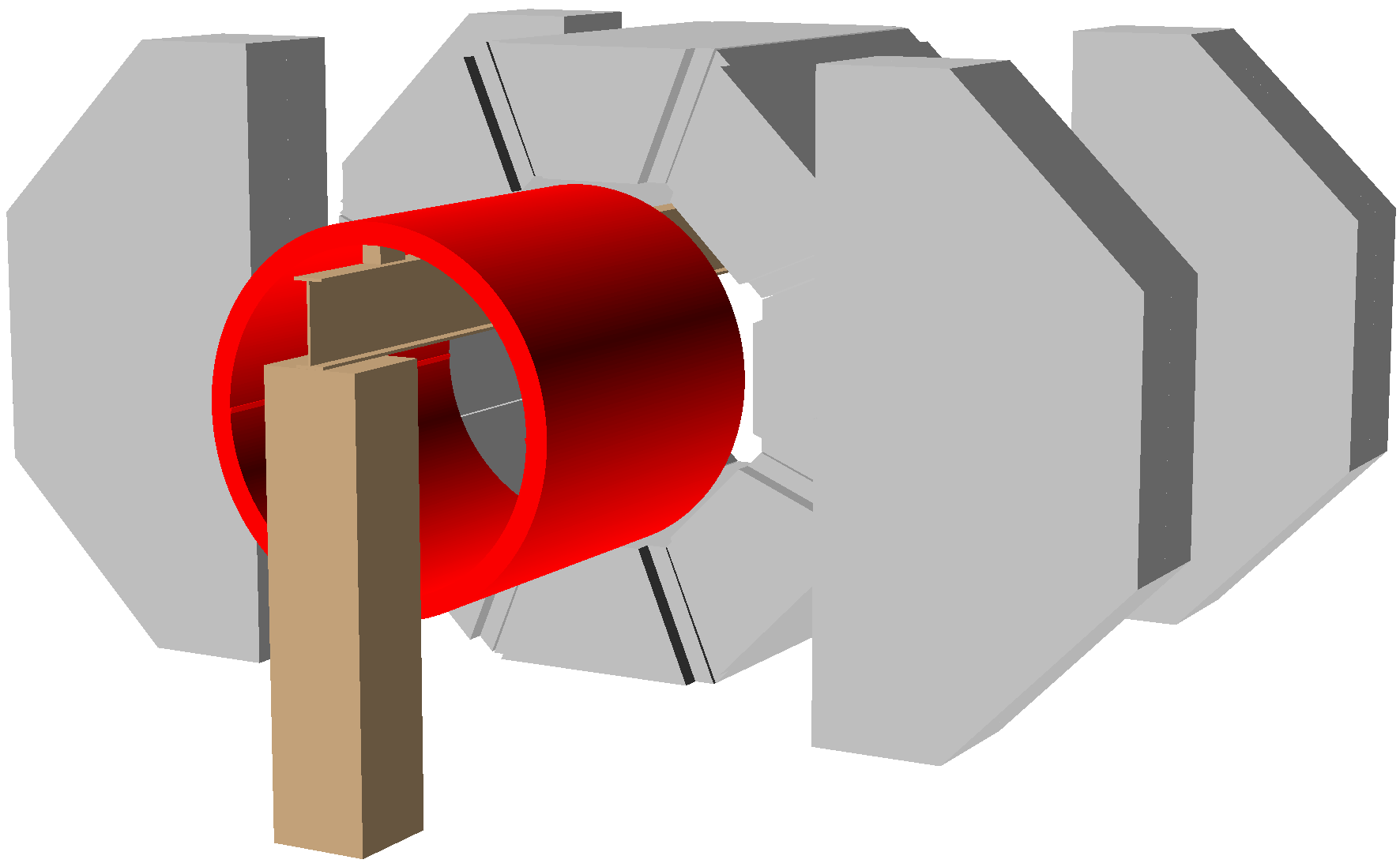}
\caption{(Var-2) The guide beam with a rail is threaded through the magnet and fixed to the pillars on both sides.
}
\label{fig_Mag_v5}
\vspace*{0.5cm}
\includegraphics[width=0.56\textwidth]{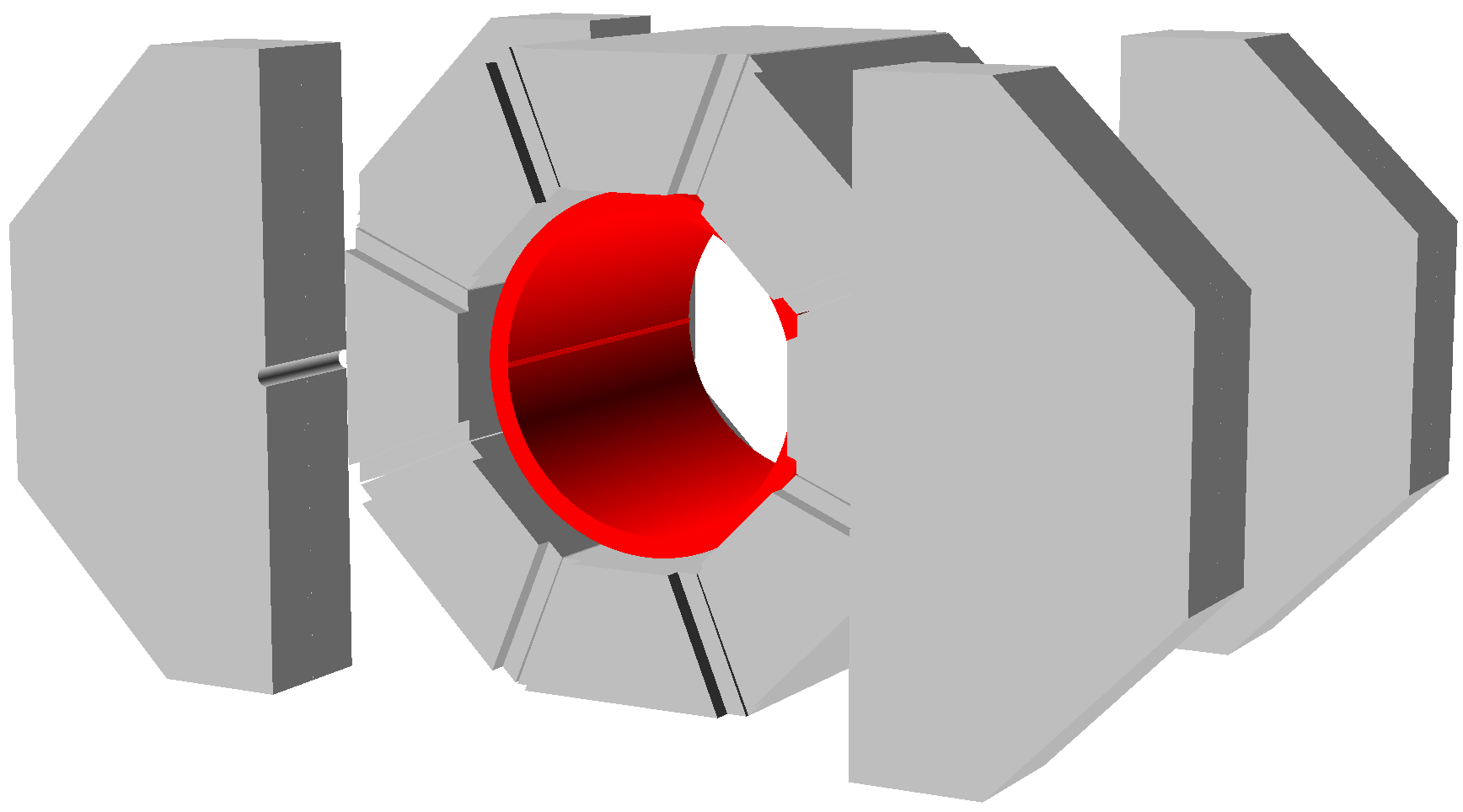}
\caption{Final configuration with the magnet installed and bolted to RS.}
\label{fig_Mag_v3}
\end{figure}

The next step is to install the superconducting solenoid. The magnet has a cylindrical shape with terminals for cable and pipeline facilities. Structurally, the magnet is divided into an external cryostat and a "cold mass" suspended inside the cryostat on stretchers. The total weight is approximately 16 tons. The magnet is delivered to the experimental site in an already assembled state. When switched on, the magnet creates a solenoid field with a magnetic induction of 1\,T along the axis. This field has a retracting effect on the RS end-caps with a force of about 300 tons per every of two sections of end-cap. Two options for mounting the magnet are considered: 
\begin{itemize}

\item  The magnet is lowered by a crane into a cradle, as shown in Fig.\,\ref{fig_Mag_v1}.
The cradle contains  lifting jacks for adjusting the height and aligning the magnet with the RS barrel.
The magnet is then transported in the axial direction to the working position.
The location of the guiding equipment (rails and rollers)  is yet to be defined.

\item  The magnet is lowered by a crane and held in the position to make it coaxial with RS.
The guide beam with a rail is threaded through the magnet and fixed to  the pillars on both sides, as shown in Fig.\,\ref{fig_Mag_v5}.
It is necessary to check how feasible this procedure is, since the length of the beam is 11\,m.
Finally, the magnet is moved on rollers along the rails to its working position. 

\end{itemize}
Once the magnet is in position, it will be bolted to the RS barrel and connected to the cryogenic equipment located on the top platform. 
The final configuration of the magnet in the RS barrel is shown in Fig.\,\ref{fig_Mag_v3}.

\begin{figure}[th]
\centering
\includegraphics[width=0.5\textwidth]{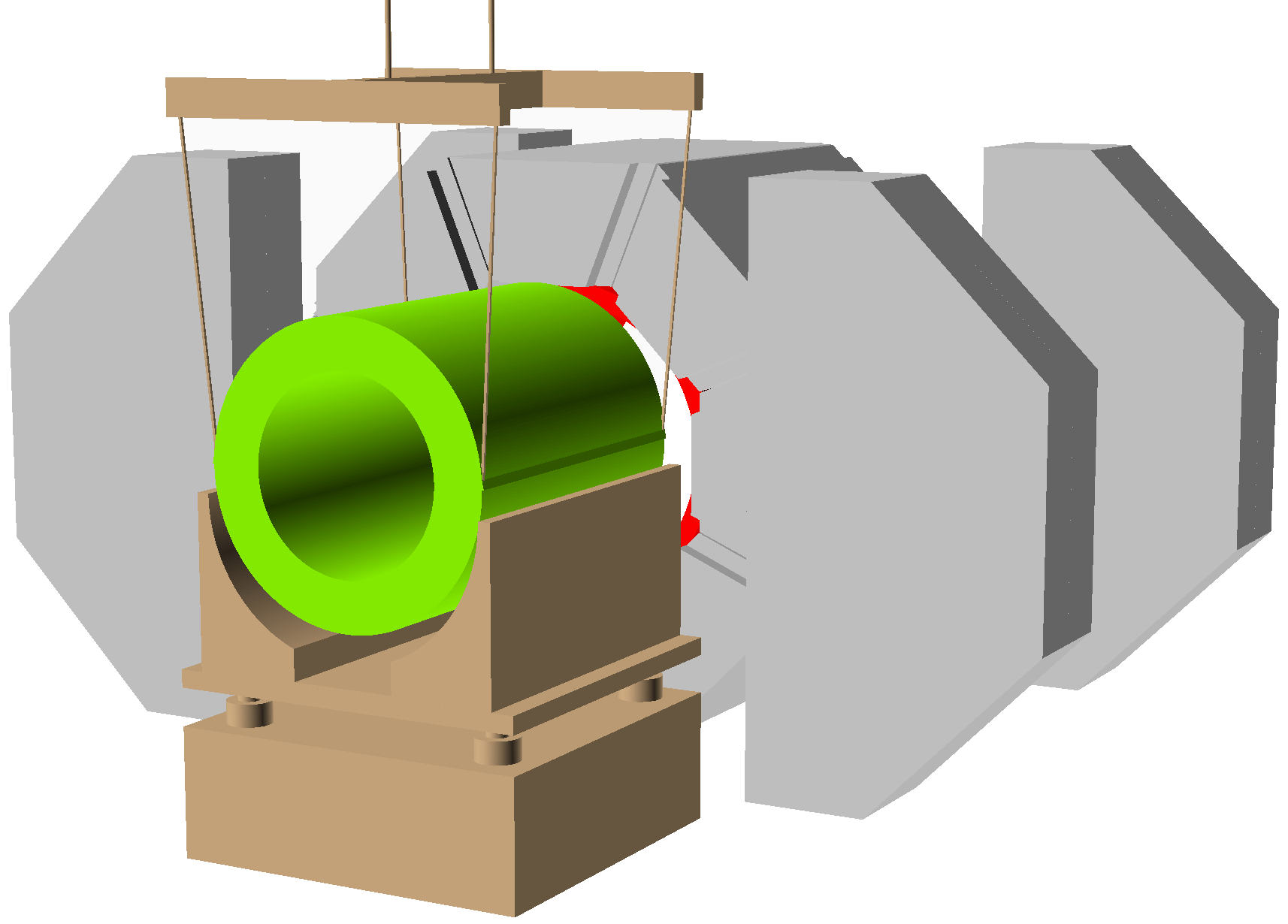}
\caption{The ECal barrel is lowered by a crane into the cradle.
The lifting platform of the cradle can be the same as for the magnet (see Fig.\,\ref{fig_Mag_v1}), but the vise mechanism needs to be adjusted to the new size. The weight of the ECal barrel is close to 40\,tons.}
\label{fig_ECalBR_v1}
\vspace*{0.5cm}
\includegraphics[width=0.55\textwidth]{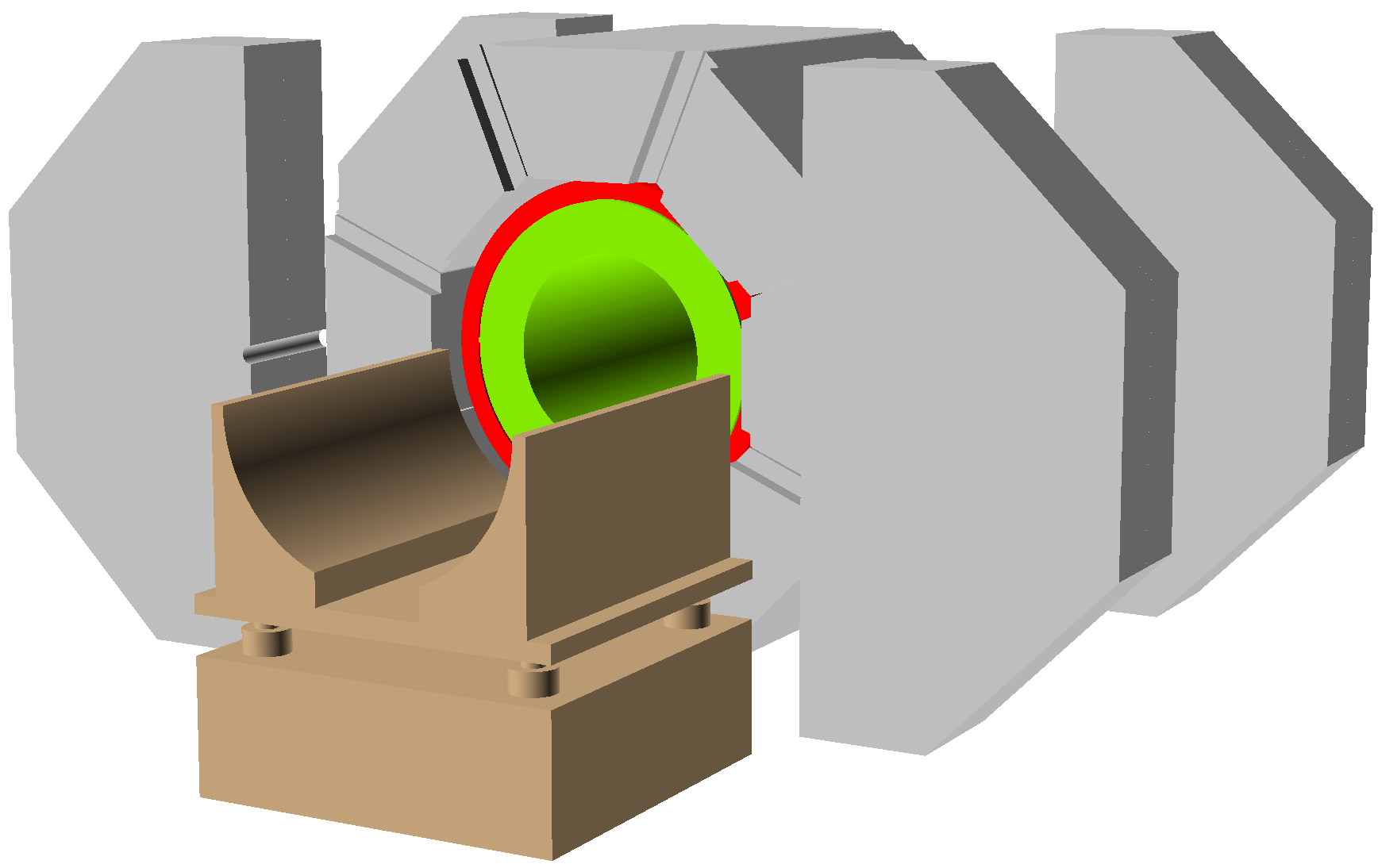}
\caption{The ECal barrel is moved along the rails to the working position.
Two fixation options are under consideration: (1) hanging ECal on the magnet,
(2) ECal is supported at its ends (fixed to RS).}
\label{fig_ECalBR_v2}
\end{figure}

The next step is to install the ECal barrel. It is lowered by a crane into the cradle, as shown in Fig.\,\ref{fig_ECalBR_v1}.
The lifting platform of the cradle can be the same as for the magnet (see Fig.\,\ref{fig_Mag_v1}),
but the vise mechanism needs to be adjusted to the new size. The weight of the ECal barrel is close to 40\,tons.
The ECal power frame will have to support the detector itself, as well as all other interior parts of the  central detectors.
Two loading and fixation options are under consideration: 

\begin{itemize}

\item  A fully or partially assembled ECal barrel is moved along the rails to the working position as shown in Fig.\,\ref{fig_ECalBR_v2}.
Cam roller guides, attached to ECal, will have a mating rail structure that will be fixed to the magnet inner shell in the median plane.
It may be necessary to thicken the walls of the inner shell of the magnet cryostat to reduce deflection. 
This mounting approach is used, for instance, to place the HCAL detector inside the magnet of CMS/LHC.

\item In this configuration, a power frame without active elements of ECal is installed  first.
The frame  is supported at its ends, which, in turn, will be fixed to the RS barrel. 
The last step will be to load the active elements of ECal.
This approach is used to mount the ECal detector inside the magnet of MPD/NICA.

\end{itemize}
Nota bene. The most likely the ECal will not be ready to the first stage of the experiment. 
However, in order to test the magnet cryostat for deflection, 
it will be necessary to make a mock ECal barrel at full weight. 
This mock-up can be a hollow steel cylinder filled with concrete. 
It can stay inside the magnet to maintain the load balance until ECal itself is installed.

\begin{figure}[th]
\centering
\includegraphics[width=0.52\textwidth]{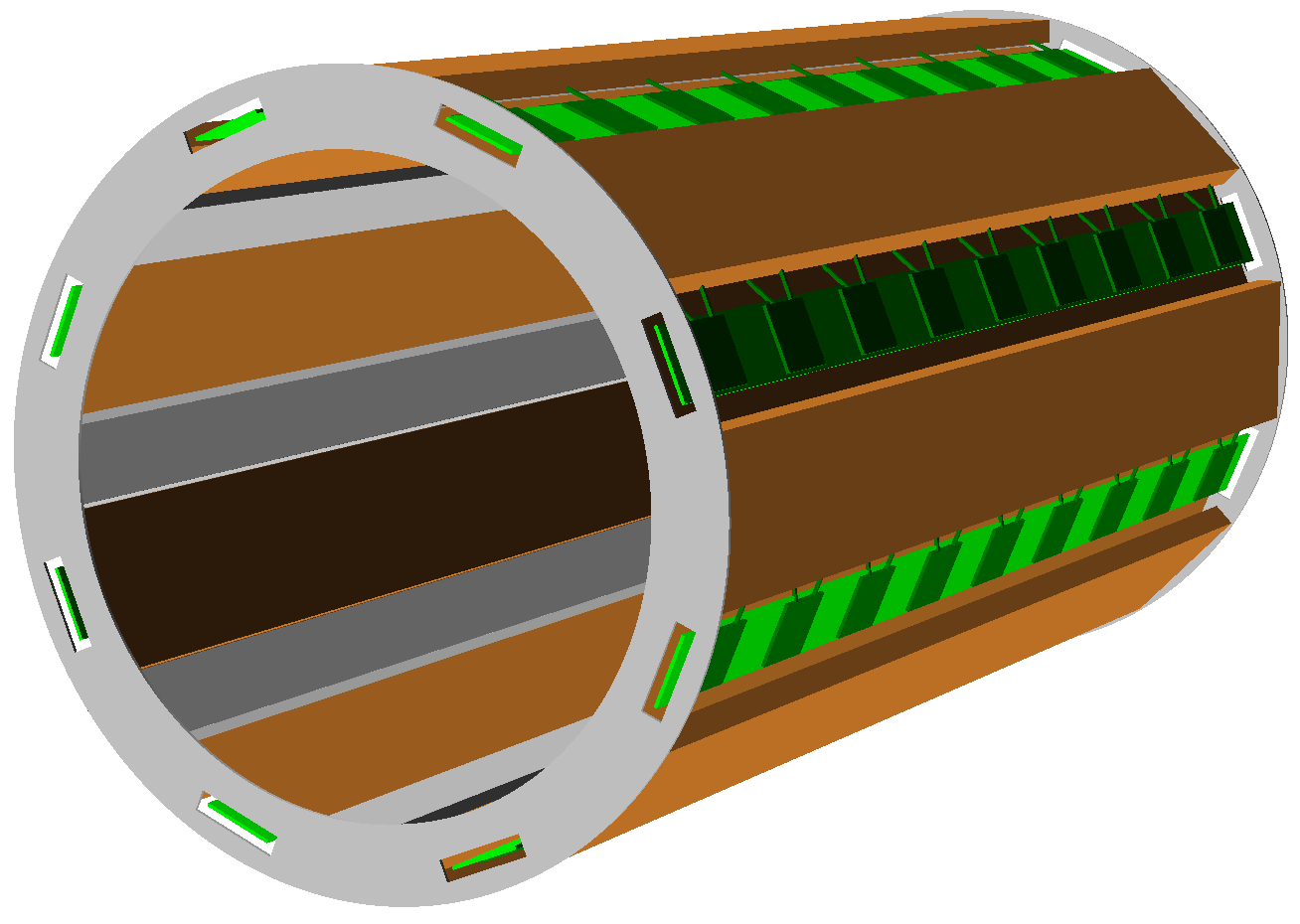}
\hfill
\includegraphics[width=0.44\textwidth]{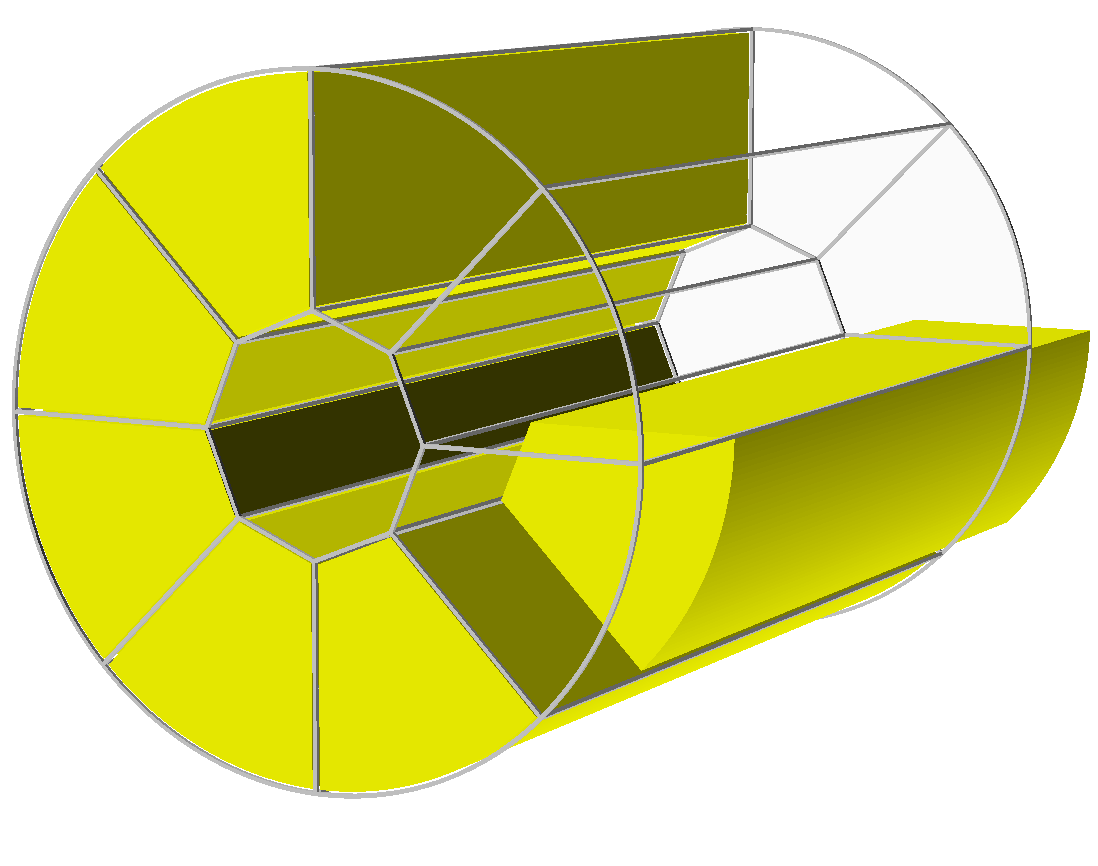}
\caption{{\it Left}: The TOF barrel with  inner and outer layers of MRPC-supermodules installed.
The expected weight is close to 4 tons.
{\it Right}: The ST frame partially loaded with the straw modules (octants). 
The weight of the ST barrel is about 120\,kg.
It will be inserted and fixed inside the TOF barrel.}
\label{fig_TOF_ST}
\end{figure}

The next step is to install the TOF barrel.
The TOF barrel consists of staggered  MRPC chambers adjacent to the ECal barrel. 
Unfortunately, the ECal barrel cannot be used as a base for mounting MRPCs, due to uncertainties associated with the time of its appearance in the setup. 
Thus, it is proposed to use a self-supporting truss structure for mounting the chambers. This truss will lean on ECal (or its mock-up) only at its ends, which will  simplify the installation. Individual MRPCs arranged in a row along the installation axis are combined into supermodules. 
The TOF barrel with  inner and outer layers of MRPC-supermodules installed is shown in Fig.\,\ref{fig_TOF_ST} (left).
The estimated weight of the entire TOF barrel, including modules, is about 3$\div$4 tons. 

Assembling of the TOF barrel is carried out outside the SPD pavilion. 
The barrel is delivered as a whole by a crane to the mounting position in front of  RS. 
The barrel is rolled inside the ECal along the guides and fixed at the ends. 
Installation of the TOF barrel can also be done after installing the ST, VD, Aerogel, and BBC detectors inside the barrel.

The next step is to install the ST barrel.
The ST barrel is divided into 8 sectors (octants). The position of each octant is determined in space using a frame. 
The central part of the frame is an octagonal prism that will house the vertex detector, 
while the ST octants are located outside this prism and fill the entire space up to the TOF barrel. 
The ST frame partially loaded with the ST octants is shown in Fig.\,\ref{fig_TOF_ST} (right). 
Due to the low weight of the octants, they can be inserted into the frame manually.
The frame will be fixed inside the TOF barrel and, in principle, can be a self-supporting structure. 
The total weight of the  fully assembled ST barrel is approximately 120 kg.

\begin{figure}[th]
\centering
\includegraphics[width=0.6\textwidth]{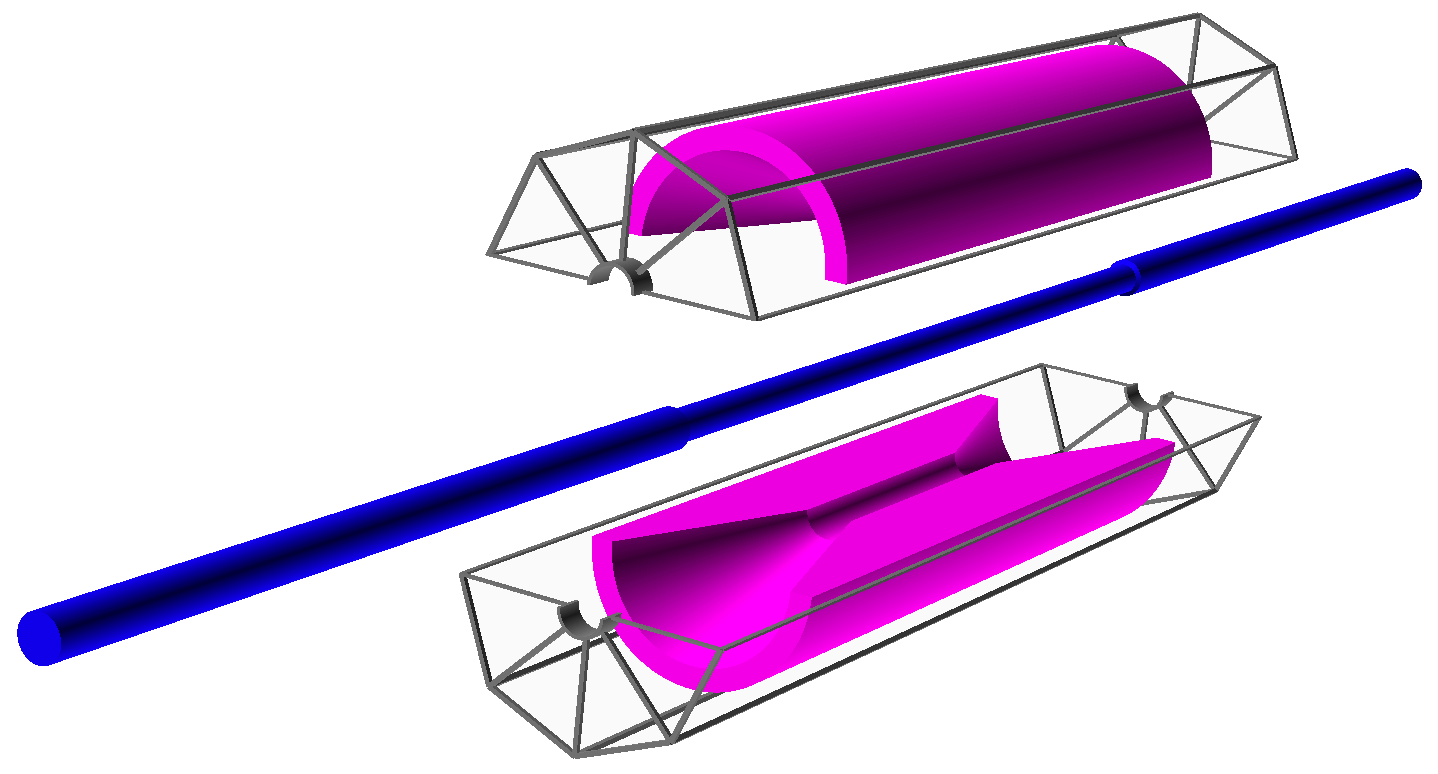}
\caption{The beam pipe and two halves of the VD detector, which are fixed on two half-cages.
These two half-cages form a load-bearing power element that carries the weight of the VD and the beam pipe.
The weight of VD in its MM option is close to 40 kg. The silicone option of VD is much lighter.}
 \label{fig_VD_v1}
 \vspace*{0.9cm}
 \includegraphics[width=0.85\textwidth]{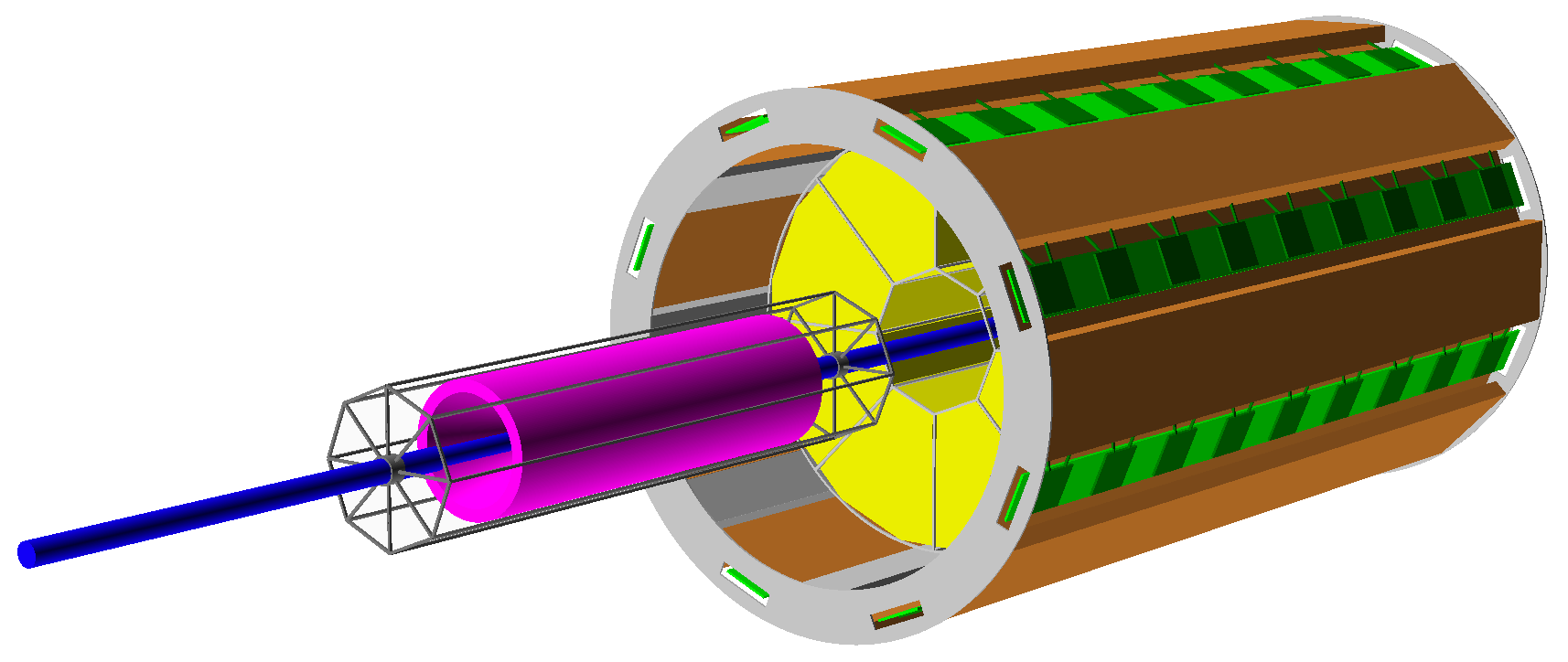}
\caption{The cage with VD and beam pipe slides into its central position inside the ST frame.}
 \label{fig_VD_v2}
\end{figure}

The next step is to install the VD with the beam pipe.
Since the beam pipe is a thin-walled tube of variable diameter expanding towards the ends, the installation procedure with passing the pipe through the VD is not feasible. 
Therefore, the detector is divided into halves, which are assembled separately and closed around the pipe during the mounting procedure. 
Thus, the VD and the beam pipe form a single module, which will be inserted into the center of the ST frame as a whole. 
The beam pipe and two halves of the VD detector in  disassembled state are shown in Fig.\,\ref{fig_VD_v1}.
These two half-cages form a load-bearing power element that carries the weight of the VD and the beam pipe.
To date, two options of the VD are being considered: MicroMegas and silicone detectors. 
The weight of VD in its MicroMegas option is close to 40 kg, while the silicone option of VD is lighter.
The cage with VD and beam pipe, sliding into the central position inside the ST frame, are shown in Fig.\,\ref{fig_VD_v2}.

\begin{figure}[th]
\centering
\includegraphics[width=0.9\textwidth]{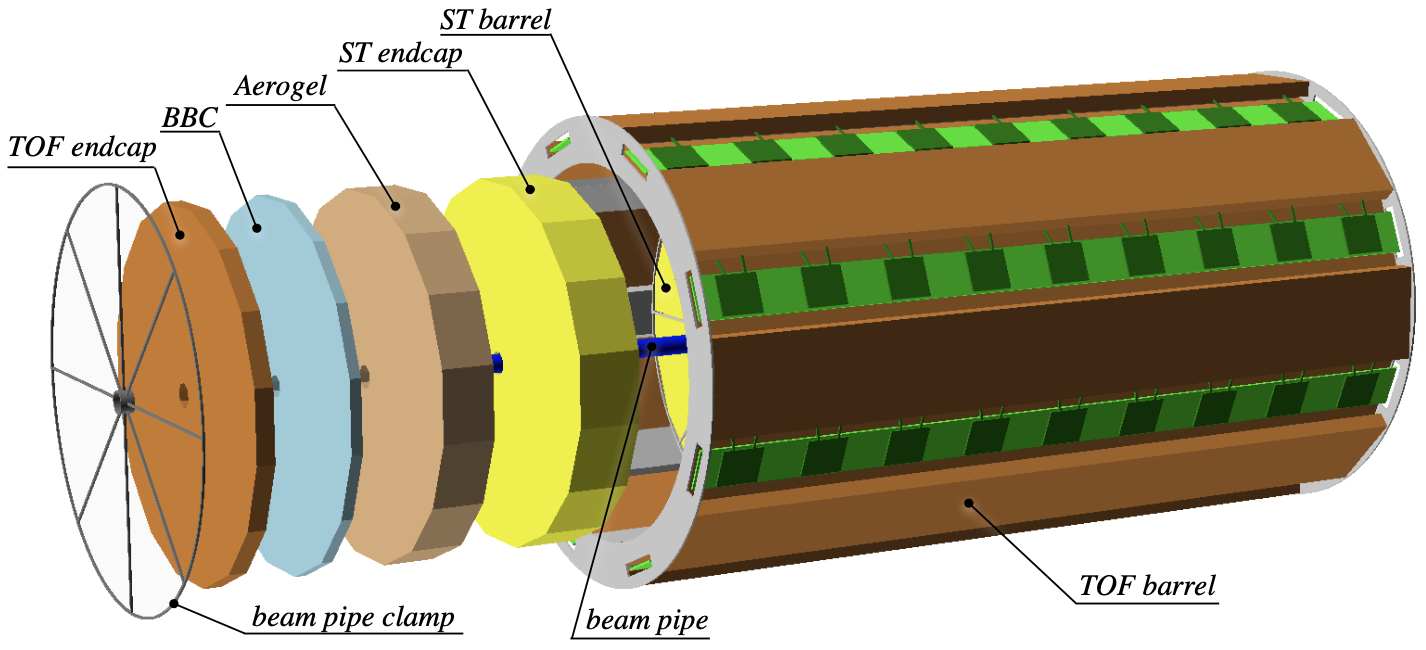}
\caption{The End-caps of ST, Aerogel, BBC, and TOF are mounted one-by-one by moving them along the beam pipe.
 The last object in this sequence is a clamp that holds the pipe. All end-caps and the clamp are braced to the TOF barrel frame.}
\label{fig_EndCap_v1}
\end{figure}

\begin{figure}[th]
\centering
\includegraphics[width=0.59\textwidth]{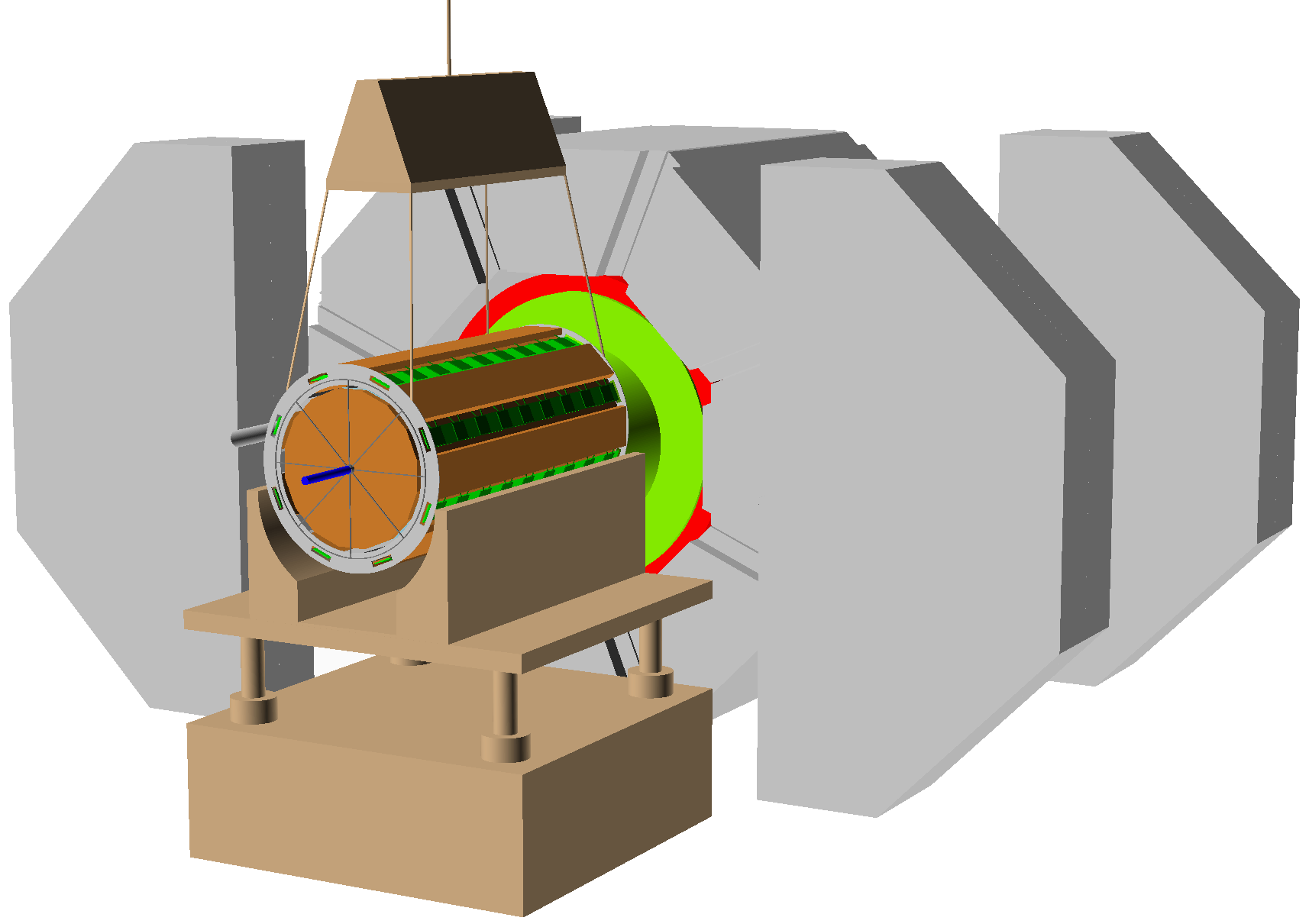}
\caption{The TOF barrel with other inner detectors (ST, Aerogel, BBC, and VD) is lowered by a crane into the cradle.
The lifting platform of the cradle can be the same as for the magnet (see Fig.\,\ref{fig_Mag_v1}) but the vise mechanism is to be adjusted to the new size.}
 \label{fig_TOF_v1}
 \end{figure}

The next step is to install end-caps of ST, Aerogel, BBC, and TOF.
All these end-caps have a cylindrical shape with a hole in the middle for the beam pipe. 
The radial dimensions of the end-caps are similar. The end-caps are edge-mounted on the TOF frame. 
They will be inserted one-by-one by moving them along the beam pipe, as shown in Fig.\,\ref{fig_EndCap_v1}.
The last object in this sequence is a clamp that holds the pipe. 
The installation procedure is carried out manually.
The assembling procedure can be carried out either directly in the experimental hall, or somewhere  outside the experimental area.
The latter case has an advantage in view of the fact that during the operation of the accelerator, access to the experimental area will be restricted.
In this case, the whole assembly  (TOF, ST, Aerogel, BBC, and VD) will be transferred by a crane to the cradle,
as shown in Fig.\,\ref{fig_TOF_v1}.
The lifting platform of the cradle can be the same as for the magnet (see Fig.\,\ref{fig_Mag_v1}), 
but the vise mechanism is to be adjusted to the size of the TOF frame.
As the last step, the assembly is moved along the rails to its working position inside ECal.

\begin{figure}[th]
\centering
\includegraphics[width=0.56\textwidth]{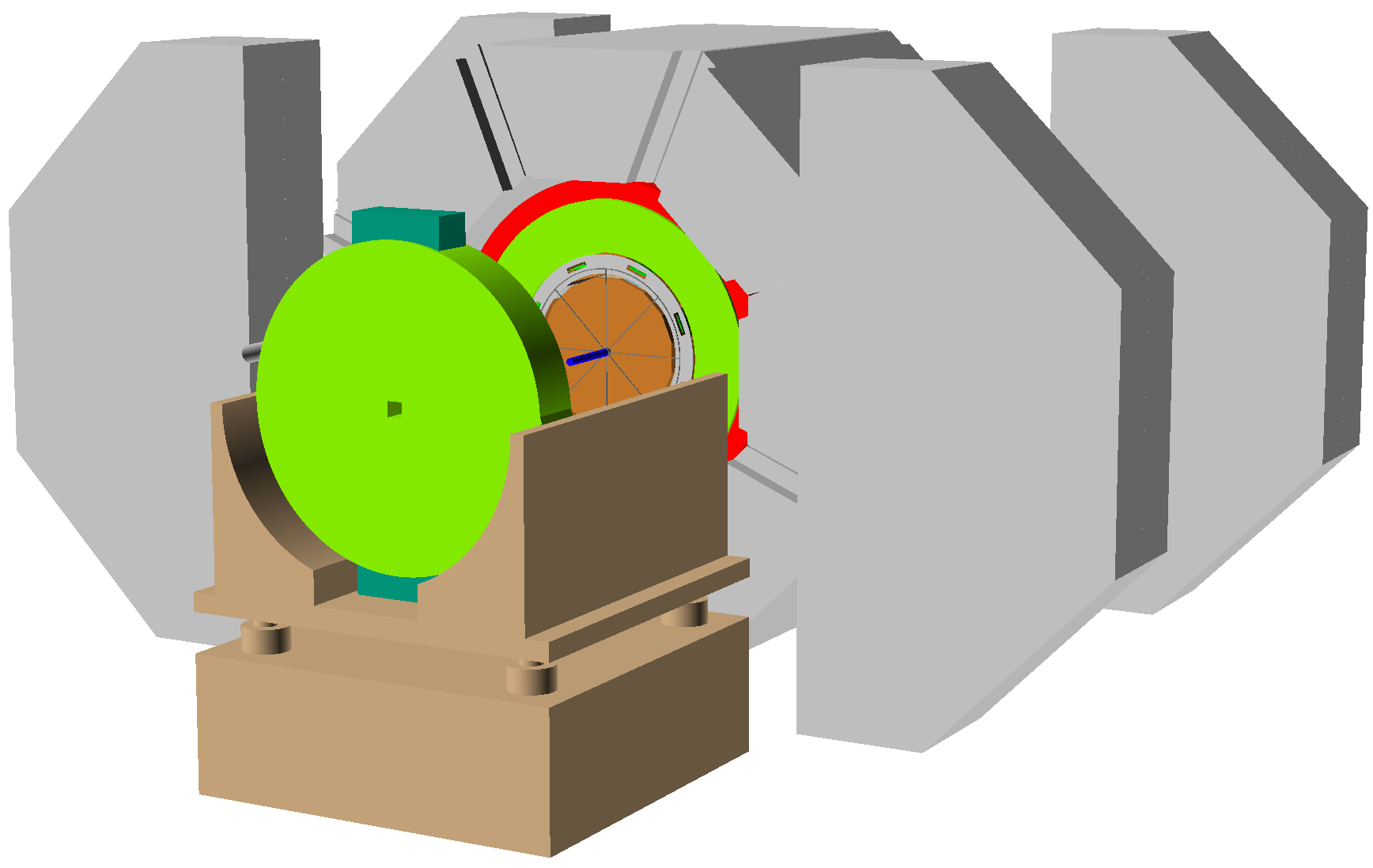}
\caption{The ECal end-cap is lowered by a crane into the the same cradle.
Assuming the same transversal dimention of the ECal end-cap as for the ECal barrel,
the same configuration of the cradle can be used as for the ECal barrel.}
\label{fig_ECalEC_v1}
\end{figure}

The next step is to install the ECal end-caps.
Assuming the same transversal dimension of the ECal end-cap as for the ECal barrel,
the same configuration of the cradle as for the ECal barrel can be used.
Therefore, the ECal end-cap will be lowered by a crane and placed into the the same cradle.
The ECal end-cap is transferred along the rails to the working position and fixed to the RS barrel.
The procedure is shown in Fig.\,\ref{fig_ECalEC_v1}.
The beam pipe passes through the central hole of the ECal end-cap, and its flange comes out. 
This flange will be used to attach the outer segment of the pipe.

\begin{figure}[th]
\centering
\includegraphics[width=0.55\textwidth]{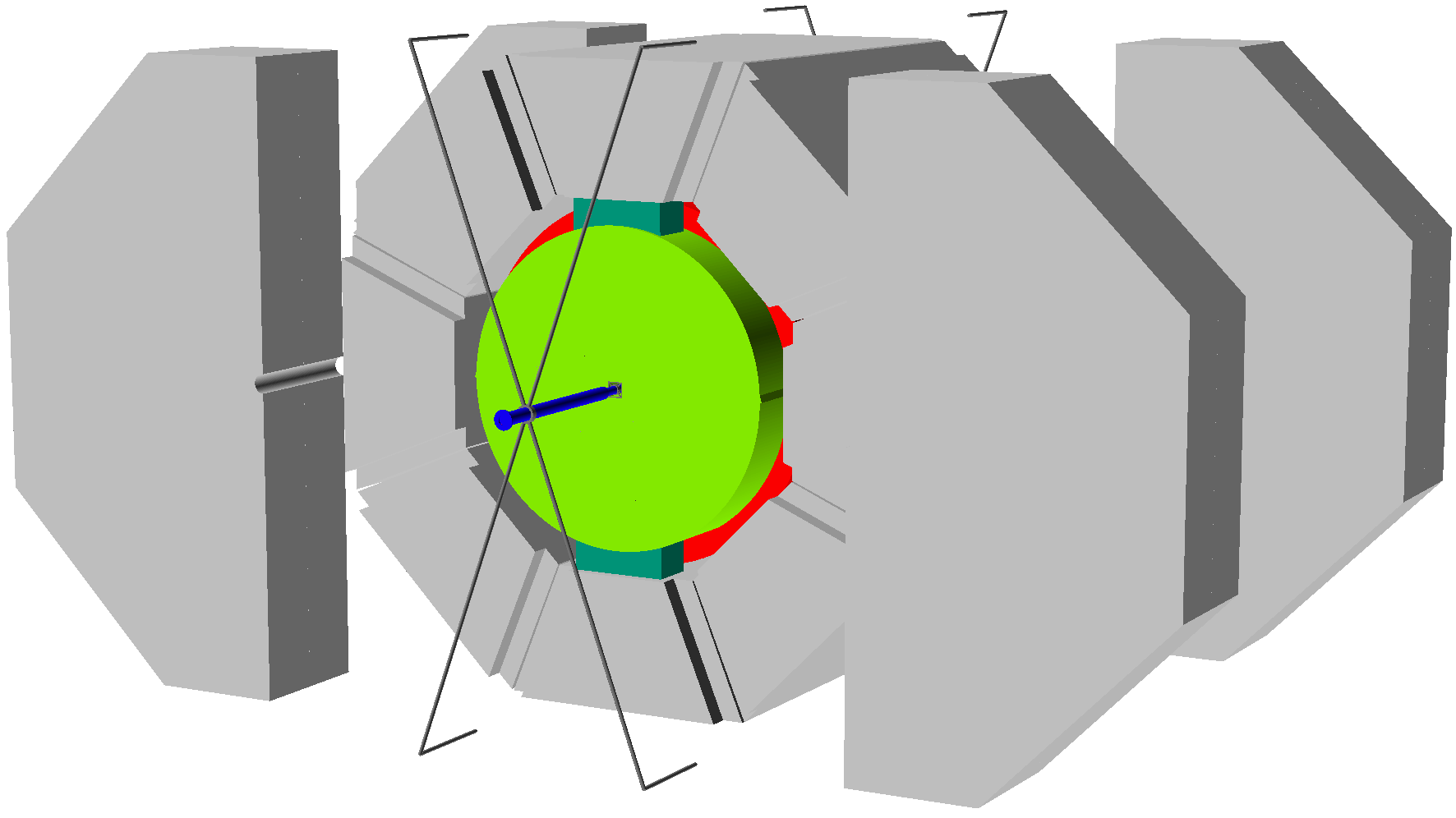}
\hfill
\includegraphics[width=0.34\textwidth]{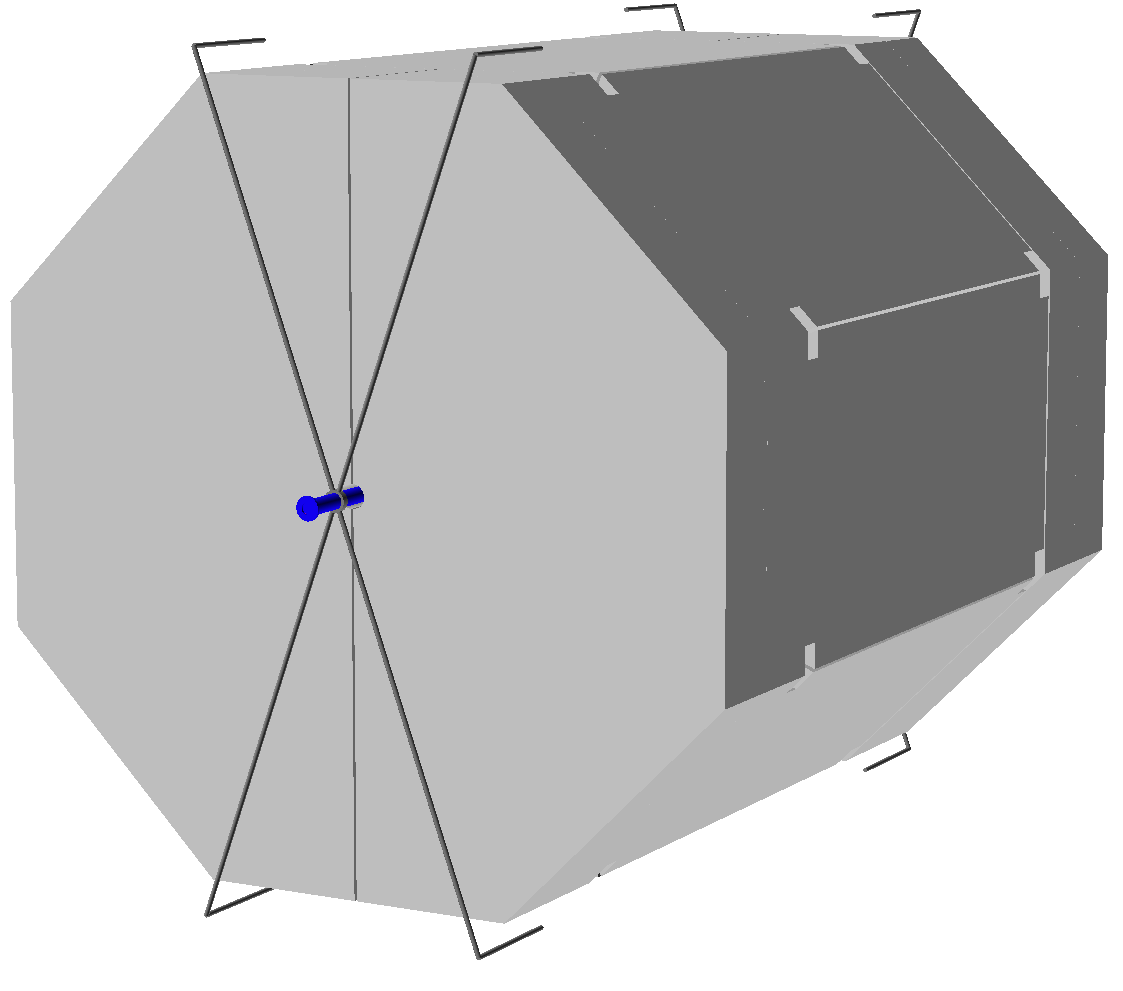}
\caption{{\it Left}: The edge section is bolted to the beam pipe at one end. 
At the outer end, the section is held by four rods, which, in turn, are attached to the lower and upper parts of the support frame.
{\it Right}: Two sections of the RS end-cap are closed. The setup is ready to be transported to the beam position.}
\label{fig_pipe_v2}
\end{figure}

The last step is to mount the outer segments of the beam pipe and to close  the two sections of the RS end-cap.
The outer segments are the heaviest part of the  beam pipe. 
Each of these segments will require two additional holding clamps. 
One clamp holds the beam pipe  in the middle of the central hole of the ECal end-cap. 
At the outer end, the segment is held by four rods, which, in turn, 
are attached to the lower and upper parts of the support frame, as shown in Fig.\,\ref{fig_pipe_v2} (left).
Finally, two sections of the RS end-cap are closed, as shown in Fig.\,\ref{fig_pipe_v2} (right), and the SPD detector is ready to be transported to the beam position.


\chapter{Detector Control System}
      The SPD detector control system (DCS) is designed to control the basic operating modes of the detector parts and the detector as a whole and to  continuously monitor  slowly changing parameters of the detector, engineering means, which provide the detector operation, and the environment. The DCS is synchronized with the basic operating modes of the NICA accelerator complex by means of a synchronization subsystem shared between the DCS and the SPD DAQ. The DCS provides parameterization of the managed object (i.e. the SPD detector), implements algorithms for normalization, parameters measurement and control, based on these parameters, and generates the necessary sets of abstractions and options for presenting them to the operator in an intuitive manner.
     Critical values of the parameters going beyond the predefined limits in predetermined situations cause emergency events and initiate procedures for handling such events, including the procedure for an automatic detector shutdown in order to prevent its damage. Parameter values, archived in a database for long-term monitoring of the detector operation, identify possible failures in the operation of the equipment and emergency situations. The configurations of the detector parameters saved in the database make it possible to start the detector promptly  with various preset parameters and in various operating modes, in accordance with the requirements of a particular physics experiment.
     
The DCS allows the autonomous operation of each detector subsystem at the stage of the initial start-up, as well as its  periodic maintenance, calibration sessions, and planned upgrades.
     The number of parameters in the system is expected to be significant, therefore, it is assumed that the system should be extendable and flexibly configurable. Architectural and software solutions, based on the event-driven model \cite{Michelson} and client-server and producer-consumer \cite{Etschberger} interaction models, should be preferred for communications when building the general DCS and the control systems of each part of the detector. Centralized systems operating in the master-slave polling mode should be avoided.

\section{DCS concept}

    Most of the high-energy physics detectors include parts consisting of similar systems, built from devices, sensors, and actuators with similar or identical functionality. This determines the parameterization of the entire detector as a managed object. Such systems include:
\begin{enumerate}
\item high voltage (HV) power supply system for powering gas detectors and light (photon) sensors (PMT and SiPM);
\item low voltage (LV)  power supplies for powering magnets, digital and analog electronics;
\item  cryogenic systems;
\item gas supply and mixing systems;
\item vacuum systems;
\item  front-end electronics LV powering control and temperature monitoring;
\item different cooling and temperature control systems;
\item DAQ system;
\item accelerator interface and synchronization;
\item general external electricity and water cooling stations, etc.
\end{enumerate}
          
\begin{figure}[ht]
\begin{center}
\includegraphics[width=1.0\linewidth]{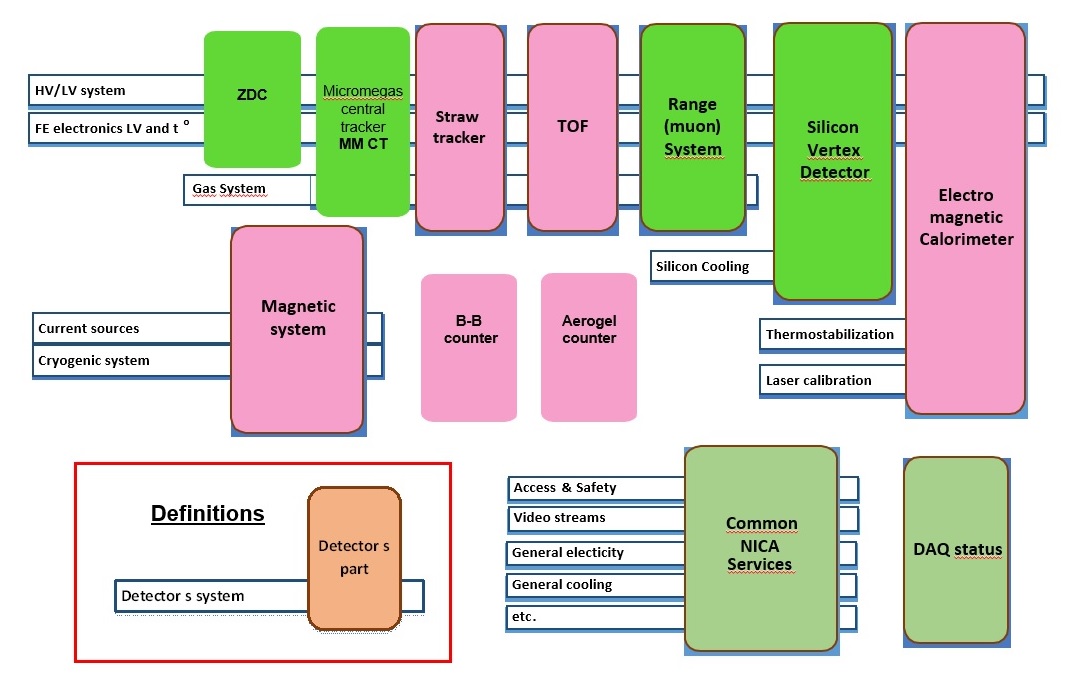}
\caption{SPD detector control system layout.}
\label{fig:dcs1}
\end{center}
\end{figure}

          The SPD detector is no exception and includes almost all of these systems spread among different parts of the detector, as shown in the layout diagram,  Fig. \ref{fig:dcs1}. Each part of the detector refers to one or more subsystems. The architecture of the systems will be refined, as the individual parts of the detector are developed.

All the systems can be similarly parameterized and shown to the operator in an intuitive presentation in order to simplify the operator's decision-making algorithm. However, the physical implementation of these elements at the hardware level may vary significantly in different parts of the SPD for the following reasons:
\begin{itemize}
\item the parts inherit the experience of their developers gained in previous experiments;
\item hardware and software components are selected based on their cost and availability;
\item parts of the detector are manufactured at different times.
\end{itemize}

Nevertheless, in order to optimize financial and human resources costs for the creation of the entire detector and the DCS, in particular, it is necessary to recommend to the developers of the detector parts to strive for standardization of the used hardware and embedded software. This will significantly reduce the efforts put into developing, deploying, and operating the detector, and will result in significant cost savings. To achieve these goals, at the stage of prototyping the detector systems, it is advisable to work out not only the detector itself, the front-end electronics, and the DAQ, but also the slow control systems. This work can be carried out in the SPD test zone, where the  slow control system must be made as similar to the final DCS version as possible.

\section{DCS architecture }
\begin{figure}[!hb]
\begin{center}
\includegraphics[width=0.85\linewidth]{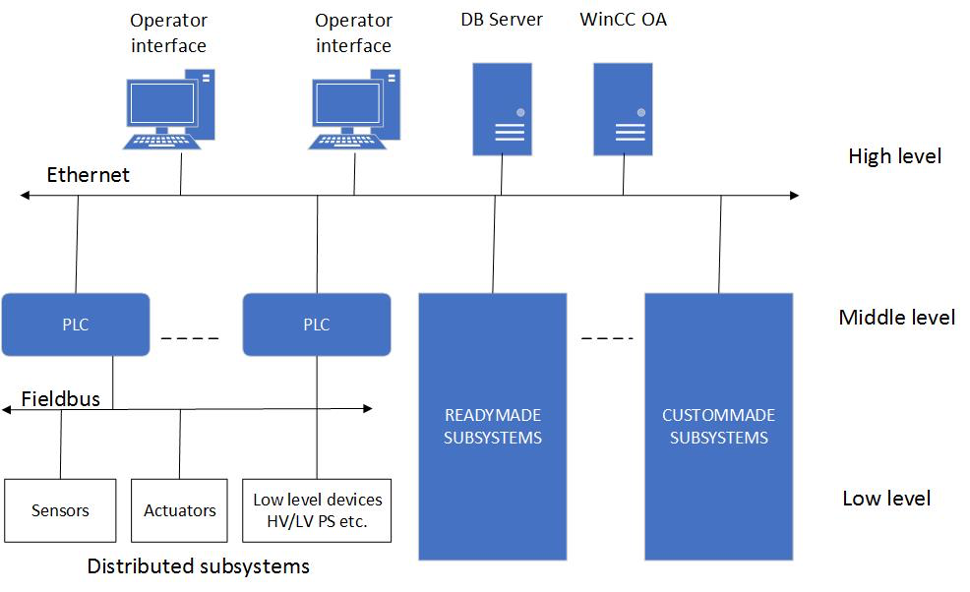}
\caption{SPD detector control system architecture.}
\label{fig:dcs2}
\end{center}
\end{figure}

The detector control system is divided into three logical levels (Fig. \ref{fig:dcs2}). 
The lower level includes measurement channels built into the front-end electronics (FEE) and data acquisition system (DAQ) of the detector parts, various stand-alone sensors, I/O devices, and low and high-voltage power supplies. The middle level is represented by programmable logic controllers and integrated ready-made and custom-made subsystems (vacuum posts, gas consoles, multichannel ready-made power subsystems, etc.). The interfaces to the FEE and DAQ that provide data for the detector control system are also on this level. The upper level is designed to provide a human-machine interface for operators, implement a database of detector parameters and configurations, communicate with the external world (accelerator, engineering support systems, access system, etc.), and implement macro-control algorithms, common for the entire detector. All these levels are connected in a hierarchical network using fieldbuses between the first and the second level, for example, a CAN-bus with a CANopen protocol. An Ethernet LAN is used between the middle and the upper levels. At the top level, special software, such as SCADA (Supervisory Control And Data Acquisition), is used, which provides control, collection, and storage of data in real time. It is proposed to use the WinCC OA system, common at  CERN, as a SCADA system. We understand that for smooth and reliable communication with the control system of the Nuclotron, a gateway to the Tango Controls \cite{tango,Gorbachev:2016yrd} system should be developed and deployed.

\section{SCADA for the DCS }
WinCC OA (ex PVSS-II) \cite{WinCC,Boterenbrood:530680} is a commercial SCADA system. It is a software component constructor that allows one to use preinstalled prototypes and templates, as well as software modules and system components, developed in C.  This system is actively used in many experiments at CERN and has support and safety certificates in the Russian Federation. The following properties make WinCC OA an attractive solution to be used in the DCS of the Spin Physics Detector:
\begin{itemize}
\item object-oriented approach, built into the system, ensures an efficient development process and the ability to flexibly expand the system;
\item	capability to create distributed systems - up to 2048 WinCC OA servers;
\item	scalability from a simple single-user system to a distributed redundant network system with $>10$ million tags (physical and synthetic parameters);
\item	platform-independent system, available for Windows and Linux;
\item	event-driven system;
\item   hot standby and 2$\times$2 redundancy (DRSystem), the required level of availability and reliability;
\item	wide range of drivers and options for communication OPC, OPC UA, S7, Modbus, IEC 60870-5-101/104, DNP3, XML, JSON, SOAP, etc.;
\item	support by major manufacturers of electronic devices for building automation systems in high energy physics.
\end{itemize}

\begin{figure}[!h]
\begin{center}
\includegraphics[width=0.85\linewidth]{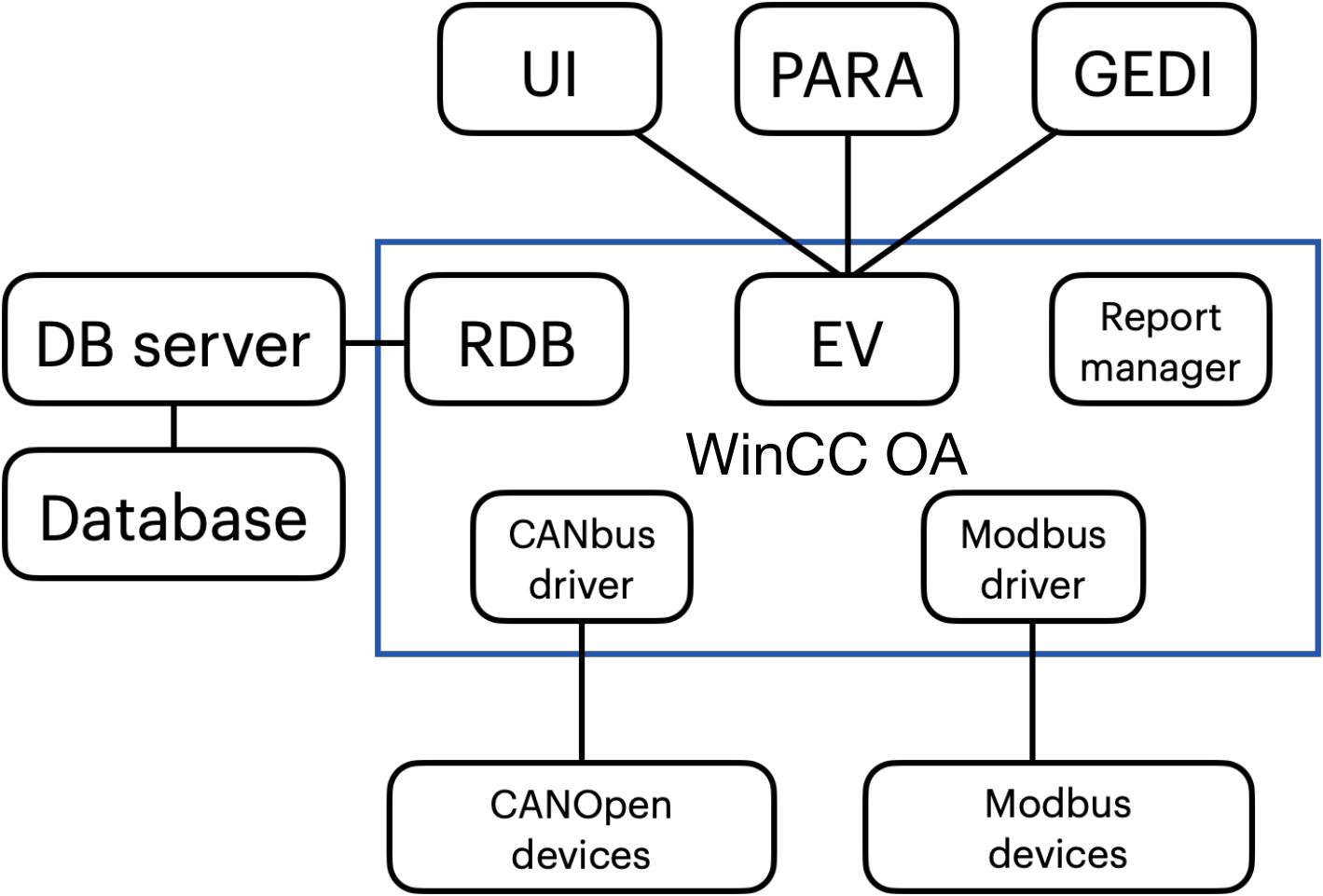}
\caption{SCADA structural scheme of the WinCC OA software.}
\label{fig:dcs3}
\end{center}
\end{figure}

Each functional unit of the system that is software implemented as a separate process is called the manager. A set of managers forms a system. Data exchange and communications between managers are done via TCP. The data is exchanged by means of passing events. The system allows parallelizing processes (managers) by running them on different computers with different OS. The system is scalable and balances the load on the control computers. The required managers start only if necessary, and multiple instances may run simultaneously. Managers can be distributed across multiple computers/servers. The WinCC OA block diagram is shown in Figure \ref{fig:dcs3}.

The main process is the Event Manager. It contains and manages the process image (current values of all process variables), receives and qualifies data (central message manager), distributes data across other managers, acts as a data server for others, manages users' authorization, as well as generation and status of the alarm messages.

The Database Manager receives data from the Event Manager and handles it according to its own algorithm. The historical database can use either a proprietary database (HDB) or an Oracle DBMS (the Oracle Real Application Clusters configuration is also supported). Parallel archiving in Oracle and HDB databases is possible. It is also possible to record user-defined data and log system events and messages in an external relational database (MS SQL, MySQL, Oracle, etc).

The WinCC OA Report Manager supports different ways of generating reports: 
\begin{itemize}
\item	in the Microsoft Excel format;
\item	in the {\it xml} format with the ability to display in any external tool for working with reports (Eclipse BIRT, Crystal Reports, SYMATIC Information Server, etc.). SOAP (Simple Object Access Protocol) protocol is also supported. 
\end{itemize}

Project development for the WinCC OA system is based on an object-oriented approach. In the WinCC OA data model, objects are represented as data points (called tags) that characterize the image of a specific physical device or process. For each data point element, properties and actions, such as signal processing (smoothing, setting limits, etc.), communication with external systems, archiving, generation of alarm messages (alarms), etc. can be defined accordingly.  Typing and inheritance are supported, therefore arbitrary hierarchical data structures can be created. Similarly, the principles of inheritance and reusability are implemented for graphical objects. The WinCC OA IDE includes the PARA configuration editor and the GEDI graphical editor of the User Interface Manager (UI) (includes a data model editor, mass configuration tools, administration tools, an interface to version control systems, a debugger, etc.).

Changes to data structures and graphics are applied without restarting the project. Writing custom scripts can be done using CONTROL++ (a programming language, the syntax of which is similar to C/C++). Such scripts can be both event handlers, associated with the elements of the graphical interface, and data processing procedures.

The system includes a standard graphical objects library; it can be extended by developing user objects or using the Qt Toolkit widgets. It is also possible to use the JavaScript libraries available on the market, or the included JavaScript scripts. Thanks to the open API (C++ / C\# API), it is possible to create managers, drivers, widgets, and CONTROL++ extensions. A new set of tools is available for the concept of High-Speed Programming implementation, which supports documentation build-up from the source code, unit testing, and autocompletion of program structures.

	It is also planned to provide data exchange between the WinCC OA and Tango Controls, which is used as the upper level of the Nuclotron control system. This can be implemented using standard OPC technologies with a client-server architecture, or it can be implemented using SQL tools, as a common database for  both SCADA systems used for the accelerator and the detector. The final choice of a suitable solution will be made at the stage of system implementation. 

\section{Cost estimate }
The cost of the DCS development and implementation includes the cost of the off-the-shelf components, customized elements, and the development of the application software including configuration of the selected SCADA. It seems to be optimal to carry out all the development of hardware and software during the first stage of the experiment, while the additional  purchase of the off-the-shelf components and additional production of custom components is reasonable to do during the second stage. In this regard, the cost of DCS implementation during the stage I is 1.5 M\$, with a total cost estimate of 2.7 M\$ for the stages I+II.

\chapter{Data Acquisition System}
\section{Introduction}

The data acquisition system (DAQ) of SPD should provide  a readout and  a transfer of data from all detector subsystems. These include:
\begin{itemize}
\item Vertex Detector (VD): a MAPS-based (or double-strip based) silicon tracking detector for precise determination of the primary interaction point and measurement of  the secondary vertices from the decays of short-lived particles;
\item Micromegas-based Central Tracker (MCT): a  tracking detector designed to improve the momentum resolution at the first stage of the experiment;
\item Straw Tracker: the main tracking detector for measurement of the primary and secondary particle momenta based on curvature of tracks in a magnetic field, and also for the ionization loss measurement;
\item particle identification detectors: i) time-of-flight resistive plate chambers (TOF) and ii) aerogel Cherenkov counters (Aerogel);
\item Electromagnetic Calorimeter (ECal) for measurement of the energy of gammas and electrons (positrons);
\item Range System (RS) for detection and identification of muons;
\item Beam-Beam Counters (BBC) for local polarimetry and monitoring of the beam collisions;
\item Zero Degree Calorimeters (ZDC) for local polarimetry using forward neutrons and for luminosity measurement.
\end{itemize}

At the maximum center mass energy $\sqrt{s} = 27$~GeV and maximum luminosity $10^{32}$~$\rm cm^2 s^{-1}$ the event rate was estimated as $3 \times 10^6$~$\rm s^{-1}$. This high rate has brought us to the decision to build a triggerless DAQ. Such DAQ systems will be created in many new experiments under preparation, in particular, in GSI (PANDA, CBM, NUSTAR), PSI (Mu3e), in future HL-LHC experiments, also foreseen for ILC, CLIC etc.

With a free-running DAQ, there is no hardware trigger system, therefore  the readout process is not controlled by  a trigger anymore.  All data that exceed the thresholds in  the front-end electronics are read out together with their timing marks. Then in DAQ the data are grouped on the basis of timing and detector affiliation, and finally are delivered to the input buffer of the  online filter in the form suitable for analysis.

The front-end electronics of the detectors should meet the requirements of a free-running DAQ:
\begin{itemize}
\item self-triggered operation;
\item digitizing on-board;
\item timestamp included in the output format.
\end{itemize}

A reasonable estimation of the total data flux in the SPD DAQ was obtained,  partly using the simulation, and partly -- the results of the beam tests of  the detectors of other experiments (MPD, PANDA, ALICE), or the parameters of the appropriate front-end electronics which  already exist or are under development.

The data flux was estimated at the maximum energy 27~GeV, maximum luminosity $10^{32}$~cm$^2$s$^{-1}$, and at the correspondent event rate of $3 \times 10^6$~s$^{-1}$. According to CDR, multiplicity value was taken equal to 8 for charged particles and 6 for neutrals. With some approximations, the total flux was found to be 20~GByte/s. This number includes transfer of the measured quantities themselves (time, amplitude, coordinate), relevant headers, also takes into account clustering of hits and partly noise counts in some detectors. The largest contributions to the total flux  are produced by VD, Straw and RS detectors, while the contributions of BBC and ZDC are almost negligible.

Note, that at the first stage of the experiment, with not all of the detectors implemented and at  a reduced energy and luminosity, the data flux could be less by up to two orders of magnitude.

A total number of channels to  be read out at the first stage will not exceed 200 thousand. For the full scale experiment this number is expected to be about 500 thousand. Some numbers for both stages are summarized  in Tables ~\ref{tab:daq-ch1} and \ref{tab:daq-ch2}. 

\begin{table}
\caption{Summary of detector outputs to DAQ at the first stage. Information type: T means time, A -- amplitude (or charge).}
\label{tab:daq-ch1}
\renewcommand{\arraystretch}{1.2}
\begin{center}
\begin{tabular}{|l|l|l|l|l|}
    \hline
    \rule{0pt}{14pt}
Sub-detector    & Information & Number & Channels& Number  \\
            & type   & of channels & per FE card& of outputs\\ \hline
Micromegas Vertex & T + A & 5384 & 128 & 43 \\ \hline
Straw tracker   & T + A  & 
25904 + 4608  & 128  & 239 \\ \hline
BBC & T + A & 
1056 &  64     &  18 \\ \hline
BBC MCP & T & 
64 &  32   &  2 \\ \hline
Range system    & T     & 137600 & 192  & 717 \\ \hline
ZDC            & T + A & 1050$\times$2  & 64 & 34 \\ \hline \hline
Total (max)     &       & 176716 & &   1053 \\ \hline
\end{tabular}
\end{center}
\end{table}

\begin{table}
\caption{Summary of detector outputs to DAQ for the full scale experiment. Information type: T means time, A -- amplitude (or charge).}
\label{tab:daq-ch2}
\renewcommand{\arraystretch}{1.2}
\begin{center}
\begin{tabular}{|l|l|l|l|l|}
    \hline
    \rule{0pt}{14pt}
Sub-detector    & Information & Number & Channels& Number  \\
            & type   & of channels & per FE card& of outputs\\ \hline
Vertex detector  DSSD &  A & 327680 & 640 & 512 \\ \hline
Vertex detector  MAPS &  A &  12000 &  &  2204 \\ \hline
Straw tracker   & T + A  & 
25904 + 4608  & 128  & 239 \\ \hline
Calorimeter     & T + A & 27168  & 64  & 425 \\ \hline
TOF         & T     & 12228 & 8 & 1529 \\ \hline
FARICH     & T &  70144  & 64 & 1096 \\ \hline
BBC & T + A & 
1056 &  64     &  18 \\ \hline
BBC MCP & T & 
64 &  32   &  2 \\ \hline
Range system    & T     & 137600 & 192  & 717 \\ \hline
ZDC            & T + A & 1050$\times$2  & 64 & 34 \\ \hline \hline
Total (vertex DSSD)     &       & 608552 & &   4572 \\ \hline
Total (vertex MAPS)     &       & 283076 & &   6264 \\ \hline
\end{tabular}
\end{center}
\end{table}
\section{DAQ structure}

Data in the DAQ system is grouped by time into parts called slices. 
A slice length is 10$\div$100~\textmu{}s. A sequence of slices forms a frame. The frame 
length is 0.1$\div$10~s. Each slice is  stamped with two numbers:
a frame number (the order of the frames inside the run) and
a slice number (the order of the slice inside the frame). The frame and slice 
numbers start from 0.

A clock called the global clock is used for synchronization in the DAQ system.
A global clock frequency is 125~MHz (a period of 8~ns). The required jitter 
for the global clock is $\leq 50$~ps.

The data of each hit must have a timestamp. Time in the SPD detector is measured from the beginning of the frame and makes sense only inside that frame. No correlation is assumed between the time in two frames. Thus, one frame is the maximum possible time interval in SPD between time-correlated events.

To control the operation of the DAQ system, the following commands are used:
\begin{itemize}
\item commands synchronous with the global clock signal;
\item asynchronous commands.
\end{itemize}

The so-called Time Synchronization System (TSS) is responsible for the distribution of the global clock signal and synchronous commands throughout the whole setup. 

Logically, the DAQ system can be divided into three subsystems: a Readout Chain, a Slice Building System, and a Time Synchronization System.

Now the interaction of the DAQ system with the front-end electronics is well developed for the already existing electronics of the Range System. The current requirements for the front-end electronics are developed assuming an easy implementation to supposed readout systems of other detectors. 

All information about the DAQ settings (slice and frame length, enabled subdetectors, etc.), configuration of the front-end electronics (calibration, the mapping of electronics channels, etc.) and experimental conditions (beam energy, polarization, etc.) will be stored in a database.
\section{Readout chain} 

\begin{figure}[htb]
\includegraphics[width=\columnwidth]{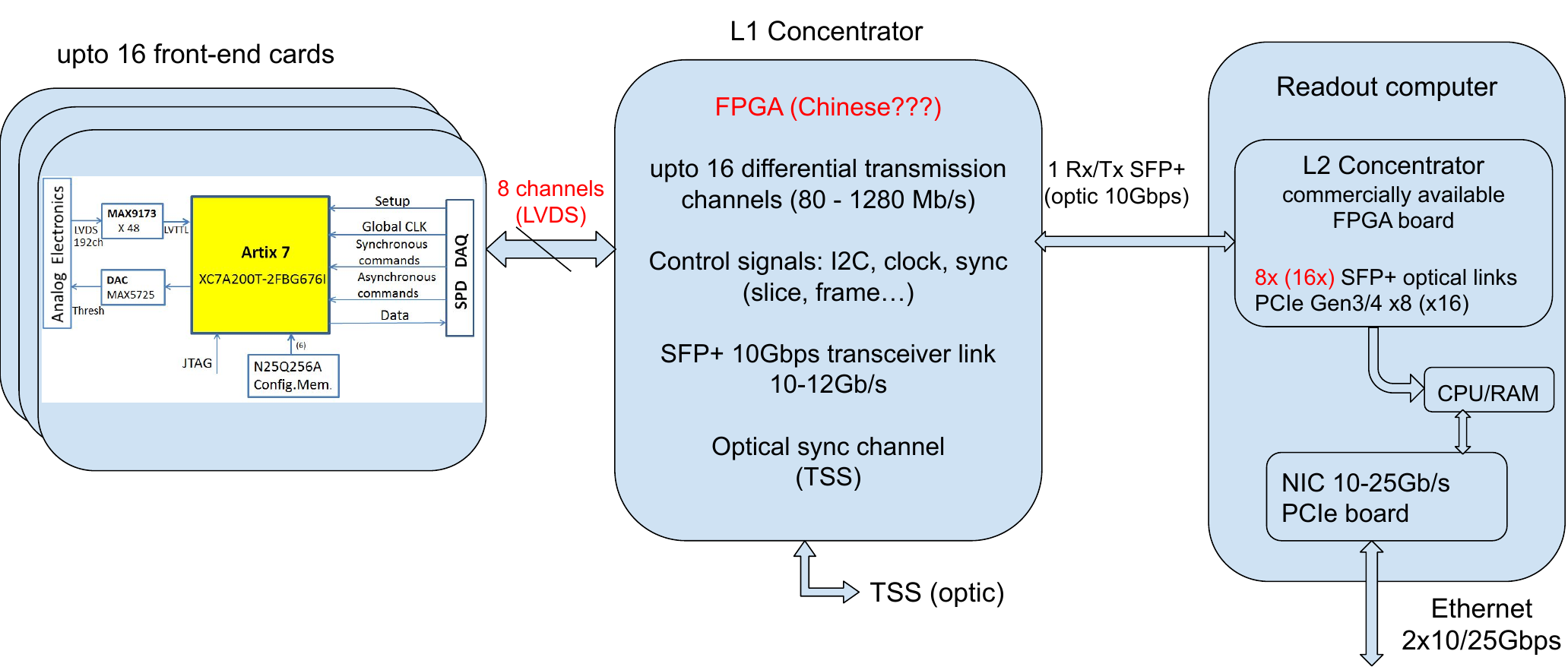}
\caption{SPD readout chain.}
\label{fig:daq-readout_chain}
\end{figure}

The general view of the readout chain is shown in Fig.~\ref{fig:daq-readout_chain}. The readout chain consists of the front-end electronic FEE cards (the example of the RS FEE cards is used at the presented scheme), a data concentrator of the 1st level (L1 concentrator) and a data concentrator of the 2nd level (L2 concentrator). 

The FEE cards are connected to the L1 concentrator using a copper cable or a flexible cable, depending on the technical design of the detectors. The length of the cables can be up to 5$\div$7 meters, depending on the data transfer rate. 

The following LVDS signals are used: data transmission channel, global clock and signals for the implementation of synchronous commands. In addition to  the LVDS signals, other signals are used: signals for implementing the I$^2$C bus (SDA, SCL) and  a reset signal.

The I$^2$C bus is used to transmit asynchronous commands. These commands are used to control and monitor FEE cards and are specific for various electronics.

The L1 concentrator board allows to connect from 8 up to 16 FEE cards, with a 
data transfer rate of up to 1.2~Gbps per channel, and to manage them. 
The L1 concentrator is based on FPGA chips similar to Intel Cyclone10GX or 
Lattice CertusPro-NX. Data for setting up and managing the FEE cards are
received from the L2 concentrator via an optical link. 
Up-level connection of the board is supposed to be high speed optical
bidirectional channel GBTx with the rate up to 10$\div$12~Gbps. The main 
limitation for the L1 concentrator is the requirement that the input data flow
rate does not exceed the transfer rate at the output of the concentrator. If
this requirement is violated (for example, all 16 links of  the FEE cards have the 
rate 1~Gb/s), then the number of the input links is to be reduced to the bandwidth 
of the output optical link. Thus, the data stream from  the FEE cards can always 
be transferred to the L2 concentrator without data loss. 

\begin{figure}[htb]
\begin{center}
\includegraphics[angle=90,width=0.7\columnwidth]{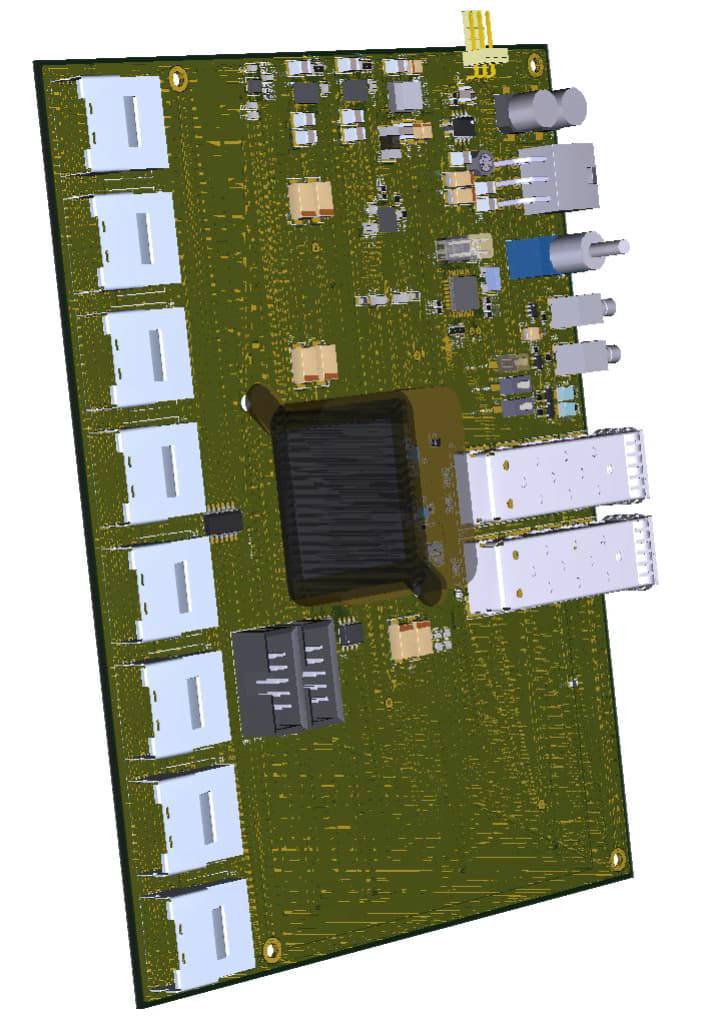}
\caption{Prototype of L1 concentrator.}
\label{fig:daq-L1}
\end{center}
\end{figure}

Now a prototype of L1 concentrator based on Intel Cyclone10GX is under development (see Fig.~\ref{fig:daq-L1}) It contains 8 Links for frontend electronics
boards connecting via miniSAS connectors, SFP+ 10Gb transceiver for data
transmission and SFP+ 10Gb transceiver for timing (White Rabbit).

The global clock signal and synchronous commands are generated and distributed through the L1 concentrators using a Time Synchronization System (TSS). An optical fiber is used to connect TSS to the L1 concentrators. The readout chain operates under the control of TSS: frames and slices begin and end in accordance with the commands coming from TSS. On the other hand, data transmission is initiated by the  front-end electronics as a reaction to the commands from TSS, but there is no strict time correlation between TSS commands and data transmission.

\begin{figure}[htbp]
\includegraphics[width=0.9\columnwidth]{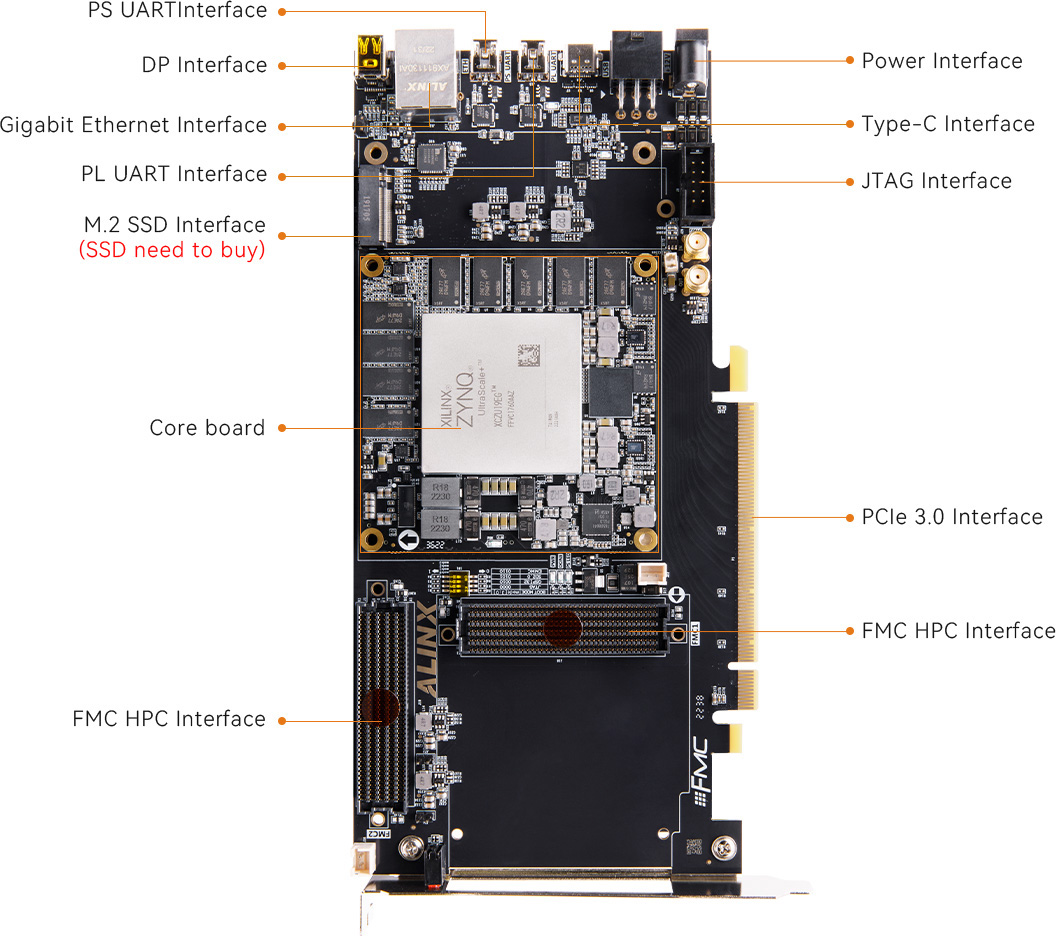}
\includegraphics[width=0.9\columnwidth]{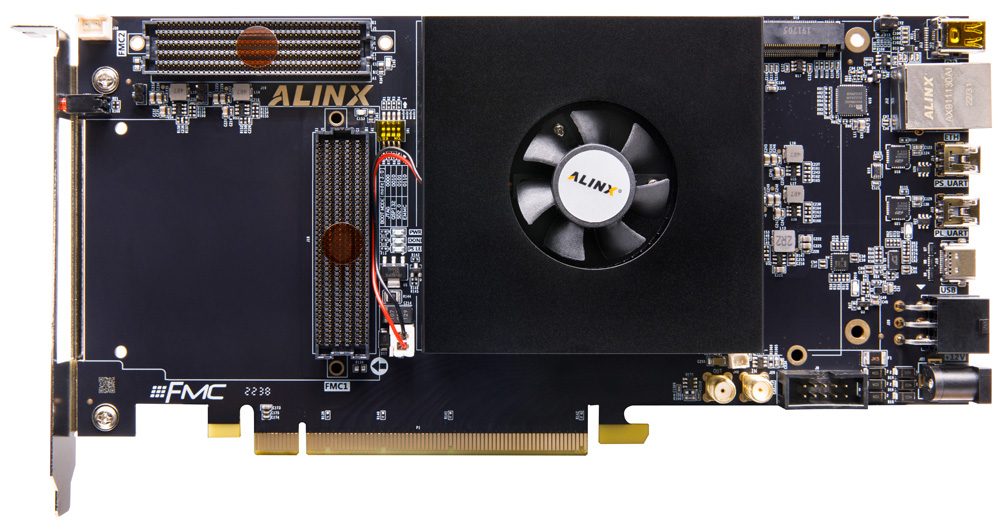}
\caption{Possible hardware for L2 concentrator.}
\label{fig:daq-l2_boards}
\end{figure} 

A commercially available PCIe card installed in the readout computer is used
as the L2 concentrator (see Fig.~\ref{fig:daq-l2_boards}). From 8 up to 16 L1 concentrators 
can be connected to one L2 concentrator (or from 64 upto 256 FEE cards). Data is 
written directly to the computer memory (RAM) at a speed of 4--8~GBps 
through the PCIe bus. A dedicated firmware need to be developed for these cards. Up to 4 such card can be installed in one readout computer.

For the 1st stage of the experiment it is assumed to install 2 L2 concentrators in each readout computer. With the minimal available multiplication of 8 for each L1 and L2 concentrators it gives 128 for the compression factor. Thus about 1000 of the detector outputs (see  Table~\ref{tab:daq-ch1}) can be read out by about 10 computers. For the 2nd stage the expected output number up to 6500 (see  Table~\ref{tab:daq-ch2}) makes the number of readout computer extremely high. More over, it assumes significant underloading of the readout computers. Thus for the 2nd stage a significant increase in compression of the readout channel numbers should be considered for each level of compression including the front-end electronics for the 2nd stage detectors. Now for the 2nd stage it is assumed that 4 L2 concentrators are installed in one readout computer.  
\section{Slice building system}

\begin{figure}[htbp]
\centering
\includegraphics[width=\columnwidth]{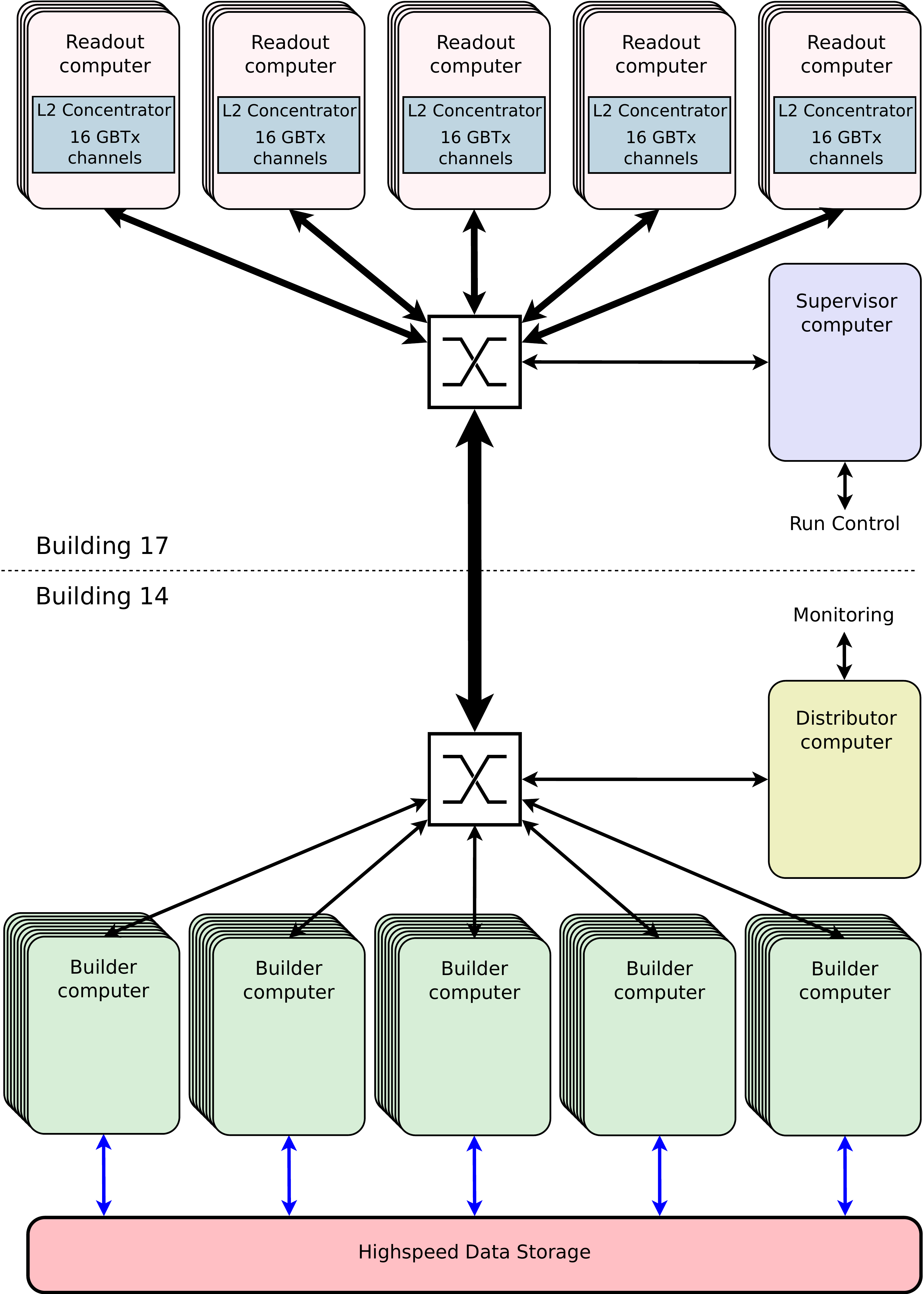}
\caption{Slice building system.}
\label{fig:daq-slice_build}
\end{figure}

As a consequence of the structure of the readout chain, the data of one slice
is distributed across all L2 concentrator. The main task of the slice building
system is to receive data from all L2 concentrators, combine them into complete
slice data and write this data to a data storage. In addition to this, a part 
of the built slices should be distributed among a group of computers with
monitoring purposes.

The data in the slice building system is processed in parts, each part covers 
a certain period of time. The natural choice for this time period is a 
frame (0.1--10~s). In some cases, the amount of data in the frame may be 
too large for convenient data processing. Therefore, the frame is 
programmatically divided into parts called chunks. A reasonable chunk length 
is about 1~s and can be chosen taking into account the actual data flow.
Splitting the frame into chunks is performed on the border of the slices,
if the frame length exceeds the specified length of the chunk. Thus, the chunk
is a data processing unit in the slice building system. The chunk is transparent
to the rest of the software.

Since different data structures of the slice building system can 
simultaneously contain slices related to different runs, 3 parameters must be 
used to address the slice: the run number, the frame number and the
slice number. This imposes the condition that the run number must be unique.
At the same time, the need to sort slices by time requires a monotonous
increase in this number. In order to leave some freedom in choosing the 
numbering of runs, for the purposes of identifying the run, the run ID will
be used --- an automatically generated monotonically increasing number
that is assigned to each run (actually, to each attempt to start a run).
Thus, three numbers will be used to address the slice: the run ID,
the frame number, and the slice number. The run number that will come from 
the run control system is still exist in various data structures, but will
only be used in text messages.

Two numbers will be used to address the chunk: the run ID and the chunk number.
The chunk number is obtained from the frame number and the slice number of 
the first slice belonging to the chunk.

The general view of the slice building system is shown in
Fig.~\ref{fig:daq-slice_build}.
The slice building system consists of readout computers, a supervisor computer,
builder computers, distributor computers and network switches.
The various components of the system are distributed across 2 buildings: 
Building 17 (NICA collider) and Building 14 (NICA Data Center).

From the software point of view, the slice building system includes the 
following main processes: reader processes, supervisor process, builder
processes and distributor processes.

All the processes mentioned above are arranged as follows. At startup, they 
read their basic configuration from a file. This basic configuration includes
only parameters that do not require frequent changes, such as: process names
(used to designate a process in text messages), parameters for connecting to
the serviced equipment (device name of the L2 concentrator for the reader
process), addresses for establishing connections with other processes of
slice building system, and parameters for establishing connections with other
parts of software (for example, to access a database). After that,
the process creates a child process and waits for it to finish.
When the child process terminates, the parent process creates a new child
process. All work is done by a child process. 

Each time a child process is started, it gets an additional configuration.
The specific way to get this configuration depends on the process. This 
configuration includes parameters that may require more frequent changes or
that must be changed consistently for the entire system. The reader and builder
processes receive such configuration from the supervisor process when they
first connect to the supervisor process. The supervisor process reads this
configuration from the database.

Once the additional configuration is received, the child process
allocates and initializes all internal data structures. Most data structures
in the child process are statically allocated, and their sizes cannot be changed
during the life of the child process. This is necessary to avoid the overhead of
allocating and deleting memory blocks, which in our case is essential.

After initialization is completed, the child process begins to perform its
main tasks. All descriptions below refer to child processes unless explicitly stated otherwise.

\begin{figure}[htbp]
\centering
\includegraphics[width=\columnwidth]{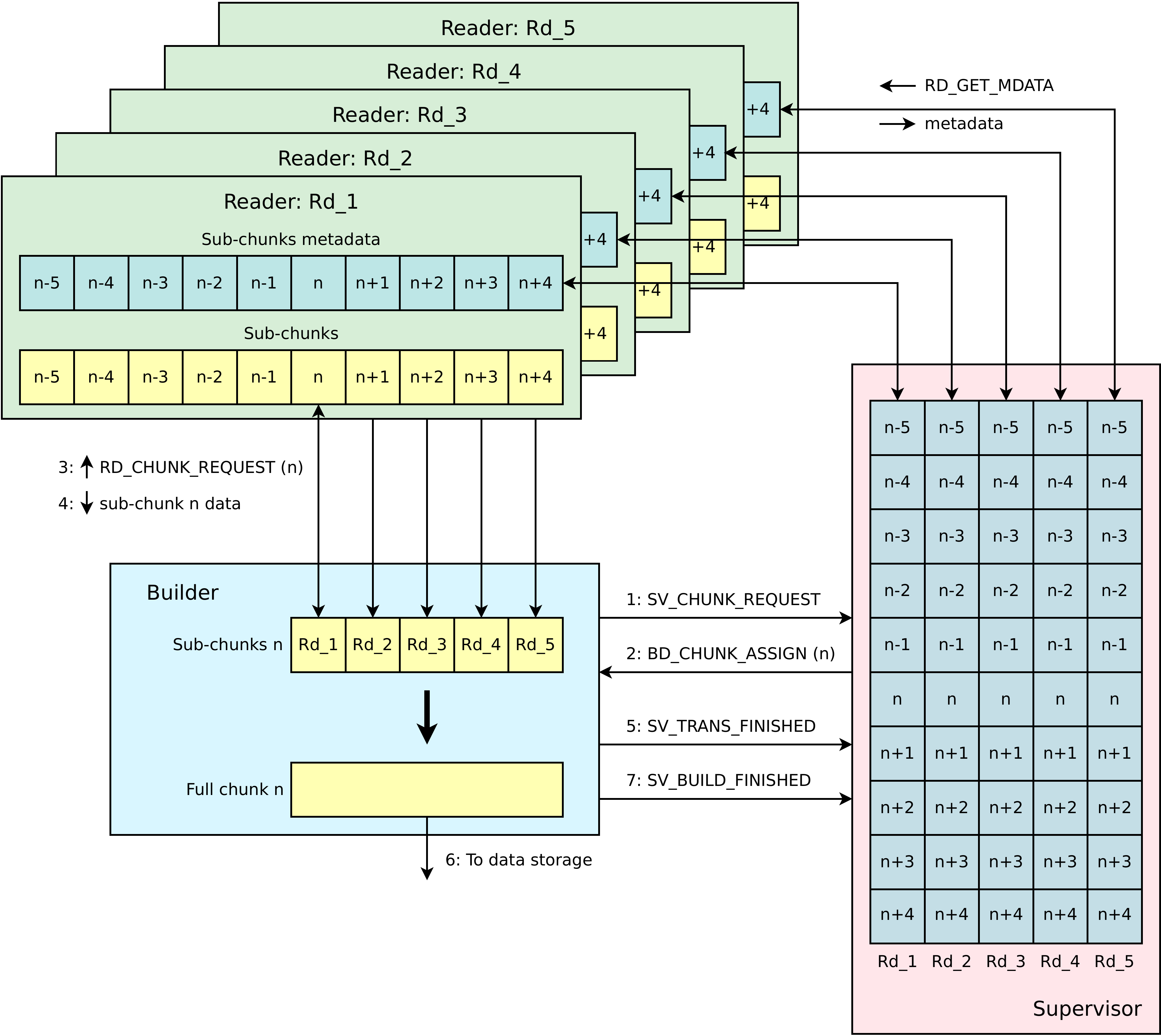}
\caption{Illustration of processing one chunk. 5 reader processes are 
taken as an example.}
\label{fig:daq-chunk_proc}
\end{figure}

Fig.~\ref{fig:daq-chunk_proc} shows the process of building a single chunk.
Below is a detailed description of the interaction of all processes involved
in this procedure.

\subsection{Reader process}

The reader processes are performed on the readout computers, one process 
for each L2 concentrator. The reader process receives data from the L2 
concentrator, buffers it in the RAM, and sends the data to the builder processes
according to their requests.

The reader process has a unique name directly associated with the ID 
of the L2 concentrator it serves.

Slice data coming from the L2 concentrator is stored in the data buffer. 
In addition, an entry is added to the corresponding list for each slice, chunk 
and run. Entries in the list of runs are created when the run start command 
is executed, entries in the list of chunks are created when the first slice
belonging to the chunk appears. As long as there is at least one slice in the 
buffer belonging to the corresponding time interval, entries about it remain
in the lists.

The reader process is a multithreaded program, main functional threads are:
supervisor thread, reader thread, builder threads and monitoring threads.

The supervisor thread reads and executes commands from the supervisor process.
At start it waits for a connection from the supervisor process, then it receives
an additional configuration from it and sends back to the supervisor an address
that will be used by the builder to connect to the reader.
When initialization is complete, the supervisor thread executes the following
supervisor commands:
\begin{enumerate}
\item \texttt{RD\_RUN\_TEST}, \texttt{RD\_RUN\_START}, 
\texttt{RD\_RUN\_STOP}:
the run control commands. The command checks whether the run can be started,
starts or stops the run, respectively. As an argument of the 
\texttt{RD\_RUN\_START} command, the reader process receives the run number and 
the run ID.
\item \texttt{RD\_GET\_MDATA}: get metadata command. In response, the
reader process sends to the supervisor process the information about the chunks
present in the chunk list. The 
information sent for each chunks includes the number of slices read from the L2 
concentrator, the number of slices lost due to lack of space in data
structures, the number of slices sent to builder process, and the amount of 
data read and sent.
\item \texttt{RD\_CHUNK\_DROP}: drop of chunk command. In response, the
specified chunk and all slices belonging to the specified chunk will be 
deleted from the corresponding data structures.
\end{enumerate}
In case if connection to the supervisor process is lost, the reader child
process is terminating.

The reader thread receives data from the L2 concentrator and buffers it in
the memory. Depending on the actual implementation of
the L2 concentrator, the data coming from the L2 concentrator may already be
formed in slices or may be independent streams from each front-end card. In
the latter case, the reader thread must reorganize the data into slices. The
reader thread stores data organized as slices in the data buffer, and
updates
data structures for chunks and slices. For testing purposes, the reader thread
can receive data from an external program or generate dummy data on command.
In case of an error during data reading, the reader thread closes the 
connection to the L2 concentrator and tries to reopen it.

The builder threads send the requested data to builder processes.
For each connection from the builder process, its own builder thread is 
created. At startup, the builder thread gets the required run ID and chunk 
number from the builder process. It sends the chunk data to the builder process
slice by slice, releases the transferred slices from the data buffer, and
updates the data structures for chunks and slices. The builder thread 
terminates when the transfer is completed or in case of an error.
Unsent slices are released from data structures.

The monitoring thread periodically sends information about the status of the 
reader process to the monitoring program. For each connection from the 
monitoring program, its own monitoring thread is created, but the information 
submitted to different monitoring programs is identical and refers to the same 
time points. The information provided includes the state of internal data 
structures and the status of the run at the time of sending, as well as 
accumulative information for the last time interval. The duration of the time
interval is set in the configuration of the reader process. Accumulative 
information consists of the slice rate, the number of slices received,
lost and sent, the number of various errors, the size of the data received 
and sent. The monitoring thread is terminated in case of a communication
error with the monitoring program.

\subsection{Supervisor process}

The supervisor process runs on the dedicated computer. The system
uses only one supervisor process, which works as a dispatcher for the entire 
slice building system. The termination of the supervisor process will result 
in the termination of all reader and builder child processes. The supervisor
process collects metadata from reader processes and chunk processing requests
from builder processes. Based on this information, it decides which builder
process should handle which chunk.

The basic configuration of the supervisor process contains the names and
addresses of configured reader processes, as well as a rather primitive
description of the network topology, in terms of connections between builder and readout processes.
Information about the reader processes partially duplicates information from 
their configurations, since both processes (the reader and the supervisor)
need this information before they establish a connection. The supervisor
process verifies the identity of the parameters in both sources at the moment
when a connection is established with the reader process.

The additional configuration contains the parameters to be transferred to the
reader and builder processes.

The main data structures of the supervisor process are: data structures for 
runs, chunks, and a ready queue. Chunk data structures contain information about 
all chunks present in the chunk list of at least one reader process. The ready 
queue contains chunks that are ready for processing by the builders.

The supervisor process is a multithreaded program, its main functional threads
are: the run control thread, the supervisor thread, the reader threads,
the builder threads, the scheduler thread and two types of monitoring threads.

The run control thread reads and executes commands of the run control system.
At startup, the run control thread waits for a connection from
the run control system. When the connection is established, the run control 
thread receives and executes following commands:
\begin{enumerate}
\item \texttt{SBS\_GET\_CONF}: get configuration command. In response, 
the supervisor process returns a list of configured reader processes.
\item \texttt{SBS\_GET\_STATUS}: get status command. In response, the supervisor
process returns the current state, which includes the run status, the data queue
flag, the run number, the run ID, the data recording flag, and the statuses of all 
configured reader processes. The run status can have one of the following values:
  \begin{enumerate}
  \item \texttt{SBS\_RUN\_STOPPED} --- the run is stopped;
  \item \texttt{SBS\_RUN\_RUNNING} --- the run is active;
  \item \texttt{SBS\_RUN\_DEGRADED} --- the connection to one or more of the enabled
  reader processes was lost when the run was in the active state;
  \item \texttt{SBS\_RUN\_UNKNOWN} --- the error was detected when
  starting/stopping the run.
  \end{enumerate}
From \texttt{SBS\_RUN\_DEGRADED} and \texttt{SBS\_RUN\_UNKNOWN} states
the system can only be transferred to \texttt{SBS\_RUN\_STOPPED} state.
\item \texttt{SBS\_READER\_ON}/\texttt{SBS\_READER\_OFF}: enabling/disabling a
reader process. If the reader process is disabled, it will be ignored
by the supervisor process, which effectively means that the corresponding
L2 concentrator will be excluded from the data acquisition process.
This commands can only be used if two conditions are met: the run must be 
stopped and the data buffers in all reader processes must be empty.
\item \texttt{SBS\_DATA\_QUEUE}: setting data queue flag. At the moment when
the flag switches to the ``unchecked'' position, all chunks from the ready 
queue are removed, and new chunks will not be added to the queue while this 
flag is unchecked. The flag can be used to clear data buffers in reader
processes during a stopped run or for testing purposes.
\item \texttt{SBS\_RUN\_TEST}, \texttt{SBS\_RUN\_START}, \texttt{SBS\_RUN\_STOP}:
the run control commands. The command checks whether the run can be started,
starts or stops the run, respectively. As parameters of the run starting
command, the supervisor process receives the run number and the data recording
flag. If the data recording flag is unchecked, the data is not written to the
data storage, but the distribution of slices for monitoring purposes continues.
When starting a run, the supervisor process generates a unique monotonically
increasing run ID.
\item \texttt{SBS\_RESTART}: restarting of slice building system.
This command terminates the supervisor child process, which leads to the 
termination of all reader and builder child processes, which, in turn,
leads to the restart of the slice building system, except the 
distributor processes.
\end{enumerate}
If the connection to the run control system is lost, the run control thread
proceeds to wait for a new connection. A lost connection to the run control 
system does not have an immediate effect on the slice building system.

The supervisor thread periodically reads metadata from all enabled reader
processes. Based on the information received, it makes a decision about the
chunks whose reading from the L2 concentrator has been completed or stopped
due to a timeout. A timeout situation is registered when data from any
reader process did not appear during the specified time after the first
data about this chunk from any other reader process was appeared.
Incomplete chunks (with missing slices or with missing chunk information 
from one or more of the reader processes) are deleted. Complete chunks are placed in a 
queue of chunks ready for building (ready queue). The supervisor thread does not
allow this queue to grow indefinitely, leaving only the specified number of
chunks in the queue and deleting old chunks. Necessity of deleting 
chunks indicates insufficient number of builder processes. 
The queue size should be chosen in such a way that full chunks are not lost
due to a situation when there are temporarily not enough builder processes,
and at the same time incomplete chunks do not occur when one reader process
receives a full chunk, and another reader process does not have space for a chunk.

The reader thread interacts with the reader process. One reader thread is 
created for each configured reader process. At startup, the reader thread
tries to connect to the reader process. When the connection is established,
the reader thread exchanges some parameters with the reader process. It
receives from the reader process the name of the reader and the
address for connecting the builder process to the reader process. In turn,
the reader thread sends the parameters of additional configuration to the
reader process. At this moment, the reader process is assigned the status 
\texttt{READY}. After that, the reader thread, waiting for commands from
the run control and supervisor threads, passes them to the reader process
and reads its response. Most commands are broadcast, and all reader threads
process them in parallel, each with its own reader process. If an error occurs
when interacting with the reader process, the reader process loses the
\texttt{READY} status, the reader thread closes the connection with the 
reader process and tries to open it again.

The builder thread interacts with the builder process. The builder thread
is created every time a connection is made by the builder process. When the 
builder thread is started, it exchanges some configuration parameters with 
the builder process. Parameters received from the builder process include
the name of the builder process, the network location of the computer that
runs the builder process, and the bandwidth of the connection. The builder
name need not to be unique. The fact is that the data structures inside
the supervisor process describing the builder process exist for some time
after the connection to the builder process is completed. Requiring a unique 
name for the builder process will block reconnection to the supervisor using
the same name when restarting the builder process. The parameters sent to the
builder process include the unique builder ID and the addresses of all
configured reader processes. Thus, to indicate the builder process in text
messages, the builder name and builder ID are used.

After initialization is complete, the builder thread runs in a loop. The cycle
starts with waiting for the \texttt{SV\_CHUNK\_REQUEST} command from the
builder process, which informs that the builder is ready to process the
chunk. When builder is selected for processing, the builder thread sends
the \texttt{BD\_CHUNK\_ASSIGN} command to the builder process with a list of
enabled reader processes, the run ID and the chunk number assigned for 
processing. After that, the builder thread waits for the
\texttt{SV\_TRANS\_FINISHED} command from the builder process, which informs
that the transfer of chunk data from the reader processes is completed.
The loop ends with the command \texttt{SV\_BUILD\_FINISHED} from
the builder process, which reports the completion of chunk processing.

The builder thread terminates in case of any communication errors
with the builder process.

When the ready queue is not empty and there is at least one request to process
a chunk from the builder, the scheduler thread chooses which builder will
process which chunk. The selection takes into account the order in which
requests are received and the network load at the time of making the decision.
To calculate the network load, the scheduler thread uses information about
the network topology, the location and connection speed of the builders,
as well as information about which builders are currently receiving data from 
reader processes. To do this, it is taken into account when the
\texttt{BD\_CHUNK\_ASSIGN} command was issued to the builder processes and
when the \texttt{SV\_TRANS\_FINISHED} response was received from the 
builder processes. The scheduler thread selects the builder that will have 
the highest network connection speed at the time of making the decision. 
Network load accounting can be useful if more than one builder process is
running on the same computer.

There are two types of monitoring threads in the supervisor process: one
provides information about the supervisor itself, and the second
one provides information about each builder process separately. For each connection 
of the monitoring program, its own monitoring thread is created, but the
information sent to different monitoring programs is identical and refers
to the same time points or time intervals. The duration of the time interval
is set in the supervisor process configuration.

The information provided about the supervisor process consists of instant
information at the time of sending and accumulative information for the last 
period of time. Instant information includes the status of internal data
structures, the number and status of configured reader processes, the number 
and status of connected builder processes, the status of the run and
the length of ready queue. The accumulative information consists of the number
of dropped chunks, the number of chunk transfers started, the number of 
completed chunk transfers, and the number of chunks built, in total, 
by all the builder processes.

The information provided about each builder process is accumulative information 
for the last time interval: the number of chunk transfer started, the number of
completed chunk transfers, and the number of chunks built.

The monitoring thread is terminated in case of a communication error with
the monitoring program.

\subsection{Builder process}

The builder process reads data belonging to one chunk from all reader
processes, creates a complete chunk, writes it to the data storage
and makes the corresponding entry in the database. In parallel with writing 
to the data storage, the builder process sends part of the chunk slices to 
the master distributor process.

Builder processes run on builder computers, which should be located
near the data storage. This determines the location of the builder computers 
in the building~14.

The builder process is a multithreaded process, the main functional
threads are: supervisor thread, reader threads and distributor thread.

The supervisor thread interacts with the supervisor process and performs 
the main functions of the builder.

At startup, the supervisor thread connects to the supervisor process and
exchanges configuration parameters with them, receiving its additional
configuration from the supervisor process, and performs initialization.

After initialization is completed, the supervisor thread runs in a loop.
When the builder process is ready to process the chunk, the supervisor thread
sends the \texttt{SV\_CHUNK\_REQUEST} command to the supervisor process.
After that, the supervisor thread waits for a response from the supervisor
process in the form of the command \texttt{BD\_CHUNK\_ASSIGN}. As a parameter
of this command, the builder process receives a list of enabled reader
processes, the run ID and the number of the chunk assigned for processing,
the data write flag and the percentage of slices that will be sent to the
master distributor process.

The supervisor thread creates reader threads, one for each enabled reader
process. The reader threads connect to the corresponding reader processes,
send them the required run ID and chunk number, receive chunk data and buffer
them in the RAM of the builder computer. All reader threads work in parallel. 
After the data transfer is completed, the reader thread ends.

When the data transfer is completed, the supervisor thread informs the 
supervisor process about it using the \texttt{SV\_TRANS\_FINISHED} command
and begins assembling the complete chunk.

When chunk assembling is finished the supervisor thread writes
the chunk data to data storage and then updates the database information.
In parallel, the distributor thread send part of slices from assembled chunk
to the master distributor process.

The loop ends with the \texttt{SV\_BUILD\_FINISHED} command, which is sent
from the supervisor thread to the supervisor process and informs the supervisor
about the completion of chunk processing.

\subsection{Distributor process}

The distributor process collects slices sent by the builder processes and
provides them to a group of processes (clients) for the purpose of data
monitoring. The percentage of slices for monitoring may vary depending on the 
type of the run and actual data flow. The maximum stream can be up to 1~GB/s.

\begin{figure}[htbp]
\centering
\includegraphics[width=\columnwidth]{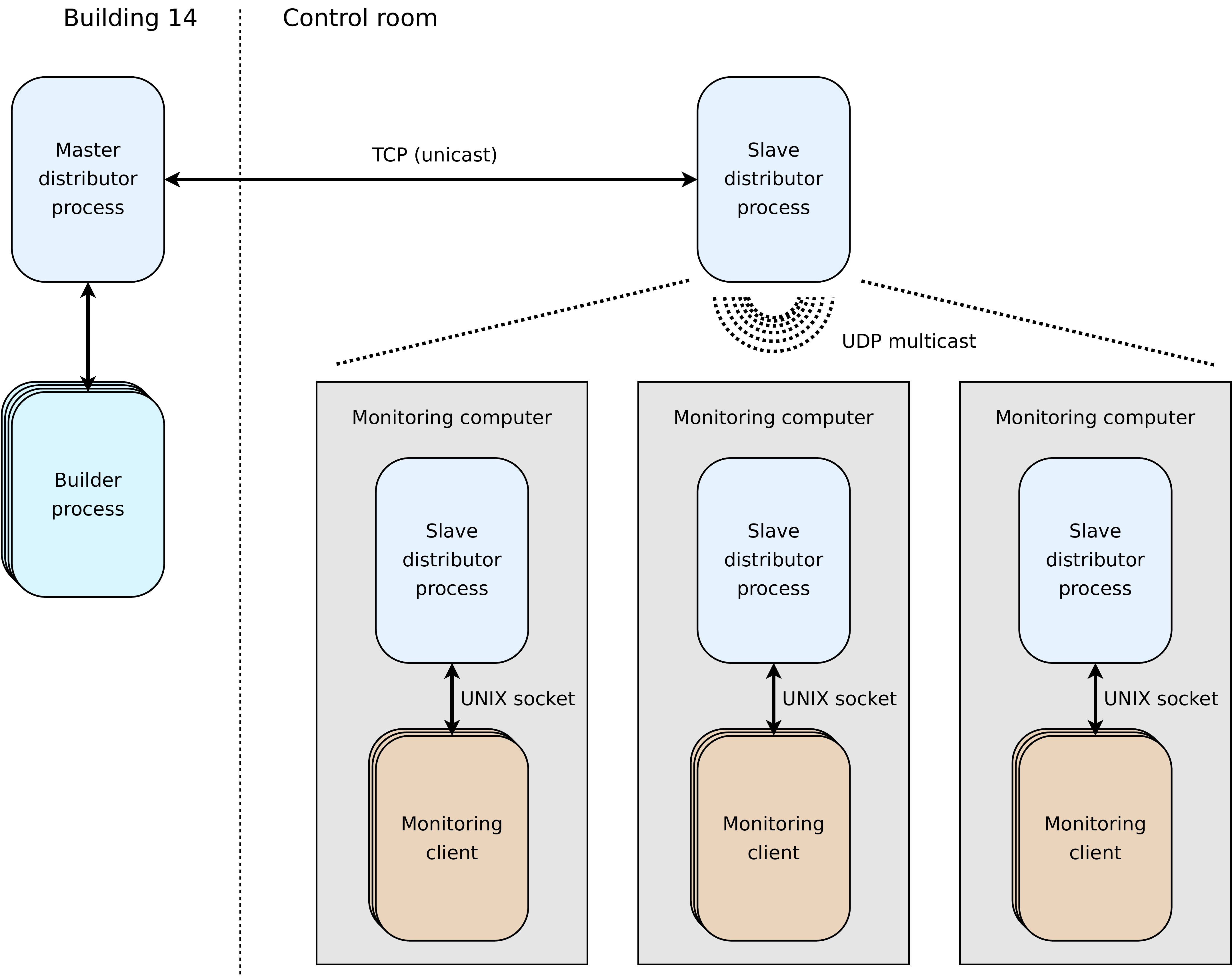}
\caption{Distribution of slices for monitoring purposes.}
\label{fig:daq-distr_chain}
\end{figure}

For the distribution of slices, several distributor processes will be organized in
a tree, as shown in Fig.~\ref{fig:daq-distr_chain}.
The process at the root of the tree is called the master distributor
process, and the others are called slave distributor processes. All
distributor processes are identical, differences in functionality are
determined by the position in the tree and are related to the way the data is
received. The master distributor process will collect data from the builder
processes, and slave distributor will receive it from the previous distributor
process in the chain, just like regular monitoring clients.

The main data structure of the distributor process is a slice buffer. 
The received slices are stored in this buffer and will be kept in it for
some time, sufficient for the client to have some freedom of action. 
Old slices are removed from the buffer. The client can read the slice only
once by requesting (explicitly or implicitly) the ``next'' slice. A 
request for a specific slice is not provided. The distributor always returns
the oldest slice presented in the buffer that has not yet been read by a particular
client. However a client that disconnected from the distributor process and 
reconnected can re-receive slices that have already been received. The
client is responsible to resolve this situation, for which
one can use the run ID, frame number and slice number identifying each slice.

The distributor process is a multithreaded process, the main functional
threads are: builder threads, client threads, broadcast thread, and
distributor thread.

The builder threads exist only in the master distributor process. One 
builder thread is created for each connection from the builder process.
The builder thread receives slices from the builder process and places them 
in the slice buffer. The slices in the buffer must be ordered by run ID,
frame number, and slice number. Since it is impossible to guarantee the
correct order of slices received from different builder processes, slices
will be available to clients with some delay relative to their appearance
in the buffer to be able to sort incoming slices. If a slice
was read from the builder too late to be inserted into the right place in
the buffer, it will be discarded.

The client thread sends slices from the buffer to the client process. One 
client thread is created for each connection from the client process. 
Two address families are supported: UNIX domain socket for local clients and 
INET (INET6) domain socket for remote clients. In both cases, a socket of the
\texttt{SOCK\_STREAM} type is used, which ensures a reliable connection between the
distributor and the client. For both address families, two interaction models 
are supported: one assumes that the client makes a request for each slice,
and the second --- slices are sent in streaming mode, without an explicit
request from the client. In any case, there is no guarantee that
the client will read all the slices received by the distributor process:
the client will read as many slices as it can.

The broadcast thread sends slices from the buffer using a multicast UDP
transport for some multicast group. Any client who has joined this multicast
group can receive this slices. This makes it possible to distribute monitoring
data between a large number of clients with minimal use of network bandwidth.
In this case, INET domain socket of the \texttt{SOCK\_DGRAM} type is used, which
provides datagrams: unreliable messages without connection. It is the client's
responsibility to ensure that all datagrams belonging to the same slice are
received. If one or more datagrams was lost, the entire slice should be
discarded by client. Since the expected slice size of about 200~kB is much 
larger than the datagram size even if a jumbo frame is used (about 9~kB),
and the expected data flow is up to 1~GB/s, it is critically important to
reduce datagram losses. To do this, the distributor and all client computers
must have the same network settings (connection speed and maximum transmission
unit). It is also proposed to place these computers (network interfaces)
into a dedicated network segment.

The distributor thread exists only in the slave distributor process.
It receives slices from the previous distributor in the chain in the same way
as a regular monitoring client. There are two options. The first one is to
connects to the previous distributor using an INET (INET6) domain socket of
the \texttt{SOCK\_STREAM} type. The second one is to join the multicast group of the
previous distributor. In this case, an INET domain socket with the 
\texttt{SOCK\_DGRAM} type is used. The received slices are placed in the slice buffer
and are immediately provided to clients due to they arrived in the correct order.
In case of an error, the distributor thread tries to reconnect to the previous
distributor or re-join the multicast group.

\subsection{Reaction to abnormal situations}

Since the real work in all the processes of the slice building system is 
performed by a child process, then if for any reason the child process is died,
a new child process will be created automatically by the parent process.
Of course, this behavior is not suitable for all cases, but in some cases 
it should make it easier to restart the system. 

The processes of the slice building system can be started in any order, but 
the situation when the server part starts after the client part will lead to 
some delays during the system start.

The supervisor process is crucial for the slice building system. A problem
with the supervisor process will interrupt the current run and cause the
restart of all reader and builder processes also. This will be accompanied by
the loss of data not recorded to the data storage at the problem appearance moment.

A problem with the reader process will not interrupt the current run, but all
chunks from the problem occurrence moment will not contain data from the
failed reader process and, therefore, will be dropped. Practically, this
means that the run will be stopped by the operator. The processing of chunks,
whose data transfer to the builder processes was completed at the
problem moment, can be continued, and the assembled chunks can be written to the
data storage.

A problem with the builder process, depending on the moment of its occurrence, 
can lead to the one chunk data loss. The run will continue.

A problem with the distribution process will lead to a temporary incapacity 
of the monitoring system, but should not affect the data taking.

\subsection{Equipment}

The number of L2 concentrators that can be installed in one readout computer
depends on the required conversion of data coming from the concentrator into 
slices, which can be a resource-intensive procedure. Therefore, we will
proceed from a data stream of 1 GB/s through one readout computer, which
gives the number of readout computers about 20.

The readout computer must have 2 network interfaces with a speed of 10~Gbit/s 
or 1 network interface with a speed of 25~Gbit/s for data transmission.
The buffering depth should be sufficient, taking into account the
architecture of the slice building system and possible emergencies,
such as delays in data transmission and recording. A reasonable buffering
time is 30--60~s, which requires 64--128~GB of RAM for readout computer.

The final specification for the readout computer can be written only after
knowledge of the design of the L2 concentrator.

The builder computers must have a 10~Gbit/s network interface for communication
with the readout computers and a separate connection to the data storage.
The amount of RAM in the builder computer should allow us to store
at least 2 full chunks in memory, which means at least 64~GB of RAM.

The final specification for the builder computer depends on the implementation
of the data storage.

Assuming that there are 20 readout computers, around 40 builder
computers must run simultaneously to ensure operations with data flows 
from each readout computer at 1~GB/s. This approximate calculation assumes 
that the speed of data reading from the readout computers is approximately 
equal to the speed of data writing to the data storage. The time of 
building a chunk is insignificant compared to the time of data reading and writing,
and the operations of reading data, building a chunk, and writing data
are  performed sequentially.

\section{Synchronization and time measurement}
\label{sect:daq-sync_and_time}

Synchronization and time measurement are based on a single clock signal, which is distributed throughout the setup. This clock is called the global clock.

Two types of commands are used to control the readout chain: synchronous with a global clock and asynchronous.

The following synchronous commands are used:
\begin{itemize}
\item \textsl{Set Next Frame} (SNF),
\item \textsl{Start of Frame} (SOF),
\item \textsl{Start of Slice} (SOS).
\end{itemize}
Synchronous commands together with the global clock are sent to the L1 concentrator from a special system called Time Synchronization System (TSS). All synchronous commands are broadcast commands. 

The \textsl{Set Next Frame} command loads the number for the next frame into the front-end electronics.

The \textsl{Start of Frame} command completes the current frame (if there was one) and simultaneously starts a new frame if the new number for the frame has been loaded.  The command \textsl{Start of Frame} marks the loaded number of the next frame as used. Before the next \textsl{Start of Frame} command, a new value for the number of the next frame must be loaded. The command starts the first slice of the frame simultaneously with the start of the frame, and simultaneously with the stop of the frame, the command stops the last slice in the frame.

The \textsl{Start Slice} command completes the current slice and starts a new one inside the frame. The front-end electronics automatically numbers the slices, the slice number is reset to 0 by the \textsl{Start of Frame} command.

The \textsl{Set Next Frame} command transmits the next frame number and therefore may require more than one clock cycle to implement. The technical implementation of synchronous commands has not yet been defined, but this implementation must ensure that the commands are atomic, i.e. if the transmission of a command takes several clock cycles, then the front-end electronics must be able to completely accept the entire command or skip it, regardless of which cycle of the transmission of the command electronics detected this transfer.

The following asynchronous commands are used:
\begin{itemize}
\item \textsl{Disarm};
\item \textsl{Arm};
\item other commands specific for various front-end electronics.
\end{itemize}

Asynchronous commands are sent to the L1 concentrator from the L2 concentrator.

Two standard asynchronous commands \textsl{Disarm} and \textsl{Arm} are used to control the reset signal line that comes from the L1 concentrator to the front-end module. These commands are executed by the L1 concentrator and are not directly visible to the front-end modules. The \textsl{Disarm} command sets the active level of the reset signal, in response to which the front-end module enters the reset process. The \textsl{Arm} command removes the active level of the reset signal. All the time when the reset signal is active, the front-end module must ignore any synchronous commands. Thus, the \textsl{Arm} command enables and \textsl{Disarm} disables synchronous commands for the front-end electronics. The \textsl{Disarm} and \textsl{Arm} commands can be addressed to a single front-end module or to all modules connected to a single L1 concentrator.

In addition to the standard asynchronous commands, various front-end electronics may have different sets of specific asynchronous commands for  their initialization and monitoring. Monitoring commands that do not change the state of the front-end electronics can be used at any time, even in parallel with the synchronous commands. These asynchronous commands are transmitted to the front-end module in the form of I$^2$C commands, and are address commands.

Synchronous commands are generated by the TSS, which in turn is controlled by the corresponding commands. A detailed description of TSS is given in section \ref{sect:daq-tss}. Here we list only two main commands:
\begin{itemize}
\item \textsl{Start of Sequence} --- upon receiving this command, TSS starts generating a sequence of synchronous commands according to the specified parameters;
\item \textsl{Stop of Sequence} --- upon receipt of this command, TSS stops the generation of synchronous commands.
\end{itemize}

\begin{figure}[htb]
\centering
\includegraphics[width=\columnwidth]{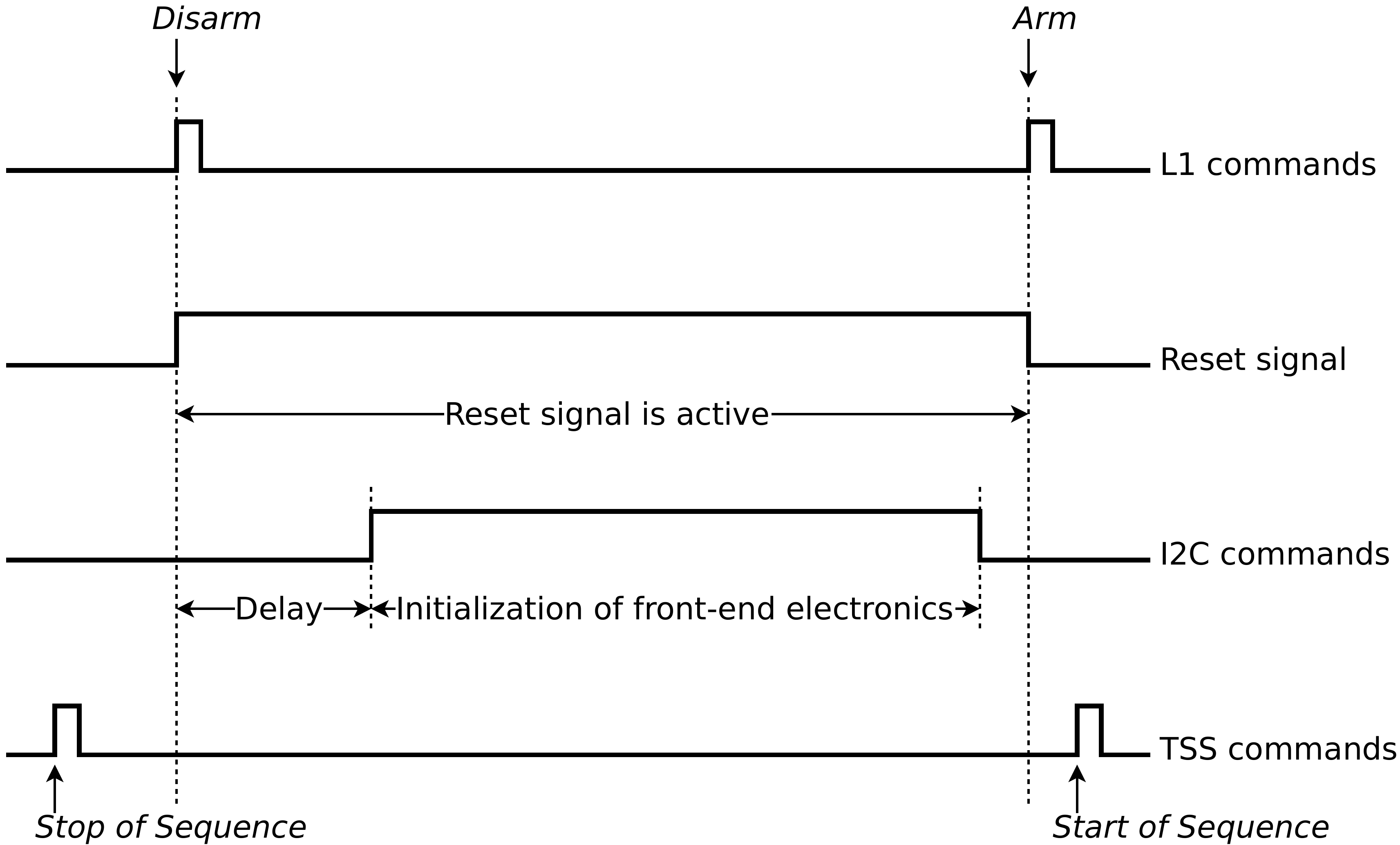}
\caption{The procedure for starting the run.}
\label{fig:daq-run_start}
\end{figure}

The start of the run procedure is shown in Fig.~\ref{fig:daq-run_start} and is as follows:
\begin{itemize}
\item The generation of all synchronous commands is disabled by issuing the \textsl{Stop of Sequence} command to TSS.
\item The \textsl{Disarm} command is issued for all front-end modules, in response to which the modules enter the reset process.
\item After some delay necessary to bring the electronics into a state of readiness to receive incoming commands, the DAQ system begins initializing the front-end electronics using specific I$^2$C commands.
\item The \textsl{Arm} command is issued for all front-end modules. Now the front-end electronics are ready to receive synchronous commands. It should be noted that the \textsl{Arm} command, as well as the \textsl{Disarm} command, is not atomic for the entire installation: it is a sequence of \textsl{Arm} (\textsl{Disarm}) commands addressed to various front-end modules.
\item The generation of  a synchronous command is started by issuing the \textsl{Start of Sequence} command to TSS.
\end{itemize}

\begin{figure}[p]
\centering
\includegraphics[angle=90, height=0.95\textheight]{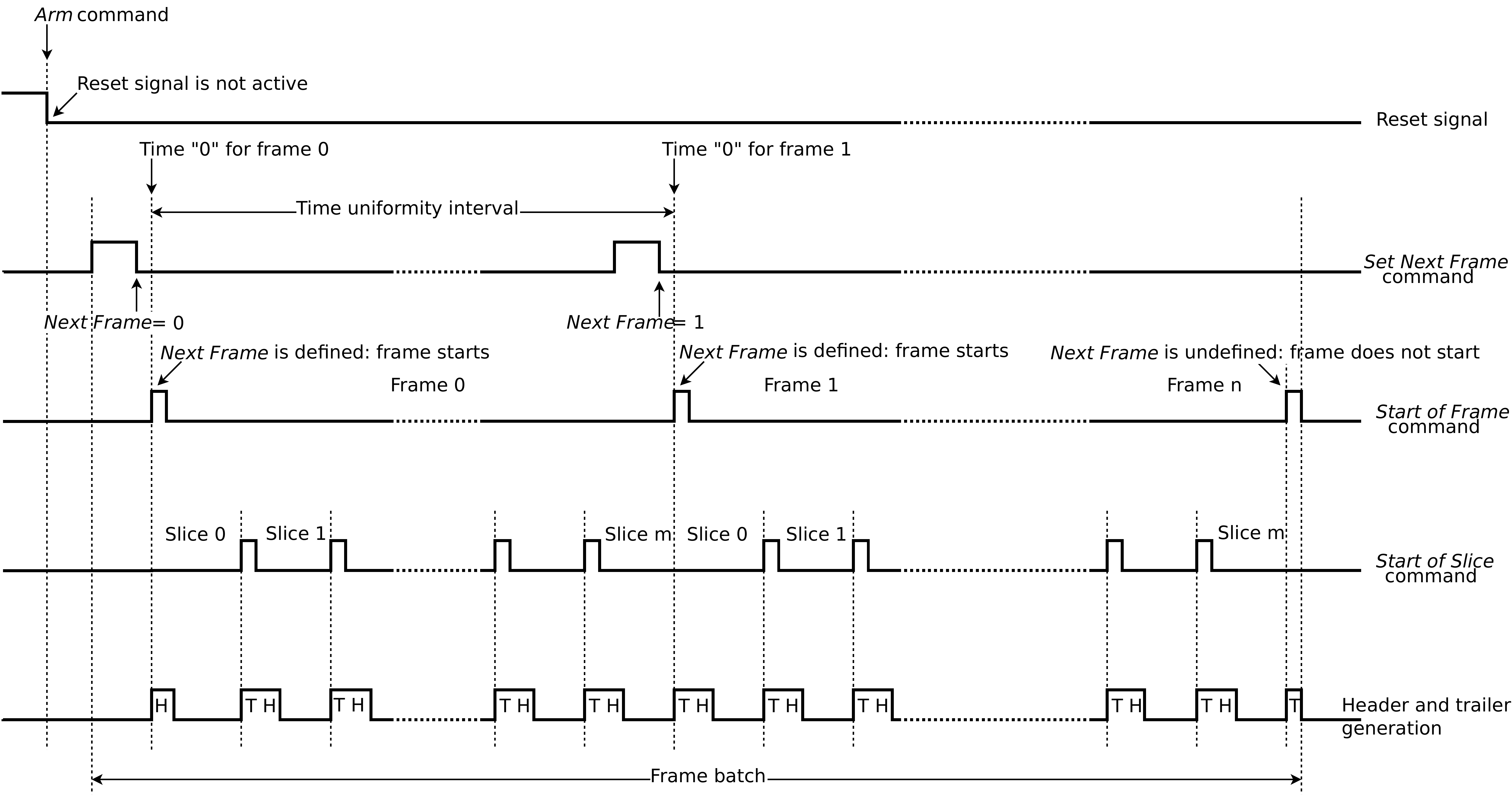}
\caption{Time structure of  a frame batch.}
\label{fig:daq-frame_batch}
\end{figure}

The structural unit of the run is a package of frames, called a frame batch. A frame batch contains a continuous sequence of frames following each other, without time intervals between frames. On the other hand, there are time intervals between frame batches that can be used by the front-end electronics to perform the necessary periodic actions, such as resetting. In the absence of such a need, one frame batch can be stretched for the entire run. In addition, the frame batch is interrupted when the run is put into a suspended state. Frames in the run have continuous numbering, independent of the grouping of frames into batches. There is no direct limit on the number of frame batches per run or on the number of frames per batch. There is only a general limit on the number of frames per run, resulting from the size of the corresponding field in the data format.

The frame batch is shown in Fig.~\ref{fig:daq-frame_batch}. Before the first frame in the batch, the \textsl{Set Next Frame} command is executed, which loads the number for the first frame in the batch into the front-end electronics. All frames in the batch, excluding the last frame, contain the \textsl{Set Next Frame} command closer to their end. There is no \textsl{Set Next Frame} command in the last frame of the batch, and therefore the last \textsl{Start of Frame} command ends the frame without starting a new one.

The frame, in turn, consists of slices. The \textsl{Start of Frame} command simultaneously with the start of a new frame starts the first slice of this frame and simultaneously with the completion of the frame completes the last slice of this frame. Inside the frame, the \textsl{Start of Slice} command has completed one slice and is starting a new one.

\begin{figure}[p]
\centering
\includegraphics[angle=90, height=0.95\textheight]{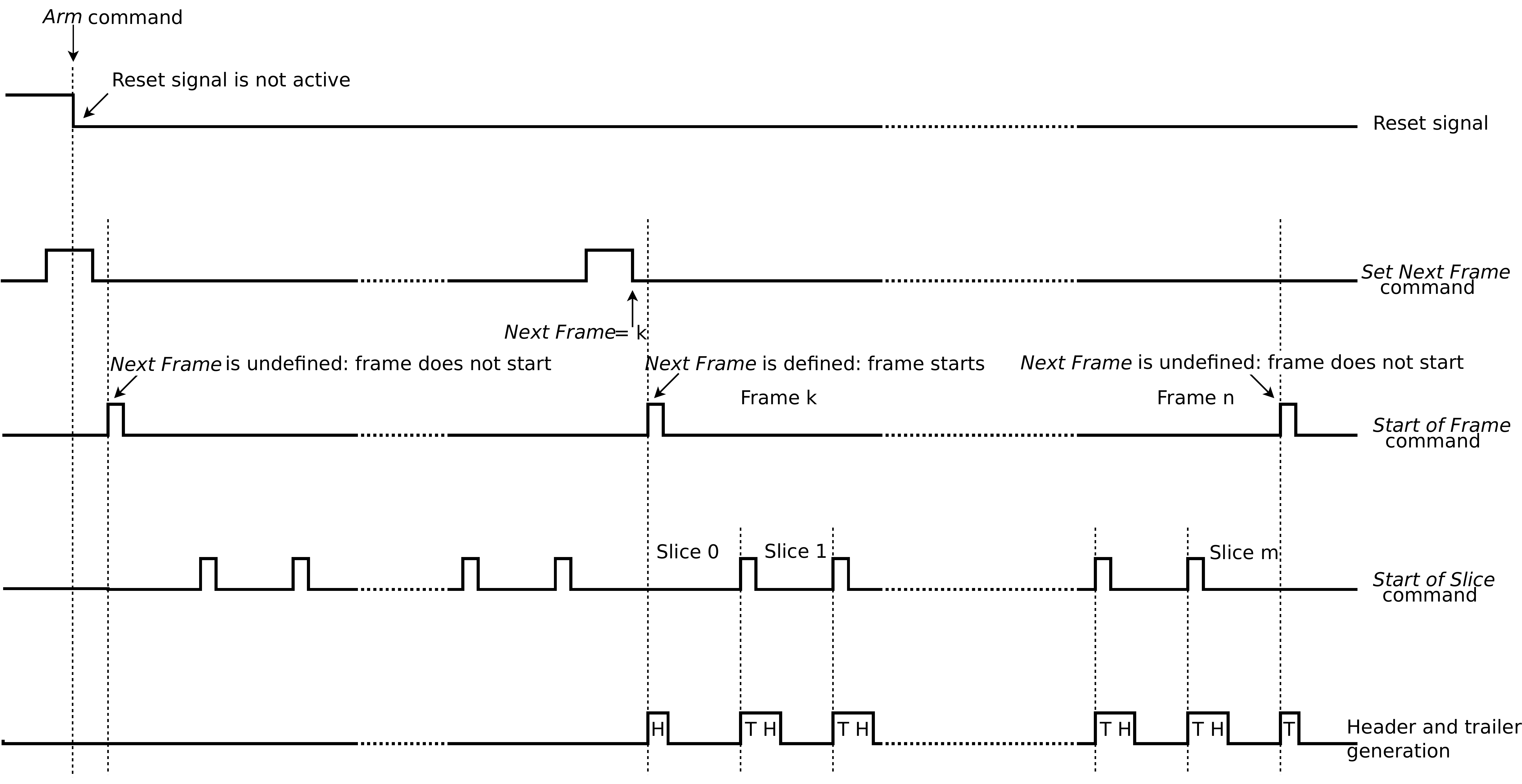}
\caption{Returning the front-end module to operation after reset.}
\label{fig:daq-reset}
\end{figure}

If necessary, the front-end electronics can be reset during the run by commands from the DAQ system in one of two ways:
\begin{itemize}
\item All front-end electronics can be reset simultaneously using a procedure similar to the run start procedure, but with the preservation of frame numbering. This involves stopping and then restarting of the frame batch.
\item Individual modules of the front-end electronics can be reset during the run without stopping the frame batch. To do this, the active level must be set on the reset signal lines of the required modules by the corresponding \textsl{Disarm} command(s). After that, the required modules can be initialized. The modules will return to operation when the active level on the reset signal lines is removed by the corresponding \textsl{Arm} command(s) (see Fig.~\ref{fig:daq-reset}).
\end{itemize}

The time in the DAQ system is measured inside the frame and from the beginning of the frame. No correlation is assumed between the time in two different frames. The time measurement must provide a continuous and uniform time inside one frame, in particular, there must be no breaks in the calculation of time within the frame. Each hit processed by the front-end electronics must have a time stamp that allows the time of the hit from the start of the frame to be calculated.

To measure the time, the front-end electronics can use various clocks,  called internal clocks. From an abstract point of view, time is measured by counting the pulses of the internal clock. It is assumed that the internal pulse counter operates in the so-called rollover mode: when the counter reaches its maximum value, it is reset to 0. The actual implementation of time measurements may be different.

Two different methods of measuring the hit time are used:
\begin{description}
\item[Reset at Start of Frame (RSOF):]
It is based on a synchronous reset performed at the start of the frame, which sets the time to zero. In this case, the number of internal clock pulses (taking into account the possible rollover of the counter) corresponds to the hit time from the beginning of the frame. Since this type of reset is  usually time-bound to the internal clock signal, in order to ensure that the reset is time-bound to the global clock signal, the internal clock must be synchronized with the global clock, and its frequency must be a multiple of the frequency of the global clock.
\item[Measuring of Start of Slice (MSOS):]
This method does not require a common zero time for the entire installation: each module of the front-end electronics can have its own zero time. It is based on the measurement of the moment of the beginning of the slice. In this case, the difference between the numbers of internal clock pulses for a hit and the start of slice (taking into account the possible rollover of the counter) corresponds to the hit time from the beginning of the slice. This method does not require a correlation between the global clock and the internal clock, provided that both clock signals are ``sufficiently'' stable.
\end{description}

A front-end electronics developer can choose one of these methods that is better suited for their electronics.

The reaction of the front-end electronics to the commands depends on which method of measuring time is used.

To describe a hypothetical front-end electronics, assume that this electronics has the following registers:
\begin{itemize}
\item \textit{Frame Number} - contains the number of the current frame inside the run.
\item \textit{Next Frame} - contains the number of the next frame, this register is loaded by the \textsl{Set Next Frame} command. In addition to the register value itself, this register has an initialization state: the initial state of this register is  `uninitialized', the \textsl{Set Next Frame} command changes the state to `initialized', and the \textsl{Start of Frame} command resets the state back to `uninitialized'.
\item \textit{Slice Number} - contains the number of the current slice inside the frame.
\end{itemize}
The contents of the \textit{Frame Number} and \textit{Slice Number} registers are used to assign hit data to a specific frame and slice. In addition, the front-end electronics stores the frame status: the frame status is `active' inside the frame and `inactive' otherwise.

The actions of the front-end electronics in response to receiving commands are described in detail below.

\begin{itemize}

\item[$\bullet$]
The \textsl{Disarm} command sets the active level on the reset signal line. This command is executed by the L1 concentrator and is not directly visible to the front-end electronics. The reset signal must have the highest priority for the front-end electronics. With this signal, the electronics must be reset to some initial state. What exactly is reset in electronics is determined by the electronics developer. But the following conditions must be met:
\begin{enumerate}
\item digitization of new hits must be stopped;
\item data transmission in the direction of the L1 concentrator  must be stopped;
\item the untransmitted data from the internal buffers must be deleted;
\item the frame status must be reset to `inactive';
\item the state of the \textit{Next Frame} register must be reset to  `uninitialized';
\item all synchronous commands must be ignored and remain ignored for as long as the reset signal is active;
\item the electronics must be ready, perhaps with some delay, to receive I$^2$C commands for initialization.
\end{enumerate}

\item[$\bullet$]
The \textsl{Arm} command removes the active level on the reset signal line. This command is executed by the L1 concentrator and is not directly visible to the front-end electronics. After removing the reset signal, the front-end electronics must start executing synchronous commands.

\item[$\bullet$]
In response to the \textsl{Set Next Frame} command, the electronics must load the number received with this command into the \textit{Next Frame} register and change the state of this register to `initialized'.

\item[$\bullet$]
The response to the \textsl{Start of Frame} command depends on two conditions: status of the frame and state of the \textit{Next Frame} register. There are four possible options which may happen:
\begin{itemize}
\item the command starts the first frame in the frame batch or the front-end module goes into operation mode after reset;
\item the command completes one frame and starts another inside the frame batch;
\item the command completes the last frame in the frame batch;
\item the interface module received this command during the transition to operation mode after reset, but did not receive the previous \textsl{Set Next Frame} command, since the reset signal was shot in between these commands.
\end{itemize}

In response to the \textsl{Start of Frame} command, the following actions must be performed:
\begin{description}

  \item[Frame status --- any,
        \textit{Next Frame} register --- `initialized':]
    This~ situation~ occurs~ when a new frame is started, regardless
    of the position of the new frame in the frame batch.
    \begin{enumerate}
    \item
      \begin{description}
      \item[RSOF method:]
        All time-related counters are reset to their initial state. This is 
        called synchronous reset and sets the start of time in the frame.
        The number of the global clock pulses required for synchronous reset must
        be fixed for a specific front-end electronics, but may vary from one 
        electronics to another, which leads to a time zero offset in different
        front-end electronics.
      \item[MSOS method:]
        The time of the global clock pulse corresponding to the SOF signal is
        measured, the data obtained must be attributed to a new slice.
      \end{description}
    \item
      The \textit{Frame Number} register is reloaded from the
      \textit{Next Frame} register.
    \item
      The status of the \textit{Next Frame} register is changed
      to `uninitialized'.
    \item
      The \textit{Slice Number} register is reset to 0.
    \end{enumerate}

  \item[Frame status --- `inactive',
        \textit{Next Frame} register --- `initialized':]
    This~ situation~ occurs~ when starting a new batch of frames or when switching the front-end module to operating mode after reset.
    The actions performed at the start of any new frame and described above
    must be supplemented with the following:
    \begin{enumerate}
    \item The frame status is changed to `active'.
    \item The front-end electronics starts digitizing new hits.
    \item As soon as the data is ready, the front-end electronics starts 
          transmitting data to the L1 concentrator.
    \end{enumerate}

  \item[Frame status --- `active', 
        \textit{Next Frame} register --- `uninitialized':]
    The~ following~ actions~ must be performed for
    stopping the current frame and frame batch:
    \begin{enumerate}
    \item The frame status is changed to `inactive'.
    \item The front-end electronics stops digitizing new hits.
    \item The front-end electronics continues transmitting the already collected  data 
          to the L1 concentrator.
    \end{enumerate}

  \item[Frame status --- `inactive', 
        \textit{Next Frame} register --- `uninitialized':]
    This situation occurs in the process of switching to the operating mode after the reset. The command must be ignored.
\end{description}

\item[$\bullet$]
 The front-end electronics responds to the \textsl{Start of Slice} command only when the frame status is `active', otherwise the command must be ignored. In response of \textsl{Start of Slice} command, the following actions must be performed:
\begin{enumerate}
\item
\begin{description}
\item[RSOF method:]
Nothing.
\item[MSOS method:]
The time of the global clock pulse corresponding to the SOS signal is measured, the data obtained must be attributed to a new slice.
\end{description}
\item The \textit{Slice Number} register is incremented by 1.
\end{enumerate}

\end{itemize}

The result of digitizing the hit time is 3 numbers: the value of the internal  clock counter, the frame and slice numbers. All 3 numbers should (ideally) correspond to a single point in time. The value of the internal clock counter is transmitted in a front-end electronics specific format in the body of the front-end data block.  The frame and slice numbers are used in order to group data by slices and are recorded in the header of the front-end data block.

The calculation of the hit time from the beginning of the frame depends on which time measurement method is used. In the case of the MSOS method, the calculation of the hit time is obvious, and requires the fulfillment of the condition that the rollover period of the internal clock counter in astronomical units is longer than the slice duration.

Consider the case of the RSOF method.

Taking into account the rollover of internal clock counter, the hit time can be calculated as:
\begin{equation}
t = (n + N R) \Delta_{\rm int},
\label{hit_time_eq}
\end{equation}
where $t$ is time from the start of the frame (in physical units), $n$ is recorded value of the internal clock counter, $N$ is a number of internal clock rollovers since the start of the frame, $R$ is the internal clock counter rollover period (in internal clocks), and $\Delta_{\rm int}$ is the internal clock period. 
The unknown value $N$ can be found knowing the slice number $m$ calculated as an integer part from dividing the time $t$ by duration of a slice:
\begin{equation}
m = \lfloor t / (S \Delta_{\rm clk}) \rfloor,
\label{slice_number_eq}
\end{equation}
here $S$ is the slice length (in the global clocks) and $\Delta_{\rm clk}$ is the global clock period.

Equations \ref{hit_time_eq} and \ref{slice_number_eq} have a unique solution only when
\begin{equation}
R \Delta_{\rm int} >= S \Delta_{\rm clk},
\end{equation}
that is, when the rollover period is equal to or greater than the length of the slice. This limits the possible length of the slice depending on the bit depth of the internal clock counter.

All of the above is the ideal case when the electronics outputs 2 matched numbers $n$ and $m$. In the case where $n$ is calculated by some ASIC, and $m$ is calculeted using an FPGA that provides a front-end card interface, it can be difficult to compute these two values in a consistent way. The FPGA, if it calculates the value of $m$ after reading hit data from the ASIC, will do it with some delay, when a new slice can already begin and the value of $m$ will differ from the correct one upwards.

In order to simplify the front-end electronics development, it is possible to require a softer condition: $m$ may differ from the correct value by no more than +1. In this case, one of the following equations must be true:
\begin{equation}
\begin{array}{ll}
m-1 = \lfloor t / (S \Delta_{\rm clk}) \rfloor,\\
m = \lfloor t / (S \Delta_{\rm clk}) \rfloor
\end{array}
\label{slice_number2_eq}
\end{equation}
and for a unique solution it is necessary
\begin{equation}
R \Delta_{\rm int} >= 2S \Delta_{\rm clk}.
\end{equation}

The algorithm for calculation of the hit time is reduced to finding such a value of $N$, in which the time $t$ calculated by formula \ref{hit_time_eq}, substituted into formula \ref{slice_number_eq}, leads to the fact that equation \ref{slice_number_eq} becomes true. The following value can be used as an initial approximation for the number $N$:
\begin{equation}
N = \lfloor (m S \Delta_{\rm clk})/(R \Delta_{\rm int}) \rfloor.
\end{equation}
If equation \ref{slice_number_eq} is not true, then the number $N$ should be increased by 1, if the value of $m$ calculated by formula \ref{slice_number_eq} is less than the required value, and reduced by 1 otherwise. Repeat this step until equation \ref{slice_number_eq} becomes true.

In the case discussed above, when the value of $m$ recorded in the data may differ from the true by 1, it is necessary to check expressions \ref{slice_number2_eq} accordingly: the algorithm stops as soon as one of the equations \ref{slice_number2_eq} becomes true. 
\section{Time Synchronization System}
\label{sect:daq-tss}

The main purposes of the Time Synchronization System (TSS) are:
\begin{itemize}
\item distribution of the global clock signal throughout the installation;
\item generation and distribution of synchronous commands throughout the installation.
\end{itemize}
At the same time, the design of this system should also provide the possibility of obtaining information about the bunch crossing and beam polarization, as well as the possibility of recording this information in the general data stream. The interaction interface with NICA and the method of obtaining such information are currently not defined.

The main part of the TSS is the so-called TSS controller. The TSS controller generates synchronous commands, which are then distributed throughout the installation. The TSS controller is managed by setting the values of its registers and issuing commands to it. In response to these commands, the TSS controller generates a sequence of frame batches separated by a certain time interval. The structural unit of the sequence is the frame batch described in the section~\ref{sect:daq-sync_and_time} and shown in Fig.~\ref{fig:daq-frame_batch}.

The TSS controller has the following read/write registers:
\begin{itemize}
\item \textit{Slice Length} --- the length of the slice, measured in the clock cycles of the global clock signal.
\item \textit{Frame Length} --- the length of the frame, measured in the number of slices.
\item \textit{Batch Length} --- the length of the frame batch, measured in the number of frames. The value 0 means unlimited batch length.
\item \textit{Sequence Length} --- the length of the sequence of batches in the number of batches. The value 0 means unlimited length of sequence.
\item \textit{Batch Interval} --- time interval between batches in the sequence, measured in the clock cycles of the global clock signal.
\item \textit{Next Frame} --- the number used as the next frame number when generating the \textsl{Set Next Frame} command. With each generated command, the contents of the register increases by 1.
\item \textit{Last Frame} --- the number of the last frame. The generation of a sequence of frame batches must be stopped after the completion of the frame with the number specified in this register.
\end{itemize}
The TSS controller implements the following commands:
\begin{itemize}
\item \textsl{Start of Sequence} --- upon receiving this command, the TSS controller must start the generation of a sequence of frame batches using parameter values loaded into its registers.
\item \textsl{Stop of Sequence} --- upon receiving this command, the TSS controller must complete the generation of a sequence of frame batches.
\end{itemize}

At the start of the run, the registers of the TSS controller are initialized. Then, at the \textsl{Start of Sequence} command, the TSS controller starts generation of a sequence of frame batches. The TSS controller completes the generation of a sequence of frame batches when one of the following conditions is met:
\begin{itemize}
\item The specified number of batches in the sequence \textit{Sequence Length} have been reached.
\item The value of the \textit{Next Frame} register has reached the value of the \textit{Last Frame} register.
\item The command to stop the sequence \textsl{Stop of Sequence} was received.
\end{itemize}
In the first two cases, the generation of a sequence of frame batches stops at the frame boundary, while the TSS controller knows about this event in advance. Stopping the generation of a sequence of frame batches with the \textsl{Stop of Sequence} command is performed on the slice boundary as soon as possible. Depending on the implementation of  the command delivery to the L1 concentrators (see below), this may take some time, because in the case of WR-based delivery, commands will be sent to  the L1 concentrators in some portions in advance and the implementation of canceling of commands already sent to  the L1 concentrators leads to unjustified complication of the L1 concentrators. To stop the sequence, the TSS controller must issue the \textsl{Start of Frame} command, provided that no \textsl{Set Next Frame} command has been issued since the previous \textsl{Start of Frame} command. If this is no longer possible, then the TSS controller starts a new frame in the usual way and stops at the end of the first slice of this frame.

Two approaches for obtaining information about bunch-crossings and beam polarizations can take place. 

The first involves the transmission of information from NICA in the form of electrical signals and the measurement of the moment of signal reception to obtain the time of bunch-crossing by the global clock. In this case, embedding the bunch-crossing data into the general SPD data stream seems to be a fairly standard task, similar to the situation with any other subsystem. The difference may lie in the fact that the spatial location of the measurement of the bunches crossing can be significantly remote from the rest of the installation and signal propagation delays can be significant. Using a White Rabbit can solve this problem.

The second approach assumes that the information from NICA comes in digital form and, accordingly, such information must have an absolute timestamp with an accuracy of 1~ns or better in the NICA clock, the global clock must be synchronized with the NICA clock, and each frame must have an absolute timestamp according to the NICA clock. With embedding this data into the overall SPD data stream there is a lot more uncertainty, because at present the interface for transmitting information from NICA is not defined. We assume that, in any case, White Rabbit will be used, even if not to control NICA, but to distribute a common clock to NICA and to SPD, and this signal will be used as the signal of our global clock.

To implement both approaches, it is proposed to use White Rabbit as the basis for the TSS system. This leads to the fact that the TSS controller will combine the functionality of the White Rabbit node and the L1 concentrator. The TSS controller will receive a global clock signal as a WR network node from the NICA master clock or from the SPD master clock, depending on which scheme for obtaining information about the bunches crossing will be implemented. In addition to generating synchronous commands, the TSS controller will timestamp each frame using the master clock and send this data to the L2 concentrator in the same way as any other L1 concentrator.

The TSS controller will be managed via the WR interface.

Currently, two approaches are being considered to implement the delivery of the global clock signal and the synchronous commands to  the L1 concentrators: using the TCS system or White Rabbit.

\subsection{TCS-based delivery}

\begin{figure}[htb]
\centering
\includegraphics[width=\textwidth]{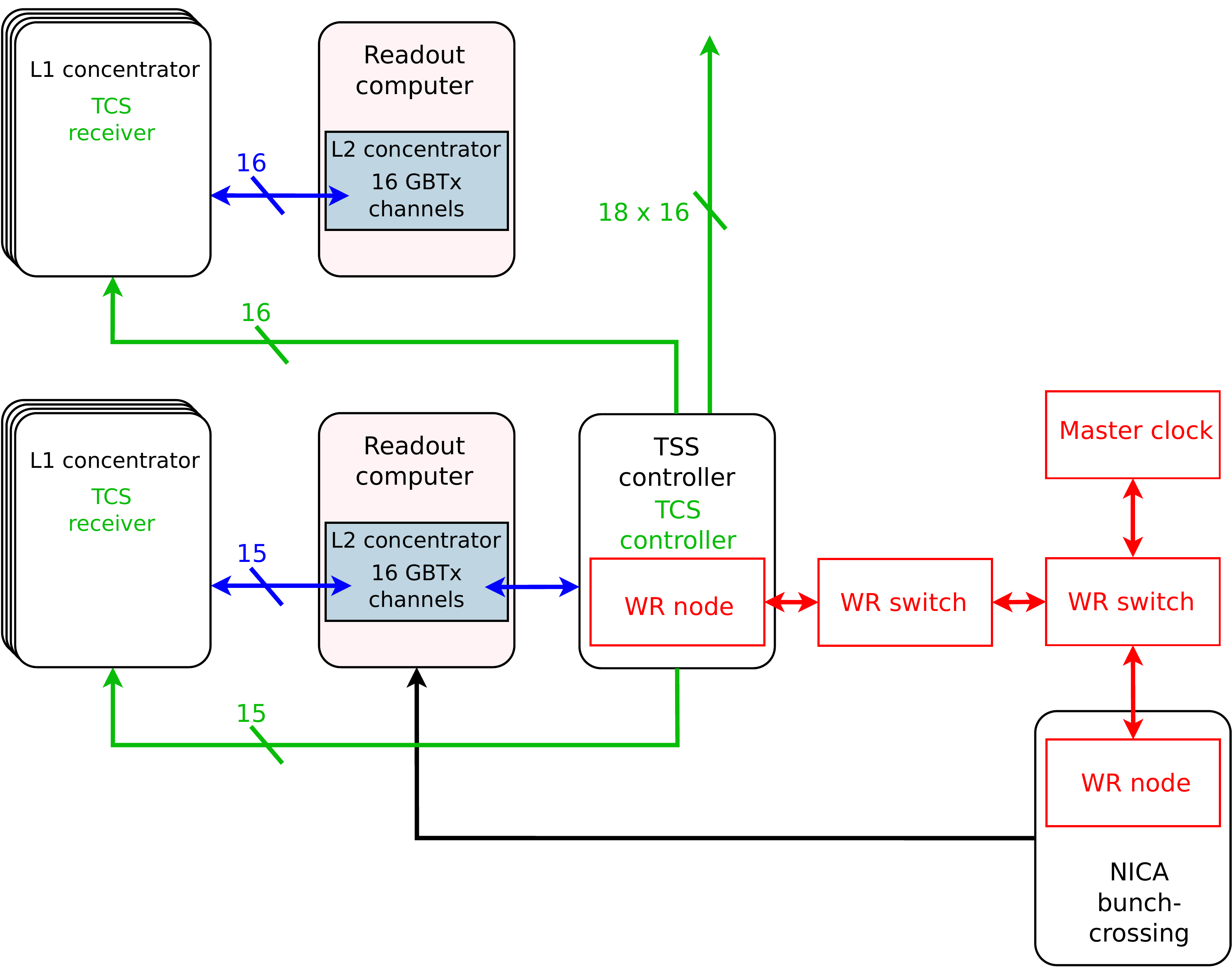}
\caption{Time Synchronization System with TCS-based delivery.}
\label{fig:daq-tcs_delivery}
\end{figure}

 The first approach (Fig.~\ref{fig:daq-tcs_delivery}) uses TCS (Trigger/Timing and Control System) to distribute the global clock signal and  the synchronous commands to the L1 concentrators. The TCS system is based on passive optical splitters. Initially, TCS was developed for the COMPASS experiment, and the modernization of this system is necessary, in particular, the frequency of the system clock should be changed to 125~MHz, which is determined by compatibility with White Rabbit.

In this case, the TSS controller, in addition to the functions described above, must have the functionality of a TCS controller. The TCS controller will generate a signal including a global clock signal and synchronous commands, which will then be delivered via optical lines to the L1 concentrators.

The main disadvantage of this approach is that in this case it is difficult to ensure the same delay of signal propagation from the TSS controller to different L1 concentrators. To do this, it will be necessary to align the lengths of the optical cables between the TSS controller and all L1 concentrators. This, in turn, may require calibration to be performed directly on the installation. Opportunities for this should be provided at the electronics design stage, for example, the possibility of applying an external signal and measuring the time of its arrival at the L1 concentrator or at the front-end electronics. At the same time, there remains the problem of phase shift with temperature.

\subsection{WR-based delivery}

\begin{figure}[b!]
\centering
\includegraphics[width=\textwidth]{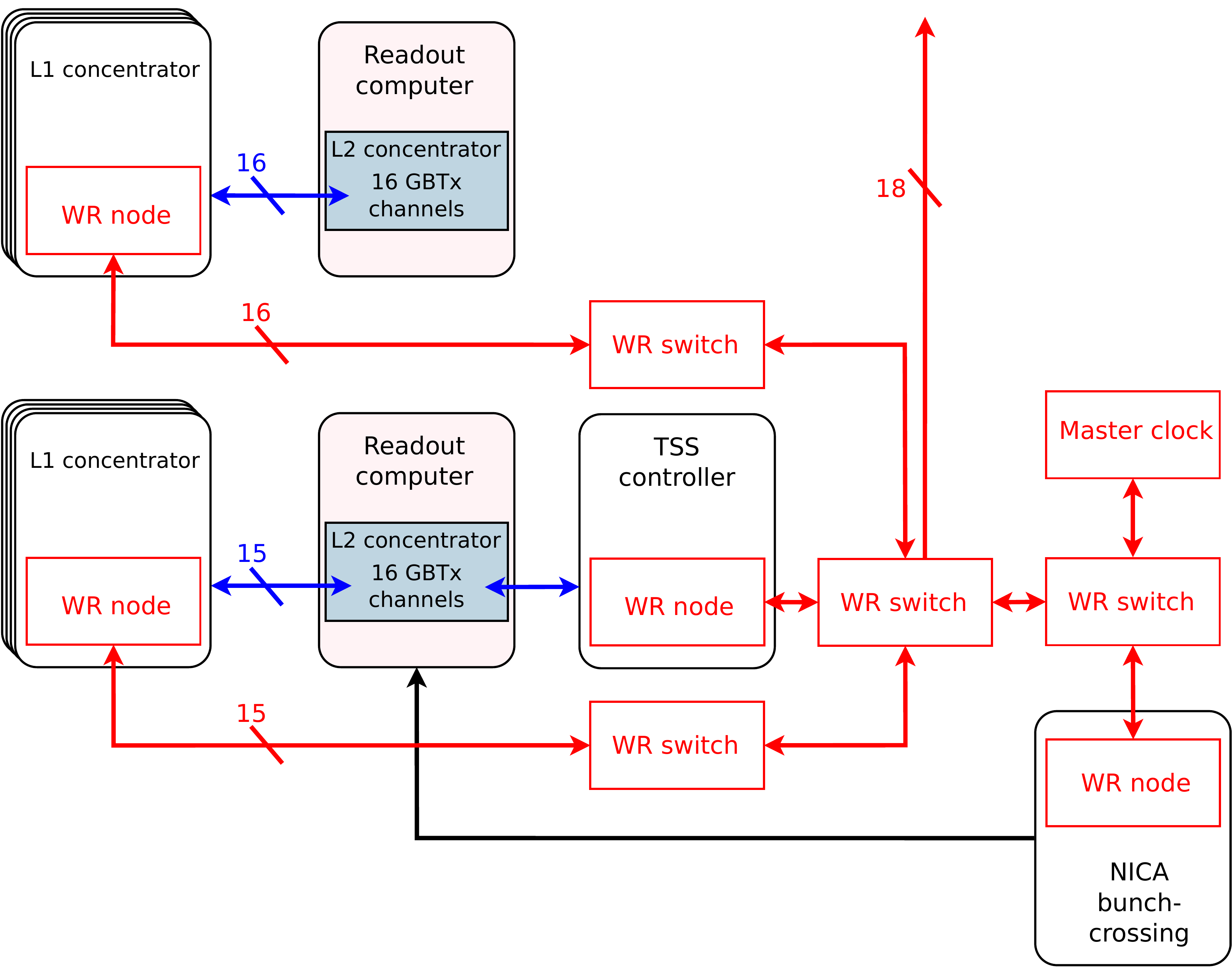}
\caption{Time Synchronization System with WR-based delivery.}
\label{fig:daq-wr_delivery}
\end{figure}

The second approach (Fig.~\ref{fig:daq-wr_delivery}) involves the integration of a White Rabbit node in each L1 concentrator. In this case, the L1 concentrators will receive the global clock signal directly from the master clock as any node of the WR network.  The L1 concentrators will receive synchronous commands from the TSS controller in the form of WR control data.

This approach has some advantages compared to TCS-based delivery:
\begin{itemize}
\item The global clock signal is automatically aligned  to astronomical time in different parts of the installation with an accuracy better than 1~ns, and, as a result, adding a new L1 concentrator is much easier. This improvement is very important, especially if the SPD detector will be created in several stages.
\item Ethernet link of  a WR node can be used to control the L1 concentrator.
\item There is no need for a dedicated powerful optical splitter.
\item WR switch is a commercially available product.
\end{itemize}

The main disadvantages of this approach are:
\begin{itemize}
\item An L1 concentrator with a WR node will be more complex, especially for a radiation-resistant implementation.
\item The development and testing time will be longer.
\item Higher  cost. The  cost of the L1 concentrator will increase slightly due to the addition of the WR node. The main increase will occur due to the need to purchase a significant number of the WR switches ($>100$~k\$).
\end{itemize}

\section{Data format}

The choice of data format is an important point in the development of a data
acquisition system. On the one hand, the data should contain as much information as
possible:
\begin{itemize}
\item  the location of the hits;
\item the hit registration time;
\item the value of the measured signal (for a certain type of detectors);
\item information about measurement and/or data acquisition errors;
\item some other information, which we may not even assume now.
\end{itemize}
On the other hand, even one extra byte per slice in data of a front-end card can
increase an overall data flux approximately by 0.5~GB/s (for 10~\textmu s slice).

Nevertheless, for the convenience of data processing and transmission, the total size
of the data and all its logical units must be aligned to the size of a 32-bits word.
Thus, if the front-end electronics data format assumes 24-bits, 16-bits, or some
other data representation, then the data acquisition system at some level will have
to automatically align them to the 32-bits word boundary. 

An important property of any data acquisition system is its flexibility for
different working conditions. Obviously, the data received during detector
commissioning should contain more information about the operation of these detectors
and electronics. Thus, we must provide the possibility of working of our data acquisition system with different data formats and have the ability to filter
debugging information. 

The unit of information in our data acquisition system is a slice. It must
contain all the necessary information for an unambiguous reconstruction of all
events received during the time of one slice. (The cases when an event  occurres on
the border of two neighboring slices and is therefore divided into two parts will
require an additional discussion and we do not consider them now.)
The data structure of a slice is quite simple, and since the cost of one
additional byte of data in a slice structure is relatively small (no more than 100~kB/s for 10~\textmu s slice), we may  not limit ourselves here.

\begin{figure}[hbt!]
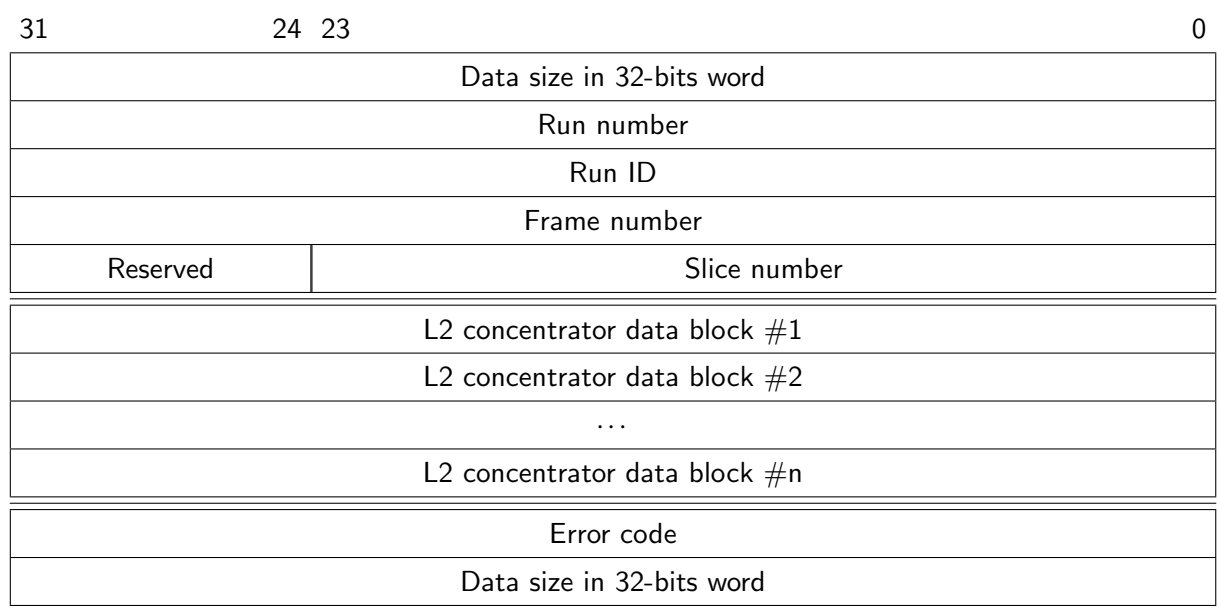

\begin{center}
\sf
\renewcommand{\arraystretch}{1.3}
\centering
\begin{tabular}{*{32}{@{}p{0.031\columnwidth}@{}}}
\multicolumn{2}{@{~}l@{~}}{31}
& & & & & 
\multicolumn{2}{@{~}r@{~}}{24}
&
\multicolumn{2}{@{~}l@{~}}{23}
& & & & & & & & & & & & & & & & & & & & &
\multicolumn{2}{@{~}r@{~}}{0} \\ \hline
\multicolumn{32}{|@{}c@{}|}{Data size in 32-bits word} \\ \hline
\multicolumn{32}{|@{}c@{}|}{Run number} \\ \hline
\multicolumn{32}{|@{}c@{}|}{Run ID} \\ \hline
\multicolumn{32}{|@{}c@{}|}{Frame number} \\ \hline
\multicolumn{8}{|@{}c@{}|}{Reserved} &
\multicolumn{24}{|@{}c@{}|}{Slice number} \\ \hline \hline \hline
\multicolumn{32}{|@{}c@{}|}{L2 concentrator data block \#1} \\ \hline
\multicolumn{32}{|@{}c@{}|}{L2 concentrator data block \#2} \\ \hline
\multicolumn{32}{|@{}c@{}|}{$\cdots$} \\ \hline
\multicolumn{32}{|@{}c@{}|}{L2 concentrator data block \#n} \\ \hline \hline \hline
\multicolumn{32}{|@{}c@{}|}{Error code} \\ \hline
\multicolumn{32}{|@{}c@{}|}{Data size in 32-bits word} \\ \hline
& & & & & & & & & & & & & & & & & & & & & & & & & & & & & & & \\
\end{tabular}
\end{center}
	\caption{Slice data block.}
	\label{fig:daq-slice1}
\end{figure}

The slice structure is presented in Fig.~\ref{fig:daq-slice1} and consists of:
\begin{itemize}
  \item 5 words of  the slice header:
  \begin{itemize}
    \item size of the slice in 32-bits words including sizes of the header and trailer;
    \item run number;
    \item run ID;
    \item frame number;
    \item 8-bits  reserved field, 24-bits slice number;
  \end{itemize}
  \item a slice body containing blocks of data received by each readout computer and written sequentially;
  \item 2 words of  the slice trailer:
  \begin{itemize}
    \item error code;
    \item size of the slice in 32-bits words including sizes of the header and trailer.
  \end{itemize}
\end{itemize}

Obviously, the data sizes in the header and the trailer of the data block must match, and that will be used to assess quickly the integrity of the data during processing. The error code will be determined by the L1 and L2 concentrators and will be used by  the data filter. 

The total size of the service information per slice will be 24 bytes (4 header words + 2 trailer words), which increases total data flux of about  2.4~MB/s for 10~\textmu{}s slice. 

\begin{figure}[hbt!]
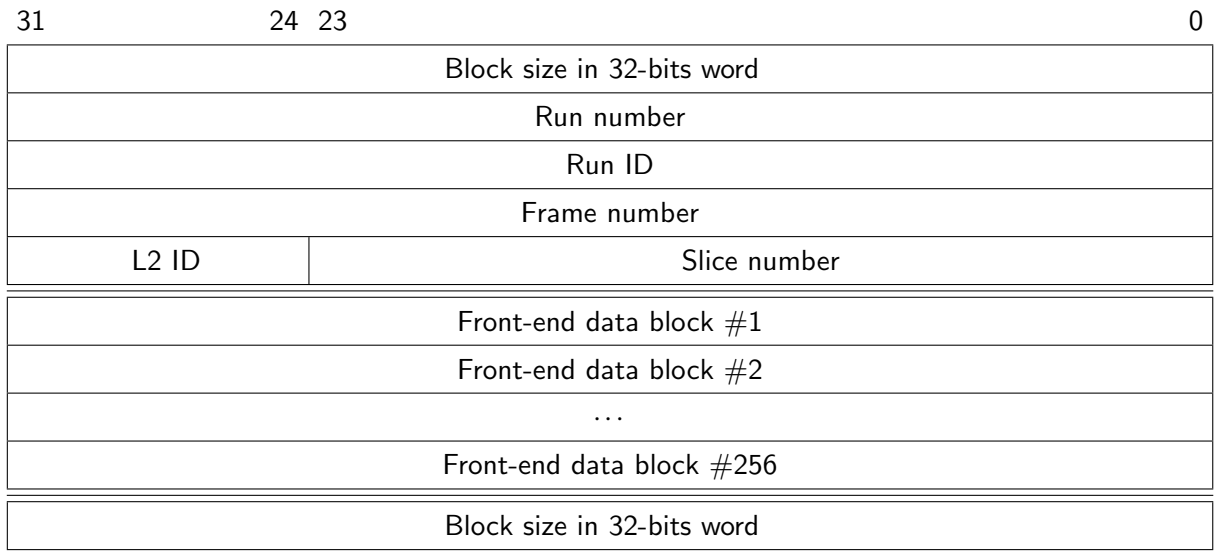

\begin{center}
\sf
\renewcommand{\arraystretch}{1.3}
\centering
\begin{tabular}{*{32}{@{}p{0.031\columnwidth}@{}}}
\multicolumn{2}{@{~}l@{~}}{31}
& & & & & 
\multicolumn{2}{@{~}r@{~}}{24}
&
\multicolumn{2}{@{~}l@{~}}{23}
& & & & & & & & & & & & & & & & & & & & &
\multicolumn{2}{@{~}r@{~}}{0} \\ \hline
\multicolumn{32}{|@{}c|@{}}{Block size in 32-bits word} \\ \hline
\multicolumn{32}{|@{}c@{}|}{Run number} \\ \hline
\multicolumn{32}{|@{}c@{}|}{Run ID} \\ \hline
\multicolumn{32}{|@{}c@{}|}{Frame number} \\ \hline
\multicolumn{8}{@{}|c@{}}{L2 ID} &
\multicolumn{24}{|@{}c@{}|}{Slice number} \\ \hline \hline \hline
\multicolumn{32}{|@{}c@{}|}{Front-end data block \#1} \\ \hline
\multicolumn{32}{|@{}c@{}|}{Front-end data block \#2} \\ \hline
\multicolumn{32}{|@{}c@{}|}{$\cdots$} \\ \hline
\multicolumn{32}{|@{}c@{}|}{Front-end data block \#256} \\ \hline \hline \hline
\multicolumn{32}{|@{}c|@{}}{Block size in 32-bits word} \\ \hline
& & & & & & & & & & & & & & & & & & & & & & & & & & & & & & & \\
\end{tabular}
\end{center}
	\caption{L2 concentrator data block.}
	\label{fig:daq-subslice1}
\end{figure}

The data blocks from  the L2 concentrators are collected by  the readout computers. Their
structure is presented in Fig.~\ref{fig:daq-subslice1} and consists of:
\begin{itemize}
  \item 5 words of block header:
  \begin{itemize}
    \item size of the data block in 32-bits words including sizes of the header and trailer;
    \item run number;
    \item run ID;
    \item frame number;
    \item 8-bits L2 concentrator ID, 24-bits slice number;
  \end{itemize}
  \item data block body, which contains a sequential record of data from all front-end cards connected to the concentrator;
  \item size of the data block in 32-bits words including sizes of the header and trailer.
\end{itemize}

Thus, with a slice length of 10~\textmu s, the overhead of each readout computer (4
words of the header plus 1 trailer word) will add approximately 2~MB/s to the total
data flux.

\begin{figure}[hbt!]
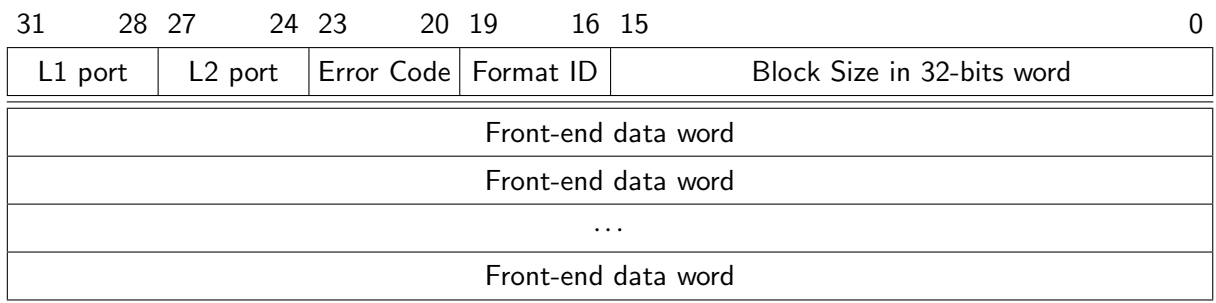

\begin{center}
\sf
\renewcommand{\arraystretch}{1.3}
\centering
\begin{tabular}{*{32}{@{}p{0.031\columnwidth}@{}}}
\multicolumn{2}{@{~}l@{~}}{31}
& 
\multicolumn{2}{@{~}r@{~}}{28}
&
\multicolumn{2}{@{~}l@{~}}{27}
&
\multicolumn{2}{@{~}r@{~}}{24}
&
\multicolumn{2}{@{~}l@{~}}{23}
& 
\multicolumn{2}{@{~}r@{~}}{20}
&
\multicolumn{2}{@{~}l@{~}}{19}
& 
\multicolumn{2}{@{~}r@{~}}{16}
&
\multicolumn{2}{@{~}l@{~}}{15}
& & & & & & & & & & & & & 
\multicolumn{2}{@{~}r@{~}}{0} \\ \hline
\multicolumn{4}{|@{}c@{}}{L1 port} & \multicolumn{4}{|c}{L2 port} &
\multicolumn{4}{|@{}c@{}}{Error Code} &
\multicolumn{4}{|@{}c@{}}{Format ID} &
\multicolumn{16}{|@{}c@{}|}{Block Size in 32-bits word} \\ \hline \hline \hline
\multicolumn{32}{|@{}c@{}|}{Front-end data word} \\ \hline
\multicolumn{32}{|@{}c@{}|}{Front-end data word} \\ \hline
\multicolumn{32}{|@{}c@{}|}{$\cdots$} \\ \hline
\multicolumn{32}{|@{}c@{}|}{Front-end data word} \\ \hline
& & & & & & & & & & & & & & & & & & & & & & & & & & & & & & & \\
\end{tabular}
\end{center}
	\caption{Front-end data block.}
	\label{fig:daq-fedata}
\end{figure}

The body of the L2 concentrator data block contains sequentially recorded information from each of the 256 front-end cards connected to the L2 concentrator. The format of the front-end data block is shown in Fig.~\ref{fig:daq-fedata}. The data coming from the front-end card will be converted to this format on the readout computer as described below.

The format of data transmission from the front-end card to the L2 concentrator depends on the electronics, an example of such format for RS is presented in Fig.~\ref{fig:daq-rsdata}.

Front-end data will be sent to the L2 concentrator sequentially, as soon as information arrives from the detector, and it is assumed that each slice will begin with a header containing the number of the current slice and frame (or just their LSB) and data format ID. And at the end of the slice, the front-end card should send a trailer containing the size of the data transferred for the slice and the error code. 

To split the data stream into slices, it is necessary to be able to quickly find the header and the trailer of the slice in the data stream. For this purpose, it is assumed that at least  two of the most significant bits of the front-end data words must contain a flag that uniquely distinguishes the header and the trailer from the data itself. This flag can later be removed from the data stream on the readout computer to reduce the overall data flow.

As mentioned above, each additional overhead byte in the front-end data block  makes a significant contribution to the overall data flow. So, we have to delete all the information that is repeated in the header of the data block of the L2 concentrator, leaving only the geographical address, error code, and front-end data format identifier in the header of the front-end data block (see Fig.~\ref{fig:daq-fedata}).

\begin{figure}[hbt!]
\begin{center}
\sf
\renewcommand{\arraystretch}{1.3}
\centering
\begin{tabular}{*{32}{@{}p{0.031\columnwidth}@{}}}
\multicolumn{2}{@{~}l@{~}}{31}
& 
\multicolumn{2}{@{~}r@{~}}{28}
&
\multicolumn{2}{@{~}l@{~}}{27}
&
\multicolumn{2}{@{~}r@{~}}{24}
&
\multicolumn{2}{@{~}l@{~}}{23}
& 
\multicolumn{2}{@{~}r@{~}}{20}
&
\multicolumn{2}{@{~}l@{~}}{19}
& 
\multicolumn{2}{@{~}r@{~}}{16}
&
\multicolumn{2}{@{~}l@{~}}{15}
& & & & & & & & & & & & & 
\multicolumn{2}{@{~}r@{~}}{0} \\ \hline
\multicolumn{1}{|@{}c@{}}{1} &
\multicolumn{1}{|@{}c@{}}{0} &
\multicolumn{1}{|@{}c@{}}{X} &
\multicolumn{1}{|@{}c@{}}{X} &
\multicolumn{4}{|@{}c@{}}{Format ID} &
\multicolumn{8}{|@{}c@{}}{LSB of Frame Number} &
\multicolumn{16}{|@{}c@{}|}{LSB of Slice Number} \\ \hline \hline \hline
\multicolumn{1}{|@{}c@{}}{0} &
\multicolumn{1}{|@{}c@{}}{0} &
\multicolumn{1}{|@{}c@{}}{X} &
\multicolumn{1}{|@{}c@{}}{X} &
\multicolumn{8}{|@{}c@{}}{Channel Number} &
\multicolumn{20}{|@{}c@{}|}{Hit Time} \\ \hline
\multicolumn{1}{|@{}c@{}}{0} &
\multicolumn{1}{|@{}c@{}}{0} &
\multicolumn{1}{|@{}c@{}}{X} &
\multicolumn{1}{|@{}c@{}}{X} &
\multicolumn{8}{|@{}c@{}}{Channel Number} &
\multicolumn{20}{|@{}c@{}|}{Hit Time} \\ \hline
\multicolumn{1}{|@{}c@{}}{0} &
\multicolumn{1}{|@{}c@{}}{0} &
\multicolumn{1}{|@{}c@{}}{X} &
\multicolumn{1}{|@{}c@{}}{X} &
\multicolumn{8}{|@{}c@{}}{Channel Number} &
\multicolumn{20}{|@{}c@{}|}{Hit Time} \\ \hline
\multicolumn{32}{|@{}c@{}|}{$\cdots$} \\ \hline
\multicolumn{1}{|@{}c@{}}{0} &
\multicolumn{1}{|@{}c@{}}{0} &
\multicolumn{1}{|@{}c@{}}{X} &
\multicolumn{1}{|@{}c@{}}{X} &
\multicolumn{8}{|@{}c@{}}{Channel Number} &
\multicolumn{20}{|@{}c@{}|}{Hit Time} \\ \hline \hline \hline
\multicolumn{1}{|@{}c@{}}{1} &
\multicolumn{1}{|@{}c@{}}{1} &
\multicolumn{1}{|@{}c@{}}{X} &
\multicolumn{1}{|@{}c@{}}{X} &
\multicolumn{8}{|@{}c@{}}{Error Code} &
\multicolumn{20}{|@{}c@{}|}{Total Number of Hits (Data words)} \\ \hline
& & & & & & & & & & & & & & & & & & & & & & & & & & & & & & & \\
\end{tabular}
\end{center}
	\caption{ Proposal of the data structure of RS.}
	\label{fig:daq-rsdata}
\end{figure}

\section{Cost estimate}
\label{sect:daq-cost}

 Based on the number of the front-end electronics to be read out at different stages of the experiment (see Tables~\ref{tab:daq-ch1} and \ref{tab:daq-ch2}), the corresponding cost estimate of the SPD DAQ system is shown in the Table~\ref{tab:daq-cost}.

\begin{table}[htbp]
\caption{DAQ cost estimate.}
\label{tab:daq-cost}
\begin{center}
\begin{tabular}{|c|c|c|c|c|c|}
    \hline
                     Item                  & Quantity & Quantity & Price per  & Price, k\$ & Price, k\$ \\
                                           &  stage 1 & stage 2 & piece, k\$ & stage 1    & stage 2 \\ \hline\hline
      L1 concentrator (vertex MAPS)     &   150    & 590 (800) &     2      &    300     &   1180 (1600) \\ \hline
      L2 concentrator (vertex MAPS)       &    25    & 100 (130) &     12     &    300     &    1200 (1560) \\ \hline
                     TSS                   &    1     & 1     &     --     &     50     &     50     \\ \hline
        Supervisor computer $+$ spare      &    2     & 2     &     4      &     8      &     8 \\ \hline
         File server (master, slave)       &    2     & 2     &     10     &     20     &     20     \\ \hline
    Database server (master, slave, proxy) &    3     & 3     &     7      &     21     &     21     \\ \hline
    Readout computer (vertex MAPS)  &    12    & 22(28) &     4      &     48     &     88 (112)    \\ \hline
            Builder network switch         &    1     & 1     &     50     &     50     &     50     \\ \hline
        Builder computer  (vertex MAPS) &    25    & 40(45)    &   5    &    125     &    200 (225) \\ \hline
            Network infrastructure         &    --    & --    &     7      &     7      &     7      \\ \hline
            General infrastructure         &    --    & --    &     25     &     25     &     25     \\ \hline
                  UPS 50~kVA               &    1     & 1     &     30     &     30     &     30     \\ \hline
             Cooling system 50~kW           &    --    & --    &     20     &     20     &     20     \\ \hline
                 Spare parts               &    --    & --    &     15     &     15     &     15     \\ \hline
                     R\&D                  &    -- &    --    &     --     &    200     &    200 \\ \hline\hline
                  Total cost (vertex DSSD)     & &          &            &    1219     &    3114 \\ \hline
                  Total cost (vertex MAPS)      & &          &            &         &    3943    \\ \hline
\end{tabular}
\end{center}
\end{table}

\chapter{Computing and offline software}

\section{Introduction}
The expected event rate of the SPD experiment is about 3 MHz (pp collisions at $\sqrt{s}$ = 27 GeV and 10$^{32}$ cm$^{-2}$s$^{-1}$ design luminosity). This is equivalent to a raw data rate of 20 GB/s or 200 PB/year, assuming a detector duty cycle of 0.3, while the signal-to-background ratio is expected to be in the order of 10$^{-5}$. Taking into account the bunch-crossing rate of 12.5 MHz, one may conclude that pile-up probability cannot be neglected.

\begin{figure}[htbp]
  \begin{center}
    \includegraphics[width=0.8\textwidth]{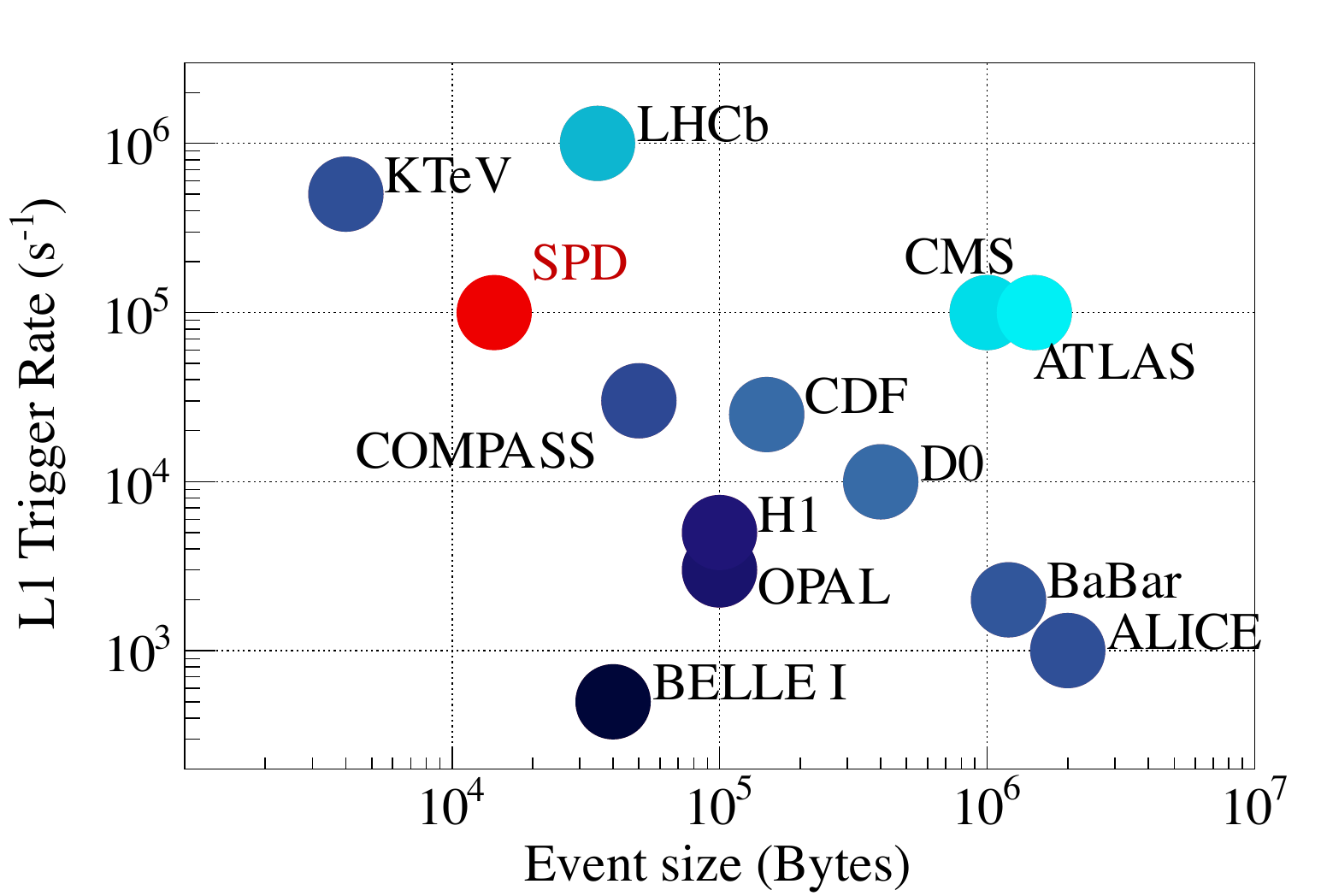}
  \end{center}
  \caption{Expected event size and event rate of the SPD setup after the online filter, compared with some other experiments \cite{Abbon:2014aex}.}
  \label{fig:rates}
\end{figure}

The key challenge of the SPD computing is the fact, that no simple selection of physics events is possible at the hardware level, because the trigger decision would depend on the measurement of momentum and vertex position, which requires tracking. Moreover, the free-running DAQ provides a continuous data stream, which requires  sophisticated unscrambling prior to building individual events. That is the reason why any reliable hardware-based trigger system turns out to be over-complicated, and the computing system will have to cope with the full amount of data supplied by the DAQ system. This makes a medium-scale setup of the SPD a large-scale data factory (Fig.~\ref{fig:rates}).

Continuous data reduction is the key point in SPD computing. While simple operations like zero suppression can still be done by the DAQ, it is an online filter that is aimed at a fast partial reconstruction of events and data selection, thus being a kind of software trigger. The goal of the online filter is at least to decrease the data rate by a factor of 20, so that the annual growth of data, including the simulated samples, stays within 10 PB. Then, data are transferred to the Tier-1 facility, where a full reconstruction takes place and the data is stored permanently. The data analysis and Monte-Carlo simulation will likely run at the remote computing centers (Tier-2s). Given the large data volume, a thorough optimization of the event model and performance of the reconstruction and simulation algorithms are necessary.

\section{SPD computing model}

\subsection{Input parameters}

The assumptions in Table~\ref{input_par} are used to calculate the storage and computing resources. At present, all processing times are higher than assumed here.

\begin{table}[ht]
  \caption{ The assumed event data sizes for various formats, the corresponding processing times, and related
    operational parameters.}
  \label{input_par}
  \begin{center}
    \begin{tabular}{lcc}
      \hline
      Item & Unit & Value \\
      \hline
      RAW event & kB & 7\\
      RECO event & kB & 15 \\
      Time for Reconstruction (1 ev) & HepSPEC & 100 \\
      Time for Simulation (1 ev) & HepSPEC & 500 \\
      Event rate at maximum luminosity & kHz & 3000 \\
      Event rate after online data filter & kHz & 150 \\
      Operation time & seconds/day & 50000 \\
      Operation time & days/year & 200 \\
      \hline
    \end{tabular}
  \end{center}
\end{table}

\subsection{Data flow and event data model}
Data processing is supposed to be done in several stages. In the first stage, data undergoes fast reconstruction and events are built from a continuous byte stream. This task will be performed online using a dedicated high-performance computing cluster (online data filter). After the fast reconstruction, the events are selected according to a set of physical criteria to suppress the contribution from background processes. Then, selected data are saved for a long-term storage and subsequent full reconstruction. Full reconstruction differs from fast reconstruction by more accurate and complex algorithms that use information from the slow control system, calibration constants, and combined analysis of information from various subsystems to identify particle types.

Two reconstruction cycles are foreseen. The first cycle includes the reconstruction of some fraction of each run, which is necessary to study the detector performance and derive calibration constants, followed by the second cycle of reconstruction of the full data sample for physics analysis. Detector simulation will run in parallel. The amount of simulated data is expected to be comparable to the amount of selected experimental data. The large amount of data, the computing resources required for its processing, and analysis within the framework of an international collaboration naturally require the development and use of a distributed data storage and processing system.

\subsection{Event building and filtering}
 
The main goal of the SPD online filter is to reconstruct events from the continuous byte stream and to suppress the background by a factor of 20 or so. The input for the online filter is the raw data files. The format of these files is determined by the front-end electronics and the DAQ system. The result of the data processing at the online filter is a set of reconstructed events, containing also raw information for the more detailed offline reconstruction. HDF5 will be used as the output format for the online filter to simplify data treatment by the offline computing system. Because of the data volume reduction, it is likely that merging several small output files to a bigger one will be necessary.

\subsection{Offline data processing}

Offline data processing includes offline reconstruction and MC simulation. Offline reconstruction starts with the output of the online filter. Detailed calibration and more precise algorithms will be used to refine the data sample and to improve the efficiency and precision of the reconstruction of physics objects. Particle identification will be made at this stage as well. The output of the offline reconstruction will be stored in ROOT trees in RECO format to allow direct use in further physics analysis.

Simulation is necessary for both the data analysis and the training of neural networks at the online filter. The latter should reproduce not only the event topology and kinematics but also the time structure similar to the one in the real data. The output of the simulation will be stored in the same format as  the reconstruction of the real data, with the addition of the MC truth information.

\subsection{User analysis}

User analysis will be done using pre-selected RECO data or derived ROOT trees.


\section{Online data filter}
\subsection{Introduction and requirements}
The SPD online filter facility will be a high-throughput computing system, which shall include:
\begin{itemize}
    \item the special heterogeneous high-performance computing cluster;
    \item the middleware complex for providing high-throughput multi-stage data processing;
    \item the software framework, which will provide the necessary abstraction, so that the common code can deliver selected functionality on different platforms.
\end{itemize}

The main goal of the online filter is a fast reconstruction of the SPD events and suppression of the background events by at least a factor of 20. This requires fast tracking and fast clustering in the electromagnetic calorimeter, followed by a reconstruction of a sequence of events from time slices and an event selection (software trigger). Several consecutive time slices shall be considered, tracker data unpacked and given for a fast tracking. The result of the fast track reconstruction is the number of tracks, an estimate of their momentum, and an estimate of the primary vertex (to distinguish between tracks belonging to different collisions). Using this outcome, the online filter should combine information from the time slices into events and add a trigger mark. The events will be separated into several data streams using the trigger tag, and an individual prescale factor for each stream will be applied.

Besides the high-level event filtering and the corresponding data reduction, the online filter will provide input for the online monitoring by the shift team and the data quality assessment, as well as local polarimetry.

\subsection{Computing system}

\begin{figure}[htbp]
  \begin{center}
    \includegraphics[width=0.8\textwidth]{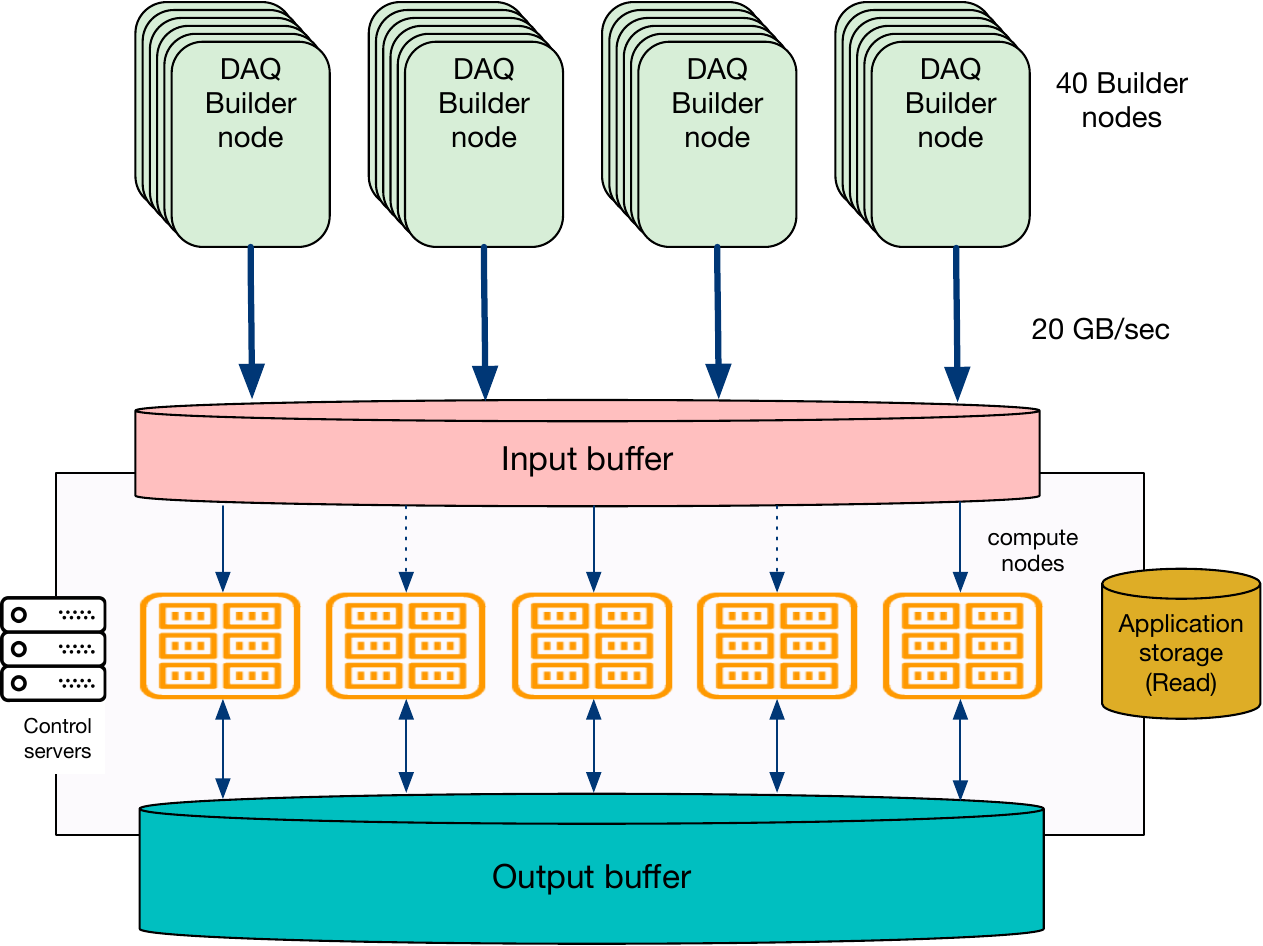}
  \end{center}
  \caption{Design of the computing cluster for SPD Online filter.}
  \label{fig:online_farm}
\end{figure}

The heterogeneous computing cluster is a hardware component of the SPD online filter (Fig.~\ref{fig:online_farm}), which includes the following components:

\begin{itemize}
    \item a high-speed storage system as an input buffer for the data received from the detector data acquisition system (DAQ);
    \item a large-volume storage system as an intermediate buffer for data produced at various stages of processing, and for fully processed data prepared for uploading to the long–term storage and further processing in the distributed offline system;
    \item a set of general-purpose multicore compute nodes;
    \item a set of multicore compute nodes equipped with hardware acceleration targeted at AI algorithms;
    \item a set of control servers for hosting middleware services that ensure the flow of high-throughput data processing.
\end{itemize}
The input buffer will be connected with DAQ (Building nodes), compute nodes, and control servers via an IP fabric which should provide the necessary peak throughput up to 400 Gbit/s. The volume of the storage system should allow storing daily data volume up to $\sim$~2 PB. The storage system will naturally be distributed. The initial implementation of such a system is supposed to use Lustre and ZFS technologies.

The intermediate buffer will be based on the same technological stack as the input buffer, but it is specifically implemented as a separate infrastructure in order to minimize possible negative effects on storage performance at the stage of recording initial raw data. The size of the intermediate buffer is expected to be on par with the input buffer.

For an accurate estimation of the required amount of compute nodes, a digital twin of the computing system is being developed, which will take into account not only available hardware technologies but also the efficiency of application software and the reliability of components.

\subsubsection{Middleware for multi-stage high-throughput data processing.}
\begin{figure}[htbp]
  \begin{center}
    \includegraphics[width=0.8\textwidth]{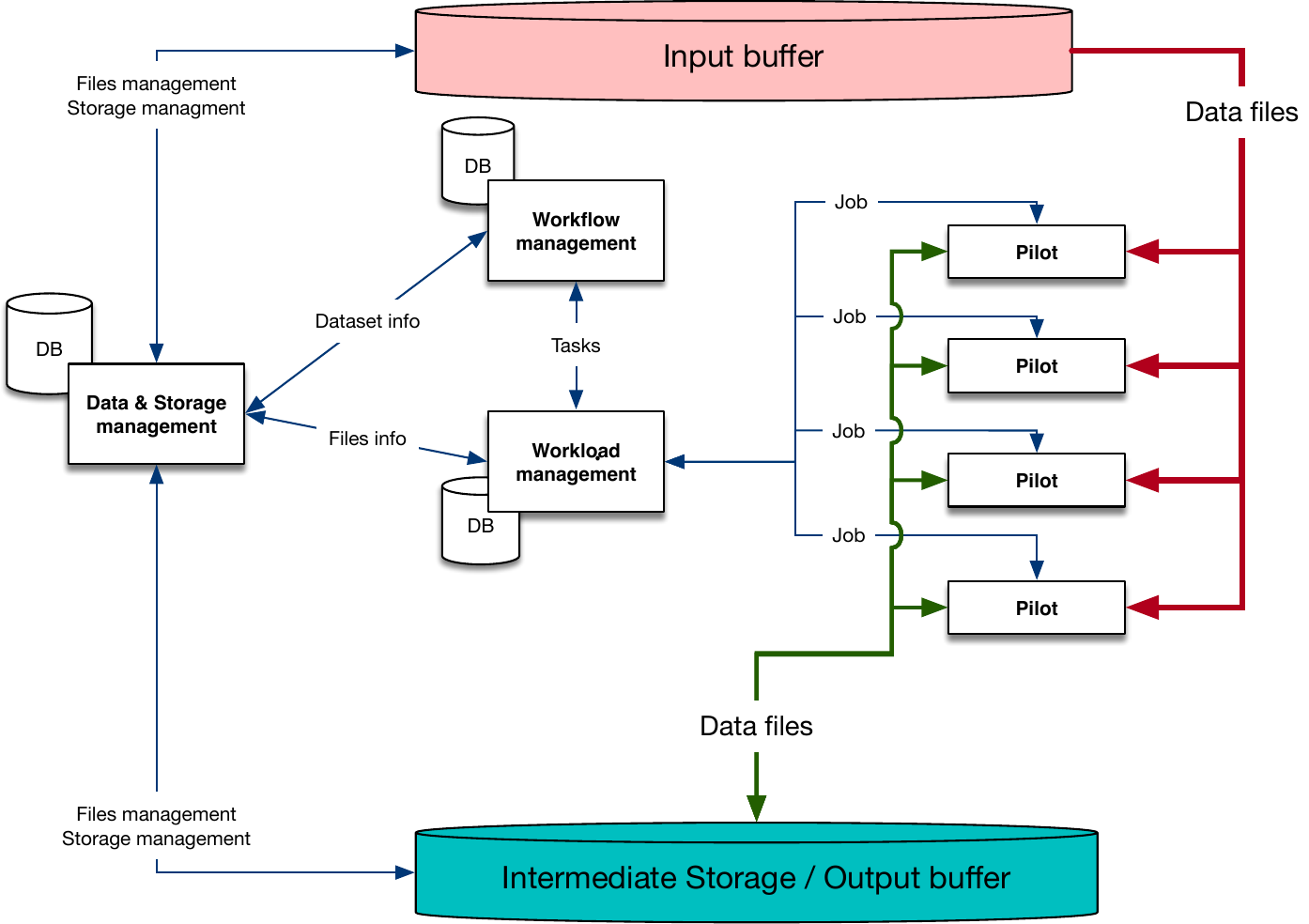}
  \end{center}
  \caption{SPD online filter middleware}
  \label{fig:online_mw}
\end{figure}

Primary data processing includes a number of workflows, each of which can be divided into atomic stages of work on homogeneous sets of data. At the same time, the processing of each individual element of such a set of data can be performed independently.
A special middleware for multi-stage data processing includes the following basic components:
\begin{itemize}
    \item data management system responsible for data catalogue, and data organization, data consistency monitoring and some control of storage;
    \item workflow management system responsible for the definition of processing pipelines and control of processing stages workflow;
    \item workload management system that is responsible for the execution of processing stages through the generation of the required number of processing jobs and efficient utilization of compute nodes.
\end{itemize}
The general scheme of interaction of the components is shown in Figure ~\ref{fig:online_mw}.

A set of sequential processing steps will be referred to as a processing pipeline. Each of the steps in such a pipeline can be defined by a predetermined template using CWL (Common Workflow Language)~\cite{commonwl}. To perform a step in the processing pipeline, a required number of tasks will be generated for the assigned data by predefined conditions (e.g. information about data, type of processing, etc.).

The unit of data processing for one task is some logical set of files - a dataset. The file, in turn, represents the amount of data for processing by one job, except for special type of tasks related to the merging of data files. The same file may belong to different datasets (at least one). Datasets in the system can be:
\begin{itemize}
    \item \textbf{Input} -- received from the data acquisition system (DAQ).
    \item \textbf{Intermediate} -- generated during processing, but not yet ready for use in the offline processing system. A lifetime of the intermediate data is coherent with the pipeline circle.
    \item \textbf{Output} -- data for subsequent processing or use outside of the functionality of the online filter.
\end{itemize}
Information about files and datasets is stored in the data catalogue. To be able to track the life cycle of data, the catalogue also stores information about their status (for example, a dataset may be open for adding files, closed, deleted, etc.).

In order for the system to be able to find out about the existence of a file on the storage, there is a file registration process that includes receiving information about the location of the file (name, physical path, metadata) and entering it into the catalogue with reference to the required dataset. 

Data consistency control refers to the regular checking of the status of files on storage (for example, their checksum), file management depending on their life cycle, as well as monitoring storage usage (for example, "dark" data - files that are on storage, but not registered in the system).

Middleware complex is expected to operate under high load. A microservice architecture built under domain-driven design is chosen, to meet scalability requirements. Microservices are as independent of each other as possible, which allows the system to be horizontally scalable.

The principles of inter-service interaction are defined, aiming at the maximization of total throughput. In cases where an instant response from the service is expected, the HTTP data exchange protocol is used. However in some other cases, this protocol could become a bottleneck for the overall performance of the system, so the AMQP protocol was chosen for asynchronous message-oriented communication.

The Workflow Management System consists of the following set of microservices, each of which has its own database:
\begin{itemize}
    \item \textbf{WfMS-Manager} -- service for interacting with the operator, which defines processing pipeline templates.
    \item \textbf{WfMS-Updater} -- data management system polling service for retrieving the datasets and sending them for processing to WfMS-Executor.
    \item \textbf{WfMS-Executor} -- service for generating a chain of tasks based on specified templates and data and then sending them for execution to the workload management system.
    \item \textbf{WfMS-Monitor} -- workload management system polling service that monitors the status of job execution and controls the life cycle of intermediate datasets by sending requests for their deletion.
\end{itemize}
The data management system also consists of its own set of microservices:
\begin{itemize}
    \item \textbf{DSM-Register} -- service that accepts asynchronous (via message queuing) requests for adding/removing data into the system.
    \item \textbf{DSM-Manager} -- service that provides a REST API for the data catalogue (CRUD operations on files and datasets, as well as sampling operations).
    \item \textbf{DSM-Inspector} -- service for running background tasks to monitor data consistency and storage health.
\end{itemize}
A unit responsible for handling a single file or other atomic processing operation, in terms of workload management, is called a job. The responsibility of job generation for performing a task, dispatch of jobs to compute nodes and job execution lies on the Workload Management System (WMS). The primary objectives of WMS are:
\begin{itemize}
    \item \textbf{Enrollment} -- submitting the necessary information and metadata for job generation.
    \item \textbf{Partitioning} -- generating the optimal number of jobs to utilize all computational resources whilst keeping a controlled job queue size.
    \item \textbf{Management} -- monitoring the state of the jobs is carried out by the pilot application, including restarting, stopping execution, ensuring load balance, and job processing rate.
\end{itemize}
The following microservices comprise the Workload Management System:
\begin{itemize}
    \item \textbf{Task manager} -- implements both external and internal REST APIs. Responsible for registering tasks for processing, cancelling tasks, and reporting on current output files and tasks in the system.
    \item \textbf{Task executor} -- responsible for the formation of jobs in the system by dataset contents.
    \item \textbf{Job manager} -- responsible for storing jobs and files metadata, as well as providing a REST API for the job executor.
    \item \textbf{Job executor} -- responsible for distributing jobs to pilot applications, updating the status of jobs, registering output files, and closing the dataset.
\end{itemize}
\begin{figure}[htbp]
  \begin{center}
    \includegraphics[width=0.8\textwidth]{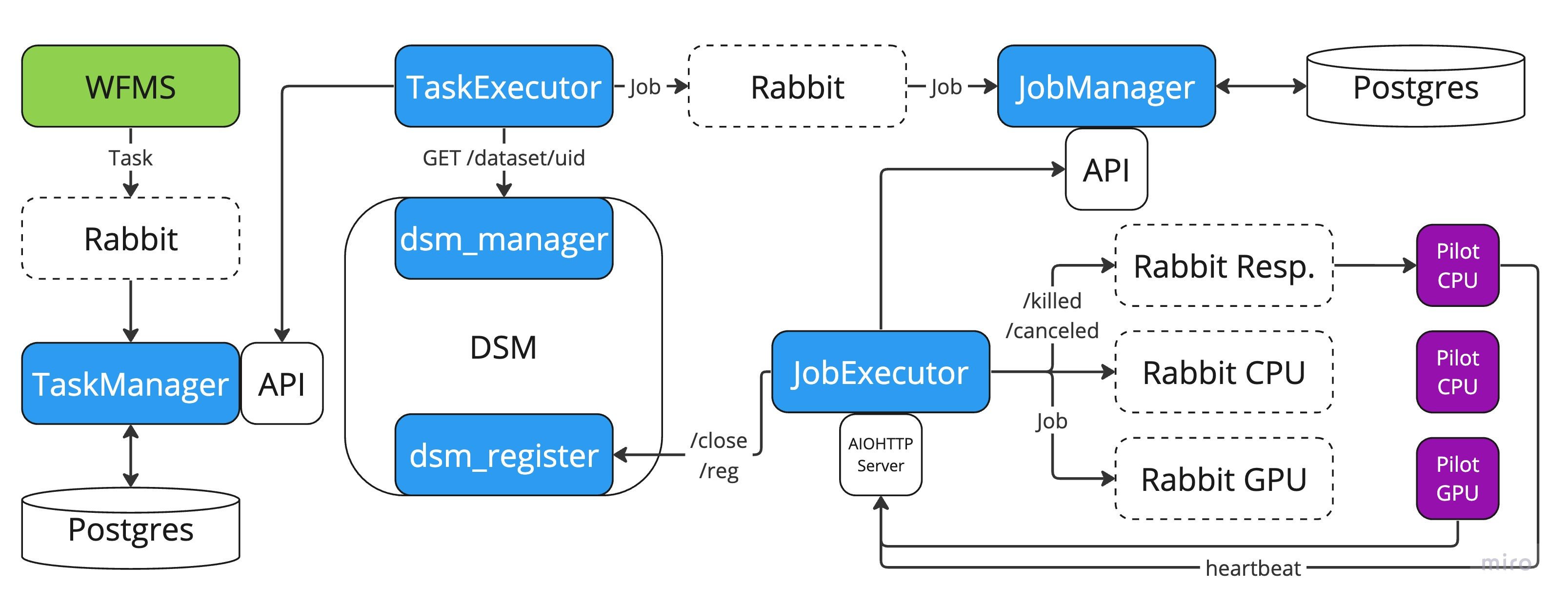}
  \end{center}
  \caption{SPD Workload Management System architecture.}
  \label{fig:online_wms}
\end{figure}

The pilot application is deployed in a container on a compute node and consists of the following two components: a UNIX daemon and the pilot itself. The UNIX daemon's objective is to run the next pilot by downloading an up-to-date version from the repository. Regardless of the presence of an error, when the pilot finishes, the UNIX daemon launches a new instance of the pilot.
Pilots are an integral part of the WMS and are responsible for executing jobs on compute nodes, organizing their execution, and communicating various information about the progress and state of the compute node to other services. Compute nodes differ only in the availability of specialized coprocessors (GPUs) and are assigned to the appropriate message broker based on the computational needs of the job (fig. ~\ref{fig:online_wms}).

\subsection{Fast event reconstruction}

\subsubsection{Fast tracking and vertex reconstruction}
The most important and complicated task in the data reconstruction chain is tracking. Traditional tracking algorithms, such as the combinatorial Kalman filter, are inherently sequential, which makes them rather slow and hard to parallelize on modern high-performance architectures (graphics processors, or GPUs). As a result, they do not scale well with the expected increase in the detector occupancy during the SPD data taking. This is especially important for the online event filter, which should be able to cope with the extremely high data rates and  fulfill a significant data reduction based on partial event reconstruction "on the fly". Parallel resources like the multicore CPU and GPU farms will likely be used as a computing platform, which requires algorithms capable of effective parallelization to be developed, as well as the overall cluster simulation and optimization.
 
Machine learning algorithms are well suited for multi-track recognition problems because of their ability to reveal effective representations of multidimensional data through learning and to model complex dynamics through computationally regular transformations, that scale linearly with the size of input data and are easily distributed across CPU or GPU cores. Moreover, these algorithms are based on the linear algebra operations and can be parallelized well, using standard ML packages. Two algorithms of track recognition in strip and pixel detectors are considered. The first algorithm, TrackNetv3, relies on the use of a Recurrent Neural Network (RNN), which allows to combine track extrapolation with testing the hypothesis that a set of points belongs to a true track and is compatible with a smooth curve. Essentially, it reproduces the idea of a Kalman filter with the difference that the physical parameters describing the track are approximated by a neural network, using synaptic weights, determined during its training. The second approach, RDGraphNet, uses a graph network and allows to implement a global search for tracks in an event, which is especially attractive when analyzing events with a large multiplicity. These approaches have already been successfully used for track recognition in the BM@N experiment at JINR and in the BESIII experiment at IHEP CAS in China~\cite{Ososkov:2020, Goncharov:2020, Shchavelev:2021, Nikolskaya:2021, Bakina:2022}. These algorithms will be adapted to find and reconstruct particle tracks in SPD data from the vertex detector and the main straw tracker. The main difficulty is the adaptation of neural networks to recover tracks in drift detectors, which requires solving the "left-right" ambiguity. To prototype neural networks and to study the quality of their work, the Ariadne software package~\cite{Goncharov:2021} will be used.

\subsubsection{ECal clustering}
 Algorithms based on convolutional networks will be developed to search for clusters in the SPD electromagnetic calorimeter and to reconstruct $\pi^0$s.
 
\subsubsection{RS clustering}
To identify muons in a muon system, the convolutional neural network will be used. Another option is to apply a simpler gradient boosting algorithm on trees.

\subsubsection{Event unscrambling}

The data stream from the DAQ will be stored in files. Each file will be processed independently from the others. Data blocks corresponding to a certain duration of data taking (from 5 to 15 $\mu$s) will be used as a reconstruction unit. Data from different subdetectors in each data block will be fed to a series of consecutive neural networks. First, vertices and track seeds will be determined, using vertex detector data. Tracks will be associated with vertices, and bunch crossing time will be determined for each vertex. Then, tracks will be reconstructed, using track seeds from the vertex detector. Hits in the straw tracker, which are not associated with any track, will be collected in a selected time window according to the bunch crossing time and attached to the events for possible use in the offline reconstruction. ECal and RS hits will be reconstructed by other neural networks and associated with vertices according to the bunch crossing time. Raw data from other subdetectors will be attached to the events according to bunch crossing time. The block of information, associated with each vertex following the procedure described above, will be called an event.



\subsection{Implementation of machine learning algorithms}

\subsubsection{Training and validation}
Training of the neural networks will be based on a dedicated large sample of simulated data.

The caution is necessary, though, to avoid possible bias due to an inadequacy of the training data to the real ones, including possible machine background and the detector noise. A dedicated workflow that includes continuous learning and re-learning of the neural network, deployment of new versions of the network, and the continuous monitoring of the performance of the neural networks, used in the online filter, will be applied.  

\subsubsection{Integration to the online data filter}
For the effective use of deep learning algorithms, besides the design of the neural networks and their training, it is necessary to solve a number of technical problems. First, open-source machine learning tools (TensorFlow, sklearn, PyTorch, etc.) tend to have Python interfaces, making them difficult to use in real-time systems. To solve this problem, it is planned to choose a package that has an advanced C++ API, and also to separate the network training (which can be implemented in any way) and its inference. Secondly, there is a risk that the neural network will malfunction if the initial data is substantially different from the data that was used for training. In a physical experiment, these differences can be caused, for example, by a fluctuation of the background from the accelerator or by a change in the detector performance or noise. Therefore, it is required to foresee a procedure for monitoring the correct operation of the neural networks. For this, it is proposed to process a certain fraction of data independently, using classical algorithms, and to compare the output with the output of deep learning algorithms. In case of large differences, one should either mark the data as questionable or retrain the neural network to work in new conditions. Third, the very application for fast reconstruction and filtering of data is proposed to be implemented in the form of a framework in order to allow flexible customization of the sequence and content of the data processing procedure.


\begin{figure}[htbp]
  \begin{center}
    \includegraphics[width=0.8\textwidth]{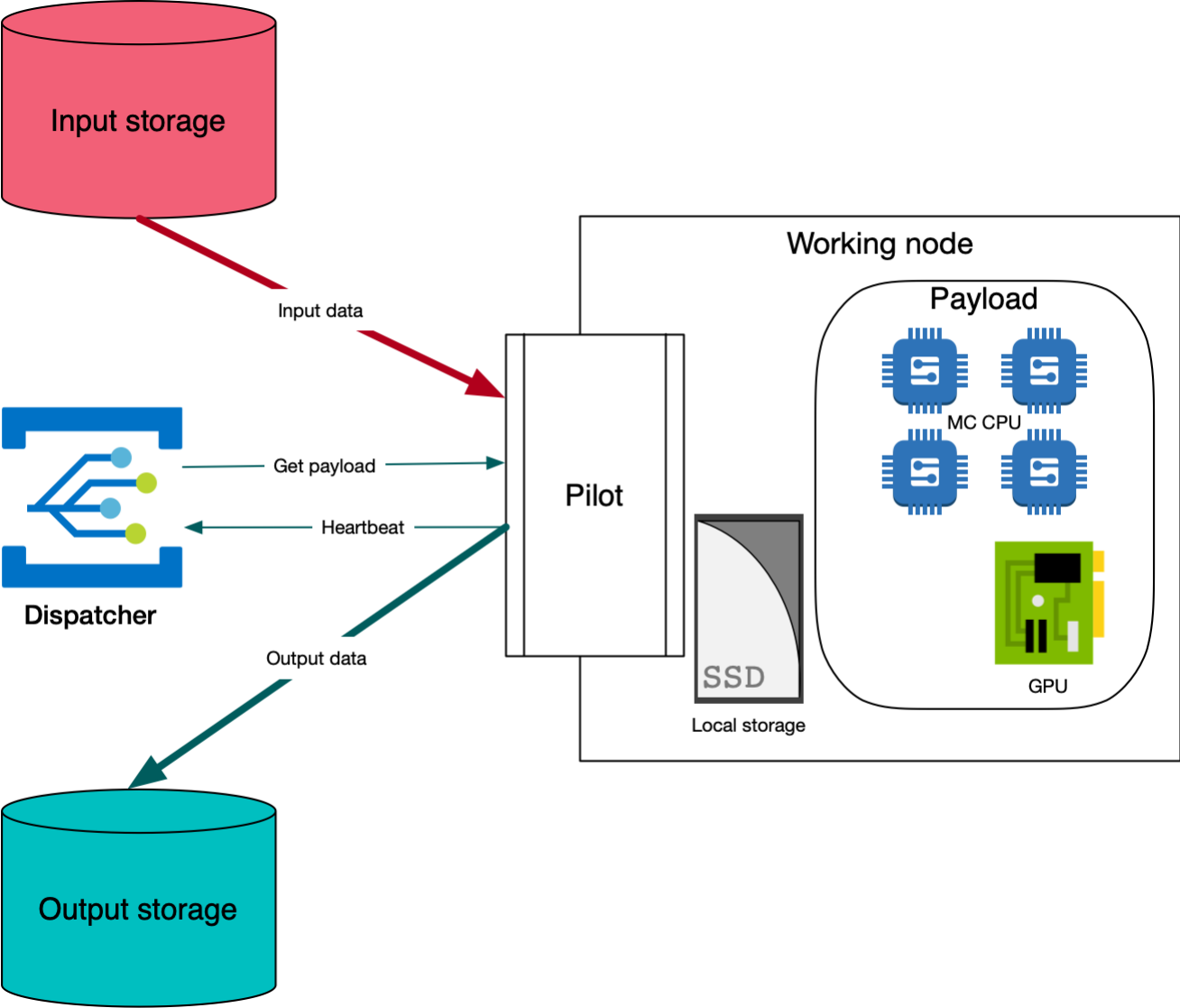}
  \end{center}
  \caption{Operation of the online data processing framework at a selected working node.}
  \label{fig:online_node}
\end{figure}


\section{Offline software}

\subsection{Introduction and requirements}
Offline software is a toolkit for event reconstruction, Monte Carlo simulation, and data analysis. Linux is chosen as the base operating system.

Recent developments in computing hardware resulted in the rapid increase of potential processing capacity from increases of the core count of CPUs and width of CPU registers. Alternative processing architectures have become more commonplace. These range from the many-core architecture based on x86\_64 compatible cores to numerous alternatives such as other CPU architectures (ARM, PowerPC) and special co-processors/accelerators: (GPUs, FPGAs, etc). For GPUs, for instance, the processing model is very different, allowing a much greater fraction of the die to be dedicated to arithmetic calculations, but at a price in programming difficulty and memory handling that tends to be slightly different with each chip generation. Further developments may even see the use of FPGAs for more general-purpose tasks.

The effective use of these computing resources may provide a significant improvement in offline data processing. The development of the concurrent-capable software framework is needed to provide the offline software for Day-1 of the SPD operation, as well as a dedicated R\&D effort to find proper solutions for the development of efficient cross-platform code.

\subsection{Choice of the framework}

\subsubsection{SpdRoot}
Currently, the offline software of the SPD experiment -- SpdRoot -- is derived from the FairRoot software~\cite{AlTurany:2012gc}. It is capable of Monte Carlo simulation, event reconstruction, and data analysis and visualization. The SPD detector description is flexible and based on the ROOT geometry package. This is the main tool to study the physics performance and to do the detector optimization during the preparation phase of the experiment.

\subsubsection{A Gaudi-based framework}
To take advantage of modern computing hardware, the offline software for the processing of experimental data should be capable of taking advantage of concurrent programming techniques, such as vectorization and thread-based programming. This is the reason why another framework based on Gaudi~\cite{gaudi} will be developed by the beginning of the SPD data taking.

\subsection{Detector description, calibration, and alignment}

The GeoModel class library~\cite{geomodel}, which is presently in use by both the ATLAS and FASER experiments, will be used for the SPD detector description. The GeoModel tool suite includes a set of tools to allow much of the detector modeling to be carried out in a lightweight development environment, outside of the offline software framework. These tools include the machinery for creating a persistent representation of the geometry, an interactive 3D visualization tool, various command-line tools, a plugin system, and XML and JSON parsers.  

Calibration and alignment constants as well as run conditions will be stored in the offline database. The use of these data in the offline algorithms will be facilitated by a dedicated service in the Gaudi framework. 

\subsection{Simulation}
Proton-proton collisions are simulated using a multipurpose generator Pythia8~\cite{Sjostrand:2014zea}. Deuteron-deuteron collisions are simulated using a modern implementation of the FRITIOF model~\cite{Andersson:1986gw, NilssonAlmqvist:1986rx}, while UrQMD \cite{Bass:1998ca, Bleicher:1999xi} generator is used to simulate nucleus-nucleus interactions. Transportation of secondary particles through the material and magnetic field of the SPD setup and the simulation of detector response is provided by Geant4 toolkit~\cite{Agostinelli:2002hh, Allison:2006ve, Allison:2016lfl}. 

\subsection{Reconstruction}
Track reconstruction uses GenFit toolkit~\cite{Rauch:2014wta}, and KFparticle package~\cite{etde_22400191} is used to reconstruct primary and secondary vertices.

\subsection{Physics analysis tools}

ROOT and Python-based tools (NumPy, SciPy, Pandas) are expected to be widely used.

A DIRAC~\cite{Stagni:2017mmz} framework could be used to run user simulation at distributed computing resources.

\subsection{Software infrastructure}
A git-based infrastructure for  SPD software development is already established at JINR~\cite{git}.

\section{Computing system} \label{sec_comp_system}

The current timeline of the evolution of the NICA complex is split into two phases according to maximum achievable luminosity. The number of events to be processed (real data and MC) in the second phase will increase by an order of magnitude in comparison to the first one: from 2$\times$10$^{11}$ to 2$\times$10$^{12}$ events per year. The expected number of events and the concomitant amounts of data, measured in petabytes, are comparable to the ones in the LHC experiments. Distributed computing systems, based on resources provided by collaborators, are proven and already well-instrumented solutions for processing such amounts of HEP data. Event-level granularity of HEP data allows independent simultaneous processing by splitting the workload among a set of jobs. Like in the MapReduce paradigm, each of these jobs will apply the same algorithm for processing a small amount of data from the same dataset. All these jobs should be performed in a controlled way to avoid duplication of data processing, and processing the whole scope of the dataset. Efficient usage of computing resources requires high-speed access to processed data. This is achieved not only by the usage of high-speed wide-area networks but also by the possibility of controlled distribution and replication of data between both intermediate and long-term storage systems, associated with data processing centers.

Already existing methods and solutions for building such distributed systems allow the achieving a sufficient level of scaling and automation of experimental data processing, along with the possibility of implementing various data processing models. Development of such systems started more than twenty years ago, and continued development allows us to meet new operational conditions and evolving computing resources. The availability of proven ready-made solutions makes it possible to significantly accelerate their commissioning while directing the main efforts at adapting the middleware components to the specific needs of the experiment.

\begin{figure}[htbp]
  \begin{center}
    \includegraphics[width=0.8\textwidth]{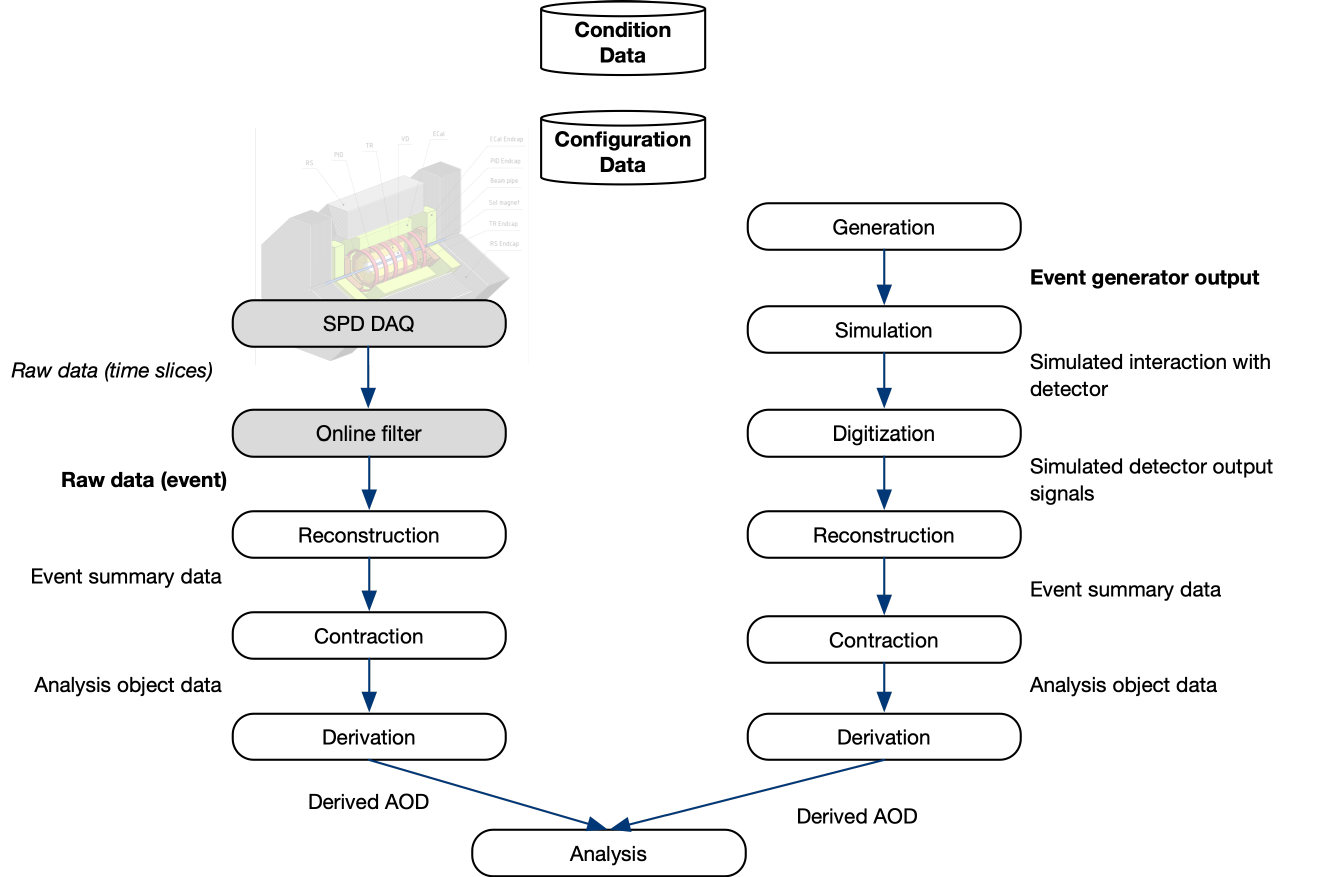}
  \end{center}
  \caption{Processing stages and data types.}
  \label{fig:comp2}
\end{figure}

Processing of large volumes of experimental data is organized in stages (Fig.~\ref{fig:comp2}). Each of the major stages will produce derived data in appropriate formats that ensure optimal data handling and storage. Derived data should be reproducible if needed. The organization of the processing will be determined in accordance with the current policy agreed upon the physical program of the experiment.

In order to avoid large-scale upgrade of the entire computing infrastructure during the transition to the second phase of the experiment, it is necessary to calculate its parameters already now, when building the model for acquisition, storage, and processing of SPD experiment data, taking into account the volumes that will be generated when operating at high luminosity of the collider. The components of the computing infrastructure should be selected based on the maximum expected load.

\subsection{Data processing workflows}

The rough estimation of a number of physical events, which should be processed per year to fulfill the physical program of the SPD experiment, after reaching the planned performance indicators of the NICA complex during the second stage, are 2$\times$10$^{12}$. It is assumed that it will take an average of 1 second to process one event on one CPU core, regardless of the type of processing. Thus, in order to cope with such a volume of calculations, the computing infrastructure of the experiment must consist of at least 60,000 processor cores. For optimal load distribution, the initial data stream should be divided into tasks that will be performed for about 8 hours, for example, 28,800 events each. If the size of one event is 15 kB, the file will be 450 MB. To achieve optimal loading of compute nodes, as well as the optimal file size for storage (on tape and disk storage systems) and for transmission over the network, it is proposed to combine such files in 16 pieces. Thus, the size of the merged file will reach 7 gigabytes, and the number of events in it will be 460,800. About 4.5 million raw data files will be created each year.

\begin{figure}[htbp]
  \begin{center}
    \includegraphics[width=0.8\textwidth]{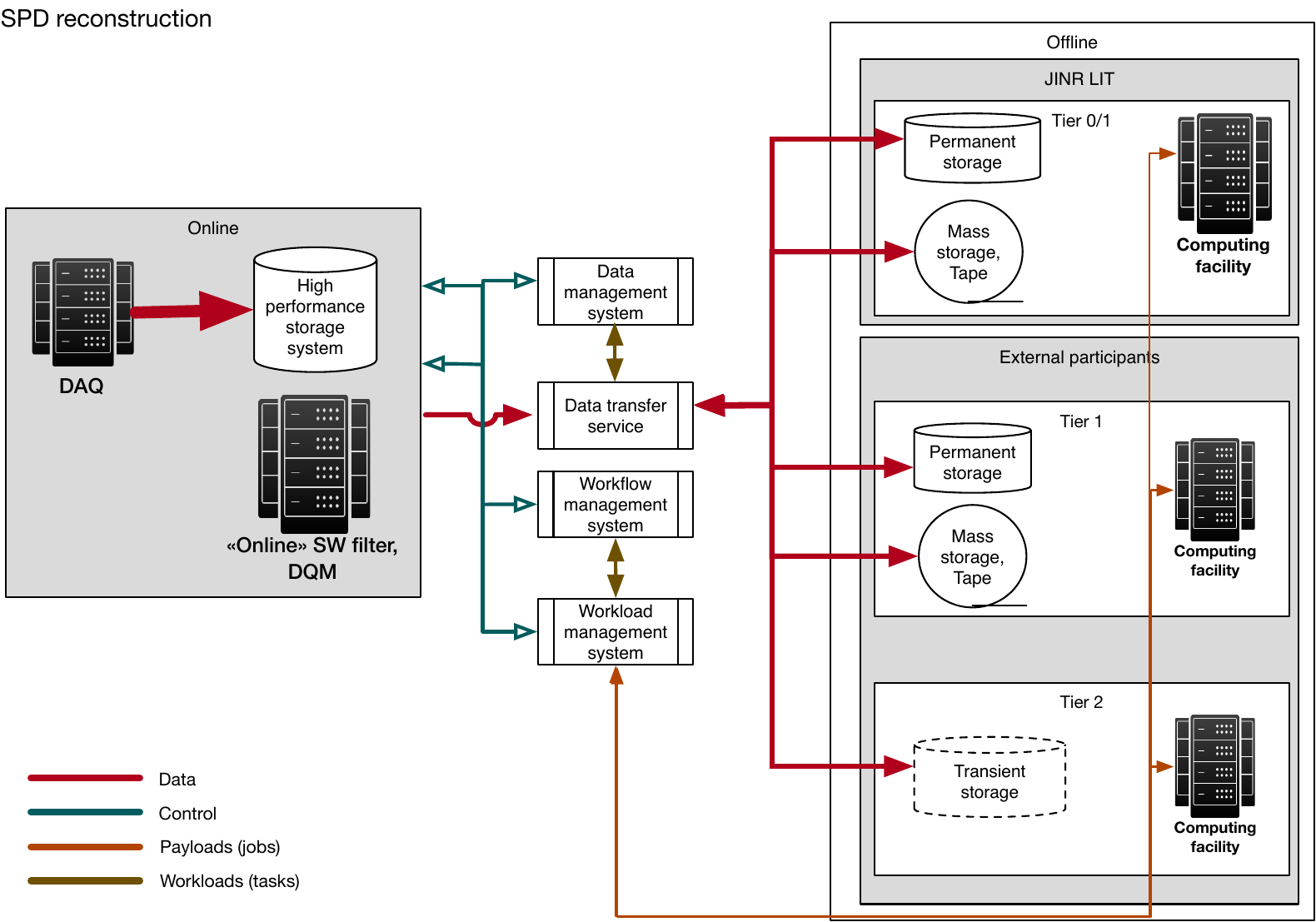}
  \end{center}
  \caption{Real data reconstruction and involved services.}
  \label{fig:comp3}
\end{figure}

The main requirement for the application software is to support multithreaded data processing. This will significantly reduce the load on the IT infrastructure by reducing the number of tasks running in the system: in the case of  single-threaded data processing, at least 60,000 jobs should be managed simultaneously. The scheme of data flow in the process of performing real data reconstruction is shown in Fig.~\ref{fig:comp3}.

\begin{figure}[htbp]
  \begin{center}
    \includegraphics[width=0.8\textwidth]{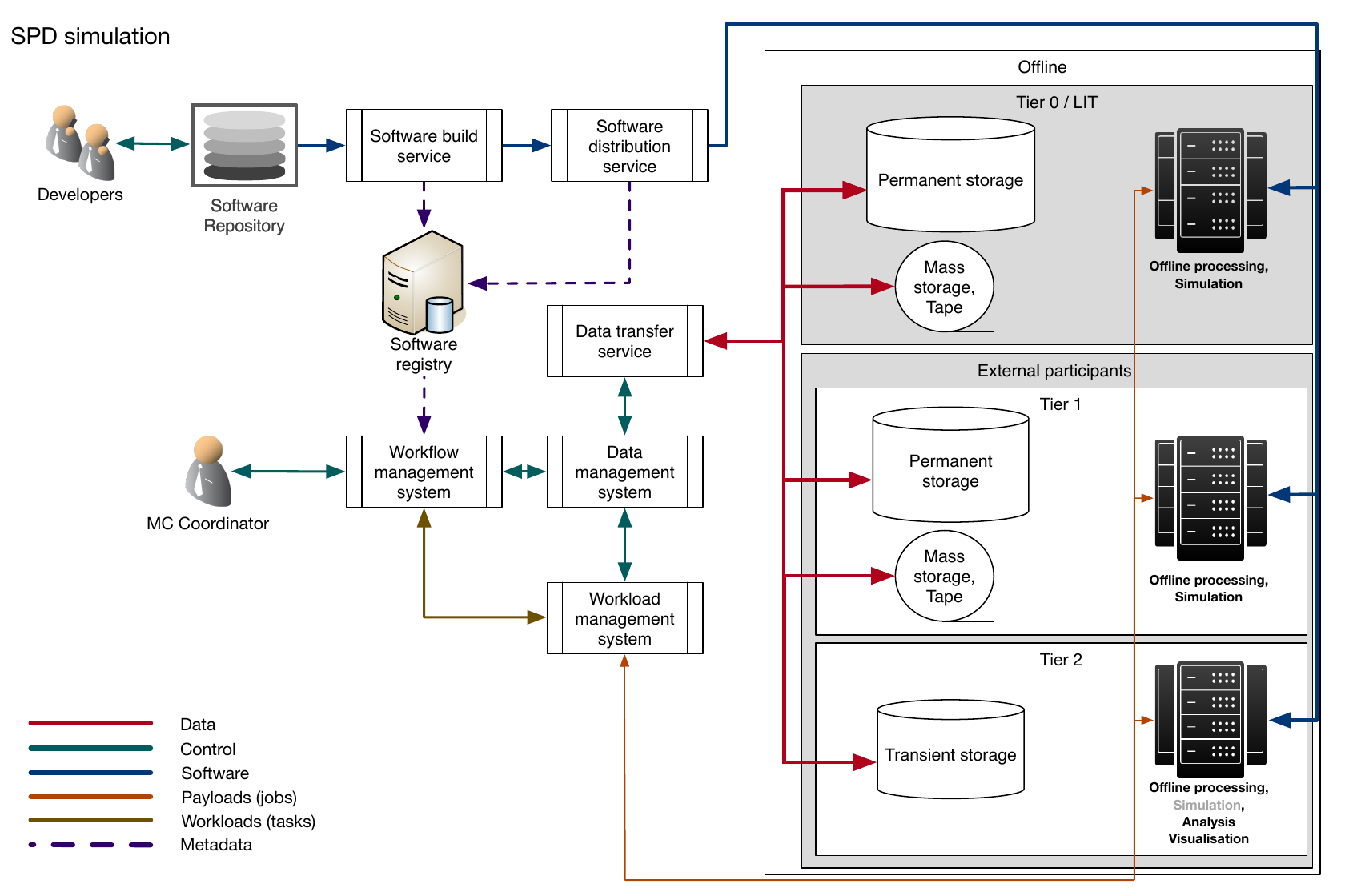}
  \end{center}
  \caption{Monte Carlo simulation and involved services.}
  \label{fig:comp4}
\end{figure}

Monte Carlo simulation includes event generation, simulation of the detector response, and digitization. Together with the following reconstruction it represents Monte Carlo data production. It is expected that the number of files generated during the Monte Carlo simulation will be comparable to the processing of data collected from the detector. Jobs of this type can be easily parallelised, so it is assumed that there will be about 4,000 tasks of the Monte Carlo simulation in the system at the same time. The flow diagram of the Monte Carlo simulation is shown in Fig.~\ref{fig:comp4}.

\subsection{Data volumes}
\label{datavolume}

Based on the aforementioned estimates, it is possible to formulate the requirements for the computing infrastructure for each of the stages:
\begin{itemize}
\item Preparation for the experiment.
Monte Carlo simulation from 2024 to 2028 will provide 2 PB per year.
Total per stage: 10 PB.

\item Stage I: running at low luminosity of the NICA collider.
Monte Carlo simulation and real data taking from 2028 to 2030 will provide 4 PB per year. Reprocessing: 2 PB per year.
Total per stage: 18 PB.

\item Upgrade of the setup for operation at high luminosity.
Monte Carlo simulation from 2031 to 2032 will provide 2 PB per year. Reprocessing: 2 PB per year.
Total per stage: 8 PB.

\item Stage II: running at maximum design luminosity of the NICA collider.
Monte Carlo simulation and real data taking from 2033 to 2036 will provide 20 PB per year. Reprocessing: 10 PB per year.
Total per stage: 120 PB.
\end{itemize}
 Total for all stages: 156 PB.

\subsection{Data processing infrastructure}
  
The experimental facility and the SPD online filter computing farm will be located at the VBLHEP JINR site, and the computing resources of the primary data processing center will be located at the DLNP JINR site. Two campuses are connected by a high-speed communication channel with a bandwidth of 400 gigabits per second. It is planned to use a wide range of storage systems for data storage and processing: disk storage systems for short-term and medium-term storage of initial and intermediate data during their processing, and tape storage systems for long-term storage of the most important data. The JINR computing infrastructure provides a number of different computing systems: CICC, the JINR Central Information and Computer Complex, which will include the SPD primary data processing center, the Govorun high-performance computing complex which is designed for artificial intelligence tasks and resource-intensive computing, a cloud computing infrastructure that can be used for user analysis.

The distributed computing infrastructure of SPD has to encompass all of the above computing systems, as well as geographically distributed resources provided by the collaboration participants. The organization of a sufficiently reliable and scalable distributed computing infrastructure for experimental data processing is carried out by a set of systems and services of an intermediate level providing safe use of resources, efficient distribution of tasks between computing resources, distribution and management of data, and a high level of automation of data processing. An additional requirement for the computing infrastructure is the ability to connect opportunistic resources that do not work within the framework of the agreements established in the collaboration. These can be supercomputer centers, commercial cloud infrastructures, and resources provided on a volunteer basis.

\begin{figure}[htbp]
  \begin{center}
    \includegraphics[width=0.8\textwidth]{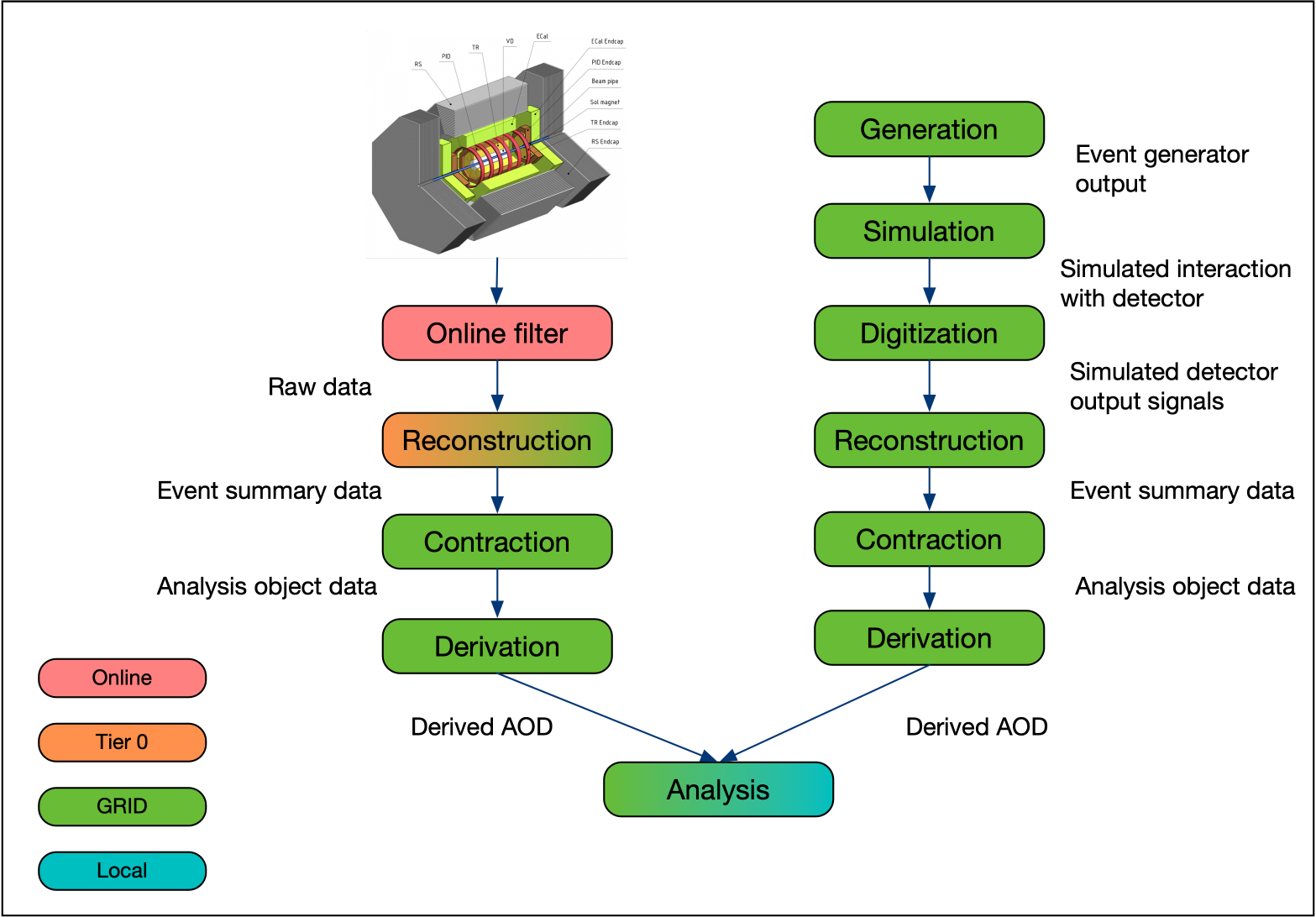}
  \end{center}
  \caption{Data processing stages and their distribution over various types of computing resources.}
  \label{fig:comp5}
\end{figure}

The two main operations that require organised data processing are Monte Carlo simulation and reconstruction of both real and simulated data. Therefore, we have to distribute the tasks to these computing resources that suit them best (Fig.~\ref{fig:comp5}). To perform such tasks on distributed resources they must be connected by communication channels with sufficiently high bandwidth. Tier-1 centers are expected to have a network bandwidth of at least 100 gigabits per second, while for Tier-2 centers wishing to participate in the Monte Carlo simulation the network bandwidth must be at least 10 gigabits per second.



\subsection{SPD production system}
To ensure a high level of automation of data processing, it is necessary to use a production system. Such a system allows one to describe the pipeline for processing a given amount of data in the form of a task graph. Each task (a graph node) is a set of jobs of the same type that process part of the initial data. Moving along the task graph to the next node is possible only after successful completion of all jobs on the current node.

The production system interacts with the distributed data management system and the workload management system, implementing a high-level control and abstraction layer on top of possibly different architectures of computing and storage systems.

The ultimate goal of the production system is to manage the mass production of data sets, combined into tasks on the computing resources available to the experiment. The Monte Carlo simulation and the reconstruction of real and simulated data should be implemented in the framework of the production system, where each operation is usually split into several basic steps, such as data preparation, processing of individual jobs, merging of the results, deleting temporary files, checking the validity of intermediate and final results, and so on.

Like in any large physics experiment, the SPD production system should be developed with an eye toward the specifics of the SPD research program and expected data volume, capable of data processing on geographically distributed resources of various types. This implies both the delivery of jobs to remote computing nodes and the relevant data management.

Quite a few software components have been developed by experiments at the LHC and are already in use by some other scientific projects dealing with high-throughput data processing on geographically distributed computing resources. These components include the PanDA workload management system and the Rucio data management system. Given that these two components are being developed and supported by the ATLAS and CMS experiments at the LHC for processing  of data volumes comparable to SPD on the widely distributed resources of the Worldwide LHC Computing Grid (WLCG), one can be reasonably confident in their suitability for the needs of the SPD project.

\subsection{Production setup and software distribution}
The CernVM File System (CVMFS)~\cite{cvmfs} provides a scalable, reliable, and low-maintenance software distribution service. It was developed to assist High Energy Physics collaborations to deploy software on the worldwide-distributed computing infrastructure used to run data processing applications. CVMFS is implemented as a POSIX read-only file system in user space. Files and directories are hosted on standard web servers and mounted in the universal namespace /cvmfs.

SPD home (/cvmfs/spd.jinr.ru/) at CVMFS is an entry point to the numbered releases of SPD software. Production setups in CVMFS are delivered to the remote compute nodes in the form of frozen sandboxes. Each new production means a new directory with all dependencies gathered in CVMFS, so that it will remain reproducible in the case of reprocessing, bug fix, or other reasons. There is also a naming convention: each production setup in CVMFS corresponds to a path with the same name in the distributed storage system. Software packages will be distributed as Docker containers to minimize external dependencies.

\subsection{Data organization and naming convention}
Due to the relatively high cost per byte of disk storage systems, it is intended to only use disk storages during data acquisition, processing or reprocessing, i.e. disk storages will only be used as a buffer. Long-term data storage will be done on tapes. EOS is currently used as a disk storage system, while CTA (CERN Tape Archive) is installed and will be used as a new interface to the tape storage.

Here is an example of the processing and data movement stages during the Monte Carlo simulation pipeline:
\begin{enumerate}
    \item preparing a processing task;
    \item start processing;
    \item request to migrate tape data to the disk storage;
    \item start jobs for processing as data moves;
    \item performing jobs;
    \item writing intermediate results to the disk storage;
    \item if there is a need to merge intermediate results, start merge tasks;
    \item writing final data to the disk storage;
    \item request to migrate disk data to the tape storage;
    \item delete all intermediate data from the disk storage.
\end{enumerate}

The most valuable data of any physical experiment are the raw data. The preservation of the raw data must be treated most responsibly: not only should they be saved, but they should be saved for as long as possible. Since it is difficult to guarantee the safety of large amounts of data within one computing center for a long period of time, there must be at least one copy of the data at some remote location. In SPD the most valuable data will be replicated from JINR to one of the remote Tier-1 centers as soon as possible.

When dealing with large amounts of data, their storage and cataloguing must be organized in such a way as to facilitate their further use. The following self-explanatory dataset description scheme is proposed in Table~\ref{dataset_naming_convention_tab}.

\begin{table}[ht]
  \caption{SPD dataset naming convention}
  \label{dataset_naming_convention_tab}
  \begin{tabular}{| l | l | l | l |}
    \hline
    Grouping & Field & Description & Example \\
    tier & & & \\
    \hline
    0 & [YEAR] & Main Scope - & 2030 \\
    & & the year of data production & \\
    \hline
    1 & [MC or DATA] & Real data or simulated data & DATA \\
    \hline
    2 & [energy][polarization] &  & 250LT \\
    \hline
    3 & [desc] & Short name of physics aim & minbias \\
    \hline
    4 & [RunNumber] &  Run number for DATA, & 27189 \\
    & & ID for MC & \\
    \hline
    5 & [data type] & EVGEN, SIMUL, RECO... & RAW \\
    \hline
    6 & [DatasetUID] & unique ID of the dataset & 636763fd78df7d \\
    \hline
    7 & [Version] & for reprocessing & 0 \\
    \hline
  \end{tabular}
\end{table}

In the text fields, whitespaces between words must be replaced by underscores. Example of the dataset name, formed using the convention scheme presented above: 
\begin{equation*}
\emph{2030.DATA.250LT.minbias.27189.RAW.636763fd78df7d.0}.
\end{equation*}

\subsection{PanDA workload management system}

Function of the PanDA (Production and Distributed Analysis system) workload management system includes the delivery of jobs to the computing nodes of remote computing centers and control over their execution~\cite{Megino:2015evs}. The system has a multi-component architecture and allows organizing high-throughput computing using heterogeneous computing resources, such as grid sites, cloud infrastructures, and high-performance (HPC) systems. The PanDA system is widely used in scientific computing, usually in projects where it is required to manage the processing of a large number of simultaneously executing jobs by distributing them across a large number of computing resources. PanDA performs job distribution by finding the most suitable computing nodes based on a large number of metrics, such as resource capabilities, current load, presence of the necessary data on the local storage, utilization of network connection, and so on. Within the PanDA system, it is possible to work not only with jobs but also with tasks, which makes it possible to achieve better system scalability and ensure optimal load on the SPD computing infrastructure. Deep integration of PanDA with the Rucio distributed data management system allows implementation of various work strategies: both delivery of tasks to data and delivery of  data by the time processing starts at one of the computing centers.

\subsection{Rucio distributed data management system}

Rucio distributed data management system~\cite{rucio} is a software platform that provides all the necessary functionality for organizing, managing, and accessing data hosted on heterogeneous, geographically distributed storage resources. Rucio is designed to manage billions of files and exabytes of data. For many scientific projects, data management is becoming increasingly challenging as the number of data-intensive instruments generating unprecedented volumes of data is growing, and their surrounding workflows are becoming increasingly complex. Their storage and computing resources are heterogeneous and can be distributed at numerous geographical locations, belonging to different administrative domains and organizations. Rucio has been built as a comprehensive solution for data organization, management, and access for scientific experiments, which can be deployed on top of the existing infrastructure and make it much easier to interact with it. One of the guiding principles of Rucio is data flow autonomy and automation.

\section{Databases}
Several databases will be developed to handle various kinds of data:

\begin{itemize}
    \item data taking conditions and calibrations;
    \item physics events, both collected from the detector and simulated;
    \item software versions and Monte Carlo configurations;
    \item information related to the detector hardware parameters and configuration;
    \item collaboration management data.
\end{itemize}

Databases are not seen as a stand-alone thing but as a part of a complex information system (IS) that includes data collection and transfer tools, APIs for access from the production and analysis software, client software including command line and GUI versions, supervisors, and monitoring. Databases and their applications should be designed aiming at scalability, long-term operation, optimized data flows, and high data rates.

Information systems are bound to each other and to the detector components, and these bindings define the priorities in development. Some of the information systems will be shared with or inherited from other experiments, some will be unique for the SPD. It is presumed that access to the information systems will be provided using the JINR SSO service. A centralized database management and hosting service will be required by the various information systems of the NICA experiments. A PostgreSQL RDBMS is being considered as the default database platform for the SPD.

A set of SPD databases will be created to store the detector hardware configuration, run information, slow control data, and calibration constants. Separate replicas should be made available for use by the DAQ, online filter, and offline computing.

\subsection{Hardware database and mapping}

A catalog of hardware components that make up the detector will be stored in the hardware database. It should contain information about the detector components, electronic parts, cables, racks, and crates, as well as the location history of all of them. It includes equipment models, vendors, parameters, and other (semi)permanent characteristics. This should help in the maintenance of the detector subsystems and will be especially helpful for the knowledge transfer between team members. A prototype system with the PostgreSQL back-end is currently being developed and will be made available via REST API from the Web interface.

The number of read-out channels of the SPD will be several hundred thousand. Signals from the detector will pass through several communication devices. It is necessary to have a mapping of the channel address on the DAQ output to the device from which this signal originated. Due to the large number of elements in the system, it is virtually impossible to maintain such a mapping manually. For the components involved in the transmission of digital signals, an automatic mapping procedure should be implemented. The component must issue a Hardware ID over the data channel in response to a distinct signal. For parts of the system that are not equipped with automatic source ID recognition, an interface must be provided that allows data entry by groups of similar components and from the configuration files.

\subsection{Conditions and calibration data}

Conditions data are  non-event data representing the detector status. It includes detector hardware and data taking conditions, calibrations and alignments, luminosity, and polarization measurements. It is used for subsystem calibration, online data processing, reconstruction, and user analysis. Conditions data contain various pieces of information heterogeneous both in terms of data type and temporal granularity. The data should be organized by "Intervals of Validity" (IOV) - the time intervals over which that data are considered valid. It can be recomputed later if the understanding of the detector response improves or the quality of the input data increases, except for the detector and trigger configurations. Careful versioning of conditions data groups for production use cases is critical to ensure reproducibility. Conditions data are typically written once and read frequently, so read rates up to several kHz must be supported for distributed computing.

\subsection{Monitoring and logging}

Monitoring information system has to be developed as detector components and other information systems become ready. It will use various sources of information from the subsystems and other databases. Data will be transferred via HTTP requests in some common format, preferably JSON with mutable schema, with only a few mandatory fields like source ID, time stamp, etc.
Time series database should be used as a back-end, with some commonly used solutions, like Grafana, for data visualization. 

\subsection{Physics metadata}

Physics metadata information system holds information about datasets and runs, provenance chains of processed data with links to production task configurations, cross-sections, and configurations used for simulations as well as online filter and luminosity information for real and simulated data. It will gather a great part of its data from other information systems and provide backlinks. The scope of this system is going to grow along with the experiment development: at first, it will include just the Monte Carlo related information, then, with the commissioning of other components, it will include more and more information. The use cases are also going to change with time. The load on the system is expected to be high and the request caching is expected to be less effective as for the conditions data. Based on the above, the Physics metadata IS has to be both flexible and efficient, making its design a challenging task.

\subsection{Event Index}

Event Index is a system designed to be a complete catalog of SPD events, real and simulated data. It will provide means of retrieving event information by indexing event data files and storing this information in the Event Index database. It will also provide access to this information through API for data processing and analysis, and through a Web-based interface for interactive user access.

An entry in the Event Index database will contain the following fields: Event ID, online filter results, UUID of the raw data file containing this event, ID of the dataset this file is included in, UUIDs of files with reconstructed events (AOD) and other important event parameters that can be used for classification and selection.

\begin{figure}[ht]
    \centering
    \includegraphics[width=0.8\linewidth]{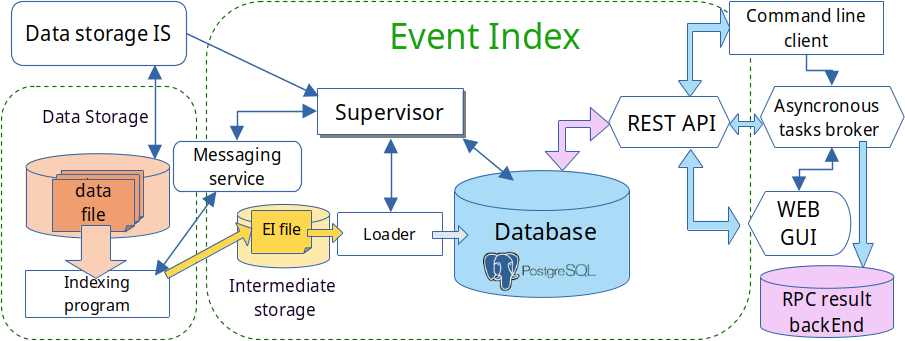}
    \caption{Event Index architecture}
    \label{fig:eiarch}
\end{figure}

A general architecture of the Event Index information system is presented in Fig.~\ref{fig:eiarch}.

The process of importing data into the database will be controlled by the Supervisor module. When a new dataset appears in the storage, a special indexing task will be launched to collect event metadata and store it in the temporary dataset. Upon successful completion of the indexing task, the production system will inform the Supervisor and the temporary dataset will be transferred to the Event Index server where it will be queued for importing into the database.

The estimated flow of data to be indexed ranges from tens to hundreds of thousands of events per second. The PostgreSQL RDBMS was chosen for index storage due to its ability to efficiently input large amounts of data, multi-threading, and high performance of bulk loading. Various optimization techniques are currently being investigated to speed up the input of data into the database.

A convenient and efficient RESTful API was developed and implemented for data exchange. The frontend part of the client interface is implemented using the Angular framework. For the server side, the fastAPI was chosen: a lightweight asynchronous RESTful framework for Python. RabbitMQ and Celery were used for asynchronous task processing, which allowed to improve overall system performance. Further development of the Event Index will be carried out in parallel with the development of other information systems.

\subsection{Collaboration management database}

This information system is needed to organize effective cooperation between hundreds of people involved in the SPD experiment. The main tasks for this system are handling of personnel, and organizations' data, support for workgroups (membership, roles, access rights, contribution accounting), and generating reports broken down by various parameters. In addition, procedures for creating, approving and editing related documents, registering, and changing collaboration members, creating and editing group and privilege lists, and managing the List of Authors are to be implemented.

An additional information system will be created for preparing and publishing the official results of the SPD experiment. This includes tracking the progress of publications, organizing communication between authors, reviewers, and curators, searching through documents, tracking  SPD publications in external information systems, and generating relevant reports.

It will also be used for organizing presentations and reports at conferences and meetings, compiling a list of conferences and available papers, organizing the call for and selection of speakers, receiving, reviewing, and approving of titles, abstracts, and slides, and tracking the publication of proceedings.

%
%
%
%
%
%
%
%

\section{Resource requirements}
The necessary amount of computing power and storage capacity will increase along the SPD experiment lifetime from the rather modest resource requirements needed for Monte-Carlo production at the preparation stage to the much larger resources needed to process the experimental data after data taking starts. 

For the online filter, we assume a processing throughput of 1'000 SPD events per CPU core per second. Thus, fast tracking requires 3'000 CPU cores simultaneously. Taking into account the additional cost of the data unpacking and event unscrambling, and recognising that actual CPU efficiency will be less than 100\%, one can derive the necessary CPU resources for the online filter to be 6'000 CPU cores. This number sets the upper limit, and the required computing power may decrease substantially if an efficient way to use GPU cores is implemented for event filtering. As for the data storage, a high-performance 2 PB input disk buffer is needed, capable of storing about one day's worth of data taking, and the output buffer of the same size.

For offline computing, the data storage capacity gradually increases from 2 PB/year at the preparation stage to 30 PB/year at running at maximum luminosity (Stage II), as detailed in section~\ref{datavolume}, reaching 156 PB of long-term storage for the whole experiment's lifetime. We assume that  half of the annual data volume ($\sim$15 PB) is kept on the disk storage, and the rest is stored on tape. The CPU resources required to process this amount of data and to run the Monte Carlo simulation are estimated at 60'000 CPU cores. The summary of the computing resources is given in Table.~\ref{cost_tab}, assuming prices at the end of 2023, i.e. 250 USD per CPU as part of the computing server, 88'000 USD per 1 PB of disk storage, and 70'000 USD per 10 PB of tape storage. The cost estimate is conservative and will be defined more accurately when detailed hardware solutions and their actual prices on the market are considered.


The estimate above is made for the maximum design luminosity of the NICA collider. For the low luminosity  stage, all numbers will be one order of magnitude smaller. More precisely, for the first stage of the experiment, it is supposed to commission 50\% of the full capacity of the SPD Online filter hardware. The required resources for offline computing should be at least 30\% of the declared resources with a gradual increase of disc and tape storage systems up to 30\% by the end of the first stage.

The burden of operating the SPD computing system shall be shared between the computing centers of the participating institutes. Tier-0 at JINR will provide about 25-30\% of all computational resources necessary for processing and storage of experimental data. Approximately the same amount is needed on Tier-1 sites, while the rest will be distributed between Tier-2s. 

\begin{table}[hb]
  \caption{Required SPD computing resources.}
  \label{cost_tab}
  \begin{center}
  \begin{tabular}{lccc}
    \hline
                  & CPU [cores] & Disk [PB] & Tape [PB] \\
    \hline
    Online filter (Stage I) & 3000 & 2 & none \\
    Offline computing (Stage I) & 20000 & 5 & 6 per year \\
    \hline
    Cost estimate (Stage I) [k\$] & 5750 & 616 & 42 per year \\
    \hline
    Online filter (Stage II) & 6000 & 4 & none \\
    Offline computing (Stage II) & 60000 & 15 & 30 per year \\
    \hline
    Cost estimate (Stage II) [k\$] & 16500 & 1672 & 210 per year \\
    \hline
    {\bf Total for Stage I: 6.4 M\$} \\
    {\bf Total for Stage II: 18.2 M\$ + 0.2 M\$ per year} \\
  \end{tabular}
  \end{center}
\end{table}



\chapter{Overall cost estimate}

 The estimated cost of the Spin Physics Detector at current prices ({\bf Nov 2023, 1 Euro = 1.08\$, 1 \$ = 92 RUB}) for two stages of the project implementation is 
 presented in Table \ref{tab:cost}. Estimates are rounded to one significant digit after the decimal point.
 The cost of the first stage is 50.0 M\$, while the full setup in the most preferable configuration is priced at 110.6 M\$. 
 This amount does not include possible R\&D for the second stage of the project.
   Any expenses related to the development and construction of an infrastructure for polarized beams at NICA are also out of the scope of this estimation.
  
 
\begin{table}[htp]
\caption{Cost estimate of the SPD setup.}
\begin{center}
\begin{tabular}{|l|l|c|c|}
\hline
\hline
 Subsystem       &  Option      &   Stage &  Cost, M\$ \\     
\hline
\hline
 SPD setup  &Vertex detector: &  & \\
                   &~~~~  -- DSSD & II &  8.8 \\
                   &~~~~  -- MAPS  & II & \cellcolor[HTML]{FFCCC9} 13.5 \\
                   &Micromegas Central Tracker  &  I & \cellcolor[HTML]{9AFF99} 0.7 \\                   
\hline
&Straw tracker    & I+II   &\cellcolor[HTML]{9AFF99} 3.7\\
\hline
&PID system:      &   &  \\
&~~~~ -- TOF     & II  & \cellcolor[HTML]{FFCCC9} 2.2\\
&~~~~ -- FARICH & II & \cellcolor[HTML]{FFCCC9} 16.7 \\
\hline
&ECal &  &   \\
&~~~~ -- mock-up      & I &\cellcolor[HTML]{9AFF99} 0.4\\
&           &II& \cellcolor[HTML]{FFCCC9} 12.0\\
\hline
&Range system  & I+II   &\cellcolor[HTML]{9AFF99} 17.3$^*$ \\
\hline
&ZDC                  & I+II &\cellcolor[HTML]{9AFF99}  0.7 \\
\hline 
&BBC + BBC MCP                 & I+II &\cellcolor[HTML]{9AFF99} 0.8 \\
\hline
&Magnetic system & & \cellcolor[HTML]{9AFF99} 9.4\\
& \& cryogenic infrastructure &  & \cellcolor[HTML]{9AFF99} 6.4\\
\hline
&Beam pipe           &  &  \\
& ~~~~ -- Al                            & I & \cellcolor[HTML]{9AFF99} 0.1  \\
& ~~~~ -- Be                            & II &  \cellcolor[HTML]{FFCCC9} 0.4  \\
\hline
\hline
General infrastructure & & &\\
&                               & I &  \cellcolor[HTML]{9AFF99} 1.8  \\
&                               & I+II & \cellcolor[HTML]{FFCCC9}  2.5  \\
\hline
Detector Control System  &  &  & \\
                                  &  & I &\cellcolor[HTML]{9AFF99}  1.5 \\
                                  &  & I+II & \cellcolor[HTML]{FFCCC9}  2.7 \\
\hline
Data Acquisition System &  & & \\
                                       &  & I & \cellcolor[HTML]{9AFF99}  1.2 \\
                                       &   & I+II & \cellcolor[HTML]{FFCCC9} 3.9 \\
\hline
Computing & & & \\
                                       &  & I &\cellcolor[HTML]{9AFF99}  6 \\
                                       &   & I+II & \cellcolor[HTML]{FFCCC9} 17$^{**}$\\
\hline
\hline
TOTAL COST & stage I & & \cellcolor[HTML]{9AFF99} 50.0  \\
		     & stage I+II & & \cellcolor[HTML]{FFCCC9} 110.6  \\
\hline
\end{tabular}
\end{center}
\label{tab:cost}
   $^*$ including 6.2 M\$ of the steel yoke of the SC solenoid \\
   $^{**}$ + 4.5 M\$  per year for tapes at the stage II \\
\end{table}%

\chapter{Summary}

We have presented the technical design of the Spin Physics Detector at NICA,
a sophisticated experimental apparatus for the study of the spin structure of the proton and deuteron, as well as the fundamental properties of the strong interaction. The building of this detector and conducting research on it are included in the JINR long-term development plan.
Although the construction of a detector on such a scale is a very challenging task -- even considering the extensive experience of the members of the SPD Collaboration -- nevertheless, we are confident that  we are able to implement all our plans within the framework of international cooperation.

\chapter*{Authors list}
\begin{center}
V.~Abazov$^1$, V.~Abramov$^{28}$, L.~Afanasyev$^1$, R.~Akhunzyanov$^1$, A.~Akindinov$^5$, I.~Alekseev$^5$, A.~Aleshko$^9$, V.~Alexakhin$^1$, G.~Alexeev$^1$,  L.~Alimov$^{11}$
% ...
\\

\vspace{0.6cm}

$^1$ Joint Institute for Nuclear Research, Dubna 141980, Russia\\
% ...
$^5$ National Research Center Kurchatov Institute, Moscow 123182, Russia\\
% ...
$^9$ Skobeltsyn Institute of Nuclear Physics of Lomonosov Moscow State University,  Moscow 119991,  Russia\\
% ...
$^{11}$ Samara National Research University, Samara 443086, Russia \\
% ...
$^{28}$ Institute for High Energy Physics of National Research Center Kurchatov Institute, Protvino 142284, Russia \\
% ...
\end{center}

\chapter*{Acknowledgements}
We thank  M.~Barnyakov, V.~Bobrovnikov,  S.~Kononov, E.~Kravchenko, A.~Kutov, I.~Ovtin, N.~Podgornov, E.~Shchavelev, A.~Tkachenko, and  E.~Tomasi-Gustafsson for their valuable contributions to the preparation of this document.

We also thank our colleagues from scientific organizations in the Czech Republic, France, Italy, Poland, Ukraine and South Africa for their contributions to the development of our project.

This work was supported by ANID -- Millennium Science Initiative Program -- ICN2019 044 (Chile);  Grant No. 1240066  FONDECYT (Chile).

The SPD Collaboration would like to thank the members of the SPD International Detector Advisory Committee
 Andrea Bressan (INFN Trieste  and University of Trieste), Pasquale Di Nezza (INFN-LNF), and Peter Hristov (CERN) who followed the SPD project in 2021--2022.

\begin{appendices}

\chapter{MDT workshop: production and test areas}
\begin{figure}[!h]
   \begin{minipage}{0.48\textwidth}
     \centering
     \includegraphics[width=0.98\linewidth]{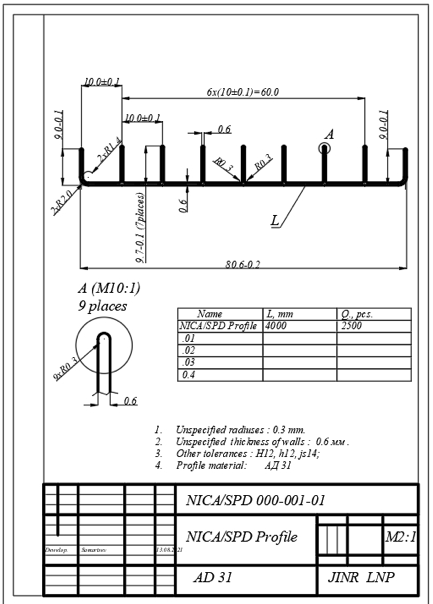} \\ (a)
   \end{minipage}\hfill
   \begin{minipage}{0.48\textwidth}
     \centering
     \includegraphics[width=1.\linewidth]{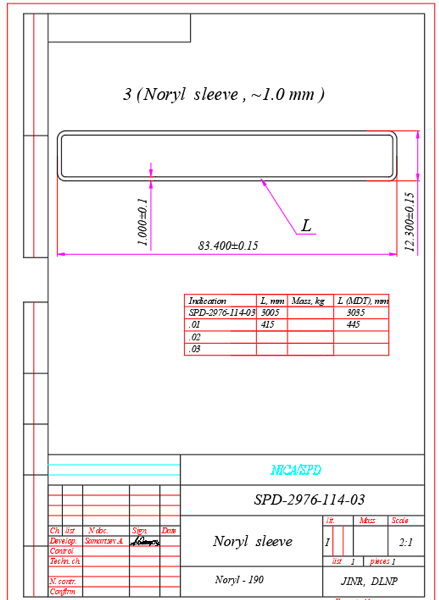} \\ (b)
   \end{minipage}
     \caption{(a) Drawing of aluminum extruded profile manufactured by AGRISOVGAZ. (b) Drawing of the plastic envelope for the MDT detector.}
      \label{fig:RS_MDT_Case}
\end{figure}

\begin{figure}[!h]
    \center{\includegraphics[width=0.8\linewidth]{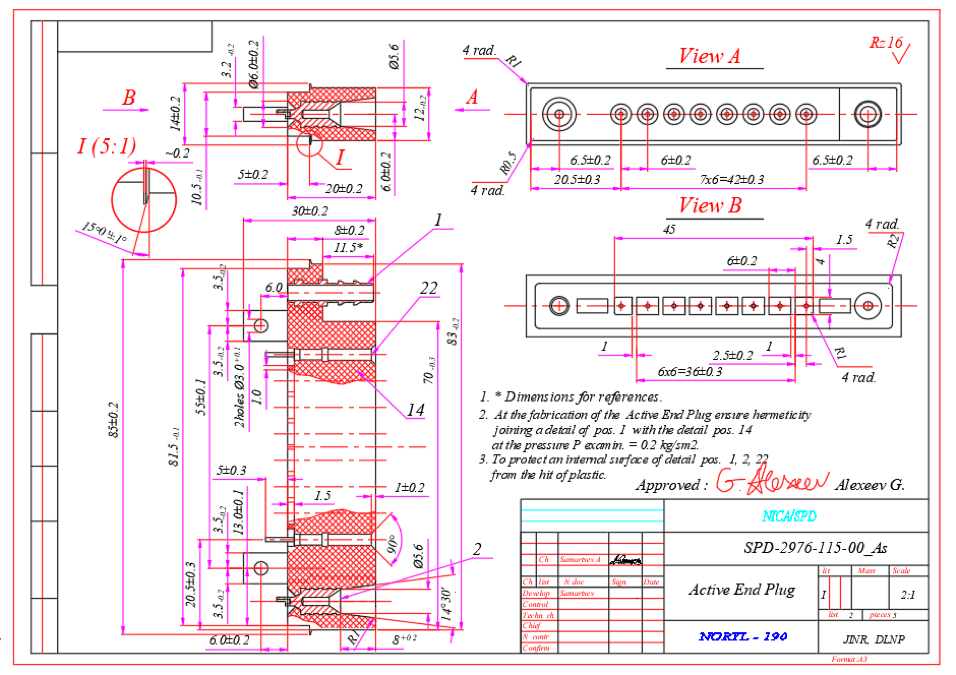}}
     \caption{Drawing of the active end-plug which contains 8 signal connectors, the high voltage and gas connectors.}
       \label{fig:RS_MDT_Active}
\end{figure}

\begin{figure}[!h]
    \center{\includegraphics[width=0.8\linewidth]{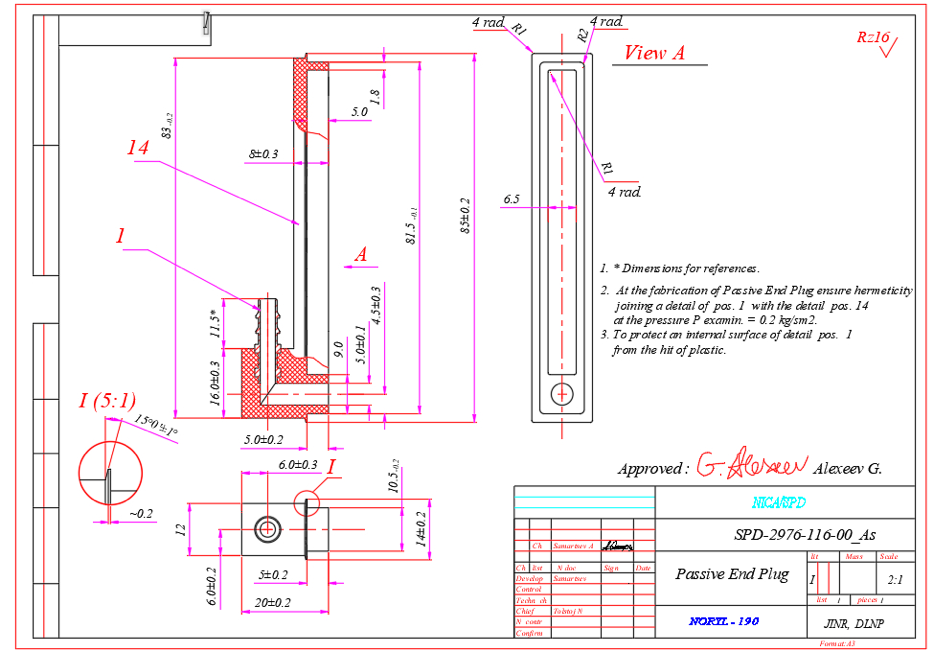}}
     \caption{Drawing of the passive end-plug which contains only the gas connector.}
       \label{fig:RS_MDT_Passive}
\end{figure}

MDT (see Fig.~\ref{fig:RS_MDT}) are gaseous detectors manufactured from thin-walled (0.6 mm thickness) 8-section aluminum profiles (see Fig.~\ref{fig:RS_MDT_Case}~(a)), with tungsten wires of a 50 $\mu$m-diameter stretched inside each section. The profile is placed in a plastic case (see Fig.~\ref{fig:RS_MDT_Case}~(b)) sealed with plastic end-plugs from both ends (see Figs.~\ref{fig:RS_MDT_Active} and ~\ref{fig:RS_MDT_Passive}). The end-plugs have special openings designed for the input and output of the working gas, as well as for high voltage supply and detector signals readout.

It is planned to organize a full cycle for the serial production of a large number of MDT detectors with length dimensions reaching up to 4.5 m in building 73.
The production cycle implies the presence of the following premises: two procurement halls, a washing room, assembly and test areas.

\section{Technological process of MDT production}
The preparation of the MDT plastic cases and aluminium profiles with the desired lengths, as well as perforation (penetration of holes for laying the bases) of profiles , are made in the procurement area. Then cases are  blown with compressed air to remove the dust and moved to a separate rack in the assembly area. Ready profiles are moved to the washing room, where they go through a washing procedure in 3 stages. At the first stage, the profiles are cleaned in hot water containing detergent powder at a temperature of about 60 $^{\circ}$C to remove possible grease stains and other contaminants. In the second and third stages, the profiles are rinsed in hot and distilled water, respectively.

The washed and dried profiles are stored on the racks in the second procurement hall. In a separate room, the selection, purging, and washing of the plastic parts (end-plugs, covers, bases, boards, support spacers) are carried out in a solution of ethyl alcohol with water; wires are soldered to the active end-plugs and matching resistors to the boards; installation and welding of bases into profiles is done; spacers are marked and laid into the profiles. Then the profiles are transferred to the assembly area.

Soldering of the 8 anode wires, laid along the profile sections on the spacers,  is made using a semi-automatic winding machine. Then the spacers are welded on the brewing machine to fix them in certain places inside the profile. After that, the wires of the active end-plug are soldered to the boards installed in the profile. Then a plastic case muffled from one end by a passive end-plug is put on the profile.

After that, the testing cycle begins on special stands for the anode wire tension test and preliminary high voltage test. In the case of a positive test result, the second end of the case is muffled by the active end-plug and welded using a brewing machine. Then the ready tube is filled with the working gas and passes high-voltage training. The last stage of testing is a gas tightness check.

Verified MDTs are stored on special racks.

\section{The terms of reference}
The terms of reference provides for:
\begin{enumerate}
\item partial reconstruction of building 73.
\item allocation of manufacturing facility (previously used at the JINR Experimental Workshop) for the MDT mass-production and testing in the announced building.
\item fulfillment of all requirements for production and auxiliary premises, workplaces, equipment, and personnel, needed:
\begin{itemize}
\item to provide  safe working conditions in the manufacture of products at all stages of the technological process of assembly and installation;
\item to manufacture the products that meet the requirements.
\end{itemize}
\end{enumerate}

\begin{figure}[!h]
   \begin{minipage}{0.48\textwidth}
     \centering
     \includegraphics[width=0.98\linewidth]{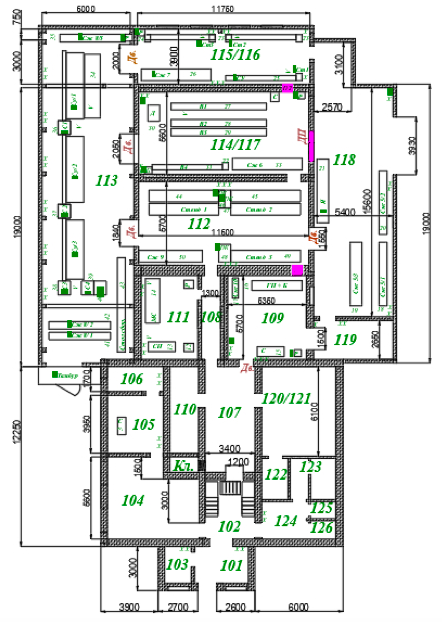} \\ (a)
   \end{minipage}\hfill
   \begin{minipage}{0.48\textwidth}
     \centering
     \includegraphics[width=1.\linewidth]{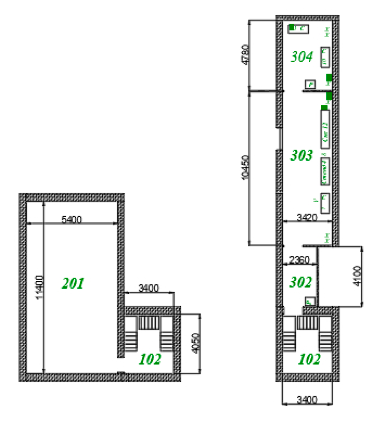}\\ (b)
   \end{minipage}
     \caption{Plan of the building  to be used for the MDT detectors mass production workshop and  deployment of the equipment for the production of auxiliary parts for the MDT detectors: (a) first (b) second and third floors.}
      \label{fig:RS_bld73_23}
\end{figure}

The building plan is given in Fig. ~\ref{fig:RS_bld73_23}, and includes the following premises distinguished by the type of work:
\begin{enumerate}
\item Service and utility rooms:
  \begin{enumerate}
  \item Room 1, 1A -- physicists rooms;
  \item Room 2 -- warehouse;
  \item Room 3 -- service room;
  \item Room 11 -- computer room;
  \item Room 11A -- service room;
  \item WC room -- bathroom and shower;
  \item Meeting and relaxation room, 3$^{rd}$ floor.
  \end{enumerate} 
 
\item Industrial premises:
  \begin{enumerate}
  \item Room 4 -- assembly room:
    \begin{enumerate} 
    \item C -- table for washing and preparation of plastic parts (end-plugs, bases, spacers, covers);
    \item MC -- assembly table for soldering wires to active end-plugs and putting on shrink tubes.
    \end{enumerate}
  \item Room 5 -- procurement area No. 1:
    \begin{enumerate} 
    \item Racks 5$_{1,2}$ -- racks for profiles and plastic cases before processing;
    \item Racks 5$_{3}$ -- rack for finished products;
    \item Electric hoist for 0.5 ton.
    \end{enumerate}
  \item Room 6 -- washing area:
    \begin{enumerate} 
    \item B1, B2, and B3 -- three baths for sequential washing of profiles. Baths B1 and B2 are filled with hot water (about 0.6m$^{3}$) from the plumbing system and maintained at a temperature of about 60 $^{\circ}$C using electric heaters. 
    Dimensions of B1 and B2: 600$\times$600$\times$6000 mm$^3$, B3: 400$\times$400$\times$6000 mm$^3$. Bath B3 is filled with distilled water (about 0.4m$^{3}$) at room temperature.
    \item D -- distiller. Overall dimensions: 700$\times$700$\times$2000 mm$^3$.
    \item B4 -- tank for checking the gas tightness of the manufactured MDT. Dimensions B4: 300$\times$360$\times$7000 mm. Bath B4 is filled with distilled water (about 0.4m$^{3}$).
    \item B -- a cylinder with N2 gas.
    \item Czh6 -- rack for preliminary drying of washed profiles.
  \end{enumerate}
 \item Room 7 -- procurement area No. 2:
     \begin{enumerate} 
     \item installation of boards, bases, spacers;
     \item Szh7 -- rack for prepared profiles;
     \item St1 and St2 -- machines for profiles trimming and perforating;
     \item St3 -- machine for cutting of plastic cases.
  \end{enumerate}
  \item Room 8 -- Assembly area:
    \begin{enumerate}
    \item C1 -- machine 1 for wire winding; 
    \item C2 -- machine 2 for spacers welding;
    \item C3 -- workstation for soldering wires of active end-plugs to boards;
    \item C4 -- machine 4 for welding of the end-plugs to the plastic cases;
    \item Collection table -- mobile table for the MDT assembly;
    \item B -- tank with cooling water for machine tool 4;
    \item Szh8 -- rack for MDTs prepared for testing.
  \end{enumerate}
  \item Room 9 -- area for testing of the manufactured MDT:
    \begin{enumerate}
    \item Stand 1 -- wire tension test bench;
    \item PC - personal computers (3 pcs.);
    \item Stand 2 -- stand for testing MDT for dark current;
    \item Stand 3 -- stand for high-voltage training of the MDT with a working gas mixture of 30\% CO$_{2}$ + 70\% Ar;
    \item Szh9 -- rack for the finished product.
  \end{enumerate}
  \item Room 10:
    \begin{enumerate}
    \item  GP -- gas control panel;
    \item Cylinders with CO$_{2}$ (1--2 pieces) and Ar (3--4 pieces);
    \item Szh10 -- rack. 
 \end{enumerate} 

  \item Room 12 -- Scintillator site:
      \begin{enumerate}
      \item Stand 4 -- stand for scintillator assembly, gluing, and taking of scintillator samples optical characteristics;
      \item Szh12 -- rack 12.
      \end{enumerate}
\end{enumerate}  
\end{enumerate}

  \chapter{List of abbreviations}
AERD --  Address-Encoder Reset Decoder\\
  ADC -- Analogue-to Digital Converter\\
  AOD -- Analysis Object Data \\
  API -- Application Programming Interface\\
  ASIC -- Application-Specific Integrated Circuit\\
  BBC -- Beam-Beam Counter\\
  BPM -- Beam Position Monitor \\
 CCD -- Charge-Coupled Device\\
  CDR -- Conceptual Design Project \\
 CICC --  Central Information and Computer Complex\\
  CF (method) -- Constant Fraction (method)\\
  CFD -- Constant Fraction Discriminator \\
  CM -- Center-of-Mass\\
  CNI - Coulomb-Nuclear Interference \\
  CPU -- Central Processing Unit \\
  CWL -- Common Workflow Language \\
  DAC --  Digital-to-Analogue Converter\\
  DAQ -- Data AcQuisition\\
  DBSCAN -- Density-Based Spatial Clustering Application with Noise\\
 DC -- Direct Current\\
 DCCT -- DC Current Transformer \\
 DCR -- Dark Count Rate\\
 DCS  -- Detector Control System\\
 DLC -- Diamond-Like Carbon\\
  DSSD -- Double-Sided Silicon Detector\\
 EC -- End-Cap\\
  ECal -- Electromagnetic Calorimeter\\
  ECS -- Electron Cooling System \\
  EES - Energy Extraction System \\
 ENC -- Equivalent Noise Charge \\
 (F)ARICH -- (Focusing) Aerogel RICH detector\\
 FAT -- Factory Acceptance Testing \\
  FBBC (monitor) -- Fast Beam-Beam Collisions (monitor)\\
  FDIRC -- Focusing Detector of Internally Reflected Cherenkov light \\
  FDM --  FPGA Digital Module \\
 FD-MAPS -- Fully Depleted Monolithic Active Pixel Sensor\\
  FDR -- Final Design Review\\
  FEA -- Finit Element Analysis \\
  FE -- Front-End\\
  FEE -- Front-End Electronics\\
  FPGA -- Field-Programmable Gate Array \\
  FSM -- Finite State Machine\\
   GPU -- Graphics Processing Unit \\
   GUI -- Graphical User Interface \\
  GWP -- Global Warming Potential \\
  HAPD -- Hybrid Avalanche Photon Detector\\
  HV -- High Voltage\\
   IOV -- Interval of Validity \\
   IP -- Interaction Point\\
   IS --  Information System\\
   I/O -- Input / Output\\
LED -- Light-Emitting Diode\\
 LV -- Low Voltage\\
 LVDS -- Low Voltage Differential Signaling\\
  MAPS -- Monolithic Active Pixel Sensor\\
  MC (simulation) -- Monte Carlo (simulation) \\
  MCP -- MicroCannel Plate\\
  MCT -- Micromegas-based Central Tracker\\
  MDT -- Mini Drift Tubes \\
  MFDM -- Muon FPGA Digital Module\\
  MIP -- Minimum Ionizing Particle \\
  ML -- Machine Learning \\
  MM (detector) -- MicroMegas (detector) \\
  MPD -- MultiPurpose Detector\\
  MPPC - MultiPixel Photon Counter\\
  MPZ -- Minimum Propagation Zone\\
  MRPC - Multigap Resistive Plate Chamber\\
  MS -- Magnetic System\\
    MWDB -- Muon Wall Digital Board \\
  NICA -- Nuclotron-based Ion Collider fAcility\\
  NPE -- Number of PhotoElectrons\\
PC -- Personal Comtuter\\
 PCB -- Printed Circuit Board \\
 PDE -- Photon Detection Efficiency\\
PEI -- PolyEtherImide \\
PET -- PolyEthylene Terephthalate \\
  PID -- Particle IDentification\\
  QPS -- Quench Protection System \\
RAM -- Random-Access Memory\\
RICH (detector) -- Ring-Imaging CHerenkov detector \\
  RMS -- Root Mean Square\\
  RNN -- Recurrent Neural Network \\
  RO -- ReadOut \\
  RRR -- Residual-Resistance Ratio \\
  RS -- Range System\\
SC -- SuperConductive \\
SAT -- Site Acceptance Testing \\
SCADA -- Supervisory Control And Data Acquisition \\
SEM -- Scanning Electron Microscope\\
  SiPM -- Silicon PhotoMultiplier\\
  SKM  -- Septum Kicker Magnet\\
  SLVS -- Scalable Low-Voltage Signaling \\
  SPD -- Spin Physics Detector\\
  SRS --  Scalable Readout System\\
    SS -- Superconducting Solenoid \\
    SSO -- Single Sign-On \\
  ST -- Straw Tracker\\
  STS -- Support and Transportation System\\
  (S)VD -- (Silicon) Vertex Detector\\
  TCS -- Trigger/Timing and Control System\\
  TMD (PDF) -- Transverse Momentum Dependent  (PDF)\\
  TOF (system) -- Time-Of-Flight (system)\\
  ToT -- Time Over Threshold (method)\\
  TPC -- Time Projection Chamber\\
  TSS -- Time Synchronization System\\
  UUID -- Universally Unique IDentifier \\
  WFD -- WaveForm Digitization \\
  WfMS -- Workflow Management System\\
  WLS -- WaveLength Shifter\\
  WMS -- Workload Management System \\
  WR -- White Rabbit \\
  ZDC -- Zero Degree Calorimeter \\
  ZFS -- Zettabyte File System \\
  
\end{appendices}

\bibliographystyle{./sty/hunsrt} 



\bibliography{./bib/case_DY,./bib/case_PPa,./bib/jpsi,./bib/DAQ,./bib/case_GPD,./bib/polarimetry,./bib/ECAL,./bib/general,./bib/summary,./bib/htp,./bib/central,./bib/csw,./bib/t0,./bib/other,./bib/quark,./bib/pid,./bib/vectors,./bib/physics1,./bib/straw,./bib/ZDC,./bib/MC,./bib/reham,./bib/slow,./bib/charmoniaMC,./bib/RS,./bib/tof,./bib/litra,./bib/RB,./bib/TDR,./bib/electromagnetic_calorimeter,./bib/main,./bib/mm,./bib/farich,./bib/straw2}


\end{document}